\RequirePackage{latex/atlaslatexpath}
\documentclass[BOOK, atlasdraft=false, texlive=2020, UKenglish, palatino]{atlasdoc}

\usepackage{atlaspackage}
\usepackage{atlasbiblatex}

\usepackage{atlasphysics}

\ifthenelse{\boolean{AtlasDraft}}{%
  \usepackage[output=true, shift=true]{atlastodo}
}{}

\graphicspath{{./logos/}{./figures/}{./figures_chapt2/}{./figures_chapt3/}{./figures_chapt4/}{./figures_chapt5/}{./figures_chapt6/}{./figures_chapt7/}{./figures_chapt8/}{./figures_chapt9/}{./figures_chapt10/}{./figures_chapt11/}{./figures_chapt12/}}

\usepackage{luxetdr-defs}
\usepackage{luxepixel-defs}

\usepackage{overpic}
\usepackage{amssymb}
\usepackage{comment}
\usepackage{makecell}
\usepackage{rotating}
\usepackage[official]{eurosym}
\usepackage{wasysym}
\usepackage{multirow}
\usepackage{longtable}
\usepackage{pdflscape}
\usepackage[symbol]{footmisc}
\usepackage{upgreek}

\PassOptionsToPackage{percent}{overpic}

\AtlasTitle{Technical Design Report for the LUXE Experiment}

\AtlasAbstract{%
  This  Technical Design Report presents a detailed description of all aspects of the LUXE (Laser Und XFEL Experiment),  an experiment that will combine the high-quality and high-energy electron beam of the European XFEL with a high-intensity laser, to explore the uncharted terrain of strong-field quantum electrodynamics characterised by both high energy and high
intensity, reaching the Schwinger field and beyond. The further implications for the search of physics beyond the Standard Model are also discussed.

}

\AtlasDate{1st August 2023}

\hypersetup{pdftitle={LUXE TDR}, pdfauthor={The LUXE Collaboration}}
\begin{document}


\frontmatter
\maketitle

\chapter*{The LUXE Collaboration}

\renewcommand{\thefootnote}{\fnsymbol{footnote}}
\begin{center}
\begin{large}

H. Abramowicz~$^{\textrm{\scriptsize 1}}$,
M.~Almanza Soto~$^{\textrm{\scriptsize 2}}$,
M. Altarelli~$^{\textrm{\scriptsize 3}}$,
R.~A{\ss}mann~$^{\textrm{\scriptsize 4}}$,
A.~Athanassiadis\footnote[1]{Also at HU Hamburg}~$^{\textrm{\scriptsize 4}}$,
G.~Avoni~$^{\textrm{\scriptsize 5}}$,
T.~Behnke~$^{\textrm{\scriptsize 4}}$,
M.~Benettoni~$^{\textrm{\scriptsize 6}}$,
Y.~Benhammou~$^{\textrm{\scriptsize 1}}$,
J.~Bhatt~$^{\textrm{\scriptsize 7}}$
% ...

\bigskip

$^{\textrm{\scriptsize 1}}$\textit{Tel Aviv University, Tel Aviv,  Israel}\\
$^{\textrm{\scriptsize 2}}$\textit{IFIC, Universitat de València and CSIC,  Paterna, Spain}\\
%
$^{\textrm{\scriptsize 3}}$\textit{Max Planck Institute for Structure and Dynamics of Matter,  Hamburg, Germany}\\
$^{\textrm{\scriptsize 4}}$\textit{Deutsches Elektronen-Synchrotron (DESY), Hamburg, Germany}\\
$^{\textrm{\scriptsize 5}}$\textit{INFN and University of Bologna, Bologna, Italy}\\
$^{\textrm{\scriptsize 6}}$\textit{INFN and University of Padova, Padova, Italy}\\
$^{\textrm{\scriptsize 7}}$\textit{University College London, London,  UK}\\
% ...





\end{large}
\end{center}

\renewcommand{\thefootnote}{\arabic{footnote}}

\chapter*{Foreword}
LUXE is going to explore quantum physics in a novel way! What could be more exciting for a physicist? It is rare that we have the opportunity to test fundamental physics in a completely new regime. Julian Schwinger showed in 1951 that at field strengths of $E=m_{e}^{2}/q_{e} \sim 1.32 \times 10^{18}$\,V/m (now known as the Schwinger field) the QED vacuum becomes unstable and decays into electron-positron pairs. Already earlier this effect was proposed by Sauter and Heisenberg and Euler also discussed it but the technology to create such strong fields was not available. 

With the recent advances in laser technology and the high energy electron beam of the European XFEL, LUXE can finally go directly into this regime and measure this effect in detail as a function of the strength of the field. The mission of LUXE is to observe the behaviour of QED in the strong-field regime, and then to also measure it in both Compton scattering and in electron-positron pair production with high precision. LUXE started in 2017 with a coffee meeting of three people, and by now consists of a collaboration of about 100 particle, accelerator and laser physicists. 

A huge thanks goes to all those who have contributed directly to this document, but also to those who gave advice and to the people and agencies who supported the project and its teams. I am excited in seeing LUXE now become a reality at DESY and the European XFEL in Hamburg, and look forward to its breakthrough observations and measurements!

Beate Heinemann\\
LUXE Spokesperson

\tableofcontents


\date{\today}
\clearpage
\mainmatter
\chapter{Introduction}
\chapterauthors{M. Altarelli \\ Max Planck Institute for the Structure and Dynamics of Matter, Hamburg (Germany)}{R.M. Jacobs \\ \desyaffil}{A. Levy \\ Tel Aviv University, Tel Aviv (Israel)}{M. Wing \\ University College London, London (UK)}
\label{chapt1}
\begin{refsection}
\vspace*{-1.5cm}

\begin{chaptabstract}
This introductory chapter outlines the genesis and early history of the LUXE project, and provides a guide to the structure of this Technical Design Report.
\newline
\newline
\newline
\end{chaptabstract}

This Technical Design Report of the LUXE project is organised in separate chapters, aiming to cover all aspects of the implementation of an apparatus for the observation of collisions between ultra-relativistic electrons extracted from the European XFEL Linac (EuXFEL) (or the high-energy $\gamma$-rays they generate in a solid target) and the photons of a high-power, short-pulse laser. The scientific ambitions of LUXE are to break new ground in the exploration of strong-field Quantum Electrodynamics, and to contribute to the search for particles and phenomena beyond the Standard Model.

The report is a collective effort of the whole LUXE collaboration, that as of now includes 23 laboratories\footnote{Four more Russian groups were originally part of the LUXE collaboration but as of February 2022 have been suspended until further notice.} in 9 countries (see Table 1.1). Needless to say, the project is developing in close consultation with the EuXFEL. 

\begin{table}[htbp]
\begin{center}

\begin{tabular}{l|l}
COUNTRY          & INSTITUTE OR GROUP\\ \hline \\
Germany   & DESY, Hamburg     \\
          & Albert-Ludwig-Universität Freiburg, Freiburg    \\ 
          & Max-Planck Institute for Structure and Dynamics of Matter, Hamburg    \\ 
          & Helmholtz Institut Jena, Jena \\ 
          & Friedrich-Schiller-Universität Jena, Jena \\ 
          & Karlsruhe Institute of Technology (KIT), Karlsruhe   \\ \hline \\
          
France    &The University of Paris-Saclay, CNRS/IN2P3, IJCLab, Orsay, Paris   \\
           &LLR CNRS, Ecole Polytechnique,  Institut Polytechnique de Paris  \\ \hline \\

 Israel    & Nuclear Research Centre - Negev, Beer Sheva \\
         & Tel Aviv University, Tel Aviv \\
           & Technion—Israel Institute of Technology, Haifa   \\
           & Weizmann Institute of Science, Rehovot  \\ \hline \\

Italy     & INFN and University of Bologna, Bologna    \\
          & INFN and University of Padova, Padova  \\ \hline \\
 Poland   & AGH Cracow (University of Science and Technology), Cracow  \\
          & University of Warsaw, Warsaw \\ \hline \\
          
Romania  & ISS (Institute of Space Science), Bucharest  \\ \hline \\

Spain    & IFIC (Instituto de Fisica Corpuscular) Valencia, Valencia \\ \hline \\

Sweden   & University of Gothenburg, Gothenburg  \\ \hline \\

United Kingdom  & John Adams Institute, Royal Holloway \\                 & Queens University Belfast, Belfast \\
               &  University College London, London  \\
               & University of Plymouth, Plymouth \\ \hline \\

Russia   & MePhi (Moscow Engineering Physics Institute), Moscow$^{*}$  \\

          & Skoltech (Skolkovo Institute of Science and Technology), Moscow$^{*}$ \\

             & National Research Tomsk State University, Tomsk$^{*}$ \\ \hline \\

 International    & Joint Institute for Nuclear Research, Dubna$^{*}$  \\ 
 Research Centers &     \\   \hline \\
 \end{tabular}
\caption{List of institutes and laboratories currently contributing to the LUXE project. The asterisk (*) denotes institutes located in Russia that originally joined the project but that  in February 2022 had their participation suspended until further notice. }
\end{center}
\end{table}

The collaboration originated in Hamburg in 2017, taking inspiration from the E144 experiment performed at SLAC in the 1990s~\cite{Bula:1996st, Burke:1997ew} and from earlier discussions about the possible use of the EuXFEL for fundamental studies (see e.g.~\cite{Ringwald:2001ib}). A major catalyst was the tremendous progress in laser technology in the preceding 20 years that allows significant expansion of the E144 results, by achieving electric fields exceeding the Schwinger critical field \cite{Schwinger:1951}, the scale for tunneling of real electron-positron pairs out of the vacuum; in addition, the high repetition rate of the EuXFEL linac, much higher than the few Hz rep rate of high power lasers, allows events to be accumulated whenever the linac is operated to produce x-rays, without any perceptible disturbance of the x-ray users, resulting in a much faster accumulation of data and better noise suppression (see Chapter~\ref{chapt2}).  The project rapidly gained momentum and support, that first resulted in the Conceptual Design Report~\cite{luxecdr} in 2021. 

The present report is organised in 12 chapters, including this introduction, and an appendix (not included in the present edition) 
describing the project planning and organisation. The following 11 chapters present all key aspects and components of the LUXE project, starting with the scientific case and objectives, and continuing with the description of the laser system and diagnostics, the set of photon, electron and positron detectors, with data acquisition, computing and simulation aspects, and concluding with a description of the technical infrastructure and of the installation plans. The chapters were formulated with the aim of being reasonably self-contained and suitable to be read independently, according to the interests and expertise of the individual readers. An effort to unify notations and definitions should nonetheless facilitate quick reference to other chapters, when required. 

The timetable for the implementation of the project has to take into account the scheduled shutdowns of the EuXFEL linac, as access for installations in the tunnels is only possible when the linac is down. The optimal time scale for the installation is 3.5\,years (see Chapter~\ref{chapt12}). Over such a time scale, some of the technical solutions envisaged in this Report may be improved, and also some of the external boundary conditions may necessitate revisions in the LUXE plans. All of these possible and even probable updates shall be inserted in the copy available in the LUXE web page, which the interested reader is encouraged to consult from time to time. In addition, the authors of each chapter are listed next to the title page, and can be contacted via e-mail for specific queries.

\section*{Acknowledgements}
We thank the DESY technical staff for continuous assistance and the DESY directorate for their strong support and the hospitality they extend to the non-DESY members of the collaboration. This work has benefited from computing services provided by the German National Analysis Facility (NAF) and the Swedish National Infrastructure for Computing (SNIC).
\printbibliography
\end{refsection}


\chapter{Overview and Scientific Objectives}
\label{chapt2}
\begin{refsection}

\chapterauthors{B. Heinemann \\\desyaffil}{B. King \\University of Plymouth, Plymouth (UK)}
\date{\today}

\begin{chaptabstract}
    This chapter presents an overview of all aspects of LUXE (Laser Und XFEL Experiment), an experiment that will combine the high-quality and high-energy electron beam of the European XFEL with a high-intensity laser to explore the uncharted terrain of strong-field quantum electrodynamics characterised by both high energy and high intensity. The chapter  also illustrates the physics performance that will be achieved using a few selected example measurements.
\end{chaptabstract}

\section{Introduction and Scientific Motivation}
\label{sec:theory}

The LUXE experiment and its scientific motivation have previously been described in a Conceptual Design Report (CDR)~\cite{luxecdr}. In this technical design report a brief recap of the scientific motivation is presented but the main focus is on the updated technical design of the experiment, including options for a staged installation. 

There are two main layouts foreseen. The first (and initial) mode will collide the electron beam of the European XFEL directly with a high-power laser. In a second mode, first the electron beam is converted to a photon beam and that photon beam then interacts with the laser, directly scattering light off light. In these interactions, photons, electrons and positrons can be produced and a set of detectors is designed to measure their multiplicities and energies. 

The main aims of the LUXE experiment are:
\begin{itemize}
    \item  measure the interactions of real photons with electrons and photons at field strengths where the coupling to charges becomes non-perturbative;
    \item make precision measurements of electron-photon and photon-photon interactions in a transition from the perturbative to the non-perturbative regime of quantum electrodynamics (QED);
   \item use strong-field QED (SFQED) processes to design a sensitive search for new particles beyond the Standard Model that couple to photons.
\end{itemize}

QED has been tested to a very high precision in systems with weak electromagnetic (EM) fields. For example, precision measurements of the electron anomalous magnetic moment \cite{Hanneke_2011} agree with the predictions of QED \cite{Aoyama:2019ryr} to an accuracy of around one part in a billion. However, in strong EM fields, QED predicts many phenomena that are yet to be confirmed experimentally. One reason is the high value of the natural EM field strength scale occurring in QED, the Schwinger limit: $\mathcal{E}_{\textrm{cr}}=m^{2}c^{3}/(e\hslash)=1.32\times 10^{18}\,\textrm{Vm}^{-1}$, which is currently orders of magnitude above any terrestrially producible field strength. However, in the rest frame of a high-energy probe charge, the EM field strength $\mathcal{E}$ is boosted by the Lorentz factor, $\gamma$, to $\mathcal{E}_{\ast}=\gamma \mathcal{E}(1+\cos\theta)$ where the collision angle at LUXE is $\theta=17.2\;\textrm{degrees}$. In this way, by colliding the $16.5\,\textrm{GeV}$ \euxfel  
electron beam with intense photon pulses produced by the laser, the electron rest-frame field at LUXE can reach and exceed the Schwinger limit and hence strong-field QED phenomena can be accessed.

The implementation of this project greatly expands the scope of the E144 experiments, performed at SLAC in the 1990s \cite{Bamber:1999zt}, leveraging on the huge progress of high-power laser and accelerator technologies in the last three decades.

QED in intense EM fields \cite{Fedotov:2022ely}, such as will be probed at LUXE, can arise in numerous settings. In astrophysics, pair creation can accompany gravitational collapse of black holes \cite{Ruffini:2009hg} and affect the propagation of cosmic rays \cite{nikishov62}. Some neutron stars are so strongly magnetised, that their magnetospheres probe the Schwinger limit \cite{Kouveliotou:1998ze,Harding_2006,Turolla:2015mwa}. In particle physics, in beam-beam collisions at future high-energy lepton colliders, strong-field effects are expected to feature prominently~\cite{yakimenko2019prospect,Bucksbaum:2020}, and probing of the Coulomb field of heavy ions can be sensitive to strong-field effects~\cite{Akhmadaliev:2001ik}. Strong fields, albeit in non-relativistic systems where the field strength scale is set by the ionisation potential, are the target of much investigation in the atomic and molecular physics communities (see e.g. \cite{ivanov}).

What separates SFQED phenomena apart from regular QED is the dependency on the dimensionless charge-field coupling. In a plane wave EM background, which approximates well the situation in a laser pulse as will be tested by LUXE, this coupling can be described with the classical nonlinearity parameter $\xi=|e|\mathcal{E}_{L}\lambdabar_{e}/\hslash \omega_{L}$: the work done by the EM field, $\mathcal{E}_{L}$, over a Compton wavelength in units of the EM field photon energy ($|e|$ is the charge of a positron, $\lambdabar_{e}$ the reduced Compton wavelength of an electron and $\hslash\omega_{L}$ the energy of a laser photon). Therefore, $\xi$ quantifies how many laser photons interact with the charge in a given QED process, with the probability of $n$ laser photons scaling as $\sim\xi^{2n}$. In weak fields, probabilities of QED events scale as $\sim \xi^{2}$, with higher-order interactions with the background being suppressed. But clearly when $\xi \sim O(1)$, this perturbative hierarchy breaks down, and all numbers of interactions between the charge and laser background must be taken into account. Hence in sufficiently strong fields, the probability of processes depend in a  non-perturbative way on the EM field. This parameter is sometimes referred to as the `intensity parameter', because it can be written $\xi = \sqrt{I_{L}/I_{\textrm{cr}}}(m_e c^{2}/\hslash \omega_{L})$, where $I_{L}/I_{\textrm{cr}}$ is the ratio of the laser intensity to the intensity of a field at the Schwinger limit and $m_e$ is the electron mass. 

\subsection{The Nonlinear Compton Process}
\begin{figure}[ht]
   \centering
   \includegraphics*[width=6cm]{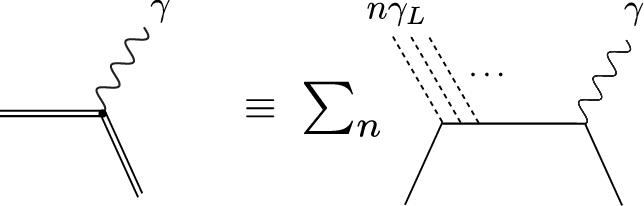}
   \caption{The nonlinear Compton process. Left: an electron `dressed' in the laser background (double solid lines) emits a single high energy photon, $\gamma$. Right: a sum of Compton processes where the electron interacts with $n$ laser photons $\gamma_{L}$ (dashed lines), and emits a single high energy photon.}
   \label{fig:NLCdiag}
\end{figure}

A key target process of the LUXE experiment is nonlinear (inverse\footnote{The process is the ``nonlinear inverse Compton process'' but it is commonly called the ``nonlinear Compton process''.}) Compton (NLC) scattering, depicted in Fig.~\ref{fig:NLCdiag}, where an electron or positron absorbs $n$ optical photons, $\gamma_L$, from the laser background and converts them into a single, high-energy gamma photon:
\[
e^\pm +n\gamma_{L} \to e^{\pm} + \gamma.
\]
This process has a classical limit. How much QED deviates from the classical limit, is in part quantified by the `quantum nonlinearity parameter', $\chi$, which for an incident electron can be written as $\chi_{e} = \mathcal{E}_{\ast}/\mathcal{E}_{\textrm{cr}}$ i.e. the ratio of the laser EM field, in the rest frame of the electron, to the Schwinger limit.  
The magnitude of nonlinear and quantum effects, can be clearly seen in the position of the `Compton edge' \cite{Harvey:2009ry} in electron and photon spectra, where $\chi$ quantifies the electron recoil when it emits a photon. The ratio, $u$, of the photon energy to the incident electron energy at the Compton edge in nonlinear QED, nonlinear classical electrodynamics and linear QED respectively, is:
\[
u_{\textsf{\scriptsize nonlin.QED}} = \frac{2\eta}{2\eta + 1 + \xi^{2}}; \quad u_{\textsf{\scriptsize nonlin.class.}} = \frac{2\eta}{1 + \xi^{2}}; \quad
u_{\textsf{\scriptsize lin.QED}} = \frac{2\eta}{2\eta + 1},
\]
where $\eta=\gamma \hslash \omega_{L} (1+\cos\theta)/m_{e}c^{2}$ is the `energy' parameter (from the definitions above, it is easily seen that $\chi_{e} = \eta \xi$, so that the preceding expressions for $u$ can be written in terms of $\chi_{e}$).  For typical LUXE parameters around $\xi=1$, the position of the edge predicted by these three theories differs by $O(100\,\textrm{MeV})$. Therefore measurement of the position of the Compton edge allows quantum and nonlinear effects to be to differentiated.

\subsection{The Nonlinear Breit-Wheeler Process}
\begin{figure}[ht]
   \centering
   \includegraphics*[width=6cm]{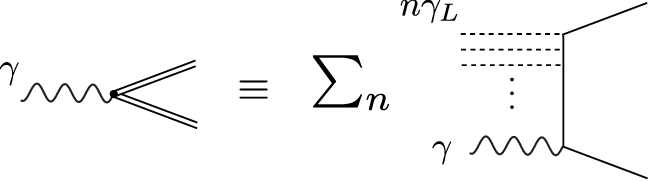}
   \caption{The nonlinear Breit-Wheeler process. Left: a high energy photon $\gamma$ produces an electron-positron pair that is `dressed' in the laser background (double solid lines). Right: a sum of Breit-Wheeler processes involving $n$ laser photons $\gamma_{L}$ (dashed lines). 
   \label{fig:NBWdiag}}
\end{figure}

Another key target process of the LUXE experiment, is nonlinear Breit-Wheeler (NBW) pair-creation:
\[
\gamma + n\gamma_{L} \to e^{+} + e^{-},
\]
depicted in Fig.~\ref{fig:NBWdiag}. This is a purely quantum effect with no classical analogue and hence requires the quantum nonlinearity parameter of the photon, $\chi_{\gamma}$, to fulfill $\chi_{\gamma}\sim O(1)$ in order to proceed, where $\chi_{\gamma} = (\mathcal{E}_{L}/\mathcal{E}_{\textrm{cr}})(\hslash \omega_{\gamma}/m_{e}c^{2})(1+\cos\theta)$; here $\hslash\omega_{\gamma}$ is the energy of the gamma photon and $\mathcal{E}_{L}$ is the laser field strength (both taken in the laboratory frame). Although linear Breit-Wheeler pair production was recently observed in Coulomb fields in heavy ion collisions \cite{STAR:2019wlg}, LUXE would provide the first observation of the nonlinear process in the non-perturbative regime, using real photons.
The non-perturbative nature of this process is exposed in the $\xi \gg 1$, $\chi_{\gamma} \ll 1$ limit, where the rate scales as:
\[
\Gamma_{\tiny\textrm{BW}} \propto \frac{\mathcal{E}_{L}}{\mathcal{E}_{\textrm{cr}}}\exp\left[-\frac{8}{3}\frac{m_e}{\omega_{\gamma}(1+\cos\theta)}\frac{\mathcal{E}_{\textrm{cr}}}{\mathcal{E}_{L}}\right],
\]
and we recall the charge-field coupling $\xi \propto \mathcal{E}_{L}$. The Breit-Wheeler rate has a similarity to the Schwinger process in a static electric field $\mathcal{E}_{S}$ (see e.g. \cite{Hartin:2018sha})
\[
\Gamma_{\tiny\textrm{S}} \propto \left(\frac{\mathcal{E}_{S}}{\mathcal{E}_{\textrm{cr}}}\right)^{2}\exp\left[-\pi\frac{\mathcal{E}_{\textrm{cr}}}{\mathcal{E}_{S}}\right].
\]

\subsection{The Nonlinear Trident Process}
\begin{figure}[ht]
   \centering
   \includegraphics*[width=3cm]{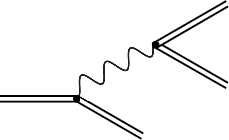}
   \caption{The nonlinear trident process. Double lines represent electrons and positrons dressed in the laser background.
   \label{fig:tridentDiag}}
\end{figure}

Another target process of LUXE is the nonlinear trident process, depicted in Fig.~\ref{fig:tridentDiag}, which combines a nonlinear Compton step and a nonlinear Breit-Wheeler pair-creation step off the radiated photon:
\[
e^{-} + n \gamma_{L} \to e^{-} + \gamma \qquad \textrm{and} \qquad \gamma + n'\gamma_{L} \to e^{+} + e^{-},
\]
where the number of background laser photons in the Breit-Wheeler step, $n'$, will in general be different to the number of photons involved in the Compton step, $n$. 
Although this process was measured in the E144 experiment \cite{Burke:1997ew}, it was done in the perturbative, multiphoton regime. Nonlinear trident in the non-perturbative $\xi \gg 1$ region has recently been measured for the first time by the NA63 experiment in the collision of 200 GeV electrons with oriented crystals \cite{nielsen2022precision}. 
LUXE will also measure nonlinear trident in the non-perturbative, all-order regime, by colliding electrons with real photons, as a test of SFQED. In order to achieve this, a new approximation framework has been developed around  locally monochromatic rates \cite{Heinzl:2020ynb}, which have been built into a bespoke numerical simulation code called \textsc{Ptarmigan} \cite{ptarmigan} that models the laser interaction point and was rigorously benchmarked with direct calculations from QED \cite{Blackburn:2021rqm,Blackburn:2021cuq}.

\subsection{Beyond the Standard Model}
\begin{figure}[ht]
   \centering
   \includegraphics*[height=4.9cm]
   {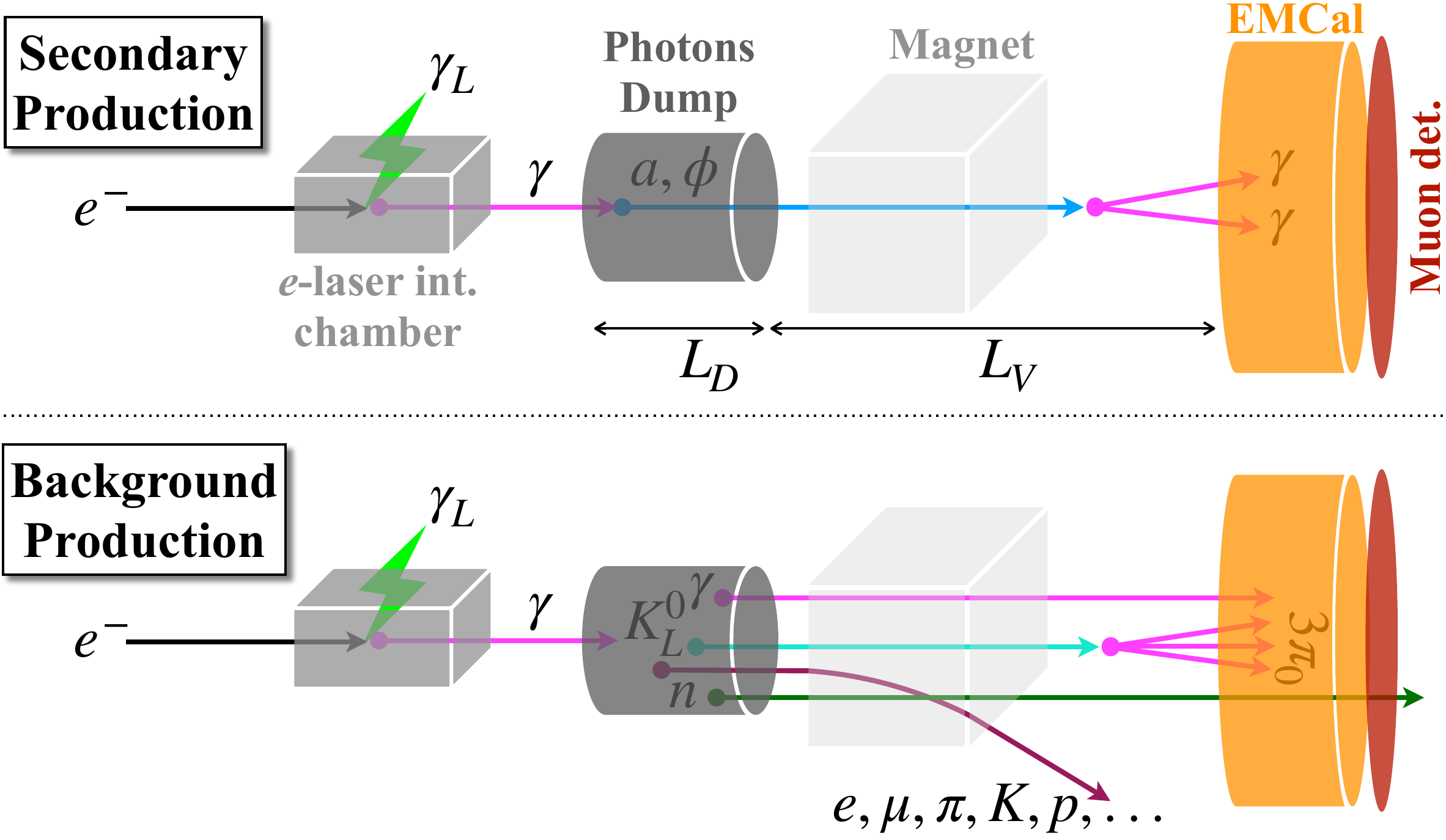}
   \caption{A schematic illustration in the LUXE-NPOD concept. Top: the secondary production mechanism realization in the experimental setup. Bottom: the relevant background topologies. The charged particles are deflected by a magnet placed right after the interaction chamber.
   (from~\cite{LUXEBSM})}
   \label{fig:BSM}
\end{figure}

There is significant experimental evidence for new physics beyond the Standard Model \cite{McCullough:2018knz,Strategy:2019vxc}. A possible explanation of these experimental signals includes the existence of new, light degrees of freedom, which are weakly coupled to the Standard Model and potentially long-lived. Axion-like particles (ALPs), which are generalisations of the posited axion that solves the strong-CP problem \cite{Peccei:1977hh, Peccei:1977ur, Wilczek:1977pj, Weinberg:1977ma}, can couple to two photons and hence produce a signal in the LUXE experiment. A `secondary production' mechanism will be used, which involves using high-energy photons generated via the nonlinear Compton process, propagating downstream to a beam dump, in which ALPs can be created (via the Primakoff effect) and then decay on the other side of the dump to two photons, see Fig.~\ref{fig:BSM}. A detector is then placed to detect the photons, reject backgrounds due to e.g. neutrons and to measure the energies and angles of the photons. This aspect of LUXE is called LUXE-NPOD  (LUXE New Physics search with Optical Dump)
and described in more detail in Ref.~\cite{LUXEBSM}. 

\subsection{Expected Particle Spectra}
\begin{figure}[ht]
   \centering
   \includegraphics*[width=0.45\textwidth]{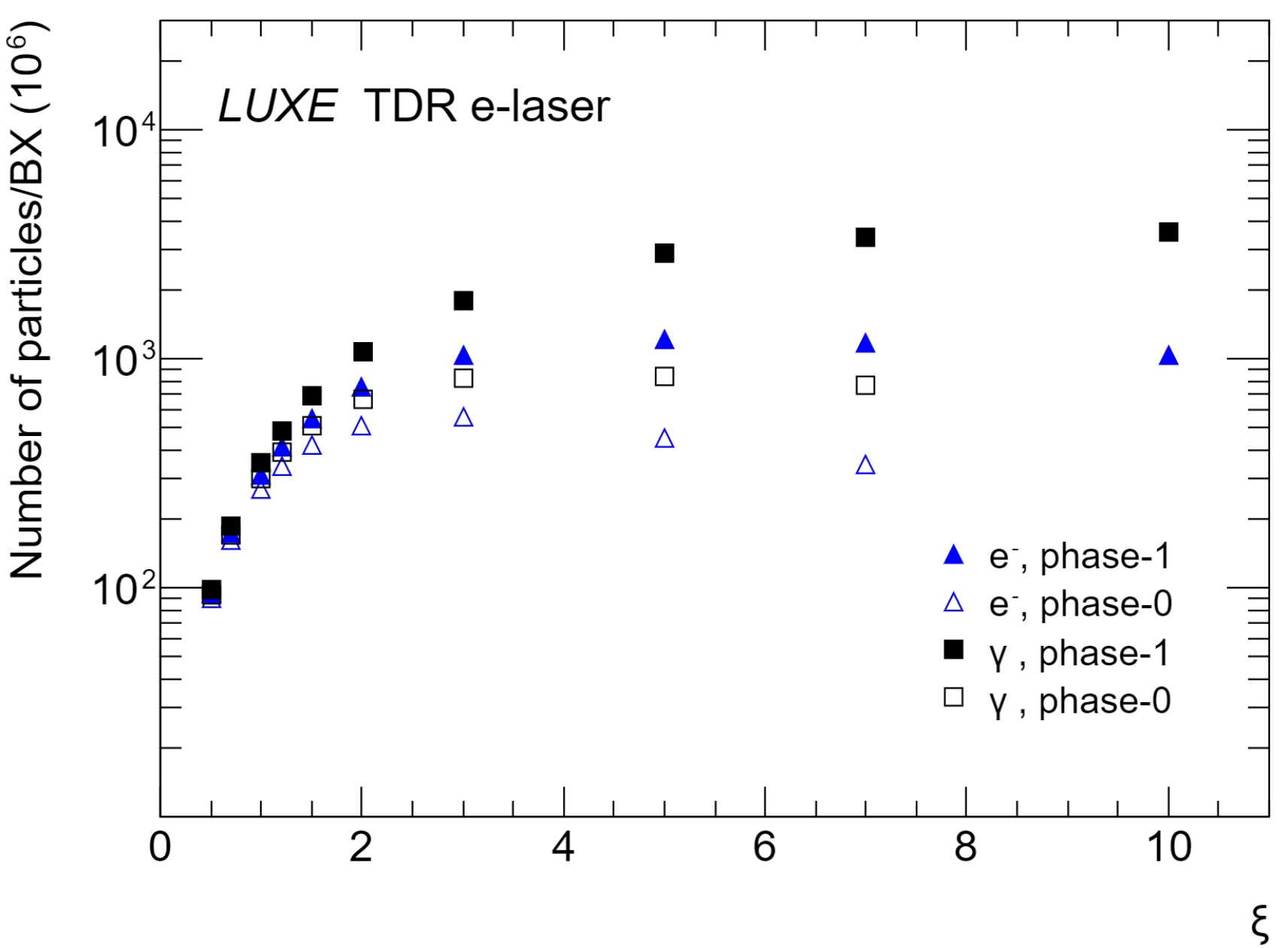}
   \includegraphics*[width=0.48\textwidth]{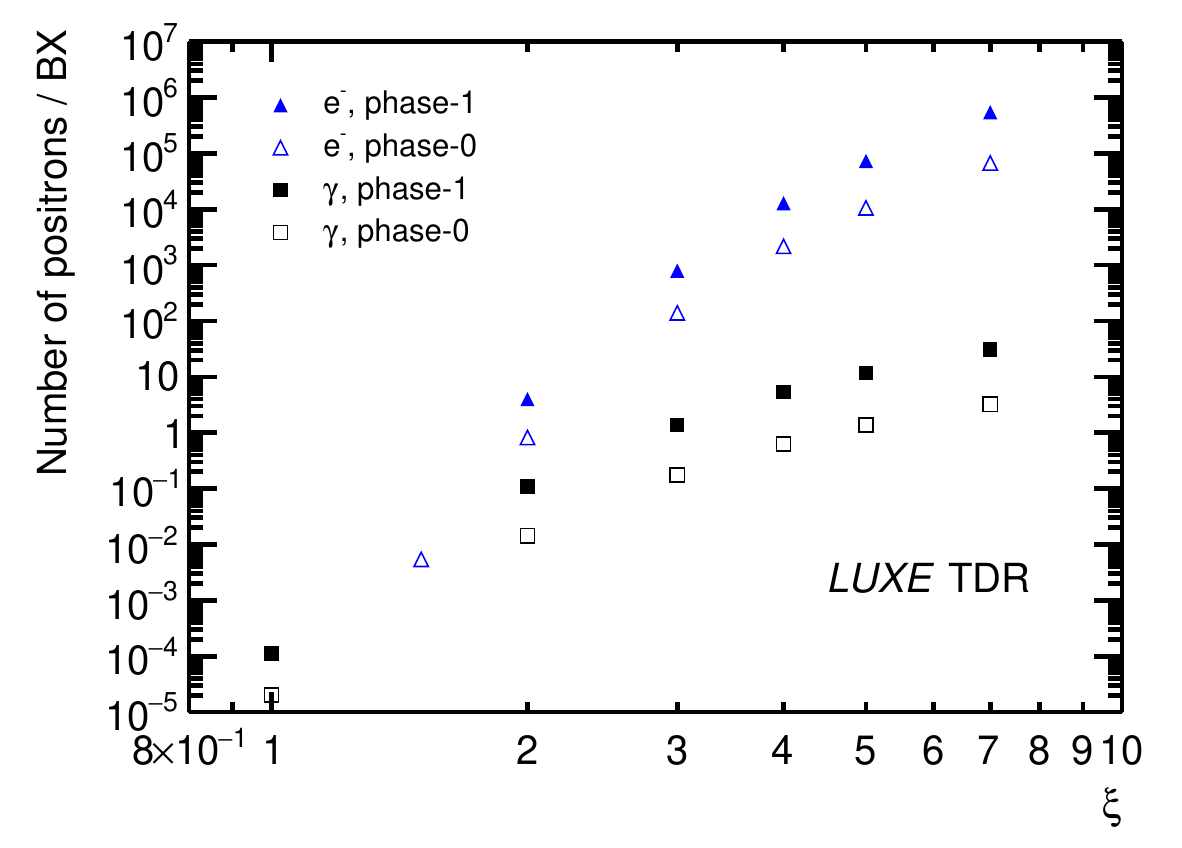}
   \caption{Left: Number of electrons and photons in \elaser collisions as function of $\xi$. Right: Number of positrons in \elaser  and \glaser collisions as function of $\xi$. Shown are the expected results based on \textsc{Ptarmigan} for two phases (\phaseone with a laser power of 40\,TW  and \phasetwo with a laser power of 350\,TW) of the experiment that differ by the laser power, see Sec.~\ref{sec:laser}.}
   \label{fig:rates}
\end{figure}

The \textsc{Ptarmigan}~\cite{ptarmigan} MC generator is used to predict particle rates and spectra for the purpose of optimising the design of the experiment. Fig.~\ref{fig:rates} shows the number of particles produced in the Compton and Breit-Wheeler processes. The parameters for the figures are the same as those in Tables~\ref{tab:xfelepara} and~\ref{tab:laser} with the exception of the bunch length which was assumed to be of 40~$\mu$m instead of $5-10$~$\mu$m. The impact on physics observables of interest is at most 25\%.

\begin{figure}[ht]
   \centering
   \includegraphics*[width=0.48\textwidth]{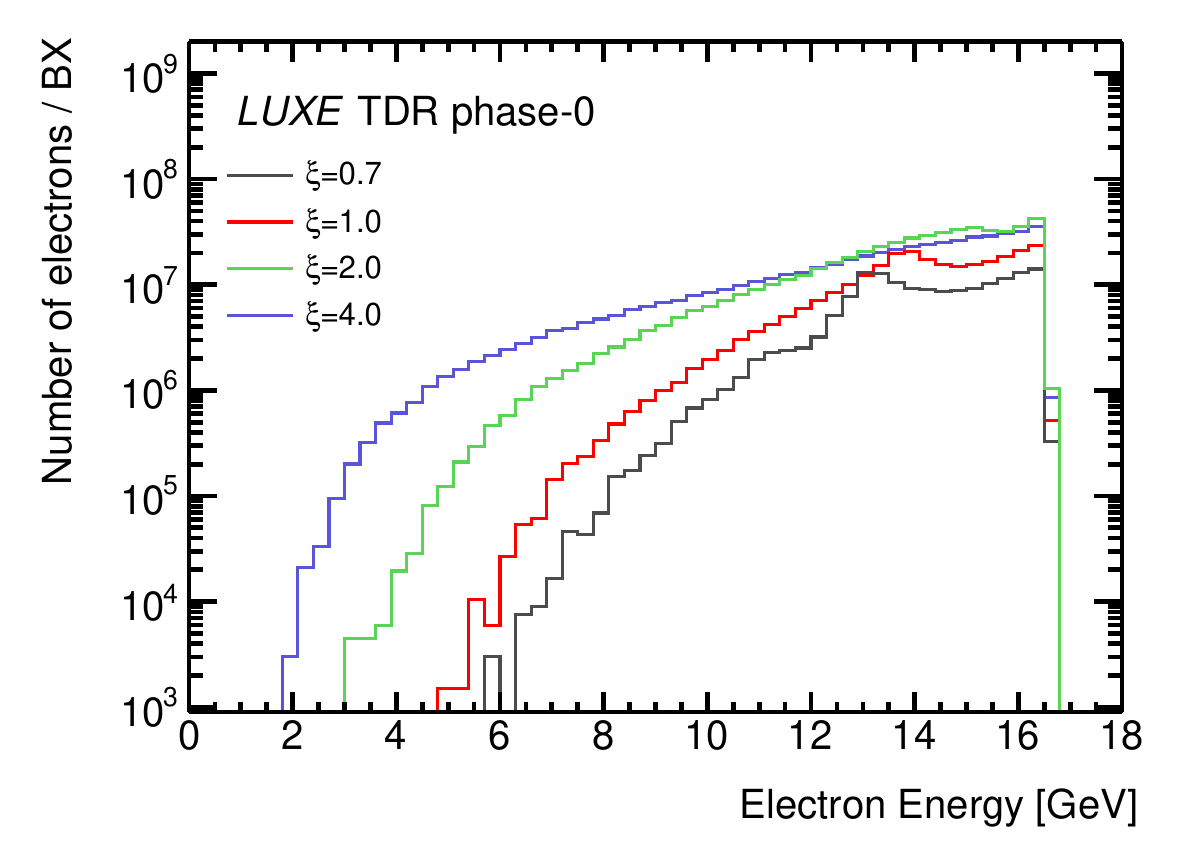}
   \includegraphics*[width=0.48\textwidth]{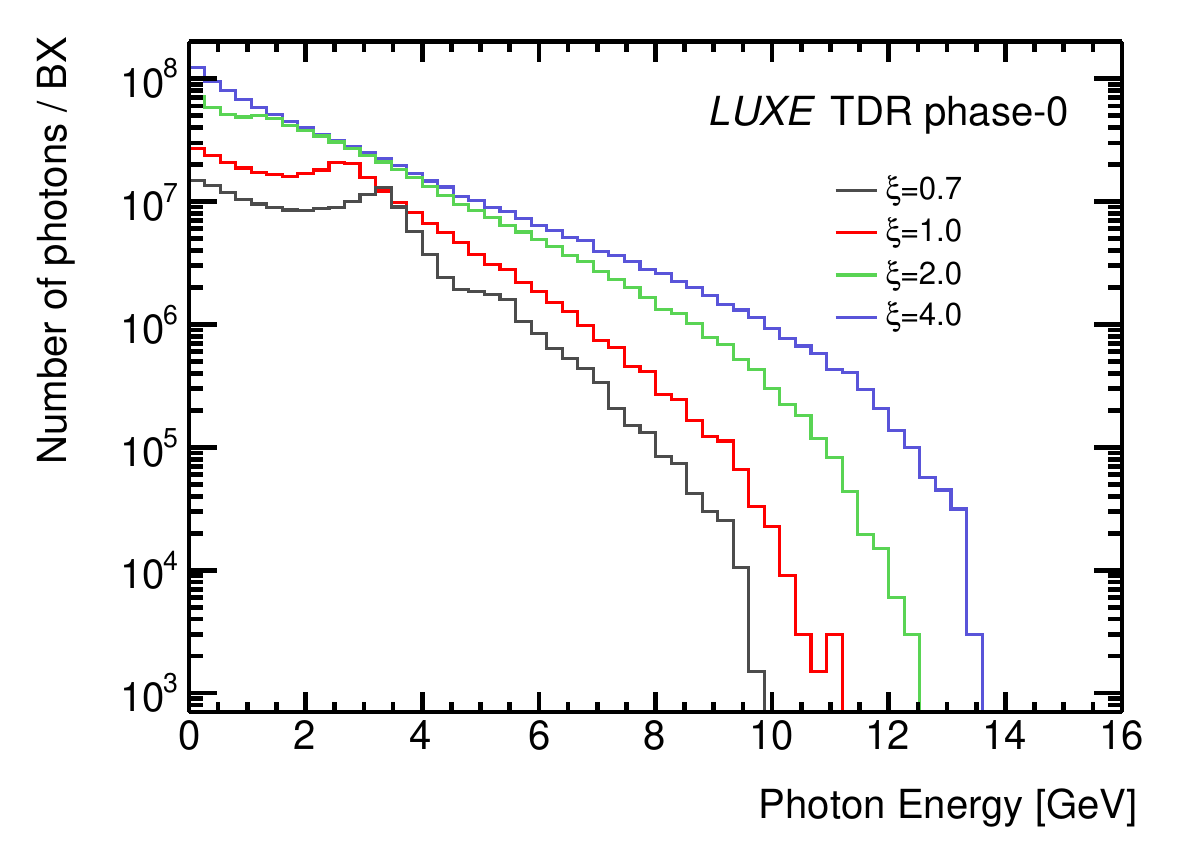}
   \caption{Electron and photon energy spectra for the Compton process in the \elaser mode for four values of $\xi$ in \phaseone. The distributions are normalised per bunch crossing.}
   \label{fig:elspec} 
\end{figure}

The energy spectra for electrons and photons predicted by \textsc{Ptarmigan} for the Compton process for the LUXE parameters are shown in Fig.~\ref{fig:elspec} for various values of $\xi$. At low $\xi$ the Compton edges are clearly visible. With increasing $\xi$ they become less and less pronounced and the electron energy spectrum extends to lower energies and, correspondingly, the photon energy spectrum extends to higher energies. The goal is to measure both the electron and the photon energy spectra and to determine the position of the edges.

\begin{figure}[ht]
   \centering
   \includegraphics*[width=0.48\textwidth]{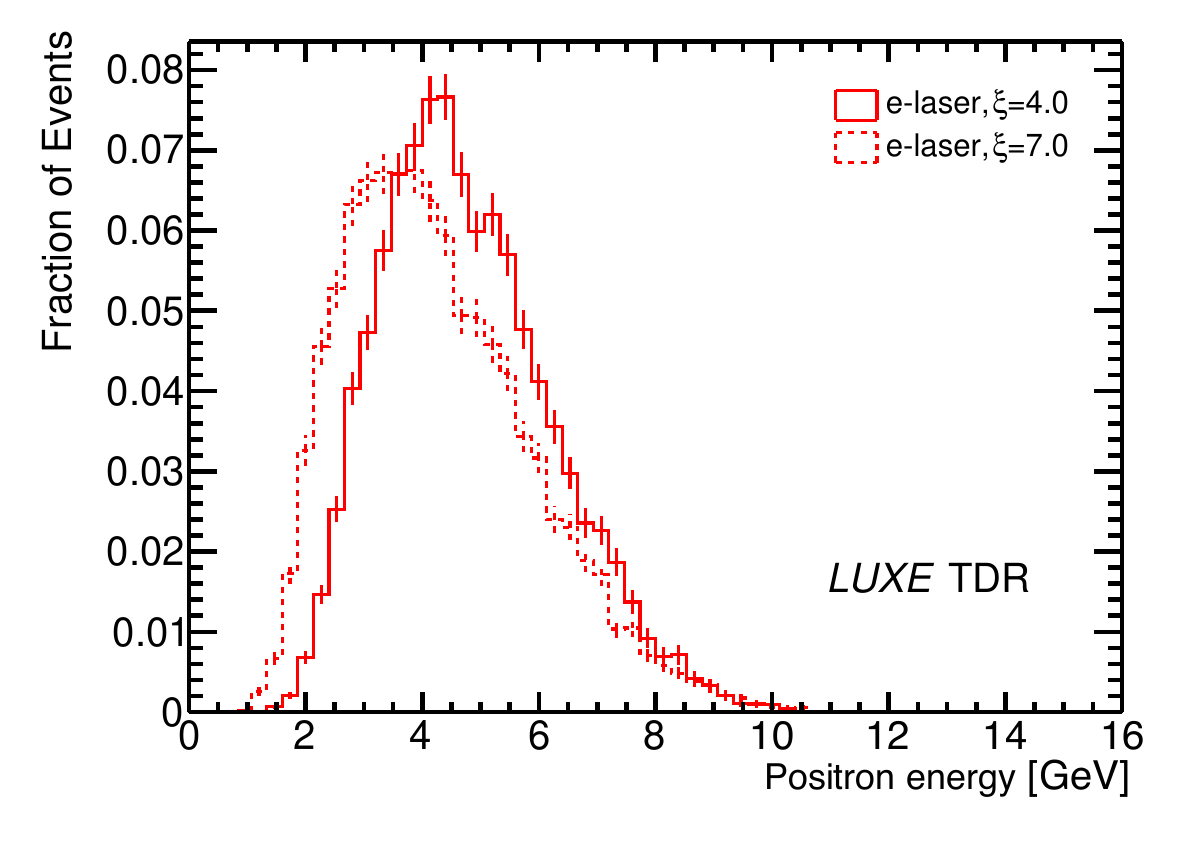}
   \includegraphics*[width=0.48\textwidth]{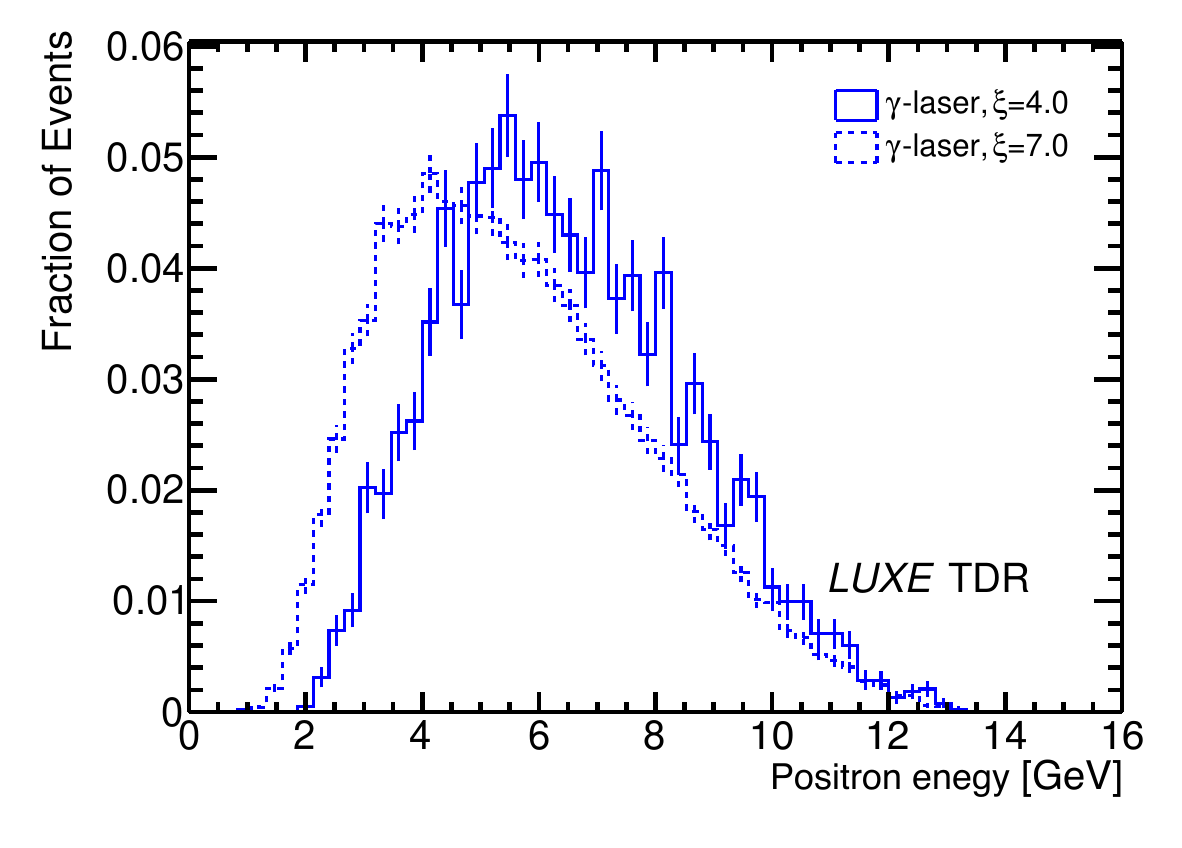}
   \caption{Positron energy spectra for the Breit-Wheeler process in the \elaser mode (left) and the \glaser mode (right) for two values of $\xi$ in \phaseone. The distributions are normalised to equal area.}
   \label{fig:posspec}
\end{figure}

Fig.~\ref{fig:posspec} shows the energy spectra of positrons, resulting from the Breit-Wheeler process, for the \elaser{} and \glaser{} modes. The detectors are designed to reconstruct these spectra with good precision.

\section{European XFEL Accelerator}
A very high-quality electron beam with a design energy of up to $\sim 17.5$~GeV is delivered by the linear accelerator of the European XFEL. The \euxfel{} has been operating since 2017. 

The beam consists of bunch trains which contain up to 2700 individual bunches, and the rate of bunch trains is normally 10~Hz. Only one of the bunches within a train will be used for the LUXE experiment as the laser operates with a frequency of 1~Hz only (the other 9~Hz will be used to measure the backgrounds). The properties of the individual bunches at the LUXE interaction point (IP) are summarised in Table~\ref{tab:xfelepara}. For the purpose of this document, electron beam energies of 16.5 and 14.0~GeV are assumed. For LUXE, the higher the energy the better but it is also interesting to perform the measurement at several energies. The accelerator is designed for a bunch charge of 1~nC but mostly operates at 0.25~nC, corresponding to $1.5\times 10^9$ electrons. 

\begin{table}[htbp]
\begin{center}
\begin{tabular}{l|r}
Parameter          & Value \\ \hline
Beam Energy [GeV]  & $16.5$\\
Bunch Charge [nC]  & 0.25   \\ 
Repetition Rate [Hz]    & 10 \\ 
Spotsize at the IP [$\mu$m]        & $5-20$  \\ 
Bunch length [$\mu$m]        & $5-10$  \\ 
Normalised projected emittance [mm mrad] & 1.4 \\
Energy spread $\delta E/E$ (\%) & 0.1\\
\end{tabular}
\caption{Electron beam parameters for the \euxfel{} as assumed for the design of the LUXE experiment. The electron beam spot size at the LUXE interaction point IP can be adjusted. The bunch length is expected to be about 5-10~$\mu$m, depending on the bunch charge and electron energy.}
\label{tab:xfelepara}
\end{center}
\end{table}

\begin{figure}[ht]
   \centering
   \includegraphics*[height=4.2cm]{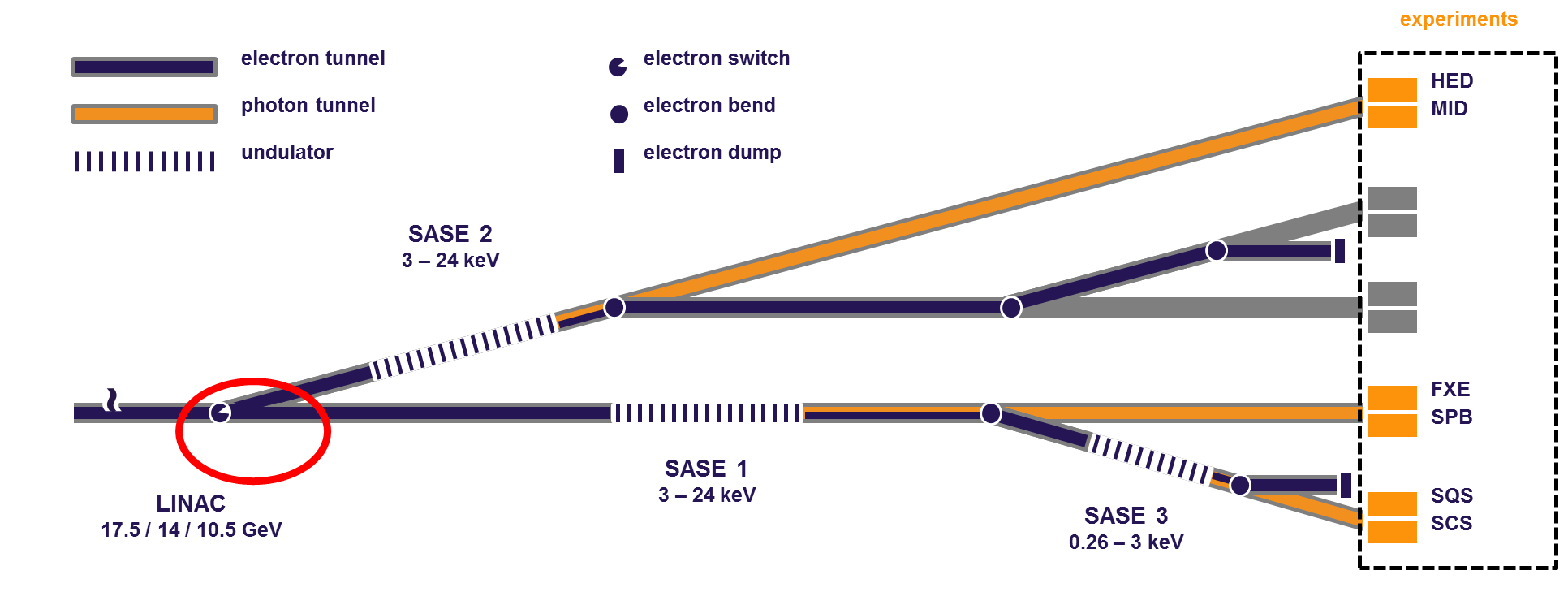}
   \caption{Schematic drawing of the \euxfel fan. The location foreseen for the extraction of the beam for the LUXE experiment is circled in red. The SASE1, SASE2 and SASE3 undulators and the experimental hall are also shown.}
   \label{fig:XFEL_sketch}
\end{figure}

The electron bunch for LUXE is extracted with a fast kicker magnet and then directed towards an annex of the XS1 building where the LUXE experiment can be housed. A sketch of the \euxfel fan and the end of the linac is shown in Fig.~\ref{fig:XFEL_sketch}. The red circle indicates the location of the XS1 building. 

\begin{figure}
    \centering
    \includegraphics*[width=0.9\textwidth]{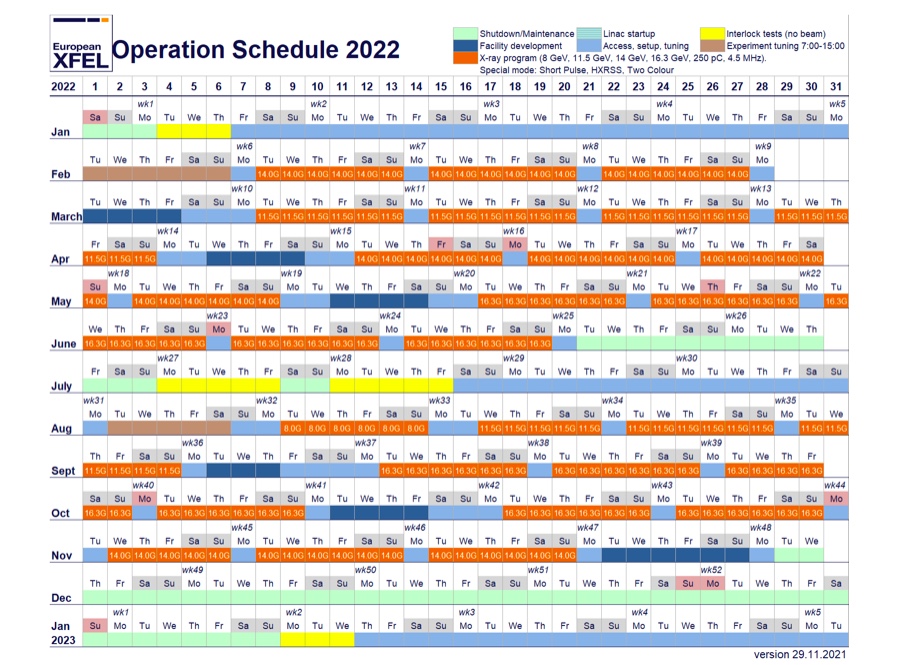}
    \caption{Operations Schedule of the \euxfel for 2022.}
    \label{fig:euxfelops}
\end{figure}

The 2022 operations schedule of the \euxfel is shown in Fig.~\ref{fig:euxfelops} for illustration. Previous years (except 2020 which was impacted by the COVID19 pandemic) were comparable. 
It is seen that the winter shutdown extends over about two months (including the break during the holiday season) and there is another 2-week shutdown in the summer during which there is access to the experimental area of LUXE. Otherwise, there are short access opportunities on Mondays. In total, there were 28 weeks of photon delivery in 2021 (the last complete year for which data are available) at three different energies: 16 weeks at $E=14$~GeV, 6 weeks each at $E=11.5$~GeV and  $E=16.5$~GeV. The schedule for 2022 is similar and results in 29 weeks of photon delivery.

\begin{figure}[h!]
    \centering
    \includegraphics*[width=0.7\textwidth]{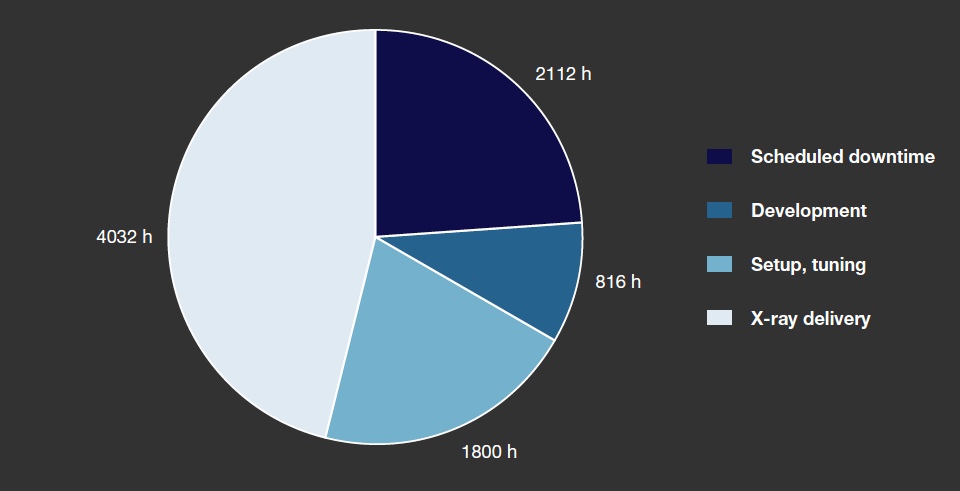}
    \caption{Operational modes of the \euxfel accelerator and the number of hours spent in each mode in 2021. Stable physics running corresponds to ``X-ray delivery''. Source: European XFEL Annual Report 2021}
    \label{fig:euxfelhour}
\end{figure}

Another aspect to consider is the fraction of time where stable physics data taking is possible. In 2021 (which again is representative also for other years) the facility delivered photons to the photon science experiments for 4032 hours ($1.45\times 10^7$s). A breakdown of the hours of the facility is shown in Fig.~\ref{fig:euxfelhour}.

Based on this past performance, we assume that LUXE will have an opportunity for physics data taking for $1.3\times 10^7$~s, and considering a data taking efficiency of LUXE of 75\% and a similar splitting between the various energies as in 2021, about $5\times 10^6$~s are provided at $E_e=14$~GeV and $2\times 10^{6}$~s at $E_e=16.5$~GeV. Data will also be taken at $E=11.5$~GeV but no  simulations of the QED processes for LUXE have yet been performed at this energy.

Furthermore we assume that normally we have a two-week access each summer and an access of 4-6 weeks each winter. Additionally, in 2025 there is a six-month shutdown planned during which a variety of work will be performed at the \euxfel{} facility. Among the plans is the construction of the beam extraction for the 2nd fan of the European XFEL, which LUXE can benefit from until the 2nd fan construction commences in about 2030. 

\section{Laser and Diagnostics}

As shall be detailed in Chapter~\ref{chapt3}, the laser will be installed in a small building at the ground level, directly above the beam extraction tunnel. A titanium sapphire laser with a wavelength of 800~nm and a pulse length of 25-30~fs will be used. Initially a laser with 40~TW power will be utilised (termed ``phase-0''), and in a later phase it will be upgraded to 350~TW (``phase-1'').

\begin{figure}[h!]
    \centering
    \includegraphics*[width=0.89\textwidth]{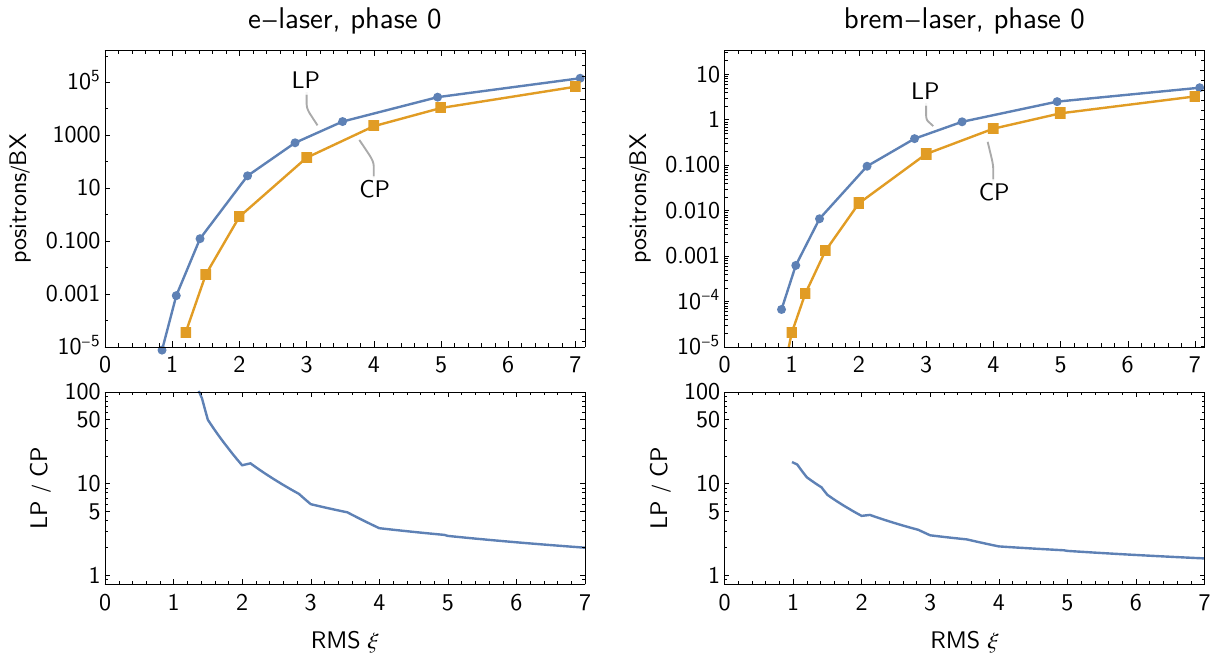}
    \caption{Comparison of positron yield per beam crossing in the e-laser (left) and $\gamma$-laser modes (right) with linear (LP) or circular (CP) polarisation. In the bottom panel the corresponding ratios are plotted. Horizontal scales: see text.} 
    \label{fig:PositronYield}
\end{figure}

The polarisation of the laser will normally be linear. However, the physics prospects shown are based on the assumption of circular polarisation unless otherwise stated. This is due to the current availability of data generated for the circular polarisation case, which is quicker to produce than the linear case. In Fig.~\ref{fig:PositronYield} we illustrate the difference in the pair yield for a circularly-polarised compared to a linearly-polarised laser, in the electron-laser and Bremsstrahlung gamma-laser set-ups. The comparison is plotted against the cycle-averaged root-mean-square intensity rather than the peak intensity, $\xi$, which in the circularly-polarised case is $\xi$ and linearly-polarised case $\xi/\sqrt{2}$.
A further discussion of the impact by using linear rather than circular polarisation can be found in Ref.~\cite{luxecdr}. 

The laser beam will be transported down to the experimental area to an interaction chamber via a 40m-long vacuum beam pipe with a diameter of 15-20~cm. Inside the interaction chamber it will be focused tightly to a waist $w_0$ (the FWHM of the laser) between 3 and 150~$\mu$m, depending on the $\xi$ value targeted. Inside the chamber the laser photons cross the electron beam at an angle\footnote{The exact angle will depend upon the choice of off-axis parabola.} of 17.2$^\circ$ and with a repetition rate of 1~Hz. 

\begin{figure}[h!]
    \centering
    \includegraphics*[width=0.75\textwidth]{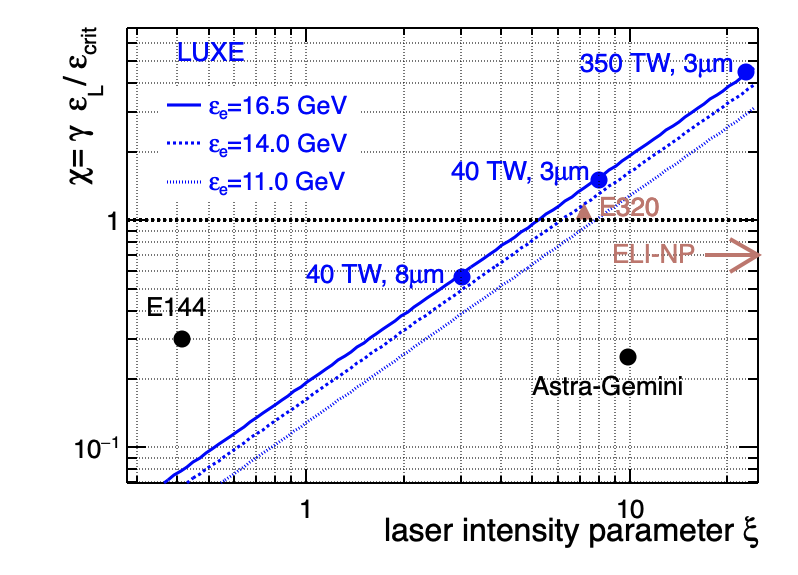}
    \caption{Quantum parameter $\chi$ versus the laser intensity parameter $\xi$. The blue lines show the values accessible at the LUXE experiment for three different energy values of the EuXFEL accelerator. The three LUXE points are for different combinations of laser power and spot size.  Also shown are previous (E144 \cite{Bula:1996st,Burke:1997ew}, Astra-Gemini  \cite{Cole:2017zca,Poder:2018ifi}) and present or planned (E320 \cite{Chen:22} and ELI-NP \cite{homma2016combined} experiments).}
    \label{fig:megaplot}
\end{figure}

A summary of the laser parameters and expected $\xi$ and $\chi$ values is presented in Table~\ref{tab:laser}. Fig.~\ref{fig:megaplot} shows the $\xi$ and $\chi$ values that can be accessed at LUXE at the various beam energies.

\begin{table}[h!]
\begin{center}
\begin{tabular}{|l|c|c|}
\hline
\textbf{Laser parameters} &  phase-0 & phase-1\\
\hline
Fraction of ideal Gaussian intensity in focus (\%) & \multicolumn{2}{c|}{50}\\   
Laser central wavelength (nm) & \multicolumn{2}{c|}{800}\\
Laser pulse duration (fs)  & \multicolumn{2}{c|}{25-30}\\  
Laser focal spot waist $w_0$ ($\mu$m)  & \multicolumn{2}{c|}{$\geq 3$} \\ 
    Laser repetition rate (Hz) & \multicolumn{2}{c|}{1-10}\\  
   Electron-laser crossing angle (rad)  &  \multicolumn{2}{c|}{0.3}\\  
   \hline
Laser energy after compression (J) & 1.2 & 10 \\
Laser power (TW) & 40 & 350 \\
Peak intensity in focus ($\times10^{20}$~W/cm$^{2}$)  & $< 1.33$ & $\leq 12$ \\  
    Dimensionless peak intensity, $\xi$ &  $<7.9$ &  $<23.6$\\   
     \hline
      \multicolumn{3}{c}{}   \\
      \hline
    \textbf{Quantum parameter} & phase-0 & phase-1\\  
 \hline
    $\chi_e$ for $E_e=14.0$~GeV  &  $<1.28$ &  $<3.77$\\ 
    $\chi_e$ for $E_e=16.5$~GeV  &  $<1.50$ &  $<4.45$\\  
    $\chi_e$ for $E_e=17.5$~GeV  &  $<1.6$ &  $<4.72$\\  
    \hline
    \hline
\end{tabular}
\vspace{1.2cm}
\caption{Laser parameters for the two phases of the LUXE experiment, and the quantum parameter $\chi_e$ is given for two different electron energies. All values are for circular polarisation with peak field parameters $\xi$ and $\chi$ increasing by a factor of $\sqrt{2}$ for linearly polarised laser irradiation. For focal spot the lowest value targeted is given, and the peak intensity ($\xi$ and $\chi_e$) values given correspond to this waist value. The laser pulse is expected to be Gaussian in all three dimensions in the focal point. The laser is expected to operate at 1 Hz repetition rate, a 10 Hz repetition rate is in principle available at lower intensities.}\label{tab:laser}
\end{center}
\end{table}

The laser will be synchronised with the electron beam via a system  ~\cite{Synchroxfel,Schulz} developed by DESY for the synchronisation of lasers used by the so-called ``pump and probe'' experiments at the \euxfel{}. The system designed at DESY has demonstrated the stable synchronisation of two RF signals to better than 13~fs, compared to a requirement of 25~fs given the electron and laser bunch lengths and their relative collision angle of $17.2^\circ$.

\begin{figure}[h!]
    \centering
    \includegraphics[width=0.9\textwidth, angle=0]{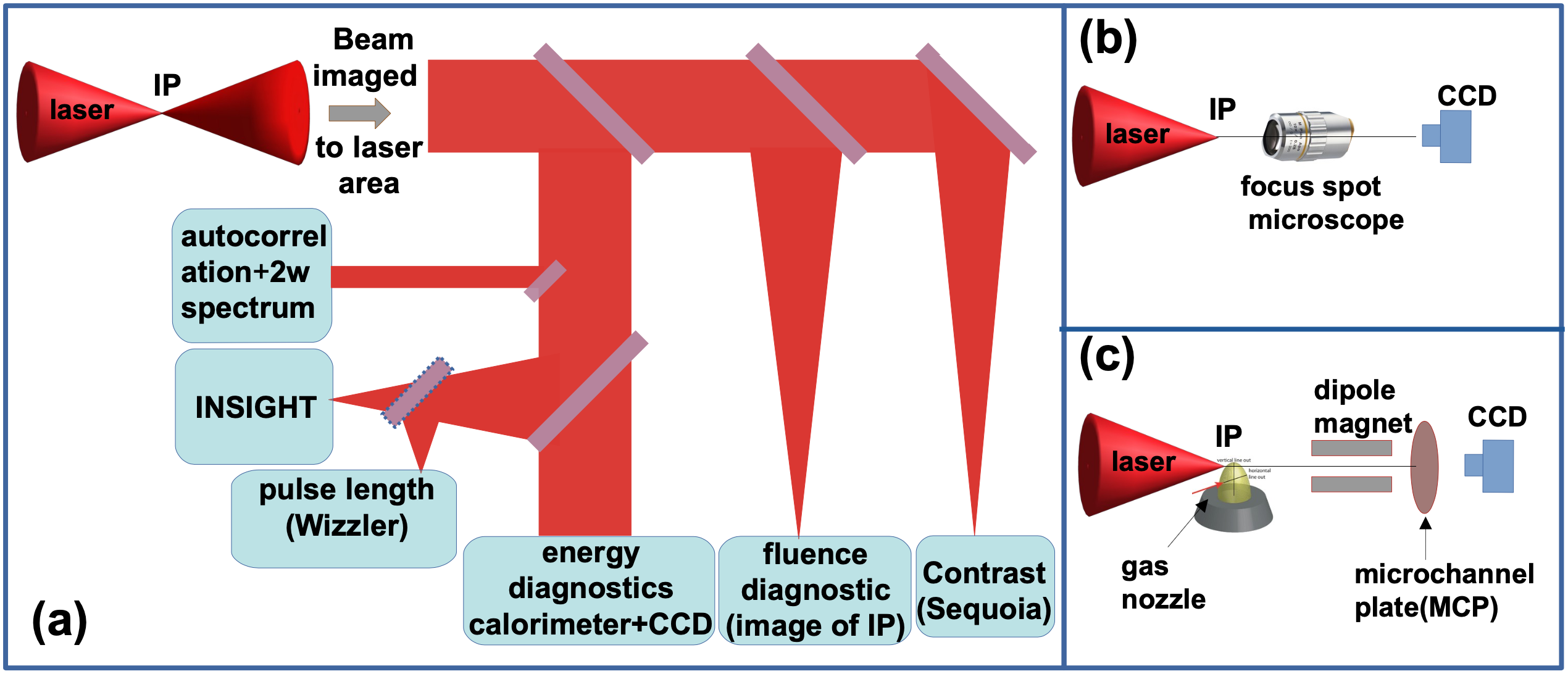}
    \vspace{1.2cm}
    \caption{Schematic diagram of the proposed intensity tagging diagnostics. The laser beam will be  attenuated and imaged on the return path to the diagnostics. (a) Layout of the laser shot tagging diagnostics in the laser area. (b) Focus diagnostic for the spot size measurement at the interaction point. (c) Set-up for an \textit{in situ} laser intensity measurement at the IP. }
    \label{fig:laserdiag}
\end{figure}

In addition, an elaborate diagnostics system has been designed with the goal to characterise the intensity to better than 1\% relatively (shot by shot) and 5\% absolutely. It relies on a variety of devices that measure the energy, fluence and other properties of the laser beam. The diagnostics is based on the beam after the interaction. A sketch of the diagnostics is shown in Fig.~\ref{fig:laserdiag}. 

\label{sec:laser}


\newpage
\section{LUXE Detectors}

The LUXE experiment is designed to analyse electron-laser and photon-laser interactions that occur at the interaction point inside the interaction chamber. Furthermore, it is also planned to use it for searches of axion-like particles resulting from photons interacting in the photon beam dump via the Primakoff effect. However, this latter part of the experiment is not yet as advanced and is considered an upgrade that will be installed at a later point, see Ref.~\cite{luxecdr}. Here, only the setup relevant to the studies of the QED interactions is discussed.

In order to fully characterise these interactions it is necessary to measure the multiplicities of particles produced as well as their energy spectra. The particles of interest are electrons, photons and positrons and the fluxes of these particles vary strongly depending on the running mode and the location. In general, the experiment is designed such that it has redundancy in many of the measurements which will allow cross-checks and cross-calibrations between the different approaches. A full simulation of the experimental area was performed with \geant to decide on the technologies for the detectors. A more elaborate discussion of the rates of signal and background can be found in Sec.~5.3 of the CDR~\cite{luxecdr} and in Chapters~\ref{chapt3} to~\ref{chapt10} of this TDR.

\begin{figure}[htbp]
\centering
\includegraphics[width=0.48\textwidth]{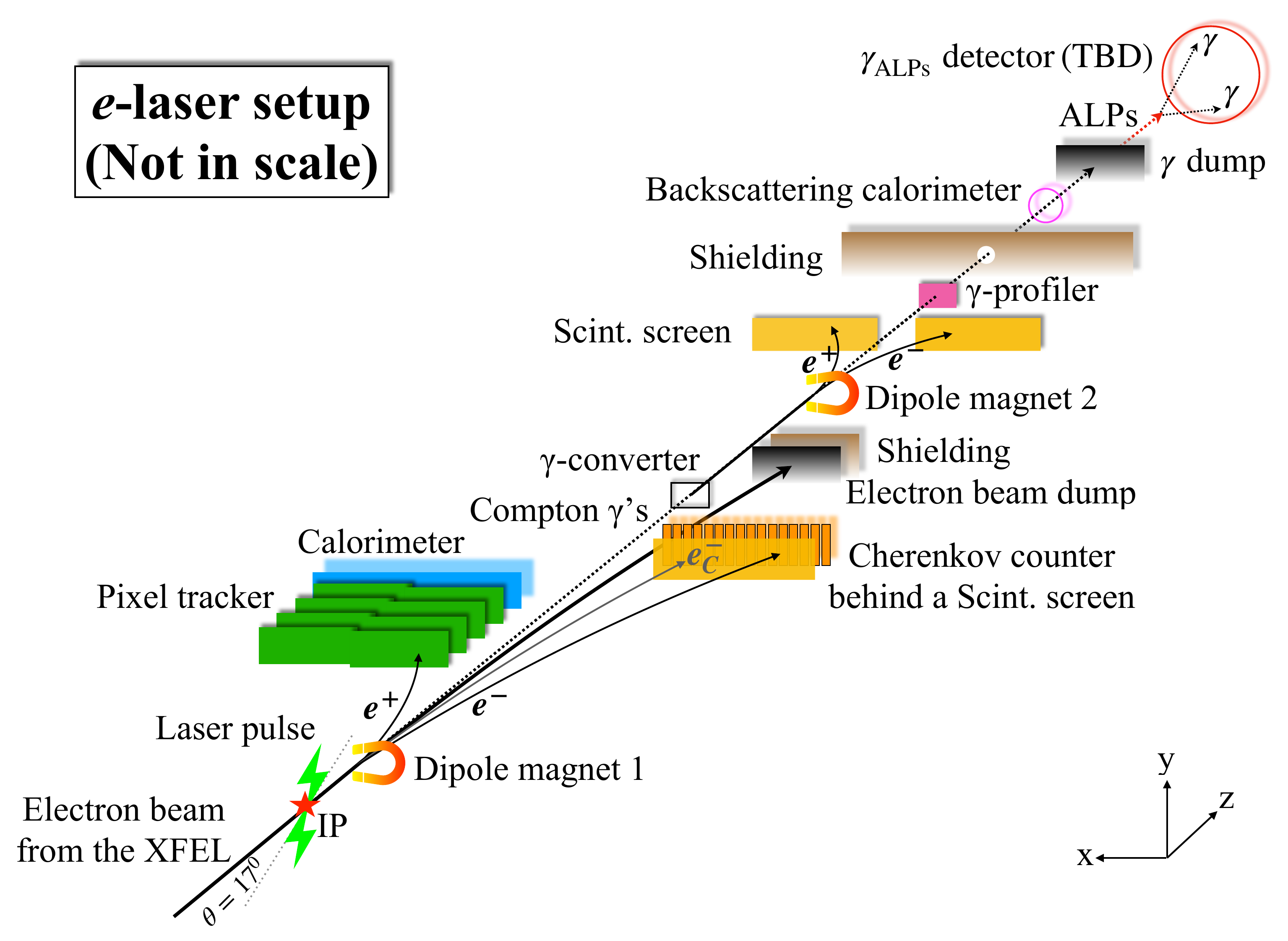} 
\includegraphics[width=0.48\textwidth]{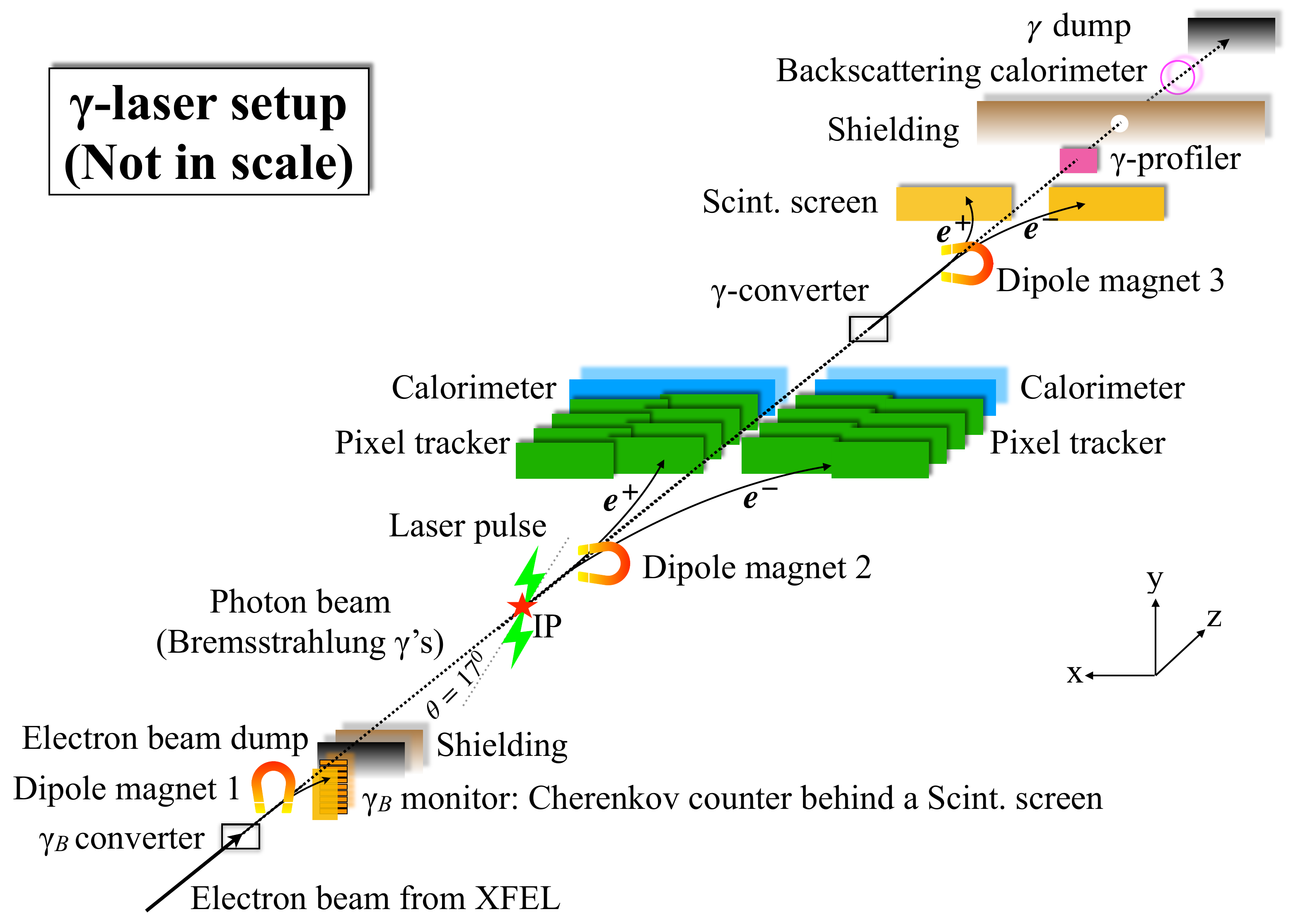} 

\caption{Schematic layouts for the \elaser{} and \glaser{} setup. Shown are the magnets, detectors and main shielding and absorbing elements.}

\label{fig:layoutdet} 
\end{figure}

Sketches of the layout of the experiment for the \elaser and the \glaser modes are shown in Fig.~\ref{fig:layoutdet}. 

There are three separate areas in the LUXE experiment that are required for the strong-field QED aspects of the research programme:
\begin{itemize}
    \item Target area: this area is only relevant for the \glaser{ } mode. A target chamber is installed for the purpose of converting the electron beam to a photon beam either via bremsstrahlung on a tungsten target or via inverse Compton 
    scattering on a low-intensity laser beam. 
    
    Behind the target chamber there is a dipole magnet to separate electrons, photons and positrons. A scintillation screen and camera and a Cherenkov detector is installed to measure the electron flux which can be used to reconstruct the energy spectrum and measure the flux of electrons after the bremsstrahlung process. In the \elaser mode the magnet is turned off and the electrons proceed to the IP. 
    
    \item IP area: at the IP, electrons, positrons and photons are expected to be produced. Behind the IP there is a dipole magnet to separate these three particles physically in the horizontal plane. Behind the magnet on the left-hand side ($x>0$, the ``positron side''), for both the \elaser and \glaser setup there is a system to measure the number of positrons and their energy spectrum. It consists of a tracker (see Chapter~\ref{chapt4}) and a calorimeter (see Chapter~\ref{chapt5}). On the right-hand side ($x<0$, the ``electron side''), in the \elaser case there is a scintillator screen with a camera (see Chapter~\ref{chapt6}) and a Cherenkov detector (see Chapter~\ref{chapt7}). The positron rates vary between $10^{-4}$ and $10^4$ per bunch crossing, and thus the technology chosen must be able to achieve an excellent background rejection (ideally $<10^{-3}$ per bunch crossing) to facilitate measurements at very low multiplicities, and achieve a good linearity at high multiplicities. A tracker followed by a calorimeter was chosen as these technologies are adequate for such particle multiplicities and are very effective at reducing backgrounds from stray secondary and tertiary particles. On the electron side the particle flux is much higher for the \elaser setup and can reach values as high as $10^9$ per bunch crossing. Therefore the technologies used here are designed to be adequate for such high fluxes. It is important that the position resolution in these detectors is high so that the energy resolution is better than 2\% in order to be able to resolve the Compton edges, see Chapter~\ref{chapt7}. This motivated the choice of  Cherenkov detectors and scintillation screens. 
    
    In the \glaser case, we expect $e^+e^-$ pairs to be produced with fluxes between $10^{-4}$ and $\sim 10$ $e^\pm$ per BX. Thus, here a tracker and a calorimeter is used on both the positron and the electron side. Due to the low multiplicity of electrons and positrons it is possible to combine the electron and positron and thus reconstruct the energy of the initial photon event by event.  
    
    \item Photon area: any photons that are produced at the IP in the \elaser mode (the Compton photons) or that fly through the IP area undisturbed in the \glaser mode end up in the photon detection system which is designed to measure the photon flux, angular and energy spectrum. The energy spectrum is determined using  a gamma spectrometer (see Chapter~\ref{chapt8}) consisting of a small tungsten target which converts photons to $e^+e^-$ pairs, a dipole magnet that separates them, and two scintillation screens (with cameras) that measure the electron and positron energy spectra. It is followed by a gamma profiler (see Chapter~\ref{chapt9}), made up of two sapphire strip detectors which are designed to measure the size of the photon beam in the horizontal and transverse direction. For a linearly polarised beam the photon spot size in the two planes can be used to get an independent measurement of $\xi$. The sapphire detector granularity is chosen such that this can be obtained for each shot with a precision of 5\%. Finally there is a gamma flux monitor (see Chapter~\ref{chapt10}), which consists of lead-glass calorimeter blocks that were originally designed for the HERA-b experiment but were never used there. It will measure the overall photon flux and also serve as a type of luminosity monitor as the photon flux is a very good measure of the overlap between the electron beam and the laser for the \elaser setup. The gamma profiler and the gamma spectrometer are also sensitive to the photon flux. 
\end{itemize}

Three beam dumps are required to dump the electron or photon beams. They are designed to reduce back-scattering in the IP and photon area sections where signal particle rates are low. Furthermore, shielding components are placed to further reduce the background radiation based on a full simulation of the experimental area with the \geant program~\cite{Allison:2006ve, Allison:2016lfl}.

A common data acquisition system is used (see Chapter~\ref{chapt11}). For 10~Hz of data taking, the data rate is about 200 MB/s and is dominated by the scintillation screens. The readout of each detector and of each laser diagnostic is controlled by a dedicated PC which is connected to a central ``Trigger Logic Unit'' (TLU), which was developed within the EU AIDA-2020 project, which receives a clock from the \euxfel accelerator and distributes the signal to the detectors and the laser. A central PC receives data from the TLU and builds the events and enables monitoring of the data. The data are then written to tape storage. 

The LUXE experiment is very modular in that it consists of a series of independent pieces which are readily accessible when access to the tunnel is allowed. It is thus possible to stage the installation, depending on when a given detector piece is needed and/or ready. For all components, it is foreseen that the installation is prepared and tested very well on the surface ahead of time so that the time needed for installation in the tunnel is minimised. 

The readout boards are mostly placed in a counting room on the 2nd floor that is accessible during \euxfel operation.

\begin{figure}[htbp]
\centering
\includegraphics[width=0.48\textwidth]{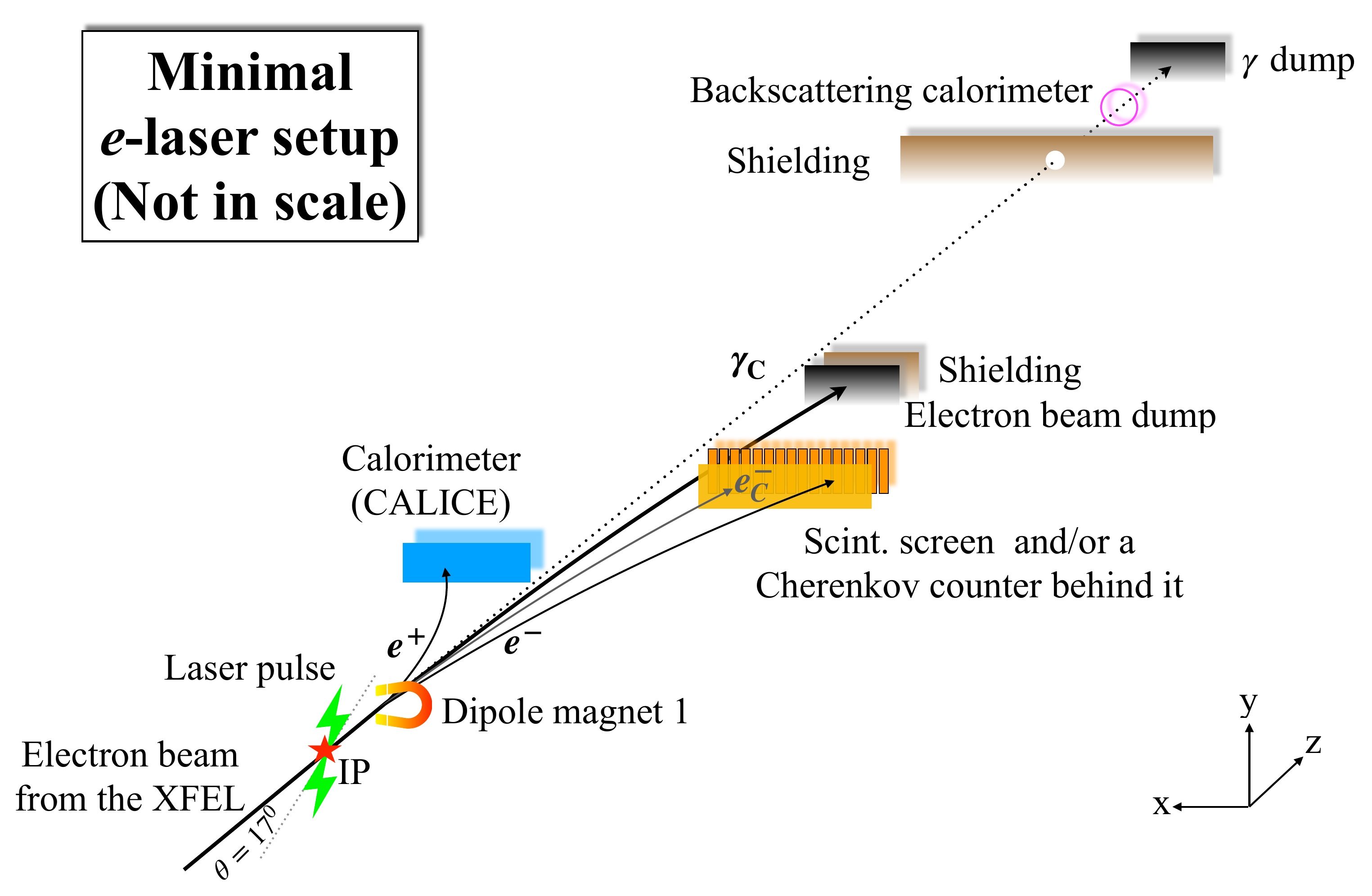} 
\includegraphics[width=0.48\textwidth]{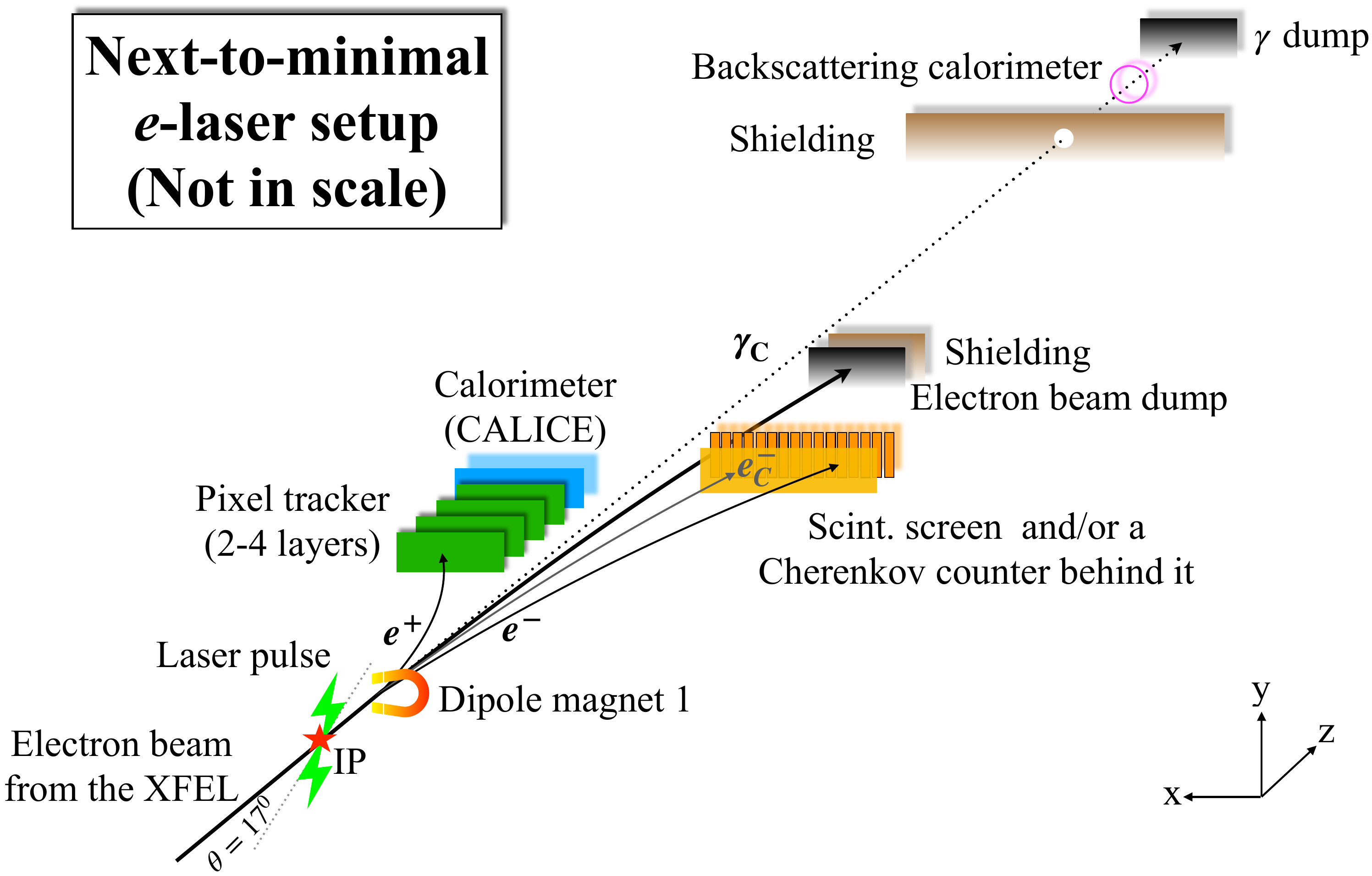} 
\caption{Schematic layouts for a minimal (left) and a next-to-minimal (right) version of the LUXE experiment for the \elaser setup. Shown are the magnets, detectors and main shielding and absorbing elements. 
} 
\label{fig:luxebare} 
\end{figure}

Given the possible uncertainties on the time scales on which all the detectors can be ready, it is prudent to consider a \textit{minimal} version of the LUXE experiment which could be ready for first data. A conceptual sketch of such a minimal version is shown in Fig.~\ref{fig:luxebare}.

This minimal version could be used for a first measurement of the Compton process via the measurement of the electron spectrum with the scintillation screen and/or Cherenkov detector. Additionally, the already existing prototype of the CALICE~\cite{Irles:2019ikz} calorimeter could be installed on the positron side and could be used to measure the number of positrons within the limited energy range of about $4-8$~GeV~\footnote{The existing prototype of the CALICE calorimeter is only 18~cm long and thus cannot cover the entire energy spectrum. However, it is planned to construct a 36~cm prototype by 2024 which could be used and would cover about 80\% of the spectrum.}. Additionally, the back-scattering calorimeter (gamma flux monitor) provides a measure of the photon flux. It is important to stress that such a minimal version is only able to cover a fraction of the LUXE physics programme but it would already allow some physics measurements, even if several detectors are delayed due to e.g.\! delayed delivery of components, missing resources or technical difficulties. However, both the physics scope and the precision will be very limited compared to the design shown in Fig.~\ref{fig:layoutdet}, as discussed in Sec.~\ref{sec:physics}.

In principle, any scenario between the bare minimum and full version should be conceivable depending on the readiness of the components. For instance, if available, tracker half-layers could be installed in front of the calorimeter (see next-to-minimal version in Fig.~\ref{fig:luxebare}).

The data taking strategy needs to be optimised to achieve the best overall precision. For the Compton process the statistical uncertainties are negligible even for a single shot, since the number of electrons and photons exceed 1 million per shot (see Fig.~\ref{fig:rates}), yielding a 0.1\% statistical uncertainty. Thus these measurements can be made within minutes of data taking once the experiment is commissioned. For the Breit-Wheeler process the particles rates are significantly lower, and are $\ll 1$ at low $\xi \leq 2$, see Fig.~\ref{fig:rates}. Furthermore, the measurements of this process will be subject to a significant systematic uncertainty of 20-40\% due to the uncertainty on the laser intensity, see Sec.~\ref{sec:physics}. Therefore it seems reasonable to take data at a given laser intensity for a time period that allows a statistical uncertainty of better than 10\% to be achieved. Assuming no background, for \phaseone (\phasetwo) this can be achieved within a single day for $\xi\geq 1.5$ ($\xi\geq 1.2$). At lower $\xi$, it is important to collect statistics for an extended period of time. For instance a 30\% uncertainty at $\xi=1$ requires at least 10 days of data taking for \phaseone.   
Based on these considerations, we assume that about 80\% of the data taking will be devoted to the low-$\xi$ region ($\xi<2$) and only 20\% to data taking at high $\xi$. 

At the highest $\xi$ values, the particle rates are actually rather too high than too low in the \elaser mode. Here, the rates can be reduced by large factors by defocusing the electron beam. For instance, by increasing the beam spot size in the transverse plane from 5 to 20~$\mu$m, at the highest $\xi$ the rate can be reduced by a factor of 10.

\section{Physics Prospects}
\label{sec:physics}

The \textsc{Ptarmigan} Monte Carlo event generator is used to assess the physics sensitivity of the experiment as described in Sec.~\ref{sec:theory}. Unless otherwise stated, it is assumed that the full experiment as shown in Fig.~\ref{fig:layoutdet} is available. 

Three important goals of the experiment are the measurement of the following observables as function of the $\xi$-parameter: i) the Compton edges; ii) the number of photons radiated per electron in \elaser collisions and iii) the rate of positrons in \glaser and \elaser collisions. 

\begin{figure}[htbp]
\centering
\begin{minipage}[c]{0.45\textwidth}
\centering
\includegraphics[width=\textwidth]{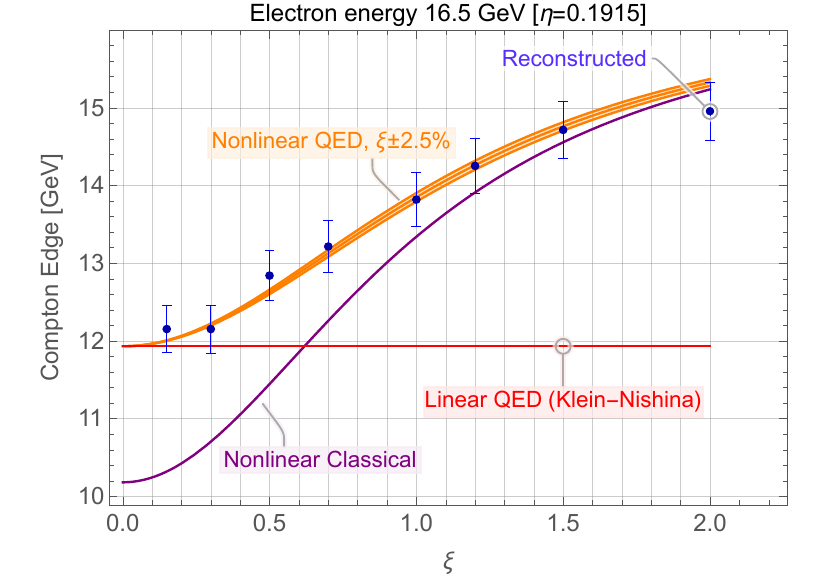}
\includegraphics[width=0.9\textwidth]{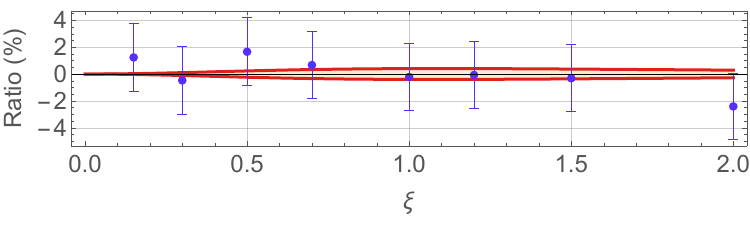}
\end{minipage}
\begin{minipage}[c]{0.45\textwidth}
\includegraphics[width=\textwidth]{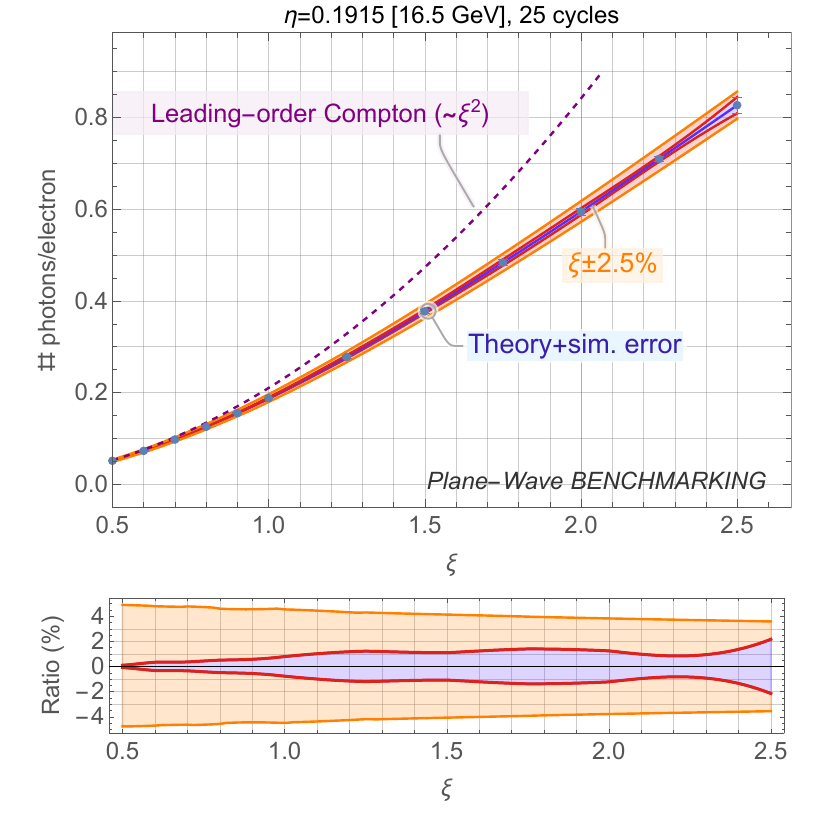}
\end{minipage}
\caption{Expected experimental results for the measurement of the Compton edge in the electron energy spectrum in \elaser  interactions (left) and the ratio of the number of photons produced per electron as a function of $\xi$ (right, where the number of cycles denotes the number of periods included by the laser pulse envelope). A 2.5\% uncertainty on $\xi$, corresponding to a 5\% uncertainty on the laser intensity, is illustrated in orange as uncertainty on the theoretical nonlinear QED prediction. The uncertainties on the data points illustrate a 2\% uncertainty on the electron energy scale. The predictions of the Klein-Nishina formula and a nonlinear classical predictions (Thomson scattering) are also shown. The bottom panels show the ratio to the central nonlinear QED prediction, and the uncertainty due to the uncertainty on $\xi$, and (on the right only) due to other uncertainties in the theoretical prediction.
} 
\label{fig:resultscom} 
\end{figure}

The Compton edges are measured in the IP area on the electron side by the scintillation screen and the Cherenkov detector, and in the photon area by the gamma spectrometer~\footnote{In the gamma spectrometer, the photon spectrum is inferred by inverting the measured energy spectra of electrons and positrons created by photon conversion.}. Examples of these reconstructed spectra are shown in Fig.~\ref{fig:elspec}. The edge is then identified based on a so-called finite-impulse-response-filter (FIR) method, which was proposed in the context of kinematic edge reconstruction at the International Linear Collider (ILC)~\cite{Tsai:1974}. The edge position as a function of $\xi$ is shown in Fig.~\ref{fig:resultscom} compared to the theoretical prediction of \sfqed, the linear (purely perturbative) QED prediction (Klein-Nishina formula) and the classical prediction (Thomson scattering). The Klein-Nishina prediction would yield a fixed value of 12~GeV for all $\xi$ while the full QED calculation increases versus $\xi$ approaching the classical prediction at high $\xi$. It is seen that a 2.5\% uncertainty in $\xi$ affects the Compton edge position by at most 0.5\%. The experimental uncertainties are expected to be about 2\% for the Compton edge position due to the uncertainty on the absolute energy scale.

Another interesting measurement is the relative fraction of photons radiated per electron versus $\xi$. As can be seen from the right-hand plot of Fig.~\ref{fig:resultscom}, whereas the leading-order perturbative result scales with $\xi^{2}$, the effect of the non-perturbative  coupling between the electron and the laser field can be seen as a suppression of the photon emission rate. Statistical uncertainties are expected to be negligible. The 2.5\% uncertainty on $\xi$ results in an uncertainty on the relative fraction of photons produced of up to 4\%. Thus the relative systematic uncertainty on the photon and electron flux normalisation should be kept below 2\% to avoid this being the dominant source. 

Another important measurement is the positron multiplicity versus $\xi$ for \glaser interactions. The goal is to observe Breit-Wheeler $e^+e^-$ production from collisions of real photons, and to measure the predicted non-perturbative effects at high $\xi$. 

\begin{figure}[htbp]
\centering
\includegraphics[width=0.48\textwidth]{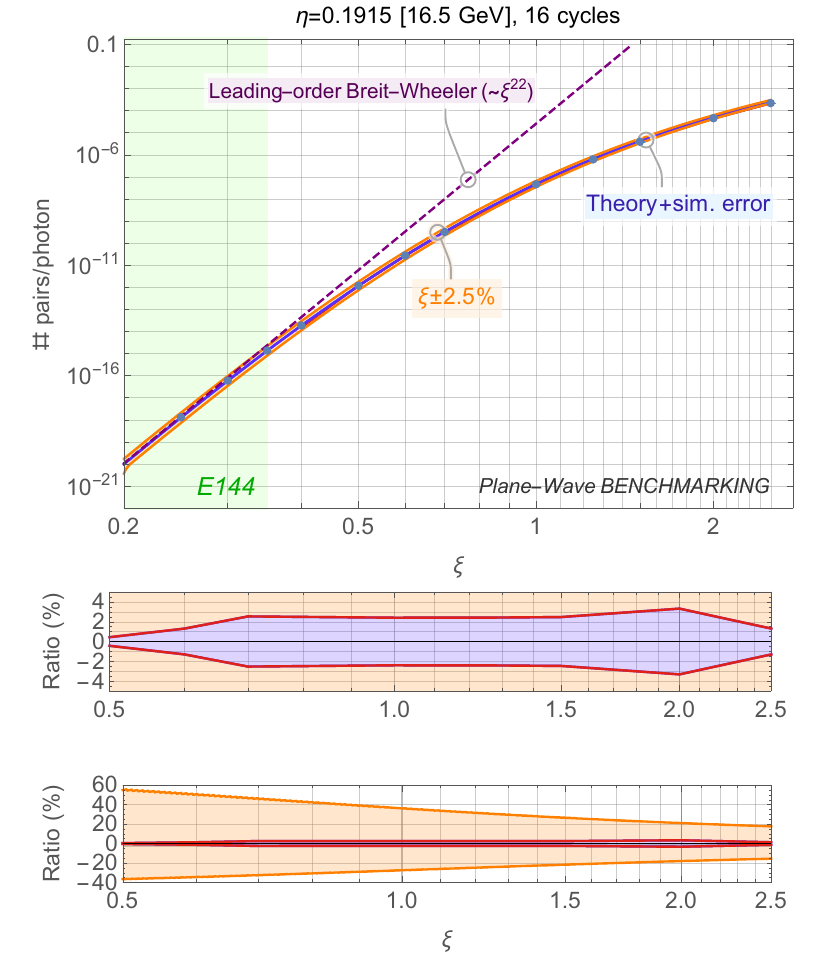}
\raisebox{+0.2\height}{\includegraphics[width=0.48\textwidth]{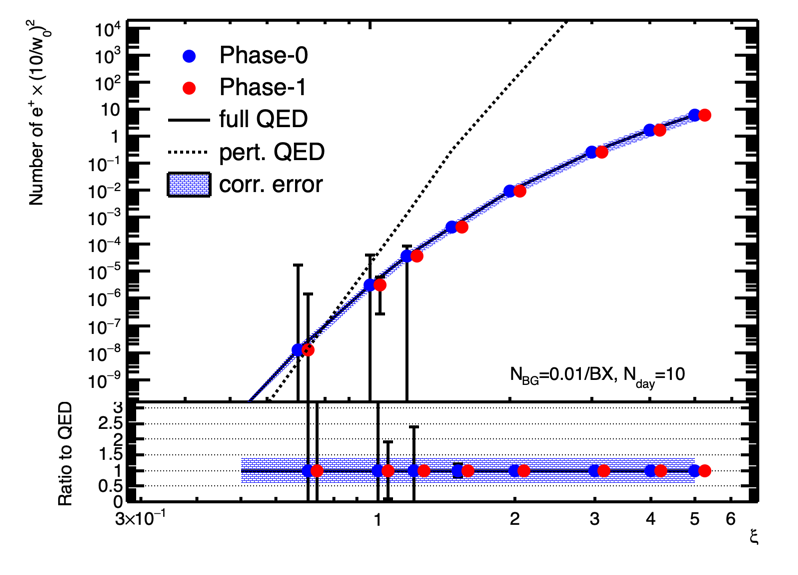}}
\caption{Left: Number of positrons produced in \glaser{} collisions as a function of $\xi$ as expected in the full QED calculation (solid line) and in the purely perturbative calculation (dashed line, in which the probability of pair creation scales as $\sim \xi^{22}$). Also shown is the impact of a 2.5\% uncertainty on $\xi$ (wide orange band). The bottom two panels show the ratio to the full QED prediction to make the uncertainties visible (the blue band is the difference between the full QED theory calculation and simulation).
Right: Number of positrons produced in \glaser{} collisions as function of $\xi$, normalised by the laser waist for phase-0 and phase-1. The data uncertainties shown represent the statistical uncertainties corresponding to 10 days of data taking and a background of 0.01 particles per bunch crossing. Also shown is a 40\% correlated uncertainty to illustrate the impact of the uncertainty on $\xi$.} 
\label{fig:resultsbw} 
\end{figure}

Fig.~\ref{fig:resultsbw} shows the expected results for the Breit-Wheeler process. The left figure shows the theoretical prediction and illustrates the importance of the uncertainty on $\xi$. Due to the very steep dependence of the $e^+$ rate on $\xi$, a 2.5\% error in $\xi$ can lead to a 20\%-60\% error on the number of positrons measured (with the error increasing, for lower positron counts). 
The uncertainties represent those anticipated if there is an irreducible background of $0.01$ particles per BX and for each $\xi$ value 10 days of data taking are used. A correlated uncertainty on the prediction arises from a 2.5\% uncertainty on the absolute value of $\xi$ (see left figure). In addition an uncertainty shot-by-shot of 1\% on the laser intensity, corresponding to 0.5\% on $\xi$, will result in an uncorrelated error between the data points of about 10\% at low $\xi$ decreasing to about 5\% at high $\xi$.

A minimal LUXE experiment, as illustrated in Fig.~\ref{fig:luxebare},  would be able to measure the Compton edge, as shown in Fig.~\ref{fig:resultscom}, although likely the uncertainty on the theoretical prediction would be larger due to a less well-known laser intensity (since the gamma profiler cannot constrain it). And, the measurement could only be performed with electrons and not with photons due to a missing gamma spectrometer. It would also be able to measure the positron rate albeit in a restricted energy range and likely with a larger uncertainty due to inferior background rejection.

\begin{figure}[htbp]
\centering
\includegraphics[width=0.48\textwidth]{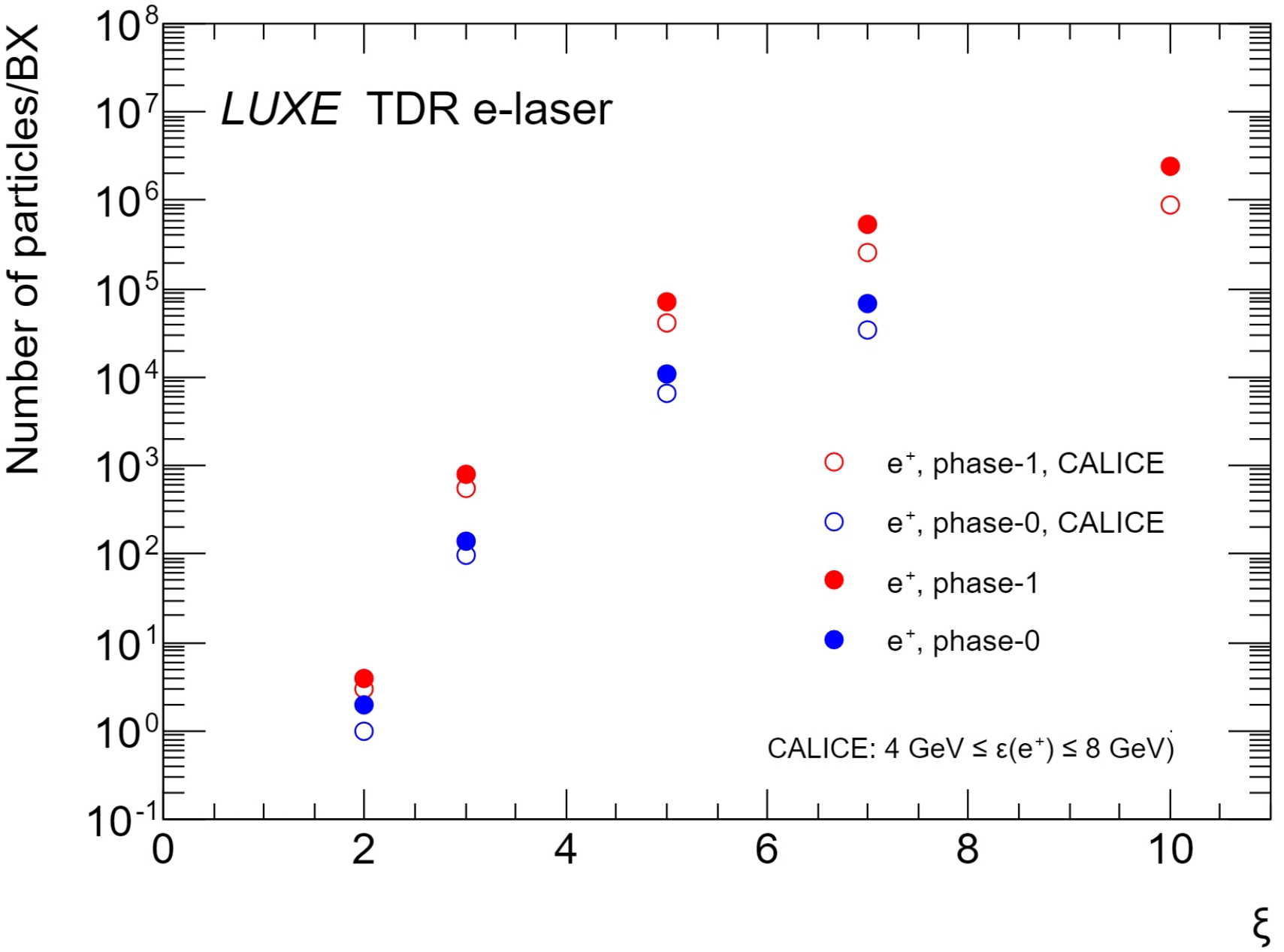}
\includegraphics[width=0.48\textwidth]{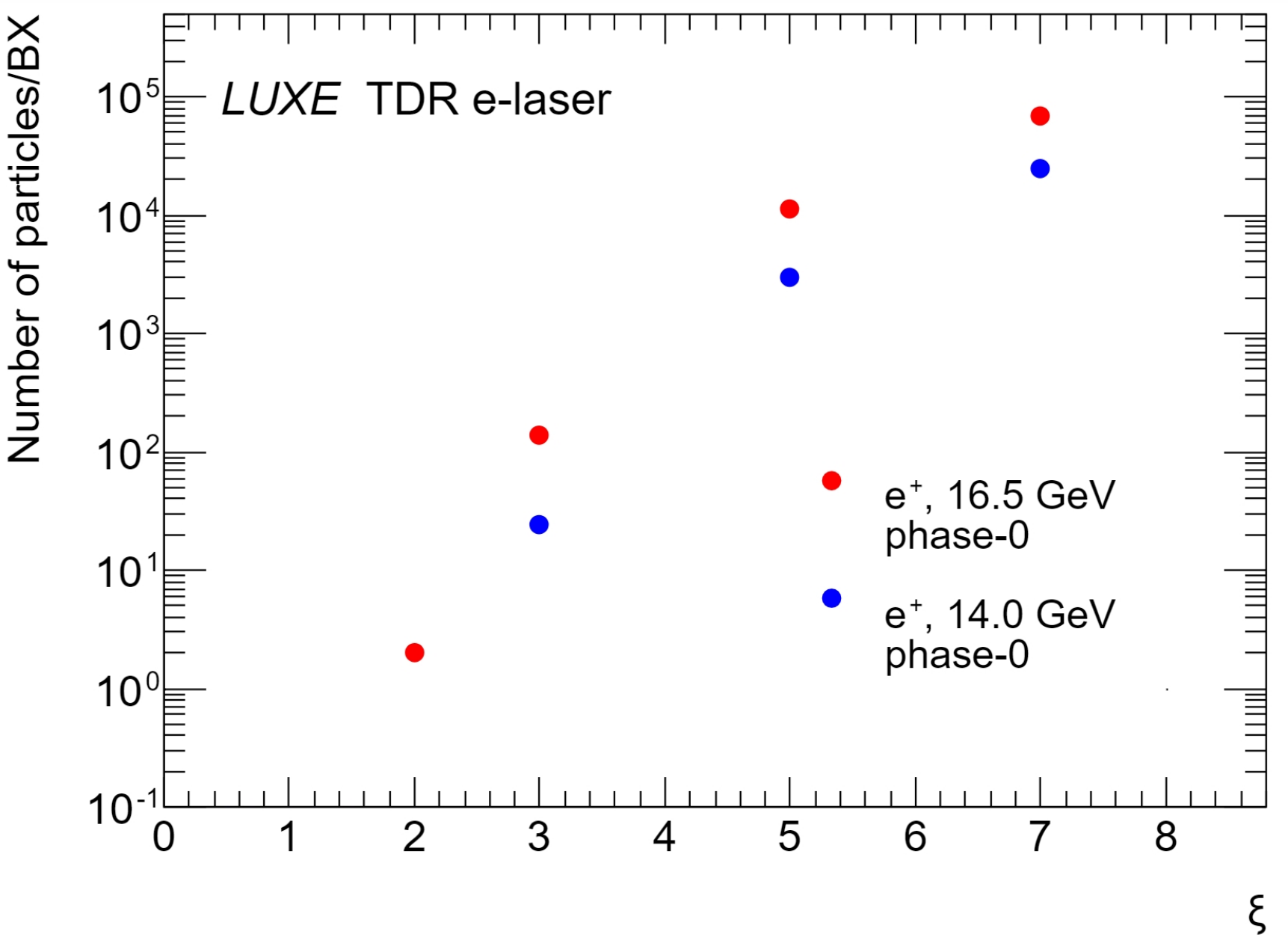}
\caption{Left: Number of positrons produced in \elaser collisions as function of $\xi$ for a restricted energy range (CALICE: 4-8 GeV) and the full energy range. 
Right: Number of positrons produced in \elaser collisions as function of $\xi$ for an electron beam energy of 16.5~GeV compared to 14~GeV. 
} 
\label{fig:resultsextra} 
\end{figure}

Figure~\ref{fig:resultsextra} shows the expected positron rates in the \elaser mode in the restricted energy range compared to the full range, and also shows a comparison of the positron rates with the electron energy of 14~GeV versus 16.5~GeV. It is seen that the shape of the distribution is not changed significantly by reducing the positron energy range. The lower electron beam energy reduces the particle rates by factors of 5-10 depending on $\xi$. 

In addition to the QED physics prospects discussed above, a first study of LUXE's sensitivity for a few scenarios of new physics is discussed in~~\cite{LUXEBSM}.
It will be sensitive to the pseudoscalar ALP couplings in the region of $10^{-5}\,\textrm{GeV}^{-1}$ for pseudoscalar masses around $\sim 200\,\textrm{MeV}$, a parameter space as yet unexplored by running experiments. It also presents a novel and complementary way compared to classic dump experiments, e.g. compared to dumping the electron beam directly this method results in significantly lower backgrounds.

\section{Conclusion}
LUXE will pioneer investigation of a region of QED and be able to measure the transition from the perturbative to the nonperturbative regime in various observables. It makes use of the existing \euxfel accelerator electron beam and a high-intensity laser to study the strong-field regime which is relevant for a variety of astrophysical phenomena, atomic physics and future high energy electron accelerators. Detectors are designed to achieve a high signal efficiency and a good background rejection. Risks due to delays of components are mitigated by planning  for redundancy in the measurements and for a staged installation.

\printbibliography
\end{refsection}


\chapter{Laser \& Diagnostics}
\label{chapt3}
\begin{refsection}

\chapterauthors{Xinhe Huang \\\desyaffil, \HIJenaaffil, \UJenaaffil}{Rajendra Prasad \\\desyaffil}{Matt Zepf \\\HIJenaaffil, \UJenaaffil}
\date{\today}


\begin{chaptabstract}
    Attainment of the non-linear QED regime by approaching the critical field strength and beyond (with $\chi \approx 1$) requires a state-of-art high intensity laser with a laser power of ultimately 350 TW. LUXE has the unique advantage of a continuous data-taking mode, with electron bunches arriving at a rate of 10 Hz, allowing the study of the non-linear QED regime with unprecedented precision.  This poses important requirements on the laser system and, in particular, its diagnostics. In  this chapter we describe the envisaged laser system and diagnostics to match the design goals of LUXE as well as aspects of the technical implementation such as beam transport and interaction chamber.
\end{chaptabstract}

\section{Introduction}

\begin{figure}[htbp]
    \centering
    \includegraphics[width=0.7\textwidth]{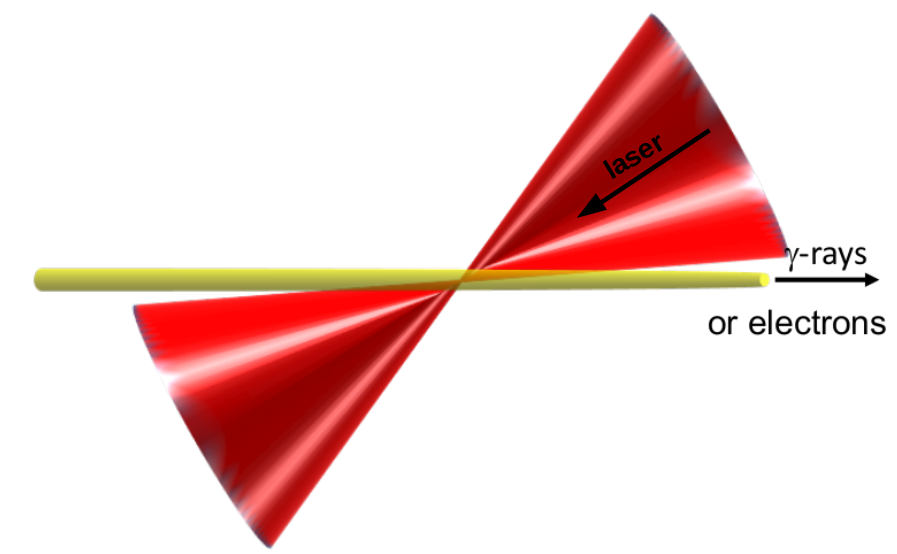}        
    \caption{The collision geometry for LUXE. Particles with GeV energy (photons or electrons) collide with a near counter-propagating intense laser pulse which can exceed the Schwinger field in the collision frame of reference.}
    \label{fig:Collision_geom}
\end{figure}

Exploring the phenomena associated with strong-field QED (\hiqed) requires reaching intensities close to and exceeding the Schwinger critical field: $\mathcal{E}_{\textrm{cr}}=m^{2}c^{3}/(e\hslash)=1.32\times 10^{18}\,\textrm{V\,m}^{-1}$. As detailed  in Chapter~\ref{chapt2}, such extreme fields are achievable in the boosted frame of a high energy electron colliding with an intense laser field. The \euxfel  accelerator provides electrons (or $\gamma$-ray photons)  in the range up to 17.5~GeV.  Reaching, and importantly, exceeding $\mathcal{E}_{\textrm{cr}}$ can be achieved using  state-of-the-art high-intensity laser with peak power of 350~TW capable of reaching intensities of $10^{21}$ W\,cm$^{-2}$. The LUXE experiment will exploit the existing European XFEL infrastructure at the Osdorfer Born facility where a high power laser will be installed in a dedicated clean-room and transported to the interaction point. In Fig. \ref{fig:Collision_geom}, the interaction point of the experiment is illustrated.

\begin{figure}[ht]
    \centering
    \includegraphics[width=0.9\textwidth]{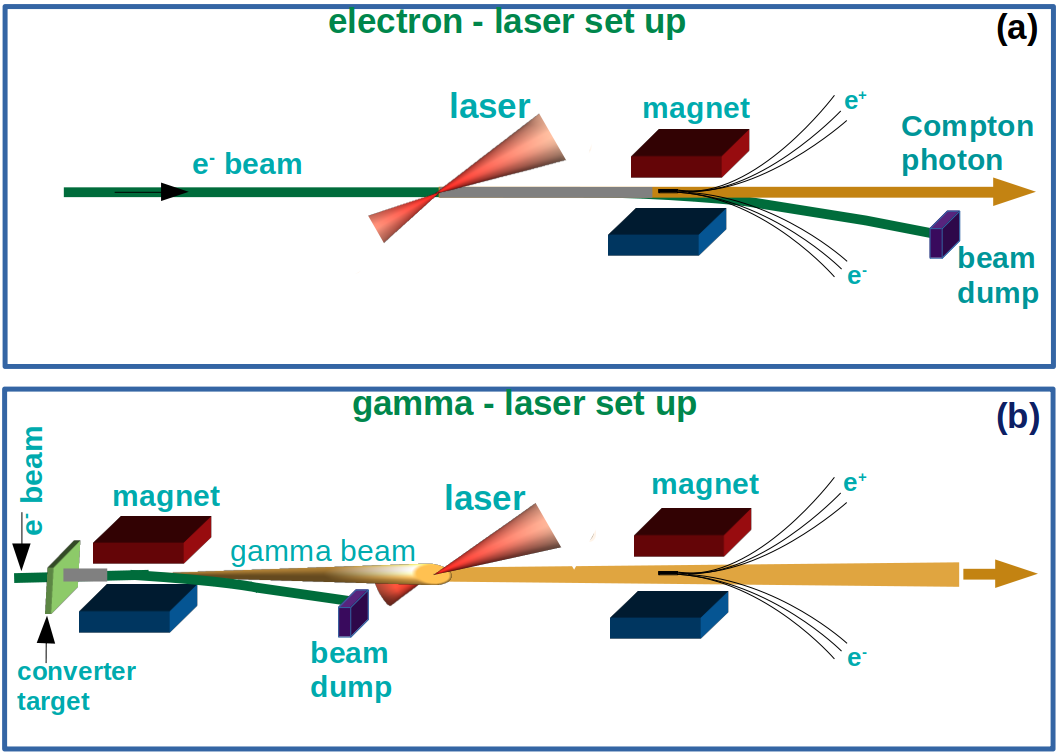}
    \caption{Interaction geometry between the high-power laser and GeV beam options of either high-energy electrons (a) or photons (b). The laser beam is focused by an off-axis parabolic mirror to a high intensity point focus and recollimated for on-shot analysis. The typical $\gamma$-ray beam diameter will be around 200 $\mu$m while the laser spot can be focused to spots as small 3~$\mu$m for the highest $\xi$. The longitudinal dimensions characterising the interactions are the electron bunch length of  $ct \sim 12$~$\mu$m and laser pulse length $ct \sim 9$~$\mu$m and Rayleigh range $z_R=35\, \mu$m.}
   \label{fig:interaction_geometry}
\end{figure}

The two distinct geometries of the LUXE interaction point (IP) are schematically illustrated in Fig.~\ref{fig:interaction_geometry} (for a more detailed description see Fig. 2.14). 
In the first case (Fig.~\ref{fig:interaction_geometry}a) the laser pulse collides directly with the electron beam. This is the so called ``e-laser'' mode in which the electron beam energy is in the 8-17.5\,GeV range by \euxfel operations (cf. CDR \cite{cdrluxe2021}). In Fig.~\ref{fig:interaction_geometry}(b), the electron beam  first interacts with a bremsstrahlung converter target (or alternatively a secondary, low intensity laser pulse to  produce $\gamma$-rays via inverse Compton scattering (ICS)). The primary electrons are then separated from the experimental axis and the high energy photons collide with the high intensity laser pulse (``$\gamma$-laser'' mode). 


These two configurations  enable studies of all foreseen physical processes at LUXE with the help of state-of-art charged particle/photon detectors situated downstream of the interaction point.  
The fundamental interaction geometry at the IP  is identical in both e-laser and $\gamma$-laser modes. The main difference in terms of laser requirements is the spot-size of the GeV particles. These vary from  10~$\mu$m for the e-laser case to $>100\, \mu$m  for the $\gamma$-laser configuration. This compares to a focal laser spot size of the order of 3~$\mu$m, illustrating that achieving and maintaining overlap is expected to be strforward in the $\gamma$-laser mode, while requiring stringent engineering and, potentially, active stabilisation in the e-laser mode. 

\begin{figure}[ht]
   \centering
   \includegraphics*[height=8.4cm]{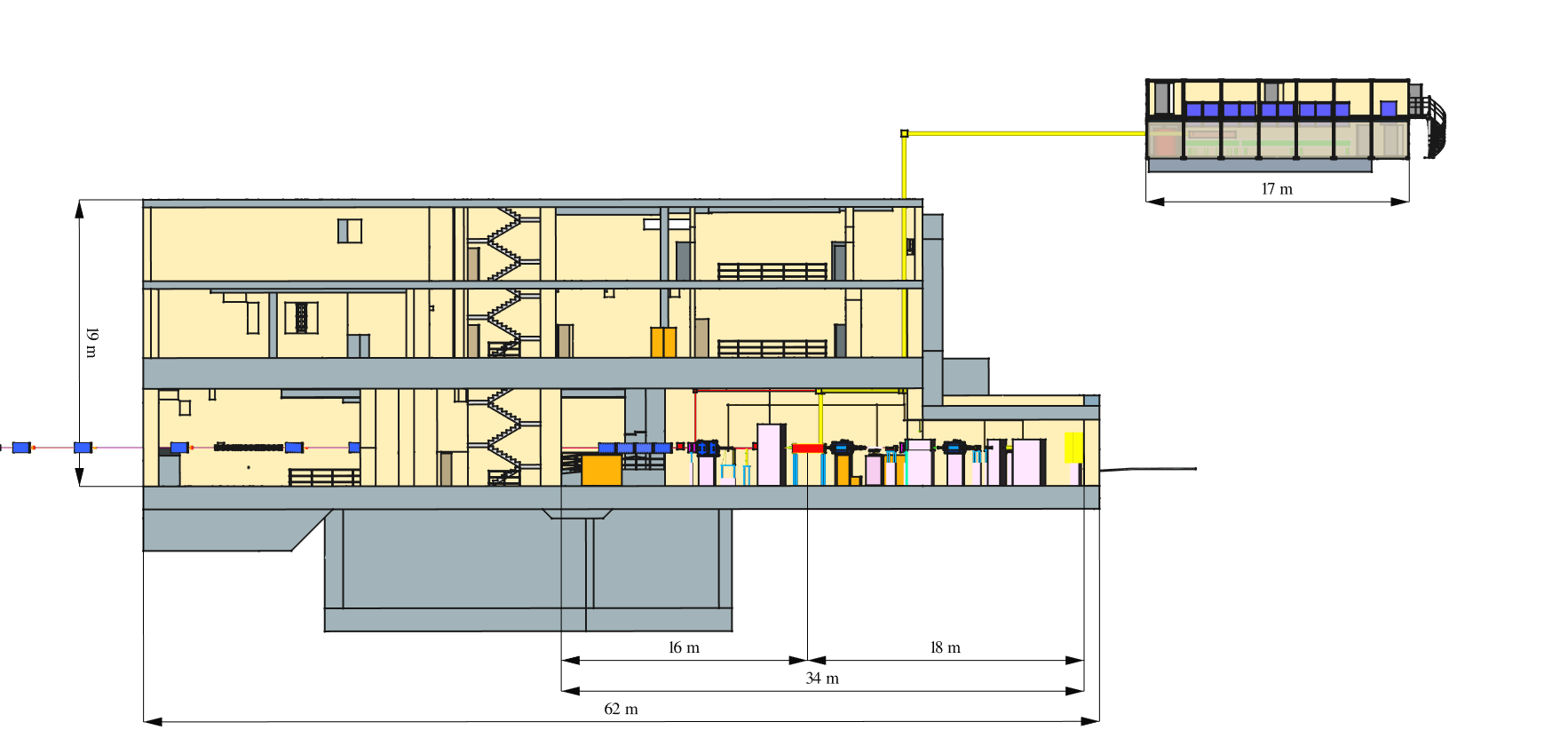}

    \caption{Side view of the installation at Osdorfer Born in the preferred configuration. The clean-room is shown in a temporary annex on the surface with the beamline in yellow running vertically from the ground level to the -3 level, where the interaction chamber is shown in red. The large grey shaded area below level -3 denotes the location of the \euxfel beam dump (for details see Chapter~\ref{chapt12}).}
    \label{fig:XS1_side_intro}
\end{figure}

In Fig. \ref{fig:XS1_side_intro} the laser clean-room and interaction area in the building at the Osdorfer Born are illustrated.

Apart from the installation at \euxfel and integration into LUXE, a significant effort is required on  precision laser diagnostics.

In this Chapter, we detail the required laser system together with the diagnostics and their implementation. The installation in the particular location of the laser clean-room and the beamline paths will be detailed in Chapter~\ref{chapt12}. Here we detail the optimal choice for the installation. 


\section{Requirements and Challenges}

The structure of this Chapter is as follows.  We first provide an overview of each area and relate the requirements to the experimental aims of LUXE. We then provide a description of the laser system options (Sections~\ref{section3-3} and~\ref{section3-4}) and describe how the requirements will be technically met in Sections~\ref{section3-5} and~\ref{section3-6}. Operational and organisational considerations are dealt with in the final sections.

\subsection{Laser System Requirements}
The headline performance requirements were scientifically motivated and detailed in the LUXE Conceptual Design Report~\cite{cdrluxe2021}, as well as in Chapter~\ref{chapt2}. These can be summarised in terms of the interaction parameters for the normalised vector potential $\xi$-values between $\sim 0.1$ and $> 20$ and  quantum parameter $\chi$ values in the range of $0.3 < \chi < 5$. These values allow the transition from perturbative behaviour to the non-perturbative regime at larger values of $\chi$ to be probed.

The peak power and pulse duration that the laser must achieve to reach these required values of $\xi$, and $\chi$ are determined by the interaction geometry and parameters of the GeV beam. The smallest required focal spot size is set by the interplay of peak focused intensity, focused e-beam size and stability requirements (the laser beam waist must be within the e-beam focus). This can be formulated by requiring that the 3$\sigma$ pointing jitter radius is not larger than the electron beam focus of 10 $\mu$m.

The pulse duration of the laser system is determined by achieving maximum overlap with the GeV beam. It optimises  for minimum laser energy when the laser pulse is matched to the pulse duration of the GeV beam. The high intensity will need to be maintained over a longitudinal extent along the experimental axis (given by the confocal parameter $b=2 Z_R=2\pi w_0^2/\lambda$, here $Z_R$ is the Rayleigh range, $\omega_0$ is the laser beam waist and $\lambda$ is the laser central wavelength) that is matched to/greater than the  GeV beam duration of $ L_\textrm{bunch}=40~\textrm{fs} \, \cdot c \sim 12\,\mu$m. This condition is met for the minimum chosen beam waist of $w_0=3\, \mu$m.

The laser system will use up to three auxiliary beams (detailed below) for timing, pointing stabilisation and inverse Compton scattering.
We note that the operating wavelength of the main laser system is uncritical so long as $\xi$ and $\chi$ requirements are met.

Table~\ref{tab:laser} details the chosen match to these requirements for using a titanium-sapphire laser system, currently the  high power laser system platform with the highest level of maturity. Typical operating wavelength of such  lasers are 800 nm with a bandwidth of $\approx$ 120 nm (Ti:Sa provides usable gain from beyond 1000 nm to almost 700 nm).

 The LUXE laser architecture is inherently modular and allows the installation to progress in two phases with the initial laser (phase-0) giving access to the regime up to $\chi \approx 1$ with the full implementation (phase-1) reaching into the $\chi > 1$ regime as shown  in Table~\ref{tab:laser}. The  laser in the first column (phase-0) will have an energy of $1.2$~J and a power of $\sim 40$~TW. During the later phase (phase-1), the laser system requires peak power of $350 $ TW to reach peak values of $\chi \sim 4$. These values in the table are based on a realistically achievable peak intensities for this class of laser. Reduction of peak intensity due to transport losses and non-ideal focal energy distribution have been factored in.

Testing QED theory in the strong field regime requires controlled parameter variation with  intensity and pulse duration scans essential to the LUXE experimental goals. While the main laser parameters of comparable systems such as JETI200 are stable to the percent level ($1-3\%$), the RMS intensity fluctuations in focus do not meet  the LUXE design goal of $<1\%$. As shown below, the dominant contribution to the intensity uncertainty are variations in the focal spot size. To achieve the required level of precision one can consider either modifying the laser to improve stability or to measure the intensity on each shot. As the perturbation of the laser beam during the interaction in the IP is negligible (less than one photon is scattered per incoming GeV particle), we will achieve the required precision by measuring the laser parameter on each shot with precision  diagnostics (see relevant sections) rather than embarking on a laser development program.

The laser repetition rate will match the \euxfel  macro-bunch structure of 10 Hz. Due to technical (grating thermal load \cite{Leroux:2020gd}) and cost optimisation constraints  full energy shots will be fired only at 1 Hz repetition rate. The remaining 9 Hz of GeV bunches will be dedicated to determining precision statistics on the background  signal. The  reduced ``on target'' repetition rate prevents  deformation of laser wavefront due to thermal heating and hence ensures stable operating conditions. The amplifiers of the laser will however operate at 10 Hz. This ensures  stable operation, full repetition rate on the timing feedback signal and ICS optimisation. 
Since the repetition rate  is thermally limited by the compressor, higher repetition rates than 1 Hz are possible at reduced pulse energy. This allows data taking at enhanced repetition rate at reduced intensity. The latter operating mode (partially) compensates for the reduced signal rate and allows good statistics at lower intensity while maintaining the spatio-temporal shape of the laser in focus.     \\

\subsubsection{Precision Laser Diagnostics}

The intensity scaling of experimental observables constitutes a key test of theory and is therefore central to the LUXE project.
 The shot-to-shot reproducibility of the peak laser intensity is limited by phase and pulse energy fluctuations in the  laser system. Precision measurements with excellent statistics require the interaction conditions to be known with high precision for each shot. For the accelerator the  reproducibility of the bunch parameters is sufficiently high, especially since the experimental rates are only linear in the bunch charge and focused beam intensity. 
By contrast, the interaction has a strong non-linear dependence on the laser intensity (see Fig.~\ref{fig:resultscom} and~\ref{fig:resultsbw}). Therefore the intensity and intensity distribution must be known on each shot with high precision. The primary source of shot-to-shot variations are fluctuations in the spatial phase across the beam. These are the result of air currents, small vibrations in the laser and fluctuations in the pump intensity distribution due to the multi-mode nature of the high energy pump lasers which results in small stochastic intensity variations of the pump profile and associated refractive index fluctuations in the amplifier.

 We will ensure that the intensity of each data shot is known with high precision by using commercial lasers with their already high quality as a basis for the LUXE laser system. We will characterise each shot to determine the actual intensity in the IP with high precision.  This approach is excellently suited to LUXE, as the laser beam itself is almost not affected at all during the interaction with the GeV beam. The goal is to have a relative precision on the shot intensity measurement below 1\%, ideally below 0.1\%. Averaging over many shots with the same nominal intensity as measured by our diagnostic will further improve the relative precision of experimental data. 

The absolute calibration of the laser intensity is another crucial element for comparing theory and experiment. Several key measurements will be reported as a function of the intensity parameter $\xi$ which depends on the square-root of the intensity. A 5\% uncertainty in the intensity corresponds to a 2.5\% uncertainty on $\xi$. Figure~\ref{fig:resultscom} and~\ref{fig:resultsbw} show three key observables LUXE aims to measure versus $\xi$: the $e^+e^-$ rate in $\gamma$-laser collisions, the number of photons radiated per electron and the Compton edge versus $\xi$ in e-laser collisions. Also shown is the impact of a systematic  uncertainty of $\xi$ by 2.5\% (absolute).

 Uncertainty in the absolute intensity results in a correlated shift of all data points (shown here for 2.5\%). For the Breit-Wheeler process the expected impact ranges between 60\% at low $\xi$ and 20\% at high $\xi$. For the Compton process the impact is only about 4\% for the ratio of photons to electrons and 0.5\% for the Compton edge position. Absolute cross-calibration can be obtained either from cross-calibration with known non-linear effects for attenuated beams or with effects such as ponderomotive scattering which provide an intensity signature at high $\xi$.


\subsubsection{Tunnel Diagnostics}
\label{sec:laser:overlap}
Achieving -- and maintaining -- spatio-temporal overlap between the laser focus and the electrons (or the bremsstrahlung or ICS photons) at the IP is an essential pre-requisite for precision data acquisition. Overlap in space requires an alignment, positioning and drift compensation system for the laser focus, to track the electron bunches. The timing system must ensure simultaneous arrival at the IP of the laser pulse (with nominal 30~fs duration) and the electron bunch (with typical length of $\sim 10$\,$\mu$m, corresponding to $\sim 40$~fs duration). Considering the geometry of the \elaser collisions at an angle about $\theta \simeq 0.3$\,rad and a spot size of $8\,\mu$m, a timing accuracy of $\sim 30$~fs is required. This overlap will be achieved with the target area diagnostics.

\subsubsection{Automation}
The goal of high precision and excellent statistics for LUXE necessitates long data taking runs,  particularly at low $\xi$ where event rates can be $\ll 1$ per bunch crossing. However, slow thermal drift  or residual vibrations may cause drift in laser beam parameters such as pointing, energy, pulse length etc. To maintain steady operation, feedback loops will be implemented to maintain laser performance in a narrow window over long periods.
Similarly, pointing drift in the beamline will be stabilised by an active system with predictive capability to maintain the laser focus in the same spatial position with high precision.
The temporal overlap will be maintained using the excellent timing system at \euxfel to maintain accurate timing between the oscillator and the accelerator. Pathlength drift will be measured with an electro-optical system coupled to a delay line with translation stage that can compensate beam pathlength drift at the 10 Hz level.



\subsection{Auxiliary Laser Beams}
In addition to the main laser beam with up to 350 TW, LUXE requires up to 3 auxiliary beams.
\begin{itemize}
    \item Timing Beam \\
    The timing beam will probe the induced polarisation rotation in the timing crystal. 
    \item Pointing Stabilisation Beam\\
    To allow for active stabilisation a CW beam with small diameter will propagate parallel to the main beam and be used for active stabilisation of the beamline if required.
    \item ICS beam \\
    The ICS beam is an option for phase-1 and will allow the bremsstrahlung source to be replaced by an inverse Compton scattering source with more precisely defined spectral characteristics. The beamline will be designed to accommodate this laser from the outset.
\end{itemize}

\subsection{Laser Beam Transport}

The laser beams will be transported in vacuum pipes, with a path optimising and minimising the required passage through the radiation shielding. The pointing jitter requirement detailed above dictates isolation of the mirrors from the beamline and the use of vibration isolation elements between mirror mount and wall mounting.

\subsection{Interaction Point Chamber and Inverse Compton Scattering Chamber}
The interaction point chamber (IP chamber) is  the central part of the experiment where beam crossings take place. The IP chamber must be large enough to accommodate the input and return beam paths and IP overlap diagnostics consisting of alignment targets and a microscope objective.
This requires ultra high vacuum (UHV) compatible translation stages and electronic vacuum feedthroughs for moving the IP diagnostics into the beam path.  The diagnostic cameras will be located outside of the chamber with windows providing optical path to the cameras. The cameras will be radiation shielded to extend their lifetime. 

The optical components inside the IP chamber will enable the focussing of the main laser as well as recollimation and attenuation of the beam on its path to the diagnostics. Further components will allow precise overlap with the electron beam for best interaction results. Vibration suppression with a breadboard decoupled from the vacuum system will be implemented. 

Vacuum levels (UHV) are matched to the \euxfel beamline. Independent access to the IP and ICS chambers necessitates gate valves isolating the vacuum system downstream of ICS chamber from the \euxfel beamline. The IP and ICS chambers will also provide a gate valve to isolate it from the laser beamline (which can also be isolated from the laser chambers in the laser area).

 Given the limited access time of the tunnel and multiple interfaces of the IP chamber with other components, it is also required that the alignment breadboard of the chamber can be taken in and out without disturbing the main chamber body alignment. Again due to vacuum contact with the electron beamline, the handling of this breadboard should be carried out in a clean environment to avoid any contamination.

\section{System Overview}
\label{section3-3}

\subsection{The Laser System}
To achieve the scientific goals of LUXE within the parameter constraints of the \euxfel beam energies and the limits set by the collision geometry (see Section 2) a laser with a power of ultimately 350 TW is required (as detailed above). Such lasers can be procured commercially from more than one supplier and one such laser is in operation at \euxfel already -- the ReLaX laser at the high energy density (HED) beamline \cite{ReLaX:2021hibef}. The headline technical specifications of the LUXE laser are detailed in the Table \ref{tab:laser}.

Beyond the headline performance parameters other parameters will need to meet the experimental requirements at LUXE. 
\begin{itemize}
    \item {Contrast}\\
    Pulse contrast is only critical with respect to ensuring sufficient precision of the measured intensity. We require the energy outside a notional transform limited Gaussian shape to amount to $< 0.1 \%$ of the total energy. This requirement corresponds to an intensity  $<10^{-8} I_0$ over the duration of the stretched pulse.
   \item {Pointing stability}\\
   The stability on the laser table should achieve a high level standard of $<0.5$ diffraction limits. Beamline jitter  and active neural network stabilisation are discussed below.
   \item Energy stability\\
   Energy stability should be $< 2.5\%$ rms over 6 hours without active feedback at 10 Hz operation.
   \item Strehl ratio\\
   The Strehl ratio (the ratio of the real beam intensity to the diffraction limited intensity) as measured on a wavefront sensor should reach 0.8. (NB. effects  due to high frequency aberrations and transport are factored into assumptions in Table \ref{tab:laser}).
   \item Pulse duration stability \\
   This is specified to be $<1\%$ rms at 30 fs.
   \item Energy tunability\\
   1-100\% remotely operable.
   \item Minimum pulse duration and adjustability\\
   This is specified to be $<25$ fs, to achieve $<30$ fs during routine operation. This is adjustable to 500 fs.
   \item Chromatic aberrations\\
   Less than 0.2 $Z_R$ focal shift for wavelengths within $1/e^2$ of spectral peak.
   \item Control system\\
   The control system is to be compatible with external hardware control (e.g.\ Tango) for automation.
   
\end{itemize}

\begin{figure}[ht]
    \centering
    \includegraphics[width=0.48\textwidth]{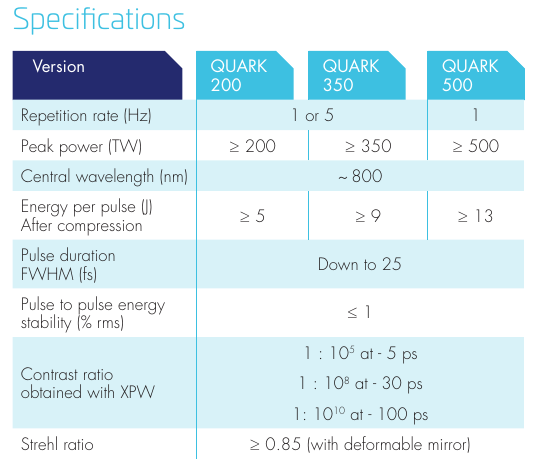}
    \includegraphics[width=0.48\textwidth]{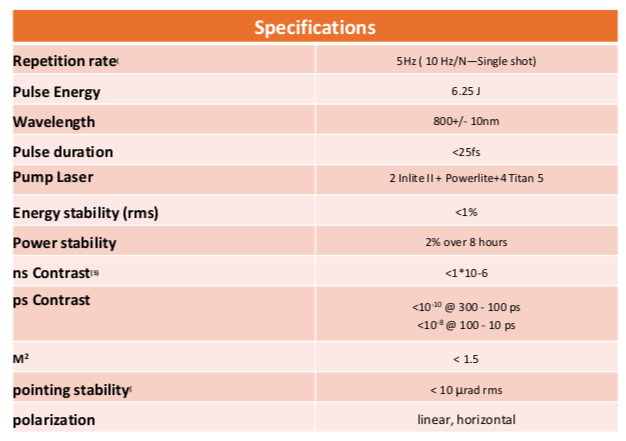}
    \caption{Performance of commercially available systems, shown as example from Thales (left) and Amplitude Laser (right) match the LUXE requirements well.}
    \label{fig:newlaser}
\end{figure}
In the following we will look in detail at our technically superior and more cost-effective option which involve the purchase of a new laser system capable of  350~TW (@30 fs) operation.  Such  a new system would operate at 40-50 TW during phase-0 (which can then be switched without affecting the installation at the IP). Figure \ref{fig:newlaser} shows the specifications of commercial laser systems available from two different suppliers.

\begin{figure}[htbp]
    \centering
    \includegraphics[width=0.7\textwidth]{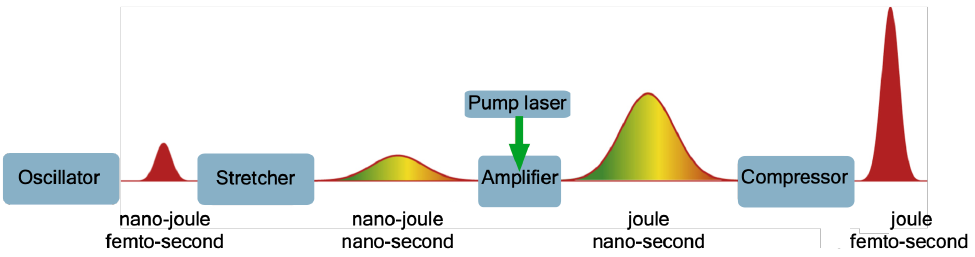}        
    \caption{Schematic depiction of the chirped pulse amplification technique.}
    \label{fig:CPA-cartoon}
\end{figure}

The architectures of such laser systems have matured and all systems under discussion follow the well-known chirped pulse amplification (CPA) concept, shown in 
Fig.~\ref{fig:CPA-cartoon}. The use of pulse stretching through spectral dispersion (chirping)  lowers the intensity in the laser chain during the amplification and significantly increases the maximum energy that can be achieved in a short pulse for a given beam diameter ~\cite{Strickland:1985gxr}.  

 The key components highlighted in the CPA schematic require adaptation for the specific requirements of LUXE.

\begin{itemize} 
\item \textbf{Front end}\newline
The laser chain starts with a commercial femtosecond oscillator which delivers a pulse train at $\approx 76$\,MHz. The important requirement for the oscillator is to produce sufficiently short pulses with high stability and quality. This requirement is easily met by modern fs-oscillators.
The specific requirement for LUXE is for this oscillator to be synchronised to the  \euxfel master clock. This is achieved by choosing  the pulse train frequency to be an exact  integer fraction of the master clock rate of 1.3\,GHz to ensure that timing is precise to within $<20$ fs rms. Maintaining this integer ratio requires the oscillator cavity length to have a feedback system to maintain the exact length using  a piezo controlled end-mirror. The  \euxfel master clock is compared to (a high harmonic of) the RF signal generated by the laser to provide the phase-locked loop (PLL) feedback for the cavity length adjustment.

\begin{figure}[htbp]
    \centering
    \includegraphics[width=.9\textwidth]{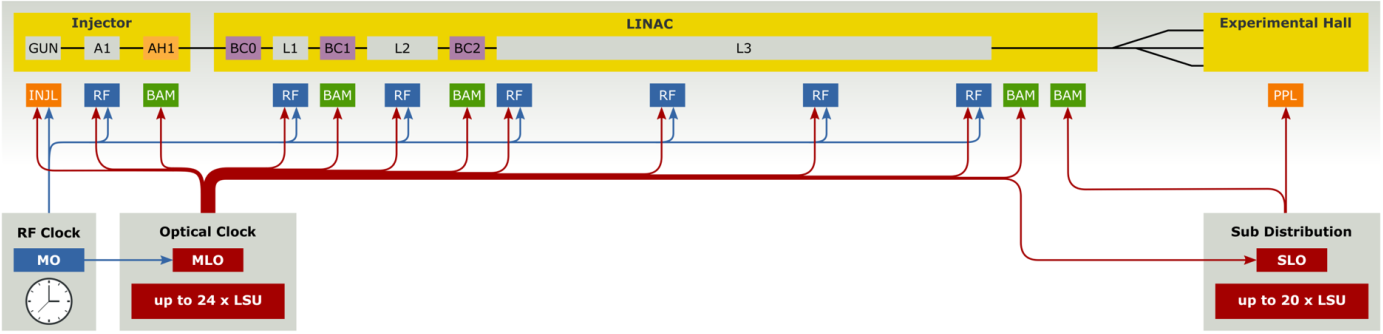}
    \caption{Schematic overview of the synchronisation system at EuXFEL. The electron bunches are produced by a laser synchronised to the 1.3 GHz master oscillator. The master oscillator signal is distributed to other systems requiring synchronisation all the way to the experimental hutches with 20 fs rms accuracy at oscillator level.}  
    \label{fig:synchro}
\end{figure}

 This requirement is not specific to LUXE and has been achieved with a variety of lasers at \euxfel including the very similar ReLaX system \cite{Schulz}. In Fig. \ref{fig:synchro} a schematic of synchronisation implemented at \euxfel is illustrated. A second element to compensate for slow drift due to changes in the beam propagation to the IP after the oscillator is discussed below in the context of the timing tool.
 
 Each of these fs pulses contains an energy of a few nJ. A pulse selector produces a reduced repetition rate pulse-train matched to the repetition rate of the front end amplifiers  (e.g. 10\,Hz)  selected  and amplified to about 0.5\,mJ in a booster amplifier. The booster is optional but present in the case of many systems.  
 
 The rejected pulses of the oscillator pulse train are available as a high repetition beamline stabilisation fiducial.

\item \textbf{Optical pulse stretcher}\newline
Following the front end, the pulses are sent into a grating-based pulse stretcher.  There they are steered to hit an all-reflective grating four times which introduces a nano-second scale delay between long and short wavelength components and stretches the pulse around $10^5$ times.  The stretcher band-pass is set to $>$100\,nm to avoid clipping effects that would adversely affect the pulse shape and will be designed to allow a narrow band component at around 900 nm to drive the ICS interaction if required.

\item \textbf{Amplification to the mJ level}\newline  
Following the stretcher, the pulses are amplified in a regenerative amplifier with a TEM$_{00}$ spatial mode and amplified in three amplification stages at 10\,Hz repetition rate allowing for stable operation. Each stage consists of a Ti:Sa crystal pumped with ns-duration green (532\,nm) laser pulses. Following each stage, the beam is expanded to remain below the damage threshold of the following optics in the optical chain. 

Note that the regenerative amplifier is a cavity amplifier and, as the final cavity, defines the spatial mode of the laser system. It can also be operated unseeded (i.e. with the input from the oscillator blocked) to produce ns-duration pulses, which will be used during alignment modes.

\item \textbf{First Ti:Sa power amplifier}\newline
The regenerative amplifier is followed by a series of power amplifiers which are in multi-pass geometry (so called bow-tie amplifiers). 
It also provides enough input energy for alignment tasks in the system (even those with strong attenuation in the beam path) and can therefore be considered part of the laser front-end.

\item \textbf{Power amplifiers}\newline
The power amplifiers – while a major cost factor of the laser system due to large crystals and pump lasers – are conceptually simple. They consist of collimated beams with increasing large apertures taking (typically) 4 passes through a pumped laser crystal. 
As long as space is available they can be added modularly one after the other to increase laser energy to a limit ultimately set by the compressor design. The beam expanders in the final amplifier stages will be all-reflective to avoid pulse distortion by radial group delay variation and the associated deformation of the laser pulse.

\item \textbf{Optical pulse compressor}\newline    
The fully amplified pulses will be expanded and sent into an optical pulse compressor which operates under vacuum.  The compressor ultimately sets the performance limit of the laser in terms of pulse energy and average power. The aim of LUXE is to stay as closely as possible with proven design concepts to minimise any risk associated with the laser to the project. The possibility of cooled gratings to maximise repetition rate was therefore rejected. The alternative to upscale the compressor to large area to reduce the fluence was rejected on cost and space (especially beamline and IP chamber) considerations.
The compressor will be based on the standard design of four reflections from two large gold-coated diffraction gratings on glass substrates. This compressor design  limits the practical repetition rate for high intensity beam crossings to 1\,Hz due to thermal effects. At higher repetition rates thermal aberrations affect the performance of the laser and higher repetition rates would require costly and complex additional features such as grating cooling or active compensation of grating aberrations. 
The compressor  will be designed to accommodate phase-1 from the outset resulting in cost savings over the duration of the project and full 10 Hz data acquisition for phase-0 power levels.

\item \textbf{Adaptive mirror}\newline    
An adaptive mirror is used to correct any wavefront distortions introduced by imperfect optical components. Using an adaptive mirror allows a Strehl ratio of typically $>0.7$ to be achieved when measured with a wavefront sensor.

\end{itemize}

The preferred location for the laser clean room environment is at  ground level at Osdorfer Born and will be discussed in Chapter~\ref{chapt12}.  Sufficient space for planned installations in phase-0 and phase-1 will be included (see Fig. \ref{fig:XS1_side_intro}).

\subsection{Laser Diagnostics}
Full characterisation of the laser with a full suite of  laser diagnostics is central to the LUXE experiment as discussed above and required to obtain well defined interaction conditions. 

To achieve the high level of precision  on a shot-by-shot basis, state-of-art diagnostics will be implemented in the laser clean room. It is crucial to have high quality diagnostics to closely monitor the effects of shot-to-shot fluctuations and long-term drifts on the precision of the data. High-power, femtosecond laser systems are precision tools and respond with significant performance changes to relatively small drifts in alignment. This is because small variations of spatial and spectral phase have noticeable effects on the peak intensity. Such variation can be caused by thermal effects, air currents and mechanical vibrations and drift.

Instead of attempting to control the laser output a dedicated diagnostic system capable of measuring the fluctuations and tagging each shot with a precise relative intensity will be set up with all the critical components in a vacuum system to prevent air currents from affecting the measurement.  A schematic of the proposed diagnostic system is shown in Fig.~\ref{fig:laserdiag}.   A nozzle valve is used to produce a  tenuous gas jet. The accelerated particle spectra can be recorded employing a magnet spectrometer and micro channel plate detectors (MCP) allowing the absolute value of $\xi$ to be constrained by comparison to 3D particle simulations using PIC codes.

  

Each laser shot will get an intensity tag from the shot-by-shot diagnostics suite, which delivers highly accurate rations among the intensities of different shots. A key factor allowing us to tag the laser parameters on a shot-to-shot basis is the fact that  the laser beam is not distorted (unlike the case, for example, in high power laser-matter interactions) by the interaction. Hence as shown in Fig.~\ref{fig:laserdiag}, the laser beam will be relay imaged to the laser area on a diagnostic table for tagging during collision. High fidelity imaging is clearly the key to the success of this scheme (see performance section).

Some general points as to the operation of the diagnostic system are important before we consider each diagnostic in turn. As can be seen in the diagnostic layout, the laser can either be imaged from the IP to the laser diagnostics or be input directly into the diagnostic chain. Each laser parameter can therefore be measured in two configurations: directly into the laser diagnostics without the laser being transported to the IP  and imaged from the IP to the diagnostics.

This is important both scientifically and operationally. Operationally, the laser system + diagnostics can be run fully independently from the LUXE experiment in the accelerator tunnel. This allows all laser work apart from that directly related to the IP to be independent of \euxfel schedule.
Scientifically, it allows the precise and
{\normalfont \itshape routine} calibration of the effects of the transport beamline on the laser performance. Any slow deterioration during LUXE operations will be instantly noticeable during routine comparison from `direct' to `via-IP' measurements of the laser performance.

Additionally to the direct and via-IP modes, the fluence distribution of the focal spot (`spot-image') can be measured at the IP directly  allowing the direct calibration of the spatial imaging fidelity.

Here we briefly describe the individual diagnostics and their role, beginning with the on shot diagnostics.

\begin{itemize}
    \item Optical Spectrometer and Autocorrelator \\
    The spectrometer will measure the laser spectrum on each shot and will alert to changes in alignment and performance. From experience, spectral changes are a sensitive signature of small alignment drifts and also important for the energy measurement. A standard autocorrelator capable of measuring longer pulses outside of the Wizzler range will be available in this channel. 
    
    \item Energy Monitor\\
    The relative energy of the laser will be monitored for each shot by a spectrally calibrated CCD camera providing a system with a high dynamic range and high precision. The on shot spectrum will be used to convert the CCD counts into a precise measurement of laser energy.
    
    \item Spot Monitor/Fluence Monitor\\
    Our data shows that the biggest contribution in focal intensity variations comes from fluctuations in the laser spot energy distribution. We will image the spot at the IP to a high dynamic range CCD diagnostic where a fluence map of the laser focus will be produced for each shot. The ratio of laser beam energy to peak fluence will be monitored closely as an early sign of changes in the laser system.
    
    \item 2$\omega$ Fluence\\
    The fluence at 2$\omega$ (or alternatively cross polarisation wave (XPW) conversion) will be measured alongside the fluence at the laser frequency $\omega$. We anticipate to be able to calibrate the system such that in the narrow range of LUXE operational parameters the ratio of the non-linear (2$\omega$ or XPW) and $\omega$ fluences provide an intensity map for each shot.
    
    \item Pulse Duration\\
    A spectral interferometry device (Wizzler) will be used to measure the pulse duration and shape on every shot. Combining this data with laser energy and spot energy distribution allows the intensity to be calculated for each shot.
    
\end{itemize}

In addition to the on-shot diagnostics we will have diagnostics to measure multi-shot averages of the laser:

\begin{itemize}
    \item $3^{\mathrm{rd}}$ Order Autocorrelator\\
    To measure the pulse intensity over nanoseconds in the $10^{-3}$ to $10^{-11}$ range inaccessible to the Wizzler. Excess low level intensity can lead to systematic errors in the peak intensity calculation. During operations this measurement will be performed on a monthly cycle.
    
    \item{Insight-Interferometer}\\
    The `Insight' device can calculate the spatio-temporal shape of a pulse in focus and show up key problems such as chromatic aberrations. This measurement will be performed daily to prevent systematic uncertainties.
\end{itemize}

With this approach, we aim to tag each shot with a relative precision of below 1\% (ultimately 0.1\% is believed to be achievable based on currently ongoing developments at the JETI laser system in Jena). The measurement accuracy of the pulse duration and systematic variations in the fidelity of the spot image being the ultimate limitation.

\subsection{IP and ICS Chambers}
The two chambers combine the GeV beams with lasers transported by the beamline and provide space for the laser diagnostics that must be in the LUXE tunnel. \\

\begin{figure}[htbp]
    \centering
    \includegraphics[width=.6\textwidth, angle=0]{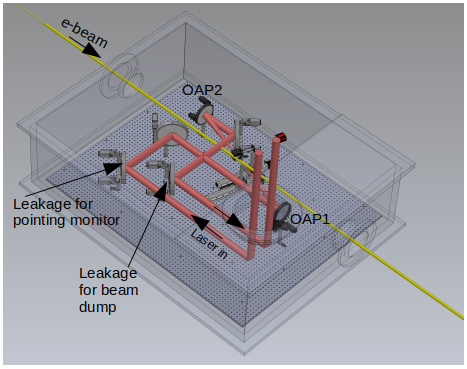}        
    \caption{ Layout of the interaction of laser pulse with the electron beam inside the IP chamber. The first off-axis parabola (OAP1) focuses and  the second (OAP2) recollimates and sends back the beam to laser area for post diagnosis. The electron beam enters from the left in the chamber.}
    \label{fig:ipchamber_only}
\end{figure}

\textbf {The IP Chamber} \\
The IP chamber is the central part of the experiment where the GeV beams and the lasers interact. It will host all the optical components required for focusing the laser beam and overlap it with the electron beam for the collision as shown in Fig.~\ref{fig:ipchamber_only}. The interaction angle is  set  to 300 mrad or $17.2^{\circ}$ at this stage. This angle is sufficiently small to make the reduction in $\chi$ due to the cosine factor in the collision equations acceptable, while providing enough distance between the parabola edge and the GeV beam axis to avoid unwanted backgrounds due to any low intensity  halo that might be associated with the GeV beams. We note that the collision angle is therefore not a critical parameter for LUXE and may vary subject to discussions with optics manufacturers.


 The IP chamber also hosts the tunnel diagnostics and alignment targets. \\

\textbf {The ICS Chamber} \\
The ICS chamber is a smaller chamber positioned upstream of the IP chamber and is separated from the IP chamber by the magnet and the first beam dump. This chamber houses the timing tool and the gamma converter target. 

For the later stages of phase-1 moving from a bremsstrahlung gamma source to  an ICS source  is planned. At slightly reduced $\gamma$-ray flux, the ICS source provides a narrow-band input for the interaction. In this scenario the collision will be at 180$^\circ$ to maximise the energy of the scattered radiation and enable long interactions lengths.

\subsection{Tunnel Diagnostics}
\begin{figure}[htbp]
    \centering
    \includegraphics[width=0.7\textwidth]{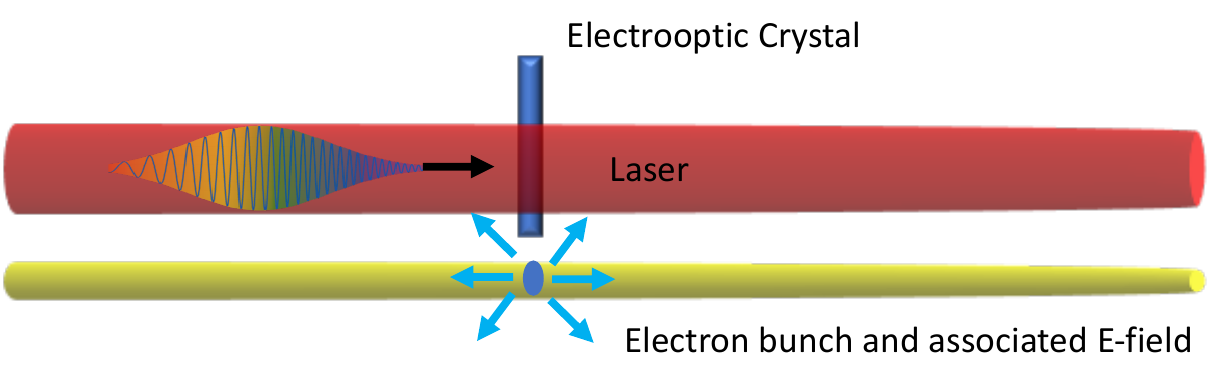}
    \caption{The laser oscillator and the \euxfel oscillator will be synchronised at the 20~fs level using the \euxfel timing system. To compensate for drift and jitter in the beam transport at low repetition rate, an electro-optical timing tool will be implemented which provides a signal proportional to the  polarisation change induced by the field of the electron bunch.}
    \label{fig:timingtool}
\end{figure}

Tunnel  diagnostics refers to the set of diagnostics placed inside the IP/ICS  chambers to allow focus optimisation of the laser beam and its precise spatial and temporal overlap with the electron beam.  These are:

\begin{itemize}
    \item The IP microscope\\ 
    This provides a high resolution image of the focal spot. This microscope will provide reference images at the beginning and end of each data run to compare with spot diagnostics in the laser area. The readout CCD is located outside the IP Chamber and radiation shielded.
    
    \item Imaging beamline calibration targets\\
    Well characterised targets (pinholes, wires) will be on a positioner to define the area of minimum aberration on the axis of  the imaging parabola. The interaction must take place within this region to avoid aberrations and systematic error (the two off-axis parabolae (OAPs) must be aligned precisely to each other in 3 space and 2 angular dimensions; imaging test targets allows the optimum position for OAP2 to be defined and characterised).  
    
    \item GeV beam alignment targets\\
    Scintillation screens and knife edges will be on the diagnostic stages to allow precise definition of the GeV beam spots and precise positioning of the laser relative to the GeV spot. The imaging OAP will be optimised to image this interaction point.
    
    \item Laser position and pointing diagnostics\\
    The  leakage through a beamline mirror in the chamber will be taken and analysed with suitable optical elements to measure the beam position on the mirror (``near-field" in laser terms) and focal spot (``far-field") and check for drift or other issues in the beamline. 
    
    \item Electro-optical (EO) timing tool \\
    An EO crystal probed by the auxiliary timing laser beam will provide an on-shot signature of the relative timing between the electron beam and laser (Fig.~\ref{fig:timingtool}). This diagnostic will be situated in the ICS chamber, where the electron beam is present in all operation modes and allow any thermal or mechanical drift in the timing to be compensated by a feedback controlled dog-leg in the laser area \cite{Steffen:2020eo}.
    
    \item Top view camera \\
    A top view CCD will allow monitoring of the moving elements in the chambers.
    
\end{itemize}

All of the diagnostics above will exist  in both the IP and ICS chambers.

All the related technical details are described in Section~\ref{section3-5}.

\subsection{Transport Beamline}

The laser transport beamline will accommodate all laser beams described in this report. Importantly, it will transport the required beams both {\normalfont \itshape to} the LUXE IP and {\normalfont \itshape back from} the LUXE IP to the laser area. This  ``two-way beamline" has major advantages. It allows all laser work to be performed in the laser clean room -- separately from any tunnel access and provides a low radiation environment for sensitive detectors, which would otherwise be damaged over time.
It also minimises the number of feedthroughs to the tunnel and uses the smallest number of remote controlled mounts possible.

 Note that the return beams will be low power with the energy attenuated by highly transmissive mirrors after the collimation parabola, with the pulse energy sufficient for the   diagnostic suite outlined above or the timing tool. The minimum laser beam size is set by the fluence limit of $<60$ mJ/cm$^{2}$ and is about 5 cm in phase-0 and 15 cm in phase-1.

\section{Expected Performance}
\label{section3-4}


\subsection{Laser System Performance}
Generally  high power laser systems (10-100s TW) operate stably for extended periods of time even without active feedback systems. We have extensive experience of operating such systems in our partner institutions. Preparatory work for LUXE has been taking place in Jena on the JETI200 system, which has comparable performance to the phase-1 LUXE laser. 

\begin{figure}[h]
	\begin{center}
		\includegraphics[width=11cm]{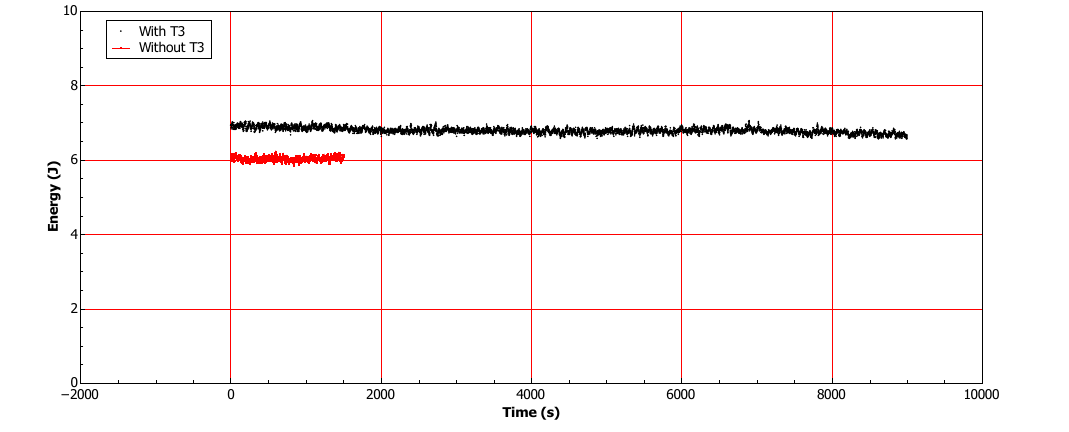}
	\end{center}
	\caption{Long term energy stability over 2.5 hrs at JETI200. Small drifts in the average  are visible and result in an increase over the short term stability 1.1 \% rms energy. Two different pump configurations are shown.}
	\label{fig:Energy_long}
\end{figure}
\begin{figure}[h]
	\begin{center}
		\includegraphics[width=11cm]{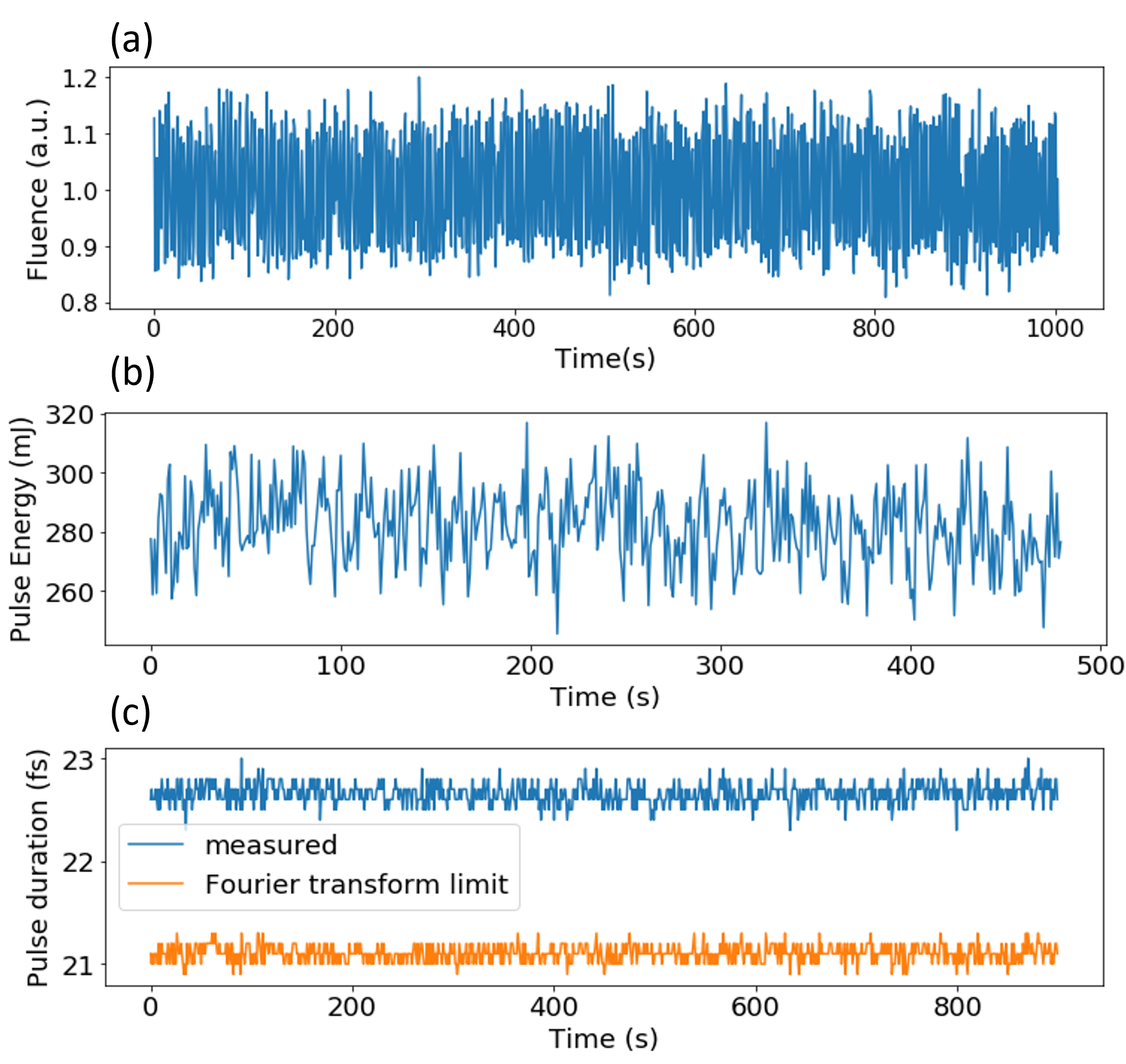}
	\end{center}
	\caption{Stability data from JETI 200. The main contribution to the intensity fluctuation in focus comes from variations in the focal spot profile (10\% rms fluence variation). The pulse duration (22.6 fs $\pm$ 0.4 \%) rms and energy $\pm$ 0.4 \% rms contribute negligibly. The laser was attenuated for the measurement.}
	\label{fig:stability}
\end{figure}

High energy Ti:Sa lasers use flash-lamp technology for the pump-lasers which have typical energy fluctuations around the \%-level rms with averaging over multiple independent pump sources, resulting in sub-\% level fluctuations shown in Fig. \ref{fig:Energy_long}.

As can be seen from Fig.~\ref{fig:stability} these  overall energy fluctuations are of similar magnitude as the fluctuations of the pulse duration measured with a Wizzler. These contributions, while  significant, are not the dominant contribution to the  shot-to-shot intensity fluctuations. The dominant contribution are the small changes in phase and amplitude across the laser beam, which result in intensity fluctuations in focus: the spot radius and the encircled energy fluctuating at the few \% level result in the 10\% level intensity fluctuations seen in the data. 

This level of stability is already fully acceptable for LUXE in combination with the on-shot diagnostic approach. Nevertheless, feedback control to maintain the laser in a stable performance envelope will be implemented. Experience in operating  JETI and commercially available systems implies that the achievement of the   performance requirements is a low risk with new lasers (cf. Fig.~\ref{fig:newlaser}) whereby the  the requirements are covered by  commercial warranty. 

\subsubsection{Synchronisation}

The synchronisation performance at oscillator level is expected to match identical systems throughout \euxfel at $\Delta t < 20$ fs rms. This level of performance meets the criteria $\Delta t \ll c Z_R$ and $\Delta t < \tau_\textrm{bunch}$ for LUXE.

\subsection{Diagnostic Performance}

The level of precision that can be obtained from the individual diagnostics is required at better than 1\% to achieve the LUXE goal of 2.5 \% relative shot-to-shot precision. In the following we consider the precision obtainable through each diagnostic.

\subsubsection{Optical Spectrometer}
The optical spectrometer will have a high dynamic range. Assuming that a 16-bit camera with many pixels well exposed per wavelength resolution element, precision at each resolution element in relative intensity of better than 12-bit is readily achieved,  provides no limitation on the overall achievable precision of relative shot-to-shot intensity.

\subsubsection{Energy Monitor}
The expanded beam will be imaged onto a CCD. A high dynamic range large area CCD can easily exceed a total of $C=10^{10} $ counts and therefore relative precision of $\sqrt{C}/C<10^{-5}$. The main limitation in the energy measurement will therefore be due to convolving the integrated CCD count with the spectrum, to account for the variation of the CCD sensitivity over the range of interest (a variation in quantum efficiency from  around 50\% @ 700 nm to 15\% @ 900 nm is typical). An overall relative precision of $10^{-4}$ is therefore achievable, well beyond the LUXE target requirement.

\subsubsection{Spot Monitor/Fluence Monitor}
The precision obtainable regarding the spatial energy distribution of the focal spot follows the same statistical considerations as the energy monitor. To determine the uncertainty in determining peak fluence of the focal spot we assume that the central 10$\times$10 pixel area is illuminated close to saturation on a 16-bit camera. This provides a statistical uncertainty of the peak fluence of $<5 \times 10^{-4}$

The imaging can introduce a systematic uncertainty. We assume that  aberrations in the imaging system from the IP to the diagnostics of $<\lambda/8$ rms are achievable.

\subsubsection{$2\omega$ Fluence Monitor}
The $2\omega$ fluence monitor (or alternatively XPW) will provide a nonlinear copy of the spot image. The ratio of these two images will provide a valuable cross check for the intensity tagging with uncertainty determined by the $N^{1/2}$ counting shot noise. The intensity ratio of the nonlinear and linear signal $R_I=I_{NL}/I_L$ will be calibrated separately over the range of laser parameters (most importantly chirp) relevant to LUXE.

\subsubsection{Pulse Duration}
As our measurements on JETI200 have shown, the combined fluctuation of laser pulse duration variations {\normalfont \itshape and} Wizzler pulse reconstruction uncertainty combines to a 0.4 \% rms variation in the measured pulse duration (see Fig. \ref{fig:stability}). Theoretical considerations show that the precision of the reconstruction depends on the complexity of the pulse both in terms of spectrum and level of chirp (deviation from Fourier transform limit). LUXE will operate with pulses close to transform limit where the theoretically predicted  precision of the Wizzler is $<0.1$ \% over the range of interest. This performance level may set the ultimate limit in determining relative intensity between individual shots and is over 10$\times$ better than the LUXE minimum requirement \cite{Jolly:2020}.

\subsubsection{`Insight' Spatio-Temporal Reconstruction}
The `Insight' Spatio-Temporal reconstruction of the 3D  pulse structure is a multi-shot measurement and does not provide any intensity information or single shot information in itself. However it demonstrates that the fundamental assumptions underlying the inferred pulse duration/intensity from the measurements above are valid and provides a valuable tool for problem shooting and validating.

\subsubsection{Contrast Measurement}
The contrast measurement using a $3^{\mathrm{rd}}$ order autocorrelator can routinely deliver a sensitivity of $>1:10^{-10}$ (ratio of peak intensity to pedestal) over temporal ranges up to 2 ns with a resolution of around 30 fs. This measurement will allow the energy outside the ideal, Fourier-transform limited pulse-shape to be quantified and any  systematic errors  to be minimised.

\subsubsection{Target Area Focal Spot Imager}
The focal spot imaging microscope in the target area will not be available on a shot-by-shot basis as it intercepts the beam in the IP. The spot image will serve as calibration for the quality of the relay system to the laser diagnostic suite.

The microscope will  also be configured to allow spot images to be taken with different levels of attenuation and allow high-dynamic range (HDR) composite images of multiple frames with differential attenuation, providing the spatial counterpart to the high dynamic 3$^\textrm{rd}$ order autocorrelation. In both cases, low level intensities distributed over large spatial (or temporal) scales can lead to a systematic over-estimate of peak intensity, which is prevented by the HDR diagnostics.

\subsection{Beamline}
The beamline fulfills the important task of linking the laser with the accelerator and is described at this point because the design detail is required in the following descriptions. The primary purpose (and challenge) is to transport the beams with minimal distortions and vibrations and to allow seamless operation of the various components.

\begin{figure}[h]
	\begin{center}
		\includegraphics[width=11cm]{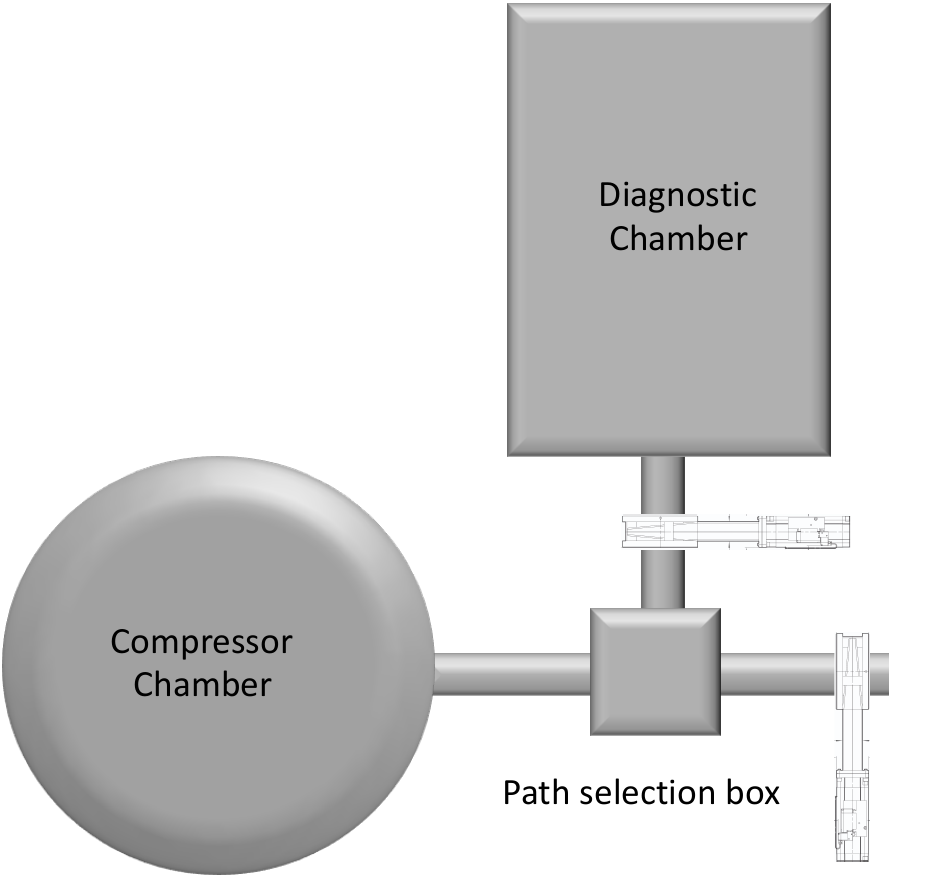}
		
	\end{center}
	\caption{Layout of the compressor and diagnostic chamber. A turning box allows either the laser to be directed to the diagnostics directly or the returning laser from the IP to be analysed. Gatevalves allow the chamber and beamline vacuum to be separated. }
	\label{fig:Cleanroom_chambers}
\end{figure}

To facilitate smooth working, the vacuum chambers will be designed to operate at a pressure of $<10^{-8}$ mbar.  To minimise pumping time for each opening of the IP or compressor chambers the beamline will be isolated by windowed gate valves at either end with a path selection box for direct injection of the laser into the diagnostic chamber or via the IP (see Fig. \ref{fig:Cleanroom_chambers}).  

To minimise motorisation and independent sources of vibration all beams will be combined in a 400 mm ID beam pipe. 

\begin{figure}[h]
	\begin{center}
		\includegraphics[width=7.5cm]{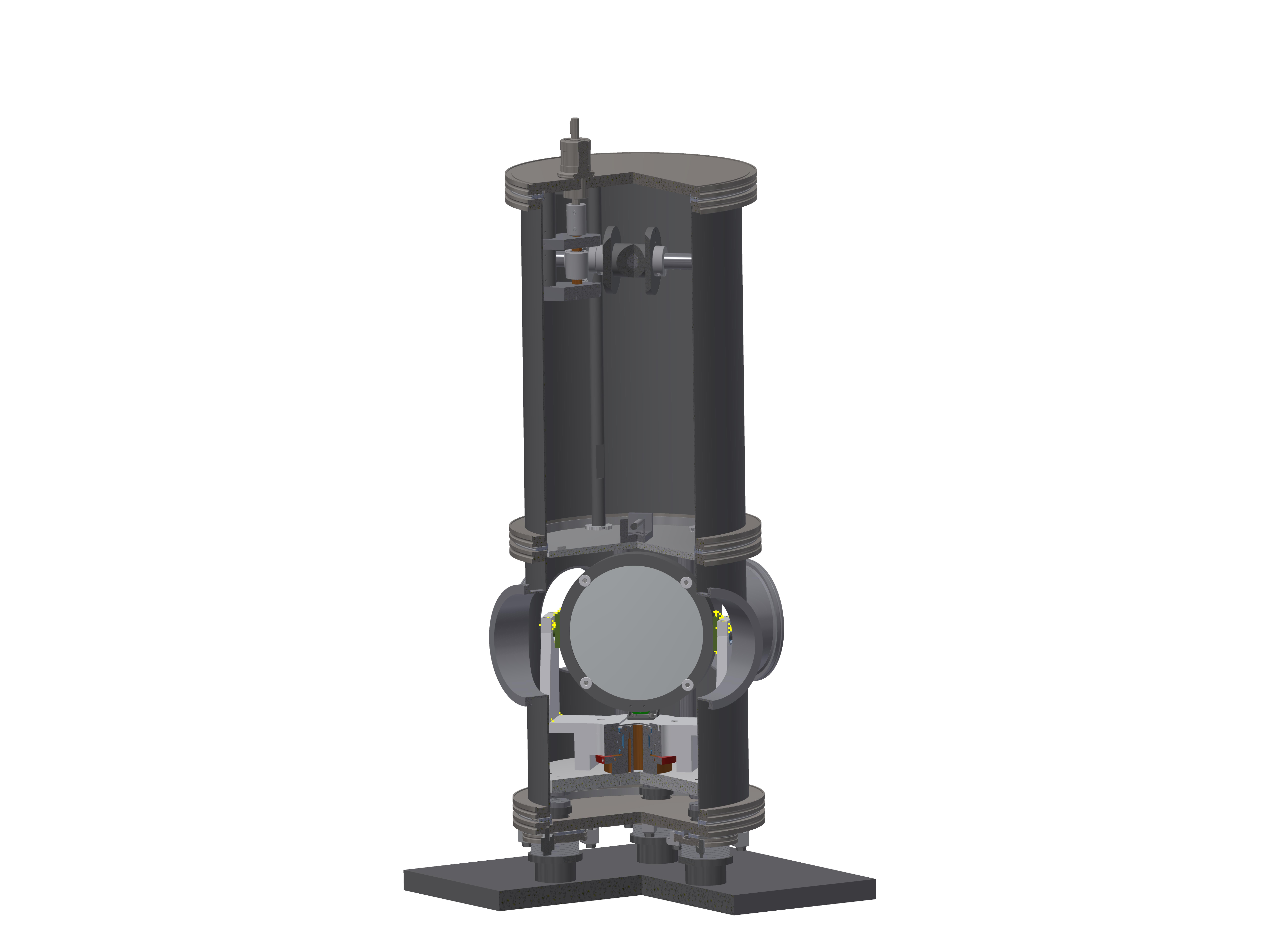}
		\includegraphics[width=7.5cm]{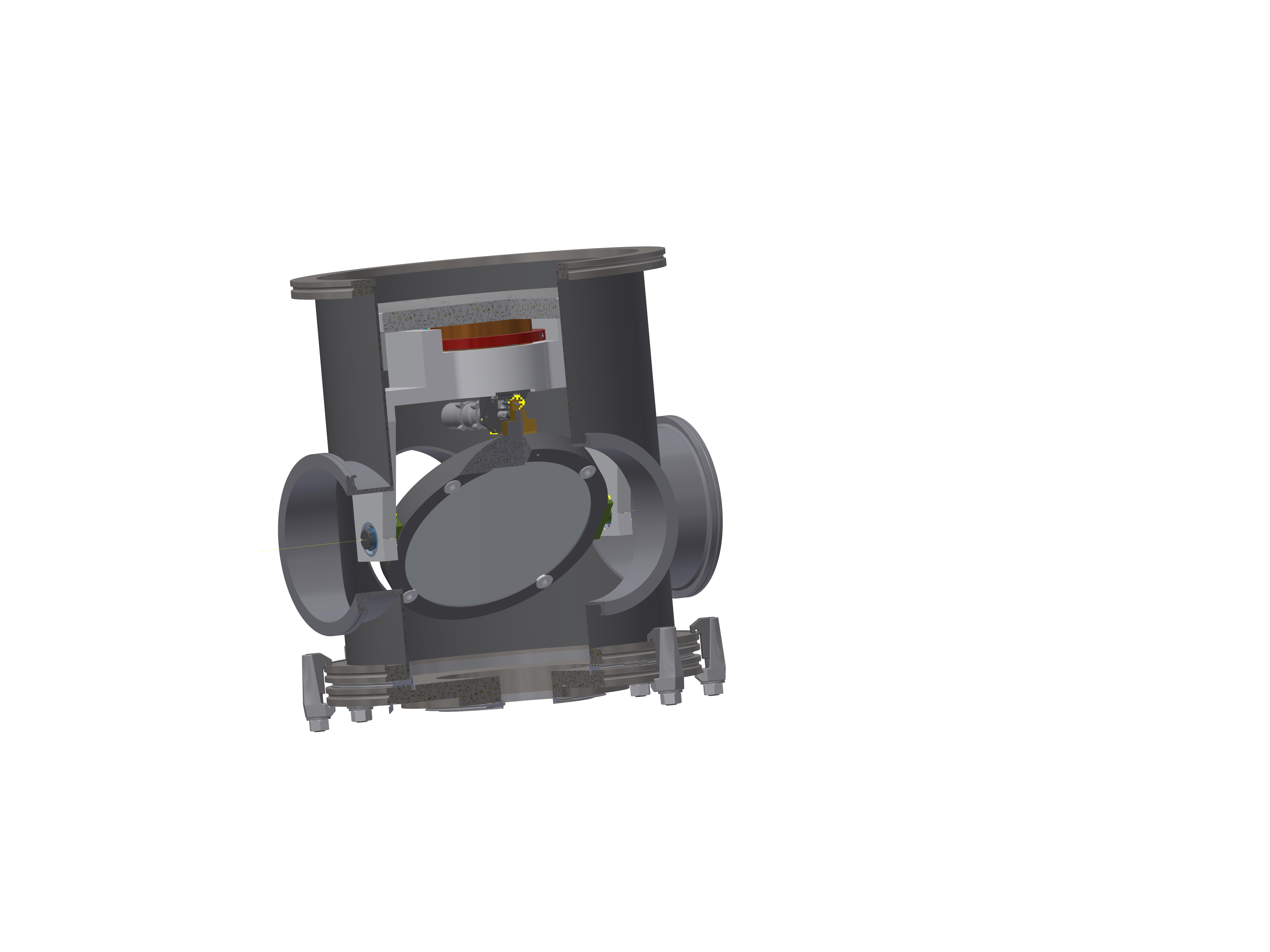}
	\end{center}
	\caption{Mirror mount design for the beamline for single circular beam for a (left) removable mirror and (right) standard mirror.}
	\label{fig:JETImirr1}
\end{figure}

There are currently two design options to transport the laser beams to and from the interaction chamber: Either all beams will be transported in a single 450\,mm diameter beam pipe, or there will be two smaller 200\,mm diameter beam pipes separating the beams going into and out of the interaction chamber. For more detail, see Section \ref{subsec:transportbeamline}.  The mirrors will be mounted separately from the vacuum tubing with a design based on the JETI200 design (Fig. \ref{fig:JETImirr1}). High precision motors will allow for fine control under vacuum to compensate any movement on pump down.
The mirrors will be connected with the building using vibration damping mounts.

\section{Technical Description}
\label{section3-5}

Additional technical aspects for the various components are considered below. 
\subsection{Laser System}

The laser system will be installed in the new, air-conditioned clean-room at the Osdorfer Born site (described in detail in Chapter 12). The clean room will have space for the full phase-1 installation and provide the required services (power, cooling) which are known from previous installations and the manufacturer. Components such as coolers, vacuum pumps and power supplies that can be housed outside the clean-room will be housed in a separate space outside the clean-room. The laser itself will be installed on a vibration isolating table system with enclosures under a flow-box hood. This provides protection against fluctuations due to mechanical vibrations, suppresses effects of air-currents and provides additional protection against dust.

Non standard components such as motorised mirrors for feedback controlled operation / automation, auxiliary beams, delay stages  and attenuators for the required operating modes will be added by the LUXE laser team. The set-up with discrete optics allows modifications to be implemented to such systems with relative ease.

\subsubsection{Synchronisation of Laser and Electron Beam}

The precision of the temporal overlap  between the 30 fs laser pulse and the 40 fs electron or photon pulse  is  determined by the collision geometry and the requirement that the laser be focused at the time of collision at the IP.  Since the laser waist varies along its longitudinal axis, $z$,  as 

\begin{equation}
    w(z)=w_0 \sqrt{1+\left(\frac{\lambda\, z}{\pi w_0^2}\right)^2}\,\,,
\end{equation}

maintaining constant intensity to within 1\% requires the collision time to occur at the laser focus with a tolerance of 30 fs peak-to-peak in the most stringent scenario of small spots and short \euxfel bunches.

 The European XFEL has developed world-leading synchronisation capabilities which have demonstrated the stable synchronisation of two RF signals to better than 13~fs even when two sources lie a few kilometres apart from each other. In the presence of active feedback to correct slow drifts, as will be implemented in LUXE, 20~fs peak to peak variations over a period of two days have been demonstrated~\cite{Schulz}. This technology will be used to synchronise the \euxfel RF master clock oscillator to the oscillator of the laser system. The repetition rate of the laser oscillators will be set to be precisely matched to a sub-harmonic of the master clock by adjusting the cavity length. This match will be maintained by fine cavity length adjustments using piezo-actuators. This method is successfully used across the \euxfel to synchronise lasers to the accelerator.
 Successful synchronisation can be verified by measuring the inverse Compton scattered photons from the main laser beam in the IP, where maximum ellipticity of the $\gamma$-beam angular distribution corresponds to best focus. A scanning measurement over many shots allows this optimum to be determined with high precision. 
 
 The above approach requires many shots for high precision, is not available for $\gamma$-laser interactions and can only operate at a maximum of 1 Hz and operates well for high intensity shots only. This necessitates an online diagnostics to provide the continuous feedback for stable operation.  The timing tool (Fig. \ref{fig:timingtool})  operates by measuring the effect of the field of the electron bunch on the light transmitted through an electro-optic (EO) crystal installed in the ICS chamber. For a slightly chirped pulse, time and laser wavelength are correlated, allowing the peak of the EO effect to be determined using a spectrometer to analyse the transmitted light with femtosecond accuracy. Such a system  has been successfully used to measure 35 fs jitter for long (400 fs) electron bunches \cite{Steffen:2020eo}, and is well suited to stabilising any drift in LUXE. Averaging over many shots will provide stable feedback data on second time-scales to maintain stable timing by moving mirrors positioned on a  timing dogleg. This can be enhanced with our neural network predictive correction system if required (see below). This timing dogleg will be inserted with small mirrors immediately after the regen amplifier. The probe beam will   be sampled after the regen amplifier  with a power of 100\,$\mu$W during operation and propagate to the ICS chamber along a separate path but sharing the same path apart from delay line mirrors on the laser table (see beamline description for transport).

\subsubsection{Active Stabilisation}

\begin{figure}[htbp]
    \centering
    \includegraphics[width=0.8\textwidth]{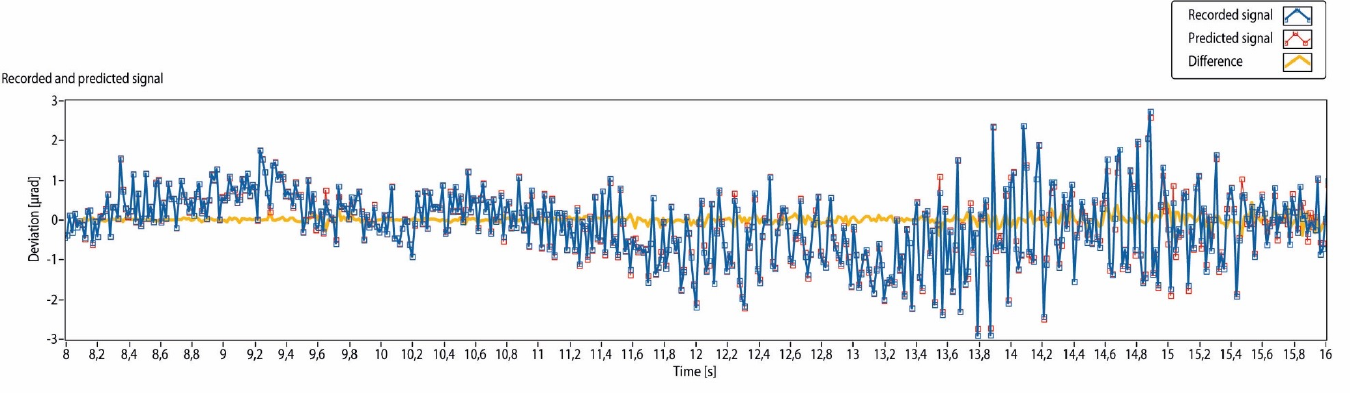}        
    \caption{Data from neural network predictions of pointing jitter. The frequency contributions to  the jitter cannot be resolved with the 5 Hz repetition rate of the data, however the neural network predicts the next position correctly, allowing predictive correction.}
    \label{fig:neural}
\end{figure}

Active stabilisation systems have been developed with LUXE in mind based on neural networks. These are designed to predict (over short timescales) the correction values for parameters such as pointing and timing based on a past data set. The system takes data and learns the patterns for the next data point and improves its estimates with increasing data. Using this method pointing and timing data that cannot resolve all timescales for Fourier-decomposition can be treated and the correct value to place the next shot exactly with the right parameter for timing or pointing can be determined.  Figure \ref{fig:neural} shows the results from JETI200 which will allow ultra-stable timing and pointing to be implemented with active, predictive feedback at LUXE.

\subsubsection{Attenuator for energy control of the pulse}

\begin{figure}[htbp]
    \centering
    \includegraphics[width=0.8\textwidth]{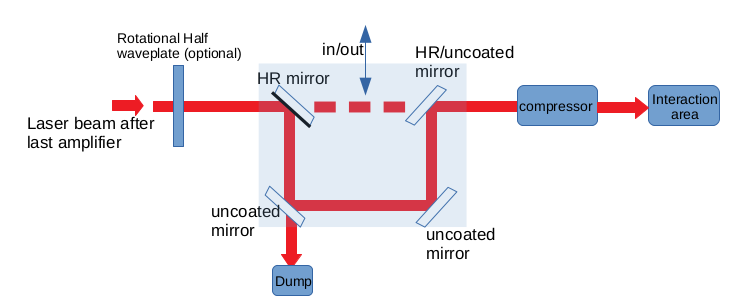}        
    \caption{Schematic of the planned attenuator based on low reflective mirrors. }
    \label{fig:attenuator}
\end{figure}

During the full amplification, the wavefront of the beam may get more deformed than the non amplified beam (e.g at front end -- mJ level). As discussed, an attenuator with an attenuation factor of up to $10^{6}$ can be inserted in the path of the laser beam after the last stage of the amplification and before the compressor using high quality ($\lambda/20$) uncoated optics (Fig. ~\ref{fig:attenuator}). This attenuates the laser without affecting other parameters and allows the focus at full energy to be measured with the IP spot diagnostic without distortion in the microscope objective. This attenuator consists of four mirrors with typically three uncoated and one high reflective (HR) mirrors.  Each uncoated mirror results in about  two orders of magnitude (0.9 \% reflectivity - BK7 @ 45 degrees) attenuation. The insertion of this attenuator will be done using a motorised translation stage. Switchable neutral density (ND) filters in front of the IP spot camera and laser pointing monitor on the laser table allow the intensity to be adjusted for best exposure. The final laser pointing monitor on the laser table will ensure that pointing is not affected by the attenuator. This attenuator provides a $\mu$J level beam otherwise identical to the full beam. Note that the main low power for alignment mode uses high energy ($>$mJ) / long pulse ($>$10 ns) duration from the regenerative amplifier to facilitate operations and alignment.

\subsection{Laser Diagnostics} 
The general  layout of the diagnostic suite has been shown in a previous section. Thanks to the availability of the JETI40, offline preparation and testing of various diagnostics  is currently in progress, which will allow faster implementation at LUXE. The laser can be fed directly into the diagnostics via an insertable mirror and focusing optics to match the laser into the diagnostics beam path described below. This allows the laser to be diagnosed and the diagnostics to be optimised separately from the operational beam path that goes via the IP in the tunnel.

\subsubsection{Imaging of the Beam from IP to Diagnostics}
The first component of the laser diagnostics chain during full LUXE operations is the imaging system from IP to diagnostics. As shown in Fig. \ref{fig:ipchamber_only} the laser is focused to the IP with the first OAP which will be optimised with IP microscope objective. Once a satisfactory focus has been reached at the IP, the IP focus must be imaged to the diagnostics for on shot measurements.

The current design of the beam path for LUXE is  40\,m long. The challenge is to transport the laser with minimal distortion to the diagnostics so that the intensity of each shot can be determined with the highest possible relative precision.\\

The first component is a collimating OAP. In the case of phase-1 this OAP will be identical to the focusing OAP and specified for minimum wavefront distortion with a surface figure of $<\lambda /15$ @ 633 nm aimed for. Parabolas of such aperture and quality have previously been successfully manufactured (e.g. by Optical Surfaces, UK). 

The second component is a wedged, uncoated, high quality reference flat to reduce the beam energy. This flat will reduce the intensity of the horizontally polarised beam by a factor of $9.2 \times 10^{-3}$ for safe transport in the beam line. The detailed alignment procedure for IP, imaging OAP and beamline  is described below.

\begin{figure}[h!]
    \centering
    \includegraphics[width=0.85\textwidth, angle=0]{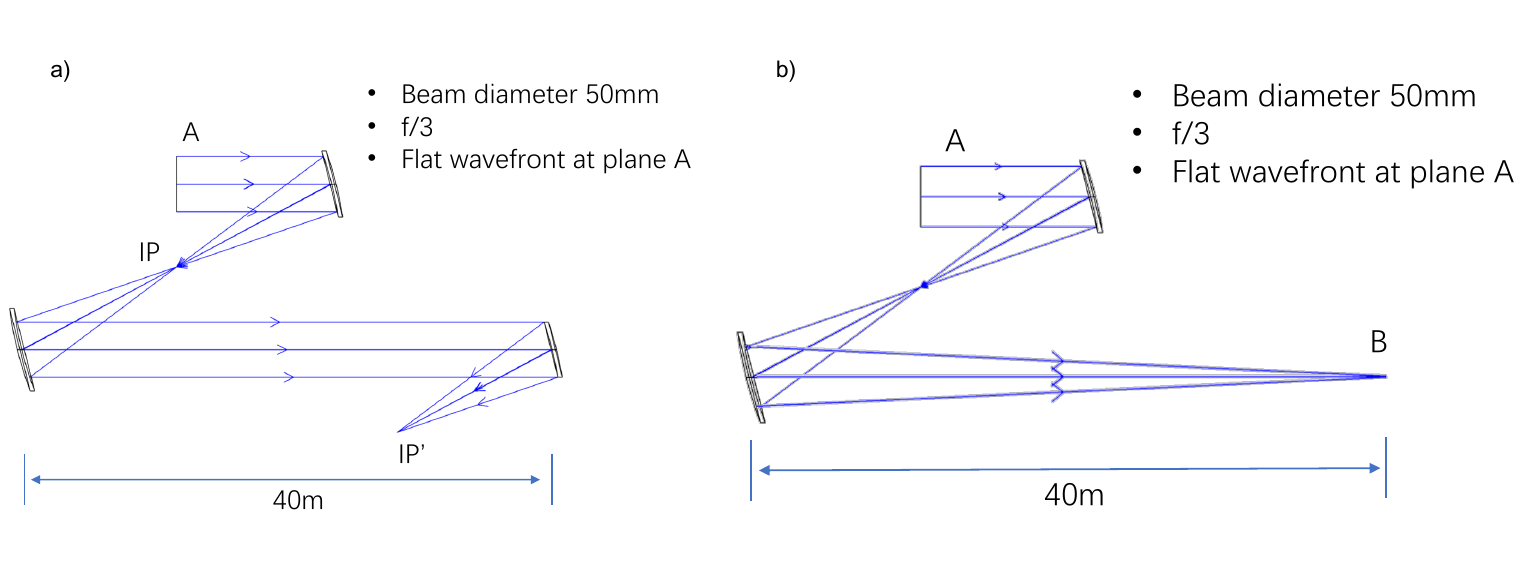}
    \caption{Imaging set-up to measure the focal spot intensity. Two basic options are considered for LUXE. In the first option (left) the first off-axis parabola (OAP) forms the IP focus. The second OAP collimates the diverging laser beam and an image of the focus is formed at the focus of a third OAP. In the alternate 'direct imaging' scheme analysed using ZEMAX (right) the OAP is used to form an image directly at a distance of 40 m from the IP.  }
    \label{fig:relay-imaging}
\end{figure}

The schematic of the imaging set up is shown in Fig.~\ref{fig:relay-imaging}.  There are two distinct options to achieve a high fidelity image. The first is to use the collimating OAP to achieve a beam with diffraction limited divergence and transport a collimated beam to the diagnostic suite and form an image of the focal spot in the focus of an OAP at the entrance to the diagnostic table.
Assuming aberration free imaging a perfect image of the IP spot is formed on the diagnostics for processing. 

This approach requires all of the (depending on final design) 11 transport surfaces and the diagnostic OAP to be of excellent quality ($<\lambda/20$ @ 633 nm) to achieve a high fidelity wavefront. These optical elements would all require full aperture.

\begin{figure}[h!]
	\begin{center}
		\includegraphics[width=0.85\textwidth]{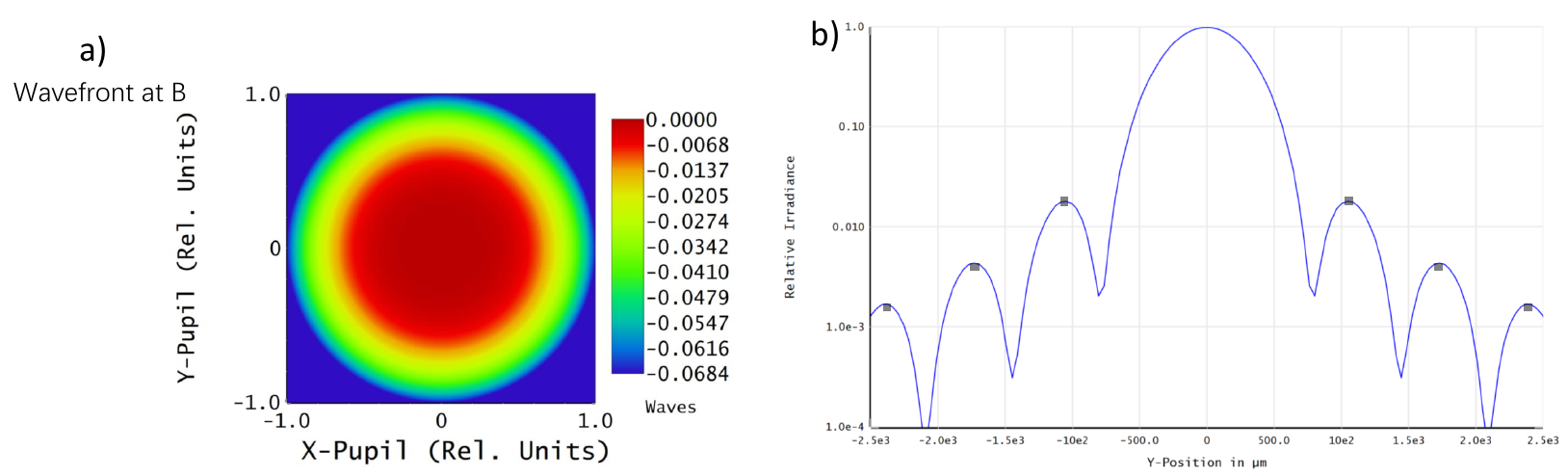}
	\end{center}
	\caption{Results of the ZEMAX simulation for the 'direct imaging' option. The additional wavefront aberrations compared to the ideal case can be seen to be only $\lambda/16$ at the beam edges. Compared to the surface accuracy of  best available OAPs ($\lambda/20$ rms wavefront distortion) the effect of using the OAP directly to image the spot is clearly acceptable. The simulation was performed for a perfect top-hat beam and the resulting intensity profile of the 'direct imaging' option is shown relative to the three OAP scheme assuming perfect optical surfaces (with side maxima peaks for this case shown as grey squares). While small deviations are visible, the fidelity is clearly excellent and meets the project goals of LUXE.}
	\label{fig:ZEMAX}
\end{figure}

A second approach is shown on the right panel of Fig.~\ref{fig:relay-imaging}. Here the collimating OAP is moved slightly from its nominal position to result in a converging beam. The large distance to the diagnostics is a distinct advantage here as we can still operate close to collimation and form an image at the desired distance. A ZEMAX simulation performed for this option  (Fig.~\ref{fig:ZEMAX}) shows the additional aberrations are very small and allow a high quality image to be formed. The additional advantage is that the size of the high quality optics in the beam path is much reduced resulting in cost savings.




\subsubsection{Focal Spot Imaging}
\begin{figure}[htbp]
    \centering
    \includegraphics[width=0.5\textwidth, angle=0]{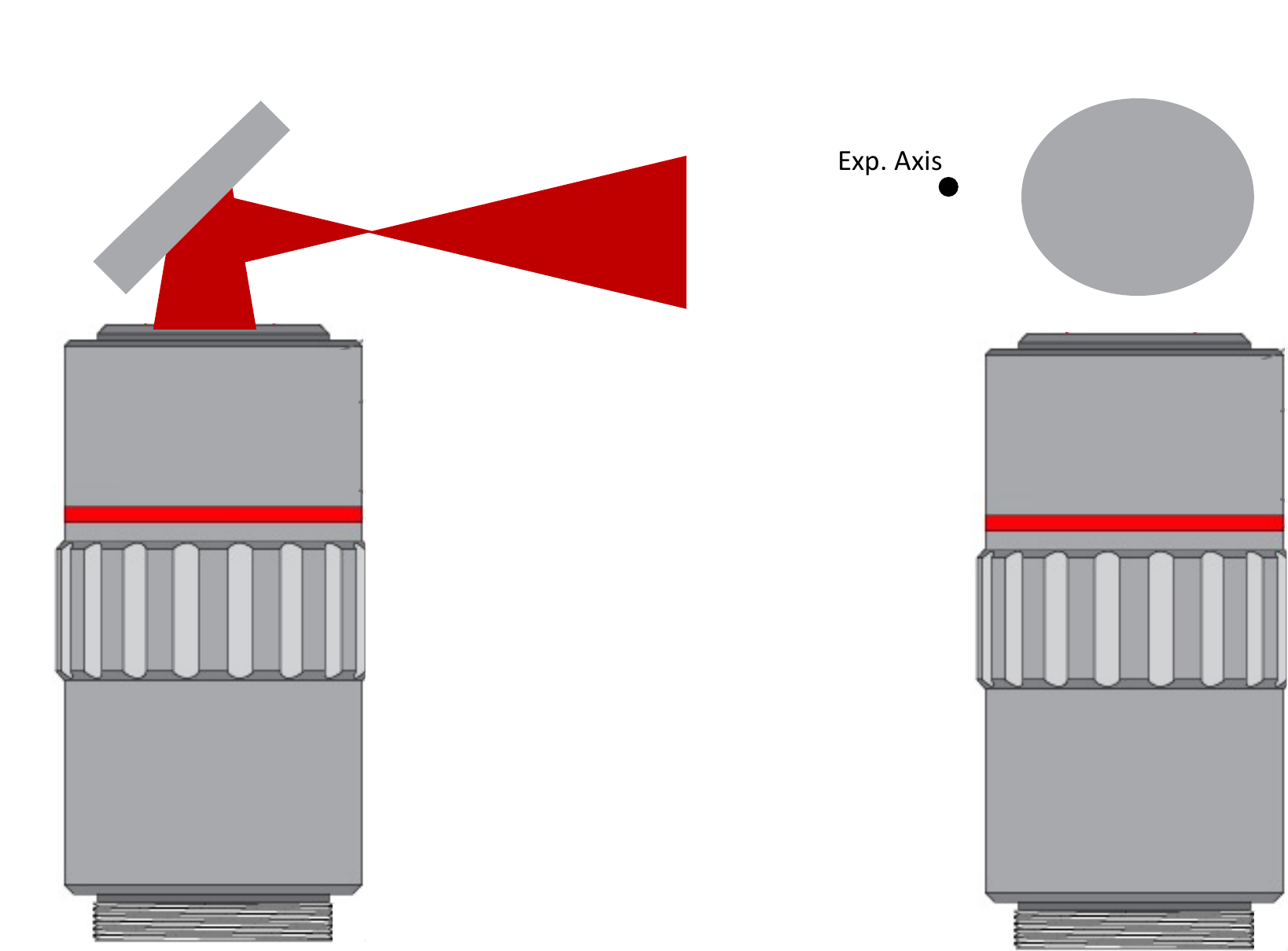}
    \caption{Detailed view of the long working range microscope objective inserted into the IP from below. A mirror enables the spot to be imaged while the  objective is mounted vertically and can be retracted vertically out for data taking mode. A view along the experimental axis is shown on the right. The spot marking the axis has a 4 mm size for clarity and does not represent the actual beam size.}
    \label{fig:microscope}
\end{figure}

 Focal spot imaging will be performed with a microscope objective that is insertable from underneath the IP and actuated through a bellows from beneath the chamber. This set-up allows the electron beam to pass past the small turning mirror mounted directly on top of the long working range objective as shown in Fig. \ref{fig:microscope}. To prevent damage to the mirror the energy of the laser is reduced to $\mu$J level using an attenuator, which allows the laser to operate at full power (described in this section) for precise measurements.

\begin{figure}[htbp]
    \centering
    \includegraphics[width=0.40\textwidth, angle=0]{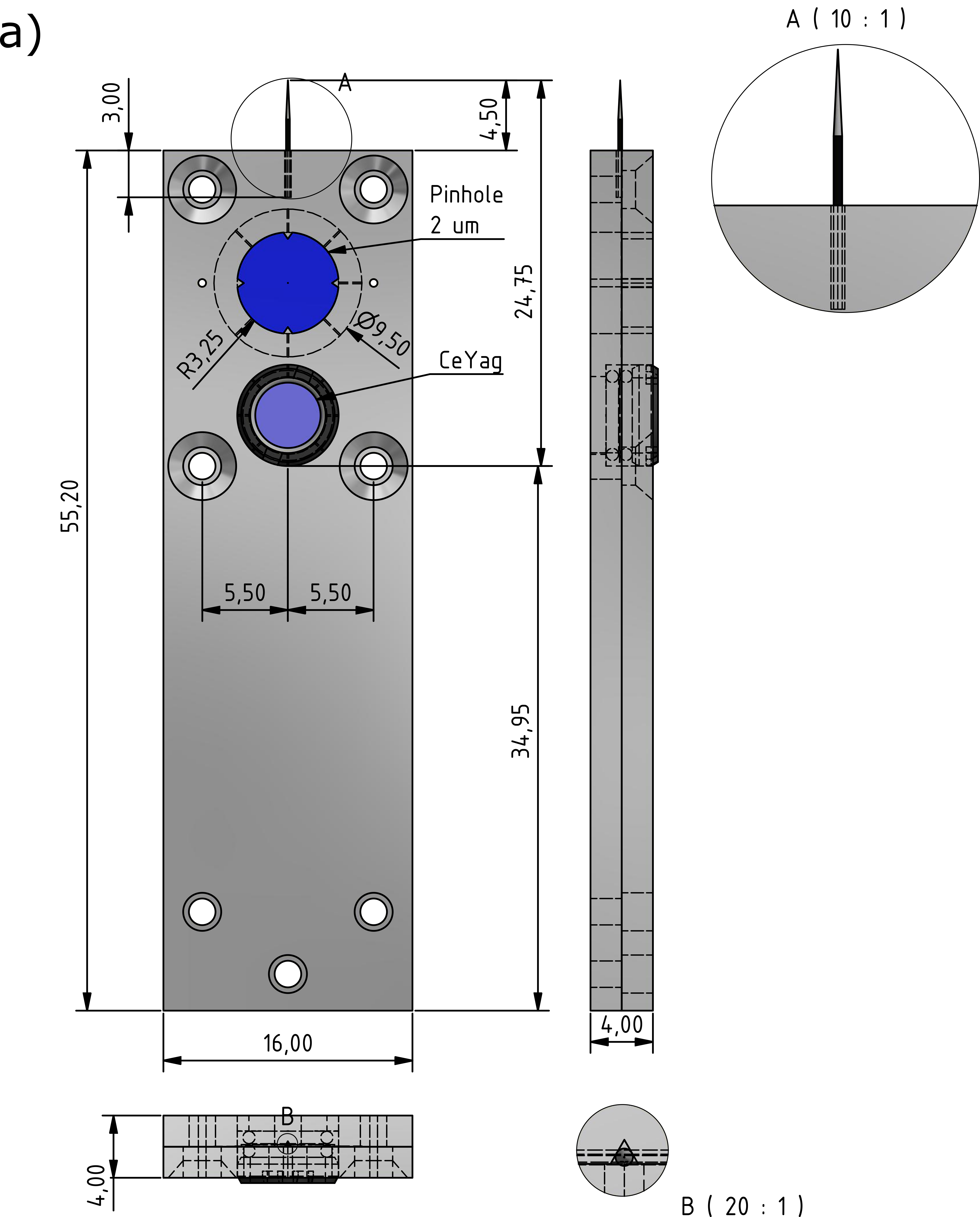}        
    \caption{ Alignment-target holder for the IP diagnostics. A pin, pinhole and Ce:YAG screen are available to determine the position of the electron beam and ensure that the imaging precision and position are accurate using a pinhole.}
    \label{fig:Targetholder}
\end{figure}

 A double OAP set-up is capable of highly accurate imaging, but only over a limited field of view. The stigmatic field of view will be checked for stigmatic imaging using test objects (pinhole), which have been separately characterised (Fig. \ref{fig:Targetholder}). The imaging system will be aligned such that the position of the focus from OAP1 corresponds to the field of view for stigmatic imaging of OAP2. (Note: a separate alignment target and microscope is required as a diffraction limited beam can be focused to a distorted spot by a misaligned OAP1 which can then be corrected by the symmetrical and identical OAP2. Imaging quality must be therefore calibrated separately with a test object in the focus of OAP2.)
 
\begin{figure}[htbp]
    \centering
    \includegraphics[width=.80\textwidth]{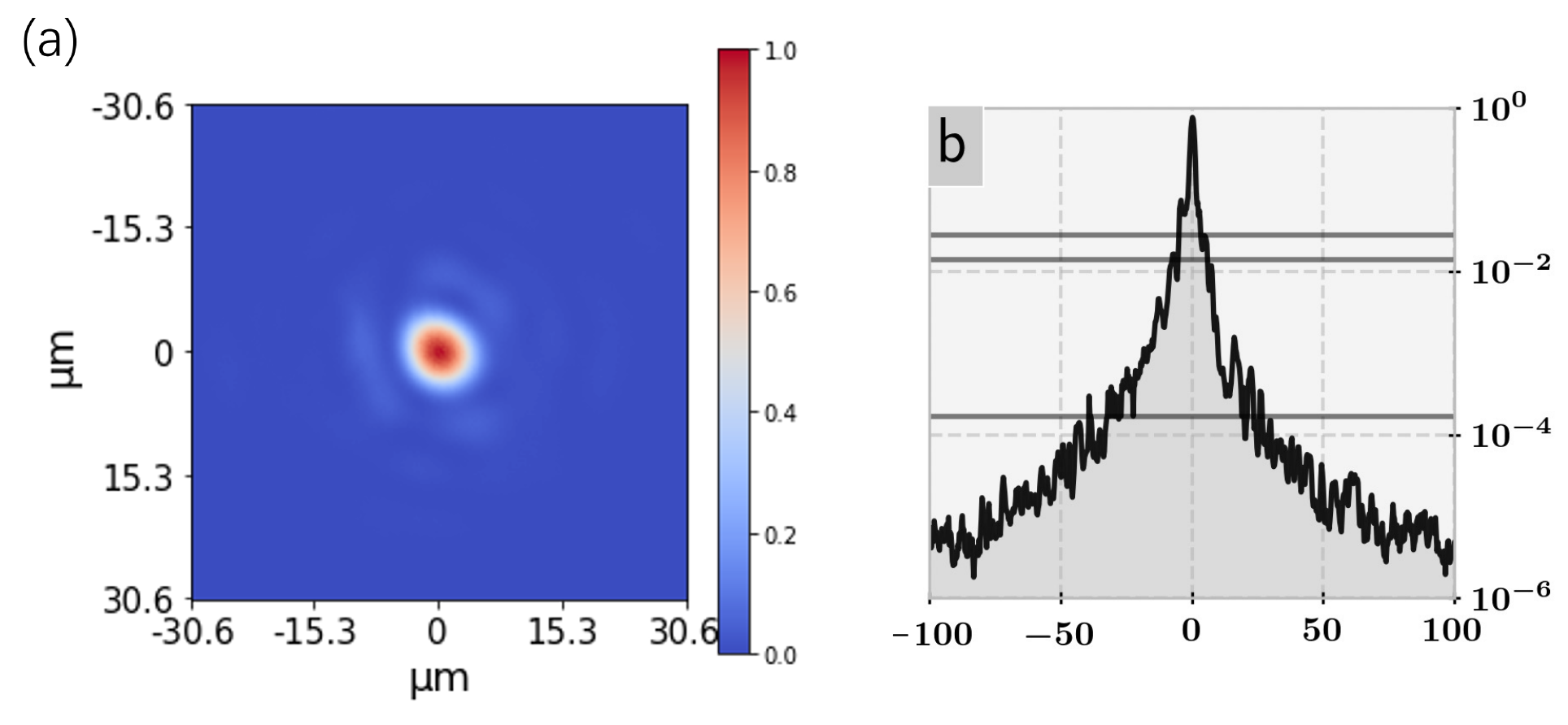}
    \caption{(a) The focus profile of a near diffraction limited spot. (b) High dynamic range  composite image with a diamond turned OAP with high scatter demonstrating that 6 orders of magnitude are feasible at LUXE.}
    \label{fig:focus}
\end{figure}

 Figure~\ref{fig:focus}(a) shows the focus profile taken with the test set-up at JETI40 with a comparable imaging set-up showing a near diffraction limited spot. An HDR image of a focus of a diamond turned OAP chosen to exemplify HDR imaging is also shown. The HDR data was obtained by combining shots with different attenuation as shown in Fig.~\ref{fig:focus}(b).  A dynamic range of $>10^6$ is obtained in our test experiments demonstrating that the focal intensity distribution can be reconstructed with high precision. In this image over 5\% of the pulse energy would be `hidden' in low dynamic range diagnostics, which is a common source of systematic laser intensity overestimate.

\subsubsection{Spatial Overlap}
The spatial overlap of the electron beam and laser focus at the IP will be carried out by using an alignment target containing a Ce:YAG scintillation screen to view the position of the focus of the electron beam with tens of $\mu$m precision. Higher precision will be achieved using a needle with a  micrometer scale tungsten tip  (Fig. \ref{fig:Targetholder}). The alignment targets will be installed on a linear translation stage actuated through a bellows from beneath the chamber. First the needle will be moved across in the electron beam and optimise the bremsstrahlung signal by downstream detectors. Then the tip of the needle at this optimised signal case will be imaged by the focal diagnostic camera placed inside the chamber. This plane will serve as a reference plane at the IP defined now by focus diagnostics. The laser focus now can be optimised to match this point in space and OAP2 can be optimised to provide a stigmatic image of test objects at this position. In turn both the electron beam and laser beam will  be spatially overlapped at the IP. The linear stages (for the needle and focus) have to be vacuum compatible as per required by the  vacuum group.

The spatio-temporal overlap will be confirmed by intersecting the laser pulse and the e-bunch. The sensitivity can be adjusted by changing pulse duration and spot-size to, for example, 500 fs and 100 µm making initial alignment straight forward. Accuracy can be then gradually increased by reducing size and duration. The gamma detectors will be able to see any overlap easily as they are sensitive to single scattered gamma rays from the e-laser interaction. By varying the time delay of the laser pulse (using a  delay stage in the laser area) the signal rate can be optimised and the aspect ratio of the angular distribution provides confirmation of the peak intensity as discussed in the context of the detectors in this TDR. 

For $\gamma$-laser mode the angular pointing from the conversion target will be confirmed by the downstream diagnostics to be on the experimental axis. The significantly larger spot ($>$200 $\mu$m) makes spatial overlap uncritical and can be aligned with the electron beam and cross-checked with the Ce:YAG screen and downstream $\gamma$-diagnostics.

\subsubsection{Pointing Monitor}
The near and far field monitors will be installed using the leakage beam from the second turning mirror (cf.  Fig. ~\ref{fig:ipchamber_only}). The leakage light will be split in two parts. One part will be used for near field and other part for far field. For the near field, the beam will be collimated down to fit onto the CCD camera. To obtain the the far field  an appropriate lens will be used and then the focus will be imaged onto a CCD camera. These setups are very compact and would fit on 50 $\times$ 70 cm$^2$ breadboard and will be attached to the isolating breadboard of the IP chamber. The accuracy of the pointing monitor will be near diffraction limited in angular resolution of the centroid, thereby providing sufficient precision to ensure high quality relative alignment of the incoming beam to the IP setup.

\subsubsection{EO Timing Tool}
The EO timing tool will be placed inside the ICS chamber (in case it is installed in phase-0) or in a simplified target chamber (only for the purpose of gamma converter and beam arrival monitor by machine group). In our case an electro-optical crystal (10 $\times$ 10 mm$^2$) sufficiently thin to ensure  femtosecond time resolution will be  placed close to the electron beam axis (about 5-10 mm). The crystal will be gallium phosphide (GaP) which offers detection of frequencies up to 10 THz for an 800 nm Ti:Sa laser pulse. The probe beam will be of diameter about 10 mm and guided through the crystal with the help of mirrors. The other remaining setup will be placed outside. The setup is rather compact and can be placed on a small (30 $\times$ 30 cm$^2$) breadboard.

\subsection{Transport Beamline} 
\label{subsec:transportbeamline}
\begin{figure}[htbp]
    \centering
    \includegraphics[width=0.5\textwidth]{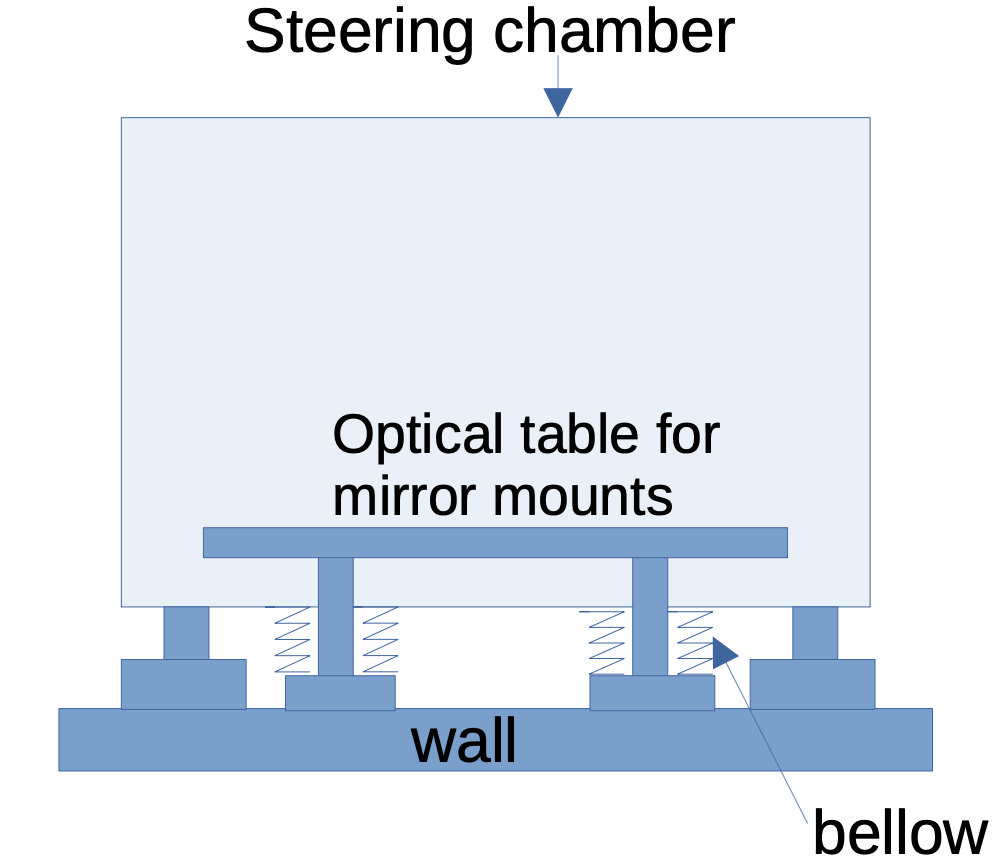}        
    \caption{Schematic showing the isolation technique for the optical breadboard from the main steering chamber.}
    \label{fig:beam_transport_isolation}
\end{figure}

The beam transport line layout for LUXE is discussed further in Chapter~\ref{chapt12}. Here we emphasise the key features. A two-way beamline is needed, to guide the beam to the IP and bring it back to the laser area for the remote diagnostics of the laser parameter on a shot-to-shot basis. In addition to the main beam (for collision at IP), auxiliary beams for the ICS and EO timing tool will also be transported through this beamline. The ICS beam would be of approximately 5 cm in size and the EO timing tool probe beam would be of 1 cm diameter. Furthermore, to avoid low frequency noise due to vacuum pumps etc., the breadboards holding the mirror mounts must be isolated from the vessel containing it, as exemplified in Fig. \ref{fig:beam_transport_isolation}. 


For the two-ways beam transport, two different solutions were explored; in the first, a single large pipe with a length of about 40 m and a 45 cm diameter would allow easy transportation of multiple beams in both directions inside smaller internal beam pipes. In the second, two smaller (about 20 cm  diameter) separate pipes would convey beams in opposite directions. The arguments in favour of this second solution are a reduction in the volume of the vessels, reducing the vacuum effort; the better flexibility in routing the transfer lines, feedthroughs, etc.; and the simpler design of the entry port in the interaction chamber.




\subsection{Vacuum Chambers}
\label{sec:vac_chamber}
\subsubsection{Interaction Point (IP) Chamber} 

\begin{figure}[htbp]
   \centering
    \includegraphics[width=0.6\textwidth, angle=270]{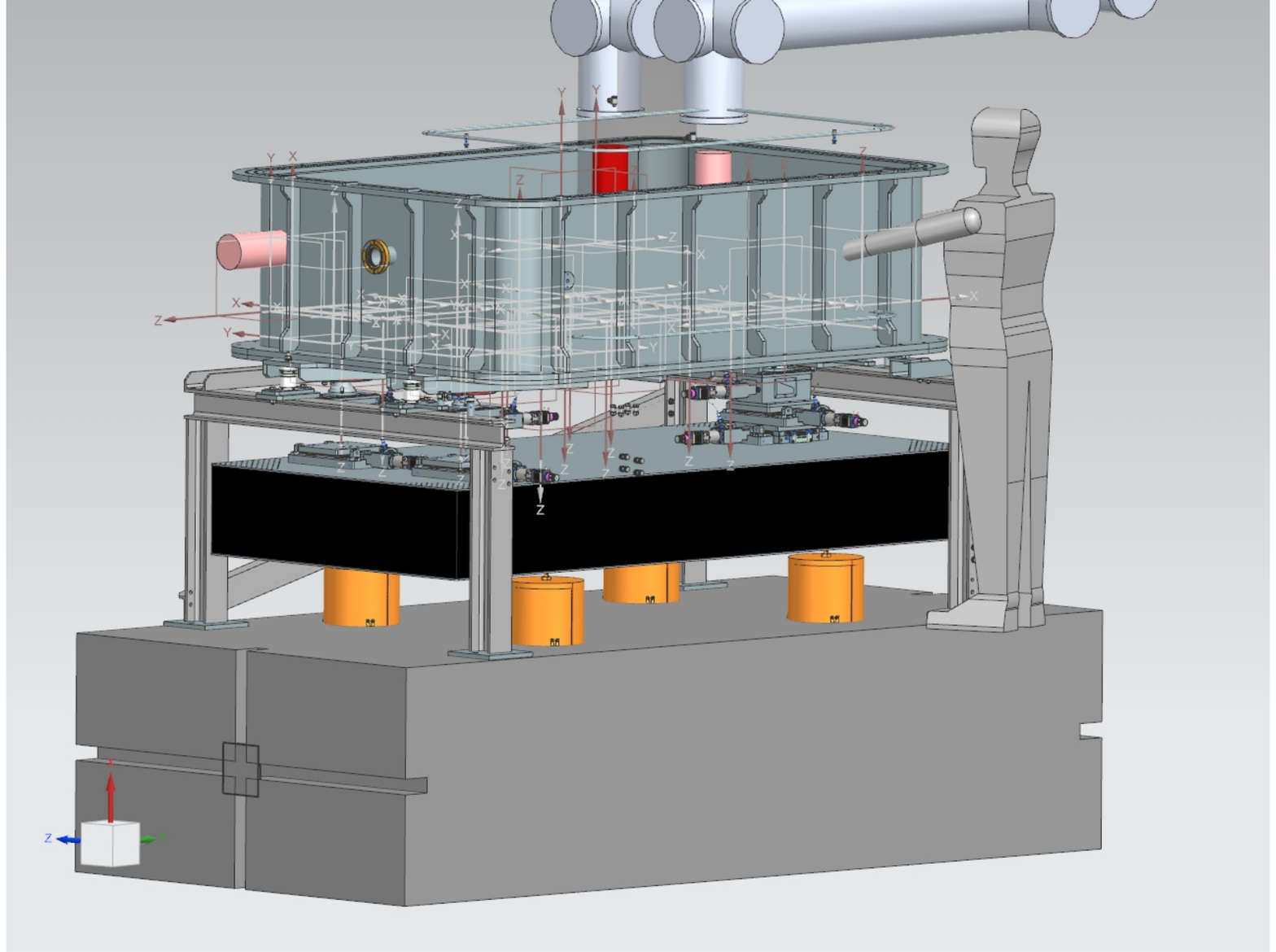}        
    \caption{ Detailed view of the IP chamber construction. The chamber has removable breadboards on kinematic locators inside for external prealignment and positioning. The optical set-up can be fully prealigned and brought to the IP chamber allowing maintenance to be performed in short access slots.  The breadboard inside the chamber will be mounted on the optical table underneath and isolated by the bellows.}
    \label{fig:ipchamber}
\end{figure}

CAD drawings for the interaction chamber are presented in Fig. \ref{fig:ipchamber}. As the height of the IP from the floor is about 2 m, large concrete blocks will be laid out underneath the chamber (the concrete block is shown in grey in Fig. \ref{fig:ipchamber}) to provide mechanical and vibration free stability. 

The incoming beam enters from the top in the chamber and  will propagate over three bend mirrors and an $f/\#=f/D_\mathrm{Laser}=f/3$  off-axis parabolic mirror to focus at the interaction point (where $D_\mathrm{Laser}$ is the laser beam diameter). 
A second parabolic mirror will re-collimate the beam and direct it to a beam-dump outside the chamber. Before reaching the beam-dump, about $1\%$ of the laser's energy will be sampled and sent back to the pulse diagnostics suite (also see  Fig.~\ref{fig:ipchamber_only}). 

All optics will be motorised for fine manipulation under vacuum. A motorised in-vacuum microscope will be used to optimise the mirror alignment, and to evaluate the laser focal spot size. We will monitor the incoming laser pointing stability by imaging the near- and far-fields of its leaking fraction behind the second bend mirror.

We will mechanically decouple the optics from the chamber's body by fixing the breadboard to an optical table beneath the chamber. The chamber itself, with its set of legs, will be set directly on the floor. Flexible bellows will form the contact surface between the vacuum vessel and the optical table similar to the concept shown in Fig. \ref{fig:beam_transport_isolation}.


Further discussion of the IP chamber body design, its integration to the other interfaces is found in Chapter~\ref{chapt12}.

\subsubsection{Inverse Compton Scattering (ICS) Chamber} 
\begin{figure}[htbp]
    \centering
    \includegraphics[width=0.8\textwidth]{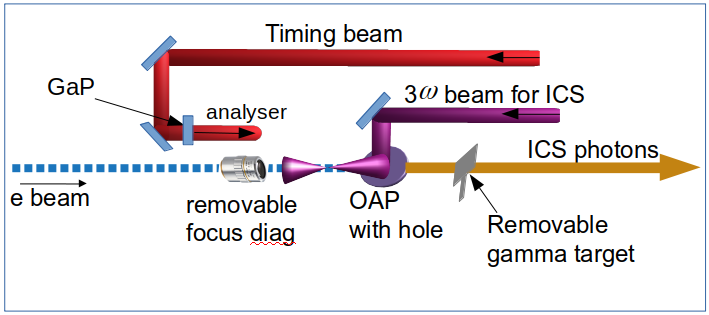}        
    \caption{Schematic of components inside the ICS chamber. It includes the jitter monitor tool and gamma converter target for phase-0. In phase-1 the ICS setup will be implemented. }
    \label{fig:ics_setup}
\end{figure}

A smaller ``target chamber" will be located at the position of the bremsstrahlung target. The ICS chamber will be a smaller scale version of the IP chamber and shares the basic features in terms of actuator design and pointing monitors for beams. It can accommodate the timing beam and the ICS beam (in a geometry closely matched to the IP chamber interaction) as shown in Fig. \ref{fig:ics_setup}. The interaction will take place at a head-on collision angle of zero degrees relative to the experimental axis. 
 
For the phase-0, this chamber will host an EO timing tool  and a gamma converter target for the $\gamma$-laser mode.  For phase-1, in ICS mode the electron beam will interact with a low intensity, narrowband $3 \omega_L$ laser pulse derived from the LUXE system. The conversion to $3 \omega_L$ and delay compensation will take place in the laser area. Both the $ \omega_L$ and $3 \omega_L$ probe will travel through the main beamline, although spatially separated. An off-axis parabola (OAP) with a hole will be used. This will allow the passing of the ICS photons towards the main interaction chamber. With a few 10s of mJ energy at $3 \omega_L$, a focus spot size of $\sim$ 5 $\mu$m and an interaction length of $2Z_R \sim 300- 400\, \mu$m, one can easily reach $\xi$ values of 0.1. The electron beam will be focused to a matching spot size as detailed in the spatial overlap criteria above.

\subsubsection{Adaptation of Interaction Geometry from Phase-0 to Phase-1}
As mentioned previously, the interaction geometry for  phase-0 and phase-1 will be the same in terms of focal spot and f-number (f/$\#$). The beamline will use the full beam size of 15 cm from the outset, requiring no further work on the beamline and IP.
For the initial installation, the smaller 5 cm beam will propagate through the beamline with full size mirror holders. A shorter focal length set of OAPs is required to maintain f/3 focusing in this case requiring them to be replaced as part of the upgrade. For the beam transport and optics inside the IP chamber, mounts would be prepared from outset to be compatible with the upgrade to phase-1.


\section{Interfaces and Integration} 
\label{section3-6}
\subsection{Interface of the Laser System}
 
The  laser system comes with its own control software for routine operation which will be interfaced for remote control and automation.  All the triggers and time delays are provided by the devices with the laser system. This includes all the Pockel's cell timing, high voltage switching, time delay on the pump laser and monitoring devices. Analogous control systems will be part of any other laser system to be purchased. 

In addition, we will implement feedback controlled systems to automate some of the routine realignment tasks or to stop data taking for manual realignment should the control parameter range be exceeded.

For this purpose parameters (such as energy, pointing, spatial profiles ) in the laser chain will be monitored by software and feedback control implemented (i.e. motorised mirrors, pump energy feedback). This will be integrated with the planned DOOCS/JDDD \cite{DOOCS} software control system used that will control/log the LUXE experiment status and data taking. 

\subsection{Data Acquisition}
The operation of the data acquisition (DAQ) system is described in Chapter~\ref{chapt11}. A full set of outputs of the diagnostics suite described previously will be saved at a minimum rate of 1 Hz. Selected outputs  will be saved at the full 10 Hz repetition rate to enable full analysis of the system performance. Precision experiments require the full data to be available for post-analysis and this informs the data strategy.  To facilitate this task, the laser data will be dealt with on the same footing as the data handling for  other parts of LUXE, via the DAQ software and control system.  This will allow coherent data storage and easy processing. More details about the architecture of the data flow to the central DAQ units are discussed in Chapter~\ref{chapt11}. 



\subsubsection{Event Based DAQ}
The event based DAQ refers to the storage of all relevant data when the collision of the laser pulse with the electron/photon beam takes place. Starting from the laser amplifier of the laser chain, the energy in the pulse will be measured using calibrated CCD camera and power meter which will be needed for the calibration of the energy diagnostics at the laser diagnostic suite. The major part of the data come from the diagnostic suite. Here for example, the energy (using a CCD camera), focus spot (16-bit CCD camera), spot in $\omega/ 2\omega$ (16-bit CCD camera) will be stored. In addition, the pulse duration based on the data format of the wizzler device will be stored.    



\subsubsection{Run Based DAQ}
 Not all relevant laser data can be measured on each shot with important examples being the IP spot monitor and the `Insight' diagnostics. Data from these will be taken at the beginning and end of each data run.
 
 Positions or co-ordinates of the motors in the compressor, IP chamber and ICS chamber will be logged  in a file and stored  in a database.


\subsubsection{Data Rate for DAQ}
Including all the  devices in the laser system, beam line, IP and ICS chamber and diagnostic suite, at present we estimate about 20 MB/s data flow towards the DAQ. The format of the stored data is 12- or 16-bit depth depending on the diagnostic system. The format (either images or data files) will also depend on the type of diagnostic .

\subsection{IP Chamber Interface and Integration}
The IP chamber is the central point of the experiment and has multiple interfaces. It has the interface with electron beamline in terms of vacuum and associated control system. The IP chamber will be designed to comply with \euxfel  accelerator vacuum requirements. Detailed information can be found in Chapter~\ref{chapt12}.

\section{Installation, Commissioning and Calibration}

In the following we outline  installation considerations for the laser, diagnostics and vacuum chambers. 

\subsection{Installation of New Laser}

A commercial laser system is the preferred option for the project and will be assembled and commissioned up to the factory acceptance test at the supplier. The full specification will be demonstrated at the factory acceptance test.

We note that this relaxes timelines for cleanroom installation. We use the JETI 200 timescales as a guide: the time elapsed from tender to factory acceptance test was 19 months; including the subsequent installation at the site, 24 months. We estimate the tendering process to require 3\,months preparation and 3\,months to the tender, i.e.\ a 30 months timescale is realistic according to experience.
The brand new laser system will be installed by the supplier in the presence of the laser team providing oversight and learning the system in the process. This sets the timeline for the completion of the laser cleanroom to avoid delays. The performance parameters of the factory acceptance test will be demonstrated again at the on-site acceptance test in the cleanroom.

The cleanroom will provide the specified services (electrical, water, cooling). The interlocks and synchronisation system will be prepared prior the installation of the laser by the LUXE laser team in collaboration with DESY/EuXFEL staff. These systems follow established precedents at comparable laser systems such as ReLaX. The installation of compressor hardware and vacuum will be carried out in conjunction with  engineering staff  from DESY.

\subsubsection{JETI40 Laser}
Previously, the installation of the existing JETI40 system has been considered. This option is still available and the JETI40 system is currently operational. The limited cost savings, higher operational risks and increased manpower requirements for upgrading JETI 40 to match the requirements of LUXE phase-0 are judged to outweigh the capital savings. 

\subsection{Installation of Laser Diagnostics}

\begin{itemize}
\item All the diagnostics will be pre-aligned using an alignment laser on individual breadboards. They will be placed on the diagnostics table according to their architecture plan.
\item The laser diagnostics will be purchased/developed and tested ready for installation at the JETI facility in Jena. It is anticipated that this process will be complete prior to the completion of the clean-room with the diagnostic vacuum hardware. 
\item Installation will progress using a low power laser system for pre-alignment of the paths. Once this is complete the LUXE laser can be directly injected into the diagnostic suite while bypassing the beamline to the IP in the target chamber. This allows the diagnostic commissioning to be fully autonomous from the installation work progress of components in the tunnel.
\item DAQ integration can proceed off-line to other processes before installation in the optical system.
\end{itemize}

It is anticipated that on site commissioning of the previously tested laser diagnostics will take 4 months at Osdorfer Born. Connection of the diagnostic output computers can be performed by the DAQ team as soon as each diagnostic is through testing in Jena.

{\normalfont \itshape Commissioning interdependence:} We stress that the full diagnostics suite is only required for precision measurements.  Commissioning of the LUXE tunnel elements  can proceed without full integration of the diagnostics suite into the laser.  First collisions with the electron beam for IP and diagnostics commissioning require only a minimal set of diagnostics consisting of energy calorimeter, autocorrelator and focal spot/electron beam overlap diagnostics in the IP chamber as well as the diagnostics required for the immediate running of the laser (not detailed as internal to the operation of the commercial laser system).

This minimal requirement is easily met and ensures that first e-laser collisions can take place as soon as delivery is complete. Full diagnostic suite commissioning can run in parallel with any use of the laser for e-laser collisions using the return beamline, ensuring the laser is fully available to commission both tunnel and laser diagnostics continuously.


\subsection{Installation of Transport Beamline}
The installation of the beamline requires major mechanical engineering support from EuXFEL/DESY. Two installation options are possible with standardised components DN400 to house both forward and return beam paths or DN250 for separate forward/return beamlines. An overview of the installation together with the ground level laser clean room was shown in Fig.~\ref{fig:XS1_side_intro}.


\begin{itemize}
    \item The hardware consists of vacuum  pipes with the mirrors installed in turning mirror chambers as shown above. Components of the piping system are largely commercial and standardised. The motorised mirror mount design already exists in Jena for JETI200.
    \item After the major hardware installation, vacuum pumps, gauges and gate valves will  be installed and the test related with vacuum level and control of valves will be performed.
    \item Installation can proceed at any time after delivery for components from just outside the cleanroom to the level above the \euxfel tunnel. Component availability is anticipated to be straightforward (not considering unpredictable geopolitical effects).
    \item Installation in the tunnel can proceed at a suitable shutdown separately from other laser related work.
    \item Alignment and testing of the beamline for stability can proceed using available low power lasers. These will be injected at the cleanroom connection point and allow beamline to be completed prior to any other work. Beamline installation is also possible without the IP chamber being fully installed.
    
    \item All mirror mounting components will be fully tested prior to installation. For multi-substrate mirror holders the relative pointing of  each mirror relative to the multi-mirror holder will be set prior to installation. Mirror mounts will be installed after vacuum testing of the beamline.
  
\end{itemize}    

\subsection{Installation  of IP Chambers}
The IP chamber optics are fully prealigned on removable breadboards, allowing work to be largely independent of access slots with breadboards removable and installable in a single short access period. This decouples IP issues from having to be resolved in a single routine access slot. Pre-assembly of the optical elements on the breadboard in a clean environment will require two weeks and a few days for fine alignment at the IP. We note that the design and integration of the hardware of the IP chamber body is included in Chapter~\ref{chapt12}.

The installation sequence will be as follows.

\begin{itemize}
    \item Confirmation that the chamber meets vacuum and mechanical specifications.
    \item Installation of the support structure  of the chamber as part of electron beamline installation.
    \item Installation of the isolation breadboard underneath the chamber.
    \item Installation of main chamber body without breadboard inside it.
    \item Connection with interfaces such as as the electron beam pipe, laser beam line, IP detector chamber. There would be gate valves installed with each interface to isolate the chamber for any work required in the air without interfering with other parts.
    \item Installation of vacuum pumps, gauges, gate valves, controllers etc.
    \item Vacuum tests.
    \item Install breadboard with optical elements pre-aligned. 
    \item Fine alignment with Ti:Sa laser beam at the IP.
\end{itemize}


\subsection{Installation of Tunnel Diagnostics}
Tunnel diagnostics will be fully tested with DAQ integration prior to installation. The final installation includes.

\begin{itemize}
    \item Installation of focus microscope system and test of actuators.
    \item Installation of alignment target and measurement of alignment targets.
    \item Installation of near and far field monitors for pointing measurements.
    \item The EO timing tool will be installed in the ICS chamber after prealignment on an external breadboard.  The final timing will be done during the commissioning as it requires the electron beam.
\end{itemize}

\subsection{Decommissioning}
The components in the laser area are modular and can easily be removed during the decommissioning phase. The IP and ICS chambers are the locations where interaction is happening and prone to radiation hazards. This will be checked with the radiation safety group and can be uninstalled as per their guidelines.



\section{Further Tests Planned}

\subsection{Diagnostic Development and Automation}
As discussed in the previous sections we are developing a nonlinear intensity tagging diagnostic. Tests of this diagnostic are currently ongoing on the JETI laser at Jena. The aim of this diagnostic is twofold. On the one hand it will allow the diagnostic system to produce an instant tag for the interaction intensity of each shot that can be included essentially `on the fly' without the need for intensive further post processing. The peak intensity is then simply a constant multiplied by the ratio of the brightest pixels in two images.

On the other hand it provides a `check-sum' type of cross-check. The detailed analysis and the intensity tagging must provide results closely matched to each other. If any of the assumptions become invalid or any diagnostic equipment ages/loses calibration due to alignment drift there will be a systematic offset between the intensity tagging and the multi-diagnostic full analysis.

Further work is ongoing regarding the alignment automation and enhanced stability. Issues that have been detected with cryo-heads are being resolved with liquid N$_2$ flow cooling instead of pumps. This reduces a significant source of noise and reduces pointing jitter. Furthermore, automation steps are being taken in the front end with pointing and energy control. All of these tests are being executed on existing systems and can be then rapidly integrated to the LUXE laser if successful.

A further area of ongoing work is the predictive stabilisation with neural networks described in this chapter. This has the potential for being a game-changer in the stabilising of the residual jitter in lasers and is being actively pursued.

\section*{Acknowledgements}
\addcontentsline{toc}{section}{Acknowledgements}
The authors wish to thank the JETI and ReLaX teams for their input and experience into the drafting of this note as well as the information provided by infrastructure staff at DESY and \euxfel. 

\newpage


\printbibliography
\end{refsection}



\chapter{Pixel Tracker}
\label{chapt4}
\begin{refsection}

\chapterauthors{A.~Santra\\ Weizmann Institute of Science, Rehovoth (Israel)}{N.~Tal Hod\\ Weizmann Institute of Science, Rehovoth (Israel)}
\date{\today}

\begin{chaptabstract}
The goal of the LUXE silicon pixel tracking detector is the precise measurement of the rate and energy of the non-linear Breit-Wheeler \ee pairs produced in the interaction between high energy electrons or photons and high intensity laser pulses expected in the experiment.

The tracker is based almost entirely on the technology developed for the upgrade of the inner tracking system of the ALICE experiment at the LHC and particularly the state-of-the-art ``ALPIDE'' monolithic active pixel sensors at its heart.
The ALPIDE pixel-size is $\sim 27\times 29~\um^2$ resulting in a demonstrated spatial resolution of $\sigma\sim 5$~\um.
A time resolution of a few~\us was demonstrated.

These features are packed in a very low material budget of 0.05\%$X_0$ per layer (sensor only).
The expected performance, quality, availability, and relatively low-cost of this technology make it particularly appealing for LUXE with minimal adaptations.

The tracker will be assembled as two arms of four layers staggered at each of the two sides of LUXE beamline downstream the electron/photon and laser interaction point and downstream a $\sim 1$~T dipole magnet.

A total detector volume of $ 50\times 1.5 \times 30~\mathrm{cm}^3$ is achieved providing a good coverage of the expected signal and a good lever arm for precise tracking.

The studies shown here indicate that it is possible to build an economic and precise structure to achieve $>90\%$ tracking efficiency for measuring the produced positrons and electrons above $\sim 2$~GeV,
with an energy resolution of $\lesssim 1\%$ and with a position resolution below $100~\um$ (depending mostly on the track inclination, the detector alignment and the reconstruction algorithm configuration).
This performance will allow LUXE to unambiguously measure signal rates spanning several orders of magnitude ($\sim 10^{-3}-10^4$ particles per bunch crossing), while efficiently rejecting most of the beam induced backgrounds.
At the highest signal rates ($>10^4$ particles per bunch crossing), tracking becomes highly challenging, but the detector can still be used in a counter mode.

The layout of the system, its components and its interfaces with LUXE are discussed in detail.

\end{chaptabstract}

\section{Introduction}
\label{sec:intro}
The Conceptual DesignRreport (CDR) of LUXE ~\cite{Abramowicz:2021zja} outlines the physics goals and the overall experimental design.
This chapter focuses on the description of the LUXE pixel tracking detector  and the non-linear Breit-Wheeler (NBW) pair production measurement.



The proposed layout of the LUXE experimental setup is shown in 
Fig.~\ref{fig:layoutdet}. 


The \elaser or \glaser collision rate is dictated by the electron-beam repetition rate of 10~Hz.
Therefore, LUXE will run with 1~Hz of collisions with the laser and 9~Hz of electron-beam only for background studies.

In the collision, the produced NBW \ee pairs are first separated according to their energy by a dipole magnet placed right after the IP with a magnetic field strength of roughly 1-1.5~T.
The separation induced by the dipole in the horizontal direction (transverse to the beamline) is expected to reach a maximum of $\sim 50$~cm on each side of the beamline, while the spread in the vertical direction (also transverse to the beamline) is expected to be smaller than $\sim 1$~cm.
It is expected that the signal \ee particles at the detector would range from $\sim\mathcal{O}(10^{-3})$ to $\sim\mathcal{O}(10^6)$ pairs per collision (denoted hereafter as bunch-crossing, BX), depending on the different experimental setup options and notably on the laser intensity at the IP.
At the same time, the number of charged background particles is expected to reach a level of $\sim\mathcal{O}(10^3-10^4)$ per BX.
This background comes mostly from the interaction of the electron beam (in the \elaser mode or photon beam in the \glaser mode) with the material of the different elements along the experiment as well as with the air.
Hence, with the signal rate (per BX) spanning 8-9 orders of magnitude and the background rate fixed at a few thousands, the requirement for a multi-layer solution with high sensor granularity is clear and simple counters cannot be used.
While the system is designed to work at both low and high multiplicities, a fallback option to reduce the signal rate at the most extreme cases (end of phase-1) is to defocus the electron beam.
This does not affect the physics, only the degree of overlap between the beams and hence helps to reduce the rates.

\begin{figure}[!h]
  \centering
  \begin{overpic}[width=0.49\textwidth]{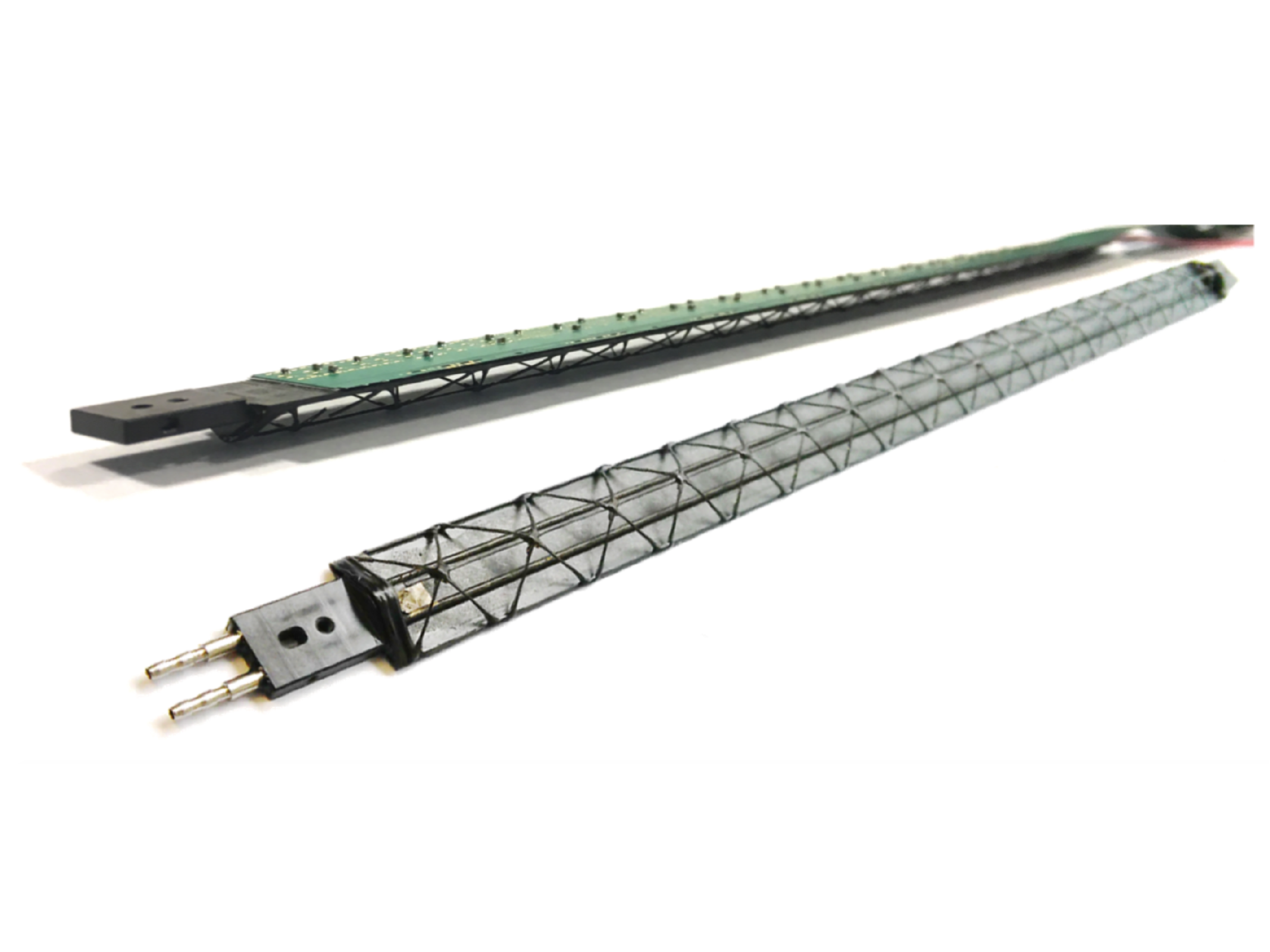}\put(20,65){\textrm{}}\end{overpic}
  \begin{overpic}[width=0.49\textwidth]{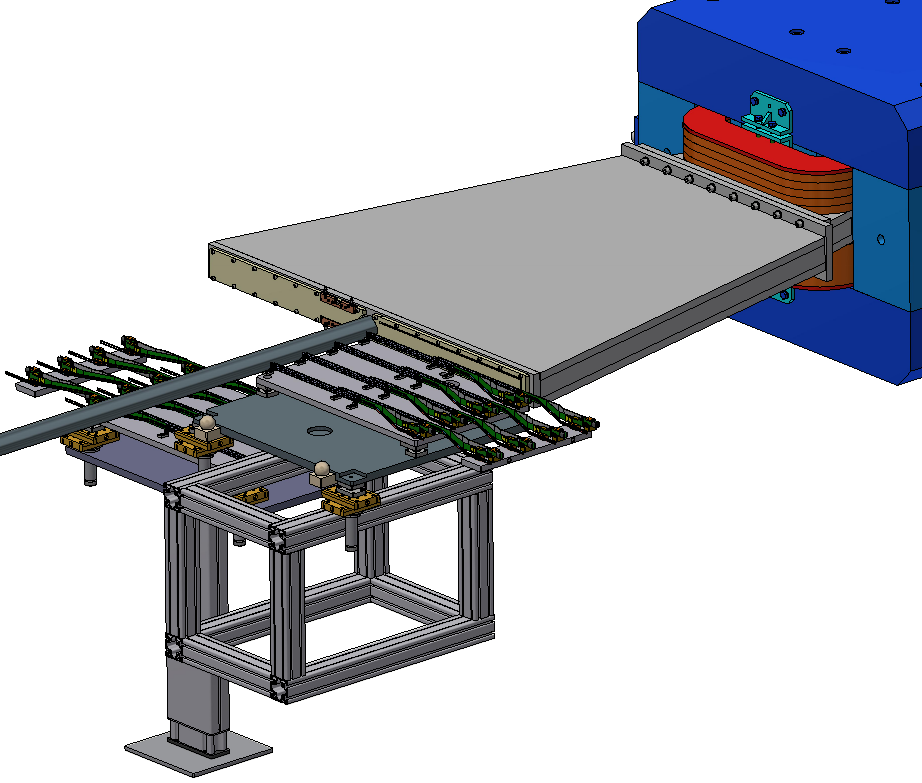}\put(20,65){}\end{overpic}
  \caption{{\footnotesize Left: an ALPIDE ``stave''~\cite{Abelevetal:2014dna,AGLIERIRINELLA2017583} including nine ($3 \times 1.5~\mathrm{cm}^2$) ALPIDE sensors flip-chip mounted on a flexible printed circuit that is glued to a carbon fibre cooling sheet and a mechanical support frame.  Right: The LUXE tracker designed at WIS, showing 8+8 staves on kinematically mounted trays in front of a vacuum chamber (grey) on the two sides of the beam. Particles are fanned by a magnet (blue/red), exit the vacuum through a thin foil window (beige) and impinge on the staves. The positron arm is not supposed to change its location during operation, while the electron arm (shown in its down position for the \elaser mode) must be brought in and out of the interaction plane, depending on the mode of operation.}}
  \label{fig:tracker}
\end{figure}

The tracking detector, shown in Fig.~\ref{fig:tracker}, 
will be built at the Weizmann Institute of Science (WIS) from state-of-the-art thin monolithic active pixel sensors (MAPS), dubbed ``ALPIDEs''~\footnote{ALPIDE stands for ``ALice PIxel DEtector''}~\cite{Abelevetal:2014dna,AGLIERIRINELLA2017583}, which integrate the sensing volume with the readout circuitry in one all-silicon chip for the first time in high energy physics (HEP) experiments.
The ALPIDE chips ($\sim 3\times1.5~\mathrm{cm}^2$) are produced by TowerJazz~\cite{TowerJazz} for the upgrade of the ALICE experiment~\cite{Collaboration_2008} at the LHC.

The basic detector element 
is called a ``stave''. 
Each stave is built from nine ALPIDE sensors lined up to cover an area of $\sim 27\times1.5~\mathrm{cm}^2$.
A photo of a stave is shown in Fig.~\ref{fig:tracker} (left). 
The LUXE tracker will be using the stave assemblies purchased from CERN. 

The LUXE tracker will be assembled as two arms of four layers, staggered at each of the two sides of LUXE beamline.
Each layer of the tracker comprises two such staves, placed one behind the other at $\sim 1.2$~cm distances along the beam direction and with $\sim 4.5$~cm of overlap in the transverse direction.
The distance between two layers is $\sim 10$~cm along the beam direction.
The detector assembly is shown in Fig.~\ref{fig:tracker} (right). 
It will be placed downstream of the electron/photon and laser IP and downstream of a $\sim 1$~T dipole magnet.
With a total detector volume of $\sim 50\times 1.5 \times 30~{\mathrm cm}^3$, good coverage of the expected signal and a good lever arm for precise tracking are achieved.

The LUXE tracker will measure positrons and electrons with a position resolution below $100~\mu{\mathrm m}$ (depending mostly on the track inclination, the detector alignment and the reconstruction algorithm configuration).
When used in conjunction with a magnetic field (the detector is situated outside the field volume), the resulting lever-arm of $\sim 30$~cm provides a momentum resolution of $\sim 1\%$, a tracking efficiency of $\sim 90-95\%$ and a background rejection power better than $10^{-3}-10^{-4}$, particularly at low signal-tracks multiplicities with $\mathcal{O}(1-10)$ signal tracks.

This excellent tracking performance is based primarily on the granularity of the sensors themselves ($\sim 27\times 29~\um^2$), but also on an algorithmic power, which enables measurement of the track momentum precisely, as well as constraining of its origin to the IP.
This power now comes from a pixel clustering algorithm, followed by a Kalman filter~\cite{KalmanFilter} track fit.

To count the \ee pairs, it is in principle sufficient to look only at the positron content of a recorded event (BX).
Clearly, measuring both positrons and electrons on the two sides of the beam provides better precision.
In low signal rates, measuring also the electrons provides event-by-event access to the parent photon.

In Section~\ref{sec:technical}, the technical aspects of the detector (services, readout, mechanics) are discussed. 
A detailed study of the tracker performance is given, based on full simulation, in Section~\ref{sec:performance}.

\section{Requirements and Challenges}
\label{sec:challenges}
\begin{figure}[!ht]
\centering
\begin{overpic}[width=0.46\textwidth]{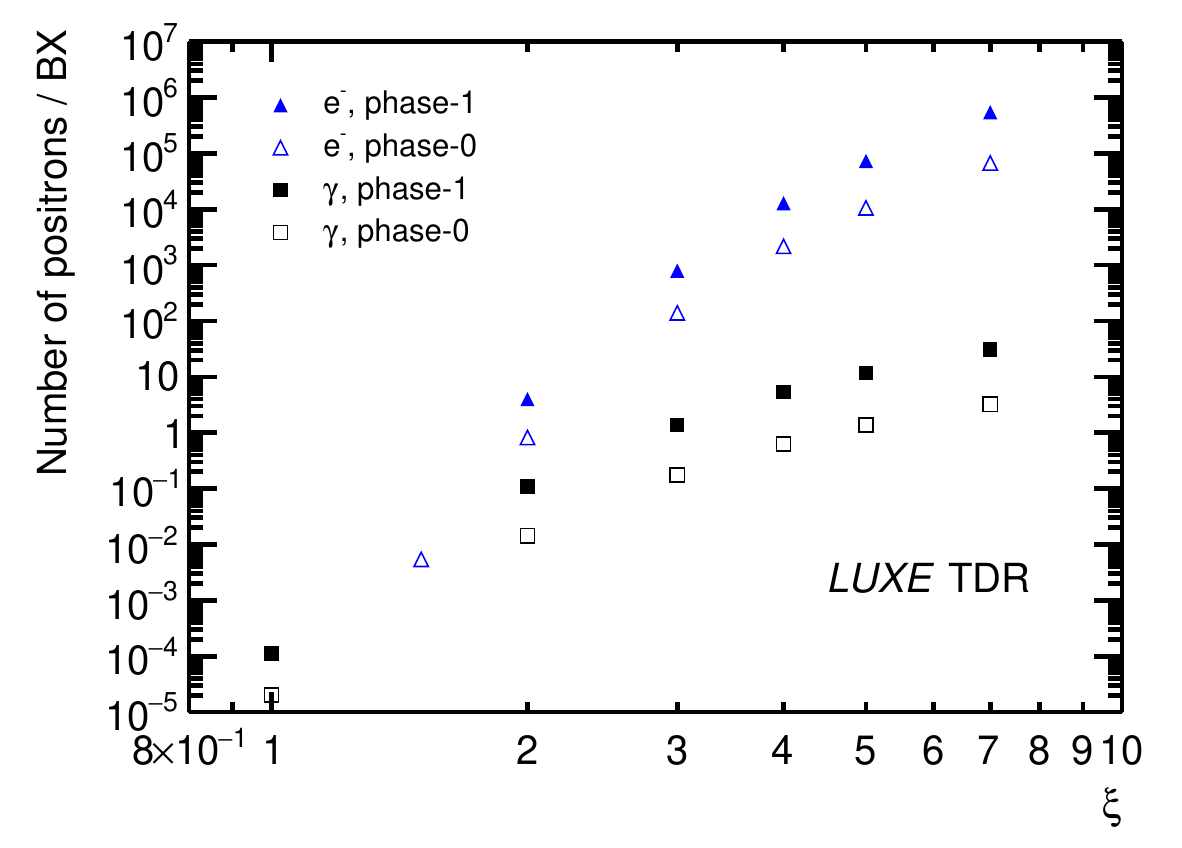}\end{overpic}
\begin{overpic}[width=0.49\textwidth]{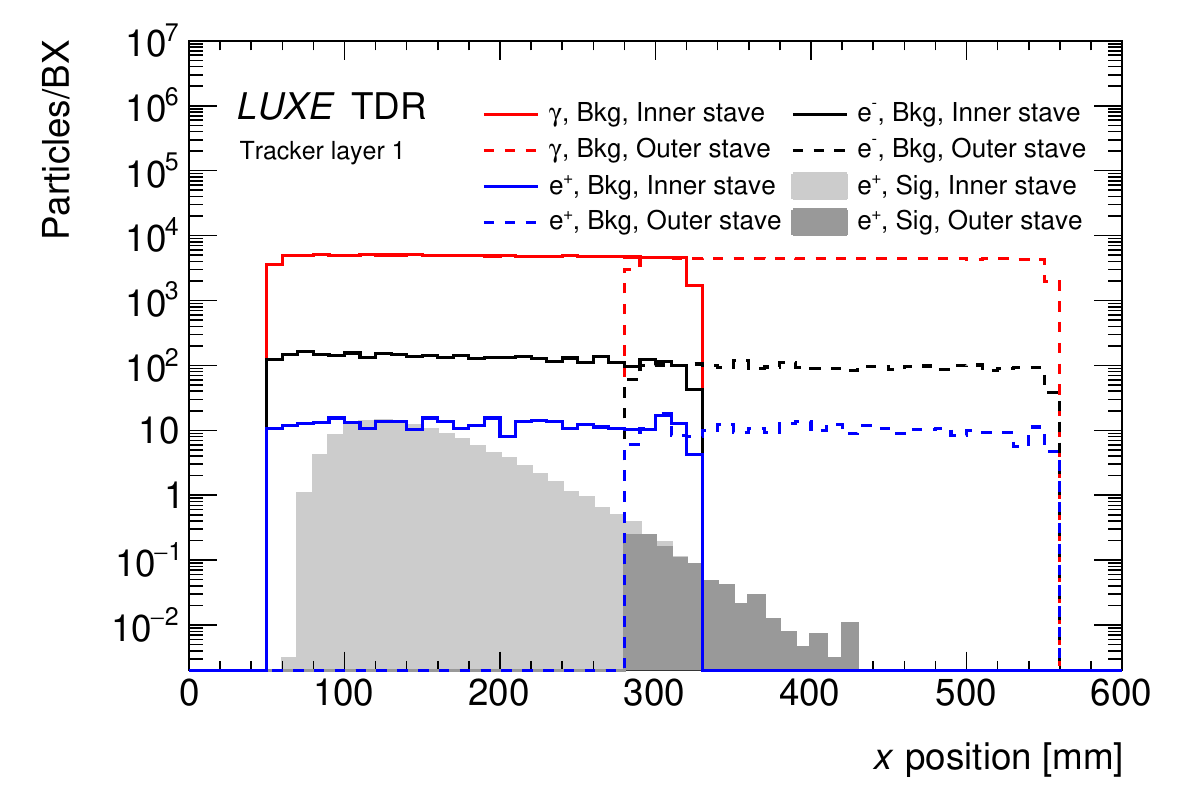}\end{overpic}
\caption{Left: number of positrons per BX in \elaser and \glaser collisions, as a function of the laser intensity parameter $\xi$. Right: the level of background (which consists of electrons, positrons and photons) compared to a typical signal in the \elaser mode, versus the $x$ position of particles for the first layer of the tracker.
All particles intersecting the first tracker layer are considered, regardless of the actual energy deposition in the sensitive volume. The solid lines (light grey fill for the signal) correspond to the tracks in the inner stave (closest to the beampipe), while the dashed lines (dark grey fill for the signal) correspond to the tracks from the outer stave (farthest from the beampipe).
The positron signal, shown in grey, corresponds to the phase-0 laser with a $\xi=3.0$.}
\label{fig:challenge_E_and_rates}
\end{figure}

A clear indication of the challenge in the measurement of NBW positrons (and electrons) can be understood from the left plot in Fig.~\ref{fig:challenge_E_and_rates} (same as 
Fig.~2.5 (right), repeated here for the convenience of the reader).
One can see the expected positron rates for different laser intensities and for the different run modes in the two phases of LUXE.
The intensities are controlled by the pulse waist\footnote{The waist is the transverse width of the laser pulse at focus.} ($w_0$) at focus, while the laser power is kept fixed.
The difference between phase-0 and phase-1 is mostly the laser power (40 and 350~TW, respectively).
The positron (and electron) detector must be capable to extract unambiguously the precise number of signal positrons, as well as their energy spectra, for both extremely low $(\sim 10^{-3})$ and extremely large $(\sim 10^{6})$ numbers of particles per BX.
If the signal is distributed uniformly across the full transverse length of one arm, the latter number corresponds to an occupancy of $\textstyle{\frac{10^6\times (27\times 29~\um^2)}{50\times 1.5~\mathrm{cm}^2}}\sim 0.1$ signal particles per pixel per BX (or $>\mathcal{O}(100)$ particles per mm$^2$ per BX).
In reality the signal is not uniformly distributed and hence the per-pixel occupancy will be even larger in some regions of the tracker (and much smaller in other regions). 
Furthermore, one particle will fire several pixels (2-3 pixels on average in the case of the ALPIDE).
Notably, it also needs to withstand the expected radiation dose ($10^{-4}$~Gy/BX and $\sim 10^3$~Gy/year assuming operation at 10~Hz with $10^7$ BXs in a year) from the runs with the largest signal multiplicities without deterioration.

On the right plot of Fig.~\ref{fig:challenge_E_and_rates}, one can see an example of how a typical signal would spatially appear ($x$ coordinate) in the detector, with respect to the expected background from electrons, positrons and photons.
Since the setup acts essentially as a spectrometer,
the $x$ coordinate has a 1:1 correspondence with the energy of the particles.
In this example, the background-particle occupancy per BX is larger by more than three orders of magnitude than the respective signal occupancy.
The detector must be capable of heavily rejecting these backgrounds.
Indeed, the energy of these particles is typically very low, while the energy of the signal positrons and electrons is at a few $(\sim 2-14)$~GeV.
Moreover, the origin of the background particles almost never coincides with the IP (as would be expected of the signal) but rather from multiple ``hot-spots'' around the experimental setup (e.g. where the electron beam exits the vacuum, the beam dumps, etc.).
Hence, if the detector can precisely measure the particles' momenta (and infer energy from it) as well as the geometrical consistency of their origin with the IP (the origin of signal particles is expected to be consistent with it), it is clear that the rejection power can be large.

The signal particles are constrained to a very small range in the vertical $(y)$ axis (a few mm), while the spread along the $x$ coordinate (where the particles will be measured after they exit the vacuum) is determined by the magnetic field.
This spread is typically constrained to $\sim 50$~cm (only signal particles with $E\leq 2$~GeV would be out of acceptance).

Since redundancy in the measurement is extremely important, the detector measurement must be as non-destructive as possible, such that another (different) detector system could be placed behind the tracker detector.
Hence, the depth of the detector and particularly its material budget also needs to be kept small enough to allow another independent measurement from a different system behind it.
Finally, while the detector needs to be efficient for charged particles (electrons and positrons), it also needs to be as agnostic as possible to background photons and neutrons, which come in large numbers from many elements along the setup.

The detector requirements can be summarised as follows:
\begin{enumerate}
    \item sensitive surface area of $\sim 1\times 50~\mathrm{cm}^2$,
    \item low material budget for a non-destructive measurement (a few 100~\um),
    \item insensitive to neutral particles (photons and neutrons),
    \item radiation tolerant to withstand the largest signal fluxes\footnote{This expected dose is very small compared to the equivalent e.g. in ALICE at the LHC} ($\sim 10^6~e^+/{\mathrm s}$) expected at $\sim 10^{-4}$~Gy/BX (and $\sim 10^3$~Gy/year assuming operation at 10~Hz with $10^7$ BXs in a year), 
    \item measure the single-particle energy ($\mathcal{O}(1\%)$) and spatial origin ($\mathcal{O}(100~\um)$),
    \item reject the background to below $\sim 10^{-3}$ particles per BX at low signal multiplicities ($\leq\mathcal{O}(10)$),
    \item fully disambiguate between particles at medium signal multiplicities ($\geq\mathcal{O}(10)$ and $\leq\mathcal{O}(10^4)$),
    \item count particles at high signal multiplicities ($\geq\mathcal{O}(10^4)$), effectively functioning as a screen (no tracking).
\end{enumerate}
These requirements impose major challenges on the choice of the technology and its physical arrangement.
Clearly, with one sensitive layer, only a subset of these requirements may be satisfied.
It is also immediately seen that a multi-layer approach with thin and highly granular sensors is necessary.
Another implicit (though very practical) requirement is that, in view of the LUXE installation timeline, the technology itself must already be ``off-the-shelf''.


With this in mind, the ALPIDE technology is a good solution, satisfying all of the requirements above.

\section{System Overview}
\label{sec:systemoverview}
As shown in Fig.~\ref{fig:tracker} (right), the tracker will be placed downstream of the exit window of the vacuum chamber.
The rectangular part of the chamber in the figure sits inside the magnet volume, while the triangular part of the chamber allows the fanned-out particles to propagate in vacuum, before they exit it and reach the detectors.
In principle, it would have been better to allow the signal particles to propagate all the way to the detectors in vacuum and also have the detectors themselves in vacuum to avoid unnecessary scattering in material or in air.
However, it was decided to leave all detectors outside vacuum due to the associated complexity with access, multiple dedicated feed-through interfaces, etc.
The exit window itself (shown in beige at the long base of the triangular chamber) is an aluminium plate with an inner part of its area $(\sim 2\times 50~\mathrm{cm}^2)$ in the middle, machined to a thickness of 300~\um.
This thickness is sufficient to maintain the required vacuum, while also minimising the interference to the signal particles on their path to the detector.
Hence, the first tracking layer sits just a few centimetres in $z$ after the window to minimise the path which the signal particles travel in air.

The tracker will have two identical arms with four layers each.
In the \glaser mode, the two tracker arms are in their nominal position.
In the \elaser mode there is only one arm since the electron side of the tracker has to be taken out of the interaction plane due to the huge radiation from the beam and electrons.
This can be achieved by lowering the hexapod table on which the electron side of the tracker will reside. 

Each layer covers an effective surface area of $\sim 50\times 1.5~\mathrm{cm}^2$ in the $x\times y$ plane.
The dimension in $y$ is dictated by the ALPIDE chip size, while the dimension in $x$ is a multiple of nine ALPIDE chips (3~cm long each) in a row, integrated on one $\sim 27$~cm long assembly as shown in Fig.~\ref{fig:tracker} (left).
At LUXE, two such $27$~cm assemblies are enough to capture most of the signal horizontally in $x$.
To provide a good lever arm for a precise measurement of the track 
position and hence the energy, the distance between the layers in $z$ is chosen to be $\sim 10$~cm.
These dimensions allow the capture of nearly 100\% of the signal particles within the tracker volume, such that at least four layers are traversed (and hence at least four hit-clusters can be expected).
Only the low-energy particles (below $\sim 2$~GeV) will experience the maximum bending in the magnetic field and will be outside of the acceptance of the tracker.
It is important to note that the signal distributions typically start at $\sim 1-2$~GeV and that, moreover, the particles with $\sim 1.5$~GeV (and below) are likely to hit the walls of the magnet.
The layers themselves are very thin ($X/X_0=0.357\%$), such that the momentum of the signal particles  does not get significantly altered when they exit the tracker volume onto the detector system behind it.

The detector services, namely the power supply (digital and analog), the readout and the water cooling come out from the side of the layers farthest from the beam axis.
These are routed to a shielded area, where the power supply system and first readout station are located, a few metres away from the detector position.

The ``bare-bones'' version of the tracker is essentially one quarter of the full tracker in the positrons side only with one 27~cm assembly instead of two.
This allows capture of the bulk of the signal needed to perform the baseline measurements in the \elaser mode alone.

A detailed description of the different components of the pixel tracking detector is given in Section~\ref{sec:technical}.

\section{Technical Description}
\label{sec:technical}
\subsection{Monolithic Active Pixel Sensors}
\label{sec:maps}
The common pixel sensor technology, used in most HEP experiments so far, has been hybrid.
These constitute a relatively large material budget, typically more than 1.5\%$X_0$ per layer (e.g. in the ATLAS Insertable B-Layer).
This thickness is distributed among the module components, the cooling and support structures.
The hybrid module production, including bump-bonding and flip-chipping, is complex and leads to a large number of production steps.
Consequently, hybrid pixel detectors are comparatively expensive.

Modern CMOS imaging sensors, instead, make use of 3D integration to combine high-resistivity and fully depleted charge collection layers with high-density CMOS circuitry, forming {\normalfont \itshape monolithic} (contrary to hybrid) pixel sensors, in which the sensor and the electronics circuitry form one entity~\cite{Garcia-Sciveres:2017ymt}.
While a combination of fully depleted high-resistivity silicon with CMOS readout is in fact a requirement of particle physics, it has also independently evolved to be the standard in commercial electronics applications like smartphone image sensors.

In Monolithic Active Pixel Sensors (MAPS), the sensing volume is effectively an epitaxial layer (epi-layer) grown on top of the lower quality substrate wafer and hosting the CMOS circuitry.
The thickness of this epi-layer typically is in the range of 1–20~\um, whereas thicker layers are often used in processes addressing CMOS camera applications.
In the first MAPS for particle detection, the charge deposited in the epi-layer can be as large as 4000$e^-$ for a typical thickness of 15~\um~\cite{Garcia-Sciveres:2017ymt}.
Since the epi-material usually has low resistivity and the allowed biasing voltages are low in CMOS technologies, the epi-layer usually is depleted only very locally around the charge collection node.
The deposited charge of a traversing particle therefore is mostly collected by diffusion rather than by drift.
This renders the signal generation relatively slow and incomplete (i.e. not all charges arrive at the collection node).
Due to the relatively longer time and path for charge collection, MAPS usually have a lower radiation hardness.
However, since the expected radiation level for HL-LHC heavy ion collisions is over three orders of magnitude lower than for LHC $pp$-experiments, the ALICE inner tracking system (ITS) upgrade~\cite{Abelevetal:2014dna,MAGER2016434} has chosen the MAPS technology based on the 180~nm CMOS node offered by TowerJazz~\cite{TowerJazz}.



For a review of both hybrid and monolithic pixel sensors technologies, see~\cite{Garcia-Sciveres:2017ymt}.

\subsection{The ALPIDE sensor}
\label{sec:alpide_sensor}
The ALPIDE chip is a $30\times 14~{\textrm{mm}}^2$ large MAPS containing approximately $5\times 10^5$ pixels, each measuring about  $27\times 29~\um^2$ arranged in 512 rows and 1024 columns that are read out in a binary hit/no-hit fashion.
It combines a continuously active, low-power, in-pixel discriminating front-end with a fully asynchronous, hit-driven combinatorial circuit. 
The ALPIDE sensors and its characteristics are discussed in detail in~\cite{YANG201561,Abelevetal:2014dna} 

The ALPIDE prototypes have been extensively tested in laboratories as well as at a number of test beam facilities in the past few years.
The chips have demonstrated very good performance both before and after irradiation; see, for example~\cite{SENYUKOV2013115, AGLIERIRINELLA2021164859,DANNHEIM2019187,MAGER2016434}. 
In~\cite{MAGER2016434}, the sensor has shown a detection efficiency above 99\%, a fake hit rate much better than $10^{-5}$, a spatial resolution of around $\sim 5$~\um and a peaking time of around 2~\us.


\subsection{Mechanics}
\label{sec:mech}

\subsubsection{Staves \& Layers}
\label{sec:staves}
We start with an assembly named ``Stave'' (in ALICE), which is built from the following elements:
\begin{itemize}
\item \textbf{Space Frame}: a carbon fibre structure providing the mechanical support and the necessary stiffness;
\item \textbf{Cold Plate}: a sheet of high thermal-conductivity carbon fibre with embedded polyamide cooling pipes, which is integrated into the space frame. The cold plate is in thermal contact with the pixel chips to remove the generated heat;
\item \textbf{Hybrid Integrated Circuit (HIC)}: an assembly consisting of a polyamide flexible printed circuit (FPC) onto which the pixel chips and some passive components are bonded.
\end{itemize}

Each stave is instrumented with one HIC, which consists of nine pixel chips in a row connected to the FPC, covering an active area of $\sim 15\times 270.8~{\textrm{mm}}^2$ including 100~\um gaps between adjacent chips along the stave.
The interconnection between pixel chips and FPC is implemented via conventional Aluminium wedge wire bonding.
The electrical substrate is the flexible printed circuit board that distributes the supply and bias voltages as well as the data and control signals to the pixel sensors.
The HIC is glued to the ``cold plate'' with the pixel chips facing it in order to maximise the cooling efficiency.
The polyamide cooling pipes embedded in the cold plate have an inner diameter of 1.024~mm and a wall thickness of 25~\um.
These are filled with water during operation.

\begin{figure}[!ht]
\centering
\begin{overpic}[width=0.6\textwidth]{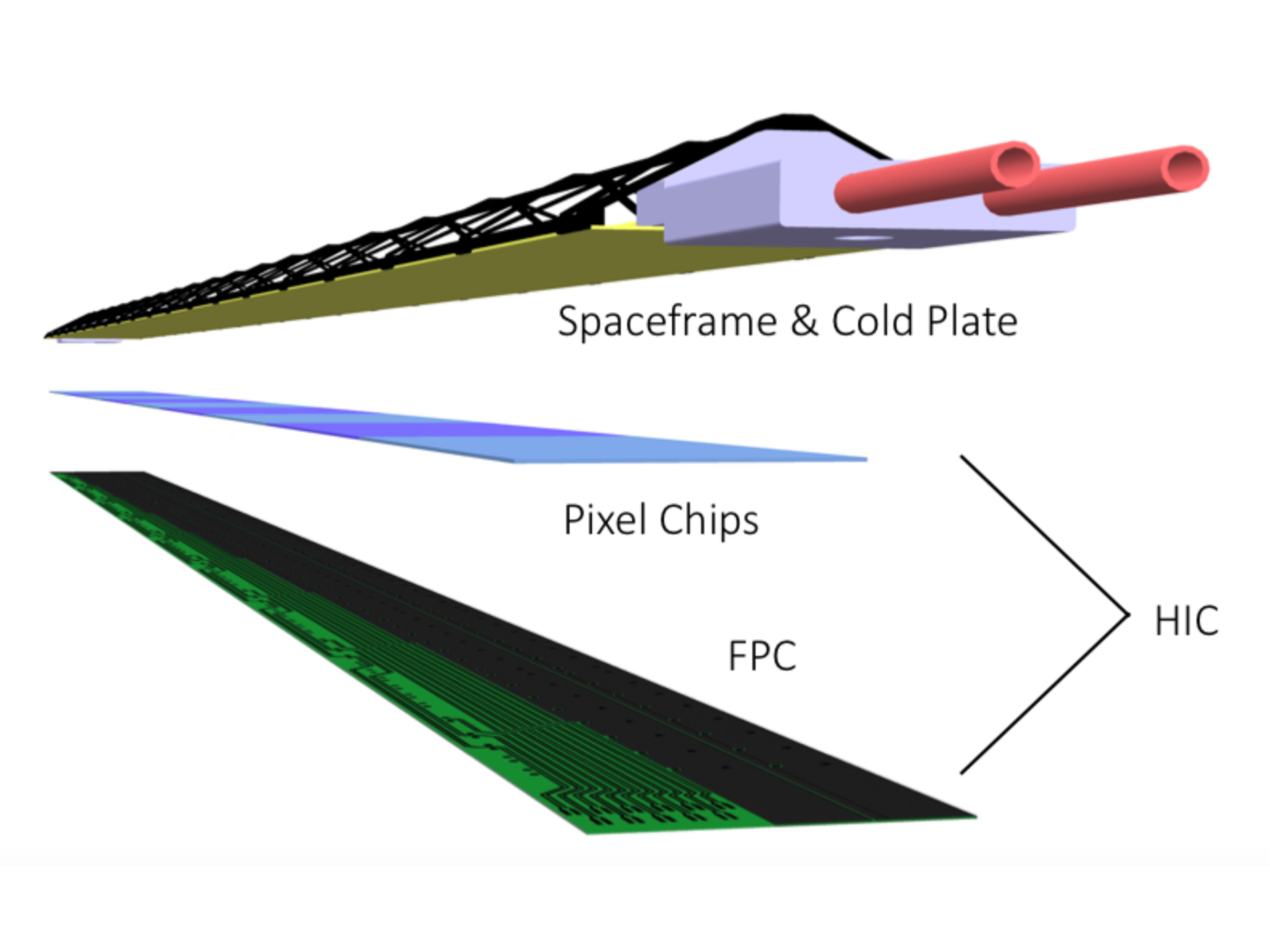}\end{overpic}
\vspace*{-0.5cm}
\caption{Schematic layout of a single stave. Nine pixel sensors are flip-chip mounted on a flexible printed circuit (FPC) to form a hybrid integrated circuit (HIC). The HIC is glued on a carbon fibre support structure (Spaceframe), which integrates a water cooling circuit (Cold Plate). }
\label{fig:stave1}
\end{figure}

A schematic drawing and a photograph of the stave is shown in Fig.~\ref{fig:stave1} and a photograph was shown in Fig.~\ref{fig:tracker} (left).
An extension of the FPC, not shown in the figure, connects the Stave to a patch panel that is served by the electrical services entering the detector only from one side.

The overall structure mean thickness is $X/X_0 = 0.357$\%, with the material budget broken down to fractions of the total width as: $\sim 15$\% silicon sensor (with absolute thickness of 50~\um), $\sim 50$\% electrical substrate (FPC) including the passive components and the glue, $\sim 20$\% cooling circuit and $\sim 15$\% carbon spaceframe~\cite{ALICE-PUBLIC-2018-013}.

In LUXE, each stave serves as a ``half-layer'' such that the total transverse length of one layer (on each side of the beam) is roughly $50$~cm.
The two staves of one layer are staggered behind each other with a small distance in $z$ ($\sim 1$~cm) and an overlap of $\sim 4$~cm in $x$.
The two staves forming one layer are staggered such that the FPC extension of the ``outer'' stave (farthest from the beam axis) does not shadow the ``inner'' stave (closest to the beam axis) and vice versa. 
Hence, the outer stave is closer to the IP in $z$ than the inner stave.

\begin{figure}[!ht]
\centering
\begin{overpic}[width=0.538\textwidth]{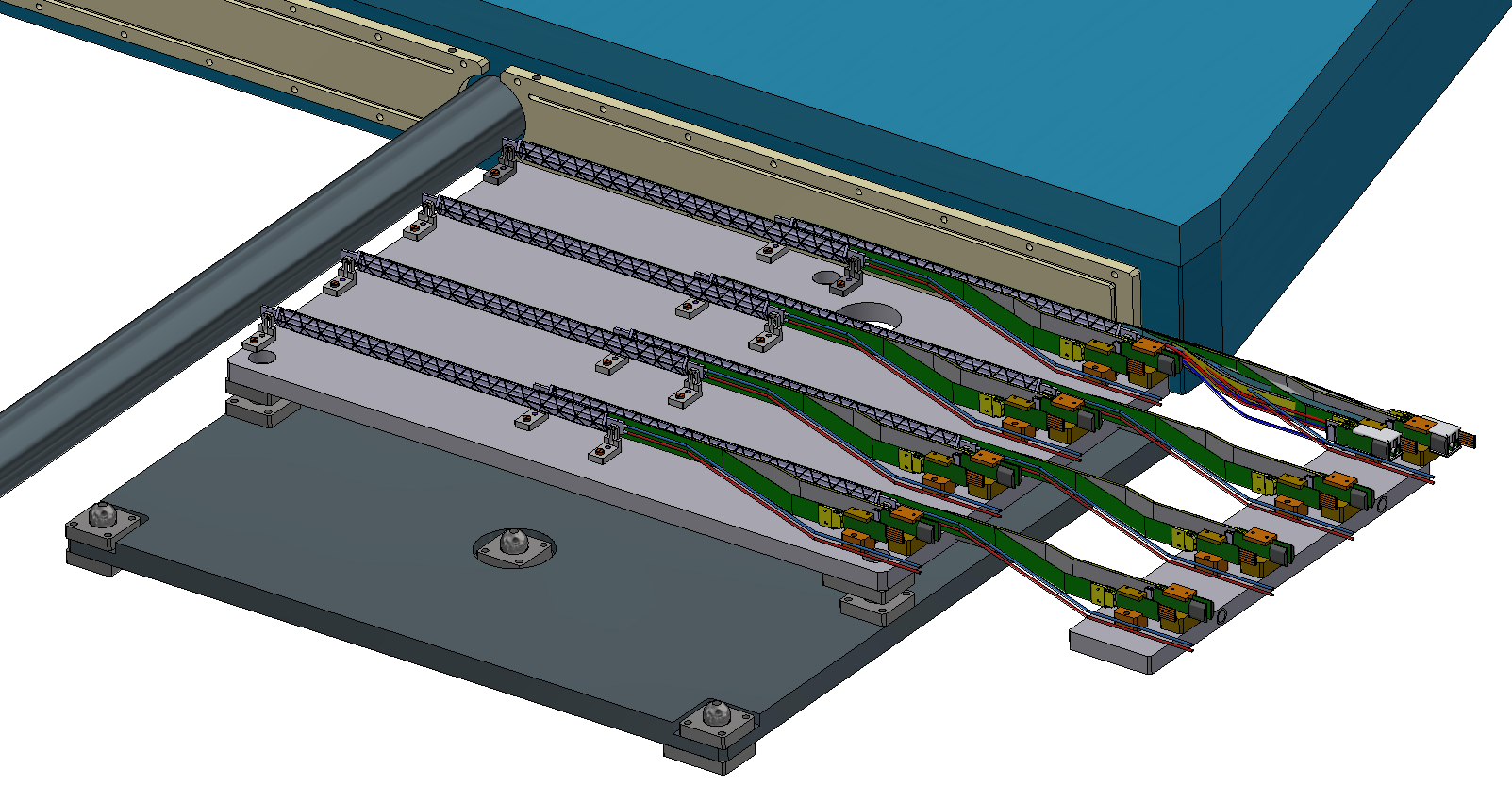}\end{overpic}
\begin{overpic}[width=0.65\textwidth]{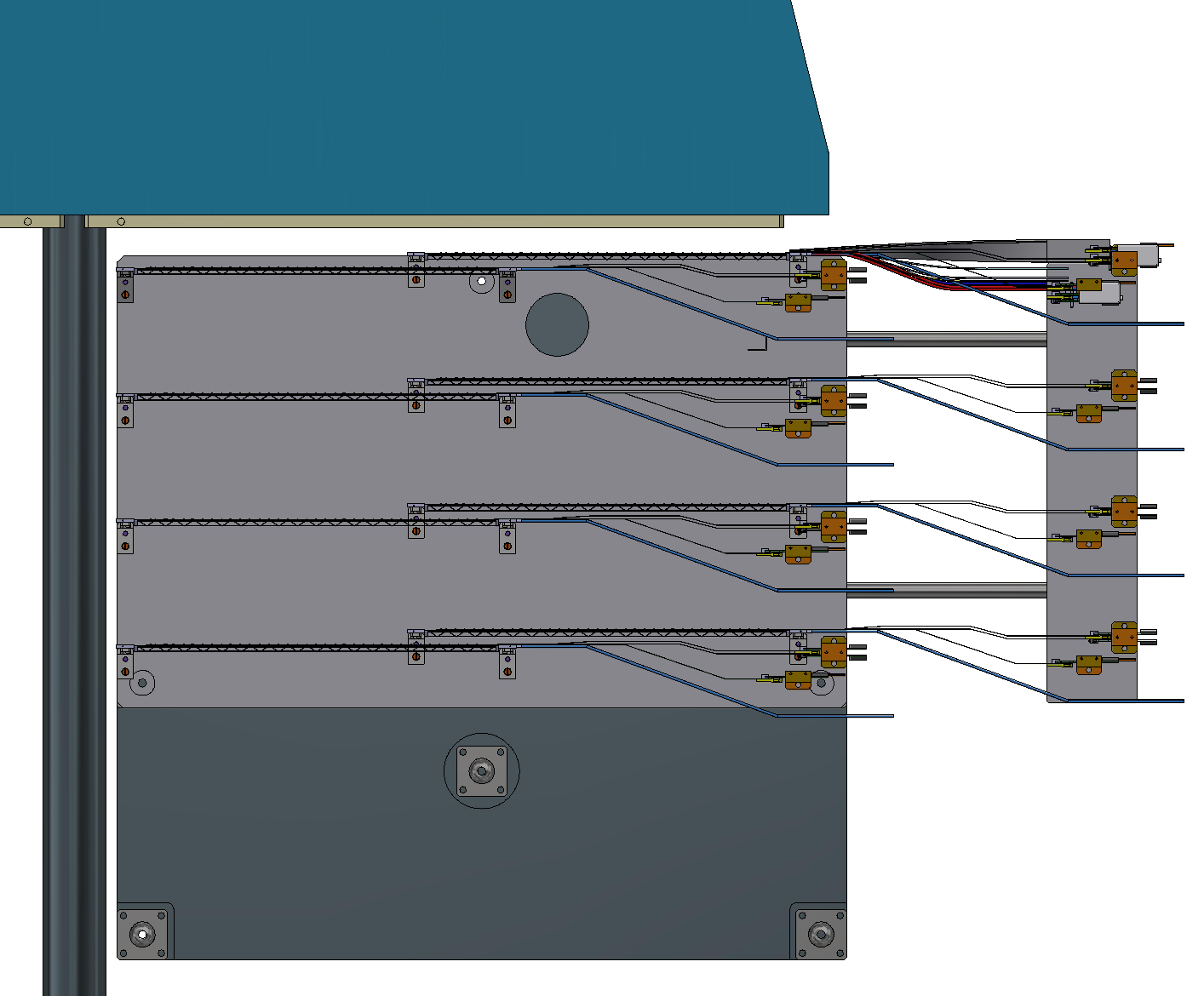}\end{overpic}
\caption{One full tracker arm showing the tray from the side (top) and from the top (bottom). The extensions (cooling pipes and flex cables) of the services are visible to the right of the tray. Also shown is the beampipe and the edge of the triangular vacuum chamber with the exit window. The base platform is shown in dark grey below the tracker tray. The free space is where the calorimeter tray is positioned.}
\label{fig:tracker_tray1}
\end{figure}

\begin{figure}[!ht]
\centering
\begin{overpic}[width=0.7\textwidth]{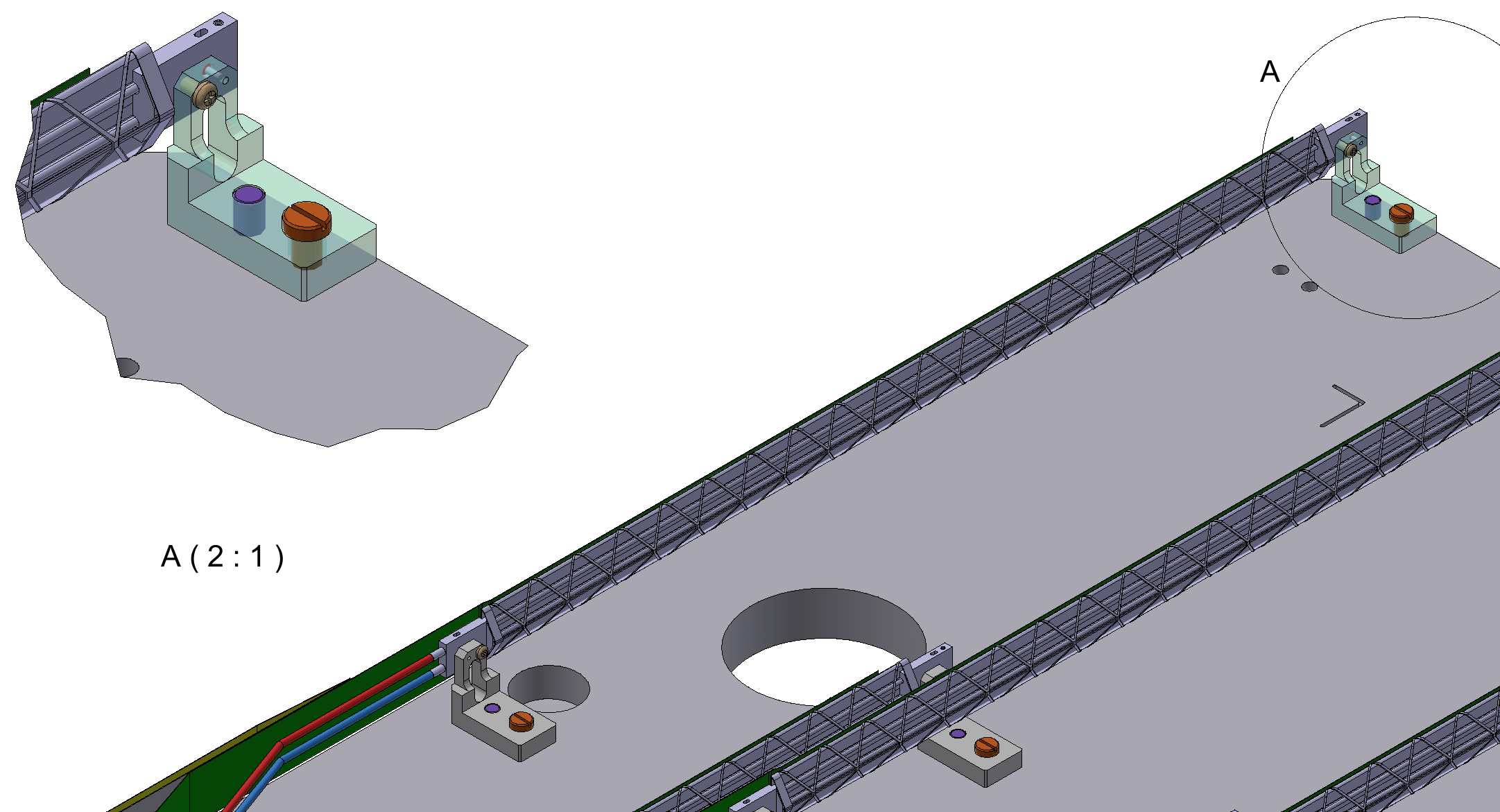}\end{overpic}
\caption{The L-shape holders which are used to fix the staves in their positions precisely, while maximising the active length of the arm and minimising the material budget of the supports.}
\label{fig:tracker_tray2}
\end{figure}

The eight staves can be fixed in space to form the four layers on each of the two sides of the beam axis, while exploiting the precision holes at the plastic edges of the space frame, as can be seen in Fig.~\ref{fig:tracker} (left).
This is done using precision pins mounted on vertical ``L-shaped'' bars which are connected to $\sim 60\times 50~{\textrm{cm}}^2$ flat trays that supports the entire structure from beneath.
The staves are surveyed by the manufacturer (ALICE at CERN) with respect to the precision holes so the position of all chips within a stave is known with a good precision.
The beampipe outer radius and the fixation of a stave on its two edges dictates the minimum distance in $x$ between the stave's active edge and the beamline axis.
This distance is $\sim 5$~cm.
The distances between subsequent layers in $z$ is $\sim 10$~cm.
This structure is shown in Figs.~\ref{fig:tracker_tray1} and~\ref{fig:tracker_tray2}.


\subsubsection{Alignment}
\label{sec:kinematicmounting}
The staves spaceframe is a rigid precision object, where the exact position is determined by only two precision holes on the spaceframe fixation ears, as can be seen in Fig.~\ref{fig:tracker_tray2}.

The staves of one tray can in principle be aligned to each other within less than $\sim 20~\um$ in all three dimensions via a precise determination of the position of the L-shapes holding the stave.

\begin{figure}[!ht]
\centering
\begin{overpic}[width=0.49\textwidth]{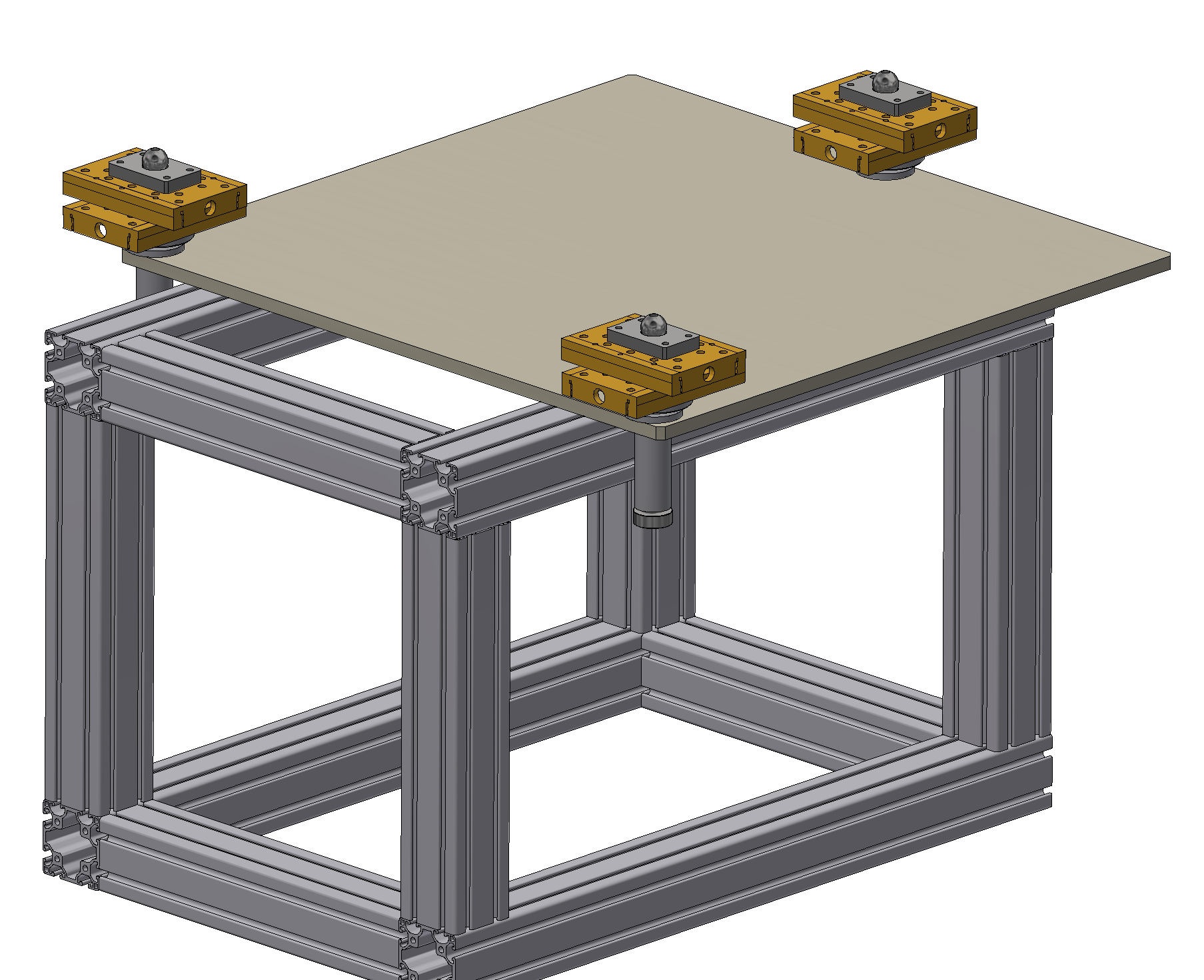}\end{overpic}
\begin{overpic}[width=0.49\textwidth]{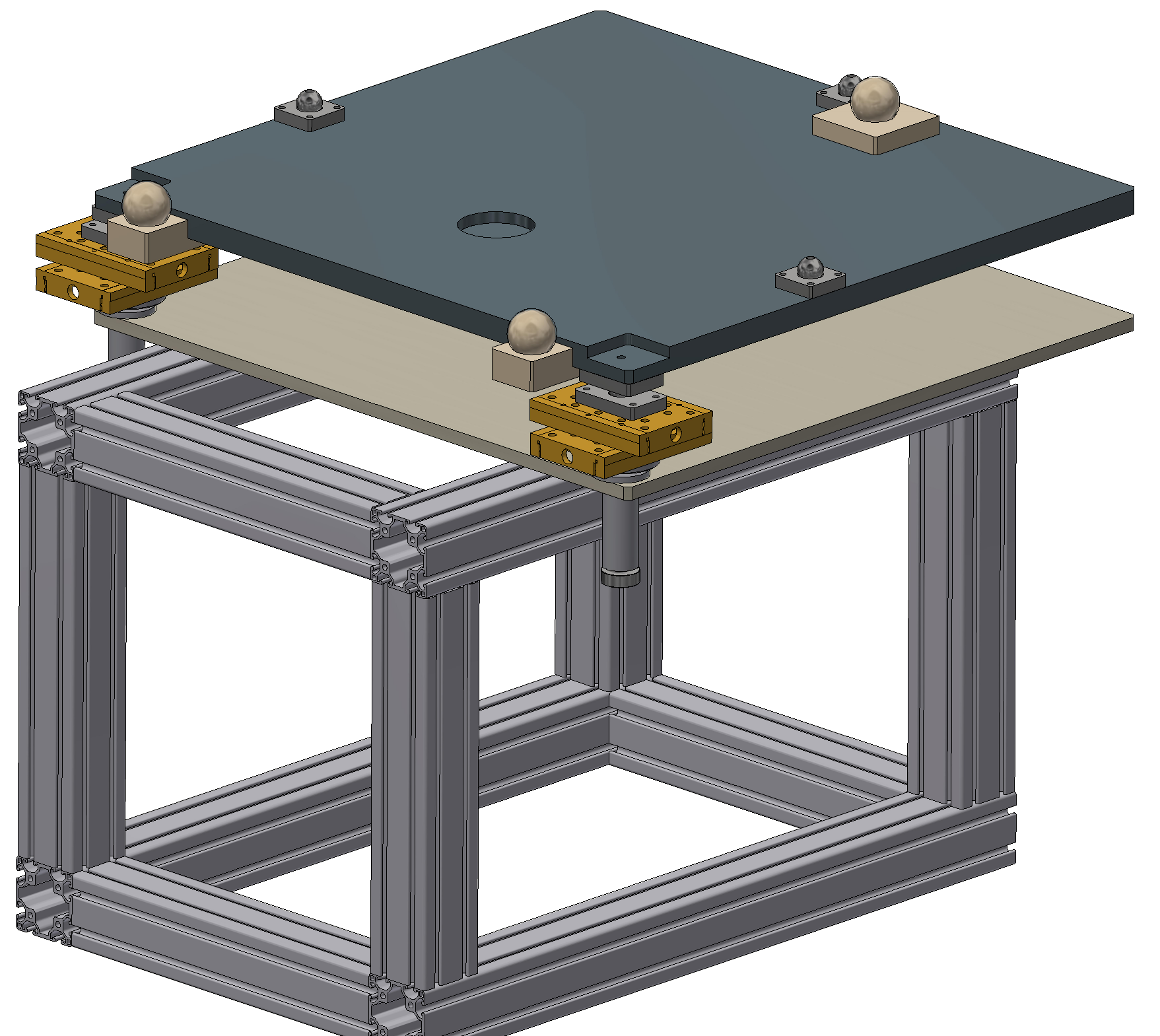}\end{overpic}
\begin{overpic}[width=0.49\textwidth]{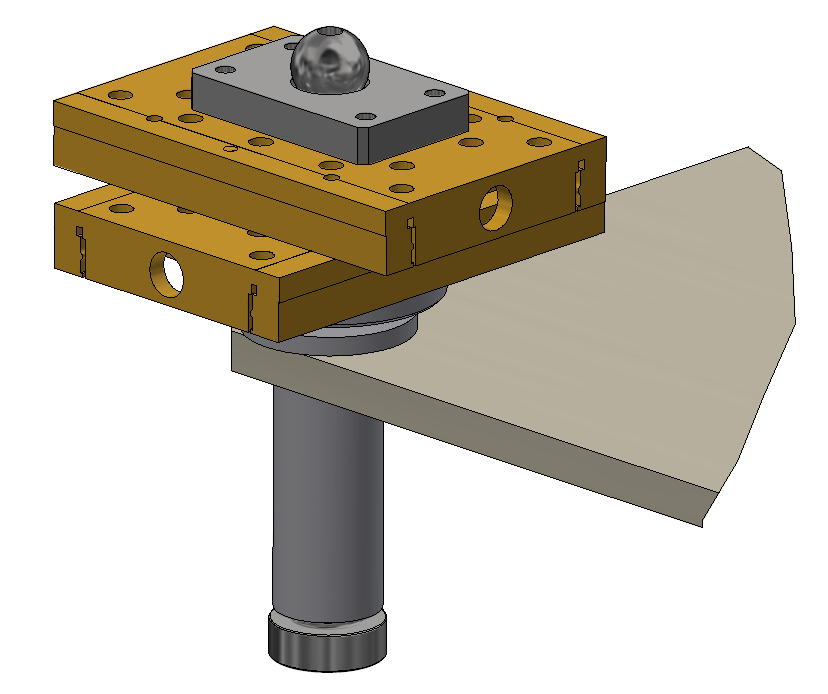}\end{overpic}
\caption{Top left: the support frame on which the base-platform (of the tracker's positron arm) is positioned.  Top right: the base platform with the kinematic mounts (grey spheres) of the tracker tray alone  and the alignment targets (light beige spheres).  Bottom: an example of one stage allowing to move the base platform in 3D.}
\label{fig:kinematicmounts_base}
\end{figure}

To align the entire structure in the experimental hall, one starts by assuming that the four layers are positioned on the two sides of the beam, stretching to $\sim 50$~cm in the transverse direction, as discussed earlier.
Next, one assumes that the staves are already surveyed and pre-aligned with respect to each other and with respect to the tray itself on the surface before mounting them in the experimental area.
The calorimeter, which sits behind the positrons tracker on a similar tray is assumed to also be pre-aligned to itself.
The two trays are positioned on the base platform (Fig.~\ref{fig:kinematicmounts_base}) through precision ``kinematic mounts'' at their bottom.

\begin{figure}[!ht]
\centering
\begin{overpic}[width=0.49\textwidth]{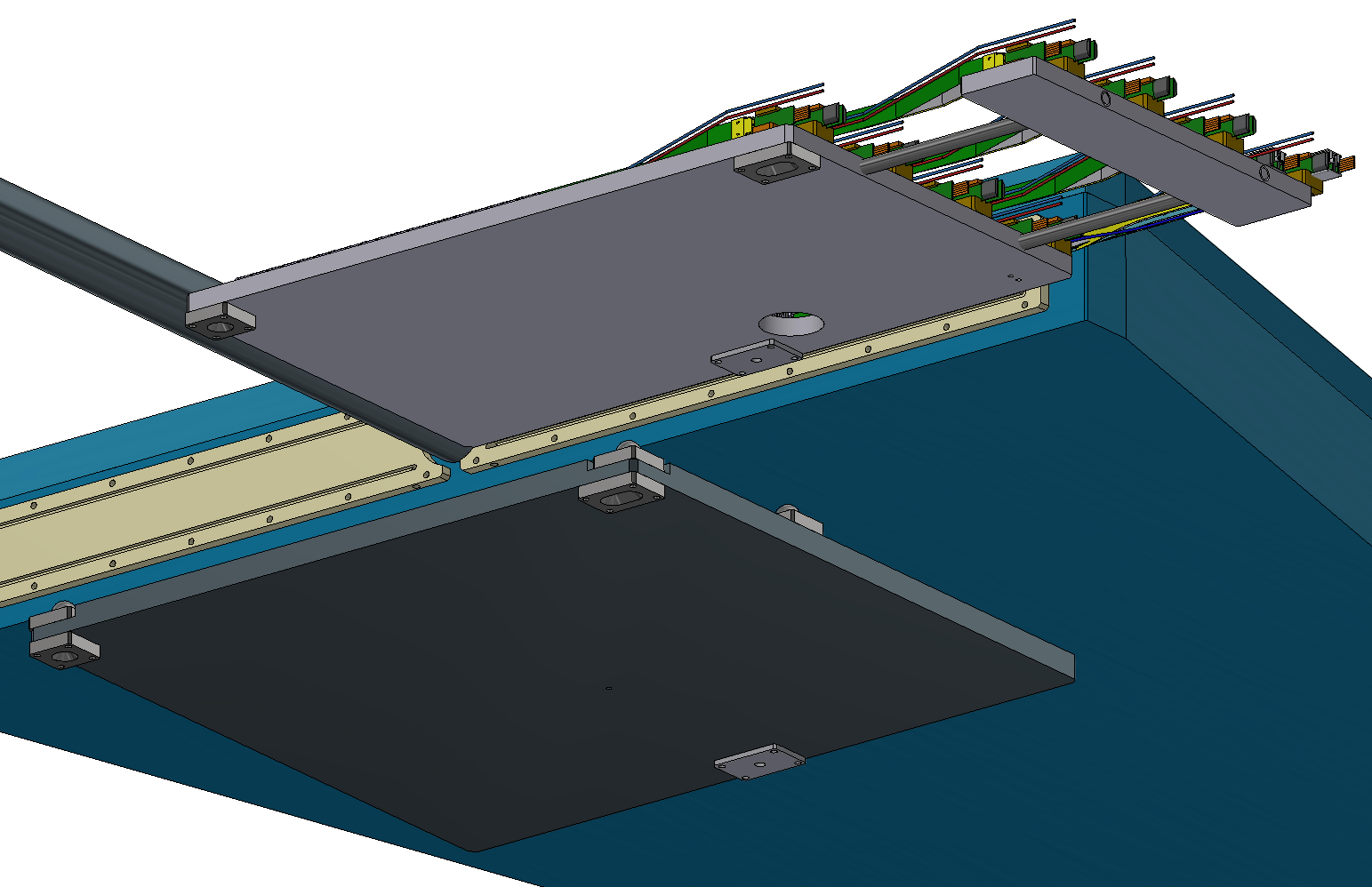}\end{overpic}
\begin{overpic}[width=0.49\textwidth]{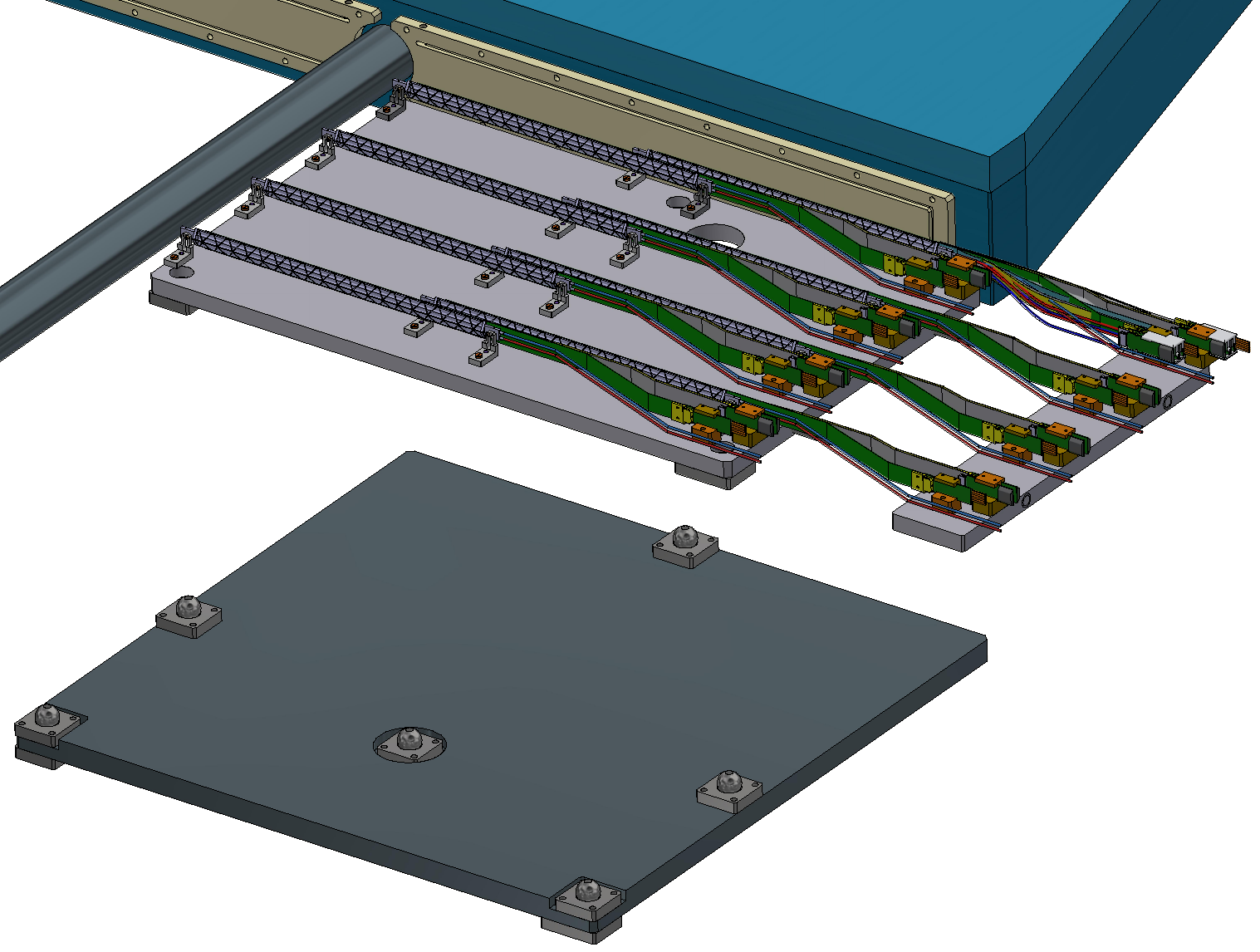}\end{overpic}
\caption{One tracker arm tray and its kinematic mounts shown separated from the base-platform from the bottom (left) and from the top (right).}
\label{fig:kinematicmounts_tray}
\end{figure}

This concept allows the lifting and re-placing of the trays without concerns about spoiling the alignment with respect to the base platform.
The precision depends on the machining process but in principle it is possible to achieve a precision of $\mathcal{O}(10~\um)$.
Furthermore, this allows to modulate the installation procedure quite flexibly.
In that respect, one assumes that the relative position of the two trays (tracker and calorimeter) on the base platform is known from an independent survey(s) outside of the experimental hall.
Fig.~\ref{fig:kinematicmounts_tray} shows one tracker arm tray and its kinematic mounts, separated from the base-platform from the bottom and from the top. The mounts allow to precisely reproduce the position of the tray on the stationary platform even if it was removed and brought back multiple times. This flexibility is important in view of the tight installation and/or technical access windows expected for LUXE.

The base platform itself is positioned on an anti-vibration table through a set of manually movable stages with three degrees of freedom per stage.
These stages allow adjustment of the platform's position and orientation with respect to the IP while performing the alignment survey (using dedicated alignment targets).

An optical alignment system from the EuXFEL is available to determine the {\normalfont \itshape in-situ} position of the base platform in the experimental hall to below 100~\um, with respect to the beam axis and the IP.

The stages holding the base-platform may be potentially motorised and remotely controlled.
However, motorised stages are important only in view of the two modes of operation (and for calibration runs).
The stages seen in Fig.~\ref{fig:kinematicmounts_base} (bottom plot) are fully manual and are only meant to be adjusted once while surveying.
As mentioned, the electron arm of the tracker has to be taken out of the interaction plane in the \elaser mode due to the high dose expected in its place stemming from the electron beam.
The limited access to the experimental area during the operation of the EuXFEL implies that the entire base platform of the electron tracker arm (at least) is required to be motorised, such that it can be brought into and out of its nominal position (e.g. tens of centimetres in $y$) depending on the run mode without physical access.
This kind of arrangement can be seen e.g. in Fig.~\ref{fig:tracker} (right plot).

\subsubsection{Dedicated shielding}
\label{sec:shielding}
\begin{figure}[!ht]
\centering
\begin{overpic}[width=0.49\textwidth]{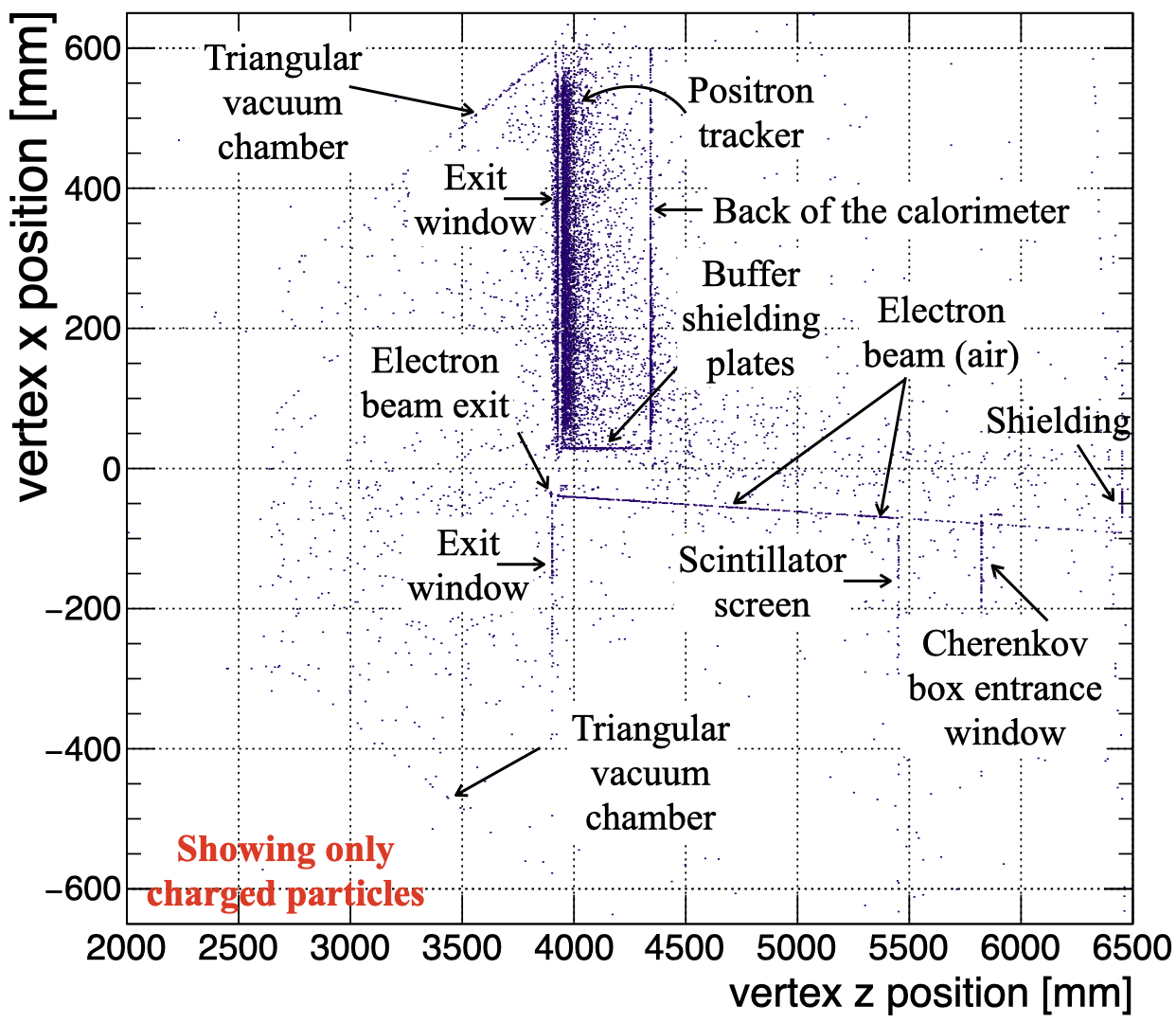}\end{overpic}
\begin{overpic}[width=0.49\textwidth]{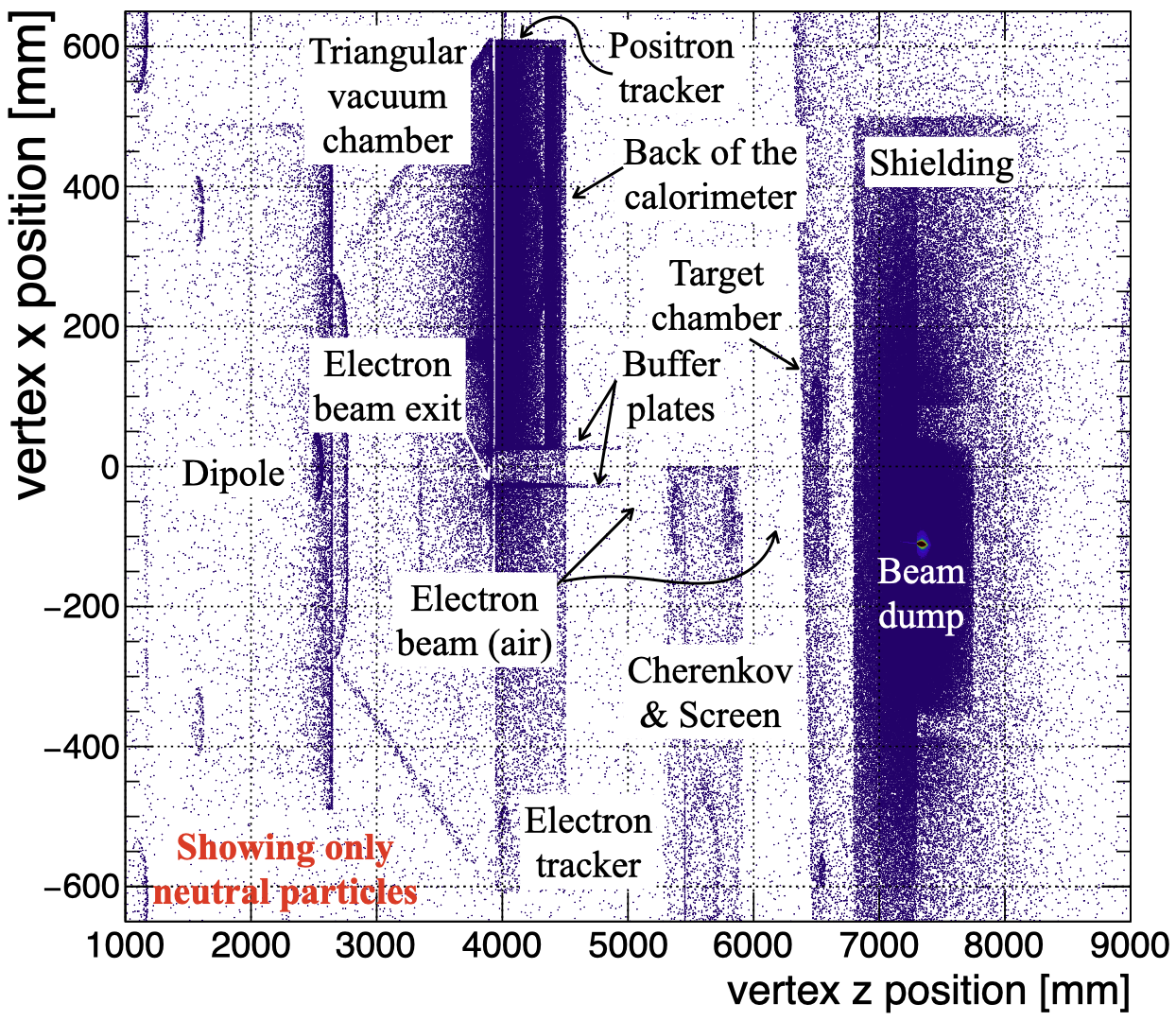}\end{overpic} 
\begin{overpic}[width=0.49\textwidth]{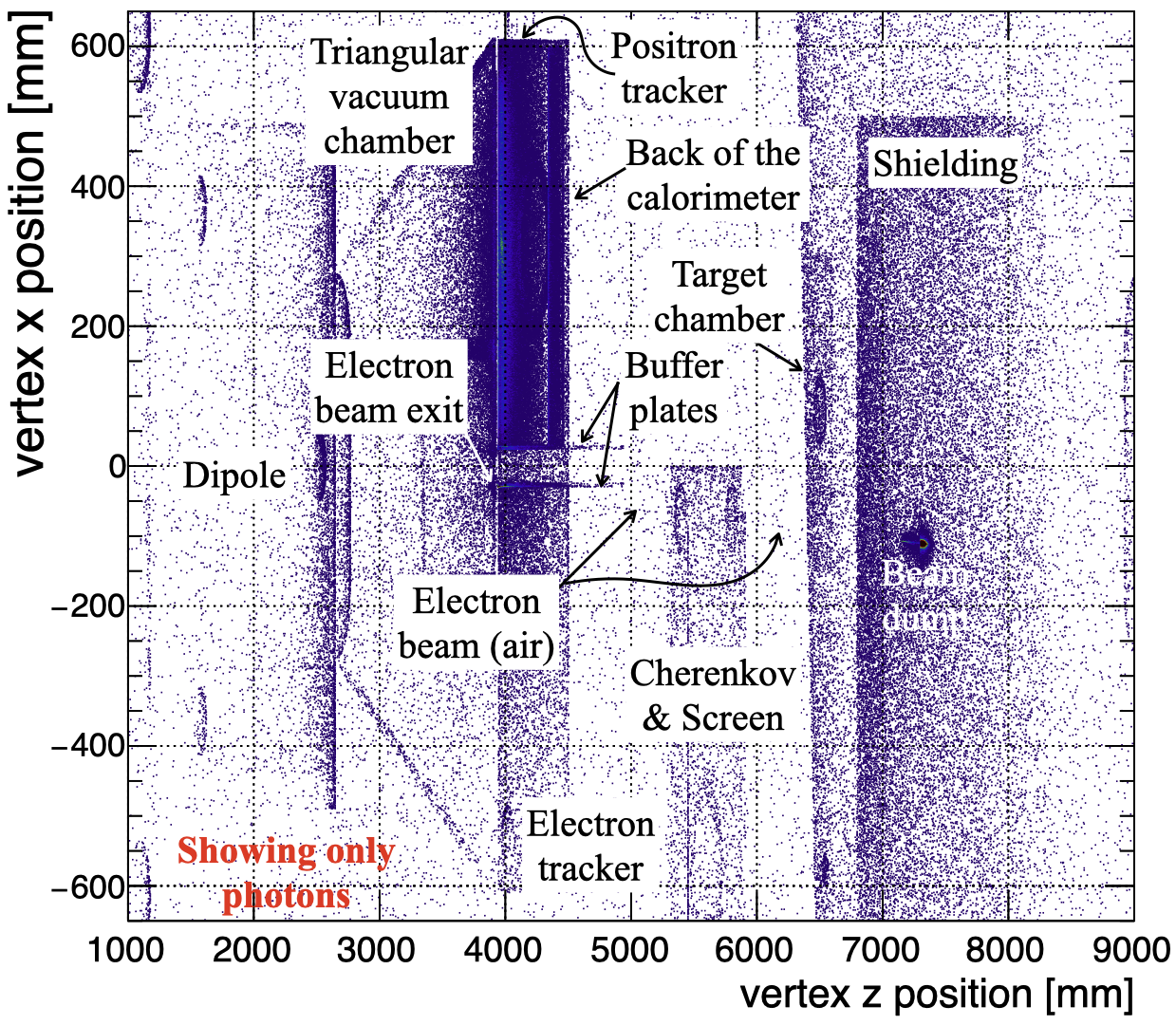}\end{overpic} 
\begin{overpic}[width=0.49\textwidth]{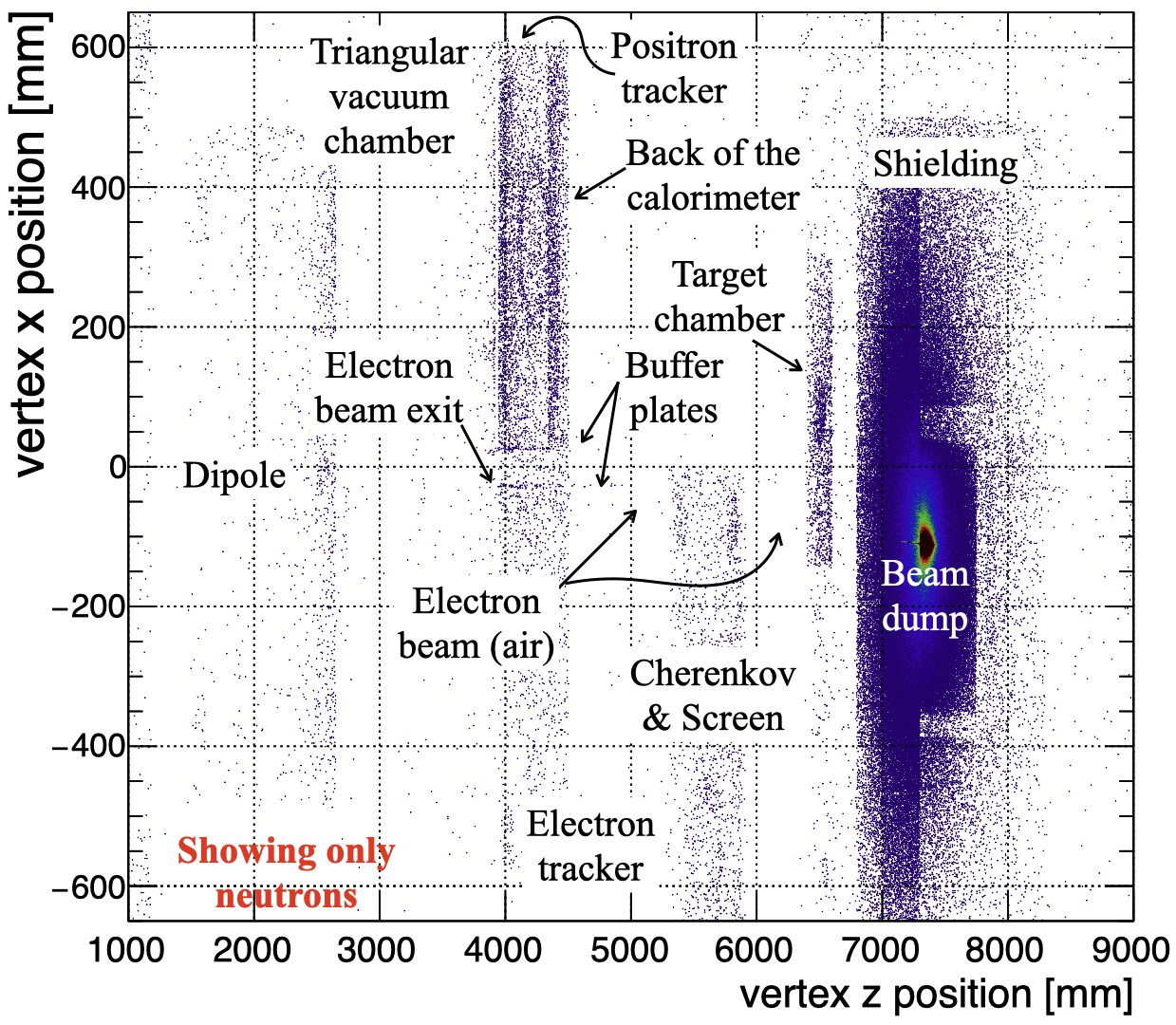}\end{overpic} 

\caption{A top view (in $x-z$) of the vertex distribution of all charged particles (top left), all neutral particles (top right),  photons only (bottom left) and neutrons only (bottom right) that reach the first layer of the positron arm of the tracker. }
\label{fig:radiation}
\end{figure}

In the \elaser mode, the background radiation is high (and higher than in the \glaser mode).
Most of it originates from the point where the electron beam exits the vacuum (300~\um thick Aluminium) as well as from the interaction of the same beam with the air on its path to the dump.
This can be clearly seen in Fig.~\ref{fig:radiation} which shows the vertex distribution of all charged particles (top left), all neutral particles (top right),  photons only (bottom left) and neutrons only (bottom right) that reach the first layer of the positron arm of the tracker. In the charged case, the distribution is dominated by electrons while in the neutral one, it is mostly photons and neutrons.  There is no filtering on the energy of the particles. The data corresponds to two BXs of beam-only (with no collision) in the \elaser setup, where the different elements in the setup around the tracker are highlighted for reference.

The electron-beam induced background can be mitigated by adding a tall 5~mm thick Tungsten plate placed vertically in the $y-z$ plane in contact with the beampipe itself to buffer between the beampipe and the positron arm of the tracker.
This problem is not present in the \glaser mode since the beam is dumped upstream, long before the tracker area, but nevertheless the shielding plates will not be moved in this mode.

A similar shielding plate will be placed above the electron arm tracker (in $x-z$) when it sits below its nominal position (in $y$, as seen in Fig.~\ref{fig:tracker}) to block the radiation coming from above.

Other than these two plates, no other dedicated shielding / casing measures are taken specifically for the tracker\footnote{Note that the calorimeter is in itself acting as a shielding for the back side of the tracker.}.


\subsection{Services \& Readout}
\label{sec:services}
\begin{figure}[!ht]
\centering
\begin{overpic}[width=0.75\textwidth]{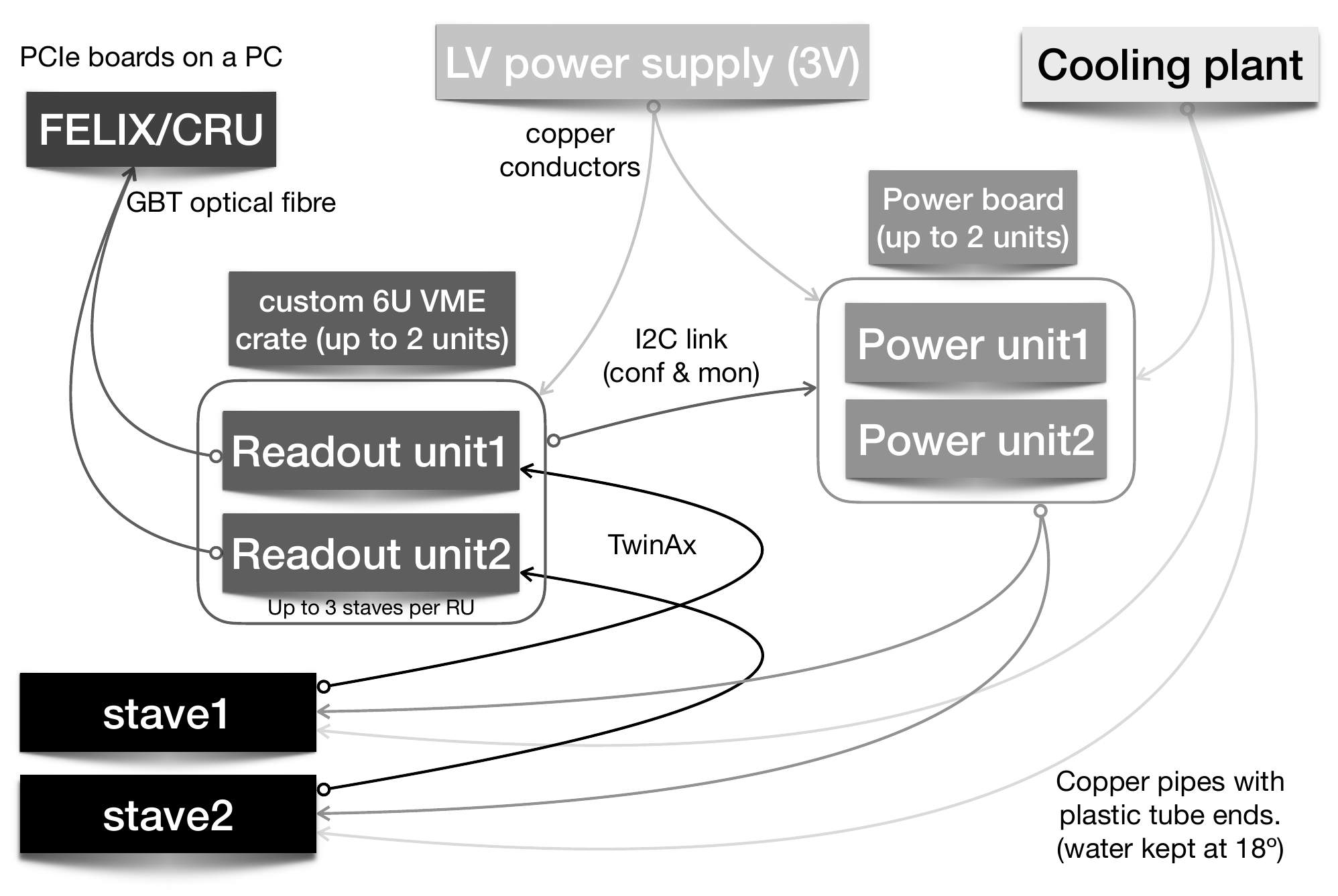}\end{overpic}
\caption{System layout, showing the staves as the end point client of all services and with the power, readout and cooling interfaces.}
\label{fig:system_layout}
\end{figure}

A schematic conceptual layout of the entire tracker system beyond the staves is shown in Fig.~\ref{fig:system_layout}.
As illustrated, one power board (PB) mounted in a custom crate has two power units supplying two staves.
One readout unit (RU) serves one stave although each RU can in principle support up to three staves as input.
In ALICE, only one stave is used per RU due to bandwidth considerations.
This is also the projected configuration for LUXE.
Each RU has up to three uplinks, and one downlink.
One FELIX (Front-End LInk eXchange, essentially a PCIe card)~\cite{Anderson_2016,FELIX1,9060037,8668437} would in principle be able to support up to 16 RUs.
The staves are connected to the RUs with relatively rigid, halogen free TwinAx cables provided and safety-approved for caverns by CERN~\footnote{The commercially available TwinAx cables are not halogen free.}.
Two such TwinAx cables are needed per stave at the length of at least 5~m in order to reach a close-by shielded rack(s), where the RUs will be positioned.
The recommended setup is to use two FELIX installed on one or two PCs, with each FELIX managing up to ten staves

\subsubsection{Cooling}
\label{sec:colling}
The cooling is provided to the staves via small pipes with outer diameter of $\sim 1.05$~mm (see also Fig.~\ref{fig:tracker_tray1}) to keep the staves at roughly below $21^\circ$C.
For that, the input needs to be at $18^\circ$C.
One or two simple water chillers will be enough to cool down all staves via a dedicated manifold. 
As in ALICE, a water-leak-less system may be foreseen.
In this case, the pressure inside the water tubes is lower than atmospheric pressure outside, such that if there is a failure, no water gets spilled outside but rather the air gets sucked inside the tubes.
The pipes are therefore subjected to a pressure from outside to inside, which may lead to buckling of the Kapton pipe wall.
The tubes have been tested up to 50~bar.

The PUs and the RUs can be cooled with proper air flow, depending on the humidity and conditions in the tunnel.

\subsubsection{Power}
\label{sec:power}
\begin{figure}[!ht]
\centering
\begin{overpic}[width=0.99\textwidth]{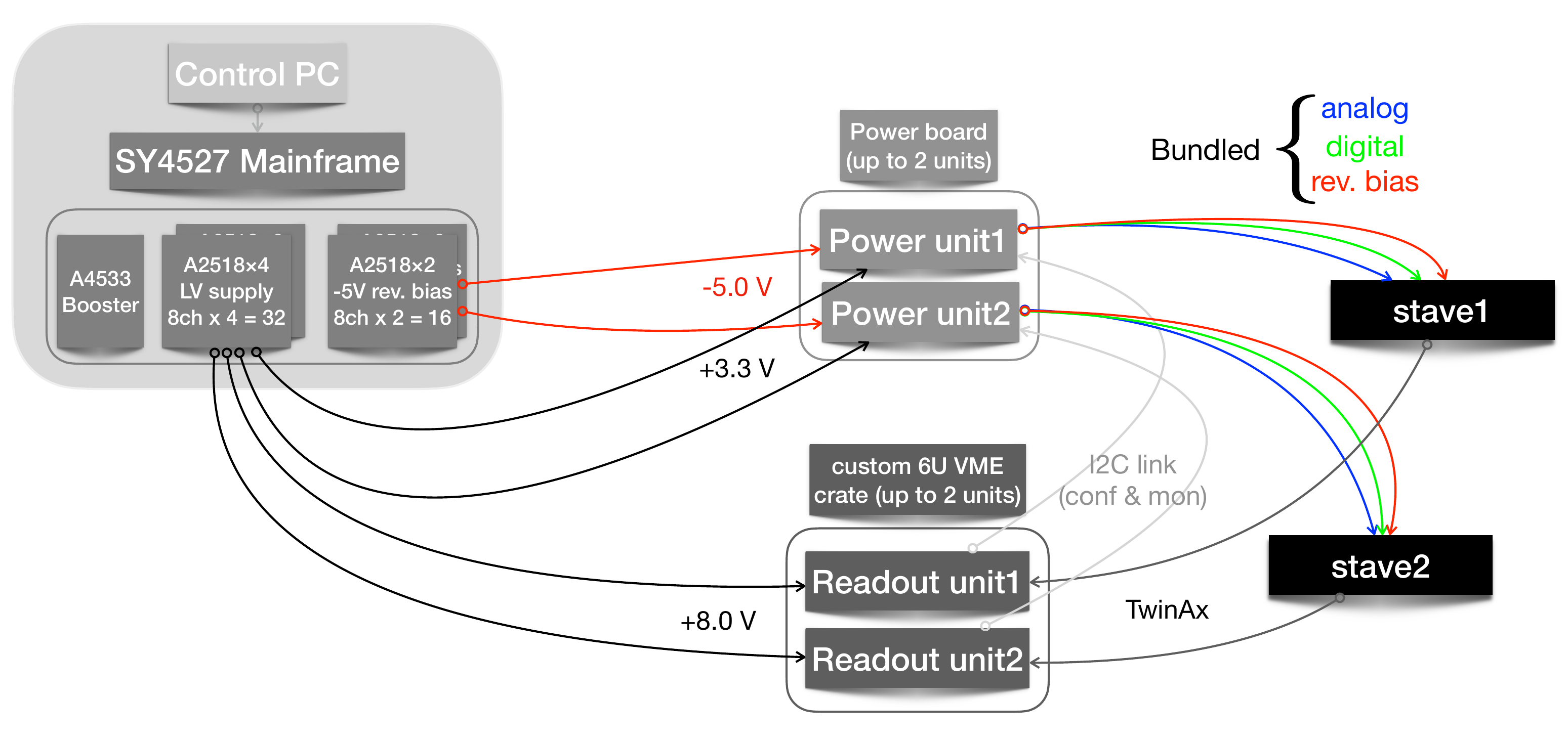}\end{overpic}
\caption{Power system layout.}
\label{fig:power_layout}
\end{figure}

A schematic layout of the tracker power system is shown in Fig.~\ref{fig:power_layout}.
It is important to stress that the components of this design cannot be placed at hostile environments in terms of radiation and they must be shielded properly in their racks.
There is an alternative power system, at a higher price tag, which may be placed in a hostile environment.
In LUXE, it is foreseen that these components can be easily shielded, while still be placed close enough to the detector ($\sim 5$~m).

A single stave has one digital low-voltage (LV) input, one analog LV input and one reverse bias supply (floating, negative with little current consumption).
One power board has 32 input channels.
The PB has a system to detect that the current does not trip on the ALPIDE chip, i.e. high charge release due to interaction between the silicon nucleus and hadrons from the collision (``latch-up'').
In that case, the PB cuts the channel.
The PB receives the reverse bias, but it only has switches to switch the on/off and not regulate them individually (the same regulation applies to all channels).
Custom control layer to control the PB, goes via the readout unit Inter-Integrated Circuits (I2C's).

One PB has 2 power units (PUs).
Each PU has 16 linear regulated (low noise in the analog supply) output channels and 2 input channels with a bank of 16 output channels, so 8 groups of 2 channels (digital+analog).
In principle, one PU can support up to eight staves but for better distributing the controls, only two staves are connected to one PU.
All PBs are tested to 10~krad and $3\times 10^{11}~1\,{\textrm{MeV}}~n_{\textrm{eq}}/{\textrm{cm}}^2$.

Each PU has a regulator of reverse bias supply with a bank of (16) switches to switch on/off the different channels.
The ALPIDE chip receives the following:
\begin{itemize}
	\item Chip needs 1.8~V
	\item PB channel: 3.3~V
	\item Reverse bias: -5~V
\end{itemize}

The power to the ITS RUs and ITS PBs is supplied by a CAEN power system of the EASY3000 type~\cite{EASY3000}, which is radiation hard.

As mentioned earlier, for LUXE this system can be of the non-rad-hard type, assuming it can be substantially shielded.
The rad-hard solution is also possible, albeit at greater cost.

The non-rad-hard system consists of a multichannel power supply mainframe mounted with boosters and LV power supply cards as shown in Fig.~\ref{fig:power_layout}.
The LV cards are used to both supply power to the RUs and the PBs and to generate the negative voltage (-5~V) for the PB reverse bias channels that regulates the biasing for the substrate of the ALPIDE chip.
Each card has 8 channels and hence, three such cards are enough for powering up the entire tracker.

Concerning the reverse-substrate bias, up to 100~krad and $1.1\times 10^{12}~1\,{\textrm{MeV}}~n_{\textrm{eq}}/{\textrm{cm}}^2$, the ALPIDE sensor can in principle be operated without reverse substrate bias without penalty in terms of detection efficiency/fake-hit rate/position resolution.
Therefore, as LUXE's application does not require it explicitly, it may be most prudent to directly connect it to analog ground on the corresponding power board.

The PUs will be manufactured in industry based on existing design files from ALICE.
Preferably the same manufacturer of the ALICE modules will be also producing the modules for LUXE.

A simpler solution may be envisioned for the ``bare-bones'' tracker based on simple power regulator board manufactured by ALICE at CERN.
This is discussed in section~\ref{sec:bareboneslayout}.

\subsubsection{Readout}
\label{sec:readout}
The FPCs of all staves are connected to the RUs such that each sensor has its own readout line with 2~Gb/s with a Samtec (TwinAx) cables transmitting the signal.
The RUs are in turn connected to the backend FELIX PCIe card with GBT optical fibres.
ALICE’s RUs are tested to 10~krad and $3\times 10^{11}~1\,{\textrm{MeV}}~n_{\textrm{eq}}/{\textrm{cm}}^2$.

The best configuration of RUs and FELIX cards would depend on the expected data rates from the staves.
The FELIX card is very similar to the ALICE CRU card; it supports up to 48 bi-directional GBT links (the CRU card is configured with only 24 bi-directional links), but features a different FPGA than the CRU card.
Both cards are PCIe cards with 16-lane Gen-3 PCIe interfaces.
In both the CRU and the FELIX card, the performance over the PCIe bus has been measured in the order of 100~Gbps, so that is the main limitation.
Each RU has up to 3 uplinks, and 1 downlink, so while the CRU supports up to 8 RU if all 3 uplinks are used, FELIX would in principle be able to support 16 RUs.
For example, for a 20-stave configuration, one can in principle attach those to 7 RUs, if one uses the 3 staves per RU configuration.
In that case all 7 RUs could be connected to one CRU or one FELIX card.

Finally, one PC can in principle handle 24 staves although some redundancy is required to be considered and hence, two DAQ PCs are foreseen in LUXE.

Like the PUs, the RUs will be manufactured in the industry based on existing design files from ALICE.
Preferably the same manufacturer of the ALICE modules will be also producing the modules for LUXE.

FELIX modules have already been purchased by WIS.
Some adaptations will need to be made for compatibility with ALICE's RUs.
This adaptation is already proven to be possible as demonstrated by the sPHENIX collaboration~\cite{KIM2019955}, which is also using the ALICE ITS staves.
This adaptation will be done in a dedicated test stand at WIS based on modules that are purchased independently.

Another solution may be envisioned for the ``bare-bones'' tracker based on existing MOSAIC~\cite{refId0} readout boards which are manufactured by ALICE colleagues in Italy. 
This is discussed in section~\ref{sec:bareboneslayout}.

\subsubsection{``Bare-bones" tracker specific system}
\label{sec:bareboneslayout}
\begin{figure}[!ht]
\centering
\begin{overpic}[width=0.8\textwidth]{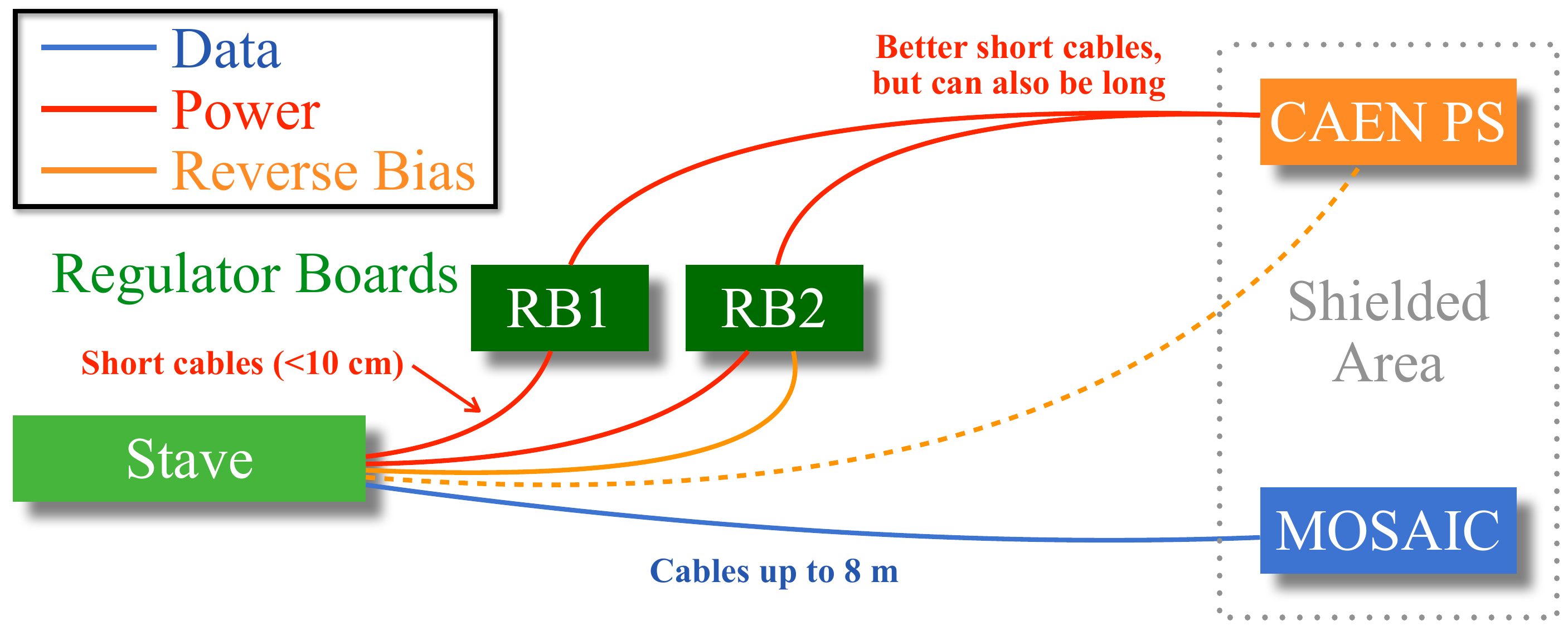}\end{overpic}
\caption{The bare-bones power and readout system layout. Note that while the MOSAIC readout boards need to be maximum 8~m away from the staves in a shielded area, the CAEN power-supply modules can be placed further away, in a different shielded area.}
\label{fig:bareboneslayout}
\end{figure}

As highlighted before, the bare-bones system can be constructed in a different, simplified way compared to the nominal system discussed so far.
For that, simple power regulator boards and the MOSAIC~\cite{refId0} readout boards will be used.
This system is depicted in Fig.~\ref{fig:bareboneslayout}.

The regulator board is basically the reference design for the Microchip MIC29302~\cite{MIC29302} with a potentiometer to adjust the output voltage.
These boards cannot be remotely controlled and the current is measured on the power supply side.
The number of regulators per CAEN channel depends on the monitoring choice, i.e., if one wants to monitor every regulator independently or not.
The recommended usage is one channel per regulator (a stave draws less than 2A).
The connection to the staves is done with a custom PCB.
The drop has to be minimized between the stave and the regulators by using short cables.
The power regulation becomes easier if the respective modules can be placed in close vicinity of the staves, to avoid the need to compensate for line losses.
Ideally, the regulators would be located at the end of the stave ($<10$~cm) and set with a simple potentiometer to a fixed voltage of 1.85~V.
This simplifies the operation massively, while compromising some controllability features.
As the regulator board stabilizes the voltage, the cables between them and the CAEN power supply modules can be made arbitrary long.
However, an adequate conductor size must be chosen to keep the voltage drop at a level which will be tolerated by the specific power supply.

One MOSAIC will replace one RU (i.e. one module per stave), with the same connectivity.
The TwinAx cables between the MOSAIC and the stave needs to be smaller than 8~m.

The regulator boards are tested to 10~krad and $3\times 10^{11}~1\,{\textrm{MeV}}~n_{\textrm{eq}}/{\textrm{cm}}^2$.
The MOSAIC board is not qualified for use in radiation environments so it must be properly shielded.

\subsection{Grounding}
\label{sec:grounding}
\begin{figure}[!ht]
\centering
\begin{overpic}[width=0.6\textwidth]{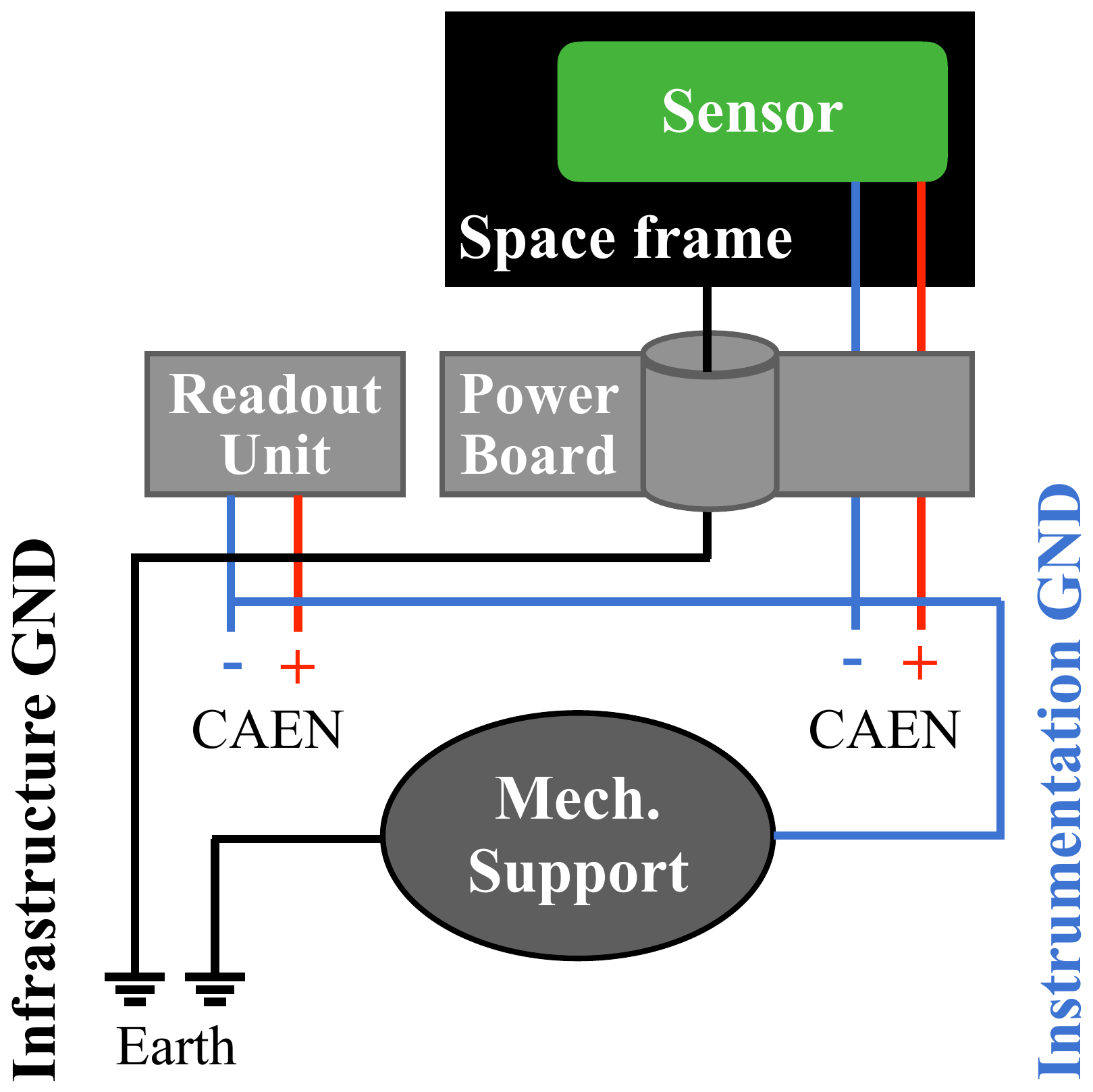}\end{overpic}
\caption{The tracker grounding scheme, following the one of the ALICE ITS upgrade.}
\label{fig:grounding}
\end{figure}

The ALPIDE chips have separate analog and digital ground connections~\cite{ALPIDEManual}.
As illustrated in Fig.~\ref{fig:grounding}, these ground lines are connected to the off-detector modules (denoted as ``Instrumentation GND''), while the staves' space-frames are connected to the infrastructure ground (denoted as ``Infrastructure GND'').
This grounding scheme follows the same scheme of the ALICE ITS.

\subsection{Data rates}
\label{sec:data}
Typically, the data rate from one tracking detector is expected to be small.
The estimation below assumes that all chips of all staves see the same (average) number of pixel hits from $N_{\textrm{ max}}({\textrm{bkg}})$ background particles at 9~Hz (no collision with the laser) and $N_{\textrm{max}}({\textrm{ sig}})$ signal particles at 1~Hz\footnote{The signal particles, on colliding with the tracker materials, can generate background too.} (from collisions with the laser).
The data rate is driven mostly by the signal multiplicity, except for the low signal multiplicity scenario where background particles drive the data rate.
It is expected to be large only under the most extreme conditions at the highest laser intensities (and under the same assumption of all chips seeing the same number of particles).
In reality, the distribution of pixel hits will not be uniform. Hence, the numbers should be seen only as an upper bound.
This bound would be relevant for a short period of the overall data taking, only in the end of phase-1, while during most of LUXE's lifetime the data rate will be much smaller.

The concepts relevant to estimate the data size and rate are discussed below
\begin{itemize}
\item {\textbf{Links:}} the link bit rate is always 1.2~Gbps (960~Mpbs with 8~b/10~b encoding overhead) and there are 9 chips in the stave, each with one link.
These 9 links need to go to some serial receiver capable of locking to 1.2~Gbps and reading and decoding in real time this stream.
If there are no trigger commands sent to the chip, this will transmit only IDLE messages on the links, meaningless data that can be discarded.
For every trigger sent to the chip, a matrix frame will be read and the chip will transmit out a chip data packet containing basic identification and timing information in the prefix CHIP HEADER and suffix CHIP TRAILER and the variable size payload matrix frame data in between.

\item {\textbf{Headers:}} a chip data packet is at least an opening CHIP HEADER (2 bytes) and a closing CHIP TRAILER (1 byte).
These 3 bytes will be transmitted for every trigger command received by the chip (and by each chip).
In a special case that the matrix frame (seen in Fig.~2.1 of~\cite{ALPIDEManual}) is empty, no pixel hits, the chip actually sends a CHIP EMPTY FRAME (2 bytes) only, no TRAILER.
This is a corner case that can be neglected (in ALICE, reducing bandwidth of the Outer Barrel case was important).

\item {\textbf{Matrix regions:}} the pixel hit data are variable length and contained in between the CHIP HEADER and CHIP TRAILER.
In this payload there is another level hierarchy, related to the regions of the matrix which is subdivided in 32 regions (seen in Fig.~2.1 of~\cite{ALPIDEManual}).
Region data packets come out in topological order (leftmost region 0 first and up to region 31).
A region with no pixel hits will not have any data in the output stream.
If there is at least one pixel hit in one given region, then a REGION HEADER (1 byte) for that region is present in the chip packet payload and this REGION HEADER marks the begin of a region data block with pixel hit information for that specific region.

\item {\textbf{Pixels:}} by default there is a compression of close-by pixel hits.
For simplicity and for being conservative, we neglect that and assume we use raw data mode: then, every single pixel hit will have a DATA SHORT (2 bytes) word in the region data block following the REGION HEADER.
Neglecting the built-in compression means DATA LONG (3 bytes) is never produced by the chip.
The DATA SHORT has the pixel hit offset inside the region.
To reconstruct the true position of the pixel one needs to combine the region address and the data of the DATA SHORT.
A single particle hit will typically generate a cluster of 2 or 3 pixel hits (average cluster size depends on many parameters - see section~\ref{sec:performance}).
\end{itemize}


To summarise, we start from the distribution of the number of pixel hits for a region (1/32 of the matrix area) and assume that all regions have at least one pixel hit and the number of pixel hits we get from our largest signal + background expectations are distributed evenly across these 32 regions.
In reality the largest number for signal particles does not distribute across all chips evenly and it is mostly centred around one half of a stave.

For each of the pixel hits in a region, the size of a given region packet can be determined (can be 0 for the boundary case of no pixel hits).
For each trigger to the chip, one needs to add to the variable size payload of the regions the 3 bytes of CHIP HEADER and CHIP TRAILER. 

\begin{figure}[!ht]
\centering
\begin{overpic}[width=0.9\textwidth]{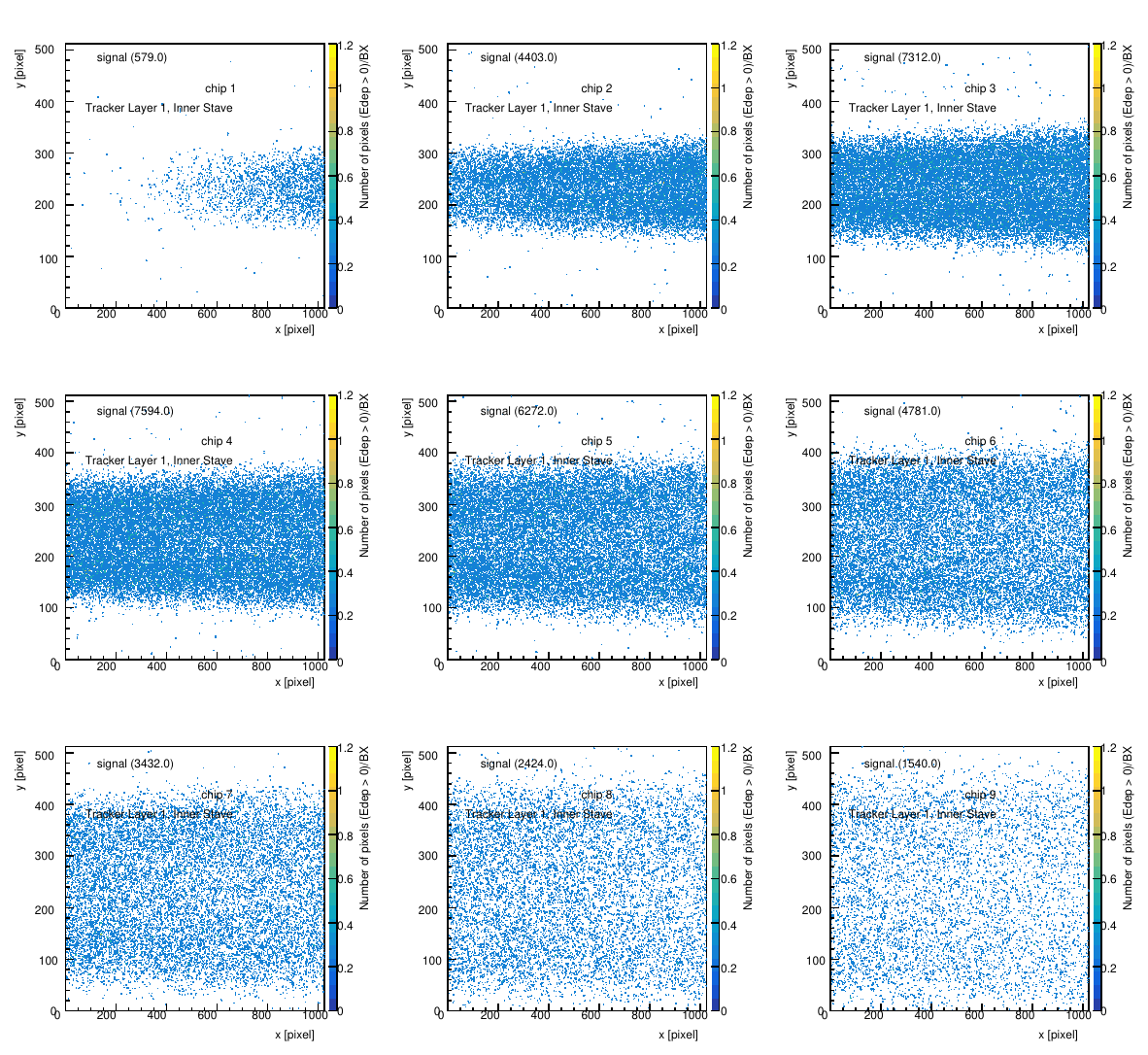}\end{overpic}
\caption{The distribution of ``hits'' (pixels with non-zero energy deposition, $E_{\textrm{dep}}>0$) from signal for the nine chips of the inner stave of the first layer of the tracker's positron arm in the \elaser mode. The binning of each sub-figure (i.e. chip) corresponds to the actual pixel matrix. The true signal particle multiplicity in the sample used is $\sim 4\times 10^4$ positron per BX. If there are several particles contributing to the energy deposited in the pixel, only one entry is filled in the histogram to avoid over-counting.}
\label{fig:hitoccupancySig}
\end{figure}

\begin{figure}[!ht]
\centering
\begin{overpic}[width=0.9\textwidth]{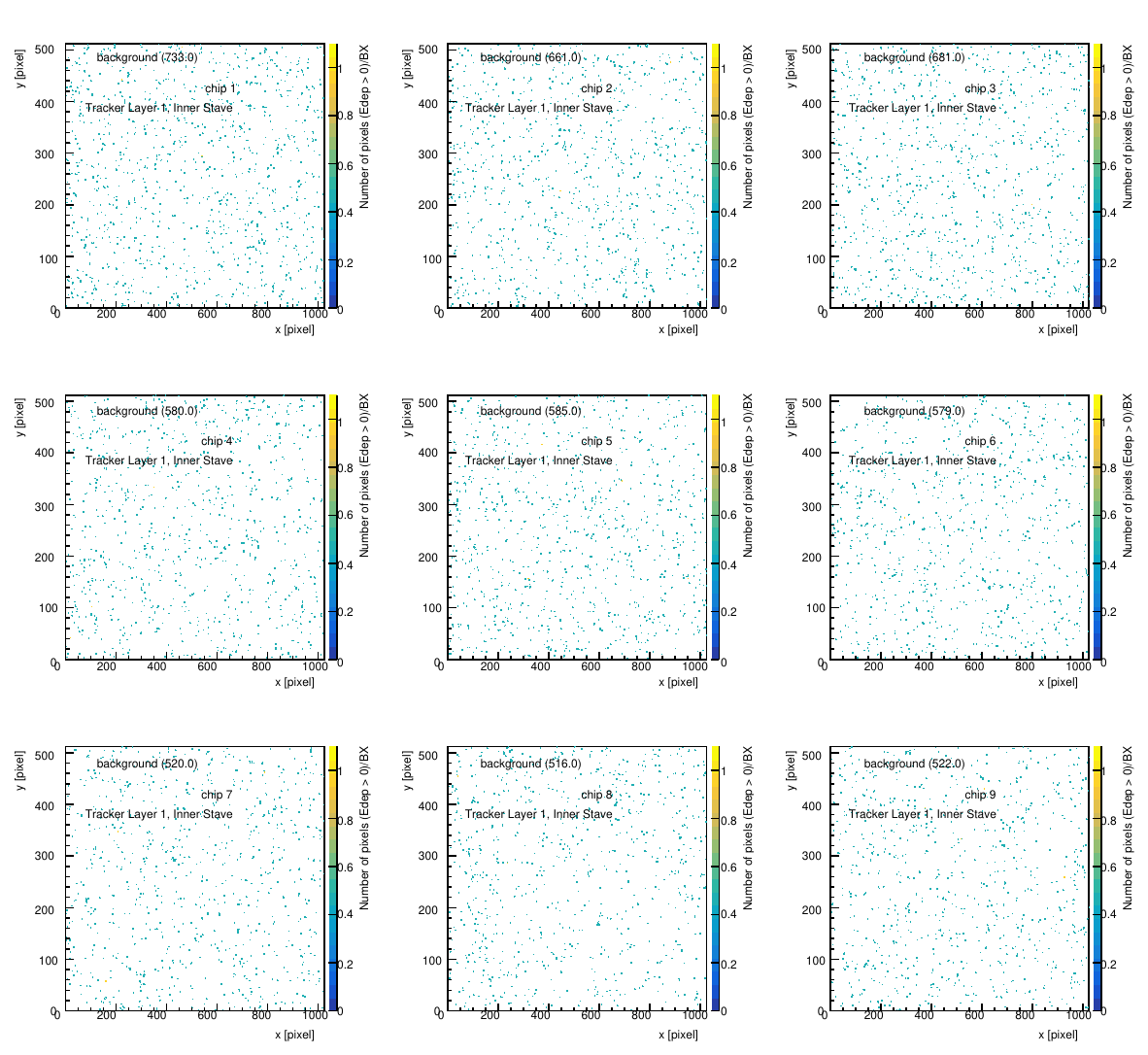}\end{overpic}
\caption{The distribution of ``hits'' (pixels with non-zero energy deposition, $E_{\textrm{dep}}>0$) from background for the nine chips of the inner stave of the first layer of the tracker's positron arm. This is shown for the \elaser mode. The binning of each sub-figure (i.e. chip) corresponds to the actual pixel matrix. If there are several particles contributing to the energy deposited in the pixel, only one entry is filled in the histogram to avoid over-counting.}
\label{fig:hitoccupancyBkg}
\end{figure}

The distribution of ``hits'' (pixels with some energy deposition, $E_{\textrm{dep}}>0$) for the case of true signal multiplicity of $\sim 4\times10^4$ particles per BX can be seen in Fig.~\ref{fig:hitoccupancySig}.
The same distribution for the background case is shown in Fig.~\ref{fig:hitoccupancyBkg}.
The two figures are shown for the \elaser mode.

The average number of hits from background is $\sim 600\,(425)$ per BX per chip in the inner (outer) stave, while for the signal shown in Fig.~\ref{fig:hitoccupancySig} it is $\sim 4300\,(500)$ per BX per chip in the inner (outer) stave.
These numbers can be used to calculate the data rates, while assuming that these average numbers of hits per chip will be seen on all chips (separately for the inner and outer staves).
The same assumption is made for the background hits.
We note that in the case of the \glaser runs, both signal and background numbers will be substantially smaller. 

Under these assumptions, the data rates are specified in table~\ref{tab:data} for three different signal scenarios.
\begin{table}[!ht]
\centering
\begin{overpic}[width=0.99\textwidth]{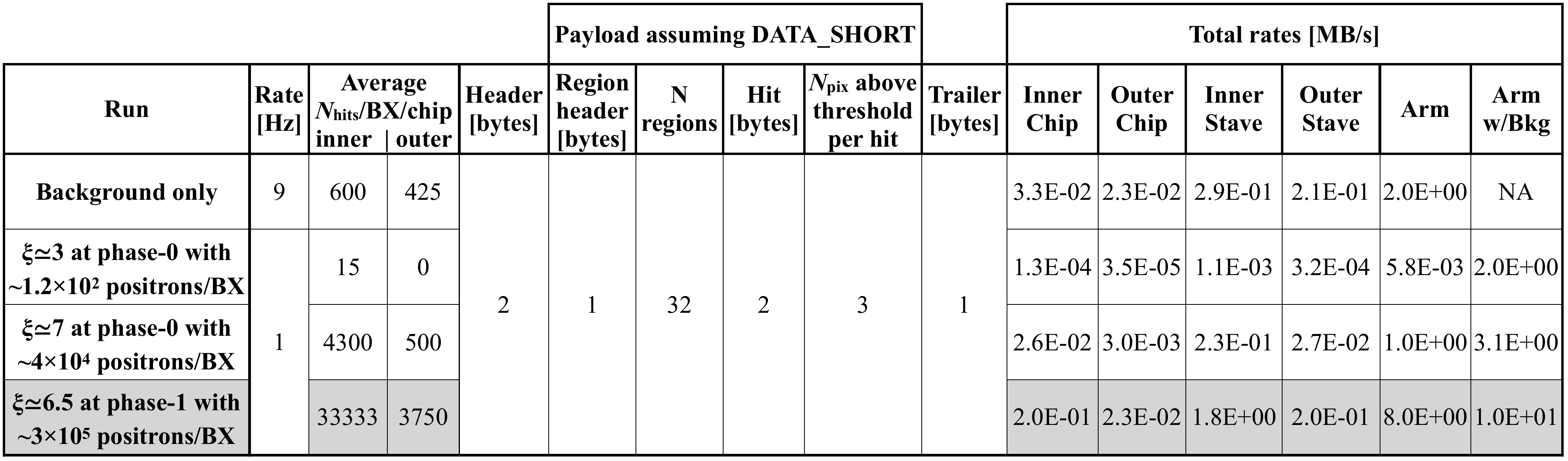}\end{overpic}
\caption{Data size and rates per chip, stave (9 chips) and arm (8 staves) for background (at 9~Hz) and signal (at 1~Hz). The rates must be divided into the inner and outer chips/staves before adding up their contributions. Recall that the same background will appear along with the signal when there are collisions and hence the total data rate of one arm is the sum of the signal and background (of that arm) shown in the last column. 
The number of pixels above threshold is assumed to scale up roughly like the number of hits times 3 pixels per hit. This assumption is discussed briefly here and in more detail later in section~\ref{sec:performance}.
The first three rows (in white) are estimated from our \geant4 simulation for ``hits'' (pixels with $E_{\textrm{ dep}}>0$). Finally, due to lack of simulation samples for the highest multiplicities, the last line (in gray) is extrapolated assuming the number of hits per chip per BX for the inner (outer) stave scales down like the number of true signal positrons per BX divided by 9 (80). This can be inferred from the ratio of the inner (outer) sub-columns of the ``Average $N_{\textrm{hits}}$/BX/chip'' column to the number of true signal positrons in the first column.}
\label{tab:data}
\end{table}
The values in the ``Average $N_{\textrm{hits}}$/BX/chip'' column of table~\ref{tab:data} are calculated from Fig.~\ref{fig:hitoccupancySig} and~\ref{fig:hitoccupancyBkg} as the average of number of hits over the 9 chips of the inner and outer stave of the first layer of the tracker arm for the background and a signal in the \elaser mode.
This is done separately for the inner and outer staves' chips.
The signal used here (third row in table~\ref{tab:data}) and in Fig.~\ref{fig:hitoccupancySig} is a typically ``large'' case, while in most of phase-0 and large parts of phase-1 the rates will be lower.
The total data size per chip is calculated by the following expression:
\begin{equation}
    {\textrm{Total}} = {\textrm{Header}}+N_{\textrm{R}}\cdot{\textrm{Header}}_{\textrm{R}}+{\textrm{Hit}}\cdot N_{\textrm{pix}}+{\textrm{Trailer}},
\end{equation}
where, if one starts from energy depositions, $N_{\textrm{ pix}}$ is the number of hits times the number of pixels above threshold per hit, i.e. $N_{\textrm{pix}} = N_{\textrm{hits}} \cdot N_{\textrm{pix/hit}}$.
An incident particle can deposit energy in one pixel, but it may induce charge generation in several adjacent pixels and hence the number of pixel per hit, taken for this calculation is empirically set to $\sim 3$.
This is illustrated later in table~\ref{tab:pixelpercluster} of section~\ref{sec:performance}.
For example, for the $\xi=3$ signal in phase-0, we have $\sim125$ positrons per BX, which lead to $\sim126$ hits and $\sim328$ fired pixels. 
Hence, $N_{\textrm{pix/hit}}\simeq3$.
This is similar for other signal points.
Finally, the number of regions is $N_{\textrm{R}}$ and ${\textrm{ Header}}_{\textrm{R}}$ denotes the region header mentioned earlier. 

In the estimation given in table~\ref{tab:data}, the data rate for one tracker arm (8 staves) is $\sim 2$~MB/s for the phase-0 $\xi=3$ signal in the \elaser mode.
In this case, the number is completely driven by the background.
In a larger signal scenario, e.g. the phase-1 $\xi=7$, the rate is driven by both the signal multiplicity and the background at $\sim 3.1$~MB/s.

The third case shown in table~\ref{tab:data} (last row, in gray) is extrapolated from the other two, fully simulated, cases due to lack of simulation samples with more than $\sim40,000$ positrons per BX.
In this extrapolation we assume that the ``Average $N_{\textrm{hits}}$/BX/chip'' number of the larger signal will scale down similarly to the existing two signals when starting from the true number of positrons per BX.
The scaling is $\sim 9$ for the inner stave's chips and $\sim 80$ for the outer ones.
Therefore, extrapolating to one of the most extreme cases for phase-1, with $\sim 3\times 10^5$ positrons per BX, the data rate becomes as large as $\leq$ 10~MB/s.

However, these runs are expected to be short and only at the very end of LUXE's lifetime.
Hence, the data rates will be considerably smaller on average throughout LUXE's lifetime and particularly in phase-0. 

This exercise is done for the \elaser mode, where the background level is higher and where the signal level is much higher than the \glaser mode.
Hence, the data rates in the \glaser mode will be considerably smaller per tracker arm (recall that there are two arms in the \glaser mode).

\section{Expected Performance}
\label{sec:performance}
Here we discuss the tracker performance, based on full simulation studies, starting sequentially from a highly detailed simulation of the electric field inside the primitive pixel cell~\cite{ALPIDE-E}, moving on to the digitisation of the energy deposition and charge generation in the pixel, continuing to the clustering of pixels and summarising with the track fitting for low to medium signal particle multiplicities, or pixels/clusters counting for large signal particle multiplicities.
The results of this chain of algorithms are discussed and quantified for the different signal and background scenarios. 

The discussion in this section is focused only on the \elaser mode with only the positron arm of the tracker.
The performance quantified for the \elaser mode can be assumed to be identical to (in the worst case) or better (in the realistic case) than the \glaser mode.
This is because (i) the background is expected to be much smaller and (ii) the signal particle multiplicity is not expected to exceed a few hundreds per BX, even at phase-1, hence it is much lower on average.
Moreover, at very low signal multiplicities the \glaser mode has the advantage of combining the information from the two arms of the tracker to further reduce the background. 

This section is organised as follows: we start by discussing the in-pixel digitisation and pixel clustering in detail.
We then move to discuss track seeding.
This part includes the data from the full LUXE simulation with the updated geometry and magnetic field.
We then proceed to discuss the track fitting.
Finally, a brief discussion is given for the ``counter mode'' operation where the number of signal particles is estimated from pixel counting.

\subsection{Digitisation of Pixel Hits}
\label{sec:digitization}
The full \geant~\cite{AGOSTINELLI2003250,ALLISON2016186,1610988} simulation~\cite{LUXETNIndico} of the signal and background particles passing through all LUXE sensitive elements provide the energy depositions as well as the truth particle information (momenta, id, timing, etc.) regardless of whether or not the particle has deposited energy.
This allows the study of both the population of particles impinging the different elements and of the subsequent electronic responses of the sensitive parts of LUXE. 

In the tracker, the energy depositions may translate into ``hits'' (pixels with $E_{\textrm{dep}}>0$), which need to be further digitised in order to be used for clustering and tracking.

The \geant toolkit can be used only up to the energy depositions and a realistic digitisation process inside the sensor needs to be done with a different tool.
In the digitisation process in silicon pixel detectors, the energy deposition is first converted into electron-hole pairs inside the silicon. 

These charges then drift or diffuse towards the electrodes (depending strongly on the electric field inside the sensor~\cite{ALPIDE-E}) before being collected at the electrode.
A hit is then formed in the front-end electronics if the collected charge is over a certain predefined threshold.

The simulation of electron-hole pair-creation, their drift or diffusion inside the active material given an electric field, the charge collection by the electrode, digitisation of the collected charge combined with the electronic noise by the front-end electronics (etc.) will be simulated with other software.

In LUXE, the digitization software used is \allpix~\cite{SPANNAGEL2018164}.
The first goal is, therefore, to simulate the effect of an energy deposition from background and signal particles in the ALPIDE sensors as realistically as possible.
\allpix can actually simulate also the energy deposition, acting effectively as a particle gun with the \geant backend handling all the particle-level simulation.
It can also start with a given geometry and energy depositions corresponding to this geometry. 
The latter is the mode of usage in LUXE since the simulation of particles is extremely extensive and is already done with the full LUXE setup in mind. 

The \allpix simulation chain is implemented using a set of configuration files and an extensible system of modules.
These modules implement separate sequential simulation steps of the digitisation process as discussed above.
These algorithms take into effect the Landau fluctuations in the energy deposition and the production of the secondary particles like delta rays, etc.
In the LUXE mode of usage, the \allpix software setup needs the energy depositions, unique particle identifier in the list of particles, particle identifiers, coordinates of the energy deposition, the geometry of the detector (position of the chip, sensor dimension, number of pixels in each chip, pixel dimensions, etc.).
This information comes from the full \geant simulation of LUXE.  In addition, we provide the \allpix software with the electric field function derived in the work done in~\cite{ALPIDE-E}.

The output of \allpix contains the fired pixels and their charge only if the pixel has charge above a certain threshold.
It also contains the truth information about the incident particle associated with the pixel. 

As it can be understood, the very detailed simulation of \allpix, essentially at the level of single charge carriers or even a group of those, can be very extensive computationally and therefore very slow.
There are ways to speed up the process but the trade-off is with accuracy.
For the time being \allpix is good enough in terms of processing time for the simulation of the tracker response.
In the future, we intend to also check the fast simulation of charge propagation using~\cite{SULJIC2020162882} which is also used by ALICE.
This software will be much faster than \allpix.

The end of the digitisation process is in principle at the level of one pixel response in terms of the collected charge.
However, the process is quantified by considering clusters of pixels\footnote{A cluster is a group of adjacent pixels, where the charge collected in each pixel is above threshold.} rather than individual pixels.
The concept of pixel clustering is introduced in detail in  Section~\ref{sec:clustering} with a simple clustering algorithm.

\subsection{Clustering of Pixels}
\label{sec:clustering}
Our 3D custom electric field map is used for the \allpix simulation.
The output of \allpix contains the index (location) of the fired pixels in the matrix, i.e. the pixels whose charge collection was more than the charge collection threshold of 120 $e$.
The output also contains the particle identifier associated with the energy deposited in the pixel.
An energetic ($\gtrsim 1$~MeV) charged particle will on average fire more than one pixel.
Therefore, to get the reconstructed position of the impinging particle one needs to average over the fired pixels.
This can be done according to the charge collected in each pixel, as long as this information exists, or simply taking the geometric average of all pixels in the cluster. 

A preliminary brute-force clustering procedure for the LUXE tracker is employed with a simple algorithm, where all adjacent pixels are grouped together and no pixel sharing is allowed between two clusters.


The algorithm flow is as follows:
\begin{itemize}
    \item loop over a list of all fired pixels;
    \item if one pixel in the list is not found to be adjacent to another pixel from the list (either in $x$-direction or $y$-direction), then that single pixel is considered a cluster and added to the list of clusters. That pixel is then removed from the list of fired pixels for future processing;
    \item if one fired pixel is found to be adjacent to another fired pixel, they are grouped together. The grouping of fired pixels continues until there is no adjacent pixel in any direction of the group. Once the grouping ends, the group of pixels is considered a cluster and added to the list of clusters. The pixels constituting the cluster are removed from the list of fired pixels to avoid double counting;
    \item the cluster position is determined from a weighted average of the constituent pixels, taking the pixels' centres as the individual positions, where the pixels' charge can be used as a weight\footnote{If the charge information is not available, a simple average with weight unity will be used.} if this information is available.
\end{itemize}
This simple algorithm is good enough for low particle multiplicities ($\leq\mathcal{O}(10,000)$ particles per BX), but it may be limited at higher multiplicities, e.g. in phase-1 of LUXE (see Fig.~\ref{fig:challenge_E_and_rates}).
In this case, a more robust algorithm may be needed, 
possibly employing machine learning techniques (see e.g.\ ~\cite{MLCLUSTERING,}).
At some very high multiplicity, the clustering of pixels becomes impossible.

In the example shown in Fig. 6 of~\cite{ALPIDE-E}, the investigator chip used for the comparison has a thickness of 100~\um, unlike the nominal ALPIDE chip which has a thickness of 50~\um (the epitaxial layer is 25~\um thick in both cases - for more information about the ALPIDE chip itself see  Sections~\ref{sec:maps} and~\ref{sec:alpide_sensor}).
The enlarged thickness of the investigator chip leads to a larger cluster size on average ($>2$) for high-energy minimum-ionising particles (MIPs).
When moving to the nominal thickness, the cluster size for high-energy MIPs becomes smaller on average ($\leq 2$).
Note that it can also become larger if the particle's energy is very low.

\begin{figure}[!ht]
\centering
\begin{overpic}[width=0.8\textwidth]{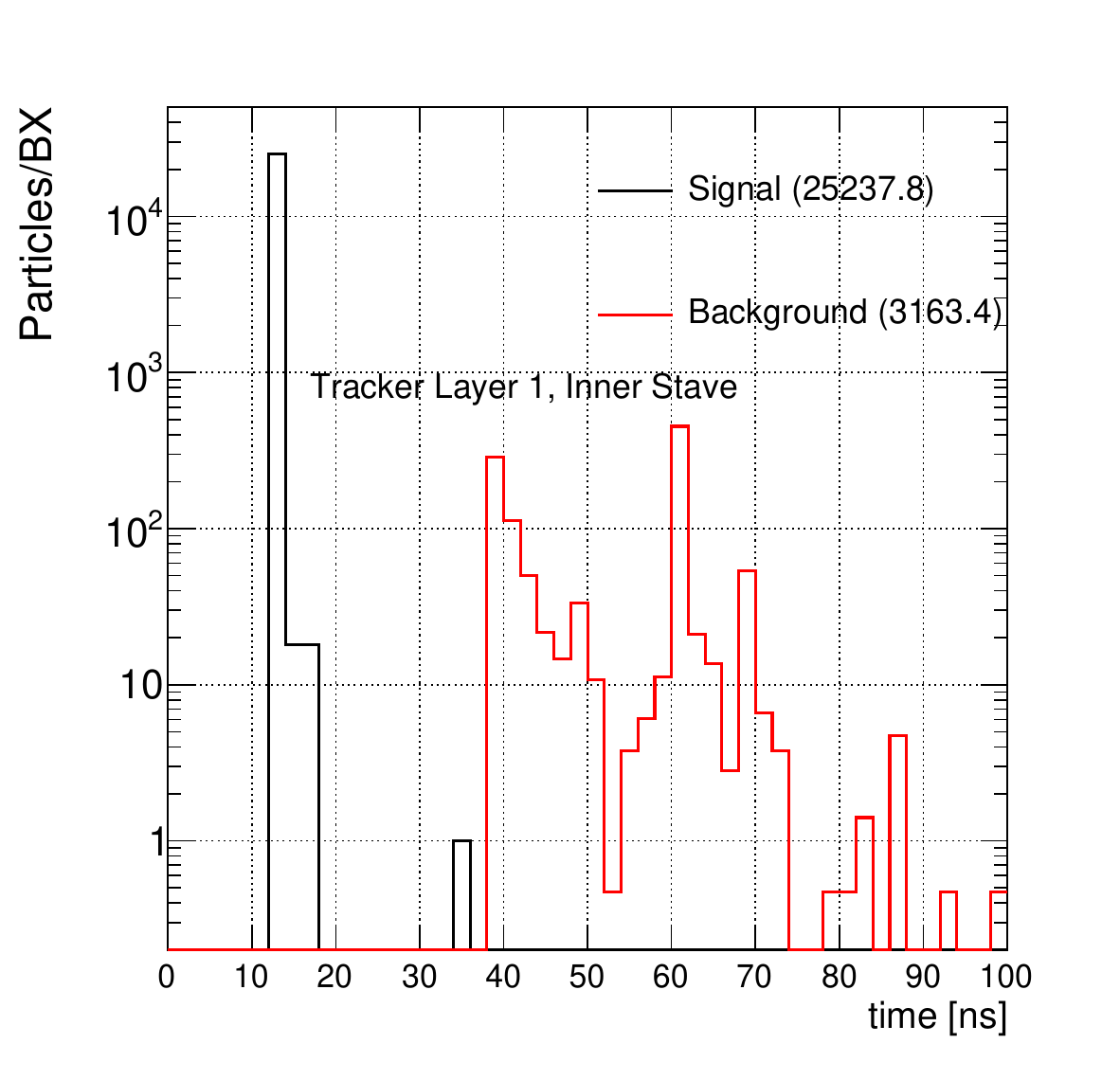}\end{overpic}
\caption{The times of arrival for signal and background particles in the full LUXE simulation integrated over all 9 chips of the inner stave of the first tracker layer.  }
\label{fig:timing}
\end{figure}

To combine the signal and background samples in the same \allpix processing, the time of arrival (from simulation) of the two species must be adjusted.
The time of arrival of particles to the first layer of the tracker is shown for signal and background in Fig.~\ref{fig:timing}.
The signal particle arrives faster than the background particles. The time alignment for background particles with the signal particle is done by subtracting 30~ns from the background particles which arrive within a 30~ns-50~ns time window. For background particles arriving after that, the time is set randomly inside the 8~ns-25~ns window. The background particle timing shown in this figure is before the time alignment.
Though the signal particles arrive faster than the background particles, the ALPIDE readout electronics is too slow ($\mathcal{O}(1~\mu {\textrm{s}})$) to discriminate based on timing, despite the sensor itself having an intrinsic time resolution of $\mathcal{O}(10~{\textrm{ns}})$.
Therefore, this feature cannot be exploited in reality and the two species of particles will appear with the same timestamp.
For the subsequent analysis we therefore align the timing of the two species of particles so that they all fall within the integration time used in the simulation (20~ns).



\subsection{Seeding of Tracks}
\label{sec:seeding}
After the digitisation and the clustering procedures discussed above, the location of the clusters in three dimensions are spatially ``embedded'' on the detector layers and the track fit may start.

The track fit algorithm takes as an input a list of the potential track ``seeds''.
A seed is characterised simply by the track's charge, the initial guess of its four-momentum and its hypothesised vertex position.

The seeds, and subsequently also the tracks,  are created by looping over all clusters in the last tracker layer (layer 4, L4), starting with the outer stave and continuing with the inner stave.

The cluster at L4 is called the pivot cluster of the seed and it can lead to up to two seeds in the list provided to the track fit algorithm.
Due to the specific staves staggering in one detector arm, depending on which stave we select the pivot cluster from, the overlap region of the other stave of the layer is ignored in the seeding (clusters in that region are taken into account for the track fit later). 
For example, if we start from the outer stave of L4 (L4O), we ignore the clusters in the overlap region of the inner stave of L4 (L4I) and vice versa. 
With this,  a pivot cluster from L4O can be matched to clusters from both the inner and outer staves of layers 1, 2 and 3 (L1I, L2I, L3I, L1O, L2O and L3O).
In contrast, a pivot cluster from L4I can only  be matched to clusters from the inner staves of layers 1, 2 and 3 (denoted as L1I, L2I and L3I respectively).
Hence, a cluster at L4 can lead to two seeds at most.

With the knowledge of both the pivot-cluster position at L4 and the map of the magnetic field strength in the three dimensions, one can in principle have a very good initial estimation of the seed four-momentum, as well as the projected position of the clusters in layers 1, 2 and 3.

The specific magnet and its field map are discussed in Chapter~\ref{chapt11}.
It is important to stress that the map is of a realistic, non-uniform field obtained from a fit to data provided by the magnet manufacturer.
The fit is modelled with a double sided Fermi function in three dimensions.
At the moment, the only component of the field considered is $B_y$ while the other two potential components are neglected as these are expected to be very small.
In that sense, the tracker simulation software is also able to incorporate fields that have three, more complex 3D components.

The prediction of the seed's four-momentum is done in a standalone run by propagating many electrons/positrons through the field volume on to the detector in a FastSim simulation setup.
The particles used are similar to those expected in LUXE in all aspects except for the energy distribution, which is flat in this case to allow coverage of the full range of the layers with enough statistics.
This is done using the specific LUXE geometry in FastSim, taking into account the IP, the vacuum volume, the vacuum exit window material, the staves' material and the air between them.
Gaussian multiple scattering and deterministic energy loss (via the Bethe-Bloch formula) are taken into account in the tracker simulation software.
For the typical energy range of the signal particles in LUXE, these effects are expected to be very small. 
After propagating the particles from the IP to the last detector layer, their position in the detector volume is registered along with their energy.

Specifically, we plot the distributions of $E(x_4)$, $\Delta x_{14}\equiv |x_4-x_1|$ and $\Delta y_{14}\equiv y_4-y_1$, where $E$ is the truth particle energy and $x_{1,4}$ and $y_{1,4}$ are the $x$ and $y$ coordinates at L1 and L4.

These plots are divided between the positron and electron arms of the tracker and further they are plotted for the \elaser mode. 
This is done separately for the \glaser mode (not shown here) due to the different geometry (two arms instead of one), but mostly due the different field strength between the two modes (1.2~T for \glaser vs. 0.95~T for \elaser).

\begin{figure}[!ht]
\centering
\begin{overpic}[width=0.99\textwidth]{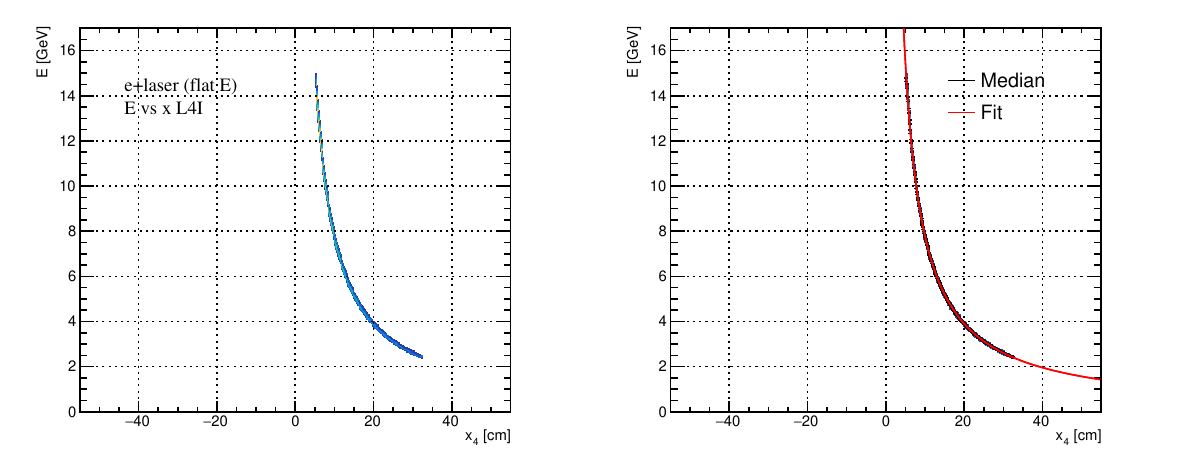}\end{overpic}
\begin{overpic}[width=0.99\textwidth]{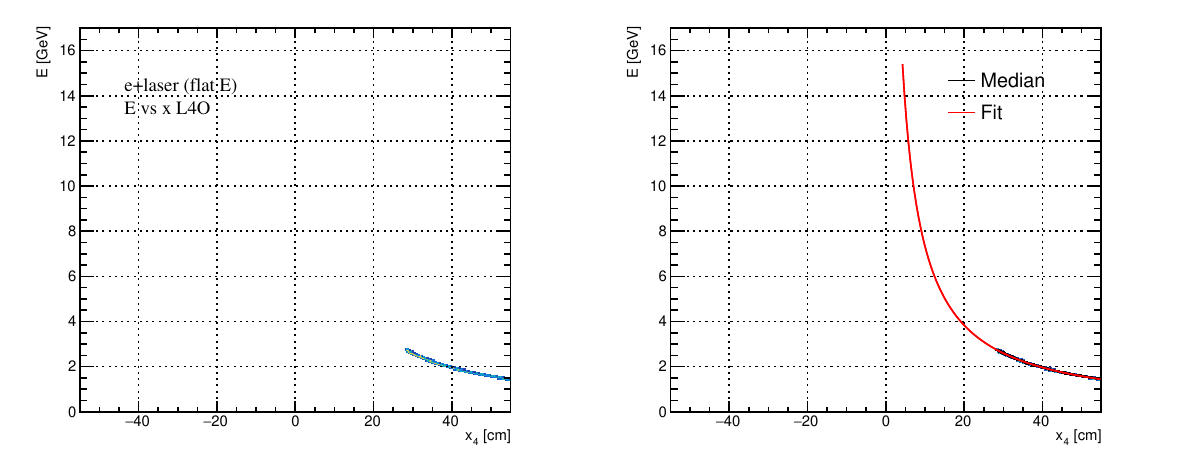}\end{overpic}
\caption{The truth particle energy as a function of its position in L4 on the left column. On the right column, the distributions' median is shown along with its fit to the function $\frac{A}{B+x_4}$. The top (bottom) plots are from the inner (outer) staves of L4. The resulting fitted functions are later used to predict the seed energy given a pivot cluster's $x$ position at L4.}
\label{fig:Eofx}
\end{figure}

\begin{figure}[p]
\centering
\begin{overpic}[width=0.99\textwidth]{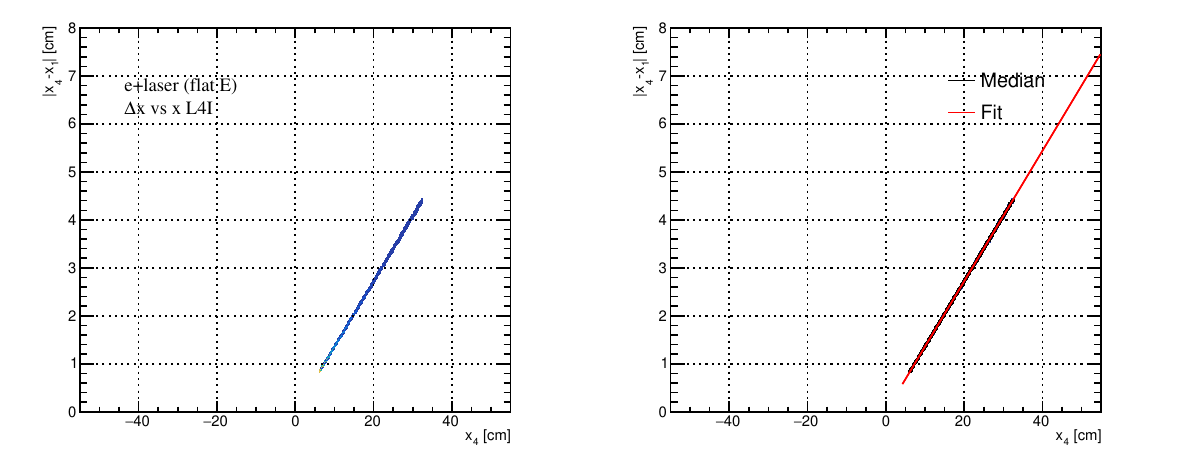}\put(10,10){II}\put(60,10){II}\end{overpic}
\begin{overpic}[width=0.99\textwidth]{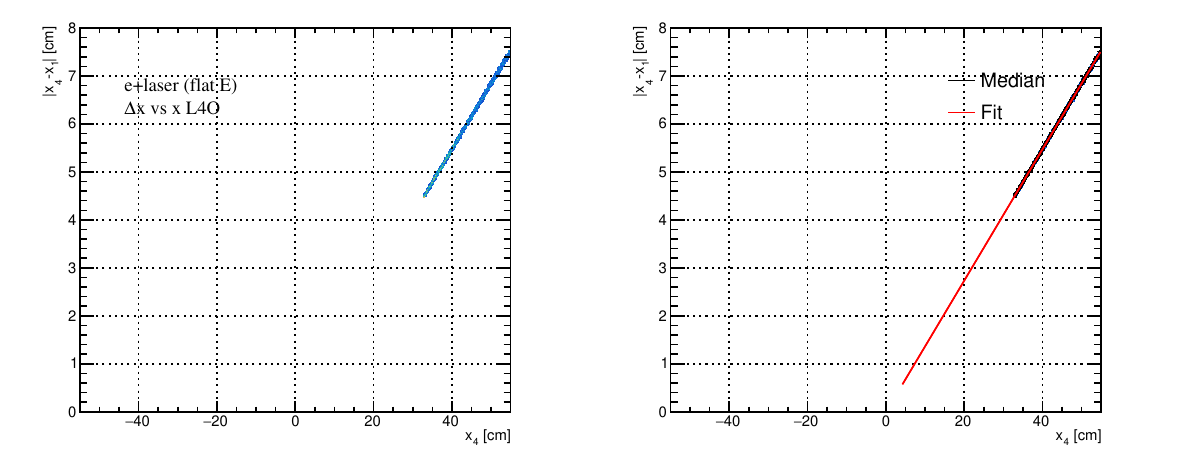}\put(10,10){OO}\put(60,10){OO}\end{overpic}
\begin{overpic}[width=0.99\textwidth]{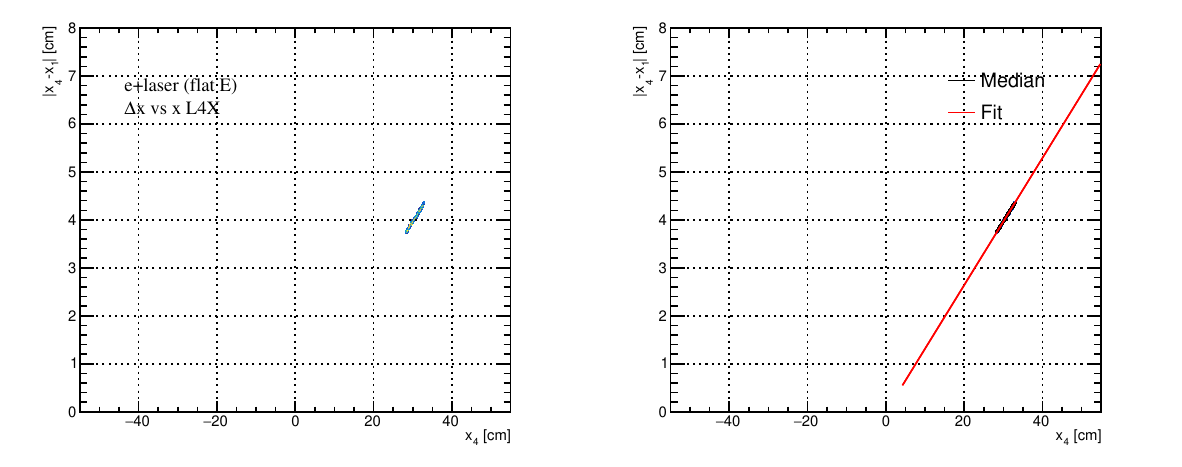}\put(10,10){OI}\put(60,10){OI}\end{overpic}
\caption{The absolute difference between the particle's position on the $x$ axis at L4 and L1 ($\Delta x_{14}$) as a function of its position along $x$ at L4 on the left column. 
On the right column, the distributions' median is shown along with its fit to a first order polynomial. The top plots are from scenario II, middle plots from scenario OO and bottom plots are from scenario OI. Scenarios OO, OI and II are described in the text.}
\label{fig:dx14}
\end{figure}

\begin{figure}[!ht]
\centering
\begin{overpic}[width=0.99\textwidth]{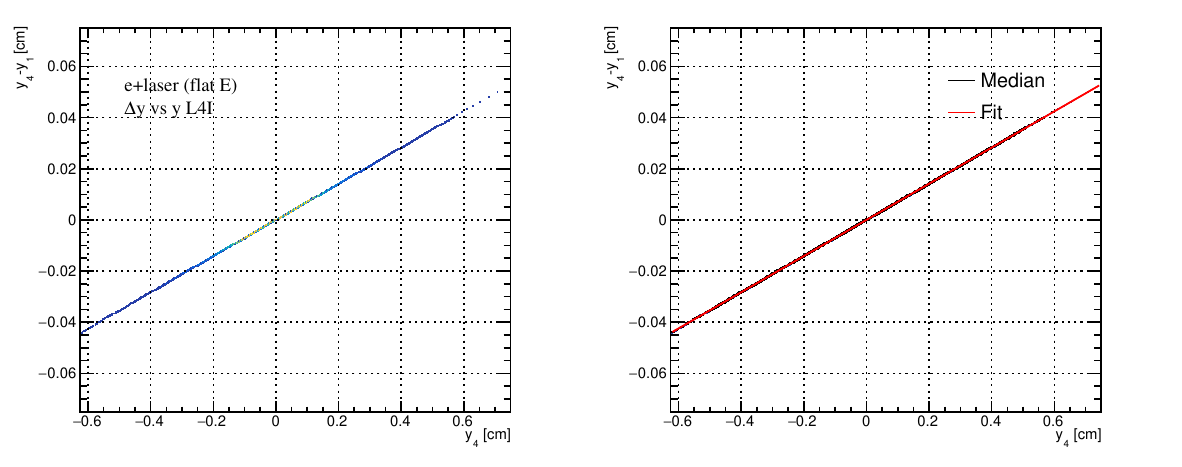}\put(10,10){II}\put(60,10){II}\end{overpic}
\begin{overpic}[width=0.99\textwidth]{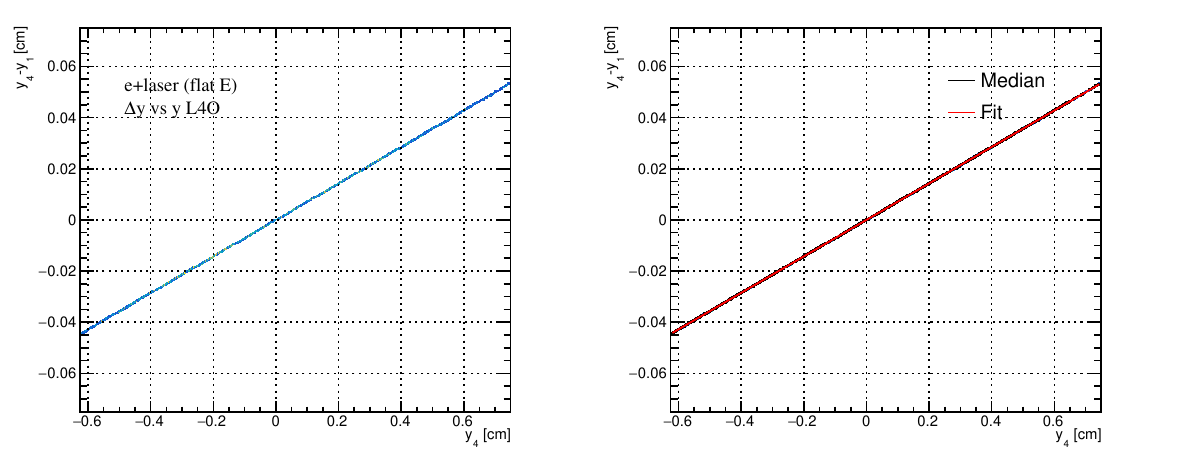}\put(10,10){OO}\put(60,10){OO}\end{overpic}
\begin{overpic}[width=0.99\textwidth]{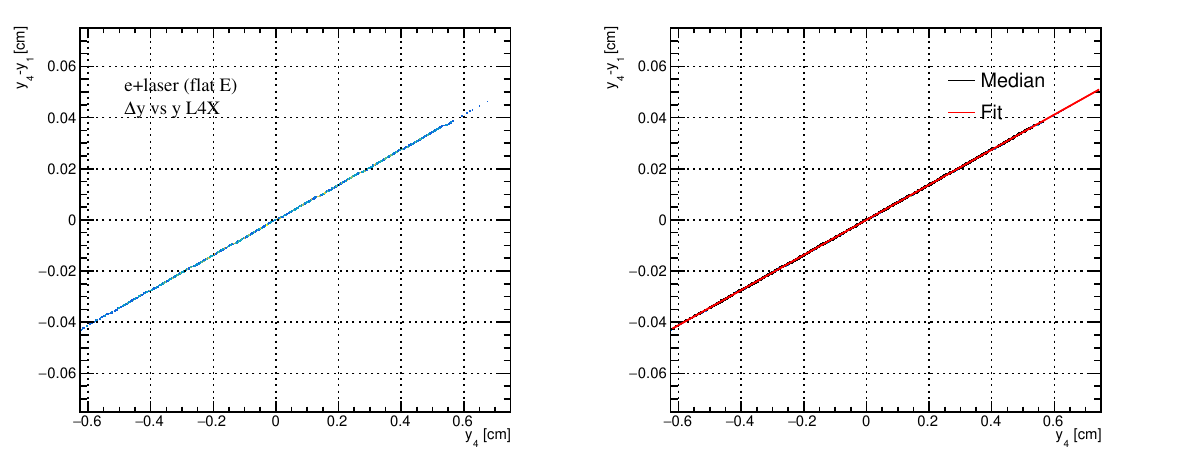}\put(10,10){OI}\put(60,10){OI}\end{overpic}
\caption{The difference between the particle's $y$ position at L4 and L1 ($y_4-y_1$) as a function of its $y$ position at L4 on the left column.
On the right column, the distributions' median is shown along with its fit to a first order polynomial. The top plots are from scenario II, middle plots from scenario OO and bottom plots are from scenario OI. Scenarios OO, OI and II are described in the text.}
\label{fig:dy14}
\end{figure}

As can be seen from Figs~\ref{fig:Eofx},~\ref{fig:dx14} and~\ref{fig:dy14}, there is a strong correlation between these quantities and the related coordinates.
These correlations are easily parameterised and used later in the remainder of the algorithm to predict the particle energy and to constrain the number of possible four-cluster combinations.
Hence, these quantities are extremely important for both the efficiency and the speed of the remainder of the algorithm.
This is done for three possible combinations of L4 and L1 clusters explained below.


The distribution of $E(x_4)$ is parameterised as $\textstyle{\frac{A}{B+x_4}}$, while the distributions of $\Delta x_{14}$ and $\Delta y_{14}$ are parameterised as first order polynomials.
In all three cases the 50\% quantile of the distribution in the vertical axis is taken as the central value, while the 33\% and 67\% quantiles are taken as the up and down uncertainties respectively.
The resulting data points and uncertainties are then fitted to the models mentioned above.

For a given cluster in L4, one can use the parameterisation of the $E(x_4)$ distribution to obtain the seed energy, $E_{\textrm{seed}}$.
Furthermore, the seed position at L1 in $x-y$ can be predicted from the $\Delta x_{14}$ and $\Delta y_{14}$ parameterisations, i.e. these two relations allow prediction of $x_1$ and $y_1$.
Knowing $\vec{r}_4=(x_4,y_4,z_4)$ and the predicted $\vec{r}_1=(x_1,y_1,z_1)$, the seed's inclination in the $y-z$ plane can be determined.
If we assume zero initial momentum (at the IP) in the $x$ direction and write down the four-momentum of the seed as
\begin{equation}
    \left|\vec{P}_{yz}\right| = \sqrt{E_{\textrm{seed}}^2-m_e^2-(p_{\textrm{seed}}^x=0)^2)},
\end{equation}
\begin{equation}
    p_{\textrm{seed}}^y = \left|\vec{P}_{yz}\right| \hat{r}_{14}^y,
\end{equation}
\begin{equation}
    p_{\textrm{seed}}^z = \left|\vec{P}_{yz}\right| \hat{r}_{14}^z,
\end{equation}
where $\vec{P}_{yz}$ is the seed momentum in the $y-z$ plane, $m_e$ is the electron mass and $p_{\textrm{seed}}^x=0$ is the initial momentum in $x$, set to zero by hand.
The unit vector $\hat{r}_{14}$ is calculated from $\vec{r}_1$ and $\vec{r}_4$ as
\begin{equation}
    \hat{r}_{14}=\textstyle{\frac{\vec{r}_{4}-\vec{r}_{1}}{\left|\vec{r}_{4}-\vec{r}_{1}\right|}}.
\end{equation}

\begin{figure}[!ht]
\centering
\begin{overpic}[width=0.75\textwidth]{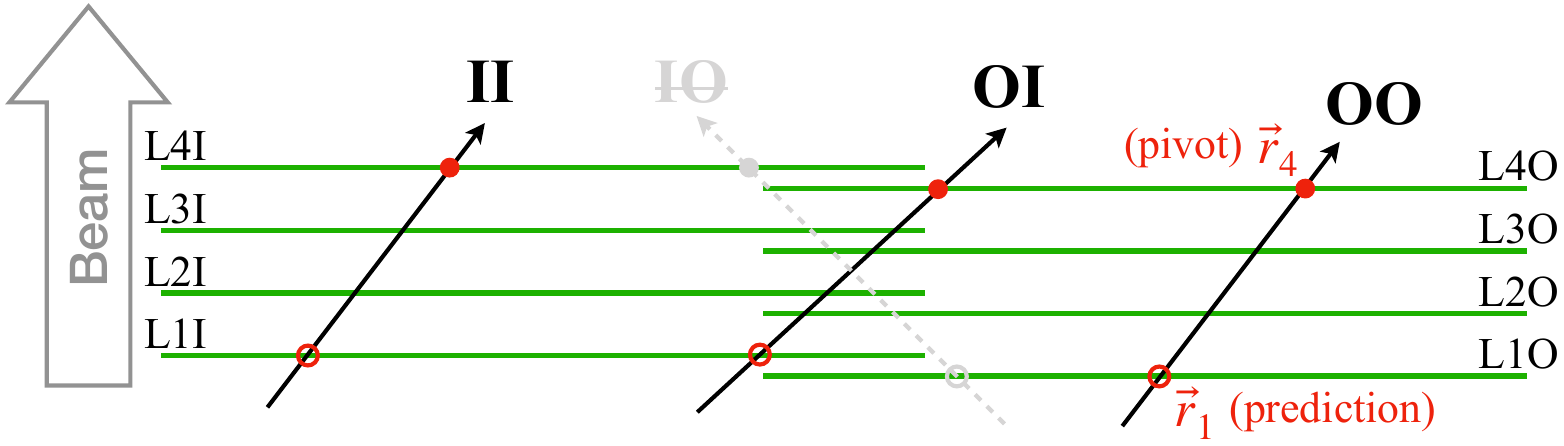}\end{overpic}
\caption{The three possible combinations defined by the pivot cluster at L4 and the respective predicted position at L1. In this illustration only the $x-z$ projection is seen. An ``impossible'' combination (IO) is shown in light grey.}
\label{fig:possibleseeds}
\end{figure}

The reason why up to two seeds may be expected per L4 cluster can be understood from Fig.~\ref{fig:possibleseeds}.
The L4 cluster may be ``linked'' to L1 in three ways, depending on its position in the layer and the position of the predicted cluster at L1 with these possible combinations:
\begin{enumerate}
    \item L4I+L1I (denoted as II),
    \item L4O+L1O (denoted as OO),
    \item L4O+L1I (denoted as OI).
\end{enumerate}

Finally, besides allowing prediction of the four-momentum of the seed, these correlations are also used for two other important tasks as discussed below.
The first task is the determination of a narrow ``road'', where clusters at L2 and L3 can be expected.
The second task is the determination whether or not one may expect to have ``holes'' (i.e. missing clusters) along the track due to e.g. passage through the gap between pixels, through the L-shape holder of the staves in the overlap region, through dead pixels, etc.

\begin{figure}[!ht]
\centering
\begin{overpic}[width=0.75\textwidth]{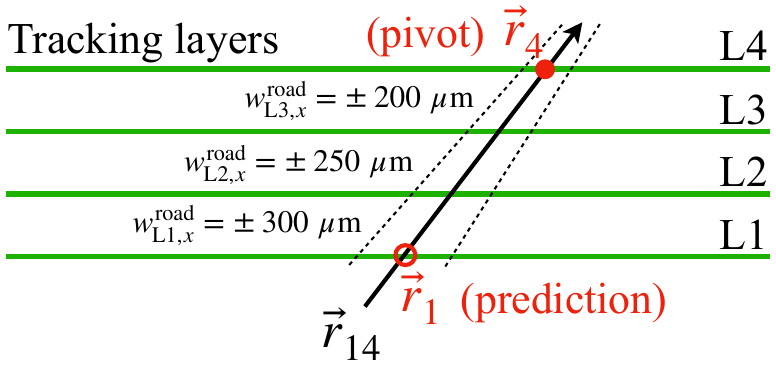}\end{overpic}
\caption{The concept of seed ``road'' defined by the pivot cluster at L4 and the respective predicted position at L1. In this illustration only the $x-z$ projection is seen and hence only the $x$-road is shown.}
\label{fig:seedroad}
\end{figure}

For the former task, the embedding of the clusters in L1, L2 and L3 is done in a selective manner inside a narrow 2D ``road'' of a few 100s of microns in $x$ and $y$ with increasing size going from L3 to L1 around the seed line $\vec{r}_{14}$:
\begin{enumerate}
    \item $w^{\textrm{road}}_{{\textrm{L1}},x}=300~\um$,
    \item $w^{\textrm{road}}_{{\textrm{L2}},x}=250~\um$,
    \item $w^{\textrm{road}}_{{\textrm{L3}},x}=200~\um$,
\end{enumerate}
where $w^{\textrm{road}}_{{\textrm{L}}j,y}$ are identical to $w^{\textrm{road}}_{{\textrm{L}}j,x}$ at the moment, with $j=1,2,3$.
These values are not yet optimised.
Later, this allows to focus the track fit algorithm only on a small number of possible combinations.
This step reduces the number of possible combinations significantly, while preserving the algorithm's accuracy and efficiency and  increasing its speed, as discussed below. 
The seed road concept is illustrated in Fig.~\ref{fig:seedroad}.

Using the latter task, we can determine how many holes are expected per layer.
With this number, and by knowing the number of clusters found in the narrow road around the seed, the track fit algorithm gets the minimum number of clusters it expects for the fit.
The road is made sufficiently wide such that this number is always larger than (high multiplicity) or equal to (low multiplicity) the expected signal clusters around the track.
At the step of the seeding, the seed is rejected if there are insufficient clusters to perform the subsequent track fit. 

\begin{figure}[!ht]
\centering
\begin{overpic}[width=0.99\textwidth]{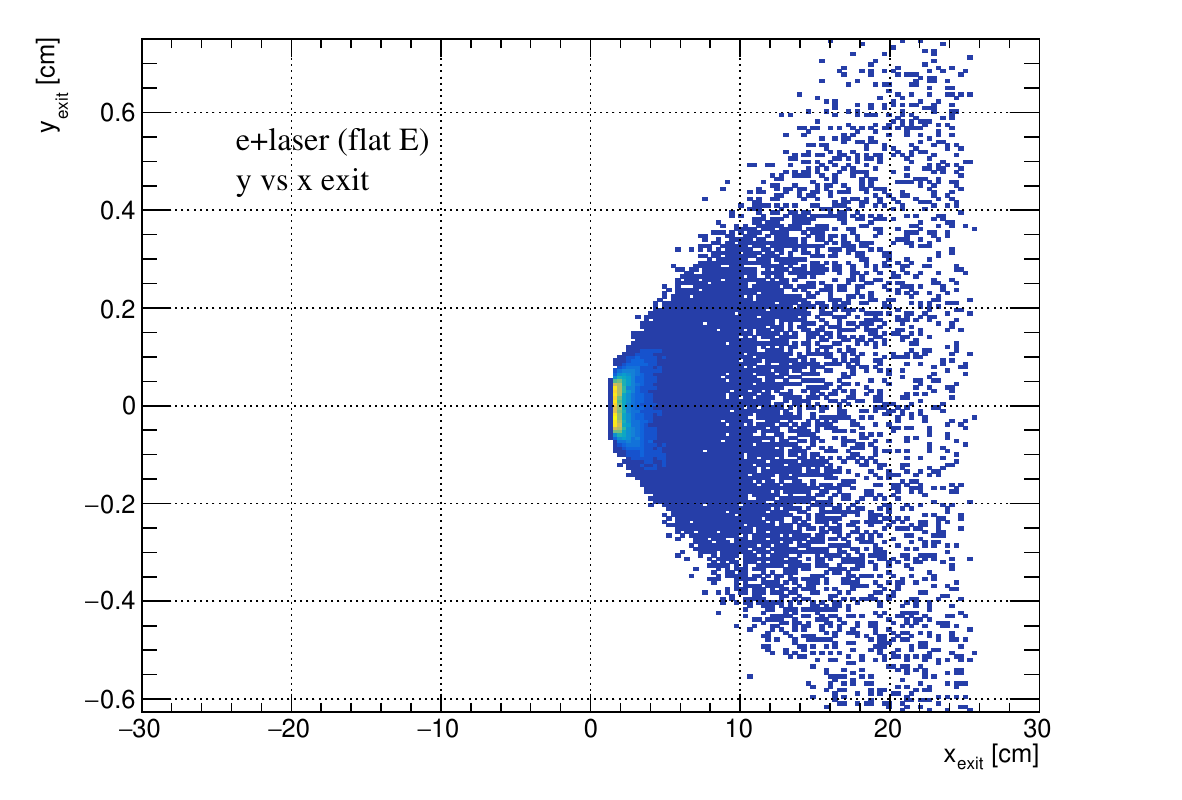}\end{overpic}
\caption{The $x-y$ distribution at the exit plane of the dipole from the flat sample. The structure is determined by the kinematics at the IP. }
\label{fig:dipoleexit_elaser}
\end{figure}

Finally, seeds which pass the above criteria
may still be rejected if the conditions listed below are met:
\begin{itemize}
    \item $E_{\textrm{seed}}<0.5$~GeV or $E_{\textrm{seed}} > 16.5$~GeV,
    \item $|y_{\textrm{exit}}| > 7.5~(6.0)$~mm for the \elaser (\glaser) setup,
    \item $|x_{\textrm{exit}}| < 18$~mm for both \elaser and \glaser setups,
    \item $|x_{\textrm{exit}}| > 300$~mm for both \elaser and \glaser setups,
\end{itemize}
where the coordinates with subscript ``exit'' denote the exit plane (in $x-y$) of the dipole volume.
For the signal particles, this area is expected to be small, while background particles may point in all directions.
This is illustrated in Fig.~\ref{fig:dipoleexit_elaser} for the \elaser setup with the flat signal sample, i.e. where the energy is uniformly distributed. If the $y$ momentum at the IP is zero the resulting distribution is constrained to $y=0$. Therefore, when making the flat samples, we are replicating the $x,y$ momenta distributions at the IP from actual signal samples.
The criteria listed above are not yet optimised and it can be seen, for example, that the rectangular cut at the dipole exit plane would perform less efficiently compared to a triangular cut.

The seeds associated with the L4 pivot cluster, which are not rejected by the above-mentioned criteria are given as an input to the track fitting algorithm as discussed in the next  Section.


\subsection{Track Fitting}
\label{sec:trackfit}
The tracking is built after the successful completion of the clustering step.
As discussed above, this prerequisite may not always be satisfied in LUXE, notably for the very large signal particle multiplicities expected.
Therefore, in this  Section we limit the discussion to the ``resolved'' scenario, i.e. for multiplicities below $\sim\mathcal{O}(10,000)$~particles/BX.
The ``non-resolved'' case where multiple signal particles may hit the same pixel is discussed in the next  Section. 

In this work, we use the Kalman-filter~\cite{KalmanFilter,BILLOIR1990219} (KF) track fitting algorithm.
It requires a set of seeds and a corresponding minimal set of clusters to start the fit procedure.
The algorithm's goals are (i) identifying the correct combination of clusters along the track's path and (ii) extracting the track's properties like momentum (and energy), the vertex, the fit quality, etc. 

However, the complexity and memory consumption increase rapidly with increasing signal multiplicities.
The background multiplicity, as large as it may get, is fixed on average.
Therefore, the embedding of the clusters in the detector layer objects of the KF algorithm must be done in an efficient way since otherwise the subsequent algorithms will become too slow and memory intensive. 

We have solved this potential problem by defining a large lookup table, 
where clusters can be selectively embedded according to their global $x-y$ coordinates.
The table is finely binned along both $x$ and $y$ per layer and all cluster objects falling in a certain bin are pre-cached into a special container corresponding to the specific bin.
Upon selecting one pivot cluster in L4, the seed line and seed road are found as discussed above.
The road edges in $x$ and $y$ are then used to find the relevant range of bins in the lookup table.
Usually there are just a few bins in the range (depending on the granularity of the table).
We then loop on the bins in that given range, while ``pouring'' the contents of the clusters' containers of each bin into the layer objects of the detector used by the KF algorithm.
The accuracy of this procedure depends on both the granularity of the table and on the seed road width.
If the road width is too small and/or the table is too coarsely binned, the wrong and/or insufficiently many clusters may be embedded. 

If there are enough clusters embedded in L1, L2 and L3 (taking into account also the expected holes), we proceed with the fit.
Otherwise, we move on to the next pivot cluster in L4. 

The KF can fit several tracks per seed, in principle.
Currently the fit requires a minimum of 4 clusters on the track, but this can be relaxed to a minimum of 3 clusters.
This may be tuned in different scenarios and also be tightened in the post-processing part.
If the fit succeeds, the ``winner'' track from the pool of fitted tracks for the specific seed is returned according to the best $\chi^2/{\textrm{DoF}}$ criterion.
The clusters associated with the winner track are removed from the detector layer objects.
These clusters are also removed from the lookup table such that there cannot be two tracks using the same cluster.
This makes the algorithm particularly fast towards the end of the loop over the pivot clusters in L4 since the lookup-table gets gradually emptier.

This behaviour can be configured differently to allow sharing of clusters.
Indeed, that depends strongly also on the specific clustering performance and for very large multiplicities ($\gtrsim 10,000$ particles per BX) this requirement must be relaxed in order to restore the efficiency.
In the results shown below, we allow the fitted tracks to share clusters only at L1, L2 and L3 for multiplicities above 10,000 particles per BX. 

To avoid accidental fit failures due to e.g. insufficient clusters in the road, excessively tight KF configuration, etc., we iteratively retry the fit, while gradually loosening some of these requirements.
There are 3-4 such iterations allowed before moving on to the next pivot cluster at L4. 

The winner track is saved for further processing before moving to the next pivot cluster at L4.
The track can be matched to the truth particle by means of matching the associated clusters with truth particles.
As it will be shown in the next section, the number (and different kinematic distributions) of the winner tracks can be tuned to resemble the number of true signal particles to a very good extent. 

Some post-processing requirements can be further imposed on the set of winner track's kinematic, origin and quality properties, as discussed in the next section.


\subsection{Reconstruction Performance}
\label{sec:trackingperf}
The performance of the tracker can be very generally gauged from three criteria, looking at several scenarios of signal in the presence of background: (i) how well we reconstruct the number of signal particles per BX, i.e.\ what is the tracking efficiency, (ii) how well the momenta (or energies) of the signal particles get reconstructed, and (iii) how well the background particles get rejected in different signal multiplicity scenarios.
We note that the clustering algorithm used here is very basic and will be replaced soon by a machine-learning-based algorithm to cope with the large multiplicity scenarios.
Hence, the clustering performance is not discussed. 

The results shown in this  Section correspond to the four BXs of signal in the \elaser mode with $\xi=3$ (7) in phase-0, leading to about 125 (40,000) positrons per BX.
The signal samples are simulated by \textsc{Ptarmigan}~\cite{ptarmigan,Blackburn:2021cuq}.
Both of these signals and the beam-induced background (no collision with the laser) are fully simulated by \geant~\cite{AGOSTINELLI2003250,1610988,ALLISON2016186} using the geometry model of LUXE.
Further, the samples are digitised (signal, background and signal + background) by \allpix~\cite{SPANNAGEL2018164,Dannheim:2020yna} as discussed in  Section~\ref{sec:digitization}.

We note that the signal positrons in the $\xi=7$ sample is mildly weighted, i.e., each positron comes with a weight of $\sim 1.8$ on average with a standard deviation of $\sim 1.3$.
However, in this  Section we quantify the detector performance for a given positron multiplicity rather than for a given $\xi$ value.
Hence, we ignore the weights in the two samples in the following discussion.
In that respect, it should be recalled that, e.g., for $\xi=7$, the actual multiplicity is $\sim 70,000$ instead of $\sim 40,000$ (see Fig.~\ref{fig:challenge_E_and_rates}, left).


For all samples, the digitisation and clustering is done for the signal alone, the background alone and the combined signal + background.
Wherever the signal is simulated, it also includes residual background due to secondary particles originating from the signal particles' interaction with the different elements of the experiment.

In addition to these two signal samples, we also draw randomly (uniformly) different numbers of signal particles per BX from the available $\xi=3$ or $\xi=7$ sample to check the linearity of the reconstruction response.
This is done with truth-particle multiplicity ranging from 1 to 10,000 per BX in the presence of background. The signal sample of $\xi=3$ was used to prepare sub-samples with truth-particle multiplicity ranging from 1 to 50, while the signal sample of $\xi=7$ was used to prepare samples with truth particle more than 50 (except for the $\sim 125$ point which is simply using the full $\xi=3$ sample).

Since the linearity study with  multiplicities larger than 50 starts from a baseline sample with high multiplicity of truth particles (40,000), it is done while ignoring the cases where the randomly chosen truth particles hit a pixel (i.e. deposit some energy in it) that is being hit by other truth particles (one or more).
This is particularly important for the low multiplicity runs, but also for the medium multiplicity runs.
The first reason for ignoring these ``overlaps'' is that this is not expected in low multiplicity cases, while it is inherently present in the baseline sample.
The second reason is that in order to quantify the performance of only the tracking part of the reconstruction chain, we need to assume that the clustering part is fully efficient, while it is clear that the simple clustering algorithm becomes quickly inefficient with increasing signal multiplicity.
In this case we already know that the clustering is not optimal and more work is needed.
Hence, for the linearity study we assume no truth particle overlaps.

For the samples used, we list for illustration in Table~\ref{tab:pixelpercluster} the total number of hits (pixels with $E_{\textrm{dep}}>0$), denoted as $N_{\textrm{hits}}$, the total number of pixels above threshold, denoted as $N_{\textrm{pixels}}$, and the total number of clusters, denoted as $N_{\textrm{clusters}}$ per BX in the inner stave of L1.
For reference, the true signal multiplicity $N_{\textrm{true}}$ is also listed in the second column.
Note that the value of $N_{\textrm{hits}}$ comes from the \geant simulation, while $N_{\textrm{pixels}}$ comes from passing those hits through \allpix.
The average number of entries per chip per BX can be obtained by simply dividing by 9.
As discussed in  Section~\ref{sec:data}, it can be seen that the scaling between pixels and hits is roughly 3 for the signal runs.
This is due to several reasons, for example the specific digitisation response and secondaries from the signal particles.
For the subsequent tracking performance discussed in this  section, only the cluster information is used, while the pixel information is used in Section~\ref{sec:countingperf}. 

\begin{table}[!ht]
\centering
\begin{tabular}{lc|ccc}
 \toprule
 Sample & $N_{\textrm{true}}$ & $N_{\textrm{hits}}$ & $N_{\textrm{pixels}}$ & $N_{\textrm{clusters}}$ \\ 
 \toprule
 Background & 0 & 5377 & 10140 & 3578 \\
 \hline
 Phase-0, $\xi=3.0$ signal-only & $\sim 125$ & 128 & 328 & 127 \\ 
 Phase-0, $\xi=3.0$ signal + background & $\sim 125$ & 5505 & 10435 & 3705 \\ 
 Phase-0, $\xi=7.0$ signal-only & $\sim 4\times 10^4$ & 38338 & 94594 & 30750 \\ 
 Phase-0, $\xi=7.0$ signal + background & $\sim 4\times 10^4$ & 43715 & 95097 & 32413 \\ 
 \bottomrule
\end{tabular}
\caption{The total number of hits (pixels with $E_{\textrm{dep}}>0$), the total number of pixels above threshold and the total number of clusters per BX in the inner stave of L1 alone for illustration.
The true signal multiplicity is also listed.}
\label{tab:pixelpercluster}
\end{table}

Below, we show a few benchmark distributions resulting from the algorithm discussed in Section~\ref{sec:trackfit} for the case of \elaser (phase-0) with $\xi=3$ having a signal multiplicity of $\sim 125$ particles per BX.
This particular multiplicity is chosen only for allowing just enough statistics to see the relevant trends.
The three samples shown include the signal alone, the background alone and the signal + background (see Fig.~\ref{fig:clustersize}).
Wherever appropriate, we also overlay the truth distribution in dashed black.

\begin{figure}[!ht]
\centering
\begin{overpic}[width=0.49\textwidth]{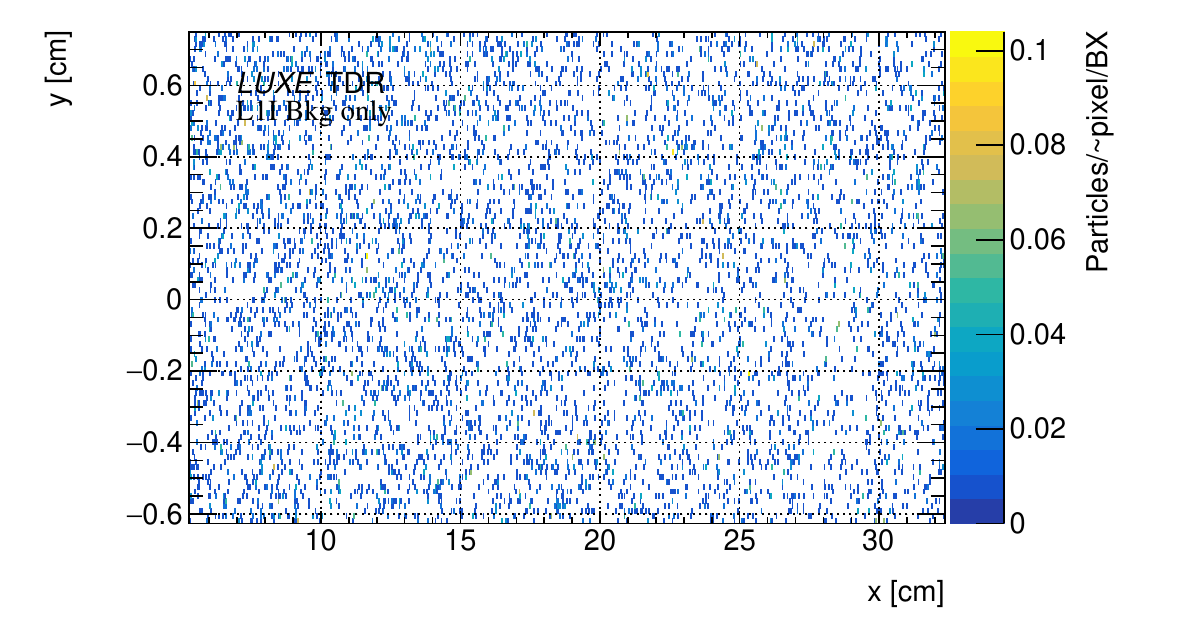} \put(18,45){\colorbox{white}{{\tiny{ \normalfont \itshape LUXE}~ TDR}}} \put(18,41){\colorbox{white}{{\tiny L1I Bkg only}}} \end{overpic}
\begin{overpic}[width=0.49\textwidth]{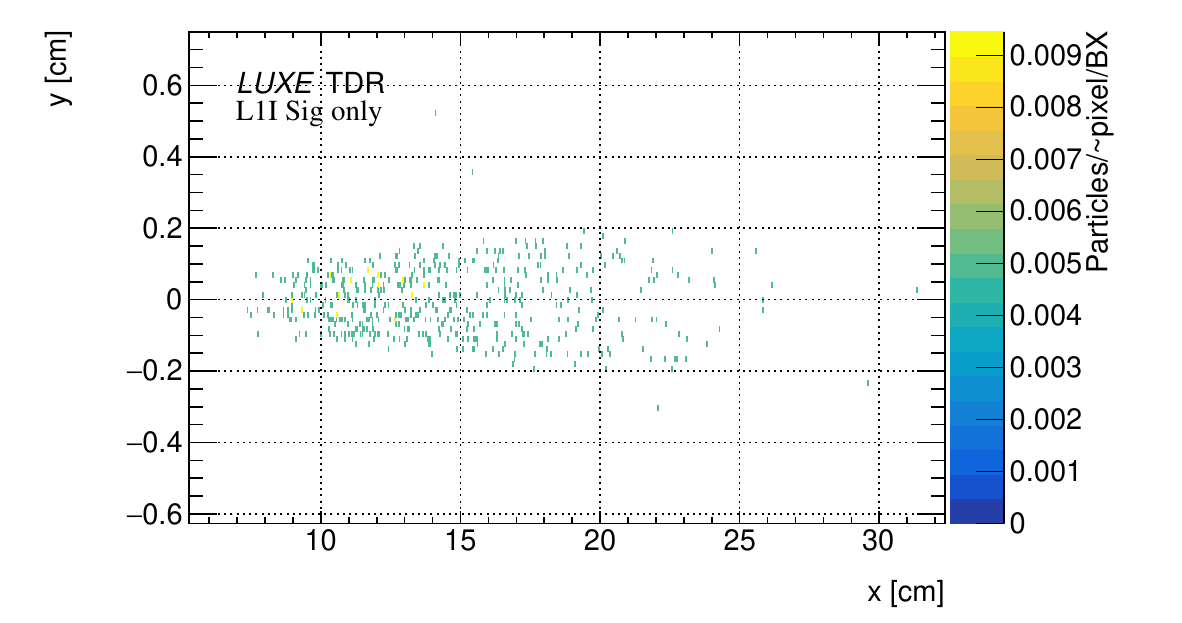} \put(18,45){\colorbox{white}{{\tiny{LUXE}~ TDR}}} \put(18,41){\colorbox{white}{{\tiny L1I Sig only}}} \end{overpic}
\begin{overpic}[width=0.49\textwidth]{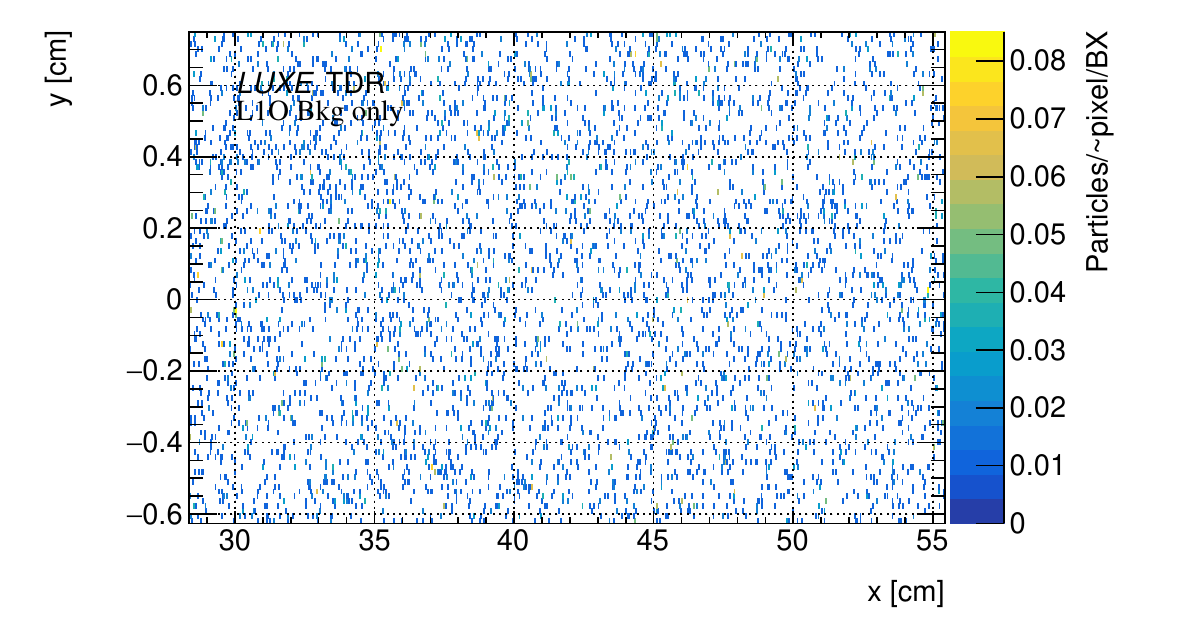} \put(18,45){\colorbox{white}{{\tiny{\normalfont \itshape LUXE}~ TDR}}} \put(18,41){\colorbox{white}{{\tiny L1O Bkg only}}} \end{overpic}
\begin{overpic}[width=0.49\textwidth]{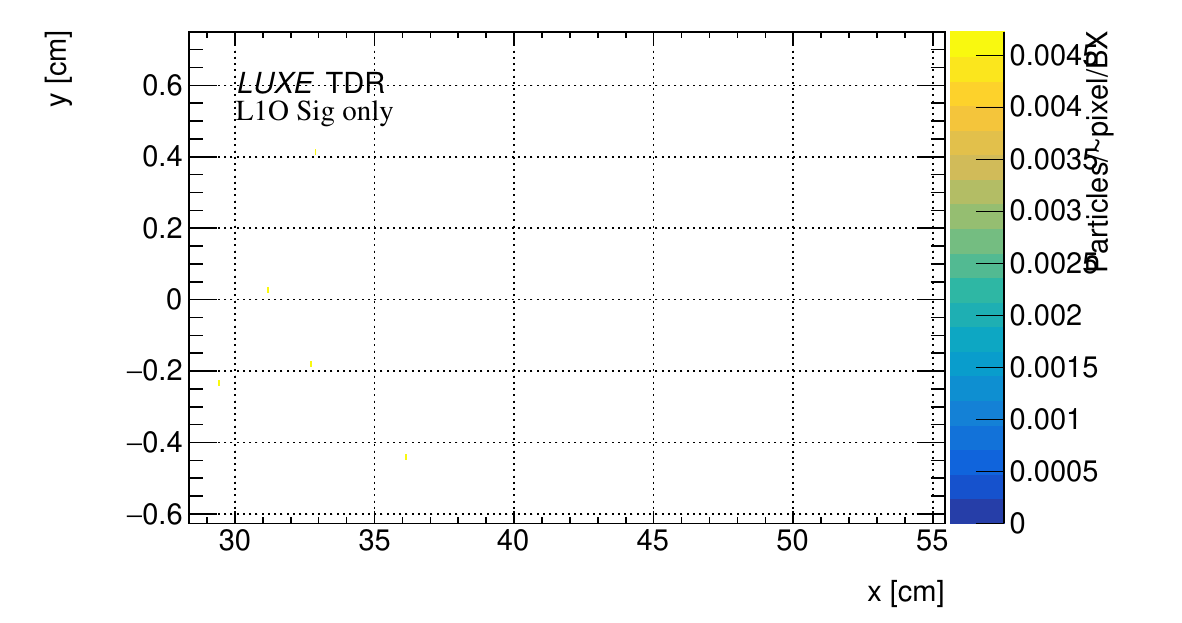} \put(18,45){\colorbox{white}{{\tiny{\normalfont \itshape LUXE}~ TDR}}} \put(18,41){\colorbox{white}{{\tiny L1O Sig only}}} \end{overpic}
\begin{overpic}[width=0.49\textwidth]{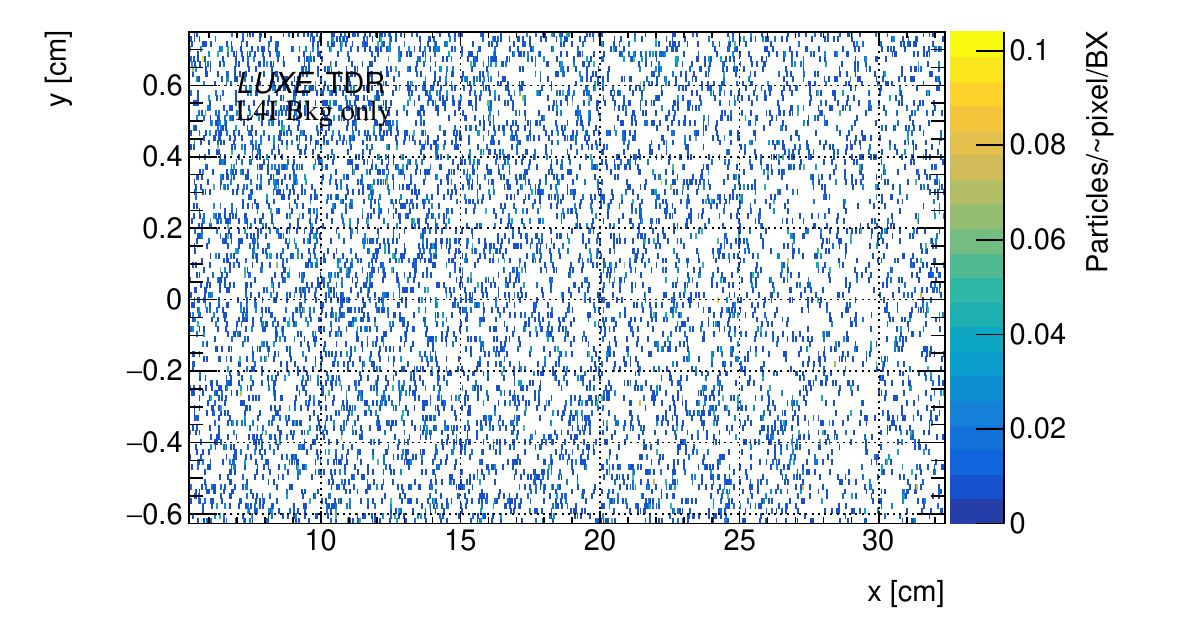} \put(18,45){\colorbox{white}{{\tiny{\normalfont \itshape LUXE}~ TDR}}} \put(18,41){\colorbox{white}{{\tiny L4I Bkg only}}} \end{overpic}
\begin{overpic}[width=0.49\textwidth]{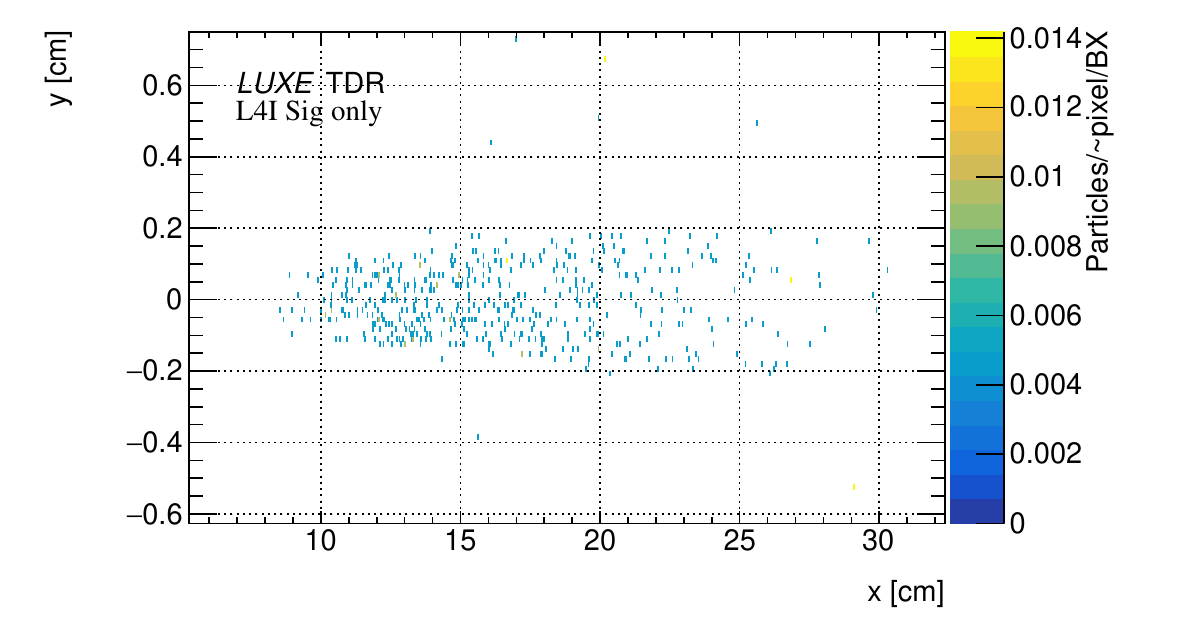} \put(18,45){\colorbox{white}{{\tiny{\normalfont \itshape LUXE}~ TDR}}} \put(18,41){\colorbox{white}{{\tiny L4I Sig only}}} \end{overpic}
\begin{overpic}[width=0.49\textwidth]{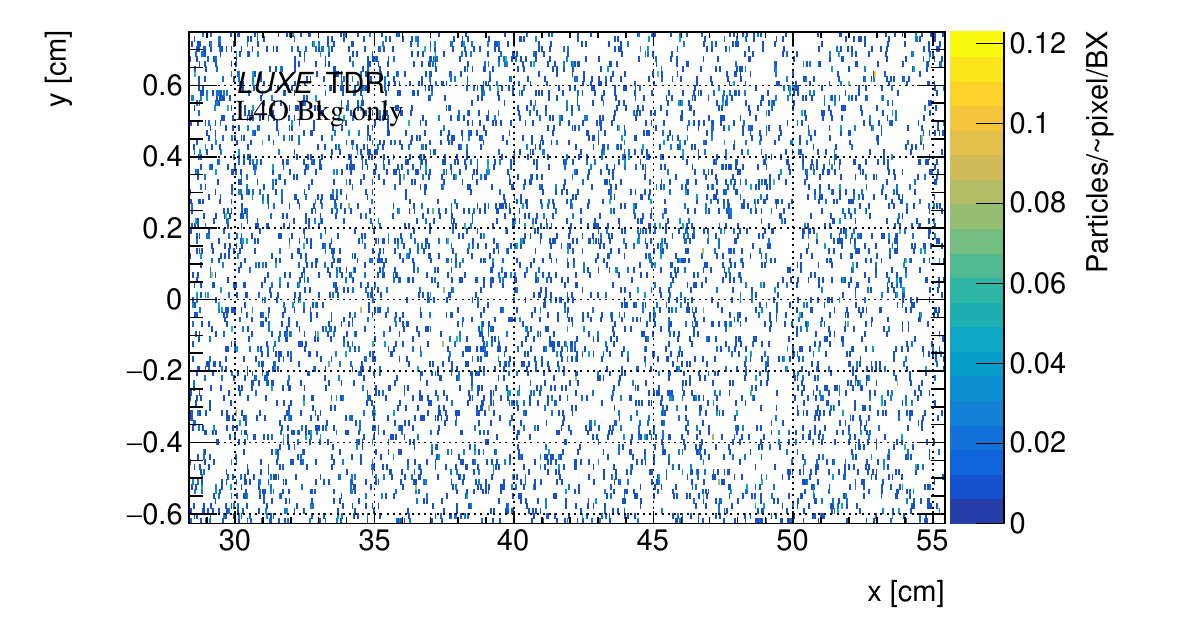} \put(18,45){\colorbox{white}{{\tiny{\normalfont \itshape LUXE}~ TDR}}} \put(18,41){\colorbox{white}{{\tiny L4O Bkg only}}} \end{overpic}
\begin{overpic}[width=0.49\textwidth]{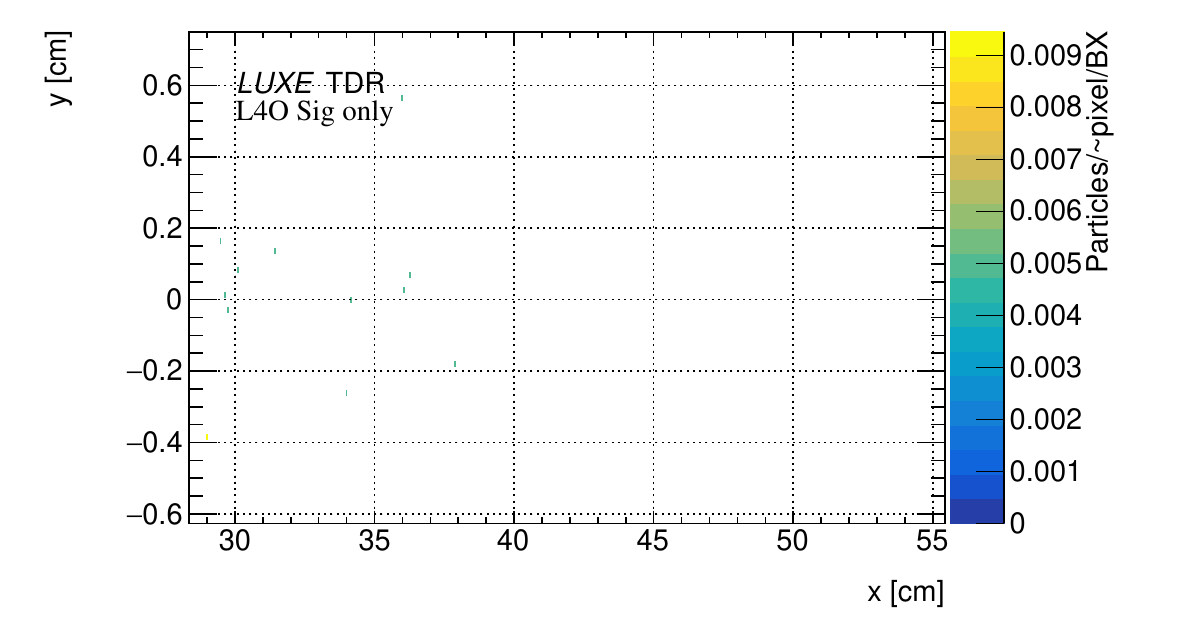}\put(18,45){\colorbox{white}{{\tiny{\normalfont \itshape LUXE}~ TDR}}}  \put(18,41){\colorbox{white}{{\tiny L4O Sig only}}} \end{overpic}
\caption{The truth particles $y$ vs $x$ position on the staves for background-only (left) and the $\xi=3$ signal only (right). }
\label{fig:occxi3}
\end{figure}

\begin{figure}[!ht]
\centering
\begin{overpic}[width=0.49\textwidth]{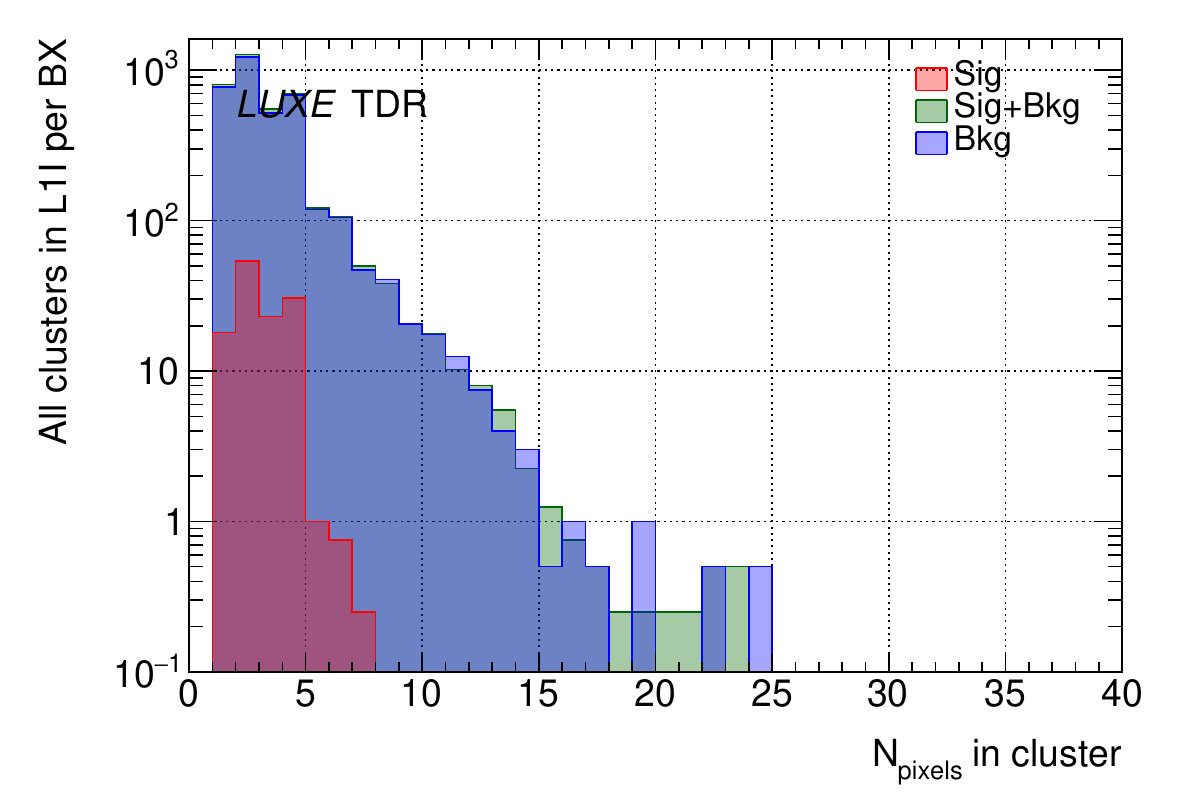}\end{overpic}
\begin{overpic}[width=0.49\textwidth]{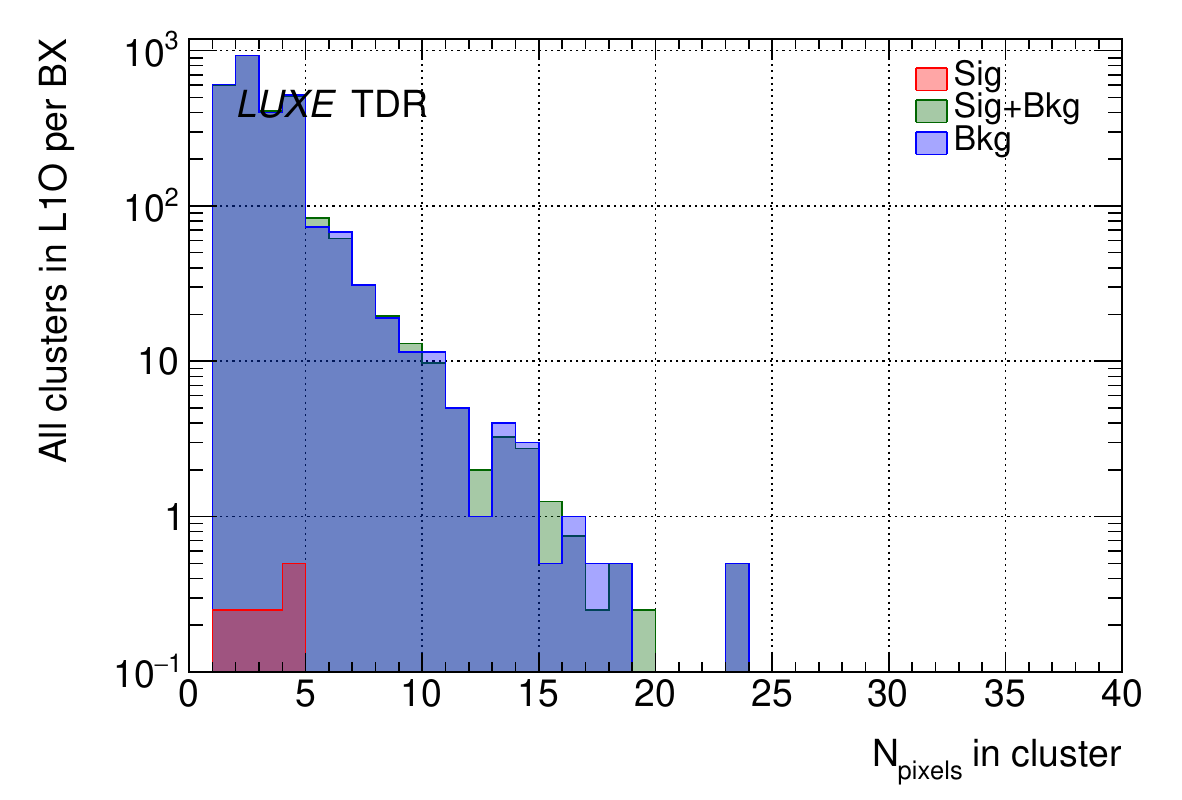}\end{overpic}
\begin{overpic}[width=0.49\textwidth]{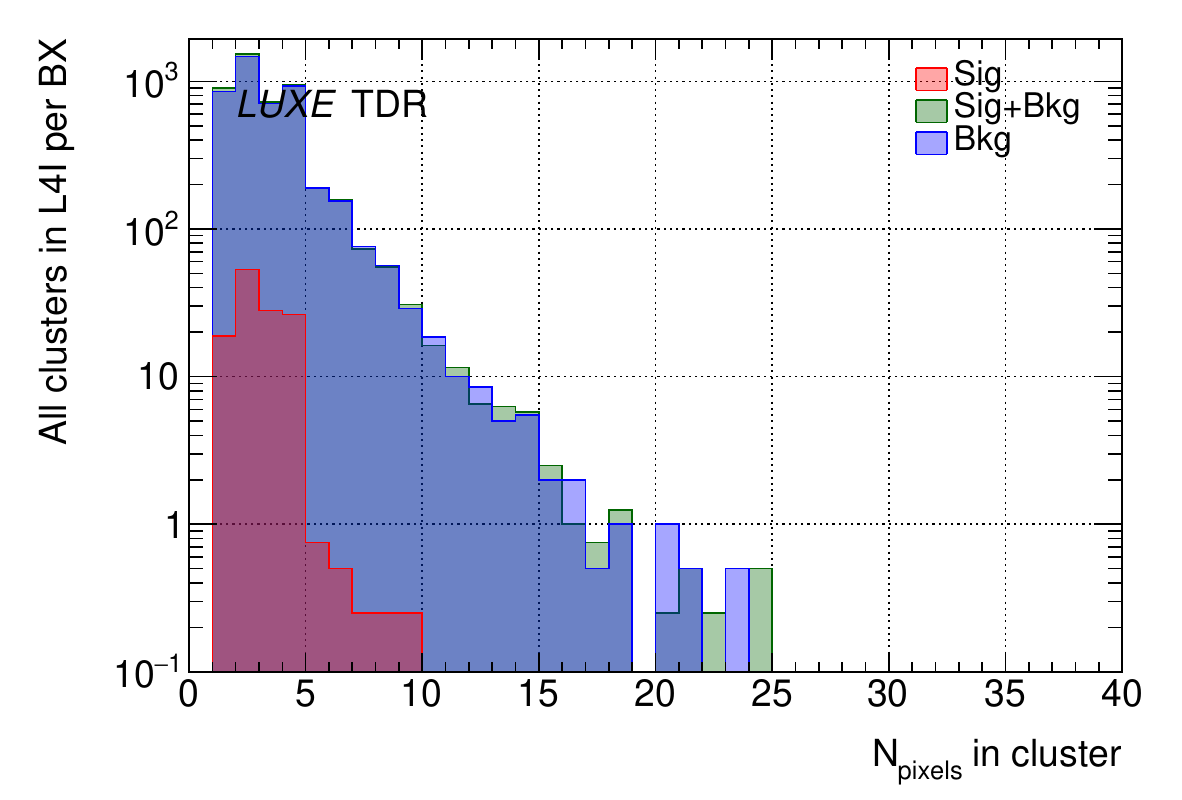}\end{overpic}
\begin{overpic}[width=0.49\textwidth]{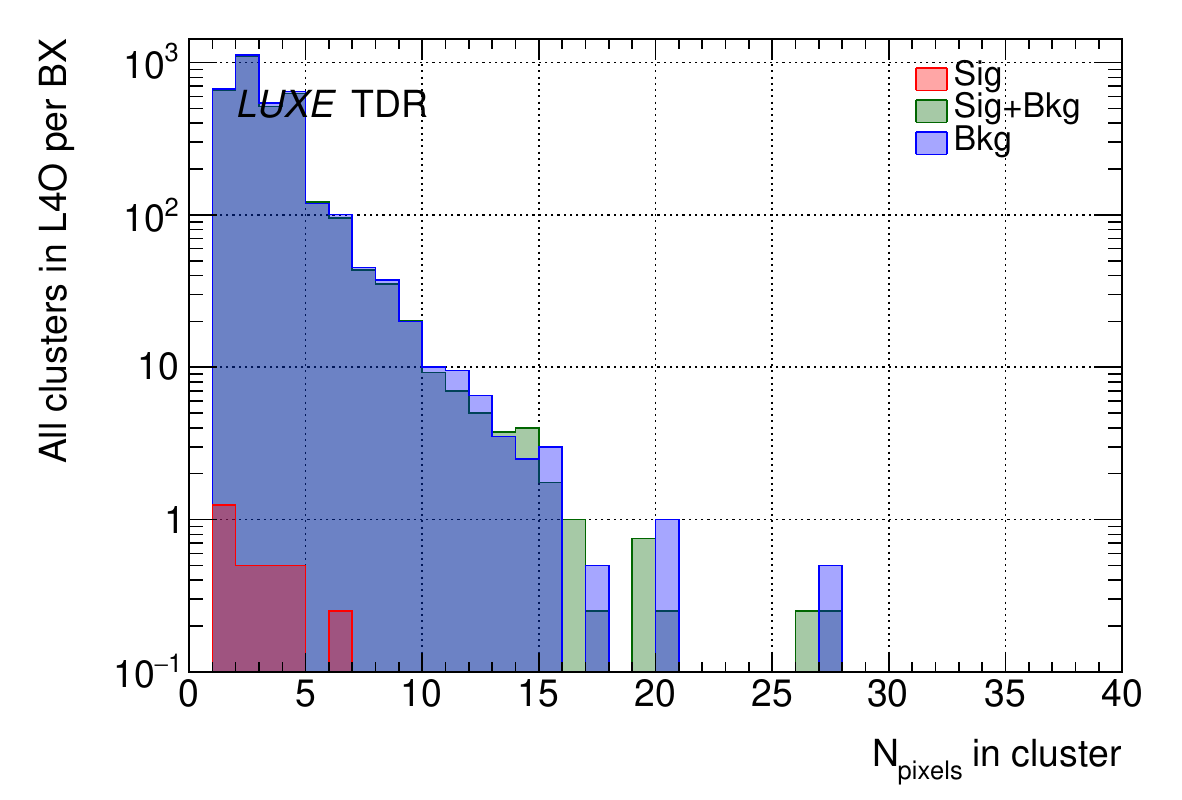}\end{overpic}
\caption{The number of pixels in a cluster from L1I (L1O) stave on the top left (right) and L4I (L4O) stave on the bottom left (right). The distributions are given for the $\xi=3$ signal, the same signal but combined with background and the background-only samples.}
\label{fig:clustersize}
\end{figure}

The truth occupancies and cluster size are shown in Figs.~\ref{fig:occxi3} and~\ref{fig:clustersize} for L4I, L4O, L1I and L1O (the behaviour in L2 and L3 can be deduced).
The distributions are drawn per pixel per BX, only for the fired pixels found in L1I/O and L4I/O. We do not filter on particle types even if more than one particle is found in the pixel. That is, all particles which have contributed some energy deposition in the pixels are accounted for. 

\begin{figure}[!ht]
\centering
\begin{overpic}[width=0.49\textwidth]{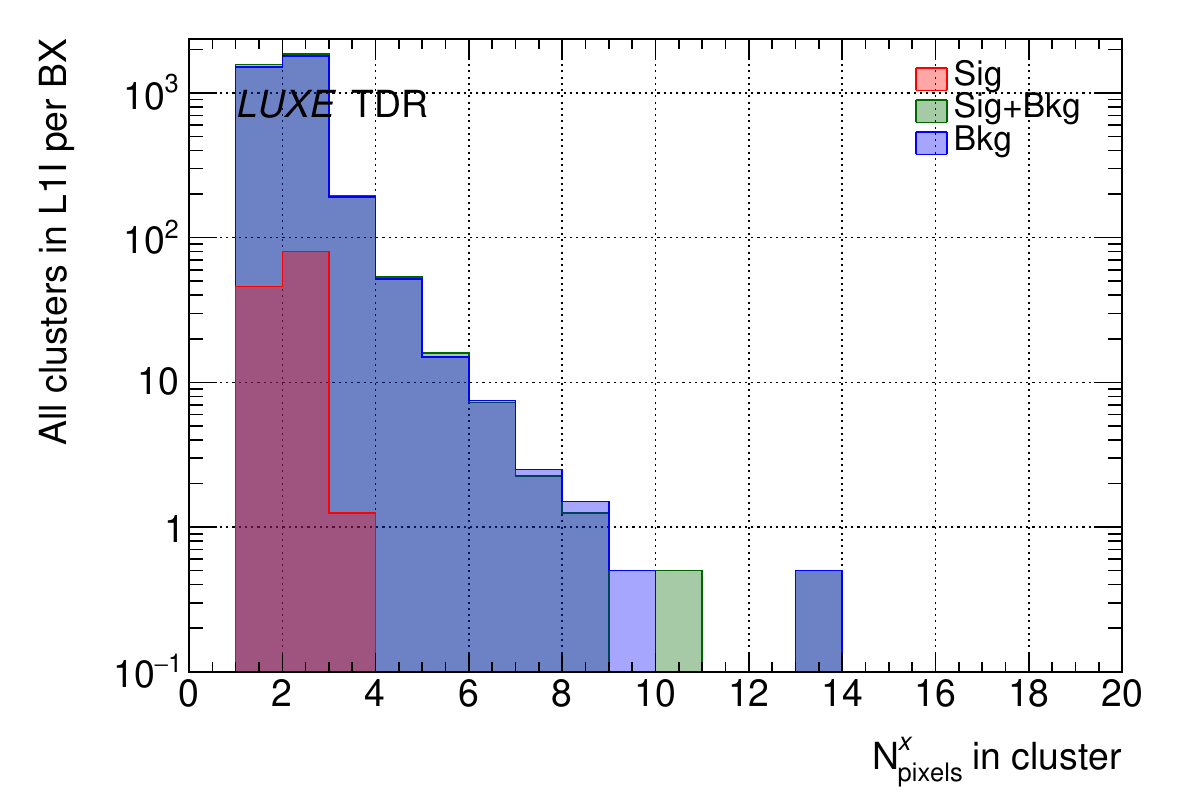}\end{overpic}
\begin{overpic}[width=0.49\textwidth]{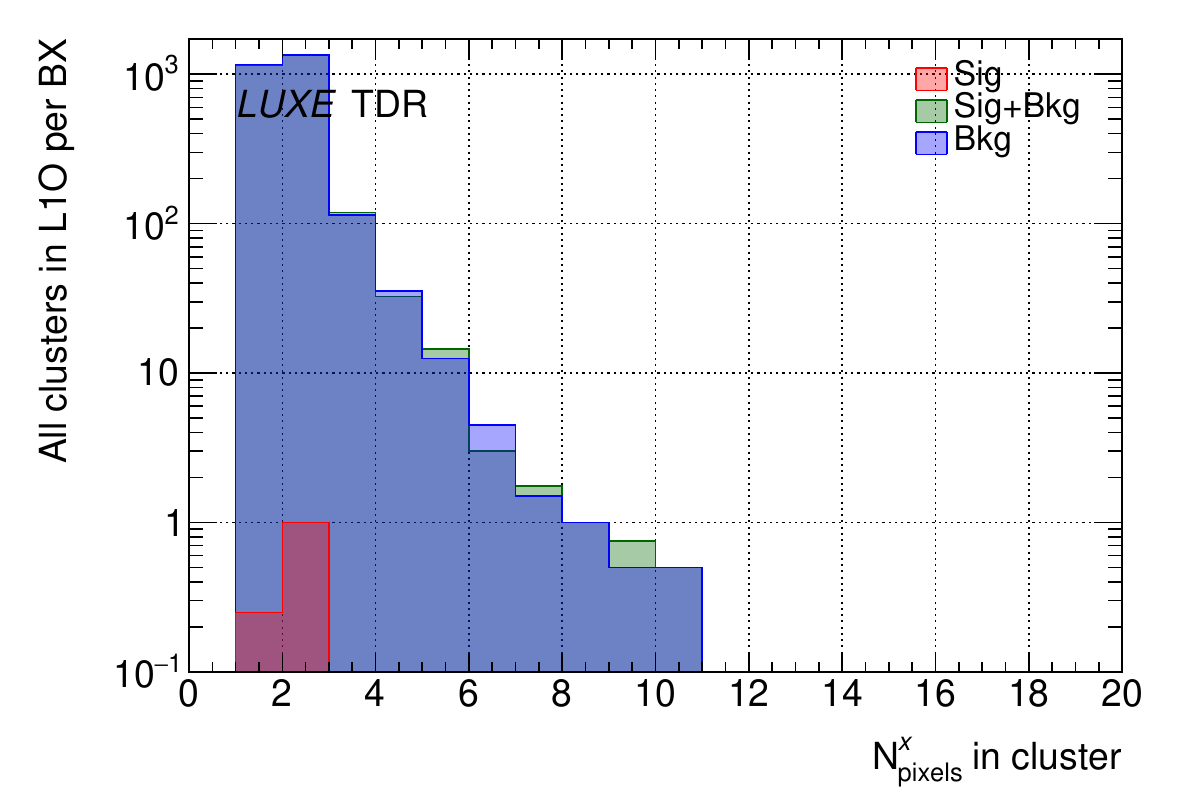}\end{overpic}
\begin{overpic}[width=0.49\textwidth]{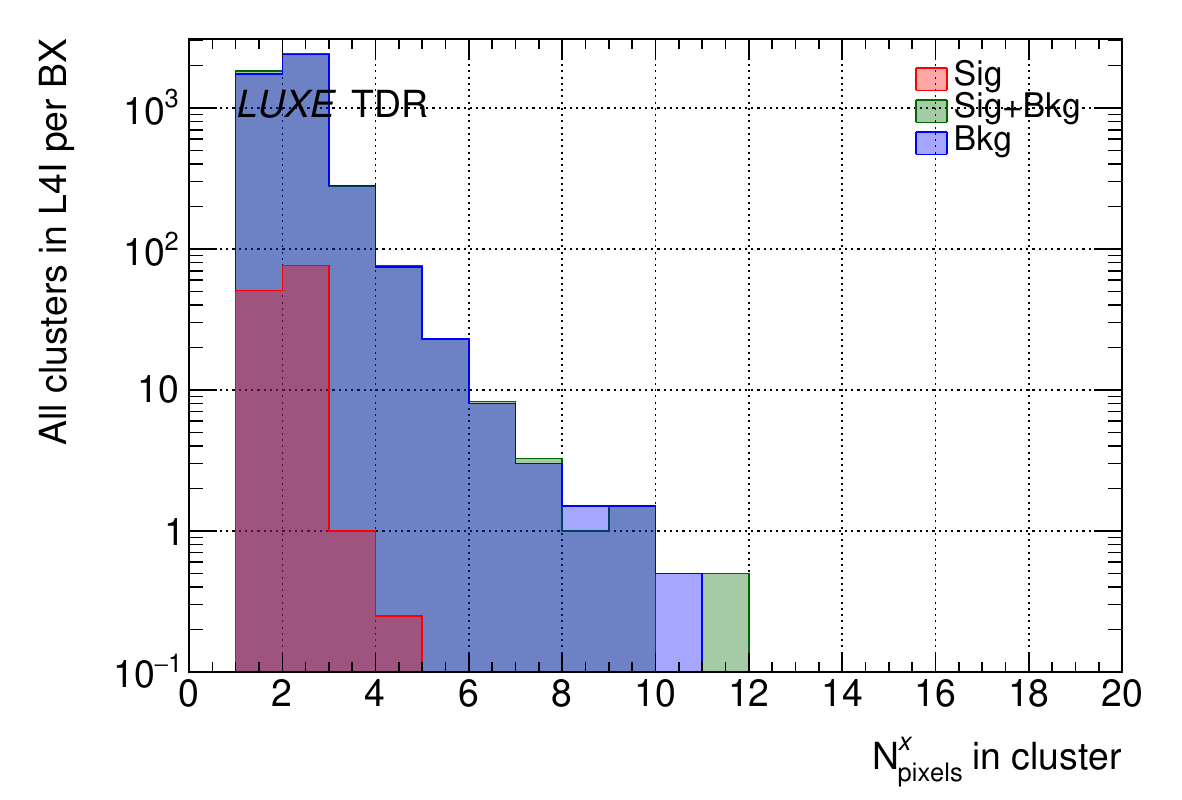}\end{overpic}
\begin{overpic}[width=0.49\textwidth]{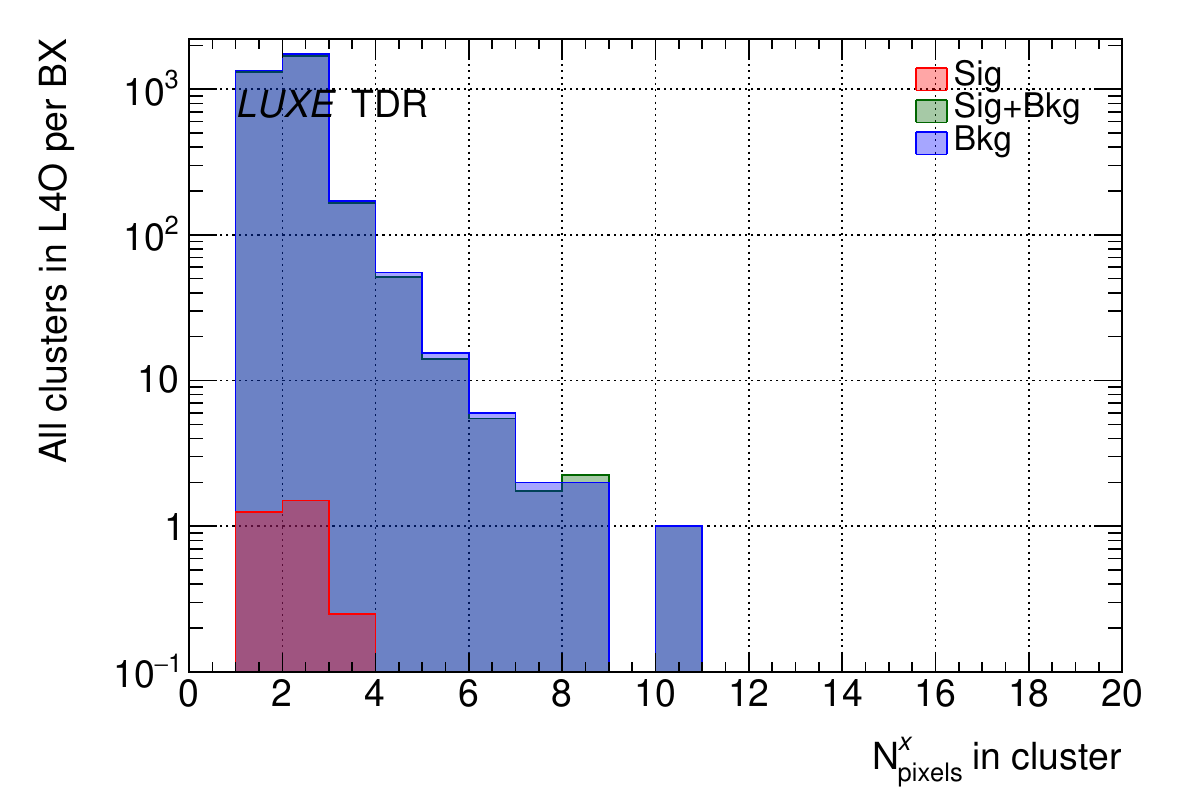}\end{overpic}
\caption{The number of pixels along $x$ in a cluster from L1I (L1O) stave on the top left (right) and L4I (L4O) stave on the bottom left (right). The distributions are given for the $\xi=3$ signal, the same signal but combined with background and the background-only samples.}
\label{fig:clustersizex}
\end{figure}

\begin{figure}[!ht]
\centering
\begin{overpic}[width=0.49\textwidth]{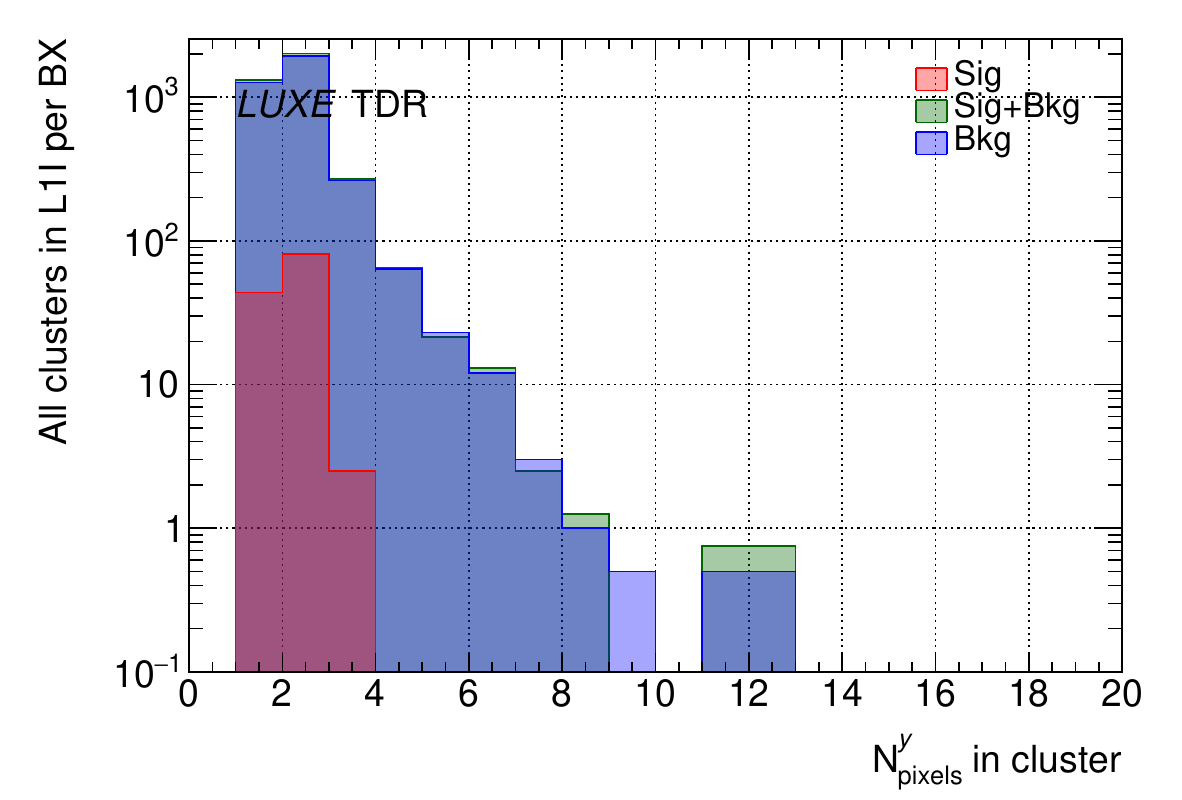}\end{overpic}
\begin{overpic}[width=0.49\textwidth]{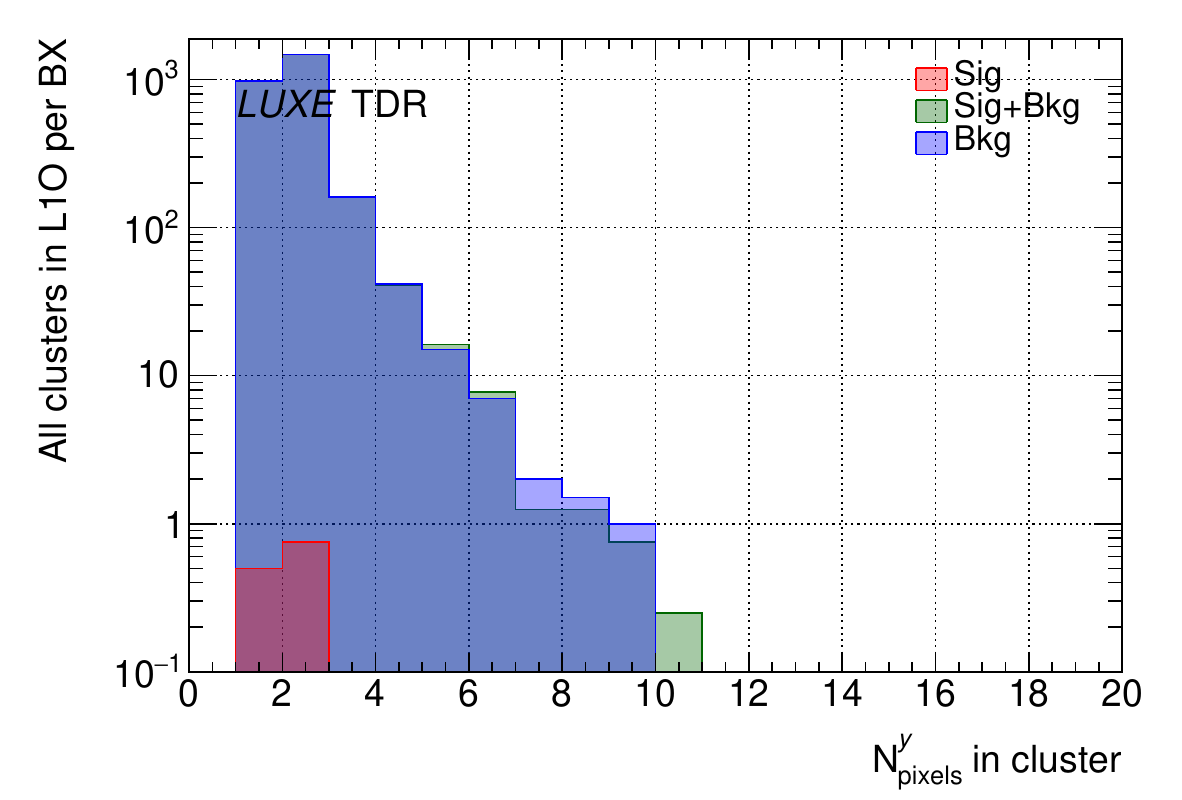}\end{overpic}
\begin{overpic}[width=0.49\textwidth]{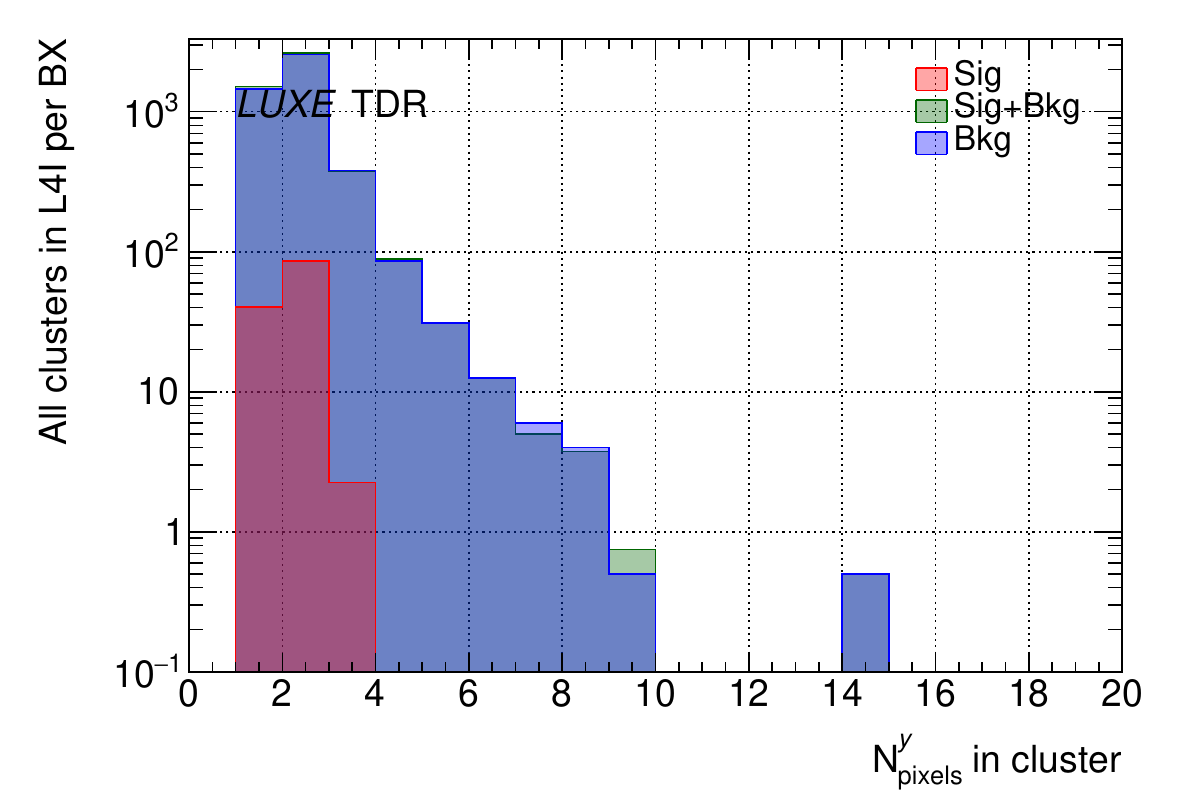}\end{overpic}
\begin{overpic}[width=0.49\textwidth]{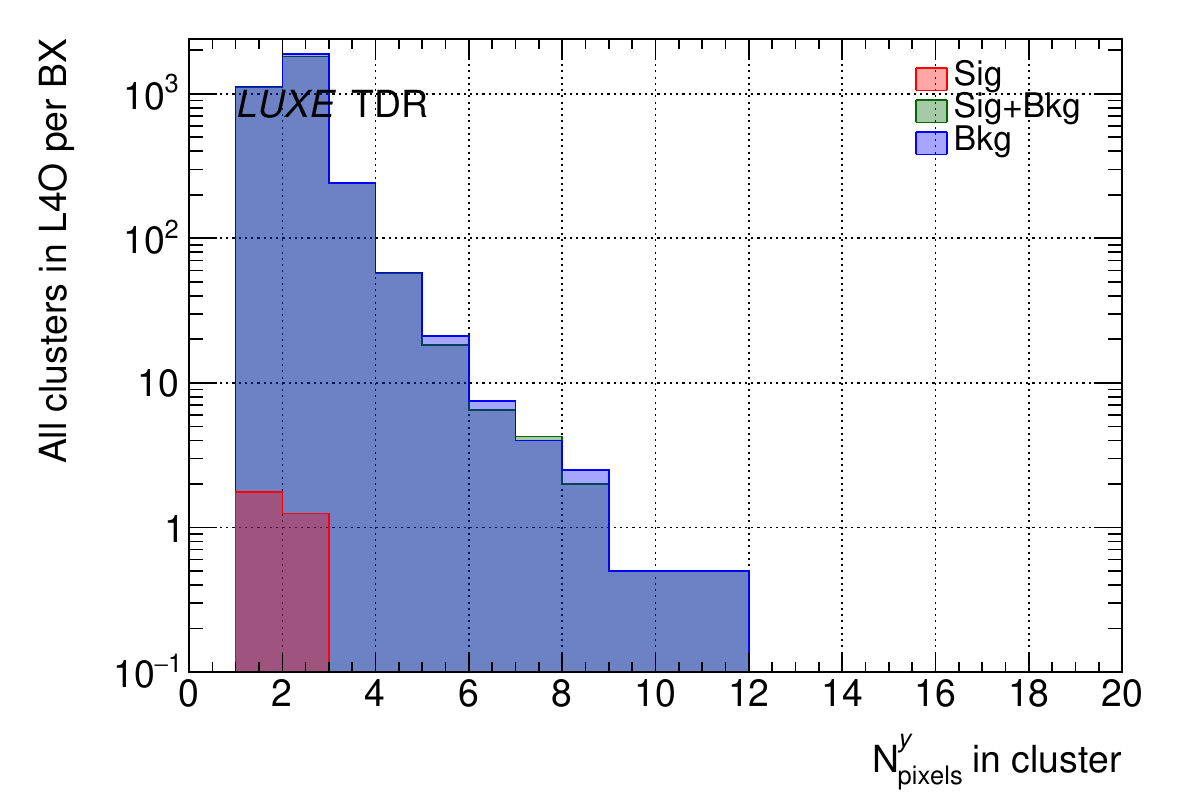}\end{overpic}
\caption{The number of pixels along $y$ in a cluster from L1I (L1O) stave on the top left (right) and L4I (L4O) stave on the bottom left (right). The distributions are given for the $\xi=3$ signal, the same signal but combined with background and the background-only samples.}
\label{fig:clustersizey}
\end{figure}

\begin{figure}[!ht]
\centering
\begin{overpic}[width=0.8\textwidth]{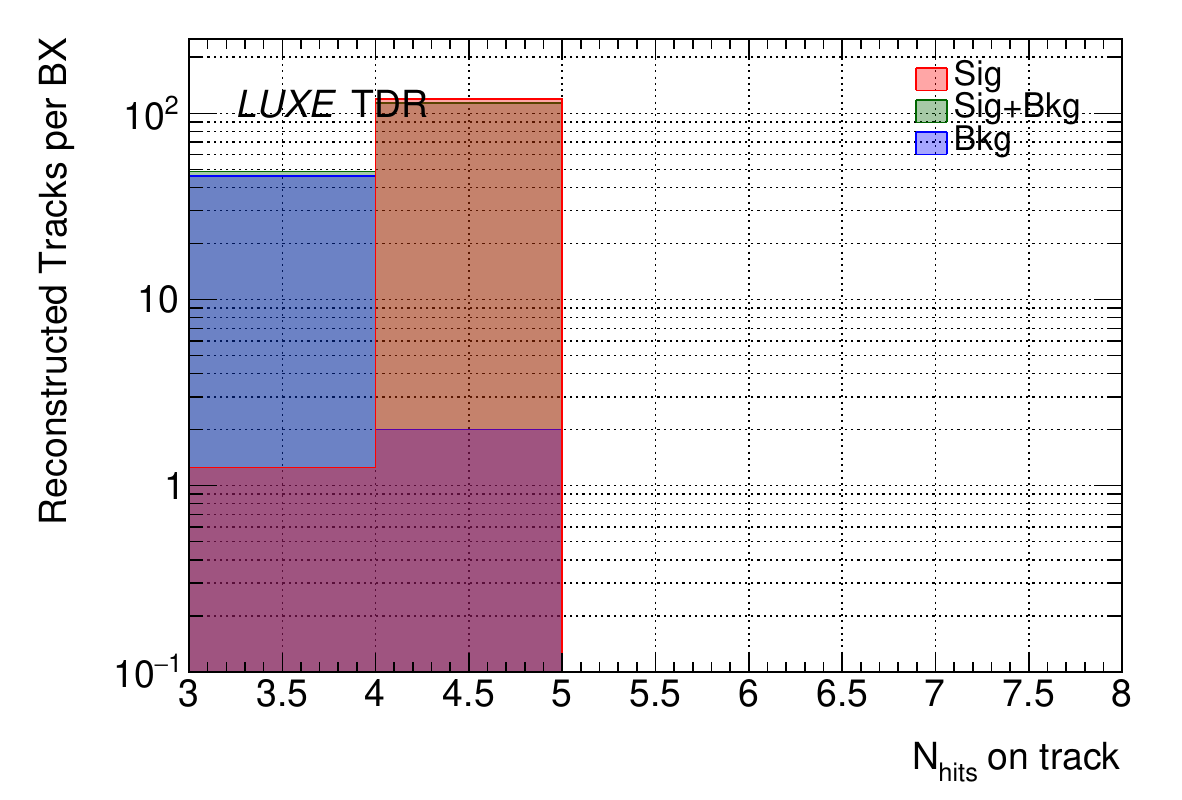}\end{overpic}
\caption{The number of clusters associated with a reconstructed track for the $\xi=3$ signal, the same signal but combined with background, and the background-only samples.}
\label{fig:nhits}
\end{figure}

\begin{figure}[!ht]
\centering
\begin{overpic}[width=0.8\textwidth]{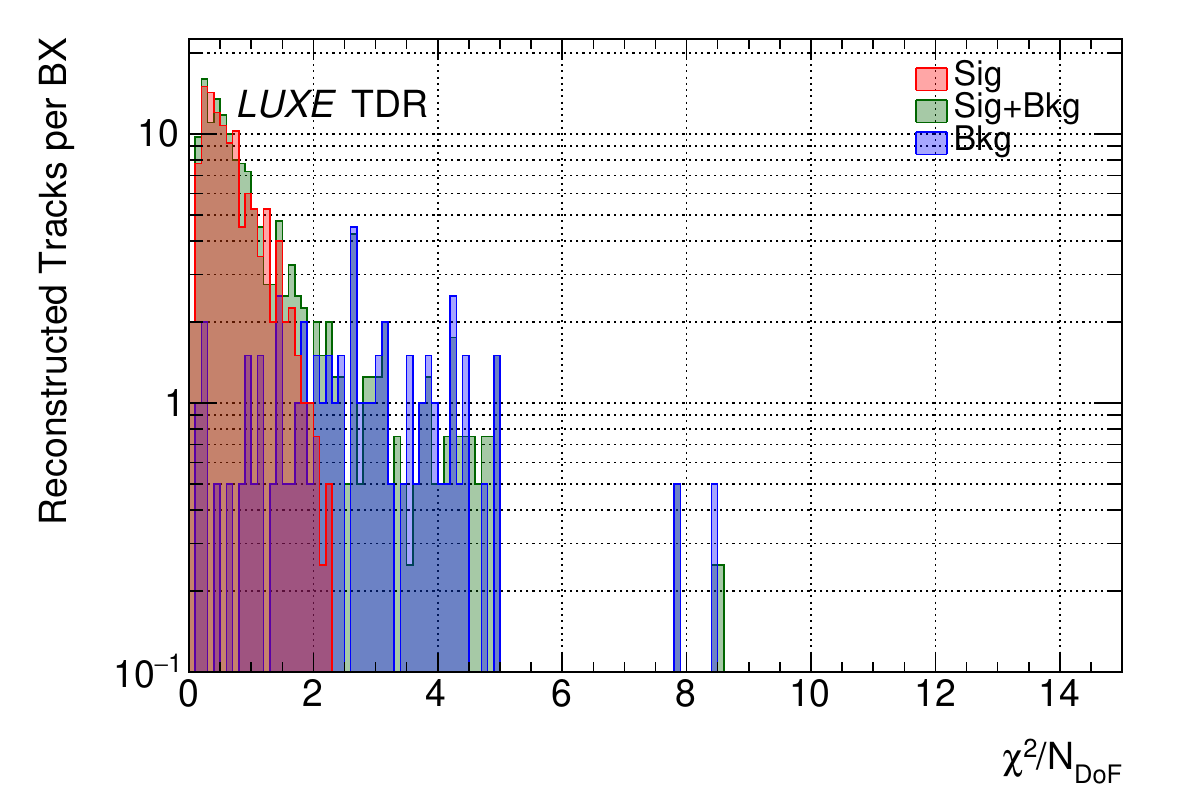}\end{overpic}
\caption{The $\chi^2/N_{\textrm{DOF}}$ of the reconstructed tracks. The distributions are given for the $\xi=3$ signal, the same signal but combined with background, and the background-only samples.}
\label{fig:trkchi2dof}
\end{figure}

Similarly, we also show the cluster size along the $x$ and $y$ axes for the same staves in Fig.~\ref{fig:clustersizex} and Fig.~\ref{fig:clustersizey}, respectively.
The number of clusters on track is shown in Fig.~\ref{fig:nhits}.
The $\chi^2/{\textrm{DoF}}$ of the winner tracks is shown in Fig.~\ref{fig:trkchi2dof}.

\begin{figure}[!ht]
\centering
\begin{overpic}[width=0.49\textwidth]{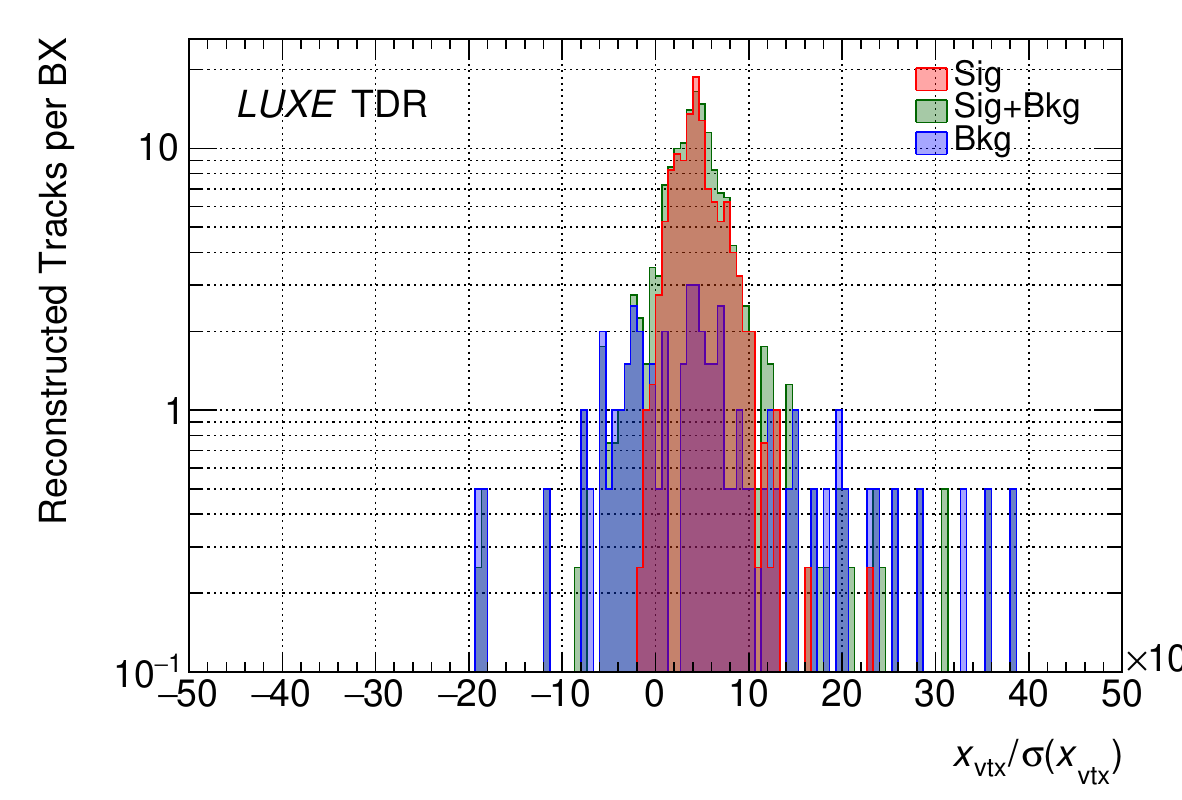}\end{overpic}
\begin{overpic}[width=0.49\textwidth]{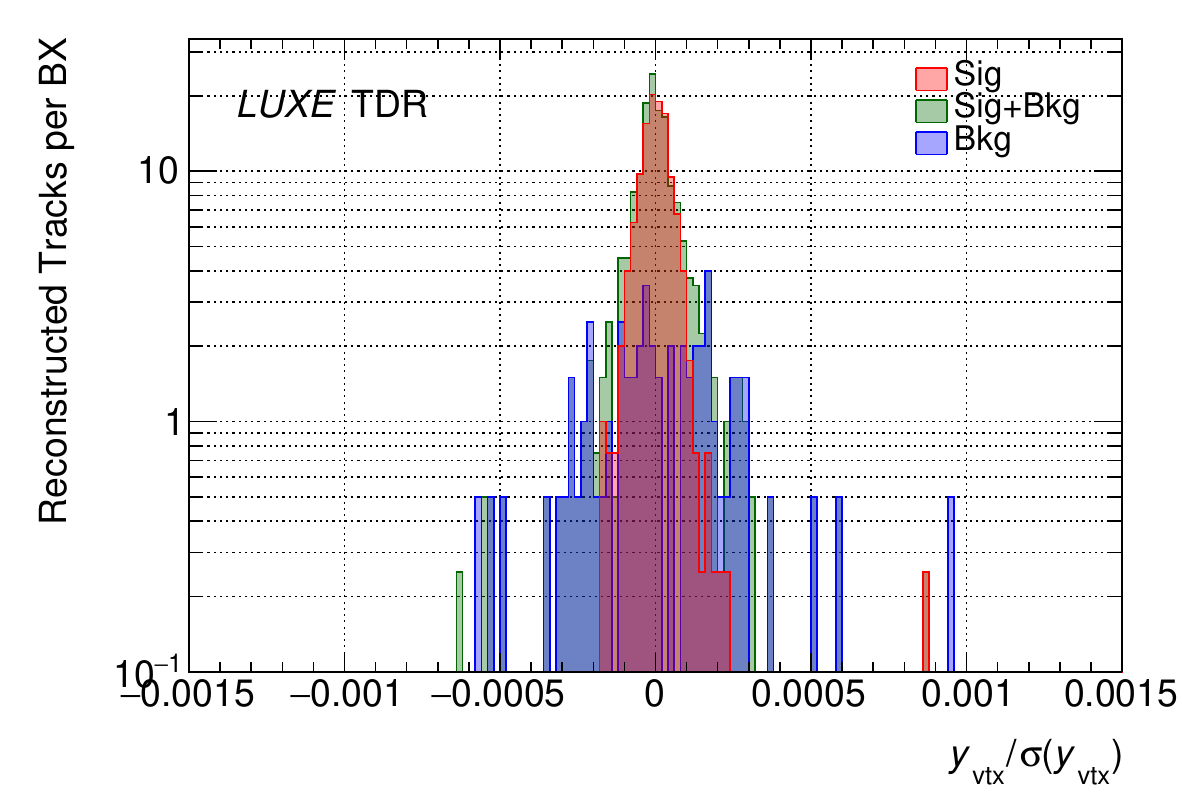}\end{overpic}
\caption{The significance of vertex position of the reconstructed tracks along $x$ (left) and along $y$ (right). The distributions are given for the $\xi=3$ signal, the same signal but combined with background and the background-only samples.}
\label{fig:vertexsig}
\end{figure}

\begin{figure}[!ht]
\centering
\begin{overpic}[width=0.49\textwidth]{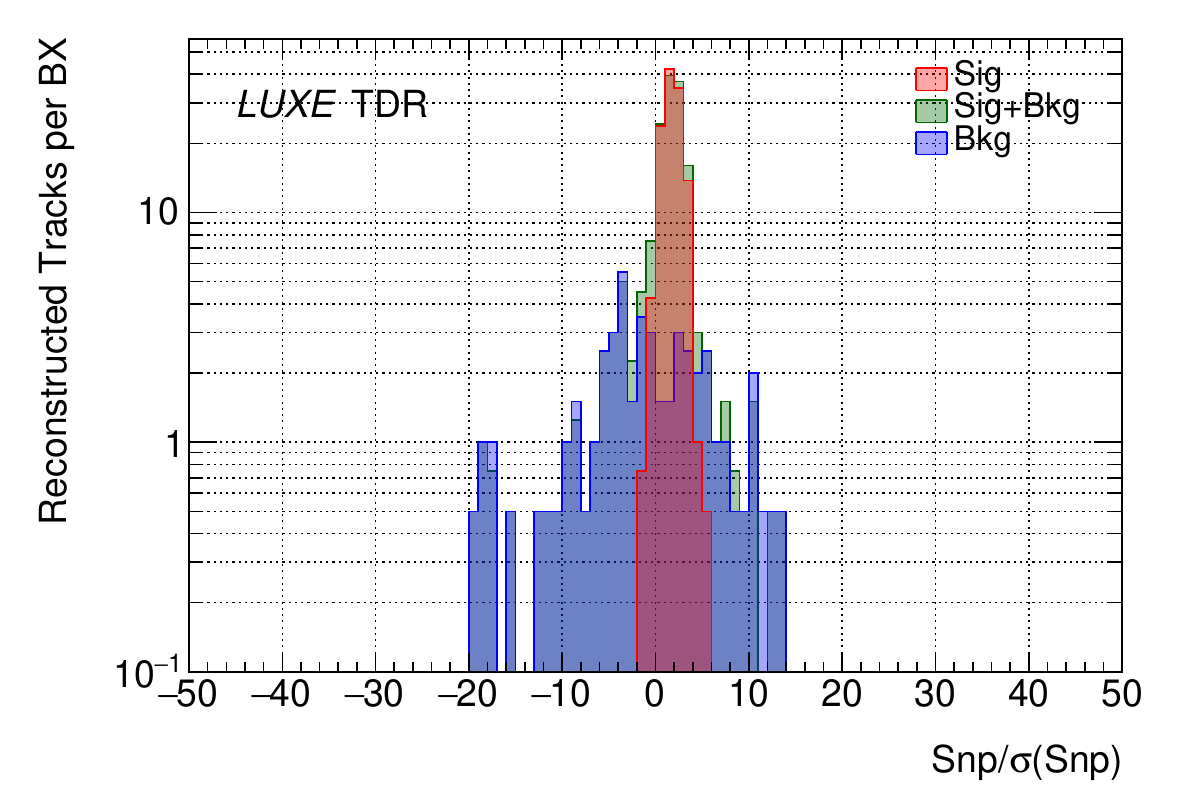}\end{overpic}
\begin{overpic}[width=0.49\textwidth]{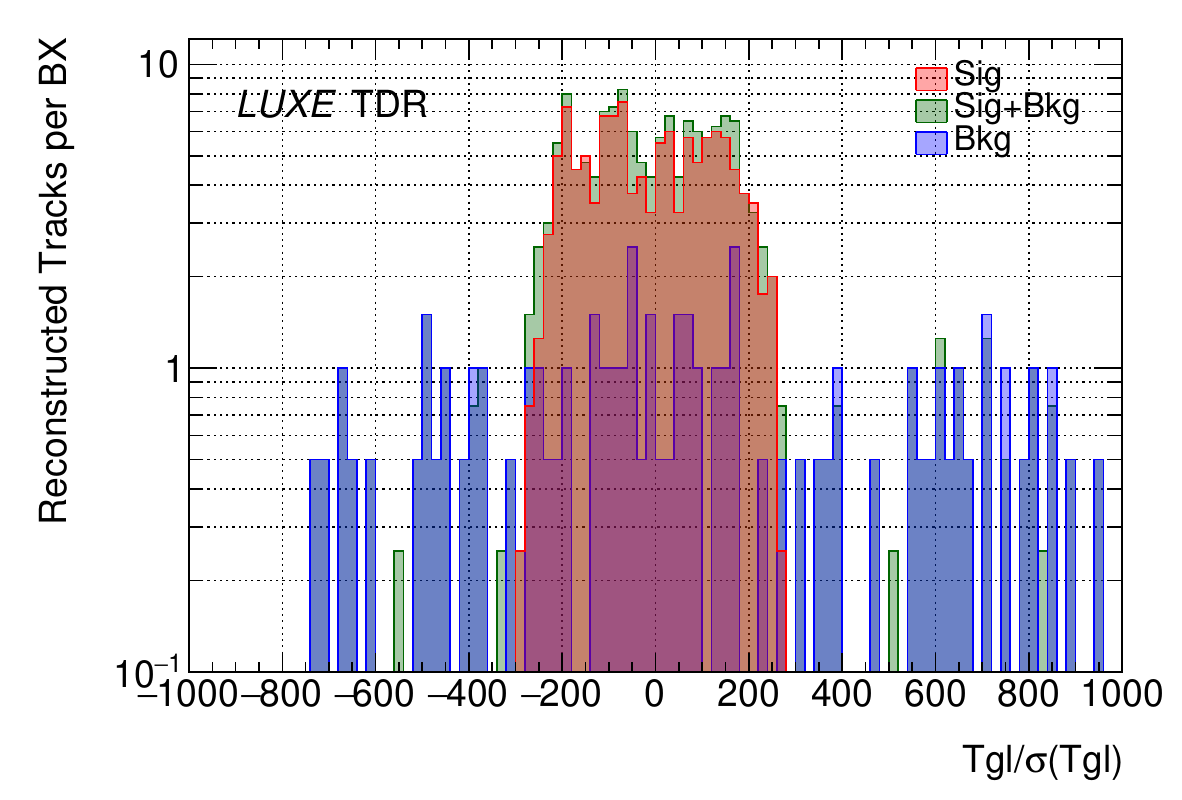}\end{overpic}
\caption{The significances of Snp (left) and Tgl (right) as defined in the text. The distributions are given for the $\xi=3$ signal, the same signal but combined with background and the background-only samples.}
\label{fig:snptgl}
\end{figure}

\begin{figure}[!ht]
\centering
\begin{overpic}[width=0.49\textwidth]{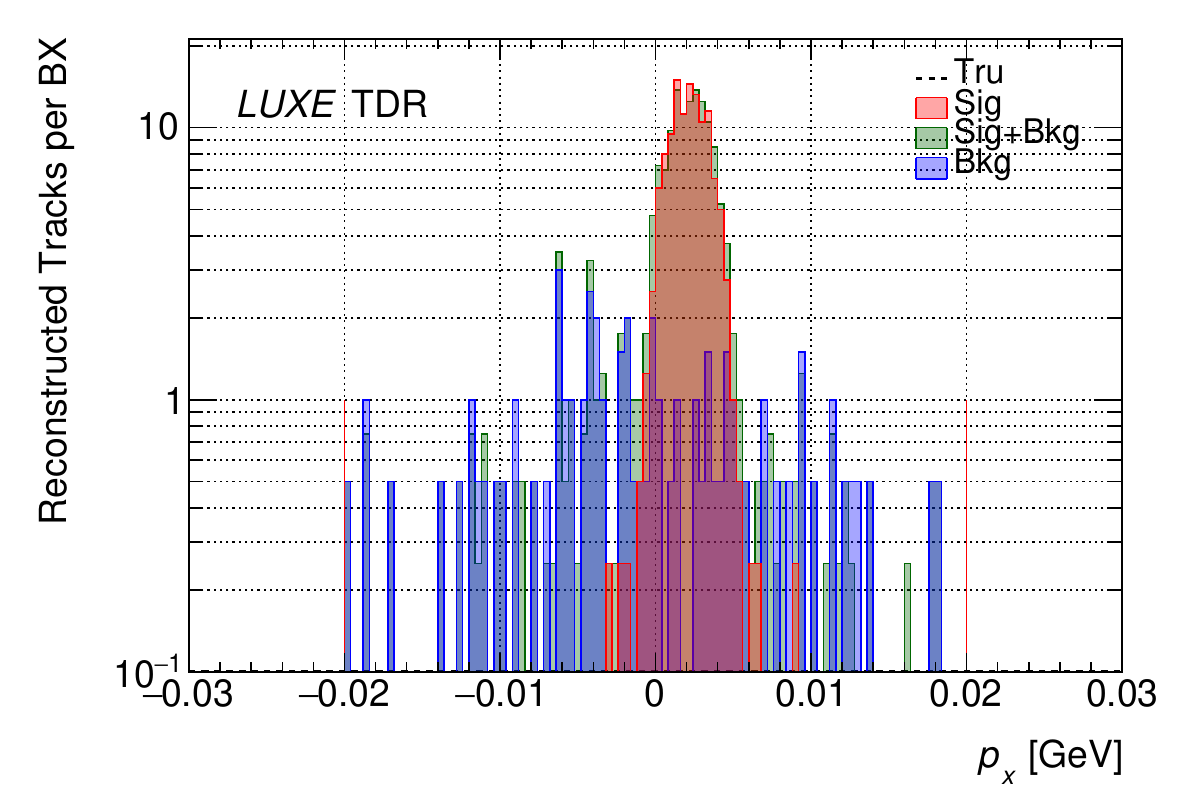}\end{overpic}
\begin{overpic}[width=0.49\textwidth]{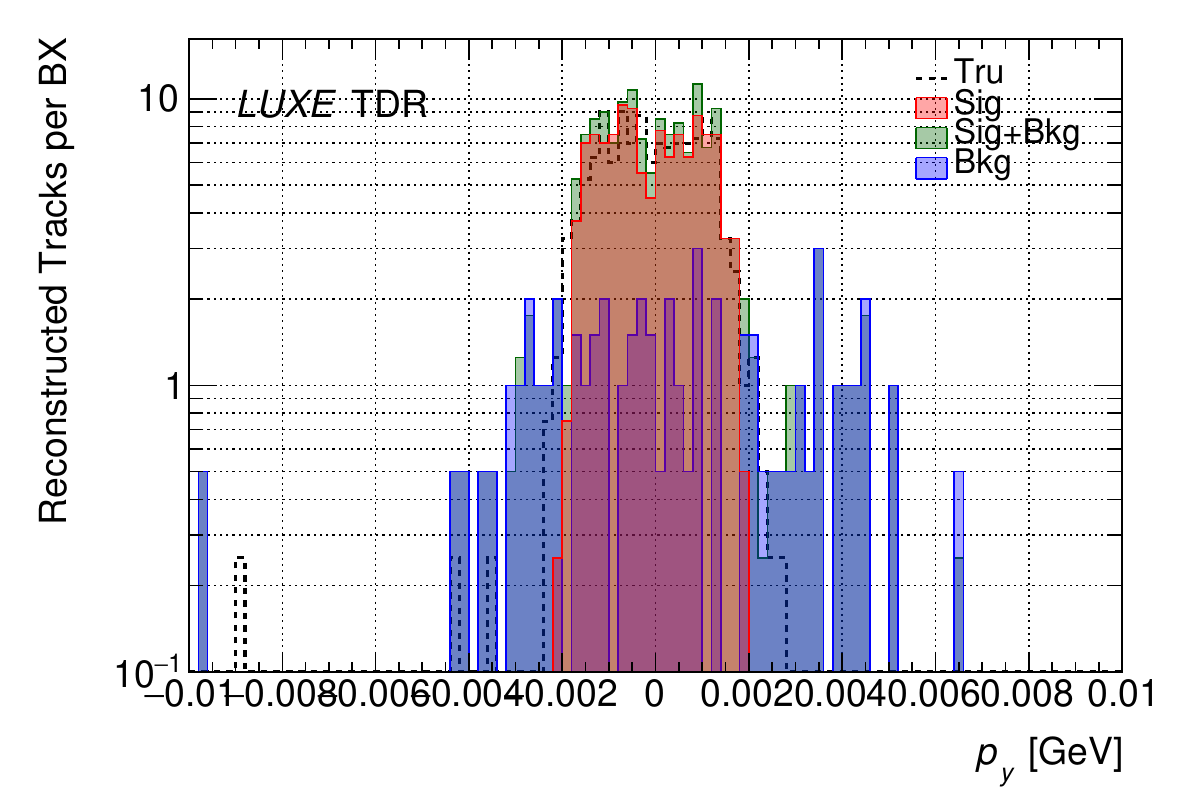}\end{overpic}
\begin{overpic}[width=0.49\textwidth]{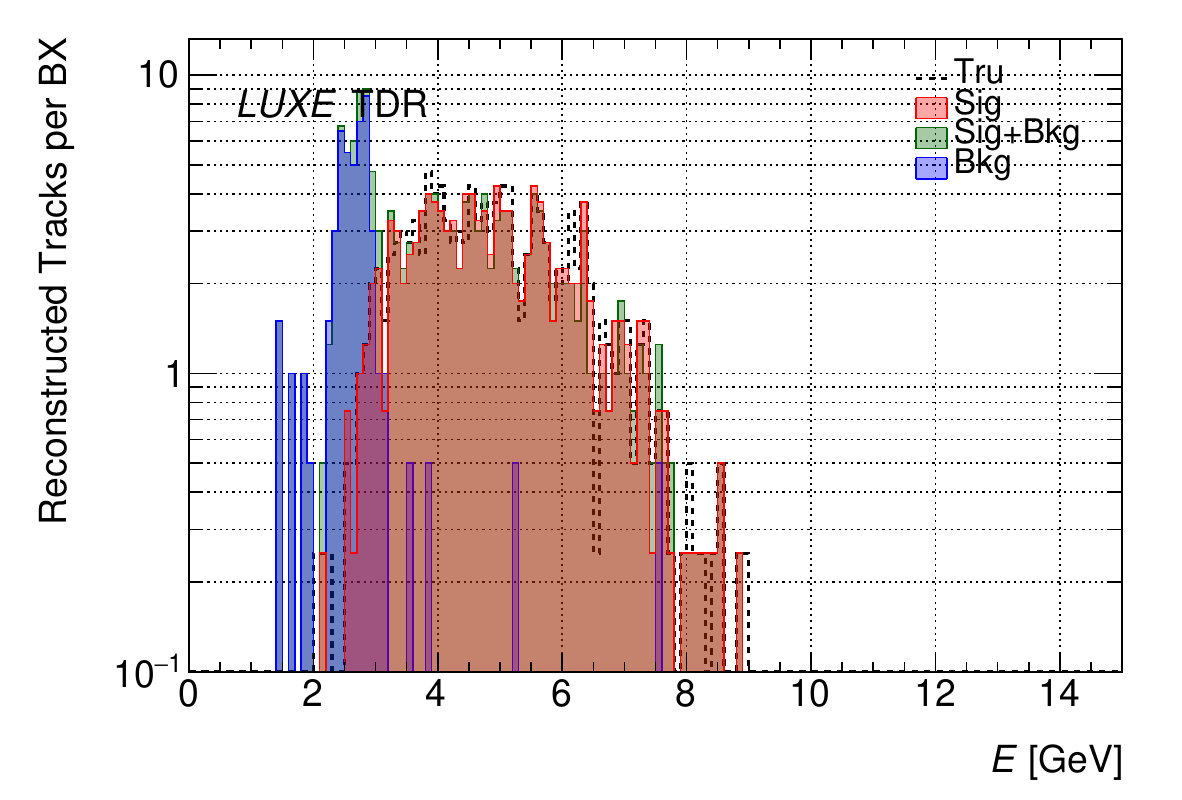}\end{overpic}
\caption{The reconstructed track momentum $p_x$ (top left), $p_y$ (top right) and the energy (bottom). Note that the truth distributions are given at the detector rather than at the IP and hence the truth $p_x$ of the particles is not distributed around zero but rather around 0.34~GeV, outside of the plot's range.
The distributions are given for the $\xi=3$ signal, the same signal but combined with background and the background-only samples.}
\label{fig:trkpyE}
\end{figure}

The track's vertex significances in $x$ and $y$ are shown in Fig.~\ref{fig:vertexsig}.
Similarly, Fig.~\ref{fig:snptgl} shows the significances in ${\textrm{Snp}}\equiv\sin\phi^{\textrm{trk}}$ and ${\textrm{Tgl}}\equiv\textstyle{\frac{p_z^{\textrm{ trk}}}{p_{\textrm{T}}^{\textrm{trk}}}}$ (the tangent of the angle with respect to the zenith), where $\phi^{\textrm{trk}}$ is the azimuthal angle in the track's frame, $p_{\textrm{T}}^{\textrm{trk}}$ is its transverse momentum in the track frame.
The transformation between the lab and track frames of some vector $\vec{v}$ is given by $v_x^{\textrm{lab}}=v_y^{\textrm{trk}}$, $v_y^{\textrm{lab}}=v_z^{\textrm{trk}}$ and $v_z^{\textrm{lab}}=v_x^{\textrm{trk}}$.
Finally, the reconstructed $p_x$, $p_y$ momenta at the IP and the energy spectra can be seen in Fig.~\ref{fig:trkpyE}.


To quantify better the quality (efficiency and resolution) of the reconstruction, the matching between the reconstructed tracks and the truth tracks must be first defined.
In this work we have two matching schemes:
\begin{itemize}
    \item {\normalfont 
    \itshape  Full matching}: requiring that all clusters participating in the track fit are associated with the same truth signal particle,
    \item {\normalfont \itshape Partial matching}: requiring that at least the pivot cluster at L4 is matched with a truth signal particle.
\end{itemize}
As is expected, the matching performance depends strongly on the truth signal particle multiplicity.
This is particularly true for the {\normalfont \itshape full matching} since the large density of clusters/tracks at the peak of the energy distribution leads to increasing ``confusion'' such that selecting at least one wrong cluster is very likely. 
In this case, the {\normalfont \itshape full matching} would return a false result, but nevertheless the fitted track may still have properties very similar to those of the truth particle. 

\begin{figure}[!ht]
\centering
\begin{overpic}[width=0.49\textwidth]{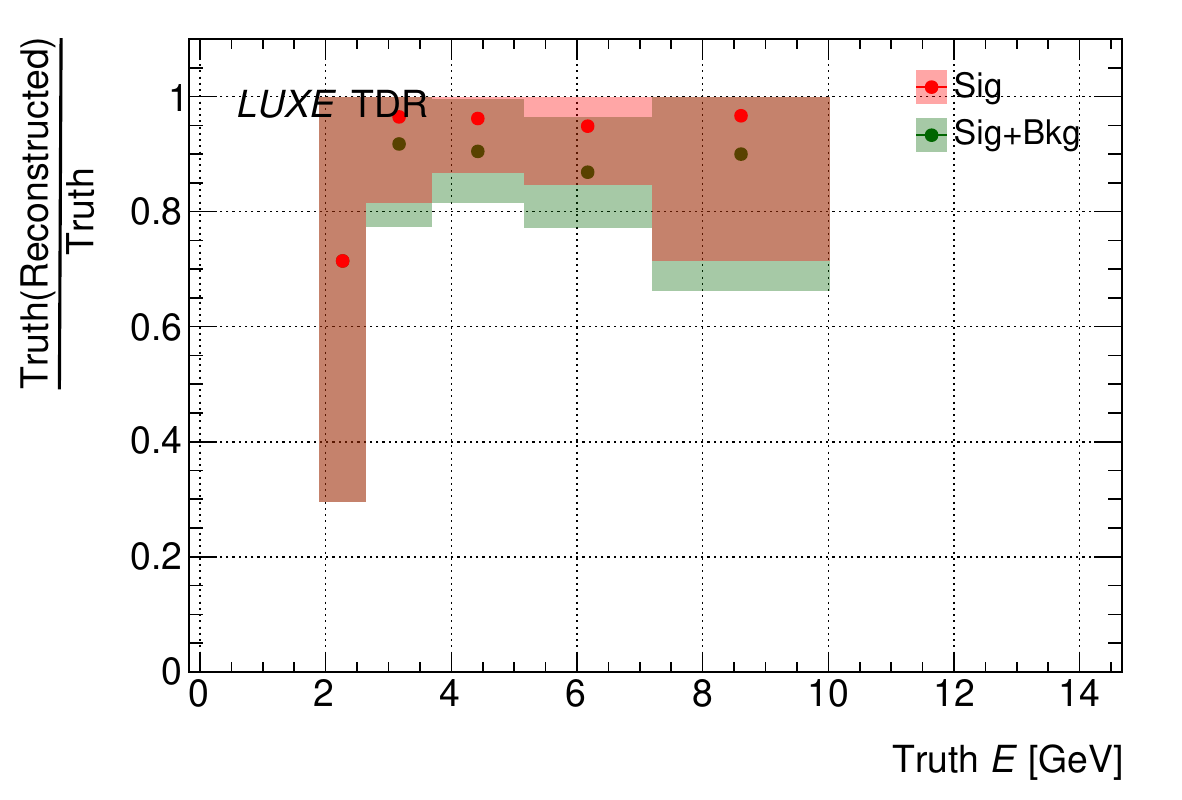}\end{overpic}
\begin{overpic}[width=0.49\textwidth]{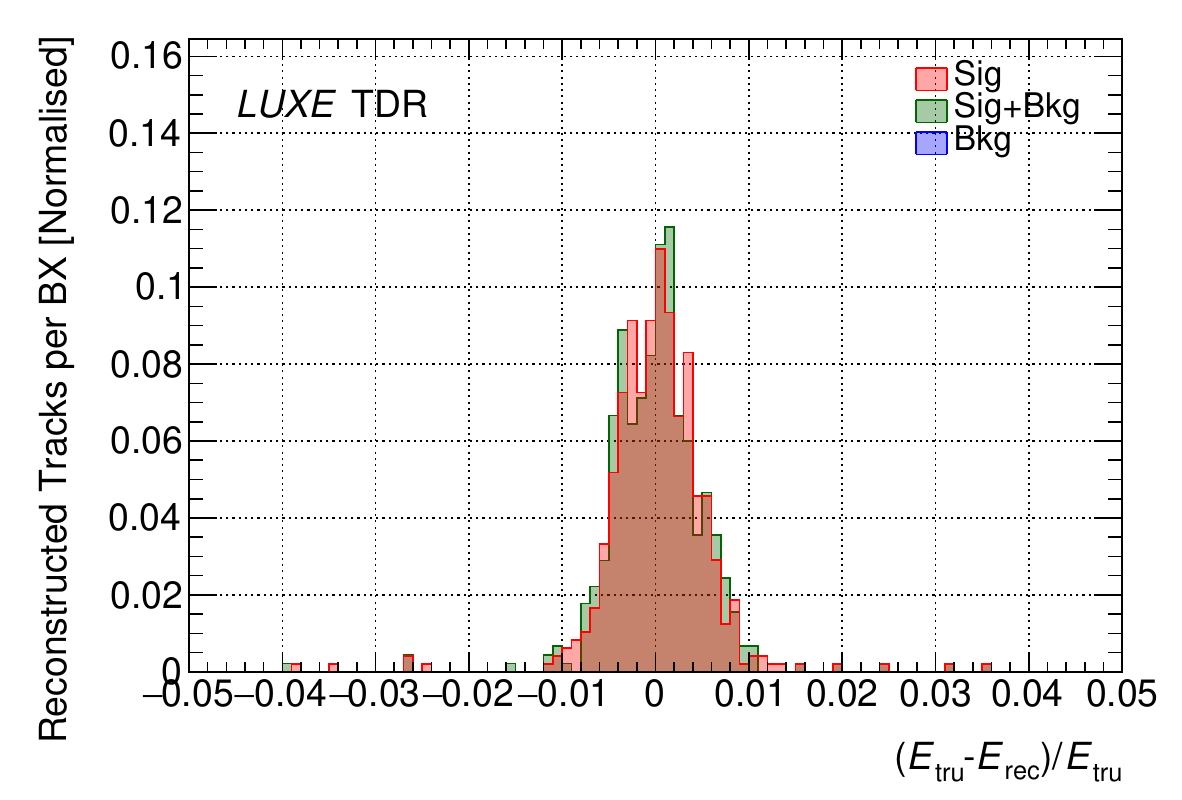}\end{overpic}
\caption{The reconstruction efficiency as a function of true energy (left) and the respective energy resolution (right) using the {\normalfont \itshape partial matching} scheme.
The distributions are given for the $\xi=3$ signal and the same signal but combined with background samples.}
\label{fig:effresE}
\end{figure}

\begin{figure}[!ht]
\centering
\begin{overpic}[width=0.49\textwidth]{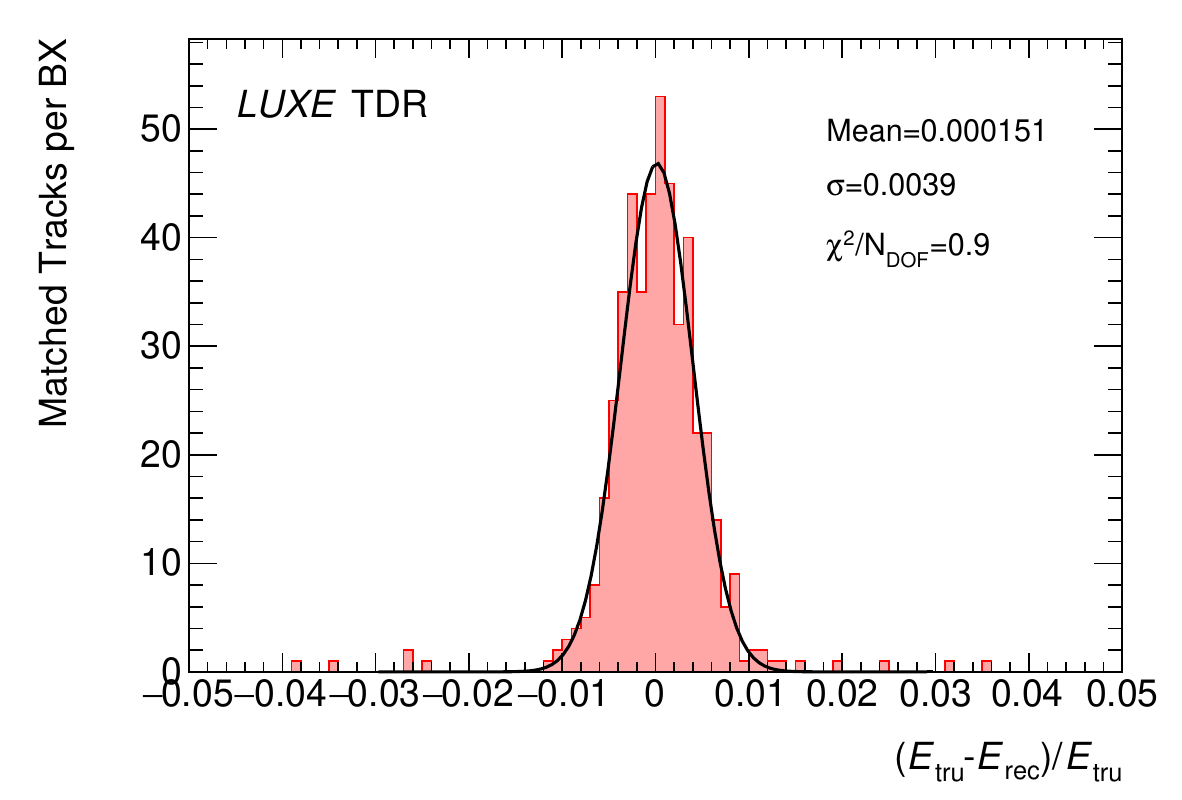}\end{overpic}
\begin{overpic}[width=0.49\textwidth]{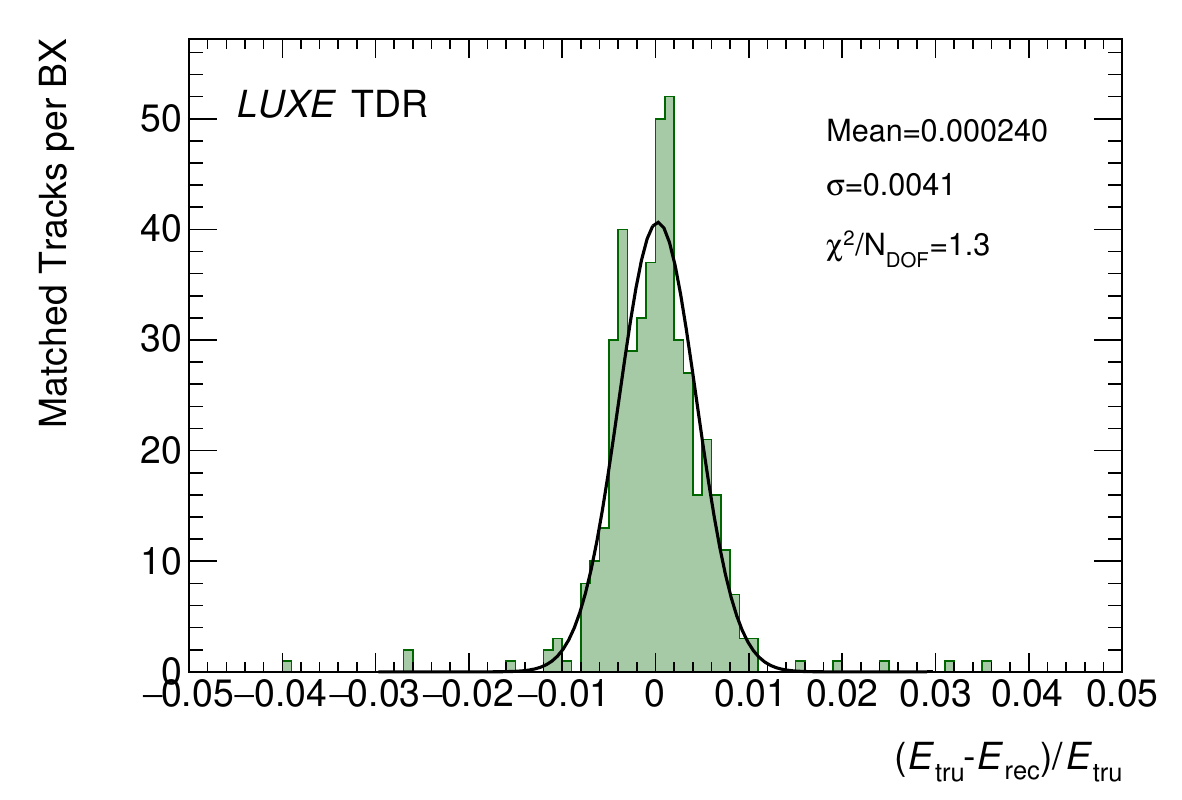}\end{overpic}
\caption{The energy resolution from signal (left) and signal + background (right) using the {\normalfont \itshape full matching} scheme.
The fits to a Gaussian are also shown.
The distributions are given for the $\xi=3$ signal and the same signal but combined with background samples.}
\label{fig:fitres}
\end{figure}

The efficiency in energy and the energy response distribution, ($\textstyle{\frac{E_{\textrm{tru}}-E_{\textrm{rec}}}{E_{\textrm{tru}}}}$), of all reconstructed tracks is shown in Fig.~\ref{fig:effresE}.
The fit of the energy response distributions for signal-only and signal + background samples is given in Fig.~\ref{fig:fitres}. These plots are using the {\normalfont \itshape partial matching} scheme.

It can be seen that the algorithm is sufficiently efficient and accurate in reconstructing the truth shapes even without imposing any special post-processing requirement.
This is evident from Fig.~\ref{fig:effresE} but also from the other figures, where the truth information can be overlaid on the reconstructed distributions.

Specifically, despite the low statistics of the $\xi=3$ sample, the efficiency is consistent with a plateau of $\sim 90-95\%$.
These efficiency curves can be later used to unfold the reconstructed energy distribution, taking into account corrections for mismodelling and uncertainties from the simulation.

The Gaussian fits to the response distributions shown in Fig.~\ref{fig:fitres} yield $\mu_E\simeq 0$ and $\sigma_E\leq 1\%$ indicating that the tracks' momenta are well matching those of the truth signal particles.
This result is consistent with runs having larger truth particle multiplicities, but unlike the efficiency, it does not depend strongly on the algorithm's tuning.

Moreover, it can be seen that in this specific case ($\xi=3$ with $\sim 125$ truth signal particles per BX), the background is very low already at the reconstruction level.
It is also clear that the residual background after reconstruction is well separable from the signal with a set of trivial post-processing selection cuts as discussed below.
It can be assumed that this would be the case for all signal track multiplicities below $\sim 125$. 

\begin{figure}[!ht]
\centering
\begin{overpic}[width=0.9\textwidth]{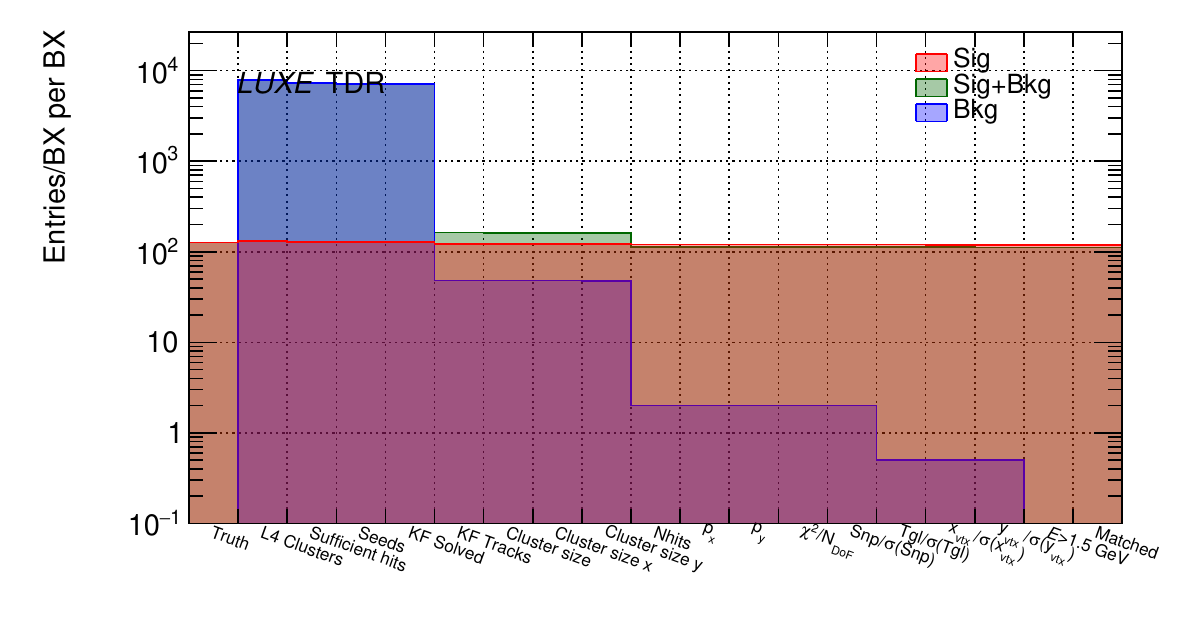}\end{overpic}
\caption{The cut-flow for the background, $\xi=3$ signal and the same signal but combined with background samples.}
\label{fig:cutflowxi3}
\end{figure}

A preliminary set of selection criteria can be applied on the reconstructed tracks' characteristics discussed and shown.
One example of the cut values used for the $\xi=3$ typical multiplicity is:
\begin{itemize}
    \item Cluster size: $N_{\textrm{pix}}<10$, $N_{\textrm{pix}}^x<5$ and $N_{\textrm{pix}}^y<5$,
    \item Minimum number of clusters on track: $N_{\textrm{clusters}}\geq 4$,
    \item Transverse momenta at the IP: $-0.003<p_x<0.008$~GeV and $-0.025<p_y<0.025$~GeV,
    \item Fit quality: $\chi^2/{\textrm{DoF}}<3$,
    \item Angles' significances: $-3<{\textrm{Snp}}/\sigma({\textrm{Snp}})<8$ and $-350<{\textrm{Tgl}}/\sigma({\textrm{Tgl}})<350$,
    \item Vertex significances: $(-3<x_{\textrm{vtx}}/\sigma(x_{\textrm{vtx}})<18)\times 10^{-9}$ and $(-2.5<y_{\textrm{vtx}}/\sigma(y_{\textrm{vtx}})<2.5)\times 10^{-4}$,
    \item Minimum energy: $E>1.5$~GeV (this corresponds to the very outskirts of the detector).
\end{itemize}
It is noted that this preliminary list of cuts is neither comprehensive nor optimised.
Nevertheless, it performs particularly well (in terms of obtaining high signal efficiency as well as achieving high background rejection) at the low multiplicity cases as discussed below.
Using this selection for the specific sample ($\xi=3$), the efficiency vs energy is effectively identical to the one given after reconstruction without selection (see left plot of Fig.~\ref{fig:effresE}). 
The respective cut-flow for the same sample is shown in Fig.~\ref{fig:cutflowxi3}. 

\begin{figure}[!ht]
\centering
\begin{overpic}[width=0.49\textwidth]{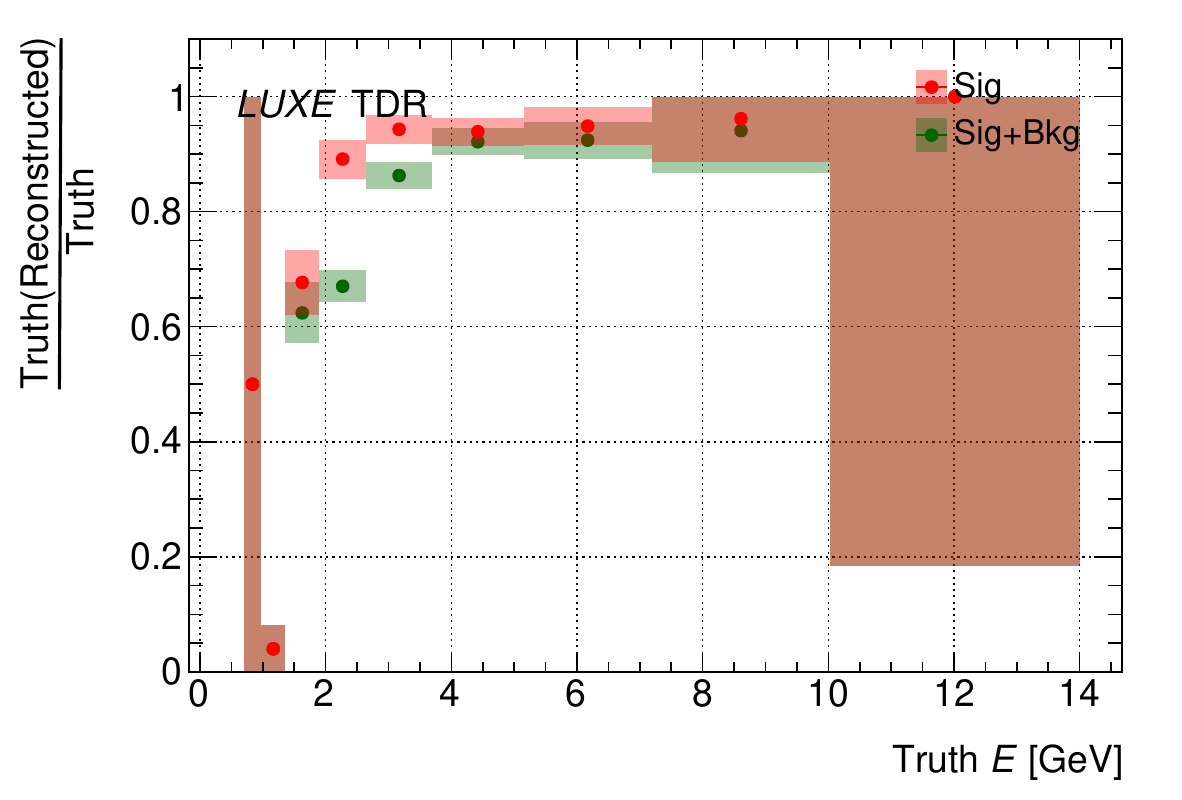}\end{overpic}
\begin{overpic}[width=0.49\textwidth]{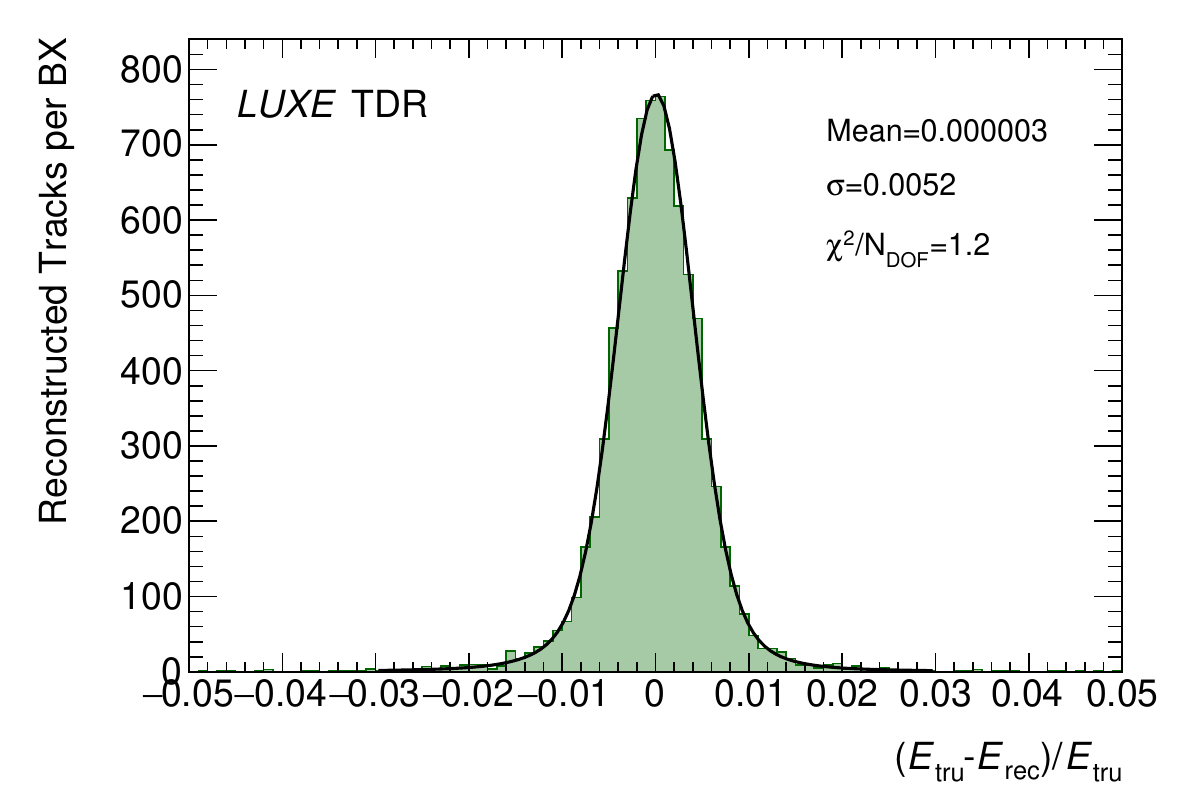}\end{overpic}
\caption{The reconstruction efficiency (left) vs the truth energy and the energy resolution of selected tracks (right) in the signal + background case alone using the {\normalfont \itshape partial matching} scheme.
The fit to a Gaussian are also shown.
The signal sample here uses 1,000 true signal particles drawn randomly from the $\xi=7$ sample per BX once for the signal sample and once combined with the background.}
\label{fig:effresE_01000}
\end{figure}

\begin{figure}[!ht]
\centering
\begin{overpic}[width=0.49\textwidth]{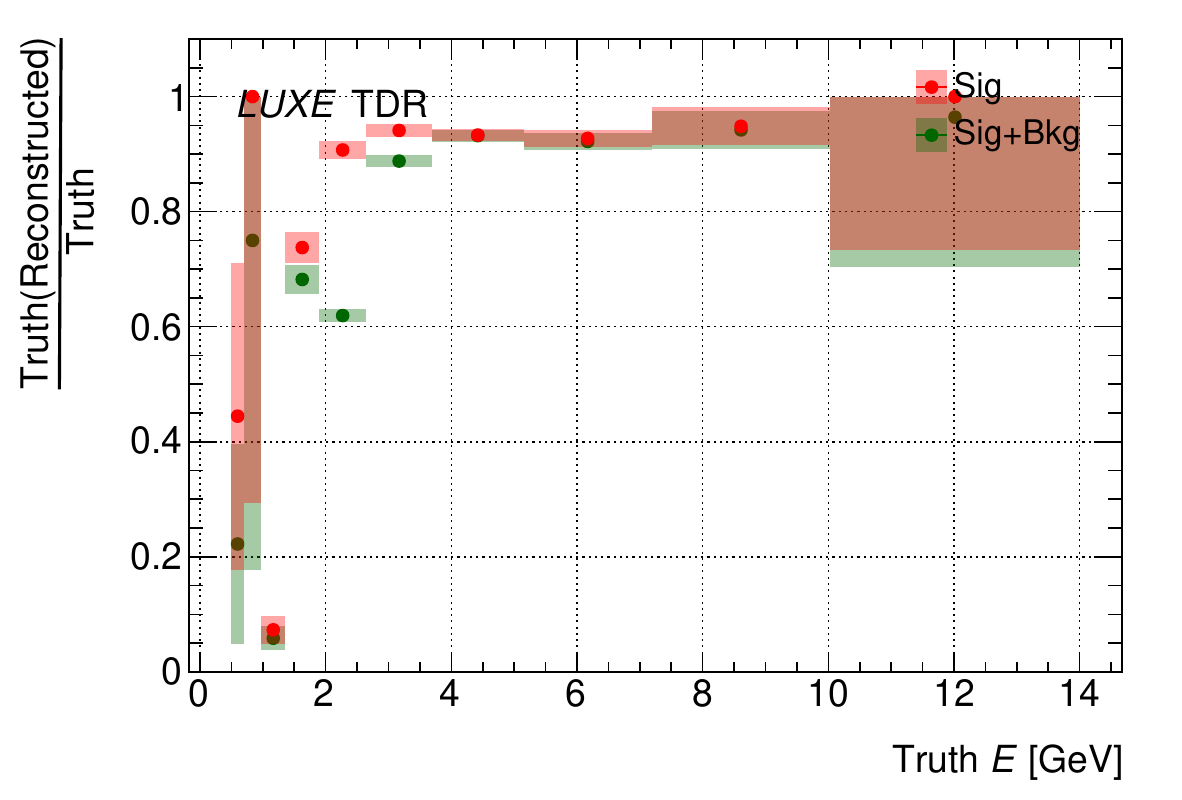}\end{overpic}
\begin{overpic}[width=0.49\textwidth]{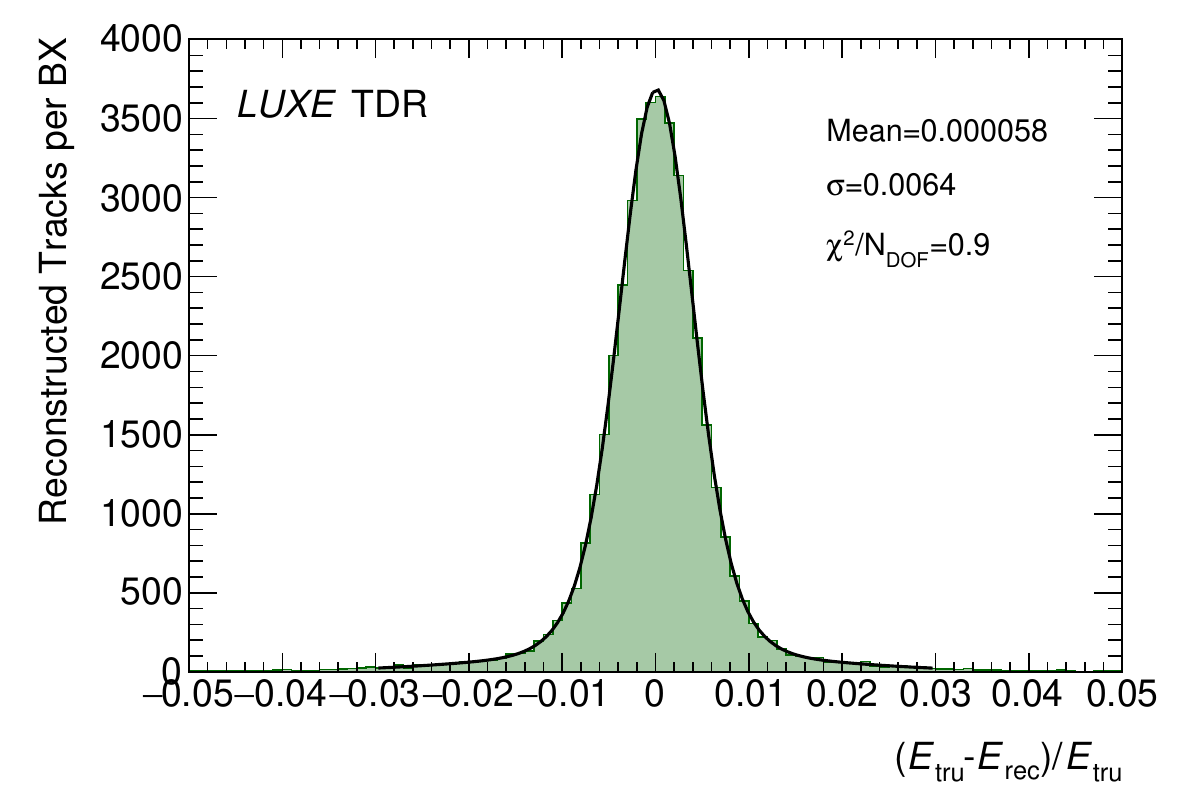}\end{overpic}
\caption{The reconstruction efficiency (left) vs the truth energy and the energy resolution of selected tracks (right) in the signal + background case alone using the {\normalfont \itshape partial matching} scheme.
The fit to a Gaussian are also shown.
The signal sample here uses 5,000 true signal particles drawn randomly from the $\xi=7$ sample per BX once for the signal sample and once combined with the background.}
\label{fig:effresE_05000}
\end{figure}

\begin{figure}[!ht]
\centering
\begin{overpic}[width=0.49\textwidth]{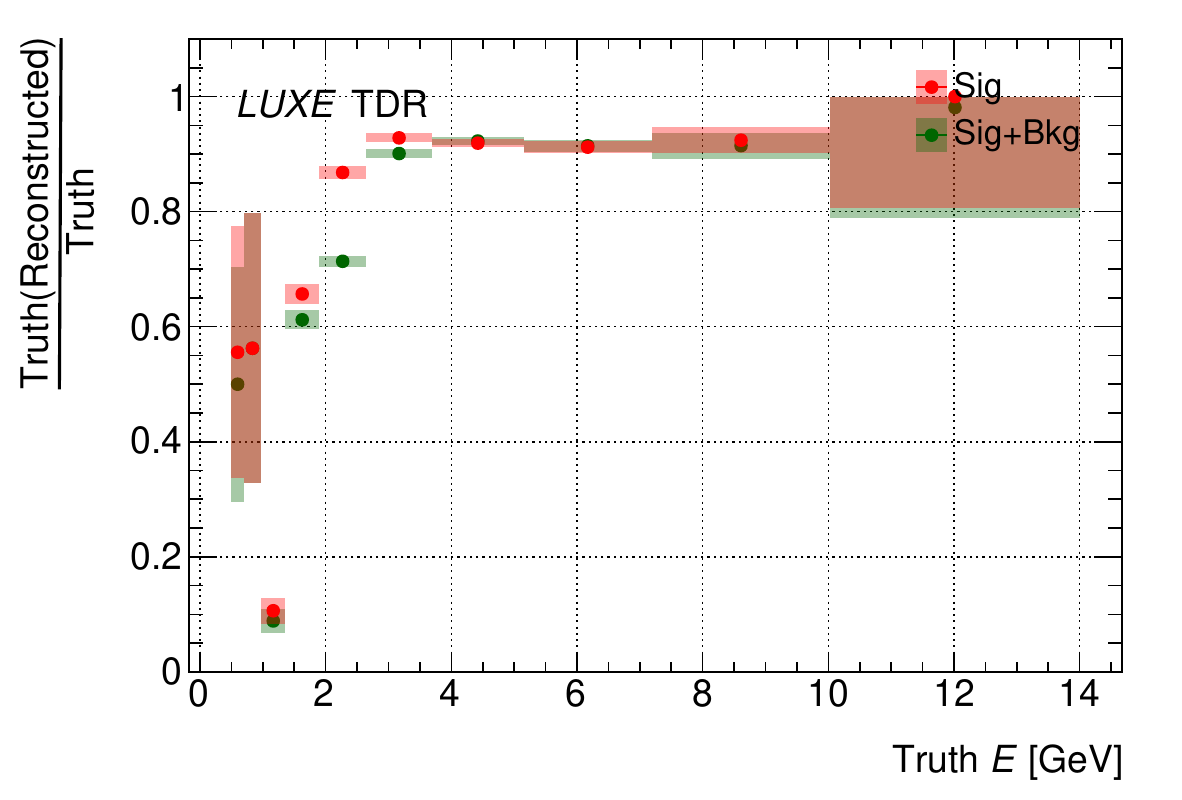}\end{overpic}
\begin{overpic}[width=0.49\textwidth]{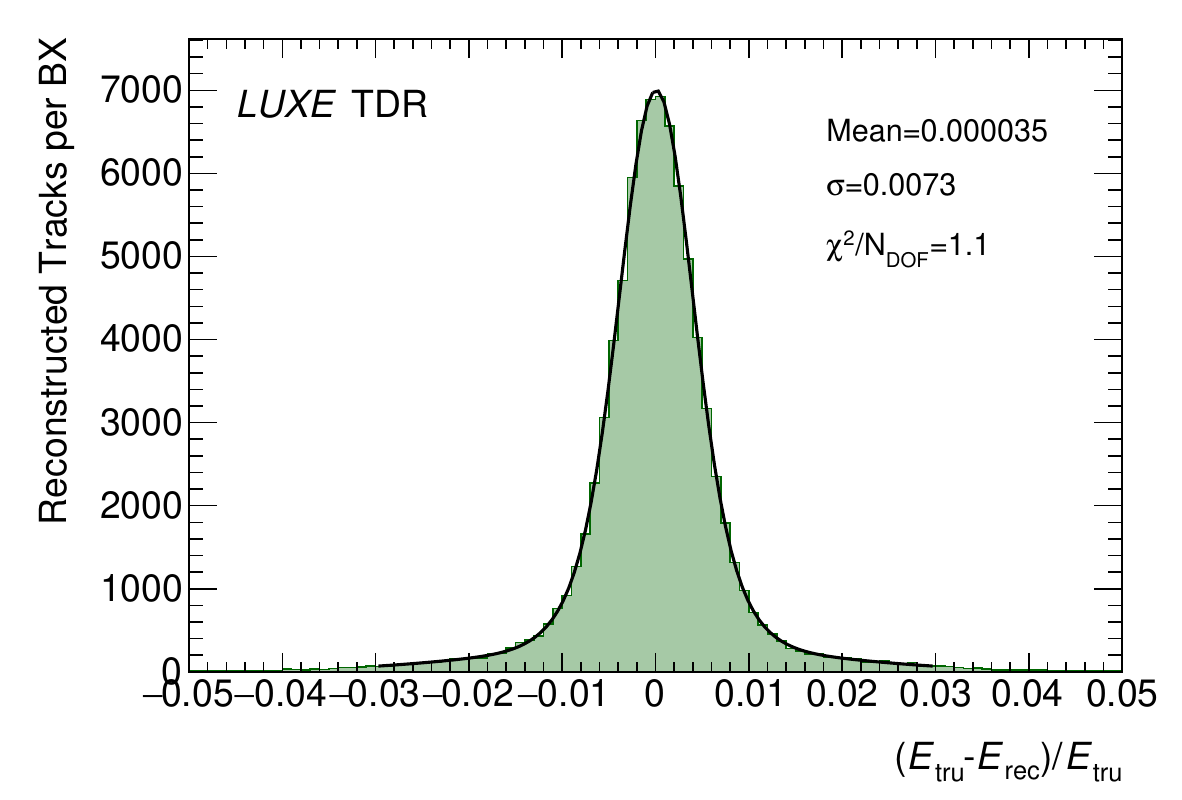}\end{overpic}
\caption{The reconstruction efficiency (left) vs the truth energy and the energy resolution of selected tracks (right) in the signal + background case alone using the {\normalfont \itshape partial matching} scheme.
Fits to a Gaussian are also shown.
The signal sample here uses 10,000 true signal particles drawn randomly from the $\xi=7$ sample per BX once for the signal sample and once combined with the background.}
\label{fig:effresE_10000}
\end{figure}

These results are consistent with runs having larger truth particle multiplicities although it may vary significantly if the algorithm is not optimally tuned for the specific multiplicity.
In Figs.~\ref{fig:effresE_01000},~\ref{fig:effresE_05000} and~\ref{fig:effresE_10000} one can see similar results for cases of 1,000, 5,000 and 10,000 true signal particle multiplicities, respectively.
While the matched (or selected) track resolution is always found to be $\leq 1\%$ regardless of the signal multiplicity, the resolution of unmatched tracks in the same samples has been quantified to be worse, at a level of $\sim$ 2\%.

\begin{figure}[!ht]
\centering
\begin{overpic}[width=0.8\textwidth]{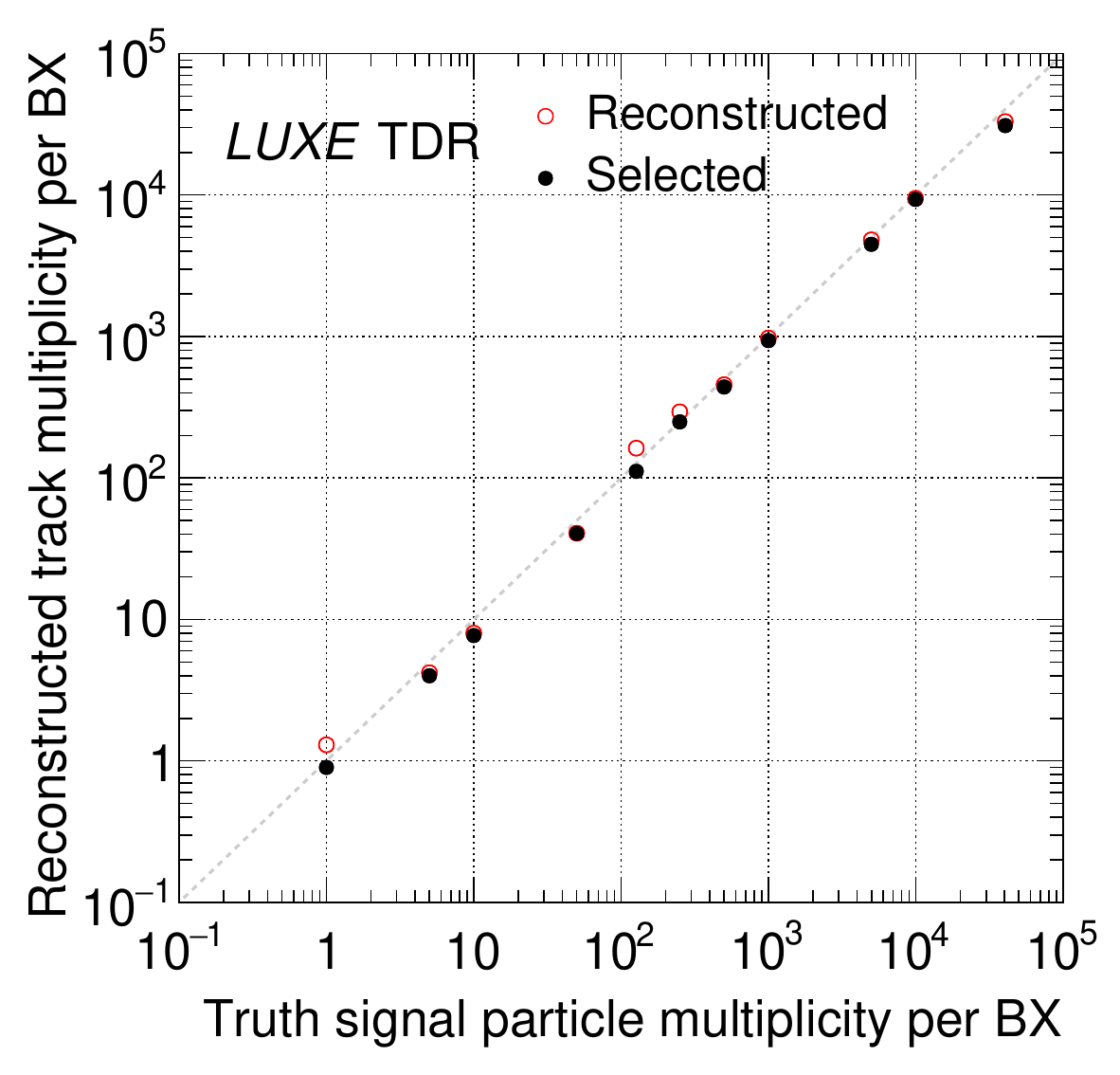}\end{overpic}
\caption{
The reconstructed track multiplicity per BX versus the true signal track multiplicity for reconstructed (hollow red) and selected (solid black) populations. The 1:1 line is given as a guidance in dashed grey.
The plot is stopped at $10^5$, the approximate range where clustering and tracking are possible.
Each point represents 10 full BXs of the background and the signal digitised and clustered together, with the appropriate number of true signal particles indicated in the $x$ axis.}
\label{fig:linearity}
\end{figure}

Finally, to quantify the linearity response of the tracker for different truth signal track multiplicities between 1 and $10^4$ (up to the approximate range where clustering and tracking are possible as discussed earlier), we show the number of reconstructed, selected and matched tracks versus the number of truth signal tracks injected in Fig.~\ref{fig:linearity}.
This quantification is done in the presence of background of about 7,000, 6,800, 6,400 and 5,800 pure background clusters in L4, L3, L2 and L1, respectively.

The study is done using the same two baseline samples of $\xi=3$ and $\xi=7$ with four BXs each.
For the injection of truth particle multiplicities below 50 (inclusive) we use the $\xi=3$ sample, while for all multiplicities above that we use the $\xi=7$ sample.
In all cases the truth particle injection is done by randomly drawing (uniformly and without repetitions) from the two samples above mentioned per BX, and redoing the digitisation and clustering together with a full BX of the background, one BX at a time.
Since there are four signal BXs and only two of the background, the latter are alternated between the former.

When selecting the truth particles randomly, we do not a priori know if e.g. a given truth particle will end up out of the acceptance or if it will overlap with the background particles (leading to biased clusters formation).

In addition to the randomly drawn samples, the full $\xi=3$ and $\xi=7$ samples themselves are also analysed.

It can be seen in Fig.~\ref{fig:linearity} that on average the points lie just below the 1:1 line.
This reflects the different trade-offs between the reconstruction and/or selection efficiency and the background rejection.
This behaviour can be tuned as needed.
Further, the preliminary response shown is not yet optimised. 

For completeness, and since this cannot be shown on the log-log plot as in Fig.~\ref{fig:linearity}, when running the algorithm with the background-only samples, i.e. without any truth signal track, we see 0.5 tracks after reconstruction and 0 tracks after selection for the two full BXs.
Moreover, this situation remains also for all runs at low multiplicity (below $\sim 125$ truth signal particles per BX) and notably for the case of 1 truth signal particle per BX.
The response is also summarised in Table~\ref{tab:linarity}.

From Table~\ref{tab:linarity} it can be inferred that for the $\xi=7$ sample, with truth signal multiplicity exceeding 40,000 particles per BX, the efficiency to reconstruct the full signal is reduced and in addition, the {\normalfont \itshape full matching} efficiency drops significantly as expected.
The former, and to a lesser extent also the latter, is largely due to the inefficiency in the simple clustering algorithm itself.
Nevertheless, even though the {\normalfont \itshape full matching} becomes much more challenging with increasing multiplicity, the track-counting efficiency is still reasonable (at $\sim 80\%$) after reconstruction and/or selection.

\begin{table}[!ht]
\centering
\begin{tabular}{l|llll}
\toprule
Truth & Seeds & Reconstructed & Selected & Matched\\
\toprule
0 (background-only) & 7103 & 0.5 & 0.0 & NA\\
\hline
1 & 6675 &1.3 (130\%) & 0.9 (90\%) & 0.9 (90\%)\\
5 & 6676.9 & 4.2 (84\%) & 4.0 (80\%) &4.0 (80\%)\\
10 & 6687 & 8.0 (80\%)	& 7.7 (77\%) & 7.7 (77\%)\\
50 & 6709.2 & 40.7 (81\%) & 40.5 (81\%) & 40.5 (81\%)\\
\hline
126.75 ($\xi=3$) & 7140.5 & 162.5 (128\%) & 111.5 (88\%) & 111.25 (88\%)\\
\hline
250 & 6920.3 & 293.3 (117\%) & 249.2 (100\%) & 211.1 (84\%)\\
499.8 & 6863 & 456.5 (91\%) & 439.8 (88\%) & 418.8 (84\%)\\
999.4 & 7670.3 & 975.3 (98\%) & 935.4 (94\%) & 828.2 (83\%)\\
4989.5 & 11506.4 & 4820.5 (97\%) & 4484.4 (90\%) & 3778 (76\%)\\
9951.2 & 16405.7 & 9478.9 (95\%) & 9317.6 (94\%) & 7499.9 (75\%)\\
\hline
40453.5 ($\xi=7$) & 38415.3 & 33033.5 (82\%) & 30951.3 (77\%) & 19824.25 (49\%)\\
\bottomrule
\end{tabular}
\caption{The reconstructed track multiplicity per BX versus the true signal track multiplicity for reconstructed, selected and truth-matched populations. Each point represents 10 full BXs of the background and the signal digitised and clustered together, with the appropriate number of true signal particles indicated in the left column. The reconstructed, selected and matched track multiplicity efficiencies with respect to the true signal track multiplicity are shown in parentheses. }
\label{tab:linarity}
\end{table}

This background behaviour at low multiplicity must be evaluated with respect to the input number of initial seeds, which lies at around 7,000 per BX for these runs (stemming from about 3,500 clusters per stave per BX since the background distributes uniformly in $x-y$).
Thus, it represents a rejection power better than $10^{-3}$ background tracks per BX.
Nevertheless, this study needs to be repeated with more BXs in order to increase the statistical significance of this statement.

In addition, there are prospects to further improve the shielding arrangements around the IP detectors to reduce the background, particularly in the \elaser setup.
This may have a dramatic impact on the background count and subsequently on the signal reconstruction.
While the work shown here illustrates it is possible to effectively reject all the background at the low truth signal particle multiplicity runs, the impact of the background in the medium/large multiplicities is not negligible as can be understood in Figs.~\ref{fig:effresE_05000} and~\ref{fig:effresE_10000}, where the efficiency starts to dip in the range where the bulk of the distribution is.
The inefficiency below $\sim 3$~GeV is mostly related to (i) the large inclination of the tracks in that region and (ii) the overlap region between the inner and outer staves and hence, the increased material (L-shapes, etc.). Nevertheless, the bulk of the signal lies above that point and hence the impact of the reduced efficiency is not substantial for LUXE.

The uncertainties of the clustering and the track fitting are not quantified here.
These will stem mostly from the limited knowledge of the alignment of the tracker with respect to the IP and the the dipole, as well as the uncertainty in the knowledge of the magnetic field map.
The alignment of the staves with respect to themselves (and their tray) is expected to be well known to $\mathcal{O}(10)$~\um.
The alignment from the EuXFEL between the IP, the dipole and the tracker tray is expected to be know to within 150~\um.
Previous studies done with a similar simulation setup, where misalignments of $\sim 200$~\um were inserted by hand systematically have shown a very small impact on the tracking performance.
The largest contribution to the uncertainty is expected to come from the magnet.
At this step we do not have enough information to estimate its impact on the tracking performance.
However, these uncertainties are not expected to have a significant impact on the tracker's main physics input to LUXE, which is the number of positrons per BX and its uncertainty.
Here, the main error is dominated by the background, i.e., the overall mitigation before tracking and the rejection after tracking.
The determination of this error requires a large-statistics background-only sample. 
The uncertainty on the tracking performance, e.g., the efficiency and subsequently the unfolding, is sub-dominant.


\subsection{Counter Mode}
\label{sec:screenmode}
Multiple particles hitting the same pixel or adjacent pixels lead to clusters being merged in a way which does not reflect the true spatial distribution of the particles.
Even if somehow the clustering can still be performed efficiently, the combinatorial complexity in the subsequent track fit becomes too large.
This again leads to a picture which is not reflecting the true spatial distribution of the particles.
Therefore, in the scenario where the signal particle multiplicity is too large, the detector can no longer be used for tracking.

As mentioned above, this scenario is only expected at the very last data-taking period in phase-1 of LUXE. 

It is noted that producing such large signal multiplicities is actually not a physics requirement of LUXE but it is just a byproduct of the laser intensity and the electron beam focusing at the IP.
In that sense, the electron beam can be defocused such that the signal rates are reduced.
This does not change the underlying physics, which depends only on the laser intensity (i.e. $\xi$) and the electron beam energy. 

Irrespective of the electron beam defocusing question, the detector can  still be used, though as a counter rather than tracker.
By using simulated signal + background samples to calibrate the observed response of the detector in terms of the number of fired pixels, one can predict the true number of particles.
Beside the integrated number of particles, a crude distribution of the energy may be obtained by slicing the pixel chips in $x$ and using the knowledge of the magnetic field map and the relation between the position and the energy as shown e.g. in Fig.~\ref{fig:Eofx}. 

A preliminary quantification of the counting performance is given below in terms of the calibration from simulation.


\subsection{Counting Performance}
\label{sec:countingperf}
\begin{figure}[!ht]
\centering
\begin{overpic}[width=0.98\textwidth]{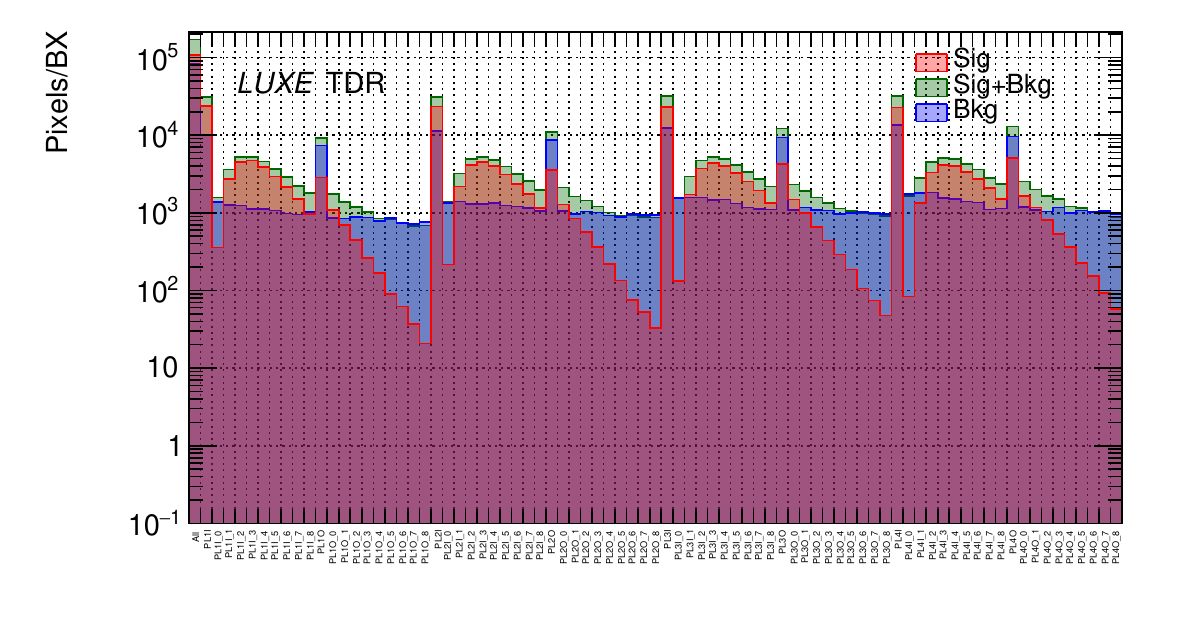}\end{overpic}
\caption{The distributions of number of fired pixels per chip for truth signal particle multiplicity of 1,000. The first bin of each plot represents the sum of all chips of the detector arm, the next bin represents the first stave and the next nine bins represent the chip of the first staves. This structure repeats itself for the rest of the staves of the detector arm along the rest of the histogram’s axis. }
\label{fig:counter1000}
\end{figure}

\begin{figure}[!ht]
\centering
\begin{overpic}[width=0.98\textwidth]{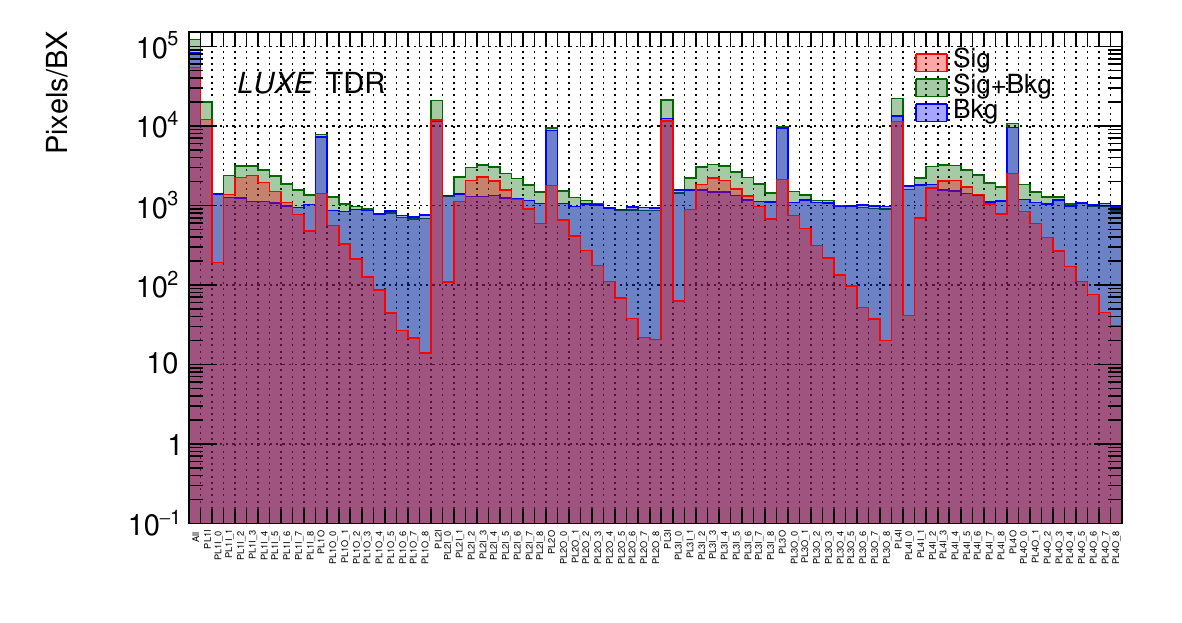}\end{overpic}
\caption{The distributions of number of fired pixels per chip for truth signal particle multiplicity of 5,000. The first bin of each plot represents the sum of all chips of the detector arm, the next bin represents the first stave and the next nine bins represent the chip of the first staves. This structure repeats for the rest of the staves of the detector arm along the rest of the histogram’s axis. }
\label{fig:counter5000}
\end{figure}

\begin{figure}[!ht]
\centering
\begin{overpic}[width=0.98\textwidth]{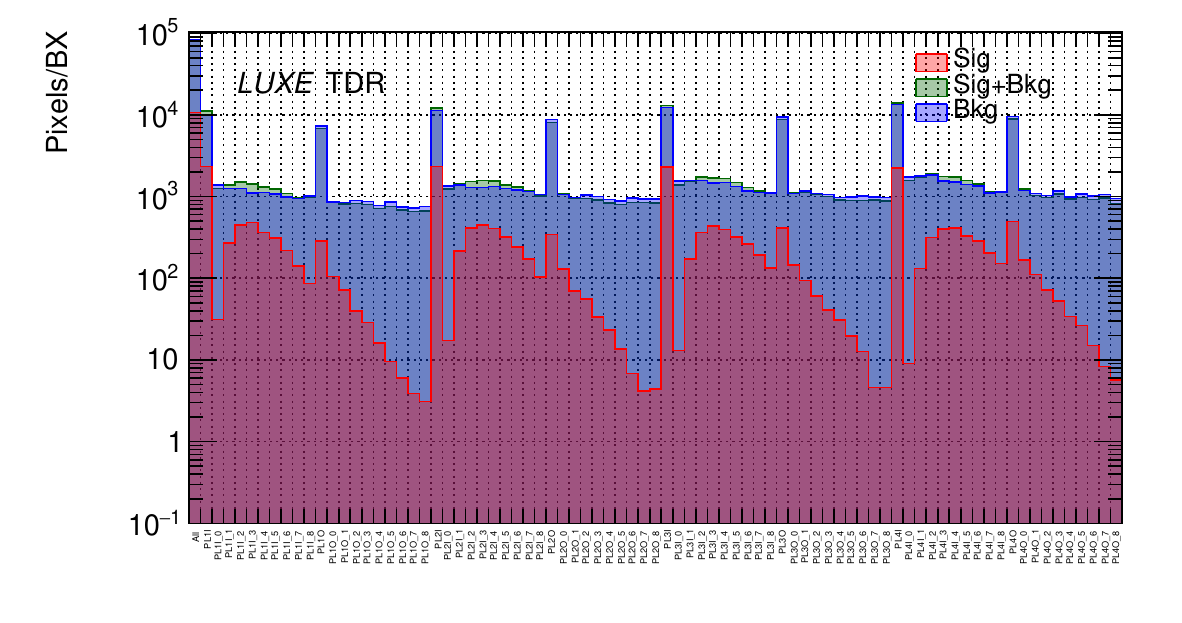}\end{overpic}
\caption{The distributions of number of fired pixels per chip for truth signal particle multiplicity of 10,000. The first bin of each plot represents the sum of all chips of the detector arm, the next bin represents the first stave and the next nine bins represent the chip of the first staves. This structure repeats itself for the rest of the staves of the detector arm along the rest of the histogram’s axis. }
\label{fig:counter10000}
\end{figure}

The distribution of the number of pixels per chip is shown in Figs.~\ref{fig:counter1000},  \ref{fig:counter5000} and~\ref{fig:counter10000} for three true signal particle multiplicities 1,000, 5,000 and 10,000, respectively.
As explained in the previous section, the particles are drawn randomly from the $\xi=7$ \elaser sample (without repetitions).
The first bin of each plot represents the sum of all chips of the detector arm, the next bin represents the first stave and the next nine bins represent the chip of the first staves. 
This structure repeats itself for the rest of the staves of the detector arm along the rest of the histogram's axis.

As can be seen, when simply counting the number of pixels above threshold, the signal starts to become visible above the expected background-only around a multiplicity of $\sim 1,000$ true signal particles per BX.
As demonstrated in the previous section, good clustering and tracking are achievable at this multiplicity and also above that.
However, clustering and tracking gradually become more difficult at higher multiplicities up to a point where it becomes impossible. 

\begin{figure}[!ht]
\centering
\begin{overpic}[width=0.8\textwidth]{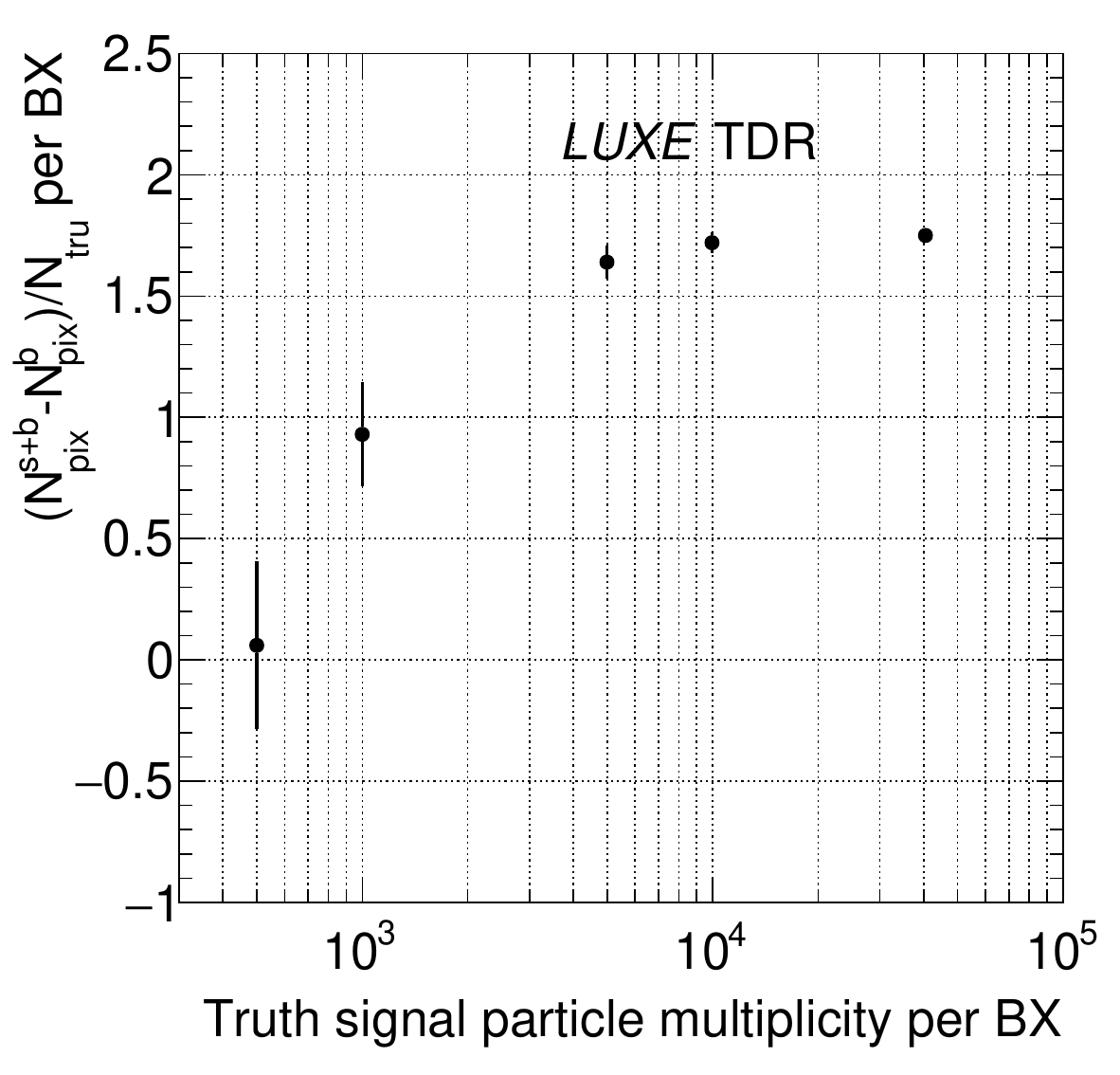}\end{overpic}
\caption{The number of fired  pixels for signal to the true signal particles ratio, $N_{\textrm{pix}}^{\textrm{s}}/N_{\textrm{tru}}$ (see Eq.~(\ref{eq:NpixS})) vs the truth signal particle multiplicity per BX. The uncertainties are statistical only.}
\label{fig:countingperf}
\end{figure}

Nevertheless, the number of truth signal particles per BX can still be obtained by looking only at chips 1-6 (in each stave the chips are counted from 0 to 8) of the inner staves alone, where the signal is expected to peak.
This number averaged over the four layers is denoted as
\begin{equation}
    N_{\textrm{pix}} = \sum_{k_{\textrm{layer}}=1}^{4}\sum_{j_{\textrm{chip}}=1}^{6} N_{\textrm{pix}}(j_{\textrm{chip}},k_{\textrm{layer}})/4.
\end{equation}
This can be obtained separately for background-only runs ($N_{\textrm{pix}}^{\textrm{b}}$) and for collision runs ($N_{\textrm{pix}}^{\textrm{s+b}}$ for signal + background).
The difference between these two estimations will yield the signal estimate alone: 
\begin{equation}
    N_{\textrm{pix}}^{\textrm{s}} \simeq N_{\textrm{pix}}^{\textrm{s+b}}-N_{\textrm{pix}}^{\textrm{b}}.
    \label{eq:NpixS}
\end{equation}
The above equation is only approximate because the background is not strictly reducible from the signal + background case and it depends on the truth signal particles multiplicity.
The summary of this procedure is given in Fig.~\ref{fig:countingperf}, where the measured value $N_{\textrm{pix}}^{\textrm{s}}$ is divided by the number of truth signal particles, $N_{\textrm{tru}}$, per BX and plotted against it.
As mentioned earlier, this can be further binned as finely as needed down to the pixel column width in principle.
Using the relation between the position $x$ and the energy, we can also produce the approximate distribution of energy.

It can be seen from Fig.~\ref{fig:countingperf} that with the given level of background, the counting approach cannot be used below a truth signal particle multiplicity of a few hundreds (e.g $\gtrsim 500$) per BX.
Irrespective of the tracking performance below a multiplicity of $\sim 10,000$ (which is very good), in order to use the counting method below a multiplicity of a few 1,000s one should fit the data and interpolate.
Above that, the trend is clear and the shape approaches saturation, as expected.
This happens at large signal rates, when the background becomes small enough compared to the signal + background count, approximately at $\textstyle{\frac{N_{\textrm{pix}}^{\textrm{b}}}{N_{\textrm{pix}}^{\textrm{s+b}}}}\leq 10\%$.
The saturation is reached at a value of $\sim 1.75$.

The values used to obtain Fig.~\ref{fig:countingperf} are summarised in Table~\ref{tab:countingperf}.

\begin{table}[!ht]
\centering
\begin{tabular}{l|ll|l|l}
\toprule
Truth ($N_{\textrm{tru}}$) & $4N_{\textrm{pix}}^{\textrm{b}}$ & $4N_{\textrm{pix}}^{\textrm{s+b}}$ & Signal-only ($4N_{\textrm{pix}}^{\textrm{s}}$) & $\frac{N_{\textrm{pix}}^{\textrm{s+b}}-N_{\textrm{pix}}^{\textrm{b}}}{{\textrm{Truth}}}$\\
\toprule
499.8	& 32592.5	& 32718.70	& 3998.10   & 0.06\\
999.4	& 32592.5	& 36292.70	& 7939.80   & 0.93\\
4989.5	& 32592.5	& 65386.30	& 40215.20  & 1.64\\
9951.2	& 32592.5	& 101098.70	& 79799.10  & 1.72\\
40453.5	& 32592.5	& 316517.75	& 317225.25 & 1.75\\
\bottomrule
\end{tabular}
\caption{The number of pixels for signal, background, signal + background, and the ratio $N_{\textrm{pix}}^{\textrm{s}}/N_{\textrm{tru}}$ (see definition in Eq.~(\ref{eq:NpixS})) for different truth signal particle multiplicities per BX. Note that the signal column is not used in the calculation of the ratio in the last column and it is just shown for reference.}
\label{tab:countingperf}
\end{table}

It is noted that this preliminary study has to be expanded to larger multiplicities, but it already provides a hint towards what can be achieved with a simple counter mode operation.

\section{Interfaces and Integration} 
\label{sec:interfaces}
The different physical interfaces of the tracker are discussed in detail in Section~\ref{sec:services}.
The system is designed to be as standalone as possible and therefore it has its own power supply and cooling.
The kinematic mount concept also ensures standalone alignment and repeatability. 

The power and temperature of the staves and the other active elements (PUs and RUs) will be monitored and controlled remotely via the data streamed to FELIX.

The independent cooling system must also be controlled remotely, assuming that the chillers are placed in the experimental hall.
In principle, the water can also be brought in and out of the experiment from outside the hall via isolated copper pipes.

The movement of the electron arm of the tracker in and out of the interaction plane must also be controlled remotely with a motorised stage.
This depends on the specific choice of the (commercial) stage solution.

The most involved interface with the rest of the experiment is the trigger and the data acquisition (DAQ).

\subsection{DAQ}
\label{sec:daq}
The EUDAQ2~\cite{Liu_2019} will be used after FELIX as a common DAQ, having user-specific interfaces for data acquisition and for data analysis.
EUDAQ2 is a generic multi-platform DAQ framework.
It is written in C++, and designed to be modular and portable.
It is planned that all systems of LUXE will be using the EUDAQ2.

The interface between the output of FELIX and EUDAQ2 has to be developed. 
It will be done in a dedicated test-stand at WIS based on modules that are purchased independently.
A full discussion of the DAQ is given in Chapter~\ref{chapt11}.

\subsection{Trigger}
\label{sec:trigger}
The trigger will be provided to FELIX by the AIDA trigger logic unit (TLU)~\cite{TLU2019}.
The TLU will provide triggers and control signals to all LUXE detectors centrally.
It will also be connected to the EuXFEL clock and to the laser.

Configuration and communication with the TLU are performed over Ethernet.
It can be employed as a standalone unit or be deployed as part of the EUDAQ2 framework discussed above, which allows it to connect to a wide range of readout systems.
The TLU can operate with a sustained particle rate of 1~MHz and with instantaneous rates up to 20~MHz, much above the rates expected in LUXE.
The unit can time-stamp incoming signals with a resolution of 1.5~ns or better.
TLUs have already been deployed successfully in beam lines at CERN and at DESY.



\section{Installation, Commissioning and Calibration}
\label{sec:installation}

\subsection{Installation}
\label{sec:install}

Most of the work on the detector itself, as well as on its services, will be performed at Weizmann to minimise the work needed at DESY on-surface and particularly underground in the experiment hall.
Ideally, this includes a system test with all components. The extent of the system test will depend on the schedule but a full readout chain with at least two staves is envisioned.

The testing step at DESY on-surface will be only for basic integrity of the different components.
This step will be made as brief as possible.

The number of people from the TC alignment survey team (TCA) is to be confirmed.
Support of $\geq 1$ TC liaison technician (TCT) will be available throughout most of the period.

There are only small differences expected between the installation procedures of the full tracker and the ``bare-bones'' tracker. 
Therefore, the full tracker scenario is taken as the benchmark for the planning here.

\subsection{Calibration Strategy}
\label{sec:calibration}
Calibration will be done based on (i) the alignment information, (ii) the in-situ magnetic field measurements at LUXE, and (iii) the ``\gneedle'' run data.

The \gneedle runs will start by converting the electron beam to a photon beam on the thin target foil upstream. This is done as planned in the \glaser runs but where the laser is off and where a thin needle is positioned in the path of the photon beam inside the IP chamber to convert it back to \ee pairs.
This process replaces the NBW pair production, but where the input photon spectrum (Bremsstrahlung) is known from simulation.

Since the \ee rate in this case is expected to be small and the background is expected to be similar to the \glaser mode (i.e.\ substantially small compared to the \elaser mode), the parent photon energy spectrum can be reconstructed from the \ee pair four-momenta.

A fit to the energy spectrum of the Bremsstrahlung photons from simulation will be done with a few free floating parameters related to the spatial alignment, the magnetic field, the energy loss in the vacuum exit window and finally, the energy scale.
The fit purpose is to reproduce the energy spectrum from simulation, which on itself has uncertainties related to the specific model used in GEANT4~\cite{AGOSTINELLI2003250, 1610988, ALLISON2016186}.

In the case where the calorimeter will be fully installed and functioning in time for the first runs and assuming it will be already calibrated, the energy measurement from the calorimeter can be potentially used to estimate the miscalibration of the tracker.

\section{Further Tests Planned}
\label{sec:future}
The period before the budget is identified is mostly driven by tests with initial modules at WIS as well as design work.
This is similar for both the full and the ``bare-bones'' tracker scenarios.

One ``silver category'' stave (each chip has more than 50 but up to 1500 dead pixels out of the 512 $\times$ 1024 pixels, so up to 0.3\% dead pixels) is purchased, along with eight individual chips on carrier boards.

The carrier boards each come with a dedicated DAQ board.
To power the stave, dedicated regulator boards (not the nominal PB) are purchased. 
Similarly, to read out the stave, three dedicated MOSAIC~\cite{refId0} boards are purchased.
The power supply system, nearly identical to the nominal one, is already available at the WIS lab.
Additionally, one FELIX~\cite{Anderson_2016} board has been purchased from CERN.    
The initial tests are started with this system.

In the case of the RU+PU solution, these units will be purchased from industry partners, based on the existing designs from ALICE.
A TLU~\cite{TLU2019} is available and the integration of the full chain with the EUDAQ2~\cite{Liu_2019} system will begin.
The goal of these tests is the evaluation of the functionality of the full chain in an identical setup  to the one foreseen in LUXE (though scaled-down).
In this way, the performance of the system up to (but not including) tracking can be quantified and feedback can be still injected to the design before orders are issued.
The performance of the ALPIDE chips themselves have been tested over multiple occasions as discussed for example in~\cite{DANNHEIM2019187,MAGER2016434,8855021} and hence this will not be part of the goals of the tests at WIS.

Besides the functionality tests and in parallel to these, the integration of the production staves with the mechanics and the cooling will also be tested.
Also here, invaluable feedback can be provided to the engineering design.

The tests foreseen will include noise runs, pulse runs, runs with radioactive sources and runs with cosmic muons.
For the latter, a scintillator trigger system will be available as well as a telescope built from the eight individual chips on carriers.
A test in another WIS lab with $\sim 1$~GeV electrons may be also carried out.

Additionally, alignment tests will be performed on a mock-up detector tray.
The cooling will be tested as well and in particular the leak-less functionality.
The latter can proceed also based on dummy pipes similar to those integrated on the production stave.


\section*{Acknowledgements}
The authors wish to thank the colleagues from the ALICE ITS project for useful discussions and help with the design of the different systems of the tracker.
We would like to thank especially Luciano Musa, Gianluca Aglieri Rinella, Antonello Di Mauro, Magnus Mager, Corrado Gargiulo, Felix Reidt, Ivan Ravasenga and Ruben Shahoyan.

We also wish to thank Simon Spannagel and Paul Jean Schutze for the detailed and dedicated help and for the useful discussions about \allpix.

The work of Noam Tal Hod is supported by a research grant from the Estate of Dr. Moshe Gl\"{u}ck, the Minerva foundation with funding from the Federal German Ministry for Education and Research, the ISRAEL SCIENCE FOUNDATION (grant No. 708/20), the Anna and Maurice Boukstein Career Development Chair, the Yeda-Sela SABRA (Supporting Advanced Basic Research) fund.

\printbibliography
\end{refsection}


\makeatletter
\def\input@path{{./chapters5/}}
\makeatother

\chapter{Electromagnetic Calorimeter}
\label{chapt5}
\begin{refsection}
\chapterauthors{H.~Abramowicz \\Tel Aviv University, Tel Aviv (Israel) }{W.~Lohmann \\Deutsches Elektronen-Synchrotron DESY, Zeuthen (Germany)}
\date{\today}

\begin{chaptabstract}
    Two electromagnetic calorimeters of the LUXE experiment, hereafter referred to as ECAL-E and ECAL-P, are designed as sampling calorimeters using silicon sensors and tungsten absorber plates. ECAL-P will measure the number of positrons and their energy in \elaser ~and \glaser ~interactions and ECAL-E, the same quantities for electrons in \glaser ~interactions. In addition, the calorimeters will be used for the cross-calibration of the energy scale with the tracker. In the following the mechanical structure, the sensors, the front-end electronics and the data acquisition are described. Monte Carlo simulations based on the \geant package are used to estimate the performance of ECAL-P in terms of energy and position resolution, as well as for the reconstruction of the number of positrons impacting the ECAL-P and their energy spectrum. Similar studies for ECAL-E are planned.
\end{chaptabstract}


\section{Introduction}


As stated in Chapter 2,
the number of electron-positron pairs and their energy distribution in \elaser and \glaser interactions are key characteristics to probe SFQED. 
In the LUXE experiment, the schematic layouts of which is shown in Fig.~{\ref{fig:layoutdet}}, positrons originating from trident production in \elaser and from the Breit-Wheeler process in   \glaser interactions will be measured in a detector system consisting of a tracker followed by an electromagnetic calorimeter, ECAL-P. The ECAL-P design is based on a technology for highly compact electromagnetic calorimeters which was developed by the FCAL collaboration~\cite{Abramowicz:2010bg,Abramowicz:2018vwb}, an R\&D collaboration to design, build and test prototypes of luminometers for future linear electron-positron colliders (ILC or CLIC). 

In the \glaser setup, also the spectra of electrons originating from the Breit-Wheeler process will be measured. For that purpose, as for positrons, the tracker, located on the opposite side of the beam, will be followed by ECAL-E, a silicon-tungsten electromagnetic calorimeter based on the technology developed by the CALICE collaboration~\cite{Kawagoe:2019dzh}. The highly granular ECAL-E is the reference design of the electromagnetic calorimeter for the International Large Detector (ILD) concept.

The trackers and both ECALs are located downstream of a dipole magnet which directs the positrons and electrons from the interaction point to the detectors. The correlation between the position and the energy of the positrons and electrons, determined by the dipole magnetic field, constitutes the basis for energy measurements in the tracker. By directly measuring the energy, the ECALs can determine whether more than one particle had almost the same path. Thus, the role of the ECALs is to determine, independently of the tracker, the number and energy spectra of eletrons and positrons. The combination will provide an independent 
{\normalfont \itshape in situ} 
energy calibration. In addition, it will ensure a good control of the beam-related background.

\section{Requirements and Challenges}

To measure the number and the energy spectrum of electrons and positrons per bunch crossing, challenges and requirements depend on two major scenarios. In the first, when the number of electrons and positrons is very small, the showers must be measured on top of low energy, but widely spread, background. In the second, when the number of electrons and positrons is large and single particle showers overlap, the number of electrons and positrons and their spectrum
can still be measured using an energy flow algorithm as explained below. However, non-linearity of the readout electronics and potentially of the signal generation in the sensor has to be taken into account.
In the first scenario, a small Moli\`ere radius and high granularity will improve the performance of the measurement. In the second, high granularity becomes essential.

\begin{figure}[htbp]
    \centering
    \includegraphics[height=0.4\textheight]{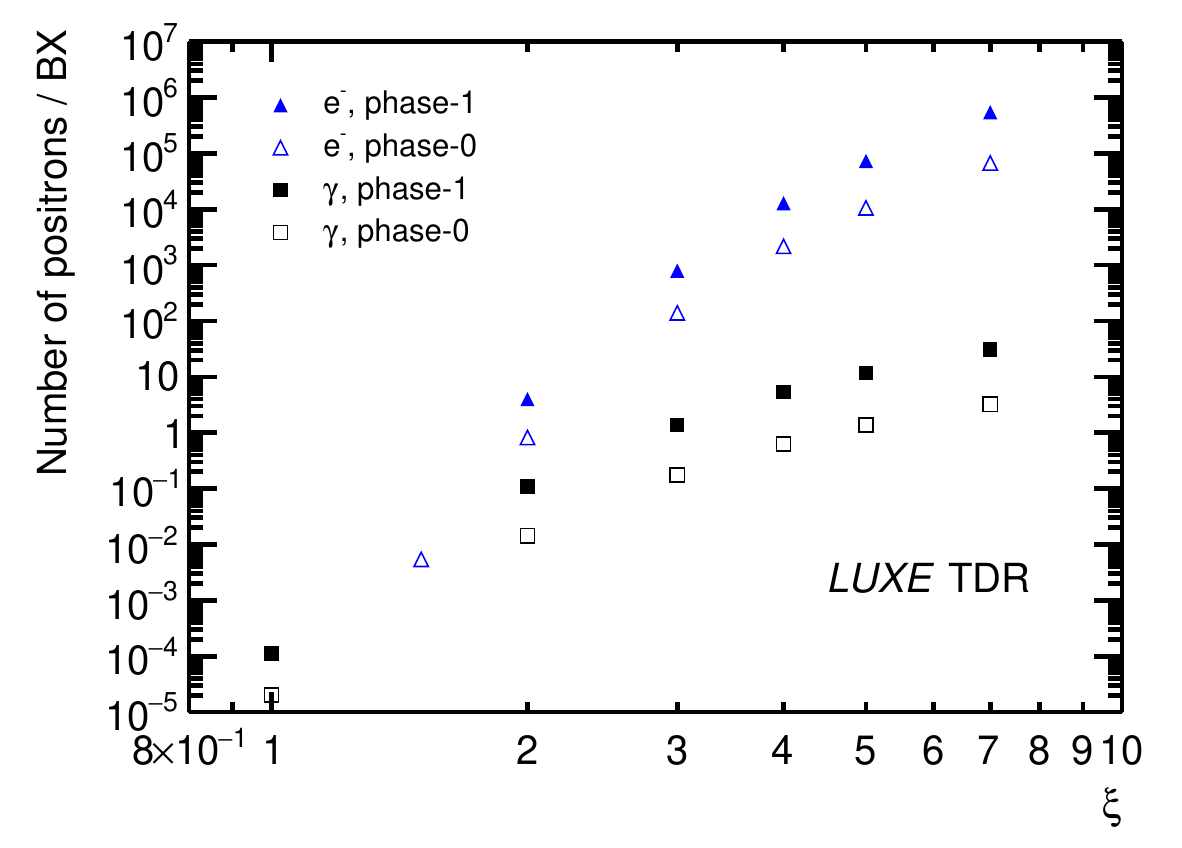}
    \caption{Number of positrons per bunch crossing produced in the \elaser and \glaser setups for \phaseone and \phasetwo, as a function of $\xi$. The electron beam energy is set to 16.5\,GeV, and the laser waist parameter varies between 100$\,\um$ and 3$\,\um$
    in the range of~$\xi$.  }
    \label{fig:posyield}
\end{figure}

\begin{figure}[htbp]
    \centering
      \includegraphics[width=0.49\textwidth]{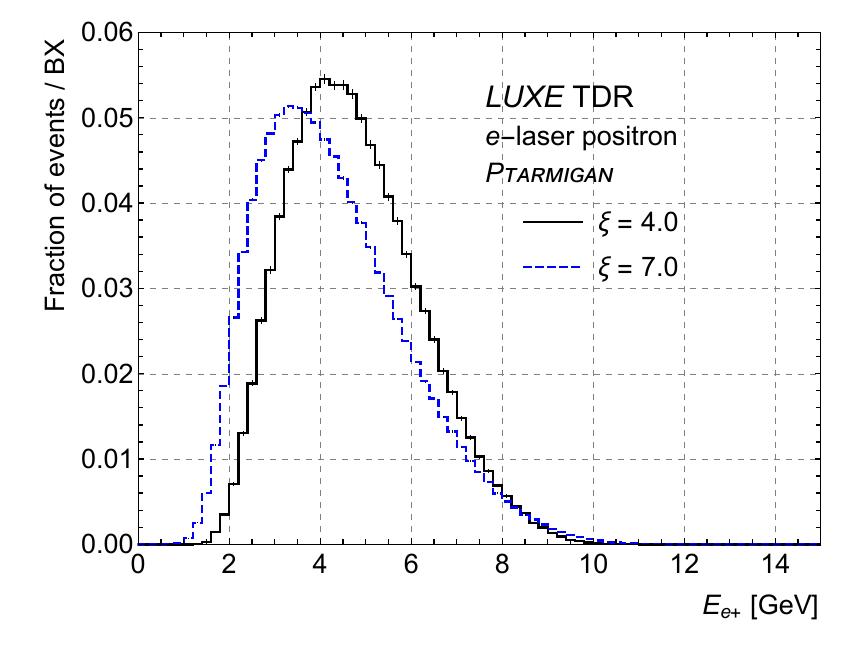}
      \includegraphics[width=0.49\textwidth]{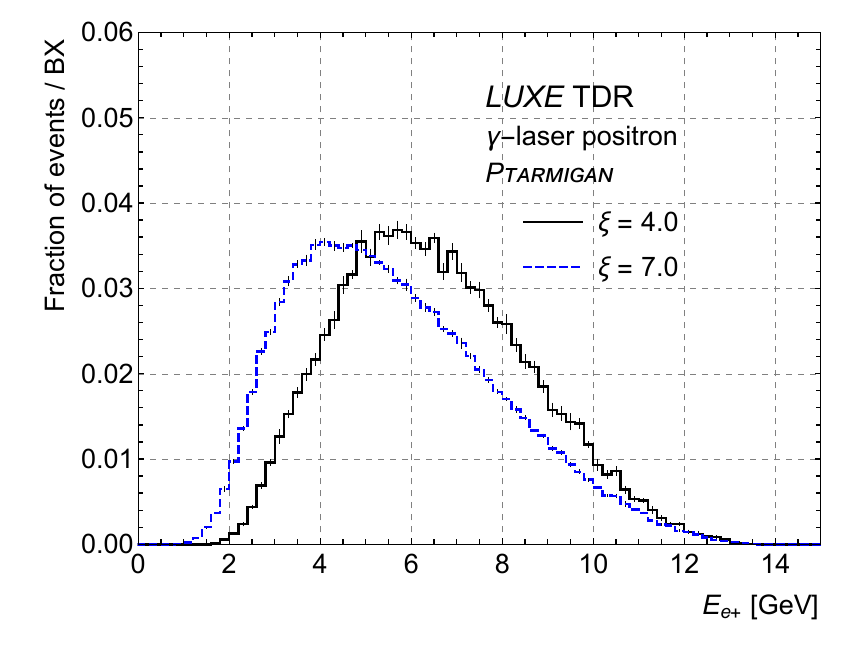}
     \caption{Positron energy-spectra for the \elaser and the \glaser setup for \phaseone for different values of $\xi$.}
    \label{fig:posspectr}
\end{figure}

The following studies are done for positrons, but they are valid also for electrons in the \glaser setup.  
To get a quantitative picture, a custom-developed strong-field QED computer code, named \textsc{Ptarmigan}~\cite{ptarmigan},
is used to simulate the strong field interactions for the relevant physics processes.
Simulations are done for different laser parameters. For \phaseone, at the start of the experiment, a laser power of $40\units{TW}$ is assumed, while in \phasetwo the laser power is increased to $350\units{TW}$.  
The positron yield per bunch crossing for \elaser and \glaser interactions are shown in Fig.~\ref{fig:posyield} as function of
the intensity parameter of the laser field
$\xi$.
This is the same as Fig. 2.15 (left), repeated here for convenience.
Different values of $\xi$ are achieved by varying the laser waist radius.
It is seen that for the \elaser mode the positron yield is larger than for the \glaser one by a factor of $\sim 10 - 10000$, depending on $\xi$. For \phaseone of the \elaser run, up to $10^5$ positrons are expected per bunch crossing, while for the \glaser running at low $\xi$, most of the bunch crossings are without a positron, and only $\sim 10^2$ are expected at the highest $\xi$. For \phasetwo the rates are roughly 10--100 times larger.\footnote{These numbers can be reduced by defocusing the electron beam.}
The expected positron spectra are typically between 2 and 14\,GeV, and are slightly softer for the \elaser mode, as can be seen in Fig.~\ref{fig:posspectr}.

Shown 
are the expected energy spectra for \phaseone for both processes for selected $\xi$ values. With increasing $\xi$ values, the maximum of the spectrum slightly shifts toward lower energies.

To keep the Moli\`ere radius small, a compact sandwich calorimeter with tungsten absorber plates interspersed with 320$\,\um$ thick silicon pad sensors
is implemented in \geant simulation for performance studies and layout optimisation.
The thickness of the tungsten plates is $3.5\units{mm}$, corresponding to one radiation length. The gap between two absorber plates is $1\units{mm}$, requiring firstly ultra-thin assembled detector planes, and secondly tungsten plates of very good flatness.
The acceptance range in the horizontal, $x$, direction  is determined by the positron energy-spectrum and the magnetic field of the dipole magnet, and amounts to about $50 \units{cm}$. The coverage in the $y$ direction is mainly determined by the Moli\`ere radius, and the estimated $5 \units{cm}$ is well within the $9 \units{cm}$ of the active calorimeter height.

Single positrons with different energies are generated by a gun for basic performance studies. For complex performance estimates,
signal and background processes of \elaser ~and \glaser ~interactions are generated with the precursor of \textsc{Ptarmigan}, \textsc{Ipstrong}~\cite{ipstrong}.\footnote{The differences between the two MC generators are mainly in the expected yields but not in the energy spectra and therefore the design of the ECAL-P will not be affected.}

\begin{figure}[htbp]
\includegraphics[width=0.49\textwidth]{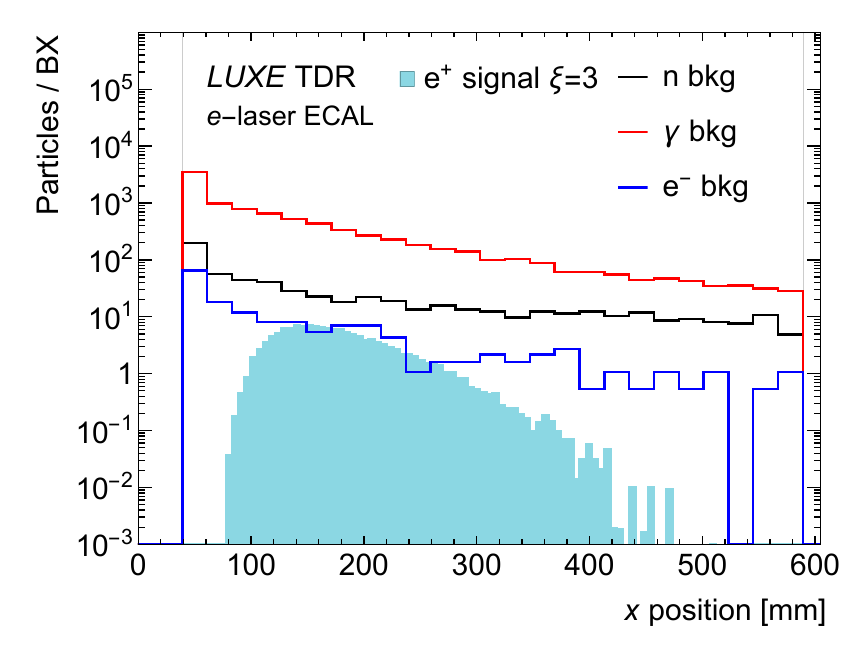}
\put(-120,10){\makebox(0,0)[tl]{\normalfont \bfseries (a)}}
\includegraphics[width=0.49\textwidth]{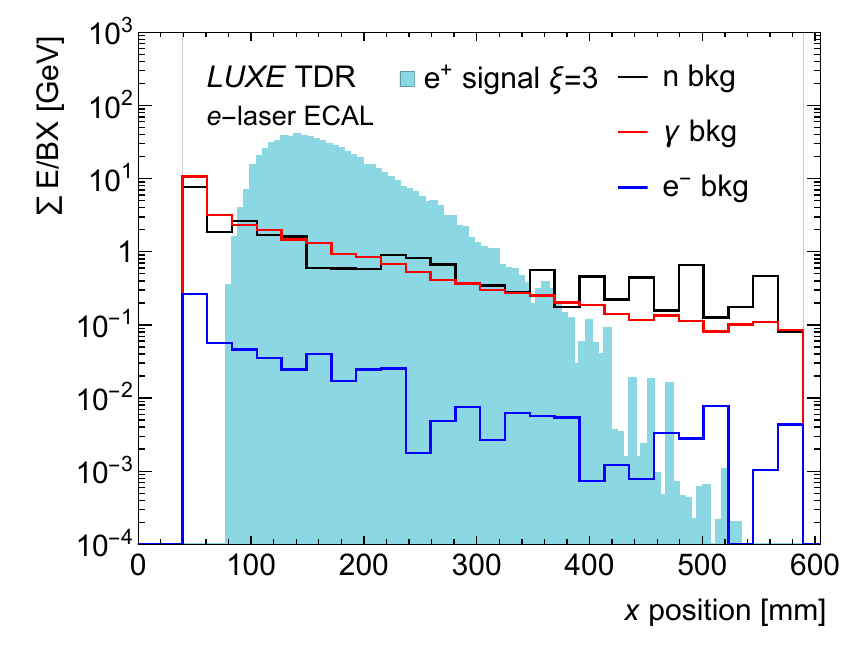}
\put(-120,10){\makebox(0,0)[tl]{\normalfont \bfseries (b)}}\\
\caption{(a) Number of particles and (b) their energy sum per BX, split into different background particle types and signal positrons as described in the legend, as a function of the $x$ position in the calorimeter for the \elaser set-up for \phaseone and $\xi=3.1$ (laser focal spot size $\omega_0 = 5\,\um$). Only background arriving within $100 \units{ns}$ is included in the plots.
}
\label{fig:ecal-elaser-bkg-rates}
\end{figure}

The ECAL-P is exposed to beam related background. 
The expected particle rates and the energy sum of particles impacting the calorimeter per bunch crossing as a function of the $x$ position in the calorimeter are shown in Fig.~\ref{fig:ecal-elaser-bkg-rates} in the \elaser set-up for the \phaseone laser and $\xi=3.1$. The background contribution is split into different particle types and compared to the expected yield of signal positrons. To reduce the background originating mainly from the neutron flux, a timing cut of $100 \units{ns}$ was imposed on the time of arrival of the signal. This substantially reduces the contribution of slow neutrons originating in the downstream beam-dump.

 Much of the background originates from a flux of very low energy secondary particles created by the beam, which will be further suppressed by optimisation of the vacuum chamber design, the shielding, and the beam-dumps. 
Once the optimisation of the beam-dumps design in particular for low neutron back-scattering is completed, extra shielding of the ECAL, for which there is ample space, will be considered. This is an ongoing effort. Finally, the remaining background depositions constitute a signal offset which will be measured in bunch crossings without laser and subsequently subtracted from signal event depositions, pad by pad.

\section{System Overview}

\begin{figure}[htbp]
\begin{center}
\vspace*{-0.5cm}
  \includegraphics[width=0.8\textwidth]{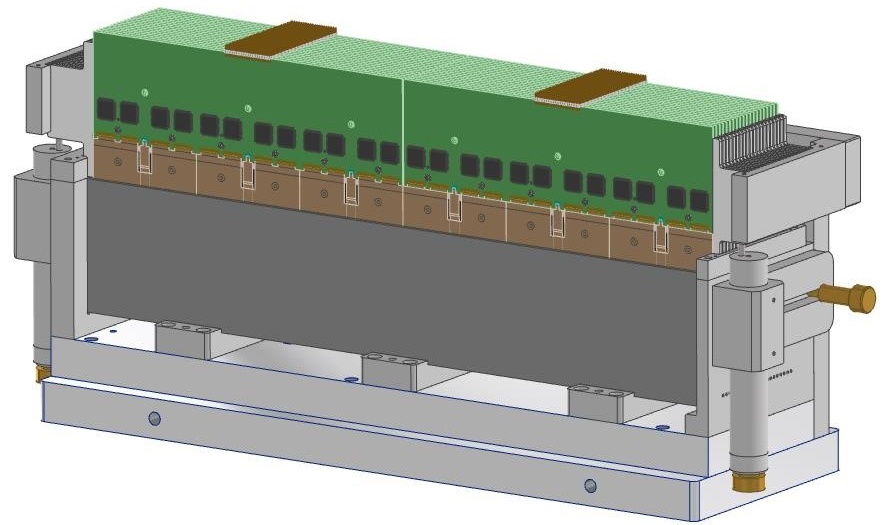}
    \caption{A sketch of the LUXE ECAL-P mechanical structure. The frame holds the tungsten absorber plates, interspersed with sensor planes. The front-end electronics mounted on the PCBs (shown in green) is positioned in the aluminium frames on top of ECAL-P and connected with sensor pads via kapton PCB foils 
    (see section~\ref{ch5_techdescr} for details).    
    } 
    \label{ECAL_layout}
    \end{center}
\end{figure}

The LUXE ECAL-P is designed as a sampling calorimeter
consisting of $21$ tungsten plates of $1 X_0$ thickness ($3.5 \units{mm}$) interspersed with $20$ instrumented silicon sensor planes, placed in a $1 \units{mm}$ gap between absorber plates. 
The whole structure will be held by an aluminium frame, with slots on top to position the front-end boards (FEB).
A sketch is shown in  Fig.~\ref{ECAL_layout}.

The ECAL-P is located $4.3 \units{m}$ from the interaction point (IP), $10 \units{cm}$ behind the tracker and $4 \units{cm}$ away from the beamline on the side towards which the positrons are deflected. A $1 \units{cm}$ thick tungsten plate, extending along the beamline from the exit of the vacuum chamber, will shield it from the side. The ECAL-P will be installed on a special optical table behind the tracker. From simulation and tests of the LumiCal prototype~\cite{Abramowicz:2018vwb} on which the ECAL-P design is based, the expected energy resolution is $\sigma/E=20\%/\sqrt{E/\units{GeV}}$ and the position resolution is about $750 \,\um$ for electrons of $5 \units{GeV}$\footnote{These values are obtained when the positrons hit the calorimeter under a small angle due to deflection in the dipole magnet.}. The position resolution translates into $\sigma/E=0.5\%$ to be compared to $\sigma/E=9\%$ from calorimetry alone.

In the current design, ECAL-P sensors are made of silicon wafers of $320\,\um$ thickness.
 Each sensor has a surface of $9 \times 9 \units{cm^2}$ and consists of 256 pads of $5.5 \times 5.5\units{mm^2}$ size ($p$ on $n$-bulk type). Each complete detector plane will consist of six adjacent sensors. The fiducial volume of the calorimeter will then be $54 \times 9 \times 9 \units{cm^3}$.

The readout will be based on the FLAME multi-channel readout system~\cite{Kulis:2012zz,Bugiel:2016bfo} which will be adapted for the needs of ECAL-P. The readout of the FEBs will be orchestrated by FPGA boards, connected with cables of a few meters length. The FPGA boards will be located in a rack near the ECAL-P. The FPGAs are connected to a computer via ethernet. The data acquisition and monitoring will be implemented in the EUDAQ system~\cite{Ahlburg:2019jyj}. 
The power dissipation of the FEBs is estimated to be less than $1 \units{W}$, hence no active cooling is needed.

\begin{figure}[!t]
\centering
\includegraphics[width=0.74\textwidth]{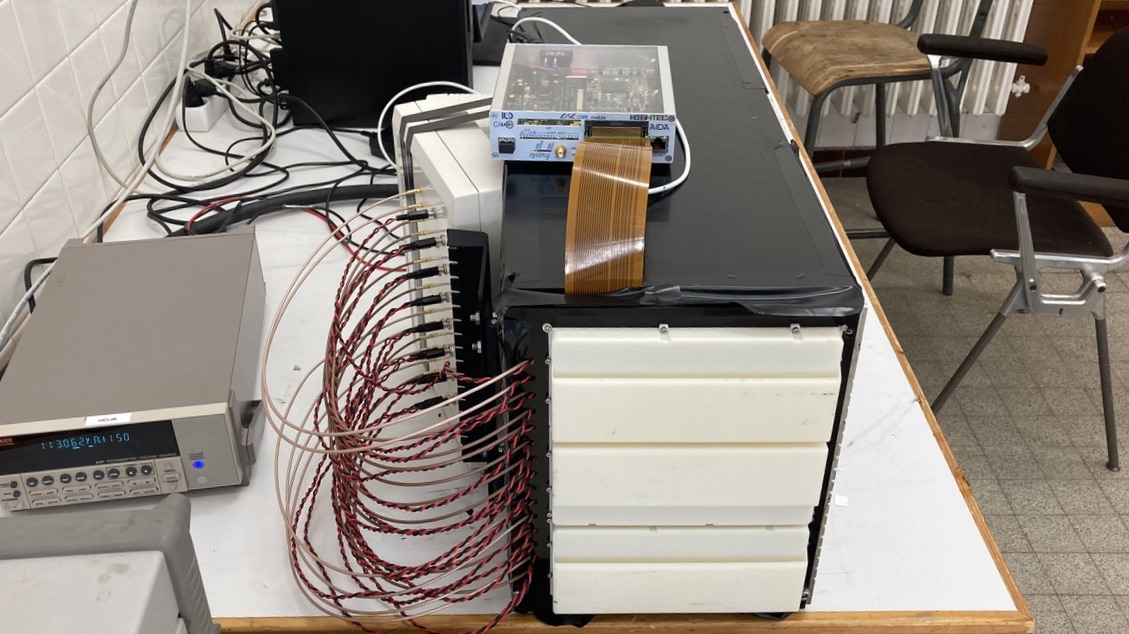}
\includegraphics[width=0.24\textwidth]{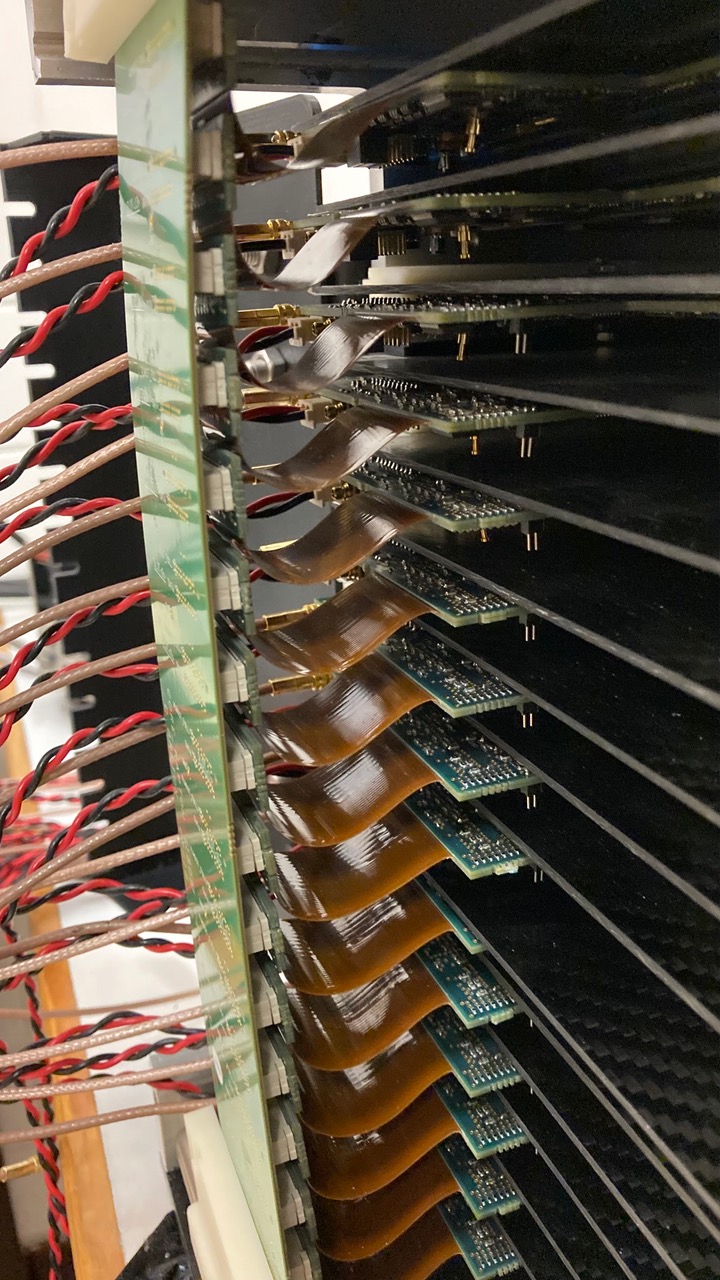}
\caption{Left: picture of the 15 layer stack of ECAL-E. Visible are the mechanical housing the HV and LV cabling and the flat Kapton cable for the readout of the 15360 pads of the stack. Right: Zoom into the extremities of the stack to appreciate the compactness of the ensemble.}
\label{fig:asu}
\end{figure}

For the \glaser setup, ECAL-E, will be installed behind the tracker on the electron side, "symmetrically" to the ECAL-P. 
ECAL-E is a sandwich calorimeter with seven tungsten absorber plates of $2.8\units{mm}$ and eight of $4.2 \units{mm}$ thickness, adding up to $15\,X_0$, interspersed with 15 silicon sensors planes.
The sensors are of the same structure as mentioned above, with pads directly connected to SKIROC2a ASICs \cite{Callier:2011zz},
embedded inside the layer structure. Four $9\times 9 \units{cm^2}$ sensors form a sensor layer of 1024 pads, read out by 16 ASICs mounted onto a PCB. The $18\times 18\,\units{cm^2}$ sensor layer, assembled with ASICs, is called an Active Sensor Unit (ASU). 
During 2021 and 2022 a stack with 15 ASUs, equivalent to 15360 readout channels, has been completed, as shown in Fig.~\ref{fig:asu}. 
The final ECAL-E will comprise three such stacks, to cover the full energy spectrum of electrons.

\section{Expected Performance}
The performance of the ECAL-P was studied with the LUXE \textsc{Geant4} simulation assuming that a sensor plane consists of $320 \,\um$ thick silicon sensor, sandwiched between two Kapton layers supported by a thin carbon envelop, with the whole structure having a thickness of $830 \,\um$.
The sensors are subdivided into pads of $5 \times 5 \units{mm^2}$ area.
For the current studies, the energy deposited in each pad is used. Signal sharing between pads and signal digitisation will be implemented in the future, based on the results of beam-tests.

In the following, we discuss the performance of the ECAL-P for single showers, with and without background, and the expected results for high multiplicity of positrons where single showers cannot be identified and the spectrum is reconstructed through an energy-flow type of approach. The latter case is studied without background implementation. For medium range of multiplicities, we are developing a clustering algorithm which should then lead to a performance similar to the one of single showers.

\subsection{Energy and Position Resolution for single positrons}
\begin{figure}[!htbp]
    \centering
    \includegraphics[width=0.45\textwidth]{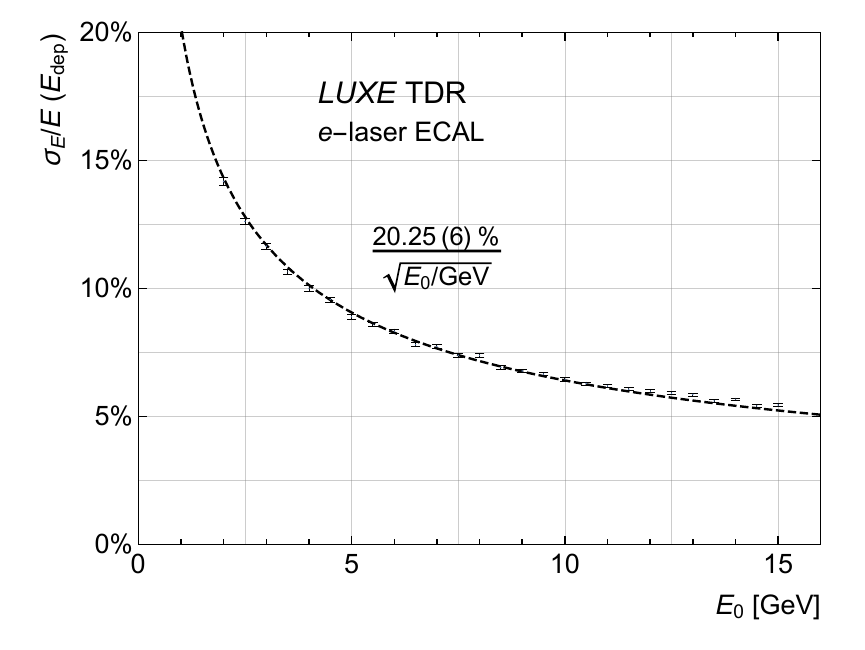}
    \includegraphics[width=0.45\textwidth]{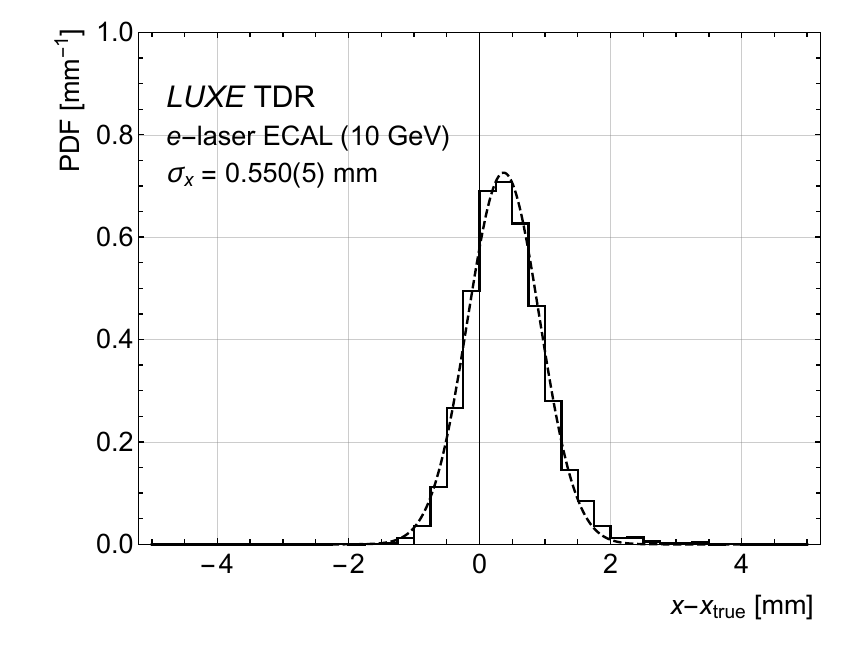}
\caption{Left: The relative energy resolution of the ECAL-P as a function of the positron energy $E_0$. Right: The probability distribution function (PDF) of the difference between the reconstructed and the generated entry point of the positron shower along the magnetic deflection ($x$) axis,  for 10\,GeV positrons.}
    \label{fig:ECAL-elaser-resolutions}
\end{figure}

The performance of the ECAL-P for single positrons was studied by generating samples of positrons in the energy ($E_0$) range between 2 and $15 \units{GeV}$ in steps of $0.5 \units{GeV}$, assuming the positrons originated from the IP. The positrons enter the calorimeter at an angle due to the deflection in the dipole magnet. After calibration, which takes into account the angular dependence of the response, the dependence of the relative resolution $\sigma/E$ as a function of the initial positron energy is shown in Fig.~\ref{fig:ECAL-elaser-resolutions} (left). A fit of the form $\sim 1/\sqrt{E_0}$ leads to
\begin{equation}
    \frac{\sigma_E}{E} = \frac{(20.2 \pm 0.1)\%  }{\sqrt{E_0/\text{GeV}}}\, .
    \label{eq:ECAL-elaser-eresolution}
\end{equation}
\begin{figure}[!htbp]
    \centering
    \includegraphics[height=0.35\textwidth]{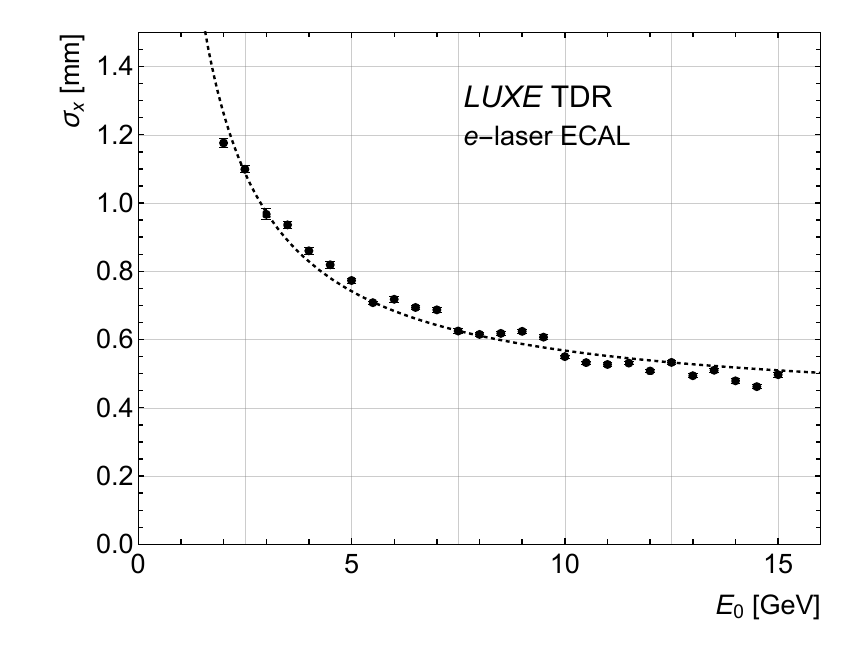}
    \includegraphics[height=0.35\textwidth]{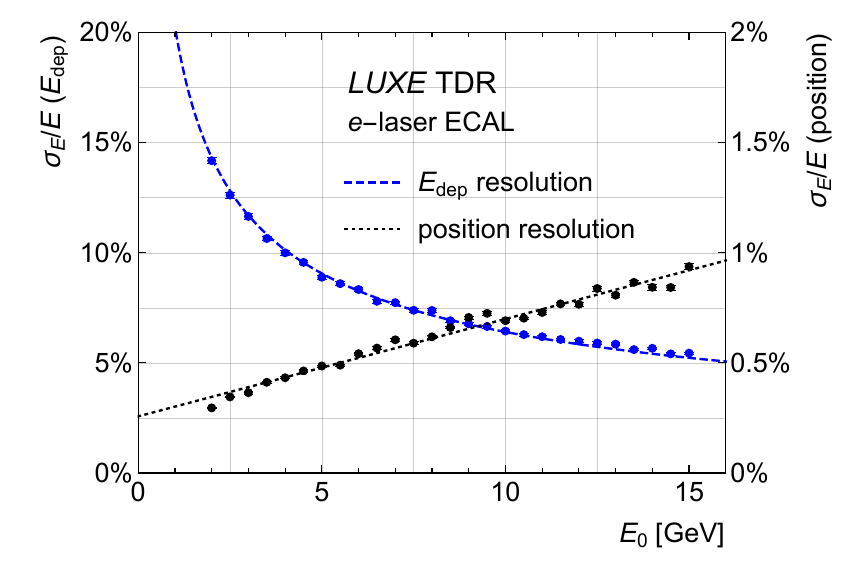}
\caption{Left: Position resolution as a function of positron energy. Right: The relative energy resolution of the ECAL-P as a function of the positron energy $E_0$ from position reconstruction (right $y$-scale) compared to the relative resolution from calorimetry (left $y$-scale).}
    \label{fig:ECAL-elaser-x-resolutions}
\end{figure}

For the position reconstruction, a logarithmic weighting of the energy deposits in the pads is applied, with a threshold optimised to give the best resolution. Along the $y$-axis, not affected by the deflection in the dipole magnet, a resolution of $\sigma_y= 460 \,\um$ is achieved (not shown). Along the $x$-axis, $\sigma_x= 550 \,\um$ at $10 \units{GeV}$ is obtained, and a mild dependence on the energy is found. However, while along the $y$-axis no bias in the position reconstruction is observed, this is not the case for the $x$ position. As shown in Fig.~\ref{fig:ECAL-elaser-resolutions} (right), for positrons of $10 \units{GeV}$, a bias of $\delta x \simeq 500  \,\um$ is generated by the fact that the positrons impact the ECAL-P under a small angle. This is not critical as this bias can be corrected for. An approach based on convoluted neural networks (CNN) was also applied to the position reconstruction. The position resolution (not shown) was not better than for the logarithmic weighting, however there was no bias in the results.

The alternative method of determining the energy for single positron showers is by making use of the correlation between energy and position of the positron entry point on the face of the ECAL-P. The relative energy resolution is then determined by the position resolution. The position resolution as a function of positron energy is shown in Fig.~\ref{fig:ECAL-elaser-x-resolutions} (left). Also shown in the figure (right) is the resulting $\sigma/E$ compared to the one from calorimetry shown in Fig.~\ref{fig:ECAL-elaser-resolutions}. Not surprisingly, the $\sigma/E$ from position reconstruction is better by factor 10 (high energy) to better than 40  (low energy) in the relevant energy range.

\subsection{Impact of background}
It is important to ensure that this performance can also be achieved in the presence of background, which is particularly sizeable near the beam-line. In this section, first the electromagnetic component of the background is considered. For the hadronic component, a preliminary result is shown in the last figure. In addition, the material of the beam exit window of the interaction chamber will be changed from Kapton to aluminium, being not yet considered in all studies. Hence, further studies 
are underway.

\begin{figure}[htbp]
\centering
    \includegraphics[width=0.49\textwidth]{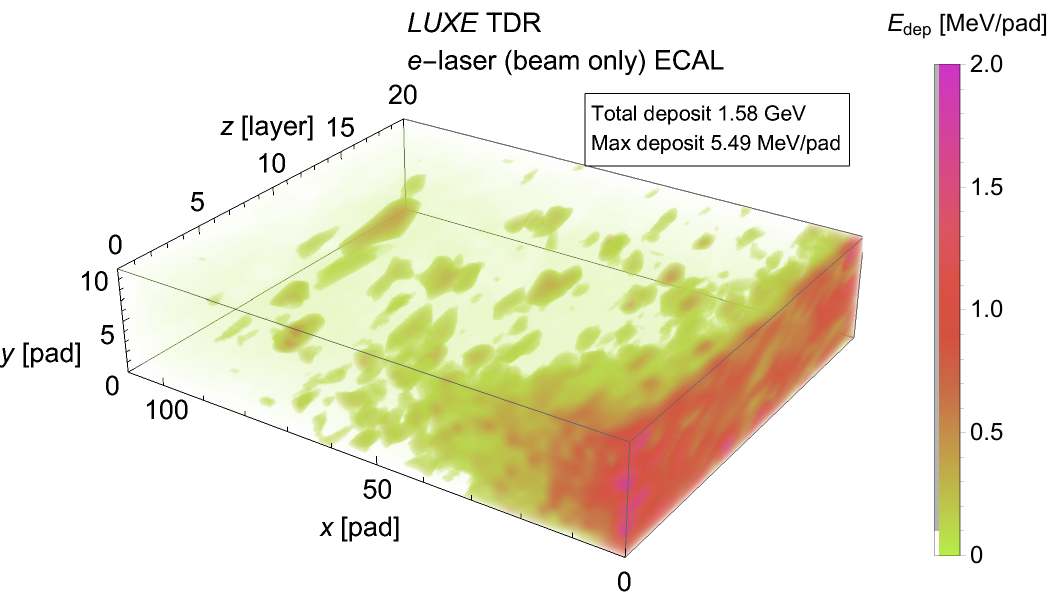}
    \includegraphics[width=0.49\textwidth]{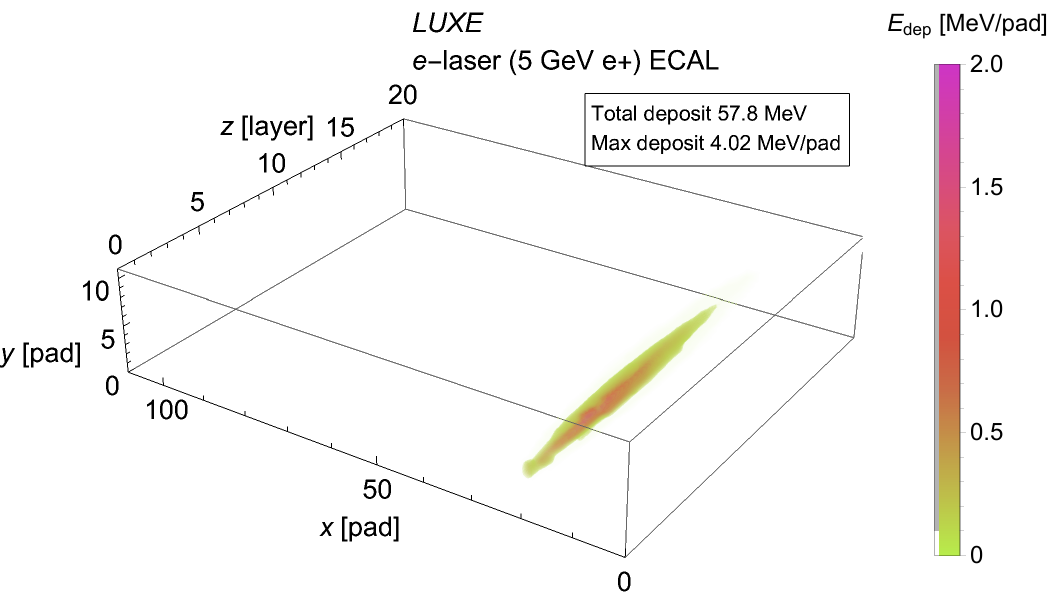}
    \caption{Left: Depositions per bunch-crossing from beam-related electromagnetic background inside the ECAL-P. The beam-pipe is located on the right side. The beam exit window at the interaction chamber is made of Kapton. Right: Depositions from a single positron of $5 \units{GeV}$.}
    \label{fig:ECAL-3D-background}
\end{figure}

The distribution of the background depositions per \elaser bunch crossing is shown in Fig.~\ref{fig:ECAL-3D-background} (left). Also shown in the figure (right) are the depositions for a single positron of $5 \units{GeV}$.
The depositions of the single positron are in a small cone around the shower axis. The background is widely spread and its contribution increases towards the beam-pipe.
\begin{figure}[htbp]
\centering
     \includegraphics[width=0.67\textwidth]{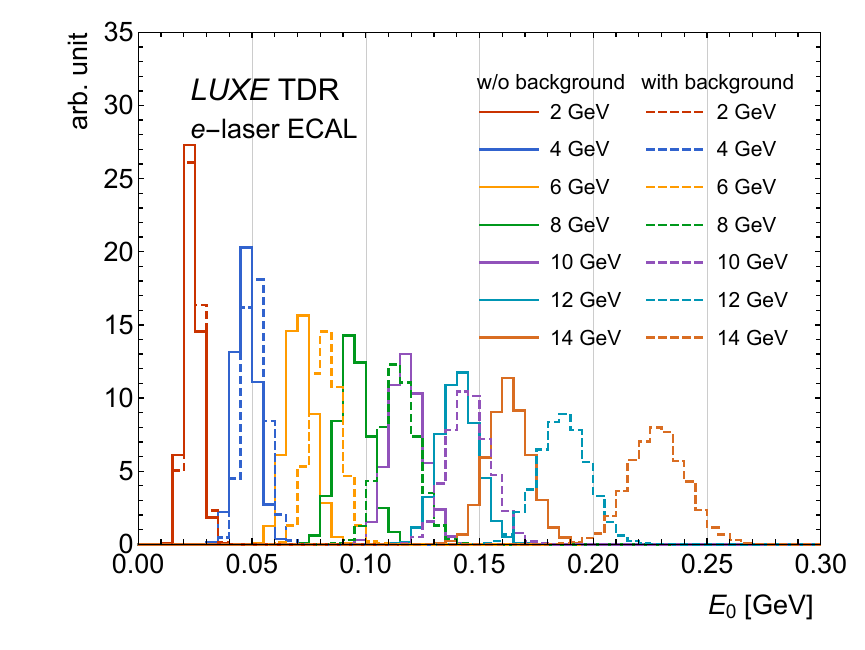}
    \caption{Distribution of total energy deposited in the sensor-pads of the ECAL-P for selected energies of positrons (full lines) and for the same positron showers with added electromagnetic background (dashed-lines) in the \elaser setup. The various energies of the positrons listed in the legend are identified by different colours. The beam exit window of the interaction chamber is made of Kapton.}
    \label{fig:ecal-singlep-edep-elaser}
\end{figure}
The effect of this background on the energy deposited in the cone around the shower axis is shown in Fig.~\ref{fig:ecal-singlep-edep-elaser} for several positron energies. The distribution of the energy deposited by positrons of a given energy is compared to the distribution obtained after adding the background to all pads containing the electromagnetic shower.
One clearly observes a shift in the total deposited energy but no significant deterioration in the resolution. As expected, the shift in energy is the more pronounced the closer it gets to the beam-line, i.e. the higher the energy of the positron. 

\begin{figure}[!hpbt]
\centering
\includegraphics[height=0.55\textwidth]{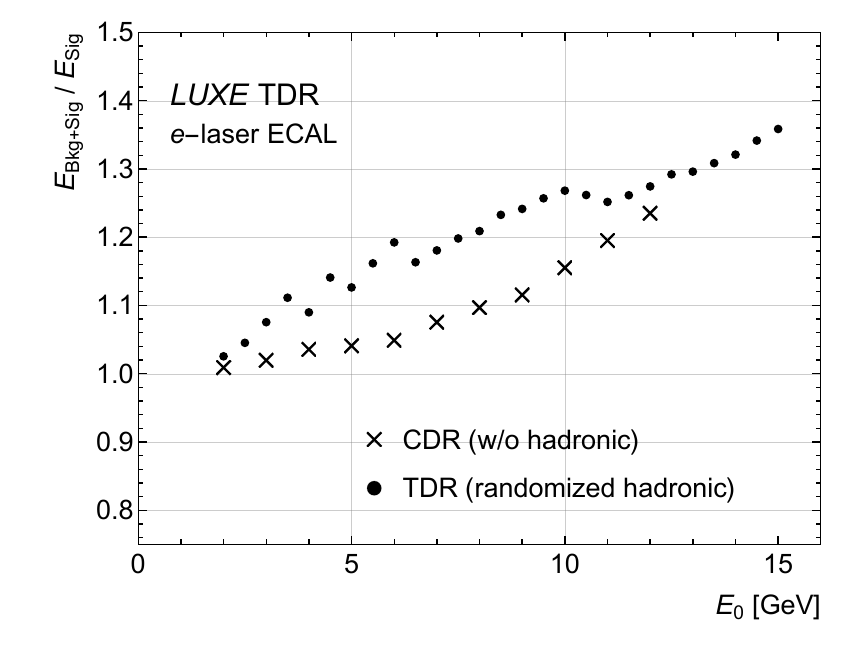}
\caption{
Ratio of the expected energy deposition in a positron shower with electromagnetic and neutron (denoted as hadronic) induced backgrounds to the one without background as a function of the positron energy $E_0$, for the \elaser setup. The distribution for the \glaser setup looks similar. The beam exit-window at the interaction chamber in these simulations is made of aluminium.}
\label{fig:ECAL-sig+bgr}
\end{figure}

In recent simulations, the depositions induced by neutrons are added \footnote{Only neutron induced depositions produced within 100 ns after the bunch crossing are considered. Depositions produced later are ignored, since they will not be registered by the read-out electronics.} and the beam exit window of the interaction chamber is made of aluminium, enhancing the electromagnetic background in the order of 10\%. 
The neutron background is estimated from a full simulation of about 2.6 \elaser bunch crossings. The majority of depositions on a pad is below one MeV. Since these depositions are smoothly distributed over the ECAL volume, their amount is normalized to the expectation of one bunch crossing and added to the signal. In a few cases, around 1\% of all hits, depositions larger than one MeV are found. These depositions are uniformly distributed over the whole ECAL volume. In the simulation of each new bunch crossing the positions of these depositions are randomly diced over the ECAL volume, and the depositions are added to the signal of the diced pads. 
In Fig.~\ref{fig:ECAL-sig+bgr} the ratio between the sum of signal and background depositions and the depositions from the signal only is shown as a function of the positron energy $E_0$ for the \elaser setup both for the electromagnetic background only, and after adding the depositions produced by neutrone. This ratio is between 1.0, at low and 1.4, at high $E_0$.
For a further suppression of the background, the shielding against photons and neutrons will be optimised. The remaining impact of the background depositions  
will finally be corrected for by direct background measurements with electron or photon bunches crossing the detector without interaction with a laser pulse.

\subsection{Reconstruction of Positron Number and Spectrum}
\begin{figure}[!htbp]
\centering
\includegraphics[height=0.35\textwidth]{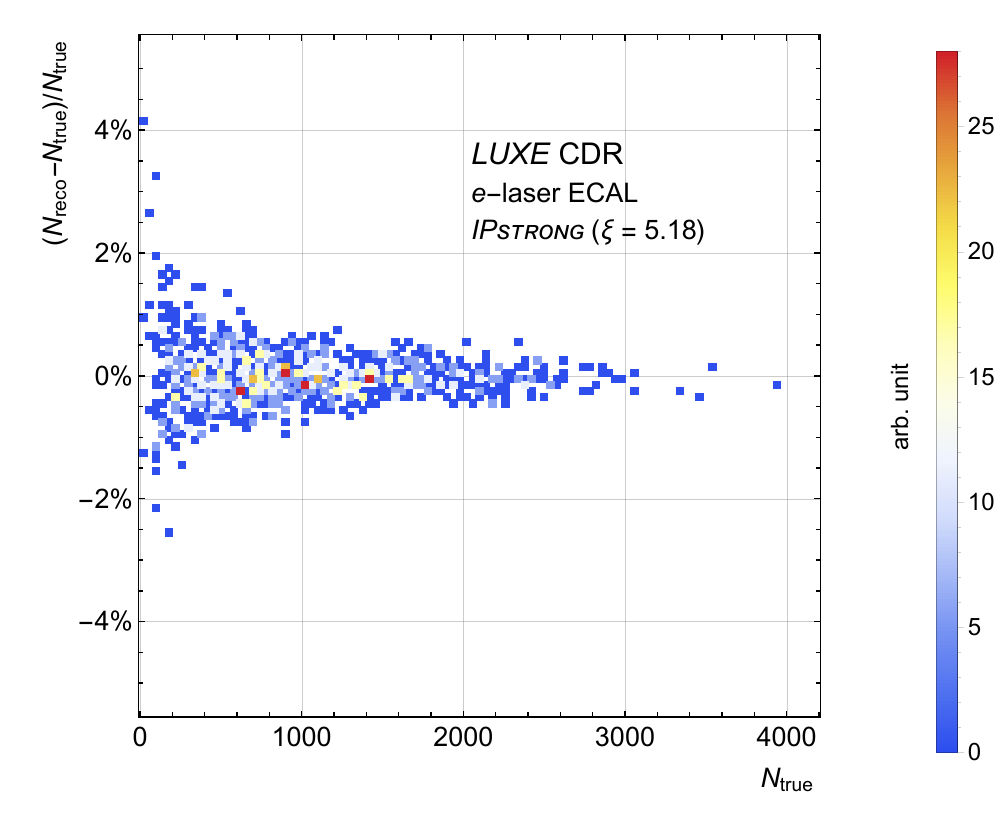}
\includegraphics[height=0.35\textwidth]{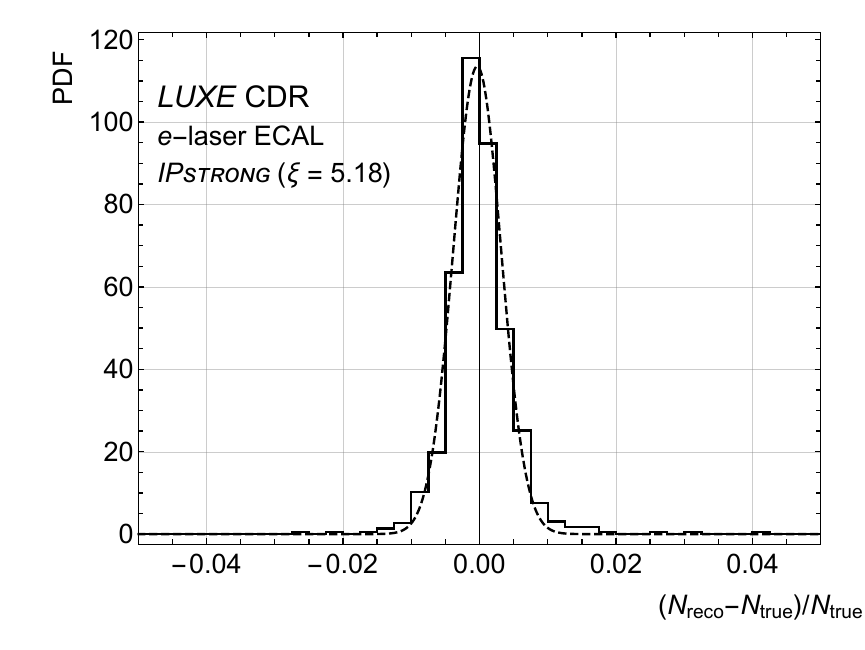}
\caption{Left: The relative spread of the number of reconstructed positrons as a function of the number of incident positrons. Right: Projection of the relative spread onto the vertical axis for an average positron multiplicity of 1000.}
\label{fig:mult-pos}
\end{figure}

\begin{figure}[!hptb]
\centering
    \includegraphics[width=0.9\textwidth]{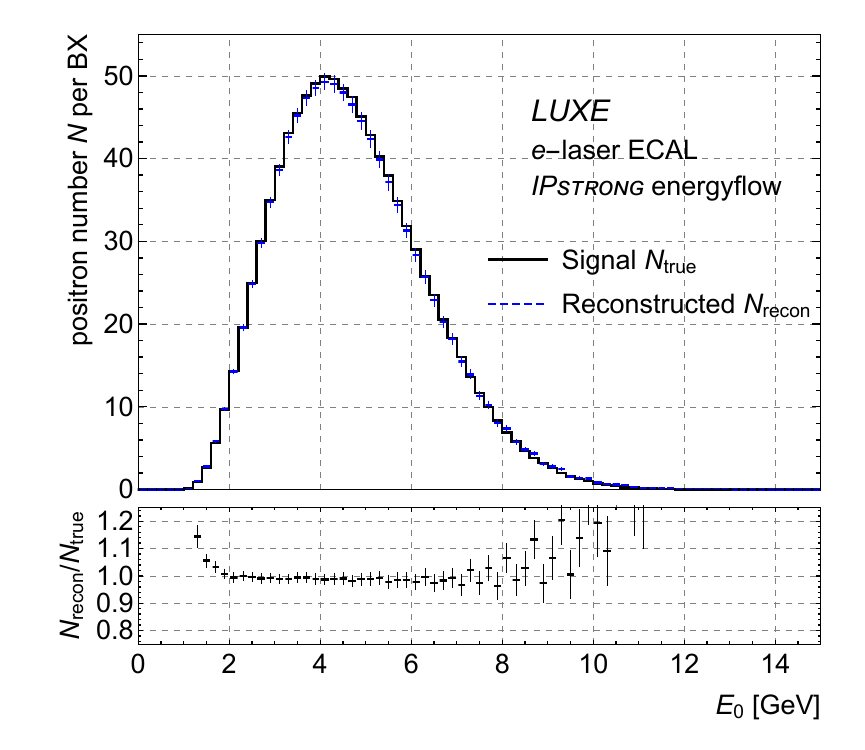}
    \caption{Comparison between the reconstructed and generated energy spectra of the positrons. The ratio of the two spectra as a function of energy is displayed in the lower panel.}
    \label{fig:ECAL-energy-spectrum}
\end{figure}

When the number of positrons entering the ECAL-P becomes large, the showers start overlapping and it becomes impossible to identify individual clusters. Instead, a method based on energy-flow is proposed~\cite{mykyta_thesis_2020}, whereby each pad contributes a fraction of the number of positrons in a given energy bin. The latter is determined from the crossing of the path connecting the centre of the pad to the IP, according to the dipole magnet field and the geometry, with the face of the ECAL-P, as if the pad was located on the axis of the shower generated by the impinging positron. At the end of this procedure, both the energy spectrum observed in the ECAL-P and the number of positrons (sum over all energy bins) are reconstructed. An overall correction of the order of 2\% has to be applied to the reconstructed number of positrons to obtain the number of generated positrons. The relative spread of the number of reconstructed positrons as a function of the number of incident positrons is shown in Fig.~\ref{fig:mult-pos} (left). For an average positron multiplicity of about 1000 the projection on the vertical axis is shown in Fig.~\ref{fig:mult-pos} (right) . The uncertainty is less than 0.5\%.
The comparison between the generated and reconstructed energy spectra is shown in Fig.~\ref{fig:ECAL-energy-spectrum}.
A good agreement between the two spectra is found, with the exception of the low-energy and high-energy tails, where leakage of the showers sets in.
The same reconstruction method was applied to multiplicities of the order of $10^5$ (not shown) and still reasonable agreement was found, though the overall correction factor increased to 5\%. However, positron multiplicities of this order very likely cannot be measured by the ECAL-P, as they would lead to distortions in the charge drift due to high charge density in the sensor and to saturation of the front-end (FE) ASICs. This problem is under study in order to determine the highest positron multiplicity that can be handled by the ECAL-P. Changing the bunch charge or the electron beam width allows at very large $\xi$ to reduce the number of positrons such that it matches the dynamic range of ECAL-P.

Studies to implement image recognition with CNN for the reconstruction of the energy spectra and multiplicities are ongoing. It turns out that with the first 10 layers of the ECAL-P one can already derive the positron multiplicity and the average energy on an event by event basis. Further studies are underway to reconstruct the full energy spectrum.

The ECAL-E has been introduced only recently into the geometry of LUXE and the study of its performance is underway. However given the similarity between the ECAL-P and ECAL-E in sampling and granularity, similar results are expected.

\section{Technical Description}
\label{ch5_techdescr}

In the following subsections the mechanical structure, the sensors, the structure of the sensor planes, the 
FE electronics and the data acquisition are described in detail.  

\subsection{Mechanical Frame and Tungsten Plates}

\begin{figure}[htb!]
    \begin{center}
    \includegraphics[width=0.8\columnwidth]{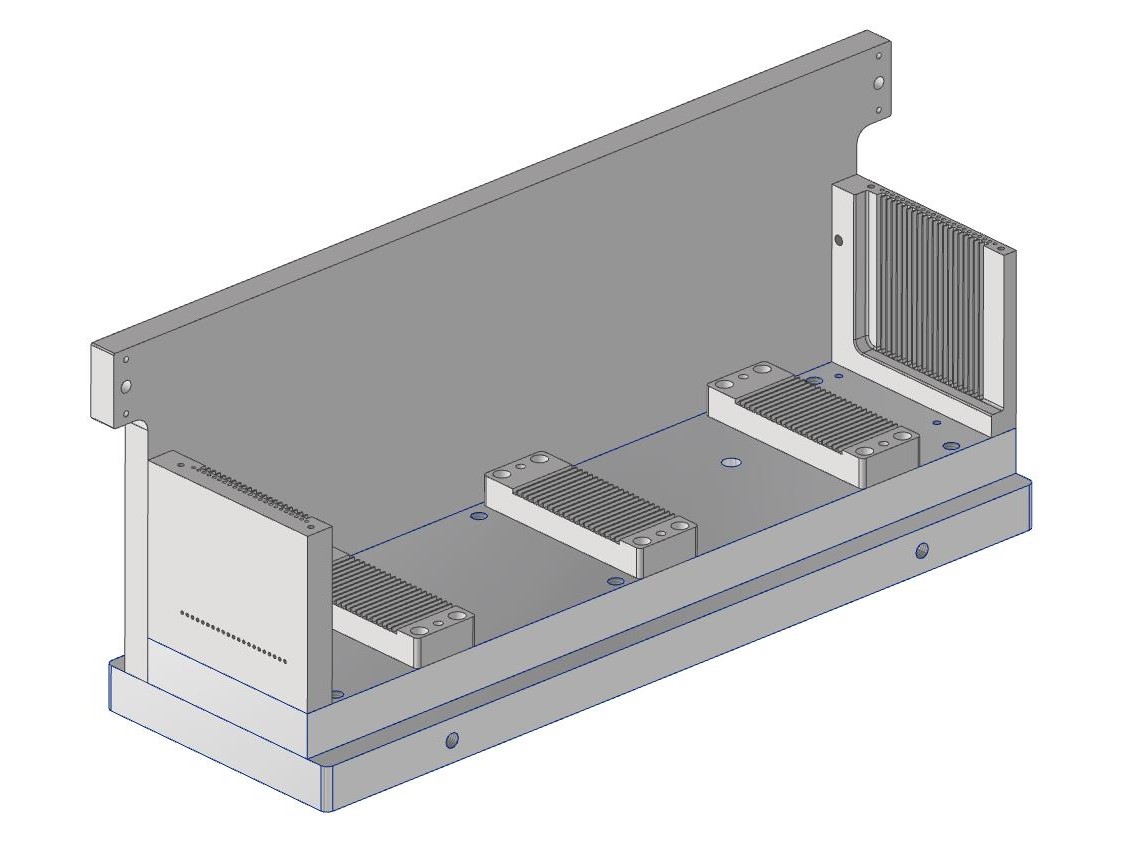}
     \end{center}
     \vspace{-1cm}
    \caption{The aluminium frame of ECAL-P with the combs for positioning the tungsten plates.}
    \label{ECAL_frame}
  \end{figure}

 Frames made of aluminium ensure the mechanical stability of the calorimeters and allow the precise positioning of sensor planes and tungsten plates. The frame of ECAL-P will be designed and constructed by the University of Warsaw. A sketch of the main aluminium frame of the calorimeter is shown in  Fig.~\ref{ECAL_frame} (see also  Fig.~\ref{ECAL_layout}).

The requirement on the mechanical frame is
to allow the insertion of 21 tungsten plates of dimensions  $550 \times 100 \times 3.5  \units {mm^3}$ parallel to each other with a gap in-between each two plates equal to $1 \units{mm}$, with a maximal deviation of $ \pm 100 \,\um $ all over the tungsten plate surface \footnote{These numbers may change slightly when the final design is completed.}. 
The positioning of the tungsten plates in the frame will be enforced by very precise aluminium combs.
The design of the mechanical structure  also includes similar combs for the positioning of the front-end electronics frames, as shown in Fig.~\ref{ECAL_layout}.
All materials used must be non-magnetic. 

The mechanical frame of ECAL-E will be designed and constructed in the IJC lab in Paris. The requirements on precision are the same as given above for the ECAL-P.

\subsection{Sensors} 
\label{Sec:sensors}

\begin{figure}[h!]
    \includegraphics[width=0.99\columnwidth]{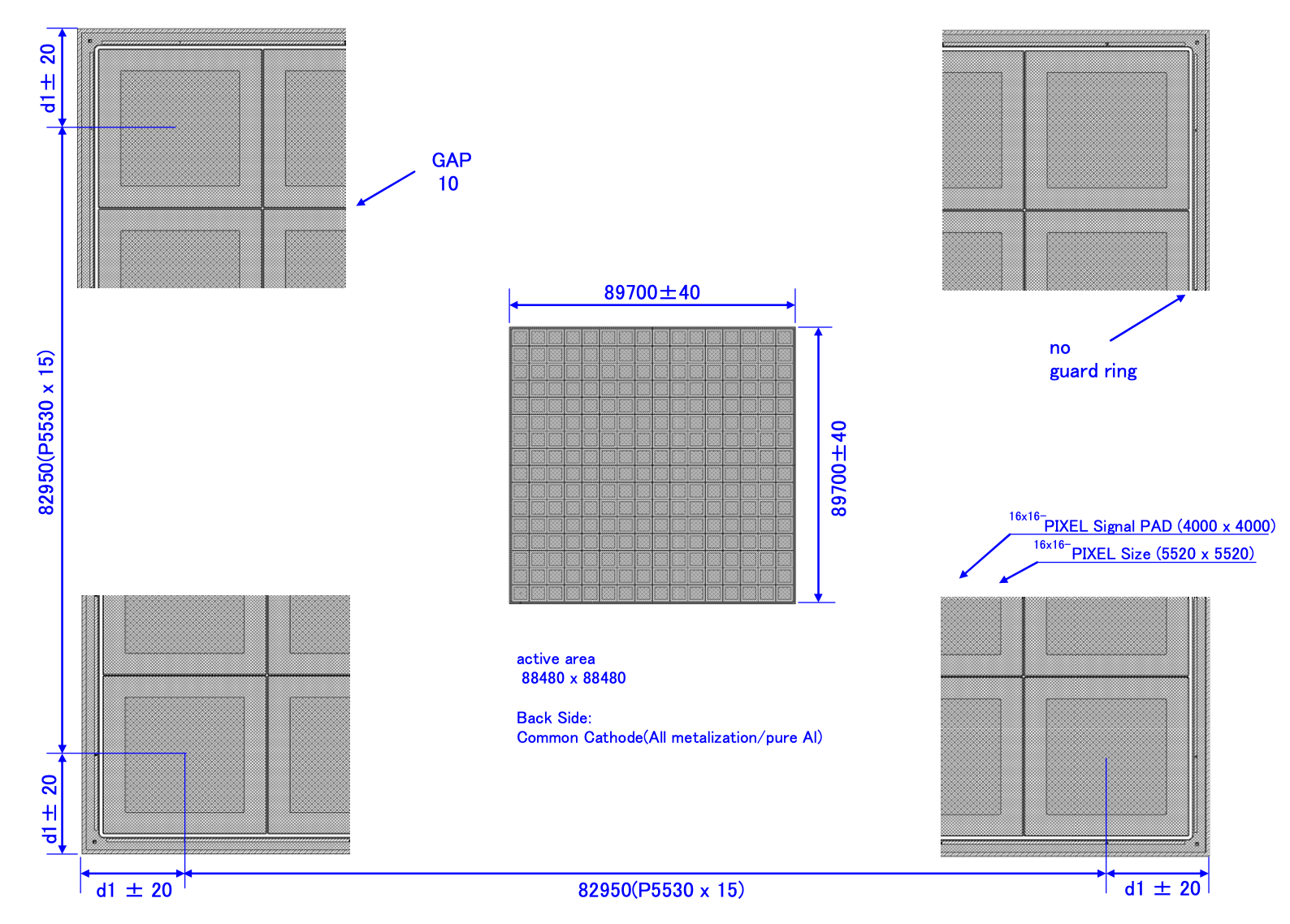}
    \caption{Details of the geometry of the CALICE sensor.}
    \label{fig:ECAL_CALICE}
  \hspace{0.025\textwidth}
\end{figure}

The silicon sensors, used both for ECAL-E and ECAL-P, are arrays of $5.5\times 5.5 \units {mm^2}$, $p$+ on $n$ substrate diodes. For ECAL-E they are made of $500 \,\um$ thick silicon with a resistivity of $3\, {k\Omega cm}$, and a reverse bias voltage of about $200\units{V}$, without guard rings. For ECAL-P sensors of $320 \,\um$
are considered.
Each sensor will have a total area of $9\times 9 \,\units{cm^2}$, corresponding to $16 \times 16$ pads. Sensors have been produced by Hamamatsu Photonics for the CALICE Collaboration, as shown in Fig.~\ref{fig:ECAL_CALICE}, and four of them were tested for ECAL-P purposes.

\begin{figure}[h!]
    \includegraphics[width=0.99\columnwidth]{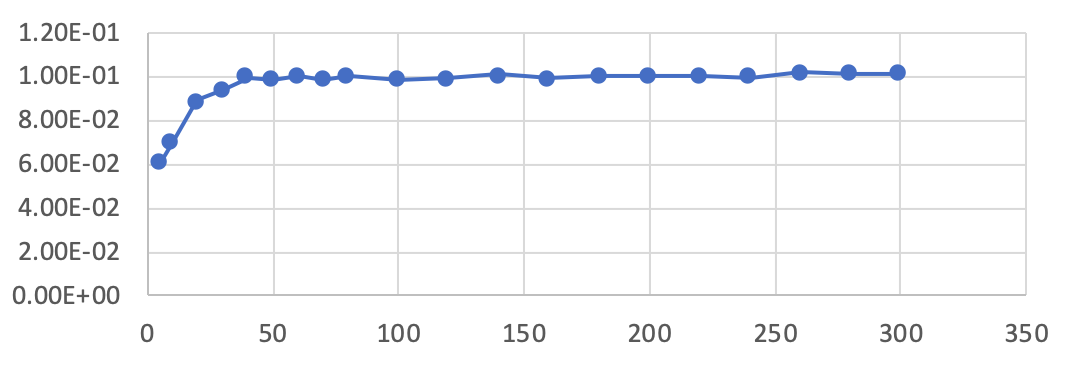}
    \caption{Leakage current (in nA) as a function of the applied voltage (in V) for a selected CALICE pad.}
    \label{fig:ECAL_CALICE_IV}
  \hspace{0.025\textwidth}
\end{figure}

 For all pads of the four sensors the leakage current was measured as a function of the bias voltage. A typical result is shown in Fig.~\ref{fig:ECAL_CALICE_IV}.


\subsection{Assembled Detector Plane}

In ECAL-P, signals from the sensor pads are routed to the FE electronics via copper traces on a Kapton PCB foil of $ 120 \,\um$ thickness. Conductive glue will be used to connect the copper traces to the sensor pads.
The bias voltage is supplied to the back-side of the sensor 
by a $70 \,\um$ flexible Kapton PCB, also glued to the sensor with a conductive glue.

\begin{figure}[htbp]
  \includegraphics[width=\textwidth]{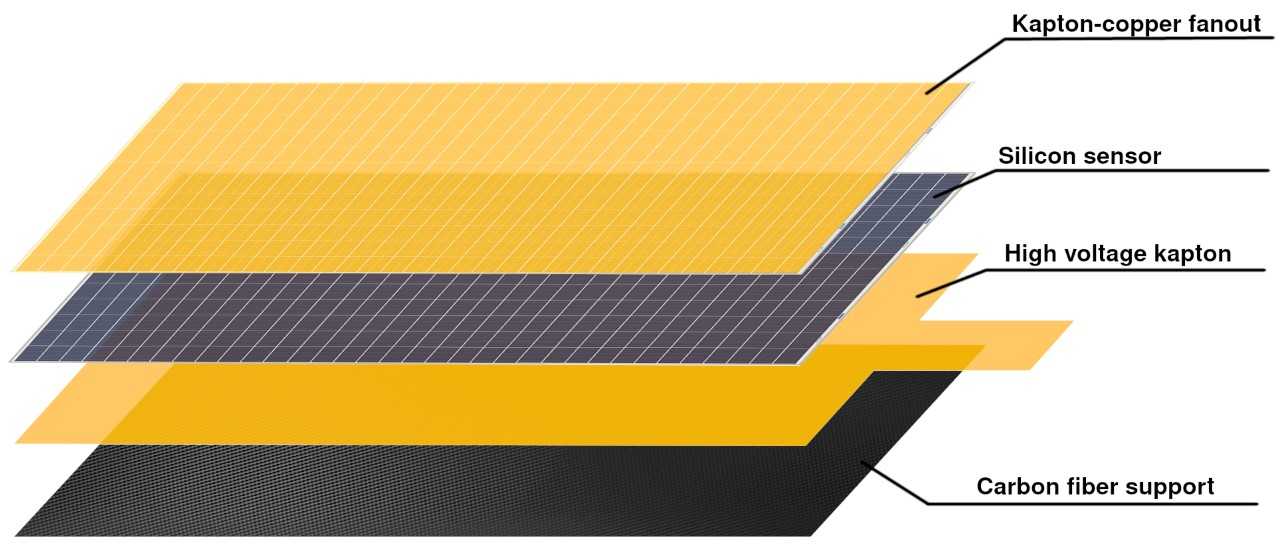}
  \caption{The structure of the assembled detector plane using a Kapton fanout for the signal routing and another PCB to supply the bias voltage. }
  \label{fig:detector_plane}
\end{figure}

 A sketch of the structure of the detector plane using Kapton PCB is shown in Fig.~\ref{fig:detector_plane}. A carbon fiber layer is used to protect the sensor.
 The total thickness of the plane is about $830 \,\um$ \footnote{To keep the detector plane thickness below 1 mm, sensors of $ 320\,\um$ are used for ECAL-P.}.
A similar technology is used for the sensor planes of the ECAL-E. However, in ECAL-E the FE ASICs are placed on the sensor plane, as described below.

\subsection{Front-end electronics} \label{sec:front_end_electronics}

In ECAL-P, each detector plane will be read out by FE ASICs called FLAXE (\textbf{F}ca\textbf{L} \textbf{A}sic for \textbf{X}FEL \textbf{E}xperiment), mounted on a PCB, denoted hereafter as front-end board, FEB, and positioned on top of the calorimeter frame.

\begin{figure}[htbp]
 \begin{center}
  \includegraphics[width=0.7\textwidth]{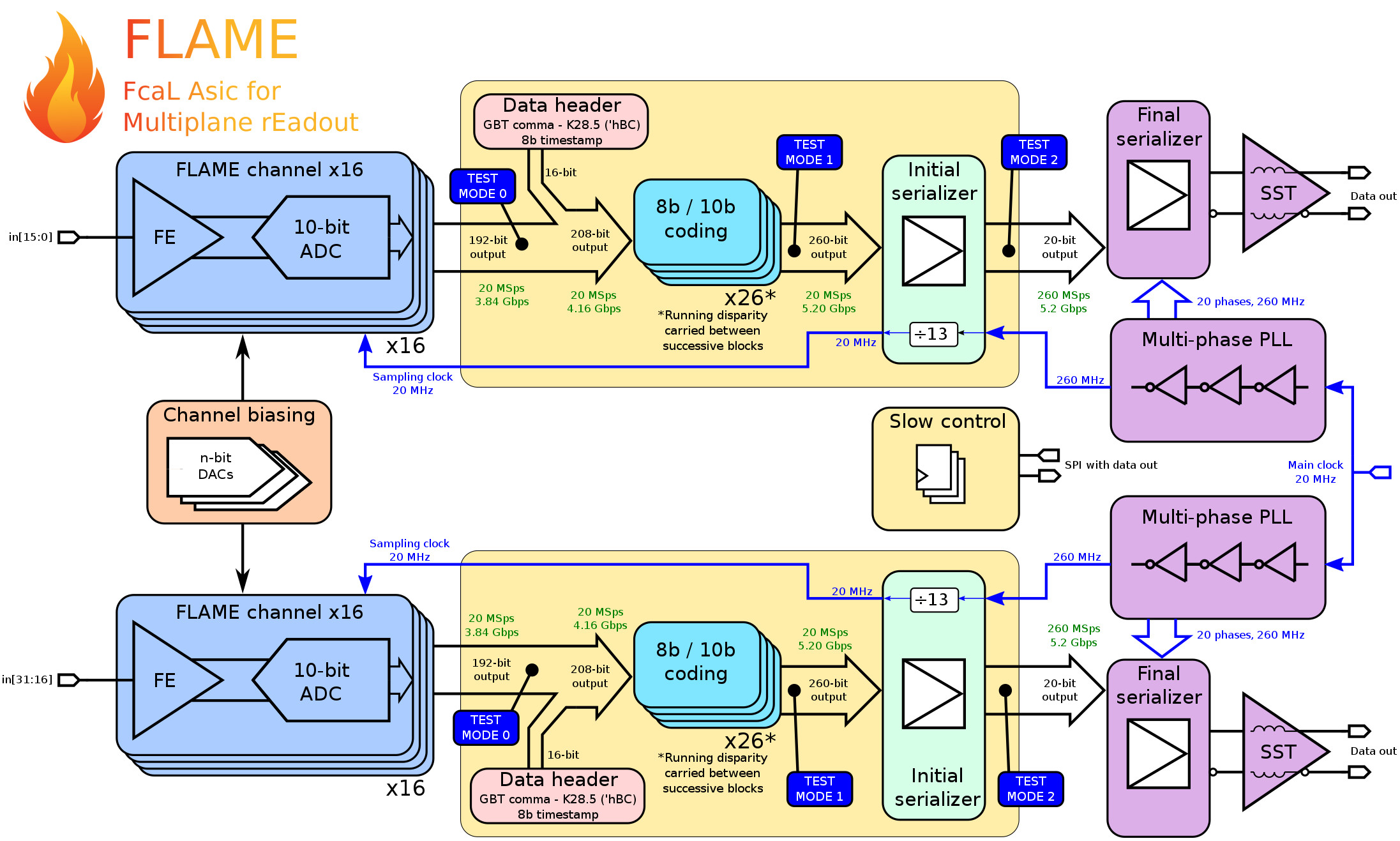}
    \caption{Block diagram of a 32-channel FLAME ASIC} 
    \label{fig:ECAL_flame}
    \end{center}
\end{figure}

On the side of the FEBs, connectors are foreseen for cables to transfer digital signals from FLAXE ASICs to the FPGAs installed at a larger distance.
The FLAXE FE electronics is based on the existing readout ASIC called FLAME (\textbf{F}ca\textbf{L} \textbf{A}sic for \textbf{M}ultiplane r\textbf{E}adout), designed for silicon-pad detectors of the LumiCal calorimeter for a future electron-positron linear collider experiment.
The main specifications of the FLAME ASIC are shown in Table~\ref{tab:flame}.
\begin{table}[htbp]
\begin{center}
\begin{tabular}{|l|l|}
\hline
Variable & Specification \\
\hline
Technology & TSMC CMOS 130~nm \\
Channels per ASIC & 32 \\
Power dissipation/channel & $\sim 2~mW$ \\
\hline
Noise & $\sim$1000 e$^-$@10\,pF + 50e$^-$/pF \\
Dynamic range & Input charge up to $\sim 6pC$ \\
Linearity & Within 5\% over dynamic range\\
Pulse shape  & T$_{peak} \sim$55~ns \\
\hline
ADC bits & 10 bits  \\
ADC sampling rate & up to $\sim 20~MSps$ \\
Calibration modes & Analogue test pulses, digital data loading\\
Output serialiser & serial Gb-link, up to 9\,GBit/s\\
Slow controls interface & I\textsuperscript{2}C, interface single-ended\\
\hline
 \end{tabular}
 \end{center}
 \caption{Summary of the specifications of the FLAME ASIC.}
 \label{tab:flame}
\end{table}
A block diagram of FLAME, a 32-channel ASIC designed in TSMC CMOS 130 nm technology, is shown in Fig.~\ref{fig:ECAL_flame}. FLAME comprises an analogue front-end and a 10-bit ADC in each channel, followed by a fast data serialiser. It extracts, filters and digitises analogue signals from the sensor, performs fast serialisation and transmits serial output data.
As seen in Fig.~\ref{fig:ECAL_flame}, the 32-channel chip is designed as a pair of two identical 16-channel blocks. Each block has its own serialiser and data transmitter so that during operation, two fast data streams are continuously sent to an external data acquisition system (DAQ). The biasing circuitry is common to the two 16-channel blocks and is placed in between. Also the slow control block is common and only one in the chip.
The analogue front-end consists of a variable gain preamplifier with pole-zero cancellation (PZC) and a fully differential CR--RC shaper with peaking time $\sim 55 \units{ns}$. 
The shaper includes also an 8-bit DAC, with 32 $\units{mV}$ range, for precise baseline setting. The analogue front-end consumes in total 1\,--\,1.5\,mW/channel.
The ADC digitises with 10-bit resolution and at least 20\,MSps sampling rate. 
The power consumption is below 0.5\,mW per channel at 20\,MSps. 
In order to ensure the linearity of the ADC, the input switches are bootstrapped, reducing significantly their dynamic resistance.

The architecture of the FLAXE ASIC will be very similar to the existing FLAME design. In particular the same analogue signal processing scheme will be kept. A few modifications are needed, mainly in the digital part.
The fast data transmission components will be replaced by a simpler and slower data transmitter to reduce the complexity and the cost of the subsequent FPGA-based back-end and DAQ system. This is possible, since the data rate and data volume in LUXE is much smaller than for LumiCal. 

The dynamic range of the ASICs can be switched between high and low gain. At high gain the response to the input charge is almost linear between deposition from 0.5 to 75  
minimum-ionising-particle (MIP) equivalent. At low gain the charge at the preamplifier input is limited to $5\units{ pC}$, corresponding to about 10000 MIP-equivalent, and the lower threshold is in the range of a few 10 MIP-equivalent.

The power dissipation of the FE electronics amounts to a maximum value of about $50 \units{W}$. However, it can be reduced by roughly a factor 10$^3$ by power pulsing, foreseen for application in LUXE. Hence for normal operation, cooling will not be needed.
However, for calibration and alignment it is planned to take data with cosmic muons. For this case, dedicated air cooling of the FEBs will be foreseen.

\begin{figure}[htbp]
 \begin{center}
  \includegraphics[width=0.4\textwidth]{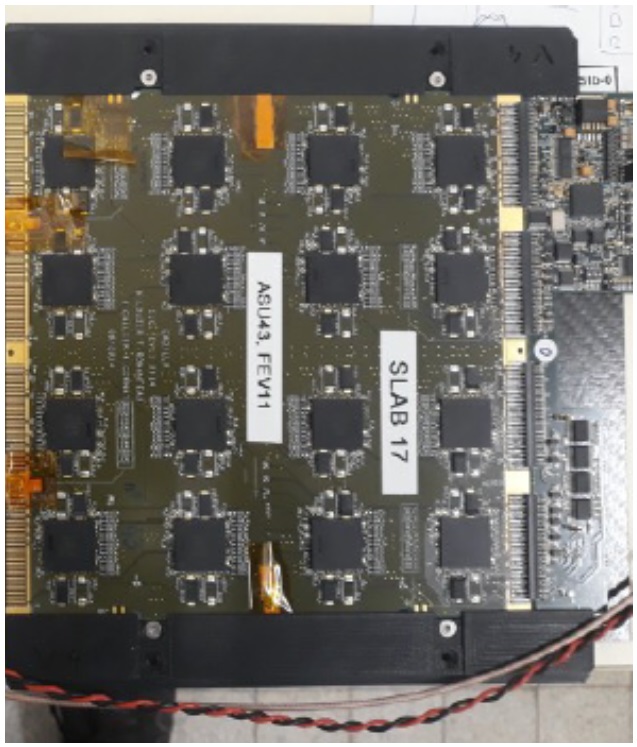}
    \caption{A fully instrumented sensor plane for ECAL-E. The SKIROC2a ASICs are positioned on the sensor plane pad sensor}.
    \label{fig:ECAL_ECAL-E_plane}
    \end{center}
\end{figure}

In ECAL-E, SKIROC2a ASICs, integrated in the sensor plane, are used. A picture of an assembled sensor plane is shown in Fig.~\ref{fig:ECAL_ECAL-E_plane}.
The main specifications of the SKIROC2a ASIC are shown in Table~\ref{tab:skiroc2}.
\begin{table}[htbp]
\begin{center}
\begin{tabular}{|l|l|}
\hline
Variable                  &           Specification \\
\hline 
Technology                &           AMS SiGe 350~nm \\
Channels per ASIC         &          64          \\
Power dissipation/channel & $\sim 6.2 ~mW$ \\
\hline
Dynamic range             & Input charge up to $\sim 10pC$ \\
                          & high and low gain             \\
Linearity                 & Within 1\% over dynamic range\\
Pulse shape               & T$_{peak}$ 30-120~ns, tunable \\
\hline
ADC and TDC bits                  & 12 bits  \\
\hline
 \end{tabular}
 \end{center}
 \caption{Summary of the specifications of the SKIROC2a ASIC.}
 \label{tab:skiroc2}
\end{table}

\section{Interfaces and Integration}
\subsection{Trigger}

\par
\begin{figure}[h!]
    \includegraphics[width=0.99\columnwidth]{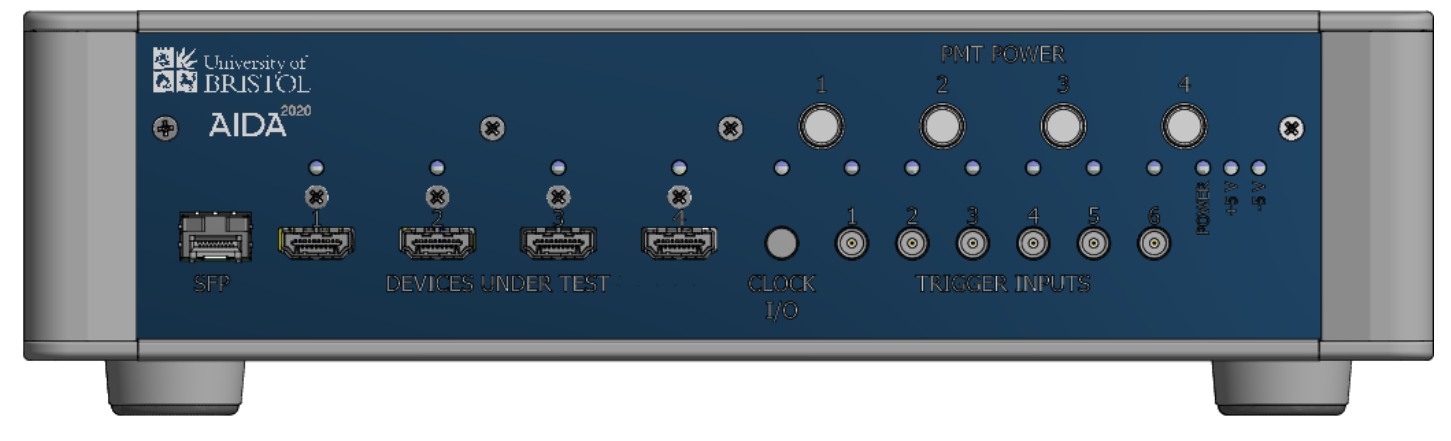}
    \caption{Front view of the TLU.}
    \label{fig:ECAL_TLU_pic}
  \hspace{0.025\textwidth}
\end{figure}

For the normal operation of ECAL-E and ECAL-P, the trigger will be provided by the Trigger Logic Unit (TLU), as shown in  Fig.~\ref{fig:ECAL_TLU_pic} and described in detail in the LUXE DAQ technical note~\cite{LUXE-DAQ}.

The TLU will distribute to all the sub-detectors the trigger signal coming from
\begin{itemize}
    \item the electron beam with a frequency of $10 \units{Hz}$,
    \item the laser beam with a frequency of $1 \units{Hz}$.
\end{itemize}
In ECAL-P, the trigger signal will be sent to the FPGA-based back-end, which will then distribute it to the FLAXE ASICs located at the FEBs via dedicated differential links. The FPGA will perform a programmable phase shift of the trigger signal in order to wake up the FLAXE ASICs just before the next bunch crossing using the power pulsing feature. Effectively, the FPGA will use the trigger originating from a particular 
bunch crossing to predict the next one, to start the analogue and the mixed-mode signal processing, as well as the data acquisition in the FLAXE ASICs. The ASICs will remain active, writing the data into an internal circular buffer until the TLU trigger signal will be received via the FPGA. The data recorded in the bunch crossing will be held in the internal FLAXE ASICs memory and successively readout by the FPGA between bunch crossings.  

During the construction of the ECAL-P at DESY, a cosmic muon trigger will be installed. It will consist of two scintillators with photomultipliers (PM). The signals from the PMs will be feed into the TLU, providing the trigger. The power pulsing will be switched off, and the ASICs will restart automatically as soon as the data is readout by the FPGA.

\subsection{Data Acquisition} 
In ECAL-P, the six silicon detector planes in each ECAL-P layer will be equipped with a single FEB, common for all sensors in the layer. All FLAXE ASICs on a FEB will be connected in parallel to the common data bus. The data bus will use a SPI-like, synchronous, serial protocol in the physical layer, with three differential links for the bus clock, SCLK, in the range of 20~--~60\,MHz, and two one-directional serial data links, providing full-duplex communication. Commands will be sent from the FPGA to the ASIC in the regime `Master Output'--`Slave Input' (MOSI) and data from the ASIC to the FPGA as `Master In'--`Slave Output' (MISO). However, in the logical layer, the bus will implement an I2C-like protocol, allowing the FPGA to address and communicate with each ASIC independently. The event data from a single ECAL-P layer will be read out from the FLAXE ASICs sequentially, one ASIC after the other. This bus will also serve the slow control and the configuration interface to the ASICs.

Up to 128 raw ADC samples will be collected in an event for each of the readout channels. With 1536 channels per single ECAL-P layer and a 10~bit ADC resolution, the event size amounts up to 1.56\,Mb. With 10\,Hz of bunch crossing rate, this results in about 15.6\,Mb of data per second read out from a single FEB. Therefore the 20\,MHz bus clock frequency provides a sufficient bandwidth for a regular operation. A higher bus clock frequency is foreseen for the data acquisition of the asynchronous muon calibration mode, in order to reduce the dead time of the system. 

The data from each FEB will be sent to the FPGA board. Since only three, relatively slow, differential links are needed for a single FEB, the data of the whole ECAL-P can be read out and processed by a single FPGA. The signals from the FEBs are fully digital, hence the FPGA board can be positioned at a larger distance from the calorimeter, e.g. in the calorimeter rack.
The FEBs will be read out in parallel via different links. 
Hence, the bandwidth which is sufficient to read out a single FEB will also be sufficient to read the whole ECAL-P between consecutive bunch crossings. 

The event will be processed by the FPGA. The signal reconstruction comprises the calculation of the deposited charge and the Time Of Arrival (TOA) from the collection of the raw ADC samples in a given channel. This procedure, combined with zero suppression, will significantly reduce the data rate. Finally the remaining event data will be sent from the FPGA board to the DAQ computer using the User Datagram Protocol (UDP) through a single 1\,Gbps Ethernet link. In parallel, the Transmission Control Protocol (TCP) over the same link will be used for the system control and monitoring.

\section{Installation, Commissioning and Calibration}

\subsection{Detector assembly}

The ECAL-E will be fully assembled and tested at LAL Orsay and shipped to DESY. After arrival, a system test will be performed and the status of all components will be documented. The time foreseen is two weeks. 

The ECAL-P assembly requires the presence of participants from four institutes engaged in building the ECAL-P:
\begin{itemize}
    \item University of Warsaw (UW) - for the mechanical frame and the tungsten plates.
    \item Tel Aviv University (TAU) - for installing the sensor planes between the tungsten plates.
    \item Akademia Gorniczo Hutnicza - University of Science and Technology (AGH-UST) - for installing the readout electronics.
    \item Institute of Space Science (ISS) - for software support and slow control.
\end{itemize}
The assembly will take place at DESY. The tasks to be performed are: 
\begin{itemize}
    \item inserting the tungsten plates, plate by plate;
    \item connecting the FEBs to the six sensors in each detector plane;
    \item inserting the detector planes;
    \item performing a cosmic run for function tests of components.
\end{itemize}
It is assumed that it will take a day to install one detector plane and perform the function tests. The detector plane will be inserted between the absorber plates positioned in the mechanical frame followed by cabling and connection of FEBs, HV and LV as well as DAQ. After performing a noise check, data from cosmics will be collected during the night. This cycle will be repeated for the 15 detector planes.
It will thus take 15 days to complete these tasks. This will be followed by a full system test for about 5 days. Another 10 days will be needed to perform the ECAL-P calibration in the DESY test-beam. The full assembly is thus expected to take about a month. With an added contingency to solve unforeseen problems, the total time estimate for ECAL-P readiness to be moved into the experiment is thus 1.5 months.

During the phase of the construction of the ECAL-P, the DAQ should be able to work in a standalone mode with the cosmic trigger. Once the ECAL-P is fully tested, the whole system is moved to the pit and the DAQ PC of ECAL-P will be included in the central LUXE DAQ.

\subsection{Installation}

For the final installation in the experimental area, it is assumed that
\begin{itemize}
\item{the tables on which the ECAL-E and ECAL-P is placed are installed;}
\item{all LV, HV and data cables are installed to connect the ECAL-E and ECAL-P subsystems with the rack.}
\end{itemize}
The installation steps, with the estimated duration and the number of participants, are as follows:
\begin{itemize}
 \item move the ECAL-E and ECAL-P to the area and place it on the table. The rough weight estimate is $50 \units{kg}$ and thus a crane will be needed.\\
 Duration: 1 day; \\
 person power: 1 technician (DESY), 2 physicists for ECAL-E and ECAL-P each;
 \item move the electronics racks to the area next to the ECAL-E and ECAL-P (time and person-power included in the previous item);
 \item perform the survey, to precisely define the position of ECAL-E and ECAL-P.\\
 Duration: 1/2 day; \\
 person-power: experts from DESY/XFEL survey, 1 physicist for ECAL-E and ECAL-P each;
 \item connect the cables between ECAL-E and ECAL-P and the rack. \\
 Duration: 1 day; \\
 person power: 1 electronic engineer, 1 technician, 1 physicist ECAL-E and ECAL-P each;
  \item Testing of HV,  LV and data connections.
  Duration: 1 day plus 2 days in reserve for potential replacements;\\ person-power: 1 electronics engineer, 2 physicists ECAL-E and ECAL-P each.
\end{itemize}

Hence between 3.5 to 5 days, each, are needed for the installation of the ECAL-E and ECAL-P in the area. The process will be performed by 2 physicists, 1 electronic engineer and 1 technician (partly provided by DESY), and the DESY survey team. 

\subsection{Commissioning}

After installation, the full system test will be repeated. Low voltage settings and currents drawn by the sensors and the FE ASICs will be checked.  Pedestal data will be collected with a special trigger, and the pedestal values and width monitored, as a proof of stability and readiness for operation. 

\subsection{Calibration Strategy}
The ECALs will be calibrated in electron beams of precisely known energy. 
The electron-beam energy available at DESY is limited to $6 \units{GeV}$. The positron/electron energy range expected in LUXE for Breit-Wheeler type of processes extends to about $15 \units{GeV}$. Therefore beam-tests at CERN will be required. 

To save time and minimise the risks of mechanical damage to the ECAL-P, the following strategy will be adopted. The existing FCAL tungsten prototype mechanical structure will be equipped with 20 sensor planes and FLAXE readout boards. It will then be tested and calibrated in the DESY and CERN test-beams. This will allow to develop the proper MC simulation with digitisation within \geant. The full ECAL-P will be exposed to the DESY test-beam and the comparison between the results obtained with the FCAL prototype and the ECAL-P
at low electron energy will then be used to project the performance of the full ECAL-P at higher electron beam energies. This approach provides more flexibility in the access to test-beams.

Tuning of the calibration after installation will be done in special runs foreseen in LUXE using a photon beam. A needle will be inserted near the IP to convert photons to electron-positron pairs. These pairs will be used to calibrate the tracker momentum by imposing the constraint that the invariant mass of the pair has to be zero. The "calibrated" tracks will then be used to cross-calibrate the ECALs. The expected momentum resolution of the tracker is within factor two the same as the one of the calorimeter when deduced from the electron or positron position reconstruction.

\subsection{Decommissioning}
The decommissioning of the ECALs will be done by the laboratories having developed and built the ECALs.
The decision about a potential further use of ECAL-E or ECAL-P, or  its dismantling, will be taken by the involved parties before the completion of the LUXE  experiment. In case of dismantling, each laboratory will take back the delivered components. 
The time needed for decommissioning depends on a potential activation of the material, e.g. near the beam line. About a potential intermediate storage at DESY, the parties will propose an agreement with DESY.

\section{ORAMS: Operability, Reliability, Availability , Maintainability and Safety}

The number of readout channels of ECAL-P and ECAL-E is about 25000, orchestrated by FPGAs. Since the readout frequency during normal data taking is 10 Hz, there should be no problem with the speed of readout and data transfer. The radiation dose for sensors and FE electronics is below a critical level, hence radiation damage is not expected. Solid state sensors are devices of high reliability. There will be no danger of over-voltage break-through or sparks. The change of the leakage current due to temperature changes will be kept small by air conditioning in the area, with a maximum temperature drift of 
a few $^{\circ}\mathrm{C}$.
The ECAL-P and ECAL-E are modular. In case of malfunctioning of a detector plane it can be replaced by a spare part within a few days of access. For the replacement of faulty FE ASICs or FPGAs, less than a day will be needed. The same holds for the Low Voltage and High Voltage power supplies.
The bias voltage for the sensors will be around 100 V. High Voltage connectors and cables will be used, matching the safety requirements.


\section{Further Tests Planned}
 
\subsection{Test of components and the whole system in the beam}

\begin{figure}[h!]
   \begin{center}
    \includegraphics[width=0.7\columnwidth]{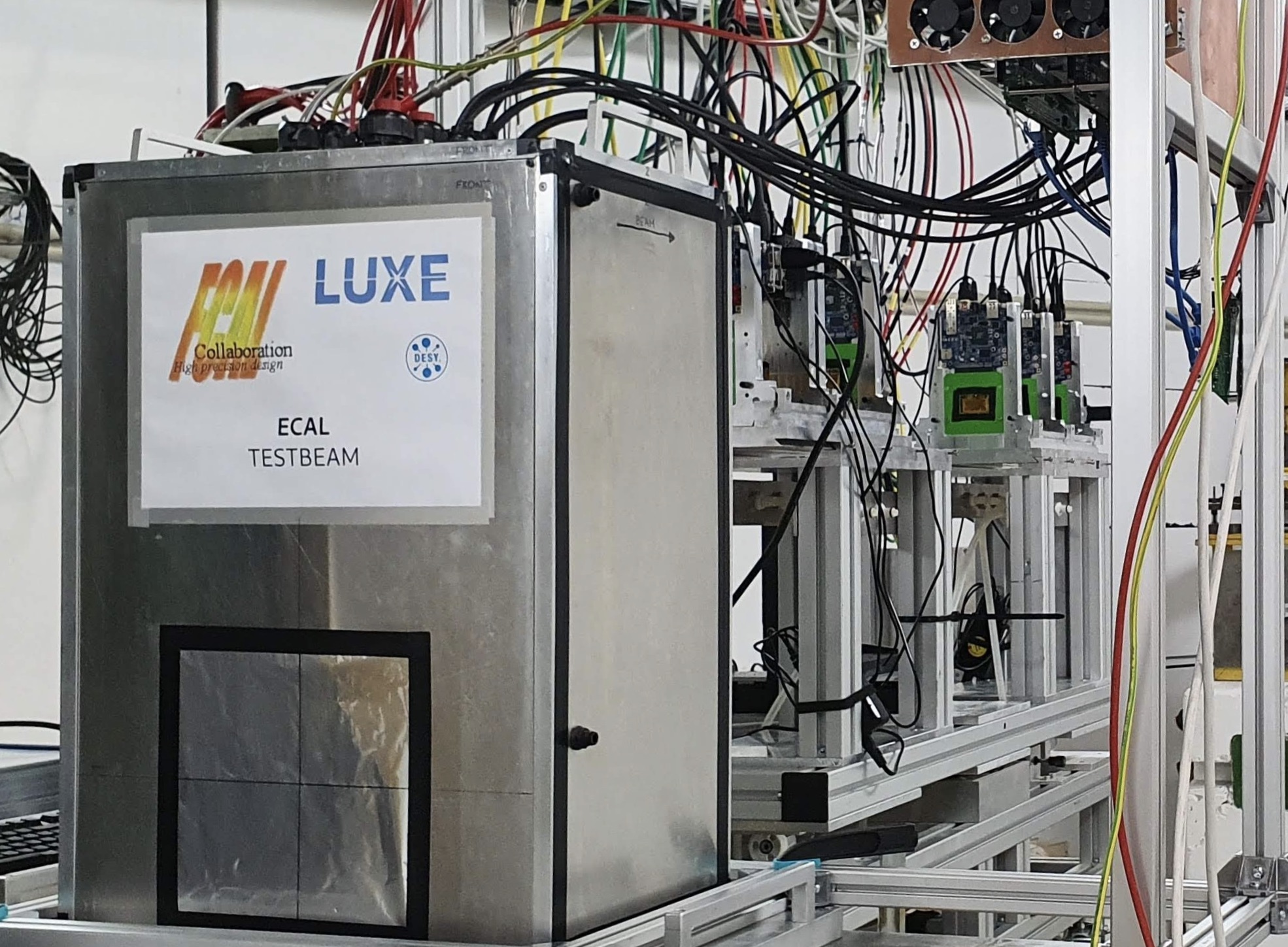}
    \end{center}
    \caption{The test-beam set-up for the measurements of the sensor response. Left in the foreground is the detector box with the sensors and FE-electronics inside. In the background on the right is the pixel telescope, used to measure the electron trajectory. }
    \label{fig:ECAL_test-beam_setup}
  \hspace{0.025\textwidth}
\end{figure}

\begin{figure}[h!]
    \centering
\includegraphics[width=0.65\columnwidth]{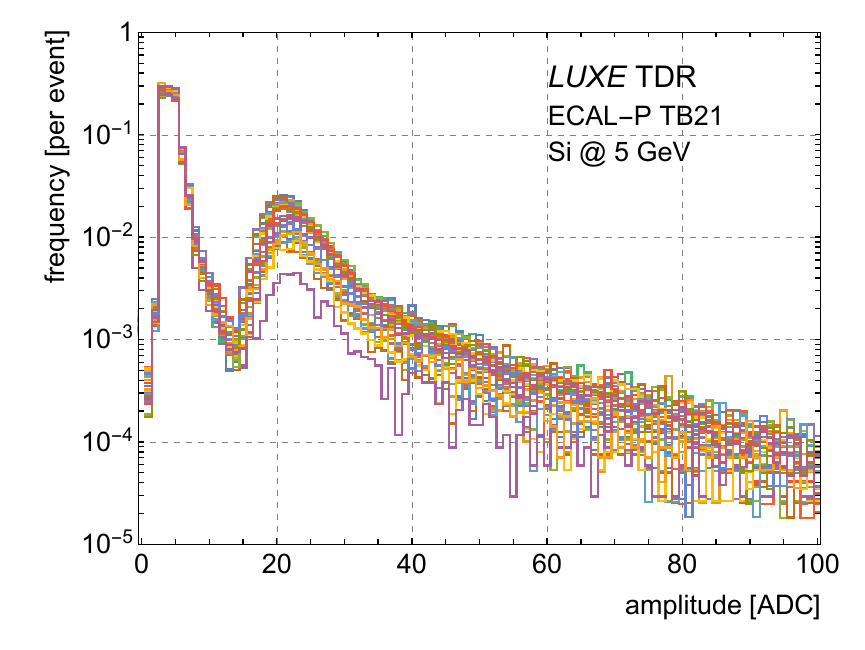}
    \caption{Distribution of the signal size in the $500 \,\um$ thick silicon sensor measured with electrons of 5\,GeV.}
    \label{fig:ECAL_sensor_response}
  \hspace{0.025\textwidth}
\end{figure}

 Studies of sensor response of a silicon pad sensor prototype to electrons with energy in the range of 1 to 5\,GeV have been performed in the DESY II electron beam at the end of November 2021. A picture of the set-up is shown in Fig.~\ref{fig:ECAL_test-beam_setup}. The detector box with the sensors under study is located downstream of the pixel telescope used to measure the beam-electron trajectory. 
Due to an unexpected electrostatic discharge event early on in the test campaign, the trigger synchronisation between the telescope and the sensors was lost and no detailed scanning of the sensor response as a function of the electron impact could be studied. The preliminary results of the signal distribution in a silicon pad sensor is shown in Fig.~\ref{fig:ECAL_sensor_response}.
As can be seen a clear signal of minimum ionising particles is visible, well separated from the noise. The size of the signal matches the expectations from the energy deposited by the relativistic electrons in the sensor material. Note that zero-suppression is applied at the readout level with a threshold of $2 \sigma$, $\sigma$ being the pedestal width, above the pedestal mean value. 
The signal distributions measured for several pads are fitted with a convolution of a Landau distribution with a Gaussian, and the most probable value, $\mu_\mathrm{Landau}$, is determined. 

\begin{figure}[h!]
   \centering
\includegraphics[width=0.65\columnwidth]{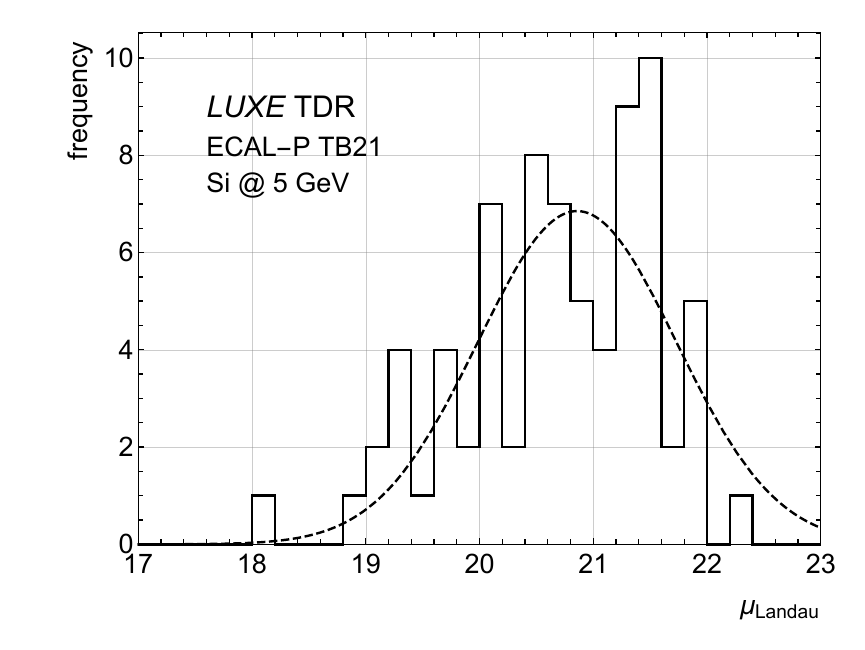}
    \caption{Distribution of the most probable value, $\mu_\mathrm{Landau}$, from fitting a convolution of a Landau and a Gaussian function to spectra measured in pads in the silicon sensor.}
    \label{fig:ECAL_sensor_response_homogeneity}
  \hspace{0.025\textwidth}
\end{figure}

The distributions of the $\mu_\mathrm{Landau}$ values for the silicon sensor, are shown in Fig.~\ref{fig:ECAL_sensor_response_homogeneity}.
The width of the convoluted Gaussian comes out to be compatible with the pedestal width (about 1 ADC count).
Within uncertainties the distribution of $\mu_\mathrm{Landau}$ as a function of the pad number is uniform.

Another campaign of beam tests was carried out in fall of 2022 to perform a detailed scanning of the sensors uniformity, cross talk and edge effects using the telescope. The data are being analysed.

The DAQ of ECAL-E and ECAL-P will be integrated into the test-beam EUDAQ system with other LUXE detectors in the test-beam.


For large values of $\xi$ where the multiplicity of positrons may reach up $10^4$ particles, there is a risk that the ECAL-P sensors or/and the front-end boards will loose their linear behaviour. The effect of these potentially high charge depositions will be studied in the recently completed DUAL HIGGINS high-power laser facility of Prof. Victor Malka at the Weizmann Institute, where up to $10^7$ electrons of $0.7 \units{GeV}$ (with 5\% energy spread) can be produced per laser shot.

\section*{Acknowledgements}
This study was partly supported by the European Union’s Horizon 2020 Research and Innovation programme under  GA no. 101004761,
by the German-Israeli Foundation under grant no. I-1492-303.7/2019, the Israeli PAZY Foundation (ID 318), the National Science Centre, Poland, under the grant  no. 2021/43/B/ST2/01107, the Romanian Ministry of Research, Innovation, and Digitalization under Romanian National Core Program LAPLAS VII – contract no. 30N/2023 and under Romanian National Program FAIR-RO - contract no. FAIR\_02/2020,  the Generalitat Valenciana (Spain) under the grant no. CIDEGENT/2020/21, and the MCIN with funding from the European Union NextGenerationEU and Generalitat Valenciana in the call Programa de Planes Complementarios de I+D+i (PRTR 2022), 
reference ASFAE$/2022/015$.

\printbibliography
\end{refsection}


\chapter{Scintillator Screens \& Camera System}
\label{chapt6}
\begin{refsection}



\newcommand{\laser}{laser~} 
\newcommand{\lasernospace}{laser}


\chapterauthors{J.~Hallford \\ University College London, London (UK)}{M.~Wing \\ University College London, London (UK)}{R.~Jacobs \\ \desyaffil}

%

\begin{chaptabstract}


This chapter describes the technical design of a scintillation screen and optical camera system intended for high-rate electron detection in the LUXE gamma beam production station of the \glaser mode and after the strong-field interaction point in the \elaser mode. The light distribution in the scintillator, imaged by remote cameras, is used to reconstruct the spatial distribution of electrons deflected in a magnetic dipole spectrometer. This allows measurement of the electron energy spectrum. The detector choice for this system is motivated by its excellent spatial resolution as well as robustness against high particle rates.



\end{chaptabstract}

%
%



\section{Introduction}
In the LUXE experiment, dedicated high-rate electron detectors are used for the detection of electrons in locations with a high particle flux, of the order of $10^3-10^8~e^-$ per detector channel. There will be two instances of high-rate detector systems in the LUXE setup, one for the detection of Compton-scattered electrons in the \elaser running mode and one for the monitoring of Bremsstrahlung production in the \glaser running mode. The two systems will be denoted as the electron detection system (EDS) and the initial Bremsstrahlung monitor (IBM). Both the IBM and EDS systems consist of a scintillating screen, imaged by an optical camera system, followed by a spatially segmented \cer detector (see Chapter~\ref{chapt7}). Figure \ref{fig:layoutdet} shows a schematic of the LUXE experimental setup in the two run modes, indicating the locations of the EDS and IBM systems~\cite{CDR}.

\begin{figure}[htb!]
\centering
\includegraphics[width=0.6\textwidth]{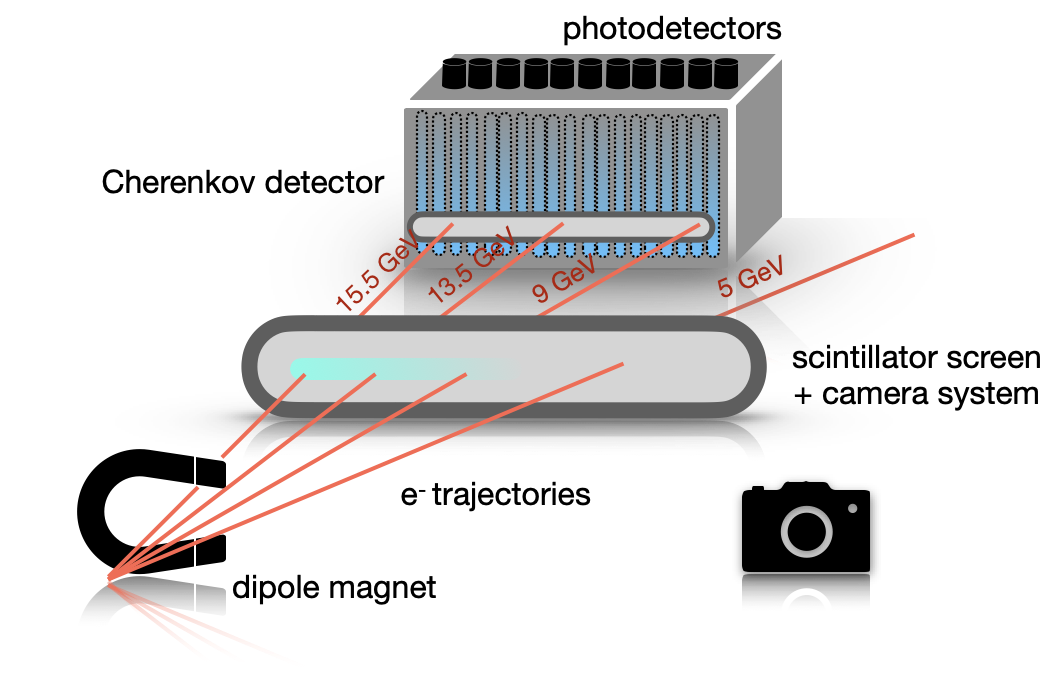}
\caption{ A visualisation of the set-up of the electron detection system. Electrons from the IP, deflected by a dipole magnet, interact first with a thin screen of scintillating material and then the \cer detector. The light profile from the screen is imaged remotely with optical cameras, and light produced in the \cer medium is reflected towards an array of photodetectors.}
\label{fig:ceds-diagram}
\end{figure}

A schematic drawing of the EDS system is presented in Fig.~\ref{fig:ceds-diagram}. The electron energy spectrum will be determined by measuring the number of electrons as a function of their displacement by a dipole spectrometer magnet. 

Two complementary technologies (\cer and scintillation screen detectors) are used to enable the possibility of cross-calibration and alignment as well as assessment of systematic uncertainties. The scintillation screen offers a simple (and inexpensive) setup with a high resolution in position. The \cer device, for an appropriate choice of refractive medium, will offer a greater resistance to low-energy non-signal particles, and also will be more robust to the challenges of high particle flux. The presence of a thin screen with a thickness of $0.5 \units{mm}$ (corresponding to $2.8\%$ of a radiation length) in front of the \cer detector will not affect the spectrum of electrons seen by the latter as the loss of the electron velocity $\beta$ is negligible. 


\begin{figure}[htb!]
\centering
\includegraphics[width=0.75\textwidth]{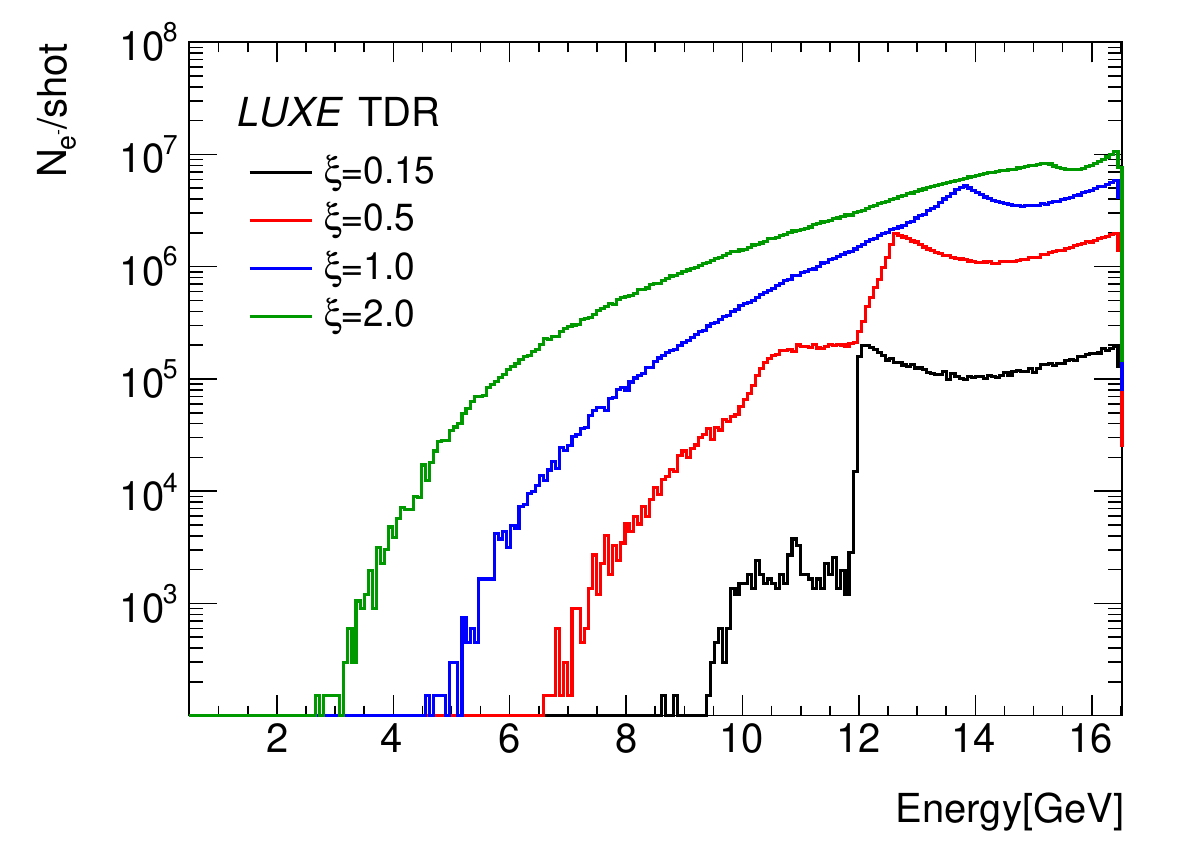}
\caption{Simulated Compton electron energy spectrum for different values of the Laser intensity parameter $\xi$. Non-linear effects increase with $\xi$.}
\label{fig:comptonspectra}
\end{figure}

The objective for the EDS system is to measure the Compton electron energy spectrum, with a particular focus on measuring the position of the first Compton edge as a function of the laser intensity parameter $\xi$. A simulation of Compton spectra for different maximum $\xi$ values using the \ptarmigan generator~\cite{ptarm} can be seen in fig. \ref{fig:comptonspectra}. It is expected that the first Compton edge is shifted to larger energy values with a functional behaviour of

\begin{equation}
\label{eq:edgexi}
 \frac{E_\text{edge}(\xi)}{E_\text{beam}}=1-\frac{2\eta}{2\eta+1+\xi^2}\,,
\end{equation}

 where $\eta=\omega_L\epsilon_e/m_e^2\cdot\beta(1+\cos\theta)$ is the electron energy parameter, with the laser frequency $\omega_L$, electron beam energy $\epsilon_e$, velocity $\beta$ and \elaser crossing angle $\theta$. For LUXE at an electron beam energy of $16.5\,\text{GeV}$ and a crossing angle of $17.2^{\circ}$, the energy parameter is $\eta \approx 0.192$. \\

The aim for the IBM is to measure the energy spectrum of electrons after the Bremsstrahlung interaction of the EuXFEL electron beam with the initial target in the LUXE \glaser collision mode. In this case the electron energy spectrum is used to monitor the energy spectrum of the produced secondary photon beam.\\

This chapter will focus on the scintillator screen and camera system. The \cer system is covered in Chapter~\ref{chapt7}.

\section{Requirements and Challenges}

The two detector locations for the \elaser EDS and \glaser IBM systems exhibit similar particle rates and have similar requirements, leading to almost identical detection systems. Although both may be installed at the same time, it is not expected that both operate concurrently. 

As shown in Fig. \ref{fig:layoutdet}, the scintillation screen and camera system will be used in these two locations. For the purposes of this document they are considered separate from the relatively similar setup in the downstream gamma spectrometer. The particular needs and flux of electrons in that system motivate the decisions made for it, laid out in chapter~\ref{chapt8}. 


\subsection{Electron Detection}

In contrast to many measurements in high-energy physics (including in other regions of the LUXE experiment, where single-particle reconstruction is performed) in the electron detection regions of LUXE only the particle flux of a single particle species is measured. The flux is used to reconstruct a distribution in energy. 

\subsubsection{\elaser IP Electron Detection System (EDS)}

In the IP of LUXE (\elaser mode) the highly focussed electron beam of \euxfel is directly collided with a \laser pulse. The beam, having interacted with the  strong field, then transforms from mono-energetic electrons to electrons following some spectrum of energy, examples of which are seen in Fig.~\ref{fig:comptonspectra}, where the particles remain largely collinear. They are then separated from photons and positrons by the dipole spectrometer magnet. This magnetic deflection is approximately inversely proportional to the energy of the charged particle, creating a plane of GeV-energy electrons. 

The high-intensity Compton scattering mechanism results in a much lower rate of low-energy electrons relative to the higher energy end of the spectrum. The acceptance of the scintillator \& camera system, constrained by the opening angle of the magnet, restricts the lower bound of any detector acceptance to the order of 2 \GeV which covers the full spectra shown in Fig.~\ref{fig:comptonspectra}, representative of all expected spectra from beam-\laser interactions in LUXE.


Achieving detection of the spectrum up to the highest energy possible -- but without running risk of intercepting the unscattered beam directly -- is required. Electrons created from the rarer trident process in the IP are also present, but are not expected to be resolvable above the signal resulting from the Compton-scattered electrons. This is because of the lower rates and greater divergence of these trident electrons, resulting in a much lower flux. The trident process is instead observed by positron detection, which are symmetrical in terms of distributions of kinematics to the electrons.

Detection is also required of the Compton edge energy, requiring not just accuracy in rates/flux, but of the energy itself. The minimum requirement of the edge detection is chosen as 2\% energy uncertainty. The problem of measuring this edge is exacerbated by the variance in $\xi$ present from the \laser focus at higher $\xi_{max}$, making the the edges appear smeared. To estimate the $\xi_{max}$ achieved Finite-Impulse-Response filters will be employed~\cite{Berggren:2020gna}. To do this as effectively as possible requires high energy (and therefore position) resolution. An analysis including this technique is available in Section~\ref{subsec:scint_fir}. 

With this in mind, not expected to be limited by statistics in measurement compared to other uncertainties, the finest energy resolution possible is desired -- using detectors which will remain tolerant to the level of radiation present, and remain cost-effective. 

\subsubsection{\glaser Initial Bremsstrahlung Monitor (IBM)}



The second site using the screen and camera system is the IBM. Monitoring the generation of both sources of the \glaser mode (inverse Compton scattering (ICS) and bremsstrahlung) can be achieved using the deflected electrons used in the generation of both. In both production modes electrons from the same \euxfel beam radiate photons. By measuring the electron energy spectrum, and with good knowledge of the beam energy, the photon spectrum can be recovered using conservation of energy: 

\begin{equation}\label{eqn:E_gamma}
\centering
E_{\gamma} = E_\textrm{beam} - E_\textrm{electron} 
\end{equation}

This is only measurable within the range of the electron beam energy minus the detector acceptance energy limits.

\begin{figure}[t]
    \centering
    \includegraphics[width=\textwidth]{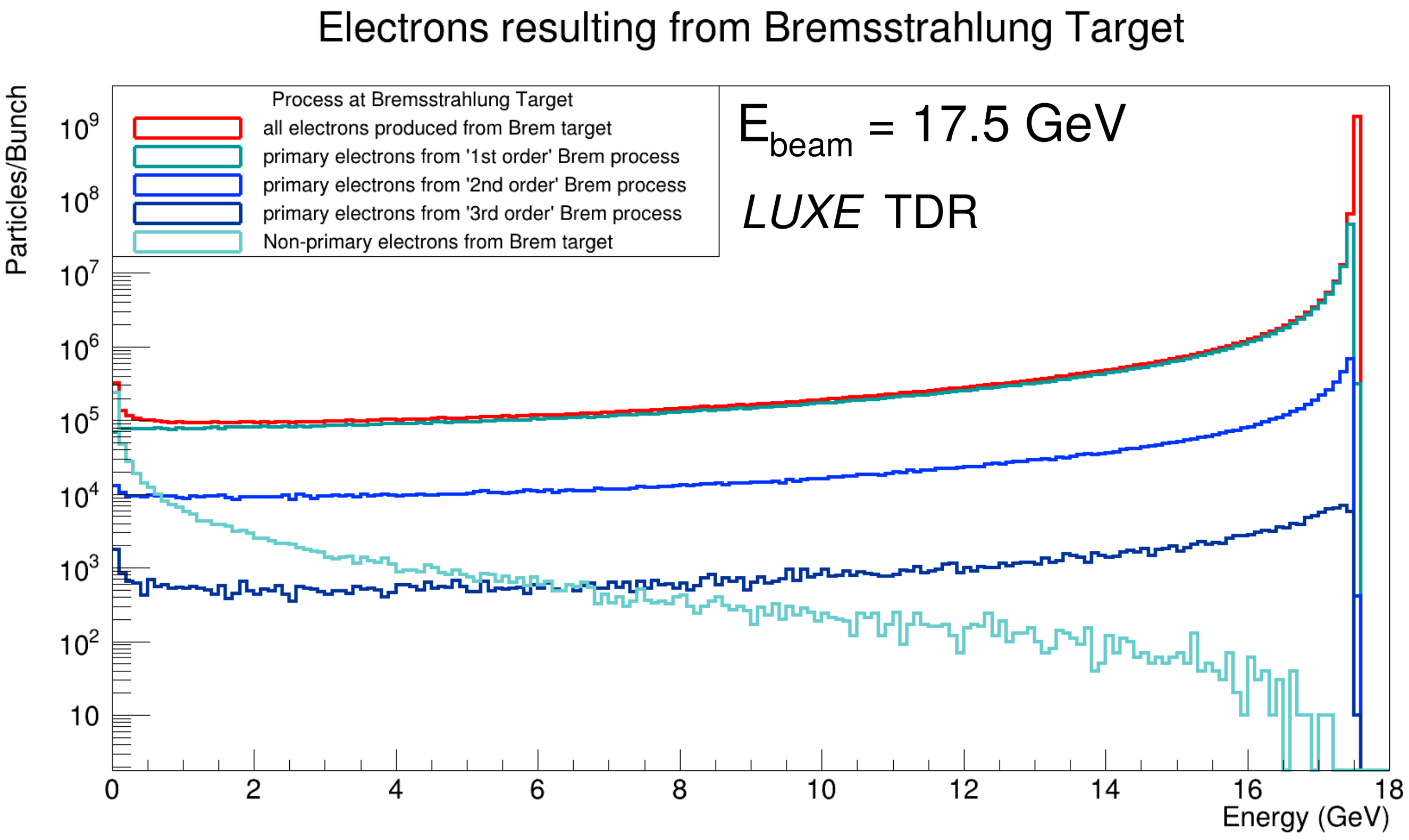}
    \caption{Energy spectrum of electrons having passed through a 35 \textmu m tungsten target simulated in \geant. This simulation uses an incident beam energy of 17.5 \GeV and bunch population of $1.5 \times 10^9$. The total spectrum is softened compared to the ideal single-emission bremsstrahlung case, because of higher-order multiple emission events. 
    }
    \label{fig:brem_spectrum}
\end{figure}

There is a subtlety in the fact that one electron can radiate multiple times. For both the bremsstrahlung and ICS modes, the distribution of these multiple-emission events and the effect they have on the final spectrum, and therefore how to unfold the electron measurement to reconstruct the final photon energy spectrum, will be achieved through simulation. For bremsstrahlung,  \geant~\cite{geant4} provides profiles of the expected rates of multiple-emission -- which comes with appreciable associated errors. For this reason the accuracy of the detection is limited, and the most precise measurement comes from the downstream suite of photon detectors, and the purpose of the scintillation screen and \cer system are as complement to the downstream detectors and to act as a shot-to-shot beam-\laser interaction quality monitor.  

The full spectrum of electrons emerging from a 35 \textmu m tungsten plate within \geant, with the comparison of the level of multiple-emission bremsstrahlung radiation events and electrons created from other sources, is shown in Fig. \ref{fig:brem_spectrum}.  


\subsection{Electron Flux and Radiation Environment}

Linearity and predictability in the response of the scintillation for a large range of fluxes, spanning several orders of magnitude, and one order of magnitude in electron energy, is essential to making a high-quality measurement. Chosen detector technologies must therefore be sensitive to the lower end of this range, but resistant to damage from the radiation environment associated with the higher end, and linear, or at least predictable, in their response throughout the range in flux.

The active elements of any detector in these regions need to be highly resistant to radiation damage, as the accumulated flux in either region is appreciable and accumulates as runs of data-taking progress. Alternatively elements at risk of damage could be replaced regularly, but only if this is enabled by the design of the detector. Observed in simulation, the dominant source of active detector element irradiation includes an approximate maximum flux of \GeV-energy electrons of $\sim 10^7\,\textrm{mm}^{-2}$ per bunch-crossing in the EDS. There is also a persistent neutron flux, the details of which are still being studied, which induces the possibility of damaging electronic elements even if they are not within the beamplane.

\section{System Overview}
\subsection{Detector}

The physical mechanism of scintillation is a response of a sensitive material to the deposition of energy by ionising radiation. Some fraction of the deposited energy is absorbed by a characteristic process, particular to each material, but consistent in the empirical consequence which is the emission of light. This release of light occurs some short time after the deposition, typically taking nano- to micro-seconds to occur. 

These properties make scintillating materials very common elements in particle detectors. For LUXE, the flux of ionising radiation is intense enough such that levels of light emitted is very high, and can therefore be imaged remotely. This offers the advantage of keeping the sensitive electronics outside the path of the direct radiation. With a relatively thin screen oriented such that the normal of the surface is aligned with the direction of ionising particles, the position of the particle within the plane of the screen is given by a near-point-like emission of light -- within the inherent position resolution of the deposition of energy and optical point spread function of the screen. 

The material chosen for the LUXE detector is the high-light-yield Gadolinium Oxysulfide doped with Terbium. The fact that the screen is thin means that the electrons typically pass through, having lost very little of their energy. This lack of impact on the electron energy spectrum means it can be placed before the \cer detector, the detection mechanism of which remains unperturbed (see chapter~\ref{chapt7}). An illustration of these two detectors and their position with respect to the deflected electrons are shown in Figure \ref{fig:ceds-diagram}. 



Using optical cameras mean this light can be used to reconstruct the energy distribution of incident particles. With good knowledge of the field map of the deflecting B-field, the energy of the particles correlating to each position is calculated and the spectrum of energy reconstructed. 

\begin{figure}[t] 
\centering
\includegraphics[width=0.65\textwidth]{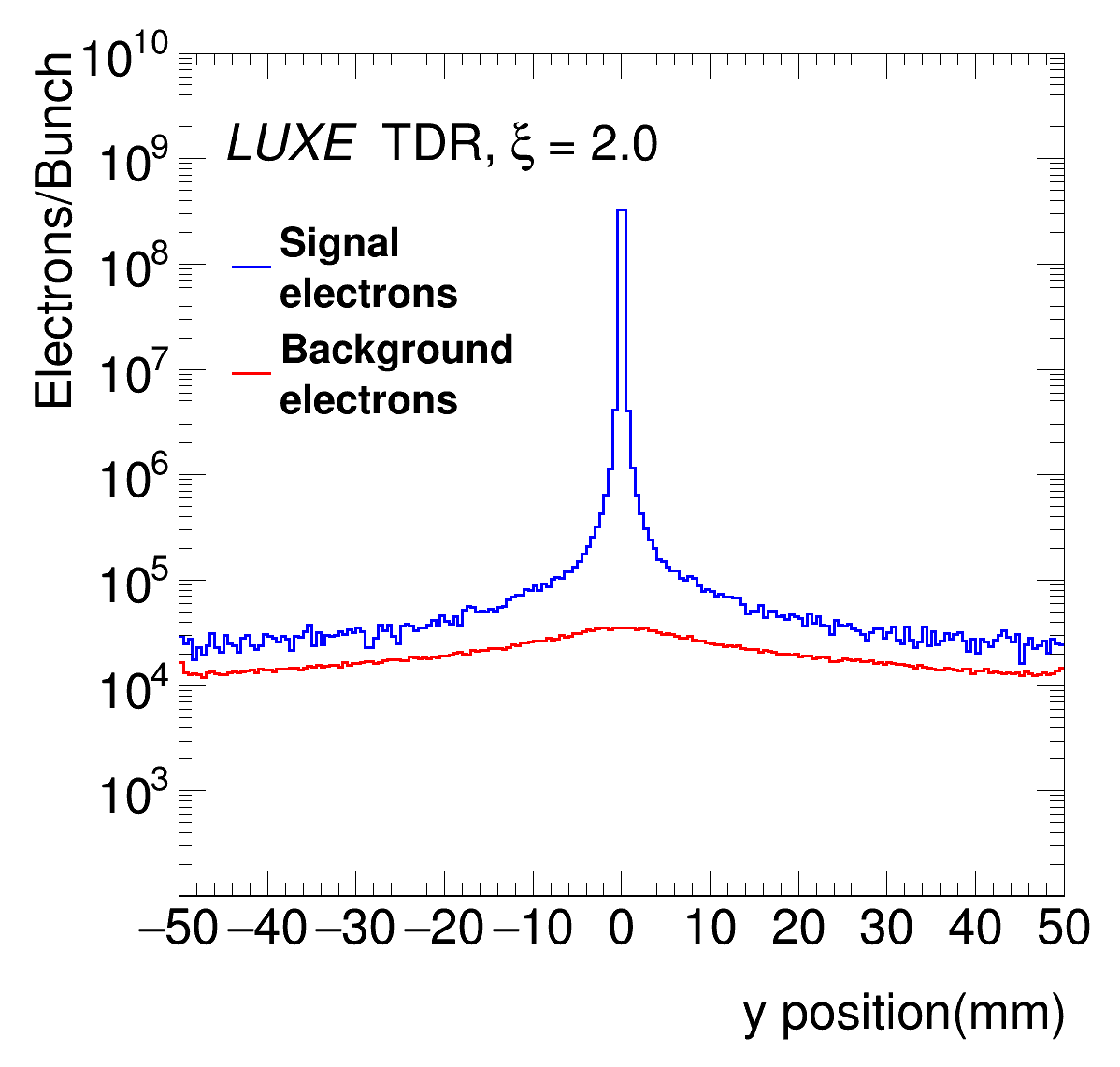}
\caption{Distribution of electrons in $y$ for the signal and background in the scintillation screen in the EDS for a sample interaction with IP $\xi = 2.0$. The background is determined from beam-only simulation. The signal measurement includes secondaries from Compton-scattered electrons within the screen's acceptance, which are taken to be unfoldable as part of signal measurement.}

\label{fig:lanex_sigbkg}
\end{figure}

In both sites in LUXE, the electrons to be reconstructed (immediately after the interaction but before the dipole) are highly collinear. Both intended sources of electrons are from the focussed \euxfel beam having undergone one or more emissions of photons. The beam energy -- typically between 14 and 16.5 GeV -- means the Lorentz factor of the beam electrons is of order $10^4$ and so their frame is highly boosted. The dispersion (orthogonal to the bending via magnetic field) of these electrons is a few millimetres (dependent on their energy) over the travel distance of $\sim$ 1-3\,m between IP/gamma target and the screens in each site. This results in a central signal band of $\sim$ 1\,cm within the screens. Extending the screen surface beyond this band in the $y$ \footnote{The general LUXE coordinate system uses the $z$ axis for the beamline, and $x$ axis is the direction in which the electrons/positrons are deflected. The $y$ axis is orthogonal to both.} dimension allows for some measurement of a more ambient background not confined to some central band. Evaluation of this ``quiet" part of the screen allows for background measurement and subsequent subtraction for clean signal measurement in this high-flux, high-statistics measurement. This is illustrated (using \ptarmigan and \geant to simulate the signal and its dispersion) in Fig. \ref{fig:lanex_sigbkg}. 

Detailed within the general LUXE documents~\cite{CDR} is the fact that signal bunch-\laser collision events  are planned with a rate of 1\,Hz, and beam-only events for a further 9\,Hz, to make a total of measurements at a rate of 10\,Hz.  The system is then designed to cater for a 10\,Hz rate. These beam-only events allow for background measurements fit for subtraction from signal collisions. The cameras chosen, technically detailed later in this document, come from the Basler Area Scan range \cite{basler_area_scan}. These present a quality position resolution, particularly choosing multiple cameras to observe one screen, a high quantum efficiency, a typical dynamic range each of $\approx 10^4$, an imaging frequency satisfactory for the 10\,Hz required, and a comparatively low cost. 

The lenses fitted depend on, and inform, the displacement between the screen and cameras. This displacement is chosen on balance between the competing needs of being close enough to gather enough light, but not suffer radiation damage and to keep the image of the screen unwarped by perspective issues (associated with lenses of small focal lengths and short working distances). 

Optical filters are fitted to the camera lenses to accept only light within a tight wave-band including the characteristic emission wavelength ($\sim 545 \units{nm}$) of the scintillator material, in order to exclude as much as possible ambient light within the experimental hall. The hall cannot be left in complete darkness, as safety lighting is necessary. These safety lights can be chosen for monochromatic emission outside the wavelength of the screen, and so optically blocked by filters chosen, available with optical transmission lower than 0.1\%
for wavelengths outside the desired waveband~\cite{edmund_filter}. 

\begin{figure}[t]
    \centering
    \includegraphics[width=0.85\textwidth]{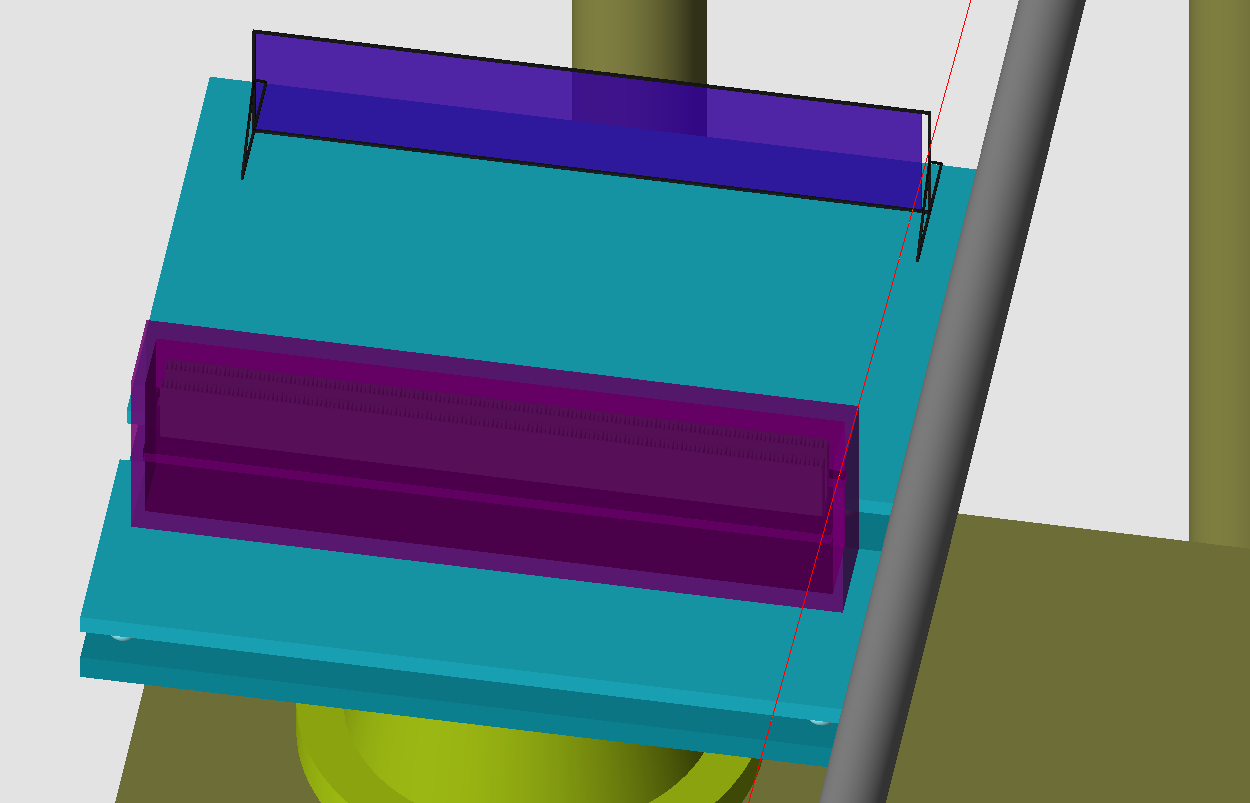}
    \caption{A \geant visualisation of the EDS scintillation screen with a traced path of electron beam in red. The beam passes through a loop in the aluminium frame.}
    \label{fig:e-laser-frame-loop}
\end{figure}

For both of the detector components (screen and camera) accurate positioning and orientation is of paramount importance. To keep each of the screens stable and free from sag, a precisely machined aluminium frame will be created to rigidly hold the screens, which are themselves fixed to an optical table alongside the \cer detector. Aluminium is chosen as a compromise between hardness and low density to minimise environmental scattering. For the EDS measurement, as the screen is fixed extremely close to the beam, a small loop of the frame is extended to enclose the beam -- this particular feature is shown in Fig.~\ref{fig:e-laser-frame-loop}.

\begin{figure}[t]
    \centering
    \includegraphics[width=\textwidth]{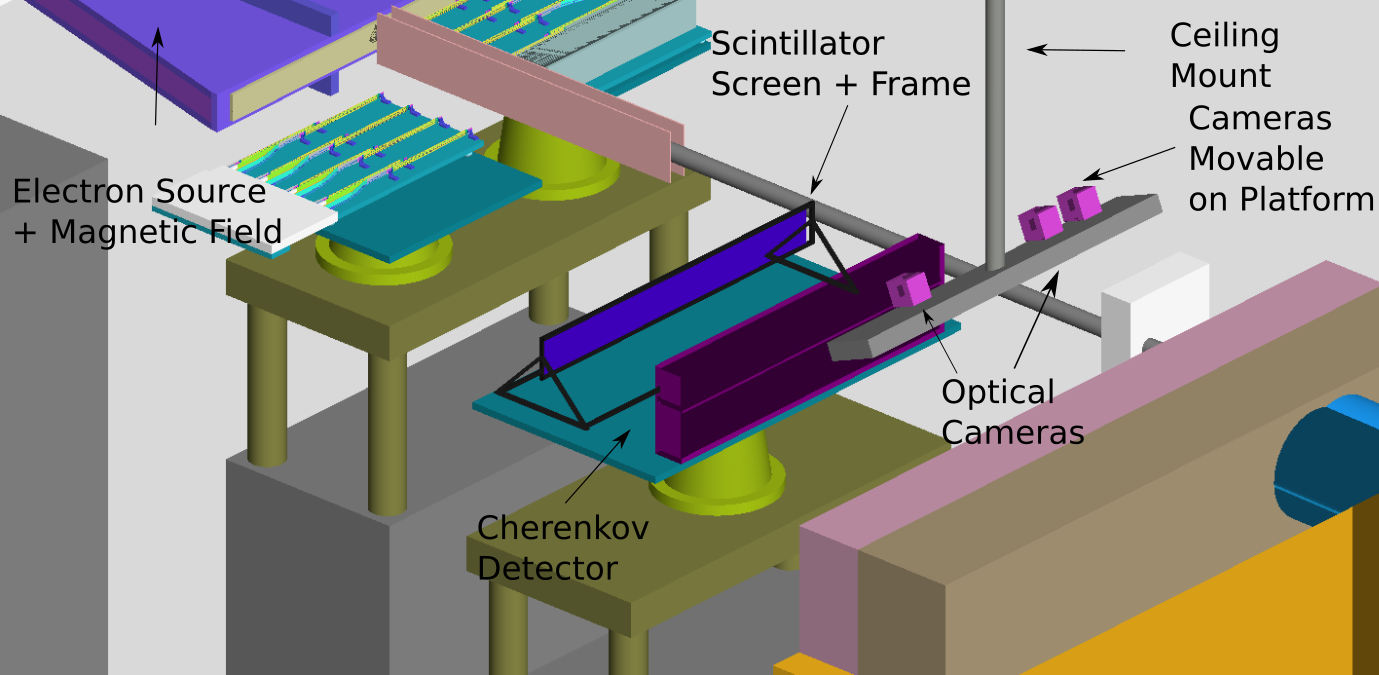}
    \caption{A \geant visualisation of the \elaser IP measurement region setup of the electron detectors, in particular the scintillation screen and cameras.}
    \label{fig:elaser_IP_visualisation}
\end{figure}

\begin{figure}[t]
    \centering
    \includegraphics[width=\textwidth]{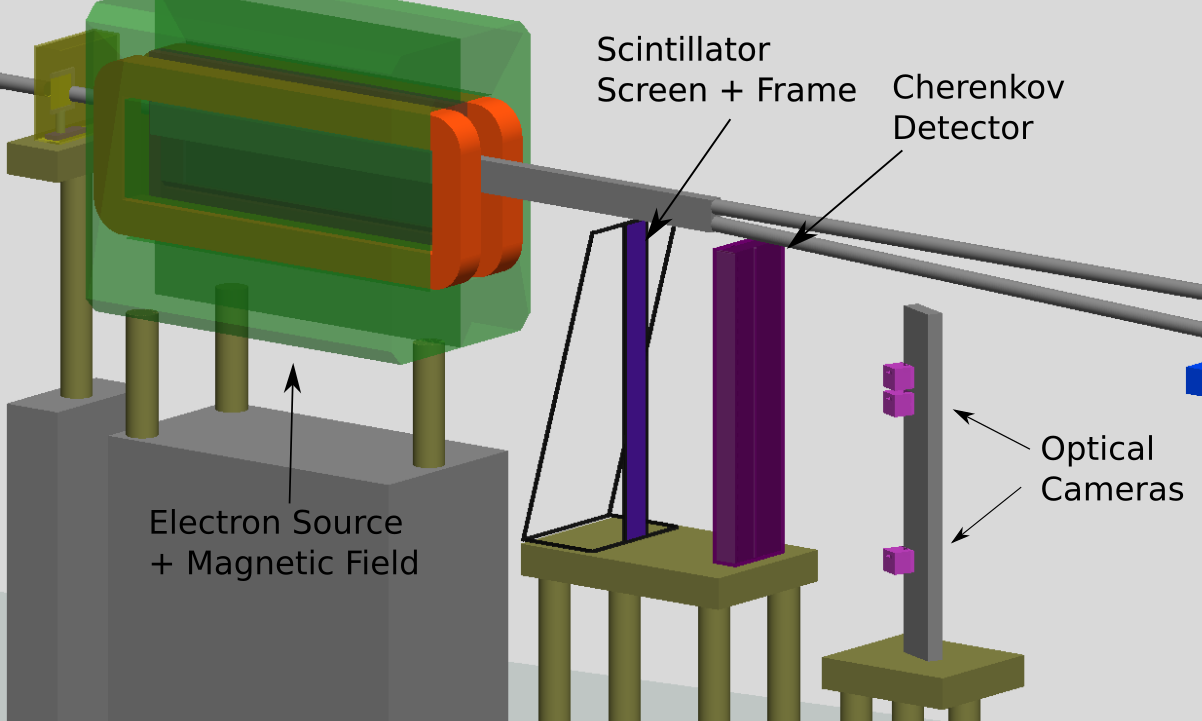}
    \caption{A \geant visualisation of the electron measurement setup of the IBM detectors, in particular the scintillation screen and cameras. As can be seen here, and in contrast to the EDS region, the deflection of electrons is into the $y$ direction of the general coordinate scheme of LUXE.}
    \label{fig:brem_visualisation}
\end{figure}

The cameras are held within lead shielding to help protect from stray radiation. This lead shielding  also assists in heat dissipation and therefore minimising electronic noise. They are secured mechanically to their mount by fixing both the lens and camera. Figures \ref{fig:elaser_IP_visualisation} and \ref{fig:brem_visualisation} show visualisations of the geometries of the screens, cameras and supporting structures in each of the two sites. 

A high linearity of response from the scintillator is expected. Although the deposition of energy within the thin screen is stochastic in nature and varies, the energy deposited per electron converges to a mean for high statistics/fluxes. This has been explicitly simulated in \geant, detailed later in this chapter, where good linearity is shown for the fluxes expected to be measurable by the screen. The linearity of the CMOS-sensor cameras, minus expected backgrounds/electronic noise, is seen in the documentation \cite{2k_camera_specs} \cite{4k_camera_specs}.

\subsection{Magnetic Action}

The setup of the electron detectors at both sites of the experiment use magnetic fields to separate the particles by charge. The magnetic deflection of these charged particles is dependent on their energy, forming a magnetic spectrometer. The relation between energy and position of particle incidence upon a screen is used to reconstruct spectra by energy. The particles' movement traces a circular path in the magnetic field with radius of curvature $R = E_{\eV}/Bc$ for field strength $B$ in Tesla,  $E_{\eV}$ in electron-volts, $R$ in m and $c$ in m/s. From this an expression for the theoretical hit point (in $x$) within a detector plane rotated by $\theta$, as seen in Fig. \ref{fig:magnet_bending} is derived:

\begin{equation} \label{eqn:magnet_bending_1}
x_1 = R(1-\cos(\phi)),\\
\end{equation}
\begin{equation} \label{eqn:magnet_bending_2}
x_2 = \tan(\phi)z_d,\\
\end{equation}
\begin{equation} \label{eqn:magnet_bending_3}
x_3 = \frac{\tan(\theta) \tan(\phi)(x_\textrm{detector} - x_1 - x_2)}{1 + \tan(\theta) \tan(\phi)}
\end{equation}

where the final $x$ position is the sum of $x_1, x_2, x_3$ and $\phi$ is equal to $\sin^{-1}(\frac{z_m}{R})$. In brief, the magnitude of the magnetic field and the distance/angle parameters are used to predict the hit point in the $x$ dimension for a particle of given energy. 

\begin{figure}[h!] 
\begin{subfigure}{0.49\hsize} 
\includegraphics[width=\textwidth]{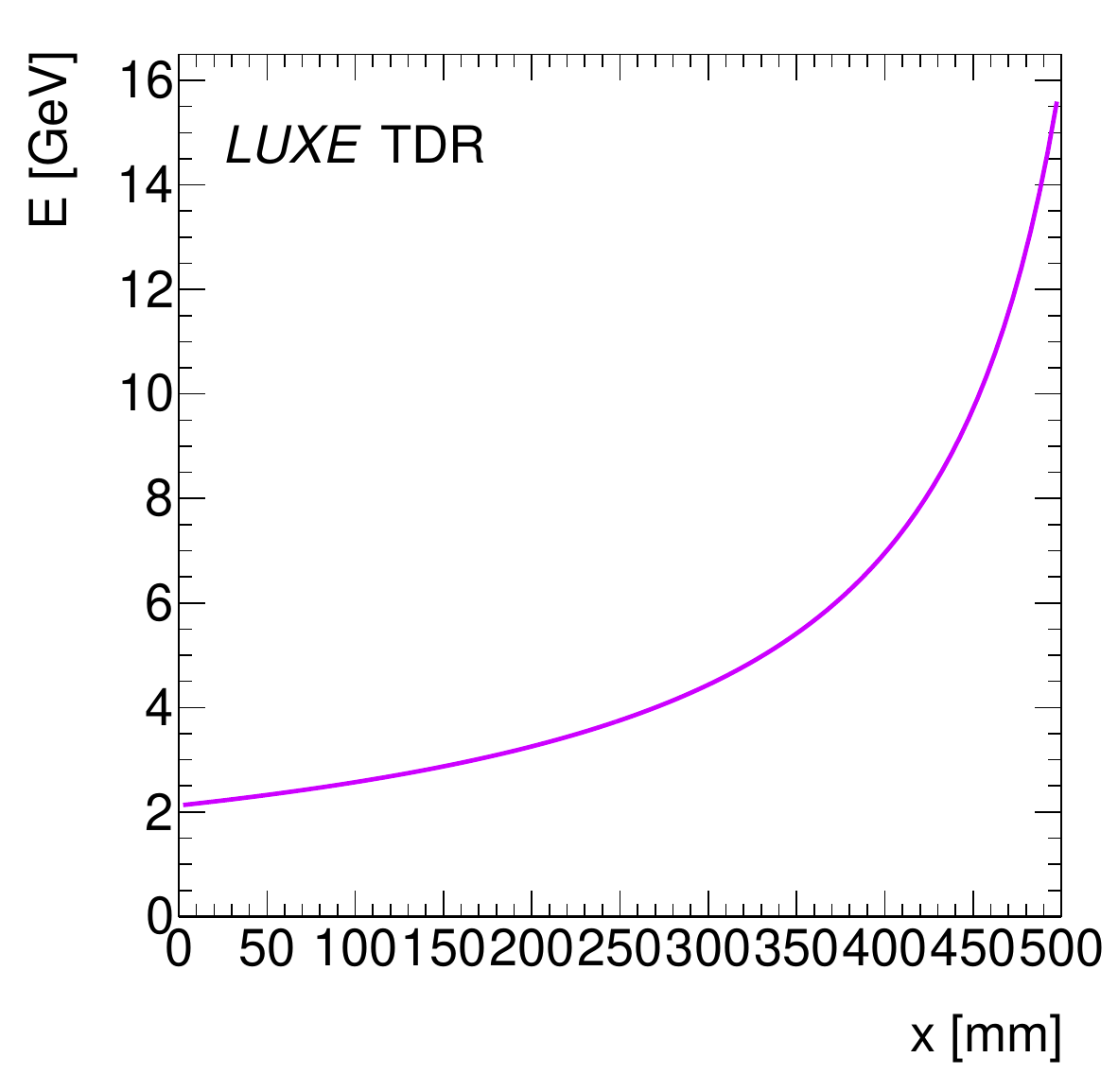}
\caption{  \label{fig:scint-light}
}
\label{fig:photon-before-IP-scint-light}
\end{subfigure}
\begin{subfigure}{0.49\hsize}
\includegraphics[width=\textwidth]{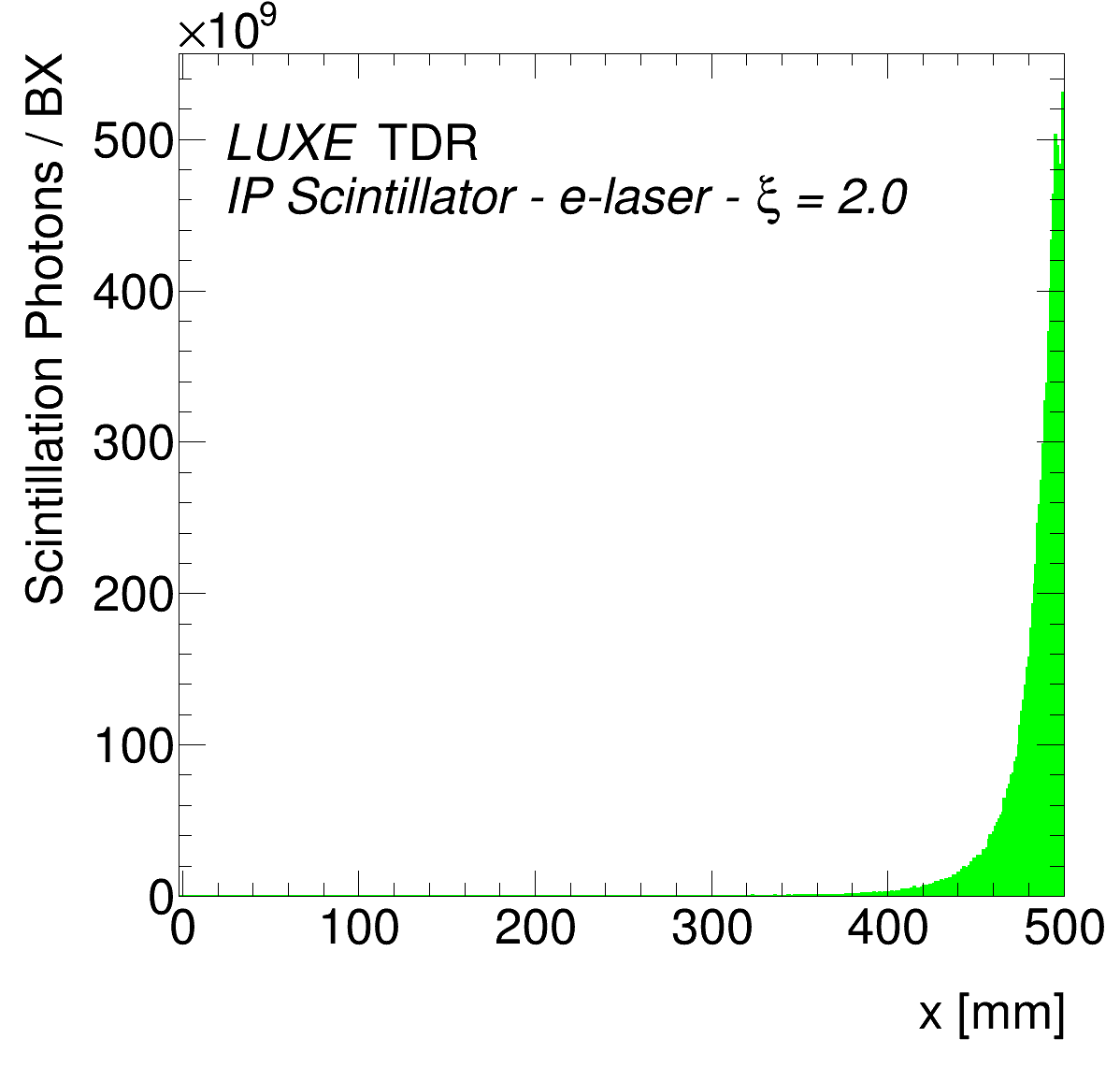}
\caption{ \label{fig:scint-recon-electron} 
}
\end{subfigure}
\caption{\label{fig:x-to-E-map} \textbf{(a)} The approximate transformation from E to $x$ plotted for the EDS scintillation screen, for a B-field of $1 \units{T}$. The zero point in the $x$ coordinate here is the low-energy edge of the screen; $x=500 \units{mm}$ is the high-energy edge. \textbf{(b)} The profile of scintillation light in $x$ position within the scintillation screen, for a $\xi = 2.0 $ \ptarmigan signal simulation, for phase-0 of the experiment. Scintillation light is simulated in \geant. 
}

\end{figure}

\begin{figure}[t]
    \centering
    \includegraphics[width=\textwidth]{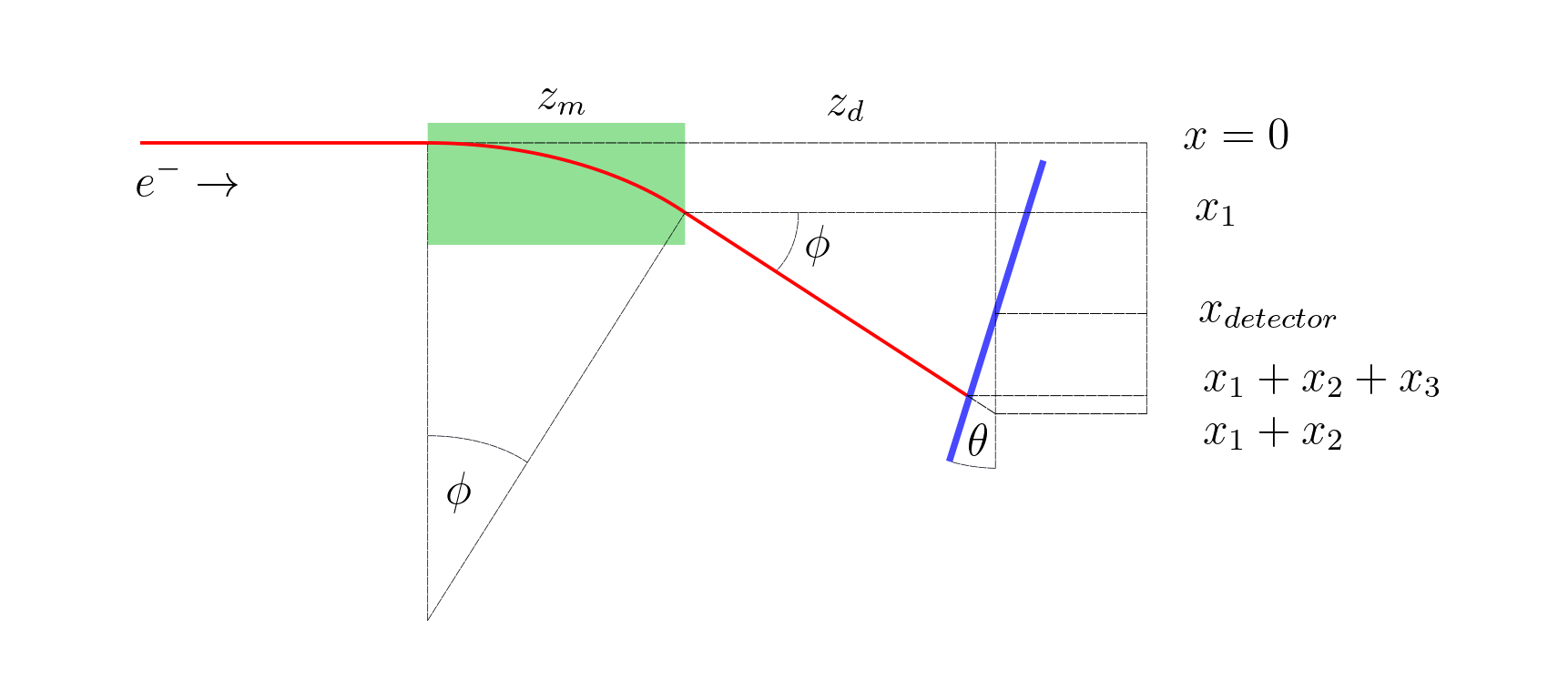}
    \caption{An illustration of the path of an electron through a box-like magnetic field. The green area represents the effective field, the red line an electron path and the blue line the scintillation screen.}
    \label{fig:magnet_bending}
\end{figure}

An approximate function of how the energy maps to the $x$ position, for some standard parameters of the EDS site, is given in Fig. \ref{fig:x-to-E-map}. Additionally, there is a plot of the scintillation light 
 profile (in $x$) in this region for a signal \elaser interaction of $\xi = 2.0$, the initial energy spectrum of which can be seen in Fig. \ref{fig:comptonspectra}.

As the electrons exit the beampipe, the beampipe wall can cause considerable scattering, especially when they approach at a shallow angle, producing background and reducing the quality of the signal. To minimise such effects, a vacuum chamber fitted with a thin beam window is utilised in the EDS, constructed of aluminium, designed to keep the electron exit at a near-normal angle to the beam window. This is largely designed and being physically tested for the EDS, and the applicability of such a window is in the process of being investigated for the IBM. 


\subsection{Reconstruction}

Reconstruction of the electron energy spectrum involves transforming the electron position map to the energy map in order to reconstruct energy for a given interval. The fidelity of these energy intervals then has a minimum dictated by the position resolution of the combined resolution of the screen and cameras. 

The conversion of light to particle flux requires the use of a light/charge calibration curve. This will be established for the screen and camera system outside LUXE, using a well-known electron test-beam of adequate energy and a high bunch population in order to measure the light emitted for some known electron flux, and ensuring its linearity.

As mentioned, the regions of the screen and image outside the central signal band, and use of beam bunches sent with no \lasernospace, allow for the creation of profiles of background to be made and subtracted from the signal measurements.  

\begin{figure}[t]
    \centering
    \includegraphics[width=\textwidth]{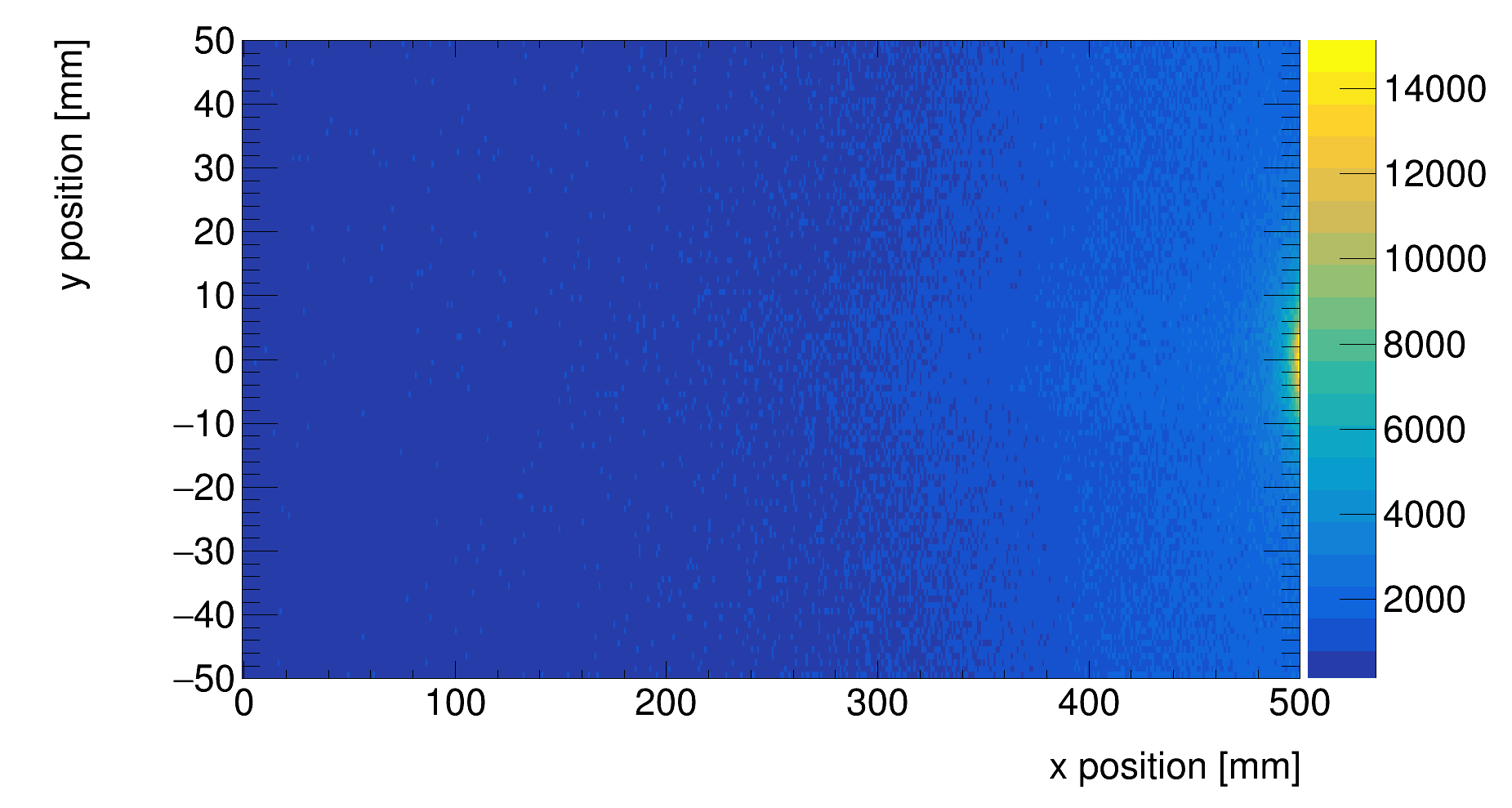}
    \caption{An indicative plot of the background levels for all particles intercepting the scintillation screen in the EDS region. Each bin represents one $\units{mm^2}$ unit of area. The $x$ and $y$ position are with respect to the centre of the screen in $y$, outermost edge in $x$, such that high $x$ points toward the beam axis. Simulation performed in \geant\cite{geant4}. 
    }
    \label{fig:bkg_plot}
\end{figure}

This background can be understood to be a composite of a largely flat distribution of scintillation-inducing particles principally coming from the respective beam-dumps, or other complex multiple-scattering events from all around the experimental hall, and a distribution of background particles reaching the screen shortly after produced from the electron beam. This latter source can then, especially in the EDS, be modelled as symmetric around the deflected electron beam axis. By measuring the background per unit area in the quiet part of the screen, for some radial distance from the modelled electron beam, this symmetry can be exploited to subtract the background from the signal band where this radial distance is preserved. 
An illustration of the interception of particles of all types from background sources is shown in Fig. \ref{fig:bkg_plot}.

Due to the inherent spread of these signal particles as they propagate through the beam exit window and the air between the screen and exit window, the point spread function of the light in the phosphor, and the optical spread of the image due to the lenses, an effective point spread function convolves the true distribution of signal electrons on the screen. This spread is dominant in terms of position resolution compared to the nominal resolution of the camera images, but the main uncertainty in energy resolution is expected due to uncertainty in the magnetic field. 


\section{Expected Performance}
\subsection{Simulation-Based Performance}
\label{subsec:scint_simperf}
The expected performance for the two sites of the scintillator and camera systems differs. Although the requirements and environments for the two sites are similar, the objectives are not quite the same: for the EDS, the energy spectrum of electrons within the detector acceptance is desired.  For the IBM, the energy spectrum of the corresponding photon beam is required. As previously noted this involves correlating the electron energy roughly to the photons by subtracting the electron energy distribution from the beam energy. Multiple-emission events obscure this by softening the electron spectrum and breaking the relation exactly. To account for this, simulation of the bremsstrahlung (and pair creation) effects in the tungsten target will be performed, and the results used to deconvolve and achieve the photon signal (using \geant \cite{geant4}). This comes with the caveat of $\pm$\,100\% errors associated for the corrections, which are expected to be appreciable for a target of 35 \textmu m. A relative level of multiple-emission events of $\sim$ 10\% is seen in Fig. \ref{fig:brem_spectrum}. In addition the region of the photon spectrum most important to characterise -- the high energy limit -- corresponds to the lowest energy electrons, which are not detectable as they do not escape the downstream dipole exit. For these reasons emphasis is placed on the `monitoring' function of the scintillator \& camera system in this region, although the quality of the straightforward electron energy reconstruction is anticipated to be similar in both regions. 

The uncertainty of measurements of rates (or fluxes) of electrons within the energy acceptances for each detector (see Tables \ref{tab:ip-el-energy-acceptance} and~\ref{tab:brem-el-energy-acceptance}) are expected to be at the few-percent level, dictated by the light--charge calibration to be performed in test-beam. Comparison can be made to Ref. \cite{Keeble:2019}, for a similar light--charge calibration which provided approximately 2\% uncertainty. 


Of great interest for the EDS, and visible in Fig. \ref{fig:comptonspectra}, is the measurement of the energy of Compton edges. The requirement for the detector system is below 2\% uncertainty on energy. The energy resolution as percentage with the magnetic spectrometer can be approximated as:

\begin{equation}
    \frac{\sigma_E}{E} = \frac{\Delta x }{Bez_m(\frac{z_m}{2}+z_d)} \times E
\end{equation}

\begin{figure}[t]
    \centering
    \includegraphics[width=0.75\textwidth]{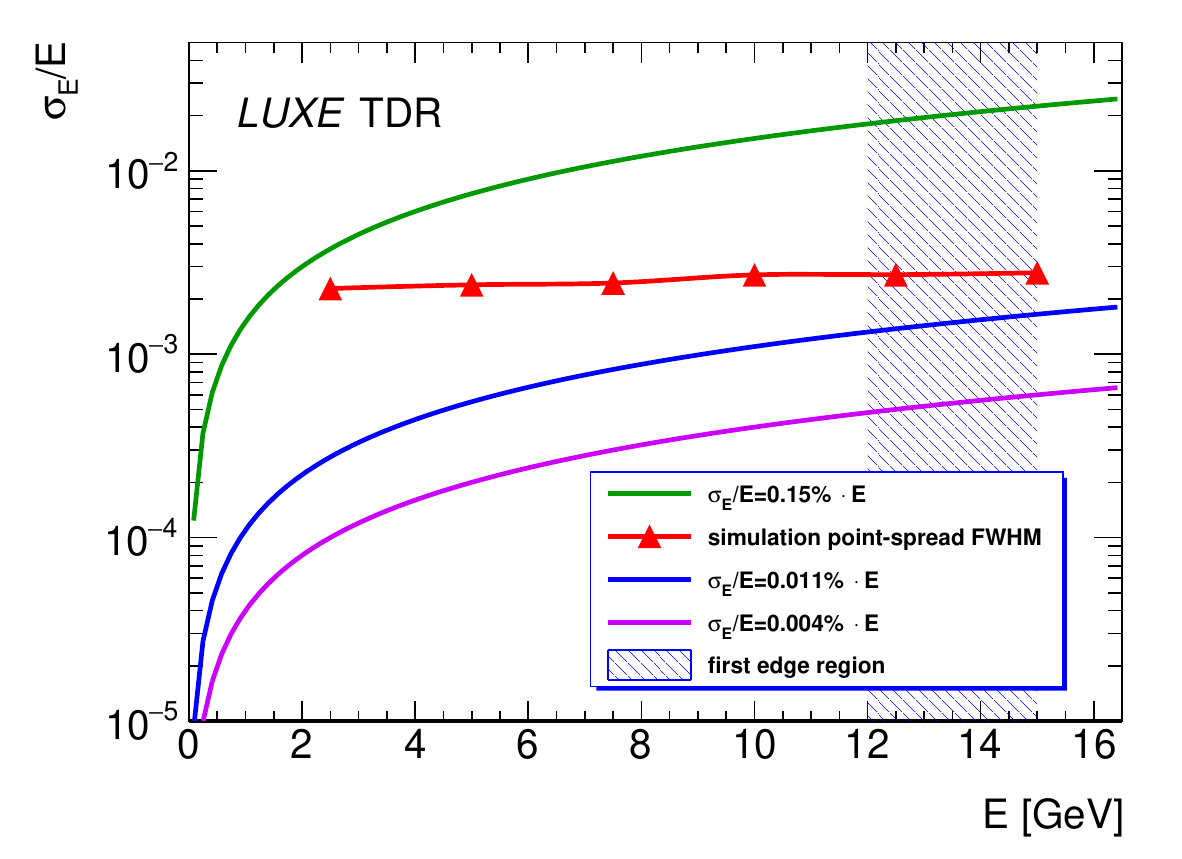}
    \caption{The functions of 0.004\% E and 0.011\% E represent the inherent position resolution of the camera setup(s). The function of 0.15\% E represents the minimum energy resolution to achieve accuracy of 2\% within the expected first edge region of Compton spectra. The dispersion of signal sources due to the geometry of the system are plotted using the full-width at half maximum of the point spread of the scintillation response of the screen from monoenergetic sources in \geant simulation. The associated trendline of these points is only empirical and not the fit of any function.}
    \label{fig:deltaEoverE}
\end{figure}

Or, condensing the expression:

\begin{equation}
    \frac{\sigma_E}{E} = \sigma_0 \times E / \GeV
\end{equation}

with the terms $z_m, z_d$ shown in Fig. \ref{fig:magnet_bending}. The expected position resolution derived for two camera models from the camera \& screen system, detailed later in Section~\ref{sec:det:scintillator}, and what they mean for the relative energy resolution, are shown in Fig. \ref{fig:deltaEoverE}. Both camera models nominally remain more than an order of magnitude below the 2\% level throughout the energy acceptance. The position resolution is also limited by the point spread of the signal through the vacuum chamber exit window and intervening air, before the energy deposition spread and optical point spread of the screen. The spread due to the environmental scattering and stochastic energy deposition within the screen has been simulated in \geant for several simulation-perfect mono-energetic signal sources from 2.5 to 15 \GeV. The full-width-half-maximum value of the resulting normal scintillation responses have been used to represent the resulting energy uncertainty due to environmental scattering and plotted too in Fig. \ref{fig:deltaEoverE}. There one can see the nominal position resolution expected from the cameras is finer than that expected of the point-spread action of the scintillation screen; exceeding the resolution requirements is useful, as it can aid in deconvolution of the smeared signal in post-analysis, and gives margin for the optical line to be redesigned without troubling the energy resolution limit or requiring a change in camera. Uncertainty in the measured B-field map will also affect this energy determination. This is expected to be at 1\% or lower, and so is actually likely to be the dominant uncertainty, and within the required 2\%.  

\begin{figure}[h!] 
\begin{subfigure}{0.49\hsize} 
\includegraphics[width=\textwidth]{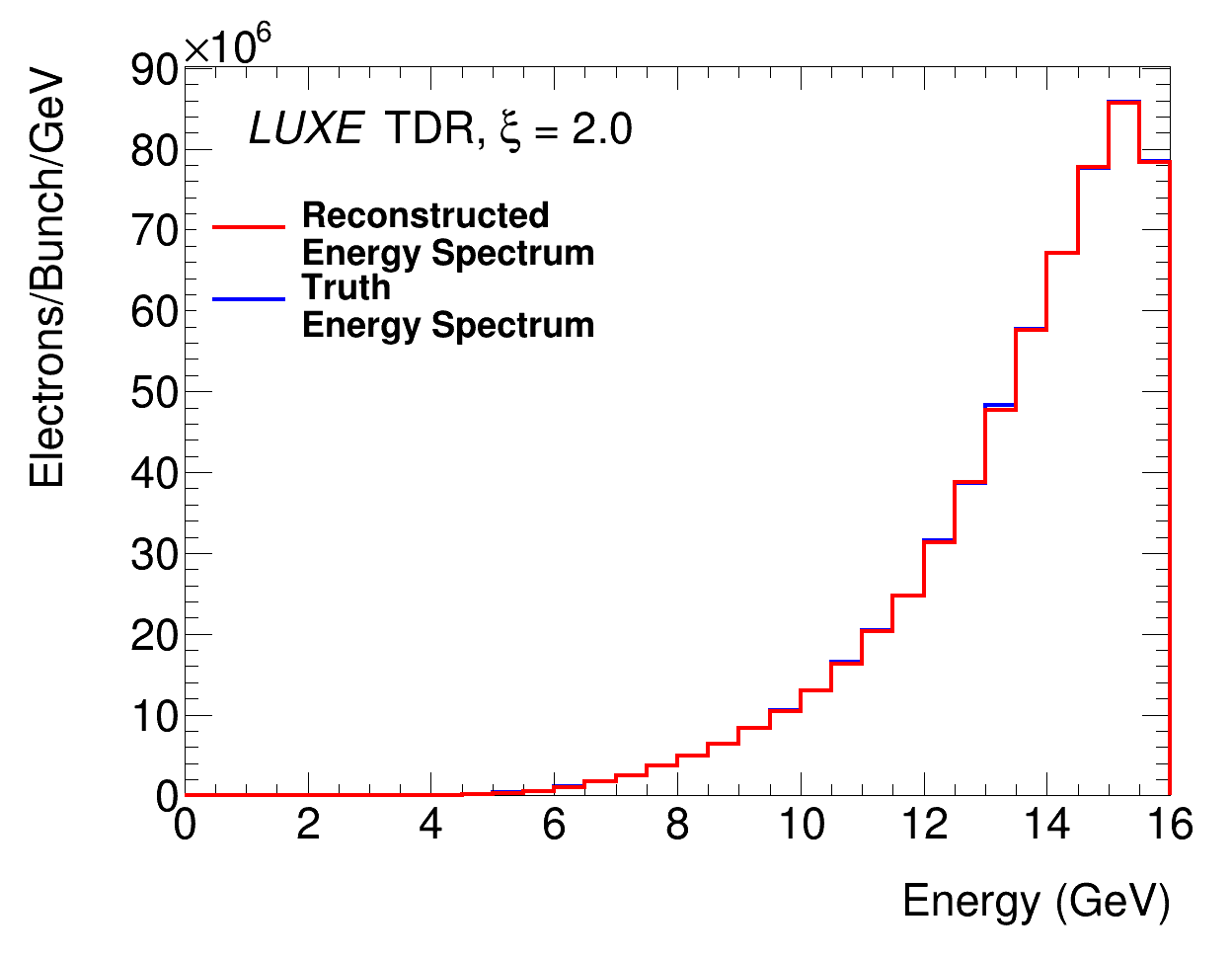}
\caption{  \label{fig:scint-light}
}
\label{fig:photon-before-IP-scint-light}
\end{subfigure}
\begin{subfigure}{0.49\hsize}
\includegraphics[width=\textwidth]{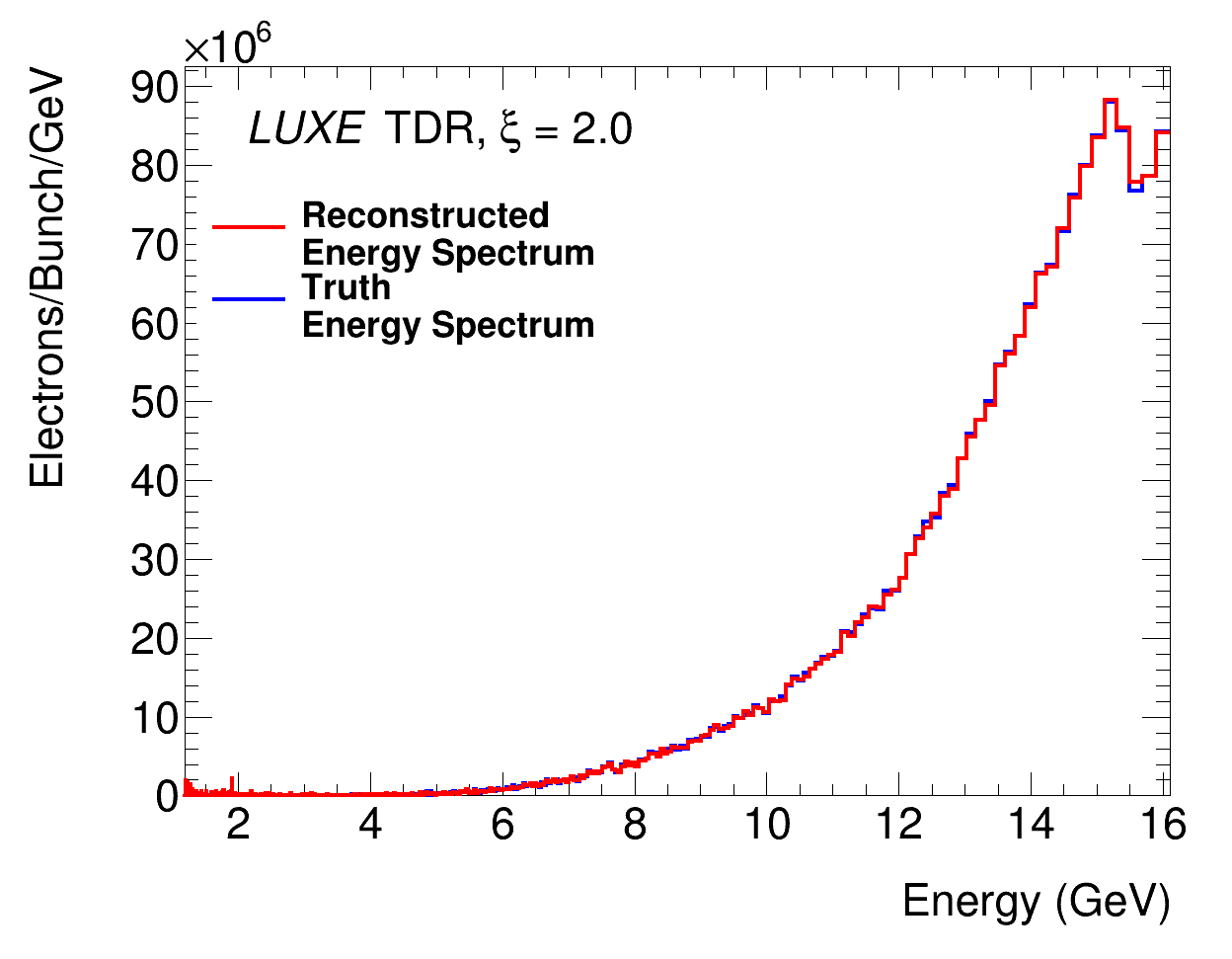}
\caption{ \label{fig:scint-recon-electron} 
}
\end{subfigure}
\caption{\label{fig:scint-recon} \textbf{(a)} A reconstruction of an example signal electron distribution (with beam) propagated through \geant. 
This represents a simulation of photons detected per electron within the active cameras, given the solid angle of the screen they cover and their quantum efficiencies, although the electronic aspects are not explicitly simulated. \textbf{(b)} Shows a reconstruction of the same data for the finer, non-regular energy binning corresponding to a position resolution of 1\,mm.
}

\end{figure}

\begin{figure}[htbp]
\includegraphics[width=0.49\textwidth]{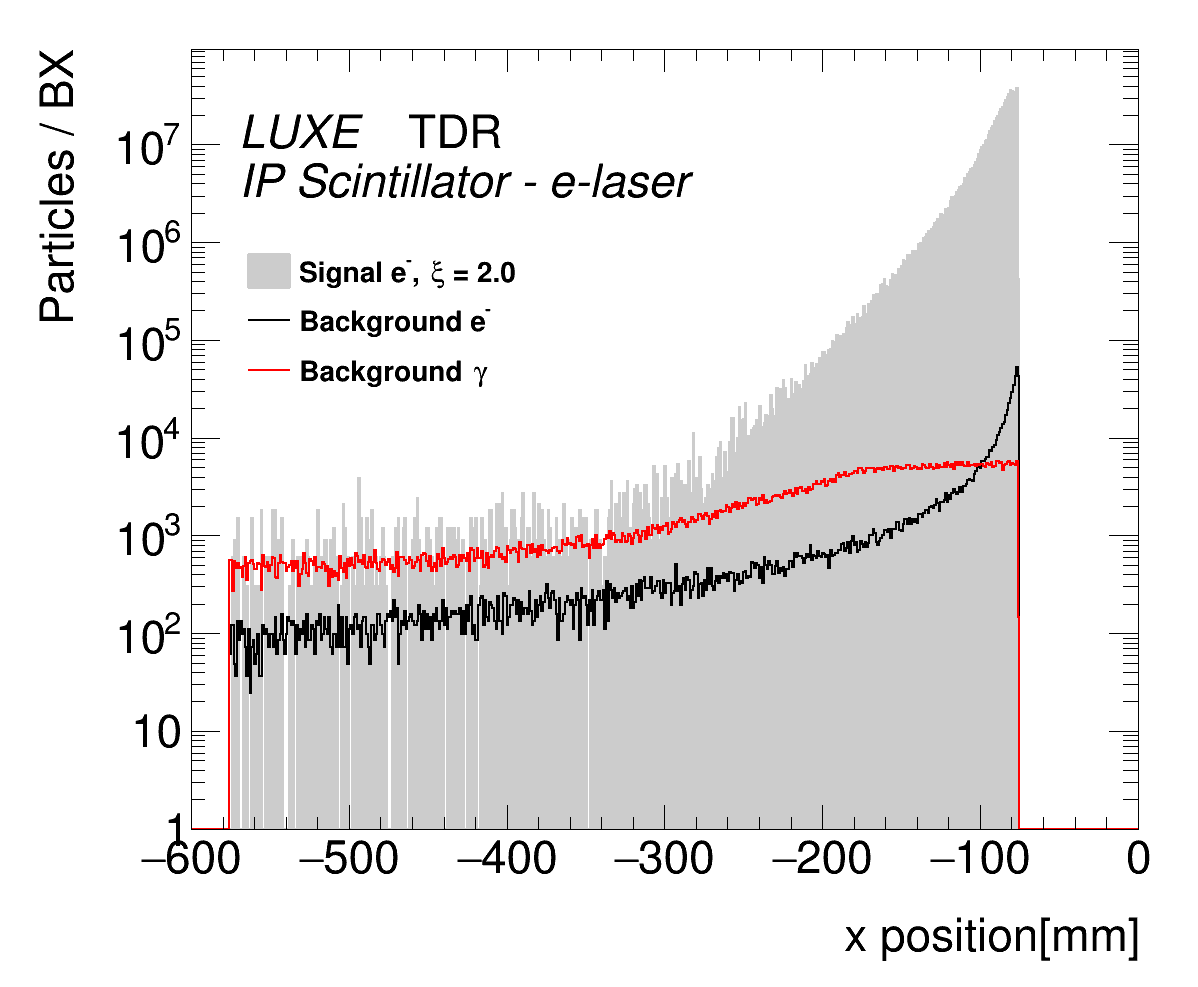}
\includegraphics[width=0.49\textwidth]{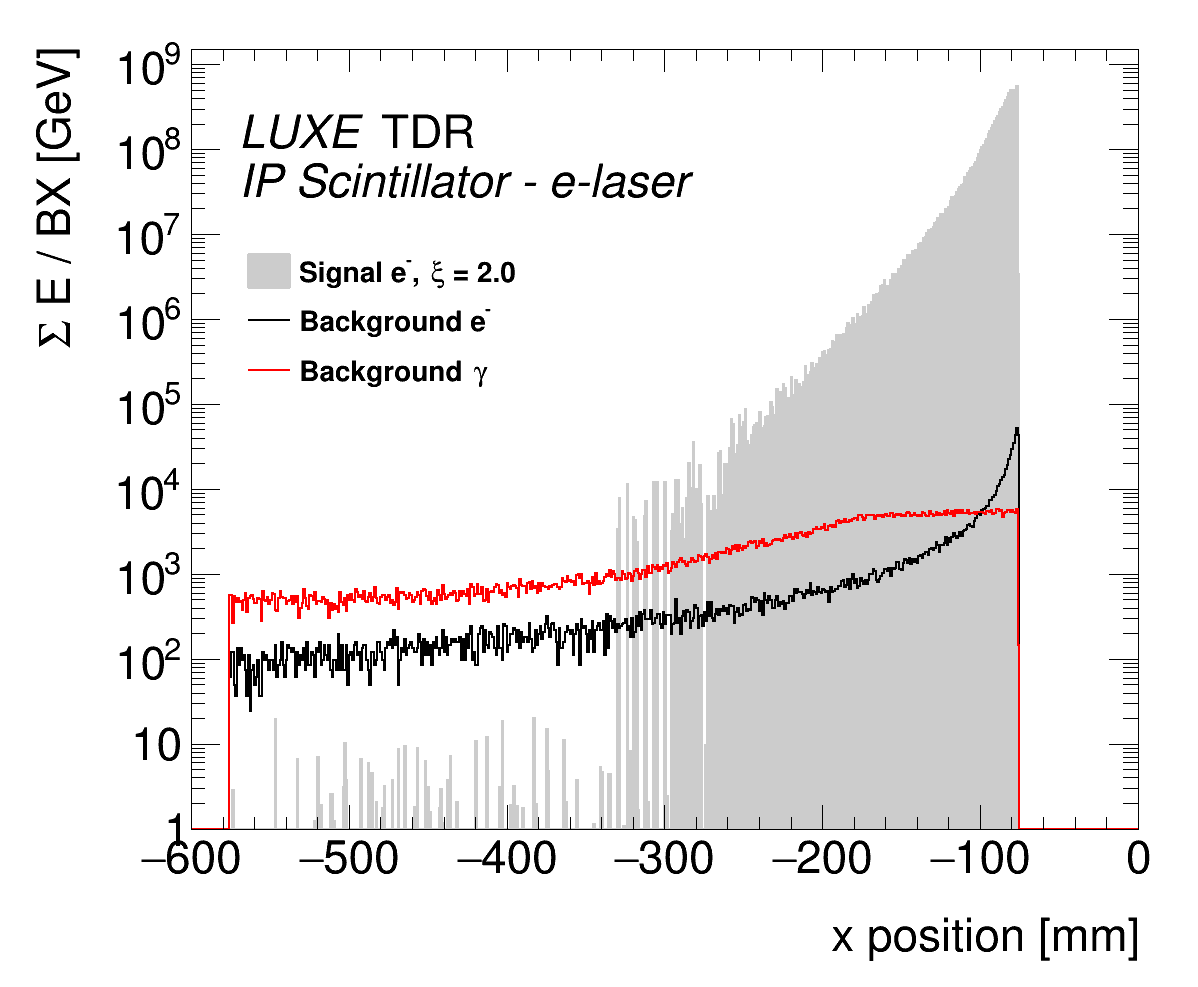}

\caption{The signal vs background rates for the EDS for phase-0 of the experiment with $\xi=2.0$. Particles per bunch crossing within mm bins is given on the left, and the integral of energy of the particles is given on the right.}
\label{fig:ip_lanex_sig_bkg}
\end{figure}

The final uncertainty for this measurement over several bunch-crossings is not expected to be dominated by statistics. The LUXE geometry has been implemented in \geant , and the scintillation physics response to signal spectra (from \ptarmigan) explicitly simulated. The light output within the simulation has been used to reconstruct the energy spectrum, using a developed reconstruction algorithm. The reconstruction in Figure~\ref{fig:scint-recon} shows a difference between reconstruction and truth spectra of <1\% for most bins. The sum of ``background" (beam-induced stray radiation), which is accounted for in these simulations and reconstructions, fluctuates between $\sim1\%$ and $\sim0.1\%$ of signal levels for relevant regions of the screen. This is seen in Figure~\ref{fig:ip_lanex_sig_bkg}. 

The final expectation for uncertainty in flux can only be given in broad terms, with precision measurements of the system in test-beam still to be performed, but reference can be made to similar experiments which have achieved uncertainty at 2\% for light-charge calibration on measurements of GeV-energy electron rates~\cite{awake_nature}. 

The dominant uncertainties then come chiefly from uncertainty associated with positioning of the components, the uncertainty of the calibration curve and the uncertainty associated with the B-field. These can all be projected to be of order 1\% or much less, so the final uncertainty of both the measurement of the total flux of electrons and the energy of features in the spectrum, are expected to remain within 2\%.


\subsection{Analysis Using Finite-Impulse-Response Filters}
\label{subsec:scint_fir}


The previously described process of simulation and energy reconstruction of the scintillation screen \& camera system uses the output of SFQED \ptarmigan \cite{ptarmigan} simulations as input electron spectrum. The ideal, nominal $\xi$ parameter of each simulation is then known, as is the distribution of $\xi$ values at the point of each instance of non-linear Compton scattering. 

As mentioned in LUXE literature~\cite{CDR}, the position of the edge feature depends on the $xi$ 
value at the point of non-linear Compton emission (see Eq.~\ref{eq:edgexi}). 
This measurement of the Compton edge becomes more challenging for the EDS detectors at higher $\xi$ as its position shifts towards higher energies.Due to the characteristics of the magnetic dipole spectrometer the energy resolution degrades as a function of energy (roughly equivalent to a constant percentage as per Fig. \ref{fig:deltaEoverE}). As can be seen in Fig. \ref{fig:comptonspectra} the clarity of the edge (above an otherwise smooth spectrum) degrades too. This is in part due to the greater range of $\xi$ values where interactions have occurred for an increasing nominal -- or maximum -- $\xi$. In order to reach a higher nominal $\xi_{max}$, the \laser must be focussed tighter, meaning a greater proportion of the finite-sized electron beam passes through a region of the 3-dimensional $\xi$ distribution lower than $\xi_{max}$. 

\begin{figure}[h!] 
\begin{subfigure}{0.49\hsize} 
\includegraphics[width=\textwidth]{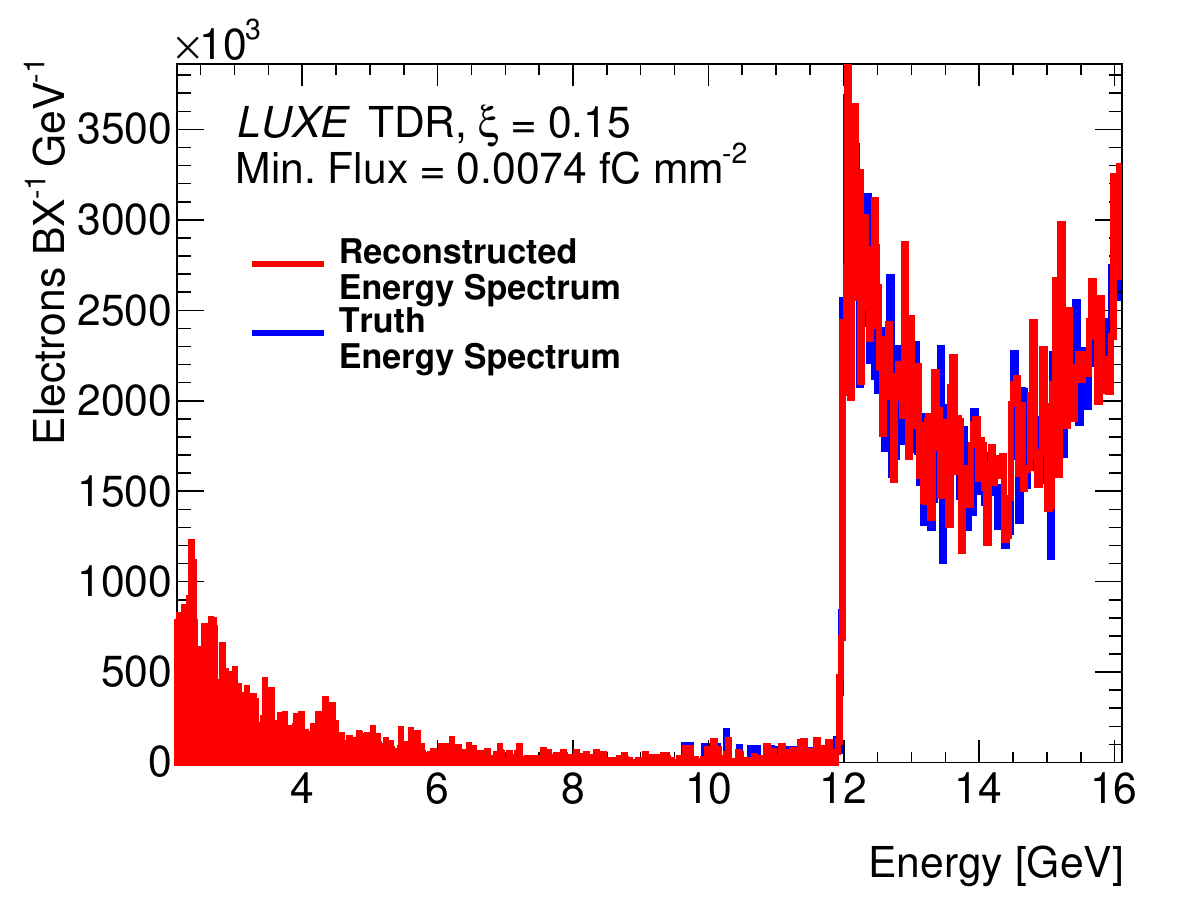}
\caption{  \label{fig:qe_2K}
}
\label{fig:photon-before-IP-scint-light}
\end{subfigure}
\begin{subfigure}{0.49\hsize}
\includegraphics[width=\textwidth]{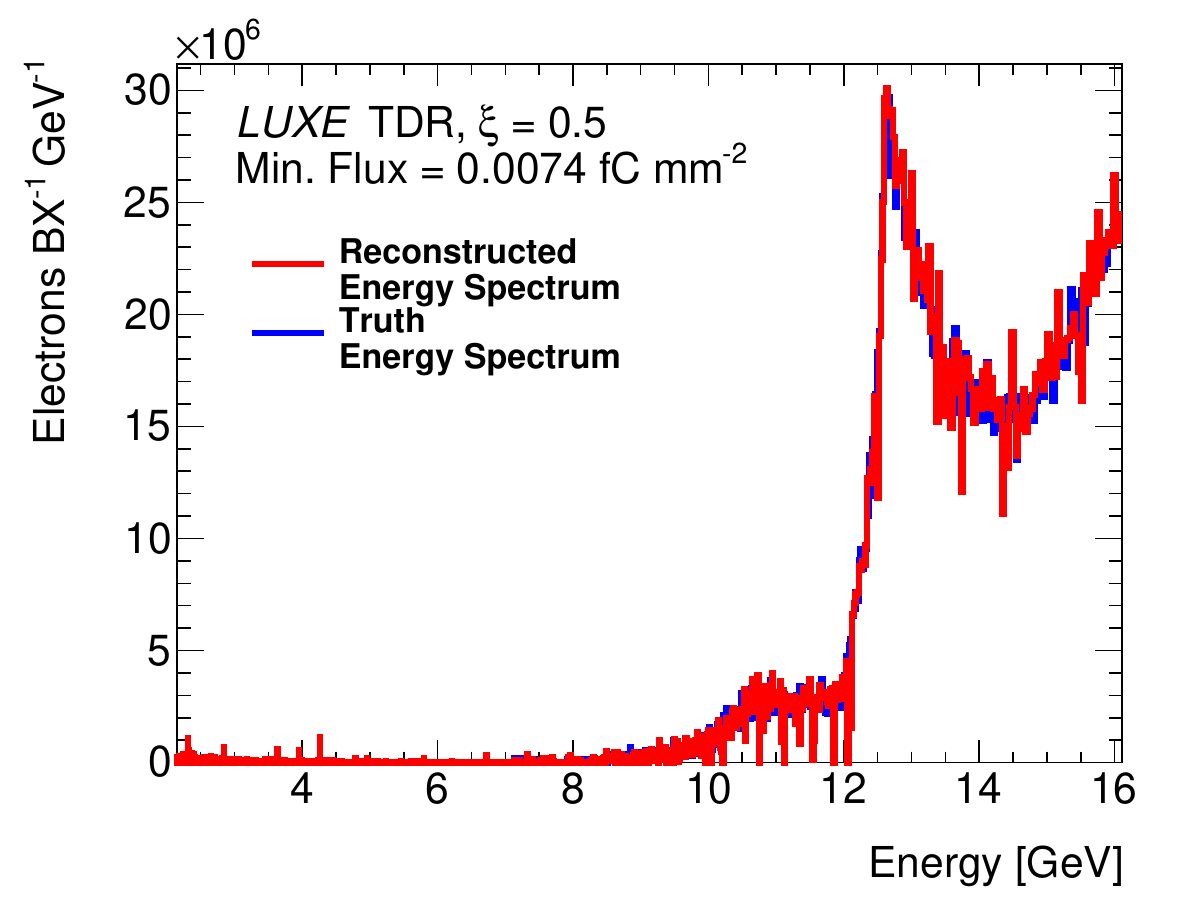}
\caption{ \label{fig:qe_4K} 
}
\end{subfigure}
\begin{subfigure}{0.49\hsize} 
\includegraphics[width=\textwidth]{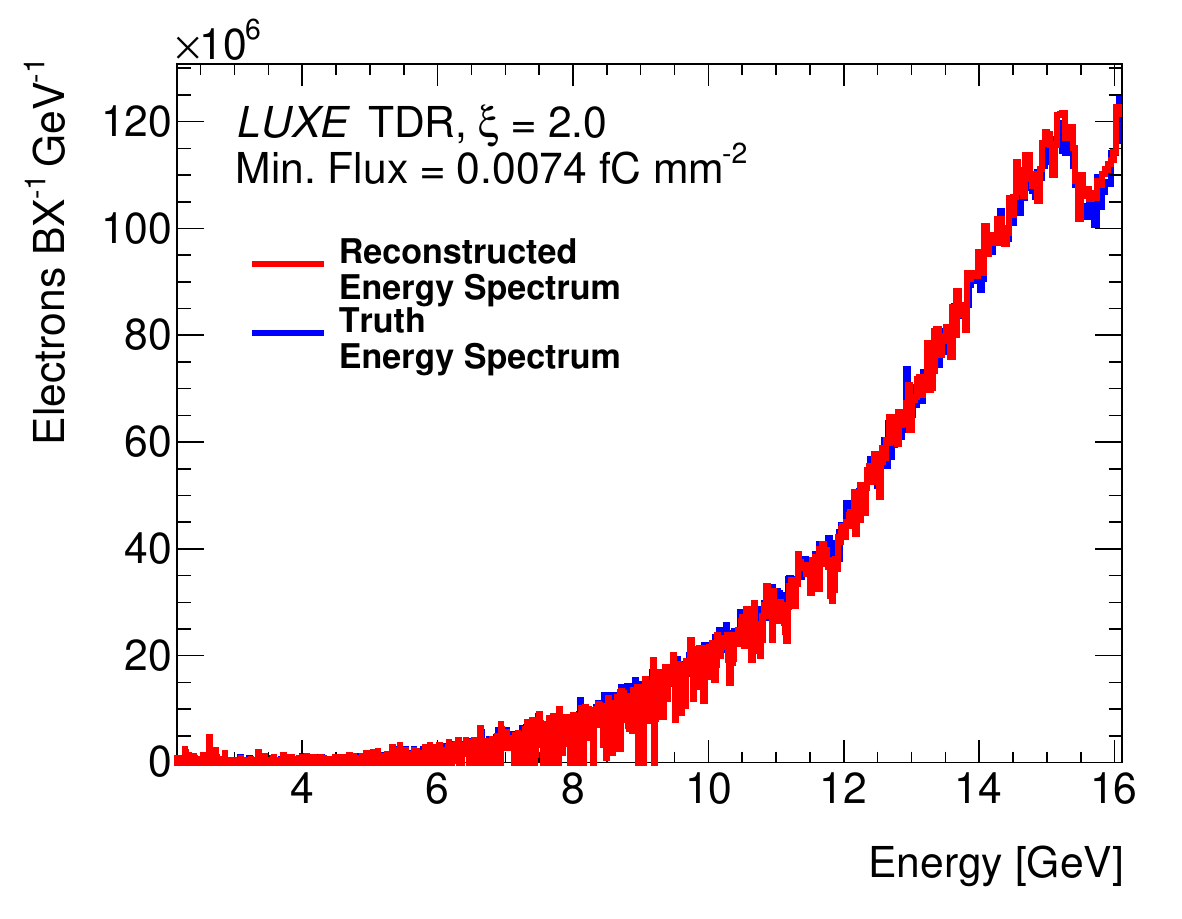}
\caption{  \label{fig:qe_2K}
}
\label{fig:photon-before-IP-scint-light}
\end{subfigure}
\begin{subfigure}{0.49\hsize}
\includegraphics[width=\textwidth]{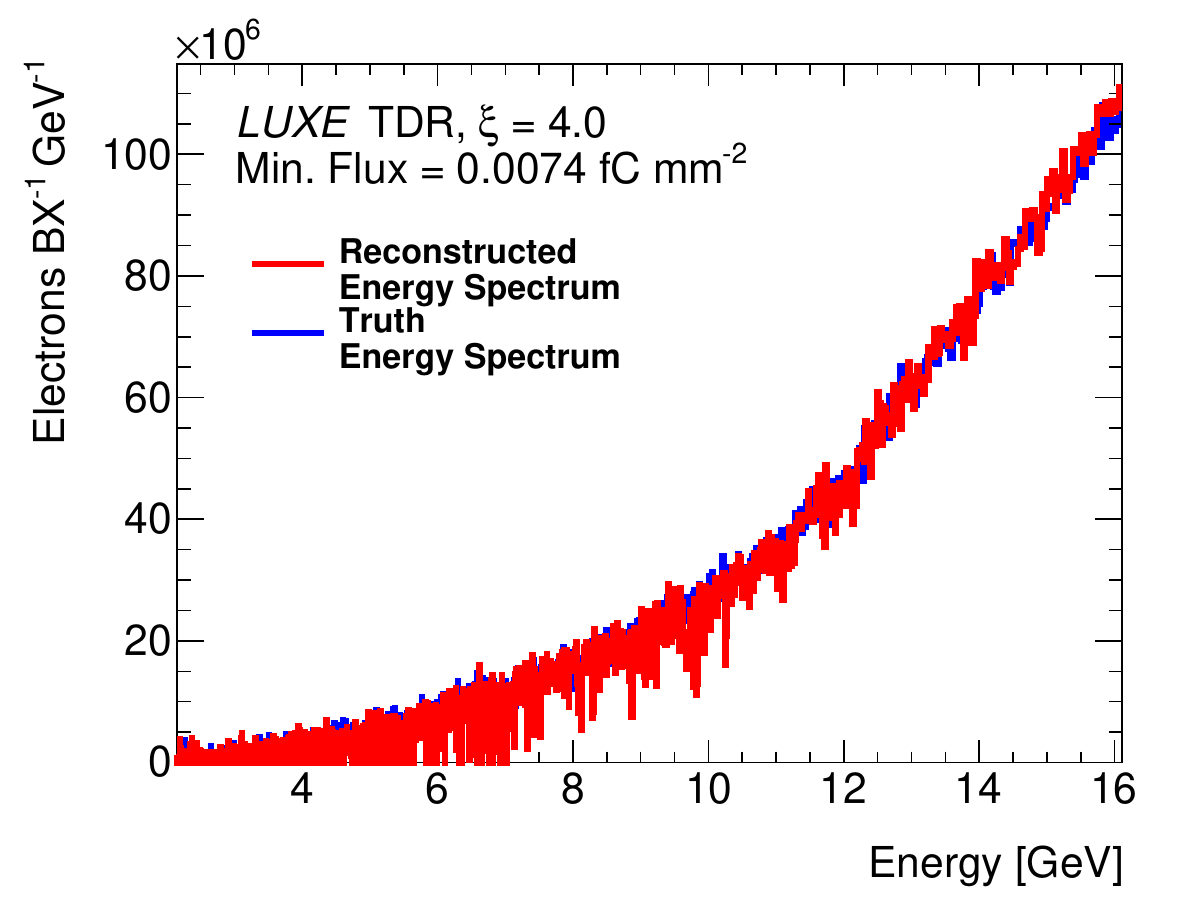}
\caption{ \label{fig:qe_4K} 
}
\end{subfigure}

\caption{\label{fig:fineresolutionsingleshots} \textbf{(a)} A reconstruction of a single-shot example SFQED electron spectrum, from \ptarmigan, with $\xi_{max}=0.15$ using the \geant simulation of the screen and camera system. This uses a threshold-of-sensitivity value (wherein no signal data at on-screen flux lower than this is used) of 0.0074 fC mm$^{-2}$ in the reconstruction. The effective resolution used for this simulation is 200 \textmu m. \textbf{(b)} Shows a similar analysis for a \ptarmigan simulation with larger value of $\xi_{max}=0.5$. \textbf{(c)} and \textbf{(d)} show reconstructions for the $\xi_{max}=2.0,~4.0$ \ptarmigan simulations respectively.}

\end{figure}

Figure \ref{fig:fineresolutionsingleshots} shows the results of reconstructions of one-shot simulations (including only the statistics one expects from one bunch-crossing), with the realistic scattering, statistics and noise characteristics of the system included. The pixel resolution here is set at 200 \textmu m, similar to the nominal resolution of the 2K camera defined in Table \ref{tab:camera_model_parameters}.

\begin{figure}[h] 

\begin{subfigure}{0.49\hsize} 
\includegraphics[width=\textwidth]{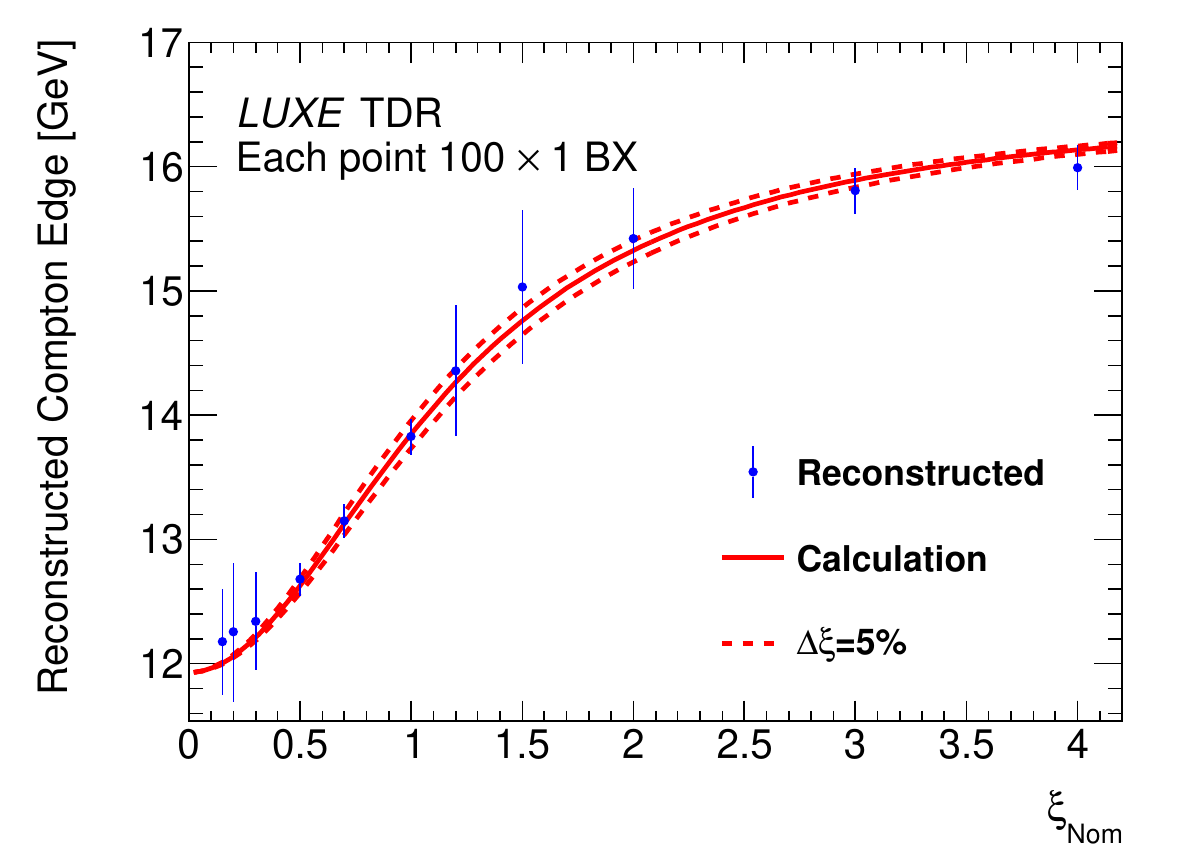}
\caption{  \label{fig:maxbinxiplotsa}
}
\label{fig:photon-before-IP-scint-light}
\end{subfigure}
\begin{subfigure}{0.49\hsize}
\includegraphics[width=\textwidth]{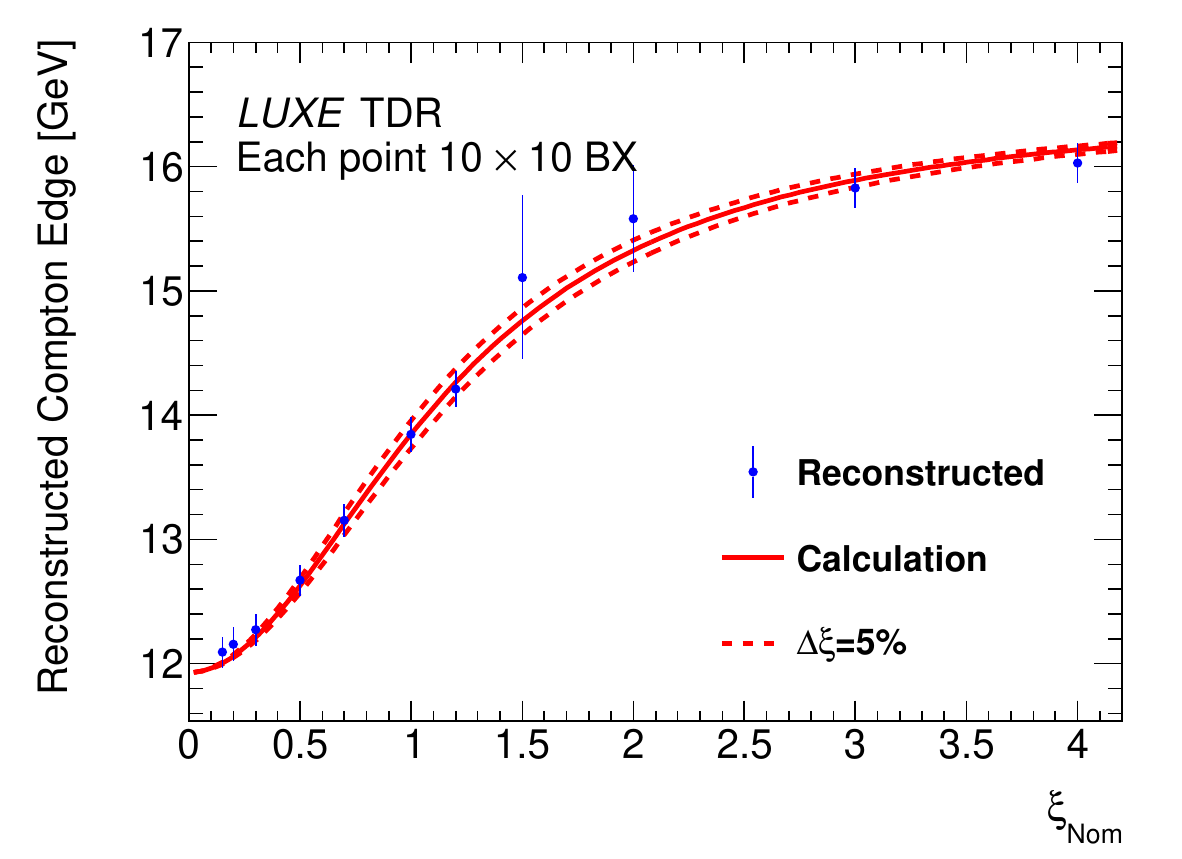}
\caption{ \label{fig:maxbinxiplotsb} 
}
\end{subfigure}

\caption{\label{fig:maxbinxiplots} \textbf{(a)} The reconstructed Compton edges, with 100 single-bunch-crossing simulations for each point in $\xi$. The reconstruction is performed with the energy of the centre of the maximum bin in the reconstructed energy spectra. \textbf{(b)} Shows a similar analysis, instead for the spectra of 10 accumulated bunch-crossings, with each point showing the mean of ten of these analyses. 
}

\end{figure}

The highest nominal $\xi_{max}$ achieved in the interaction point is reconstructed using the position in energy of these Compton edges. This can be simply analysed as the local maximum of the energy distribution. This is then compared to the expected Compton edge, calculable from the $\xi_{max}$ known from the \ptarmigan simulation using equation (\ref{eq:edgexi}). In Fig. \ref{fig:maxbinxiplots} one can see the result of this simple analysis, for a range in $\xi_{max}$ values between 0.15 and 4.0. Above this value of $\xi = 4.0$, the Compton edge is shifted beyond the acceptance of the scintillator screen.  Each point is the mean of 100 (Fig.~\ref{fig:maxbinxiplotsa}) or 10 (Fig.~\ref{fig:maxbinxiplotsb}) reconstructions from independent simulations. This is compared to the theoretical ideal, plotted alongside upper and lower bounds for $\pm 5\%$ uncertainty in $\xi$. The error bars of the reconstructed points are the sum in quadrature of a 1\%  systematic energy uncertainty estimate and the standard deviation of the distribution of reconstructed edge positions. Figure \ref{fig:maxbinxiplotsa} shows the result, for each value of $\xi$, of the mean of 10 samples of the sum of 10 bunch crossings. Figure \ref{fig:maxbinxiplotsb} shows the analysis for 100 different single-bunch-crossing energy spectra. As one might expect, the results for only one bunch-crossing are poorer than the sum of ten, represented in the typically larger errors. 

The best single-shot resolution of this edge is desirable to be able to discriminate between bunch-crossings and therefore shot-to-shot variations of the $\xi$ distribution (due to possible timing or spatial inconsistency). Fluctuations due to statistics and noise inhibit these reconstructions. One technique to recover subtle edges in kinematic distributions, particularly above noise, is the finite-impulse-response (FIR) of a matched filter function. This has been already studied in the context of the \cer detector (see Chapter~\ref{chapt7}), but eminently also has use in this context with the scintillation screen \& camera detector system. 

The application of various FIR filters is exemplary in Ref. \cite{Berggren:2020gna}. Elucidated in this publication is the use of convolutions between a discrete observed distribution, and a discrete matched filter, in order to recover any non-continuous or sharp rises or falls in the distribution, known as a kinematic `edge'. The location of edges can be obtained by finding the local extrema of the derivative of the filter convolved with the desired kinematic distribution. Consequently, for a finite edge feature, the end of the edge may be recovered from the zero-crossing in the response immediately after the local maximum. The response can be expressed in the continuous form with:


$$
R_{1}(x) = \frac{d}{dx} (f(x) * g(x)) = (f'(x) * g(x)) 
$$

Since $(f*g)' = f'*g = f*g'$, from the properties of the convolution. 

As is typical in physics, the real data expected is in a discrete form. The equivalent discrete form is:

$$
R_1 (i) = \sum_{k=-N}^{\infty} \left[ \frac{h_d(k+1) - h_d(k-1)}{2 \Delta k} \right] \cdot g_d(i-k)
$$

Within the aforementioned reference \cite{Berggren:2020gna}, it is seen that the matched filter distribution -- above denoted with $f'(x)$ -- of the first derivative of a Gaussian (also written as FDOG) provides the best results. For the analyses contained in this work, this FDOG construction of a matched filter is used. This function has a discrete form:

$$
h_d(k) = -k ~ \exp \left(- \frac{k^2}{2\sigma^2} \right)
$$

for integer $k$. If the edge has a finite, non-zero width $W$, then a characteristic scale factor $\sigma$ (of course this is the standard deviation of the original Gaussian form) similar to $W$ gives the best results in terms of filter efficiency.



\begin{figure}[h] 
\begin{subfigure}{0.49\hsize} 
\includegraphics[width=\textwidth]{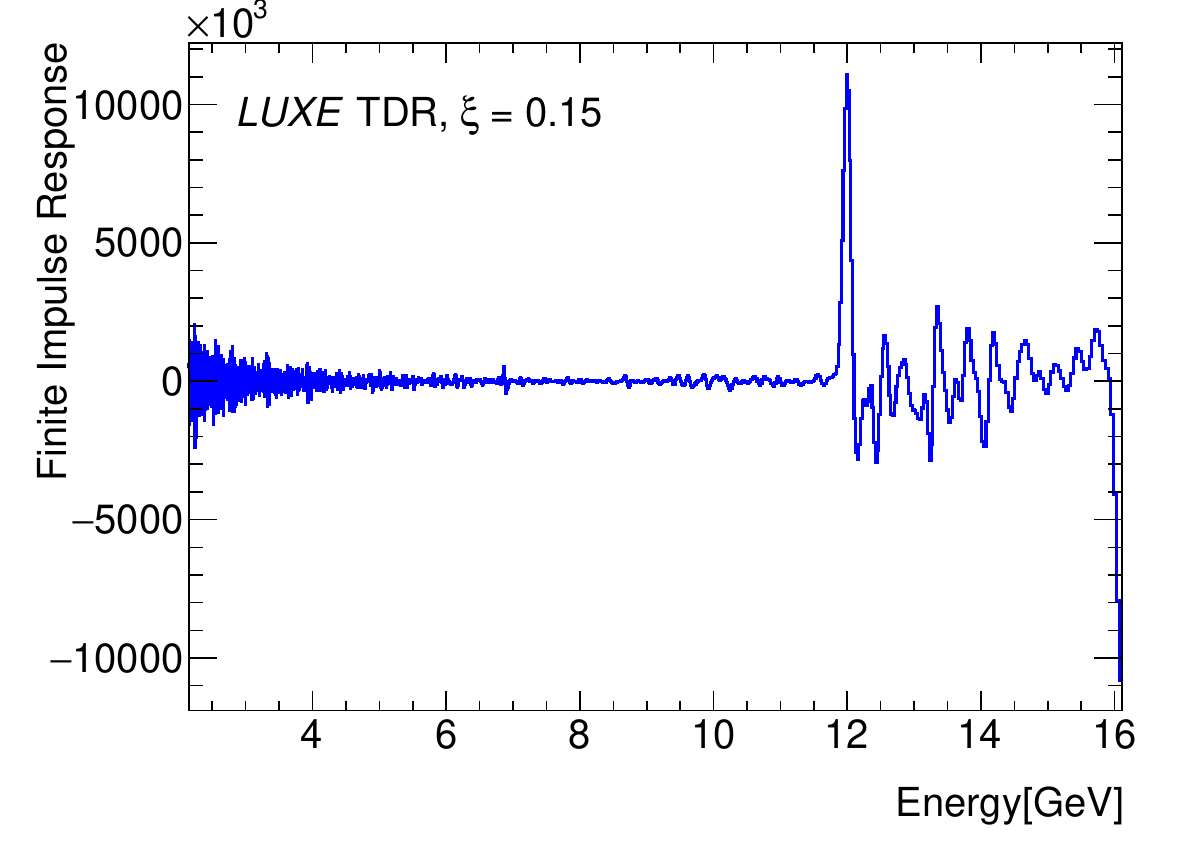}
\caption{  \label{fig:qe_2K}
}
\label{fig:photon-before-IP-scint-light}
\end{subfigure}
\begin{subfigure}{0.49\hsize}
\includegraphics[width=\textwidth]{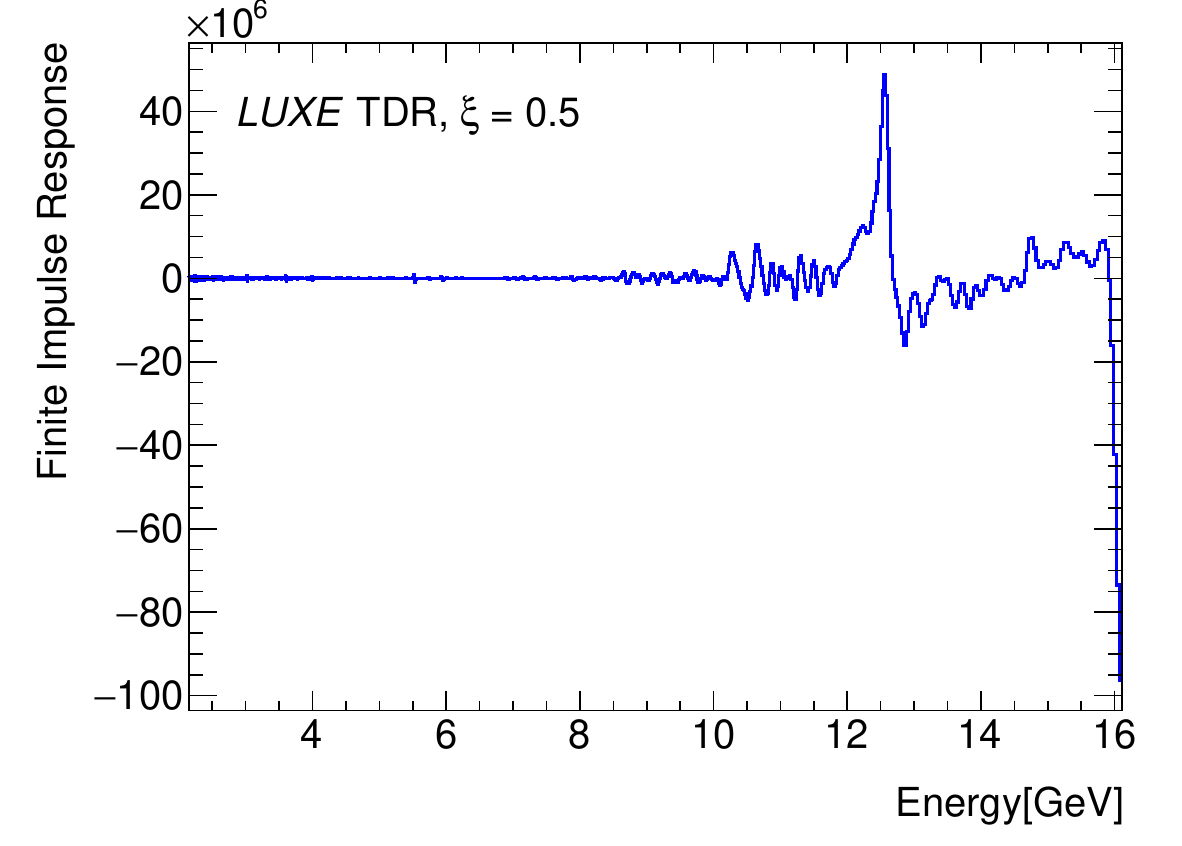}
\caption{ \label{fig:qe_4K} 
}
\end{subfigure}
\begin{subfigure}{0.49\hsize} 
\includegraphics[width=\textwidth]{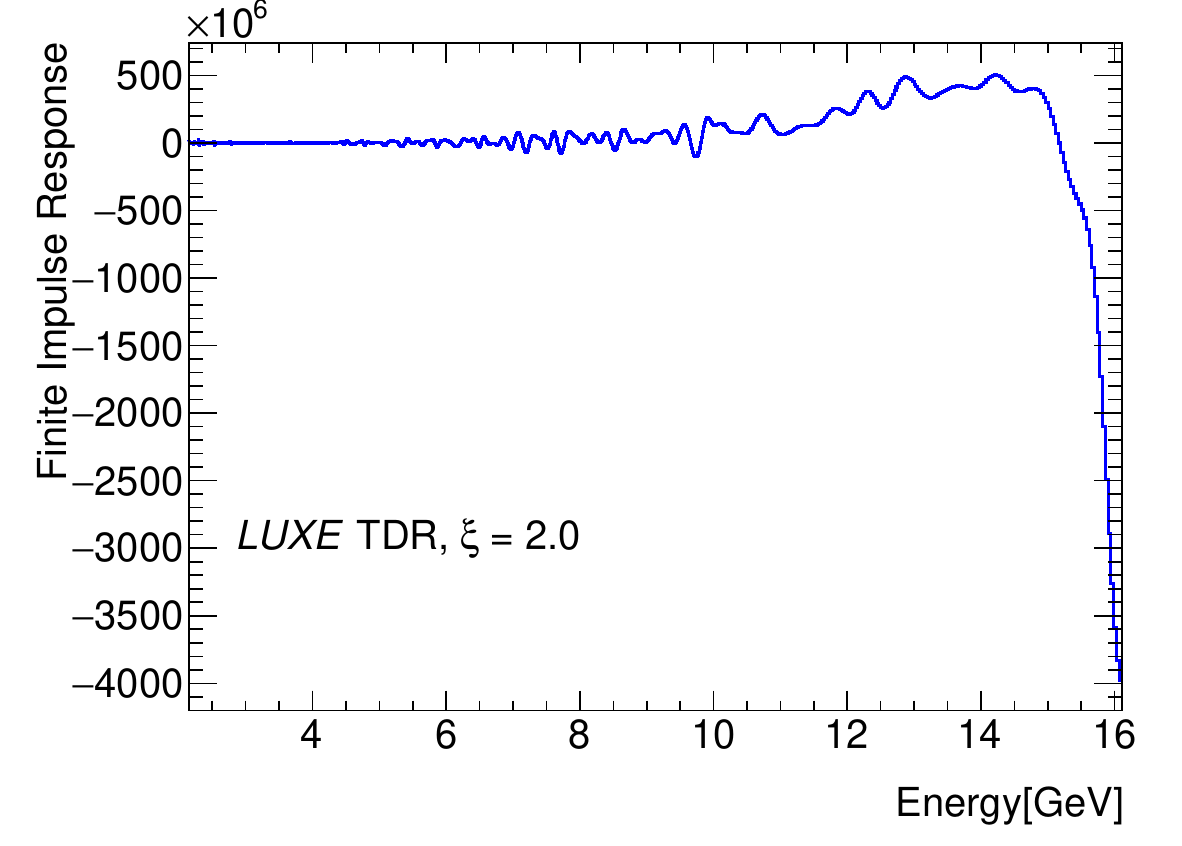}
\caption{  \label{fig:qe_2K}
}
\label{fig:photon-before-IP-scint-light}
\end{subfigure}
\begin{subfigure}{0.49\hsize}
\includegraphics[width=\textwidth]{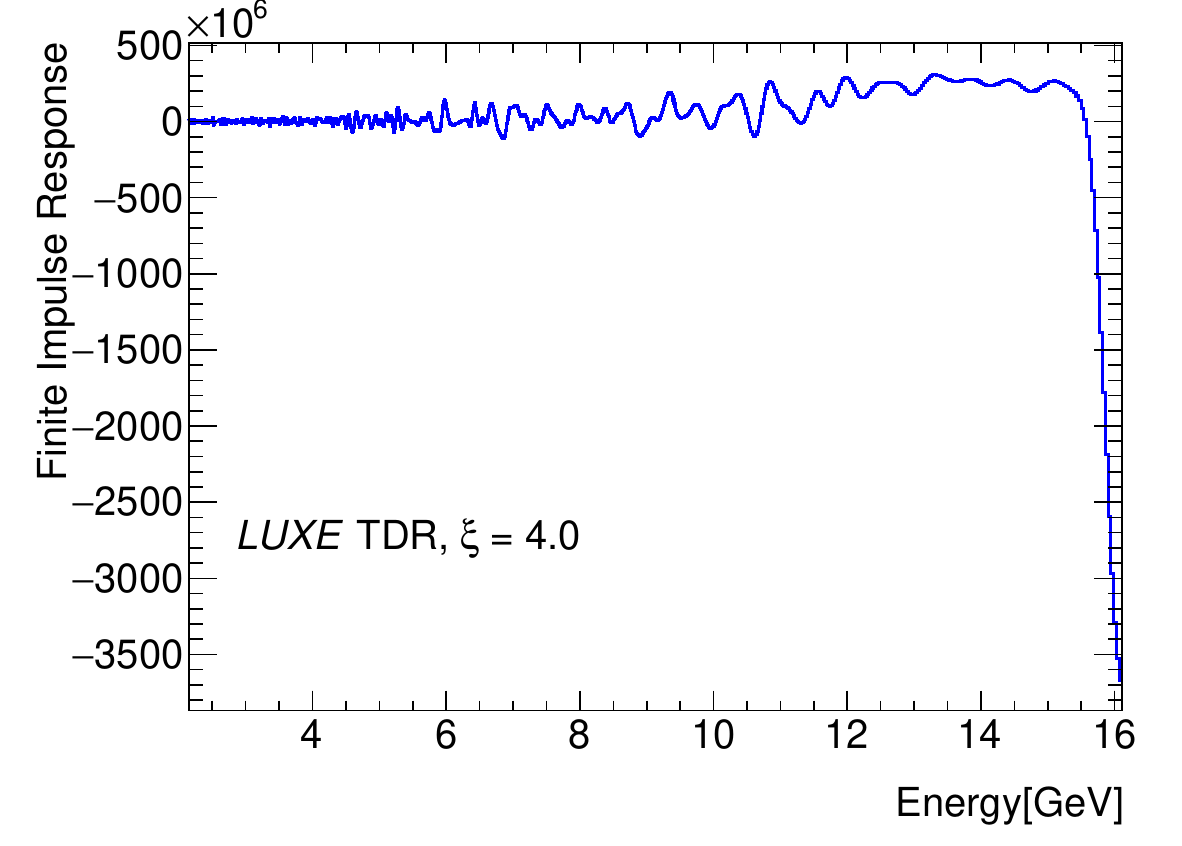}
\caption{ \label{fig:qe_4K} 
}
\end{subfigure}

\caption{\label{fig:firresponseplots} \textbf{(a)} The FIR for spectrum \textbf{(a)} for Fig. \ref{fig:fineresolutionsingleshots}. The y axis in each of these plots is arbitrary. \textbf{(b)}, \textbf{(c)} and \textbf{(d)} similarly show responses for the same filter of spectra \textbf{(b)}, \textbf{(c)} and \textbf{(d)} of Fig.  \ref{fig:fineresolutionsingleshots} respectively.}

\end{figure}

The FIR for each of the spectra in Fig.~\ref{fig:fineresolutionsingleshots} are shown in Fig. \ref{fig:firresponseplots}. The clarity of the primary peak is shown as degrading with increasing $\xi$, as explained. To find the maximum point of the finite edge, and therefore the $\xi_{max}$, the point that these distributions cross the x axis after the maximum in the spectra is used as the Compton edge position. 


The results of the reconstruction of the nominal $\xi$ using the FIR technique, again for a range of nominal $\xi$, are shown in Fig. \ref{fig:firresponsexiplots}.  Left 
shows the analysis for the application of the filters to 100 different single-bunch-crossing energy spectra.  Shown right are  
the results of the mean response of 10 samples, each of the sum of 10 bunch crossings. Compared to the simple reconstructions of nominal $\xi$ in Fig. \ref{fig:maxbinxiplots} the 10\,BX spectra results are similar, with a slight systematic underestimate coming from the FIR analysis at higher $\xi$ (notable at $\xi = 3.0, 4.0$). At the same time, for the single-shot analysis, the application of FIR filters has markedly improved the stability and mean accuracy of the low-$\xi$ (and so higher relative noise) reconstructions. These analyses are for particular values of the characteristic width of the matched filter $\sigma$, which have been iteratively optimised for each point in $\xi$. The error bars are again the sum in quadrature of a 1\%  systematic energy uncertainty and the standard deviation of the reconstructed edge positions.  

\begin{figure}[h] 

\begin{subfigure}{0.49\hsize} 
\includegraphics[width=\textwidth]{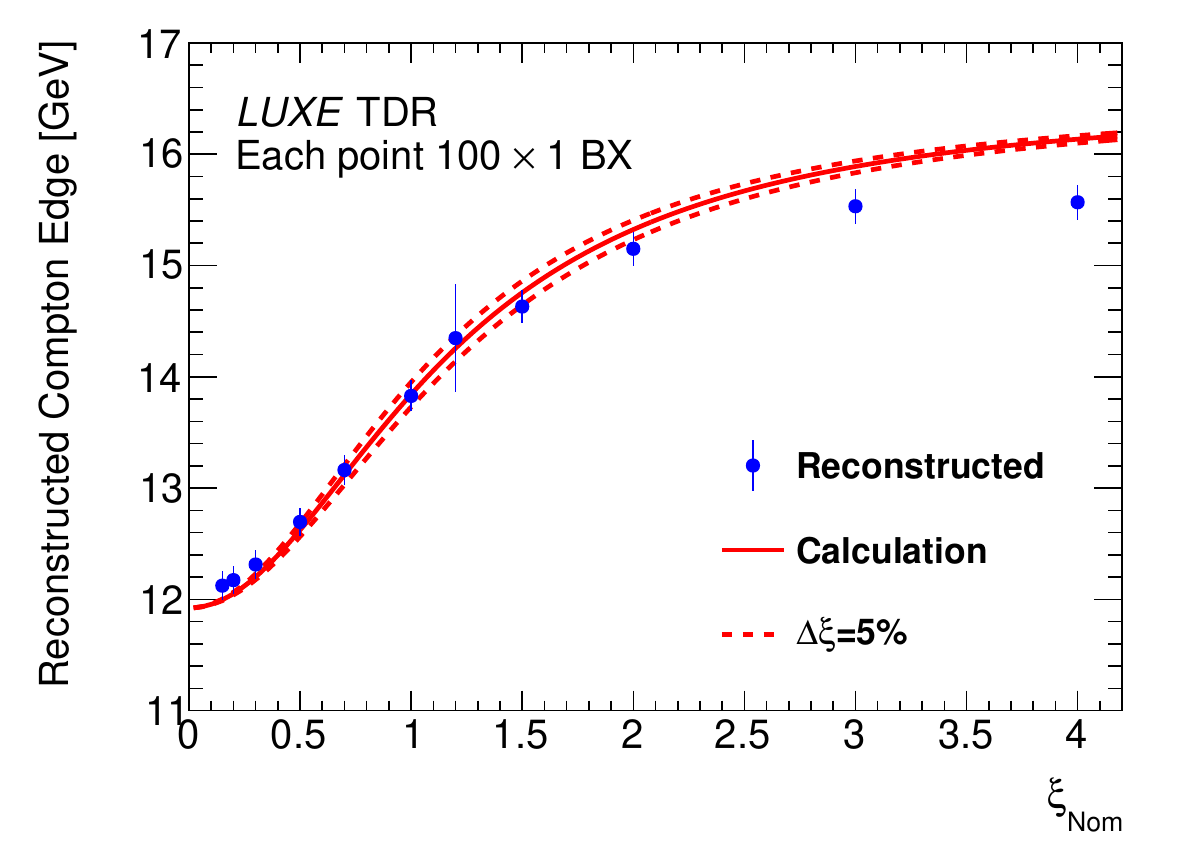}
\caption{  \label{fig:firresponsexiplots}
}
\label{fig:photon-before-IP-scint-light}
\end{subfigure}
\begin{subfigure}{0.49\hsize}
\includegraphics[width=\textwidth]{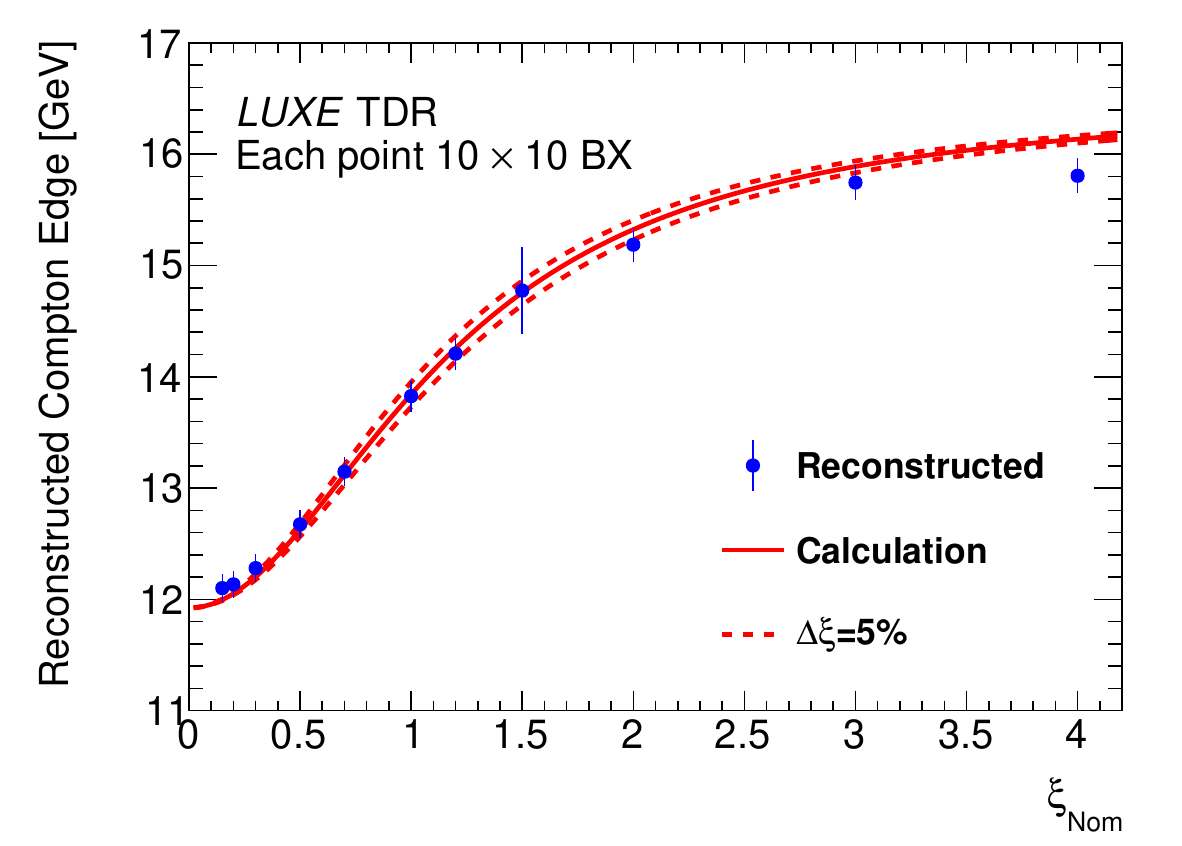}
\caption{ \label{fig:firresponsexiplots} 
}
\end{subfigure}

\caption{\label{fig:firresponsexiplots} \textbf{(a)} The reconstructed $\xi_\textrm{Nom}$ values for 100 single-bunch-crossing reconstructed spectra for each point in $\xi$. The reconstruction here is performed with the FIR technique. This is compared to the theoretical ideal alongside upper and lower bounds for $\pm 5\%$. \textbf{(b)} Shows a similar analysis, instead for the spectra of 10 accumulated bunch-crossings, with each point showing the mean of ten of these 10$\times$10\,BX spectra. 
}

\end{figure}

\begin{figure}[h] 
\centering
\includegraphics[width=0.8\textwidth]{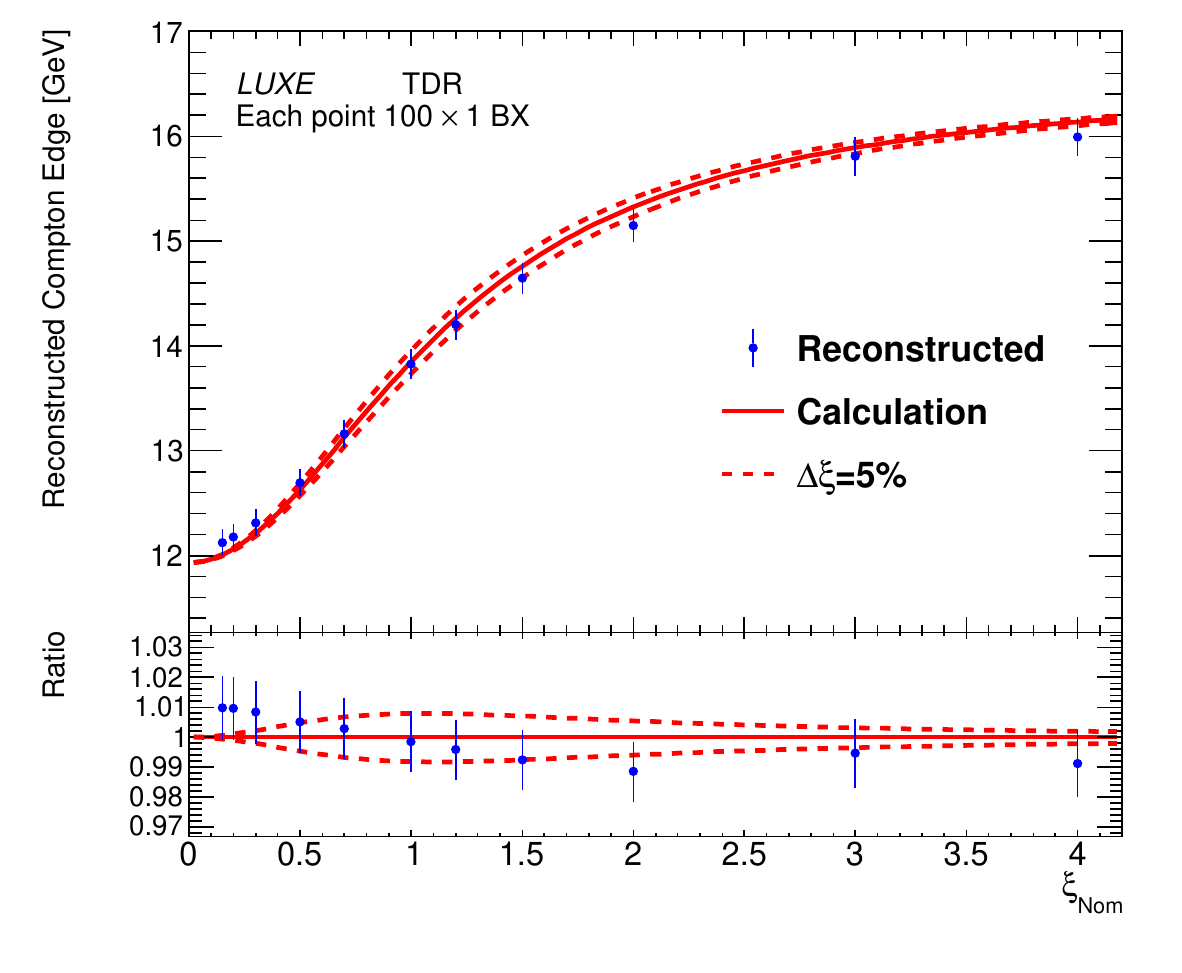}
\caption{ \label{fig:firoptimised} Optimised reconstructions of Compton edges with comparison to theoretical expectation for each $\xi_\textrm{Nom}$. Each point represents the analysis of 100 single-shot simulations.}
\end{figure}

With the use of these FIR filters for $\xi$ smaller than 2.0, and a simple maximum of the energy spectra above, the nominal $\xi$ of the simulations are reconstructed to the quality seen in Fig. \ref{fig:firoptimised}. For each of the points, the reconstructed edge is within (or very nearly within) one percent of the theoretical expectation from the simulations. Accounting for the given uncertainties, any single measurement of just one bunch-crossing is very likely to be made to an accuracy of 2 percent, satisfying the 2\% energy uncertainty requirement for the case of reconstructing Compton edges from single bunch-crossings. 

The expected position of the Compton edge at the value of $\xi = 4.0$ is at the limit of the scintillator detector acceptance, seen to be within 400 \textmu m from the edge of the screen. The results for the $\xi = 4.0$ points are therefore not meaningfully different from using the high-energy limit of the reconstruction in the case of a smooth, edge-less, rising energy distribution. In practice, we can expect to lose the ability to identify the Compton edge for some $\xi_{max}$ between 3.0 and 4.0 simply due to its shifting too close to the limit of the detector acceptance. 


\subsection{Laser-Plasma Accelerator Test-Beam}
\label{subsec:scint_beamtest}

\begin{figure}[h] 
\centering
\includegraphics[width=0.8\textwidth]{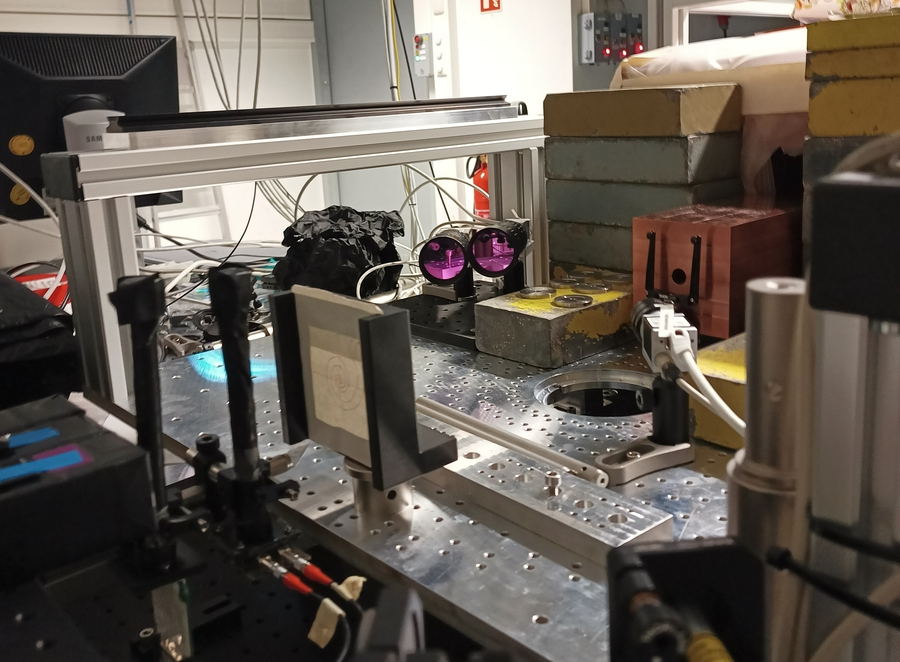}
\caption{ \label{fig:28m-testbeam-photo} A photograph of the setup in the DESY \lasernospace-plasma accelerator BOND lab. On the right hand side lies a 1\,cm-diameter collimator, from which the beam emerges, before reaching the screen in a black plastic frame, in front of Cherenkov detector straws (see chapter~\ref{chapt7}) and lead-glass calorimeter blocks (see chapter~\ref{chapt10}). Two cameras, triggered by the accelerator and fitted with filters which appear purple, collect the light.}
\end{figure}

\begin{figure}[h] 
\begin{subfigure}{0.49\hsize} 
\includegraphics[width=\textwidth]{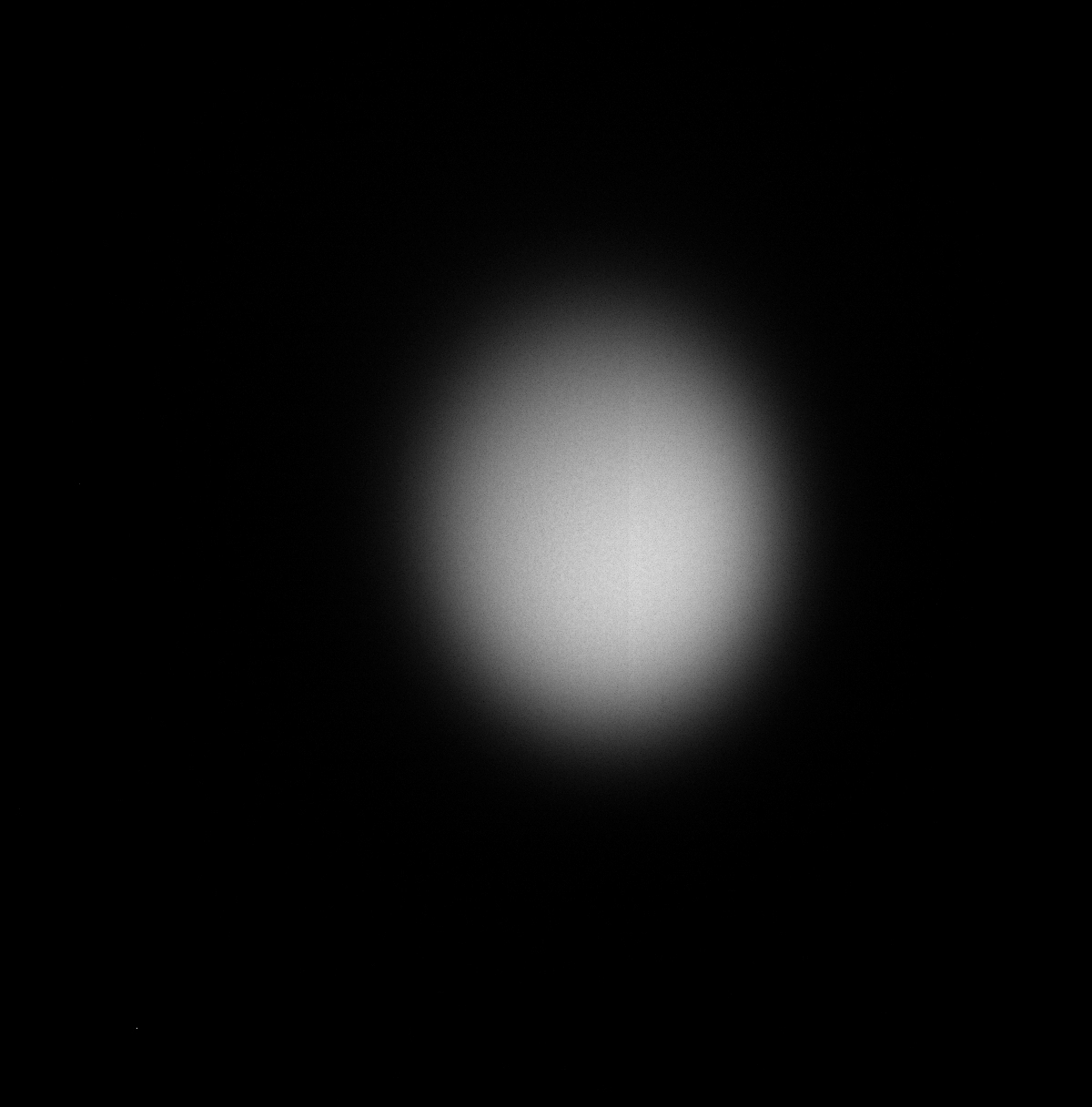}
\caption{  \label{fig:qe_2K}
}
\label{fig:photon-before-IP-scint-light}
\end{subfigure}
\begin{subfigure}{0.49\hsize}
\includegraphics[width=\textwidth]{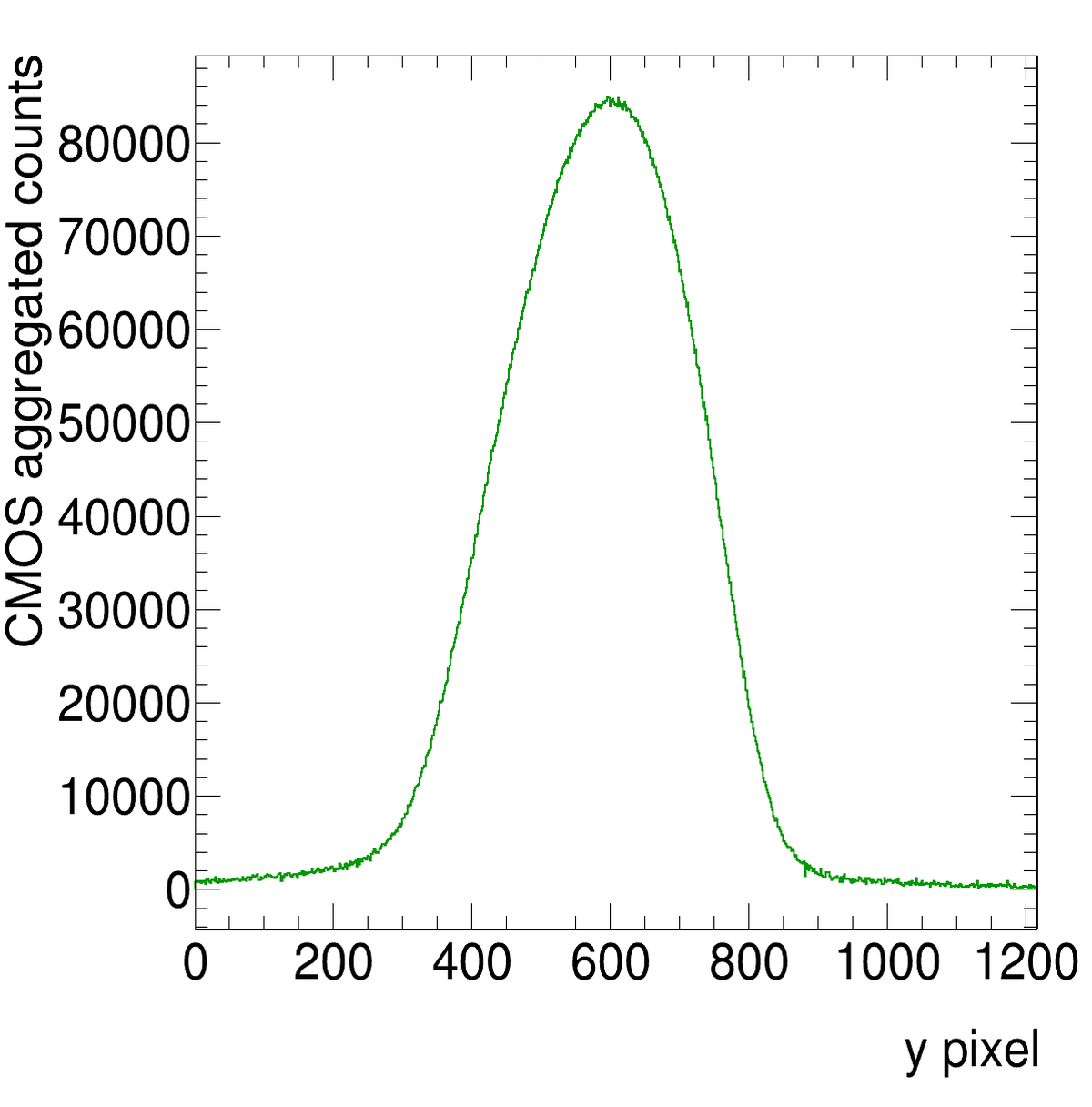}
\caption{ \label{fig:qe_4K} 
}
\end{subfigure}
\caption{\label{fig:28m-testbeam} \textbf{(a)} An accumulated image for 3000 beam bunch events, having subtracted dark-noise images, as taken by the 2K camera. The intensity scale is arbitrary for visibility. \textbf{(b)} The aggregated CMOS counts in each pixel of the same data set in the y dimension, having subtracted a background profile.}

\end{figure}

A photograph of the experimental setup within the \lasernospace-plasma facility -- alongside other detectors sharing the same beamtime -- is given in Fig. \ref{fig:28m-testbeam-photo}, and an example of accumulated signal image and a profile in one dimension are given in Fig. \ref{fig:28m-testbeam}. 

The bunches from this beam exhibited some variance in energy between 60-70 MeV, but the average response of the screen is not expected to depend on energy in this range (explicitly tested in simulation in \geant). The bunch-to-bunch charge varies much more, and the screen response then varies linearly with the incident charge. 

In the beam area two charge diagnostics were running, allowing reconstruction of the charge of individual bunches. The technology chosen is Damon Dark-Current detection, which measures the charge non-destructively, and is located within the beampipe upstream of the collimator \cite{damon}. The measurement consists of two antennae, one configured to measure, in principle, charges above 1 pC; and a second antenna intended for detection below this value. Reconstructions of the charge from the raw current data from the antennae have some significant tension; in the region where both antennae are sensitive, the reconstructions from the antenna sensitive to lower charges presented a roughly 60\% higher value. One or both of the two reconstructions then has drifted from their once-accurate calibration parameters. Hereafter the calibration correlating to the higher-charge-range-antenna is used for calculations, as this exhibited a saturation effect at an absolute value as expected around 0.5 pC (and therefore less likely to have suffered major drift in its detection response). This tension still represents a large qualitative uncertainty in the charge reconstructed bunch-to-bunch. This is in addition to the considerable shot-to-shot variance of the measurement itself, and the loss of some unknown fraction of the charge sum to the edges of the collimator. In light of these compounding effects, an effective upper and lower bound are given to the calculation of values dependent on the bunch charge. These are chosen as $\pm 50\%$. This is most illustrative below when we choose each of the three values (central, lower bound, and upper bound) of minimum detectable flux and analyse their effects on the system in LUXE, using the \geant simulation. 

\begin{figure}[h] 
\begin{subfigure}{0.49\hsize} 
\includegraphics[width=\textwidth]{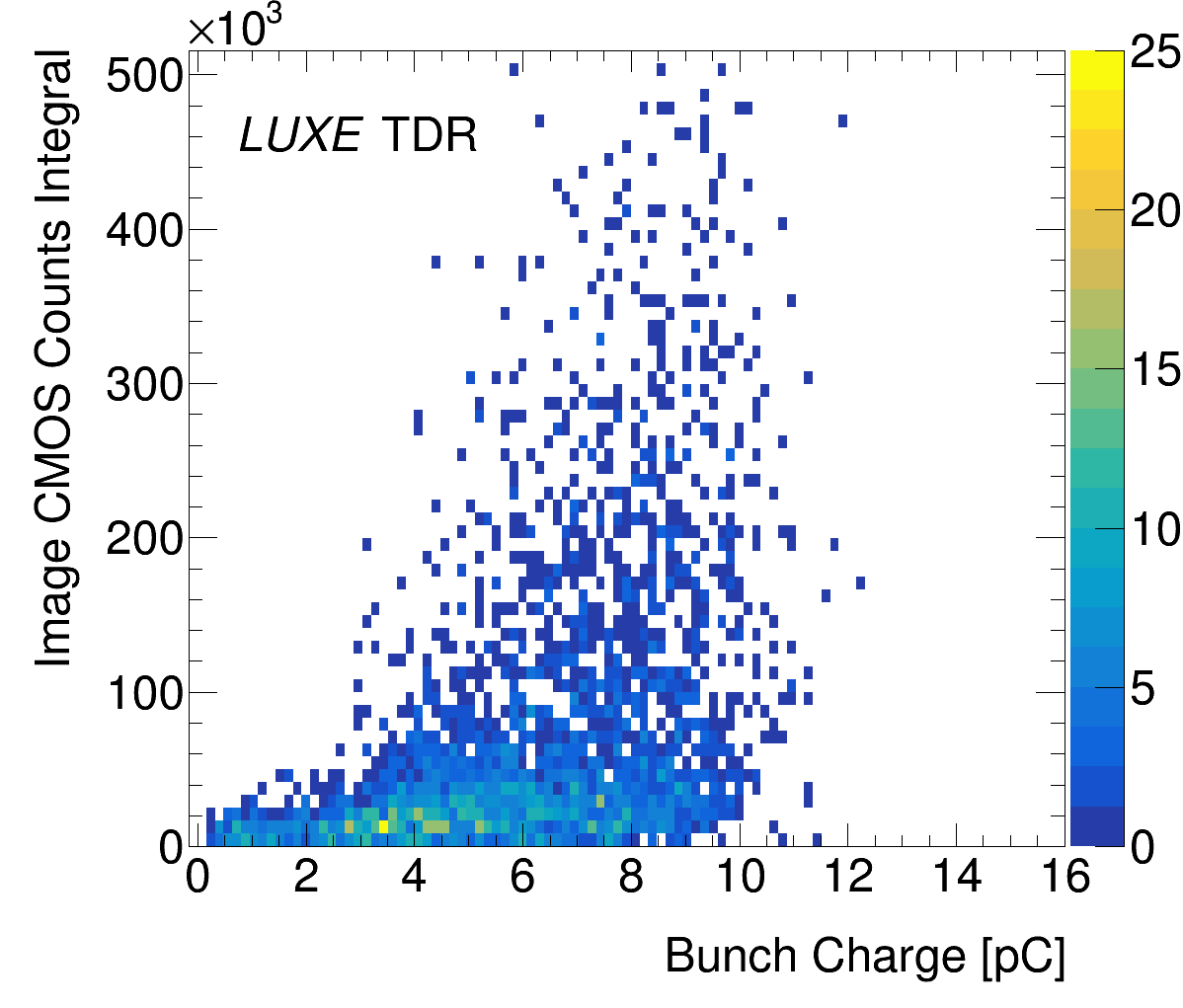}
\caption{  \label{fig:qe_2K}
}
\label{fig:photon-before-IP-scint-light}
\end{subfigure}
\begin{subfigure}{0.49\hsize}
\includegraphics[width=\textwidth]{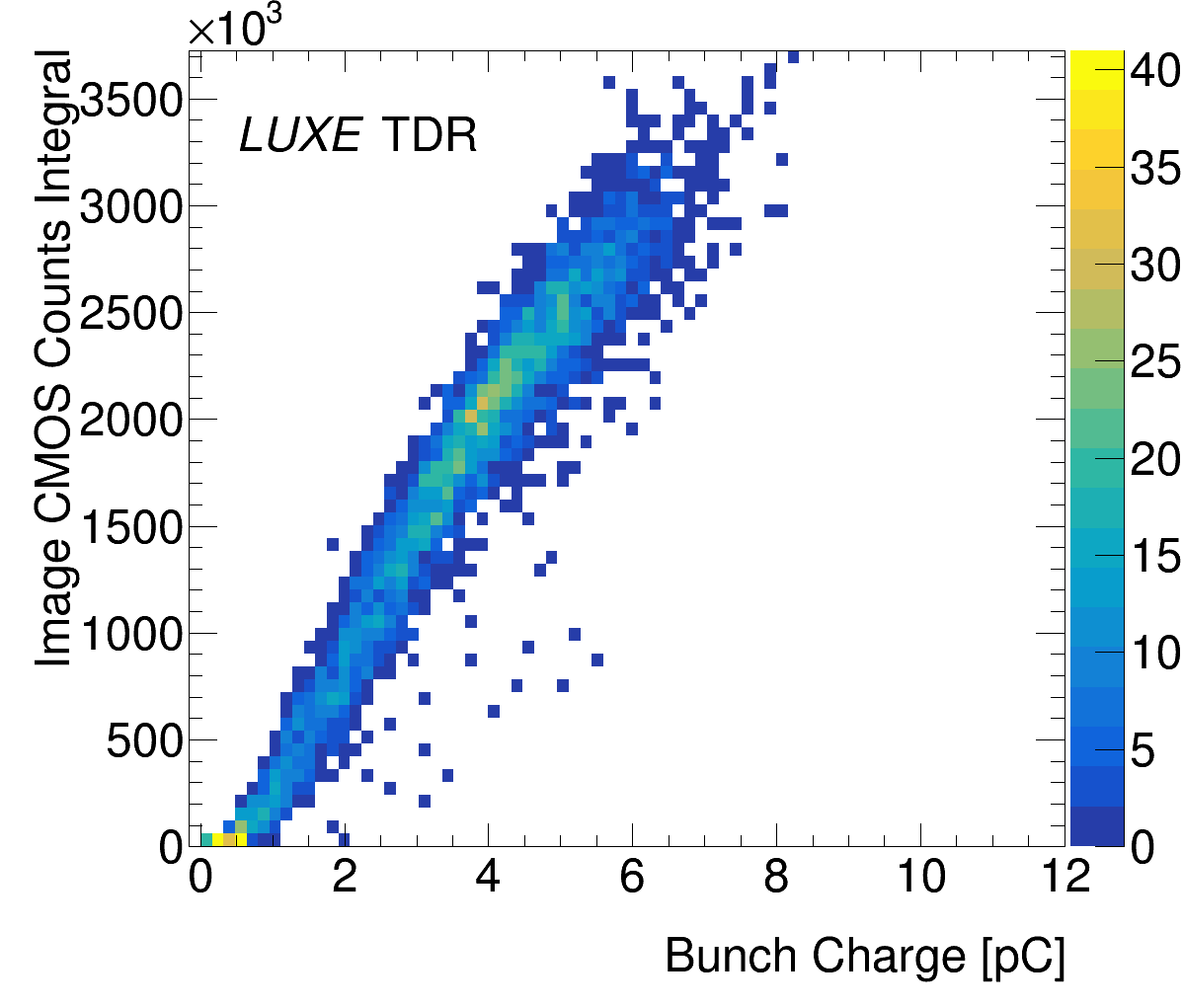}
\caption{ \label{fig:qe_4K} 
}
\end{subfigure}
\caption{\label{fig:plasmalens} \textbf{(a)} The correlation between the integral of digital signal, in terms of CMOS counts within the 2K camera, with the reconstructed incident bunch charge. This run does not use the focussing plasma lens element. \textbf{(b)} shows the data from an identical run, in terms of camera settings and the optical line, enabling the plasma lens within the accelerator.}

\end{figure}

The pointing of the laser-plasma accelerated beam was unreliable. Because the beamline is outfitted with a collimator of diameter 1 cm, the on-screen charge was highly dependent on this pointing, and had an inextricable effect invisible to the charge diagnostics due to the fact they were positioned before the collimator. The laser-plasma acceleration pointing problem was mitigated in some of the runs performed, using a plasma lens within the accelerator to better align the beam. This came at the drawback of a lower maximum bunch charge. The improved reliability of the pointing of the beam proved to be be essential, and so only data taken when using the plasma lens was chosen for analysis. Two representative plots showing the correlation between the screen and camera system signal versus the charge diagnostic are shown in Fig.~\ref{fig:plasmalens}.

The equation for integrated CMOS counts per pC on-screen charge is calculated by fitting slices of the data in the y axis, or the bunch charge, with a Gaussian function. Given the pointing of the beam will lead to the loss of some portion of the charge, likely proportional to bunch size, an estimate of the signal/charge is taken slightly above the mean of the distribution. In particular, the best estimate for the true CMOS-counts-integral for some on-screen charge is measured as the mean of the Gaussian plus one standard deviation. These values are fit with a linear expression. This is then compared to the variance of the integrated CMOS count of the area of interest (AOI) in dark images. The required charge to equal the standard deviation of the integrated CMOS counts (i.e. the point at which SNR=1) is then calculated. This is divided by the screen area covered by the AOI to retrieve an estimate for the flux at threshold of sensitivity. 

\begin{figure}[h] 
\centering
\includegraphics[width=0.8\textwidth]{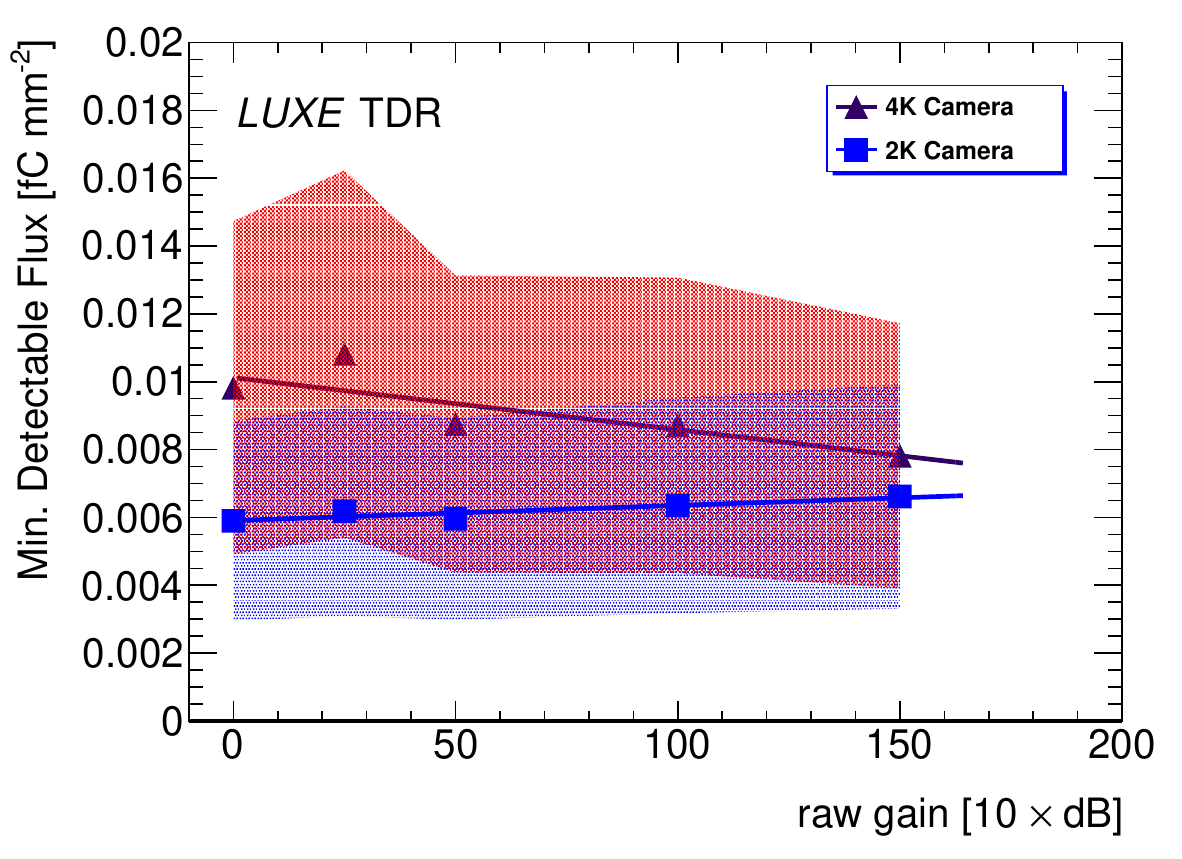}
\caption{ \label{fig:minimum-sensitivity} Plotted values for the reconstructed values of minimum detectable on-screen charge flux. This is performed with the plasma-lens-enabled data set, for varying values of electronic gain. The `raw gain' value is equivalent to the power-gain in dB multiplied by ten.}
\end{figure}

The reconstructed values for the threshold flux are shown in Fig. \ref{fig:minimum-sensitivity}, with the associated $\pm 50\%$, for various values of electronic gain. It is expected that the trend through the gain is relatively flat. The linear trendlines in Fig. \ref{fig:minimum-sensitivity} shown follow this understanding approximately, but the limited data available does not allow for accurate enough measurements to observe this in both cameras. Upper and lower bounds on the threshold-of-sensitivity flux (the mean, calculated for all values of gain) are also given in Table \ref{tab:mindetflux}. The 2K camera and its optics is shown to be more sensitive in this respect.

\begin{table}[h]
    \centering
    \caption{Reconstructed minimum detectable flux for each camera.}
\begin{tabular}{c|c|c|c}
         Minimum Detectable Flux & Upper & Central & Lower\\
         fC mm$^{-2}$ & Bound & Estimate & Bound\\
         \hline
         4K Camera & 1.398 $\times 10^{-2}$ & 9.32 $\times 10^{-3}$ & 4.662 $\times 10^{-3}$ \\
         2K Camera & 1.11 $\times 10^{-2}$ & 7.397 $\times 10^{-3}$ & 3.698  $\times 10^{-3}$ \\
         
         \end{tabular}
         \label{tab:mindetflux}
\end{table}

A simple comparison of the use of the optical filters is made. The result of the calculated value of the integrated CMOS counts per pC charge is compared for: the high-quality optical filter chosen for this system in LUXE at central wavelength 543\,nm; a less expensive filter of the Machine Vision filter \cite{machinevision} range from Edmund Optics; and no filter fitted. Unlike the absolute values of these reconstructions, when comparing the runs of differing filters relative to one another the problem of a consistent unknown charge calibration factor effectively cancels. The relative comparison is given in Table \ref{tab:filters}. 

\begin{table}[h]
    \centering
    \caption{Relative camera CMOS counts per pC for optical filter variations.}
\begin{tabular}{lccc}
         Signal & 543 nm Filter & Machine Vision Filter & No Filter\\
         \hline
         4K CMOS counts per pC & 6.98 $\times 10^{5}$ & - & 1.67 $\times 10^{6}$ \\
         2K CMOS counts per pC & 5.46 $\times 10^{5}$ & 6.96 $\times 10^{5}$ & - \\
         Relative Signal & 1 & 1.27 & 2.4 \\
         
         \end{tabular}
         \label{tab:filters}
\end{table}

By using the calculated CMOS counts per pC charge, and incorporating the optical parameters listed in Tables \ref{tab:camera_model_parameters} and \ref{tab:lens_model_parameters}, the effective upper and lower detectable flux at the limits of each camera's dynamic range are calculated. The ranges of the two cameras, which invariably share most of one another's flux range, can be combined now into one system of effective dynamic range greater than the 12 bits associated with one camera. Dynamic ranges of the system for cameras at several gain values (where the 2K and 4K camera share a raw gain value) are given in Table \ref{tab:dynamicranges}, alongside the absolute upper and lower limits of flux in the range. These tests were performed without any neutral density filter, mentioned in the technical description section~\ref{subsec:scint:opticalfilters}. The use of this filter with the 4K camera can multiply the dynamic range with an effective factor 3.163, equivalent to adding 5 dB. 

The final entry gives an extreme extrapolated example of the maximum achievable dynamic range, where the 4K camera is at full gain, and the 2K at the lowest setting. Each of these values, similar to the threshold-of-sensitivity, are dependent on the charge reconstruction, and so have the associated $\pm 50\%$ uncertainty.

\begin{table}[h]
    \centering
    \caption{Reconstructed dynamic ranges, upper and lower detectable limits of flux within the range, for values of electronic camera gain.}
\begin{tabular}{c|c|c|c}
         Raw Gain & Dynamic Range  & Flux Lower Limit & Flux Upper Limit\\
         $10\times$dB & dB & fC mm$^{-2}$ & fC mm$^{-2}$ \\
         \hline
         0 & 44.18 & $2.911\times 10^{-1}$ & $7.615\times 10^3$\\
         50 & 43.63 & $1.653\times 10^{-1}$ & $3.816\times 10^3$\\
         100 & 43.33 & $9.914\times 10^{-2}$ & $2.136\times 10^3$\\
         150 & 42.68 & $5.808\times 10^{-2}$ & $1.077\times 10^3$\\
         Max. Range & 55.85 & $1.978\times 10^{-2}$ & $7.615\times 10^3$\\

         \end{tabular}
         \label{tab:dynamicranges}
\end{table}

\begin{figure}[h] 
\begin{subfigure}{0.49\hsize} 
\includegraphics[width=\textwidth]{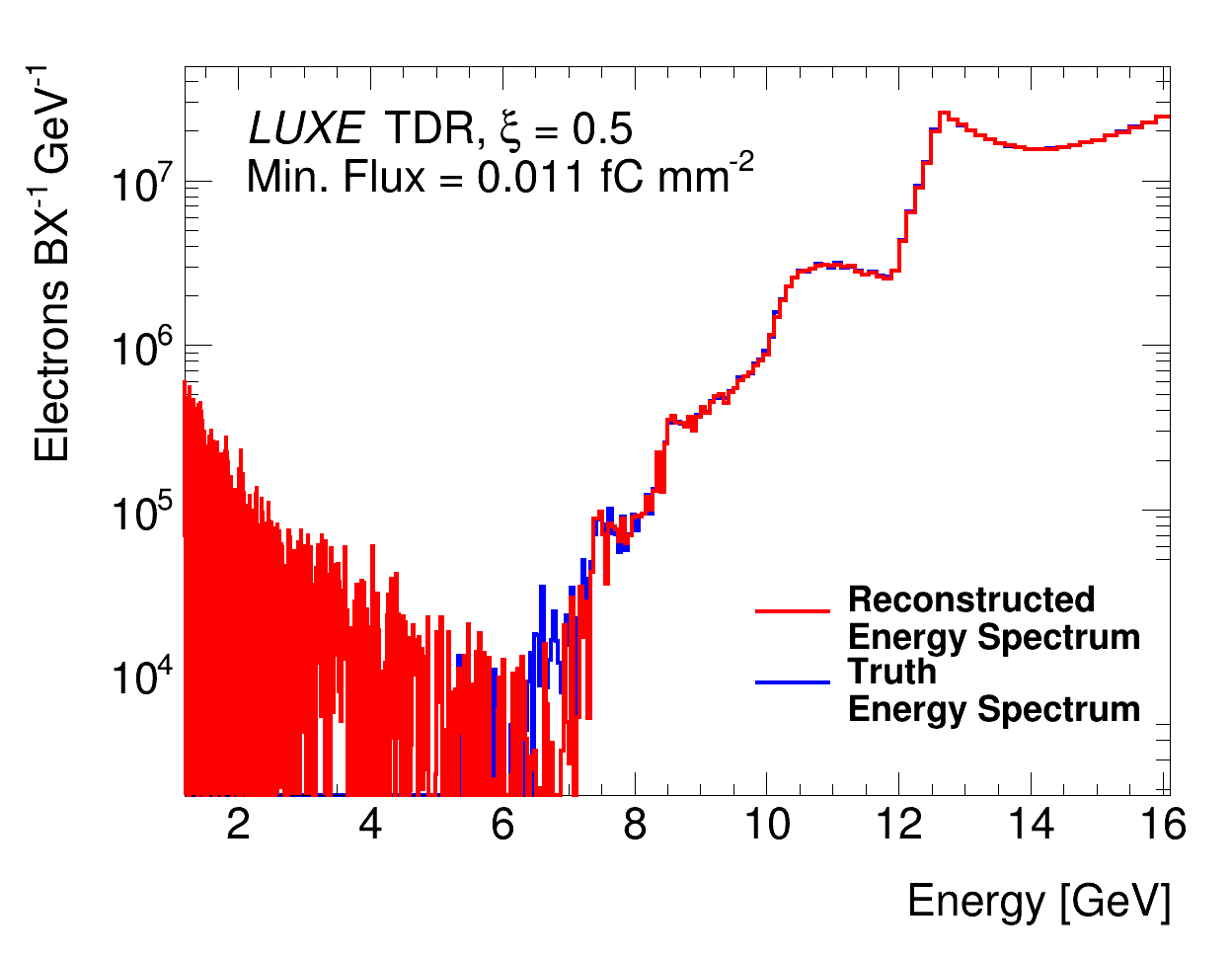}
\caption{  \label{fig:qe_2K}
}
\label{fig:photon-before-IP-scint-light}
\end{subfigure}
\begin{subfigure}{0.49\hsize}
\includegraphics[width=\textwidth]{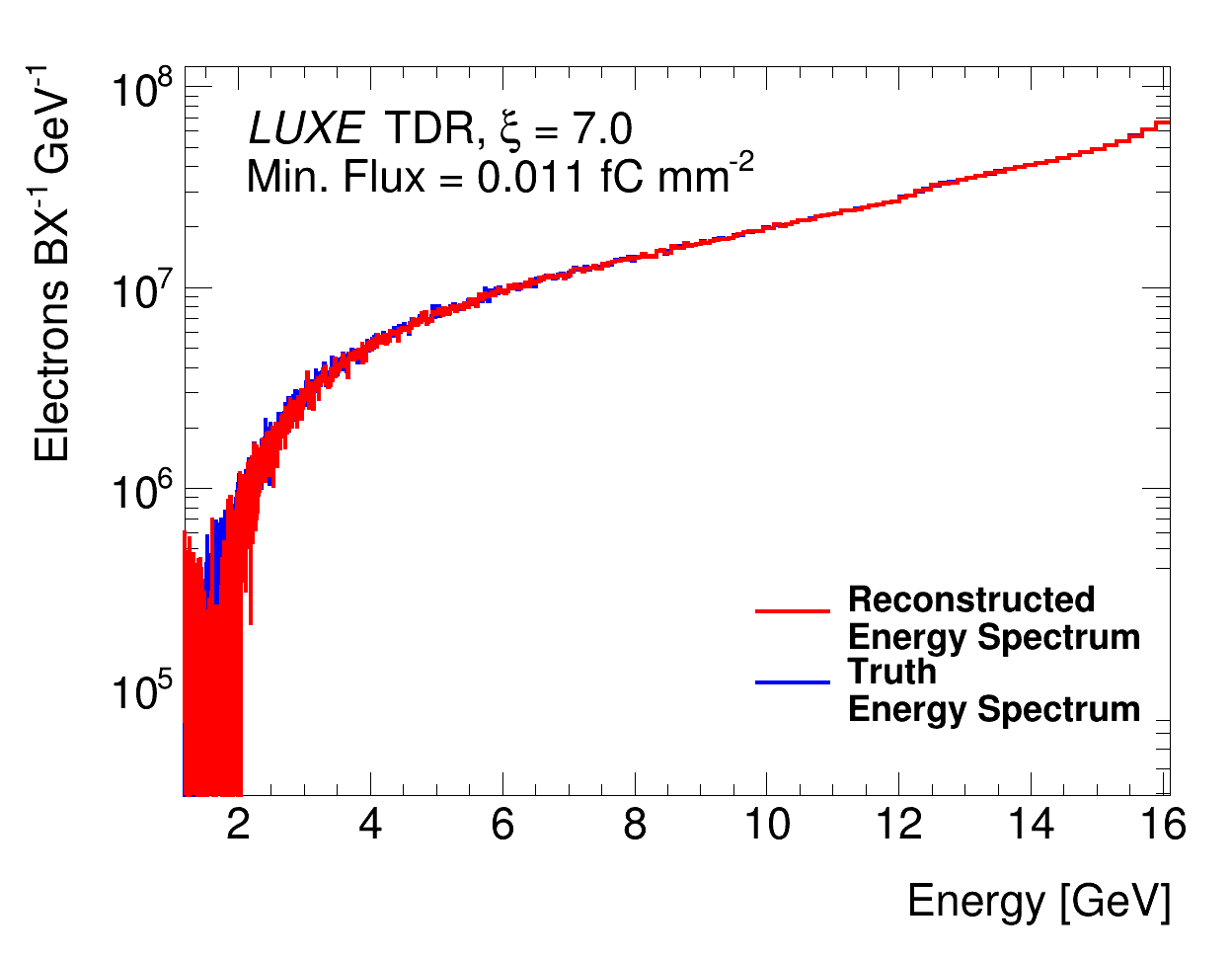}
\caption{ \label{fig:qe_4K} 
}
\end{subfigure}
\begin{subfigure}{0.49\hsize} 
\includegraphics[width=\textwidth]{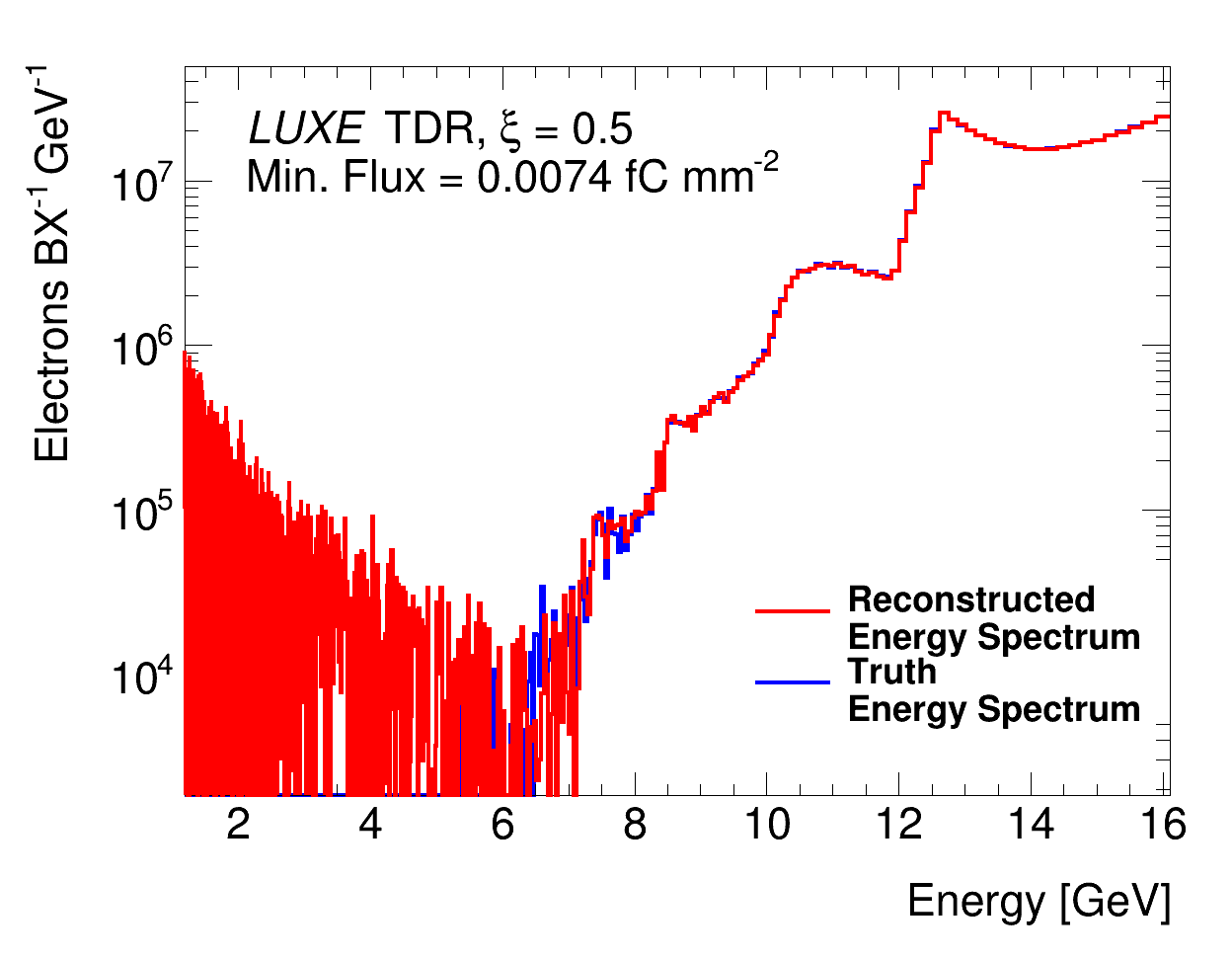}
\caption{  \label{fig:qe_2K}
}
\label{fig:photon-before-IP-scint-light}
\end{subfigure}
\begin{subfigure}{0.49\hsize}
\includegraphics[width=\textwidth]{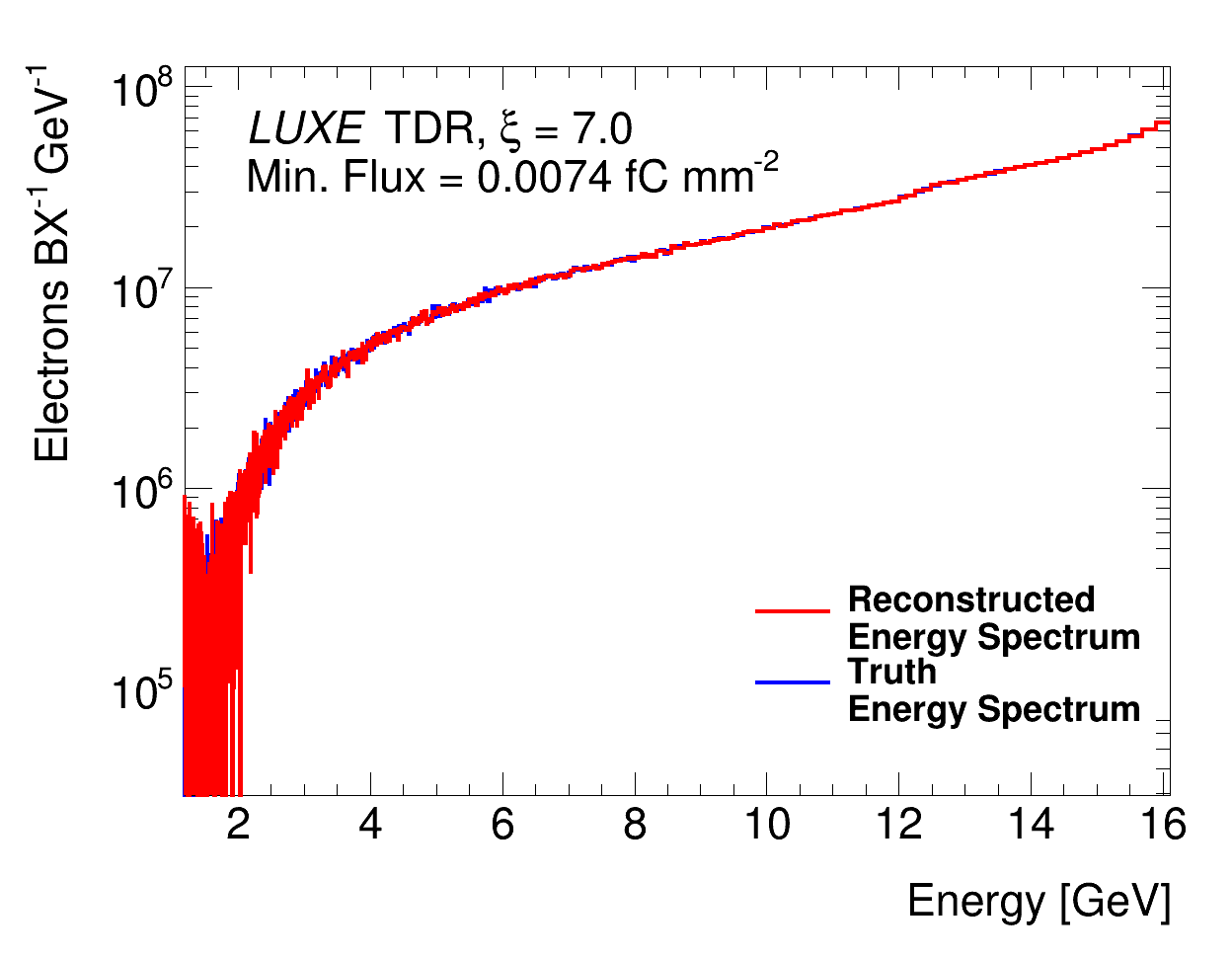}
\caption{ \label{fig:qe_4K} 
}
\end{subfigure}

\caption{\label{fig:mindetfluxgeant} \textbf{(a)} A reconstruction of an example SFQED electron spectrum, from \ptarmigan, with $\xi_{max}=0.5$ using the \geant simulation of the screen and camera system. This uses a threshold-of-sensitivity value (wherein no signal data at on-screen flux lower than this is used) of 0.011 fC mm$^{-2}$ in the reconstruction. This value is used to scale a noise map, taken from real dark images. The dynamic range in flux for raw gain of 100 is applied. \textbf{(b)} Shows a similar analysis for a \ptarmigan simulation with larger value of $\xi_{max}=7.0$. \textbf{(c)} and \textbf{(d)} show reconstructions for the $\xi_{max}=0.5,~7.0$ \ptarmigan simulations respectively, with a threshold-of-sensitivity value of 0.0074\,fC mm$^{-2}$.}

\end{figure}

The results from this data, in particular the threshold-sensitivity flux and dynamic range(s) calculated, are applied to a simulation of the scintillation screen and camera system. The resulting loss of signal possible due to fluxes being outside the detection threshold are shown for a selection of SFQED spectra, for each upper, central and lower estimate. This is performed for $\xi$ values of 0.5 and 7.0, each for the upper bound and central estimate of threshold flux of sensitivity, shown in Fig. \ref{fig:mindetfluxgeant}. This is performed with a raw gain value of 100, and the appropriate dynamic range in flux used. For the upper/lower bound on the threshold-sensitivity flux, the dynamic range in flux is transformed correspondingly. A noise map is taken from real, dark images from the prototype in test-beam. The noise from one single image is added, and the normalised sum of 100 of these images is subtracted, to emulate the realistic noise characteristics in the process of reconstructing one bunch crossing. The absolute percentage difference between integral of the reconstruction and truth are shown in Table \ref{tab:spectrum-loss}. In each case, the discrepancy between reconstruction and truth due to threshold-of-sensitivity / dynamic-range is low, below 1\%. The system is then proven to be suitable for the environment, in terms of electron signal expected, of the electron wing of the IP region of LUXE. The loss of some portion of the spectrum due to acceptance, dynamic-range/sensitivity, and statistical limitations in one shot, are seen as small compared to the uncertainty of the magnetic field measurement and screen \& camera system light/charge calibration curve (these uncertainties are not modelled in simulation).  

\begin{table}[h]
    \centering
    \caption{Percentage absolute difference between reconstructed spectra integral and truth integral.}
\begin{tabular}{l|c|c|c}
         \ptarmigan  & Upper & Central & Lower\\
         Parameters & Bound & Estimate & Bound\\
         \hline
         $\xi = 0.5$ & 0.134\% & 0.006\% & 0.41\%\\
         $\xi = 2.0$ & 0.187\% & 0.14\% & 0.077\%\\
         $\xi = 7.0$ & 0.28\% & 0.15\% & 0.021\%\\  

         \end{tabular}
         \label{tab:spectrum-loss}
\end{table}


\section{Technical Description}
\subsection{Scintillation Screen}
\label{sec:det:scintillator}
Scintillation screens, in comparison to the segmented approach of the \cer device, offer a measurement of the same electrons with a superior position resolution. The large flux of ionising radiation expected in the Compton process will induce enough light in a scintillating material to be imaged by a remote camera, while the electrons will pass through a thin screen mostly unperturbed on their path to the \cer detector.

There is little practical difficulty or high cost associated with extending the range in the $x$ direction, so the scintillation screen can be engineered to cover the entire possible energy spectrum of electrons exiting the magnetic field (which is expected to be constrained by the size of the aperture of the preceding magnet). However, at present, the design does not attempt to extend as far as possible, as the magnetic action results in the low-energy limit of the spectrum becoming increasingly spread, and given the low rates below 2.5 \GeV, the on-screen flux drops to the point of difficulty of detection. The lower energy limit is chosen more pragmatically. In Table~\ref{tab:scintdimensions} the  dimensions and key parameters of the screen are listed.

\begin{table}[h]
\caption{Key parameters of the scintillation screen used in the electron detection systems. The screens are foreseen to be identical at each site~\cite{mcio}.}
\label{tab:scintdimensions}
\centering
\begin{tabular}{|l|c|}
\hline
\hline
Parameter & Default Value \\
\hline
\textbf{Scintillation screen} & \\
Length ($x$) & $500$~mm \\ 
Height ($y$) & $100$~mm \\ 
Protective Layer PET thickness & 6~\textmu m \\
Phosphor Layer GadOx thickness & 310~\textmu m \\
Support Layer Plastic thickness & 188~\textmu m \\ 
Total thickness & 507~\textmu m \\
Thickness in Radiation Length ($X_0$)& 2.8 \% \\
Typical Phosphor Grain Size & 8.5~\textmu m\\
\hline
\hline
\end{tabular}
\end{table}

\begin{figure}[t]
    \centering
    \includegraphics[width=\textwidth]{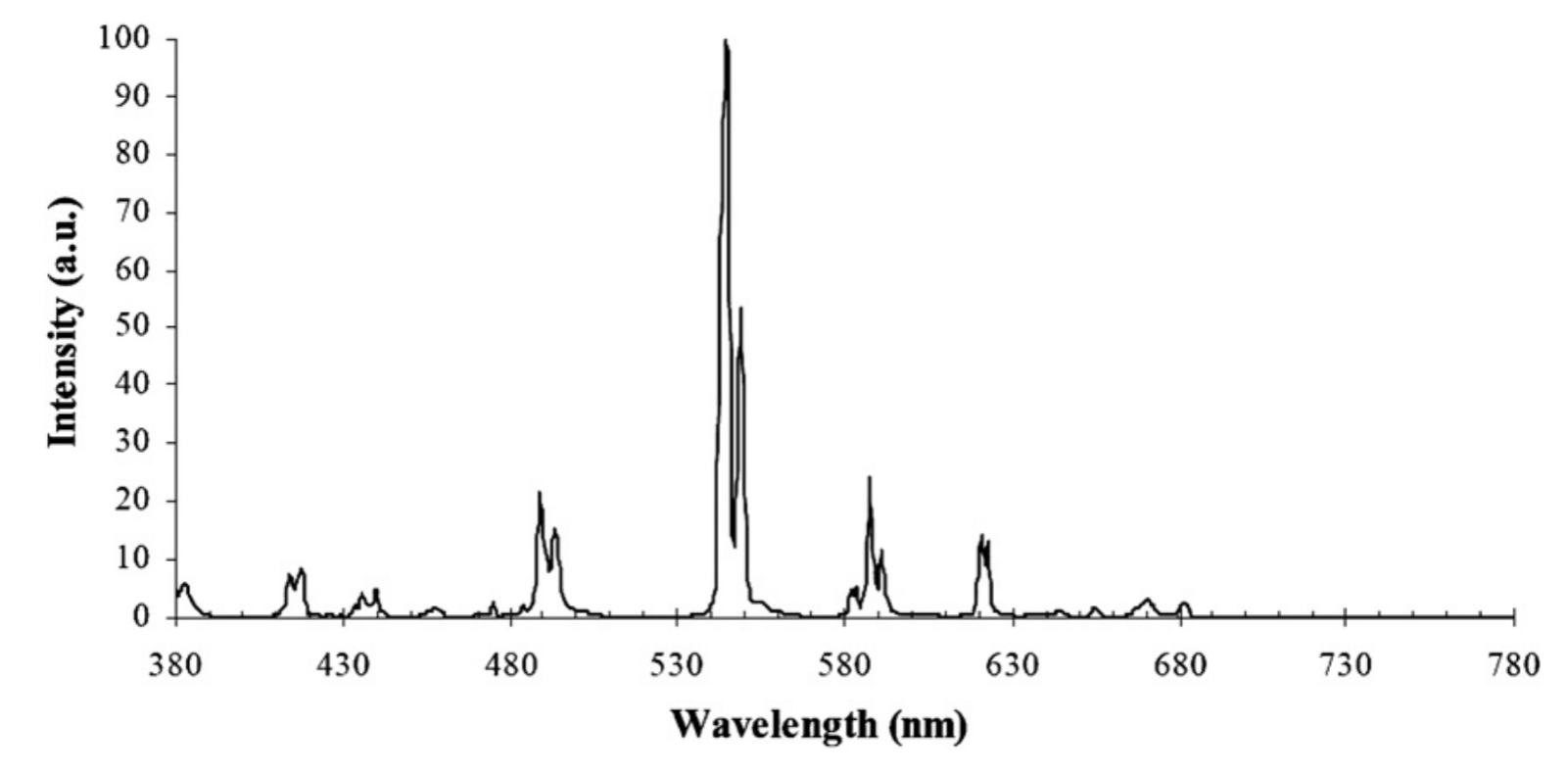}
    \caption{Emission wavelength profile from a Phosphor Technology scintillator of chemical composition $\mathrm{Gd}_2\mathrm{O}_2\mathrm{S:Tb}$~\cite{gadoxwavelengthprofile}~\cite{phosphortechnology}.}
    \label{fig:gadox_wavelength}
\end{figure}

The scintillation material Terbium-doped Gadolinium Oxysulfide ( $\mathrm{Gd}_2\mathrm{O}_2\mathrm{S:Tb}$, GadOx) is chosen for the screen. Analysis of the material's properties ensures it is suitable for use in this detector: 
\begin{itemize}
  \item \textbf{Emitted light wavelength profile:} Best chosen to be within the optical wavelength band of non-specialised optical equipment (mirrors, cameras). The characteristic light emission wavelength of GadOx doped with Terbium is around $545\units{nm}$, which easily fulfils this requirement, and filters exist which are optimised for this fluorescent wavelength. While there are secondary emission peaks, this primary emission peak contains the majority of visible light output from GadOx, visible in Figure~\ref{fig:gadox_wavelength} \cite{gadoxwavelengthprofile}.  
  \item \textbf{Light emission time profile:} The maximum bunch frequency expected at LUXE is $10 \units{Hz}$. The relatively long emission evolution of GadOx compared to most common materials -- with decay constant $\tau \approx 600$ \textmu s -- is still comfortably within $100\units{ms}$ \cite{mitsubishi_chemical}.
  \item \textbf{Scintillation photon yield:} Greater light output gives more statistical significance in general and above the ambient light background, and GadOx offers some of the highest outputs in terms of the number of photons for energy deposited \cite{MORLOTTI1997772}. The yield of the scintillator is in principle susceptible to saturation, or a quenching of the light yield for high fluxes, but the fluxes of particles to measure in LUXE will not be intense enough to observe these effects \cite{Keeble:2019}.   
  \item \textbf{Thickness:} Given the density and corresponding radiation length of GadOx, and the requirement that the screen perturb the electron energy spectrum as little as possible before measurement by the \cer device, the screen must be adequately thin. The screen's thickness as given in radiation lengths, using parameters defined in Table \ref{tab:scintdimensions}, is calculated as 2.8\%. The manufacturing of the screen involves crushing the GadOx into a fine powder, which is then suspended in a plastic base. This ensures consistency in thickness and optical properties across the surface compared to a continuous crystal.  
  \item \textbf{Radiation hardness:} Scintillators which are damaged by radiation can suffer a local decrease in scintillation efficiency. GadOx screens are radiation-hard, which is important due to the high-flux region(s) at LUXE. 
  A material's response after deposition of energy from ionising radiation is always specific to the particle type and energy profile of the radiation, as well as to the flux and frequency of exposures. This makes it difficult to quantify the radiation hardness with absolute confidence. The material's response to electrons of order $10 \units{keV}$ in the context of electron microscopy has been studied \cite{ScintiMaxdatasheet} and a threshold of noticeable change in the scintillation efficiency defined. This is calculated in terms of absorbed dose of around $10^8 \units{Gy}$. Within a simulation in \geant, fluxes of up to $1 \times 10^7 \units{mm^{-2}}$ are recorded in the \elaser mode, leading to a local maximum dose per bunch of $\sim 0.3 \units{Gy}$. For the calculated maximum dose, this allows up to $\sim 3 \times 10^8$ bunches or operation in LUXE up to 30 years. 
  
\end{itemize}

From the Bethe--Bloch formula for energy deposition over distance travelled in matter, the magnitude of signal response in this detector is related to the velocity of the particle, $\beta$, 
\begin{equation}
\frac{dE}{dx} \approx \frac{n e^6}{4 \pi m_e c^2 \beta^2 \epsilon_0^2}\left[ \ln\left( \frac{2 m_e c^2 \beta^2}{I(1-\beta^2)} \right) - \beta^2 \right] ,
\end{equation}
for electron number density $n$ and mean excitation potential $I$, both constants specific to a material. 
Like in the case of the accompanying \cer detector, always twinned with the scintillation screen in LUXE, the response of the screen is dependent on the $\beta$ of the electrons, rather than energy, and thus the detection efficiency is relatively flat over the \GeV \, energy range. The reconstructed energy is solely determined by where the particle hits the screen, i.e.\ by the spatial distribution of the scintillation light observed by the camera. 
Unlike the \cer device, the scintillator is somewhat sensitive to low-energy particles as in this regime $\frac{dE}{dx} \propto 1/\beta^2$. 

 The background can, however, be determined in-situ via two methods. The first method uses the $9\units{Hz}$ of electron-beam data where there is no \laser shot, using the fact electron bunches are foreseen to be used at rate 10\,Hz but the \laser pulses at only 1\,Hz. The second method uses the data from the  non-central bands of the scintillation screen. The scintillation screen covers $10\units{cm}$ in the vertical plane and the signal is confined to just the most central $1\units{cm}$, while the background is mostly flat in that plane, see Figure~\ref{fig:lanex_sigbkg}, with some contribution symmetric around the beam axis, and so is accounted for and subtracted. 
 
 The linearity of the scintillation response to incident electron flux is dependent on the energy deposition, which is a stochastic process, and so this deposition varies around a mean. The high fluxes and statistics measured here ensure that using the mean value is suitable in reconstruction. This process has been explicitly included in the \geant simulations included in the performance. Non-linearity in scintillators are expected in the case of saturation at extremely high fluxes, where the light response is quenched. At high fluxes it was demonstrated by the AWAKE experiment that GadOx has a linear response up to charges of 350\,pC~\cite{BAUCHE2019103}, i.e. also adequate for the primary electron beam, and so no saturation is expected. 
 
 Tables \ref{tab:ip-el-energy-acceptance} and \ref{tab:brem-el-energy-acceptance} give the position and energy acceptances for the EDS and IBM regions respectively, assuming a dipole field of $B=1\,\mathrm{T}$. 

\begin{table}[h]
\caption{Physical position and energy ranges within the acceptance of the scintillation screen in the EDS. These are calculated with the assumed \euxfel beam energy of 16.5 GeV.
$z$ denotes the position in the z dimension downstream from the IP. $z_d$ is the approximate distance from the end of the B-field, and $x$ the distance from the high-energy edge of the screen from the beam axis. The aluminium frame is characterised by the thickness of the wire of the frame, and the displacement of the loop of free air to allow the beam through unperturbed.}
\label{tab:ip-el-energy-acceptance}
\centering
\begin{tabular}{|c|c|}
\hline
   Scintillator 
   Parameter & $B=1$\,T \\
\hline

$E_\textrm{max}$ (GeV) & $15.04 $\\
$E_\textrm{min}$ (GeV) & $1.832 $\\
$z$ (m) & $4.85$ \\
$z_m$, approx. (m) & 1.2 \\
$z_d$ (m) & $2.8$ \\ 
$x$ (mm) & $66$ \\ 
Frame thickness (mm) & $10$ \\ 
Frame loop extension (mm) & $15$ \\
\hline
\hline
\end{tabular}
\end{table}

\begin{table}[h]
\caption{Physical position and energy ranges within the acceptance of the scintillation screen in the IBM. These are calculated with the assumed \euxfel beam energy of 16.5 GeV.
$z$ denotes the position in the z dimension downstream from the IP. The negative value implies the screen position is \textit{upstream} from the IP. $z_d$ is the approximate distance from the end of the B-field, and $y$ the distance from the high-energy edge of the screen from the beam axis.}
\label{tab:brem-el-energy-acceptance}
\centering
\begin{tabular}{|c|c|}
\hline
   Scintillator 
   Parameter & $B=1.5$\,T \\
\hline

$E_\textrm{max}$ (GeV) & $7.5$\\
$E_\textrm{min}$ (GeV) & $1.3 $\\
$z$ (m) & $-5.465$ \\
$z_m$, approx. (m) & 1.2 \\
$z_d$ (m) & $0.685$ \\ 
$y$ (mm) & $95$ \\ 
Frame thickness (mm) & $10$ \\ 
\hline
\hline
\end{tabular}
\end{table}

\subsection{Cameras}

The Basler camera models acA1920-40gm and acA4096-11gm are chosen~\cite{Baslercamera2K}~\cite{Baslercamera4K}. Each come from the Basler Area Scan Range, employing CMOS sensors to take a monochromatic image. The resolution of these cameras (hereafter referred to as ``2K" or ``4K" models, given the pixel resolution in the $x$ dimension) are $1920\times1200$ and $4096\times2160$ respectively.

In both sites 3 cameras will be used -- two of the 2K model, and one of the 4K model. The 2K models will cover each one half of the screen, with the images exceeding the screen by 5\,mm in both dimensions and 6mm overlap between in the middle. This assists with alignment.

The purpose of using an additional 4K camera is to increase positional resolution in the critical high-energy region of the image. As a result of the magnetic spectrometry mechanism, the high energy signal is more tightly bunched here in the $x$ direction which motivates the resolution needs. Using the finer resolution, as well as a lens with longer focal length, also gives the 4K camera a lower per-pixel signal as each pixel covers a smaller section of the screen. This means the 4K camera sensor saturates at a higher scintillator light flux, effectively boosting the dynamic range of the combined camera system.  

For the two 2K cameras, their view is centred on the centre of the screen in $y$ and at positions one quarter, and at three quarters along the length of the screen ($x$). 

\begin{figure}[t] 
\centering
\includegraphics[width=0.85\textwidth]{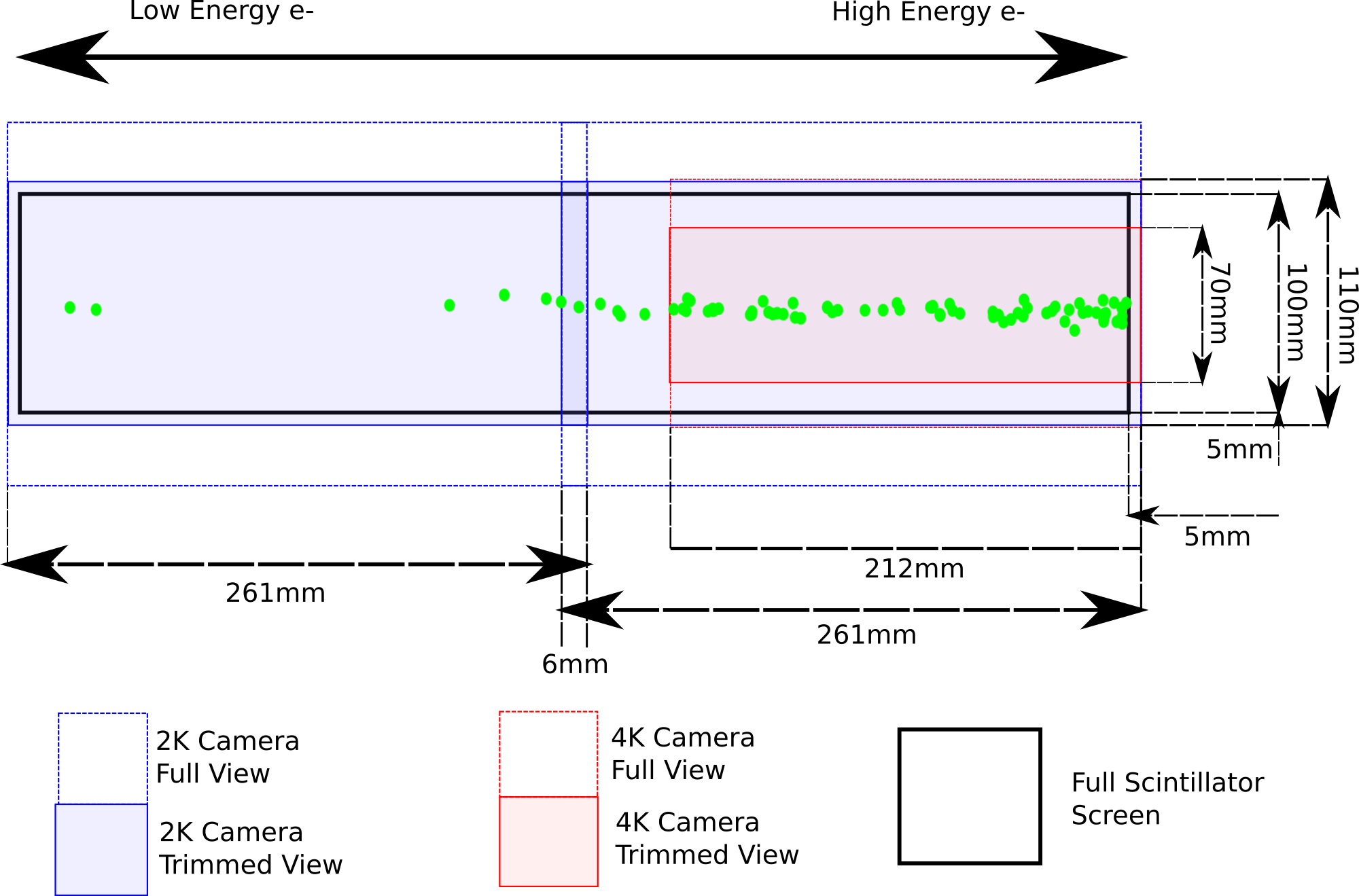}
\caption{An illustration of the coverage of the scintillation screen from each camera. Valid for both regions of the experiment.}

\label{fig:camera_coverage}
\end{figure}

The 4K camera at each site is positioned such that the image overlaps the screen in $x$ at the high energy side by 5\,mm. Again this image is aimed centrally in $y$, and 101\,mm from the side in $x$. An illustration of this scheme of coverage is shown in Fig. \ref{fig:camera_coverage}.

These cameras will be operated with 12-bit ADC pixel depth. In order to preserve the 10\,Hz functionality for the 4K camera, as for this choice of bit depth and full resolution the interface bandwidth is insufficient, the image is cropped at top and bottom. 

Both models have a global shutter, required for this imaging technique to ensure consistency in timing pixel-to-pixel. Additional technical details for both models are given in Table~\ref{tab:camera_model_parameters}. 

\begin{figure}[t] 
\begin{subfigure}{0.49\hsize} 
\includegraphics[width=\textwidth]{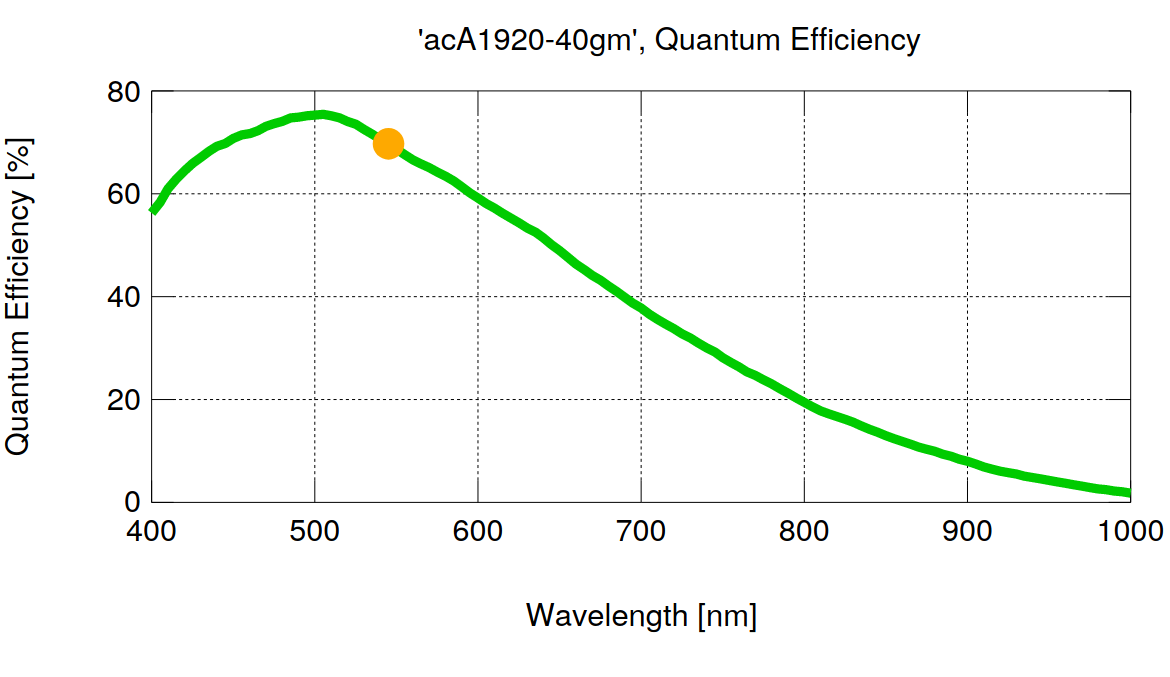}
\caption{  \label{fig:qe_2K}
}
\label{fig:photon-before-IP-scint-light}
\end{subfigure}
\begin{subfigure}{0.49\hsize}
\includegraphics[width=\textwidth]{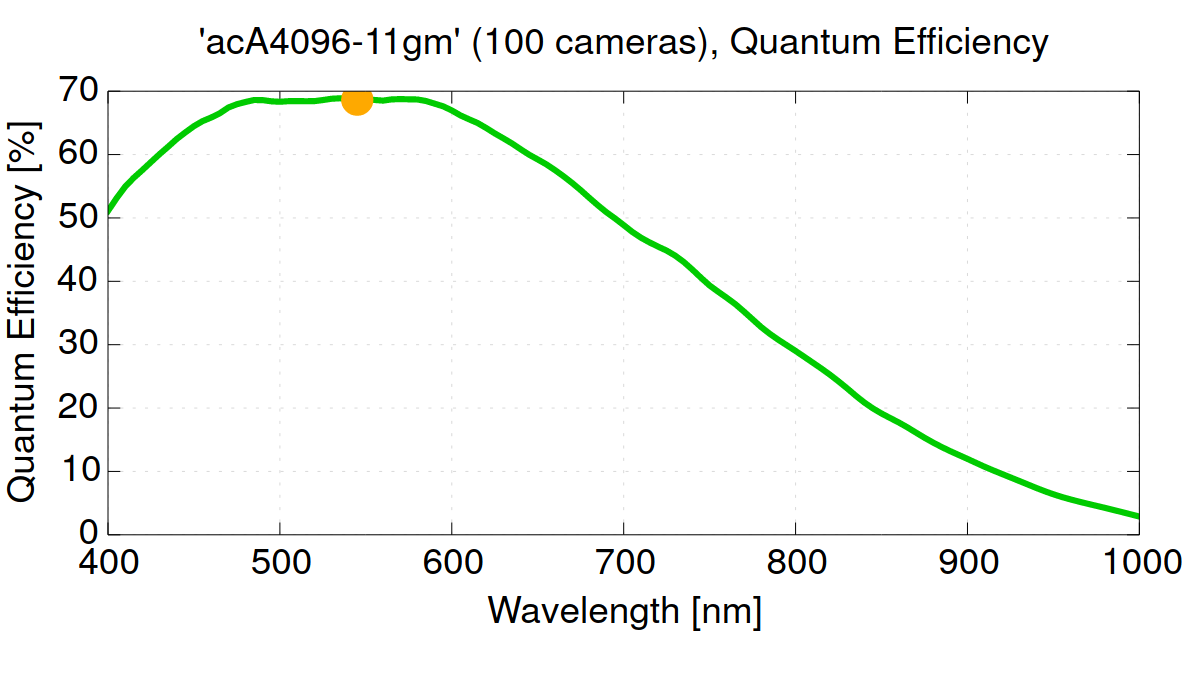}
\caption{ \label{fig:qe_4K} 
}
\end{subfigure}
\caption{\label{fig:quantum_efficiency} \textbf{(a)} The measured quantum efficiency for the 2K camera for photons of varying wavelength.  \textbf{(b)} The quantum efficiency profile for the 4K camera. Both exhibit a peak very near the target wavelength of 543 nm. $\lambda = 545$nm is marked in yellow~\cite{4k_camera_specs} \cite{2k_camera_specs}. }

\end{figure}

In recent years, the CMOS pixel technology present in these cameras has become broadly equivalent to the quality of more typical CCD devices for scientific purposes, while still preserving a low cost. The profile of relative quantum efficiencies for each model of camera is given in Fig. \ref{fig:quantum_efficiency}. This peaks for both at around 545 nm, close to the GadOx characteristic emission wavelength, at a value of $\sim 70\%$. 

The linearity in response, in terms of ADC pixel grey value per photons/pixel, of each camera model, is shown in Fig. \ref{fig:camera_sensor_linearity}. The response of the combined scintillation screen and camera system will be measured specifically in high-rate test-beam campaigns in the future. 

\begin{figure}[t] 
\begin{subfigure}{0.49\hsize} 
\includegraphics[width=\textwidth]{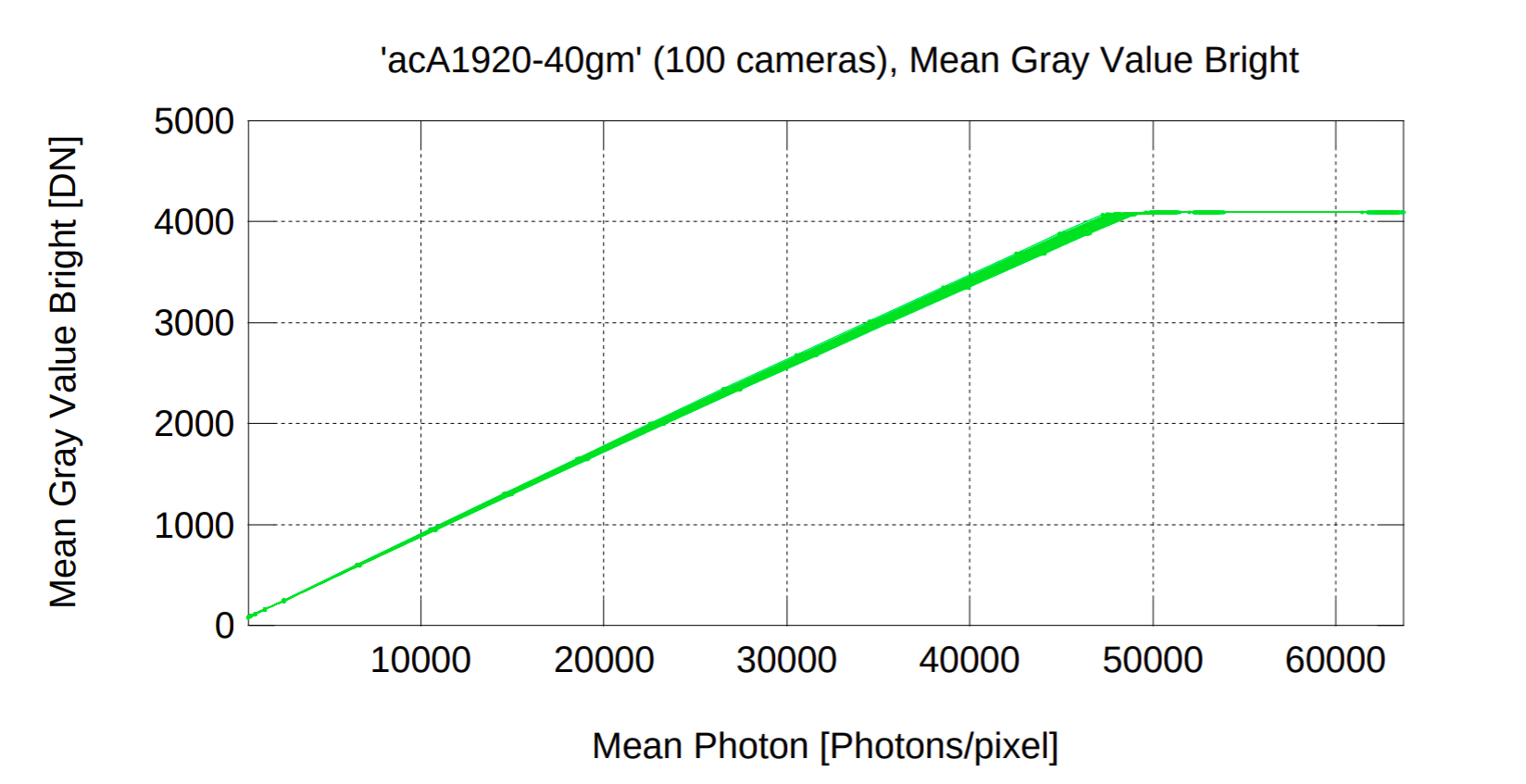}
\caption{  \label{fig:camera_sensor_linearity}
}
\label{fig:camera_sensor_linearity}
\end{subfigure}
\begin{subfigure}{0.49\hsize}
\includegraphics[width=\textwidth]{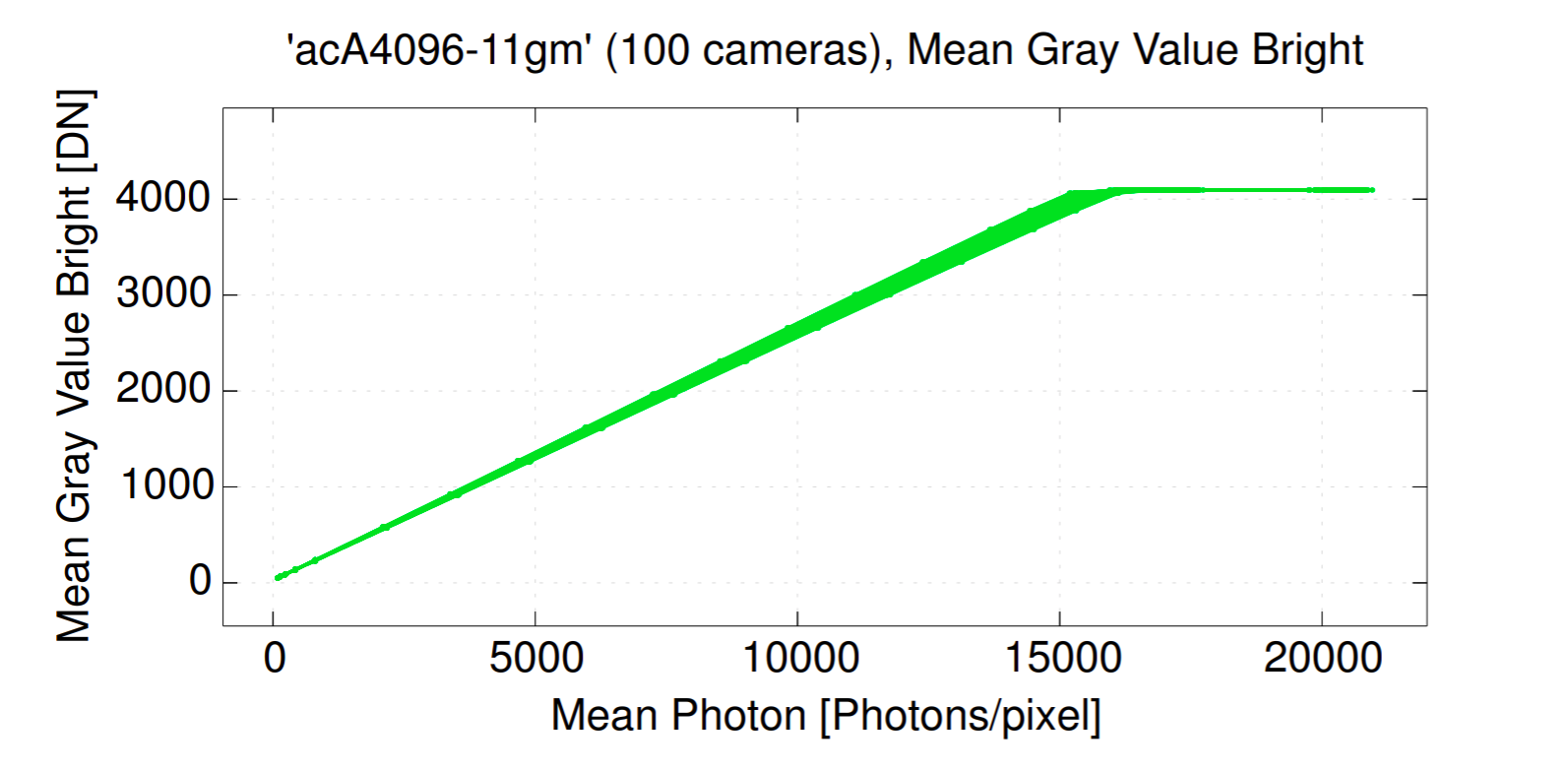}
\caption{ \label{fig:qe_4K} 
}
\end{subfigure}
\caption{\label{fig:camera_sensor_linearity} \textbf{(a)} The measured grey value in digital ADC pixel response for the 2K camera for varying intensities of photons/pixel. 100 cameras of the same model are plotted. \textbf{(b)} The measured grey value for the 4K camera~\cite{4k_camera_specs} \cite{2k_camera_specs}. }

\end{figure}

\begin{table}[h]
\caption{Key camera model parameters.
The nominal position resolution is true for both the vertical and horizontal dimensions.}
\label{tab:camera_model_parameters}
\centering
\begin{tabular}{|c|c|c|}
\hline
Camera Parameter & 4K camera & 2K camera \\
  & $f$=75\,mm lens & $f$=50\,mm lens\\
\hline
Nominal positional resolution (\textmu m) & 51.8 & 135.9 \\
Vertical view angle (deg) & 5.7 & 8.2 \\
Typical quantum efficiency at $\lambda$ = 543 nm & ~69\% & ~70\% \\
Typical sensor SNR & ~ 40 dB & ~40 dB\\
Typical power requirement (W) & 2.7 & 2.7 \\
Pixel size (\textmu m$^2$) & $3.45\times3.45$ & $5.86\times5.86$ \\
Total sensor size (mm$^2$) & $14.1\times7.5$ & $11.3\times7.1$ \\ 
Typical read noise ($e^-$) & 2.3 & 6.7 \\
\hline
\hline
\end{tabular}
\end{table}

\subsection{Lenses}

The angle of view of the cameras is dictated by the sensor size and the characteristics of fitted lenses. For the fixed position, a fixed-focal length lens can be used. 

Lenses with a focal length of $f$ = 50 mm suit the parameters well, where one 2K camera is trained upon an $x$ measurement of $\sim 260\,\mathrm{mm}$. A lens with $f$ = 75 mm fitted to the 4K camera allows for a tighter zoom in the interest of finer position resolution. Both these fixed focal lengths are readily available and models for use have been identified, from Edmund Optics and Basler~\cite{edmundlens75mm}~\cite{baslerlens50mm}.

Each of the lenses have variable focus, from a minimum working distance of 250 mm and 200 mm for the $f$ = 75 mm lens and $f$ = 50 mm lens, to a maximum infinite working distance respectively. They also have variable aperture sizes, for which closing can in principle minimise the effect of optical disturbances/dust upon the lenses. Keeping the apertures maximally open maximises total light collection, and so this setting will be used. Further technical information is present in Table \ref{tab:lens_model_parameters}.

\begin{table}[h]
\caption{Chosen lens model characteristics.}
\label{tab:lens_model_parameters}
\centering
\begin{tabular}{|c|c|c|}
\hline
Lens coverage & 4K camera & 2K camera \\
  & $f$=75\,mm lens & $f$=50\,mm lens\\
\hline
Working distance (mm) & 1200 & 1200 \\
Horizontal view angle ($^{\circ}$) & 10.8 & 12.9 \\
Vertical view angle ($^{\circ}$) & 5.7 & 8.2 \\
Horizontal image width at screen (mm) & 212 & 261 \\
Vertical image height at screen (mm) & 112 & 164 \\
Trimmed vertical image height (mm) & 70 & 110 \\ 
C-mount diameter (mm)  &  25.4 & 25.4 \\
Typical optical distortion & <0.1\% & 0.06\% \\
Mass (g) & 436 & 277.5 \\
Image circle (mm) & 21.6 & 17.6 \\
\hline
\hline
\end{tabular}
\end{table}

\subsection{Optical Filters}
\label{subsec:scint:opticalfilters}

\begin{figure}[t] 
\centering
\includegraphics[width=0.8\textwidth]{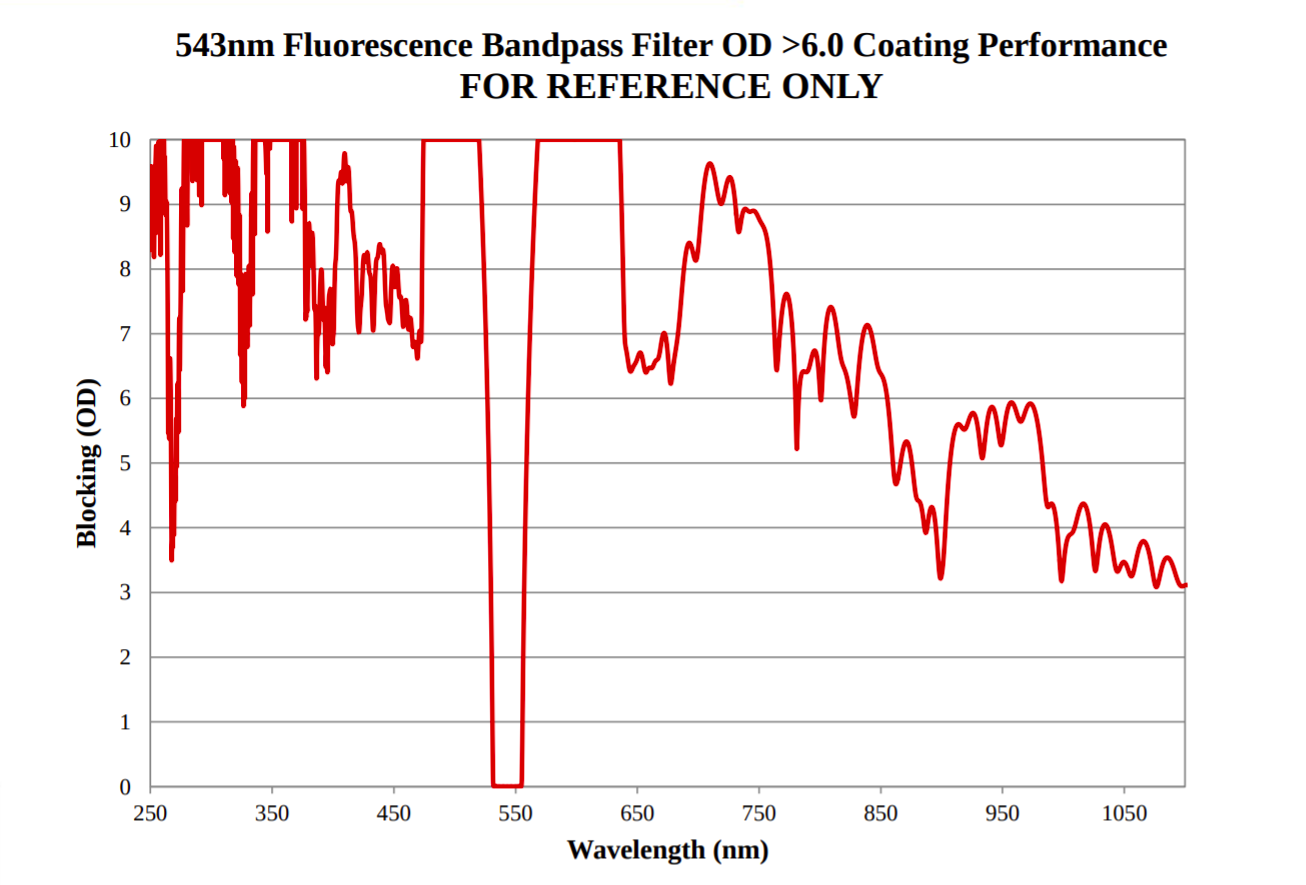}
\caption{Optical Blocking OD for the chosen optical filter. Outside the desired waveband, the blocking is greater than 3, meaning transmission is less than 0.1\%~\cite{edmund_filter}.}

\label{fig:fluorescence_blocking}
\end{figure}

The optical filter chosen also comes from Edmund Optics \cite{edmund_filter}. Its diameter of 50\,mm easily encompasses the diameter of both chosen lenses. A filter holder is required to secure them to either lens. The filter is optimised to accept light of central wavelength 543\,nm, including a full-width-half-maximum of the peak of 22\,nm. This encloses the main emission peak seen in Fig.~\ref{fig:gadox_wavelength}; the optical density as function of wavelength of the filter is given in Fig.~\ref{fig:fluorescence_blocking}.

For certain parameters of $\xi$, the spectrum of electron flux may induce an intensity of light too bright to remain within the dynamic range of the cameras at the EDS site. Obvious solutions to reduce the light levels each have drawbacks: shortening the exposure time and closing the aperture, make the measurement more susceptible to temporal jitter and are difficult to precisely control, respectively. The solution chosen for these runs is to outfit the 4K camera with an additional neutral density filter. This effectively increases the total dynamic range and will be implemented when needed (when approaching the problematic running $\xi_\textrm{max}$) with a quick exchange of the filter during short accesses to the hall. The exchange can be performed mechanically by attaching the filter to the end of the band-pass filter (or its holder). A suitable filter is identified in reference \cite{neutral_density_filter}. Its optical density of 0.5 implies a fraction of light transmitted compared to light incident of 0.316. 
 
\section{Interfaces and Integration} 
\subsection{Network}

Each of the described cameras uses ethernet to transmit data, and are powered through the same ethernet connection (power over ethernet). Similar cameras with alternate interfaces exist, offering in the case of USB or CameraLink a greater maximum bandwidth. Ethernet is chosen as the interface for data transmission given its universality, cost-effectiveness and crucially the possible length of the cable -- up to $\sim 100$\,m while USB and CameraLink are known to exhibit problems at lengths greater than 10\,m. The ultimate cable length for each camera-to-DAQ-PC link will for both sites exceed 10\,m. This is shown in Table~\ref{tab:TC:lengthService}.

\begin{table}[htbp]
\caption{Distance separating the service room to the relevant experimental areas.\label{tab:TC:lengthService}
}
\begin{center}
\begin{tabular}{|l|c|}
	\hline
	
Area & Length \\	\hline

Bremsstrahlung target & 22\,m \\
Interaction point  & 10\,m\\
	\hline

\end{tabular}
\end{center}
\end{table}

Data acquisition and treatment is handled by a C++ program making use of the Basler Pylon software \cite{basler-pylon}. An acquisition program using this basis has already been developed for a working camera \& screen prototype.  

To assist in monitoring, the data acquisition and treatment programs will send a small number of histograms to the central LUXE DOOCS \cite{DOOCS} slow-control, wherein the software parameters of the cameras will be alterable. These histograms are, per camera, one sum of the pixels in the x dimension and one sum in the y dimension, and possibly even simple live energy reconstructions. 

\subsection{Data Streaming and Storage}

The images themselves are 12-bit in bit depth, and as noted this actually limits the transmission of the 4K camera to smaller than the full resolution at 10\,Hz. The rest of the screen is covered by the 2K cameras. The vertical range of the 2K cameras too are trimmed down to the screen plus some 5\,mm buffer. This slightly reduces data bandwidth requirements. 

The anticipated data streaming bandwidth requirements are estimated using this bit-depth, image size and frequency: each 2K camera then will transmit at $\sim$\,185\,Mb/s or $\sim$\,23\,MB/s. The 4K camera in each instance transmits nearer the bandwidth of the ethernet interface, using the parameters in Table \ref{tab:lens_model_parameters}, at $\sim$\,83\,MB/s. This still preserves bandwidth to ensure safety in the quality of streaming. Summed together for one site, this is 128 MB/s of data-streaming capacity needed.

This offers an effective upper limit, but it remains likely the 4K image can be binned (data from two neighbouring pixels combined into one) within the y dimension. Since the fundamental resolution of the GadOx screen is of order $\sim 100$ \textmu m, the nominal resolution of 50 \textmu m for the 4K camera represents finer fidelity than needed. Although for deeper analyses it may be well-motivated to retain this resolution, a binning function to combine 2 pixels to one in either the y dimension, or both x and y dimension, can reduce data rates of the 4K camera by a factor of two or four respectively. These options would provide a total $\sim 87.5$ MB/s or $\sim 67$ MB/s of data rates per site respectively. 

This will be retained locally in the control room at the experimental hall location from run to run, until flushed to the file-systems at DESY and other institutes between runs. The two sites of this detector system are not expected to run concurrently and so the total required write speed remains equal to that of one site. The relatively high rate for the 4K camera motivates the need for more than a standard hard disk drive, as it exceeds typical reliable HDD write speeds alone, and so saving data to one disk per camera is ruled out. The excessive cost of solid-state drives (which feature higher write speeds) mean this technology is not planned to be used. Instead HDDs will be written to in parallel, in a RAID 1+0 array, increasing the write speed by a factor of the number of drives used. Since this comes with the drawback of one drive failing resulting in the corruption of all the data, each of these drives are mirrored. RAID6 could prove suitable too, resistant to 2 drive failures with a more efficient use of space, but is inferior with respect to the high read-speeds (multiplied by a factor of all drives present) available with RAID1+0.

\begin{table}[htbp]
\caption{Estimates for the data streaming/storage requirements of the scintillator \& camera system (one site). \label{tab:data-storage}
}
\begin{center}
\begin{tabular}{|l|c|c|c|c|}
	\hline
	
Quantity & no 4K & 2-pixel & 4-pixel & Unit \\	
 & binning & 4K binning & 4K binning & \\
\hline

Data Streaming Rate & 128 & 87.5 & 67 & MB/s \\
Approx. On-site Storage Rate & 12.8 & 8.8 & 6.7 & MB/s \\
Max. Single Run Storage & 2.2 &  1.5 & 1.15 & TB \\
Typical one-year Storage & 64 & 44 & 34 & TB \\
8-year LUXE lifetime Storage & 512 & 352 & 272 & TB \\
	\hline

\end{tabular}
\end{center}
\end{table}

The sum total cost of long-term storage becomes large over the lifetime of the experiment. Compression techniques can quickly help to mitigate the high volumes of data. In particular, the content of many images can be aggregated into a single image of higher bit depth. Conversion from images of 12-bit depth to 32 bits offers a higher dynamic range by factor $10^6$, allowing this number of images to be aggregated together. This is appropriate for the background measurements of the EDS and measurements at the IBM site, but for the signal measurement at the EDS the shot-to-shot variation is of interest, giving reason not to aggregate them entirely, and keep the individual images. For a run lasting 24 hours, this sums to $\sim$\,1.1\,TB of signal images, where the non-signal measurements data requirement is small in comparison. The aggregation of the background images is performed immediately, so the minimum value of 2.2\,TB per run (a reasonable estimate for the maximum run time is 48 hours) is the requirement for on-site storage for this detector. Regular flushing of the data off-site is possible throughout the run, so there is little to no down-time post-run.

It has been observed in tests that simple desktop PCs can process the treatment of the camera's images (one camera being processed by a single CPU thread), so there are no concerns with processing power when the final experiment runs.  

It is anticipated that standard lossless compression algorithms will reduce the long-term storage requirements further by at least 50\%. To run and gather data at the target rate of $10^7$ events per year, this gives 64 TB in long-term off-site storage per year, not counting backups. This is the largest data requirement of any detector at LUXE but is taken to be manageable. This reduces to 44 TB or 34 TB using the 4K camera images, binned in y or x \& y respectively. 

The images are used for reconstruction as described previously. Once the parameters of the magnetic action have been computed, the algorithm of reconstructing the energy spectrum from the image data is relatively fast. This is likely to be analysable to return shot-to-shot spectra at a rate of 1 Hz, or at least 0.1 Hz, suitable for live monitoring. The post-run analysis therefore is not expected to take additional time, excepting new techniques of analysis from the raw data. The cost of data storage of these reconstructed spectra is again small compared to the raw image data. The lossless algorithmic compression of the raw data is not likely to take excessive processing time. As comparison, compression of raw image data from the detector prototype into a zip file is processed at 11 MB/s with a single desktop PC thread. This compression can be performed then, essentially at pace with the 1\,Hz signal images. Table \ref{tab:data-storage} shows a summary of the data streaming/storage estimates.  

\subsection{Triggering}
Trigger comes from one of the TLUs used in LUXE (see chapter~\ref{chapt11}). Each trigger comes from the accelerator clock -- so images are triggered with every bunch. Triggers generated from the \laser at the same time for beam-\laser events, delivered to the DAQ PC, can be used to then validate those images which have recorded a \laser interaction, and categorised correctly, separating signal events from beam-only events.

Triggering is handled for these camera models by Hirose 6-pin connections. These connections are only used for the Line 1 connection and common ground in order to trigger exposure start. The other pins are foreseen to remain unused. 
Trigger is a typical TTL signal. Actual activation of the sensor happens roughly 5 \textmu s after the trigger is applied. Shot-to-shot jitter of this quantity is lower, at $\sim$\,100\,ns \cite{basler_io_timing}. This is of little concern as to losing much of the scintillation light. As explained the time constant of the scintillation material is relatively long at $\sim 600\,$\textmu s \cite{mitsubishi_chemical}. No significant portion of the light is lost, and the calibration of the screen/camera system takes this into account within the light--charge calibration curve measurement, which replicates the final experimental setup as much as possible. 

\subsection{Cabling}

The installation of cabling requires one gigabit ethernet (Cat6 and above) and one coaxial cable connection between the service room and each camera. Keeping the cables out of the beam-plane helps to protect the cables from electromagnetic interference, and they are shielded in addition. The cables must also be halogen-free in any plastic coating for safety. These requirements are modest, and available commercially at low cost. 

In the EDS region the ceiling mount allows the cables to be routed through the supporting tube, and thereafter attached to the ceiling before entering the service room.

\section{Installation, Commissioning and Calibration}
\subsection{Installation and Commissioning}

The installation and commissioning should be comparatively straightforward, although there is complication in securely fastening the ceiling mount planned for the EDS region. The main challenge is the physical alignment of 4 total components on two platforms at both sites. The envisioned setup requires both accurate alignment of the screen with respect to the beamline and B-field, and equally the cameras and their field of view to the screen(s).

Limited access to the experimental hall at installation means the shortest time possible may be required. The following estimate assumes as much preparation as reasonably possible is performed beforehand, including practice installations above ground. Work on the two sites is possible in parallel, shortening the total time, or since the \elaser mode is envisioned to run first, the IBM site can be postponed until a later access window. 

\vspace{2mm}

\textbf{Item-by-Item Installation Plan:}
\begin{enumerate}
    \item Installation of ceiling mount infrastructure -- 3 days -- technicians/engineers required
    \item Placement/alignment of IBM region components including cabling -- 3 days -- technicians/engineers required
    \item Placement/alignment of EDS region components including cabling -- 5 days -- technicians/engineers required
    \item DAQ stress tests of data streaming, triggering and running temperature (in conjunction with other detectors) -- 3 days
    \item Final calibration of imaging parameters due to light levels -- 3 days
\end{enumerate}

This gives an estimate of 8 days of access required to the experimental hall, for items 1 and 3, and item 2 performed in parallel (or at a later date), and then roughly 6 days of testing and commissioning for the last two items. The majority of this testing and commissioning comes previous to the installation.  

An anticipated possible shutdown within the winter of 2024, lasting between 4 and 6\,weeks, would prove adequate time to install, align and commission this detector in each location. This is only true under the assumption that the last two items can be controlled remotely, which is in principle true, but leaves open the risk of incomplete commissioning if a mistake is made within the installation window.  

The possible installation within the summer of 2025 of only 2 weeks is in principle possible. In order to ensure the installation is completed with a large contingency, the choice of effectively removing one of the alignment procedures from each site could be employed. This possibility requires a redesign where the position of the cameras are rigidly connected to the screen within the same, large aluminium frame. This would remove the ceiling mount for the EDS, and remove one half of the positional alignment, at the cost of material budget within the beam-plane in the region and also alignment performed during less critical time outside the experimental hall. 

\subsection{Calibration Strategy}

Calibration and tests of the scintillation response of the screens and detection characteristics of the cameras and associated optical components can be completed outside the experimental hall. The main challenge when calibrating within the experimental hall comes from alignment -- the position of both the screen and camera components must be independently accurately aligned and securely fixed. 

The physical alignment of the screen can be tested for a short time using any beam-time from the \euxfel below 16.5 \GeV, for example 11.5 GeV, using a magnetic field tuned for 16.5 GeV. The beam will then directly impinge the screen, and its location/reconstructed energy can be compared to the well-known beam parameters.  

The light-charge calibration curve will be measured with the geometry as close to that of the final experiment as possible at a high-flux test-beam. A standardised well-known source of light -- a calibration lamp -- will be used to further calibrate and test the quality of the optical line. An example product can be seen in reference \cite{lamp}. With an optical pattern of regular line spacings the modulation transfer function independent of the resolution of the screen and its scintillation function can be ascertained. This can be done with or without the calibration lamp to help illuminate. Only one instance of each of these are required between the two sites. The lamp must output light within the wavelength range accepted by the optical filter ($\sim545$ nm). 

\subsection{Decommissioning} 

The components of this system are not heavy, toxic or dangerous. The cameras will have to be safely removed from the ceiling with DESY technical help, at which point they can be returned to the institution which owns them, if this is not DESY. The screens will directly be irradiated, and the accumulated dose for any single screen will dictate if it too is returned to the owning institute or disposed of via DESY Radiation Protection. 


\section{ORAMS: Operability, Reliability, Availability, Maintainability, Safety}
\begin{itemize}
    
\item Operability -- The system is proven to be operable and not prohibitively difficult to operate \cite{awake_nature}. It is highly automatable, requiring only typical monitoring. The scintillating detector element is fully passive and the cameras standard, proven commercial equipment. The system requires a consistent, reliable, strong magnetic field to operate, but once set this field strength is unchanged during a run.  

\item Reliability -- The system is proven to be reliable within similar experiments \cite{awake_nature}, but there is always a small chance of radiation or mechanical damage. For the case of radiation damage replacements are possible. Small disruptions in position, alignment or scintillation efficiency are detectable using outside correlation with other detectors. It has been noted that scientific cameras can develop `hot pixels', that is a greater electronic read noise within a single pixel, as a result of stochastic stray particles in high-radiation environment \cite{neutron_cameras}. These hot pixels can be permanent, but consistent, and therefore mitigated when subtracting noise. Too many of these, particularly in the central signal bands, would motivate an exchange of cameras. The probability they will need to be replaced, or the frequency they are, is being investigated. 

\item Availability -- Every piece of equipment is available commercially within a period of several months, or is custom-made at DESY from standard materials, in particular aluminium for the supports. 

\item Maintainability -- Each sensitive component is cheap and easily replaceable. Replacements for the mechanical supports will also be constructed, in the rare case of mechanical damage to these supports. Damage to these may require a full re-alignment, however, requiring up to several days to do so. Replacement cameras can be attached to the same lens, fixed in place and correctly set for aperture and focus, meaning an exchange is quick and easily performed within a one-day access period, which is typically available once per week \cite{executive-summary}. The design of the screen support will also allow for easy exchange of the screen(s), which are manufactured to be similar as possible, and again this swap can be performed within one day. 

\item Safety -- Over some long integrated time the scintillation screens, like all components in the area, may absorb enough dose to be regarded as activated. To better make an estimate of this requires detailed radiation maps. If the components become activated, like the rest of the experiment, they would then require handling and disposal at the end of the experiment or the end of use of the specific screen, along guidance and consultation with the DESY D3 Radiation Protection group \cite{D3-radioprotection}. There are no hazardous materials or high-voltage equipment.


\end{itemize}


\section{Current \& Future Tests}
Test-beam campaigns have been completed using a prototype for the screen and camera system at the test-beam-facility DESY-II at DESY Hamburg~\cite{DESYTESTBEAM} and a non-user \lasernospace-plasma accelerator facility also at the DESY campus. This involved the creation of a data acquisition (C++, pylon \cite{basler-pylon}) program for the DESY II test-beam, and the use of a DOOCS-integrated data acquisition \cite{DOOCS} for the \lasernospace-plasma facility. These experiences provide a base to write further acquisition programs and proof the system is integrable within the DOOCS framework, envisioned for use in LUXE (see Chapter~\ref{chapt11}). 

In addition to these recent beam-tests is the required specific light--charge calibration curve. To achieve this, measurement must be made with a well-known, accurate, high-flux, bunch-based electron source, to calibrate the system's response to increasing flux. Candidates for suitable user-facility test-beams to perform this measurement include CLEAR~\cite{clear} at CERN and ELBE at the Forschungszentrum Rossendorf~\cite{elbe}.

\section*{Acknowledgements}
\addcontentsline{toc}{section}{Acknowledgements}
The authors JH and MW acknowledge DESY, Hamburg for their support and hospitality.
%
%
%
%
%
%
%
%

\printbibliography
\end{refsection}


\chapter{\cer Detector}
\label{chapt7}
\begin{refsection}
\chapterauthors{A.~Athanassiadis \\ \desyaffil}{L.~Hartman \\ \desyaffil \& \'Ecole Polytechnique F\'ed\'erale de Lausanne, Switzerland}{L.~Helary \\ \desyaffil}{J.~List \\ \desyaffil}{R.~Jacobs \\ \desyaffil}{E.~Ranken \\ \desyaffil}{S.~Schmitt \\ \desyaffil}
\date{\today}


\begin{chaptabstract}
    One of the experimental goals of LUXE is the measurement of non-linear Compton scattering in $e$-laser collisions, particularly the position of the Compton edge as a function of laser intensity. The challenge in this measurement lies in the wide range of electron rates of $10^3-10^8~e^-$ per event and detector channel. This note discusses the technical design of detectors for the measurement of high electron rates in LUXE. Segmented air-filled \cer detectors are used in a dipole spectrometer to reconstruct the electron energy spectrum. The choice of \cer detectors is motivated by their robustness against high particle rates and their good intrinsic background suppression. The detector segmentation is achieved via small-diameter straw tube light guides reflecting  the \cer light onto a grid of photodetectors. 
\end{chaptabstract}


\section{Introduction}
In the LUXE experiment, spatially segmented, gaseous \cer detectors are used for high-rate electron detection in two locations: In the electron-laser collision mode they are used to detect electrons from non-linear Compton scattering (EDS). In the photon-laser collision mode they are used to monitor the energy spectrum of the photons produced in the Bremsstrahlung solid-state target, or the inverse linear Compton scattering. The \glaser monitoring system is denoted IBM. The \cer detectors are part of a combined high-rate electron detection system together with a scintillator screen read out by an optical camera. For more details on the high-rate electron detection systems, and the scintillator screens, see Chapter~\ref{chapt6}. 

The combined use of two detector technologies (scintillator screens and \cer detectors) for high-rate electron detection is motivated by their complementarity, where the scintillator screen and camera system provides excellent position resolution, whereas the \cer detector adds tolerance of a large dynamic range of electron rates as  well as built-in low-energy background rejection through the \cer threshold.

The operating principle of the \cer detector is as follows: Charged particles traversing an active medium of refractive index $n$ at a speed faster than the speed of light in this medium emit \cer photons in the optical range according to the \textit{Frank-Tamm-Formula}~\cite{FrankTamm}:

\begin{equation}
    \label{eq:franktamm}
    \frac{d^2N_\gamma}{dz\,d\lambda}=\frac{2\pi\alpha}{\lambda^2}\sin^2\theta_c,
\end{equation}

where $\theta_c$ is the \cer angle, defined as $\cos\theta_c=1/\beta n$, $\lambda$ is the wavelength of the \cer photons, $\beta=v/c$ the speed of the charged particle, $z$ its path length in the active medium and $\alpha$ the fine-structure constant.\\

\begin{figure}
    \centering
    \includegraphics[width=0.6\textwidth]{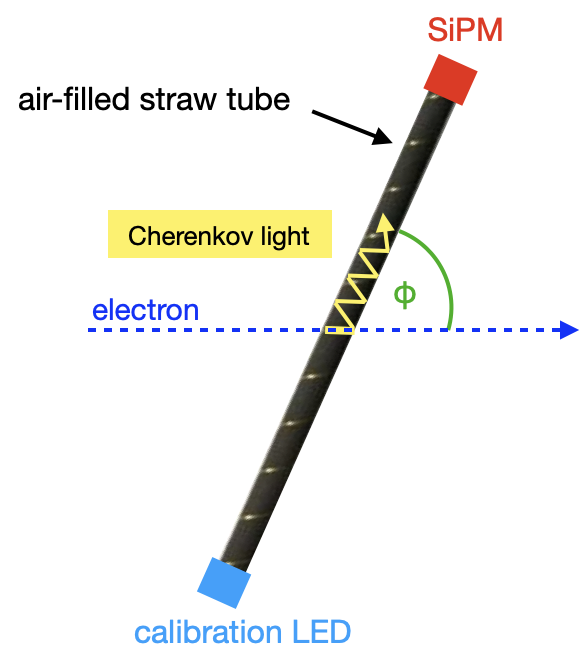}
    \caption{Visualisation of the production of \cer radiation and the light path in an air-filled reflective straw tube used as light guide.}
    \label{fig:sketch_straw}
\end{figure}

 Since the electron rates for the \cer detectors reach up to the order of $10^8$ per channel and laser shot, the straw diameter traversed by the incident electrons in the active medium produces a sufficient number of \cer photons for detection. The \cer medium used in the straws is air, which, despite its refractive index being low compared to other gases ($n_{\text{air}}\approx1.0028$) produces a sufficient number of \cer photons for the electron rates in LUXE. Fig.~\ref{fig:sketch_straw} shows a sketch of the light path in the straw light guide. Within the straw the \cer photons are reflected multiple times at the straw walls and are guided into the photo-detector placed at the end of the straw. 

The following list summarises the modifications of the \cer detector system design with respect to the LUXE CDR \cite{luxecdr}:

\begin{itemize}
    \item \textbf{Channel design:} Vertical Straw tubes instead of metal U-shaped channels (\textit{eases finer segmentation}).
    \item \textbf{Photodetectors:} SiPMs and/or APDs instead of gaseous PMs (\textit{eases finer segmentation, allows to measure larger dynamic range of electron rates}).
    \item \textbf{Active medium:} Air instead of argon (\textit{cost reduction}).
\end{itemize}

\section{Requirements and Challenges}
\label{sec:requirements}

The main requirements for the \cer system concern the energy resolution, linearity and stability of the detector response to the number of primary electrons, as well as the background rejection. They will be discussed in the following, focusing on the EDS. The requirements for the IBM are similar to those of the EDS. The IBM particularities are discussed in Section \ref{subsec:IBMreq}. 

\subsection{EDS Requirements}
The detector requirement are closely linked to the dipole spectrometer properties. The energy of an electron traversing a magnetic dipole field of length $z_m$ and strength $B$ can be reconstructed from its transverse $x$ deflection measured by  a detector placed at a distance $z_d$ from the dipole field. Approximating for a large bending radius $\rho\gg z_m$ yields:

\begin{equation}
    E(x)=Bez_m\cdot \left( \frac{z_m}{2}+z_d \right)\frac{1}{x}+ \mathcal{O}\left(\left(\frac{z_m}{\rho}\right)^3\right),
\label{eq:dipoleeofx}
\end{equation}

where $e = \sqrt{4\pi\alpha} = 0.30282$ is the dimensionless elementary charge,  $E$ is expressed in units of GeV, $B$ in T and $x$ in m. 

In order to resolve the position of the edges in the nonlinear Compton spectrum, the \cer detectors are spatially segmented. The same holds for the initial Bremsstrahlung monitor, where the energy spectrum of the electrons after Bremsstrahlung is used to extract information about the produced photons. The necessary segmentation depends on the dimension and field strength of the dipole spectrometer, as well as on the distance between the dipole magnet and the detector plane. Based on equation~\ref{eq:dipoleeofx}, one can express the energy resolution $\sigma_E/E$, achieved with a spatial segmentation $\Delta x$ of the \cer detectors as:

\begin{equation}
    \frac{\sigma_E}{E}=\frac{\Delta x}{Bez_m\cdot(\frac{z_m}{2}+z_d)} \cdot E,
\label{eq:eresosegm}
\end{equation}

\begin{figure}[htbp!]
    \centering
    \includegraphics[width=0.75\textwidth]{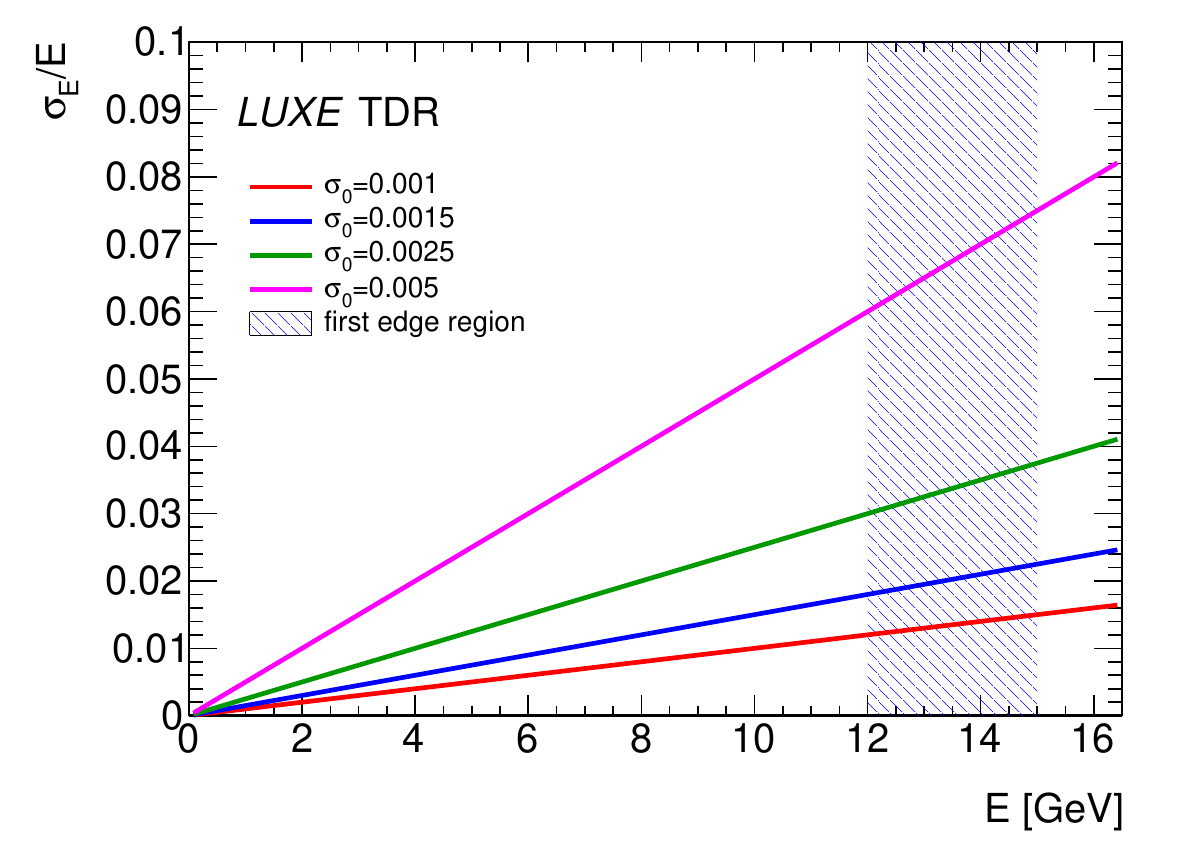}
    \caption{Relative electron energy resolution as a function of energy for different values of $\sigma_0$. The shaded blue area corresponds to the energy region in which the first Compton edge lies for different values of $\xi$. }
    \label{fig:eresocere}
\end{figure}

For reasons of simplicity one can express the resolution as:

\begin{equation}
    \frac{\sigma_E}{E}= \sigma_0 \cdot E/\text{GeV},
\label{eq:eresosegmsimp}
\end{equation}

where $\sigma_0$ denotes the dipole parameters and the spatial segmentation and position of the detector. The linear dependence of $\sigma_E/E$ on $E$ is typical for a dipole spectrometer. Figure~\ref{fig:eresocere} shows the relative electron energy resolution as a function of energy for different values of $\sigma_0$. The aim is to achieve below $2\%$ energy resolution in the first edge region, which is possible with a $\sigma_0=0.15\%$. With the LUXE dipole configuration $(z_d=3.2\,\text{m},\: z_m=1.2\,\text{m}, B=1.6\,\text{T})$ this corresponds to a spatial segmentation of $\Delta x=3\,\text{mm}$.

\begin{figure}[htbp!] 
\centering
\begin{subfigure}{0.75\hsize} 
\includegraphics[width=\textwidth]{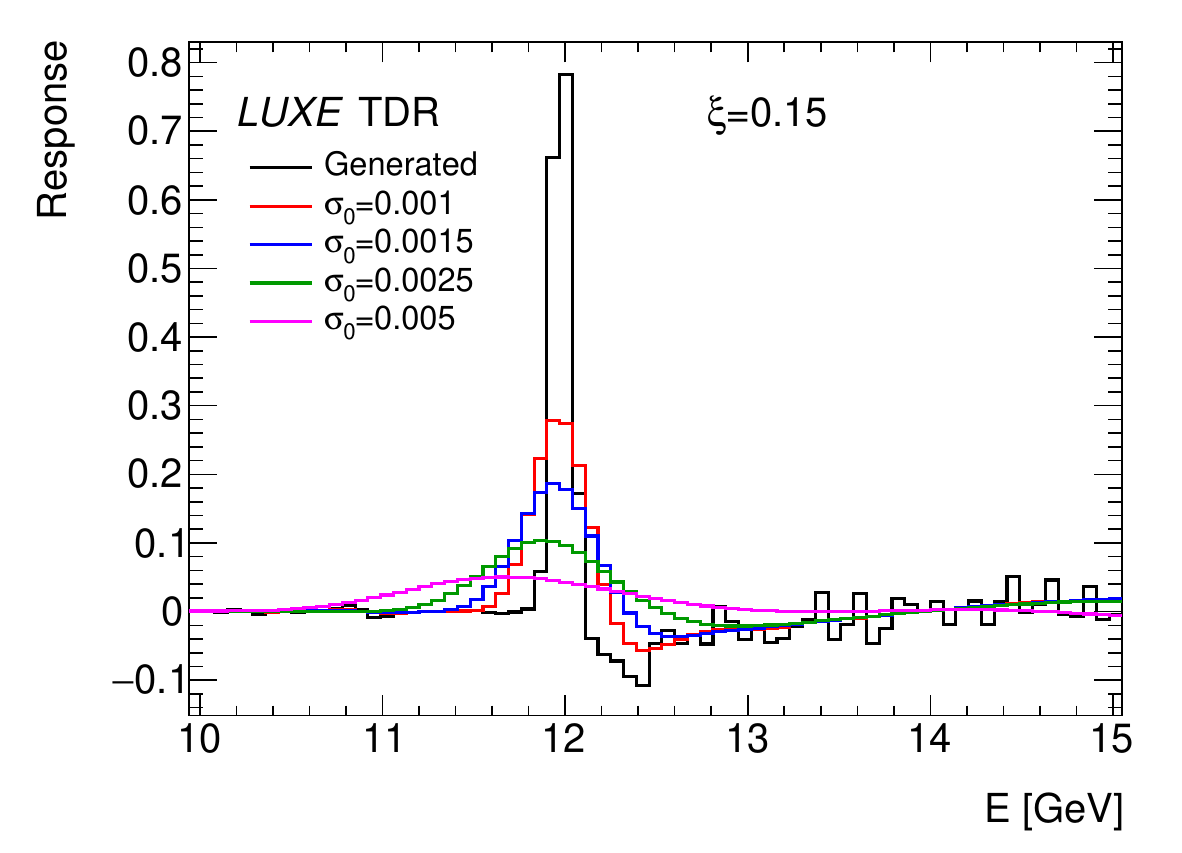}
\caption{ }
\end{subfigure}
\begin{subfigure}{0.75\hsize} 
\includegraphics[width=\textwidth]{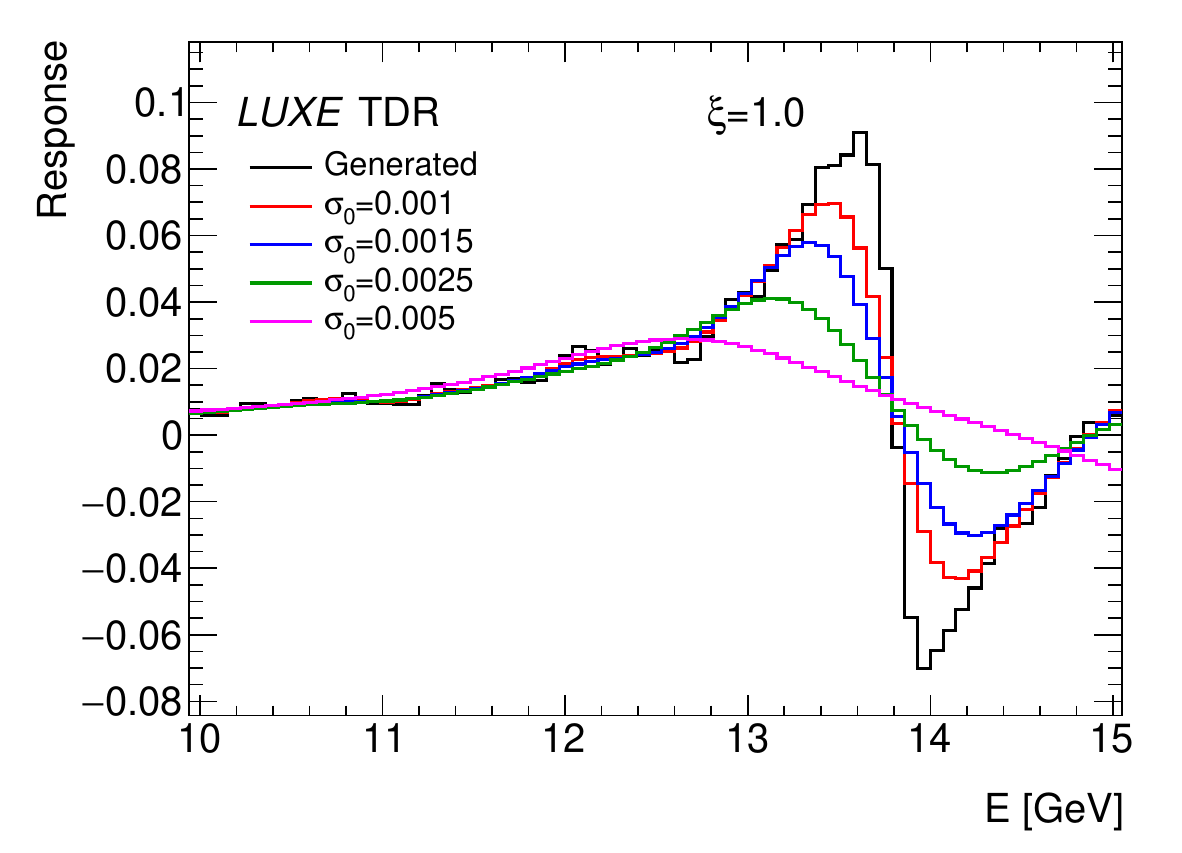}
\caption{ }
\end{subfigure}
\caption{Simulated FIR response for different energy resolution factors $\sigma_0$ for a) intensity parameter $\xi=0.15$ and b) $\xi=1.0$ }
\label{fig:response}
\end{figure}

The impact of the energy resolution on the reconstruction of the Compton edge position is demonstrated by smearing the electron energy spectrum according to equation~\ref{eq:eresosegmsimp} with different values of $\sigma_0$ and studying the impact on the edge finding algorithm performance. The reconstruction of the Compton edges is performed with a finite impulses response filter method (FIR)~\cite{Berggren:2020gna}. More detail on the FIR method is given in Section~\ref{subsec:scint_fir}. Figure~\ref{fig:response} shows the filter response for the first Compton edge for different values of $\sigma_0$ and laser intensity $\xi$. The zero-crossing of the filter response corresponds to the so-called kink-position, which is the edge position of the highest $\xi$ in the laser pulse and the main feature of interest in the Compton edge analysis. It is clearly visible that the coarser the energy resolution the more diluted the edge becomes. 

The spatial segmentation of the \cer detector is achieved by using reflective straw tubes as light-guides. The minimal spatial segmentation is limited by the minimal active area of the photodetector and the minimal spacing between photodetectors on the carrier PCB. Also, the cost of the detector is directly proportional to the number of channels needed to cover the full Compton spectrum. This means that the choice of segmentation of the \cer detector is a trade-off between a fine enough energy resolution to resolve the Compton edges and the technical and cost constraints. The channel segmentation of $\Delta x=3\,\text{mm}$ meets these requirements. 


\begin{figure}[htbp!]
    \centering
    \includegraphics[width=0.75\textwidth]{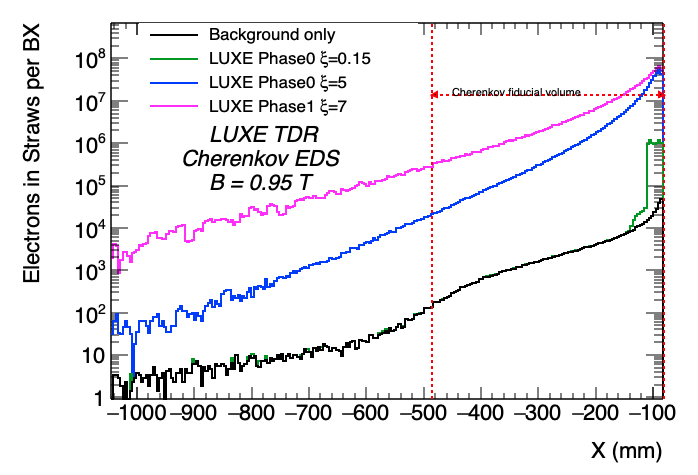}
    \caption{Simulated distribution of arriving signal and beam background particles in the EDS \cer detector ($E>20\,\text{MeV}$ \cer threshold applied) for different values of $\xi$. The binning corresponds to the EDS straw channel segmentation. Indicated by the red dashed line is the acceptance area of the EDS \cer detector.}
    \label{fig:sigbg}
\end{figure}

Figure~\ref{fig:sigbg} shows the expected signal and background distributions for the EDS system for the \cer detector simulated by \geant. Only charged particles are shown for the signal and background because the \cer detector is not sensitive to direct photon backgrounds. In addition, low-energy charged particle backgrounds are suppressed by the $20\,\text{MeV}$ \cer energy threshold in air. 

\begin{figure}
    \centering
    \includegraphics[width=0.75\textwidth]{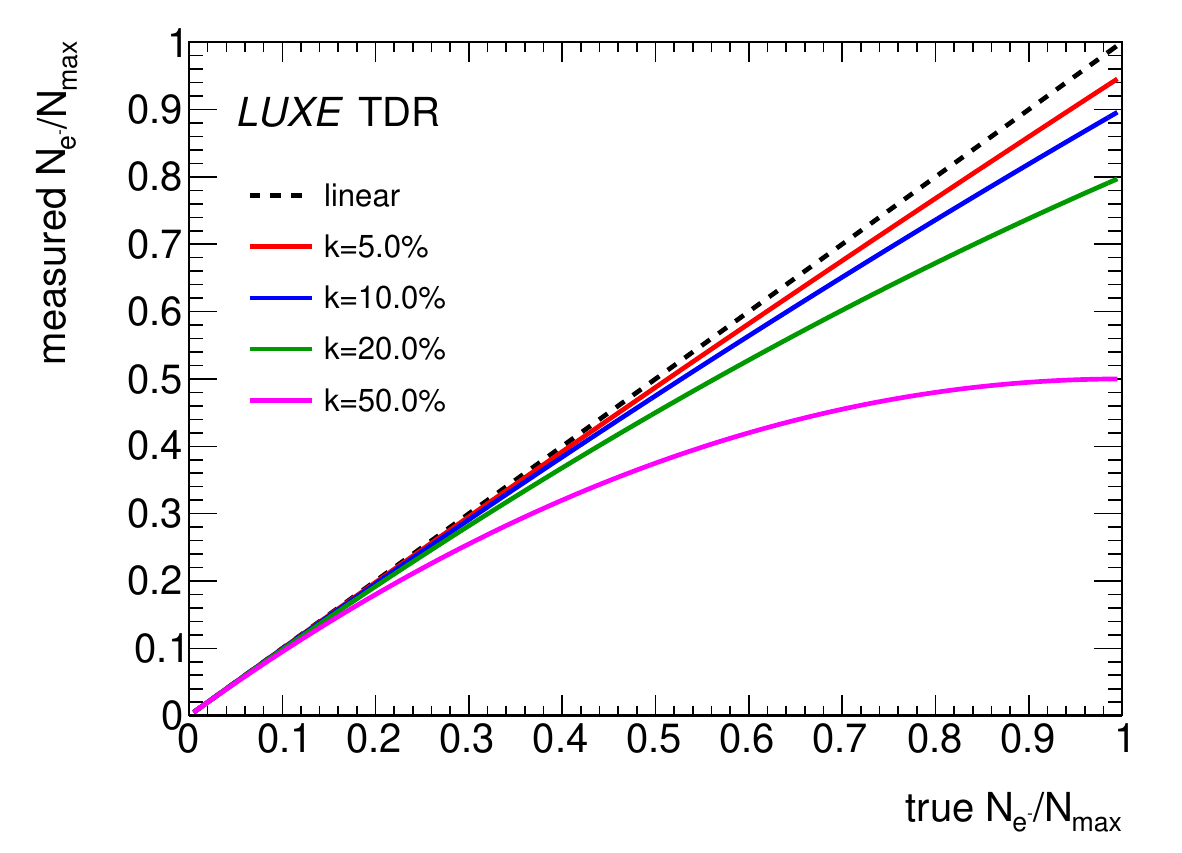}
    \caption{Non-linear distortion factor with quadratic correction to estimate the impact of detector or readout non-linearities on the EDS \cer measurement.}
    \label{fig:nonlincere}
\end{figure}

One important aspect of measuring the energy spectrum is the linearity of the photodetector and readout electronics with respect to the rate of incident electrons.  The impact of possible detector or readout non-linearities is estimated by applying a non-linear distortion factor with a quadratic term to the generated spectrum and recalculating the FIR filter response. The number of electrons in the distorted spectrum $f(N)$ is calculated according to:

\begin{equation}
    f(N)=N\cdot(1-kN),\; \text{where:} \; k=\frac{\delta_{\text{max}}}{N_{\text{max}}}
    \label{eq:nonlin}
\end{equation}

where the maximal non-linearity $\delta_{\text{max}}$ is the deviation from linear for the maximal electron count $N_{\text{max}}$ in the spectrum. The distortion factor is shown in Fig.~\ref{fig:nonlincere}.

\begin{figure}[htbp!] 
\begin{center}
\begin{subfigure}{0.75\hsize} 
\includegraphics[width=\textwidth]{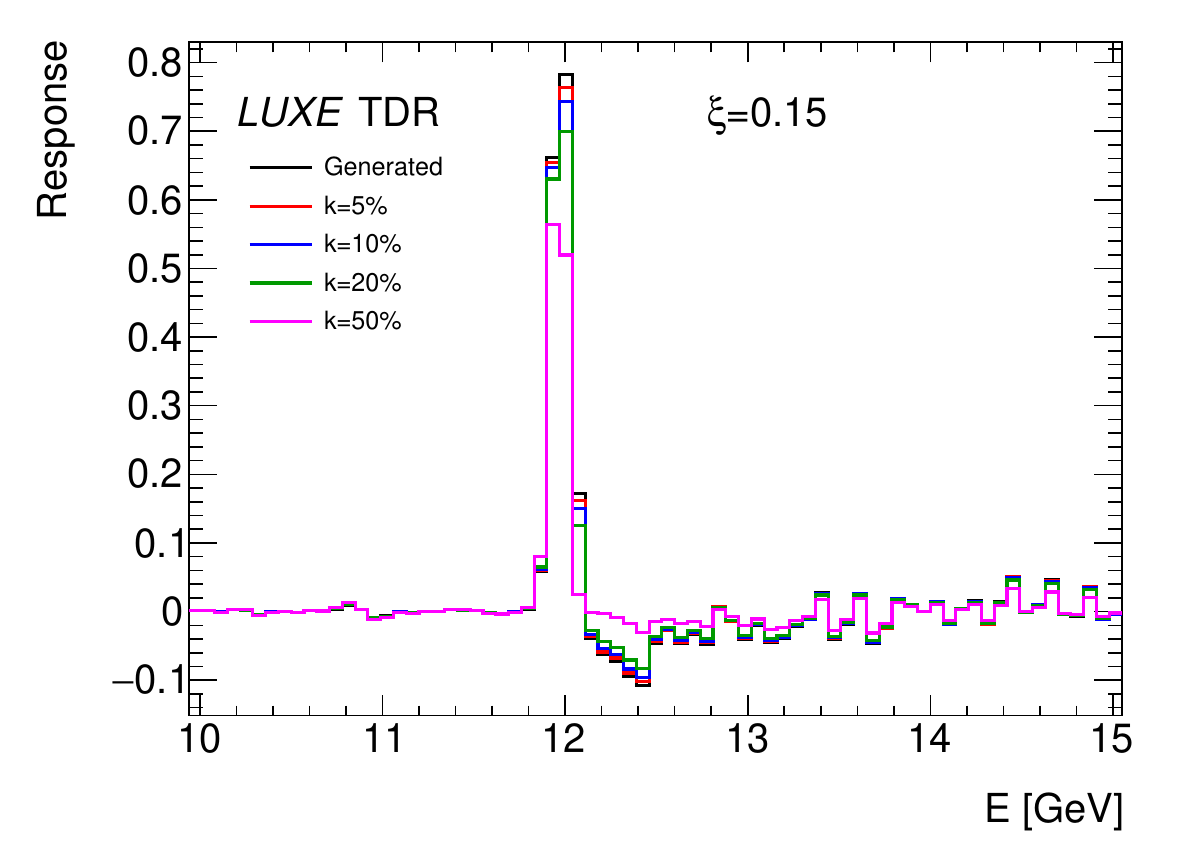}
\caption{ }
\end{subfigure}
\begin{subfigure}{0.75\hsize} 
\includegraphics[width=\textwidth]{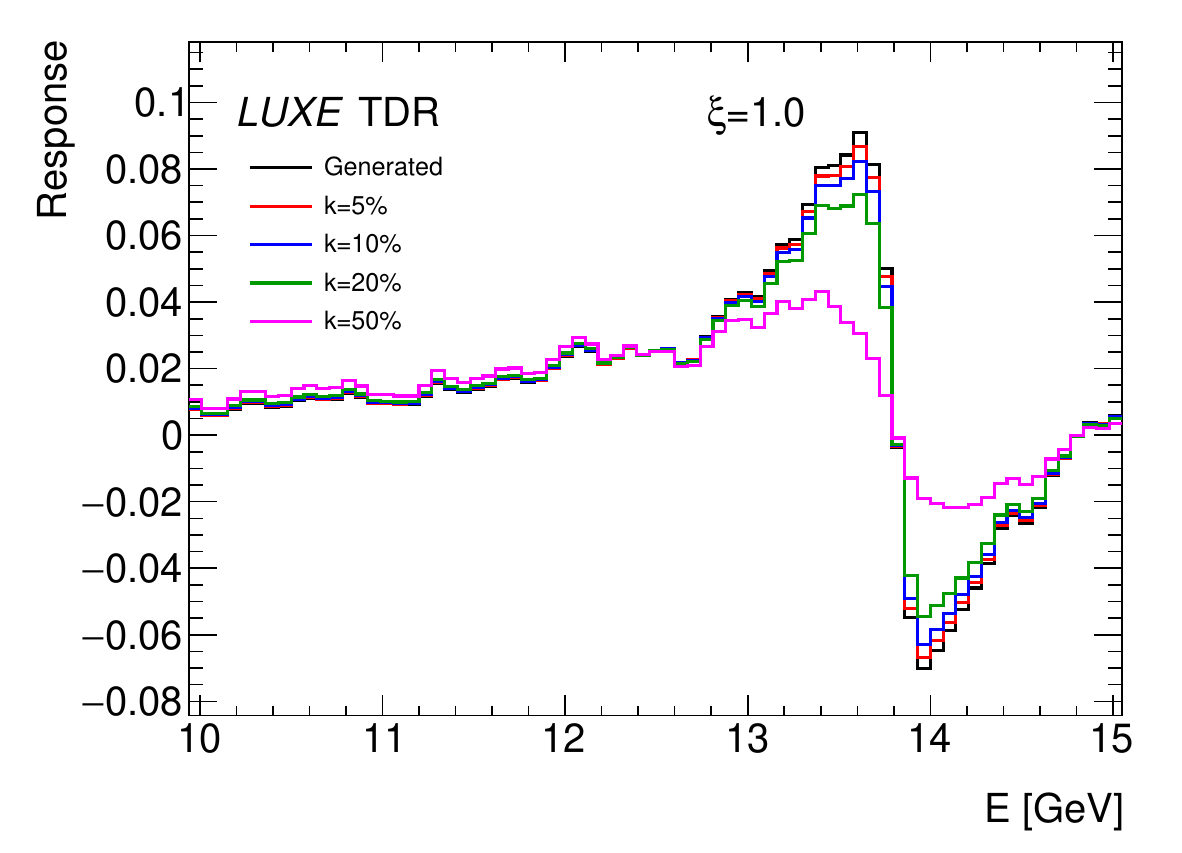}
\caption{ }
\end{subfigure}
\caption{Impact of a non-linear distortion on the FIR function of the Compton electron spectrum for a) $\xi$=0.15 and b) $\xi=1.0$ }
\label{fig:nonlin}
\end{center}
\end{figure}

The impact of the non-linear distortion factor with different $\delta_\text{max}$ on the FIR is shown in Fig.~\ref{fig:nonlin} for two different laser intensities. It is clearly visible that the response zero-crossing of the first Compton edge is robust against detector and readout non-linearities up to $\delta_\text{max}=20\%$. 

\begin{figure}
    \centering
    \includegraphics[width=0.75\textwidth]{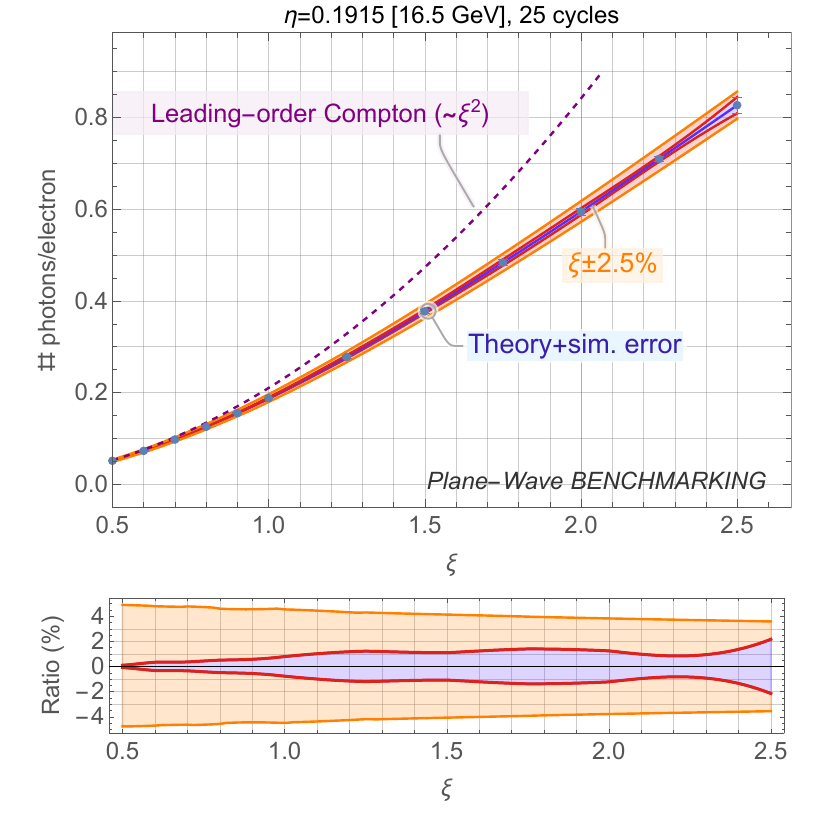}
     \caption{Predicted ratio of the number of photons and electrons produced in non-linear (solid) and linear (dashed) Compton scattering as a function of the Laser intensity $\xi$. Shown are the systematic uncertainties from theory and simulation (blue band) and from a $2.5\%$ uncertainty on $\xi$ (yellow band).}
    \label{fig:gammaeratioCompt}
\end{figure}

Another quantity of interest for the EDS \cer system is the total number of Compton-scattered electrons per event. This number is an input to calculating the ratio between the number of Compton photons and electrons, $R_{\gamma/e}=N_\gamma/N_{e^-}$, which, as a function of $\xi$, is one of the observables sensitive to the transition between the perturbative and non-perturbative QED regimes (see Fig.~\ref{fig:gammaeratioCompt}, which is the same as Fig.~\ref{fig:resultscom}, repeated here for convenience). For more details see Chapter~\ref{chapt2}.

The requirement for the EDS linearity for the total electron rate is derived as follows: The systematic uncertainty on the predicted behaviour of the Compton photon-electron ratio $R_{\gamma/e}$ as a function of $\xi$ is driven by the uncertainty of $\xi$ itself (see the yellow band in the ratio in Fig.~\ref{fig:gammaeratioCompt}). The associated relative uncertainty is estimated to be $\Delta R_{\gamma/e}/R_{\gamma/e}\approx4\%$. The total detector-related systematic uncertainty of the ratio should be smaller than this value. Translated into a requirement on the relative uncertainty on the total number of electrons $\Delta N_{e^-}/N_{e^-}$, this means that $\Delta N_{e^-}/N_{e^-}\lsim\Delta R_{\gamma/e}/R_{\gamma/e}/\sqrt{2}\approx2.8\%$.

\begin{figure}
    \centering
    \includegraphics[width=0.75\textwidth]{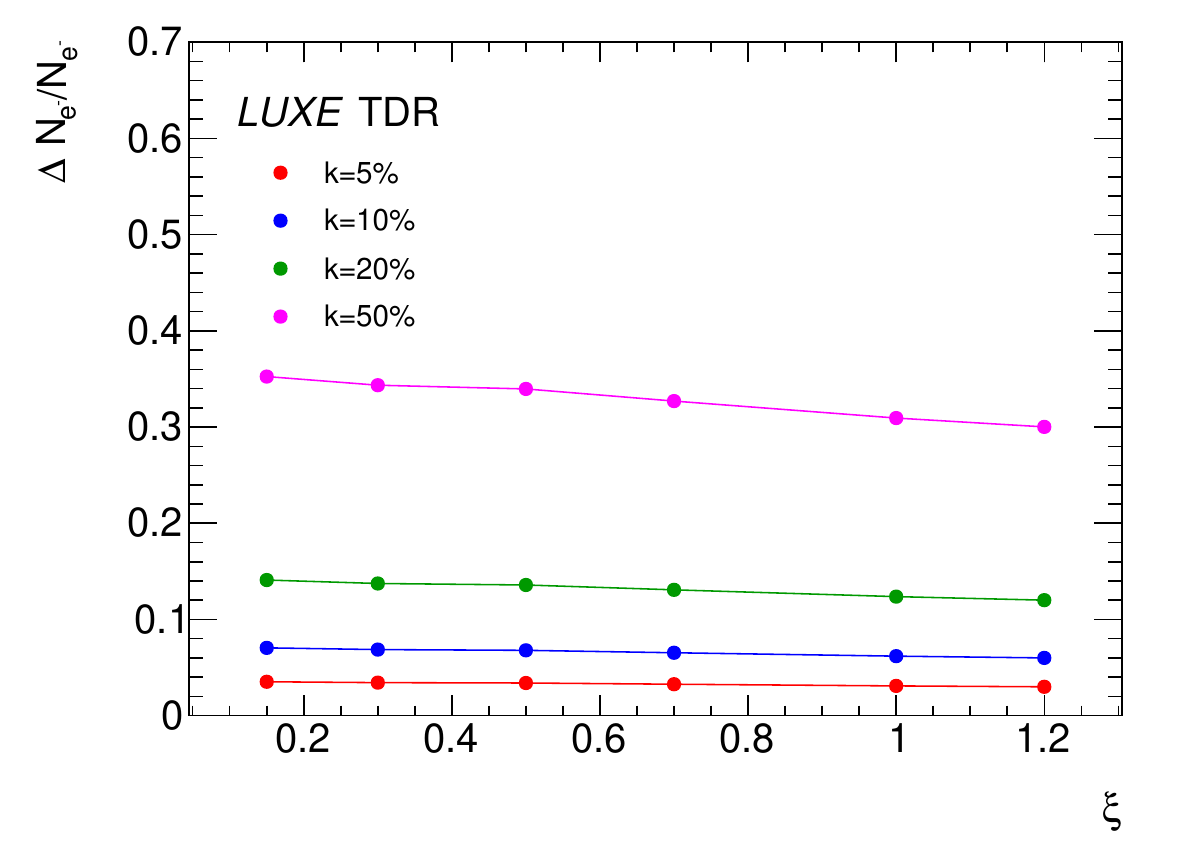}
    \caption{Impact of a non-linear response with quadratic correction on the measurement of total number of Compton electrons.}
    \label{fig:nonlinnorm}
\end{figure}

Figure~\ref{fig:nonlinnorm} shows the impact of the non-linear distortion (equation~\ref{eq:nonlin}) on the measurement of the total number of electrons. The quantity shown is the relative difference between the total number of electrons measured with and without presence of a non-linearity assuming different $k$ parameters. It is visible that the requirement on $k$ is much more stringent for the measurement of the total number of electrons, where $k=5\%$ corresponds to a variation of the total electron number of about $3\%$.

Finally the \cer detector is required to give a stable response for a given number of incident primary electrons for every bunch interaction. This includes stability as a function of the detector channel being hit, as well as stability over time. For the stability as a function of channel it is important that each photodetector and the individual channel characteristics, such as gain, channel reflectivity and optical filter efficiency are known precisely in advance. To ensure the stability of the detector over time it is important that the individual light guides are mounted such that their position is stable over time. Other possible sources of time-dependent instabilities are air temperature and pressure variations which affect the refractive index \cite{Ciddor:96,Edl_n_1966}. Using equation~\ref{eq:franktamm} and Gaussian error propagation, one can see that a change in the refractive index $n$ affects the number of \cer photons per primary charged particle $\Delta N_\gamma$ as follows:

\begin{equation}
    \label{eq:nvariation}
    \frac{\Delta N_\gamma}{N_\gamma}=\frac{2\Delta n}{n^3-n}
\end{equation}

Following equation~\ref{eq:nvariation}, and using air refractive index data from \cite{AirRefIndex}, the expected change in the number of \cer photons produced due to a temperature-induced variation of the refractive index is estimated to be $\Delta N^{\text{T}}_\gamma/N_\gamma\approx0.3\%/^\circ$C. For a pressure variation the expected change in the number of produced \cer photons is $\Delta N^{\text{P}}_\gamma/N_\gamma\approx0.1\%/\text{mbar}$. To comply with the requirement of linearity at the percent level, the temperature must therefore be controlled to within $3^\circ$C, while the gas pressure must be controlled to within $10\,\text{mbar}$. It should be noted that temperature or pressure will be monitored such that variations during the physics data-taking can be corrected offline. Both temperature and pressure-dependence of the \cer light will also be measured ahead of installation in a testbeam campaign. Finally, the optical properties of the straw light guide materials must not deteriorate over time. To ensure this, radiation-tolerant materials will be chosen for the straw light guides, and their properties investigated in an irradiation test.

\subsection{IBM Requirements}
\label{subsec:IBMreq}

The aim of the IBM system is to monitor the spectrum of electrons after the initial Bremsstrahlung conversion target in the photon-laser mode. It employs a dipole spectrometer with the same type of magnet as the EDS system. The IBM system complements the forward gamma ray detection systems that monitor the energy distribution of the downstream photon beam. 

\begin{figure}[htbp!]
    \centering
    \begin{subfigure}{0.75\hsize}
    \includegraphics[width=\textwidth]{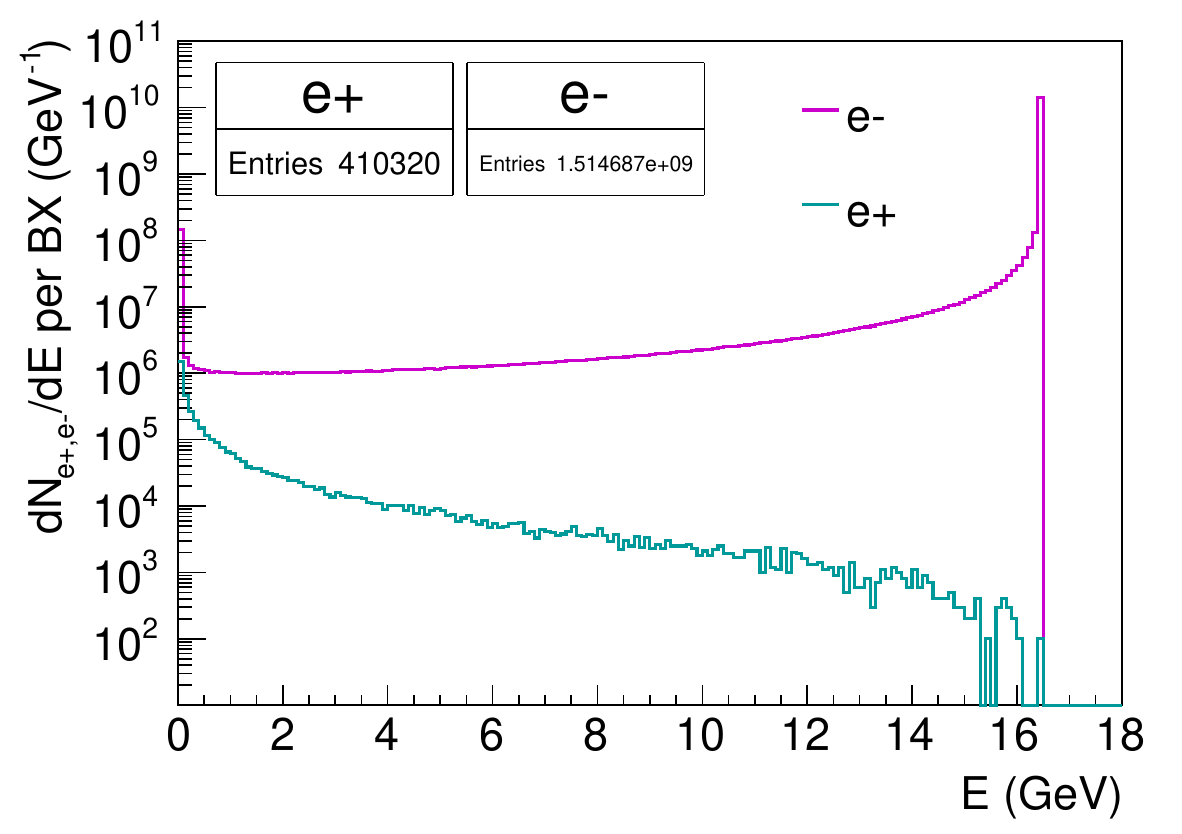}
    \caption{ }
     \label{fig:brem_el}
    \end{subfigure}
    \begin{subfigure}{0.75\hsize}
    \includegraphics[width=\textwidth]{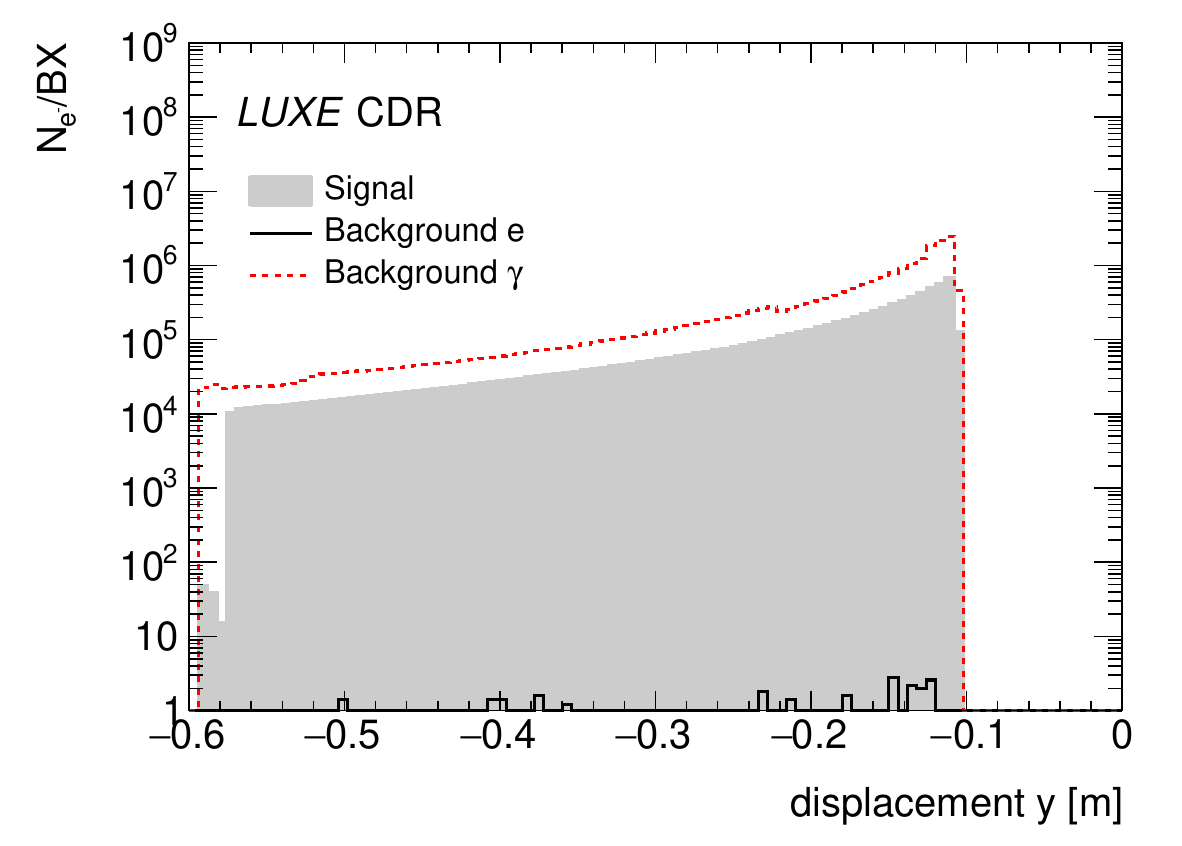}
    \caption{ }
    \label{fig:brem_sigbg}
    \end{subfigure}
    \caption{(a) Energy spectra of electrons and positrons produced in the tungsten target~\cite{luxecdr}. A tungsten target with a thickness of 35\,$\mu$m (1\%$X_{0}$) is simulated. (b) Occupancy per mm for the \cer detector after the bremsstrahlung target for \glaser scattering~\cite{luxecdr}.}
    \label{fig:ibmsystem}
\end{figure}

Figure~\ref{fig:brem_el} shows a simulated spectrum of the electrons after the Bremsstrahlung target. Since the Breit-Wheeler pair production rate increases with photon energy, the relevant part of the photon spectrum are high-energy photons, corresponding to low-energy electrons.

The main goal for the IBM \cer system is to have the largest possible acceptance towards low energies. The energy resolution according to equation~\ref{eq:eresosegm} grows linearly with the energy, therefore a coarser segmentation than the EDS system ($10\,\text{mm}$) is acceptable for the IBM detector. With a magnetic field of $B=1.6\,\text{T}$, $z_\text{m}=1.2\,\text{m}$ and $z_\text{d}=0.5\,\text{m}$, the resulting energy resolution factor (see eq.~\ref{eq:eresosegmsimp}) is $\sigma_0=0.016$, resulting in an energy resolution of $2\%$ for $1.2\,\GeV$ electrons at the lower energy acceptance limit. Furthermore, the expected electron rates are an order of magnitude lower than for the EDS, which makes a longer electron path length in the active medium acceptable without saturating the IBMs photodetectors. With fifty straws it is possible to cover the electron energy range between $1.2$ and $16\,\GeV$.  

Figure~\ref{fig:brem_sigbg} shows the simulated signal and background spectrum of the IBM system as a function of the displacement in $y$. It should be noted that the IBM spectrometer dipole magnet is tilted by $90^\circ$, therefore the displacement of charged particle tracks occurs in the vertical direction. In Fig.~\ref{fig:brem_sigbg} it is visible that the electron rate of the order of $10^4-10^5$ is on average lower than the rates at the EDS system. This allows the use of straws with larger diameter without saturating the detector, despite the longer path length of the primary electron resulting in a higher \cer light yield. In addition one can see that the main background source in this location are photons, which the \cer detector is not sensitive to.

The linearity requirement for the IBM system is still under study, but is expected to be less stringent that for the EDS system, since it only monitors the Bremsstrahlung electron energy spectrum without attempting a precision measurement of the total rate.

\begin{table}[]
    \centering
    \begin{tabular}{|c|c|c|}
         \hline
         \textbf{Parameter} & \textbf{Requirement} & \textbf{Commment}  \\ \hline
         Detector acceptance & $[5,16]\,\text{GeV}$ & coverage of first edge region \\ \hline
         Energy resolution & $\sigma_0 <0.0015$ & $2\%$ in first edge region \\ \hline
         Linearity factor k & $<0.5\%$ & total norm uncert. $<2.8\%$\\ \hline
    \end{tabular}
    \caption{Requirements for the EDS \cer detector.}
    \label{tab:req_eds}
\end{table}

\begin{table}[]
    \centering
    \begin{tabular}{|c|c|c|}
        \hline
         \textbf{Parameter} & \textbf{Required limit} & \textbf{Commment}  \\ \hline
         Detector acceptance & $[1.2,16]\,\text{GeV}$ & coverage of Brems spectrum \\ \hline
         Energy resolution & $\sigma_0=0.016$ & $2\%$ at lower limit $1.2\,\text{GeV}$ \\ \hline
         Linearity & tbc & less stringent than for EDS\\ \hline
    \end{tabular}
    \caption{Requirements for the IBM \cer detector.}
    \label{tab:req_ibm}
\end{table}

Tables \ref{tab:req_eds} and \ref{tab:req_ibm} summarise the aforementioned requirements for the EDS and IBM system.

\section{System Overview}
\begin{figure}[htbp!]
\centering
\includegraphics[trim={10cm 2cm 10cm 2cm},clip,width=\textwidth]{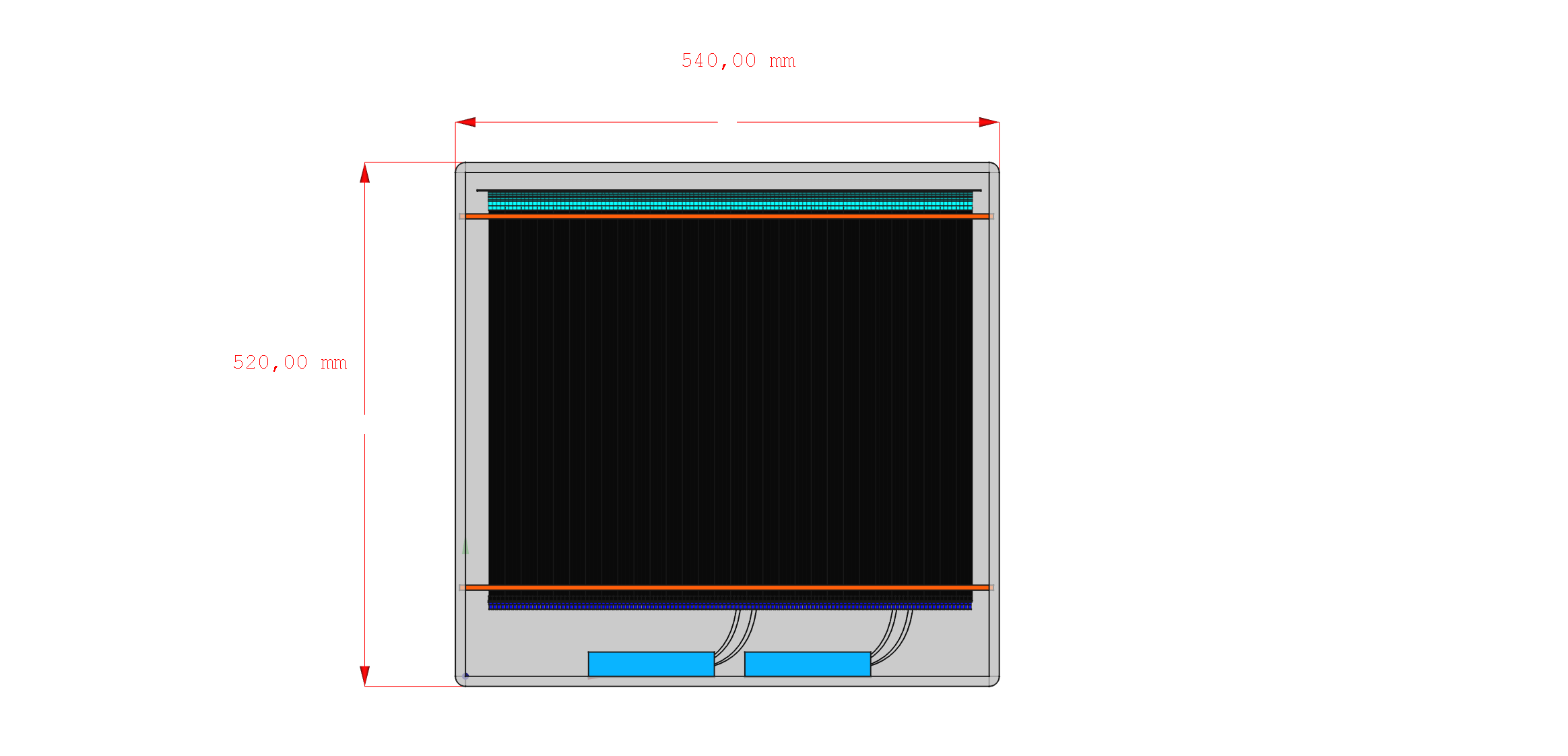}
\caption{CAD drawing of the straw \cer detector seen from the front (XY-plane) shown without front box wall, showing the straw tube light guides (black, turquoise), mounting rails (red) and LED calibration box (blue) with fibres.}
\label{fig:tech_cadfront}
\end{figure}

\begin{figure}[htbp!]
\centering
\includegraphics[trim={5cm 2cm 15cm 2cm},clip,width=\textwidth]{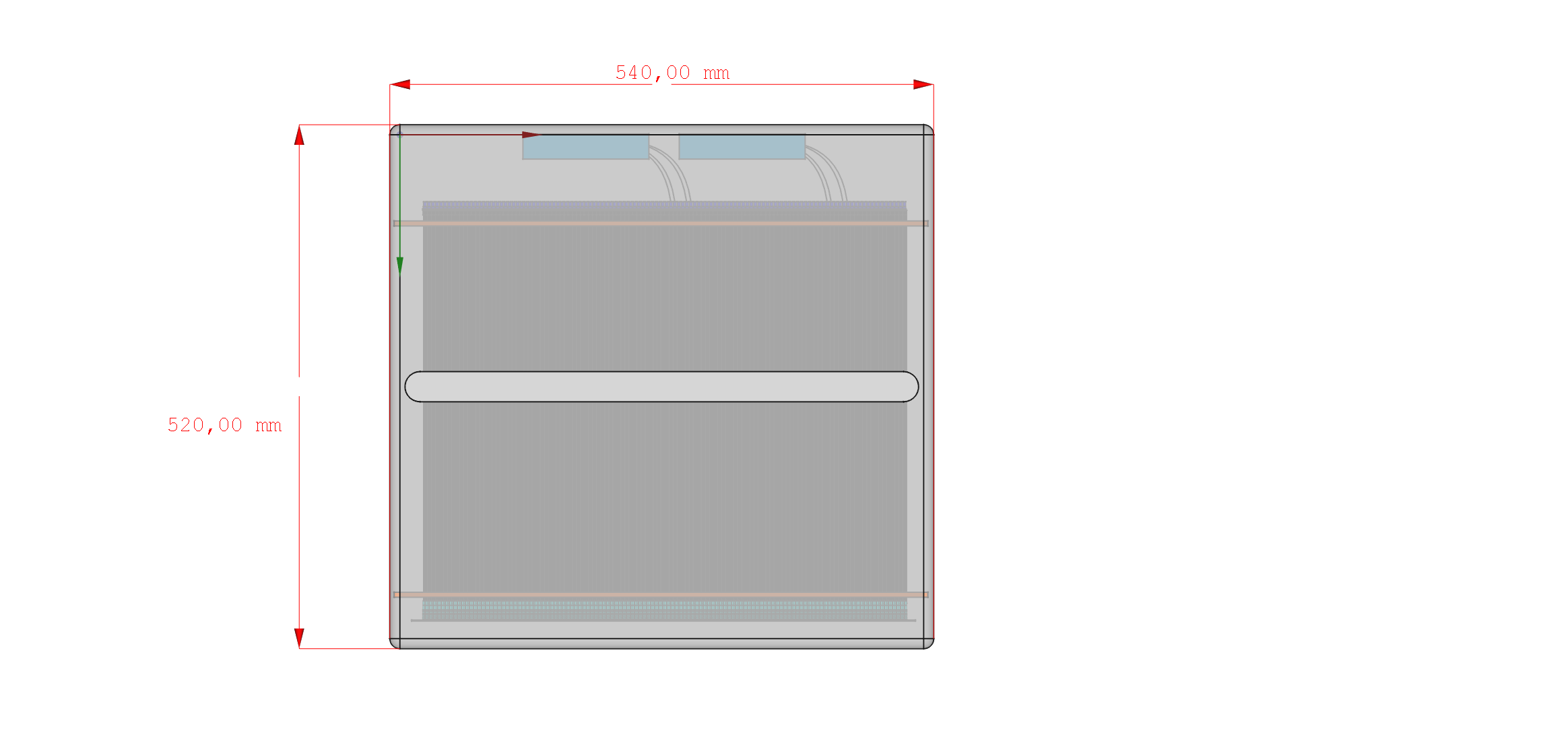}
\caption{CAD drawing of the straw \cer detector seen from the back (YX-plane, identical to front box wall), showing the electron entrance/exit window).}
\label{fig:tech_cadback}
\end{figure}

The LUXE \cer detectors consist of reflective straw tube channels filled with air as an active medium. Figures~\ref{fig:tech_cadfront} and \ref{fig:tech_cadback} show a CAD drawing of the detector design.

The straw light guides are mounted in a gas-tight box, such that the electrons pass through a beam window, perpendicularly traversing the straws and exiting the gas-tight box through an exit window. The box is filled with air at $1.1\,\text{bar}$ and is sealed to ensure a constant gas pressure.

On the bottom of the light-tight box, a LED calibration system is mounted, consisting of a pulsed UV LED and an optical fibre, which guides the LED calibration light pulses directly into the straw tubes. For more information on the LED calibration strategy, see section \ref{subsec:calib_strategy}.

\begin{figure}[htbp!]
    \centering
    \includegraphics[trim={20cm 10cm 20cm 10cm},clip,width=\textwidth]{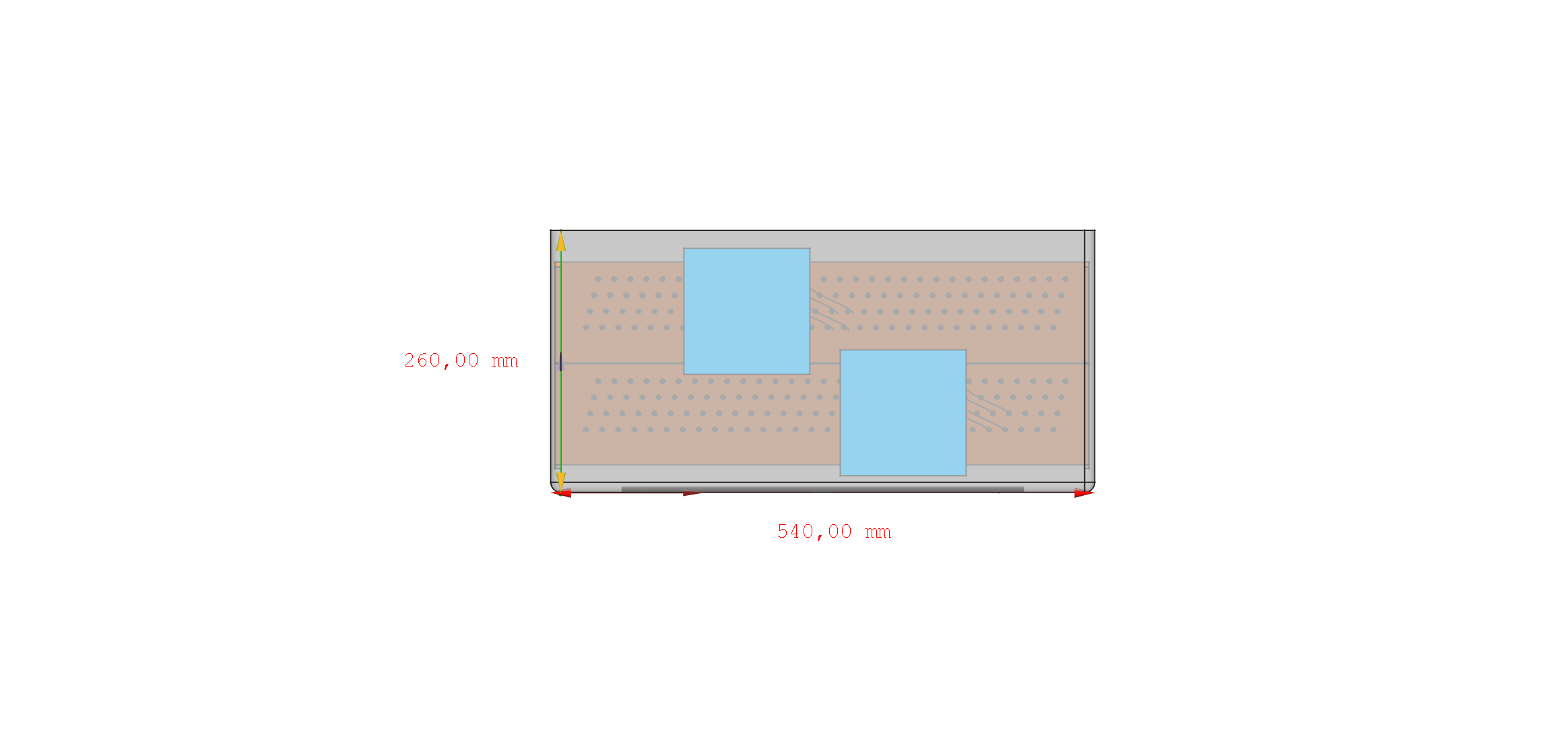}
    \caption{CAD drawing of the straw \cer detector seen from the top (XZ-plane), showing the mounting rail (red) and two super-layers (each consisting of four layers of staggered straw channels),  for APD readout and for SiPM readout. Shown in light blue is the outline of the LED calibration box with optic fibres leading to the straw channels.}
    \label{fig:tech_cadtop}
\end{figure}

Figure~\ref{fig:tech_cadtop} shows the straw configuration inside the box as seen from the top. The straws are mounted in staggered layers, to maximise the distance between the photodetectors on the PCB in order to simplify the PCB design and mounting, while retaining a fine horizontal segmentation needed to achieve an energy resolution below $2\%$ in the Compton edge regime. To allow the required large dynamic range of electron rates to be measured, two types of photodetectors will be used to read out the \cer light simultaneously, therefore two super-layers of staggerred channels are foreseen, namely one per each type of photodetector. 

The straw tubes consist of foil which is reflective on the inside (coated with a thin metal layer) and light-tight to the outside of the straw. Alternatively, aluminium straws of the same diameter could be used. The reflectivity of the straw has a significant impact on the light yield at the photodetector end. One foreseen option to optimise the detector dynamic range is to use differently reflective straw materials for the two photodetector types. Several high-energy experiments use similar straw tubes as cathodes for gas ionisation tracking detectors with an anode wire in the middle of the straw, for example the ATLAS TRT detector~\cite{CERN-LHCC-97-016} and the NA62 experiment~\cite{Hahn:1404985}. In LUXE, only the cathode tubes without the anode wire will be used to guide the \cer light created by electrons traversing the active medium inside the tube towards the photodetector placed at the end of the tube. The design is inspired by gas-filled \cer detectors for electron beam polarimetry at the ILC~\cite{Bartels:2010eb}. A similar technology also already exists for the monitoring of nuclear reactor core activity using a metal tube to guide \cer light created by a high rate of electrons to a photodetector at the end of the tube ~\cite{BENTOUMI201886}.

\begin{figure}[htbp!]
    \centering
    \includegraphics[width=\textwidth]{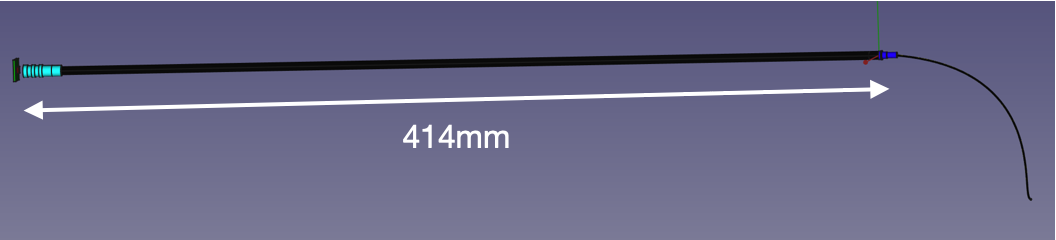}
    \caption{CAD drawing of a single straw and its connection to the PCB and fibre to the calibration LED.}
    \label{fig:tech_cadstraw}
\end{figure}

Figure~\ref{fig:tech_cadstraw} shows a CAD drawing of a straw and its connection to a photodetector on the one side, and the optical fibre leading to the UV LED on the other end of the straw. The straw tube length for the LUXE \cer detectors will be of the order of $40\,\text{cm}$, to ensure that the electrons passing in a plane in the middle of the straw do not hit the photodetectors or the LED calibration system directly. To further shield the photodetectors and to reduce backgrounds, tungsten plates will be mounted in front of the straws above and below the beam windows.

\begin{figure}[htbp!] 
\centering
\begin{subfigure}{0.75\hsize} 
\includegraphics[width=\textwidth,trim={2cm 0 2cm 0},clip]{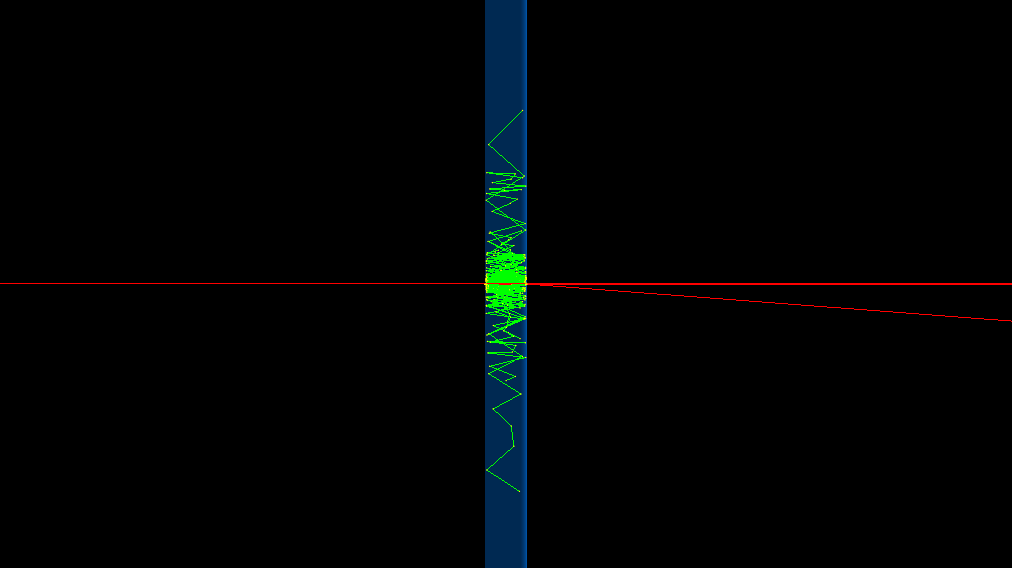}
\caption{ }
\end{subfigure}
\begin{subfigure}{0.75\hsize} 
\includegraphics[width=\textwidth,trim={2cm 0 3cm 0},clip]{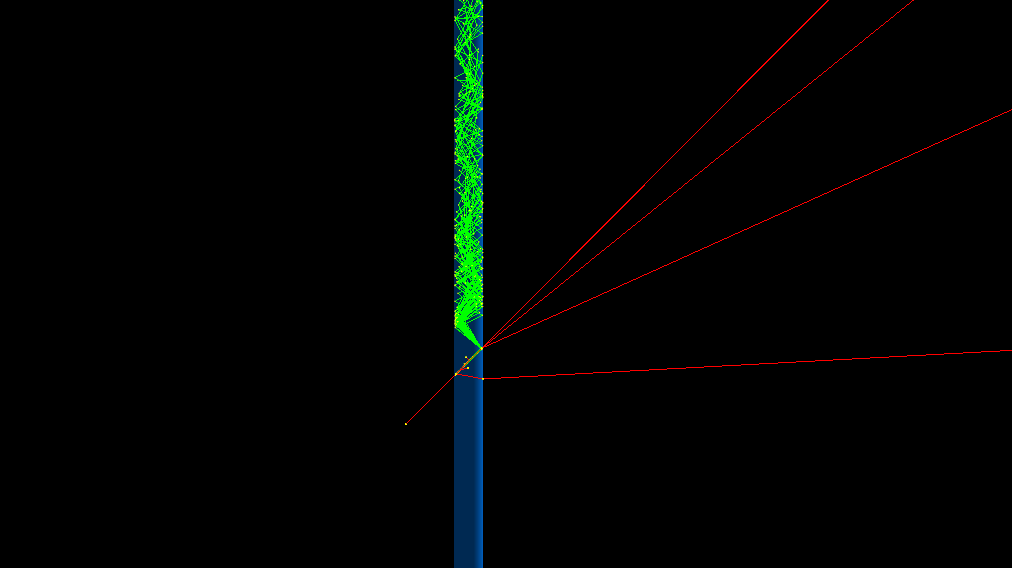}
\caption{ }
\end{subfigure}
\caption{Simulated (\geant) \cer light production in an air-filled $4\,\text{mm}$ diameter cylindrical, $85\%$ reflective polished aluminium light guide for a $5\,\text{GeV}$ electron bunch (100  $e^-$) impinging at (a) $90^\circ$ angle and (b) $45^\circ$ angle.}
\label{fig:g4visstraw}
\end{figure}

Figure~\ref{fig:g4visstraw} shows a visualisation of the simulated production of \cer photons in an air-filled $4\,\text{mm}$ diameter straw light guide using \geant. The light guide material assumed is polished aluminium with $85\%$ reflectivity. A bunch of 100 electrons with an energy of $5\,\GeV$ is impinging on the straw. The \cer photons are produced mostly collinearly to the electrons, as the \cer angle in air is $\theta_C\approx 1.4^\circ$. If the beam impinges perpendicularly on the straw, the light gets reflected diffusely (assuming an imperfectly polished surface), while in case of a $45^\circ$ angle between the electron beam and the straw, the light is reflected mostly towards the upper end of the straw.

 \begin{figure}[htbp!]
    \centering
      \includegraphics[width=0.75\textwidth]{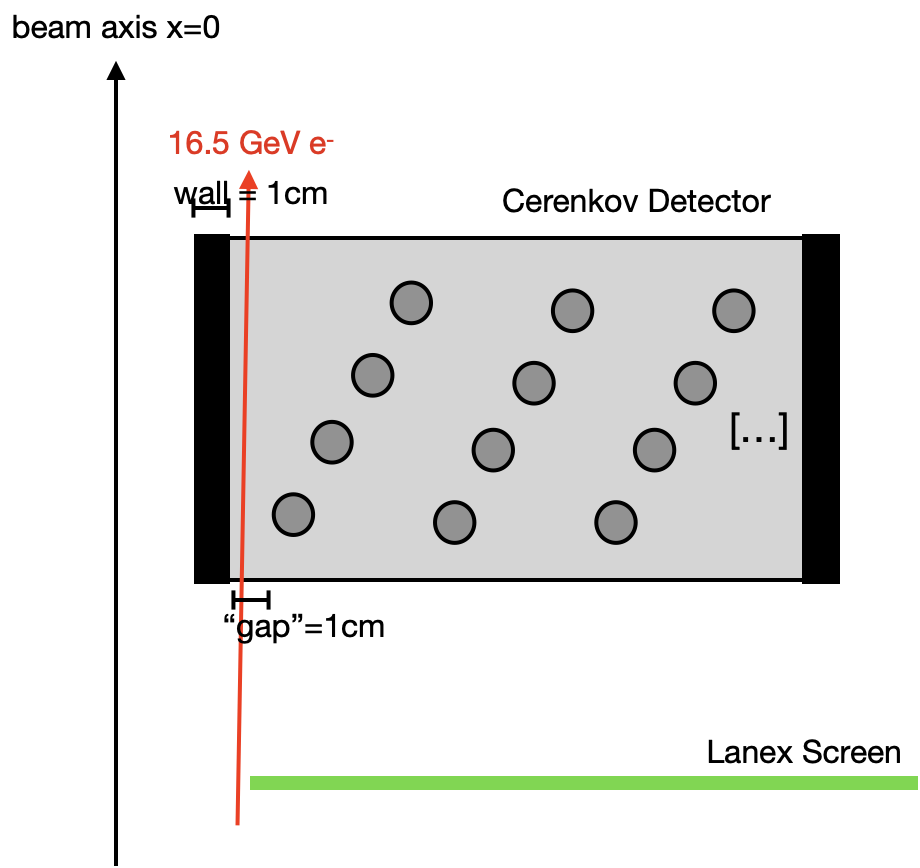}
      \caption{Sketch of the electron detection system: shown are the locations of the \cer detector and the scintillation screen with respect to the nominal beam axis. The path of beam electrons is also shown. The scintillation screen extends further to the right, away from the beam axis.}
      \label{fig:cere:positionbeam}
 \end{figure}

One important aspect for the positioning of the \cer detector is its placement with respect to the $16.5\units{GeV}$ electron beam trajectory. To reduce backgrounds from scattering of beam electrons in the \cer detector material, the amount of material crossed should be minimal. The main scattering was shown to originate from the aluminium side-wall of the gas-tight volume. To mitigate this, the \cer sensitive channels within the box should be placed according to Fig.~\ref{fig:cere:positionbeam}, leaving a gap between the wall of the gas-tight box and the first sensitive channels for the beam to pass through. The beam electrons then only pass the entrance and exit windows, avoiding the thicker outer wall of the box. The \cer detector will be placed on a moveable stage, which can be moved to align the system in all three spatial dimensions.

\section{Expected Performance}
\subsection{Simulation-Based Performance}
\label{perf:sim}

 For the analysis of the Compton edge shift according to equation \ref{eq:edgexi} the FIR filter method, described in Section \ref{subsec:scint_fir}, is used. In the following, only the spectra as reconstructed by the \cer detector are analysed to estimate the performance of the \cer detector. In the final LUXE analysis, a combination of the information from both the scintillator screen and the \cer detector is expected to yield the best result. 

 



\begin{figure}[htbp!]
\centering
\includegraphics[width=0.75\textwidth]{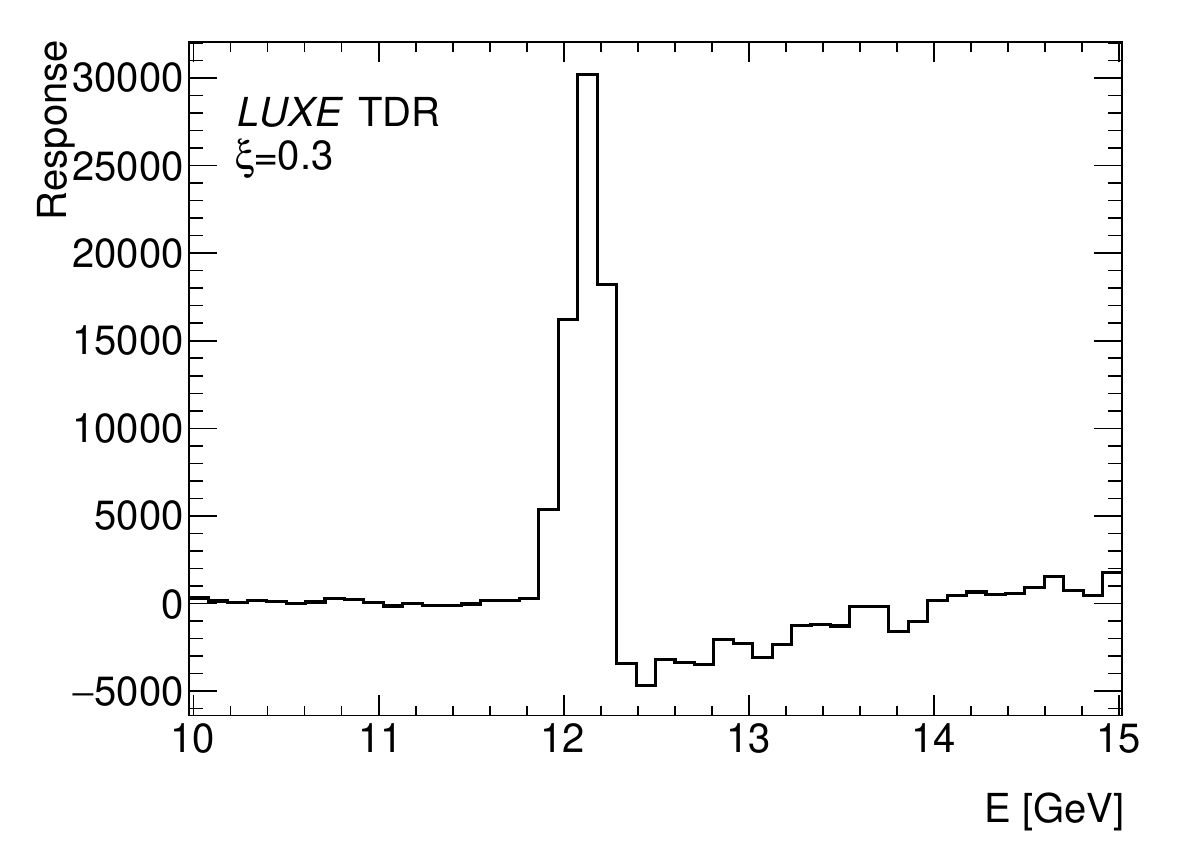}
\includegraphics[width=0.75\textwidth]{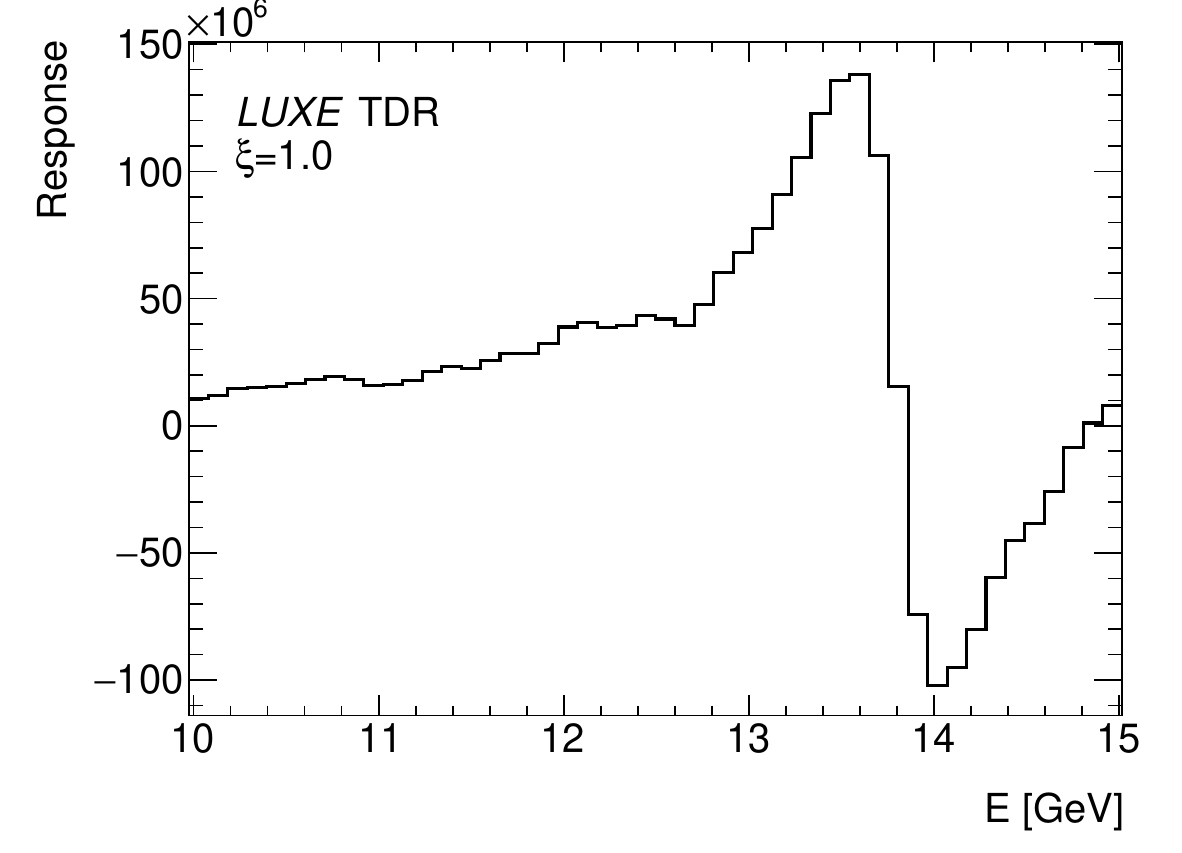} 
\caption{
Discrete response function for Compton edge finding for $\xinom=0.3$ (top) and $\xinom=1.0$ (bottom).} 
\label{fig:xiresponse} 
\end{figure}

Figure \ref{fig:xiresponse} shows the FIR distributions ($\mathbf{R_d(i)}$) corresponding to the Compton electron energy distributions ($\mathbf{g_d(i)}$) shown in fig.~\ref{fig:comptonspectra} for two different values of $\ximax$. The response is calculated for an FDOG filter with $\sigma=1$ and $N=10$. The correspondence between the edge location and the maximum of the response function is clearly visible. The choice of filter and parameters presented in this section showed acceptable performance and was not optimised. A more thorough optimisation study is expected to improve the performance of the filter method.

Due to the Gaussian shape of the laser pulse in both transverse and longitudinal coordinates  the spectra shown in fig.~\ref{fig:comptonspectra} correspond to a distribution of different true $\xi$-values with a maximum value of $\ximax$. Here, the position of the edge of the $\ximax$-value in the laser pulse is taken as a reference point. This position corresponds to the upper-energy limit of the edge, where the slope becomes negative due to the shape of the Compton edge. This position is referred to as the upper kink position. It corresponds to the first zero-crossing of the discrete response function after the response maximum.
 
\begin{figure}[htbp!]
\centering
\begin{subfigure}{0.75\hsize} 
\includegraphics[width=\textwidth]{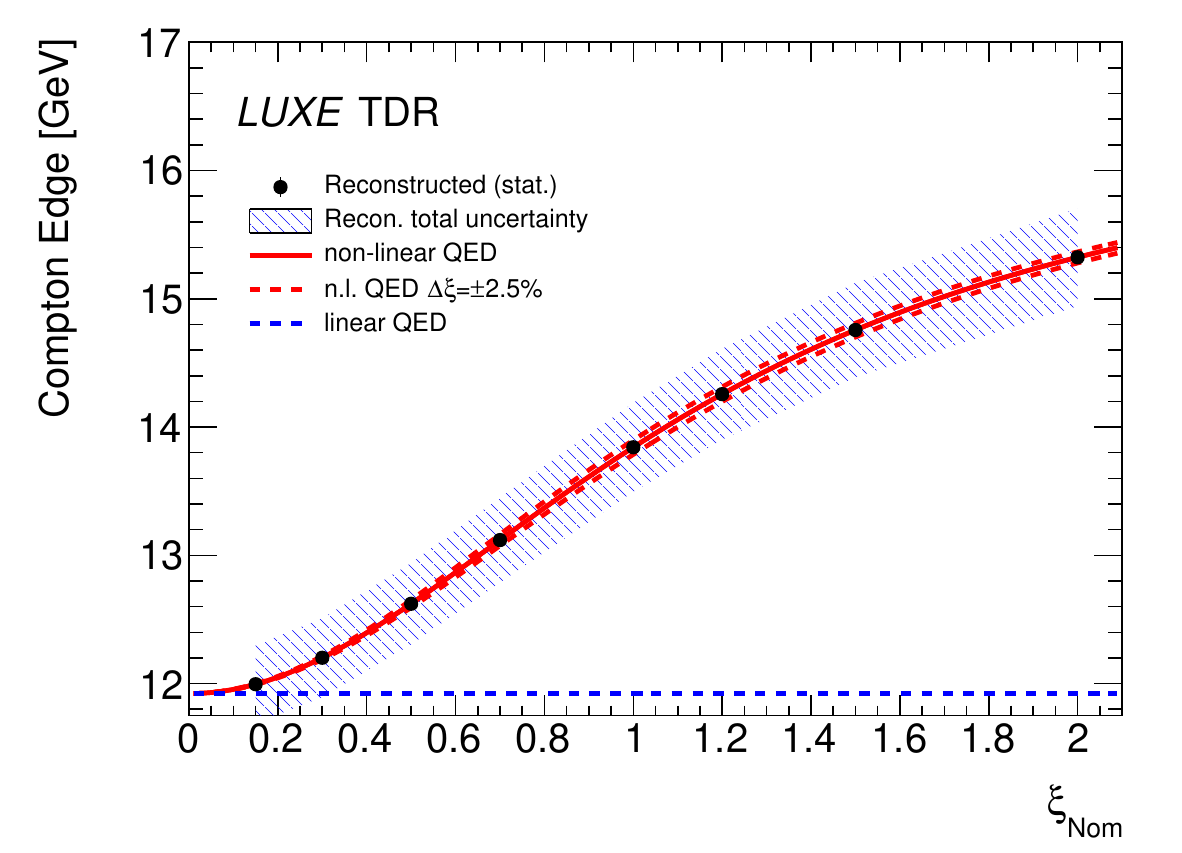} 

\caption{\label{fig:cerdet:comptresultedge}}
\end{subfigure}
\hspace{0.5cm}
\begin{subfigure}{0.75\textwidth} 
\includegraphics[width=\textwidth]{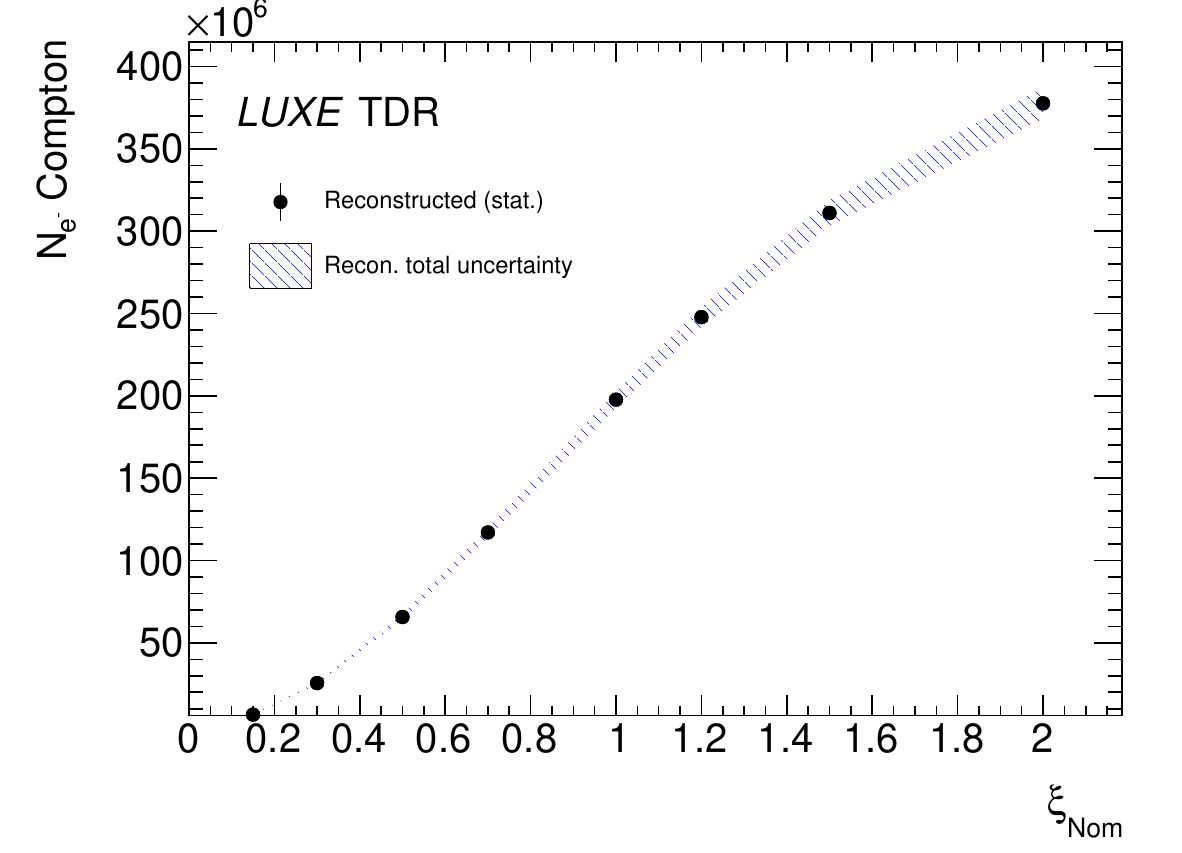}
\caption{\label{fig:cerdet:comptresultnorm}}
\end{subfigure}
\caption{a) Compton electron edge upper kink position from simulated analysis as a function of the nominal $\xi$ value. Red solid line: theoretical prediction with 2.5\% uncertainty on the laser intensity (dashed lines). Black points and error bars: reconstructed Compton Edge (shifted to theory prediction) with statistical uncertainty corresponding to 1\,h data-taking. Shaded blue area: total uncertainty on the reconstructed edge position, dominated by energy scale uncertainty of 2.5\%. b) Number of Compton electrons in fiducial volume $8.4<\varepsilon_e<15.5$~GeV as a function of $\xinom$. Black points and error bars: reconstructed number of electrons with statistical uncertainty for 1\,h of data-taking. Shaded blue area: total uncertainty, dominated by a systematic uncertainty on the response calibration of 2.5\%.}
\label{fig:ce} 
\end{figure}

In Fig.~\ref{fig:cerdet:comptresultedge}, the Compton edge upper kink position is shown as a function $\ximax$, compared to the theoretical prediction based on equation~\ref{eq:edgexi}. It can be seen that statistical uncertainties are negligible already after one hour of data-taking and the total uncertainty is dominated by a $2.5\%$ energy scale uncertainty. 

Figure~\ref{fig:cerdet:comptresultnorm} shows the total number of Compton electrons within the \cer detector acceptance as a function of $\ximax$. Since the energy spectrum gets skewed towards lower electron energies at high $\xi$, which are outside of the \cer acceptance, the total number of electrons reaches a maximum and then decreases. The fiducial region assumed in Fig.\ref{fig:cerdet:comptresultnorm} ($8.4<\varepsilon_e<15.5$~GeV) is slightly smaller than that expected for the final EDS detector geometry described in this note ($5<\varepsilon_e<16$~GeV), but it can be used to estimate an upper limit on the expected uncertainties.

It should be noted that the edge reconstruction performance estimate presented here is based on signal truth simulation with \ptarmigan and only contains a simple first-principle parametrisation of the detector acceptance as well as a simple estimate of the uncertainties via toy Monte Carlo and error propagation. To estimate the \cer detector performance in more detail, the same study should be repeated using the full \geant detector simulation, which includes the full detector geometry, multiple scattering, particles impinging on the channels at an angle and beam backgrounds. In this case a more complicated reconstruction algorithm is needed to recover the Compton spectrum from the individual channels.

\subsubsection{Multi-layer Straw Reconstruction}
\label{sec:mlstrawreco}

\begin{figure}
    \centering
    \includegraphics[width=0.75\textwidth]{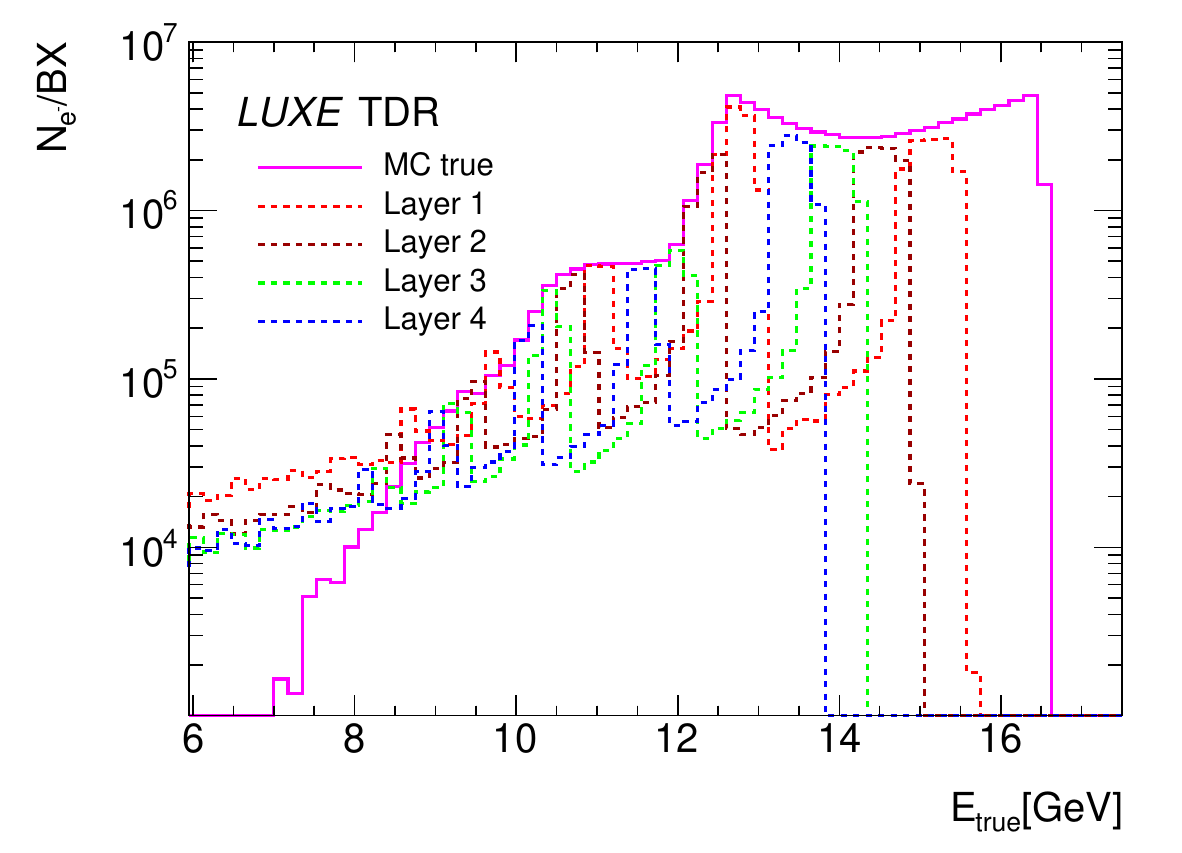}
    \caption{\geant signal and beam-induced backgrounds simulation of true energy spectra of electrons impinging on a grid of four staggered layers of straw tubes, overlaid with the true simulated spectrum (magenta).}
    \label{fig:staggered_straws}
\end{figure}

Figure \ref{fig:staggered_straws} shows the energy distributions of electrons impinging on the \cer detector with four layers of staggered straws, as shown in Fig.~\ref{fig:geooptlayouts} (a), simulated in \geant. The simulation contains both signal and beam-induced backgrounds. The structure of the straw layers is clearly visible. The long tails of the distributions towards low energies in the spectrum of each straw are contributed by beam-induced background electrons (see also fig.~\ref{fig:sigbg}). Furthermore, there is a slight energy overlap between adjacent straws, e.g.~due to the fact that the electron tracks are bent by the dipole magnet and therefore impinge on the \cer detector at an angle, crossing several straws at the same time.

A simple matrix-based technique is used to reconstruct the electron energy spectrum from the number of counts measured per straw channel, using a \geant simulation of the straw channel geometry in the dipole spectrometer. A transformation matrix $\mathbf{T}$ is derived, which transforms the vector of counts per straw $\mathbf{S}=\{s_0,\dots,s_{N_{\text{straw}}}\}$ into a binned energy histogram $\mathbf{E}=\{e_0,\dots,e_{N_{\text{bins}}}\}$:

\begin{equation}
    e_i=\sum^{N_{\text{straw}}}_{j=0} T_{ij}\cdot s_j .
\end{equation}

The matrix normalisation is chosen such that:

\begin{equation}
\frac{1}{s_j}\sum^{N_{\text{bins}}}_{i=0} T_{ij} = 1.
\end{equation}

The transformation matrix is derived using a \geant simulation of the channel geometry with a particle source producing a flat electron energy spectrum, in order to have an equal number of events for each electron energy. For several layers of straws, it can happen that for a single incident electron, multiple straws are hit, which must be considered in the reconstruction. Furthermore, since there are several layers of straws, a reference plane is identified, for which the electron energy spectrum is reconstructed. This reference plane is called the \textit{virtual detector plane}. Straw hits are then matched to true electron hits on the virtual detector plane by extrapolating the trajectories inside the straws back to the virtual detector plane and matching to the closest electron hit. It should be noted that, due to scattering that can occur upstream of the virtual detector plane, the electrons hitting the virtual detector plane are not equivalent to the primary electrons. This correspondence has to be taken into account by a further simulation-based correction factor. The transformation matrix then relates the energies of the electron virtual front plane hits to the straw hits. Multi-straw hits for a virtual front plane hit $i$ are weighted according to the path length $l_{ij}$ travelled through straw $j$, divided by the sum of all path lengths travelled through all straws matched to the virtual front plane hit $i$:

\begin{equation}
    w_{ij}=\frac{l_{ij}}{\sum^{N_{\text{match}}}_{k=0} l_{ik}}.
\end{equation}

\begin{figure}[htbp!]
    \centering
    \includegraphics[width=0.75\textwidth]{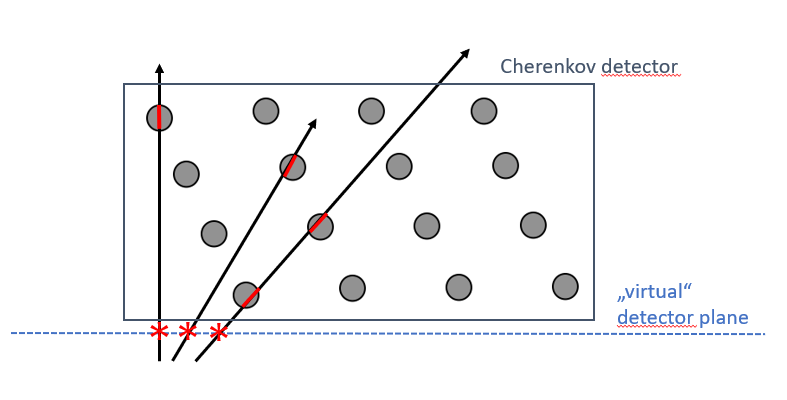}
    \caption{Electrons impinging on the virtual detector plane and the \cer channel geometry.}
    \label{fig:cherreco}
\end{figure}

Figure \ref{fig:cherreco} illustrates the incidence of electrons on the \cer detector channel geometry and shows the location of the virtual front plane hits, as well as multiple straw passes per incident electron.

The transformation matrix, by construction only contains entries for successfully matched frontplane hits and straws. In order to correct for acceptance deficits, where there is a frontplane hit, but no straw hits (e.g. electrons passing through gaps in the straw geometries), an additional acceptance correction is derived, by applying the derived transformation matrix onto the vector of straw hits and comparing the obtained energy spectrum with the front plane hits energy spectrum. The efficiency then is defined as:

\begin{equation}
    \epsilon(E)= \frac{N_{e,\mathrm{reco}}(E)}{N_{e,\mathrm{true}}(E)}.
\end{equation}

To reconstruct the signal spectrum in simulation, the transformation matrix derived in the flat energy sample is applied to the vector of counts per straw obtained from the signal sample. Furthermore the acceptance correction factor is applied. For real data-taking, the reconstruction method is the same as described above. The transformation matrix derived from the flat energy simulation, as well as the efficiency correction are applied to the measured vector of counts per straw ID. On top of this a chain of calibration and correction factors is applied, to correct for differences between the simulation and the real detector performance.

\subsubsection{Geometry Optimisation}

\begin{figure}[htbp!]
    \centering
    \includegraphics[width=0.5\textwidth]{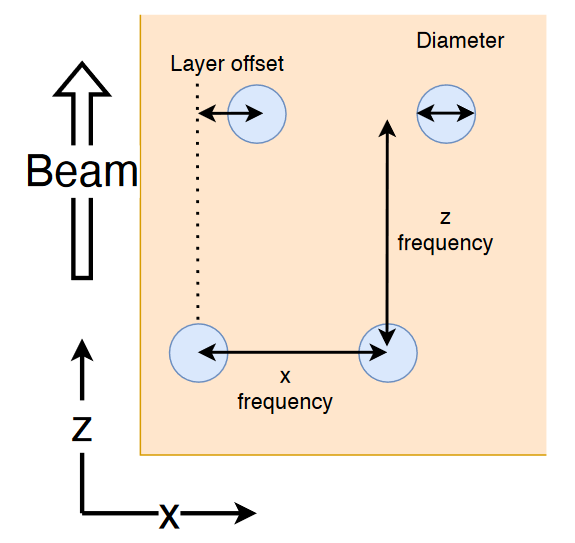}
    \caption{Relevant parameters for the \cer channel geometry optimisation}
    \label{fig:geooptpar}
\end{figure}

An optimisation of the straw tube channel geometries was performed based on the \geant simulation. Figure \ref{fig:geooptpar} show the relevant straw geometries parameters that were adjusted during the optimisation. The parameters are denoted as follows: straw diameter, $d$, layer offset, $\Delta x_{\mathrm{offs}}$, straw frequency in $x$, $f_x$, straw frequency in $z$, $f_z$. \\

The figure of merit of the geometry optimisation is the reconstruction efficiency, minimising the probability that an incoming electron misses a straw channel entirely. Another optimisation criterion was to retain an energy resolution below $2\%$. All optimisations are constrained mainly by cost-effectiveness of the detector. This mainly poses a limit on the total number of channels, and hence photodetectors, which are the main cost drivers.

\begin{figure}[htbp!]
    \centering
    \begin{subfigure}{0.49\textwidth} 
    \begin{overpic}[width=\textwidth]{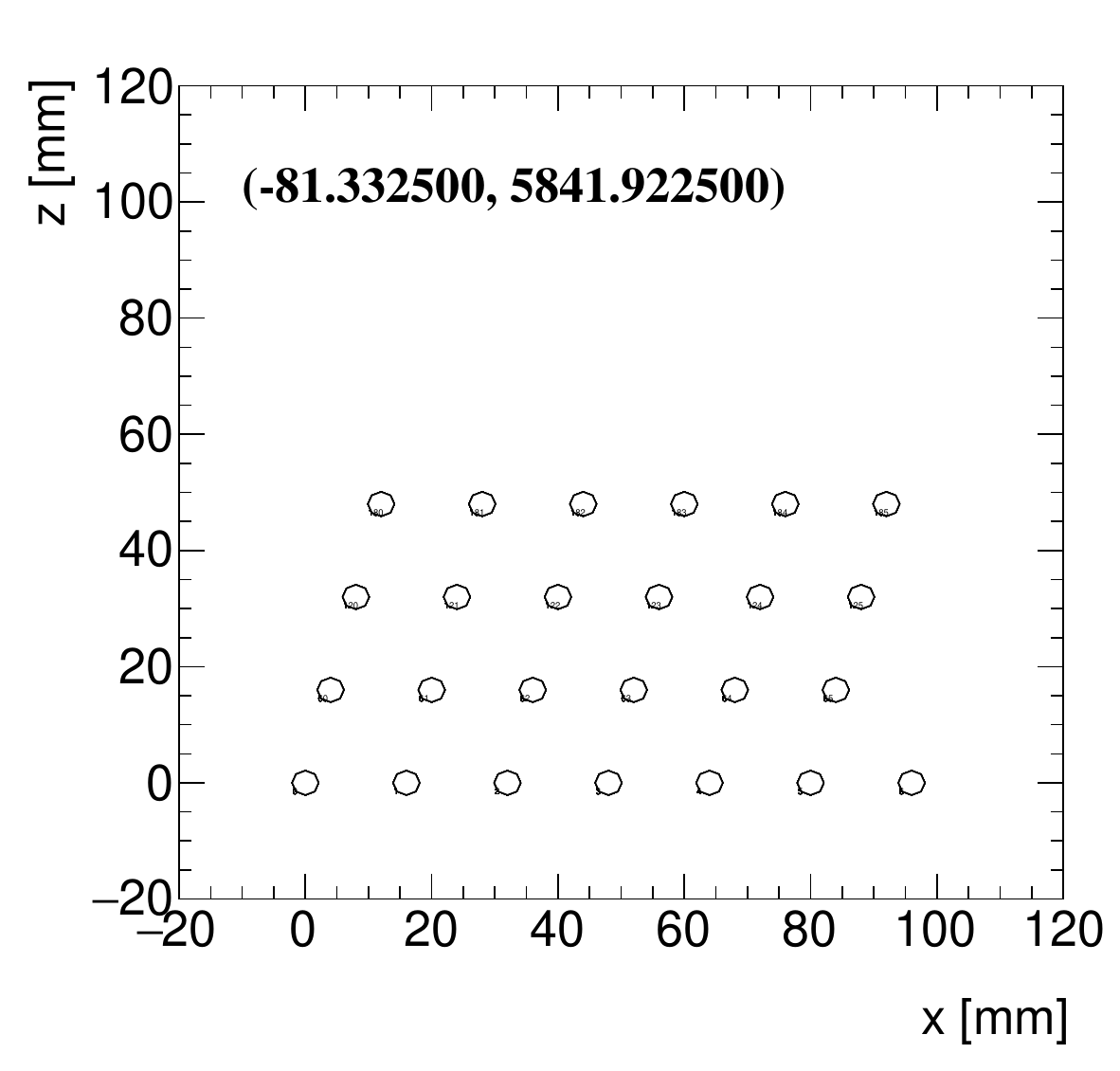}
        \put (25,70) {\fontfamily{phv}\selectfont LUXE TDR}
    \end{overpic}
    \caption{ }
    \end{subfigure}
    \begin{subfigure}{0.49\textwidth} 
     \begin{overpic}[width=\textwidth]{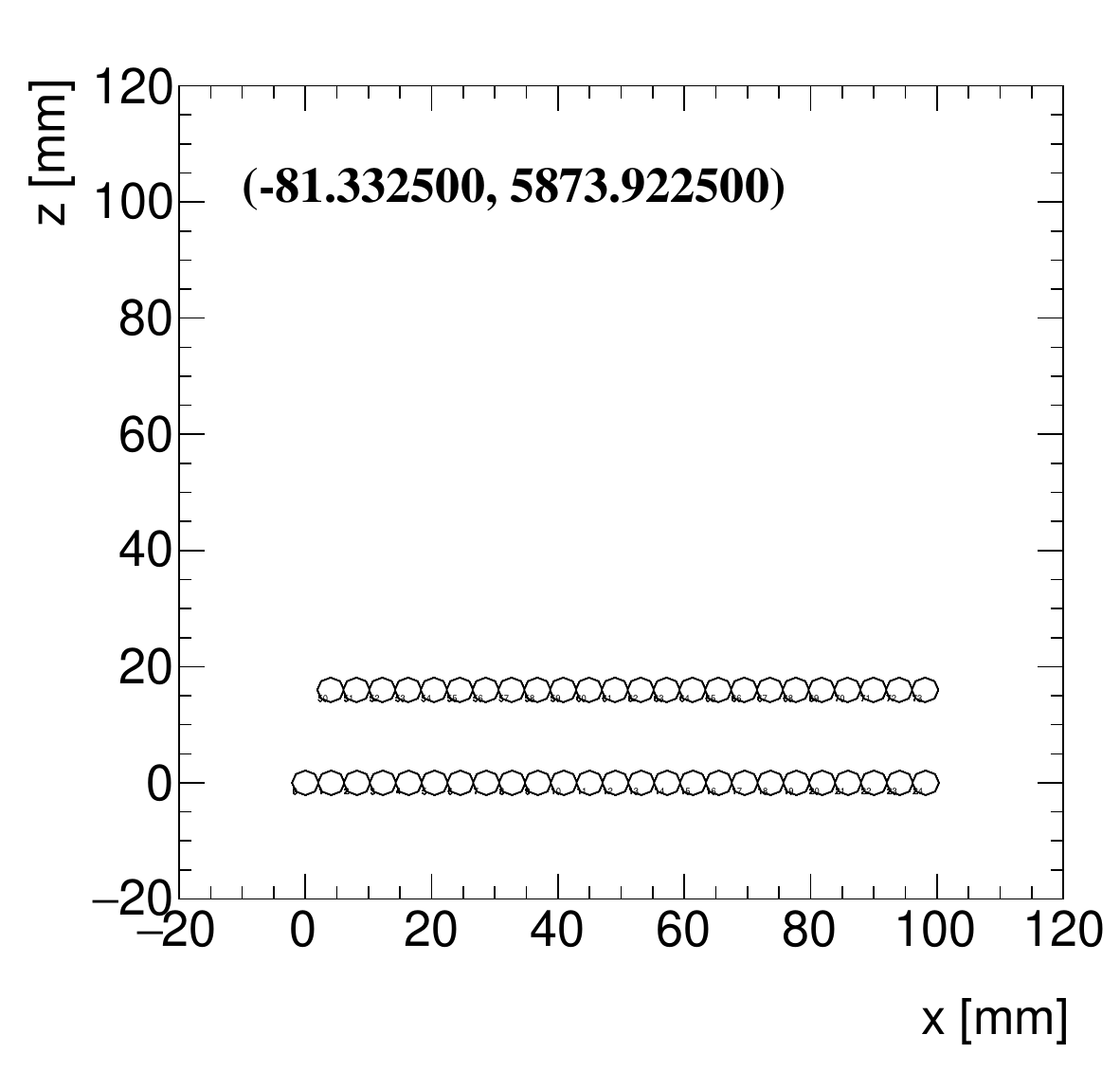}
         \put (25,70) {\fontfamily{phv}\selectfont LUXE TDR}
    \end{overpic}
    \caption{ }
    \end{subfigure}\\
   
    \caption{Geometry for a) the default design, b) the optimized 2-layer design (2L). }
     \label{fig:geooptlayouts}   
\end{figure}

In the following, two different \cer detector channel layouts will be discussed, namely the default geometry ($d=4\,\mathrm{mm}$, $\Delta x_{\mathrm{offs}}=4\,\mathrm{mm}$, $f_x=20\,\mathrm{mm}$, $f_z=20\,\mathrm{mm}$), as well as an optimised denser geometry with fewer channel layers ($d=4\,\mathrm{mm}$, $\Delta x_{\mathrm{offs}}=4\,\mathrm{mm}$, $f_x=4\,\mathrm{mm}$, $f_z=20\,\mathrm{mm}$), denoted "2L". The two channel layouts are shown in Fig. \ref{fig:geooptlayouts}.

\begin{figure}[htbp!]
    \centering
    \begin{subfigure}{0.725\textwidth} 
    \begin{overpic}[width=\textwidth]{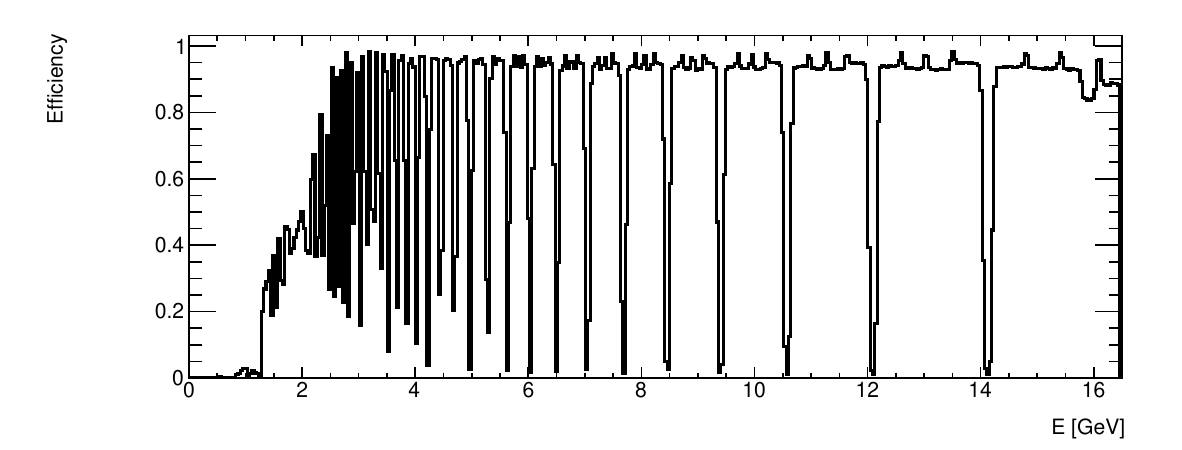}
        \put (20,30) {\fontfamily{phv}\selectfont \scriptsize LUXE}
        \put (20,26) {\fontfamily{phv}\selectfont \scriptsize TDR}
    \end{overpic}
    \caption{ }
    \end{subfigure}\\
    \begin{subfigure}{0.725\textwidth} 
    \begin{overpic}[width=\textwidth]{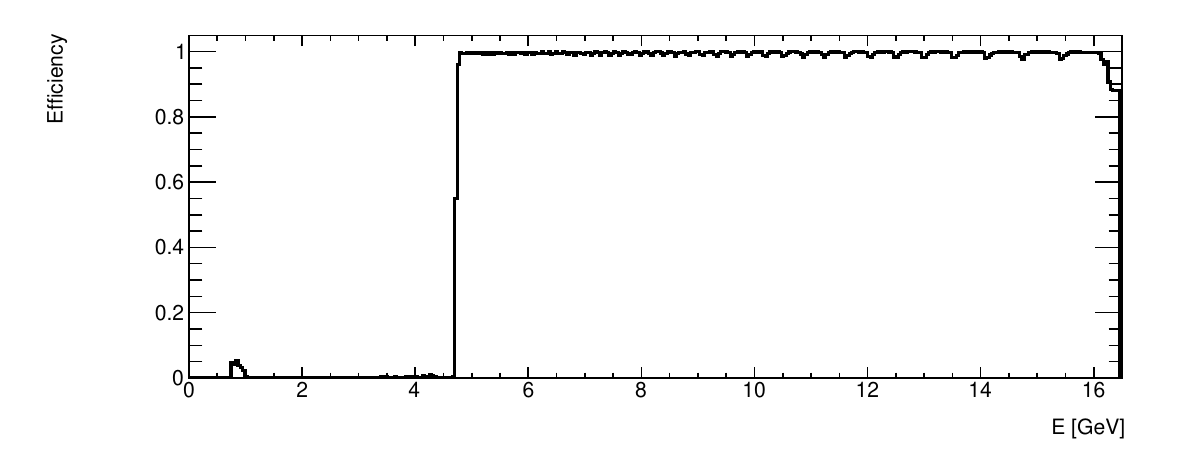}
        \put (20,30) {\fontfamily{phv}\selectfont \scriptsize LUXE TDR}
    \end{overpic}
    \caption{ }
    \end{subfigure}\\
        \begin{subfigure}{0.725\textwidth} 
     \begin{overpic}[width=\textwidth]{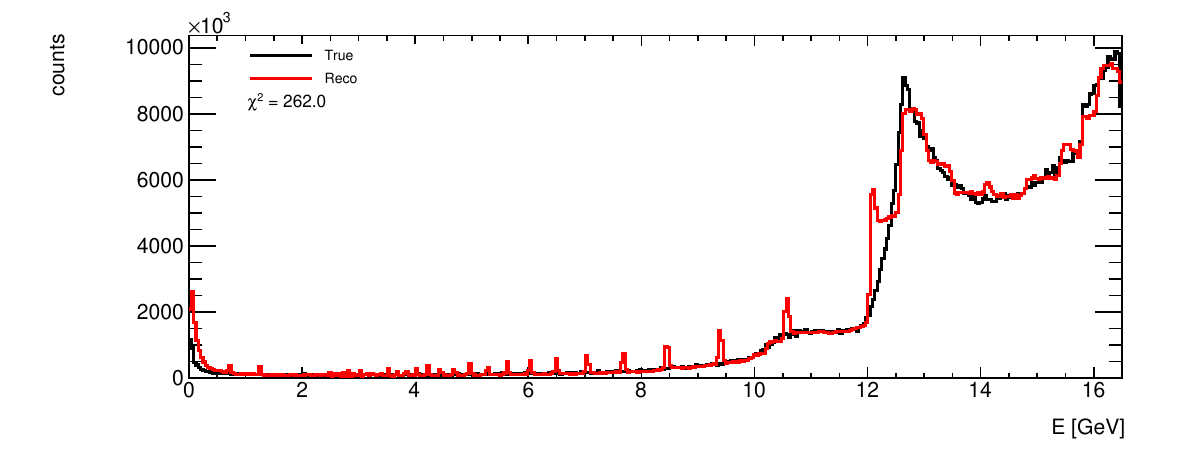}
        \put (20,25) {\fontfamily{phv}\selectfont \scriptsize LUXE TDR}
    \end{overpic}
    \caption{ }
    \end{subfigure}\\
    \begin{subfigure}{0.725\textwidth} 
    \begin{overpic}[width=\textwidth]{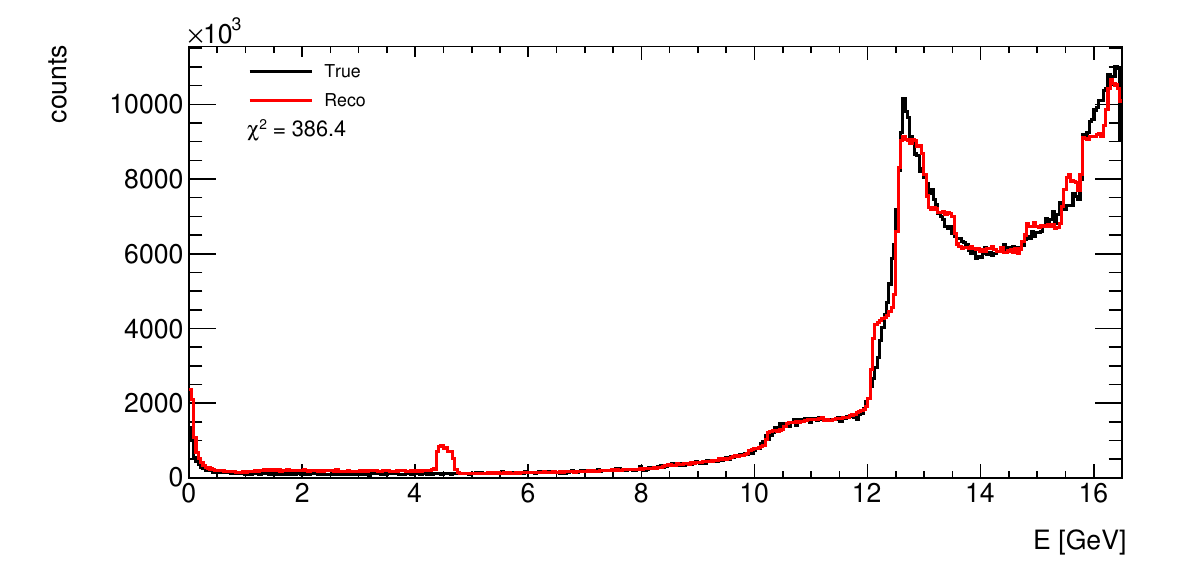}
        \put (20,33) {\fontfamily{phv}\selectfont \scriptsize LUXE TDR}
    \end{overpic}
    \caption{ }
    \end{subfigure}\\
    \caption{Results of the \cer channel geometry optimisation. Reconstruction efficiency for a) default, b) 2L as well as reconstructed signal spectrum for c) default, d) 2L. }
     \label{fig:geooptresult}   
\end{figure}

Figure \ref{fig:geooptresult} shows the reconstruction efficiency as well as the reconstructed signal spectrum for the default layout and the "2L"-layout. In the default geometry, a periodic structure of low-efficiency is visible at certain values of the energy. This is due to the significant gaps between the straw channels which leads to electrons at certain incidence angle missing all straw layers. This is alleviated in the 2L design with layers being closer together and redundancy of the two overlapping straw layers. The improvement in efficiency is reflected in the reconstructed energy spectrum, where low detection efficiency areas cause spikes in the reconstructed spectrum. These spikes no longer occur in the 2L design. The difference between the reconstructed and true electron spectrum at high energies is due to the limited resolution resulting from the straw diameter and is expected. Because of the linear behaviour of the relative energy resolution in a dipole spectrometer (see eq. \ref{eq:eresosegm}), the resolution effect of the nonzero straw diameter is dominant at high energies.

It should be noted that the channel geometry optimization discussed in this section mainly focuses on improving the detection efficiency. In order to estimate the dependence of the systematic uncertainties of the energy measurement on  the channel layout, and improving its robustness, more detailed studies are required.

\subsection{Prototype Test at DESY II Testbeam}
\label{subsec:testbeam}

To test the feasibility of the \cer straw tube detector, a proof-of-principle prototype was constructed and tested at the DESY II test beam facility \cite{DESYTESTBEAM} in November 2021.

\begin{figure}[htbp!]
    \centering
    \begin{subfigure}{0.75\textwidth} 
    \includegraphics[width=\textwidth]{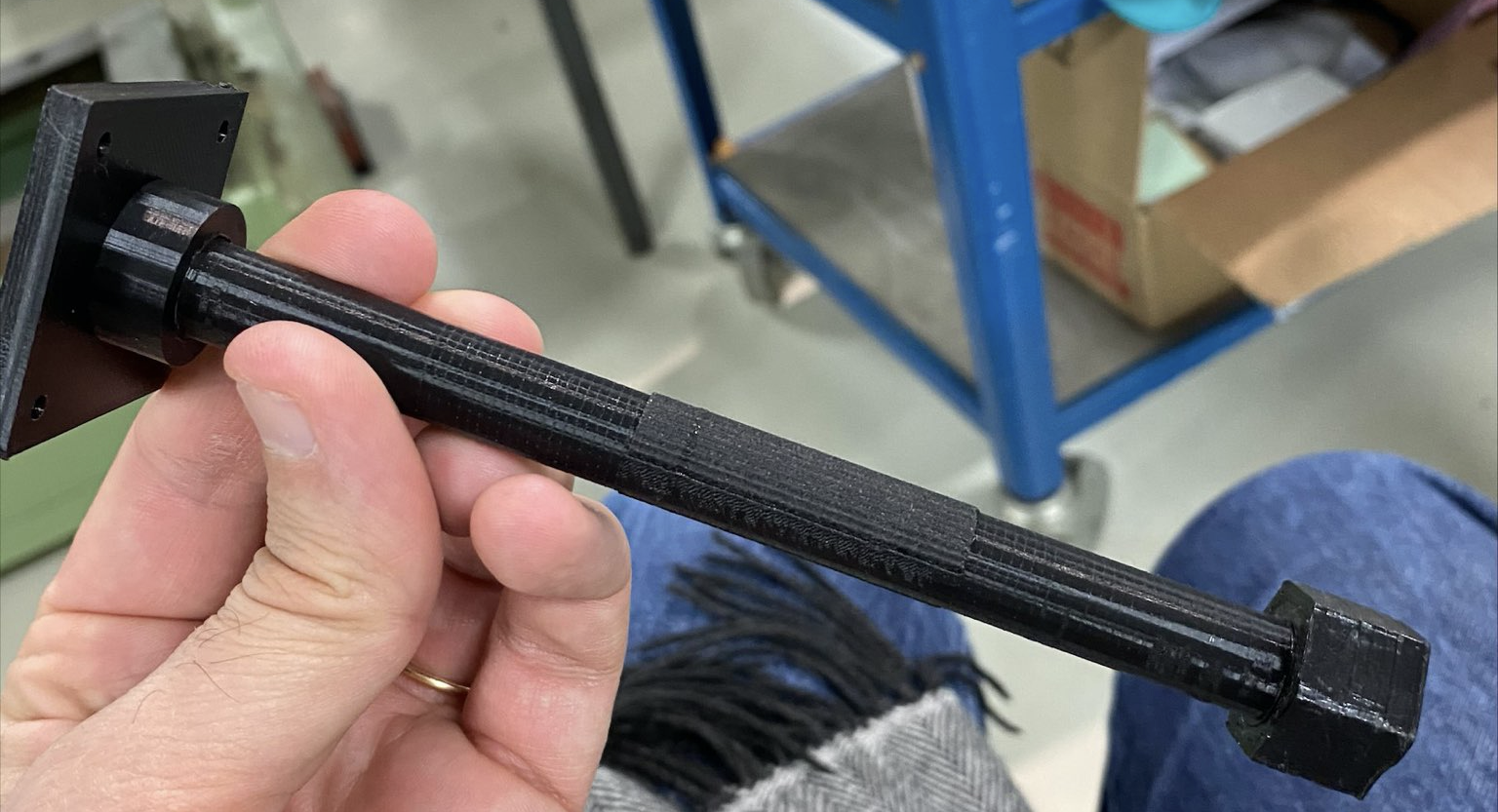}
    \caption{ }
    \end{subfigure}
    \begin{subfigure}{0.75\textwidth} 
    \includegraphics[width=\textwidth]{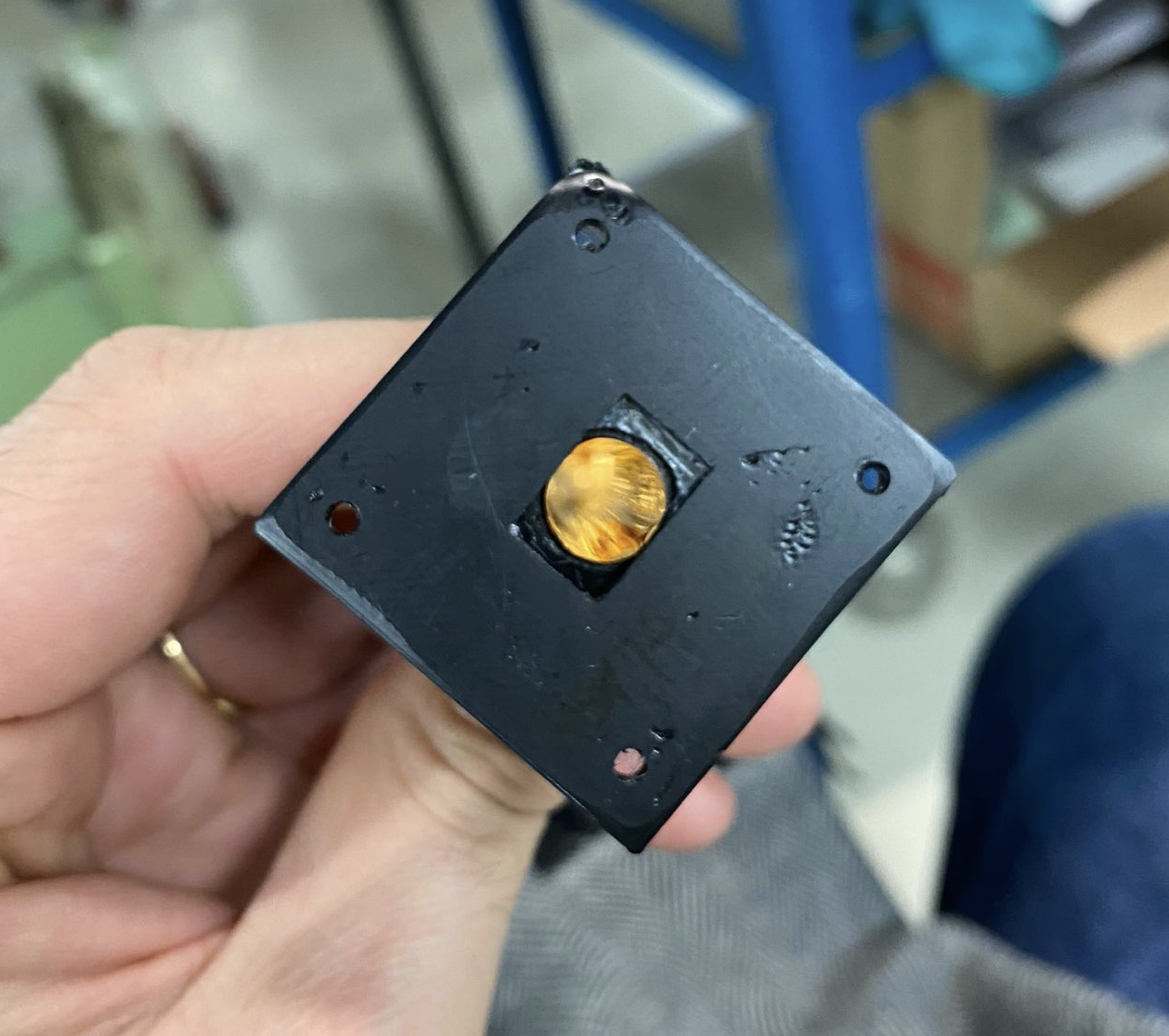}
    \caption{ }
    \end{subfigure}
    \caption{a) Side view of 3D-printed straw container holding the straw and paraffin oil. b) View into the straw container with NA62 straw as light guide. }
     \label{fig:strawprototype}   
\end{figure}

The prototype consists of a single straw tube, attached to a silicon photomultiplier (SiPM) using a 3D-printed mounting tube holding the reflective straw, attached to an end-cap with an exit window and a flange to connect to the SiPM. A photograph of the single-straw prototype is shown in Fig.~\ref{fig:strawprototype}.

Since the DESY II testbeam produces mostly single electrons between 2 and 6 GeV at a low rate (kHz) compared to LUXE, it was decided to use heavy paraffin oil with a high refractive index ($n=1.47$) as active medium in order to produce enough \cer light to generate a signal in the SiPM.

\begin{figure}[htbp!]
    \centering
    \begin{subfigure}{0.75\textwidth} 
    \includegraphics[width=\textwidth]{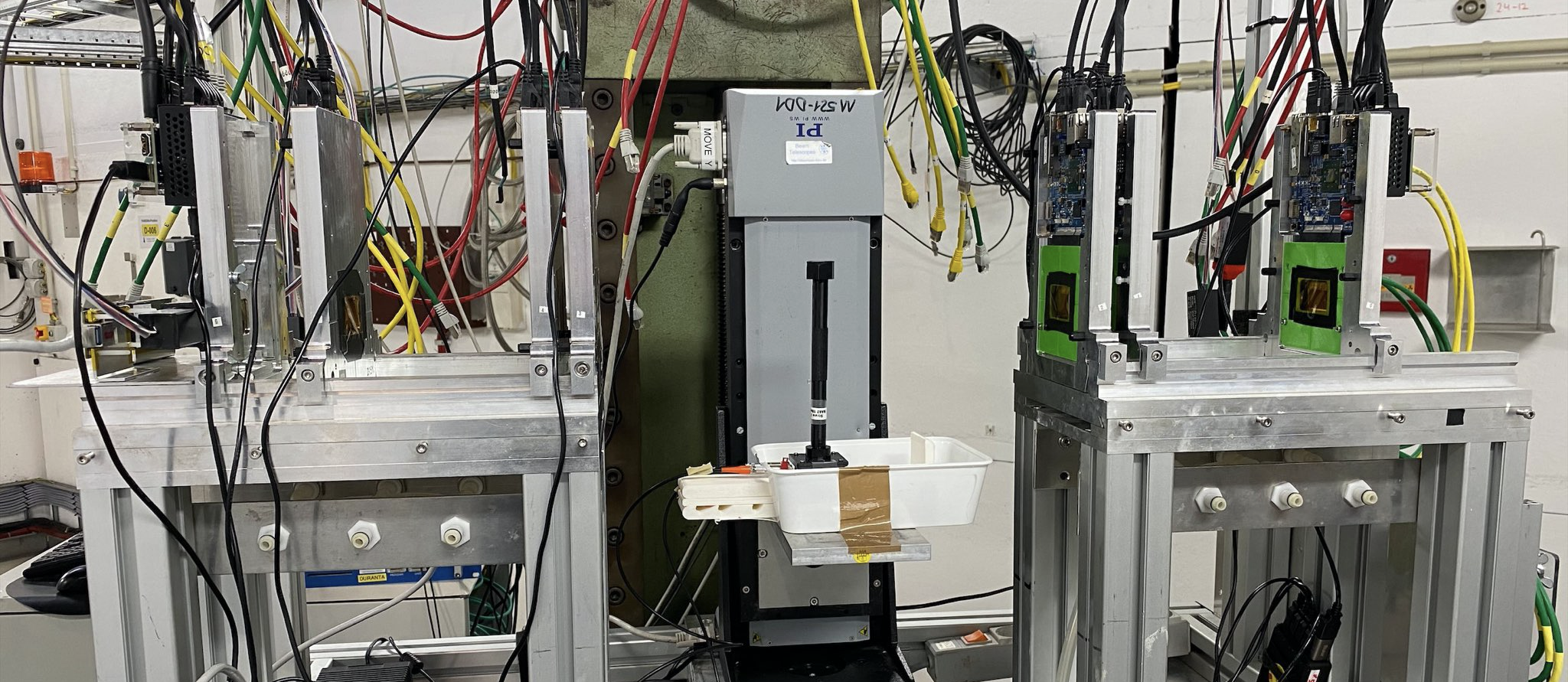}
    \caption{ }
    \end{subfigure}
    \begin{subfigure}{0.75\textwidth} 
        \includegraphics[width=\textwidth]{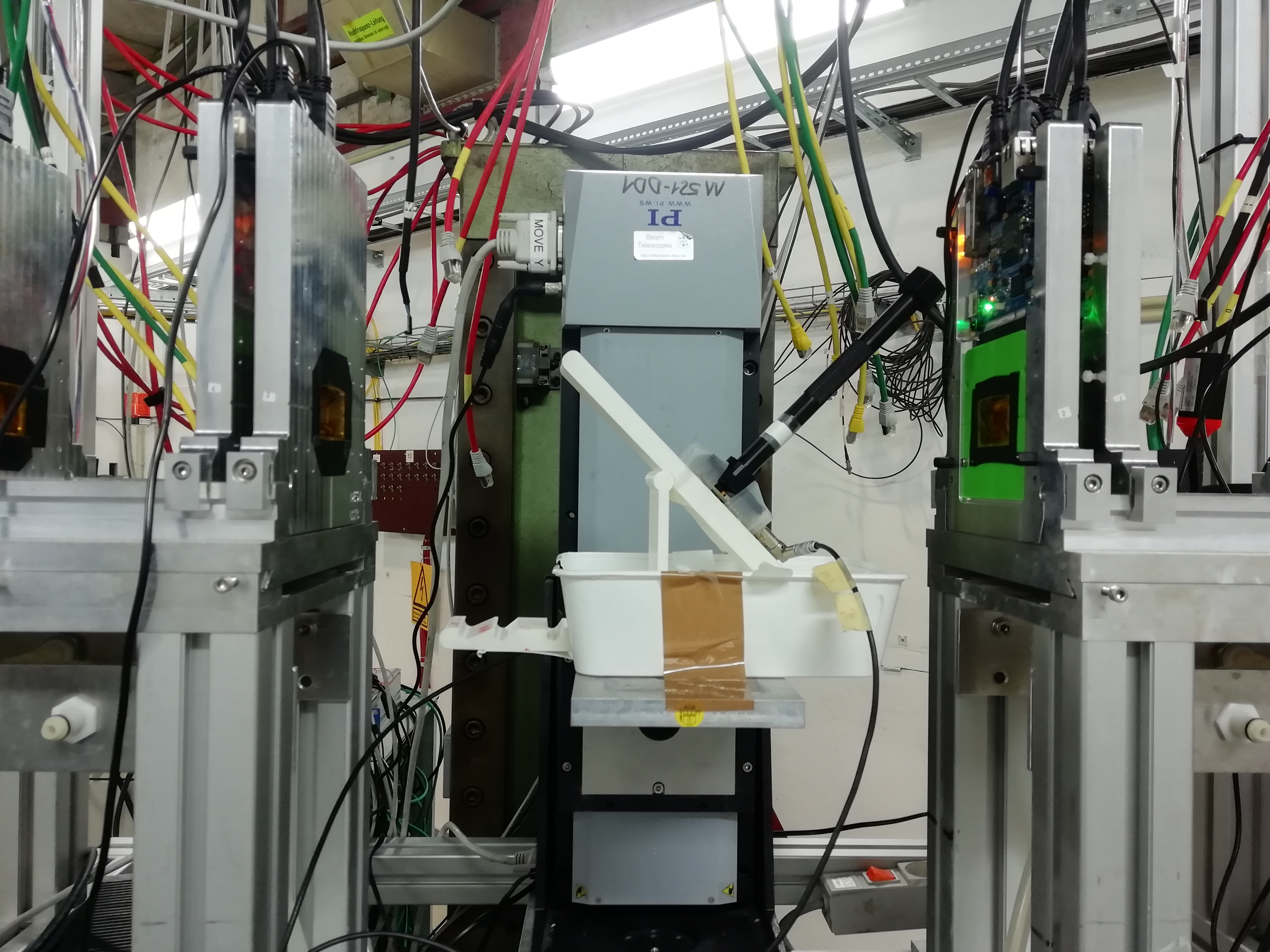}
         \caption{ }
    \end{subfigure}
    \caption{Beam test setup with single straw prototype at DESY II. a) Straw at $0^\circ$, b) Straw at $45^\circ$.}
    \label{fig:tb_setup}
\end{figure}

In order to increase the electron path length in the active medium, the straw tube was mounted at a $45^\circ$ angle with respect to the incoming particle beam. Figure~\ref{fig:tb_setup} shows a photograph of the straw tube in the beam line.

The data acquisition consists of a CAEN DT5702 SiPM readout board that is triggered externally via two scintillator fingers at the end of the setup. To obtain precision tracking information on the incoming electrons the straw tube is placed in the middle of six layers of a beam telescope. 

\begin{figure}[htbp!]
    \centering
    \begin{overpic}[width=\textwidth]{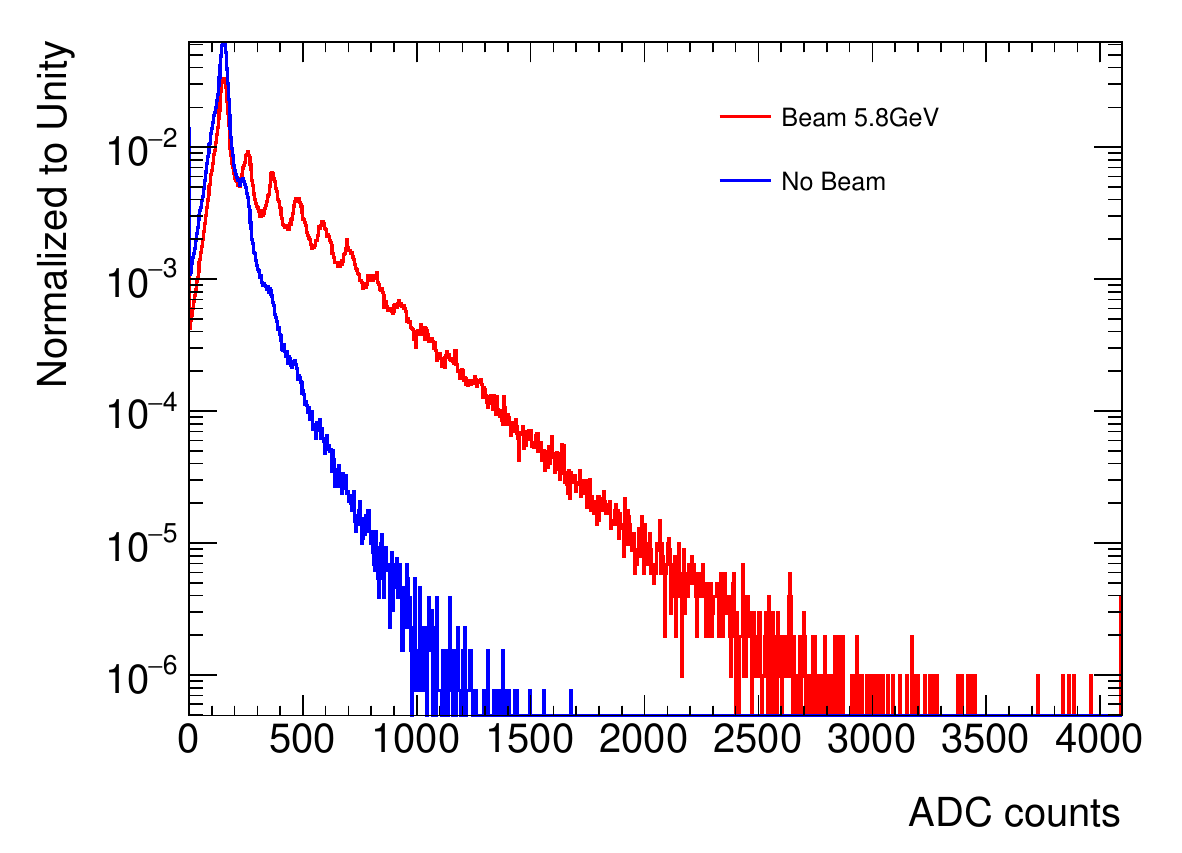}
        \put (30,60) {\fontfamily{phv}\selectfont \large LUXE TDR}
    \end{overpic}
    \caption{Photodetector pulse height spectrum for 10\,mm NA62 Straw and SiPM prototype filled with paraffin oil, with a $5.8\,\GeV$ electron beam (red) traversing the prototype and without beam (blue). }
    \label{fig:tb_sig_noise}
\end{figure}

Figure~\ref{fig:tb_sig_noise} shows the measured SiPM signal pulse height spectrum in ADC counts, which corresponds to the measured charge. The red curve corresponds to the \cer signal in the presence of $5.8\,\GeV$ electrons passing through the straw. The blue curve shows the "No Beam" spectrum when the shutter is closed and no beam electrons traverse the straw. The signal from electrons initiating \cer radiation is clearly visible. The peak structure is characteristic for the SiPM and corresponds to $N$ number of  pads firing in the presence of $N$  photons. The first maximum corresponds to the noise pedestal. The smaller higher-order maxima in the "No Beam" spectrum originate from cross-talk, where secondary electrons from an avalanche created in one SiPM pad initiate another avalanche in a neighboring pad. The "No Beam" spectrum in Fig.~\ref{fig:tb_sig_noise} shows that cross-talk plays a minor role for the SiPM operation settings used.

The number of \cer photons produced by a primary electron is given by the Frank-Tamm formula (eq.~\ref{eq:franktamm}), which, assuming highly relativistic electrons traversing a medium of constant refractive index $n$ and length $l_z$, can be expressed as:

\begin{equation}
    \frac{dN}{d\lambda}=\frac{2\pi\alpha}{\lambda^2}\left(1-\frac{1}{n^2}\right)l_z.
\end{equation}

It is visible that the total amount of \cer light created per primary electron follows a $(1-1/n^2)$ characteristic, where $n$ is the refractive index of the \cer medium, also it is proportional to the length travelled in the active medium, which is related to the straw diameter, depending on where the electrons traverse the cylindrical straw.

One of the main goals for the design of the \cer detectors is to ensure that the signal measured by the photodetector remains within the dynamic range of the readout electronics at high rates. The total charge signal from the photodetector depends on several factors related to the straw and photodetector: 

\begin{itemize}
    \item \textbf{Light guide efficiency:} The inner surface of the straw reflects the \cer light several times to guide it towards the photodetector. Depending on the reflectivity of the straw wall, more or less light is absorbed or diffused during the reflection process. The light pattern on the photodetector is also influenced by the round straw geometry and the electrons incidence angle, which makes it difficult to simulate with high accuracy. The effective reflectivity can be determined by comparing testbeam data to an optical simulation. Another contribution to the light guide efficiency would be the efficiency of an optical filter (e.g.\ a neutral density filter) used to lower the \cer light levels such that they are within the dynamic range of the chosen photodetector. The optical filter efficiency can be wavelength dependent, even for a neutral density filter. In the low-rate testbeam at DESY II, no additional optical filter was used, so the light guide efficiency purely depends on the straw reflectivity.
    
    \item \textbf{Photon detection efficiency (PDE):} Every photodetector has a limited efficiency to detect incoming photons. This efficiency typically depends on the operating voltage, temperature and the photon wavelength. For the SiPM used, the maximal PDE is $50\%$ at the peak wavelength of $420\,\text{nm}$.
    
    \item \textbf{Gain:} A SiPM has a built-in charge amplification factor. For the SiPM used, the maximal gain is $6.3\times10^6$. The gain varies as a function of the SiPM operating voltage and temperature. 
    
\end{itemize}

At the DESY II test beam, the main goal is to demonstrate the basic relationships predicted by the Frank-Tamm formula, such as the dependence on different active media (variation in refractive index $n$), straw diameters (variation in $l_z$ and light pattern) and different straw materials (variation in light guide efficiency). It should be noted that already with a single fixed-diameter straw the $l_z$ dependence is relevant, because the particle beam is much wider than the straw diameter. As a result, the beam electrons traverse the straw not only in the centre of the cylinder, but also at the sides, where they travel a shorter distance in the active medium.

\begin{figure}[htbp!]
    \centering
    \begin{overpic}[width=\textwidth]{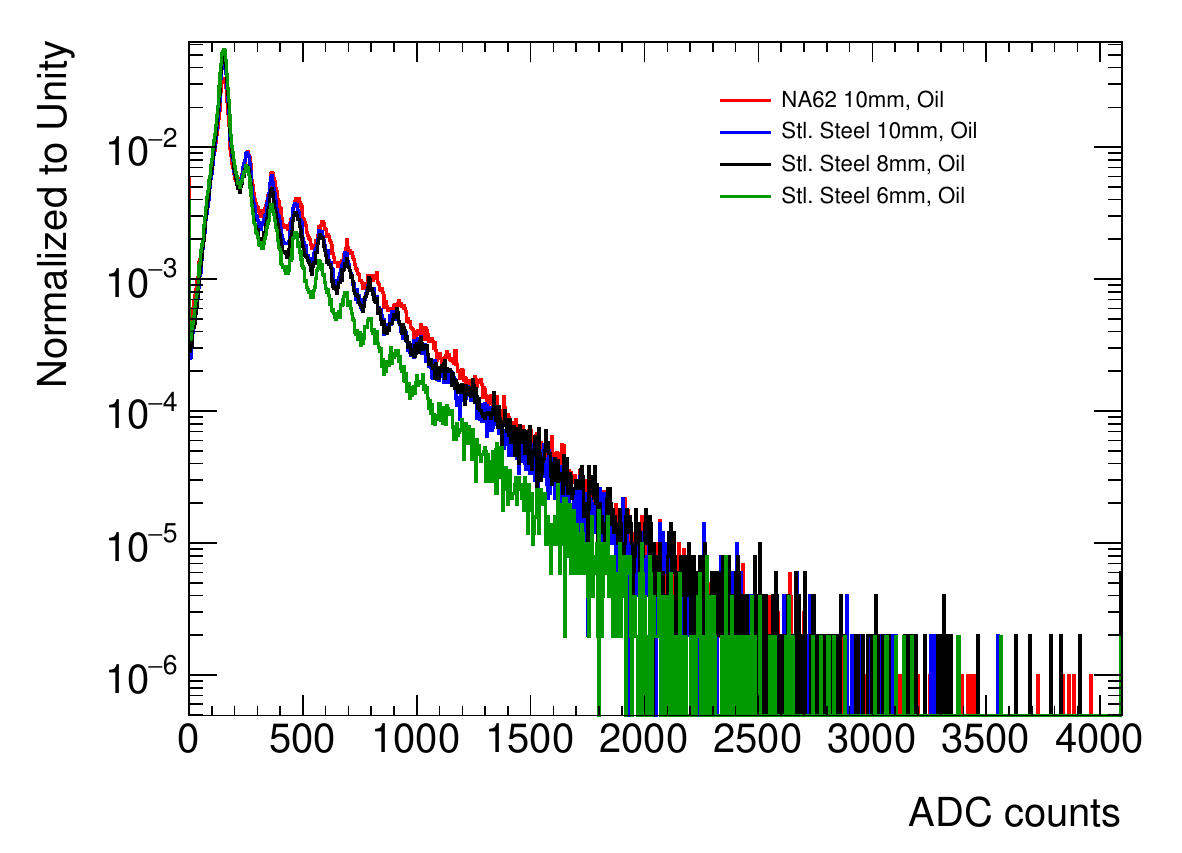}
        \put (30,60) {\fontfamily{phv}\selectfont \large LUXE TDR}
    \end{overpic}
    \caption{Photodetector Pulse height spectrum for different straw diameters and materials and SiPM prototype filled with paraffin oil, with a $5.8\,\GeV$ electron beam. }
    \label{fig:tb_strawcomp}
\end{figure}

Figure \ref{fig:tb_strawcomp} shows the pulse height spectra for a set of straws with different materials (NA62: Au+Cu coated kapton, stainless steel), and straw diameters ($10\,\text{mm}$, $8\,\text{mm}$, $6\,\text{mm}$). It is visible that the difference in reflectivity between the coated kapton and stainless steel is small. With smaller diameter, the measured light intensity decreases due to the shorter path length of the electrons in the \cer medium. 

\begin{figure}[htbp!]
    \centering
    \begin{overpic}[width=\textwidth]{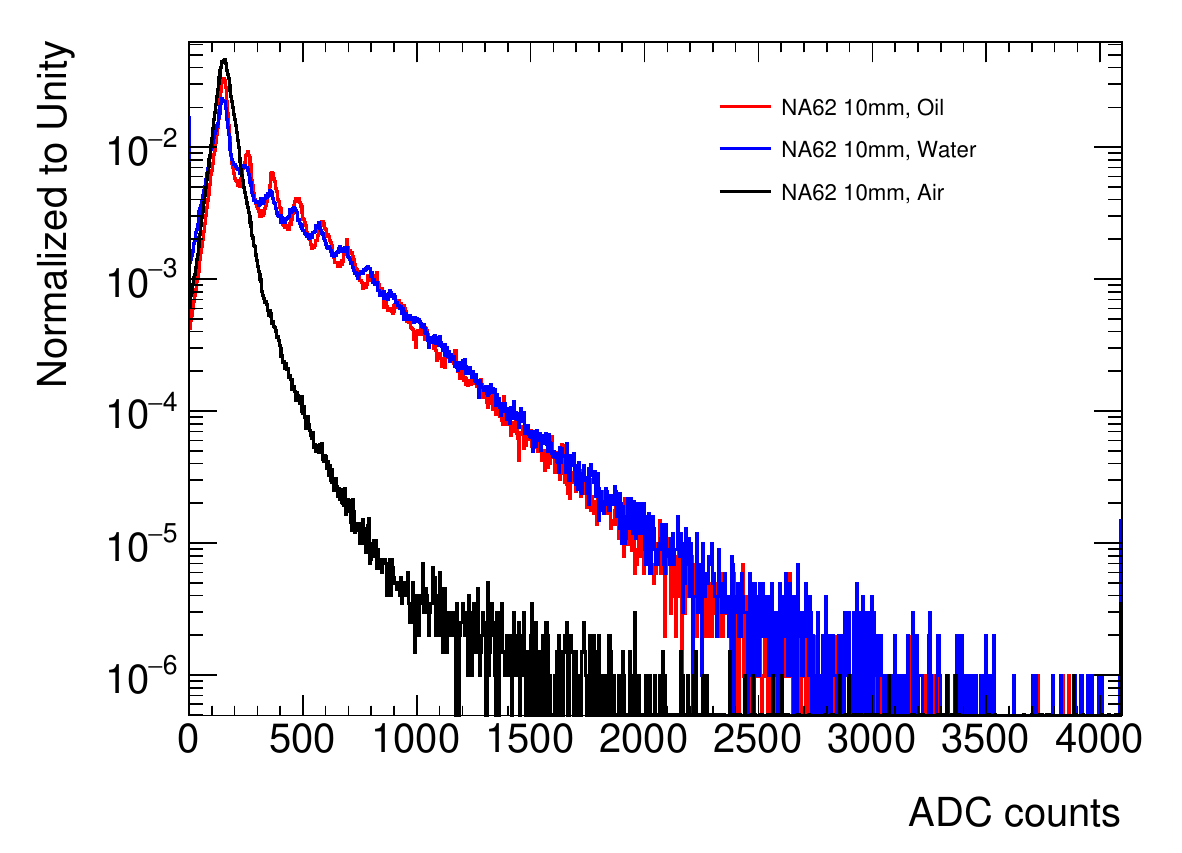}
        \put (30,60) {\fontfamily{phv}\selectfont \large LUXE TDR}
    \end{overpic}
    \caption{Photodetector pulse height spectrum for 10\,mm NA62 straw and SiPM prototype filled with paraffin oil (red), water (blue) and air(black)  with a $5.8\,\GeV$ electron beam.}
    \label{fig:tb_mediumcomp}
\end{figure}

Figure \ref{fig:tb_mediumcomp} shows a comparison of the same straw (10\,mm NA62 straw) filled with different active media, namely paraffin oil ($n=1.47$), water ($n=1.33$) and air ($n=1.00028$). The straw with water shows a similar characteristic as the straw with oil. One source of difference between oil and water is the additional creation of fluorescence photons in oil, while for water only \cer radiation is produced. In air, due to the low refractive index and the resulting low number of produced \cer photons per primary, no conclusive difference between the spectrum with and without beam could be determined for single electrons. In the LUXE environment with up to $10^8$ primary electrons impinging on a straw, the low refractive index of air will help to keep the \cer light yield low enough to avoid saturation of the photodetector and readout electronics. 

\subsection{Prototype Test at LEAP Laser Plasma Accelerator Facility}
\label{subsec:testbeamleap}

In February 2022 a test campaign using the single straw prototype described in Section~\ref{subsec:testbeam} was performed at the LEAP laser plasma accelerator experiment facility at DESY Hamburg. The LEAP accelerator supplies a bunched electron beam with an electron energy around $60\,$MeV, with a bunch charge varying around $6\,\text{pC}$. This corresponds to an electron rate of the order of $10^7e^-/$shot, which lies within the rate regime expected for LUXE.

\begin{figure}[htbp!]
    \centering
    \includegraphics[width=0.5\textwidth]{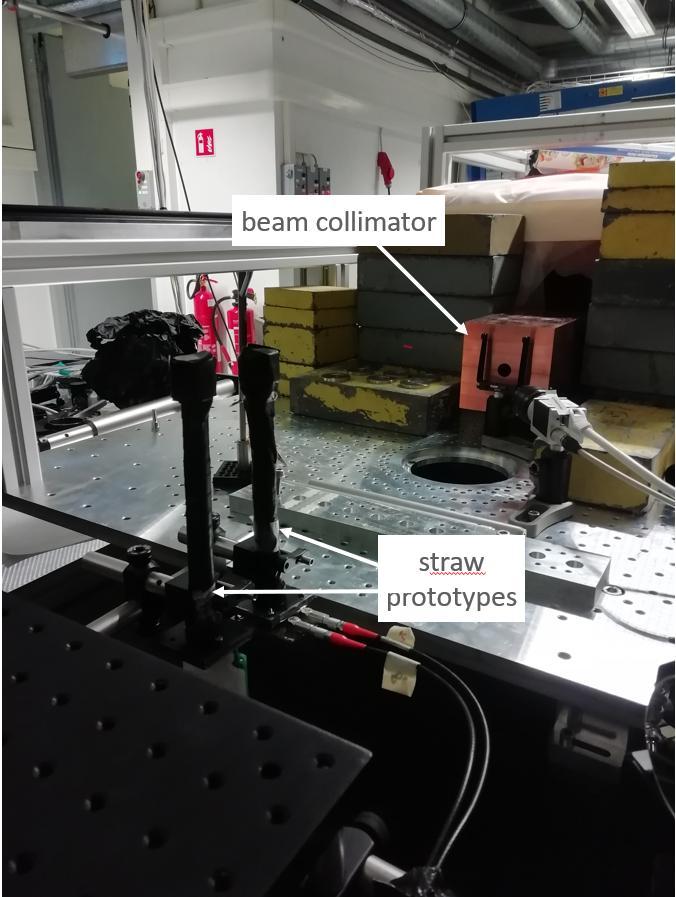}
    \caption{Straw prototypes in the LEAP laser plasma accelerator experiment setup. The $60\,$MeV electron beam emerges from the collimator (bronze) and traverses the two straw prototypes mounted one after the other.}
    \label{fig:leaptbsetup}
\end{figure}

The experimental setup at LEAP is shown in Fig.~\ref{fig:leaptbsetup}. Two single-straw prototypes are mounted consecutively in the beamline. Both are filled with air as active medium. The first straw is read out by a SiPM, the second by an APD. The SiPM-straw uses a reflector from the NA62 experiment, while the APD straw uses a reflective wrapping foil from the CALICE HGCAL scintillator tiles. Both the APD and SiPM are read out simultaneously by the DT5702 board. The readout circuit is the same stand-alone circuit as was used in the DESY II testbeam campaign, however the trigger signal to the DT5702 is supplied by the laser trigger driving the laser plasma accelerator. This trigger signal can be delayed arbitrarily with respect to the arrival of the electron beam and the delay is optimised such that the APD and SiPM signal arrive $50\,$ns after the trigger signal. It should be noted that the electron beam from the LEAP accelerator impinging on the straw is several cm wide, such that not the entire bunch charge is injected into the prototypes.

\begin{figure}[htbp!]
    \centering
    \begin{overpic}[width=\textwidth]{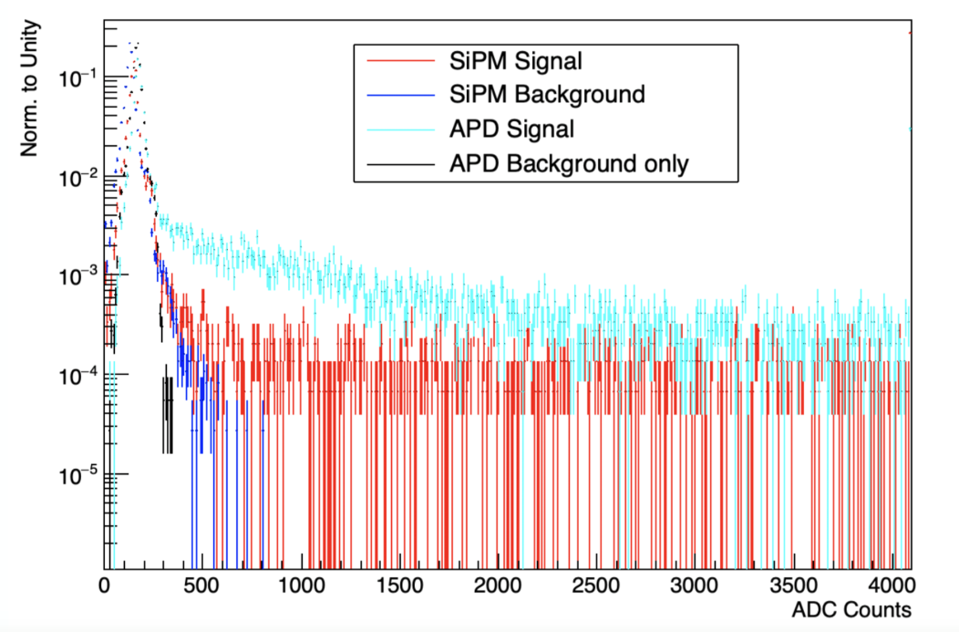}
        \put (20,55) {\fontfamily{phv}\selectfont \large LUXE TDR}
    \end{overpic}
    \caption{Photodetector pulse height spectrum for APD and SiPM air-filled straw prototypes at LEAP facility. Red: SiPM straw with electron beam, Blue: SiPM straw without beam (background), Cyan: APD straw with beam,  Black: APD straw without beam (background).}
    \label{fig:leapspectrum}
\end{figure}

Fig.~\ref{fig:leapspectrum} shows the photodetector pulse height spectrum for the SiPM and APD readout with and without the electron beam. A clear signal can be observed for the data with beam, both with the APD and SiPM, establishing the functionality of the air-filled \cer straw tubes in the LUXE electron rate regime. Furthermore, it is visible that the SiPM signal frequently saturates the ADC (entries in overflow bin 4096), while the APD signal remains within the dynamic range of the readout ADC. It was observed that lowering the preamplifier gain for the SiPM signal channel reduced the frequency of the saturating events. Based on the variations performed in this test, it is deemed possible to keep the signal within the readout dynamic range using the methods described in Section~\ref{subsec:dynrange}. A more in-depth analysis of the dataset collected at the LEAP facility is ongoing. To study the detector response in quantitative detail, a high-statistics test beam with a variable, precisely controllable bunch charge is required. Further test beam campaigns using electron bunches with bunch charges in the LUXE regime are in planning, with ongoing investigation, which facilities fulfil the beam requirements mentioned above.

\section{Technical Description}
\subsection{Detector Dimensions}

The dimensions of the \cer detector are driven by the covered energy range $[E_{\text{min}},E_{\text{max}}]$ and the dipole spectrometer parameters, namely the magnetic field strength $B$, the length of the magnetic field $z_m$ and the distance between the magnet and the detector. The approximate functional behaviour of the energy and the displacement is described by equation \ref{eq:dipoleeofx}.

The segmentation $\Delta x$ determines the energy resoluton, as discussed in section \ref{sec:requirements}. 

The parameters of the EDS system are summarised in table \ref{tab:dimensions}, the parameters of the IBM system in table \ref{tab:dimensions_ibm}. The parameters listed in tables \ref{tab:dimensions} and \ref{tab:dimensions_ibm} are determined such that they match the requirements discussed in section \ref{sec:requirements}, as well as comply with technical boundary conditions, given by e.g.~sizes of commercially available photodetectors and straw tubes, while remaining at acceptable cost.

\begin{table}[h!]
    \centering
    \begin{tabular}{|c|c|c|}
    \hline
    \textbf{Parameter [Unit]} & \textbf{Value} & \textbf{comment}  \\\hline
     $E_{\text{min}}\,[\text{GeV}]$ & 5 & minimal covered energy \\\hline $E_{\text{max}}\,[\text{GeV}]$ & 16 & maximal covered energy \\\hline
     $\Delta x\,[\text{m}]$ & 0.003 & channel segmentation \\\hline
     $N_{\text{ch}}$ & 100 ($\times2$) & number of channels (per photodetector) \\\hline
     $\sigma_0$ & 0.0014 & resolution parameter \\\hline
    $x_{\text{min}}\,[\text{m}]$ & 0.137 & minimal channel position \\\hline $x_{\text{max}}\,[\text{m}]$ & 0.44 & maximal channel position \\\hline
    $x_{\text{beam}}\,[\text{m}]$ & 0.132 & beam position \\\hline
    box size $ x\times y \times z [\text{mm}^3]$ & $540\times520\times260$ & box dimension \\\hline
    \end{tabular}
    \caption{Geometrical parameters of the EDS \cer detector.}
    \label{tab:dimensions}
\end{table}

\begin{table}[h!]
    \centering
    \begin{tabular}{|c|c|c|}
    \hline
    \textbf{Parameter [Unit]} & \textbf{Value} & \textbf{comment}  \\\hline
     $E_{\text{min}}\,[\text{GeV}]$ & 1.2 & minimal covered energy \\\hline $E_{\text{max}}\,[\text{GeV}]$ & 16 & maximal covered energy \\\hline
     $\Delta x\,[\text{m}]$ & 0.01 & channel segmentation \\\hline
     $N_{\text{ch}}$ & 50 & number of channels \\\hline
     $\sigma_0$ & 0.016 & resolution parameter \\\hline
    $x_{\text{min}}\,[\text{m}]$ & 0.025 & minimal channel position \\\hline $x_{\text{max}}\,[\text{m}]$ & 0.55 & maximal channel position \\\hline
    $x_{\text{beam}}\,[\text{m}]$ & 0.023 & beam position \\\hline
    box size $ x\times y \times z [\text{mm}^3]$ & $600\times520\times260$ & box dimension \\\hline
    \end{tabular}
    \caption{Geometrical parameters of the IBM \cer detector.}
    \label{tab:dimensions_ibm}
\end{table}

\subsection{\cer Light Detection}

The \cer light created in the active medium is guided by the straw light guides onto a photodetector placed at the upper end of the tube.

\subsubsection{Photodetector Choice}
\label{subsubsec:pdchoice}

Currently two types of photodetectors are foreseen for use in the LUXE \cer detector, namely SiPMs and avalanche photodiodes (APDs). Table \ref{tab:photodetectors} compares the parameters of the APD and SiPM candidates for the LUXE \cer detectors, namely the ON Semiconductor MicroFJ-30035 SiPM \cite{ONSemiJSeriesSiPM} and Hamamatsu S5344 Si APD \cite{Hamamatsu5344APD}.

\begin{table}[h!]
    \centering
    \begin{tabular}{|c|c|c|c|}
    \hline
    \textbf{Parameter [Unit]} & \textbf{\makecell{SiPM \\ (MicroFJ-30035)}} & \textbf{\makecell{APD \\ (Hamamatsu S5344)}} & \textbf{comment}  \\\hline
     $U_{\text{max}}\,[\text{V}]$ & 30.7 & 200 & max. operation voltage \\\hline 
         gain & $6.3\times10^6$ & 50 & gain \\\hline 
       $[\lambda_{\text{min}},\lambda_{\text{max}}]\, [\text{nm}]$ & 200 to 900  & 200 to 1000 & spectral range \\\hline 
       $\lambda_{\text{peak}}\, [\text{nm}]$  & 420 & 620 & peak sens. wavelength \\\hline 
   photosensitive area  & $3\times3\,\text{mm}^2$ & $\diameter:\: 3\,\text{mm}$ &  \\\hline 
      number of pads  & 5676 & 1 &  \\\hline
    \end{tabular}
    \caption{LUXE \cer photodetector candidates parameters.}
    \label{tab:photodetectors}
\end{table}

The advantage of the SiPMs is their relatively low operating voltage and their lower peak sensitive wavelength, compared to the APD. For the high-rate application in the LUXE \cer detectors, the advantage of APDs is their much lower gain factor. Furthermore, the APD has a continuous response to the impinging light level. For high light levels,  this is an advantage, since the maximal number of photons that can be detected simultaneously by a SiPM is limited by its number of pads. If a significant fraction of pads is fired at the same time, the SiPM signal saturates. This leads to a limited dynamic range of the SiPM compared to the APD. \\

Based on these considerations, the APD is the preferred choice for the high-electron rate environment of LUXE. However, since the electron rates per channel in the Compton spectrum strongly vary as function of energy and $\xi$, SiPMs is a preferrable choice in the regime with moderate rate. A detector design incorporating both photodetector technologies is explained in more detail in section \ref{subsec:dynrange}. \\

Ongoing are studies using a \geant simulation of the creation of \cer photons in air and their reflection in the straw light guides. Using this simulation one can predict the number of optical photons impinging on the photodetectors, as function of their wavelength $\lambda$ for every $\xi$, and thereby study in more detail the range of $\xi$ in which the SiPM readout or APD readout is applied.\\

\subsubsection{Photodetector Mounting}

The photodetectors are mounted on a PCB in an array such that they can be flanged directly onto the straw light guides. The array is designed such that there are four layers of straws, in order to increase the $x-y$-distance between the detectors on the PCB (see Fig.~\ref{fig:tech_cadtop}). The straws and photodetector are staggered such that they are adjacent in $x$. The photodetector PCB should contain an RC-terminated power line for the supply voltage, as well as shielded lines for the signal output. It is connected via a flatband bus cable to the frontend board (see Section \ref{subsec:frontend}).

\subsubsection{Dynamic Range}
\label{subsec:dynrange}

The \cer detector in LUXE will be exposed to a wide range of particle rates from $10^3-10^8$ electrons per channel and event. To ensure that the photodetector and read-out electronics will not saturate at the highest rates it is necessary to limit the amount of \cer light created by the electrons. In the LUXE \cer detector design the light-level reduction through design choices is accomplished in several ways:

\begin{itemize}
    \item{\textbf{Active medium:} The number of \cer photons per primary electron, $N_\gamma$, scales as a function of the refractive index of the active medium as $1-1/n^2$. Choosing as active medium a gas with a very low refractive index (e.g.\ air with $n=1.00028$) reduces $N_\gamma$ by a factor of 5 compared to $C_4F_{10}$ with $n=1.0014$. An additional advantage of air is its slightly higher \cer threshold at $20 \units{MeV}$ which reduces the low-energy background.} 
    \item{\textbf{Active medium length:} $N_\gamma$ scales linearly with the length $l_z$ of the active medium traversed by the primary electron. The small diameter of the straw ensures a small $l_z$. } 
    \item{\textbf{Foil reflectivity and straw length:}
    Within the straw tube, the \cer photons are reflected several times on the straw walls. Since the straw wall reflectivity factor is smaller than one, the light level at the photodetector will be reduced proportionally to the number of reflections on the straw walls. The further away from the photodetector the electrons pass through the straw, the higher will be the light reduction due to additional reflections. The impact of the channel reflectivity will be studied in detail using an optical simulation, which is still under development.}
    \item{\textbf{Neutral density optical filter:} A neutral density optical filter inserted in front of the PD can further reduce the amount of \cer light. Neutral density filters have a flat transmission spectrum and are available for different transmission coefficients. Dependent on the final choice of straw and the reflectivity of the material, an appropriate attenuation factor can be chosen.}
    \item{\textbf{Photodetector choice:} Choosing a photodetector with a lower gain with respect to a SiPM (gain $\mathcal{O}(10^6)$), such as an APD (gain $\mathcal{O}(10^2)$) reduces the signal charge by a factor of $10^4$ to avoid saturating the readout electronics. For a more detailed discussion see section \ref{subsubsec:pdchoice}.}
 \end{itemize} 
 
 As can be seen in Fig.~\ref{fig:sigbg} the electron rates vary by 4 orders of magnitude between the highest and lowest energy in the Compton spectrum as well as 3 orders of magnitude between the lowest and highest laser intensity $\xi$. It is therefore necessary to adapt the channel response to the expected electron rate not only through fixed detector design choices as listed above, but also during the operation of the detector.
 
 The LUXE \cer detector design foresees use of SiPMs and APDs at the same time, with one staggered super-layer of straws for each photodetector type. The different gain of the two technologies (see table \ref{tab:photodetectors}) enables a dynamic range over four orders of magnitude, meaning that if the SiPM is saturated for high rates, the APD is still sensitive, and if the light levels are too low for the APD, the SiPM  is sensitive. In the intermediate rate regime, the two  detector  technologies can be used to cross-calibrate. One possible option to vary the light levels at each end of the straw is to change the angle of the light guide with respect to the beam. Since the Cherenkov angle in air is small, most optical photons will be created close to the beam axis. If the straw light guide is tilted, fewer reflections are needed for the light to get into one of the photodectors, whereas less photons are reflected into the detector at the other end of the straw. Furthermore, the use of two different light guide material with different reflectivity, each optimised for the corresponding photodetector type, helps to further adjust the dynamic range.
 
 Further detector modifications to adjust the dynamic range are possible in between LUXE runs with different $\xi$. It is possible to change the active medium, to replace optical filter with one of more or less transmissivity, or to adjust the photodetector operating voltage to vary the gain. These measures require an intervention in the LUXE experimental area, however since different $\xi$ values will be measured in different runs, they can be planned in advance and carried out with minimal access time. The final detector design must particularly also be chosen such as to make any of the aforementioned interventions as easy as possible. 
 
 The final configuration of photodetectors in combination with the aforementioned mitigation techniques to adapt the detector dynamic range for different run periods of LUXE will be determined and calibrated in a test beam with variable electron bunch charge in the regime expected in LUXE.

\subsection{Straw Light Guide Mounting}
The straw tube light guides are mounted inside the \cer detector gas-tight box using two mounting rails that ensure that the straw tubes remain in place and retain their structural integrity throughout the LUXE data-taking time. Stability of the straw tube position over time is directly related to the total energy scale uncertainty on the measured energy spectrum. Different strategies exist for mounting and stabilizing foil straws for gaseous ionisation detectors which can similarly be applied to the LUXE \cer straws~\cite{Ahdida:2704147}. 

For the metal tube alternative, the mounting and stabilisation requirements are not as critical due to the greater stiffness of the material. The two mounting rails are sufficient to keep the tubes in place. However, when using metal tubes it is important that no mechanical stress is applied to the tubes during the detector assembly procedure to avoid bending. 

\subsection{DAQ and Frontend Electronics}
\label{subsec:frontend}

\begin{figure}[htbp!]
    \centering
    \includegraphics[width=0.9\textwidth]{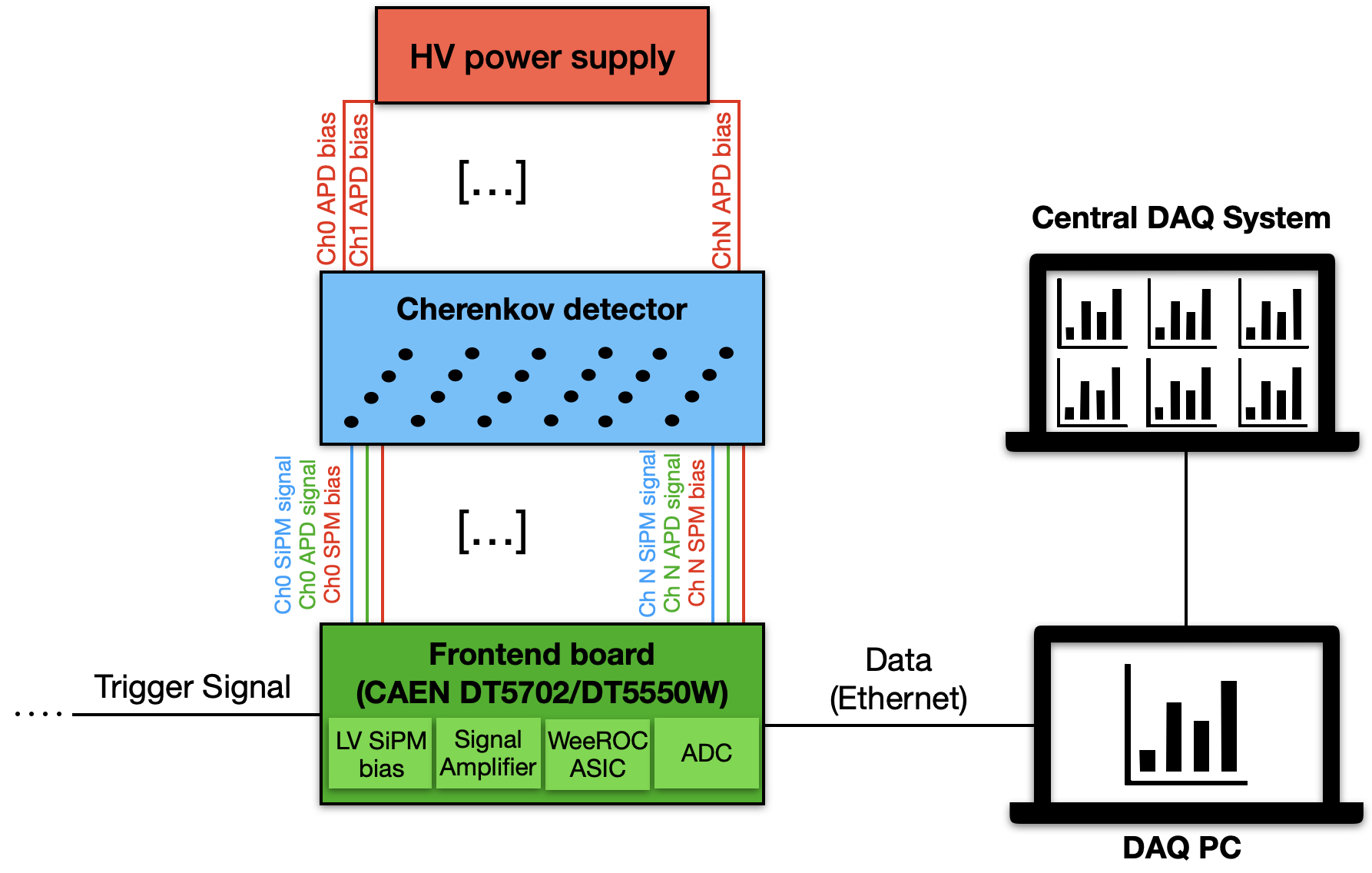}
    \caption{Block diagram of the Cherenkov detector readout and data acquisition chain.}
    \label{fig:blockdaq}
\end{figure}

Fig.~\ref{fig:blockdaq} shows a block diagram depicting the Cherenkov detector readout and data acquisition chain. A commercially available 32-channel SiPM frontend board (CAEN DT5702/DT5550W \cite{CAENDT5702,CAENDT5550W}) based on the Weeroc Citiroc/Petiroc readout chip is used to read out the photodetector signal. For each channel,  the frontend board has an integrated adjustable gain and rise-time amplifier and a 12-bit analogue-to-digital converter to digitise the peak level of the SiPM/APD signal pulse. Since the shape and amplitude are similar for APD and SiPM signal pulses, the same board type is used for SiPM and APD. Given an event trigger, the digitised signal peak value is saved for all output channels. The recorded events are saved in a ring buffer and are directly transferred to the local DAQ PC via an ethernet cable. Several boards can be daisy-chained to read out the required number of \cer detector channels. In addition the frontend boards provide one adjustable stabilised bias voltage per channel which can be used to power the SiPMs. Since the APDs require high voltage, they  must be biased by an external power supply. Compared to the DT5702, used for the first testbeam prototype, the DT5550W has a fully customisable firmware, which makes it the preferrable choice for use in the final \cer detector.

The final configuration of the frontend board parameters such as the amplifier gain, rise-time etc.~has not been optimised yet. The optimal choice depends on the channel geometry and reflectivity and the required range of particle rates to be covered.

The DAQ software foreseen in LUXE is EUDAQ2 \cite{EUDAQ2}. In the case of the \cer detectors the local DAQ PC runs an EUDAQ producer which communicates via network with the central DAQ PC. For more details, see Chapter~\ref{chapt11}.

\subsection{LED Calibration System}
\label{subsec:ledcalib}

To monitor the stability of the \cer detector channels, a calibration system based on a pulsed UV LED is installed in the detector box. The system is based on the CALICE ~\cite{6154596} calorimeter calibration source and was developed in \cite{Vormwald:168227}. It provides a very short ($\mathcal{O}(10\,\text{ns})$ long) stable light pulse with a variable intensity. The intensity is steered by an external voltage and the frequency is provided by a trigger pulse. The LED light pulse is distributed via optical fibres into each \cer detector channel. 

The LED calibration will be performed in-situ during data-taking. As the rate of \euxfel electron bunches is only $10\,\text{Hz}$ and the laser repetition rate is $1\,\text{Hz}$, the LED pulse is triggered in between the electron bunch crossings. This enables an in-situ monitoring of photodetector gain variations that are expected to occur due to fluctuations in temperature and operating voltage.

\subsection{Stabilisation and Slow Control}

The response of SiPMs and APDs are dependent on temperature and bias voltage variations. For example, the breakdown voltage of the MicroFJ-30035 SiPM \cite{ONSemiJSeriesSiPM} typically varies by $21.5\,\text{mV}/^\circ \text{C}$ as a function of temperature, whereas the Hamamatsu S5344 APD varies by $140\,\text{mV}/^\circ \text{C}$, which, due to the higher breakdown voltage of the APD corresponds to a similar relative change.  Slow variations such as temperature can be monitored and corrected using the in-situ pulsed LED calibration, however the pulsed LED intensity is temperature-dependent as well. To monitor temperature changes in order to correct the calibration, standard temperature sensors are mounted on the sensor PCBs and the LED driver. Since the LED driver is spatially separated from the photodetectors, their temperature will vary mostly independently.

The bias voltage supplied by the CAEN DT5702 and DT5550W frontend boards is stabilised, in case of power supply used for the DT5550W, the voltage ripple is quoted as being smaller than $0.1\,\text{mV}$ peak-to-peak \cite{CAENA7585}, which means the relative gain variations ($\sim0.36/\text{V}$) for the MicroFJ-30035 SiPM are small. For the APD bias, a low-noise multi-channel power supply will be used.

The air pressure in the box is monitored using one digital pressure monitor per box. After the light-tight box is filled with air at $1.1\,\text{bar}$ it is sealed. According to the discussion in Section~\ref{sec:requirements}, the pressure must be controlled to within $10\,\text{mbar}$, or monitored to this level of precision to correct the measurement offline. The same applies to the overall air temperature, which must be known within $3^\circ$C. This can be achieved using the readout of several temperature sensors in different positions on the gas-tight box.

\subsection{Future Design Decisions}

The most important design decisions still to be taken for the \cer detector are:

\begin{itemize}
    \item{\textbf{Photodetector choice:} The decision which photodetector or combination of photodetectors to use will be taken after a testbeam campaign  with variable bunch charges in the regime expected for LUXE. Most likely, both SiPM and APDs will be used simultaneously to adjust to the very wide range of electron rates covered throughout the LUXE lifetime.}
    \item{\textbf{Optimised channel geometry:} The layout of straw channels in the LUXE \cer detectors is very flexible. To determine the optimal layout, studies based on \geant simulations will be performed, studying unfolding reconstruction techniques, energy resolution as well as channel overlap.}
     \item{\textbf{Material of straw tubes:} The straw tube material and manufacturing technique will be chosen based on a high-rate test beam that will allow the required effective reflectivity to be measured. A simulation of \cer photon generation and an optical simulation of the reflection within the straw can give a first hint how this will affect the signal dynamic range.}
\end{itemize}

\section{Interfaces and Integration} 
The \cer detectors will be integrated in the central LUXE DAQ system (EUDAQ). The frontend electronics firmware will be designed such that it can be triggered and sychronised to a common LUXE event trigger.

The \cer detectors will record data with the electron beam at $10\,\text{Hz}$, of which there will be collisions with the laser pulse with a rate of $1\,\text{Hz}$. Since the signal rate is comparatively low, it is possible to take in-situ LED calibration data in between the electron beam data, as well as dark noise data, using a stand-alone trigger outside of the beam bunch arrival or LED pulse. This will enable constant monitoring of the linearity of the detector.

Since the data rate for the \cer detector is small and the readout is fast, it can easily provide information to the LUXE online data quality monitoring.

\section{Installation, Commissioning and Calibration}
 \label{sec:instcommcal}
 \subsection{Installation}
 
The two straw \cer detectors (EDS and IBM system) are almost identical, therefore the installation process for the two systems will be very similar. Both systems will be fully assembled above ground before installation. 

 The EDS \cer detector enables LUXE to measure non-linear Compton scattering even in a very reduced "bare-bones" scenario in its first e-laser collision runs, therefore the priority is to install this system in the experimental area at the earliest possible time. The IBM system needed for $\gamma$-laser collisions can be installed during a later intervention. Since the EDS and IBM system are almost identical, the installation of the IBM will profit from the experience during the installation of the EDS system. A staggered installation also gives more time for the assembly of the IBM detector. 

The EDS \cer detector is fully assembled in the laboratory in advance of the installation and is lowered in one piece into the LUXE experimental cavern. The main work to be performed in the experimental area is connecting power and readout cables to the channels, using pre-assembled patch panels, as well as the gas system. Finally, system tests are performed with intermittent access to the detector in the experimental area.

Table~\ref{tab:cere_installation} shows the necessary steps to be taken during installation of one \cer system (identical for EDS and IBM) in the control room (CR), and in the experimental area (EA), as well as the needed personnel-hours for each step. Central DESY services are mainly needed for lifting the detector in place by crane as well as the mechanical alignment. The weight of the fully assembled detector will be about $40\,\text{kg}$. The total estimated time required for the installation of the \cer system is about 2-3 weeks (assuming some tasks in table~\ref{tab:cere_installation} can be parallelised). In this case it is possible to install the \cer detector in a shorter (e.g. end-of-year) shutdown of the EuXFEL.  

\begin{table}[]
    \centering
    \makebox[\linewidth]{  
    \begin{tabular}{|c|c|c|c|c|c|c|}
        \hline
        \multirow{2}{*}{}\textbf{Activity} & \textbf{Duration} & \textbf{Access} & \textbf{ Personnel} & \multicolumn{3}{c|}{\textbf{Person-days}} \\\cline{5-7}
         & & & & PH & EN & TE \\ \hline
         \makecell{Inspection before \\ installation } & 1 day & EA, CR & \makecell{3PH, 1MEN, \\ 1EEN} & 3 & 2 & \\\hline
         \makecell{ Installation of LV \\ power supply} & 1 day & CR & 2TE, 1EEN & & 1 & 2  \\\hline
         \makecell{Test LV remote \\ control and \\ SW procedures} & 1 day & CR & 1PH, 1TE & 1  && 1 \\\hline
         \makecell{DAQ dry \\ integration tests } & 3 days & CR & 1PH, 1TE & 3 && 3\\\hline
         \makecell{Detector and \\ associated equipment \\ mechanical installation, \\test of motion control \\system, activation \\ of gas supply line } & 1 week & EA, CR & \makecell{2PH, 1MEN,\\ 1TE} & 10 & 5 & 10  \\\hline
         \makecell{Mechanical alignment \\ of detector supports} & 1 day & EA & \makecell{1MEN, 1TE, \\ 1PH} & 1 & 1 & 1 \\\hline
         \makecell{Connection and \\ test of the LV} & 2 days & EA, CR & \makecell{1EEN, 1TE, \\ 1PH} & 2 & 2 & 2  \\\hline
         \makecell{Connection and \\ test of the DAQ \\ data path} & 1 day & EA, CR & 2PH & 2 & & 1 \\\hline
         \makecell{Test and \\ commissioning of \\ stand-alone readout and \\ calibration software} & 1 day & EA, CR & 2PH & 2 & &  \\\hline
         \makecell{Detector tests and \\ noise optimization \\ with calibration signals} & 1 week & EA, CR & 3PH, 1TE & 15 & & 5  \\\hline
         \makecell{Integration and \\ commissioning \\ with central DAQ \\ and slow control \\ system, dry runs} & 1 weeks & EA, CR & 3PH, 1TE & 15 & & 5 \\\hline\hline
         \textbf{TOTAL} & 2-3 weeks & & & \textbf{54} & \textbf{11} & \textbf{30}  \\\hline
         \textbf{\makecell{TOTAL FTE \\ (1 year = 225 \\ working days}} & & & & \textbf{0.24} & \textbf{0.05} & \textbf{0.13}  \\\hline
    \end{tabular}
    }
    \caption{Summary of steps needed to complete installation of one LUXE \cer detector (EDS or IBM) and associated personnel-hours. PH = Physicist, EEN = Electrical engineer, MEN = Mechanical engineer, TE = Technician.}
    \label{tab:cere_installation}
\end{table}

\subsection{Commissioning}

The commissioning of the \cer detector will consist of the laser-based detector alignment (see table \ref{tab:cere_installation}) as well as a series of electron beam-only runs to test the timing and integration of the \cer detector readout within the LUXE DAQ system.

Especially the DAQ integration with the scintillator and camera system can be excercised in beam tests and dry runs before the installation in the LUXE area.

Furthermore a run with varying dipole magnetic field strength and varying bunch currents will be used to do a final alignment of the \cer channels. Finally a run with a Bremsstrahlung target in the IP region will be used to perform an in-situ alignment of the straw tube channels in the \cer detector, by comparing the observed spectrum to a MC prediction (see also Section~\ref{subsec:calib_strategy}).

\subsection{Calibration Strategy}
\label{subsec:calib_strategy}

The response of the individual channels of the \cer system can be pre-calibrated in a high-rate testbeam facility where a bunched electron beam is impinged on the straw channels.

The relative response of the photodetectors to a pulsed UV LED will be determined before and during detector assembly, to account for possible differences in photodetector operation points and light guide characteristics.

The stability of the charge response over time during data-taking can be achieved by using a UV LED calibration system, discussed in Section~\ref{subsec:ledcalib}. The LED calibration data-taking is expected to take place in between the arrival of beam electron bunches in the LUXE setup at $10\,\text{Hz}$. 

The absolute energy calibration of the full \cer system can be achieved by a dedicated commissioning run with the electron beam at variable bunch charge and low energy, varying the dipole magnetic field to guide the electron beam into the straw channels.

For low laser intensities the well-known linear Compton spectrum can be used to calibrate the energy and rate measurement of the EDS system, as well as correct for possible misalignment effects. Furthermore, the EDS system could be calibrated by inserting a target in the interaction region and studying the resulting spectrum of the electrons after undergoing the Bremsstrahlung process. However, the modelling of the Bremsstrahlung spectrum from the target showed a significant dependence on the physics processes included in simulation. This means that the Bremsstrahlung modelling must be studied in advance, e.g.\ using testbeam data, making the linear Compton-scattering calibration preferrable. The calibration runs should be repeated frequently during data-taking, to monitor the alignment of the \cer straw channels with respect to each other.

Finally, the laser-off data where only the electron bunches traverse the LUXE experimental setup is used to estimate the background levels from scattering of the beam particles.

\subsection{Decommissioning}

No particular challenges are foreseen for the decommissioning of the LUXE \cer detector. The active material is air, which can be released without the need for filtration. The overall material budget is low and the expected activation dose is low enough for a normal disposal of the detector material. 

\section{ORAMS: Operability, Reliability, Availability, Maintainability and Safety}
The \cer straw detector is a comparatively simple and robust system in the LUXE environment. The system uses a well-known detection principle and standard readout electronics. The expected data rate from the system is small ($\mathcal{O}(40\,\text{kByte/s})$). From simulation studies it has been shown that the impact of e.g.\ detector or readout non-linearities is small.

The main challenge in terms of operability, maintainability and reliability is to ensure the stability of the straw tube position in the detector over time. In addition it is important that the straw material can sustain the high electron fluxes present in LUXE. Furthermore, due to the number of channels, of the order of 100, it is important to make the exchange of electrical components such as photodetectors as easy as possible in case of failure. 

The \cer design uses mainly standard components, which are readily available for purchase. The source of straw tubes still needs to be determined, however industrially produced metal tubes have been shown to be a viable option.

The \cer detector system only uses non-toxic gas (air) at slight over-pressure of $100\,\text{mBar}$. The SiPMs use an operating voltage of $30\,\text{V}$. For the APDs, a higher voltage of $150\,\text{V}$ is needed.


\section{Further Tests Planned}
The most important test to be still performed in the future for the \cer detectors in LUXE is a high-rate beam test with electron bunches of $10^8$ impinging on the detector channels. This is necessary to determine the set of parameters defining the channel efficiency (channel reflectivity, use of optical filters, choice of photodetectors and optimisation of readout system). Furthermore a calibration run across the full dynamic range of electron rates for different $\xi$ values (see fig. \ref{fig:comptonspectra}) is necessary to ensure that the readout electronics is not saturated and sensitive enough to detect particles also at lower $\xi$ values. It should be noted however, that expecially for low-intensities the scintillation system provides additional information, so the focus of the \cer system must be on the high-rate regime. 

Another important aspect still to be studied is the reconstruction algorithm to obtain the non-linear Compton spectrum from the distribution measured charge per \cer channel. Using simulation, the aim of this study is to develop a robust unfolding technique either using standard matrix inversion methods or multivariate algorithms.

\section*{Acknowledgements}
\addcontentsline{toc}{section}{Acknowledgements}
This work was in part funded by the Deutsche Forschungsgemeinschaft under Germany's
Excellence Strategy – EXC 2121 “Quantum Universe” – 390833306. This work has benefited from computing services provided by the German National Analysis Facility (NAF). Parts of the measurements leading to these results have been performed at the Test Beam Facility at DESY Hamburg (Germany), a member of the Helmholtz Association (HGF). The authors would like to thank the SHiP team at University of Hamburg as well as the ATLAS TRT group for providing straw tube samples for the test beam campaigns. 



\printbibliography
\end{refsection}


\chapter{Gamma Ray Spectrometer}
\label{chapt8}
\begin{refsection}

\chapterauthors{K.~Fleck \\Queen's University Belfast, Belfast (Northern Ireland)}{N.~Cavanagh \\Queen's University Belfast, Belfast (Northern Ireland)}
{E.~Gerstmayr \\Queen's University Belfast, Belfast (Northern Ireland)}
{M.~Streeter \\Queen's University Belfast, Belfast (Northern Ireland)}{G.~Sarri \\Queen's University Belfast, Belfast (Northern Ireland)}
\date{\today}

\begin{chaptabstract}
In this chapter, a detailed description of a gamma ray spectrometer to be implemented in the LUXE experiment is provided. The instrument is designed to reconstruct the spectrum of high-flux GeV-scale gamma ray beams from the measurement of the spectrum and yield of electron--positron pairs generated during the propagation of the gamma ray beam through a thin foil. The electron--positron pairs are measured using the techniques of magnetic spectroscopy in combination with high granularity scintillator screens. The gamma ray beam will be produced either via inverse Compton scattering during the propagation of the EuXFEL electron beam through the focus of an intense laser, or via bremsstrahlung during the propagation of the same electron beam through a thin converter foil; as such, the instrument is expected to operate during both the electron--laser and gamma--laser configurations of the experiment. Numerical and analytical modelling of the spectrometer, corroborated by preliminary experimental tests, indicates energy and yield resolution at the percent level, in an energy range from 1 to 20\,GeV.  
\end{chaptabstract}

\section{Introduction}
\label{gs:intro}

The present chapter provides a technical description of a gamma ray spectrometer to be installed in the LUXE experiment. In a nutshell, the instrument aims at providing high-resolution spectral information of high-flux gamma ray beams with photon energy in the multi-GeV range, as to be obtained either from inverse Compton scattering (ICS) of the EuXFEL electron beam during its propagation through the focus of a high-intensity laser pulse or from bremsstrahlung generated during the propagation of said electron beam through a thin converter target. In its current inception, the spectrometer mainly comprises a thin converter foil, a lead collimator, a dipole magnet, and two scintillator screens imaged by CCD cameras. It will be installed downstream of the electron and positron detectors, in conjunction with a gamma ray profiler that is discussed in Chapter~\ref{chapt9} (see Fig.~\ref{fig:layoutdet}).  The design of the gamma ray spectrometer follows the general principles of operation discussed in Ref. \cite{Fleck:2020}.

\section{Requirements and Challenges} \label{sec:gs:requirements}

\begin{figure}[h!]
    \centering
    \includegraphics[width=\textwidth] {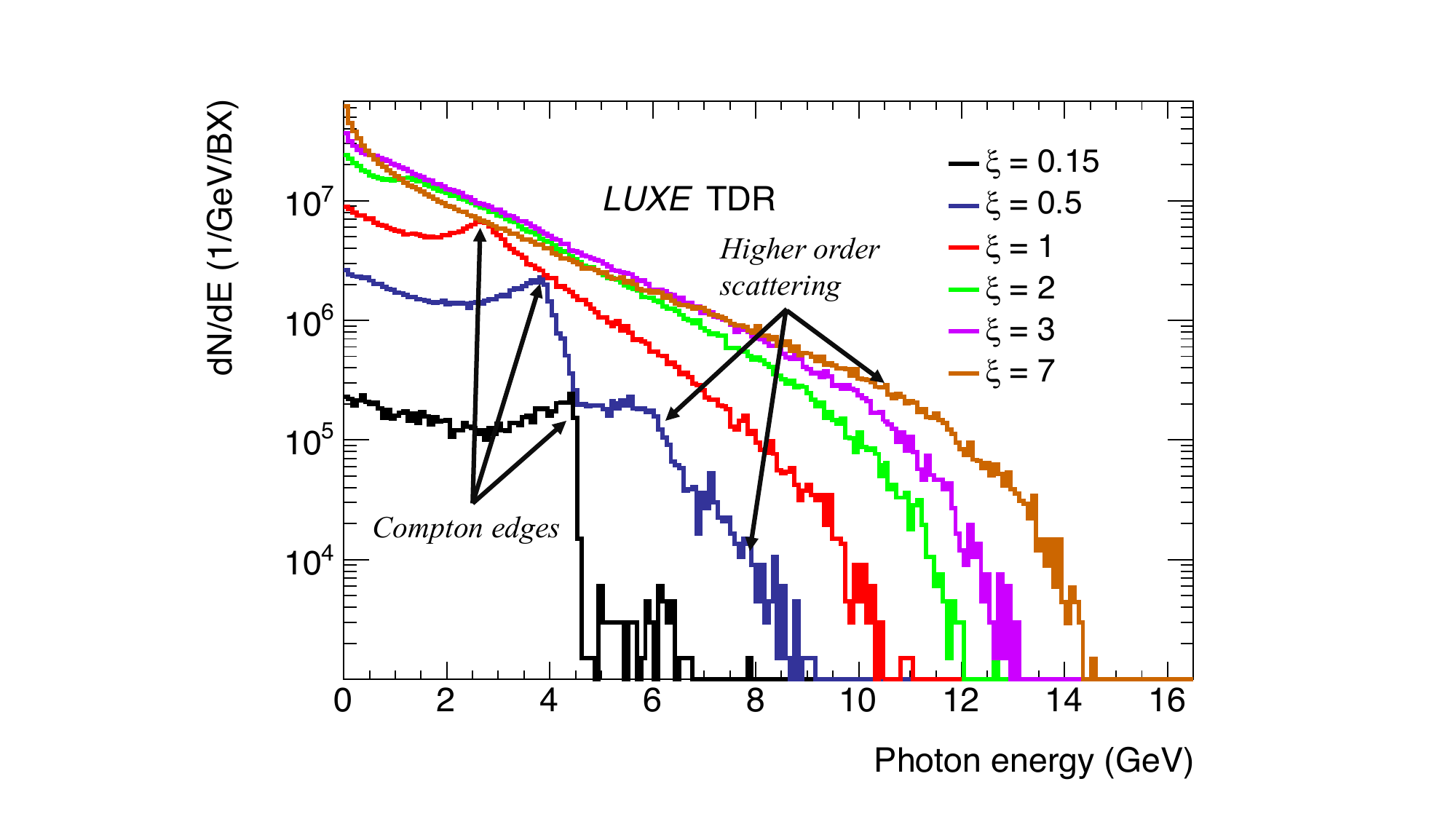}
    \caption{Simulated photon spectra from ICS of a 16.5\,GeV electron beam in the focus of a 40\,TW laser for different laser dimensionless intensities. Highlighted are some examples of the positions of Compton edges and the contributions from higher order scattering.}
    \label{Compton_spectra}
\end{figure}

The instrument will be used in both $e$-laser and $\gamma$-laser modes of operation for LUXE. In the $e$-laser configuration, the spectrometer is required to provide the spectral distribution of the ICS photon beams generated during the interaction between the electron beam and the focus of the high-intensity laser. Several distinctive features can be identified and studied in the spectrum of the photon beam generated by ICS in the laser field: $e^-$ $+$ $n\gamma_L$ $\rightarrow$ $e^-$ $+$ $\gamma$, where $\gamma_L$ is a laser photon, $n$ is the number of laser photons absorbed by one electron in a formation length, and $\gamma$ is the gamma ray photon. Primarily, the spectrometer aims at determining the position of Compton edges in the photon spectrum and the extent of nonlinear contributions in the scattering process. Figure~\ref{Compton_spectra} shows predicted ICS photon spectra to be generated at LUXE for different laser intensities.

In a linear regime (valid for 
a laser dimensionless intensity parameter $\xi\ll1$), the energy of the ICS photons can be estimated as $\hbar\omega_\gamma = 4\gamma_e^2\hbar\omega_L$, where $\gamma_e$ is the Lorentz factor of the electron at the moment of emission and $\omega_L$ is the laser frequency. This is obtained under the assumption of a maximum of one laser photon absorbed per electron at each emission event. For example, an electron energy of 16.5\,GeV ($\gamma_e = 3.23\times10^4$) and a laser photon energy of 1.2\,eV (central laser wavelength of 1\,$\mu$m) correspond to a photon energy of $\hbar\omega_\gamma= 4.8$\,GeV.
In this regime, while multi-photon absorption might still occur, it can be treated perturbatively, as experimentally verified at the E144 experiment \cite{Bula:1996st, Burke:1997ew} and more recently in fully laser-driven experimental configurations \cite{Sarri:2014}. 

 In a non-linear regime (valid for a laser dimensionless intensity parameter $\xi\geq1$), the position of the Compton edge shifts due to the higher dressed mass of the electron in the laser field, by a factor $[1+(\xi/\eta)^2]^{-1}$, where $\eta = 1$ ($\eta =\sqrt{2}$) for circular (linear) polarisation of the laser.  Moreover, multi-photon absorption becomes a dominant process, enhancing non-linear contributions to the spectrum of the ICS photons.  

A key experimental measurable of the experiment is thus the spectrum of the ICS photons for a series of reasons: first, it allows the position of the Compton edge to be identified and, thus, to experimentally observe electron mass dressing in the laser field; second, it provides information on the level of non-linearity in the Compton process, i.e., the number of photons absorbed per emission event, together with the associated probability; finally, it provides useful information on the characteristics of the gamma ray beam that produces electron--positron pairs in the laser field, via the non-linear Breit--Wheeler process. 

The spectrometer can also be used during the $\gamma$-laser collision phase of the LUXE experiment, since it represents an on-shot diagnostic of the properties of the bremsstrahlung photons to be used in modelling this class of interactions.

Ideally, one would require a spectrometer able to measure photon energies between 1 and 16.5\,GeV with a precision at the percent level. Providing on-shot, detailed spectral measurements of high-brightness and high-energy gamma ray beams is however a non-trivial task. Different systems have been proposed: methods based on pair production in a high-Z target \cite{Schumaker:2014,Barbosa:2015,wistisen:2018} are able to detect high-energy photons but are currently not designed to work at a high flux. Similarly, methods based on measuring the transverse and longitudinal extent of cascading in a material are designed to work only at a single-photon level, or present limited energy resolution for high fluxes \cite{Behm:2018}. Compton-based spectrometers (such as the one in Ref. \cite{Corvan:2014}) do work at high-fluxes but can only meaningfully measure spectra up to photon energies of a few tens of MeV. Cherenkov radiation is also used \cite{McMillan:2016} but, again, it is best suited to perform single-particle detection. Large-scale detectors such as the EUROBALL cluster \cite{Wilhelm:1996} and the AFRODITE germanium detector array \cite{Lipoglavsek:2006} can also resolve up to 10--20 MeV but their significant size make their implementation in many laboratories infeasible.

In our proposed detector, we aim at extracting the spectrum of the gamma ray photons by measuring the spectrum of electron--positron pairs generated during the propagation of the photons through a thin converter target, in a setup similar to that discussed in Ref.~\cite{Fleck:2020}. 

\section{System Overview}

\begin{figure}[h!]
    \centering
    \includegraphics[width=\textwidth] {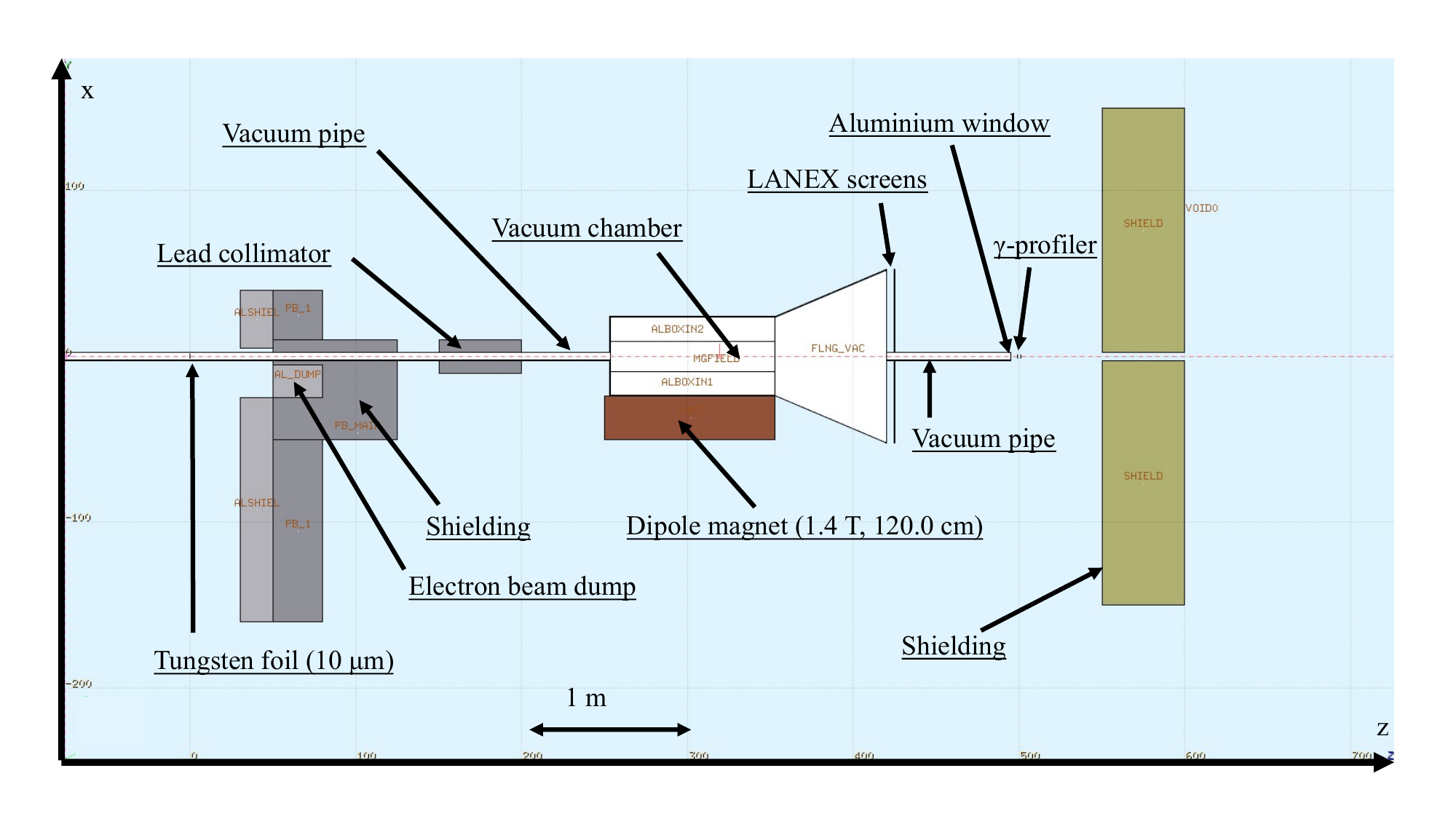}
    \includegraphics[width=0.92\textwidth] {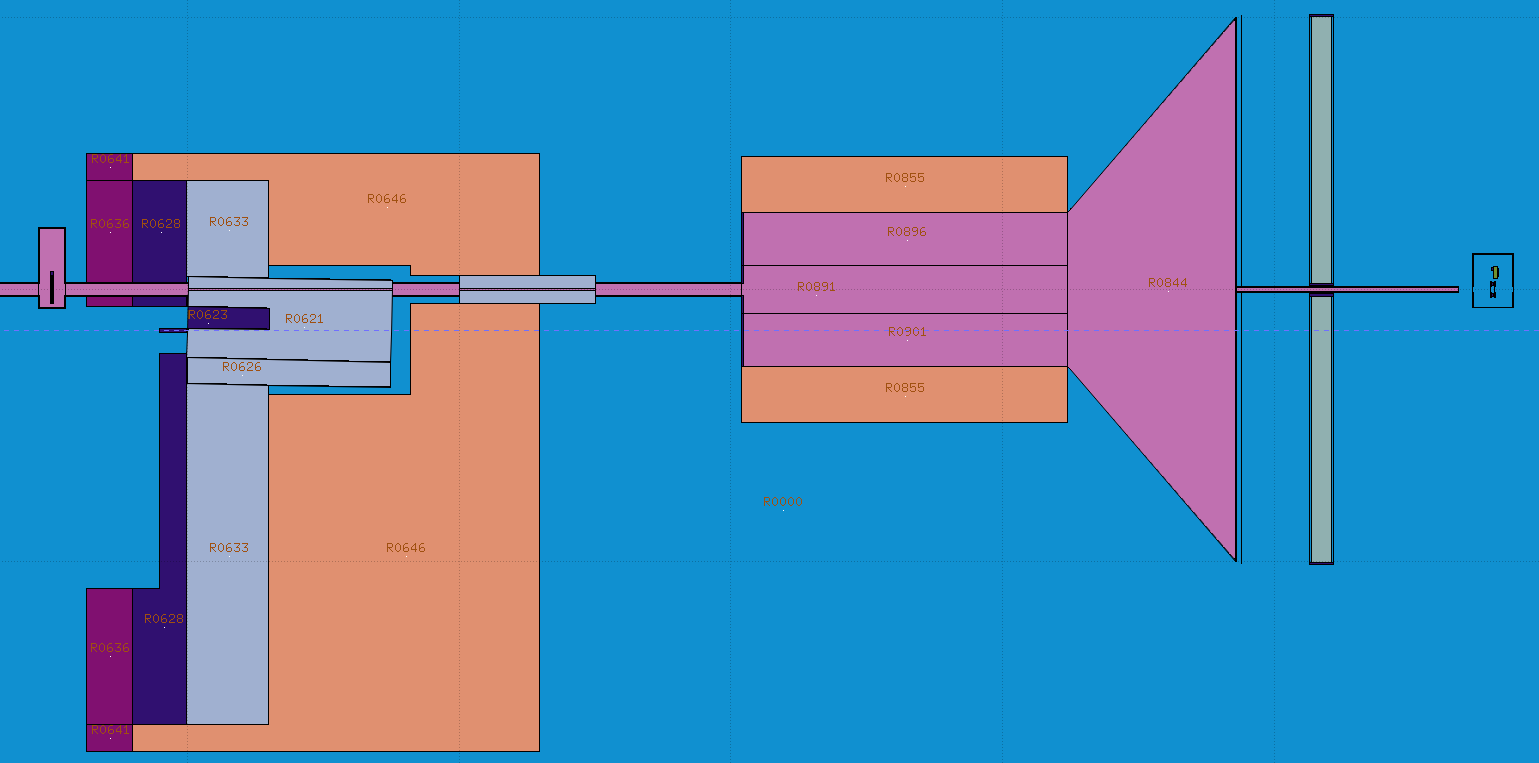}
    \caption{(Top) Top-view (in the $x-z$ plane) in-scale drawing of the gamma ray spectrometer, as used as input for Monte Carlo simulations, detailing all the main constituents of the instrument. Coordinates are given relative to the converter target in cm.
    (Bottom) The gamma ray spectrometer as implemented in fuller detail within the full-scale simulation geometry.}
    \label{setup}
\end{figure}

An in-scale drawing of the current incarnation of the gamma ray spectrometer, as implemented in the full-scale \geant  \cite{Agostinelli:2002hh,Allison:2006ve} simulations of the whole experiment, is given in Fig.~\ref{setup}. In essence, the spectrometer measures the spectrum of electrons and positrons generated during the propagation of the gamma ray beam through a thin converter target. A deconvolution algorithm (discussed in Section~\ref{sec:gs:performance}) can then be used to retrieve the spectrum of the primary photon beam from the spectrum of the electron--positron pairs. The converter, a 10\,$\mu$m thick tungsten foil, is placed 6\,m downstream of the interaction point and is followed by a shielding configuration consisting of a first structure containing the electron beam dump and a second lead collimator. The collimator is 50\,cm long with a circular aperture in the centre with a diameter of 4.8\,cm, resulting in an acceptance angle from the tungsten foil of 19\,mrad. Via Monte Carlo modelling, this configuration has been found to minimise noise arising upstream of the setup and from low-energy particles generated during the conversion process through the tungsten foil (discussed in Section~\ref{sec:gs:performance}). A vacuum pipe on-axis guarantees unaffected propagation of the gamma ray beam and generated secondary particles up to a vacuum chamber with a length of 1.6\,m and a maximum width of 1.6\,m, expected to operate at an air pressure $<10^{-3}$\,mbar. The first section of the vacuum chamber is surrounded by a dipole magnet with a field of 1.4\,T and a length of 120\,cm. Two scintillator screens  are placed outside of the vacuum chamber on either side of the main axis, 64\,cm from the magnet, to provide spatially resolved detection of the electron--positron pairs generated in the converter. The electrons and positrons will reach the scintillators through 500\,$\mu$m thick aluminium windows. A second vacuum pipe extends from the rear of the vacuum chamber up to a gamma ray profiler, discussed in Chapter 9. We propose to use two LANEX (Gd$_2$O$_2$S:Tb, GadOx) scintillator screens (main properties are given in Table~\ref{Table:1}), each imaged onto a cooled and amplified CCD camera with an interference filter centred at 545\,nm, the peak wavelength in the scintillation curve of a LANEX screen. As will be discussed in more detail in the following, amplified CCD cameras are needed in order to clearly record relatively low signal levels (see Fig.~\ref{dep} for an example of it). Assuming two 16-bit cameras with 2048 $\times$ 512 pixels and operations at 10\,Hz, the spectrometer is expected to generate a maximum of 40 MB/s of data to be stored by the centralised data acquisition system. Thanks to the long scintillation time of a LANEX screen, we aim at triggering the CCD cameras at least 1\,$\mu$s after the laser shot, in order to minimise optical noise arising from scattered laser photons.  
As the location of the CCD camera will be several metres away from the LANEX (see Section~\ref{sec:gs:installation} for possible locations currently being investigated), an optical image relay system will be required.
Commercial SLR-type wide angle lenses will be used to create an intermediate de-magnified image of the screen.
This will then be transported to the camera location by using either a $4f$ lens-based imaging system or by using a flexible imaging fibre-bundle (e.g.\ Schott IG-163).
The latter option would provide better performance and convenience but at higher cost.
Therefore, both configurations are currently being designed, with the option to upgrade to the fibre-bundle system when it is confirmed that the fibres would not be damaged by the radiation environment.

\begin{table}[b!]
\centering
\begin{tabular}{|c|c|}
\hline
Optical Property & \\
\hline
\hline
Light Yield (@ $\sim60-80$keV) & $6\times10^4$  ph/MeV\\
Decay Time & $\sim 3\, \mu$s\\
Afterglow after 3 ms & $< 0.1$\%\\
Peak Emission Wavelength & $545$ nm\\
\hline
\end{tabular}
\caption{Main scintillation properties of the LANEX scintillator screens. Taken from Ref. \cite{Eijk:2002}}
\label{Table:1}
\end{table}

\begin{figure}[h!]
    \centering
    \includegraphics[width=0.8\textwidth] {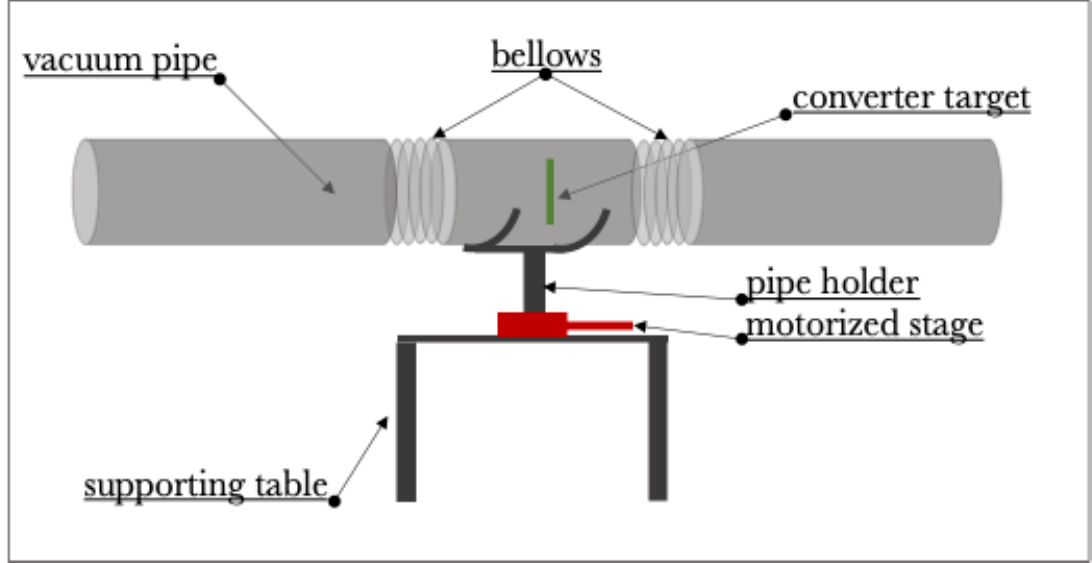}
    \caption{Artistic impression of the pipe section containing the converter target, attached with bellows to the main vacuum pipe.}
    \label{translation}
\end{figure}

As discussed in more detail in Section~\ref{sec:gs:performance}, the recorded electron and positron spectra are deconvolved to extract the spectrum of the gamma ray beam. The deconvolution algorithm requires an exact knowledge of the thickness of the converter target. While in principle gamma ray irradiation might result in target ablation and, therefore, reduction of the target thickness, extensive numerical work (detailed in Section~\ref{sec:gs:performance}) and preliminary experimental tests using the Astra-Gemini laser at the Central Laser Facility in the UK and the Apollon laser facility in France (results are currently being prepared for publication) confirm that gamma ray irradiation will remain consistently below the threshold fluence for tungsten, indicating durability of the performance of the spectrometer even at a 1\,Hz operation. The option of translating the converter target along the two transverse directions must be anyway considered, both for alignment purposes and in the unlikely event that damage would occur locally to the target. Due to the small cross section of the vacuum pipe, the position of the converter will be controlled by having it in a separate section of the pipe connected to the rest of the main pipe by bellows. A 2-D motorised stage will then be used to move the specific pipe section with respect to the rest of the pipe, a solution already adopted in several large-scale accelerators (see Fig.~\ref{translation} for an artistic impression of it).

Once the electron--positron pairs are generated in the converter, their arrival location, $x$, on the scintillator screens is directly related to their energy by the relation:
\begin{equation}
    x = \rho - \frac{\rho^2 - z_mz_d - z_m^2}{\sqrt{\rho^2 - z_m^2}}.\label{exact}
\end{equation} 
Here $\rho = E/eB$ (with $c=1$ assumed) is the radius of curvature in the ultrarelativistic limit for a particle of charge $e$ and energy $E$; $B=$ 1.4\,T is the strength of the magnetic field in the dipole; $z_m = 120$\,cm and $z_d=64$\,cm are the length of the active magnetic field and the distance between the exit aperture of the magnet to the plane of the scintillator, respectively. 

\begin{figure}[t!]
    \centering
    \includegraphics[width=0.49\linewidth]{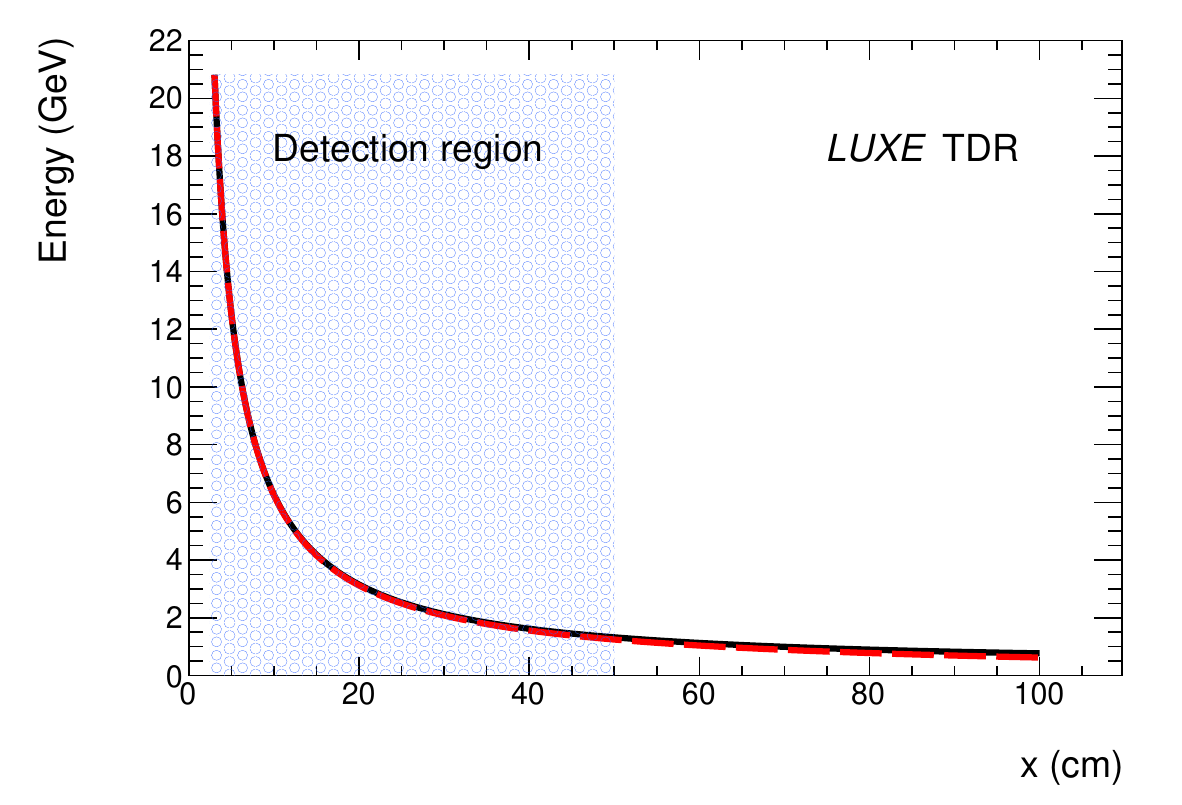}
    \includegraphics[width=0.49\linewidth]{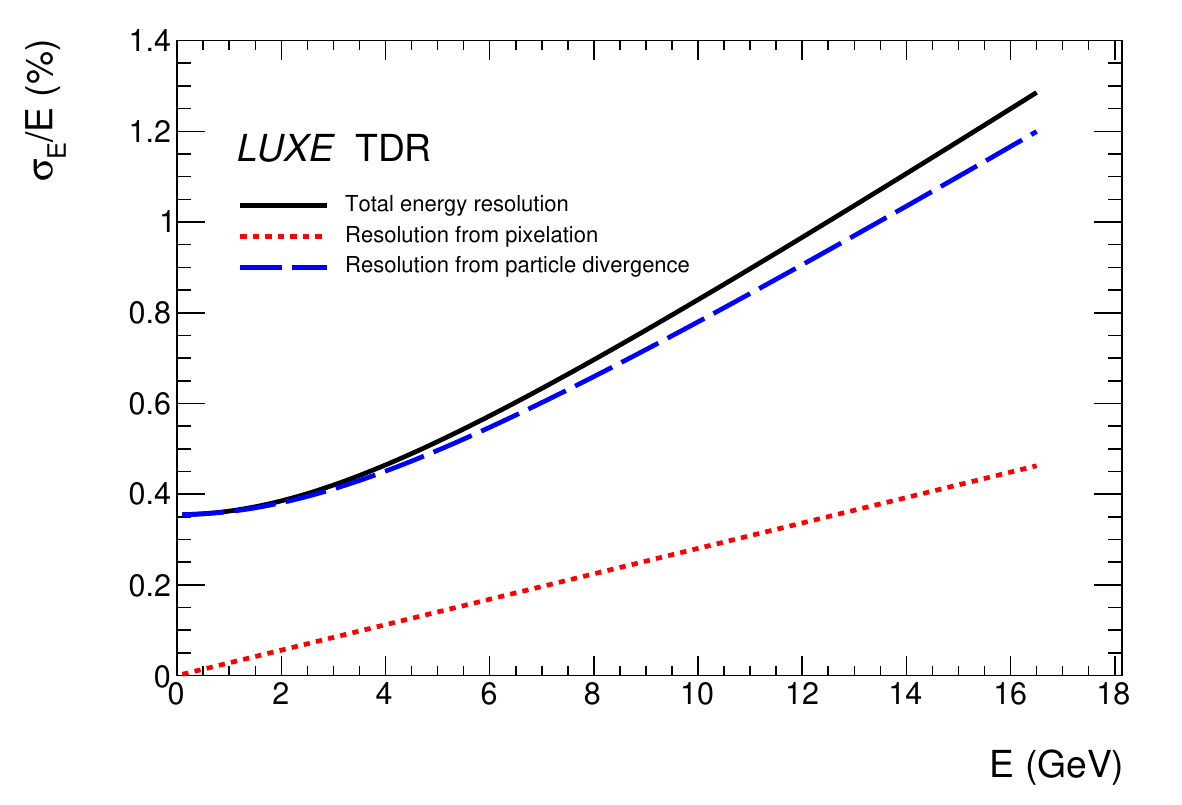}
    \caption{(Left) Exact (black, from Eq.~(\ref{exact})) and approximated (red, from Eq.~(\ref{approx})) particle energy as a function of the arrival position on the LANEX screen. Highlighted in blue is the region covered by the scintillator screen. (Right) Predicted resolution (from Eq.~(\ref{res_tot})) in measuring the spectra of electrons and positrons as a function of particle energy, showing better than 1\% energy resolution from the spectrometer up to a particle energy of approximately 14\,GeV. The red and blue lines depict the components of the resolution due to pixelation and particle divergence, respectively. Shown in black is the total resolution.}
    \label{res}
\end{figure}

The solution in $\rho(x)$, is non-trivial analytically and a root-finding method may be more appropriate. Alternatively, if $\rho \gg z_m$, which at LUXE is equivalent to $E \gg 400$ MeV, an approximation of the solution to Eq.~(\ref{exact}) is
\begin{equation}
    E \simeq eB \cdot \frac{z_m}{x}\left(\frac{z_m}{2} + z_d\right). \label{approx}
\end{equation}
This approximation is accurate to within $0.1\%$ below $1$\,GeV and improves with increasing energy (increasing radius of curvature). In Fig.~\ref{res}, the calculated particles' deflection at the scintillator plane as a function of their energy is plotted, showing particle energies between 1 and 16.5\,GeV contained in approximately 50\,cm of scintillator.

The energy resolution of the spectrometer is mainly dictated by two quantities: the energy-dependent divergence of the particles to be detected, and the binning size of the detection system. The inverse power relation between the energy of the incident electron/positron and its arrival position on the scintillator causes a decrease in resolution with increasing energy, assuming a detector that is linearly segmented. It can be shown that for a detector with uniform pixel size $\delta x$ along the dispersion plane, the resolution in energy is
\begin{equation}
    \frac{\sigma_\mathrm{pix}(E)}{E} \simeq \frac{\delta x}{z_m(z_m + z_d)}\cdot\frac{E}{eB}. \label{eq:res}
\end{equation}
If we assume 50\,cm of scintillator to be imaged onto a camera chip with a length of 2.6\,cm (2048 pixels each with a size of 13\,$\mu$m), the optical system for the scintillation light must have a magnification of $M\approx1/20$, implying a spatial resolution of the order of $\delta x=260$\,$\mu$m. This induces an energy-dependent spectral resolution that is $<0.5$\% across the energy range of interest (see red fine dashed curve in Fig.~\ref{res}\,(right)). 

The divergence of the electron/positron beam after conversion also affects the resolution in energy. If the divergence of the photon beam leaving the IP and impacting the converter target is $\theta_\gamma$, then the divergence of the electrons and positrons leaving the converter can be approximated, in the ultra-relativistic limit, as $\Theta_S \sim \sqrt{\theta_\gamma^2 + 1/\gamma^2}$, where $\gamma = E/m_e$ is the Lorentz gamma factor of the electron or positron. For particles of energy $E > 0.5$ GeV, $1/\gamma < 1$ mrad. This divergence of the electrons/positrons leads to an energy resolution of the order of 1\% (blue dashed curve in Fig.~\ref{res}\,(right)): 
\begin{equation}
 \frac{\sigma_\theta (E)}{E} = \frac{(z_S + z_m +z_d)\Theta_S}{(z_d+z_m/2)z_m} \cdot \frac{E}{eB}. 
\end{equation}

These two contributions to the energy resolution are independent and can thus be summed in quadrature to give the total spectral resolution as:
\begin{equation}
    \frac{\sigma_E(E)}{E} = \frac{\delta x}{z_m(z_m + z_d)}\cdot\frac{E}{eB} \oplus \frac{(z_S + z_m +z_d)\Theta_S}{(z_d+z_m/2)z_m} \cdot \frac{E}{eB},\label{res_tot}
\end{equation}
where $z_S = z_m + z_d$. 

An example of the achievable spectral resolution is given in Fig.~\ref{res}\,(right), showing better than 1\% energy resolution across the energies of interest for the experiment. As can be seen in the figure, the spectral resolution is dominated by the divergence of the electrons and positrons, relaxing the requirements on the pixelation of the detector. It is worth noting that for GeV-scale particle energies and a predicted divergence of the gamma ray beam $\theta_\gamma < 1$\,mrad, the divergence of the electron and positron pairs is much smaller than the acceptance angle of the spectrometer (19\,mrad), which thus has a negligible effect on the transverse distribution of the particles.

The energy deposition by the electrons/positrons in the scintillator screen must also be known to determine the number of particles impacting the screen. A simple \geant simulation of a pencil beam of electrons of various energies $\ge$ 1\,GeV interacting with a $1$\,mm thick plane of Gd$_2$O$_2$S was performed to calibrate the energy deposition within the material and compare it to theoretical predictions. The material parameter is defined as \cite{Landau:1965, Bischel:1988}
\begin{equation}
    \xi_L = 2\pi N_A r_e^2 m_e^2 \left\langle \frac{Z}{A}\right\rangle\frac{\rho L_x}{\beta^2} = 0.154 \left\langle \frac{Z}{A}\right\rangle\frac{\rho L_x}{\beta^2}\, \mathrm{MeV},
\end{equation}
where $N_A$ is Avogadro's number, $r_e$ is the classical electron radius, $\langle Z/A\rangle$ is the average atomic number--mass ratio, $\rho L_x$ is the density of the scintillator in $\mathrm{g/cm^2}$ and $\beta$ is the velocity parameter of the incident electron. 

\begin{figure}[h!]
    \centering
    \includegraphics[width=0.8\linewidth]{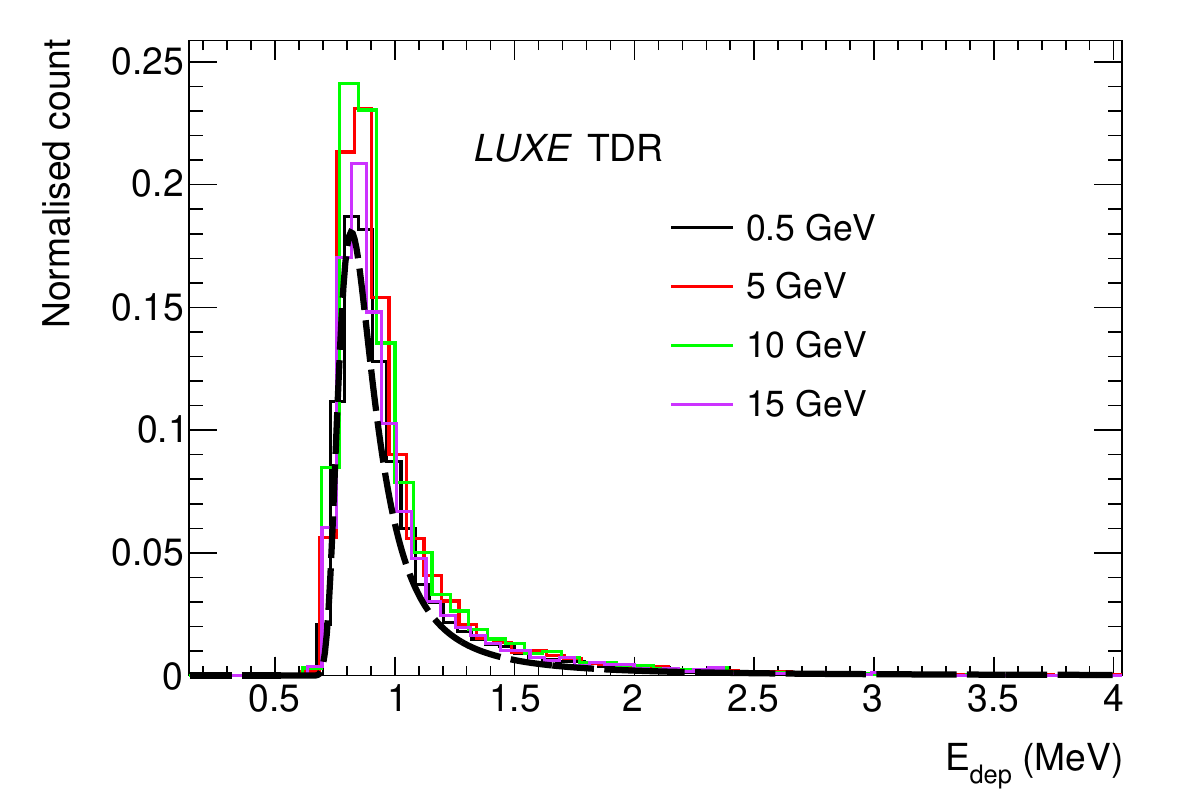}
    \caption{Normalised plot of the energy deposition spectrum in a 1.0\,mm thick Gd$_2$S$_2$O screen due to electron beams of different incident energies. Shown in black (dashed) is the theoretical Landau distribution for the deposited energy in the screen for an incident energy of 0.5\,GeV, following the formula of Ref.~\cite{Koelbig:1984}.}
    \label{fig:3}
\end{figure}

The Vavilov parameter $\kappa = \xi_L/E_{max} \sim 10^{-5}$ for $L_x = 1$\,mm of Gd$_2$O$_2$S, hence the energy deposition in the scintillator is expected to follow a Landau distribution. In the ultrarelativistic limit, the most probable value (MPV) of this distribution is given by \cite{Workman:2022}
\begin{equation}\label{landau_mpv}
    \Delta_{MPV} = \xi_L\left[\ln\frac{2m_e\xi_L}{\omega_p^2} + 0.200\right],
\end{equation}
with $\omega_p$ the plasma energy. The parameter $\xi_L$ also gives the width of the Landau distribution. Use of Eq.~(\ref{landau_mpv}) for the different values of incident energy show that the energy deposition in the scintillator reaches a Fermi plateau (where energy deposited in the scintillator no longer increases with increasing incident particle energy) of $0.82$\,MeV, in good agreement with the modal values seen in Fig.~\ref{fig:3}.

In a first approximation, it can thus be assumed that incident electrons of energy $> 1$\,GeV deposit constant energy in the detector, irrespective of the particle initial energy. Assuming a light yield as stated in Table \ref{Table:1}, the scintillation yield per incident electron is thus approximately $5\times10^4$ optical photons per particle. The scintillation yield is virtually isotropic and we can assume that light emitted in approximately 0.4\,srad can be collected and imaged, i.e.\ approximately 3\% of the total scintillation yield. With an ideal transmission of the optical system and a typical quantum efficiency of intensified CCD cameras in the optical region of the order of 10\%, we would then expect a signal of $\approx 10^2$ photons per incident particle at the camera. As discussed in the next section, while this represents a measurable photon yield, it is a relatively low signal, requiring the use of amplified CCD cameras.

\section{Expected Performance}
\label{sec:gs:performance}

The gamma ray spectrometer aims to reconstruct the energy spectrum of an incoming gamma ray beam by providing a high-resolution measurement of the spectra of electron and positron pairs generated during the propagation
of the gamma ray beam through a 10\,$\mu$m thick tungsten target. The system design results in only a small fraction ($\leq$ 0.1\%) of the gamma ray photons interacting to produce electron--positron pairs via the Bethe--Heitler process while the majority of the beam is left unchanged, to allow for further gamma ray diagnostics downstream of the spectrometer.

If we consider a photon beam with energy spectrum $f(\omega) = \frac{dN_\gamma}{d\omega}$ that collides with a converter target of thickness $\Delta z$, electrons and positrons are produced with energy spectra, $g(\epsilon_{\mp}) = \frac{dN}{d\epsilon_\mp}$, given by
\begin{equation}
    g(\epsilon_{\mp}) = \mathcal{A}\int d\omega\, \frac{d\sigma}{d\epsilon_\mp} f(\omega) \, , \hspace{5mm} \mathcal{A} = \frac{\rho N_A \Delta z}{A} \, , \label{eq:e-spectra}
\end{equation}
where $\frac{d\sigma}{d\epsilon}$ is the energy differential cross section of the pair production process, $\rho$ is the material density, $A$ is the atomic mass of the converter and $N_A$ is Avogadro's number. 
In the ultra-relativistic limit, where complete screening of the nucleus applies (roughly $> 1$\,GeV) \cite{Tsai:1974, Klein:2006}, the cross section for pair production in the field of a nucleus is given by
\begin{equation}\label{pp-cross-section}
    \frac{d\sigma}{d\epsilon_\mp} = \frac{A}{N_A X_0}\left[1 - \frac{4}{3}\frac{\epsilon_\mp}{\omega}\left(1-\frac{\epsilon_\mp}{\omega}\right)\right]
\end{equation}
where $\epsilon_{\mp}$ are the electron and positron energies, respectively, such that $\epsilon_- + \epsilon_+ = \omega$. From this point, we assume natural units.
\begin{table}[]
    \centering
    \begin{tabular}{|c|c|c|}
    \hline
    $\xi$ & $\mathcal{N}_\gamma$ ($10^6$/BX) & $\mathcal{N}_{e^\mp}$ ($10^6$/BX)  \\
    \hline
    \hline
     $0.15$ & $8.6$ & $0.2$\\
     $0.5$ & $93.5$ & $2.1$\\
     $3.0$ & $823.1$ & $18.3$\\
     $7.0$ & $757.6$ & $16.8$\\
     \hline
    \end{tabular}
    \caption{Approximate values for the expected number of photons to be produced at the IP for various values of the laser intensity parameter, $\xi$, assuming circular polarisation, and the corresponding number of converted $e^-e^+$ pairs from a 10\,$\mu$m tungsten foil.}
    \label{xi_vs_ngammas}
\end{table}

Before discussing a more rigorous procedure for the retrieval of the gamma ray spectrum, it is useful to provide first estimates of the number of pairs produced in the gamma ray spectrometer. As an example, simulations of the $e$-laser interactions for an intensity $\xi = 0.5$ indicate that $\sim 9.4\times10^7$ Compton-scattered photons are produced (values for a range of different $\xi$ can be seen in Table~\ref{xi_vs_ngammas}). Upon reaching the converter, a fraction of these photons will be converted to electron--positron pairs to be measured in the spectrometer. Integrating Eq.~(\ref{eq:e-spectra}) over all $e^{\mp}$ energies gives the total number of pairs produced
\begin{equation}
    \mathcal{N}_{e^{\mp}} = \frac{\rho N_A\Delta z}{A}\int d\omega\, \sigma(\omega)f(\omega).
\end{equation}

\begin{figure}[h!]
\centering
	\includegraphics[width=0.55\textwidth]{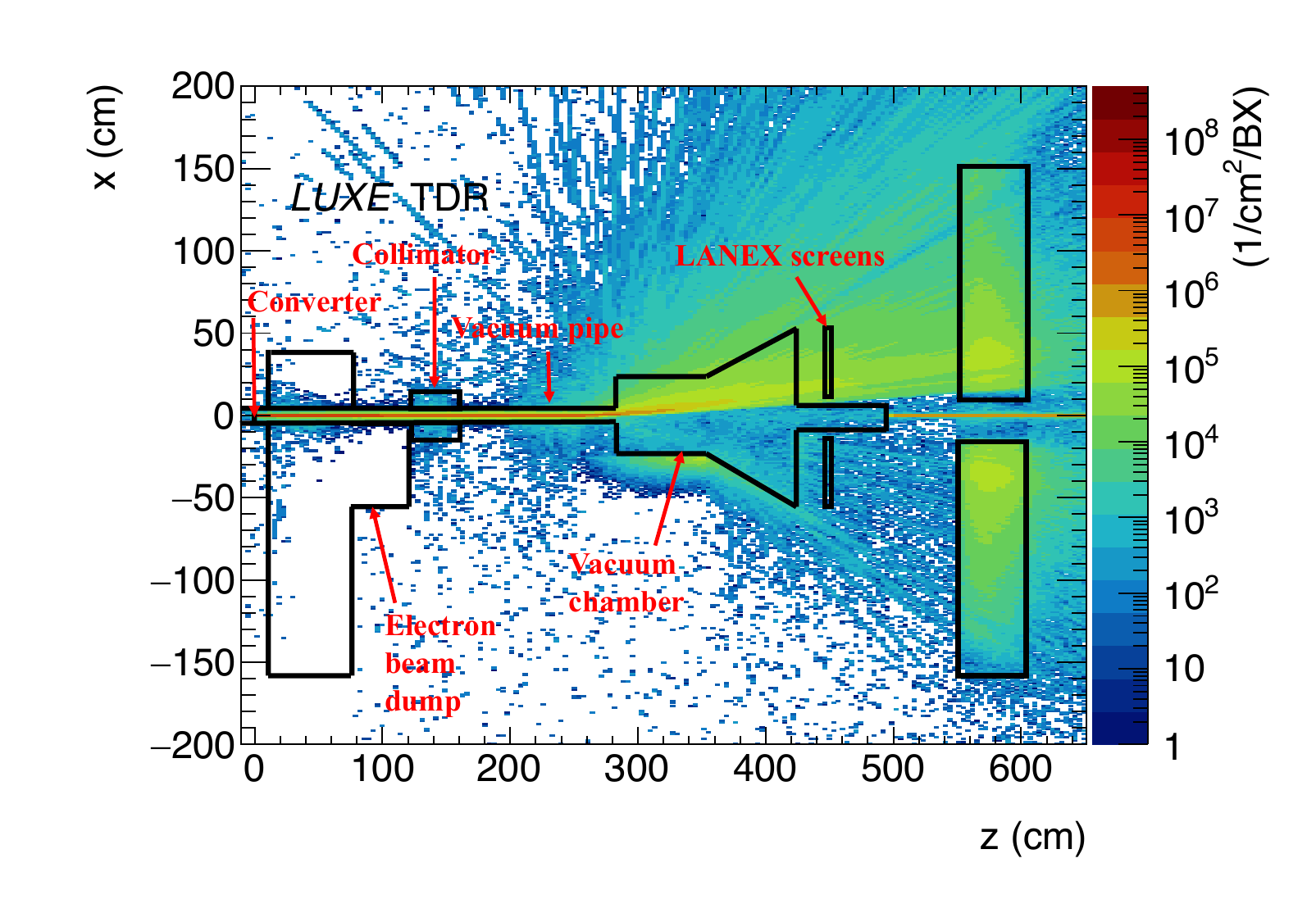}
	\includegraphics[width=0.55\textwidth]{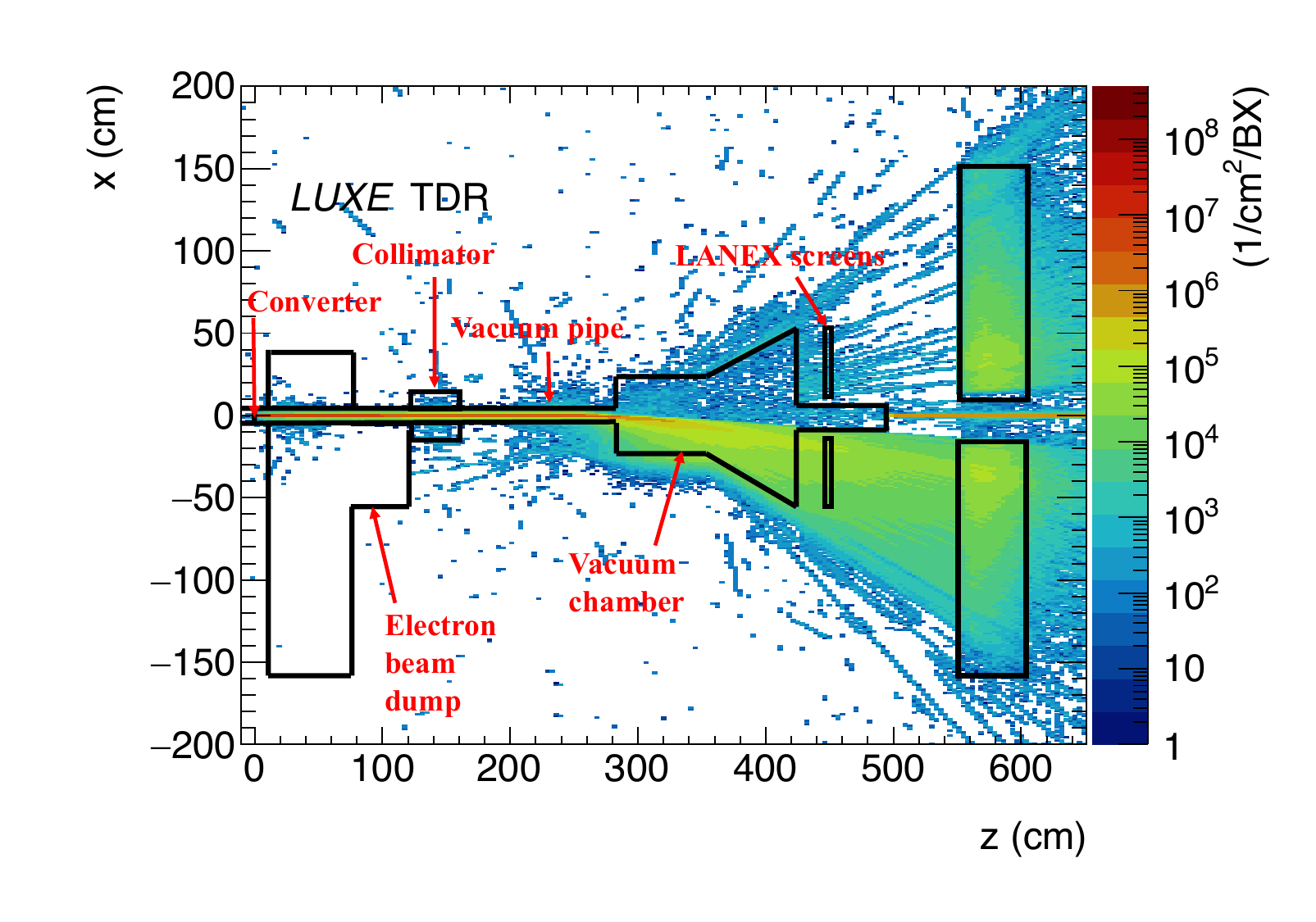} 
	\includegraphics[width=0.55\textwidth]{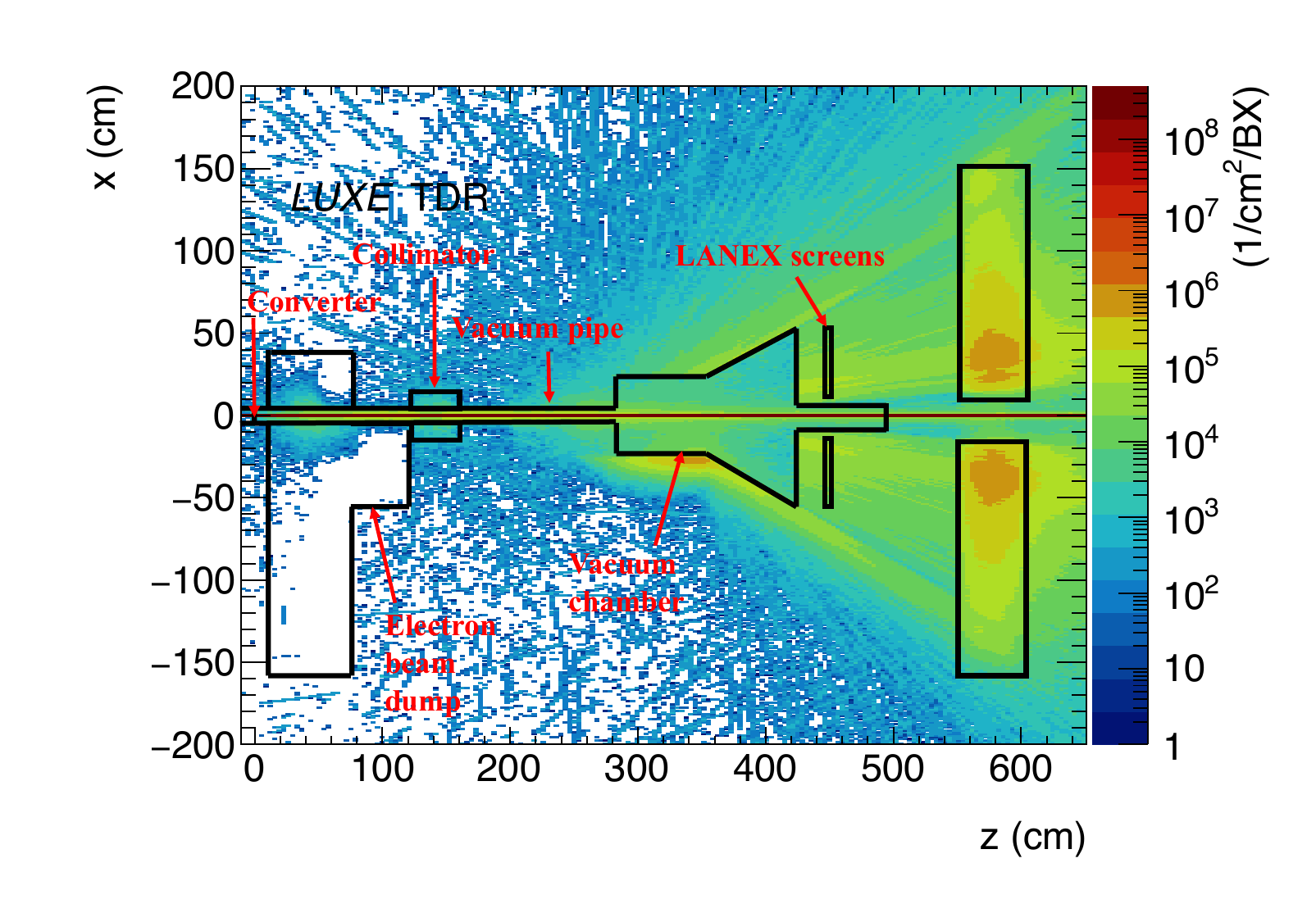}	
\caption{Simulated (top) positron, (middle) electron and (bottom) photon spatial distribution inside the spectrometer assuming a primary inverse Compton scattering photon beam generated during the interaction of the 16.5\,GeV electron beam with the focussed laser with an intensity $\xi=0.5$ (blue line in Fig.~\ref{Compton_spectra}).}
\label{maps}
\end{figure}

This can be approximated by integrating Eq.~(\ref{pp-cross-section}) to give the total pair production cross section: $\sigma = 7A/9N_AX_0$ where $X_0$ is the target density in $\mathrm{g/cm^2}$. This can be rewritten as:
\begin{equation}
    \mathcal{N}_{e^{\mp}} = \frac{7\rho \Delta z}{9X_0} \mathcal{N}_\gamma = \frac{\Delta z}{\lambda_{PP}}\mathcal{N}_\gamma,
\end{equation}
where $\lambda_{PP}$ is the mean free path of the pair production process in a particular material. For a $10\,\mu$m tungsten foil, we therefore expect  $\sim 2\times 10^6$ $e^{\pm}$ pairs to be produced (approximately 2\% of the photons are converted to pairs). This relatively low signal is the main reason why amplified CCD cameras are required to image the scintillators.

\begin{figure}[h!]
    \centering
    \includegraphics[width=0.8\textwidth]{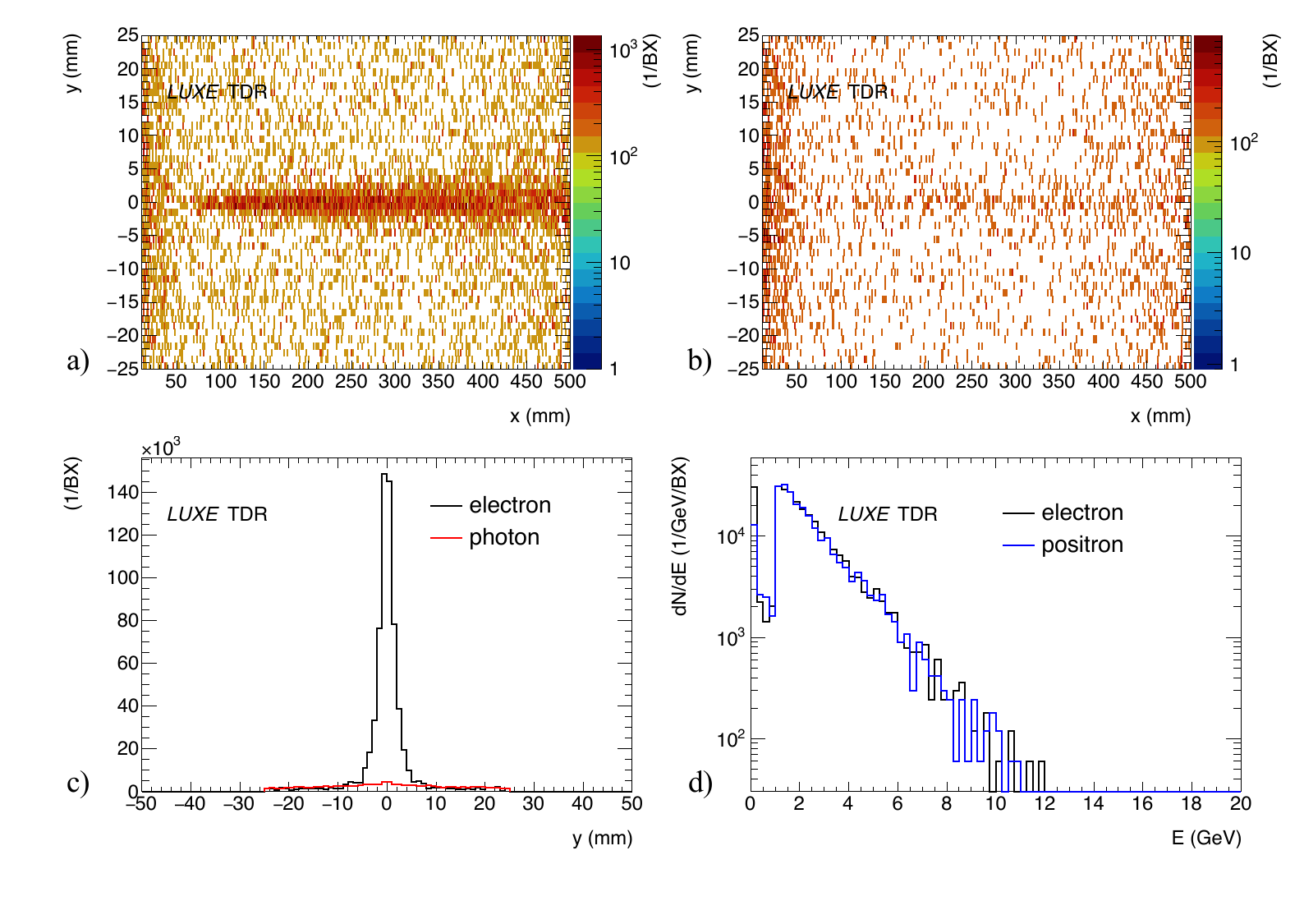}
    \caption{Particle number for (a) electrons and photons and (b) photons only per bunch crossing on the upper LANEX screen ($x-y$ plane). (c) number of electrons (black) and photons (red) from frames (a) and (b) integrated along the $x$ axis. (d) spectrum of all (signal and background) electrons impinging onto the right LANEX screen (black) and all positrons impinging onto the left LANEX screen (blue) per bunch crossing. Results shown here are for $\xi = 3$. The spectra have been linearly binned in energy according to the resolution available at $10$\,GeV from Fig.~\ref{res}\,(right). }
    \label{dep}
\end{figure}

These are the electron--positron pairs that, once dispersed by the dipole magnet, will impinge onto the two LANEX scintillator screens. As an example of the particle spatial distribution in the spectrometer, we show in Fig.~\ref{maps} the (top) positron, (middle) electron and (bottom) photon distributions obtained from Monte Carlo simulations assuming the primary inverse Compton scattering photon beam generated during the interaction of the 16.5\,GeV electron beam with the focussed laser with an intensity $\xi=0.5$ (blue line in Fig.~\ref{Compton_spectra}). These distributions are integrated along the transverse dimension $y$; in order to get a more precise estimate of the signal-to-noise on the scintillators, we plot in Fig.~\ref{dep}, the number of (a) electrons and photons and (b) photons only impinging onto the LANEX scintillator in one bunch crossing. By integrating along $x$ (the dispersed axis in the spectrometer) these particle numbers, one obtains an average signal-to-noise ratio in the region of interest exceeding 10 (see (c)). The electron and positron spectra extracted from Fig.~\ref{dep}(a) are shown in Fig.~\ref{dep}(d). The gamma ray spectrum can be retrieved by inverting Eq.~(\ref{eq:e-spectra}) using the measured electron or positron spectrum. 

The inversion of Eq.~(\ref{eq:e-spectra}) can be performed by first defining the kernel integral operator: 
\begin{equation}
    \mathcal{K} \rightarrow \mathcal{A}\int d\omega\, \frac{d\sigma}{d\epsilon}\, \circ ,
\end{equation}
Deconvolution can then be posed as the inversion of the linear operator problem
\begin{equation}\label{OperatorProblem}
    g(\epsilon) = \mathcal{K} f(\omega).
\end{equation} 
This problem can in principle be solved by direct back substitution \cite{Press:2007} as the kernel is triangular ($\mathcal{K}\rightarrow 0$ if $\omega < \epsilon$); however, this method would lead to a significant uncertainty in the retrieved photon spectrum (not discussed here for the sake of brevity). However, combined with a statistical approach based on a Bayesian interpretation of regularisation \cite{Piana:1994, Djafari:1993, Djafari:1996}, it can provide an accurate reconstruction of the original photon spectrum.
To pose the deconvolution problem in a statistical manner, a noise term is introduced to the operator equation
\begin{equation}\label{operator_problem}
    g(\epsilon) = \mathcal{K}f(\omega) + \eta(\epsilon).
\end{equation}

A statistical distribution is then applied to the noise -- this can be influenced by empirical measurements; however, a common choice is Gaussian noise that is independent of the signal, i.e.\ $\eta(\epsilon_i) \sim \mathcal{N}(0, \sigma_b^2)$ where $\sigma_b^2 = \frac{1}{\beta}$ is the variance in the noise and $\beta$ is the equivalent precision. The noise is also assumed to be centred with mean zero. From Eq.(\ref{operator_problem}),
\begin{eqnarray}
    & & g(\epsilon) - \mathcal{K}f(\omega) \sim \mathcal{N}(0, \sigma_b^2I) \nonumber \\
    & & \Rightarrow \, \pi(g| f, \beta, \mathcal{K}) = \frac{1}{Z_1(\beta)}\exp\left(-\frac{1}{2}\beta|| g(\epsilon) - \mathcal{K}f(\omega)||^2_2\right).
\end{eqnarray}
Here, $\pi(g| f,\beta, \mathcal{K})$ is the likelihood function of the problem and $Z_1(\beta) = \left(\frac{2\pi}{\beta}\right)^{\frac{n}{2}}$ is the normalisation factor with $n = \dim g(\epsilon)$, and $I$ is the identity matrix. To apply Bayesian methods, a prior distribution must be stated; we make only very general assumptions about our Bayesian prior, namely that the values of the spectrum cannot be negative, i.e.\ $f(\omega) \in [0,\infty)$. As such, the maximum entropy prior satisfying this condition is chosen: a normal distribution with unknown variance to be inferred, and mean determined by the back substitution solution to Eq.~(\ref{OperatorProblem}). This ensures that we maintain generality and prevent bias towards a desired solution, making the spectrometer useful for measuring any general gamma ray spectrum.
An application of empirical Bayes inference which involves maximising the marginal likelihood function can then be used. 

\begin{figure}[h!]
    \centering
    \includegraphics[width=0.8\textwidth]{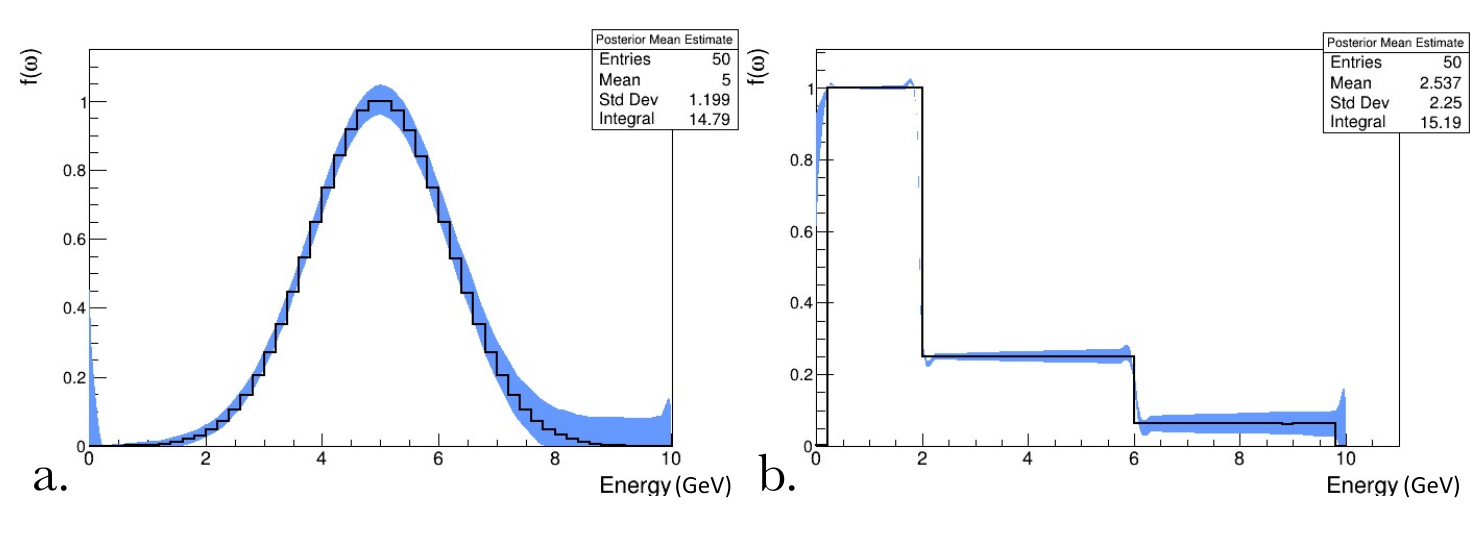}
    \caption{Original (black) and retreived (blue) photon spectrum in the case of a (a) Gaussian  or a (b) step-like gamma ray spectrum.}
    \label{deconv}
\end{figure}

The error associated with the reconstruction of the spectrum is represented by the highest posterior density (HPD) interval about the solution. The $100\,(1-\alpha)$\% HPD interval for $x$ is a particular $100\,(1-\alpha)$\% level credible interval defined as follows: let $\mathcal{C} \subseteq X$ be a subset of the support, satisfying $\mathcal{C} = \{x\in X : \pi(x) \geq k\}$ where $k$ is the largest value such that
\begin{equation}
    \int_{\mathcal{C}} dx\, \pi(x) = 1-\alpha.
\end{equation}
An example of the performance of the deconvolution algorithm is given in Fig.~\ref{deconv}. In these tests, a Gaussian or a step-like function have been assumed for the spectrum of the gamma ray beam, which was then propagated through the spectrometer. The resulting electron and positron spectra were then deconvolved using the Bayesian approach outlined above, to give the retrieved gamma ray spectrum, with associated uncertainties (blue bands in Fig.~\ref{deconv}). As can be seen from these two examples, the spectrometer is able to reconstruct the gamma ray spectrum with a high fidelity, and able to precisely identify the edges in the spectrum. For example, the edge at 2\,GeV in Fig.~\ref{deconv}(b) is identified with an uncertainty of $\pm$\,120\,MeV ($\approx$ 6\%), while the uncertainty in identifying the edge at 6\,GeV is of the order of $\pm$\,170\,MeV ($\approx$ 3\%). It must be noted that these uncertainties, associated with the level of confidence in retrieving the gamma ray spectrum, are much larger than the intrinsic resolution of the spectrometer (see Fig.~\ref{res}), and are thus the dominant factor in estimating an uncertainty in reconstructing the gamma ray spectrum from the detector.
A higher statistics retrieval of gamma ray spectra expected to be generated in LUXE (Fig.~\ref{Compton_spectra}) could then be achieved by accumulating over several bunch crossings. 

The reconstruction ability of the Bayesian deconvolution algorithm and the accuracy of edge location has also been tested for exemplary gamma ray spectra expected in the $e$-laser configuration of the LUXE experiment, using the \textsc{Ptarmigan} MC data for two $\xi$ values: $0.5$ and $7.0$ of the scattering laser. The photon spectrum produced at the IP, $f_0(\omega)$, was convolved with the Bethe--Heitler cross section, reproducing the effect of the converter target in producing $e^+e^-$ pairs, to produce the `true' electron spectrum, $g_0(\epsilon) = \mathcal{K}f_0(\omega)$. 

\begin{figure}[h!]
\centering
\includegraphics[width=0.8\textwidth]{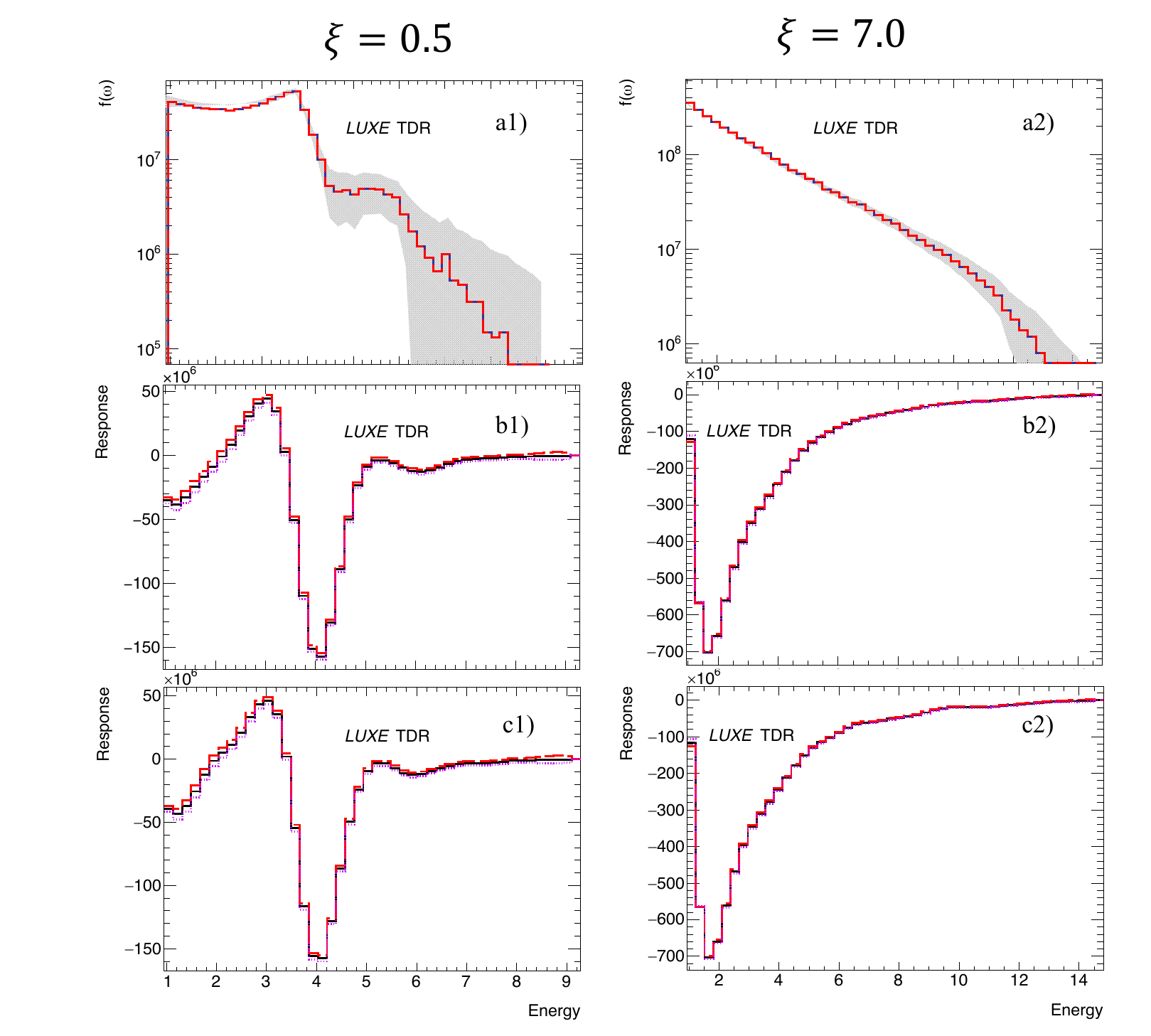}
\caption{Plots showing: (a1) the deconvolution of $g(\epsilon)$ (blue, solid) with the original photon input $f_0(\omega)$ (red, dashed) and the 95\% HPD interval (grey) for $\xi = 0.5$ and $\ell = 1$. Also shown is the response of the deconvolved spectrum (blue) and the lower and upper HPD bounds (red, dashed and purple dotted, respectively) to the FIR filter for (b1) $\ell = 1.0$ and (c1) $\ell = 10^6$. The equivalent for $\xi = 7.0$ are shown in (a2), (b2) and (c2), respectively.}
\label{deconv_analysis}
\end{figure}

A perturbed spectrum was then used as the actual input for the deconvolution algorithm:
\begin{equation}
    g(\epsilon) \sim \mathcal{N}[g_0(\epsilon), \ell^2 g_0(\epsilon)],
\end{equation}
where $\ell$ is a parameter controlling the size of the bin error. Two exemplary cases will be considered; for $\ell = 1$, the large-number limit of the uncertainty in a Poissonian process is applied. On the other hand, $\ell=10^6$ would correspond to an extremely noisy electron spectrum.  The deconvolution algorithm was then applied to $g(\epsilon)$ to retrieve the solution to Eq.~(\ref{OperatorProblem}). Similar to the analysis of the IP high-rate electron detectors, an FIR (finite impulse response) filter (see section \ref{subsec:scint_fir}) was used to determine the position of the Compton edge -- this corresponded to the minimum in the filter response. The filter was applied to both the solution and the lower and upper bounds of the HPD interval describing the uncertainty in reconstruction, giving extremely similar results.

Examples of the results obtained are shown in Fig.~\ref{deconv_analysis}, where the first column depicts results for $\xi=0.5$ and the second for $\xi=7$. For each column, the first frame shows a comparison between the original gamma ray spectrum (red dashed line) and that obtained from deconvolving the perturbed electron spectrum (blue solid line), together with the bands associated with 95\% HPD interval. As one can see, there is high fidelity in reproducing the spectrum at relatively high fluxes, with error bands increasing at the high-energy end of the spectrum. The large uncertainty at the high-energy end of the spectrum is mainly due to the low particle flux in that region. However, one can see a precise determination of the first Compton edge, for example at approximately 4\,GeV for $\xi=0.5$. This precise determination is further elucidated by applying the FIR filter to the signal (second and third frame in each column of Fig. \ref{deconv_analysis}). The two cases in each column correspond to $\ell=1$ and and $\ell=10^6$. In both cases, minima in the function identify the position of edges in the spectrum. For example, for $\xi=0.5$, the FIR filter identifies a Compton edge at $4.1 \pm 0.1$\,GeV (relative uncertainty of 2.4 \%) for $\ell=1$. Even in the extreme case of $\ell=10^6$, the FIR filter still identifies a Compton edge at $3.9 \pm 0.1$\,GeV (relative uncertainty of 2.6 \%).

It is interesting to note that the FIR is able to identify edges in the spectrum even in the case of $\xi=7$, where any edge might not be apparent by naked eye (second column in Fig. \ref{deconv_analysis}). In this case, an edge is identified at $1.8 \pm 0.1$\,GeV (relative uncertainty of 5.5 \%), where the uncertainty in energy mainly arises from the size of the energy bin used.

As a final remark, it has to be noted that the deconvolution algorithm outlined above relies on a precise knowledge of the thickness of the converter target for each bunch crossing. However, the gamma ray irradiation might in principle degrade the target in time, by causing surface ablation. To check this possibility, we have carried out Monte Carlo simulations of the dose deposited in the target during each bunch crossing for the whole range of expected gamma ray parameters. The highest gamma ray yield is expected for $\xi=7$ (see Table~\ref{xi_vs_ngammas}), with $N_\gamma \simeq 10^9$. Monte Carlo simulations indicate an average energy deposited per photon of $E_{\mathrm{dep}} = 23$\,keV over an area of $3\times10^{-2}$\,mm$^2$. This results in a maximum fluence on the target surface of $10^{-2}$\,J/cm$^2$. This fluence is much smaller than the threshold fluence of tungsten, which is of the order of 0.4\,J/cm$^2$ for a single irradiation and goes down to a minimum of $0.2$\,J/cm$^2$ for sustained irradiation at 1\,Hz. As such, it is predicted that the target will be able to withstand sustained gamma ray irradiation without noticeable degradation, consistently with preliminary tests carried out at the Central Laser Facility and the Apollon laser facility. During the initial phases of calibration and alignment, the resistance to radiation of the converter target will be monitored closely, to confirm the numerical expectations and experimental tests. This can be done by directly imaging the front surface of the converter during irradiation, as discussed in more detail in the previous section.

\section{Installation, Commissioning and Calibration}
\label{sec:gs:installation}

\begin{figure}[h!]
    \centering
    \includegraphics[width=0.8\textwidth]{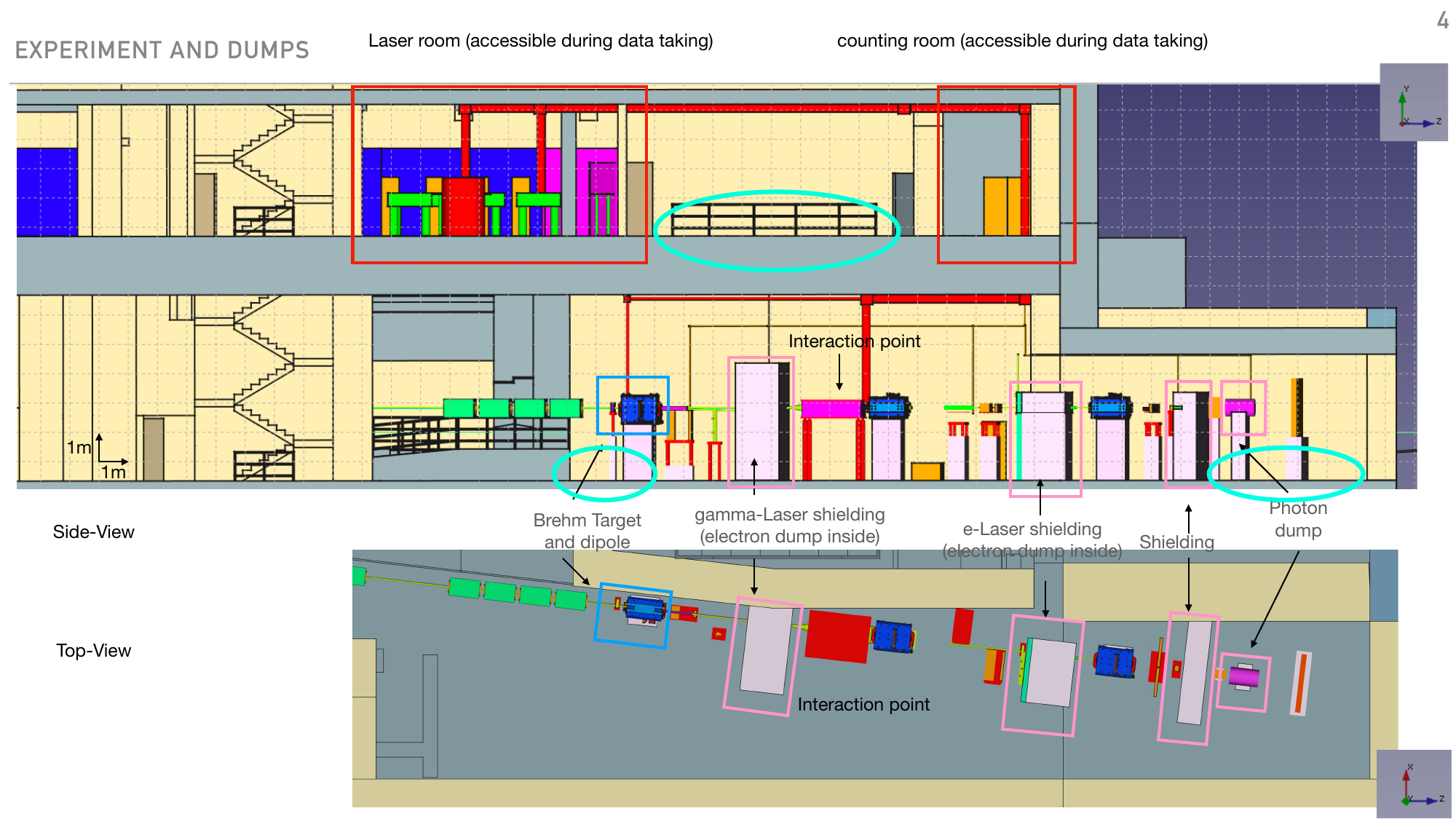}
    \caption{Schematic diagram of LUXE at the EuXFEL showing the experimental chamber and surrounding rooms. The potential locations of the cameras for imaging the scintillator screens are marked in cyan. }
    \label{luxe_ccd_camera_locations}
\end{figure}

Here, we provide preliminary estimates of the time and support required for the installation, calibration, and testing of the gamma ray spectrometer in the LUXE experiment. We expect two weeks to be required to install the vacuum chamber, detectors and  optics in the area and to connect the CCD cameras to the main data acquisition and triggering systems. Three potential locations for these cameras are highlighted in Fig.~\ref{luxe_ccd_camera_locations}. Two areas within the experimental chamber that do not have a large presence of background are: in front of the bremsstrahlung target and dipole magnet and behind the photon dump located at the rear of the experiment. Additionally, there is a counting room located above the experimental hall with suitable holes already drilled on its floor. This room  could also host the CCD cameras, even though at a cost of a longer optical beamline for the scintillation light. The determination of the ideal placement for the cameras is currently under investigation, even though the preferred location will be behind the photon dump, with an optical path from the scintillators to the cameras of approximately 5\,m.

Ideally, the lead collimator and the detectors should be installed on translation stages for fine alignment, which should take approximately one week of dedicated access.  Two weeks of dedicated access should then be allocated for testing and calibration.
For both the alignment and calibration, we aim to use a bremsstrahlung photon source obtainable by inserting a thin converter upstream of the interaction chamber. During alignment, a scintillator screen can be placed downstream of the collimators, on axis; best alignment will then be achieved by maximising the photon signal on the scintillator. Given the high flux of bremsstrahlung photons, no amplification is needed for the camera used during this alignment procedure. Once the system is properly aligned, the same bremsstrahlung source can be used to calibrate the spectrometer, by recording the electron and positron signal on the detectors arising from a known source. During this time, we will also confirm experimentally the numerical expectation that sustained gamma ray irradiation will not result in a noticeable degradation of the converter target.

Given the affinity of this detector with the gamma ray profiler, some of the calibration and testing of these two detectors can be carried out in parallel. Overall, we would thus require approximately five weeks of access to install, test, and calibrate the whole system. This is a conservative estimate, which would mostly depend on the time required to install and test large components of the spectrometer, such as the vacuum chamber and the dipole magnet. During this whole period, we would require specific technical support from the facility, especially  engineering aspects such as handling and installing heavy elements, connecting the chamber to a vacuum pump system, and connecting the spectrometer to the general triggering system and the data acquisition system. 

Finally, decommissioning of the system is expected to be straightforward, with iCCD cameras to be returned to QUB for further use and the vacuum chamber and dipole magnet to be kept at DESY.

\section*{Acknowledgements}

The work presented in this document has been partially supported by the Engineering and Physical Sciences Research Council (grant No.: EP/T021659/1 and EP/V049186/1).

\printbibliography
\end{refsection}


\chapter{Gamma Beam Profiler}
\label{chapt9}
\begin{refsection}

\chapterauthors{M. Bruschi, 
A. Sbrizzi, F. Lasagni Manghi \\
INFN and University of Bologna, Bologna (Italy)\\ \\
M. Benettoni, S. Bonaldo, F.~Dal~Corso, U. Dosselli, M. Giorato, P. Giubilato, P. Grutta, S. ~Mattiazzo, M. Morandin, A. Paccagnella, D.~Pantano, G. Simi, S. Vasiukov \\
INFN and University of Padova, Padova (Italy)\\ \\
N. Cavanagh, K. Fleck, G. Sarri\\
School of Mathematics and Physics, The Queen’s University of Belfast, Belfast (UK)}\\
\date{\today}

\begin{chaptabstract}
The present chapter describes a gamma ray beam profiler to be installed as part of the diagnostic suite of the LUXE experiment.  The instrument is designed to record the transverse profile of high-flux and high-energy gamma ray beams, with a predicted spatial resolution of 5\,$\mu$m. The gamma ray beam will be produced in two modes. The first is via inverse Compton scattering during the propagation of the EuXFEL electron beam through the focus of an intense laser. In this case, the profiler is implemented in order to provide precise on-shot information about the laser intensity at the interaction point and, therefore, about the quality of the collision. The profiler can also be used to characterise bremsstrahlung photons during the propagation of the same electron beam through a thin converter foil, providing pivotal information to model the interaction in the gamma ray laser configuration.
\end{chaptabstract}

\section{Introduction}
\label{gbp:intro}


In Fig.~\ref{fig:layoutdet}, a scheme for the LUXE detectors is shown. Several diagnostics will be implemented in the experiment. As well as the detectors described in the other chapters, it is necessary to measure the transverse angular distribution of the gamma ray photons, which will yield precious information on the interaction, such as the  quality of overlap and intensity of the laser with the electron beam at the interaction point. For this purpose, a gamma beam profiler (GBP) detector is proposed. The expected high flux of photons per bunch crossing (BX) implies that such a detector must be radiation hard at the level of several MGy per year, together with the capability of reconstructing the beam profile position and width with a spatial resolution of the order of $5\,\mu$m.

It is important to mention that the GBP will be \emph{the only LUXE detector} able to provide information on the spatial distribution of the produced gamma-ray photons. For this reason, it is planned to have a very reliable detector capable of  providing the beam profile information to the experiment from the very beginning, during the first commissioning, when the optimisation of the $e$--laser interaction will be of utmost importance to understand the quality of the interactions themselves and the shot-by-shot stability. Moreover, the GBP information should be continuously available  during the running periods, and the possibility to lose data due to detector failures should be minimised. For this reason, we are planning to install (possibly from the very beginning) two identical detector stations, along the photon beam and close to each other. In this way, the following goals will be achieved:

\begin{itemize}
    \item minimise the possibility to lose data for detector failures. The running of the EuXFEL machine will allow only two accesses per month to the experimental area, and this could cause important data loss to the GBP, in case of some failure in the detectors;
    \item improve up to a factor $\sqrt 2$ the precision in the beam profile reconstruction;
    \item since we are proposing that the detectors are mounted on movable tables having micro-metric step precision (see Section \ref{gbp_subsec:mechanics}), the possibility to displace one detector with respect to the other will allow assessment of the detector precision in reconstructing the beam profile during the special calibration runs proposed in Section \ref{gbp_subsec:calibrationstrategy}. 
\end{itemize}

The GBP was already introduced in the LUXE CDR~\cite{luxecdr}. The present document will present an enlarged description of that detector adding many Monte Carlo (MC) studies not available at the time of writing the CDR. The main substantial differences of the present version with respect to the CDR are shortly reported in the following section.

\subsection{Main Differences with respect to the CDR}

In the CDR, the possibility to have the exit window of the beam-pipe at around 11\,m from the IP was still under discussion. In the CDR the proposed beam line  ended at about 6\,m from the IP. Under this assumption, two options were considered: either placing the GBP stations 6\,m away from the interaction point outside of the vacuum region, or close to the photon beam dump, at a distance of 11\,m. The former solution minimised multiple scattering and conversions in air, but had the drawback of narrower beam profiles, which are harder to reconstruct. In the present LUXE design the exit window of the photon beam-pipe is at 11\,m and the detector position is therefore ideal to optimise all the above mentioned effects.

Besides this, in the CDR the detector design was mostly based on a simplified standalone MC simulation, while in the present version realistic beam profile distributions and detector digitisation have been implemented (see Section~\ref{sec:gbp:performance}). 

Finally, Compton scattering in the laser field was modelled in the CDR using the so-called local constant field approximation (LCFA), which assumes that the photon emission occurs at a constant electric field perpendicular to the electron momentum. Due to the oscillatory nature of the laser electric field, this approximation is valid only as long as the photon formation length $\ell_\gamma$ is much smaller than the laser wavelength $\lambda$, a condition that translates in assuming that the laser normalised vector potential is much larger than unity. In this document, a more accurate approximation has been adopted, the so-called local monochromatic approximation (LMA), which only assumes a slowly varying laser envelope, a condition that is well respected over the whole range of intensities to be accessed in the LUXE experiment. These new more accurate simulations of the gamma-ray beam profiles evidence a clearer relation between the laser intensity and the angular distribution of the Compton-scattered photons, allowing a more precise reconstruction of the laser interaction at the interaction point.

\section{Requirements and Challenges} \label{sec:gbp:requirements}

In the following, a first evaluation of the performances of the GBP in reconstructing the beam profile and the $\xi$ parameter are described.

 \subsection{Beam Profile Reconstruction}
 
The GBP will be able to record the charge delivered to each one of its readout channels for each BX. This will allow precise measurement (at percent level) of all the main features of the beam profile distribution (in both the $x$ and $y$ directions): the amplitude ($A_x$ and $A_y$), the position ($x_0$ and $y_0$) and the rms value ($\sigma_x$ and $\sigma_y$). This could be performed either with a fit to the distribution (or to the main part of it, excluding the tails, for instance) or by direct calculation using the charge deposited in each strips values (centre of gravity type algorithms).

Data quality plots reporting $A_x$ vs $A_y$, $x_0$ vs $y_0$ and $\sigma_x$ vs $\sigma_y$, for instance, will be a straightforward and effective way to monitor precisely the shot-by-shot stability of the interactions of the electron or gamma beams with the laser light and to provide valuable and unique online information both in the LUXE commissioning and running phases. An example of the GBP performances in reconstructing the beam profile is reported in Section~\ref{sa_simulation}.

\subsection{Reconstruction of the $\xi$ Parameter}

For an electron in an external laser field, the nature of Compton scattering is governed by the relativistic invariant normalised vector potential (also referred to as laser intensity parameter in the following): 

\begin{equation}
\xi = eE_L/m_e\omega_L c\,,   
\end{equation}
where $E_L$ and $\omega_L$ are the laser electric field and carrier frequency, respectively. For $\xi\ll 1$, the electron oscillations along the laser polarisation are non-relativistic, and thus magnetic contributions to the electron motion can be neglected. In this case, the electron will then oscillate in phase with the laser, at the frequency $\omega_L$. If the electron in the laboratory frame has a Lorentz factor $\gamma_e$, this oscillation will be accompanied by emission of photons at a frequency, in the laboratory frame, of $\omega_{ph} = 4\gamma_e^2\omega_L$ within a cone angle $\theta = 1/\gamma_e$ around its instantaneous velocity. At higher laser intensities ($\xi>1$), the electron oscillation becomes relativistic and magnetic contributions in the electron dynamics cannot be neglected. Quantum mechanically, this can be interpreted as the incident electron absorbing more than one laser photon in one formation length. This is a situation equivalent to a wiggler with a strength parameter $K>1$.

For $\xi>1$ and neglecting for now radiative energy losses for the electron, the photon emission becomes asymmetric, with a cone angle in the direction perpendicular (parallel) to the laser polarisation axis of $\theta_\perp = 1/\gamma_e$ ($\theta_{||} = \xi/\gamma_e$). The above assumes a perfectly collimated (no divergence) primary electron beam, a reasonable assumption for the LUXE experiment that has been confirmed by numerical simulations (see Section~\ref{sec:gbp:performance}). In this case, the ellipticity of the photon beam ($\eta_{ell}$) would thus represent a direct measurement of the normalised intensity of the laser beam at the moment of emission: $\eta_{ell} = \theta_{||}/\theta_\perp = \xi$. 

If the non-zero divergence ($\theta_e$) of the electron beam is taken into account, the emission angles can be, in a first approximation, expressed as:
\begin{equation}
\theta_\perp = \sqrt{1/\gamma_e^2 + \theta_e^2} ,  \hspace{5mm}
\theta_{||} = \sqrt{\xi^2/\gamma_e^2 + \theta_e^2}.
\end{equation}

This treatment is valid in a plane-wave approximation and neglecting energy loss of the electron during its propagation through the laser field. However, for a more realistic treatment of the problem, one must take into account the focussed nature of the laser field and the relatively long duration of the laser pulse (more than 10 laser periods for the first phase of experiments at LUXE). We can assume the laser field in focus to have a Gaussian transverse profile, with a standard deviation that is smaller than the electron beam diameter at the interaction point. In this case, each electron in the beam will experience a different local value for the laser intensity, resulting in a different emission pattern. Moreover, one must take into account that the electron is likely to lose a significant portion of its energy to radiation during its propagation in the laser field. As such, there is a high probability of several emissions from the same electron, each time at a different electron Lorentz factor. This will result in a different energy distribution of the photon beam, which can be modelled as~\cite{Blackburn:PRAB2020}:

\begin{equation}
 \xi^2 = 4\sqrt{2} \zeta \langle\gamma_i\rangle\langle\gamma_f\rangle(\theta_{||}^2-\theta_\perp^2), \label{a0}  
\end{equation}
where $\langle\gamma_i\rangle$ ($\langle\gamma_f\rangle$) is the average Lorentz factor of the electron beam before (after) the interaction with the laser field and $\zeta$ is a geometrical factor that takes into account the relative size of the electron and laser beam. For Gaussian beams, $\zeta$ can be expressed as:
\begin{equation}
 \zeta =  \left(\sqrt{\frac{1+4\rho^2}{1+8\rho^2}}\exp\left(\frac{-\alpha^2}{(1+4\rho^2)(1+8\rho^2)}\right)\right)^{-1}, \label{beta}
\end{equation}
where $\rho = r_b/w_0$ is the ratio between the radius of the electron beam ($r_b$) and the laser focal spot size ($w_0$), and $\alpha = x_b / w_0$ is the ratio between the transverse misalignment between the electron and the laser ($x_b$) and the laser focal spot size. 

Assuming detailed knowledge of the factor $\zeta$ (as obtainable from different diagnostics in the experiment), the uncertainty in retrieving the laser normalised intensity $\xi$ can be expressed as:
\begin{equation}
\frac{\delta \xi}{\xi} = \sqrt{\frac{1}{4}\left(\frac{\delta k}{k}\right)^2+\frac{\sigma_{||}^2+\sigma_\perp^2}{(\sigma_{||}^2-\sigma_\perp^2)^2}(\delta\sigma)^2}, \label{error}
\end{equation}
where $k=4\sqrt{2}\zeta\langle\gamma_i\rangle\langle\gamma_f\rangle$ and $\delta k / k\approx 1$\%. 
However, for a meaningful modelling of the detector, one has to consider that high-energy photons (energy per photon ranging from a few MeV up to a few GeV) will have an energy deposition in a thin low-$Z$ material that is roughly independent of the initial photon energy. As such, the detector described here will be predominantly sensitive to the number of photons per unit area, rather than the energy deposited. As one might expect, the photon number distribution slightly differs from the energy distribution, and it is the former that will be considered hereafter. As shown later in the chapter, the width of the distribution of the number of photons reaching the detector still has a clear and monotonic dependence on the laser intensity $\xi$, which will be used for the intensity retrieval. 

We aim at determining the laser intensity at the interaction point with a maximum relative uncertainty better than 5\%, which in turn implies a relative uncertainty in $\xi$ of $\delta\xi/\xi\leq 2.5$\%. As justified in more detail in Section~\ref{sec:gbp:performance}, a gamma-ray profiler placed 11.5\,m away from the interaction point must then have a spatial resolution of the order of 5--10\,$\mu$m, in order to achieve the required precision in retrieving the laser intensity at the interaction point. 

It must be noted here that other diagnostics will be deployed to estimate the laser intensity on a shot-to-shot basis. For instance, the laser duration and energy per shot will be measured at the laser bay -- i.e., before the laser enters the interaction chamber -- and the laser focal spot distribution will be measured on a shot-to-shot basis. However, one must still take into account that the obtained measurement of the intensity is not necessarily representative of the laser intensity experienced by the electron at the interaction point. Even in the case of perfect overlap between the electron beam and the laser focus, the pointing of a laser of this kind is expected to fluctuate, on a shot-to-shot basis, with a Gaussian distribution centred around the electron beam axis and with a standard deviation that is typically of the order of a few $\mu$rad. This implies slightly different effective laser intensities experienced by the electron beam on each shot. As such, the gamma-ray profiler will not only represent a key diagnostic in the experiment for retrieving the laser intensity at the interaction point, but it will also provide key information on the overlap and synchronisation of the electron beam with the laser, providing simple and reliable discard criteria for sub-optimal events.

\section{System Overview}

The LUXE beam profiler must operate in a stable and reliable way in the presence of a very intense high energy gamma ray flux.
The typical photon energy is of the order of GeV, and in the following it will be assumed that electron--positron pair production in the detector is the dominant process.
\begin{figure}[!htb]
\center
\includegraphics[width=\textwidth]{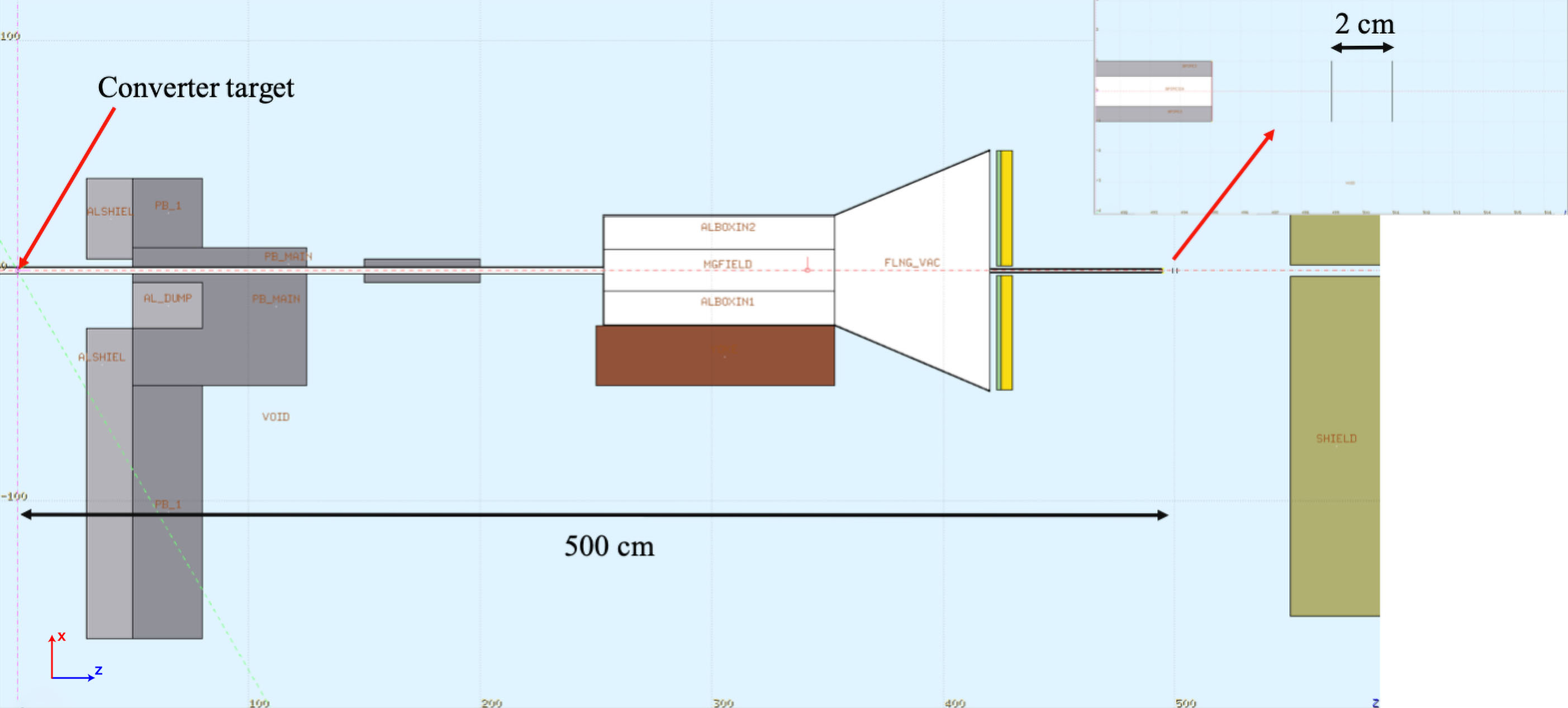}
\caption{The geometry implemented in the FLUKA simulation showing the two planes of sapphire strips placed at 5\,m from the converter target and 11.5\,m from the IP. The red arrow points to a zoom of the planes area. }
\label{fig:geometry}
\end{figure}
In the following, we will adhere to the axis convention adopted in LUXE, where $z$ represents the main beam propagation axis, and the $x$ and $y$ axes are parallel and perpendicular to the laser polarisation, respectively.

The expected beam spatial distributions have been simulated using FLUKA. The geometry implemented in the FLUKA simulation is shown in Fig.~\ref{fig:geometry}. Two $x$ and $y$ planes are placed 2\,cm distance from each other at 11.5\,m downstream of the LUXE IP.

\begin{figure}[!htb]
\center
\includegraphics[width=0.49\textwidth]{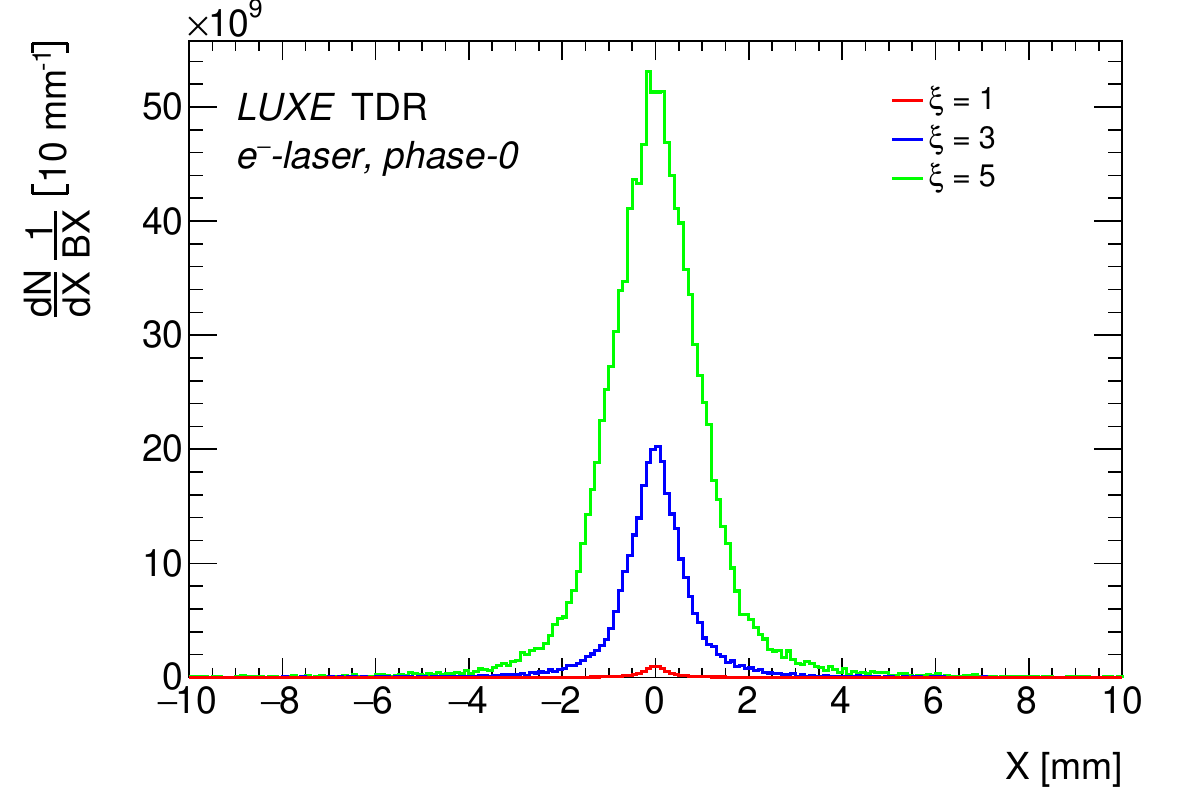}
\includegraphics[width=0.49\textwidth]{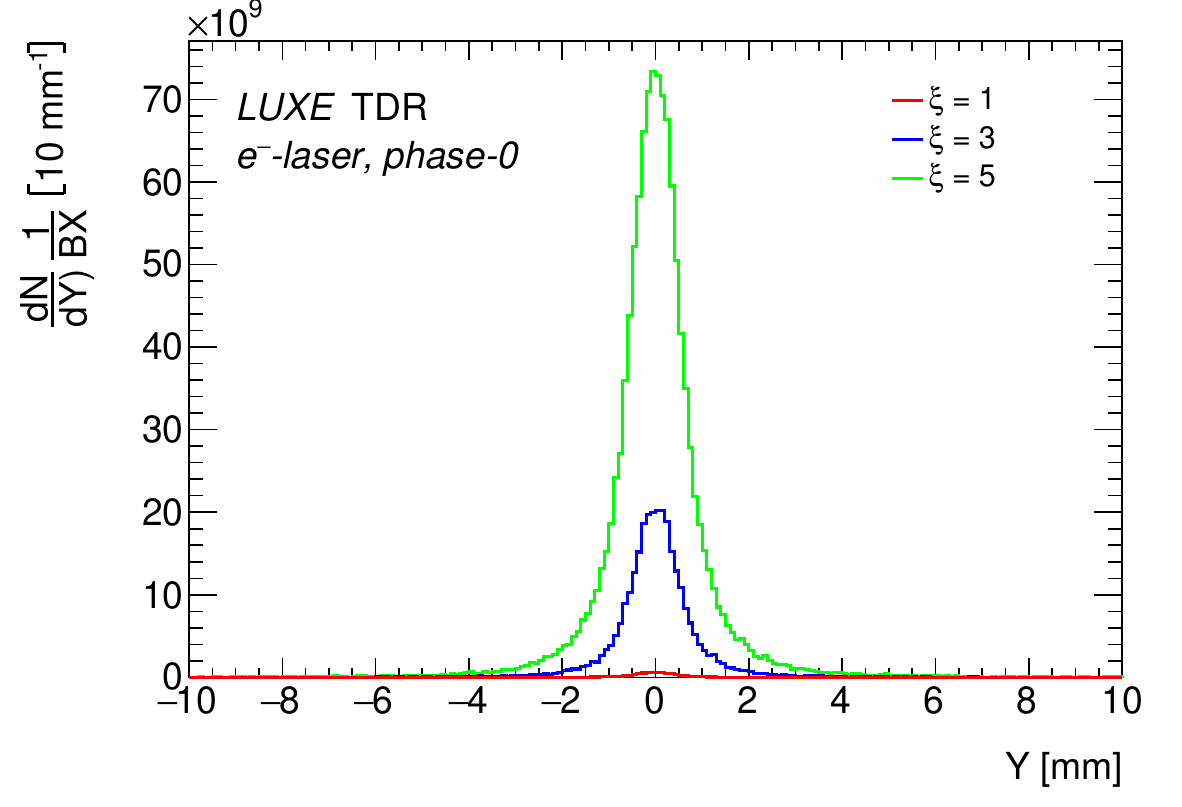}
\caption{Beam profile for $\xi=1$, $\xi=3$ and $\xi=5$ along the $x$-direction (left) and $y$-direction (right) at the entrance of the GBP in the $e$--laser phase-0 LUXE setup.
}
\label{fig:beam_profile}
\end{figure}

The simulation (see Section~\ref{sec:gbp:fullmc} for details) shows that the beam width increases with $\xi$, that beam profile projections are characterised by long tails and most of the profile is contained in a $2\times 2$\,cm$^2$ square as can be seen from Fig.~\ref{fig:beam_profile}.

\begin{figure}[!htb]
\center
\includegraphics[width=0.8\textwidth]{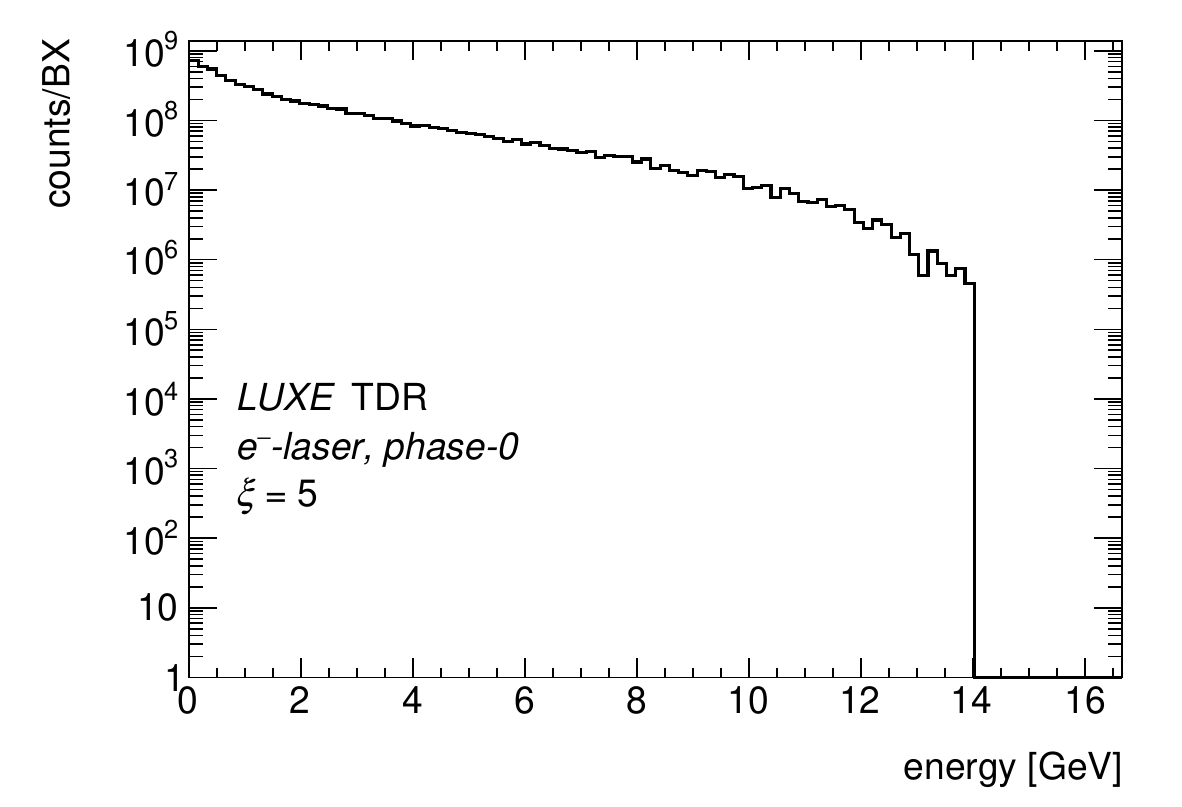} 
\caption{Energy distribution per BX of particles at the entrance of the GBP detector ($e$--laser mode, phase-0, $\xi=5$). }
\label{fig:energy_distribution}
\end{figure}

The minimum beam width which will be measured by the profiler will be about 70\,$\mu$m. The maximum intensity per bunch crossing (BX) of the gamma beam is $N_\gamma\simeq5\cdot10^{8}$ photons/BX in the $e$--laser mode. In Fig.~\ref{fig:energy_distribution}, the Compton photon energy distribution per BX at the input of the GBP detector is shown (\mbox{$e$--laser} mode).

Given the very high dose that the detectors will withstand, a sapphire strips detector is proposed, as this material is very resistant to the radiation dose (up to 10\,MGy~\cite{Karacheban:2015jga}).
The choice of a strip readout will allow the readout electronics to be kept in a low irradiated area. The use of two successive strip detectors (with strips placed perpendicular between themselves) will cause a beam broadening due to the Coulomb scattering in the first detector the beam will hit, but this has been proven to be negligible in MC simulation which assumed a distance between the two detectors of about 2\,cm. Also, the second detector will be submitted to an approximately double dose and will have higher signal amplitude, but the benefit to save the readout electronics by high irradiation level will compensate these drawbacks which are anyway expected to be tolerable.

The detector should be as thin as possible to minimise the dose absorbed by the more downstream plane but still preserving an acceptable signal level.
The minimum detector thickness that can be manufactured still having a good mechanical stability is about 100\,$\mu$m.
In the following a sapphire detector having $2\times 2$\,cm$^2$ area, $dz=100$\,$\mu$m thickness and $\Delta x$ $(\Delta y)=100$\,$\mu$m strip pitch will be considered. It must be mentioned here that in the present time sapphire samples from different companies are under test, and some company could only manufacture wafers having 150\,$\mu$m thickness (which has been shown in some studies not reported here still acceptable).

Given the very high beam intensity and occupancy per strip, the beam position and the beam width can be reconstructed with the desired precision by using an analog readout and a charge centre-of-gravity (COG) and/or fit to spatial distributions algorithms. A total of about 400 channels will be needed to readout two orthogonally  placed strip detectors.

\subsection{Sapphire Detector Characteristics} \label{sec:sapphire}

Artificial sapphire (aluminum oxide) is a wide band gap ($9.9$\,eV) insulator material which gained some interest in high energy detector physics in recent times~\cite{Karacheban:2015jga}.  It has excellent mechanical and electrical properties and it is produced industrially in large amounts. In Table~\ref{tab:sapphire} the main characteristics of sapphire with respect to silicon and diamond are reported. The band gap of sapphire is larger than for diamond, resulting in a factor of two larger energy needed to create an electron--hole pair. The energy loss of a charged particle moving in sapphire is, however, larger than in diamond. The amount of charge carriers generated per unit length in sapphire is about 60\% of the amount generated in diamond.

 It was shown recently~\cite{Karacheban:2015jga} that optical grade sapphire can be used directly as a detector material even without extra purification with the advantage of low cost and superior radiation hardness.

\begin{table}[htbp]
  \centering
    \begin{tabular}{l c c c}
    \hline \hline
    Material properties & \multicolumn{1}{|l|}{sapphire} & \multicolumn{1}{l|}{diamond} & \multicolumn{1}{l}{silicon} \\
    \hline
    density [g/cm$^3$] & 3.98 & 3.52 & 2.33 \\
    \hline
    bandgap [eV] & 9.9 & 5.47 & 1.12 \\
    \hline
    mean energy to create an eh pair [eV] & 27 & 13 & 3.6 \\
    \hline
    dielectric constant & \multicolumn{1}{c}{9.3-11.5} & 5.7 & 11.7 \\
    \hline
    dielectric strength [MV/cm] & 0.4 & 1.0 & 0.3 \\
    \hline
    resistivity [Ohm cm] at 20$^\circ$C & 1.0E+16 & 1.0E+16 & 1.0E+05 \\
    \hline
    electron mobility [cm$^2$/(V s)] at 20$^\circ$C & 600 & 2800 & 1400 \\
    \hline
    MIP eh created [eh/$\mu$m] & 22 & 36 & 73 \\
    \hline \hline
    \end{tabular}%
     \caption{Overview of sapphire material characteristics compared to diamond and silicon.}
  \label{tab:sapphire}%
\end{table}%

Signals collected from sapphire detectors are limited by a relatively small amount of 22 electron--hole
(eh) pairs produced per micron of MIP track and by the charge collection efficiency that has been measured to range 
between 2\% and 10\% in 500\,$\mu$m thick detectors~\cite{Karacheban:2015jga}.
This intrinsic small MIP signal amplitudes should still be suitable for the GBP case given the large fluxes of particles simultaneously hitting the detector.
Extremely low leakage current ($\sim$pA) at room temperature even after high dose
irradiation makes sapphire detectors practically noiseless.

A typical detector design includes a thin sapphire plate with continuous metallisation on one side and a pattern of electrodes (pads or strips) on the opposite side. 
Aluminum is proven to be a good material for metallisation, the usual thickness is a few microns.
The operation voltage of the detector is a few hundred volts, the signal duration is several
nanoseconds, the charge collection is completely dominated by electrons drifting in the electric field.

Taking into account the characteristics described above, a sapphire plate strip detector is considered as a suitable candidate for the LUXE profile monitoring system.

\section{Beam Tests on Sapphire Detectors}

\begin{figure}[!htb]
    \center
    \includegraphics[width=0.49\textwidth]{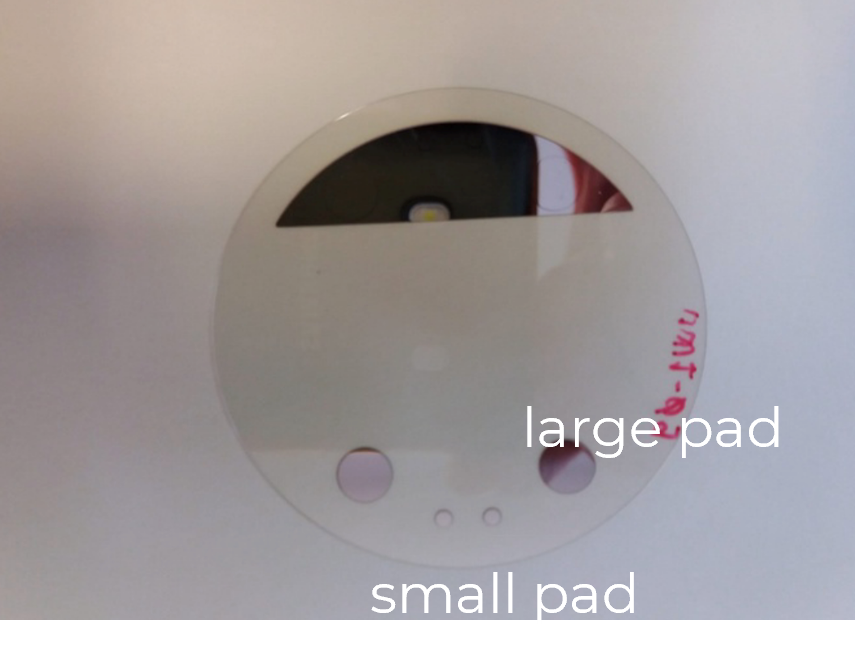}
    \includegraphics[width=0.49\textwidth]{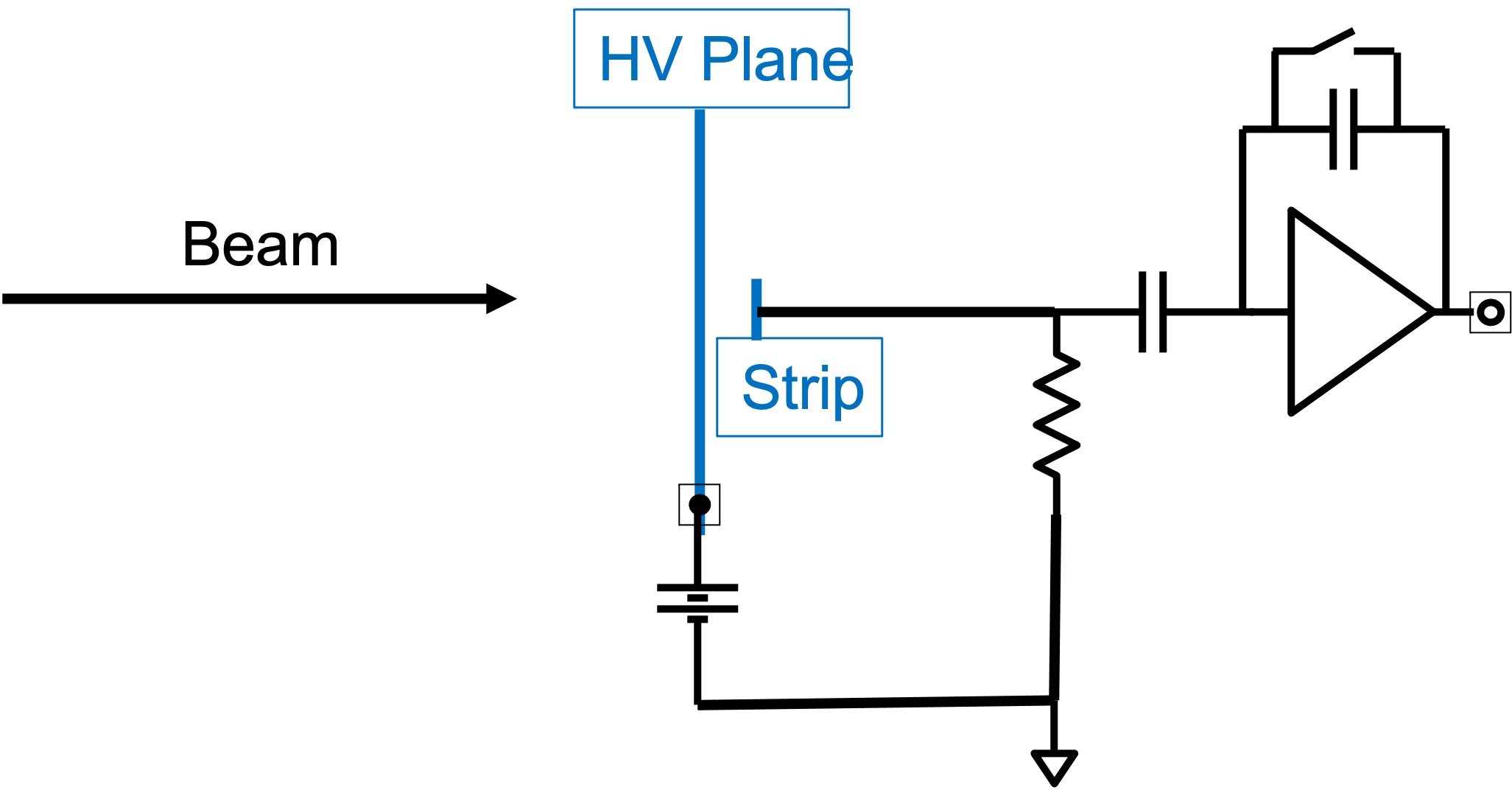}
    \caption{Left: the sapphire wafer of the `pad-sensor' before the assembly. Four large and small pads are present on the uppermost surface, and the `half-moon' shaped metallisation on the back. Right: the simplified electrical diagram of the sensor board.}
    \label{fig:testbeam_detectors}
\end{figure}

An experimental campaign has been planned to investigate the response of sapphire sensors to incident radiation. The main purpose of the first investigation was to determine sapphire charge collection efficiency as a function of the external biasing voltage and of the beam bunch charge. The experiments have been conducted at the Beam Test Facility (BTF) at the Frascati INFN national laboratories~\cite{dafne-btf}.
The 'pad-sensor' prototype shown in the Fig.~\ref{fig:testbeam_detectors} has been tested. It is made of a 2-inch sapphire disk where two metal-plated pads (with radius $r_{\small \mathrm{SP}}=0.8\,{\mathrm{mm}}$ and $r_{\small\mathrm{LP}}=2.75\,{\mathrm{mm}}$) have been deposited on the top surface. On the bottom, there is a backside metallisation with the shape of a half-moon which provides the ground plane.

\begin{figure}[tb]
    \centering
    \includegraphics[width=0.46\textwidth]{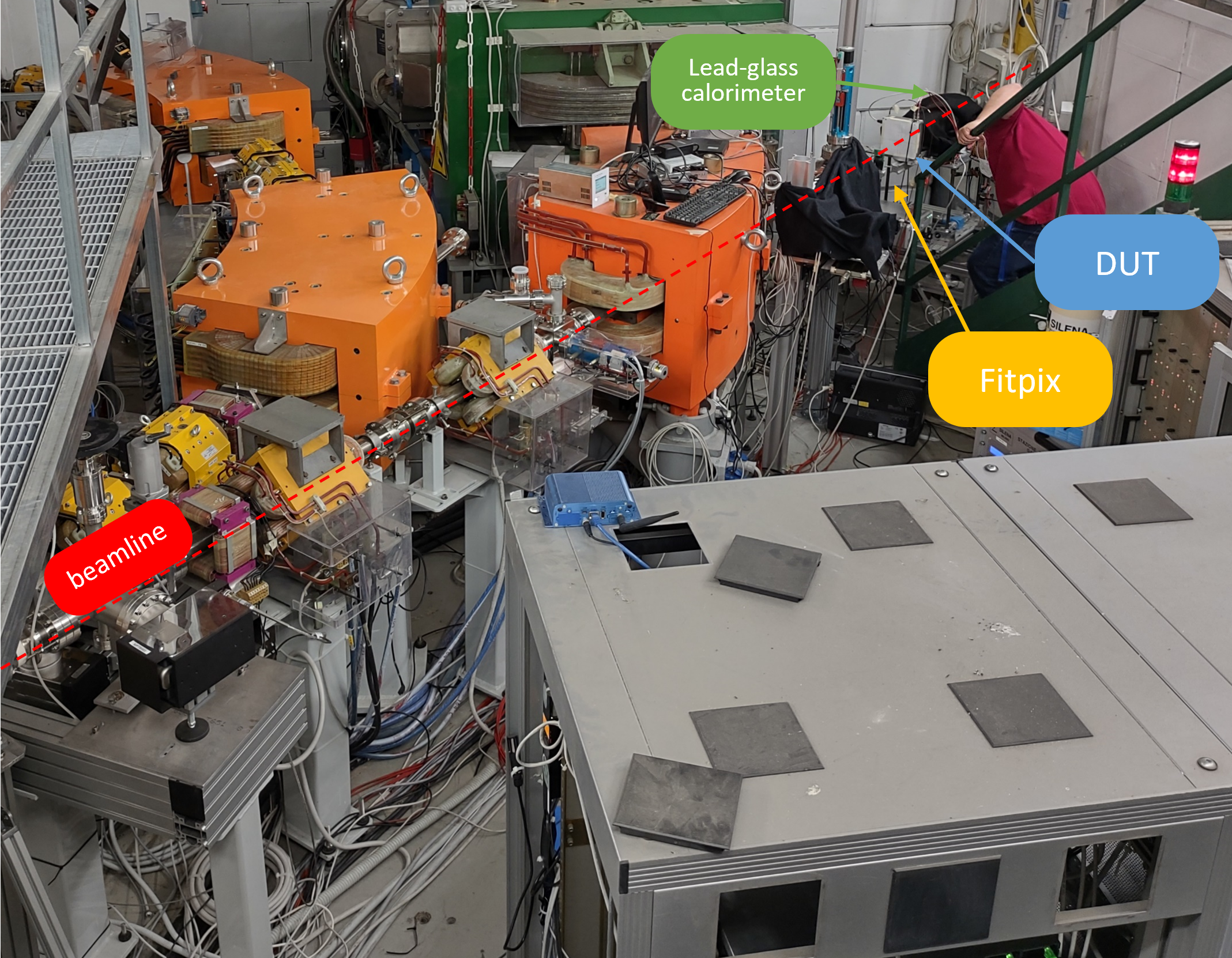}
    \quad
    \includegraphics[width=0.46\textwidth]{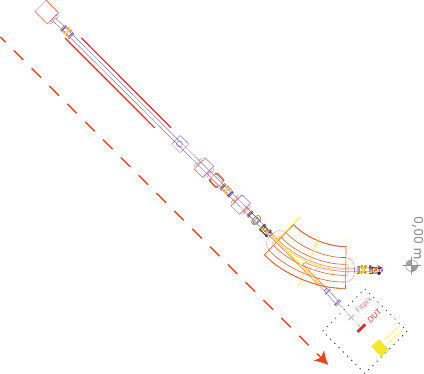}
    \caption{Beam test facility at Frascati. Left: wide angle picture of the beam-pipe and detectors. Right: in scale top-view of the beam line and detector placement.}%
    \label{Fig:BTF_setup}
\end{figure}

The behaviour of two sapphire detectors of thicknesses $d$ of 110\,$\mu$m and 150\,$\mu$m has been investigated. The experiment was performed at BTF running in parasitic mode, using a 300\,MeV electron beam with a bunch repetition rate of a $\mathrm{\sim Hz}$. Electron bunches had an intensity in the range $1-10^5$ electrons/bunch, and a length of approx.\ 10\,ns. The beam profile and position were measured in the proximity of the sensors by means of a 400\,$\mu$m-thick silicon pixel detector (Fitpix) and the beam charge by a lead-glass calorimeter. The final part of the beam line is shown in Fig.\ref{Fig:BTF_setup}: the electron beam -- which exits the vacuum pipe by crossing a 50\,$\mu$m titanium exit-window -- travels in the air for 33\,cm. It crosses the Fitpix detector, the device under test (DUT) with our sensors located 22\,cm downstream and is finally absorbed in the calorimeter.

\begin{figure}[bht]
    \centering
    \includegraphics[width=0.45\textwidth]{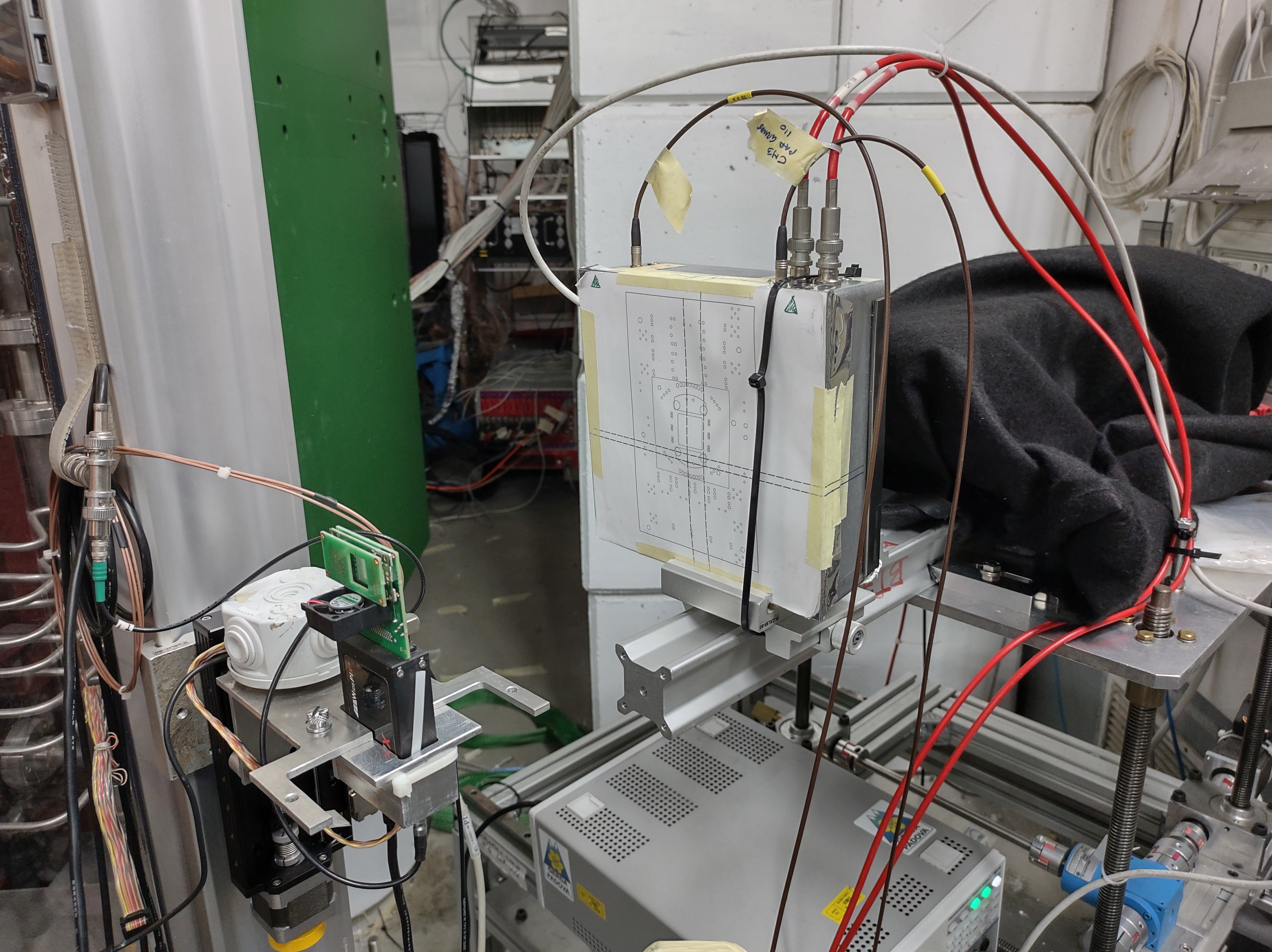}
    \includegraphics[width=0.45\textwidth]{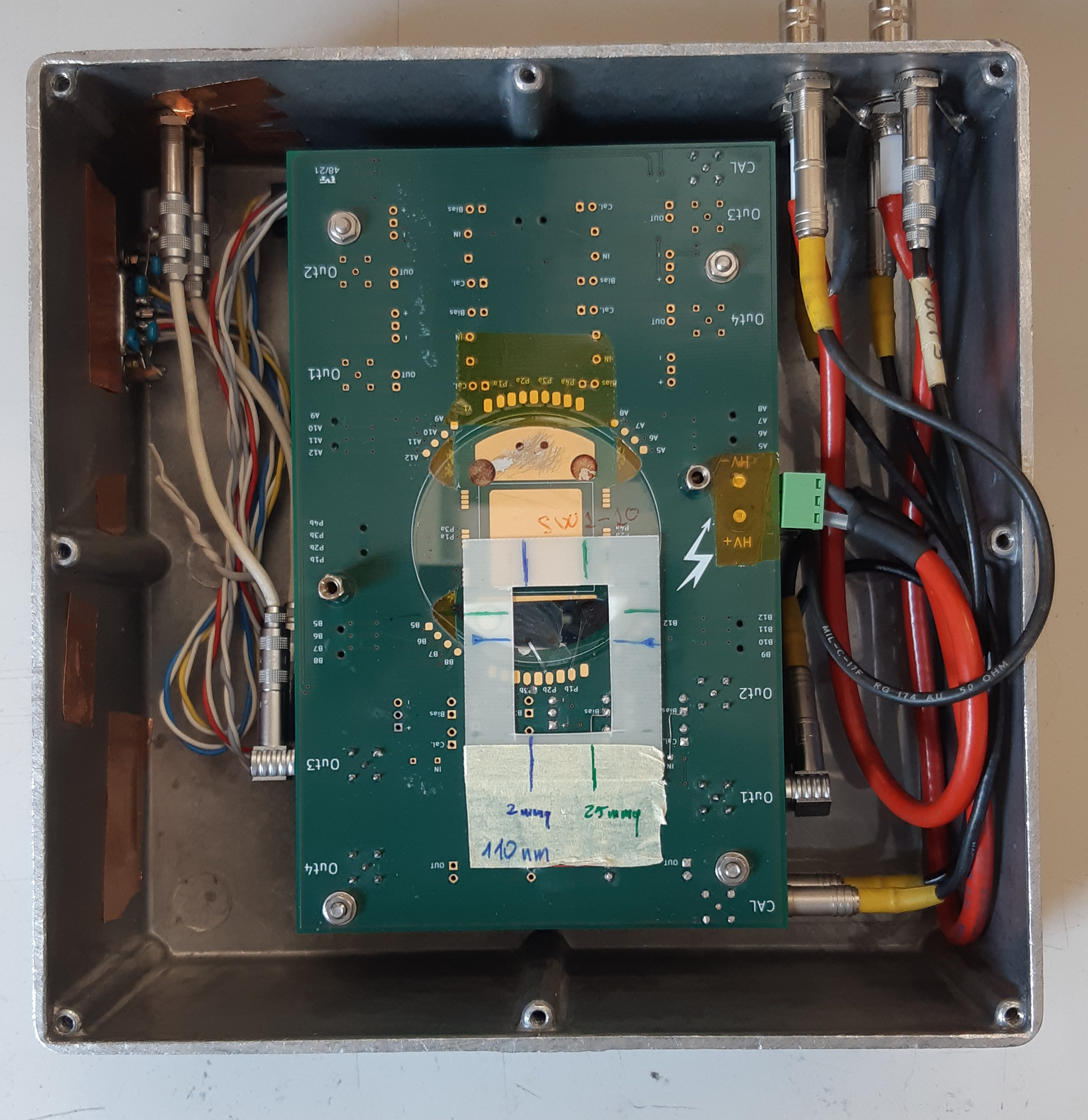}
    \caption{Left: the PCB hosting the $110\,\mu$m thick sapphire sensor. Right: the DUT assembly in its final position on the movable table.}%
    \label{Fig:dutbox_open}%
\end{figure}

The DUT is placed on a movable table capable of sub-millimetre displacements. It consists of a stack of two detectors, mounted on identical PCBs and separated 2\,cm apart. They are contained inside a 3\,mm thick metallic box, which provides electromagnetic shielding. The box (shown in Fig.~\ref{Fig:dutbox_open}) is closed with a thin 0.3\,mm thick Al foil in the beam line. The sensors are placed in the box in such a way that the electron beam first intercepts the $110\,\mu$m \emph{upstream} sensor and then the $150\,\mu$m \emph{downstream} sensor. The schematic of each circuit board is shown in Fig.~\ref{fig:testbeam_detectors}: the signal from small/large pad is routed to a dedicated charge sensitive amplifier with a gain of $\sim$200\,mV/fC.

The four outputs of the amplifiers (two pads by two sensors) are connected to a LeCroy Waverunner 640zi 4-ch. 40\,GS/s 4\,GHz oscilloscope in high-impedance AC-coupling mode, which is placed close to the detector and remotely controlled. The signal is sampled at 100\,MS/s and digitally processed online by the scope itself. 

The charge collection efficiency is defined as the ratio between the charge collected at the pads of the detector and the total charge generated by the passage of a ionising particle within the active material, that is
\begin{equation}
	{\mathrm{CCE}} \;\equiv\; \frac{Q_\textrm{col}}{Q_\textrm{dep}}
	\label{eq:cce}
\end{equation}
The collected charge, appearing at the numerator, is directly measured as described above. The charge produced by ionisation, at the denominator, is obtained by dividing the energy deposited, estimated via a \geant simulation of the experimental setup, by the average eh pair creation energy (see Table~\ref{tab:sapphire}). A correction is applied to subtract the contribution to the signal due to ionisation charges generated outside the cylindrical volume defined by the pads.
The CCE was estimated as a function of the biasing voltage $ V_{\mathrm{bias}} $ (both forward and reverse configuration) and of the beam intensity $ N_{e} $ (at fixed $ V_{\mathrm{bias}}\,=\,100\,$V). By convention, the forward bias is the configuration where the pad (on top of sapphire) is at a positive potential +V with respect to the backside metallisation.\\

An acquisition typically occurs in about a minute leading to a statistics of about 800--1000 bunches, corresponding to a negligible statistical error when compared with systematic uncertainties.

\paragraph{Systematic uncertainties}
Contributions to the systematic uncertainties come from the uncertainties of the electron beam position and profile in the transverse plane. This was investigated in detail by means of MC simulations of the beam line and the total systematic uncertainty on the CCE measurements turned out to be in the range between 3\% and 6\%.

\paragraph{Fit model} If we assume the following conditions to hold:
\begin{enumerate}
    \item the detector is planar (thickness $ d $ is negligible compared to other dimensions). The electric biasing field is uniform $ E_0\equiv\frac{V}{d} $ directed perpendicular to the pads;
    \item the transport properties and the electric field are uniform over the whole volume of the sensor;
    \item the free charge in steady state conditions is negligible with respect to the charge inducing the electric field and its generation, due to a photon (or a particle) absorption, is instantaneous;
    \item diffusion and de-trapping phenomena are negligible and the density of charge carriers while drifting decreases with time as $\sim e^{-t/\tau_{e,h}} $.
\end{enumerate}
It is straightforward to compute the charge collected at an electrode due to the motion of a charge carrier starting at the position $y_0$.

\begin{figure}[h]
    \centering
    \includegraphics[width=0.95\textwidth]{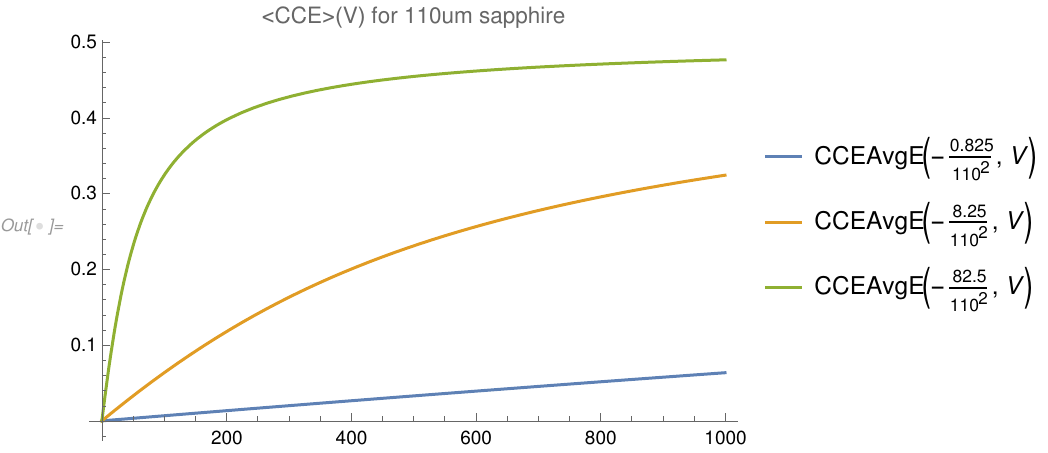}
    \caption{Plot of  Eq.~\ref{eq:fitequation} for a 110~$\mu$m planar detector with $ -(\mu\tau)_e $ products of 0.825\,$\mu$m$^2$\,V$^{-1}$, 8.25\,$\mu$m$^2$\,V$^{-1}$ and  82.5\,$\mu$m$^2$\,V$^{-1}$.}
    \label{Fig:fitequations}
\end{figure}

In the experiment, a minimum ionising electron track crosses the detector along its thickness creating distributed charge carriers along the way. Furthermore, in sapphire the main role in signal generation is played by electrons only \cite{Karacheban:2015jga}. Therefore, neglecting holes contribution to charge collection, our fit model is described by the equation

\begin{equation}
        \textrm{CCE}(V) = k\,V \left[1 + k\,V\left( \exp\left(-\frac{1}{k\,V}\right) - 1 \right) \right]
    \label{eq:fitequation}
\end{equation}
with $k \equiv -\frac{\mu_e\tau_e}{d^2}$.
The behaviour of Eq.~\ref{eq:fitequation} is shown in  Fig.~\ref{Fig:fitequations} for different values of the parameter $k$.

\begin{figure}[htb]
    \centering
    {\includegraphics[width=0.49\textwidth]{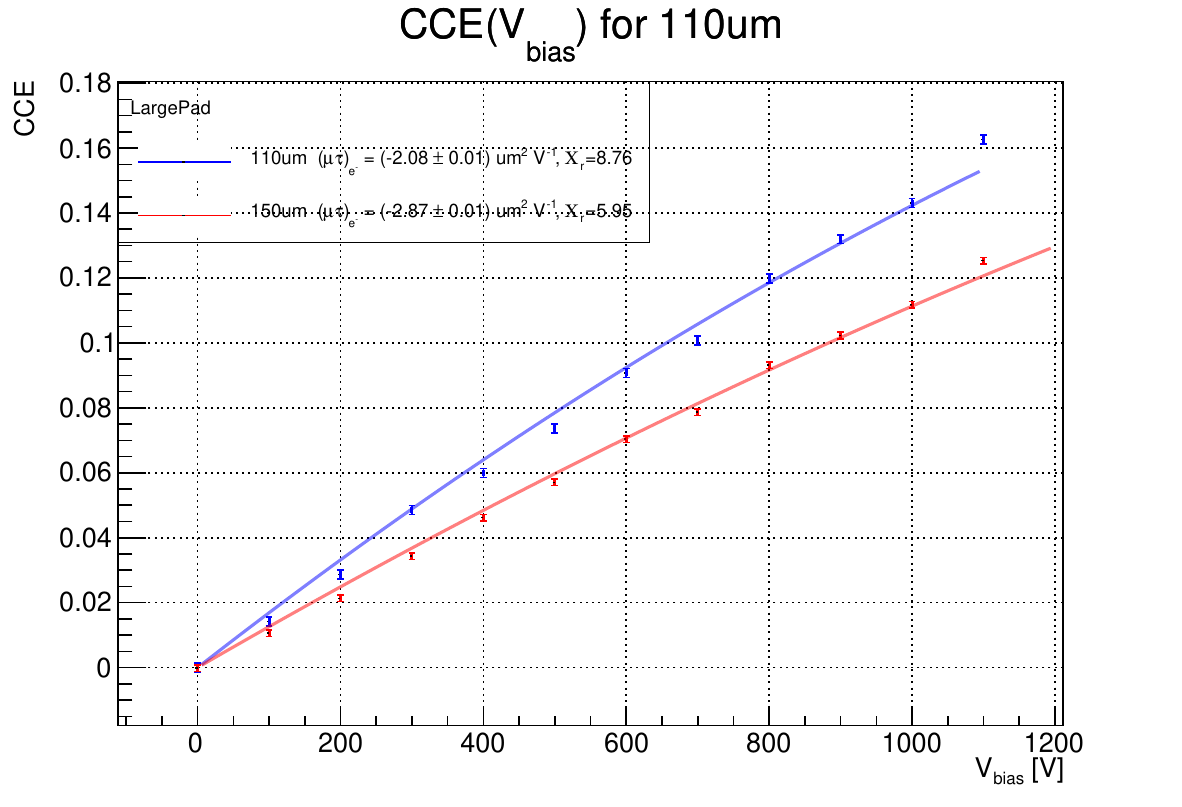}}
    {\includegraphics[width=0.49\textwidth]{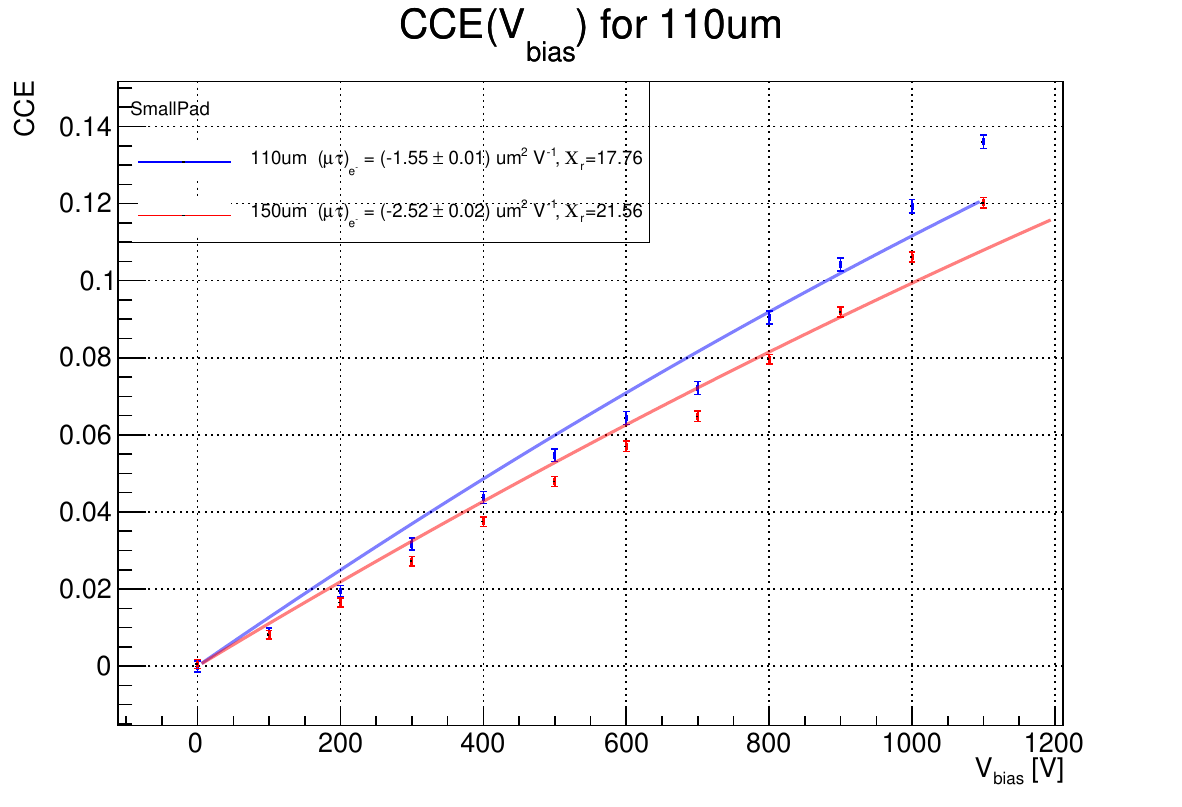}}
    \caption{Scan of the charge collection efficiency as a function of the forward biasing voltage $ V_{\mathrm{ bias}} $. Statistical uncertainties are attached to data points. Left: CCE for the large pad (r=2.75mm). Right: CCE for small pad (r=0.8mm). Blue (red) line for the 110~$\mu$m (150~$\mu$m) wafer.}
    \label{Fig:cceVSvbias_forward}
\end{figure}

\paragraph{Results. CCE(V)} The charge collection efficiency as a function of the HV is shown in Fig.~\ref{Fig:cceVSvbias_forward}. A typical efficiency of 14\% (12\%) is observed when operating the 110\,$\mu$m (150\,$\mu$m) sapphire sensors at 1\,kV. The $(\mu\tau)_{\mathrm{ e}}$ product is extracted from fit parameters of the charge collection efficiency curves, using the fit, Eq.~\ref{eq:fitequation}. Such (model dependent) extractions of $(\mu\tau)_{\mathrm e}$ gave typical values in the range $\mathrm 2-3\;\mu$m$^2$V$^{-1}$ for the two sapphire samples.

It is worth stressing a few points: (a) if the typical drift distance ($d_{\mathrm{ CCE}}\sim{\mathrm{ CCE}}\times d$) is small with respect to the sensor thickness $d$, the charge collection efficiency is proportional to the sensor thickness $d$; (b) the signal generated in the sensor is given by the product ${\mathrm{ CCE}}\times Q_{\mathrm{ dep}}$, which in the above condition does not change with thickness -- i.e.\ the increase in the charge deposited is cancelled by the decrease in the collection efficiency; (c) the two sensors of different thicknesses investigated here are from different manufacturers, and the dependence of the CCE from thickness cannot be extracted from the data. 

\begin{figure}[!htb]
    \centering
    \includegraphics[width=0.6\textwidth]{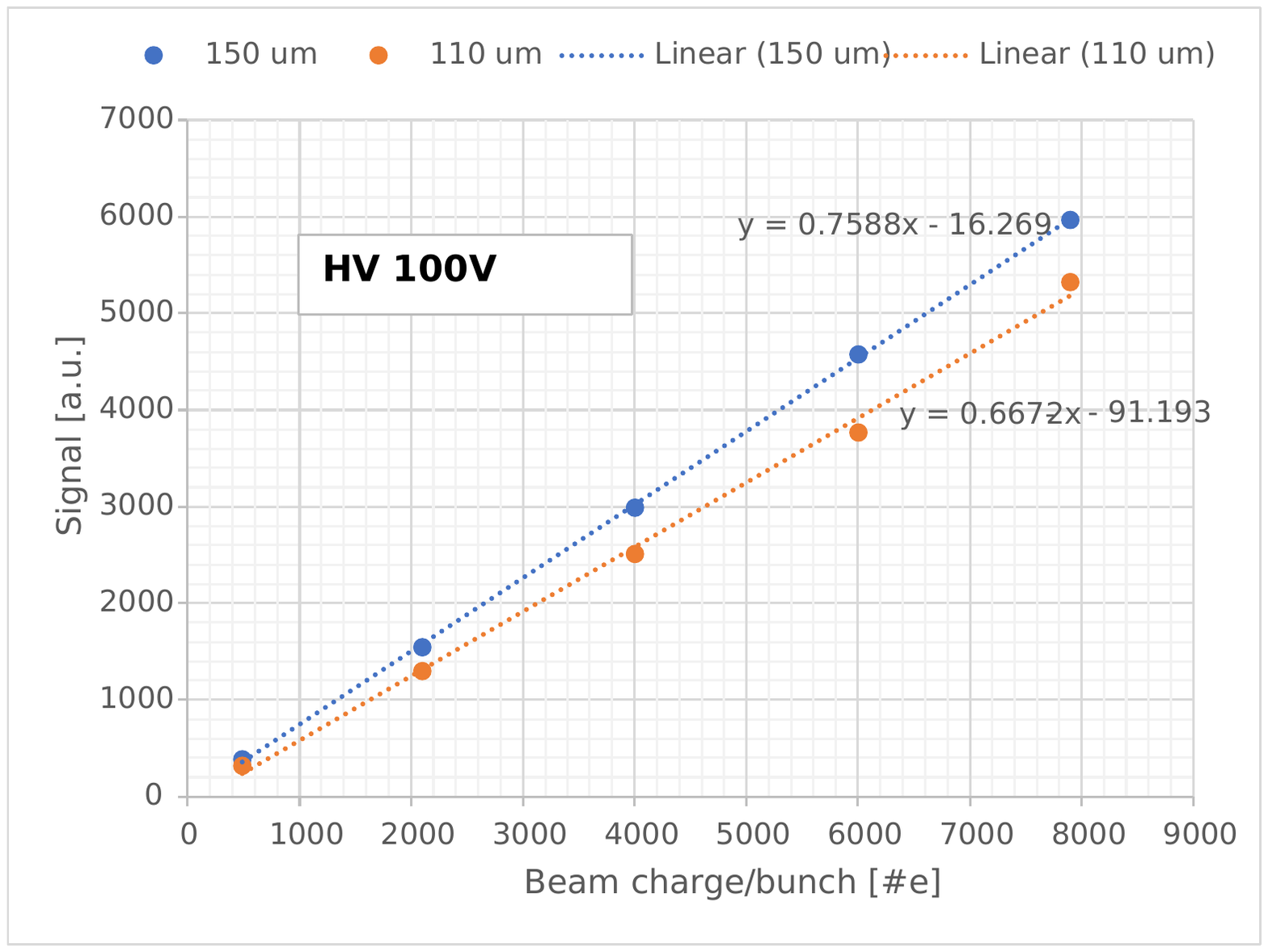}
    \caption{Behaviour of the collected charge from the large pad as a function of the beam multiplicity, at a fixed bias voltage of $V_{\mathrm{ bias}}\,=\,100\,\mathrm{ V}$.}
    \label{Fig:chg_linearity}
\end{figure}

\paragraph{Results. CCE(Q)} The detector response as a function of the beam charge is investigated using the large pad, by keeping the biasing voltage fixed and varying the beam charge from 1 to 8k electrons/bunch. The signal amplitude is measured as a function of beam multiplicity, and reported in arbitrary units in Fig.~\ref{Fig:chg_linearity}.

It is worth noting that both the 110\,$\mu$m and 150\,$\mu$m detectors show a linear dependence with the beam charge up to 9k electrons/bunch. The slope of the two linear fits are within 13\% of each other and the projected offsets are consistent with amplifier's RMS noise ($\sim \mathrm{ 30\;mV}$), as expected.

\section{Expected Performance}
\label{sec:gbp:performance}

The proposal is to build a LUXE beam profiler station with two sapphire strip detectors of $2\times2$ \,cm$^2$ area, thickness $100\,\mu$m, strip pitch $100\,\mu$m, placed at about 2\,cm distance from each other along the beam direction, with the $y$-strip detector upstream and the $x$-strip detector downstream. The first upstream detector will be at 11.5\,m from the IP. 

As mentioned in the introduction, given the importance of the GBP for monitoring the beam conditions of LUXE, the intention is to install two such stations. 
In the following we will refer to the more upstream station as the first station. The second station is in all effects similar to the first, so that the main conclusions drawn for the first detector will broadly apply also to the second.

The rates of gammas impinging on the GBP will depend on the LUXE running setup ($e$--laser or $\gamma$--laser) and on $\xi$, the laser intensity parameter. A precise reconstruction of $\xi$ from the beam profile, as described later in the section, will be possible only in the $e$--laser mode.  In this mode, the expected number of Compton gammas per BX produced for several values of the laser intensity parameter $\xi$ is shown in Table~\ref{tab:rates}. From this table, one can infer that the Compton gamma rate seen by the GBP will span the range $10^6-10^9$ being almost constant at the maximum level for $\xi >3$. Since LUXE will operate for most of its time for $\xi >1$, the best precision on reconstructing $\xi$ will be needed for values largely exceeding unity. As a consequence, unless differently stated, the detector design parameters which will be reported in this section have been fixed considering a typical running condition around $\xi=5$ which is expected to be reached also in phase-0.

\begin{table}[htbp]
  \centering
    \begin{tabular}{ c c }
    \hline
    $\xi$ & $N_\gamma$ ($10^6$/BX) \\
    \hline
    0.15 & 4.6 \\
    \hline
    0.5 & 47.3 \\
    \hline
    1.0 & 154 \\
    \hline
    2.0 & 354 \\
    \hline
    5.0 & 458 \\
    \hline
    7.0 & 428 \\
    \hline
    10.0 & 366 \\
    \hline
    15.0 & 327 \\
    \hline
    \end{tabular}%
  \caption{Expected rates of produced Compton gammas per BX for several values of the laser intensity parameter $\xi$ in the LUXE $e$--laser running mode with linearly polarised laser light.}
  \label{tab:rates}%
\end{table}%

The relevant quantities to characterise the detector performance can be calculated using the standalone MC simulation described later in the section. It will be shown there that the more upstream detector will receive a dose of about 0.28\,Gy/BX and the collected maximum charge in the central strip of the detector of about 47\,pC (assuming conservatively a charge collection efficiency of 10\%). The MC simulation also shows that the peak absorbed dose increases by roughly a factor 1.5 for each successive detector layer added, while the charge increases by roughly a factor 2.2. 
As a consequence, the more downstream layer in the second station will absorb about 0.42\,Gy/BX and will have a peak charge of about 107\,pC. We will therefore assume conservatively, as design parameters, that the detector planes should withstand a maximum dose of about 1\,Gy/BX and the front end electronics should be able to read efficiently signals that span the range 0.05--100\,pC. The latter range corresponds to the amplitude swing of a Gaussian profile from the maximum to a 3 standard deviations value.
These design parameters values are summarised in Table~\ref{tab:output}, and the value of the strip geometrical capacitance is also reported. 

\begin{table}[htbp]
  \centering
    \begin{tabular}{cccl}
    \hline
    \hline
    Quantity & Value & Units & Description \\
    \hline
    $D_{MAX}$ & $\sim 1$ & Gy/BX & Maximum dose absorbed \\
    \hline
    $Q$ & $\sim 0.05-100 $ & pC/BX & Range of collected charged \\
    \hline
    C &  $\sim 5$ & pF & Strip capacitance \\
    \hline
    \end{tabular}%
      \caption{Typical relevant parameters assumed for the detector and electronics design. The maximum dose is evaluated at the more downstream plane in the second station. The typical charge is evaluated at the more upstream plane of the first station, which corresponds to the lowest level of collected charge expected.} 
  \label{tab:output}%
\end{table}%

Thanks to the COG algorithm, a spatial resolution of about $5\,\mu$m in a single BX can be achieved with a detector resolution in charge reconstruction of $\simeq1\%$.

Since sapphire detectors are supposed to work properly up to a dose of about 10\,MGy, the maximum dose reported in Table~\ref{tab:output} ($D_{MAX}$) corresponds to a detector lifetime of about $10^7$s, about 1 year of running. This means that in general the detectors should be changed once per year.
However, the real performance of the sapphire is under evaluation; its use in the GBP has still to be measured as a function of the absorbed dose.
In order to mitigate possible performance worsening for doses absorbed in running periods shorter than one year, the detector supports will be provided with two independent sub-millimetre movements allowing regular displacement from the most irradiated areas.
This approach is possible with sapphire detectors (not with silicon) since the leakage current of the irradiated sensor remains at negligible levels (few pA) after intense irradiation.
The possibility to have movable support for the detectors will also be fundamental for the calibration strategy (see Section~\ref{sec:gbp:installation}).

\subsection{Standalone Simulation}
\label{sa_simulation}

A standalone \geant MC simulation was set up in order to investigate the detector response in detail. The present detector design (see Section~\ref{sec:sapphire_detector}) was implemented: there are two sapphire microstrip sensors, separated from each other by 2\,cm and placed 10\,cm downstream of the beam pipe window, which is implemented as 5\,cm $\times$ 5\,cm $\times$ 200\,$\mu$m kapton. Any other volume is filled with air. The simulation starts with initial particles (mostly Compton gammas) imported from the \geant MC simulation of the full LUXE experiment. A detailed amount of information -- i.e.\ energy deposited from each particle at each step of the particle tracking -- is recorded inside the sensitive volumes of the upstream/downstream detector, giving the possibility to study the profile reconstruction after the beam particles have interacted with the sapphire  and to assess the absorbed dose and the possible background contributions. However, for implementing the generation of the electric signal and its digitisation, a specific tool developed for silicon pixel detectors (Allpix2) has been adopted to simulate the charge generation and propagation within the sapphire, and to find the induced signal on the electrodes.

The study of the GBP performance has been carried out by taking the high intensity Compton photons at the IP, as resulting from the most recent LMA simulations of the $e$--laser interaction. An idealised electron beam (e.g.\ zero divergence, transverse beam radius of 0.1\,$\mu$m and charge of 240\,pC) has been used to simulate the IP. The main conclusions are unaffected with respect to the case where realistic electron beam at IP is used. The high intensity Compton tracks are propagated from the IP downstream for 11.5\,m, where they are generated in front of the 200\,$\mu$m kapton beam pipe exit window in the standalone simulation for the profiler. A few cm downstream, a gamma beam profiler station is instrumented with two orthogonal sapphire detectors.

\begin{figure}[htb]
    \center
    \includegraphics[width=\textwidth]{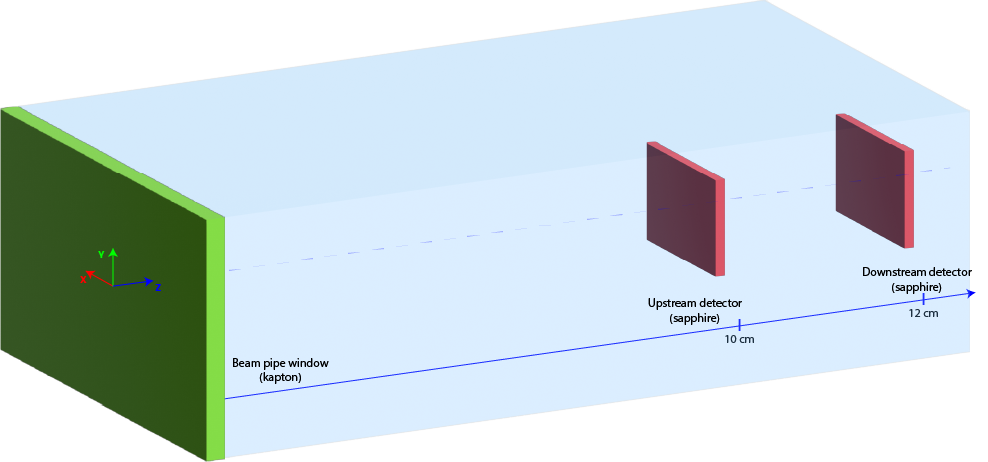}
    \caption{Illustration of the geometry for the \geant standalone MC simulation for the GBP. In order from left to right: the beam pipe kapton window (green), the upstream and downstream detectors of the first station.}
    \label{fig:GBP_volume}
\end{figure}

The geometry is illustrated in  Fig.~\ref{fig:GBP_volume}. The upstream (downstream) sensor has strips parallel to the $y$ ($x$) axis, measuring the photon profile along the parallel (perpendicular) laser polarisation axis.

The typical number of Compton photons produced in a BX for the phase-0 $e$--laser $\xi=5$ configuration is $\mathcal{O}(5\times10^8)$.
At present, only $\mathcal{O}(10^6)$ $e$--laser interactions have been simulated, resulting in a statistics of $\mathcal{O}(5\times10^6)$ Compton photons. The standalone GBP simulation picks the initial photon tracks from this set, and it simulates the statistics of a typical $\xi=5$ BX by artificially increasing the number of primary Compton photons.
A summary of the type and number of particles entering the GBP volume is given in Table~\ref{tab:PartOverview}.

\begin{table}[!hb]
\begin{center}
\begin{tabular}{cc}
        \hline
        \hline
        type & total number \\
        \hline
        $\gamma$ & $1.02 \times 10^{8}$\\
        \hline
        $e^-$ ($e^+$) & $1.47 \times 10^{5}$ ($1.41\times 10^{5} $) \\
        \hline
        $n$ ($p$) & $320$ ($252$) \\
        \hline
        $\pi^{\pm},\,{\mathrm{ etc.}}$ & $184$ \\
        \hline
    \end{tabular}
    \caption{Summary of the particles at the entrance of the upstream GBP sensor, listed by type, from the $\xi=5$ simulation with $1.03\times10^9$ Compton photons.}
    \label{tab:PartOverview}
\end{center}
\end{table}

\begin{figure}[!hb]
\center
\includegraphics[width=.49\textwidth]{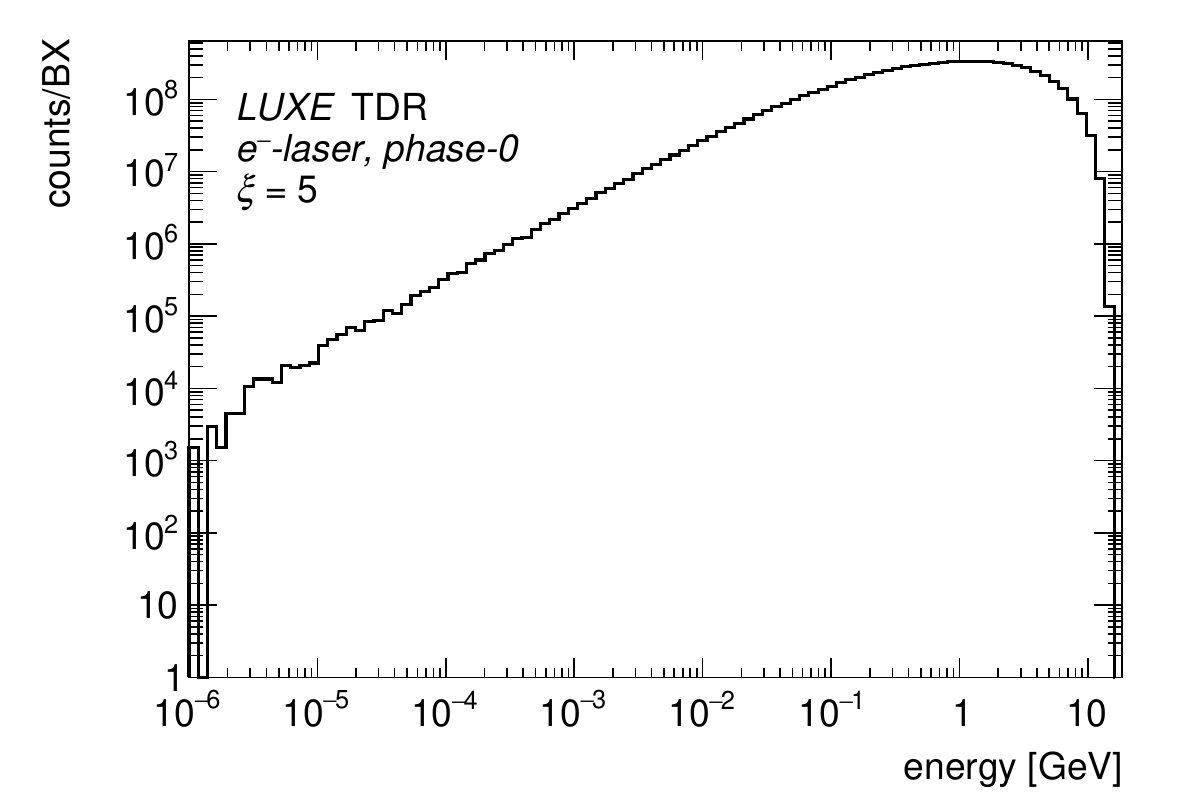}
\includegraphics[width=.49\textwidth]{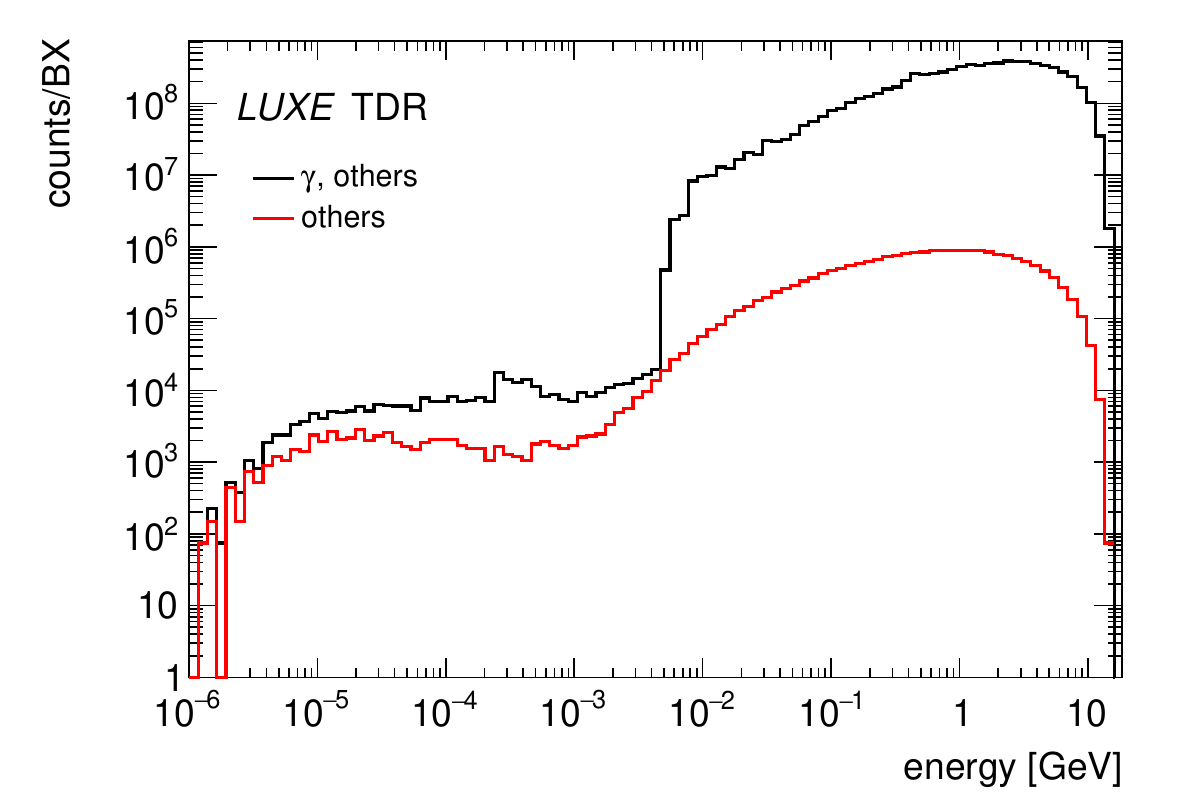}
\caption{Energy spectrum of the Compton photons at the IP (left) and at the entrance of the upstream GBP sensor (right). ($e$--laser, phase-0, $\xi=5$ setup.)}
\label{fig:GBP_primaryeSpectr}
\end{figure}

The energy spectrum of these particles at the entrance of the upstream profiler is shown in Fig.~\ref{fig:GBP_primaryeSpectr}. The kapton window generates a particle background mostly composed of $e^+e^-$ pairs from gamma conversion, while the energy of the Compton beam is mainly contained in the range $1\;{\mathrm{ MeV}}\,\leq E_\gamma \leq 16\;{\mathrm{ GeV}}$.

\begin{figure}[htb]
\center
\includegraphics[width=0.49\textwidth]{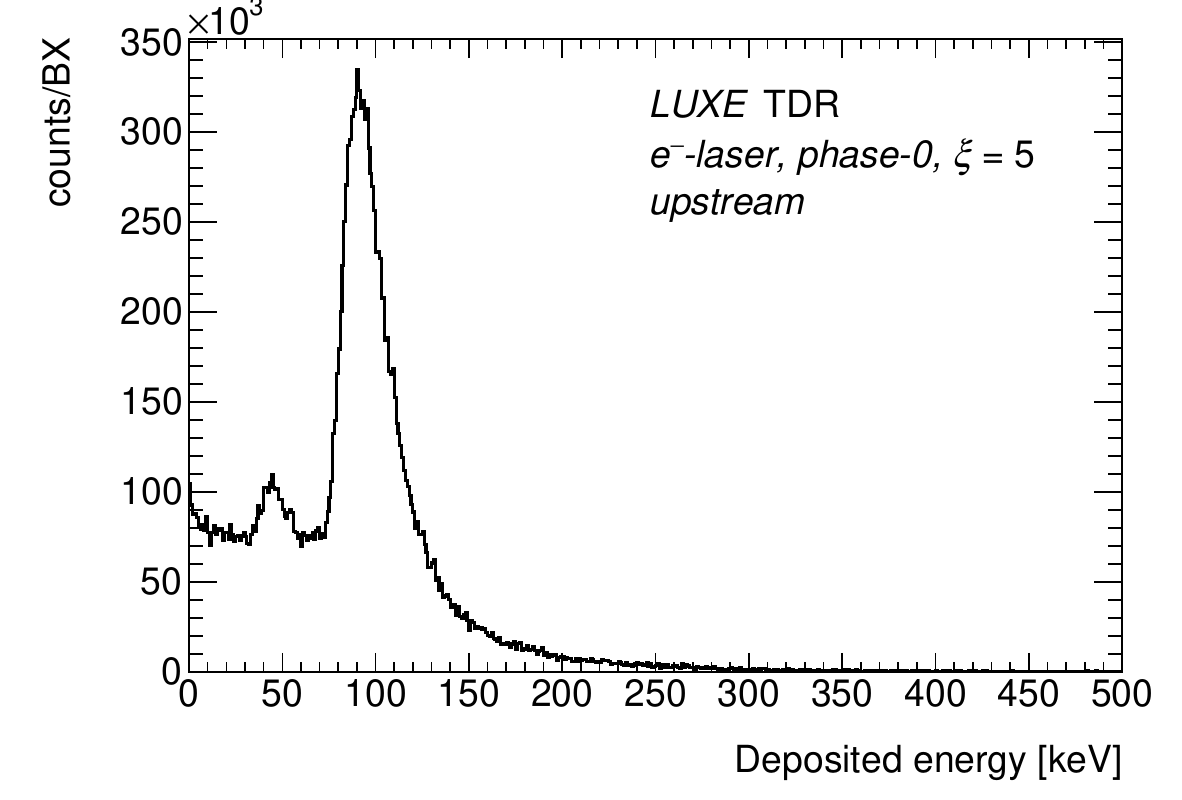}
\includegraphics[width=0.49\textwidth]{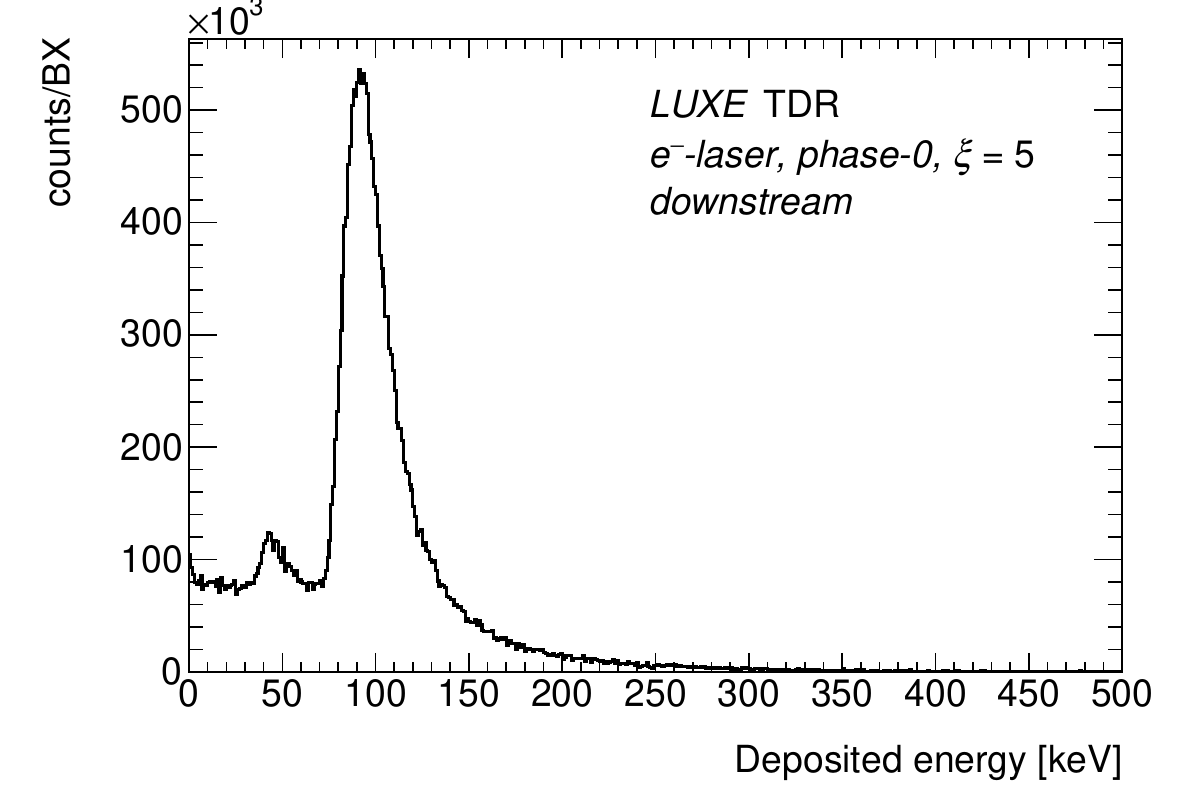}
\caption{Spectrum of energy deposits by signal particles in the sapphire upstream (left) and downstream (right) profilers.
}
\label{fig:GBP_edepSpectrum}
\end{figure}

The simulation of a BX is made by generating about $N=1.9\times10^9$ Compton photons at the position of the kapton beam-pipe window -- i.e.\ the green box in the left of Fig.~\ref{fig:GBP_volume} -- according to the spatial and momentum distributions of the Compton photons in the initial imported beam for $\xi=5$ $e$--laser phase-0 with a pencil electron beam.
The spectrum of the energy depositions in the sapphire detectors is shown in Fig.~\ref{fig:GBP_edepSpectrum}. Most of the energy depositions have energies $E\leq200\,{\mathrm{ keV}}$, although long tails up to $\sim30\,{\mathrm{ MeV}}$ are populated with a few events (less than 0.2\% total.)

A minimum ionising particle deposits in the sapphire an energy which is peaked at $E_{\small \mathrm{ MIP}} \simeq 48\,{\mathrm{ keV}}$. Two peaks can be identified in the deposited energy spectrum: one in correspondence to the MIP signal, and one at twice the MIP energy.
In this energy range, each primary photon can interact with the detector via two main mechanisms: Compton scattering or pair production, with the latter being the dominant process. The electron and positron resulting from the gamma conversion will each deposit an energy of about $48\,{\mathrm{ keV}}$, hence the peak in the spectrum at $96\,{\mathrm{ keV}}$. These conversions take place in the sapphire detectors, but also in the beam pipe exit window and in the air. The less populated peak at $48\,{\mathrm{ keV}}$ is instead predominantly due to Compton scattering processes within the detector. Also, secondary production of $e^+e^-$ pairs eventually reaching the sapphire sensor contributes to this peak.

\begin{figure}[!htp]
\center
\includegraphics[width=0.49\textwidth]{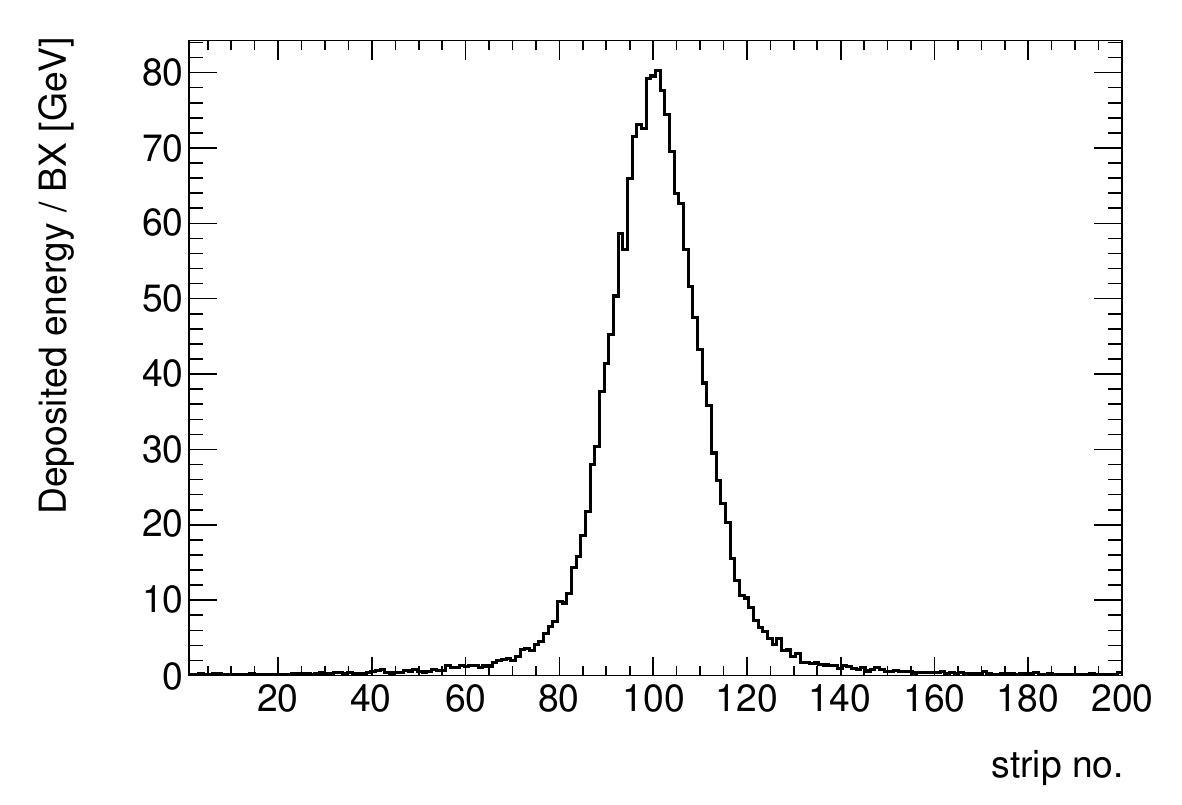}
\includegraphics[width=0.49\textwidth]{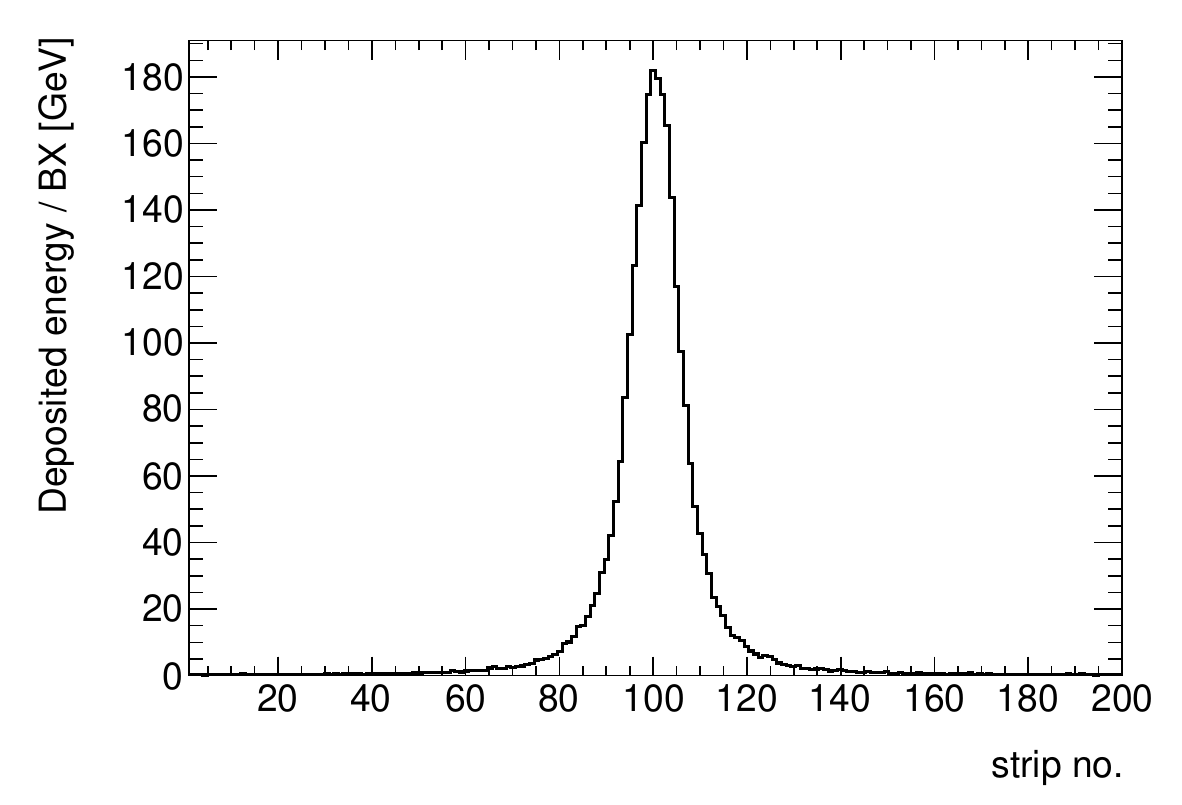}
\caption{Energy deposited in the strips for the upstream (left) and downstream (right) detectors, from the Compton gamma beam signal.}
\label{fig:GBP_edepStripChg}
\end{figure}

The profile of energy deposited in sapphire is shown in Fig.~\ref{fig:GBP_edepStripChg}. The highest amount of energy is released in the central strips of the detector. The maximum energy deposition in the strips is about $80\,{\mathrm{ GeV}}$ upstream, and $180\,{\mathrm{ GeV}}$ downstream. The amount of charge generated in a strip can be calculated using the mean energy to create an eh pair (see Table~\ref{tab:sapphire}): in the central strip we have $474\,{\mathrm{ pC}}$ upstream and $1067\,{\mathrm{ pC}}$ downstream. The charge collected should then be reduced accordingly to the CCE of the detector, which is expected to be at the level of about 10\%~\cite{Karacheban:2015jga}. 

\begin{figure}[!hbt]
\center
\includegraphics[width=0.49\textwidth]{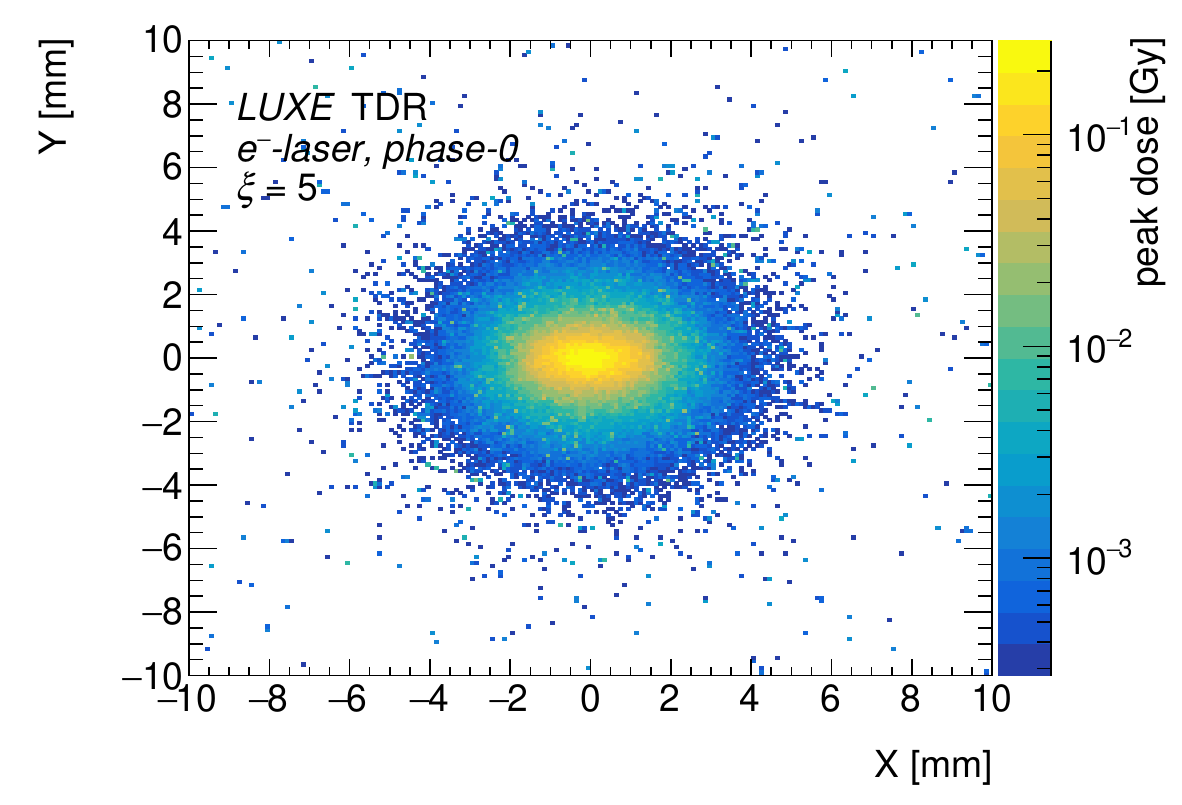}
\includegraphics[width=0.49\textwidth]{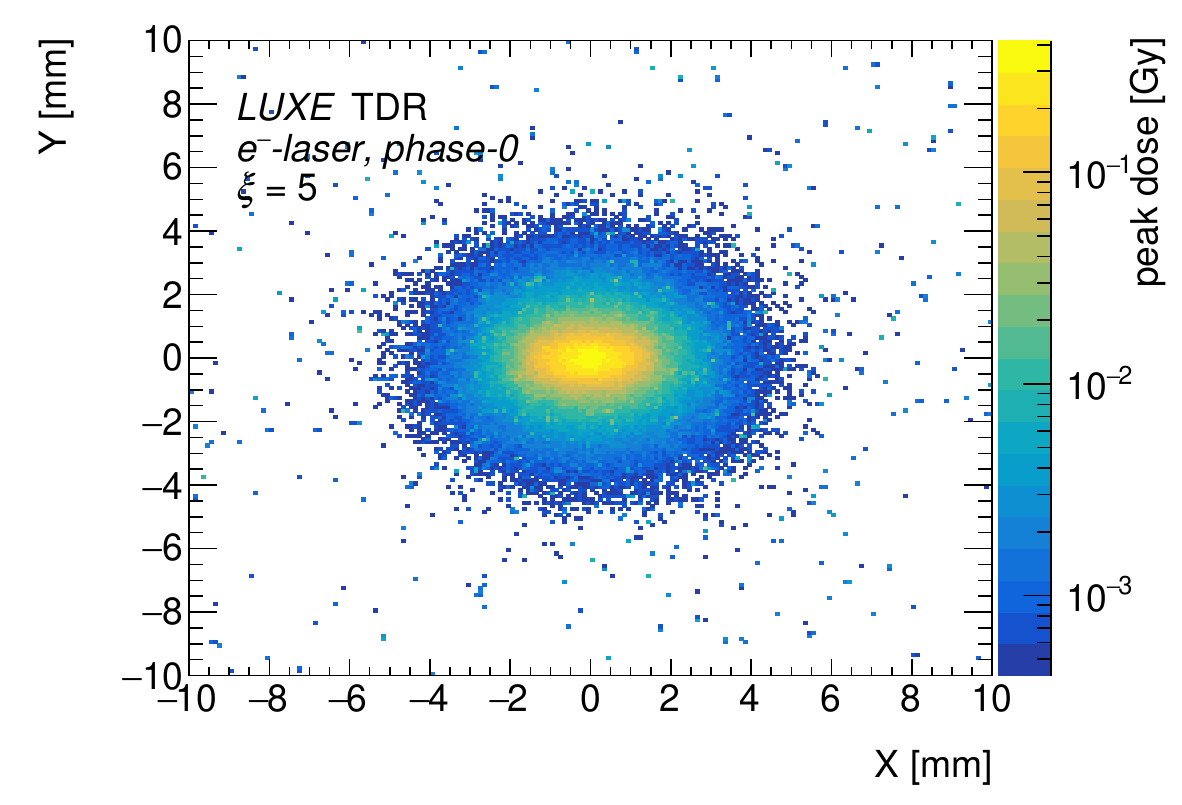}
\caption{Dose distribution profile for the upstream (left) and downstream (right) detectors. Energy depositions from the Compton beam signal are considered. Points where no depositions occurred are blank.}
\label{fig:GBP_dose}
\end{figure}

The radiation damage is determined by calculating the \emph{peak dose}, which is determined by looking at the highest amount of energy deposited in a cubic volume of conventional size $L=100$\,$\mu$m. From the map of energy depositions shown in Fig.~\ref{fig:GBP_dose} we estimate the peak dose for the upstream and downstream detectors to be $0.278\,{\mathrm{ Gy/BX}}$ and $0.418\,{\mathrm{ Gy/BX}}$, respectively.
These values are within the maximum dose estimated in Table~\ref{tab:output}, which uses as upper limit the extrapolated dose absorbed per BX by the most downstream plane of the second station.

\subsubsection*{Background Study}

The simultaneous arrival of background particles with the signal may affect the ability of the GBP to reconstruct the spatial distribution of the Compton photons. The particles produced in the beam pipe exit window should not be considered background, because their spatial distribution is strongly correlated with Compton gammas from the IP. However, there is a very small number of particles which are beyond secondary generation particle (e.g.\ a secondary produced by an IP Compton) and are not correlated with the photon beam. For example, neutrons and other particles produced in the beam dumps belong to this category. The main conclusions, which have been drawn using the LCFA simulation of the $e$--laser interaction, still hold for the recent simulations of the $e$--laser interactions with LMA. The main results are collected below for completeness.

The fraction of `background particles' is of the order of $f<0.02\%$ of all the initial Compton photons produced in a BX. The contribution of the background to the reconstructed profile is completely negligible and, analogously, the effect on the accumulated dose. In fact, the total dose the detectors absorb during the operation must account also for the 9\,Hz non-interacting electron bunches. 

\subsection{Full MC Simulation}
\label{sec:gbp:fullmc}

\begin{figure}[t!]
\center
    \centering
    \includegraphics[width=\linewidth]{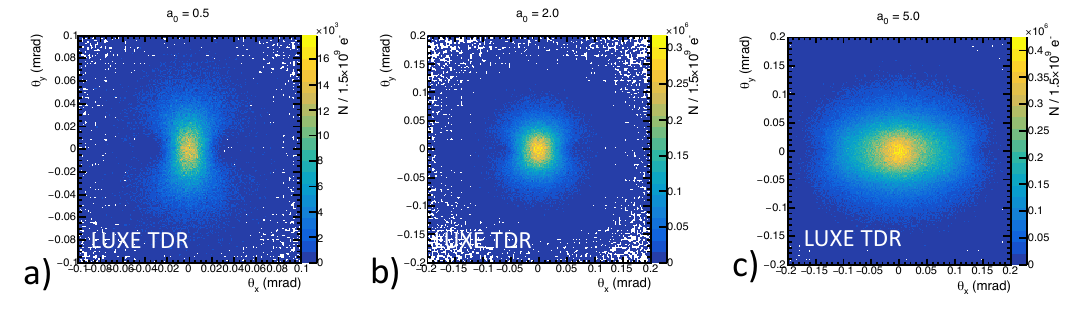}
\caption{Examples of the angular distributions of the number of Compton-scattered photons at the interaction point for a) $\xi=0.5$, b) $\xi=2$, and c) $\xi=5$. The laser polarisation is along the horizontal axis. As justified in more detail later, we show here the distribution of photons with an energy exceeding 1\,MeV.}
\label{profile_xi}
\end{figure}

The spatial distribution of the Compton-scattered photons hitting the detector has been simulated in FLUKA, assuming the full experimental geometry and the spectral and spatial distribution of the Compton-scattered photons obtained from \ptarmigan~\cite{ptarmigan}. A set of simulations were performed, using laser intensities in the range $0.1\leq\xi\leq14$ (see examples in Fig.~\ref{profile_xi}) and either an idealised electron beam (e.g.\ pencil-like and monochromatic) or a realistic beam (e.g.\ expected electron energy distribution and opening angle). Nonetheless, results shown in the following indicate that the electron beam characteristics have a  negligible effect on the photon angular distributions and can thus be neglected in the modelling and data analysis.

\begin{figure}[htbp]
\begin{center}
\includegraphics[width=\linewidth]{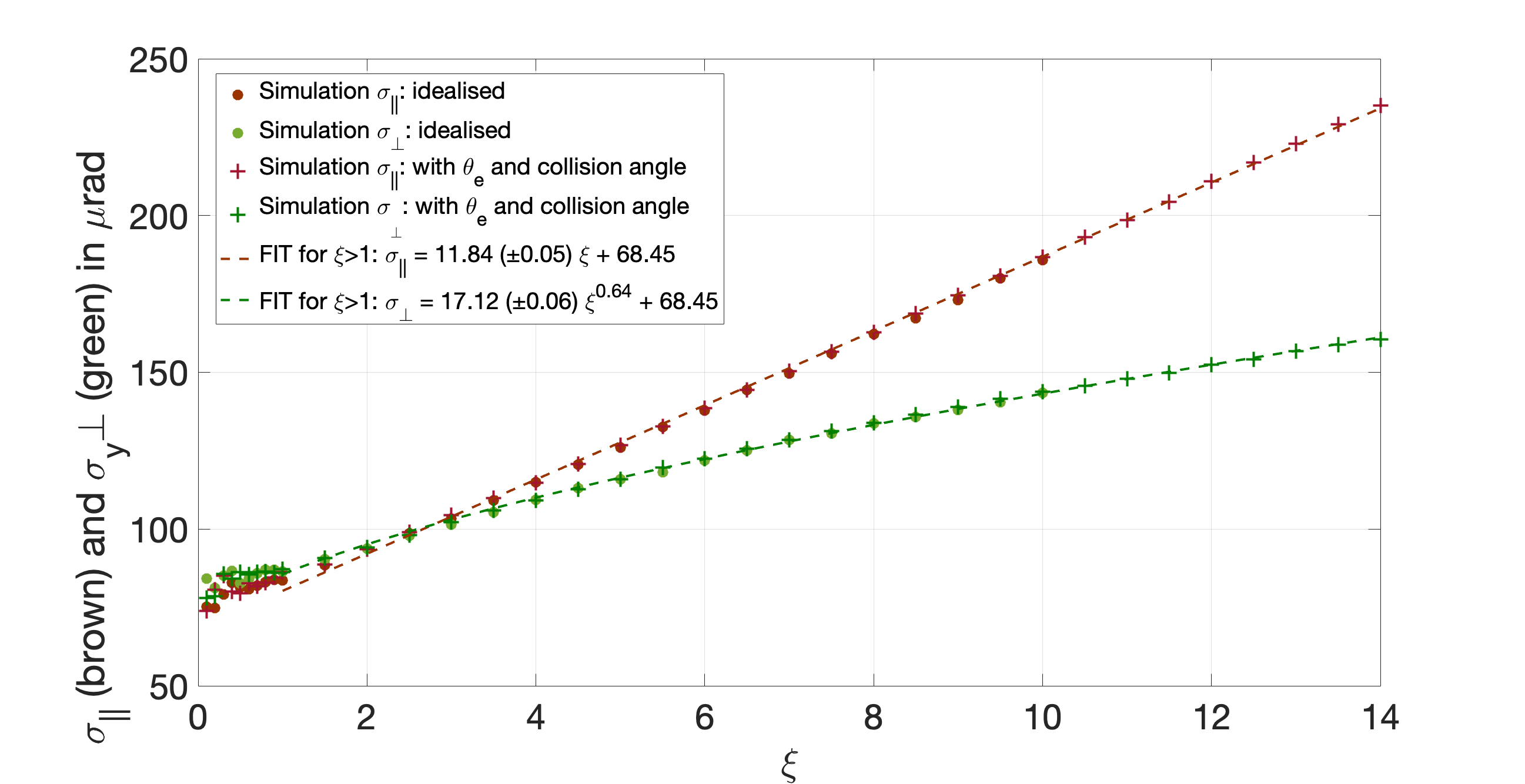}
\caption{Width of the Compton-scattered photon beam (energy per photon $>$ 1\,MeV) at the generation point along the direction parallel (brown) and perpendicular (green) to the laser polarisation. Circles and crosses indicate results from simulations assuming either an ideal or a realistic electron beam, respectively.}
\label{widthatgeneration}
\end{center}
\end{figure}

Qualitatively, at low laser intensities ($\xi<1$) the electron dynamics in the field is linear and thus the photon emission pattern follows the expected dipolar emission for a charge in harmonic motion. At higher laser intensities, the electron dynamics starts to be significantly affected by the laser magnetic field and thus results in a widening of the distribution, especially along the laser polarisation. From a quantum point of view, these two situations correspond to linear Compton (one laser photon absorbed per electron) and non-linear Compton (more than one photon absorbed per electron) scattering, respectively.
For each distribution, the numerical standard deviation has been calculated and used as a representative quantity for the width of the angular distribution along both axes. The extracted widths of the photon beam at the generation point (for photons with energies $>$ 1\,MeV, which is justified later) as a function of the laser intensity parameter $\xi$ is shown in Fig. \ref{widthatgeneration}.

Several clear features can be identified. First, the inclusion of realistic collision parameters (such as a non-collinear geometry and non-zero divergence electron beam) induces a negligible change in the photon distributions and can thus be neglected hereafter. The minimum photon width is of the order of $\sigma_{min} \approx 60 - 65$ $\mu$rad, which corresponds to the minimum opening angle induced by the relativistic beaming of the radiation ($\sigma_{min} \approx 2/\gamma_e$, with $\gamma_e = 3.2\times10^4$ the initial Lorentz factor of 16.5 GeV electrons). For $\xi<1$, while the width of the photon beam appears to depend on the laser intensity, the dependence is not clear and, most importantly, not monotonic. It is important to note that, in this intensity range, the laser intensity can be directly inferred by simply looking at the integrated number of photons emitted, which is directly proportional to the physical laser intensity ($I_L\propto \xi^2$). For $\xi>1$, the width of the photon beam parallel to the laser polarisation is, as expected, linear with $\xi$. The behaviour along the axis perpendicular to the laser polarisation is more complex. Even though the behaviour is still monotonic with intensity, it appears to follow a trend of the kind $\sigma_{\perp} \propto \xi^{2/3}$. By fitting the simulated data, the relations between $\sigma_{||},  \sigma_{\perp}$ (in $\mu$rad units) and $\xi$ are found to be:

\begin{eqnarray}
\sigma_{||}(\xi) & = & 11.84 (\pm0.05) \xi + 68.45 \nonumber \\
\sigma_{\perp}(\xi) & = & 17.12 (\pm0.06) \xi^{0.64} + 68.45 
\label{eq:widths}
\end{eqnarray}

\begin{figure}[htbp]
\center
\includegraphics[width=0.8\linewidth]{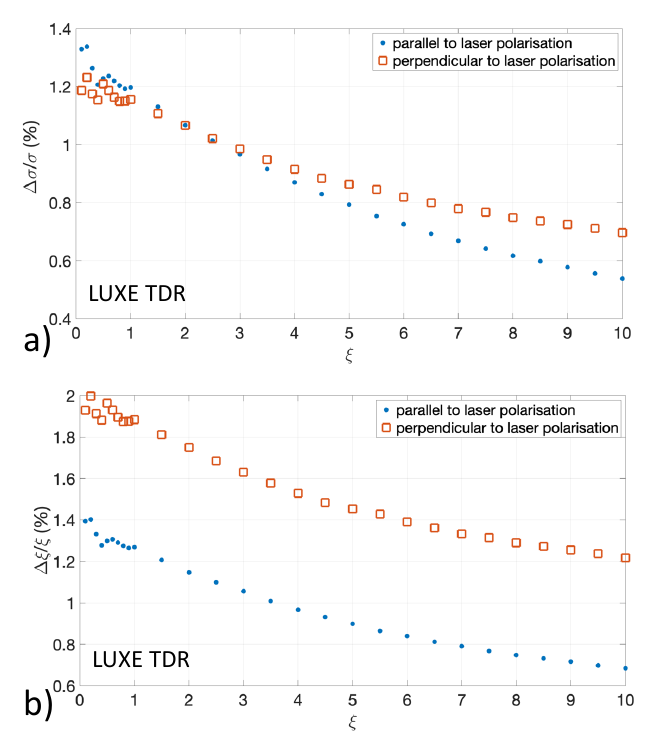}
\caption{a) Relative uncertainty in the width of the gamma-ray beam conservatively assuming a spatial resolution of the detector $\Delta x = 10$ $\mu$m. b) Corresponding uncertainty in the extracted value of $\xi$. In both panels, blue circles and orange squares correspond to distributions parallel and perpendicular to the laser polarisation, respectively.}
\label{fig:error}
\end{figure}

While qualitatively expected, a quantitative theoretical explanation of these trends is currently under development and will be presented elsewhere. Nonetheless, the fitting curves can be used for the laser intensity retrieval. The spatial resolution of the detector and the uncertainty in the fitting parameters allow the extraction of a precision in retrieving the intensity parameter $\xi$ from the widths of the gamma ray beam. Following Eqs.~\ref{eq:widths} and conservatively assuming a spatial resolution of $\Delta x = 10$ $\mu$m, the uncertainty in $\xi$ can be estimated as:

\begin{eqnarray}
\frac{\Delta\xi_{||}}{\xi_{||}} & = & \sqrt{\left(\frac{\Delta\sigma_{||}}{\sigma_{||}}\right)^2 + \left(\frac{\Delta a} {a}\right)^2} \\
\frac{\Delta\xi_{\perp}}{\xi_{\perp}} & = &  1.56\sqrt{\left(\frac{\Delta\sigma_{\perp}}{\sigma_{\perp}}\right)^2 + \left(\frac{\Delta a} {a}\right)^2},
\end{eqnarray} \label{uncertainty}
where $\xi_{||}$ and $\xi_{\perp}$ refer to whether the intensity is retrieved from the distribution
parallel or perpendicular to the laser polarisation, respectively, and $a$ is the multiplying coefficient in Eqs.~\ref{eq:widths} (i.e. $a = 11.84\pm0.05$ and $a = 17.12\pm0.06$, respectively). The relative uncertainty expected for $\xi$, together with the relative uncertainty in the widths of the distribution is shown in Fig.~\ref{fig:error}. As one can see from the figure, the uncertainty in $\xi$ is of the order of (if not less than) 1\% for the whole intensity range of interest.
Degradation of the precision due to systematic effects has not been studied so far.  A sub-par performance of the accelerator could for example result in a higher electron divergence, which can affect the precision in reconstructing the laser intensity parameter. While this can be easily factored in the analysis, we assume in this document nominal performance of the accelerator.

\begin{figure}[htbp]
\center
\includegraphics[width=0.8\linewidth]{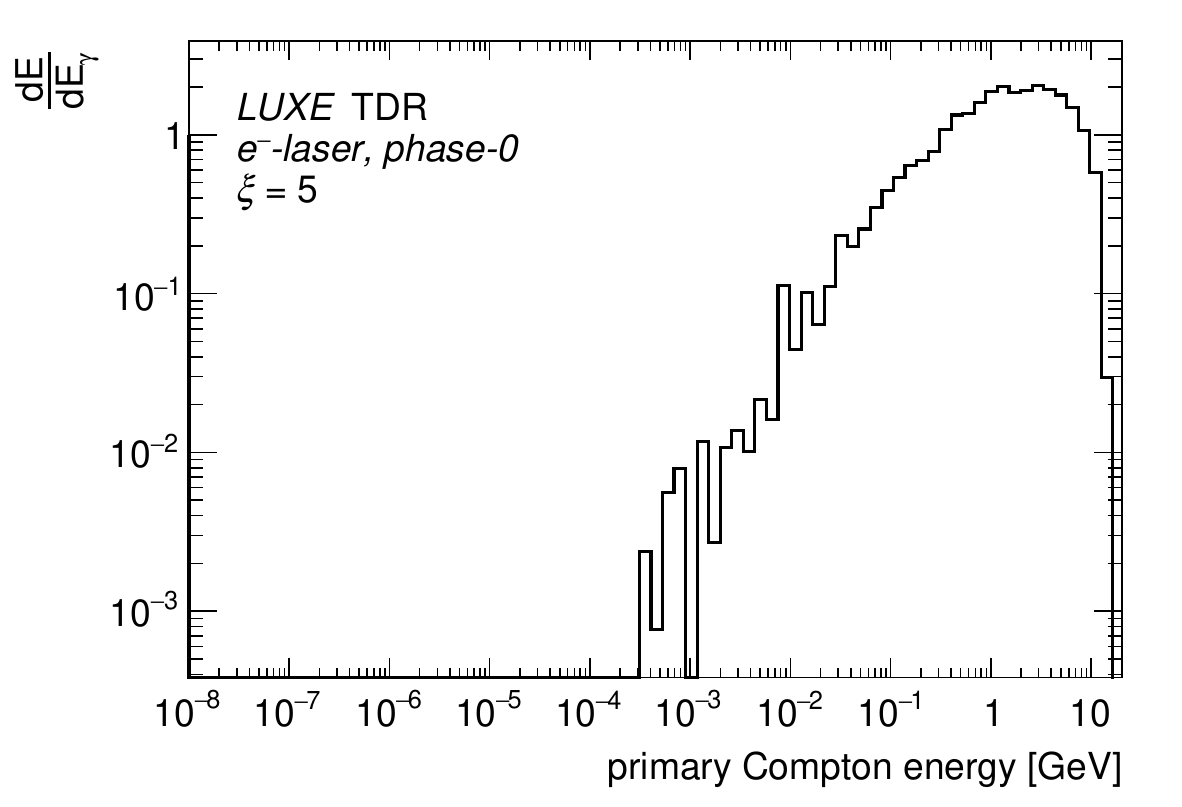}
\caption{Energy deposited in the upstream sapphire sensor as a function of the primary Compton energy. ($e$--laser (LMA), phase-0, $\xi=5$ setup.)}
\label{spectrum_detector}
\end{figure}

This treatment is valid for the gamma ray beam characteristics at the generation point. However, the beam will have to travel approximately 11.5\,m before reaching the detector, passing through the converter of the gamma ray spectrometer, a kapton window ending the vacuum region in the experiment, and a certain length of air. Monte Carlo simulations have thus been performed to check on the characteristics of the gamma ray beam at the detector plane. As shown in Fig.~\ref{spectrum_detector}, the contribution to the energy deposited and, therefore, to the signal, is negligible for photons with energy below 1\,MeV, thus justifying the assumption of studying the beam distribution only for photons with energies exceeding 1\,MeV.

\begin{figure}[htbp]
\center
\includegraphics[width=0.8\linewidth]{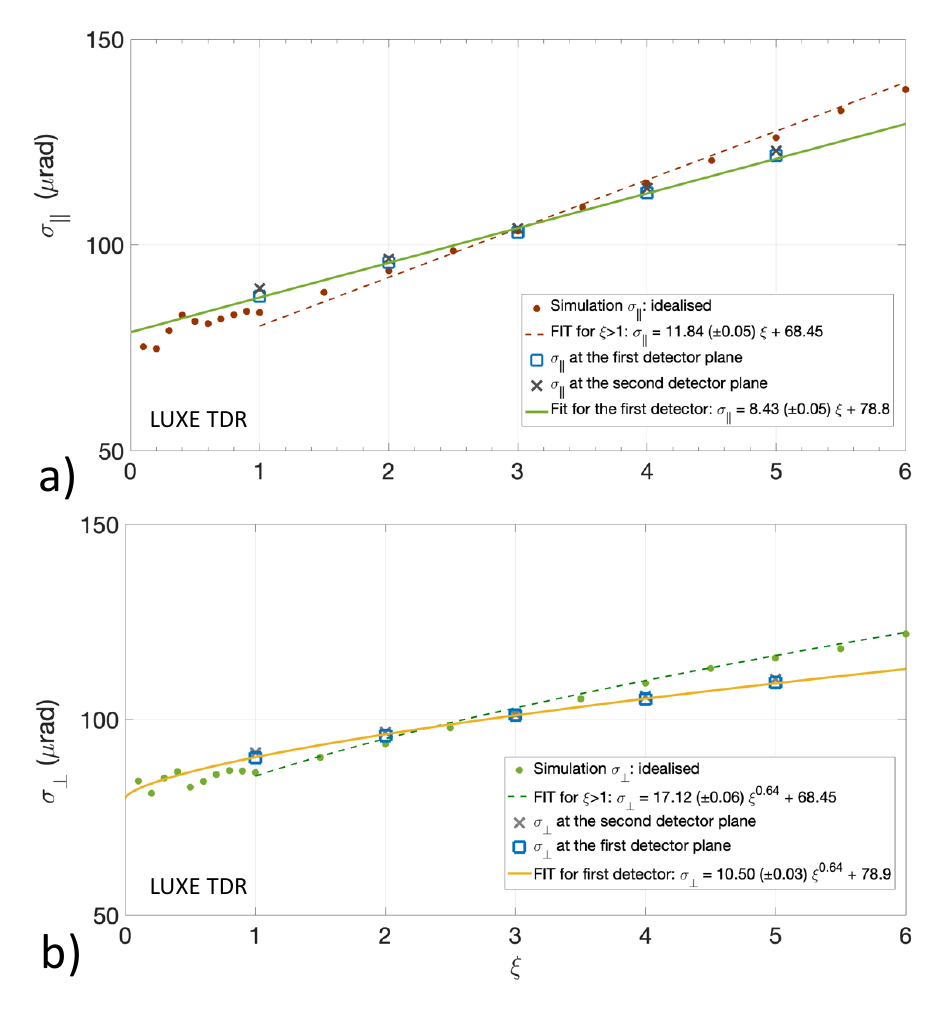}
\caption{a) Comparison between the width of the gamma ray beam parallel to the laser polarisation at the generation point (brown circles with corresponding fit) and at the plane of the first (blue squares) and second (grey crosses) detector. The green line represents a linear fit of the data at the first detector plane. b) Comparison between the width of the gamma ray beam perpendicular to the laser polarisation at the generation point (green circles with corresponding fit) and at the plane of the first (blue squares) and second (grey crosses) detector. The yellow line represents a linear fit of the data at the first detector plane with the same functional dependence on $\xi$.}
\label{at_detector}
\end{figure}

The propagation, however, also affects the angular distribution of the photons at the detector plane. To illustrate this effect, we show in Fig.~\ref{at_detector} a comparison between the simulated results at the generation point and at the detector plane for the directions parallel and perpendicular to the laser polarisation. It is important to note that, while the propagation does induce a slight flattening of the dependence of the width on the laser intensity, it does not change the functional dependence. The same fitting curves can then in principle be used to extract the laser intensity, although with slightly different coefficients; similar curves can also be easily obtained once the whole experimental setup has been installed.
Finally, the distributions at the first and second detector plane appear virtually identical, indicating a negligible effect of the propagation through the first detector on the angular distribution of the photons (Fig.~\ref{at_detector}).  

\section{Technical Description}
\label{techdescr}

\subsection{Sapphire Detector}
\label{sec:sapphire_detector}

\begin{figure}
      \centering
      \includegraphics[width=1\linewidth]{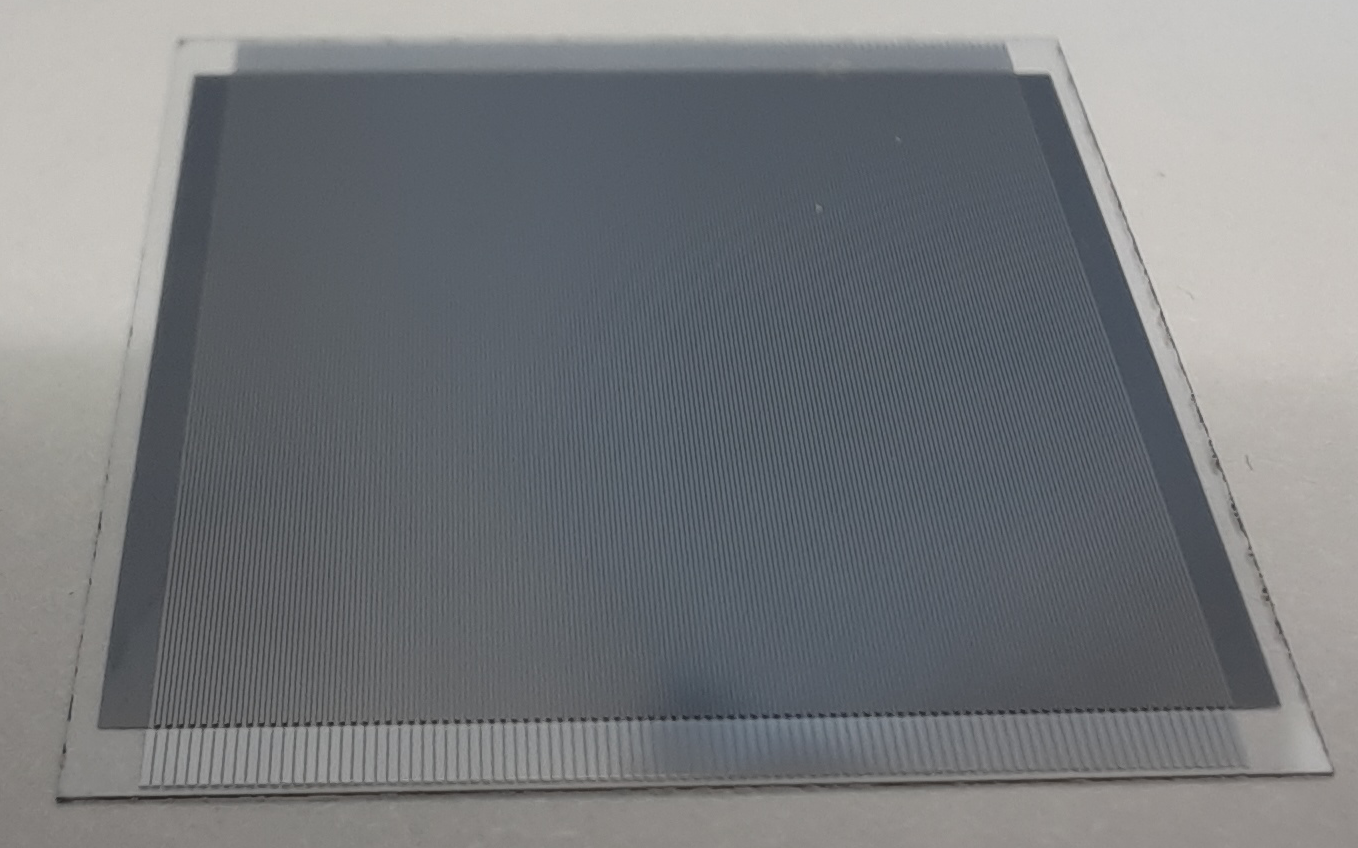}
      \includegraphics[width=1\linewidth]{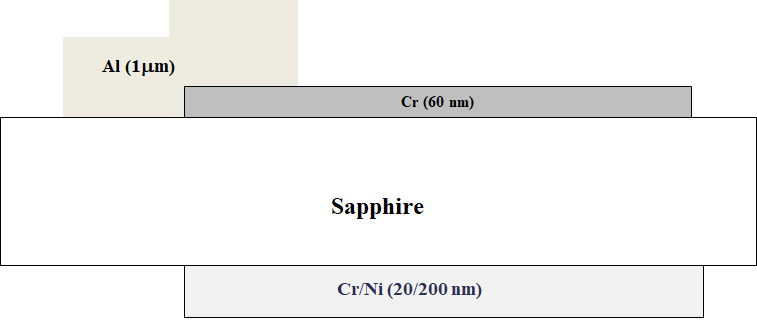}
    \caption{ (Top) 192 microstrips sensor. (Bottom) Sapphire sensor contact structure.}
    \label{fig:192_microstrips_sensor}
\end{figure}

\begin{figure}[htp]
    \centering
    \includegraphics[width=0.8\textwidth]{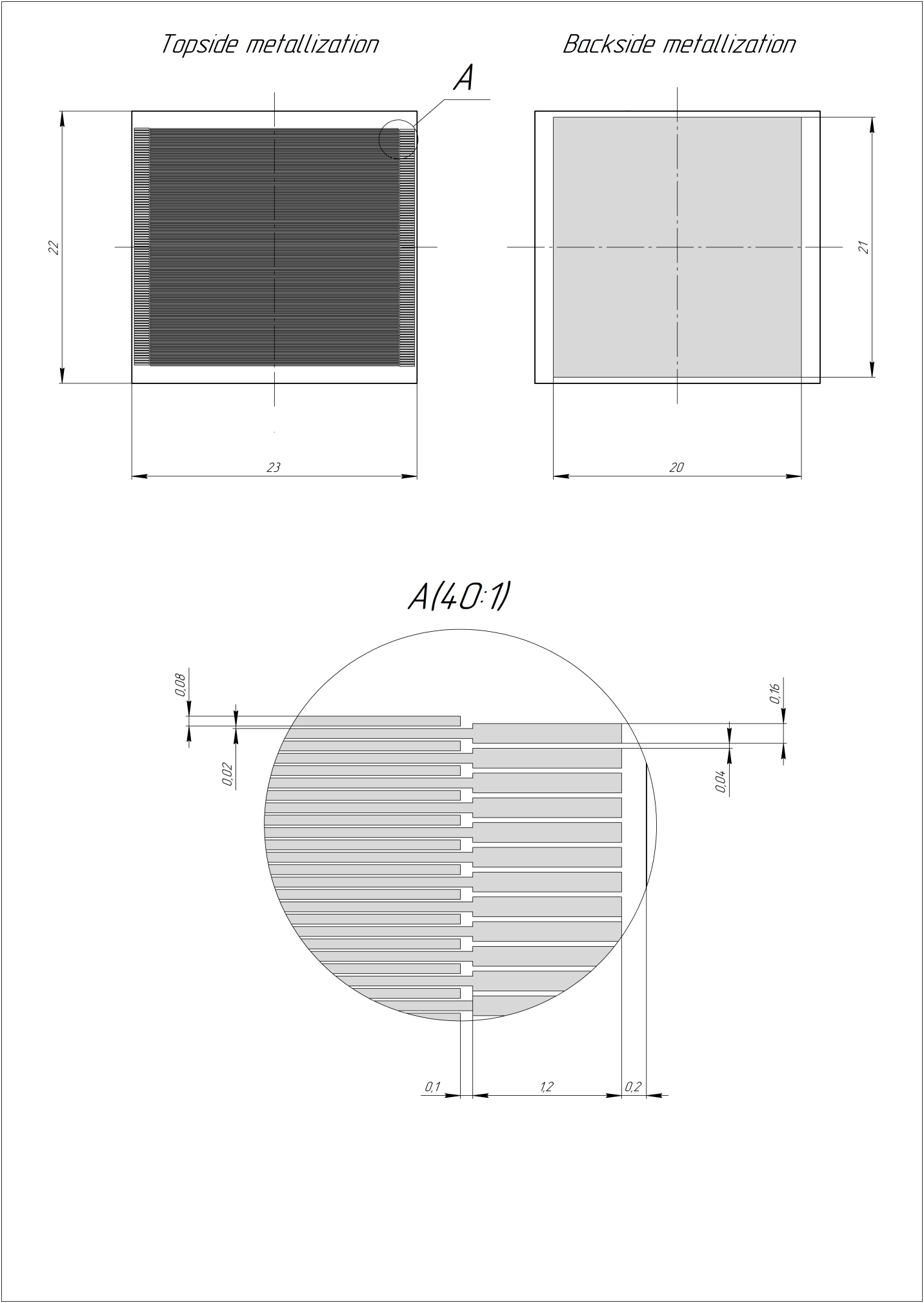}
    \caption{Sapphire sensor contact topology (scale 4:1).}
    \label{fig:sapphire_sensor_contact_topology}
\end{figure}

Sapphire microstrip sensors are made of sapphire single crystals. Microstrips of $0.08\times20$\,mm$^2$ sizes, 0.1 mm pitch as well as backplane contact are made of 1\,$\mu$m thick Al layers by means of magnetron sputtering technology. The sensors are 100--170\,$\mu$m thick with overall sizes of $22\times23$\,mm$^2$. The active area is $20\times19.2$\,mm$^2$ (192 microstrips). A 192 microstrips sensor is shown in Fig.~\ref{fig:192_microstrips_sensor}. There is no guard ring. The backplane contact is intended for high voltage (HV) bias and has a size of $20\times21$\,mm$^2$. The sensor layout is shown in Fig.~\ref{fig:sapphire_sensor_contact_topology} and Table~\ref{tab:sapphire_sensor_specification}.

\begin{table}[ht]
    \centering
    \begin{tabular}{l|r}
        Sapphire thickness, $\mu $m & 100--170 \\
        \hline
        Contact pad & Al\\
        \hline
        Metallization thickness, $\mu $m & 1.0 \\
        \hline
        Pad size, µm & $1200\times160$ \\
        \hline
        Bonding type & Left-right \\
        \hline
        Number of strips & 192 
    \end{tabular}
    \caption{Sapphire sensor specification.}
    \label{tab:sapphire_sensor_specification}
\end{table}

\begin{figure}
    \centering
    \includegraphics[width=0.6\textwidth]{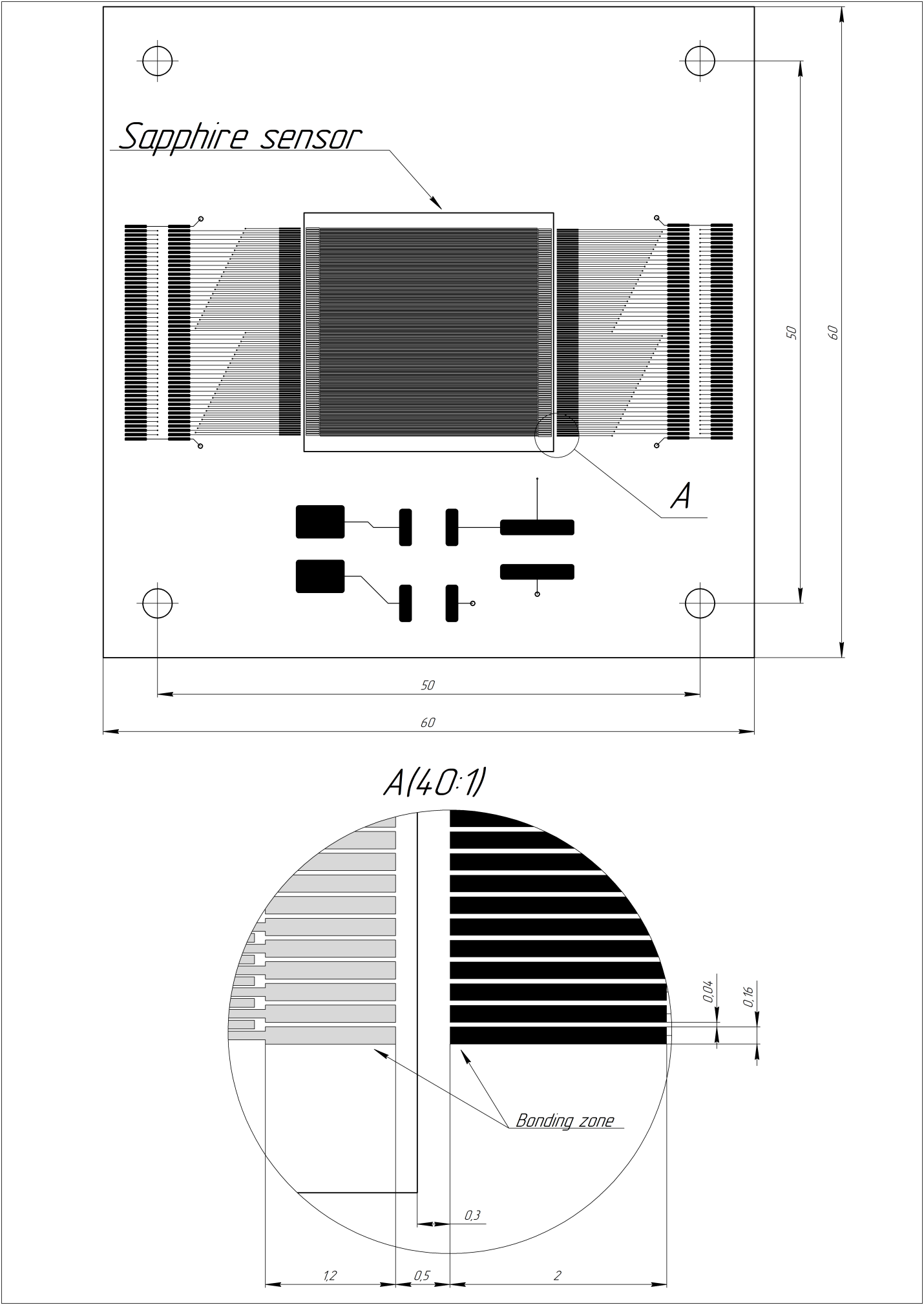}   
    \caption{Placement of the sensor on printed circuit board (scale 4:1).}
    \label{fig:sensor_placement}
\end{figure}

The sensor is assembled on a printed circuit board (PCB) by epoxy glue. The PCB has RC elements for HV filtering as well as fan-out for electrical connections with sensor microstrips with a 25\,$\mu$m diameter Al wire. The overall size of the PCB is $60\times60$\,mm$^2$. The PCB layout is shown in Fig.~\ref{fig:sensor_placement} and Table~\ref{tab:PCB_specification}. The signals from strips via fan-out are delivered to 1.27\,mm pitch SAMTEC connectors (not shown in Fig.~\ref{fig:sensor_placement}).

\begin{table}[ht]
    \centering
    \begin{tabular}{l|r}
        PCB Size, mm & $60\times60$ \\
        \hline
        Thickness, mm & 1.6\\
        \hline
        Number of layers & 4 layers \\
        \hline
        Cu thickness, $\mu $m & 18 \\
        \hline
        Type metallization & ENIG \\
        \hline
        Length of aluminum wire, mm & 0.7--1 \\
        \hline
        Wire diameter, $\mu $m & 25 
    \end{tabular}
    \caption{PCB board specification.}
    \label{tab:PCB_specification}
\end{table}

\subsubsection{Photo-Current Mapping}

The photo-current mappings were carried out to investigate the suitability of sapphire wafers for producing micro-strip sensors. The $1.8 \times 1.8$\,mm$^2$ or $4.5 \times 4.5$\,mm$^2$ pixels were formed on one of wafer surface by means of magnetron sputtering of an Al target followed by photoresistive mask formation and selective etching of the Al layer. The opposite sides of sapphire wafers had continuous 1\,$\mu$m thick Al contact. 

\begin{figure}
    \centering
    \includegraphics[width=0.6\textwidth]{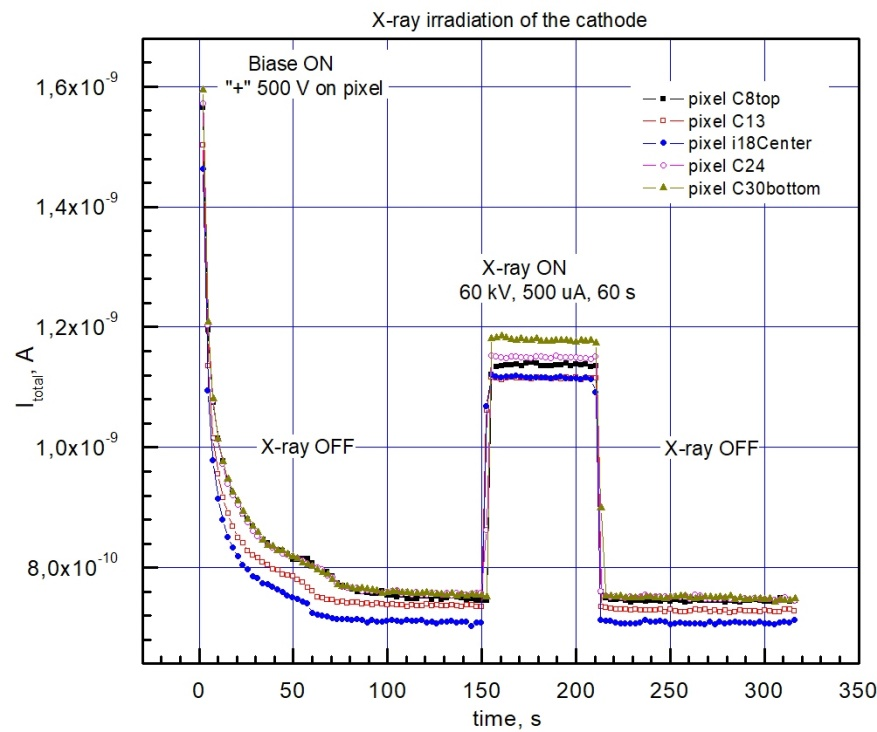}   
    \caption{Temporal response of leakage current and photo-current of $1.8\times1.8$\,mm$^2$ pixels under X-ray irradiation. X-ray pulse duration is 60\,s and 500\,V bias on pixels.}
    \label{fig:temporal_response}
\end{figure}

\begin{figure}[htp]
    \centering
      \includegraphics[width=0.8\linewidth]{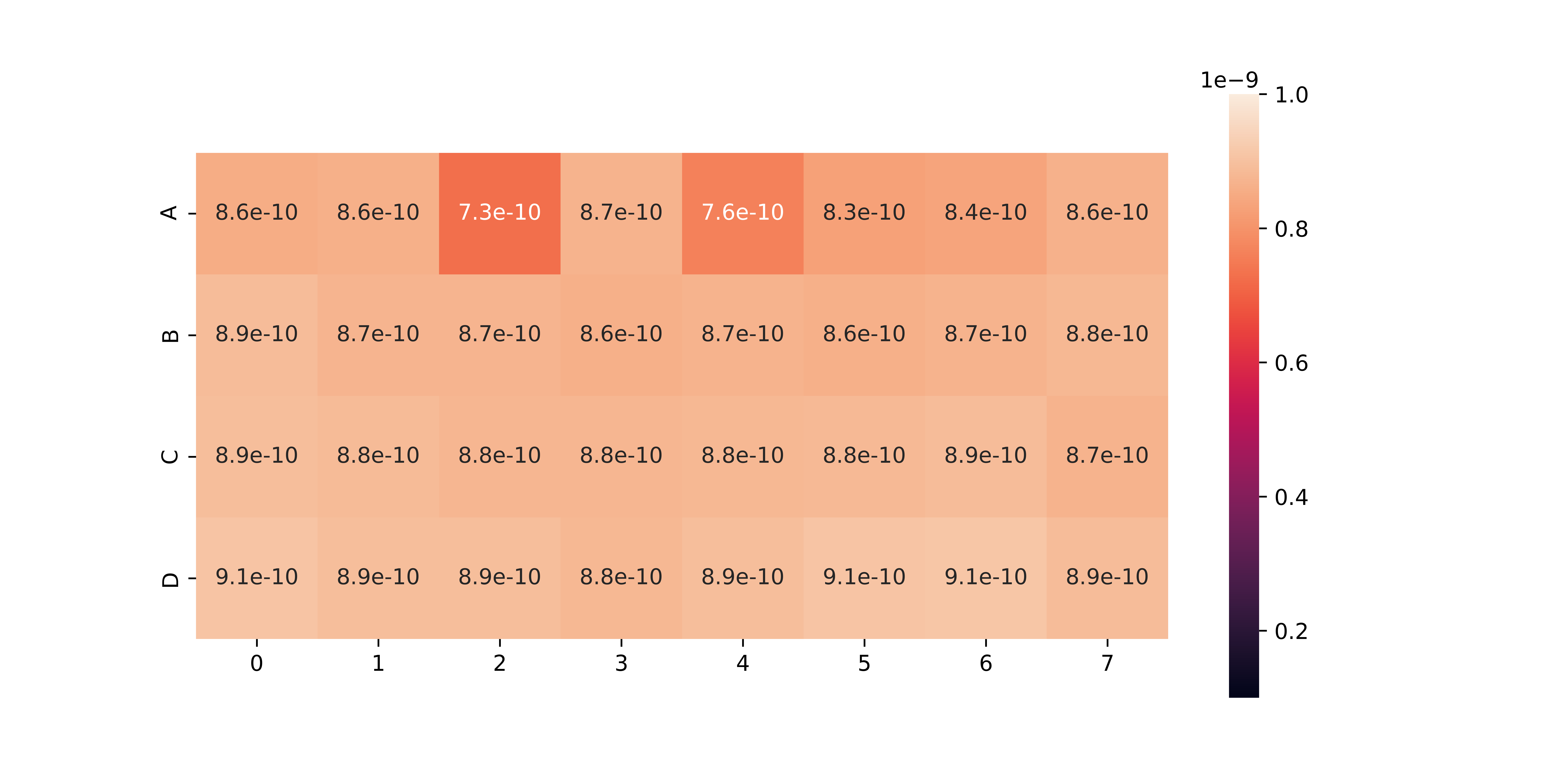}
      \includegraphics[width=0.8\linewidth]{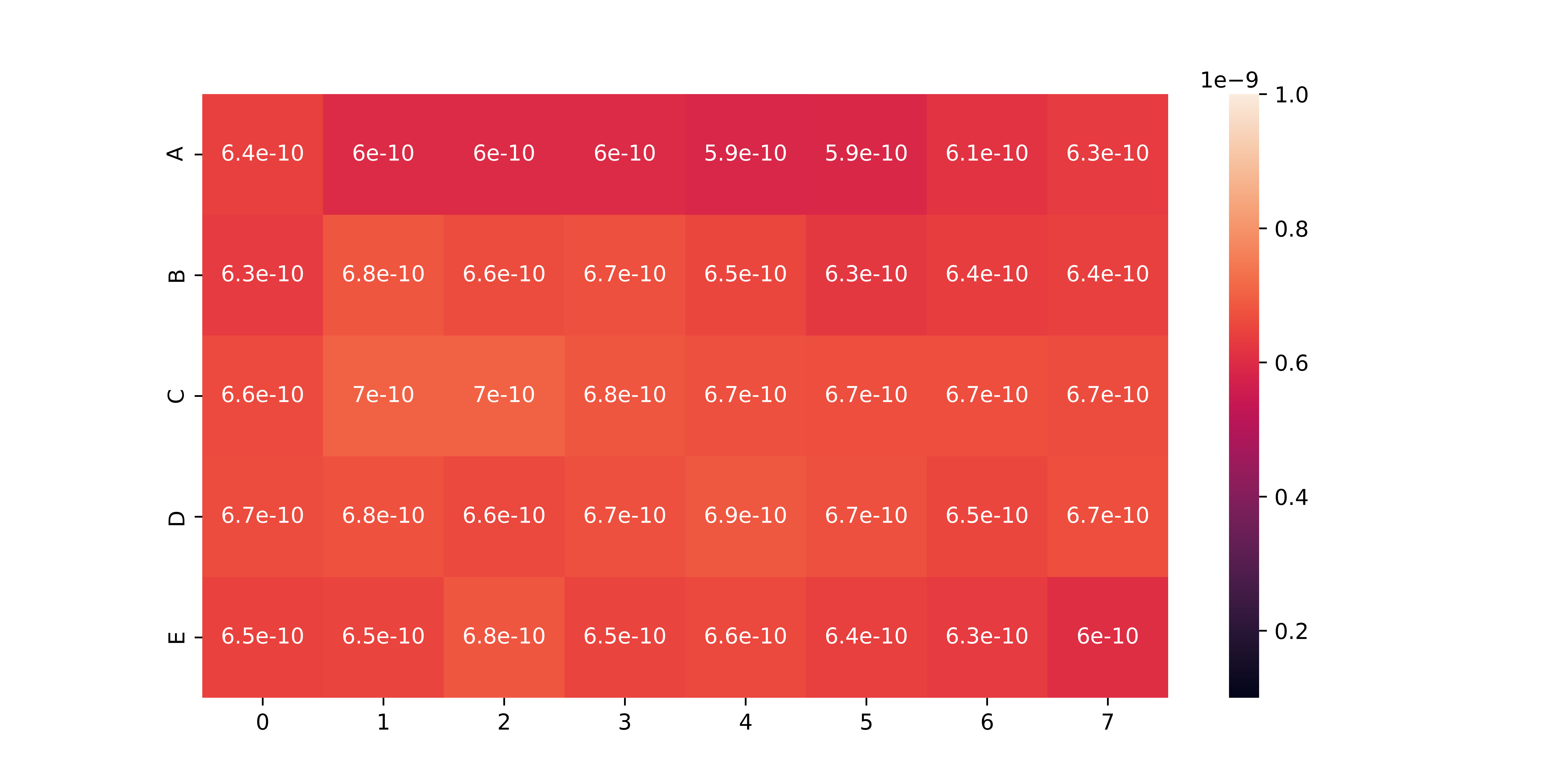}
      \includegraphics[width=0.8\linewidth ]{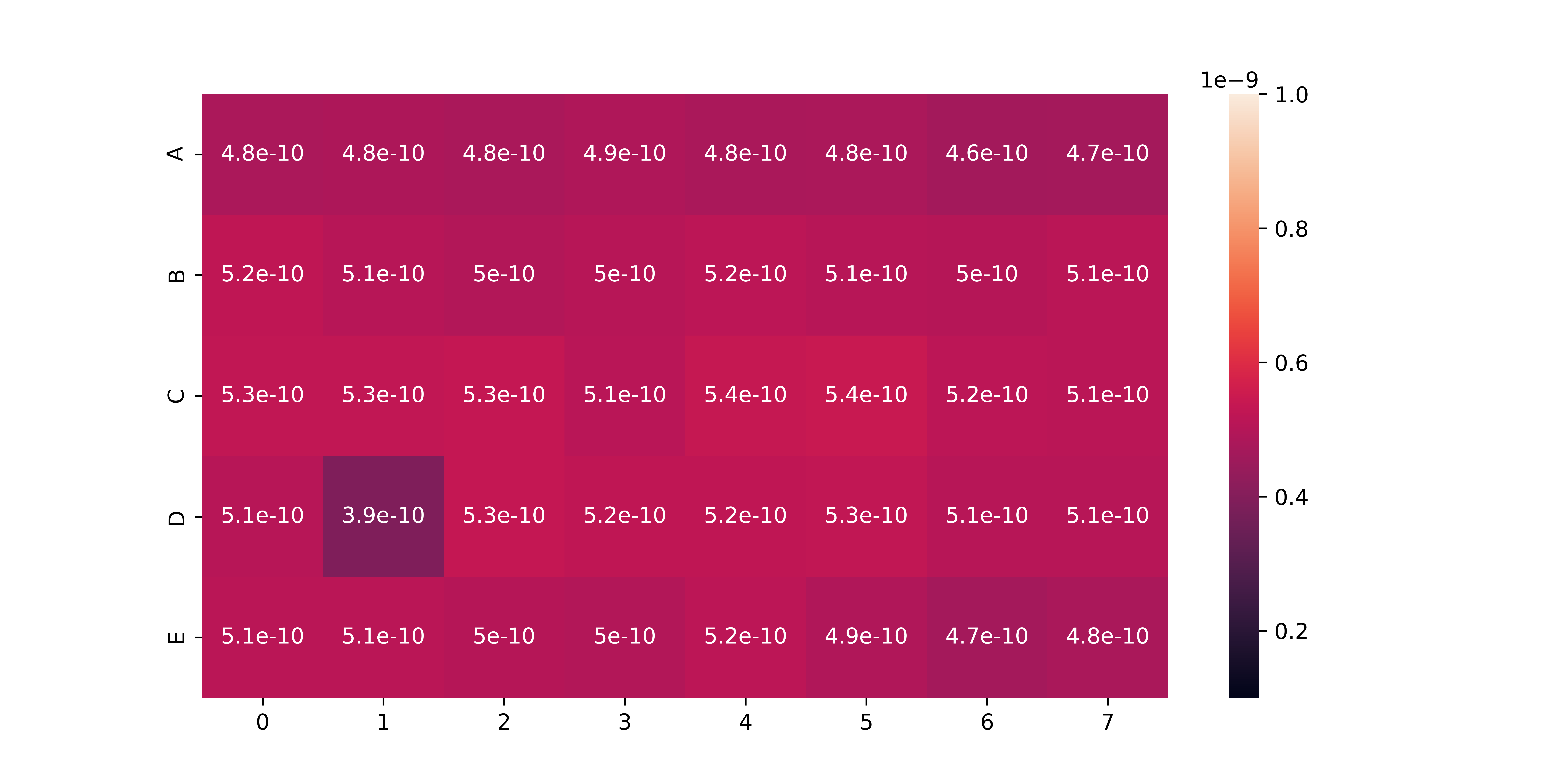}
    \caption { Photo-current (in A) mappings for three sapphire wafers from different producers: (top) Producer \textnumero 1 , $30\times48$\,mm$^2$ and 0.15\,mm thick; (middle) Producer \textnumero 2, 50\,mm diameter, and 0.1\,mm thick; (bottom) Producer \textnumero 3, 50\.mm diameter, and 0.15\,mm thick.}
    \label{fig:photocurrent_mappings}
\end{figure}

The photo-current was generated with X-ray irradiation of a W anode X-ray tube at 60\,kV bias and 500\,$\mu$A current. The pixel photo-currents were measured at 500\,V bias on pixels. Figure~\ref{fig:temporal_response} shows temporal response of pixel photo-current for pixel with sizes of $1.8 \times 1.8$\,mm$^2$. Photo-current mappings of the wafers from different producers (in the following named as \textnumero 1, 2 and 3) are shown in Fig.~\ref{fig:photocurrent_mappings}. For that case $4.5 \times 4.5$\,mm$^2$ pixels were used. Nine sapphire wafers were tested in total.

It can be seen that the photo-current distribution within each wafer is fairly uniform. Thus, the entire area of the wafers can be used to manufacture microstrip sensor (Figs.~\ref{fig:192_microstrips_sensor} and~\ref{fig:sapphire_sensor_contact_topology}).

\subsubsection{Simulation of Charge Collection Efficiency}
\label{sec:cce_simulation}

\begin{figure}
    \centering
      \includegraphics[width=1\linewidth]{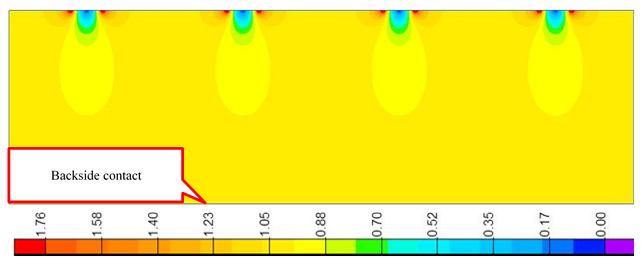}
      \includegraphics[width=1\linewidth]{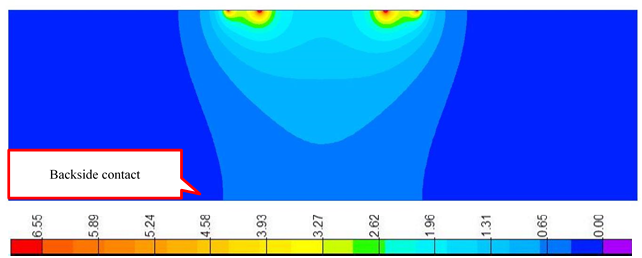}
    \caption {Simulations of 2D electric field (a) and “weight” field (b) distributions into the volume of 100\,$\mu $m thick sapphire micro-strip sensor. Pixel bias: 100\,V (a) and 1\,V (b). Electric field strength in V/m, “weight” field strength in 1/m.}
    \label{fig:sapphire_microstrip_simulation}
\end{figure}

Simulations of 2D electric field and “weight” field distributions into the volume of sapphire micro-strip sensor were carried out. The simulations are based on solving 2D Laplace equation for the given sensor topology. The input data are thickness, pitch and strip width of the sensor. The results are shown on Fig.~\ref{fig:sapphire_microstrip_simulation}.

\begin{figure}
    \centering
      \includegraphics[width=0.65\linewidth]{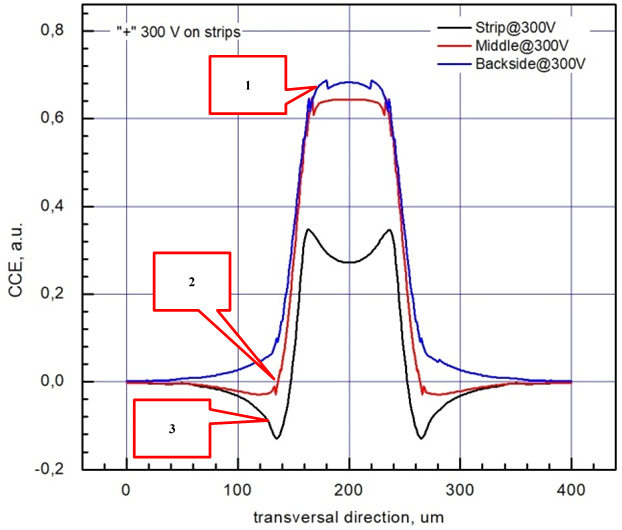}
      \includegraphics[width=0.65\linewidth]{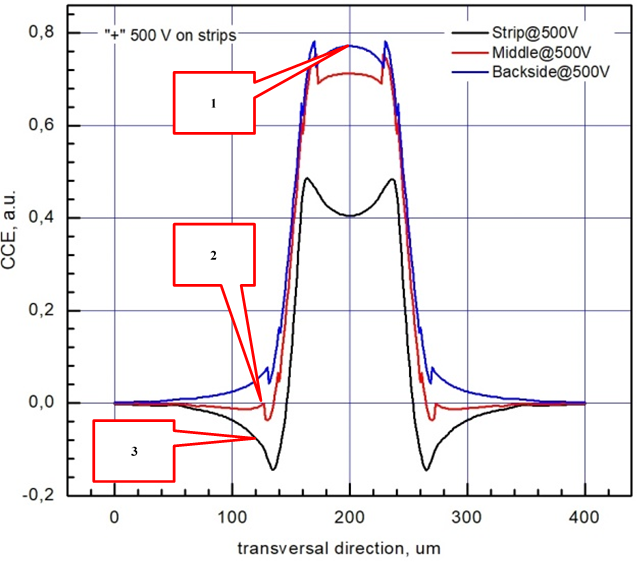}
    \caption { Simulated 2D CCE profile for X (width) direction at different depths into volume of 100\,$\mu$m thick micro-strip sapphire sensor: 1,  5\,$\mu$m above backside contact; 2, at half the sensor thickness; 3, 5\,$\mu$m below micro-strip surface. Sensor bias are “+” 300\,V (a) and “+” 500\,V (b) on microstrips.}
    \label{fig:simulation_2d_cce}
\end{figure}

Simulation of 2D charge collection efficiency (CCE) distribution into the volume of sapphire micro-strip sensor was done. The simulation is based on calculation of 2D induced current in accordance with the Ramo-Shockley theorem~\cite{Shockley:1938itm}. The input data are the 2D electric and “weight” fields, the charge carrier life times and mobilities. The results are presented in Fig.~\ref{fig:simulation_2d_cce}.

\subsection{Electronics}
The main characteristics to take into account for the design of the detector electronics are the following:
\begin{itemize}
    \item 
    the central strips will deliver plenty of signal, of the order of tens of pC;
    \item
    a dynamic range of 12--13 bits will  be  more than sufficient in order to reconstruct well the tails of the gamma beam profile distribution and to measure the beam intensity with a precision at percent level;
    \item
    the acquisition rate will not be higher than 10\,Hz.
\end{itemize}
Given the signal abundance and the low acquisition rate, the design of the electronics is considered to be quite straightforward. The CAEN company has developed the high-scalable system FERS-5200~\cite{fers} which matches all these requirements and this is proposed as the baseline readout system for the GBP.
The FERS-5200 system consists of:

\begin{itemize}
    \item an A5202 front-end card, reading 64 channels at time. Three front-end cards will be therefore needed to read out one plane of 192 strips;
    \item a DT5215 FERS data concentrator, which can connect and manage up to 16 A5202 units in a daisy chain for each of its 8 TDlinks inputs.
\end{itemize}

The FERS-5200 system has been designed for the readout of large detector arrays, such as SiPMs, multianode PMTs, silicon strips detectors and many others, and it can be adopted also for the GBP sapphire strips detector readout.
A description of each of these modules is reported in~\cite{fers_manual}.


\subsubsection{The A5202 Front-End Card}

\begin{figure}
    \centering
    \includegraphics[width=\textwidth]{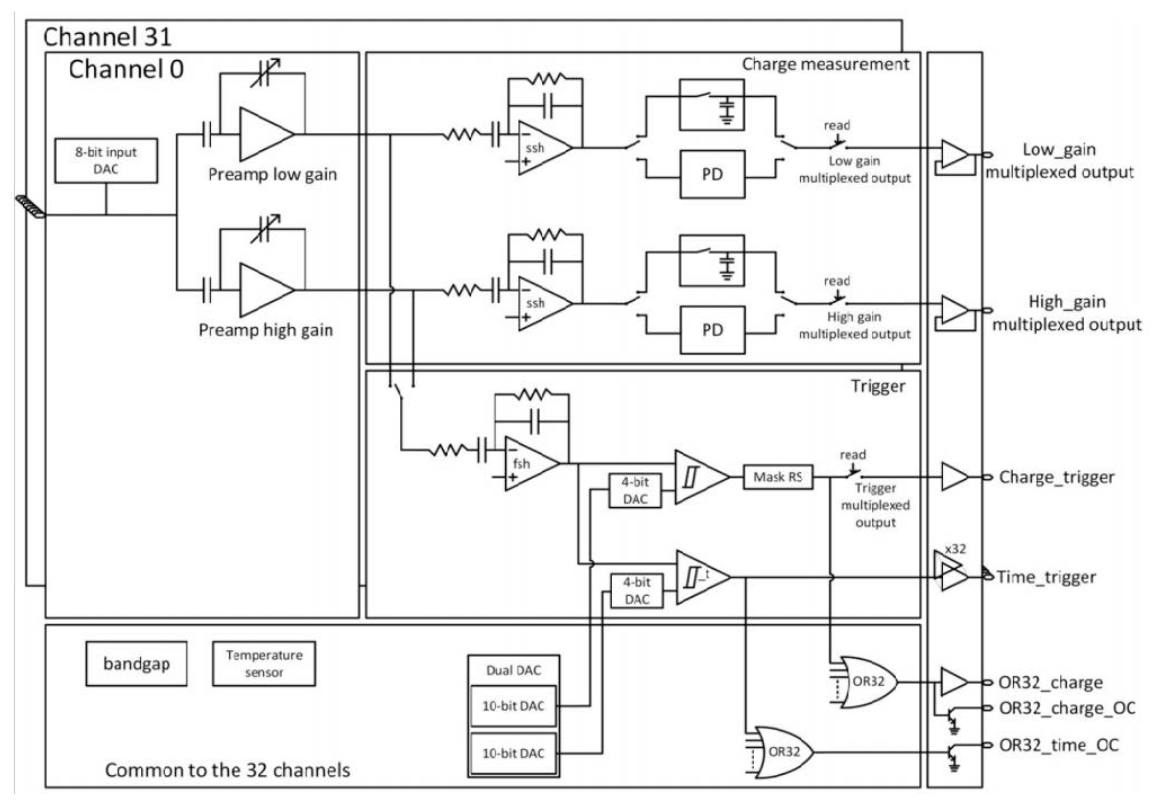}
    \caption{General ASIC block scheme.}
    \label{fig:asic}
\end{figure}

The front-end card is based on the Citiroc 1A ASIC~\cite{citiroc} whose block scheme is shown in Fig.~\ref{fig:asic}.

One ASIC is able to read 32 channels. Citiroc 1A is compatible with positive-charge signals. Two measurement
lines with different gain (1 to 10 ratio) are working in parallel in order to maximise the dynamic range of
Citiroc 1A. For each channel, two parallel AC coupled voltage preamplifiers are embedded to ensure a dynamic 
range from about 10\,fC to about 100\,pC (the full range could actually be 400\,pC), well suited to accommodate the 
different amount of collected charge from different strips. Each preamplifier is followed by a variable 
shaper with an adjustable time constant (from 12.5\,ns to 87.5\,ns with a 12.5\,ns pitch) to filter the signal 
bandwidth and optimise the signal-to-noise ratio (S/N). The signal from the two shapers can be sampled using either a sample \& hold controlled by an external signal (all the 32 channels are held at the same time) or by a peak-detector disabled by an external signal.
The GBP will use the former of these two features since the charge integration will  achieve a better S/N. The stored charge information is read out in serial using two analogue multiplexed outputs, one for low gain and one for high gain. The multiplexed outputs are controlled by a read shift-register. Both information on low gain and high gain of the same channel are present at the same time. Along with the two charge measurement information, a hit-register provides a trigger status information. This trigger status information will be used for debugging purposes in the lab or in the test beam, but will not be relevant for the LUXE data acquisition mode which is planned to record all the BX data.

\begin{figure}
    \centering
    \includegraphics[width=\textwidth]{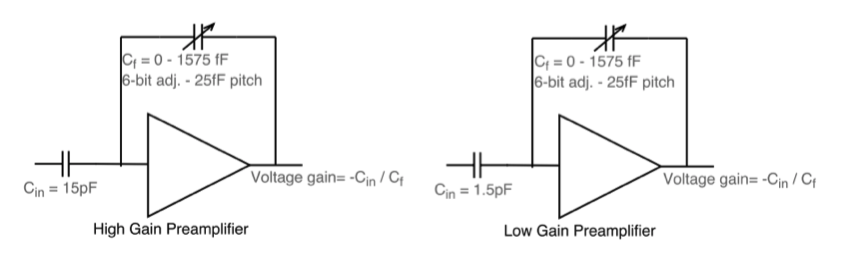}
    \caption{High gain and low gain voltage sensitive preamplifier.}
    \label{fig:amplifier}
\end{figure}

Each channel of Citiroc 1A embeds two channel-by-channel independent programmable variable-gain preamplifiers ensuring a wide coverage of the dynamic range depending on the application (see Fig.~\ref{fig:amplifier}). Both low gain and high gain preamplifiers can be tuned on 6 bits (Cf can be tuned from 0 to 1575\,fF with a step of 25\,fF). The voltage gain is given by Cin/Cf, with Cin\,=\,15\,pF for high gain and 1.5\,pF for low gain.
Any channel preamplifier can be shut down by a slow control bit to disconnect any noisy channel from the measurement chain.
A calibration input is included in each channel and can be enabled individually by slow control parameters. In order to have a good precision, only one calibration input should be allowed at any given time. The injection calibration capacitance value is about 3\,pF.

Two CRRC slow shapers are connected on the two preamplifiers outputs for each channel. The peaking time of each slow shaper can be tuned from 12.5\,ns to 87.5\,ns with a 12.5\,ns pitch. The peaking time is set commonly for all the 32 channels, however it can be different between low gain shaper and high gain shaper.
A 15 ns peaking-time fast shaper can either be connected to the high gain preamplifier or to the low gain. That connection is set by slow control. The fast shaper peaking time and gain are not programmable. The fast shaper is used to provide trigger output which is of limited interest for the LUXE running mode.

Each A5202 front-end card will contain two Citiroc 1A ASICs, so allowing the readout of 64 channels. The board logic will allow  configuration of the readout channels,  transmission of the readout data to the DT215 data concentrator, and 13 bit conversion to be made for each analog level from the track and hold capacitors.

\subsubsection{The DT5215 Data Concentrator}

One DT5215 (FERS data concentrator) can manage up to 8 TDlinks (optical links, 4.25\,Gbps each), each connected up to 16 FERS units in a daisy chain: in the case of the A5202 FERS unit, this sums to 8192 readout channels. 
The data concentrator is connected to a Host computer through a 1/10\,Gb Ethernet or USB 3.0.
A Linux-based single board computer is embedded in the concentrator board. It manages the data readout from the network of FERS units and the event data building according to the time stamp and/or trigger ID of the event fragments acquired by each unit. Sorted and merged data packets are then stored to the local memory and finally sent to the host computers through a fast 10\,GbE or USB\,3.0 link. The decision about which of the two options to adopt, has still to be taken. Custom algorithms for data processing and reduction will be developed and uploaded into the embedded CPU to fully integrate the system in the LUXE data acquisition system. 

\subsubsection{Readout System Description}

Each sapphire strip detector will need three A5202 front-end cards to be read out, for a total of 192 readout channels per detector. The three cards will be plugged in a patch panel collecting the signals from the one detector (fed into kapton flat cables) and distributing them to the A5202 front-end cards. The patch panel will be mounted in the platform supporting the detector in the proximity of the two detectors themselves.
In this way, the front-end will be placed very close to the detector, in a configuration which should optimise the obtainable S/N. One patch will be needed to read out a station of two $xy$ strips detectors, so accommodating the insertion of 3+3 front-end cards.
The three cards reading out one strip detector will be daisy chained together via short optical fibres, and the overall output fed in a ``long" (about 20\,m) fibre connecting via a fast TDlink (4.25\,Gb/s) the front-end to one input of the DT5215 data concentrator box placed in the experiment counting room. 
The system is modular, and a second station with another pair of $xy$ sapphire strips detector can be easily added in the future.

\begin{figure}
    \centering
    \includegraphics[width=0.5\textwidth]{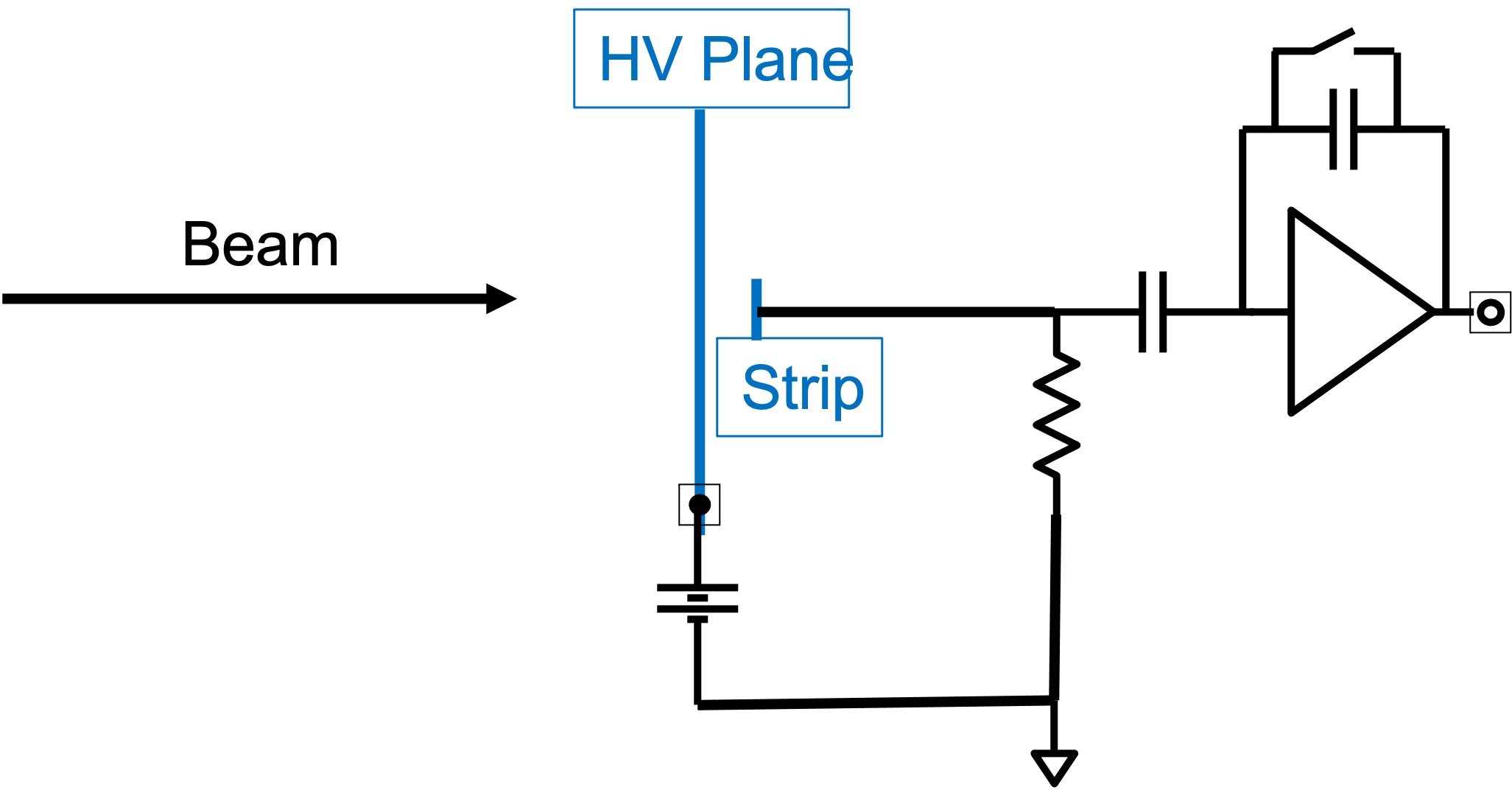}
    \caption{Connection of the detector to the HV.}
    \label{fig:HV_plane}
\end{figure}

The way the detector will be connected to the high-voltage (HV) system and to the front-end card is sketched in Fig.~\ref{fig:HV_plane}. Due to the very different mobility of electrons and holes in sapphire, the proposed connection will yield positive input signals (as required by the A5202) and full collection of charge also for ionisation created by gammas which convert to electrons in the proximity of the detector strip.

\subsubsection{High- and Low-Voltage Systems}

For powering the detector and the associated electronics, the SY5527LC modular system developed by CAEN is proposed~\cite{sy5227}. The system is based on a main crate (SY5527LC) which will host one low-voltage (LV) card (A2519CA) to power the electronics and one HV card (A1561H) to power the detectors. The crate, with its HV and LV cards will be placed in the experiment counting room at about 20\,m distance from the detector.
One HV card and two LV cards will be sufficient to power two detector stations (4 sapphire strip planes). The system will be fully configurable and controlled remotely.
The HV card (A1561H) will provide up to 1\,kV/20\,$\mu$A per channel with the possibility to read out the current with default precision of 50\,pA. The CAEN company could improve the latter parameter if necessary. The board is designed in order to provide the HV with a common floating return which is a very important option to possibly decouple the grounding of the detector from the grounding of the HV.

\subsection{Mechanics}
\label{gbp_subsec:mechanics}
The GBP consists of two identical $xy$ detectors, positioned at the end of the beam line before the beam dump.
Each $xy$ detector is composed by two sapphire sensors, both mounted on a $xy$ moving stage through two PCBs acting as mechanical and connection interfaces. One sapphire sensor will measure the beam $x$ coordinate and the second sensor the $y$ coordinate. 

Each pair of sensors are mounted on a $xy$ controlled stage that allows measurement of the beam profile using different portions of the sapphire detectors. 

The stages will be remotely controlled in order to displace the sensors of the needed amount in both $x$ and $y$ directions. The strokes of both $x$ and $y$ axes are 25\,mm, thus allowing to fully cover the sensor sensible surface of $20 \times 20$\,mm.

\begin{figure}
    \centering
    \includegraphics[width=1.0\textwidth]{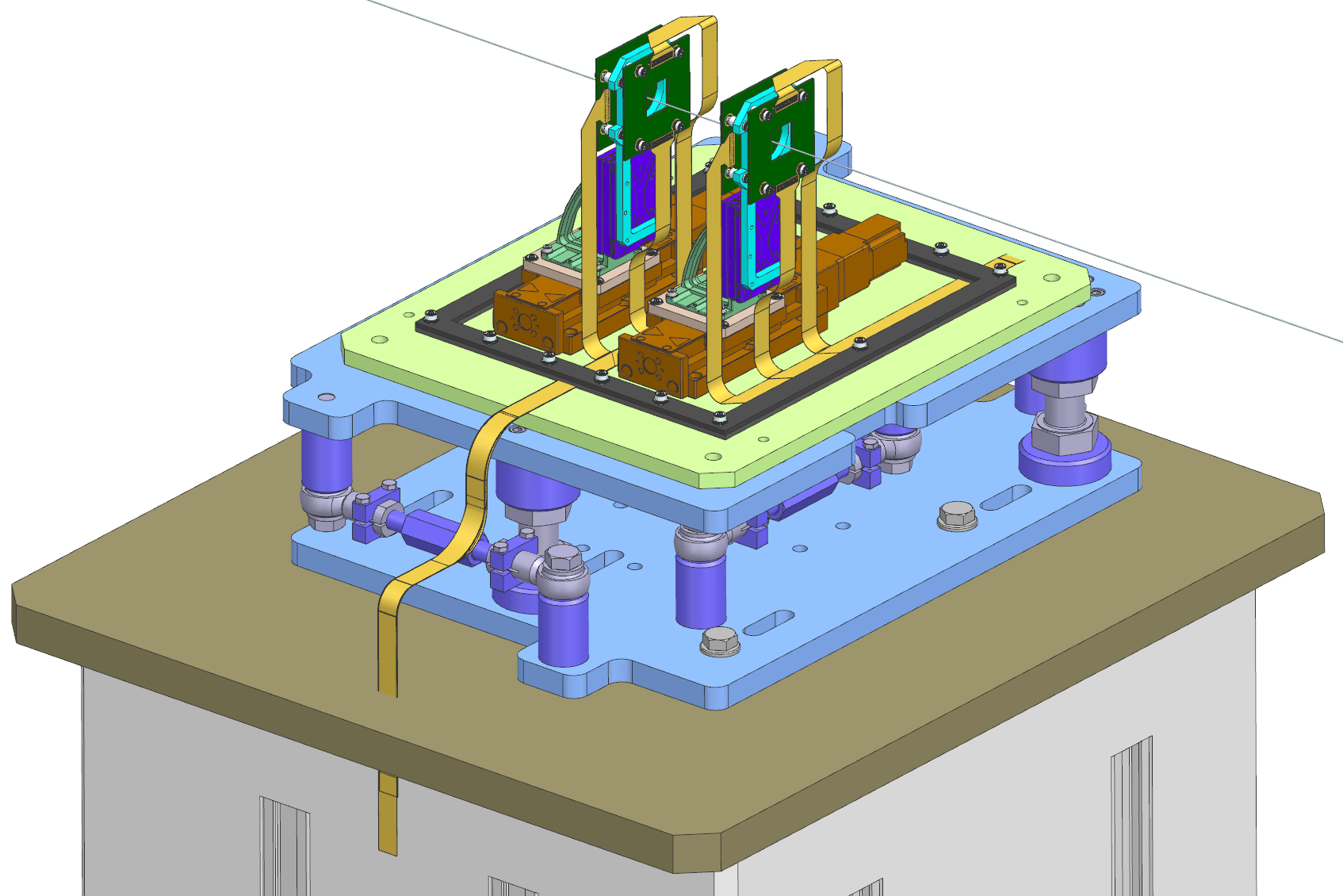}
    \caption{GBP station overview (brown: $x$ stages; violet: $y$ stages.)}
    \label{fig:Luxe2A_9}
\end{figure}

The system is represented in Fig.~\ref{fig:Luxe2A_9}: the $x$ axis (brown stages) are mounted on the support platform. The $y$ axis stages are the violet  prism, mounted on the $x$ axis by means of precise 90$^\circ$ brackets. The light blue brackets are mounted on the movable part of the $y$ axis and supports the four green boards which represent the PCBs on which the four sapphire sensors are mounted.

The accuracy and precision required for detector positioning is of the order of 10\,$\mu$m overall on a stroke of a few millimetres, in both $x$ and $y$ directions. 
The definition of the final specifications will be assessed by the construction and test of the prototype, when the maximum deviation of the detector positioning and the precision of the full mechanism chain will be measured.

The stages of the two $x$ and $y$ axes shall cope with the following requirements:

\begin{itemize}
\item high precision and accuracy inside a small stroke range;
\item adequate stiffness compared to the applied load;
\item lightweight;
\item self-locking capability;
\item low values of  pitch, yaw and roll;
\item low straightness deviation;
\item compact dimensions, mostly for the $y$ stage.
\end{itemize}

In particular the pitch of the horizontal axis and yaw of the vertical axis can have a large impact on the accuracy of the detector positioning, as these angular deviations are amplified by the relatively large lever-arm of the detector w.r.t.\ the trolley of the stage.  
Also the deviation from straightness of each single axis, in both  vertical or horizontal  directions, can have an impact on the detector position. 

In the current design the use of PI Physik Instrumente stages have been selected, in particular an L-836 universal linear stage for the horizontal movement and  a Q-545 Q‑Motion stage for the vertical movement. These linear stages match quite well the specifications listed above, where the choice of the stage is driven mainly by the requirement of adequate stiffness for the $x$ axis, carrying more load, while for the $y$ axis stage the choice is driven by the requirement of lightness, as it is carrying a smaller load, and self locking capabilities. 

The construction of the prototype with two different types of linear stages  will allow  the behaviour of both types of stages to be tested in order to qualify them for the given purpose, and also to test the adoption of the same type of linear stage for both $x$ and $y$ axes. In case this can bring some advantage in terms of simplification, e.g.\ the use of a unique control interface. 

The coupling of axes will be assembled through precisely machined coupling brackets in order to realise the perpendicularity of the two movements.

All the mechanical parts mounted on the system will be designed in order to be adequately stiff and light, as reduction of weight and maximising of stiffness of each component of the moving mechanism will contribute to reaching adequate precision of the detector positioning  and to minimise the effect of possible vibration sources.

In and out flat cables will be  supported and routed in such way to minimise forces and loads transmitted from them to the $xy$ movement.

Commercial drivers/controllers and control software and interface will be procured with the linear stages. The system will be chosen among adequately tested, reliable and user-friendly commercial products. 

\begin{figure}
    \centering
    \includegraphics[width=1\textwidth]{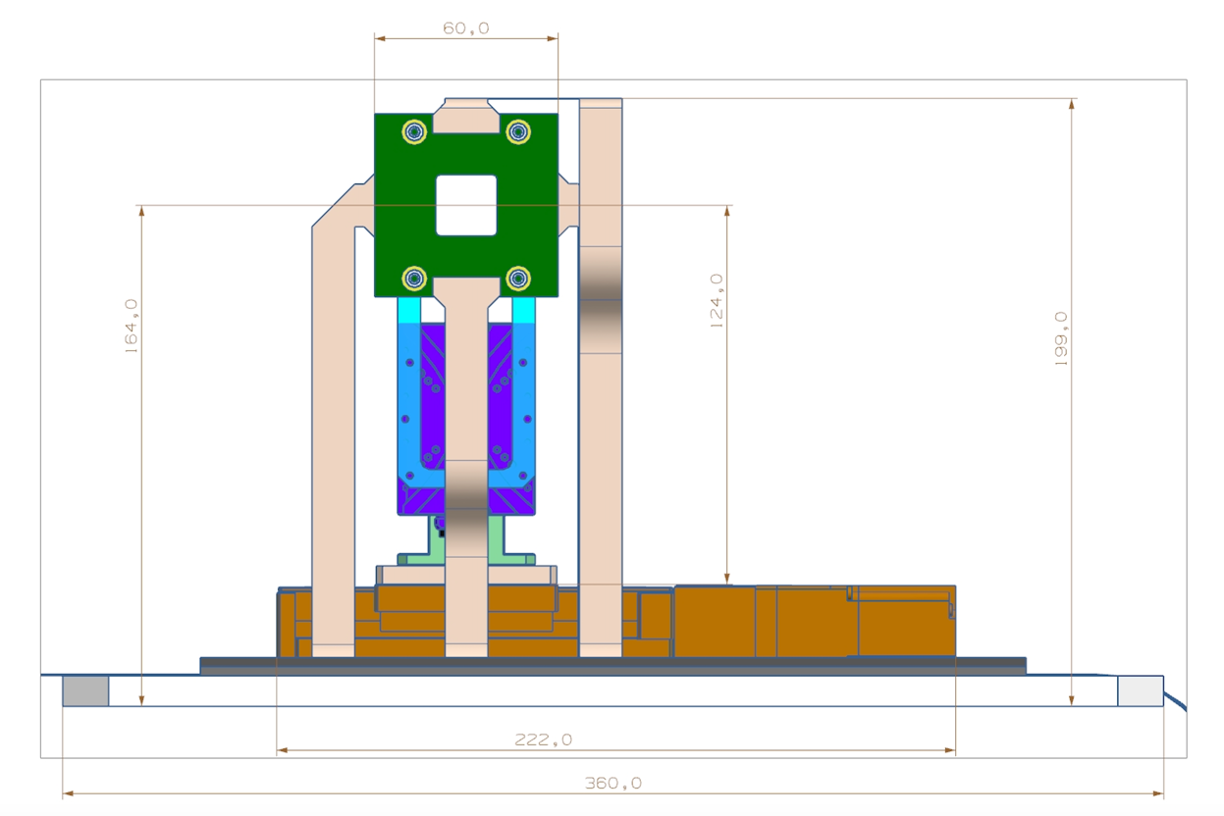}
    \caption{$xy$ stages front view. Detail of the station main dimensions. Support plate dimensions are about $35 \times 35$\,cm$^2$.}
    \label{fig:Luxe2A_4b}
\end{figure}

\begin{figure}
    \centering
    \includegraphics[width=0.8\textwidth]{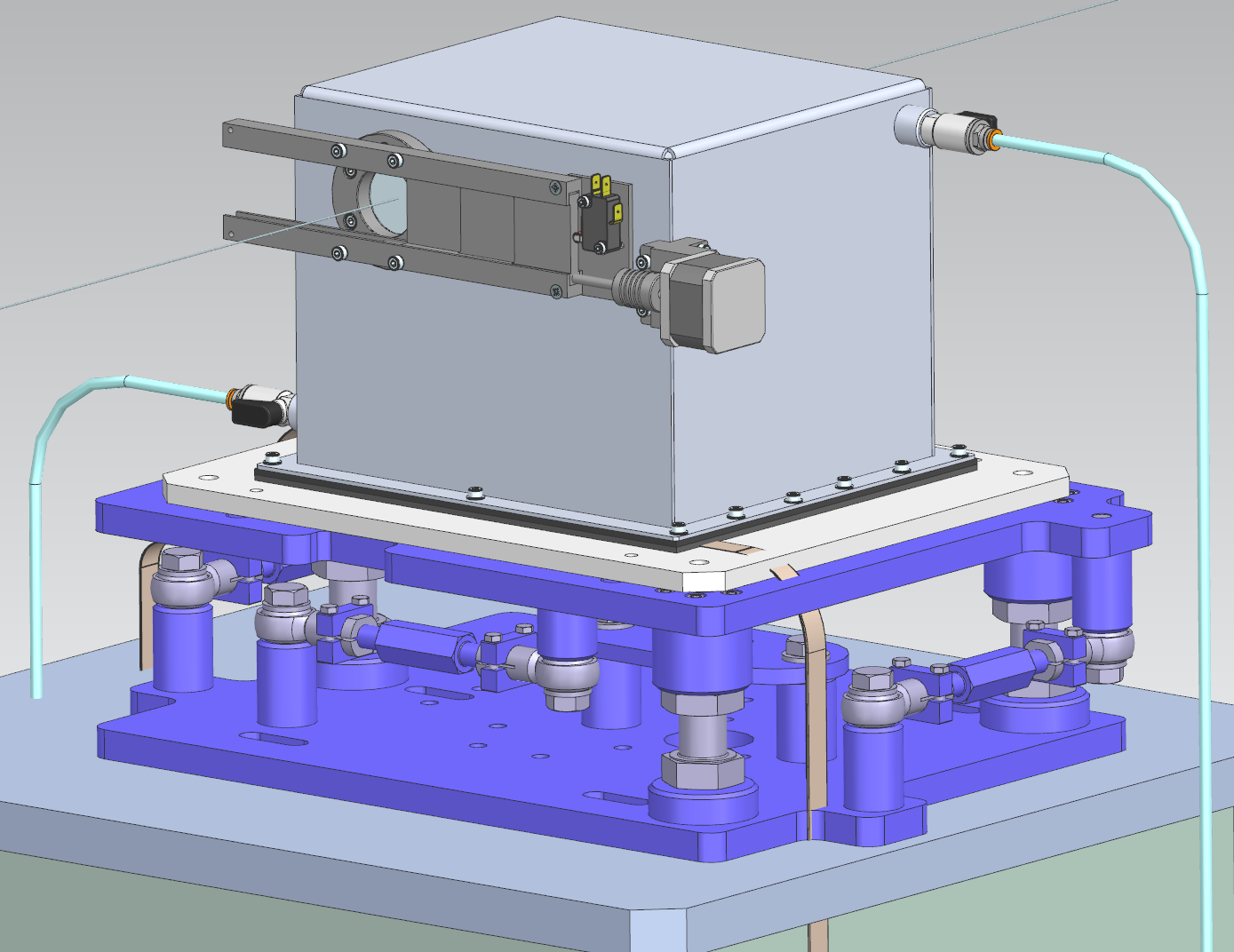}
    \caption{Faraday cage and gas enclosure with windows.}
    \label{fig:Luxe2A_1}
\end{figure}

Both $xy$ detectors will be positioned along the beam line on a kinematic mount (see Figs.~\ref{fig:Luxe2A_4b} and~\ref{fig:Luxe2A_1}) provided by DESY, that allows to centre and align the initial  position and tilting of the stages with respect to the beam axis. The size of the table is about $35 \times 35$\,cm$^2$.

The kinematic mount allows the position to be adjusted and fixed over all degrees of freedom, in particular, the initial positioning along $x$ and $y$ can be adjusted by the relevant adjusting screws, and the tilt can also be controlled and corrected by fine-tuning the vertical adjusting screws.

The detectors will be enclosed in a Faraday shield, see Fig.~\ref{fig:Luxe2A_1}, acting also as a gas enclosure, that will be flushed with dry air or nitrogen to preserve cleanliness.
The shield will have entry and exit windows aligned with the beamline, sealed by a thin film of polyimide or polyester. These windows may be aluminised in order to preserve the electromagnetic tightness of the Faraday cage. Various thicknesses are available, as low as 13\,$\mu$m from different producers.
Each window will be clamped by flanges  sealed by O-rings and grounded through the contact with the clamping flanges.
On the front face of the cage an absorber plate will be mounted if necessary. It will be movable, with sectors having different thicknesses in order to allow different levels of gamma conversion. It will be driven by a simple stepper motor so that it can be controlled remotely to move the plate to different positions. Tests and simulations will be carried out to define the actual thickness and material needed.

\begin{figure}[htbp]
    \centering
    \includegraphics[width=0.8\textwidth]{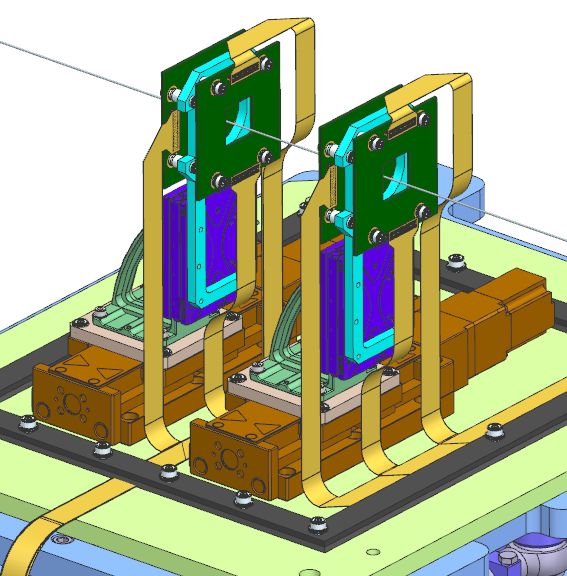}
    \caption{Signal flat cables preliminary routing.}
    \label{fig:Luxe2A_0}
\end{figure}

\begin{figure}
    \centering
    \includegraphics[width=0.8\textwidth]{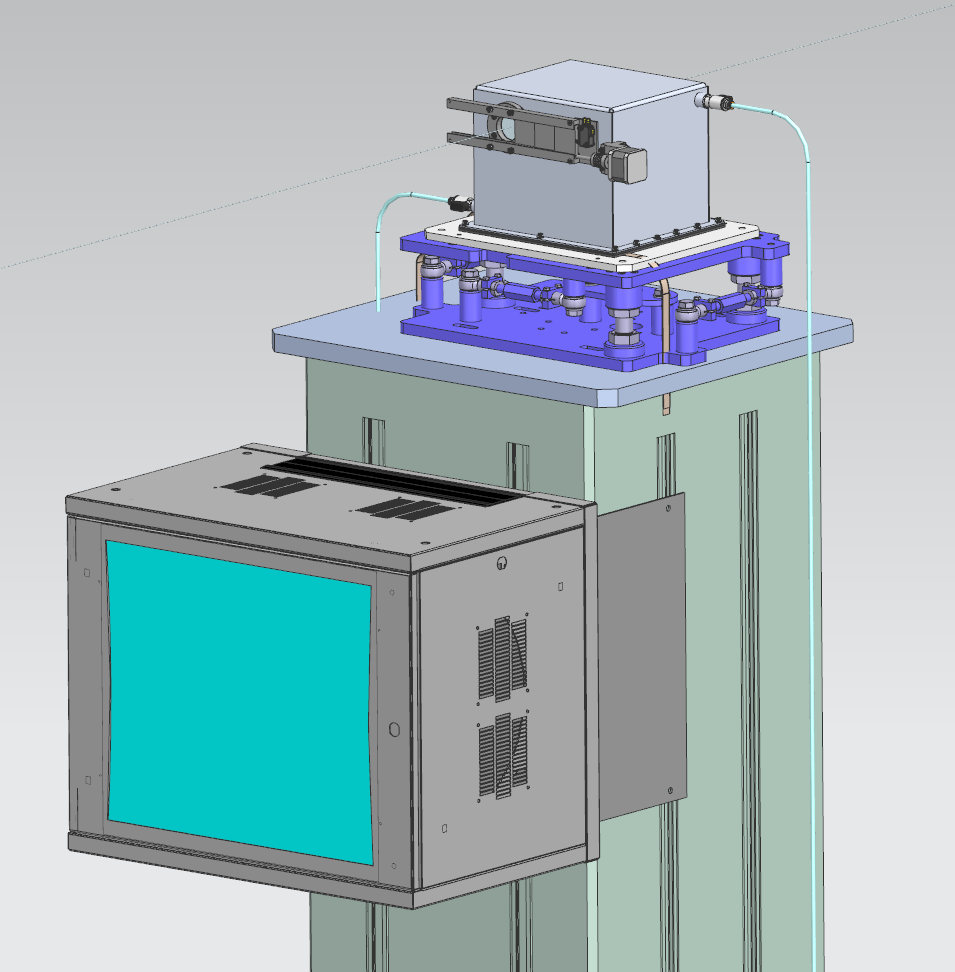}
    \caption{Overview of the GBP station on the support pillar and the rack.}
    \label{fig:Luxe2A_2}
\end{figure}

Power and signal cables will pass through a rubber sealing of the gas enclosure, embedded in the support plate of the detector (Fig.~\ref{fig:Luxe2A_0}).
Cables will be supported and routed to the related electronics by means of simple supports, cable trays and cable harnesses.  On the other side, they will be routed towards the rack containing the electronics, which will be attached to the support concrete pillar (see Fig.~\ref{fig:Luxe2A_2}).

\section{Interfaces and Integration}
\label{interfaces}

\subsection{The Data Acquisition System and Trigger System}

Given the very low maximum data taking frequency of 10\,Hz, the LUXE data acquisition (DAQ) system is in its nature quite simple and robust.

A general scheme of the system is discussed in Chapter~\ref{chapt11}. The system, based on the EUDAQ2~\cite{Liu:2019wim} software, has been intensively used as standard for test beams at CERN and DESY and is based on a trigger logic unit (TLU) distributing event readout trigger and synchronising the PCs performing the event readout of the LUXE detectors.
The DT5215 module can be connected to an external clock (NIM/VSDL) so it could accept the TLU signal. There are two ways to communicate with the DT5215 module: ethernet or USB. The former communication will be adopted  since it is easier to be integrated in the EUDAQ framework. 
It is not expected that the GBP produces any signal trigger to be used by LUXE.

\subsubsection{Data Structure and Computing Aspects}

The GBP event size will be quite small (less than 10\,kbytes/s per detector plane) and one PC will be sufficient to collect this amount of data from the CAEN DT5215 data concentrator. All the BX data will be kept, and there will be no physics trigger.
The software to analyse and reconstruct the data from the GBP must still be developed. For each detector plane, the main quantities that should be reconstructed for the beam profile are its width, position and intensity. The details about how these quantities will be reconstructed are still not defined: one possibility is given by fits to the distributions in case a good modelling of the beam profile is found, or FWHM algorithms, or a combination of the two methods.
Given the small amount of data to process for each BX and the hopefully low level of complication of the reconstruction algorithms, a small CPU time for processing one BX is probably needed.

\subsection{The Slow Control System}

The LV and HV power supplies will be integrated with the LUXE slow control  system.  Standard ethernet connections will be used via standard protocols.

\subsection{Services}
The GBP will require the availability of sockets for AC power, ethernet connections and an outlet for the nitrogen gas.

\section{Installation, Commissioning and Calibration}
\label{sec:gbp:installation}

Installation of the GBP should be rather straightforward and the physical placement of the components will not require the use of cranes or special handling equipment.  A detailed planning of the activities has been drafted, showing that all components (detectors, FE electronics, HV and LV power supplies, etc.) could be physically put in place in a couple of weeks. 

The interconnections with the DAQ system and the LUXE slow control system should then be completed and extensively checked in a few days.  Integration of the GBP system within LUXE and commissioning will then proceed without requiring access to the experimental area.   

Although thorough testing of the GBP system is planned to take place beforehand in the laboratory, different environmental conditions might affect the expected performance of the detector.   In particular, the noise level can be higher in the experimental area than in the laboratory once all the other detectors are in place.  So further access  to the detector, for short periods of time,  might be required after the installation, to optimise the noise shielding or the grounding of the detector.

The personpower to be provided by the GBP group has been estimated by means of a resource loaded schedule and a summary is outlined in Table~\ref{tab:GBP_install_mp}.

\begin{table}[h!]
  \begin{center}
    \begin{tabular}{l|c|c} 
      \textbf{Personnel involved} & \textbf{No. of working days} & \textbf{No. of personnel units} \\
      \hline
      Physicists (30\% Students)  & 82 & 4 \\
      Mechanical Engineers & 5 & 1 \\
      Electrical Technicians & 26 & 2 \\
      Mechanical Technicians & 7 & 1 \\
    \end{tabular}
    \caption{Estimated personpower for installation and initial commissioning with no beams.}
    \label{tab:GBP_install_mp}
  \end{center}
\end{table}

In this schedule it is assumed that the cables and optical fibers carrying power and digital signals will be laid down before installation starts and that AC power and LAN sockets will be available both in the experimental area and in the control room.    

\subsection{Commissioning}

After the initial integration of the GBP in LUXE, the commissioning will be possible as soon as a gamma beam impinging on the detector becomes available. In the $e$--laser setup, if the laser pulses are not available, a thin converter could be placed at the IP to generate the $\gamma$ beam.

\subsection{Sensor Replacement}

Although sapphire offers the possibility of displacing the sensors to expose them in multiple areas up to the most tolerable doses, it is foreseen that, at most every few months, the sensors will have to be replaced due to radiation damage.
This operation will require an access of a few hours to remove the Faraday cage, disconnect the sensor support PCBs, replace them with the spares equipped with the new sensors,  reconnect the cables, put back the Faraday cage in position and finally test the electrical connections.    
It is expected that it will be possible to plan this intervention during the periodic technical stops of the EuXFEL accelerator without normally requiring a dedicated access to the experimental area.

\subsection{Calibration Strategy}
\label{gbp_subsec:calibrationstrategy}
Calibrations aimed at evaluating and monitoring detector response and electronic noise will be routinely carried with data collected in normal running conditions.  Electronic channel calibrations will be carried out periodically with self-generated pulses.   In addition, a calibration with alpha sources will be performed when needed by running the front-end boards in continuous ADC conversion mode.

\subsubsection{Calibration in Special Runs}

The possibility to calibrate the GBP in LUXE special runs of short duration is under study. During these special runs a converter target material (a tungsten wire, for instance, as it is planned for the LUXE $\gamma$--laser running mode) should be inserted in the electron beam line in proximity to the IP. In this way, a bremsstrahlung photon beam will be produced whose profile will be recorded by the GBP stations. It will be possible for instance to monitor the stability of the detector response with time, assuming regular periodic runs of this type. The duration of such data taking should be of the order of a  couple of minutes, allowing scans for instance of ten different positions in $x$ and ten in $y$ for the detectors of each station. The acquisition rate during this special run will be 10\,Hz implying 1\,s of data taking for each scan setting. 

This calibration mode will need an integration of the station's movable table's slow control with the LUXE DAQ system, but the details of this can not be provided at this early stage of the project. It is worth noting that, due to the affinity of this detector with the gamma-ray spectrometer, both could be calibrated simultaneously during the same dedicated runs.

\subsection{Decommissioning}

As for installation, decommissioning of the GBP will be rather simple and quick.  Equipment in the control room will be disconnected and dismounted and the detectors with the associated patch panels and cables removed from the experimental area.  Precautions could be necessary for the sensors if they are activated due to absorbed radiation.

\section{Further Tests Planned}
\label{sec:gbp:future}

A first test to evaluate the radiation resistance of sapphire
specimens supplied by various manufacturers was carried out using three 4-strip sapphire sensors at the CLEAR electron linac at CERN in September 2022.  However the measurements of the charge collection efficiency during the irradiation were affected by instabilities in the reference detector used to estimate the beam charge and therefore the test will have to be repeated at CLEAR or at the BTF in Frascati in 2023.

Therefore, we plan to combine the measurement of the radiation resistance with the evaluation of response uniformity and the detector performance in a single campaign, since, meanwhile, 196-strip detectors have been produced.

The test will include detailed scans in direction perpendicular to the strips direction in order to reconstruct the beam parameters.  This test will last about 1\,week and will require the final CAEN readout system and the movable table which will support the GBP in the LUXE experiment in its final version. This test can be considered as a final validation test for the GBP in all its parts.

\section*{Acknowledgements}

Part of the work reported in this chapter of the TDR was carried out by our colleagues M. Skakunov, A. Tyazhev, A. Vinnik, A. Zarubin, whose contribution we would like to acknowledge.

\printbibliography
\end{refsection}


\chapter{Gamma Flux Monitor}
\label{chapt10}
\begin{refsection}

\chapterauthors{M.~Borysova \\\desyaffil}{O.~Borysov \\\desyaffil}
\date{\today}

\begin{chaptabstract}
    In the high-flux photon region of the LUXE experiment, a photon detection system (PDS) is designed based on three complementary technologies: the gamma ray spectrometer (GRS) in which energy spectra of converted electron--positron pairs are used to determine the photon flux, the gamma beam profiler (GBP) which determines the spatial profile of the photons, and the gamma flux monitor (GFM) which is sensitive to the overall photon flux. The GFM measures a signal proportional to the energy of particles back-scattered from the photon beam dump. This chapter presents the details of the GFM design, performance and construction plans.
\end{chaptabstract}

\section{Introduction}
\label{gfm:intro}

One of the characteristics of the LUXE experiment is that the number of particles per bunch crossing (BX) substantially varies depending on the LUXE run mode and detector-subsystem location. There are several subsystems where this number exceeds $10^{5}$.
For some parts of the setup the overall flux of photons, as for example created in the $e$-laser interactions for laser intensities $\xi > 1$, exceeds $10^{8}$. This large number makes counting of photons and measuring their spectrum challenging. 

A solution proposed here is based on measuring the energy flow of particles back-scattered from the dump that absorbs the photon beam at the end of the beamline. 
This detector provides a simple and robust way to monitor the variation in the photon flux with time. It may also be used at the beginning of the run to optimise the collisions. It plays a role similar to a luminosity monitor in collider experiments.

\section{Requirements and Challenges} \label{sec:gfm:requirements}

The precise knowledge of particle fluxes in the LUXE experiment is very important for physics measurements. To measure the absolute number of photons on an event-by-event basis is crucial for spectra normalisation. The measurement goal of the gamma flux monitor (GFM) is to precisely monitor the rate and possibly to determine the angular distribution of photons originating from high-intensity Compton scattering. The latter could indicate, for example, if the beam accidentally shifts from its nominal position. The magnitude of the shift would be determined through non-uniform distributions of energy deposited in the GFM modules. Dedicated simulations will be performed to get a quantitative assessment of this potential measurement.

Due to the high charged-particle fluxes created at the interaction point and lack of photon detectors able to measure such fluxes, the usage of the energy flow of back-scattered particles~\cite{dudar_master} from the beam dump at the end of the beamline detected with a Cherenkov-type calorimeter is proposed. Detectable Cherenkov light is produced whenever a particle traverses a transparent medium with a speed $v> c/n$, where $c/n$ is the speed of light in that medium and $n$ is the refractive index of the medium. Cherenkov light is emitted on the surface of a cone centred on the particle trajectory with half-angle $\theta_{C}=\arccos(c/nv)$. Dielectric materials with $n > 1$ are good candidates for Cherenkov detectors.

Lead-glass is cheap and easy to handle, and therefore has been widely used in the past for high-energy physics applications, for example, in the NOMAD~\cite{NOMAD:1997pcg} neutrino experiment at the CERN SPS, in the OPAL~\cite{OPAL:1990yff} experiment at LEP or in HERMES~\cite{Avakian:1998bz} at DESY. One drawback of lead-glass is the poor radiation resistance -- a significant deterioration of the light output for some types of crystal is observed for doses larger than 100\,Gy~\cite{Kobayashi:1994kf}.

\begin{figure}[htbp]
\centering
\includegraphics[width=0.99\textwidth]{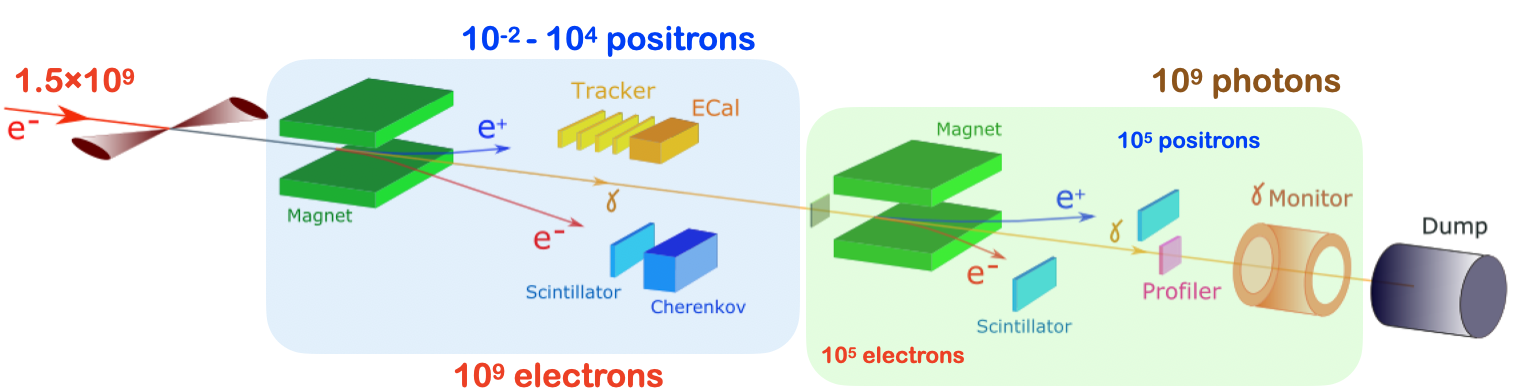}
\caption{Schematic experimental layout for the $e$-laser setup. Two main detector subsystems are highlighted: the electron/positron spectrometer, in blue, and the photon detection system, in green. Shown are the magnets, detectors and absorbing elements.} 
\label{fig:layout} 
\end{figure}

Conceptually the LUXE setup contains two detector subsystems: an electron/positron spectrometer (shaded in blue in Fig.~\ref{fig:layout}) and a photon detection system (PDS) (shaded in green in Fig.~\ref{fig:layout}). 
The GFM is placed at the end of the beam line before the photon beam dump depicted in black in Fig.~\ref{fig:layout}.

\section{System Overview}

\begin{figure}[ht!]
  \begin{center}
    \includegraphics[width=0.75\textwidth]{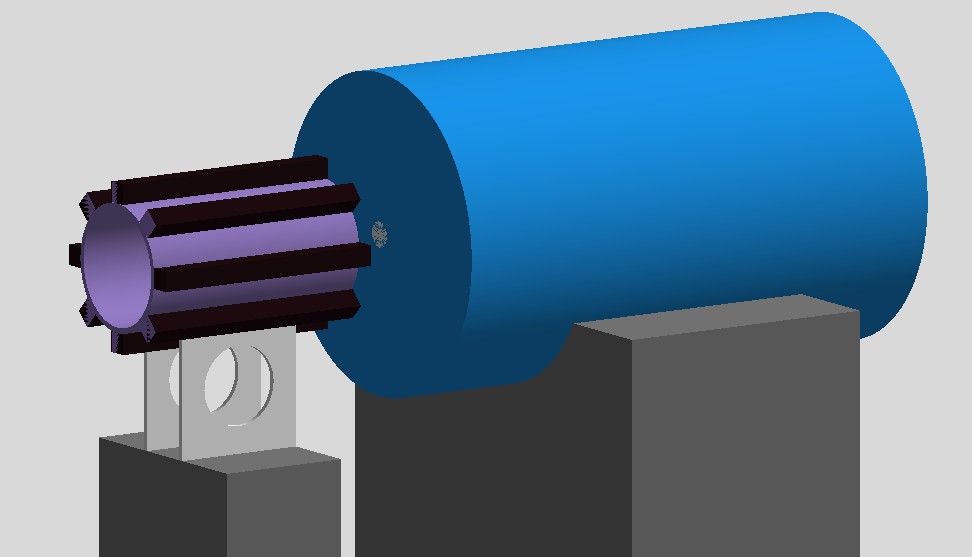}
      \end{center}
    \caption{{\geant} rendering of the back-scattering calorimeter followed by the beam dump. }
    \label{fig:GM_conf}
\end{figure}

The GFM is assembled from eight blocks of lead-glass arranged cylindrically and symmetrically around the beam axis, at a distance of $12\,\mathrm{cm}$ radially from the beamline and $10\,\mathrm{cm}$ upstream of the beam dump, as shown in Fig.~\ref{fig:GM_conf}. This configuration
is selected to ensure that the radiation load is tolerable. Also such an arrangement of the eight blocks provides sensitivity to the location and transverse dimension of the photon beam. The blocks consist of $3.8 \times 3.8 \times 45\,\mathrm{cm^3}$ TF-101 (or TF-1) type lead-glass modules. Each module is $18\,\mathrm{X_0}$ deep. 
The TF-101 lead-glass is quite similar to TF-1 (see Table~\ref{tab:LGproperties_tf} for physical properties of the lead-glass and see Table~\ref{tab:LG_Chemproperties} for the chemical composition) except for the improved radiation hardness of TF-101 which is beneficial in the presence of high fluxes. 
Measurement of the radiation hardness of TF-101 blocks with $\gamma$-rays~\cite{Kobayashi:259413} and high energy hadrons~\cite{Inyakin:1981tg} have shown that an accumulated dose of 20\,Gy produces a degradation of transmittance of less than $1\%$. Thus TF-101 is 10 to 50 times less susceptible to radiation damage than the TF-1 glass.

\begin{table}[h!]
  \begin{center}
    \caption{Physical properties of the lead-glasses,  types TF-1 and TF-101.}
    \vspace{0.2cm}
    \begin{tabular}{|l|c|c|}
      \hline
      \textbf{Properties }  & \textbf{TF-1} & \textbf{TF-101}\\
      \hline

       Radiation length (cm) &2.50 &2.78 \\
       Density ($\mathrm{g/cm^3}$) & 3.86 & 3.86\\
       Critical energy (MeV) & 15.57 & 17.97\\
       Refraction index     & 1.6476 & 1.65\\
       Moliere radius (cm)   & 3.5 & 3.28  \\
      Thermal expansion coefficient ($K^{-1}$) &  & $8.5\cdot 10^{-6}$\\
      \hline     
    \end{tabular}
     \label{tab:LGproperties_tf}
  \end{center}
\end{table}

\section{Expected Performance}

\subsection{Simulations}
\begin{figure}[ht!]
  \begin{center}
    \includegraphics[width=0.48\textwidth]{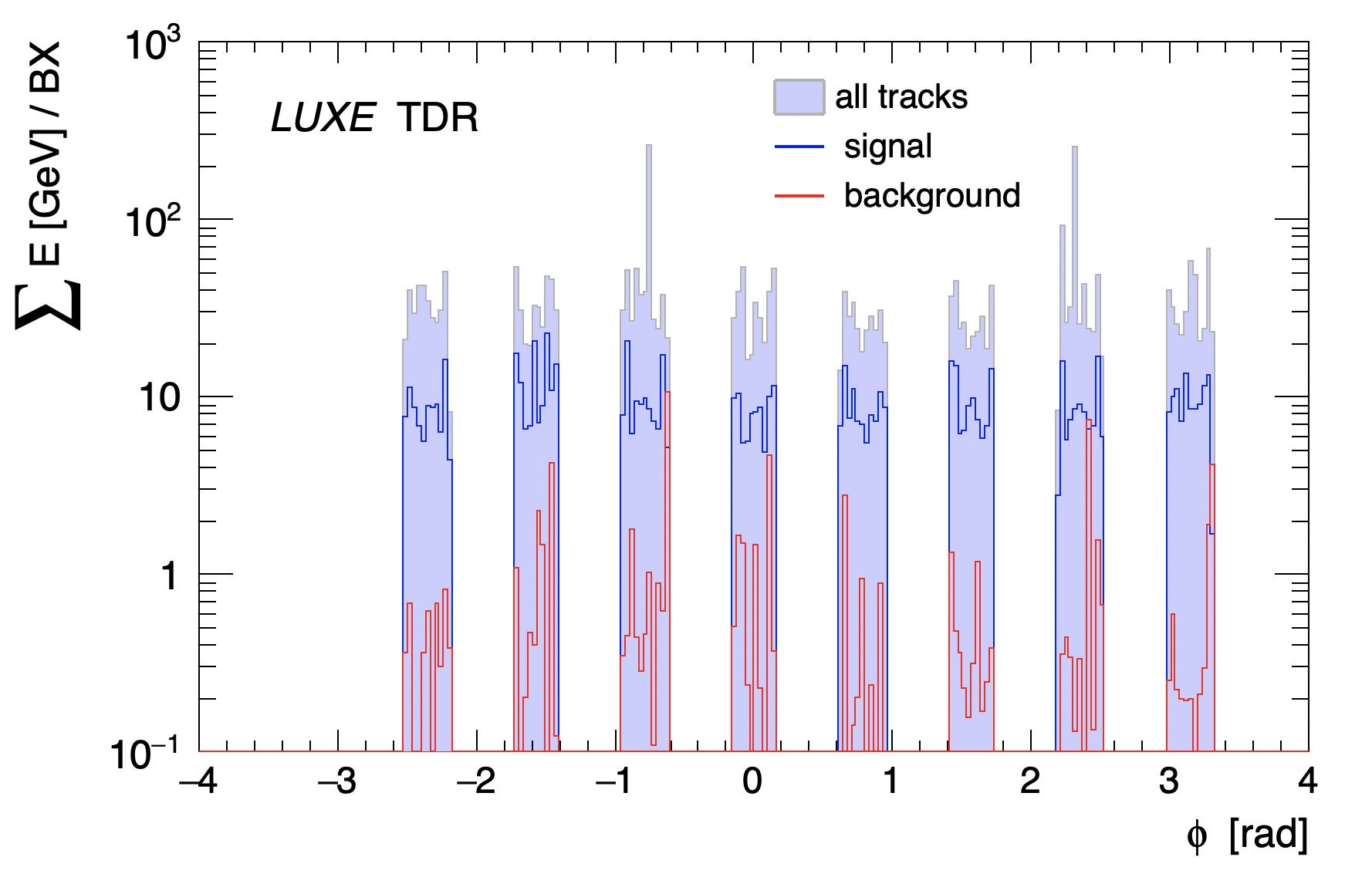}
    \includegraphics[width=0.48\textwidth]{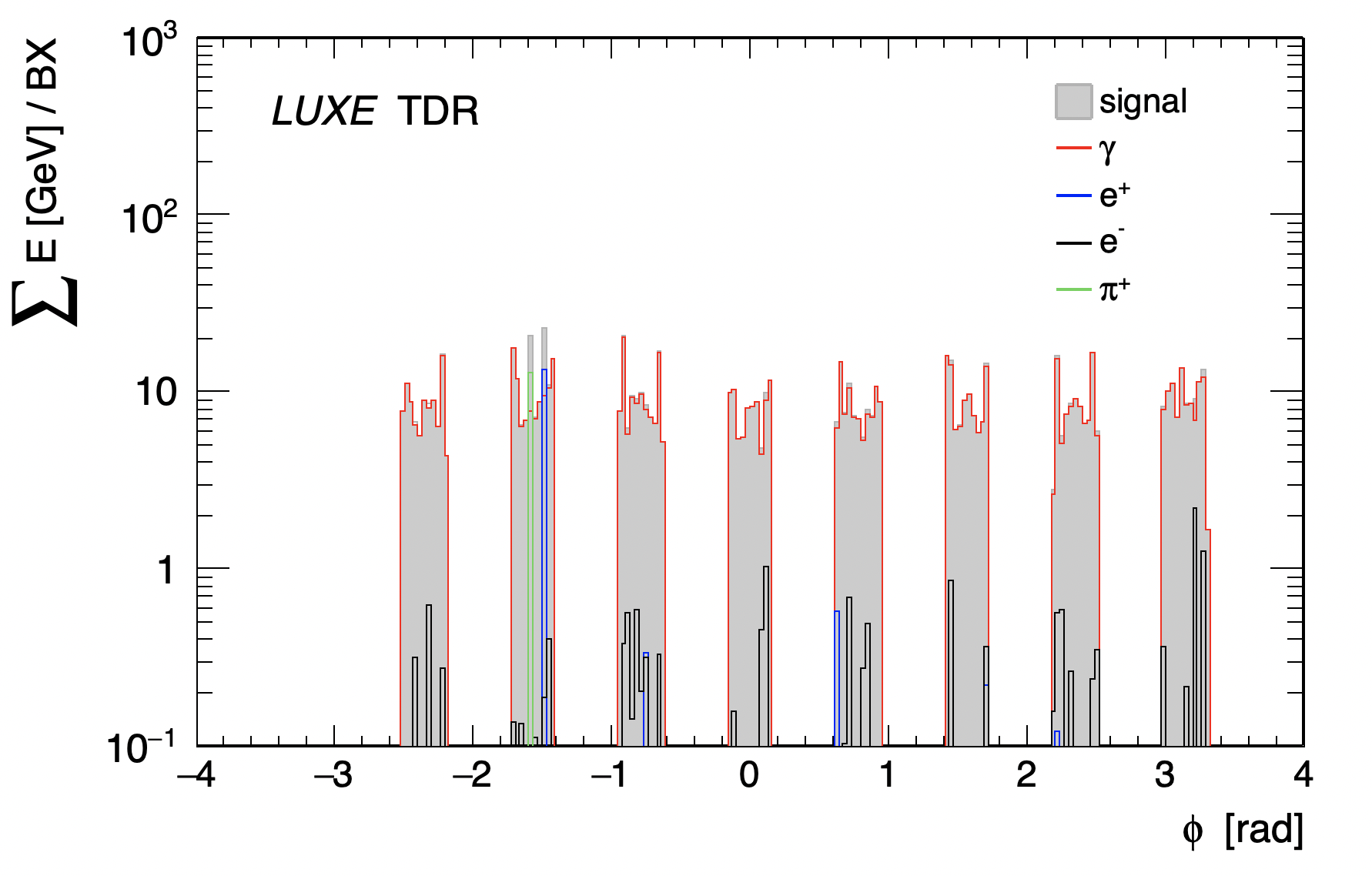}
      \end{center}
    \caption{Left: Sum of energies of signal and background particles per bunch crossing as a function of the azimuthal angle of tracks in the GFM. Signal particles correspond to those originated from the beam dump. All particles  intersecting  the GFM are  considered  regardless  of  the  actual  energy  deposition  in  the sensitive volume. Right: Sum of energies of signal particles of different species per bunch crossing as a function of the azimuthal angle of tracks in the GFM. The signal and background particles are generated in the simulation of the $e$-laser setup with the JETI40 laser (using $\xi=0.5$) and with the electron beam at 16.5 GeV.}
    \label{fig:GFM_perform}
\end{figure}

To study and optimise the performance of the GFM detector, the experimental layout is implemented in detail in the \geant~\cite{Allison:2016lfl,Allison:2006ve,Agostinelli:2002hh} geometry model. Simulations of the signal and beam induced background are studied for various configurations of the planned measurement programme to estimate the performance of the detector. Through an iterative process the layout was optimised to reduce backgrounds and radiation doses to a tolerable level. Figure~\ref{fig:GFM_perform}, left, shows the sum of energies of  all tracks, signal and background particles per bunch crossing for each lead-glass crystal of the back-scattering calorimeter. This also includes muon and electron neutrinos and their antiparticles and neutrons which do not produce signal in lead-glass. The signal consists of the particles that can produce Cherenkov light in lead-glass crystal and represents electrons, positrons, photons and pions. The signal to background ratio is better than $ 10$. In Fig.~\ref{fig:GFM_perform}, right,  the sum of energies of different particle species which compose the signal is shown and it is seen that the energy is dominated by photons.

\begin{figure}[ht!]
  \begin{center}
    \includegraphics[width=0.48\textwidth]{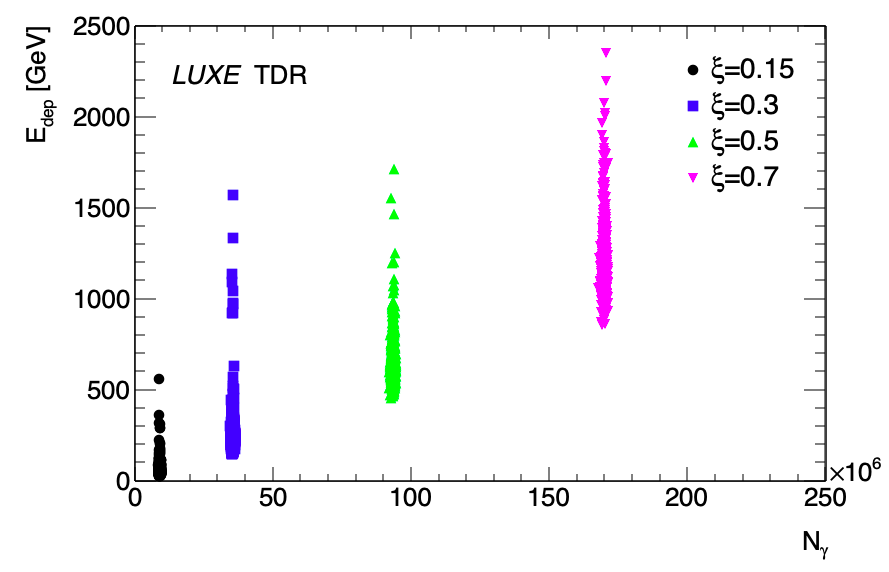}
    \includegraphics[width=0.48\textwidth]{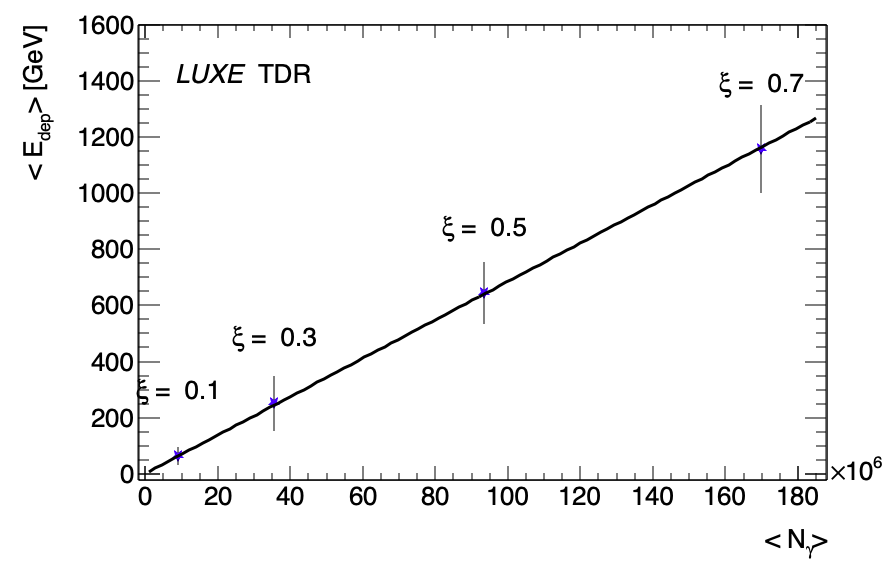}
      \end{center}
    \caption{Left: Total deposited back-scattered energy, $E_\mathrm{dep}$, as a function of the number of photons, $N_\gamma$. Each point corresponds to one bunch-crossing. Right: Average energy deposition versus the mean number of photons.
    }
    \label{fig:GM_perform}
\end{figure}

In order to study the expected performance of the GFM, the back-scattered energy deposit, $E_\mathrm{dep}$, is studied as a function of the number of dumped photons, $N_{\gamma}$. The correlation is shown in Fig.~\ref{fig:GM_perform}, for different laser intensities. A clear dependence of the deposited energy on the number of incident photons is observed. 

The precision with which $N_\gamma$ can be determined is estimated as
\begin{equation}
\label{eq:deltaN}
\Delta N_\gamma = \frac{\partial N_\gamma}{\partial E_\mathrm{dep}} \Delta E_\mathrm{dep} \, ,
\end{equation}
 where $\frac{\partial N_\gamma}{\partial E_\mathrm{dep}}$  is the value of the slope in Fig.~\ref{fig:GM_perform} (right) and $\Delta E_\mathrm{dep}$ is the measure of the inherent fluctuations of the back-scattered particle flow.
The uncertainty on the number of measured photons 
is in the range of 5 to 10\%, decreasing with increasing photon flux.
To achieve this level of precision, a system of monitoring light sources with high pulse stability will be implemented to track the stability response of the crystals as some transmission losses with time may be expected due to exposure to high radiation doses.

The radiation dose deposited in each lead-glass crystal per bunch-crossing  for phase-0 of the experiment (at $\xi=3$)  is estimated from simulations to be $\sim 1.3\cdot\,10^{-7}\,\mathrm{Gy}$. For the phase-1 at laser intensity of  $\xi=10$  the expected dose is $\sim 1.3\cdot\,10^{-5}\,\mathrm{Gy}$.
For one year equivalent of $10^7$\,s this corresponds to an annual dose of $1.3$\,Gy or $130$\,Gy, respectively. 
According to Ref.~\cite{Kobayashi:1994kf} a dose of $20$\,Gy can be tolerated with  less than $1\%$ degradation for a device of 2.8\,cm length made of TF-101 type lead-glass. For the present 45\,cm long crystal, if we accept the decrease of the transmission over the detector depth to be $1/e$, the tolerable dose is 320\,Gy. Thus the detector can operate during the LUXE life-time but continuous monitoring is important. 

\subsection{Lab Tests and Beam Tests}
\label{sec:tests}
The first prototypes for the GFM detector modules were tested in the Electronics Laboratory (ELab) at DESY and at the Laser Plasma Accelerator (LPA) facility with a high intensity electron beam. The electrical and optical tests were carried out in the ELab facility and are described in Section~\ref{gfm:app2}.
Light transmittance with different light source positions, several models of photomultiplier tubes (PMTs) and different types of wrapping materials were studied. The performed tests showed that the available  lead-glass crystals are fully functional.

The first beam-test campaign took place in November 2021 at the LPA at DESY and is described in detail in Section~\ref{gfm:app2}.  Two GFM module prototypes were tested in the electron beam of 60\,MeV with a bunch charge up to 25\,pC (\mbox{$\sim 10^8\,e^-$}) at 3.5\,Hz. Each module consists of one lead-glass block wrapped in one layer of aluminium foil and two layers of vinyl foil with the PMT attached to the end. One module was equipped with a one-anode Hamamatsu R1398 connected to a custom made voltage divider and the second with a four-anode Hamamatsu R5900-03-M4 with a PMT socket Hamamatsu E7083. Different positions and orientation of the detector modules in the beam were studied.

\begin{figure}[ht!]
  \begin{center}
    \includegraphics[width=0.8\textwidth]{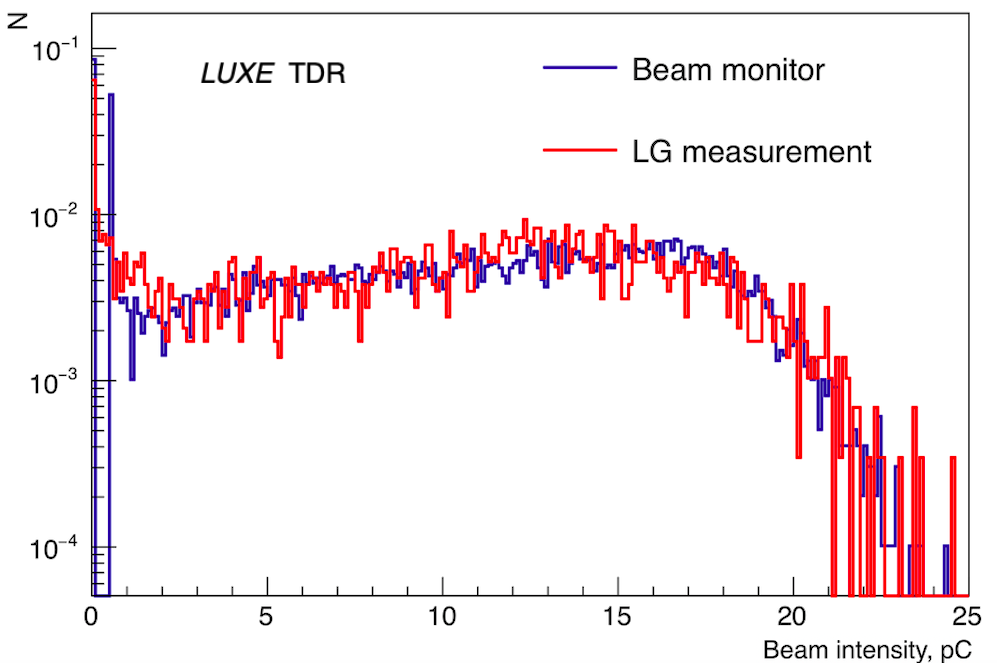}
      \end{center}
    \caption{Comparison of the spectra from the beam monitor (blue line) and the GFM prototype equipped with a four-anode PMT.
The red curve is the sum of signals from the four-anodes.}
    \label{fig:GFM_BT_result}
\end{figure}

Figure~\ref{fig:GFM_BT_result} compares spectra from the beam monitor and the GFM prototype equipped with the four-anode PMT. The electron beam spectrum is well reproduced by the measurements in the lead-glass detector modules.

\section{Technical Description}

In the following, technical details of the GFM system, including mechanics, readout on and off detector, hardware calibration system are discussed.

\subsection{Mounting Structure}
The GFM consists of eight blocks of lead-glass read by PMTs and mounted on a cylindrical pipe, made of stainless steel or copper, around the beam axis.
The pipe has a radius of 12.5\,cm, its thickness is 1\,cm  and it is located at a distance of 10\,cm upstream of the beam dump. The choice of material will be driven by the need to minimise activation.
The supporting structure serves as a holder of the lead-glass crystals as well as a shielding to reduce their radiation load.
For the supporting structure the following aspects need to be considered:
\begin{itemize}
    \item an exchange-friendly design of the clips;
    \item alignment marks for crystal mounting;
    \item mount for cable holders.
\end{itemize}

\subsection{Lead-Glass Crystals}
Approximately 50 lead-glass crystals are currently available at DESY. They are spare blocks that were to be tested for radiation hardness for the HERA-B experiment but were never installed in the experiment.
These are new crystals with dimensions of $ 3.8 \times 3.8 \times 45\,\mathrm{cm^3}$  each, made of either TF-101 or TF-1  lead-glass type.
The type of lead-glass and therefore the respective radiation hardness is not known due to lack of documentation. 
The radiation hardness of TF-101 is 20 times better than that of TF-1. In the LUXE environment TF-1 crystals would degrade faster and would needed to be replaced while with TF-101 running through the full lifetime of the LUXE experiment is possible.
The radiation damage can cause degradation of the transparency of the crystal to Cherenkov light and ultimately the glass can become visibly brown.
It would thus be beneficial to test one of the crystals under
realistic irradiation conditions and measure its transparency before and after irradiation.

\subsection{PMTs}
The Cherenkov light emitted by electrons and positrons of electromagnetic showers in lead-glass is detected by PMTs. The PMT output is directly connected  to an ADC analogue input.  The preliminary choice of PMTs is head-on Hamamatsu R972 or Hamamatsu R821. For Cherenkov light detection, a UV-transparent window is required, typically made of radiation hard quartz (silica). The R821 model has a fused silica window and the R972 window is made of $\mathrm{MgF_2} $. Both windows are sufficiently radiation hard.  The radiant sensitivity of R972 ranges from 115\,nm to 200\,nm with the  peak wavelength of 140\,nm. For R821 the spectral response is within 160--320\,nm and the wavelength of maximum response is 240\,nm.  Both ranges correspond to spectra of Cherenkov light in lead-glass crystal.  The final choice of PMTs will be made after analysis of beam test results and future study of the market.

\subsection{Readout}
\begin{figure}[ht!]
  \begin{center}
    \includegraphics[width=0.8\textwidth]{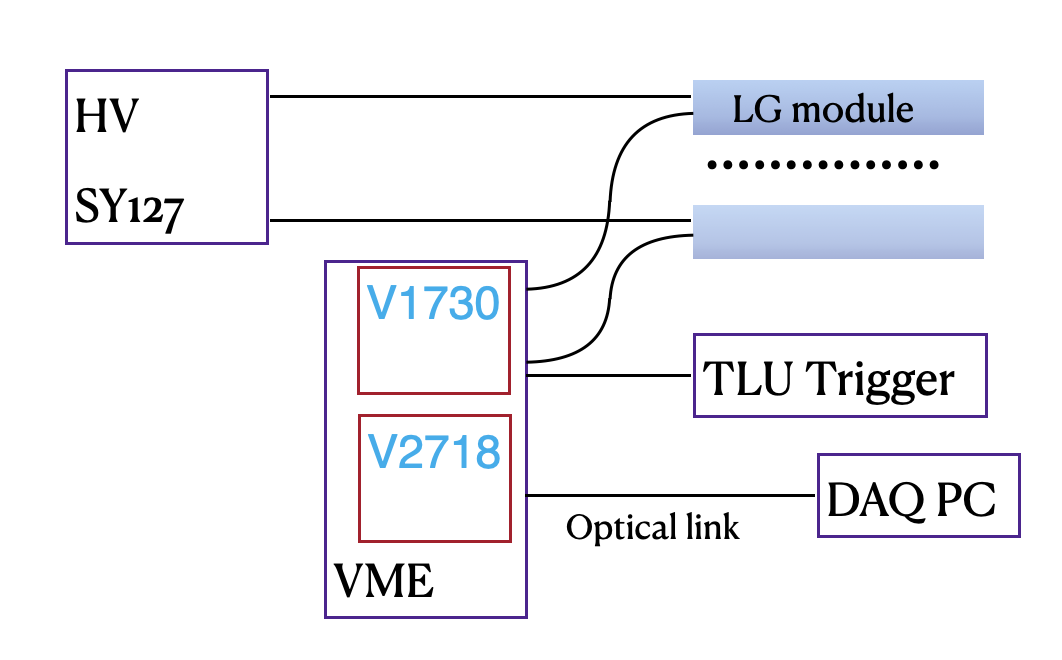}
      \end{center}
    \caption{Readout scheme for the GFM.}
    \label{fig:readout}
\end{figure}

The PMT signal is observed as a short 20\,ns pulse with an amplitude of up to several Volts when connected to a typical $50\,\Omega$ input of the processing unit (see Fig.~\ref{fig:GM_BT_signal}). A CAEN V1730SB ADC module with 14-bit resolution and sampling rate of 500\,MS/s will provide a sufficiently detailed measurement of the signal shape. This board will be tested with the detector prototype. It has 16 input channels with two selectable input dynamic ranges and programmable DC offset adjustment. A VME to PCI optical bridge can be used to interface it with the DAQ PC for module configuration and data recording. Considering the readout with two channels for each lead-glass detector module at 10\,Hz the expected data rate is 115\,kb/s including additional trigger information combined with the payload data.
The readout scheme is presented in Fig.~\ref{fig:readout}. The DAQ PC and HV will be placed in the control room; the location of the VME crate is still to be decided. A HV127 is available and was tested in the lab (see Section~\ref{gfm:app2}). During the beam-tests the HV was supplied from a CAEN Mod VME unit.
Solutions based on the microTCA technology recommended at DESY are also being investigated.

\subsubsection{Monitoring System}
\begin{figure}[ht!]
  \begin{center}
    \includegraphics[width=0.8\textwidth]{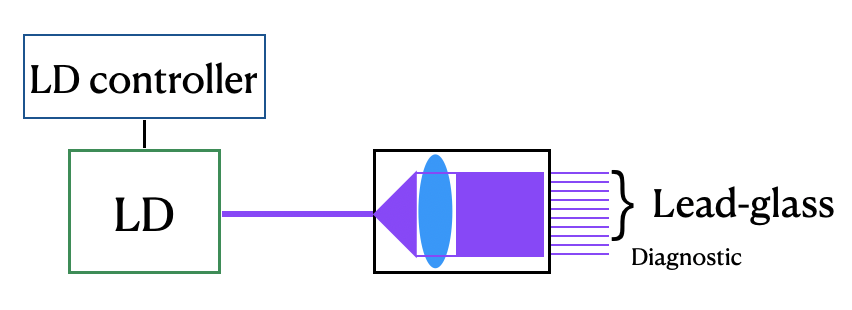}
      \end{center}
    \caption{Monitoring system for the GFM.}
    \label{fig:monitoring}
\end{figure}

The performance of each lead-glass detector module will be monitored continuously throughout the long periods between electron bunches as well as in dedicated calibration runs. The monitoring system is responsible for monitoring of the readout of all channels by sending short light pulses during and between beam bursts. The linearity of the response should be checked over the full dynamic range. For channels with non-linear response, appropriate corrections should be introduced.
The monitoring system could follow either that used in the SELEX experiment at Fermilab~\cite{SELEX:2004wtx} or in the CMS experiment at CERN~\cite{CMS:2008xjf}, down-scaled to the much lower number of channels in LUXE. In both cases it consists of a laser diode, light source and a light distribution system but uses different stabilisation methods for the light pulses.
The envisioned scheme for monitoring the dynamic range and response function for each channel of the GFM is shown in Fig.~\ref{fig:monitoring}.
The monitoring system is based on a stable (over long periods of continuous operation) light source, laser diode (LD), a light distribution system and optical fibres. 
The LD controller receives a signal from the LUXE trigger system and generates the current pulse through LD. The LD light is transmitted through an optical fibre to the experimental area and expanded through a distribution box over 10 fibres, eight of which inject the light into lead-glass crystals and the remaining two are used to monitor the light power for each trigger and to normalise the response of the PMTs.

\subsubsection{Triggers for Monitoring Procedure}
The accelerator supplies beam particles at 10\,Hz while the laser pulsing rate is 1\,Hz. Triggers without collisions can be used for background measurements and monitoring. The time between accelerator bunches is used for detector monitoring. At interspill time, a special trigger generates three outputs. 
Output one provides the trigger for the ADC pedestal measurement, while output two is split into two signals: the first is used to trigger the LD controller and the second to generate the trigger measuring the detector response.
There is a single LD system, which produces light pulses and distributes them over the GFM detector modules. 
Output three is used for the LD measurement. It sends the trigger signal to the laser and at the same time produces the trigger to measure the pedestals. The laser trigger signal is sent over a 50\,m cable to the LD controller. On this trigger, the laser light-pulse is transported back to the experimental hall via an optical fibre and delivered to the GFM modules. 

\section{Interfaces and Integration} 

This subsection discusses the integration of the GFM with the rest of the LUXE experiment, e.g.\ the DAQ and the timing synchronisation.

The ADC calibration monitoring program (CMP) for the GFM is used to record basic parameters of the detectors.  The program runs independently on the data quality monitoring (DQM) computer. For every ADC channel connected to the detector, CMP
computes the following values:

\begin{itemize}
   \item the average ADC response;
   \item the RMS of the external light signal generated by LD system;
   \item the ADC response without any light signal (pedestal).
\end{itemize}

The program estimates the quality of these values and compares them with those obtained earlier, which are considered as a stable reference. These earlier data are stored in a work database. CMP also produces an output file used to evaluate the performance of detector modules and files with configuration parameters for the ADC front-end board.

During data taking, the DQM  will also store:

\begin{itemize}
   \item the deposited energy in each lead-glass module and all channels;
   \item the history of monitoring the light intensity;
   \item the response of the detector in case of physics or monitoring event.
\end{itemize}

The slow control (SC) will include:

\begin{itemize}
   \item HV power supply current (graph as function of time);
   \item Voltage current responsible for light pulse generation.
\end{itemize}

\section{Installation, Commissioning and Calibration}

The installation will proceed in two locations: in the experimental area (EA), where GFM is positioned on the beamline, and in the control room (CR).  The list of parts that need to be installed (with the location in brackets) is: 
\begin{itemize}
    \item Detector with mechanical support, front-end electronics, HV patch panel (EA);
    \item HV power supplies (CR);
    \item HV cables (from EA to CR);
     \item LV cables (from EA to CR); 
    \item LV power supply (EA); 
    \item VME crate (EA);
    \item Light source, optical fibers (EA);
    \item DAQ (CR); 
    \item DAQ cables (from EA to CR). 
\end{itemize}

Cables run from the EA to the CR and are laid down by LUXE/DESY technical services before installation starts.  The AC 240\,V one-phase sockets and sockets to connect devices to the LAN are needed in the EA and in the CR. Sockets to connect the DAQ system and the LUXE slow control system will be needed in the CR.  The local cabling in the EA will be done by the GFM team.  Some of the activities can be carried out in parallel. For the moment the GFM team has no particular constraints regarding the installation dates therefore this will be decided on the basis of an overall optimisation of the LUXE installation schedule.

\begin{table}
\begin{minipage}{\textwidth}
    \centering
\caption{Summary of the installation and commissioning of the LUXE GFM.}
    \vspace{0.2cm}
    \begin{tabular}{p{4.cm}|p{1.5cm}|p{1.0cm} |p{1.8cm}|p{1.2cm}|p{1.5cm}}
\hline
Activity &	Duration & Access & People\footnote{PH = Physicist, PD = Postdoc, EN =  Engineer, TE = Technician} involved  & Person-days &			LUXE/ DESY technical services involved \\
\hline
\hline
Inspection before installation &	1 day&	EA, CR&	1 PH, 1 PD, 1 EN,  1 TE	& 1 &	Yes\\\hline
Installation of HV and LV power supply &	2 days &	CR	&1 TE, 1 EN	&	2&	Yes\\\hline
Test of HV and LV remote control &	1 week &	CR	&1 PD, 1 TE &	5  &	Yes\\\hline
Installation of DAQ and DAQ dry integration tests&	1 week&	CR&	1 PH 1 TE &	5&	Yes\\\hline
Detector and associated equipment mechanical installation&	1 week&	EA, CR&	1 PH, 1 PD, 1 EN, 1 TE	&	5&	Yes\\\hline
Mechanical alignment of the detector supports&	1 day&	EA&	1 EN, 1 TE, 1 PD&	1&	Yes\\\hline
Connection and test of the LV and HV systems&	3 days&	EA, CR &	1 TE, 1 PD, 1 PH&	1 (PH),
2 (PD),		3 (TE) &	Yes\\\hline
Connection and test of the DAQ data path&	2 day&	EA, CR&	1 PH, 1 PD 1 TE	&2&	Yes\\\hline
Test and commissioning of the stand-alone GFM&	2 days&	EA, CR&	1 PH, 1 PD	&2	&		\\\hline
Detector tests and noise optimisation with calibration signals and light source&	2 weeks	&EA	&1 PH, 1 PD&	10&	\\	\hline
Integration and commissioning with central DAQ and slow control system; dry runs &	1 week&	EA, CR	&1 PH, 1 PD &	5&	\\		
        \hline
    \end{tabular}
    \label{tab:instalation}
\end{minipage}
\end{table}

\begin{table}
    \centering
    \caption{Summary of the total FTE for the installation and commissioning of the LUXE GFM.}
    \vspace{0.2cm}
    \begin{tabular}{l|r|r|r|r}
\hline
&  PH &  PD & EN & TE\\ \hline
TOTAL (Person-days)		&31.5 &		31.5 &	9&	24    \\    
TOTAL (FTE-years) 1 year = 225 working days	&	0.14&	0.14 &	0.04&	0.1	\\
							
       \hline
    \end{tabular}
    \label{tab:FTE-years}
\end{table}

\subsection{Installation}
The necessary actions and their duration along with the people involved are given in Table~\ref{tab:instalation} and the summary of needed FTE integrated over one year is given in Table~\ref{tab:FTE-years}. The crane and technical help will be needed to place the detector in the cavern.  The detector installation and cabling will require DESY support. The alignment will be provided centrally. During the installation the access to the EA will be needed for up to six weeks which is in line with the overall schedule of the experiment.

\subsection{Commissioning}
Overall, approximately 4 to 6 weeks of access will be required to install, test and calibrate the whole system.  The commissioning will be carried out by a postdoc and a senior scientist and will take up to four weeks, assuming availability of photon and electron beams hitting the dump. This also includes tests and commissioning of the detector in stand-alone mode, detector tests and noise optimisation with calibration signals and light source and also integration and commissioning with the central DAQ and LUXE slow control system.

\subsection{Calibration Strategy}
Precise calorimeter response monitoring is vitally important for interpreting the data.  Changes in the temperature of the PMTs and the electronics, as well as glass transparency degradation due to radiation damage, can cause long- and short-term variations of the signal amplitudes read out from the lead-glass detector modules. Therefore, calibration constants of each lead-glass detector module must be monitored continuously during data acquisition with special triggers.  Optimal monitoring requires injection into each crystal of a number of photons that corresponds to the signal generated by the typical energy to be measured by each GFM lead-glass channel.  The monitoring light wavelength should be close to the Cherenkov light spectrum. 

Another calibration concept includes calibration with bremsstrahlung photons.  Two weeks of dedicated access should  be allocated for testing and calibration, using bremsstrahlung photons from a thin converter target placed upstream of the interaction chamber. The usage of targets of different thicknesses in the electron beam allows a different number of bremsstrahlung photons to be generated. The energy deposited in the GFM would then be correlated to the number of bremsstrahlung photons by counting the scattered electrons. This idea needs to be verified with simulation to reproduce the measurements and to correct for the Compton spectra.  The calibration also could be performed with dumped electrons; to measure the effect when primary electron beam is dumped and to compare to the simulation model.

\subsection{Decommissioning}

The detector units made from lead-glass could be used after decommissioning  beam-tests or other projects like Beam4School (at DESY). The crystal quality could be restored after the completion of the experiment via an by exposure to UV light (including sunlight) according to Ref.~\cite{Kobayashi:1994kf}.

\section{ORAMS: Operability, Reliability, Availability, Maintainability and Safety}

The design reliability of the GFM to perform under LUXE conditions was studied in simulations with a close-to-final design. The operational reliability of the detector is under study and will be further extended through extensive testing in the lab and in beam-tests. To ensure proper functioning of the detector, well tested existing technologies will be used whenever possible. An exchange-friendly design will ensure the ability to replace faulty elements of the detector within the short access time granted every week or so. There are no particular safety considerations for the GFM design.

\section{Further Tests Planned}

Future tests and studies still needed before being sure that the system will meet the requirements and can be installed are the following:

\begin{itemize}
    \item Beam-tests to study the radiation tolerance of the lead-glass blocks at the GIF facility at CERN or/and the DESY R-weg planned facility;
   \item Test the CAEN V1730SB ADC board with the detector prototype;
    \item Lab tests of the light monitoring system.
\end{itemize}

\newpage
\clearpage

\section{Supplementary Technical Information}

\subsection{ Chemical Composition of Lead-Glass}

\begin{table}[h!]
  \begin{center}
    \caption{Chemical composition of the TF-1 and TF-101 lead-glasses.}
    \vspace{0.1cm}
       \begin{tabular}{|p{1.5cm}|p{4.0cm}|p{2.5cm} |p{2.5cm}| }
      \hline
      Crystal type &Chemical composition & \% composition  &Fractions atomic units\\
      \hline
      TF-1& $\mathrm{PbO}$   &  51.2 &	Pb - 0.082232  \\
       &$\mathrm{SiO}_{2}$  & 41.3	 &  O - 0.608358  \\
       &$\mathrm{K}_{2}\mathrm{O}$   & 3.5   &  Si - 0.246406 \\ &$\mathrm{Na}_{2}\mathrm{O}$   & 3.5   &  K - 0.038057 \\
       &$\mathrm{As}_{2}\mathrm{O_{3}}$   & 0.5   &  Na - 0.023135 \\
       &  &    &  As - 0.001812 \\

       \hline
       TF-101&$\mathrm{Pb}_{3}\mathrm{O}_{4}$   &  51.23 &	Pb - 0.0795  \\
       &$\mathrm{SiO}_{2}$  & 41.53	 &  O - 0.6223  \\
       &$\mathrm{K}_{2}\mathrm{O}$   & 7.0   &  Si - 0.2450 \\ 
       &$\mathrm{Ce}$  &  0.2 &	K - 0.0527  \\
       &  &   &	Ce - 0.0005  \\
            \hline

    \end{tabular}
     \label{tab:LG_Chemproperties}
  \end{center}
\end{table}

\subsection{ Lab Tests}
\label{gfm:app2}

The lab tests were carried out in the ELab facility. The light-tight box developed for the beam polarisation measurements~\cite{Vormwald2014} was used to study different wrapping materials, light transmittance and photomultiplier tube (PMT) response. For the operation of the PMTs, a VME crate is used. VME is a standardised multi-purpose data bus, which is widely used in high-energy physics.  The VME based DAQ system  used in lab tests comprises a QDC (CAEN Mod. V965/V965A 16/8 channels Dual Range QDC) for the PMT readout, a high voltage source for the PMT, and a VME/PCI bridge, which establishes the data link between the VME bus and the PCI bus of the steering computer via an optical link.  For the high voltage source, the VME module CAEN V6533 has been tested as well as CAEN SY127. 

In the lab tests two types of available PMTs were studied. First, one one-anode round Hamamatsu R1398 PMT, that was powered with a custom made voltage divider. And  second, a Hamamatsu R5900-03-M4 with a PMT socket Hamamatsu E7083. The latter PMT features ten dynode stages and a four-fold segmented anode readout.

The PMT signal was studied with the help of a light source -- an LED driver that was used previously in beam polarisation studies and was developed based on the calibration light source of the CALICE tile hadron calorimeter~\cite{Reinecke:2011ina}. 

It is equipped with two UV-LEDs (LEDTRONICS SML0603-395-TR~\cite{led}) with peak intensity at $\lambda = 395$\,nm,  and a rather large spectral width of a few tens of nanometres. 

\begin{figure}[ht!]
  \begin{center}
    \includegraphics[width=0.65\textwidth]{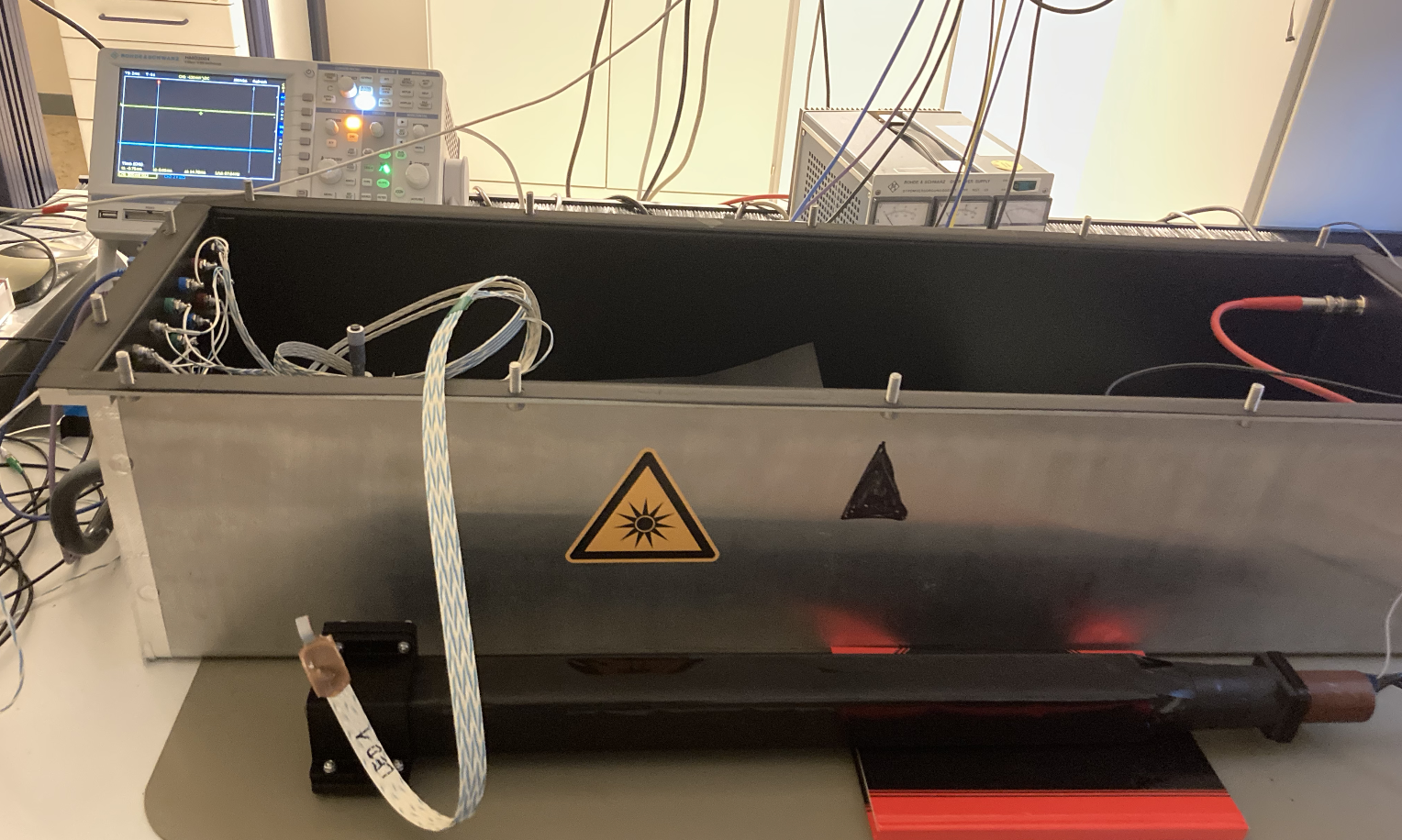}
      \end{center}
    \caption{The detector prototype in the ELab.}
    \label{fig:GM_Elab}
\end{figure}

A picture of the laboratory setup is shown in Fig.~\ref{fig:GM_Elab} where the detector prototype is wrapped in aluminium and two layers of vinyl foils exposed to environment light.

\begin{figure}[ht!]
  \begin{center}
    \includegraphics[width=0.65\textwidth]{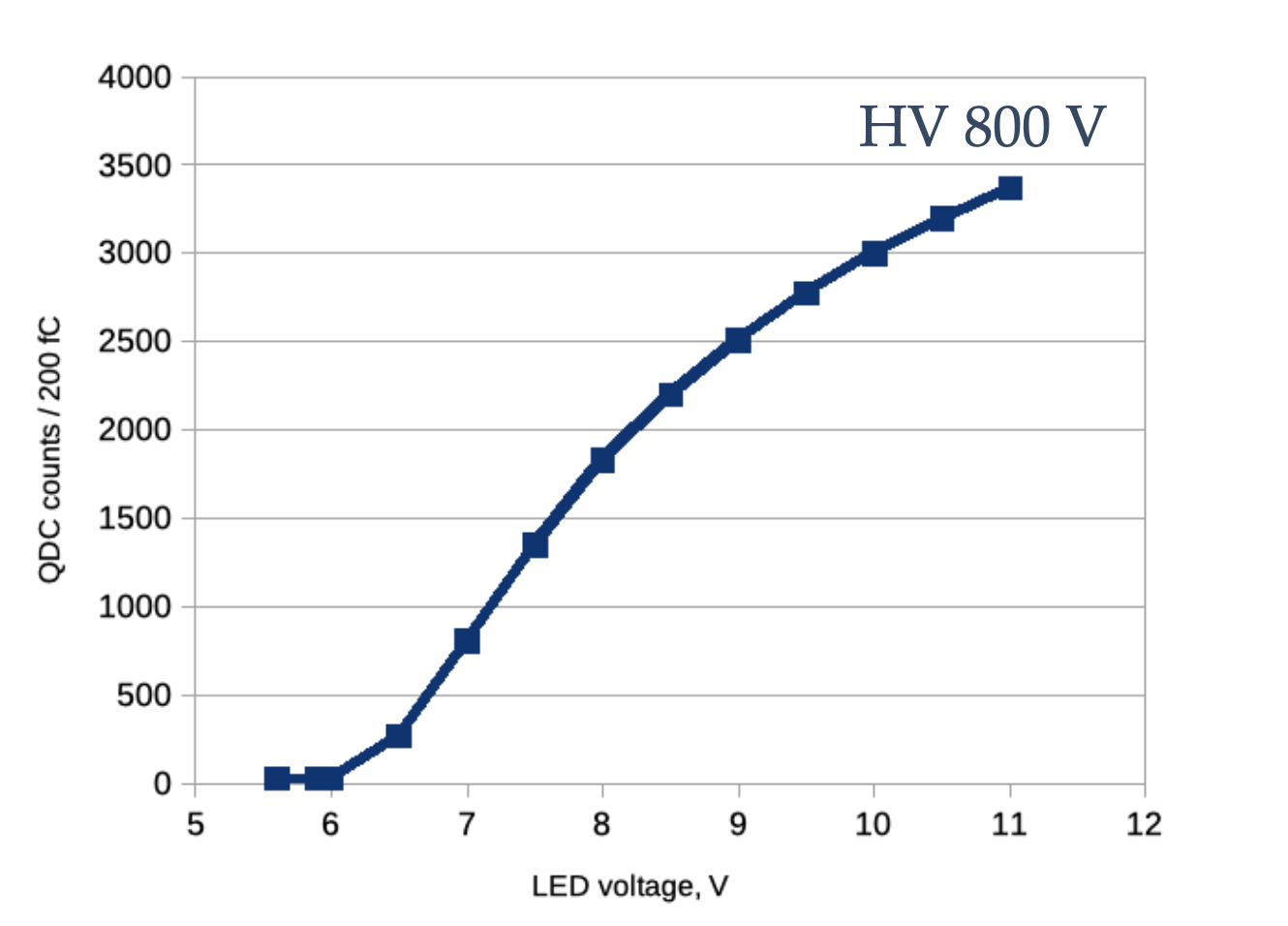}
      \end{center}
    \caption{Dependence of the PMT signal on the voltage applied to the LED.}
    \label{fig:PMT-response}
\end{figure}

Figure~\ref{fig:PMT-response} shows the PMT response at 800\,V as a function of the LED steering voltage in the range of $5.5 - 11$\,V which corresponds to $0.9 - 4.0$\,mW. 
The sensitivity is not linear especially at low light intensity and the maximum of light intensity is below the PMT saturation level.

\begin{figure}[ht!]
  \begin{center}
    \includegraphics[width=0.65\textwidth]{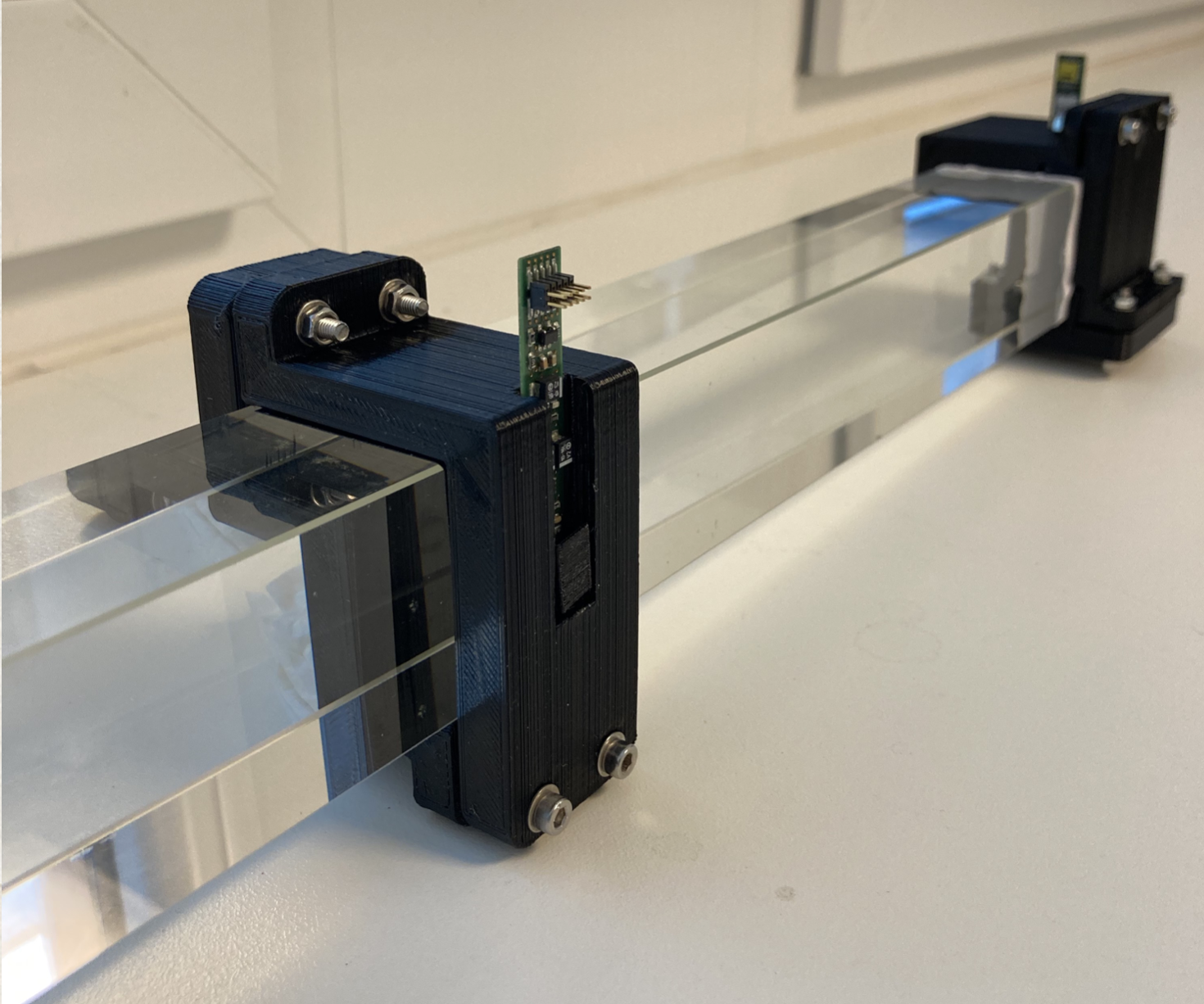}
      \end{center}
    \caption{LED driver in supporting structure for detector response studies.}
    \label{fig:GM_Led_holders}
\end{figure}

The first GFM detector unit prototype consists of one lead-glass block  with a Hamamatsu R1398 PMT, that was powered with a custom made voltage divider.
Different holding structures for the PMT  and the LED driver board have been designed and 3D-printed (see Fig.~\ref{fig:GM_Led_holders}). This allowed the dependence of the crystal response on the light source position to be studied.

\begin{figure}[h!]
  \begin{center}
    \includegraphics[width=0.7\textwidth]{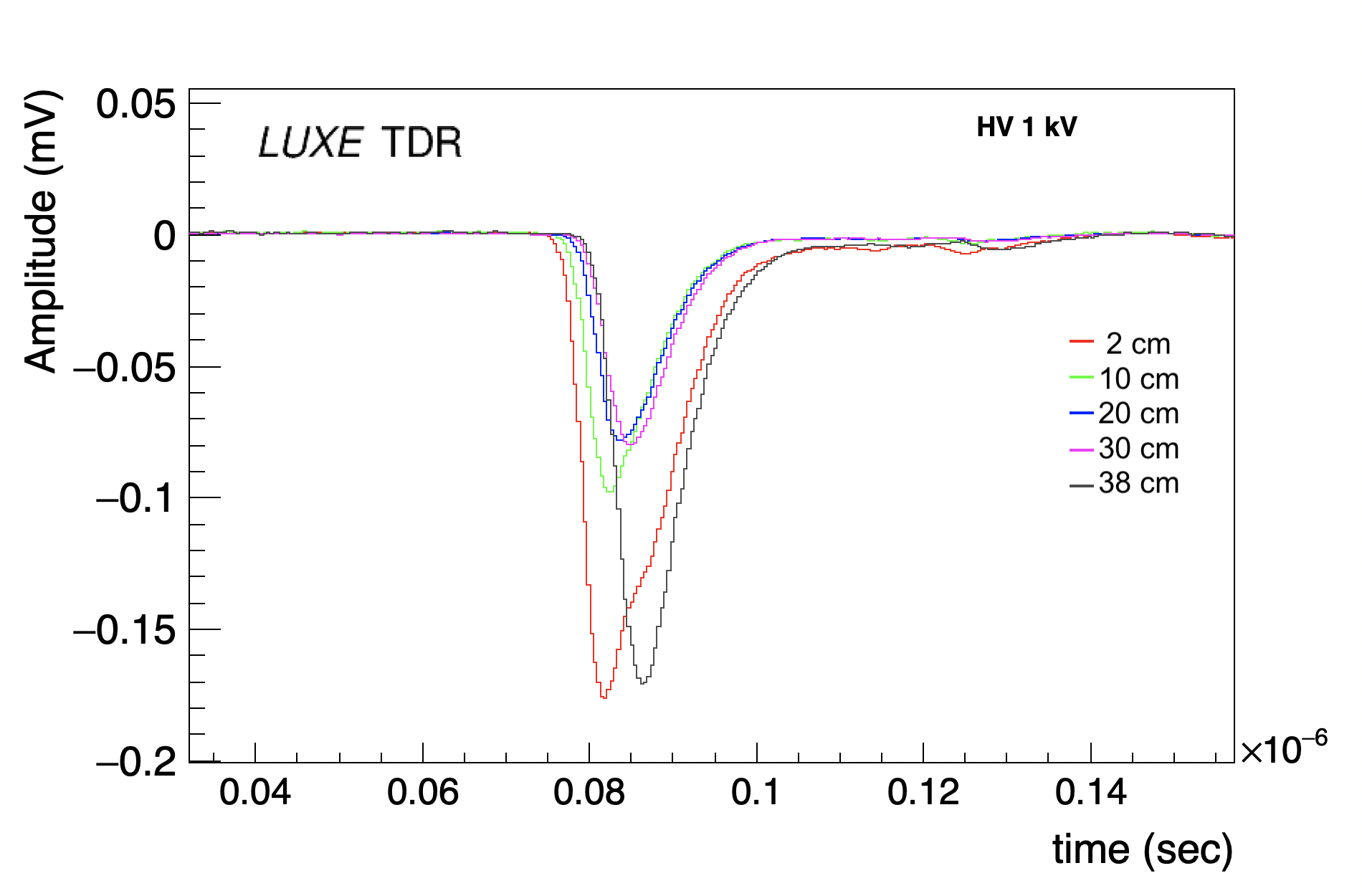}
      \end{center}
    \caption{The signal depending on LED position along the crystal.}
    \label{fig:GM_Led_position}
\end{figure}

The signal shape was measured for different LED intensities and positions along the crystal. The displayed pulse shapes in Fig.~\ref{fig:GM_Led_position} are obtained from averaging 50 recorded waveforms per LED voltage. Testing the response depending on the LED position was done at five locations: close to the PMT, 10, 20, 30 and 38\,cm from the PMT. 

The measurements with and without reflector wrapped around the crystal did not show significant differences when shining LED through the crystal’s end which is explained by the total internal reflection (lead-glass refraction index is 1.65 (Table~\ref{tab:LGproperties_tf}), which corresponds to the angle of 37$^\circ$ or 53$^\circ$ with respect to the surface.)

\subsection{Beamtests}

The first beam-test campaign took place in November 2021 at the Laser Plasma Accelerator (LPA) at DESY.  The electron beam has energy of 60\,MeV, a bunch charge 20\,pC ($\sim 10^{8}\,e^{-}$) and operates at 3.5\,Hz. The number of electrons is higher than expected for the GFM in LUXE.

\begin{figure}[h!]
  \begin{center}
    \includegraphics[width=0.55\textwidth]{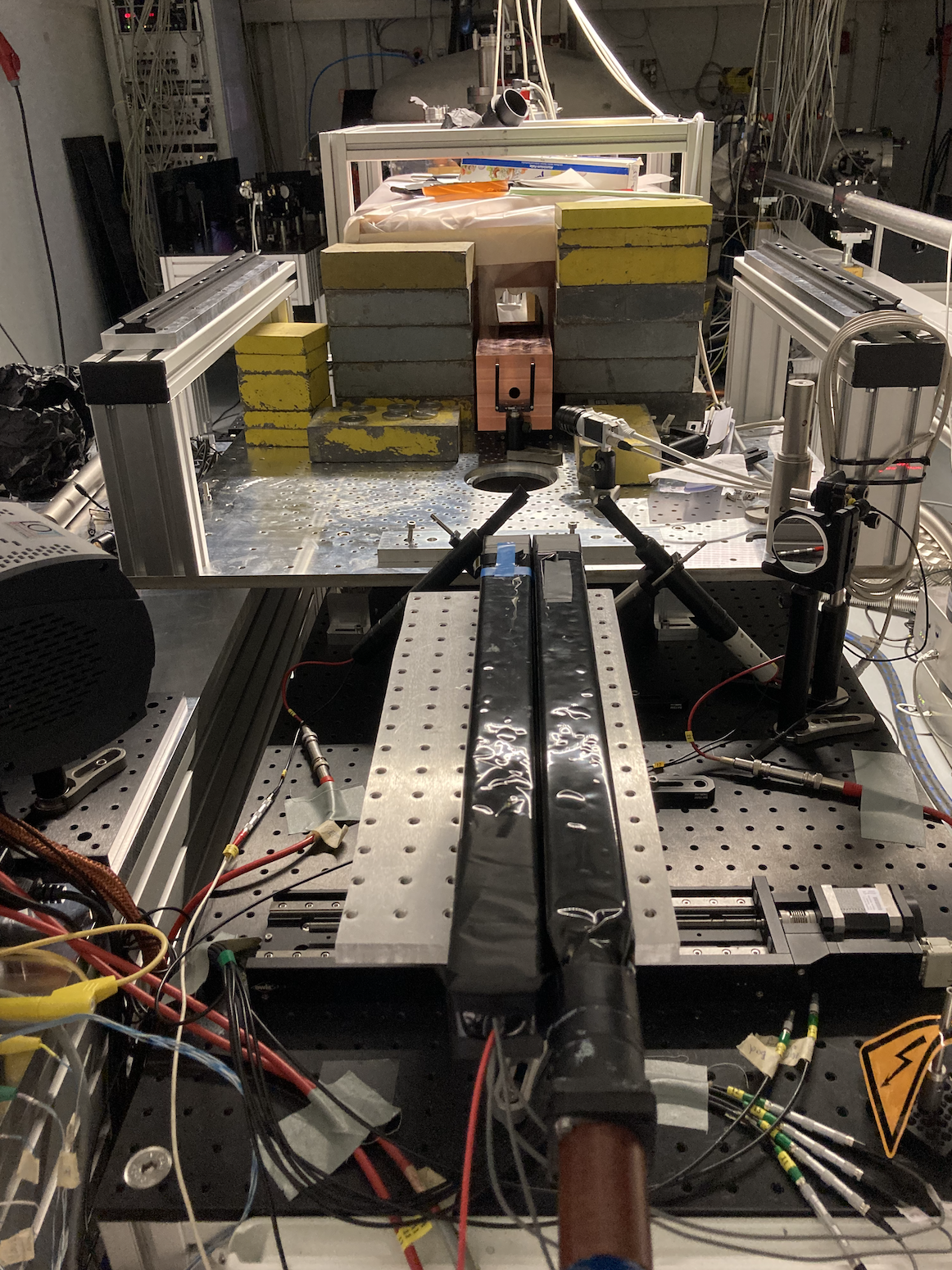}
      \end{center}
    \caption{Picture of the GFM test setup in the LPA laboratory.}
    \label{fig:GM_LPA}
\end{figure}

Two GFM module prototypes were tested in the beam. Each of them consists of one lead-glass block wrapped in one layer of aluminium foil and two layers of vinyl foil with the PMT attached to the end. One module was equipped with a Hamamatsu R1398 connected to a custom made voltage divider and second module with a Hamamatsu R5900-03-M4 with a PMT socket Hamamatsu E7083. The latter PMT features ten dynode stages and a four-fold segmented anode readout.  A picture of the measurement setup is shown in Fig.~\ref{fig:GM_LPA} where two GFM modules are placed in the beam on a moving stage. The scintillators for trigger and beam monitoring are also shown.

The trigger was provided from the laser and data were collected for several positions and orientations of the lead-glass blocks with respect to the beam.

\begin{figure}[ht!]
  \begin{center}
    \includegraphics[width=0.65\textwidth]{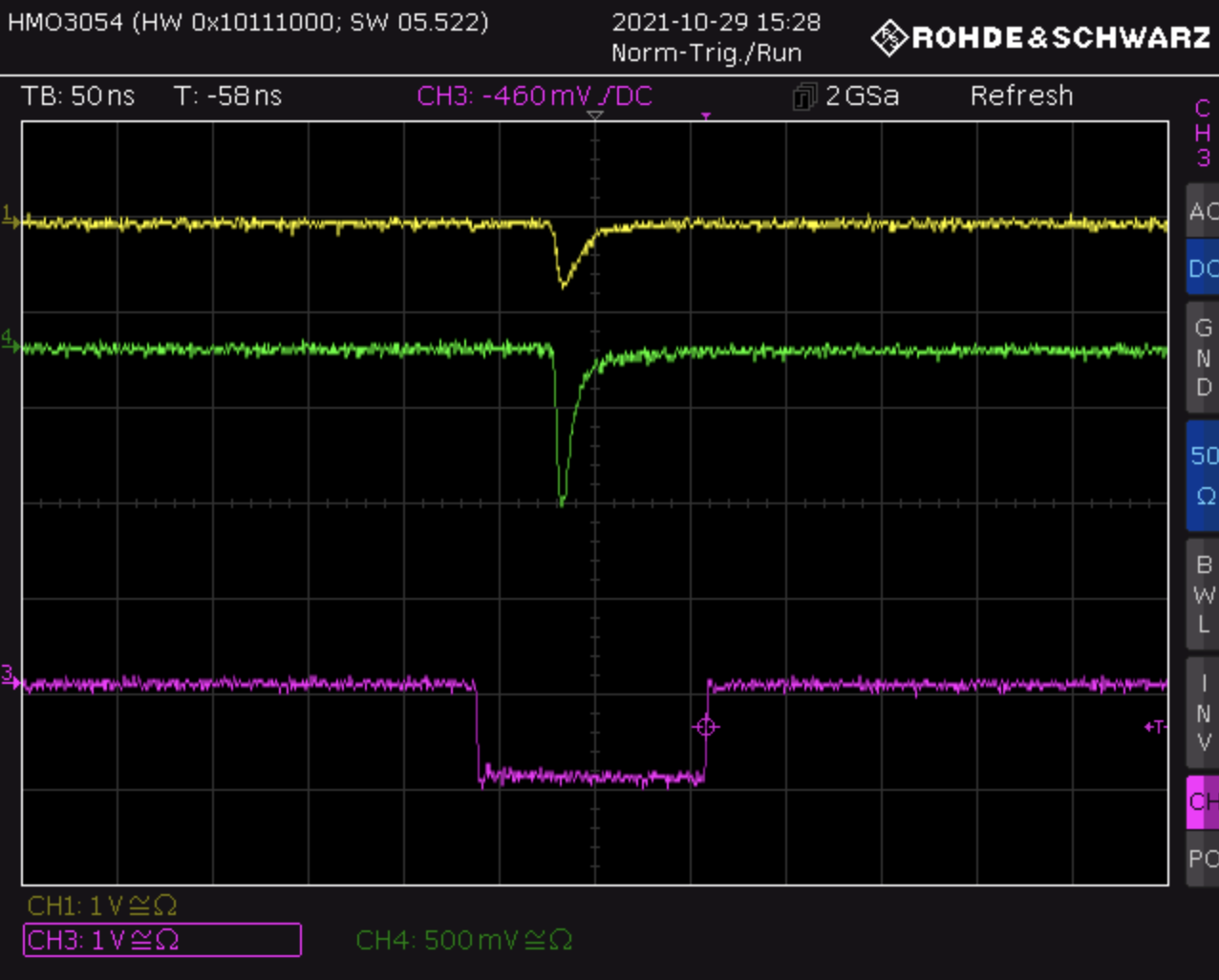}
      \end{center}
    \caption{Signal from  one block of GFM (green) exposed to electron beam of 60\,MeV at LPA shown along with the signal from the scintillator counter (yellow), which is installed in front of the lead-glass, downstream of the collimator, and the gate for charge integration (QDC) (purple). }
    \label{fig:GM_BT_signal}
\end{figure}

A high voltage scan was performed to determine the best value to fully use the dynamic range of the QDC for a fluctuating electron beam for which the number of particles per bunch can vary from zero to maximum. In this setup, a voltage of 350\,V was applied to the R5900-03-M4 and 600\,V to the R1398 PMTs. These voltages are about $20\% $ lower than the lowest required to sense LED light.  An example of the signal (green line) coming from the prototype seen on the scope is shown in Fig.~\ref{fig:GM_BT_signal}. The yellow line corresponds to the signal from the scintillator counter, and the purple to the signal from the gate for charge integration (QDC). 

\begin{figure}[ht!]
  \begin{center}
      \includegraphics[width=0.60\textwidth]{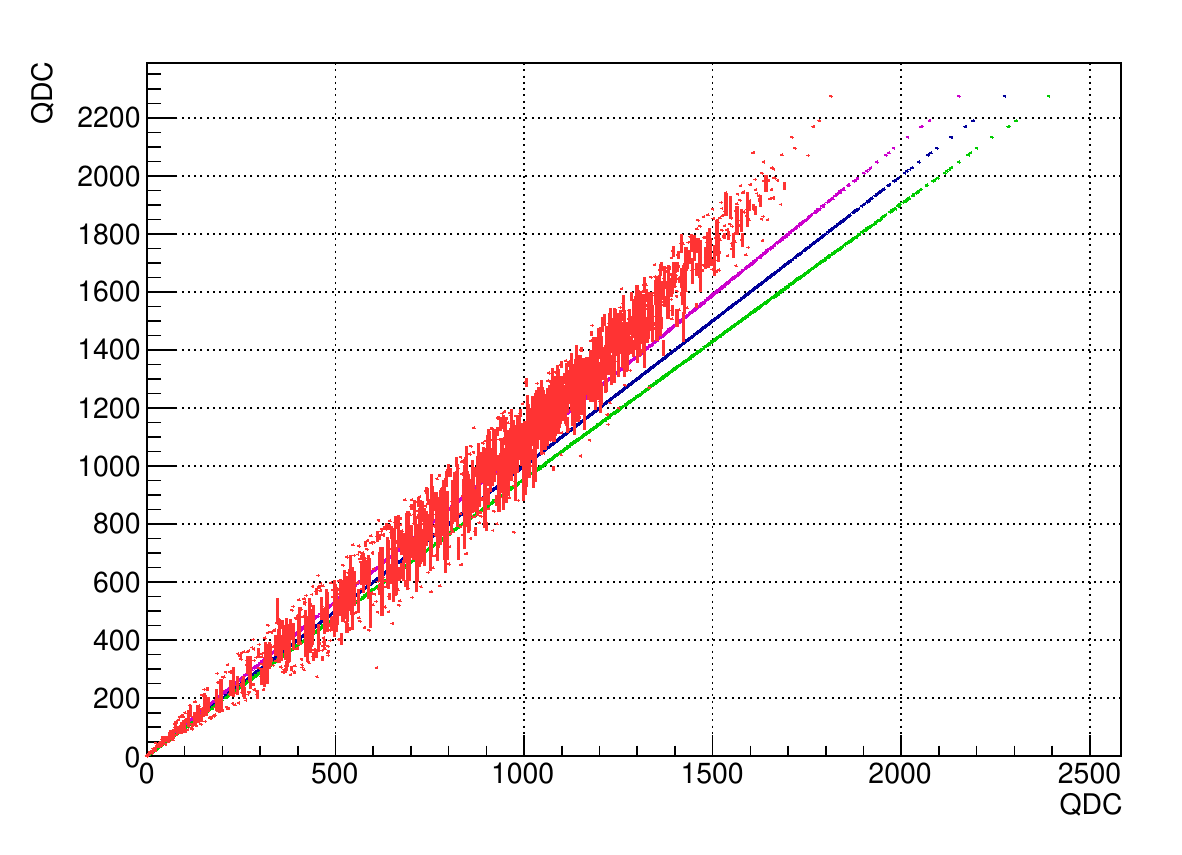}
      \end{center}
    \caption{Correlations between signals obtained from the one-anode PMT (red) and the four-anode PMT. For the four-anode PMT, the first anode was used as monitoring input for the scope, the second anode was used as a reference (blue line), and the third and fourth anodes were used as signal (purple, green).}
    \label{fig:GFM_BT_results}
\end{figure}

Figure~\ref{fig:GFM_BT_results} shows the correlations of signals between different PMT channels.  The one-anode PMT was placed at 40\,mm off the beam axis, while the module with the four-anode PMT was in the beam.  The $y$-axis in Fig.~\ref{fig:GFM_BT_results} shows the signal from one of the four-anode PMT channels which is taken as a reference signal.  The $x$-axis corresponds to the signals observed in other channels.  The thin lines which correspond to the other three channels of the four-anode PMT demonstrate good linear dependence with only a small difference in individual gain.  The signal from the one-anode PMT is linear up to 1000\,QDC, and shows slightly non-linear response for higher values. This can be attributed to a slightly wider signal waveform of the one-anode PMT for high signal amplitudes which was not taken into account when the gate was configured. For high values, part of the signal was not integrated and this lost fraction increases as the signal amplitude increases. The signal from the one-anode PMT also has a wider distribution which can be explained by fluctuations in the angular distribution of the off-beam particles and intrinsic fluctuations of light production in the lead-glass.

\printbibliography
\end{refsection}


\chapter{Data Acquisition, Computing and Simulation}
\label{chapt11}
\begin{refsection}

\chapterauthors{
H. Abramowicz\\ Tel Aviv University, Tel Aviv (Israel)}
{O.~Borysov\\ Weizmann Institute of Science, Rehovoth (Israel)}
{B. Heinemann \\ \desyaffil}{ L.~Helary \\ \desyaffil}
{F.~Meloni \\\desyaffil}{M.~Wing \\University College London, London (UK)}
\date{\today}

\begin{chaptabstract}
    This chapter describes common aspects of LUXE regarding simulation, data acquisition, data quality and slow control monitoring, and computing. 
\end{chaptabstract}

\section{Introduction}
\label{common:intro}

This chapter covers aspects of LUXE related to software and data flow that are general in nature and not specific to a given detector or system.  In Section~\ref{sec:simulation}, the  Monte Carlo program, called \textsc{Ptarmigan}, used to generate the expected strong-field QED processes is briefly described, with more information given in Chapter~\ref{chapt2} and elsewhere~\cite{luxecdr,ptarmigan}.  The \geant simulation of the beam-line, detectors and other aspects of the experimental set-up is then described.  This highlights the implementation of the experiment's geometry, the special case of the $\gamma-$laser setup and the generation of signal and background samples.  In Section~\ref{sec:daq}, the design of the data acquisition (DAQ) system is described.  This includes estimates of the data rates from all detectors and the resulting DAQ architecture, the timing and control system, the DAQ software, and the current system status.  First ideas of the slow control and data quality monitoring are discussed in Sections~\ref{sec:sc} and~\ref{sec:dqm}, respectively.  The slow control system used by the EuXFEL accelerator can also be used for the LUXE experiment, whereas the data quality monitoring will be integrated into the LUXE DAQ software.  The final common infrastructure topic, computing, is discussed in Section~\ref{sec:computing}.  Issues addressed are data volume and storage, based on the DAQ estimates, and processing.  Common software and a software framework along with organisation of the activities are also discussed.  As physics data taking is expected to proceed with a rate of 1\,Hz, cutting-edge new developments for the general hardware and software systems presented here will not be required.  We will therefore frequently rely on hardware, software and tools that already exist, and suited to LUXE, in order to minimise the risk. In Section~\ref{sec:further-tests}, some near term plans will be discussed that should reduce the risk of implementation in the final experiment. 


\section{Simulation} \label{sec:simulation}

The simulation is based on a dedicated Monte Carlo (MC) generator for the primary physics processes that will be studied by LUXE. The active and passive elements of LUXE are described in the \geant simulation package, and the interactions of the particles with the various detectors are also simulated by \geant. This section describes both aspects. 

\subsection{MC Sample Generation}
A custom-built strong-field QED MC computer code, named \textsc{Ptarmigan}~\cite{ptarmigan}, is used to simulate the strong field interactions for LUXE for the relevant physics processes, as described in Ref.~\cite{luxecdr}.
The task of the MC generator is to simulate the physics processes in a realistic scenario with the laser focusing and electron beam sizes taken into account properly. 

The MC has been used to generate datasets for all foreseen LUXE experimental configurations. These include the non-linear Compton process, the two-step trident process, and the non-linear Breit-Wheeler pair production process. The results of the MC are validated by comparing with analytic calculations (see also Refs.~\cite{luxecdr,Blackburn:2021rqm}).

\begin{figure}[htbp]
  \centering
  \includegraphics[width=0.49\textwidth]{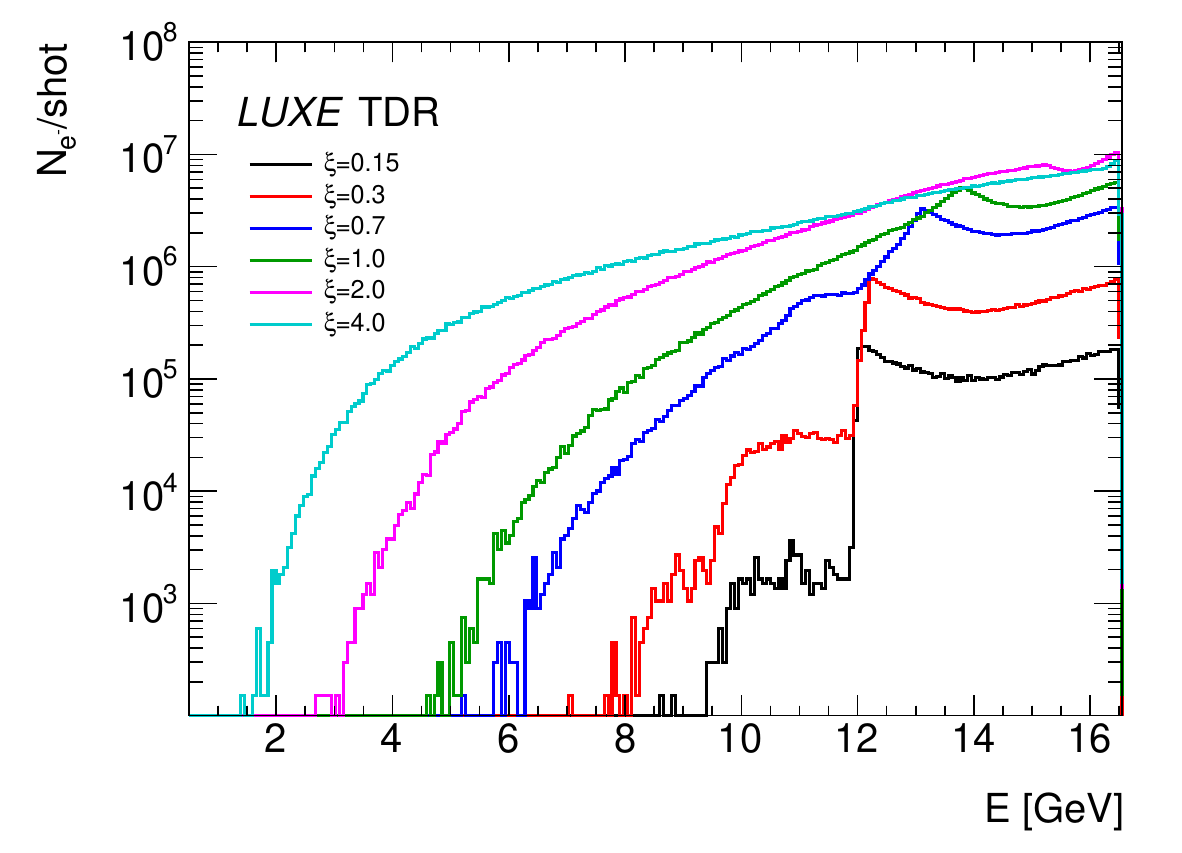}
  \includegraphics[width=0.49\textwidth]{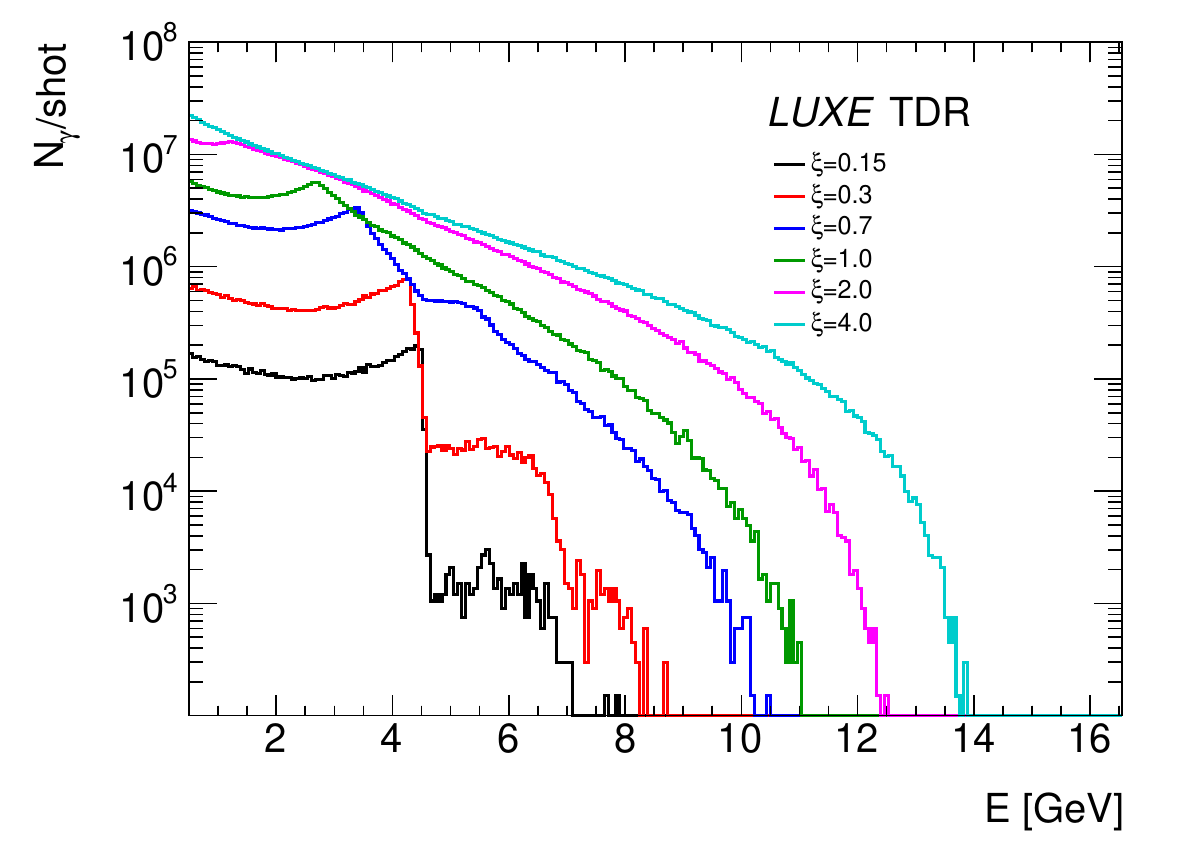}
  \includegraphics[width=0.49\textwidth]{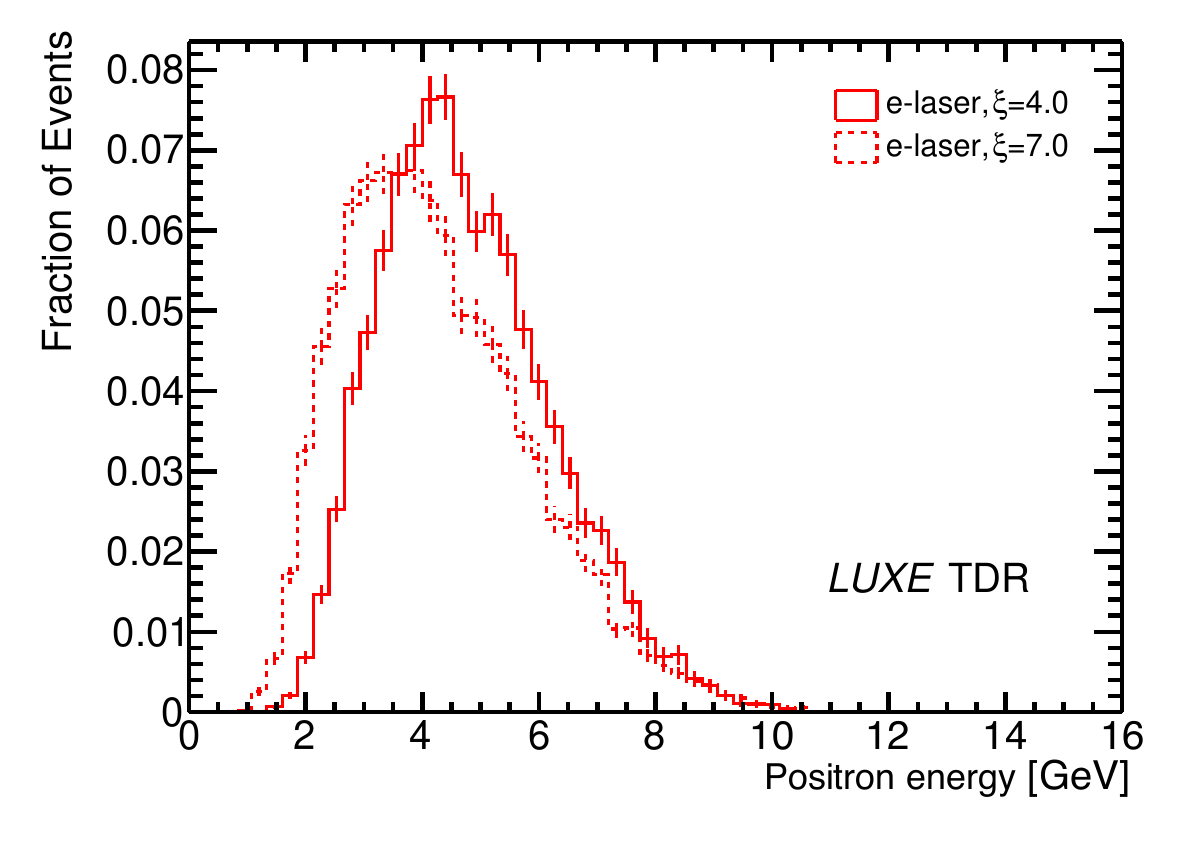}
  \includegraphics[width=0.49\textwidth]{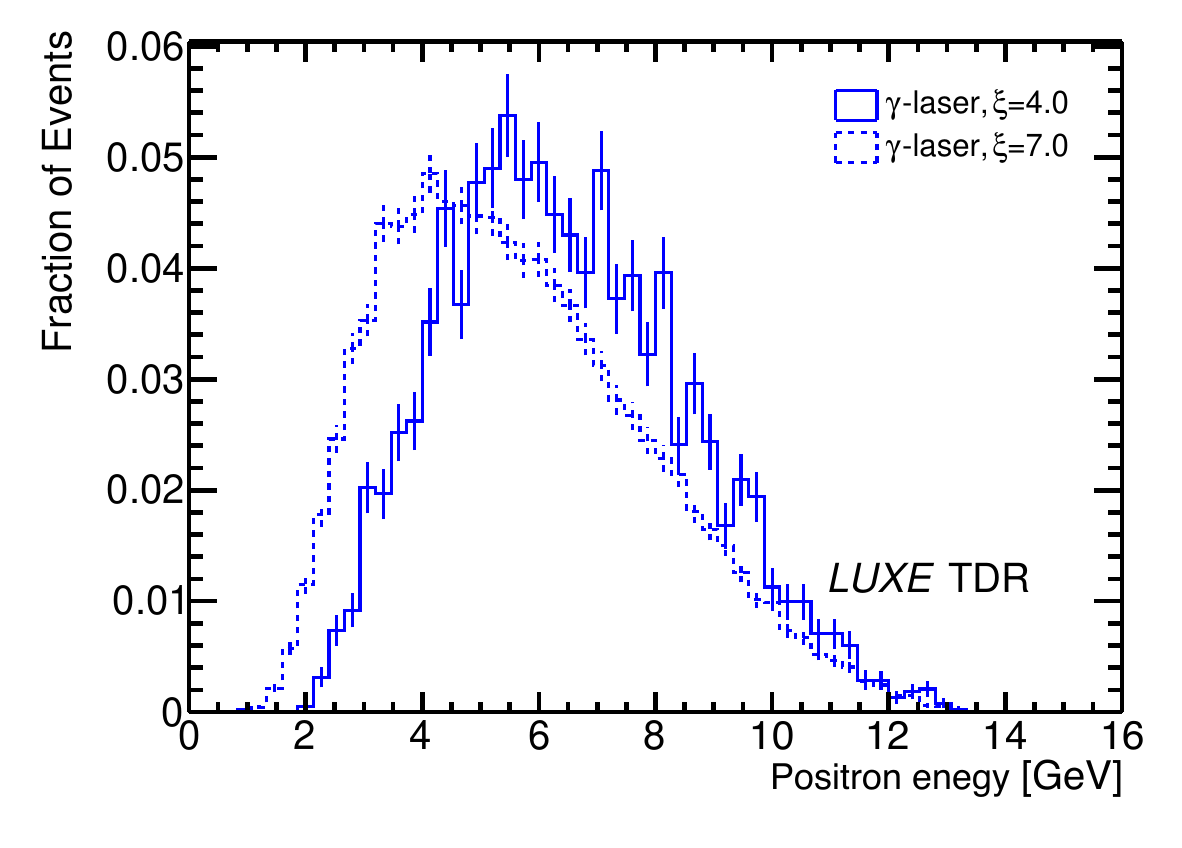}
  \caption{Top: Energy spectra of (left) electrons and (right) photons for various values of $\xi$ expected in the \elaser mode. Bottom: Energy spectra of positrons for two values of $\xi$ expected in the (left) \elaser and (right) \glaser mode. }
  \label{fig:ptarmigan}
\end{figure}

As examples, Fig.~\ref{fig:ptarmigan} shows the expected energy spectra from \textsc{Ptarmigan} for different $\xi$ values for electrons, photons and positrons.

\subsection{\geant Simulation}
\label{sim:geant4}
In this section, an overview of the common simulation and geometry is provided. The details of the individual detectors and their response to particles is discussed in the chapters on the detectors, and the beamline is discussed in Chapter~\ref{chapt12}.

A simulation of particle fluxes considering all components of the experiment is performed to estimate the rates of signal particles from strong-field QED interactions and the background due to secondary particles from the beam. Through an iterative process the layout was optimised to reduce these backgrounds. The simulation is also used to estimate the ionisation dose and to aid the development of parametric fast emulation of the response of the detectors to the various particles.

\subsubsection{Geometry of the Experiment}
\begin{figure}[htbp]
  \centering
  \includegraphics[width=0.95\textwidth]{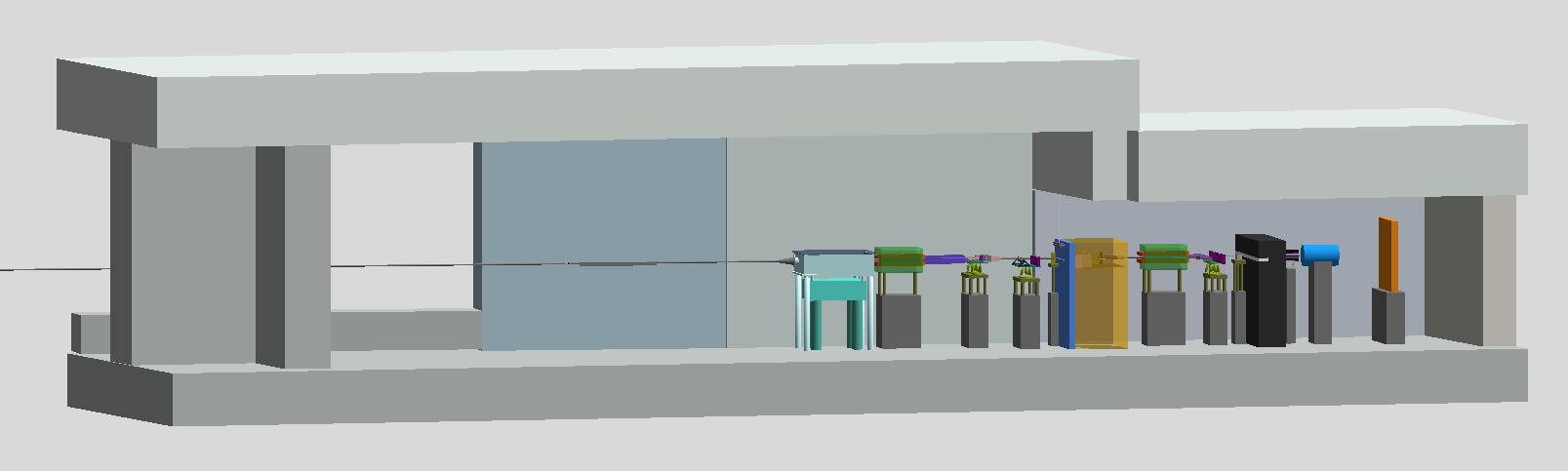}
  \includegraphics[width=0.95\textwidth]{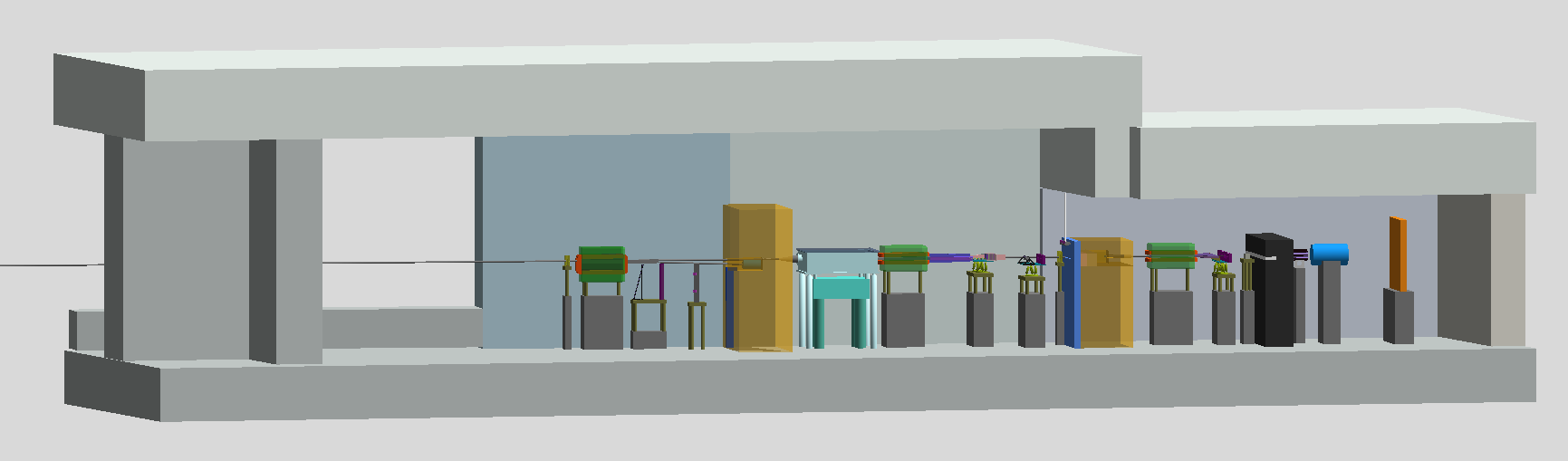}
  \caption{The layout of LUXE as modelled in the \geant simulation for the \elaser (top) and \glaser (bottom) set-up based on the CSG approach (see text). Dipole magnets after the tungsten target and the interaction chamber are shown in green, the yellow and black blocks are shielding elements. Behind the shielding in light blue is the interaction chamber.}
  \label{fig:luxe_geant4_geom}
\end{figure}

A sketch of the experiment is shown in Chapter~\ref{chapt2}. 
The geometry model of the LUXE experiment is implemented in \geant~\cite{Allison:2006ve,Allison:2016lfl} version 10.06.p01 using the QGSP\_BERT\_HP physics list. A right-handed coordinate system is used with the the $z$-axis being defined by the nominal beam direction, the $y$-axis pointing towards the sky (referred to as vertical direction) and the positive $x$-axis (transverse direction) is to the left of the beamline from the perspective of the beam. Figure~\ref{fig:luxe_geant4_geom} shows a general view of the LUXE 
simulation model for the \elaser and \glaser modes which implements the proposed layout of LUXE\footnote{The simulation software is available at~\cite{luxe-software}.}. It includes beam instrumentation components, detector systems and infrastructure of the XS1 cavern. 
Initially it was imported as a collection of tessellated objects from the existing 3D CAD model of the building. Later it was implemented using the standard \geant constructive solid geometry (CSG) approach, which provides faster simulation and requires substantially less memory. The CSG implementation matches the original 3D CAD model within a few millimetres in the areas along the LUXE experimental setup and was used in simulation studies.

Magnetic fields are approximated with the following function
\begin{equation}
\label{eq_magnet_field}
B_{i}(x,y,z) = B_{0i} f_{ix}(x)f_{iy}(y)f_{iz}(z),
\end{equation}
\begin{equation}
\label{eq_magnet_approx}
f_{ij}(x_{j}) = \frac{1}{\left(1+e^{\frac{p_{0}-x_{j}}{p_{1}}}\right)\left(1+e^{\frac{x_{j}-p_{2}}{p_{3}}}\right)},
\end{equation}
where $i$ indicates field vector components $(B_{x}, B_{y}, B_{z})$ and $j \subset \{x, y, z\}$, $B_{0i}$ is the amplitude of the field and $p_{0}$, $p_{1}$, $p_{2}$, $p_{3}$ are parameters whose values are determined by fitting the function~(\ref{eq_magnet_approx}) to the measurements of the field in transverse directions in the middle plane of the magnet. In the $y$ direction the field is considered to be uniform.

The geometrical models of the interaction and target chambers, beam dumps and detector support structures are implemented according to the design envisaged for LUXE and described in Chapters~\ref{chapt3} and~\ref{chapt12}. The locations of the key components are given in Table~\ref{tab:sim_keypar}.

\begin{sidewaystable}[htbp]
    \centering
    \resizebox{0.8\textwidth}{!}{%
    \begin{tabular}{|l|c|c|c|r|}
    \hline
     \textbf{Component}    & distance   &  transverse & transverse & purpose  \\
        & from IP (m) & size $x$ (cm) & size $y$ (cm) &  \\\hline
     \multicolumn{5}{|l|}{\textbf{Bremsstrahlung system area} ($\gamma$-laser mode only)}\\\hline
     $W$ target & $-7.47$ & small & small & produce $\gamma$\\
     Dipole magnet ($B=1.5$~T, $y$ deflection)& $[-7.22,-5.78]$ & $[-35,35]$ &  $[-49,+49]$ & deflect $e^\pm$ \\
     Scintillator & $[-5.59,-5.34]$ & $[-5,5]$ &$-[13,110]$ &  measure $e^-$\\
     \cer & $-4.79$ & $[-14,9]$ & $-[15,110]$ & measure $e^-$ \\ \hline
     \multicolumn{5}{|l|}{\textbf{IP area} ($e$-laser and $\gamma$-laser modes)}\\ \hline
     Shielding & $[-2.98,-1.48]$ & $[-239,+62]$ & $[-248,+152]$ & protect IP \\ 
     IP & $0.0$ & $0.0$ & $0.0$ & interactions\\
     Interaction chamber & $[-0.87,+1.25]$ & $[-101,+61]$ & $[-34, +27]$ & final steering and focus of laser\\
     Dipole magnet ($B=1$~T, $x$ deflection) & $[1.33,2.77]$ & $[-49,+49]$ & $[-35,35]$  & deflect $e^\pm$ \\
     Pixel tracker & $[3.95,4.26]$ & $[5,55]$ & $[-1,+1]$ & measure $e^+$ (and $e^-$ for \glaser case) \\
     Calorimeter & $[4.34,4.43]$ & $[4,59]$
     & $[-3,+3]$ & measure $e^+$ \\
     Scintillator & $5.45$ & $[-108,-8]$ & $[-5,+5]$ &  measure $e^-$ for \elaser \\
     \cer (Ar gas) & $5.87$ & $[-104,-8]$ 
     & $[-14,+9]$ & measure $e^-$ for \elaser (**)\\ 
     \cer (Quartz) not implemented & &  & & measure $e^-$ for \glaser\\
     Gamma target & $6.5$ & small & small & photons to $e^+$ $e^-$ conversion\\
     Beam dump & $[7,7.75]$ & $[-25,+5]$ & $[-15,+15]$ & dump primary $e$ beam\\
     Shielding & $[6.8,8.28]$ & $[-170,+50]$ &  $[-248,+52]$ & reduce background in PDS \\
     \hline
     \multicolumn{5}{|l|}{\textbf{Photon Detection System (PDS)} ($e$-laser and $\gamma$-laser modes)}\\\hline
     Dipole magnet ($B=1.4$~T, $x$ deflection) & $[8.92,10.36]$ & $[-49,+49]$ & $[-35,35]$ & deflect $e^\pm$ \\
     Scintillators & $10.88$ & $[-101,+101]$ & $[-5,+5]$ & measure $e^\pm$ \\
     Gamma profiler & $11.8$ & small & small & measure $x-y$ profile of $\gamma$s \\
     Backscattering calorimeter & $[13.06,13,51]$ & $[-16,+16]$ & $[-16,+16]$ & backscatter of $\gamma$s \\
     Beam dump & $[13.63,14.63]$ & $[-30,+30]$ & $[-30,+30]$ & dump photons \\\hline
     BSM detector & $[16.0,16.2]$ & $[-100,+100]$& & detect photons \\\hline
     Final wall & $17.4$ & large & large &-- \\\hline
    \end{tabular}}
    \caption{Table of the key components simulated. Given are the locations and purposes. The transverse direction here is the plane in which the charged particles are bent by a dipole magnet. In cases where the objects are sizeable, intervals are given. When a sign is given before the interval it applies to both values. The coordinate system is right-handed, and defined by $+z$ being the direction of the beam, $+y$-direction is vertically up (away from the centre of the earth) and $x$ is the horizontal direction.
    }
    \label{tab:sim_keypar}
\end{sidewaystable}

In order to save CPU resources some simplifications have been made for supporting structures in the areas which are far away from the path of particles from both the main beam or the expected physics interactions.

For both the \elaser and the \glaser studies, the primary electrons are generated in accordance with the \euxfel beam parameters (see Chapter~\ref{chapt2}).

For the \elaser simulation, the primary electron beam directly enters the IP, and after the IP it is deflected by a magnet towards a dump. Additionally, when the laser is fired, a large rate of \elaser interactions occurs resulting in a large number of lower energy electrons and photons. The photons continue in the \beampipe towards the end of the cavern, where three detection systems are placed to measure various properties of the photon flux. Charged particles produced in the \elaser interaction are deflected in the horizontal plane by a dipole magnet, and detectors are installed to measure their energy spectrum and flux.

\subsubsection{Bremsstrahlung Simulation for the \glaser Setup}

For the \glaser setup, the simulation of the bremsstrahlung process at the target is important. The energy spectrum of the bremsstrahlung photons can be approximated using the following formula for thin targets~\cite{Tanabashi:2018oca}: 
\begin{equation}{
   \frac{dN_{\gamma}}{dE_{\gamma}} = \frac{X}{E_{\gamma} X_{0}} 
                     \left( \frac{4}{3} - \frac{4}{3} \frac{E_{\gamma}}{E_{e}} + \left(\frac{E_{\gamma}}{E_{e}}\right)^{2}\right) \, ,
  }\label{eq_bremsstrahlung_pdg}
\end{equation}
where $E_{e}$ is the energy of the incident electron, $X_{0}$~the radiation length of the target material and $X$~is the target thickness. Tungsten was chosen as a target material due to its high melting point temperature, high thermal conductivity and high sputtering resistance. The target thickness was chosen to be 1\%$X_0$, corresponding to 35\,$\mu$m. A FLUKA simulation was used to check that it can stand the thermal stress exerted by the radiation from the primary \euxfel beam in a year. It also provides a sufficiently high photon yield. By increasing the target thickness the photon yield could be increased by up to a factor of two, but it would stand less radiation and thus pose a higher risk. 

\begin{figure}[htp]
    \includegraphics[width=0.5\textwidth]{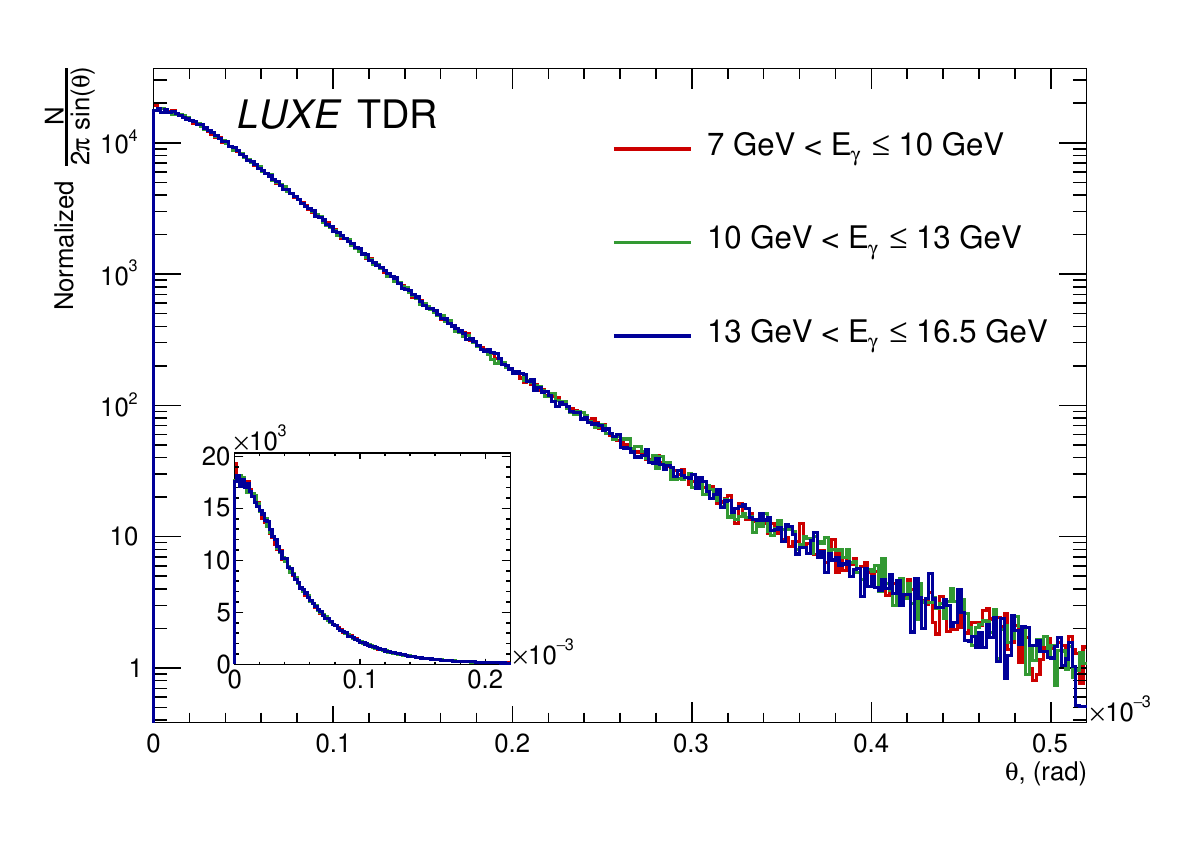}
    \put(-145,120){\makebox(0,0)[tl]{(a)}}
    \includegraphics[width=0.5\textwidth]{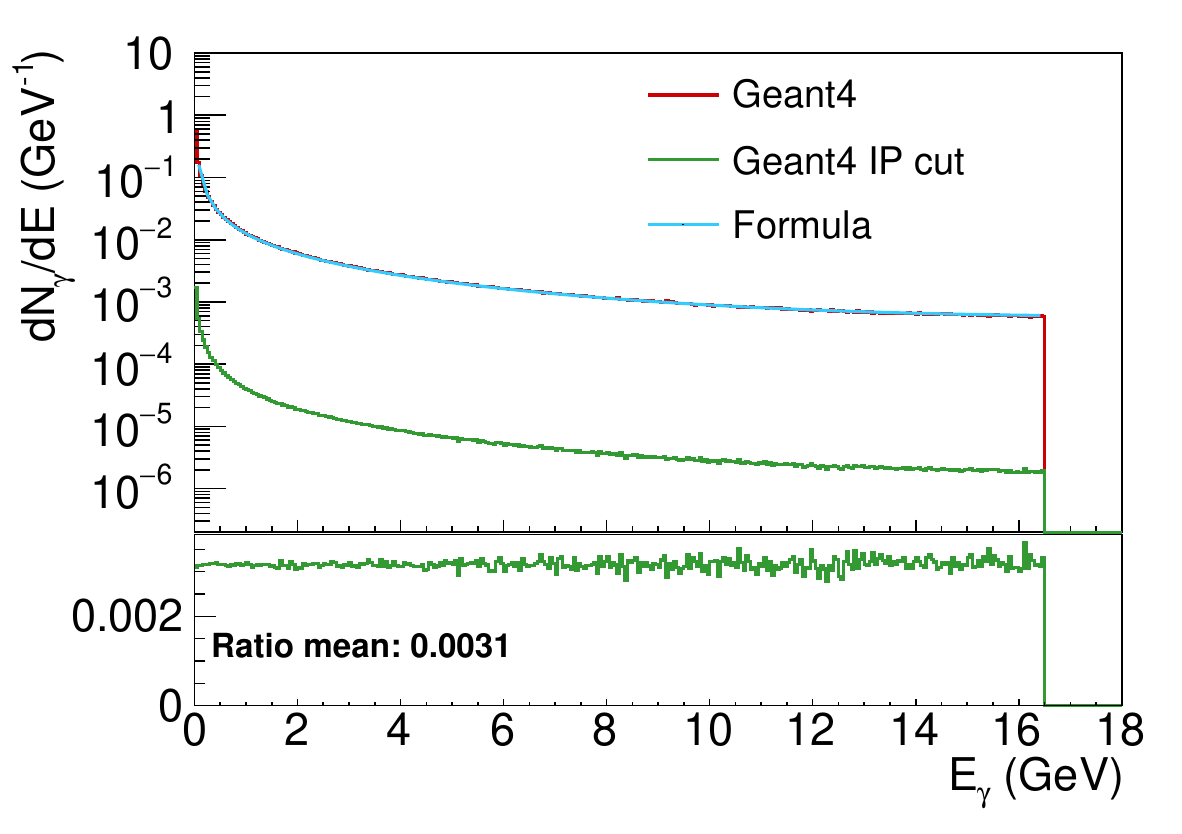}
    \put(-145,120){\makebox(0,0)[tl]{(b)}}
\\
    \includegraphics[width=0.48\textwidth]{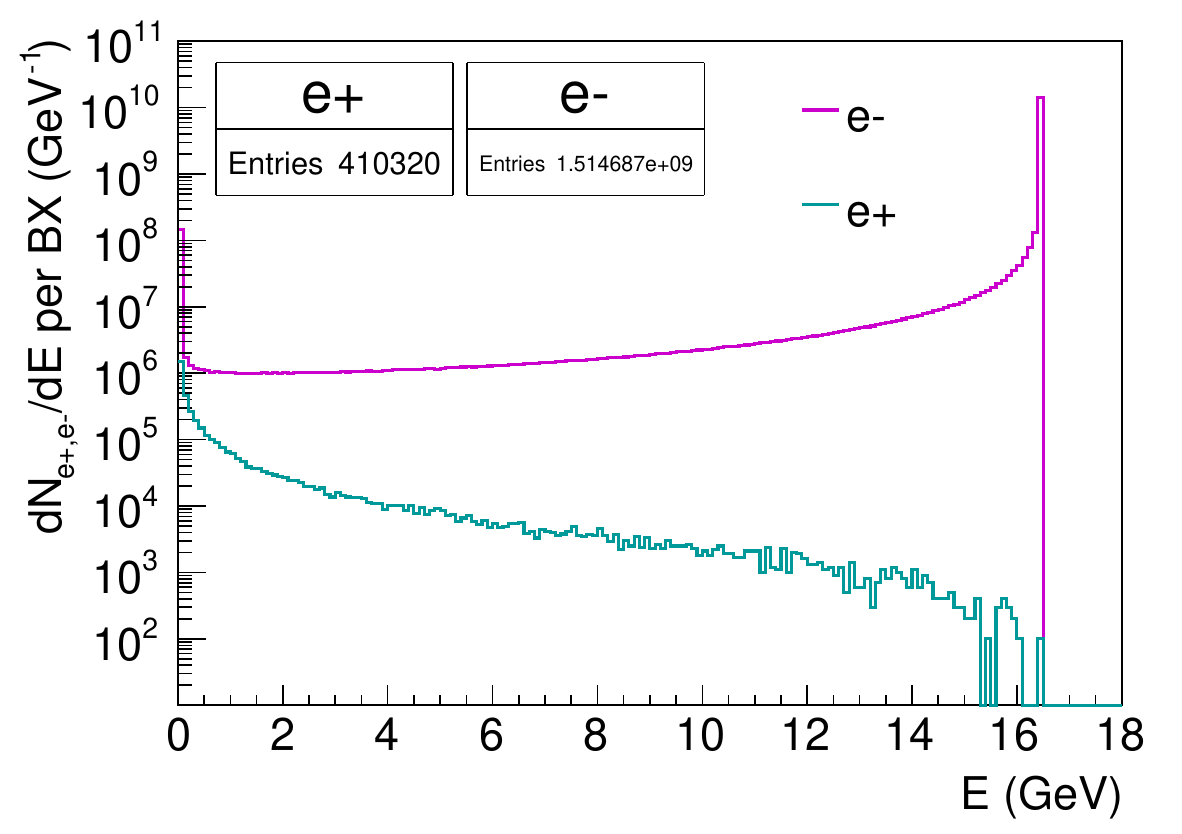}
    \put(-145,105){\makebox(0,0)[tl]{(c)}}
\caption{
(a) Polar angle distribution of bremsstrahlung photons in different energy ranges for one BX normalised by area.
(b) Energy spectrum of bremsstrahlung photons obtained by analytical calculation~(Eq.~(\ref{eq_bremsstrahlung_pdg})) and by \geant simulation. The green line shows the $\gamma$ spectrum after imposing limits on position in the transverse plane to~$\pm$25\,$\mu$m. The bottom plot shows the fraction limited by the interaction area. 
(c) Energy spectra of electrons and positrons produced in the tungsten target. In all cases, a tungsten target with a thickness of 35\,$\mu$m (1\%$X_{0}$) is simulated. 
}
    \label{fig:sim:brem}
\end{figure}

The angular spectrum of the photons is expected to follow $1/\gamma$, i.e.\ about $\sim 30$\,$\mu$rad for $\gamma=16.5~\textrm{GeV}/m_e$. It is independent of the photon energy as seen in Fig.~\ref{fig:sim:brem}(a). As the IP is at a distance of 7.5\,m from the target, the beam size at the IP is about 230\,$\mu$m, much larger than the envisaged laser spot size. 

Figure~\ref{fig:sim:brem}(b) shows the spectrum of bremsstrahlung photons produced in the \geant simulation for a 35\,$\mu$m (1\%$X_{0}$) thick tungsten target. The simulation agrees well with the calculation based on Eq.~(\ref{eq_bremsstrahlung_pdg}). Also shown is the fraction of photons that are within $\pm 25$\,$\mu$m in both the $x$ and $y$ direction at the IP. 
The fraction of bremsstrahlung photons that are within $\pm 25$\,$\mu$m at the IP is 0.3\%. This fraction depends quadratically on the laser spot size and the distance between the target and the IP, e.g.\ for a $5\,\mu$m laser beam spot reduces to 0.01\% (and this is taken into account in the simulations). 

For an electron bunch containing $1.5\cdot 10^9$ electrons, the average number of bremsstrahlung photons produced per bunch crossing (BX) is about \mbox{$1.45 \cdot 10^8$}.
Even though the target is thin there is a finite chance that these photons interact again in the target which gives rise to  production of $e^+e^-$ pairs at a rate of \mbox{$4.1 \cdot 10^5$} per BX. 
The spectra of positrons and electrons and their average numbers for one BX are presented in Fig.~\ref{fig:sim:brem}(c) (same as Fig.~\ref{fig:brem_el}(a), repeated here for convenience).
It is seen that the secondary positron rates are small, particularly in the high-energy region. Based on the measurement of the electron rate the photon energy distribution can be determined, as discussed in Chapters~\ref{chapt6} and~\ref{chapt7}. While it would be useful to measure also the positron rate, it is not critical, and so at present no detectors are foreseen for this purpose.

In summary, with the present configuration, about $5\cdot 10^5$ photons with $E>7$~GeV per BX arrive at the IP within $\pm 25$\,$\mu$m, and $1.5\cdot 10^4$ arrive within $\pm 5$\,$\mu$m. 

\subsubsection{Signal and Background Simulation} 
The technologies of the detectors were chosen based on the signal and background rates determined using this full simulation. A more detailed discussion of the rates was presented in the simulation chapter of the CDR~\cite{luxecdr}. The background arises mostly from stray particles that are produced when primary beam particles hit some material, in particular the electron dump. This type of background will be measured \textit{in situ} using the 9 Hz electron beam bunches that will pass through the experiment without the presence of a laser shot.  

The output of the simulation in the active detector elements is estimated based on the hit properties the simulation estimates and the response of the given detector. This is done separately for each of the detector technologies and discussed in the chapters on the detectors. 

\section{Data Acquisition System} \label{sec:daq}

As the maximum data-taking frequency will be 10\,Hz and all detectors are relatively small, current data acquisition (DAQ) solutions are appropriate for LUXE.  Calibration data will also be needed, e.g.\ when there is an electron bunch but no $e$–laser events or when there are no electron bunches, in order to measure pedestals, noise and backgrounds. When running with electron bunches only, the well-known energy of the electrons and charge of the bunch will allow their use for detector alignment, cross checks of the magnetic field and detector response.  These calibration and alignment data again should not yield large data rates or high data volume.  The DAQ system will need to be bi-directional as control data will need to be sent to the detectors, e.g.\ to distribute timing information, to control motorised detector stages, etc.

\subsection{Data Rates and System Overview}

The \euxfel operates bunch trains at a frequency of 10\,Hz, with each train containing 2700 individual bunches of electrons.  The LUXE experiment will require one of these bunches typically at a rate of 1\,Hz due to the limitations of high-power lasers that can maximally operate at this rate.  Therefore collision data (\elaser or \glaser) will occur typically at 1\,Hz.  Detectors will be read out at 10\,Hz, matching the bunch-train frequency of the \euxfel as calibration data will be acquired as well as collision data.

The data rate for each detector has been estimated based on a 10\,Hz frequency.  The values are shown in Table~\ref{tab:data-rate} for each sub-detector, including the laser diagnostics.  As can be seen, the rates are small for many detector systems, of order $10-100$\,kB/s.  Some systems, however, have expected rates of $O(10\,\mathrm{MB/s})$ and up to $O(100\,\mathrm{MB/s})$ for scintillation screens imaged by cameras.  The rates are relatively high, although it may not be necessary to store all 10\,events per second.  An explanation of the larger numbers is outlined in the following, with more details found in the respective chapter on the sub-detector system.

The scintillation screen system has the largest data rate as follows.   
Assuming a camera used to image a scintillation screen has 9\,Mpixel with 12\,bits, then at 10\,Hz, reading out all pixels gives a rate of 130\,MB/s. This can be trimmed by removing pixels well away from the signal region, reducing the rate to 83\,MB/s.  Two further, lower-resolution cameras image the same screen and as they have 2.3\,Mpixel, contribute a data rate of 23\,MB/s each.  Therefore the sum of the three cameras leads to a data rate of 128\,MB/s.

The gamma ray spectrometer similarly uses a scintillation screen imaged with a camera.  Here a camera with 1\,Mpixel and 16\,bits is expected giving a total data rate of 40\,MB/s for the two stations.

The other large rate comes from the diagnostic systems to characterise the laser.  This is calculated for a frequency of 1\,Hz as data will only be kept when the laser in on.  The value of 20\,MB/s consists of several systems to characterise the laser which typically contribute 2\,MB/s, being megapixel CCD cameras (see Chapter~\ref{chapt3} on the laser for more details).

\begin{table}
    \centering
    \caption{Data rates in megabytes per second by sub-detector.  A frequency of 10\,Hz is assumed, which includes collision as well as calibration data.  More details of the origin of these values is given in the respective sub-detector chapter.}
    \begin{tabular}{l|c|l}
        Sub-detector  & Data rate (MB/s) & Comment \\ \hline
        Scintillation screen & 128 & \\
        Tracker & 10 & Upper bound \\
        Calorimeter & 0.1 & \\
        Cherenkov detector & 0.04 & \\
        Gamma ray spectrometer & 40 & For two screens\\
        Gamma ray profiler & 0.04 & \\
        Gamma flux monitor & 0.01 & \\
        Laser diagnostics & 20 & 1\,Hz rate\\
        \hline
    \end{tabular}
    \label{tab:data-rate}
\end{table}

\begin{figure}[hbt]
\centering
    \includegraphics[width=0.99\textwidth]{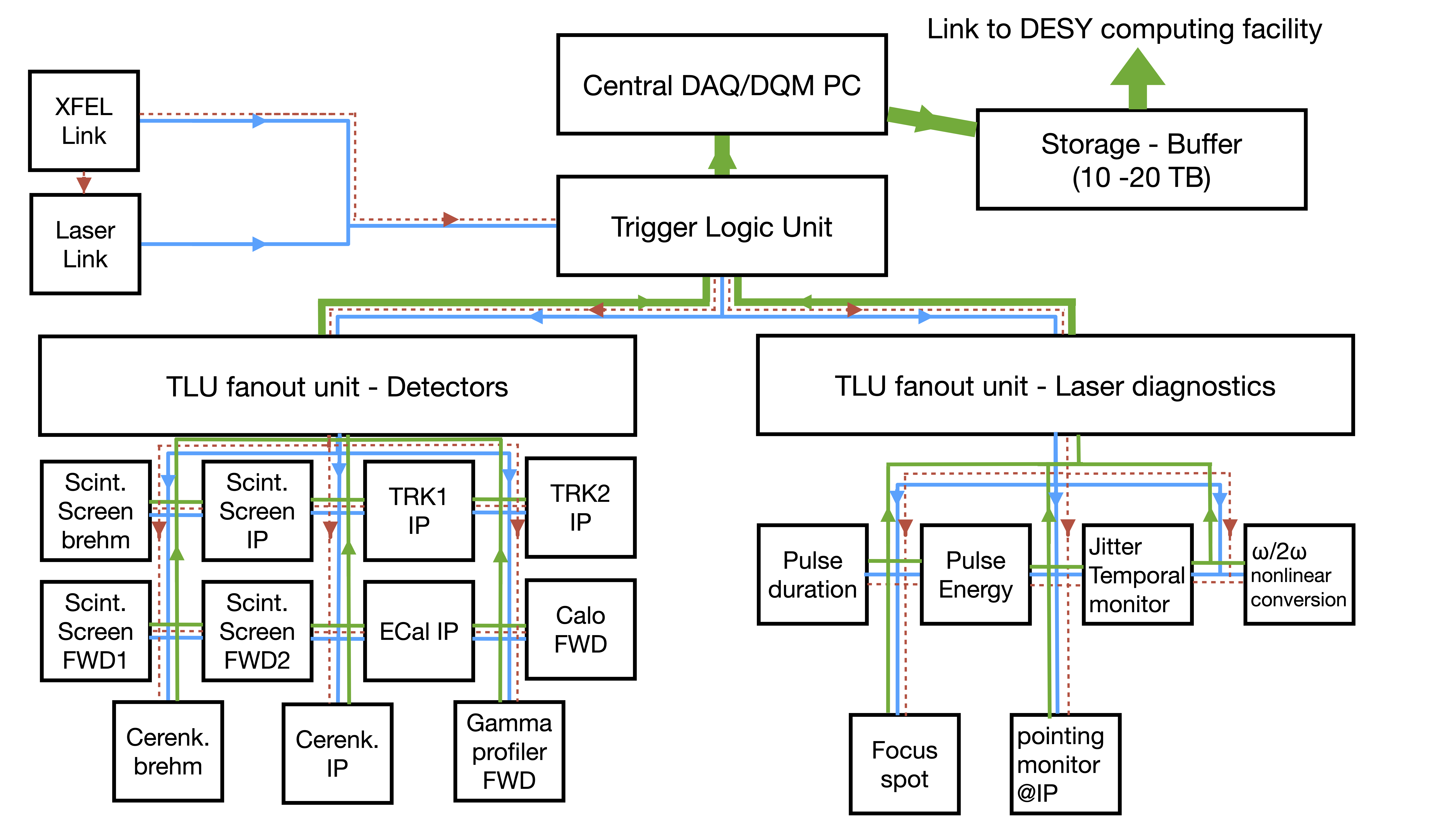}
    \caption{Proposed architecture of the data acquisition.  The central DAQ PC and Trigger Logic Unit (TLU) control the flow of data in the system.  The DAQ PC will also provide DQM checks and links to data storage, primarily used as a buffer, before transfer to the DESY computer centre.  A machine clock is received by the TLU from the \euxfel accelerator and distributed to the detectors (red line).  A clock is also provided to the laser and this sends a trigger (blue line) to the TLU which along with the machine clock confirms that a collision occurs.  Signals between the TLU and the detectors occur through a fanout so as to multiply the number of possible connections.  This fanout is shown here as two separate devices, one for the detector systems and one for the laser diagnostics, but one physical unit may be used.  Data collection is shown as the green line.}
    \label{fig:DAQ-diagram}
\end{figure}

Therefore each detector component, e.g.\ the silicon tracker after the IP or scintillation screen-cameras system, can be read out and controlled by one front-end computer (PC).  Note, more than one PC may be used for contingency, although the actual data rates would not require this. Each of the detector PCs will then be connected to a central DAQ PC, based in the LUXE control room, as well as a trigger and control system. Further PCs will be required in the control room to display detector information for e.g.\ data quality (see Section~\ref{sec:dqm}) and slow control monitoring (see Section~\ref{sec:sc}) purposes. Given these moderate constraints from the data rate, the proposed architecture of DAQ is shown in Fig.~\ref{fig:DAQ-diagram}.

Some detectors will require calibration data that is not synchronous with the \euxfel machine clock, such as cosmic-ray data.  The, e.g., calorimeter requires such data and this will require a trigger from a scintillator placed behind the calorimeter.  Such a configuration is used as standard in beam tests and so compatible with the trigger logic unit (TLU), see next section.  A separate TLU dedicated for such calibration may be used rather than the central DAQ TLU.

\subsection{Control System}

The Trigger Logic Unit (TLU)~\cite{Baesso:2019smg,TLU:www} developed most recently within the EU AIDA-2020 project has been designed to be a flexible and easily configurable unit to provide trigger and control signals to devices employed during test beams and integrating them with pixel detectors.  It has been used extensively at DESY, as well as CERN, beam lines by a number of different detectors.  

The unit can accept signals from some detector and can generate a global trigger for all LUXE detectors to indicate the start and end of data taking.  The unit can also act as a master clock unit, receiving synchronisation and trigger commands from the software DAQ, as well as a precise clock reference. Therefore the machine clock from the electron bunch  will be fed to the TLU, along with a trigger from the laser when in operation, which will then synchronise the LUXE detectors.  The TLU has been used previously to provide the clock and synchronisation for the ProtoDUNE-SP tests at CERN~\cite{CUSSANS2020162143}. 

The TLU can also mix time-stamping and triggering as it records both.  The detectors can also send signals back to the TLU to indicate that they are busy and, for example, request data taking to be paused until the busy is removed.  

Communication between the DAQ system and the TLU uses the IPBus protocol, a well-established and reliable protocol widely used in the CMS and ATLAS experiments at CERN.  The TLU is also integrated with the EUDAQ2 software (see Section~\ref{sec:eudaq}).

\begin{figure}[hbt]
\centering
    \includegraphics[width=0.8\textwidth]{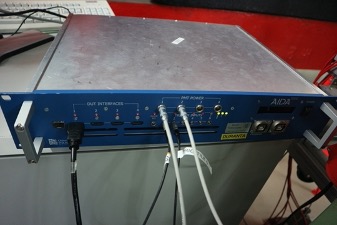}
    \caption{Photograph of the TLU (Credit: D. Cussans).}
    \label{fig:tlu}
\end{figure}

The production version of the TLU is shown in Fig.~\ref{fig:tlu}.  This is available in a small desktop case or in a rack-mount case (19-inch rack mount 2U units).  The unit has 6 trigger inputs and 4 device under test connections.  Active fanout units are available and have already been used in other TLU deployment~\cite{CUSSANS2020162143}; they are capable of providing 8 outputs, each of which can be split 8 ways with an optical splitter.  The TLU can provide a clock with a jitter of $\sim 10 - 100$\,ps.

Recently a production of 30 TLUs was made by DESY and shipped to users.  LUXE currently has access to TLUs for tests in DESY but will need to order a batch to cater for use at individual labs as well as for the final experiment.  An estimated 15 TLUs will be needed to cover all lab tests,  experimental needs and spares.

\subsection{DAQ Software}
\label{sec:eudaq}

A DAQ software will be required which could be used for individual components or, if they come with their own software, a central DAQ software will be needed to interface to the detector software.  Many different DAQ softwares exist with different levels of complexity and scale, with some designed for specific experiments and others as generic developments. At LUXE, we propose to use the EUDAQ2~\cite{EUDAQ2:www,Liu:2019wim} software which has been developed for tests of high energy physics prototype detectors (in beam tests), a similar setup to LUXE.  It can cope with different triggering and readout schemes and has been used by several different detectors and projects.  The software can be used on Unix, Windows and Mac, although Unix is generally the operating system of choice of the various users.

\begin{figure}[hbt]
\centering
    \includegraphics[width=0.99\textwidth]{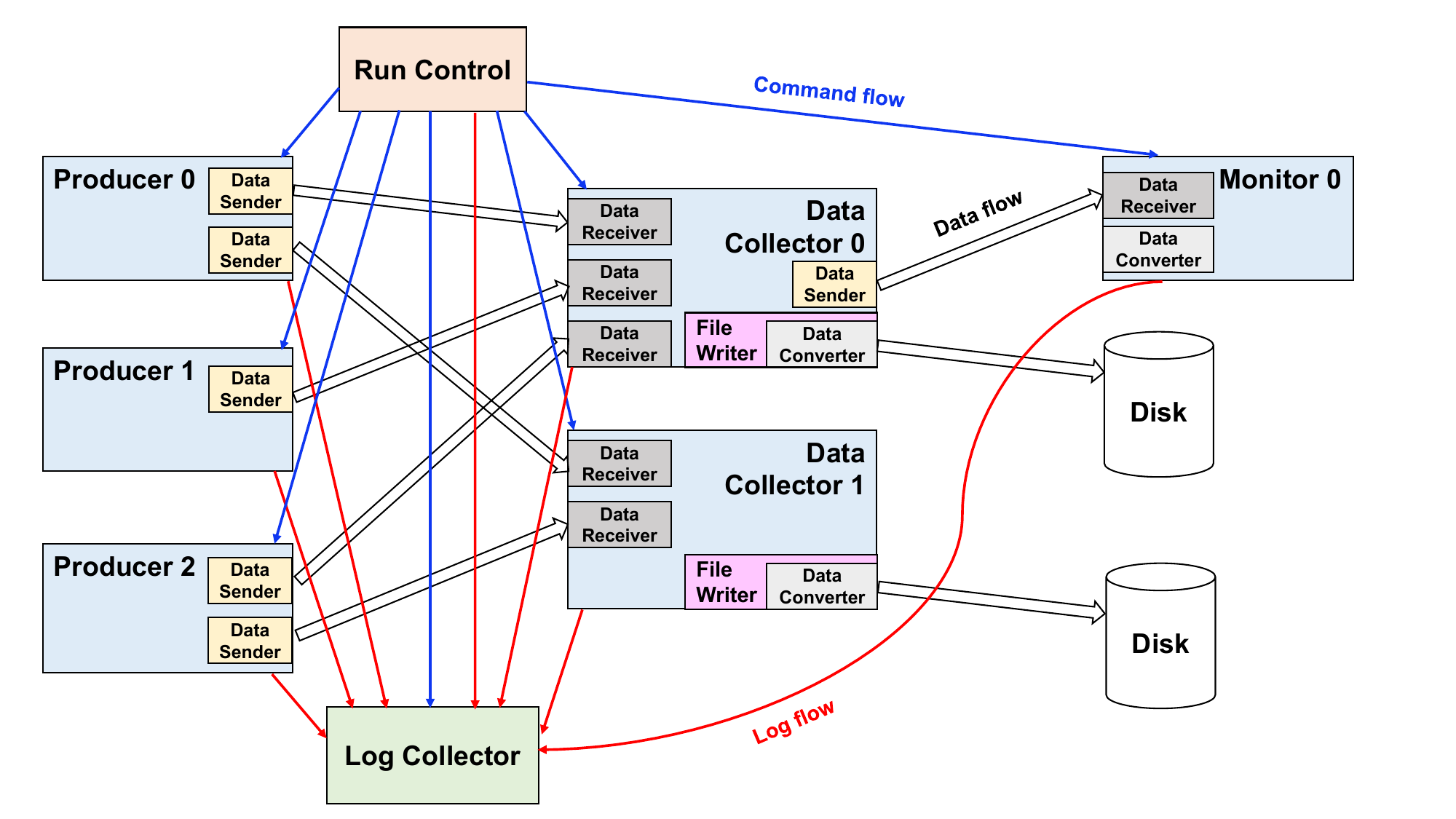}
    \caption{A schematic view of the EUDAQ2 architecture.  From~\cite{Liu:2019wim}.}
    \label{fig:EUDAQ}
\end{figure}

An overview of the EUDAQ2 software is shown in Fig.~\ref{fig:EUDAQ}. One central instance of the Run Control steers all other components on the network. Each physical device, e.g.\ detector, is implemented as a producer; the delivery of a producer is the main action required of each detector responsible to interface to EUDAQ2.  A finite state machine defines the behaviour, such as starting and stopping a run in which the state is changed from e.g.\ configured to running or running to idle.  Data can be sorted by trigger ID or event number or can be written directly to data storage.  A central instance collects log messages, errors, etc.

The DQM within EUDAQ2 is rather rudimentary but could be adapted and 
extended for use at LUXE.  Again, there are no real specific challenges with the LUXE DQM needs (see Section~\ref{sec:dqm}) and so simple solutions should be sought.

EUDAQ2's flexibility and applicability to many different types of detectors has been demonstrated~\cite{Liu:2019wim}.  
Specifically, EUDAQ2 has been used in beam tests to read out several detectors like those to be used in LUXE, such as calorimeters and pixel detectors~\cite{Liu:2019wim}, including the ALPIDE sensors.  EUDAQ2 is also fully compatible with the TLU and the two have been used as a basis for DAQ systems in several situations.  The software has been used extensively in DESY test beams, as well as other facilities.  DESY is also the lead developer of the EUDAQ2 software and so we will have access to in-house expertise.

\subsection{System Status and Availability}

The hardware, firmware and software designs of the TLU~\cite{TLU:www} and the EUDAQ2 software~\cite{EUDAQ2:www}
are freely available.  These are both accompanied by a manual as well as documentation in the published papers~\cite{Baesso:2019smg,Liu:2019wim}.  Work can therefore start quickly on the implementation of the TLU and EUDAQ2 in LUXE.

\section{Slow Control} \label{sec:sc}

The LUXE experiment will be composed of a laser system, multiple laser diagnostics, a variety of vacuum chambers and vacuum pipes to propagate the laser and the electron beams, and about 10 detector systems. Each of these systems will need to be powered, controlled, read out and monitored. And this will have to be done while being able to cross-correlate the information between them but also the information coming from the \euxfel accelerator.

Contrary to the data stored to disk using EUDAQ2, as explained in the previous section, the slow control data, or control system data, corresponds to all the information deemed of interest to be saved for long term usage, but that will not be synchronised with the electron or laser pulses. These data will typically be saved at a maximum of a few Hz and contains essential information about the data-taking environment. While not synchronised, these data could still be needed for calibration or correction purposes if drifts of some physical quantities, linked to the laser position for instance, are detected after the data-taking has taken place.
 
For all these reasons, the LUXE control system will be developed with the idea to fit in the \euxfel ecosystem, and will rely on the already well established Distributed  Object  Oriented  Control  System (DOOCS) environment~\cite{DOOCS}.
This will allow the retrieval and easy integration of the information from the \euxfel accelerator, while publishing in a coherent way the information of interest for LUXE. DOOCS has several APIs allowing the development of control and readout software in a variety of programming languages (C++, Java, Matlab, Python), while being a cross operating platform.

Given the widespread adoption of DOOCS in the different DESY projects, the collaboration will profit from a large community support but also from the pre-integration of multiple hardware systems (e.g.\ power supplies, digitisers, etc.), that will allow time to be saved in the development of the applications needed to run the experiment.

\begin{figure}[hbt]
\centering
    \includegraphics[width=0.99\textwidth]{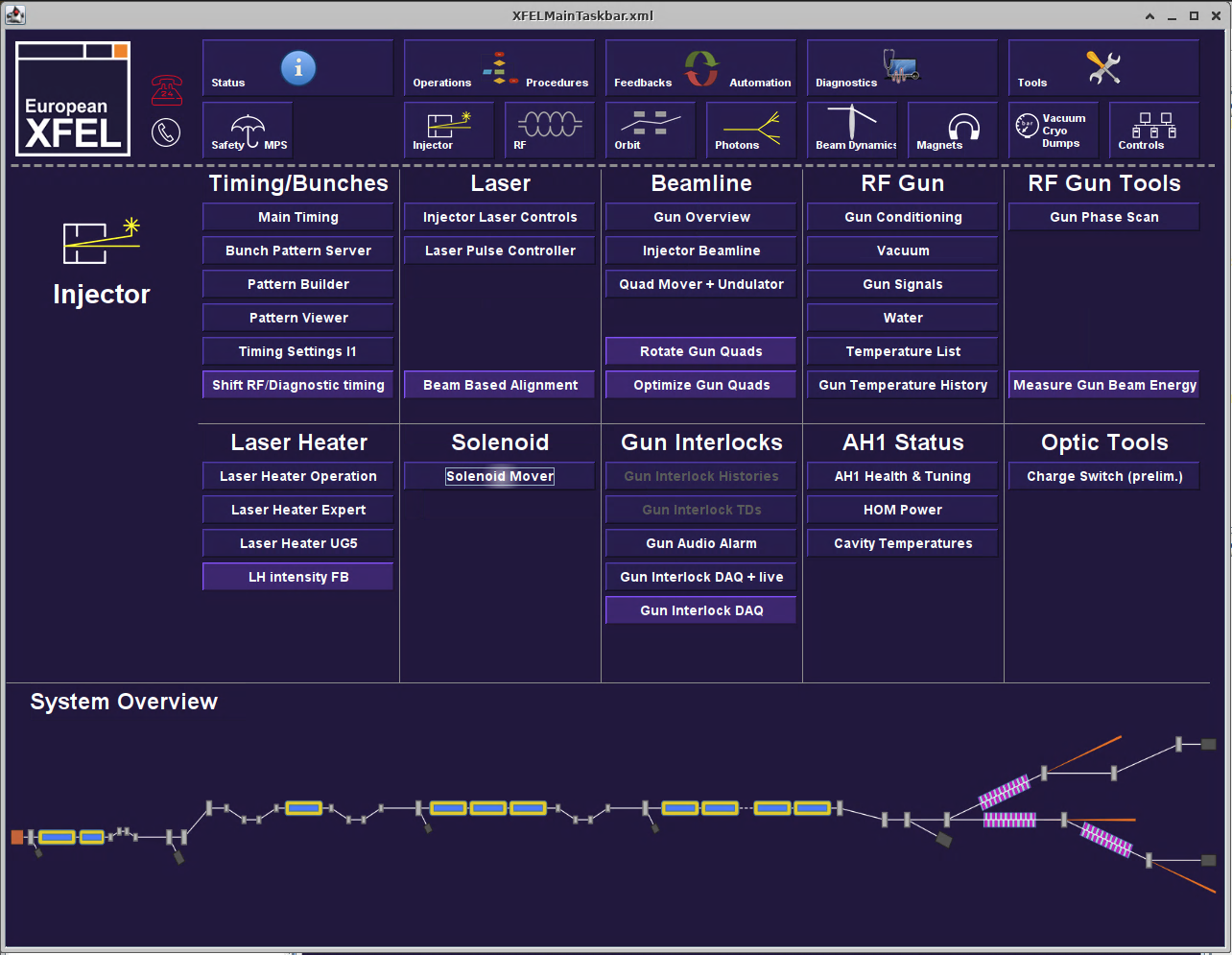}
    \caption{Main \euxfel control panel developed in the JDDD GUI, showing the possibility of the DOOCS environment.}
    \label{fig:JDDD-XFEL-panel}
\end{figure}

Finally, DOOCS also comes with a default Graphical User Interface named JAVA DOOCS DATA DISPLAY (JDDD)~\cite{Sombrowski:82503}, which allows easy access to the data collected, worldwide. As an example, the main monitoring control panel of the \euxfel is shown in Fig.~\ref{fig:JDDD-XFEL-panel}. It is also planned to develop such an interface in the future to display the control panels and data of the experiment.

\section{Data Quality Monitoring and Calibration} \label{sec:dqm}

It is critical that the quality of the online data is controlled during data taking to detect e.g.\ incorrect settings of a subdetector or miscalibrations etc. If major issues are found online, the goal is to correct them so that the following data do not suffer from that problem. 

Furthermore, the data quality needs to be assessed offline after the final offline calibration has taken place.

\subsection{Online Data Quality Monitoring}
The EUDAQ2 system allows for simple histograms to be made while the data are collected. This feature will be used to perform sets of control histograms, per detector component, that can alert the shift crew if something goes amiss. Two sets of histograms will be generated, for background (laser-off) and for laser-on data. Typically, these control histograms can be compared to a set of default histograms, previously selected by experts, to quickly identify any arising problem. 

A list of control histograms presently considered by the component experts is given in the appendix. These histograms are built from raw data. Whether simple reconstruction based on raw data will be possible remains to be investigated.

\subsection{Offline Data Quality Monitoring}
Offline monitoring will be performed on reconstructed data and the main role will be to check the consistency and robustness of the recorded data, including laser information. A predefined set of plots will be generated for inspection by expert crews. This procedure will certainly be used frequently during commissioning. It may be expected that during regular data taking, it will be a daily process.

It is also important to have methods of calibrating the detectors \textit{in situ}. For this purpose it is envisaged to install one or more thin tungsten foils in the interaction chamber which can be moved into the beamline, so that the primary beam interacts in this foil instead of with the laser. For the electron beam this will result in bremsstrahlung events and for the gamma beam in $e^+e^-$ events. The electrons (photons) in bremsstrahlung events can be used to calibrate the high-rate electron (photon) detectors. The electrons and positrons in $\gamma$-laser events can be used to align and calibrate the tracking detector and the calorimeters. Such runs can be taken regularly as needed. 

\section{Computing} \label{sec:computing}

This section describes the plans for the computing aspects of the LUXE experiment. These include: the choice of a data model for the persistence of the data, an estimate of the computing resources required for the processing and storage of the data, and the initial survey of the software requirements for the experiment.

\subsection{Storage and Computing Infrastructure}

The data associated with a single bunch crossing is called an “event”. Raw data will be delivered by the detector for reconstruction, and will consist of a serialisation of detector readouts and metadata that will be stored in a byte stream format. Full-detector events are stored at a rate of 10\,Hz (9\,Hz of this are dedicated to detector calibration and \textit{in-situ} background estimation and will not make use of the camera systems), with optional data streams dedicated to specific sub-detector calibration possibly stored at a higher rate. The estimated raw event size varies strongly with the laser intensity but is not expected to exceed 50\,MB (1\,MB for the events with empty BX). With the expected duty cycle of the LUXE detector and the \euxfel, LUXE anticipates recording between 400 and 500\,TB of raw data per year.

The expected reconstructed events size can go up to 40\,MB per event, the breakdown of which is dominated by the camera images and the number of reconstructed positron candidates (i.e.\ tracks matched with calorimetric showers). The possibility of producing lightweight samples from the reconstructed data aimed at specific physics studies is foreseen (e.g.\ samples dedicated to the secondary production of BSM particles).

The simulated data is expected to use the same data formats, with a digitisation step replicating the raw data coming from the detectors. Additional information, such as the event description at generator level will be retained in the subsequent steps. The size of this additional information also depends strongly on the laser intensity. If no reduction scheme is applied to the truth record, this can go up to about 500\,MB per event in $e$-laser interactions during LUXE phase 1. Work is required to define and implement a reduction scheme for the truth information. In the following, a minimal reduction of the truth record to a size of 30\,MB per event is assumed.

Additional information regarding the detector geometry, operation conditions, data quality, and other database needs is not expected to require significant storage resources, especially in comparison to the LHC experiments. It is currently foreseen to use a combination of versioned plain text files (potentially in JSON format) and dedicated ROOT-based databases to store this information, with more powerful SQL-like databases still being evaluated as a backup option.

The \euxfel beam condition data, available via the DOOCS system, is expected to be small (of the order of 4\,kB per event). The uncompressed laser diagnostics data is expected to dominate the raw file size (20\,MB per event).

\subsubsection*{Storage}
The expected needs for the storage of data and MC simulated events is summarised in Table~\ref{tab:computing_storage}, assuming no compression for the data from the cameras and laser diagnostics.

These needs are computed assuming $10^7$\,s of data-taking per year. The raw data is expected to be stored on backed-up tape resources in the DESY computing centre. 
The data is assumed to be processed through reconstruction about twice per year, with the two most recent versions to be kept on disk for analysis. Furthermore, we assume to store on disk at least as many simulated signal events as those collected during the data taking.

In order to reduce the risk of data loss, at least two replicas of the data will be stored at different institutes. The use of grid resources for the management and replication of the data via either the RUCIO or DIRAC backends is currently being explored.

\begin{table}[h]
    \centering
    \caption{Size estimates for the various data formats per year of operation, assuming robust disk- or tape-based storage and the maximum event size for data events at a single site.}
    \def\arraystretch{1.2}
    \begin{tabular}{|l|c|}
        \hline
        Type of data                   & Data Size \\
        \hline
        Raw data (filled BX)           & 0.46 PB \\
        Raw data (empty BX)            & 0.01 PB \\
        Reconstructed data & 0.40 PB \\
        \hline
        Simulated data (1:1 filled BX) & 0.54 PB \\
        \hline
        Total                          & 1.42 PB \\
        \hline
    \end{tabular}
    \label{tab:computing_storage}
\end{table}

\subsubsection*{Processing}
Although a combined reconstruction software is not available at the moment, it is expected that the average per-event processing time will be of the order of one minute. For example, the track reconstruction, which is expected to be the leading contribution to data reconstruction time, of an event with 40000 tracks takes on average 130\,s to be completed. Future optimisations are expected to decrease this time further. Assuming an average reconstruction time of one minute per event, approximately $10^{6}$ CPU-hours per year would be required to reconstruct all the data from the detector (twice) and the simulated bunch crossings.

The simulation of a signal bunch crossing can take up to $7 \cdot 10^{6}$\,s. While the simulation is significantly more CPU expensive than reconstructing the events, it is expected to generate only a moderate sample of events, limiting the overall processing needs. 
It is expected that above 90\% of the data-taking time will be spent at low values of the field intensity (i.e.\ $\xi < 2$). In this scenario, the signal positron multiplicity is expected to be negligible compared to the beam-induced backgrounds, electron and photon rates. The CPU budget will then be dominated by the simulation of samples which will be used to model the high particle multiplicity regions of the experiment. A budget of $10^{6}$ CPU-hours is foreseen for the simulation of these samples. The simulation of the signal events assuming high-intensity conditions (i.e.\ $\xi > 2$) is expected to require $10^{6}$ CPU-hours, to satisfy a global target of simulating at least as many bunch crossings as those recorded by the detector. Samples dedicated to specific sub-detectors are foreseen, but are expected to require a negligible amount of resources.
An additional budget of $2 \cdot 10^{6}$ CPU-hours is also foreseen as contingency and to account for the digitisation of the detector inputs.

The overall CPU requirements are considered to be modest (about 5\% of the total yearly CPU time provided by the DESY NAF) and it is expected to be able to satisfy the computing requirements of the collaboration using exclusively opportunistic resources at DESY and at other involved institutes.

\subsection{Software}

Event generation, simulation, reconstruction, and analysis are currently implemented in standalone programs.  Each of the LUXE detectors will provide the dedicated software to reconstruct the collected data, or digitise the simulated hits from the \geant simulation. Depending on the detector, the reconstructed data can go from information with a per-particle granularity (e.g.\ tracks and calorimeter clusters) to spectra (e.g.\ from the Cherenkov detector).

A shared event reconstruction software will be needed to correlate the measurements made in the tracker and calorimeter systems. No other detectors are expected to require a common reconstruction software at the time of writing. 

A common software framework with the ability to steer the reconstruction of the data from every LUXE detector, as well as to manage the Monte Carlo programs dedicated to the event generation, simulation and digitisation, is desired. Although not strictly required, such frameworks ease the production of new simulated data for specific studies, as well as allow for a centralised management of the data reconstruction campaigns. 
In order to maximise the accessibility of the data, as well as wider support from the particle physics and science community, the event data model (EDM) adopted for the reconstructed and simulated data should support access via ROOT and, possibly, other modern data science tools. 
Given the limited resources available for computing developments, the development of new non-LUXE-specific software should be kept to the minimum.
A survey of the available open source solutions has been performed, with EDM4hep and the Key4HEP software stack arising as possible candidates. This software stack, or the closely related ILCsoft, would provide a thoroughly tested and validated baseline to handle the data reconstruction and analysis.

\subsubsection*{Track Reconstruction with ACTS}
One notable example of use of open source software is the planned adoption, in the medium term, of the ACTS (“A Common Tracking System”) tracking suite. This change will allow the  leveraging of a fully-featured, open-source reconstruction framework based on the ATLAS tracking software and the deployment of more sophisticated pattern recognition, extrapolation, and track-fitting techniques.
ACTS is a collaboratively developed and maintained track reconstruction software, with substantial commitment of effort from external collaborations such as ATLAS or FASER. A preliminary integration of ACTS within the ILCsoft framework is publicly available\footnote{At \url{https://github.com/MuonColliderSoft/ACTSTracking}} and could be adopted to integrate this step into the LUXE software stack.

\section{Summary and Outlook} \label{sec:further-tests}

The LUXE simulation framework consists of a physics generator where the final-state particles are passed through a \geant simulation of the experimental set-up.  Both of these will continue to be refined as more strong-field QED processes become available and the experimental layout is updated.  Better simulation of the detector signals will also be worked out in collaboration with the individual detector groups.

The DAQ hardware (TLU) and software (EUDAQ2) have been defined.  They will undergo continual integration and testing, both in home laboratories with individual detectors and in beam tests involving single or multiple detectors.  This will reduce the risk for the final integration before data taking.  This will involve production of a new batch of TLUs for use in laboratories, which is currently being planned.  A crucial extension of the EUDAQ2 software is the DQM functionality and this will be an initial focus of work.

The slow control software (DOOCS) has also been defined.  Important tasks for both the DQM and slow control are to define the quantities to be recorded and implement these in the relevant software.  This work can also be carried out during laboratory tests and beam tests of one or more detectors.  Again, the development is to ensure that final integration has reduced risk.

Current computing solutions can cope with the data rates, volume and processing needs.  However, work will continue to optimise these such as reduction in data size for the scintillation screens.  Although several software frameworks should be able to meet the needs of LUXE, the framework to be used needs further assessment and then chosen, as again current solutions should be sufficient without the need for a bespoke software.

\section{Appendix: List of Online DQM Histograms}
\begin{table}[!h]
    \centering
     \caption{Examples of control plots for online data quality monitoring, assuming that only simple manipulations on the data are  possible within EUDAQ.}
    \label{tab:my_label}
    \begin{tabular}{l|l}
  Tracker       &  \\
  \hline
         &  2D chip-wise hit occupancy \\
& 1D chip-wise number of hits \\
& 1D chip-wise number of clusters \\
& 1D chip-wise cluster size \\
& 1D average cluster size \\
& 1D number of track seeds \\
\hline
 ECAL       &  \\
  \hline
& 1D distribution of pads number frequency \\
& 1D projection of pads signal on the $x$-plane\\
& 1D individual plane/towers on the edge to check for excess background\\
\hline
Scintillating Screen & \\
\hline
&1D $x$ histogram, up to 4000 bins \\
&1D $y$ histogram, up to 2000 bins \\
\hline
Cherenkov & \\
\hline
&1D distribution of the charge ADC counts per channel number\\
& Projective occupancy plot \\
& Event timing information and trigger rate \\
\hline
Gamma spectrometer & \\ \hline
&Camera images of the two scintillators at either end of the spectrometer \\
\hline
Gamma beam profiler & \\ \hline
&1D $x$ charge distribution per BX \\
&1D $y$ charge distribution per BX \\
&2D Fitted amplitude in $y$ vs $x$, one data point for each BX \\
&2D Fitted mean in $y$ vs $x$, idem \\
&2D Fitted $\sigma$ in $y$ vs $x$, idem\\
\hline
Gamma flux monitor &\\ \hline
&Deposited energy in each of the eight channels\\
\hline
    \end{tabular}
\end{table}

\section*{Acknowledgements}
\addcontentsline{toc}{section}{Acknowledgements}

We thank the following for useful discussions: D. Cussans, L. Huth and M. Stanitzki.
Author MW acknowledges DESY, Hamburg for their support and hospitality.
\printbibliography
\end{refsection}


\chapter{Technical Infrastructure}
\label{chapt12}
\begin{refsection}

\chapterauthors{L.~Helary \\\desyaffil}{S.~Boogert \\ University of Manchester, Manchester (UK), and Cockcroft Institute, Daresbury (UK)}
{R.~Jacobs \\\desyaffil}


\begin{chaptabstract}
This chapter summarises the technical infrastructure of LUXE, from the electron extraction and  transport beam line, to the laser clean room, the interaction chamber, the detectors and the dumps. It also outlines the main aspects and steps of the experiment installation that is foreseen for LUXE, and of the project organisation.
\end{chaptabstract}



\section{Introduction}
LUXE, as outlined in Chapter 2, is an experiment aiming at the investigation of  Strong Field Quantum Electrodynamics (SFQED) by colliding the high-quality electron beam from the accelerator of the European X-ray Free-Electron Laser (European XFEL, denoted \euxfel in the following) with a state-of-the-art multi-terawatt titanium–sapphire laser pulse.





It is planned to install the LUXE~\cite{luxecdr} experiment in the XS1 annex which is an appendix of the XS1 access shaft, located in Osdorfer Born, that has already been foreseen for a possible extension of the distribution fan (termed "2nd fan") of the \euxfel~\cite{XFELTDR}. The annex is about 35\,m long, 5.4\,m wide and 5\,m high. XS1 is located at the end of the main Linear Accelerator (LINAC) of the \euxfel which is installed in the XTL tunnel (see Fig.~\ref{fig:XFEL_sketch}.). 

\begin{figure}[htp]
   \centering
   \includegraphics*[height=4.2cm]{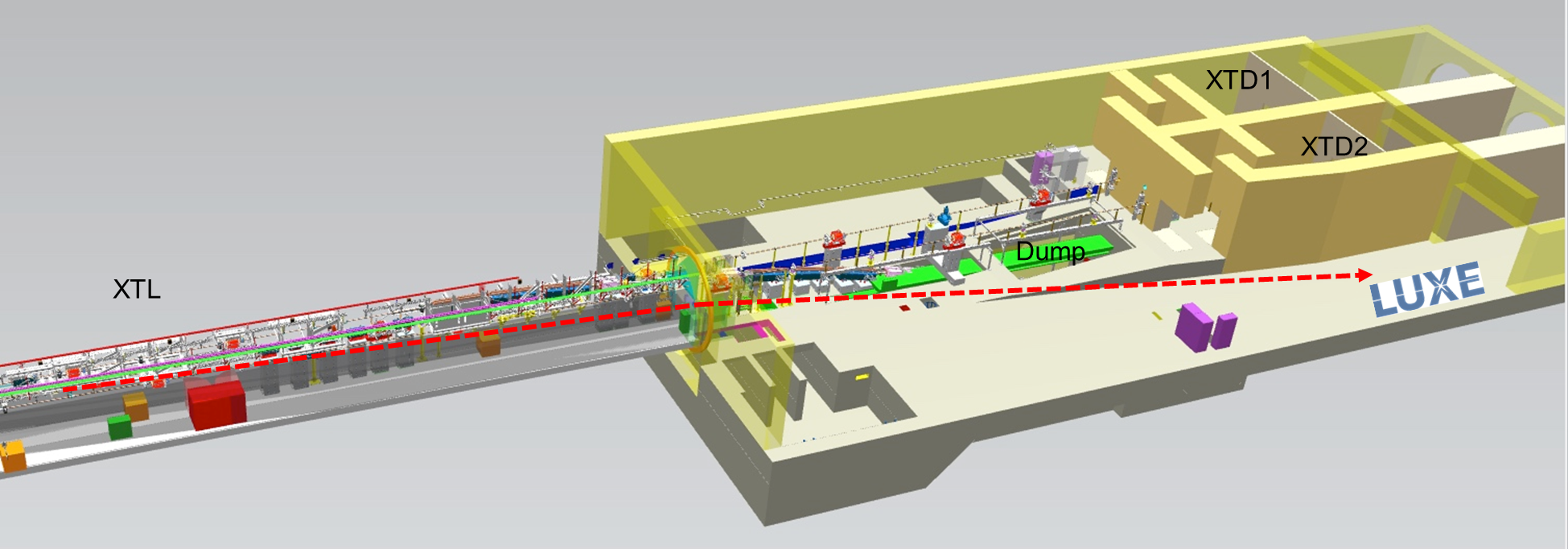}
   \caption{CAD model of the end of the \euxfel accelerator tunnel and the shaft building with the two existing
beamlines XT1 and XT2 to the undulators (SASE1 and SASE2) and the XS1 annex, where the LUXE experiment can be installed. The beam extraction and the beam line towards the experiment is sketched with the dashed line. }
   \label{fig:LUXE_cad}
\end{figure}

The beam path to the XS1 annex, required for LUXE, and the XTD1 and XTD2 tunnels are also schematically shown in Fig.~\ref{fig:LUXE_cad}. About 40\,m of installations in the XTL tunnel are needed to kick out a bunch and guide it towards the XS1 annex in a beamline portion denoted XTD20. Inside the XS1 shaft building additional components are needed to guide and focus the beam to the experiment. The length of this installation is about 50\,m. A more detailed description of the new required extraction line is given in Ref.~\cite{beamlinecdr}.

\begin{figure}[htp]
   \centering
   \includegraphics*[height=8.4cm]{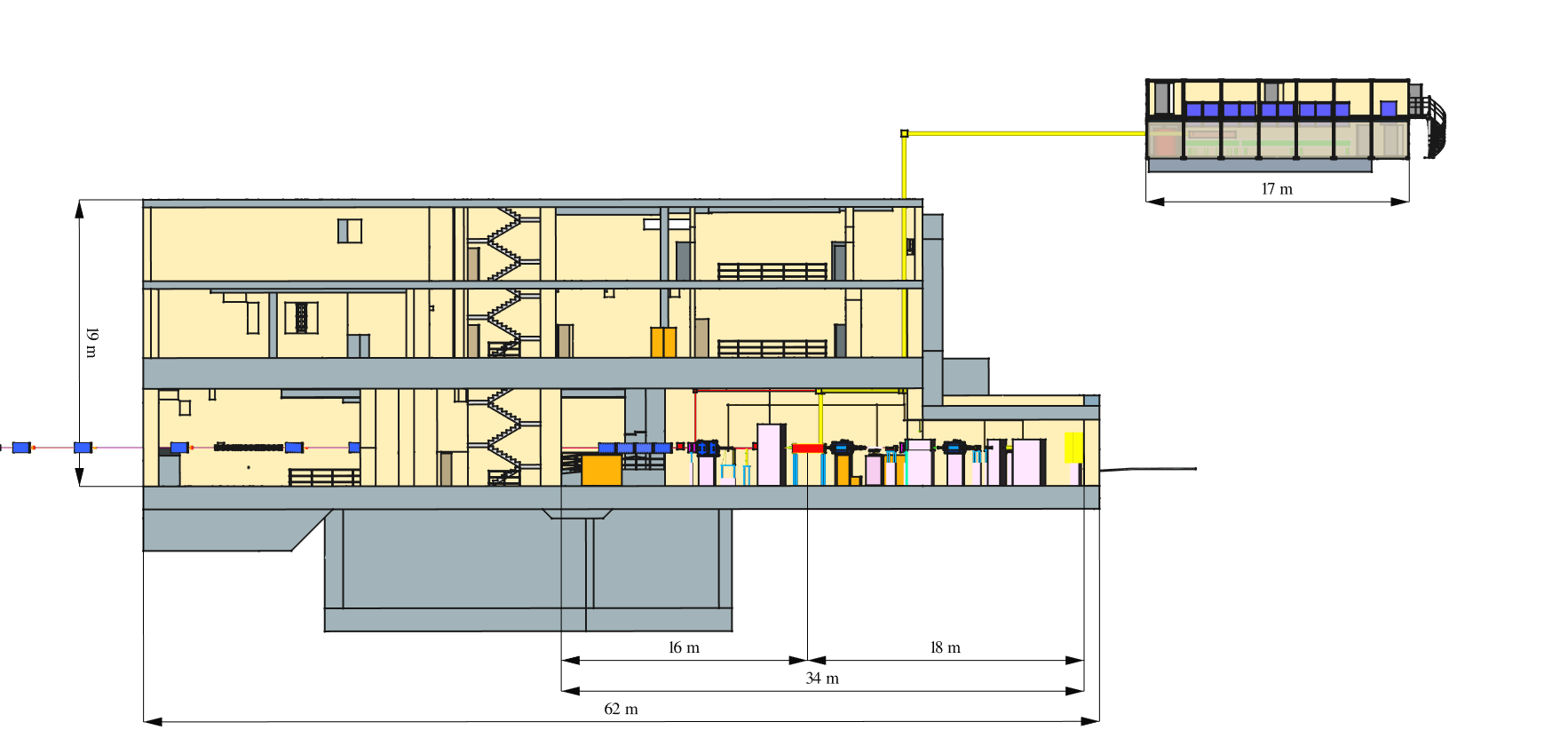}
   \caption{Side-view CAD drawing of the XS1 access shaft building, including the LUXE experiment. The laser clean room and the control room will be placed in a temporary building above ground. The XS1 building is shown in beige and grey, while the different experimental elements are shown coloured. The laser and electron beam pipes are shown in red, the supporting elements and shielding elements in concrete are shown in pink, the supporting structure are shown in light blue, the magnets are shown in dark blue, the different active components (interaction point chambers (IP), and detectors) are shown in yellow, and the services are shown in orange.
   \label{fig:LUXE_CAD_Building}}
\end{figure}

Figure~\ref{fig:LUXE_CAD_Building} shows a 2D view of a 3D CAD model of the XS1 shaft building and various components. The conversion of the experimental components from the \geant simulation described in Chapter~\ref{chapt11} to the XS1 CAD model was done using the software described in Ref.~\cite{keith_sloan_2020_4008390}. The experimental area where the electron-laser or photon-laser collisions will happen is located in underground level --3 (UG03), with all the other \euxfel beamlines passing through XS1. This area is not accessible during \euxfel operations due to radiation protection regulations. 
The laser and most services will be located in a new temporary building structure located in the surface, to allow access at any time, in particular also during \euxfel operations.

The installation of the elements that are located in UG03 have to be done while the accelerator is not running. While a 6 months shutdown of the facility is anticipated to take place in the coming years, in order to avoid relying on any specific down-time of the facility, the preliminary installation schedule was shown to be possible in the regular running cycle of the \euxfel, so assuming a 2 weeks shutdown during the summer and a 6 weeks shutdown during the winter. The other elements, which will be mostly located on the surface can be installed in parallel to the accelerator regular operations.

In this chapter, the plan for the construction, installation and operation of the LUXE experiment will be presented. In all cases the procedure is designed to minimise as much as possible any impact on the \euxfel operations.


\section{Experimental Setup}

\subsection{Electron Beamline}
 \subsubsection{System Overview}

The electron beam line is required to safely transport the 16.5\,GeV \euxfel beam to the LUXE target and focus the electron bunch to a two-dimensional Gaussian shape of variance ($\sigma$) parameter about $10\; \mu {\textrm m^{2}}$ suitable for the physics goals of LUXE. In addition to obtaining a suitable beam for LUXE experiment, the charged particle background, position and angle jitter need to be understood. The beam also needs to be safely transported to either electron beam dump, for e-laser and $\gamma$-laser modes.

\subsubsection{Beam Optics}

\begin{figure}[ht]
   \centering
   \includegraphics*[height=8cm]{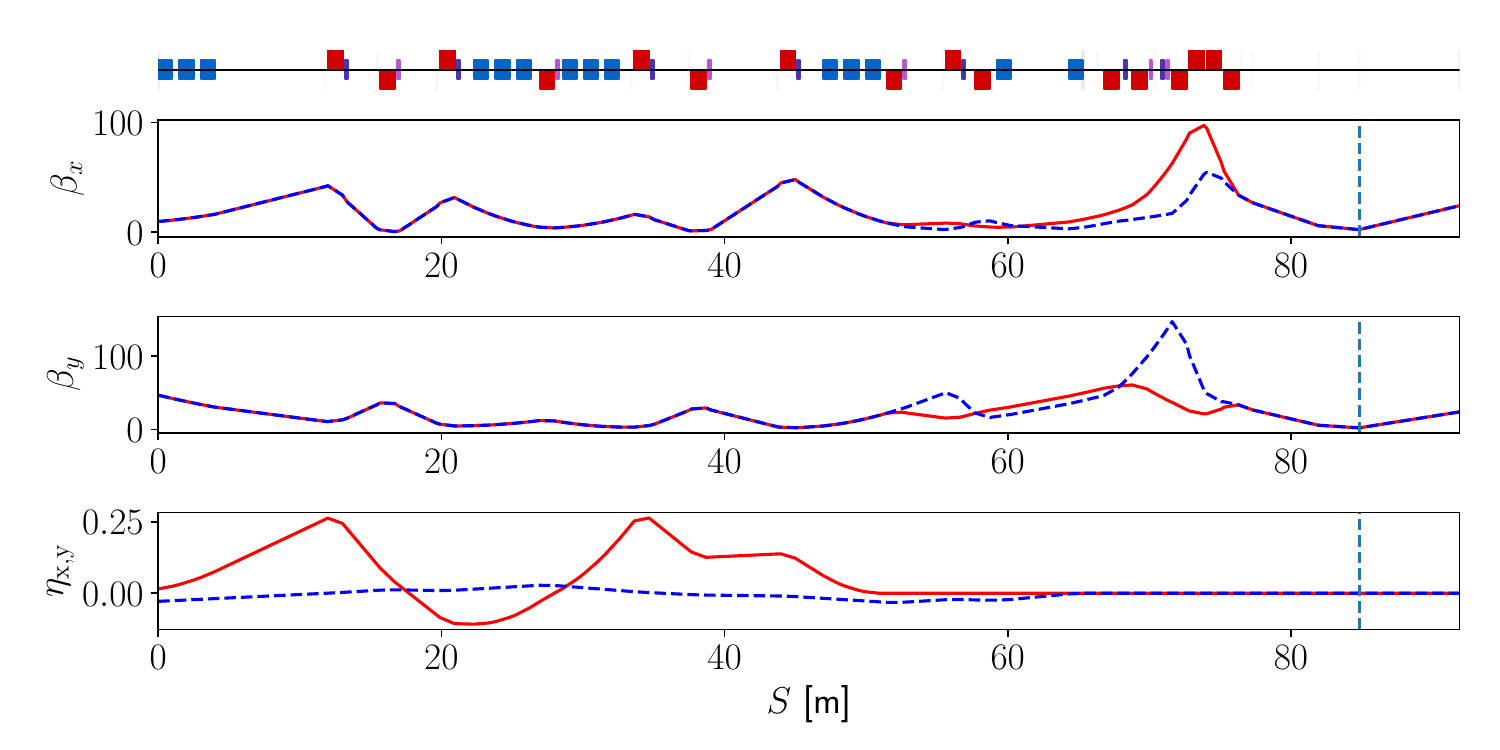}
   \caption{Beta functions (top two panels) and dispersion (bottom panel) for the T20 beam-line for the final focus magnet (red solid line) and arc-magnets (blue dashed line) optical configurations. The vertical dashed line is the IP location. Above the plots is a schematic of the beam line (blue: sector bends and red: quadrupoles).  
   \label{fig:T20-optics}}
\end{figure}

The beam line from the \euxfel to LUXE is designated T20. To reach the LUXE experimental area the \euxfel beam needs to be extracted and deflected by 6.74 degrees with respect to the linac axis. The beam is extracted from the XTL though a 84.82\,m long beam line (from sector bend BZ.2.T20 to the LUXE interaction point (IP)). The beam potentially needs to be focused at two locations: the bremsstrahlung target and the IP. In total four optics configurations have been simulated, two focus locations (at the IP and at the bremsstrahlung target) and two matching configurations (final focus quadrupoles and using the quadrupoles in the arcs). Figure ~\ref{fig:T20-optics} shows the beta functions and dispersion for a configuration with an electron beam focus at the IP. Focusing at the IP is the default for both operation modes.

The optics configuration with a round electron focus at the bremsstrahlung target (7.51\,m upstream from the IP) is similar to that presented in Fig.~\ref{fig:T20-optics}. Optics configurations have also been developed to achieve a $5\,\mu{\textrm m}$ beam. In general the T20 can be optimised for a large range of beam sizes and focus locations using the last 9 quadrupoles of the T20 line. The dispersion introduced by the arcs in the focusing section and throughout the LUXE experiment is zero, and minimal contributions to the beam size and position-angle jitter are expected from beam energy spread and variation, respectively. The largest beam size in the T20 line is in the final quadrupole quartet and for all four optical configurations the variance is less than 60\,$\mu$m. In general the beam size of the electron beam at the electron dump in e-laser mode is approximately 30\,$\mu{\textrm m}$ and it is about 15\,$\mu{\textrm m}$ in $\gamma$-laser mode.

\subsubsection{T20 Beam Line Magnets and Beam Instrumentation}

The magnets proposed for the T20 beam line are of the same type as those used in the \euxfel. For a given magnetic element the beam pipe apertures are taken from the default \euxfel component database. Given the projected beam sizes in the T20 LUXE optics, no special apertures or transitions are envisioned beyond that at the start of the LUXE experiment, just upstream of the bremsstrahlung target.  

The T20 will be instrumented with beam position monitors, wire scanners (and/or screens), bunch charge monitors and background/radiation detectors \cite{Nolle:2009}. As with the magnets, standard \euxfel beam instrumentation devices are envisioned, currently no specialised or custom beam instrumentation devices are planned. 

There are four types of beam position monitor (BPM) currently installed in the \euxfel, "cold" re-entrant buttons, "warm" button, large, and small diameter cavities \cite{Keil:2010}. To monitor beam position jitter of order a few microns (the IP beam size) transducers with intrinsic single bunch resolution of 1\,$\mu$m are required, which implies the small diameter re-entrant cavities are the only BPMs which can be employed for IP beam position variation. The \euxfel weekly stability requirement is that drifts should be less than the device resolution. Given the acceptance of the detector systems, there is a weaker requirement for the IP beam direction variation. This can be monitored with two or more BPMs located at the correction phases of the T20 beam line.   

Given beam sizes of the order of $10\,\mu\textrm{m}$, single bunch operation and a lower concern for coherent synchrotron radiation (CSR), a single optical transition radiation (OTR) screen placed at the IP position can be used to monitor and coarsely tune the IP beam size. The wire scanners for the \euxfel emittance measurement are designed for beam size measurement in a single pulse and likely be an expensive option for LUXE.

A detailed beam tracking study, which includes beam-position, size and charge monitor performance (resolution), coupled to the strong field QED Monte Carlo is required to fully determine the impact of the electron beam on the LUXE physics goals. A simple MAD-X/PTC \cite{JACOW} simulation can also be used to optimise the location of beam instrumentation devices. All the optical configurations have a dispersion free section just before the final focusing quadrupoles which is suitable for installation of beam instrumentation.   

\subsubsection{Beam Induced Backgrounds}

It is not expected that beam induced backgrounds are a potentially significant risk for the LUXE experiment, but it is still important to compute the expected background rates in the experiment. Important backgrounds to be considered are electron beam-gas interactions, synchrotron radiation from the T20 dipoles and quadrupoles, and beam halo interactions with the beam-pipe aperture. To investigate beam-gas backgrounds a small stand-alone (independent of the LUXE detector Monte Carlo) Geant4~\cite{Agostinelli:2002hh} simulation was performed. The simulation consisted of a beam pipe filled with a representative gas mixture with a density of $\sim 1\times 10^{-13}$\;\textrm{g/cm$^3$}, corresponding to $10^{-6}\,$mbar. The first electron beam-gas interaction cross-sections were scaled by a factor of $10^9$ and $10^7$ beam electrons were simulated. Although the simulated number of particles was significantly less than the typical bunch population of $1.5\times 10^{9}\; e^-/\textrm{bunch}$, the cross-section biasing makes the simulation of a sizeable number of electron beam-gas interactions possible. Scaling the simulated number of particles to the bunch population gives a rate of less than $1\times 10^{-4}$ beam-gas interactions at high energy (particle energies $>10$\,GeV) and up to $0.1$ at lower energy (particle energies $<10$\,GeV) per bunch. Synchrotron radiation is expected to have a negligible effect. Finally, beam-halo backgrounds are currently being simulated in \geant/BDSIM~\cite{Nevay:2018zhp} from just after the \euxfel collimation section (where there are existing measurements) and transported through the entire beam line simulation including realistic apertures.      


\subsection{LUXE Magnets}
\subsubsection{System Overview}

Up to three dipole spectrometer magnets will be used in the two data-taking phases of the experiment. In the $e$-laser collision mode, the electron beam arrives at the IP where it interacts with the laser beam. Just after the IP,  a first dipole magnet (denoted IP) is placed, that guides the electrons that have not interacted with the laser beam to  the dump. This magnet will also induce a spread of the electrons and positrons that have been created in the $e$-laser collisions, followed by the IP detectors systems (Cherenkov, scintillation screen, tracker and calorimeter). A second magnet is placed downstream of the forward $\gamma$-spectrometer (or photon detection system (PDS)) to allow the detection of electrons and positrons that will be created after the $\gamma$ beam interacts with a target.

In the $\gamma$-laser collisions, photons are created upstream of the IP, either via bremsstrahlung or via inverse Compton scattering. The electrons from the beam that have not interacted are dumped in the shortest possible distance,  using another magnet (denoted Brem) which is placed before the IP. As in the $e$-laser setup, the other two magnets are then used as spectrometer to measure the energy of electrons and positrons that are created in the setup.

\subsubsection{Technical Description}

The main characteristics that the three magnets used in the experiment must achieve are summarised in Table~\ref{tab:TC:Magnets}. 

\begin{table}[htbp]
\begin{center}
\begin{tabular}{|l|c|c|c|}
	\hline
Characteristic  & Brem & IP & PDS\\
 
\hline
min $e^{-}$ energy [GeV] & 1.5 & 2 & 2 \\
max $e^{-}$ energy [GeV] & 16.5 & 16.5 & 12 \\
$Z$ coordinate to object [mm] & 3950 & 4450 & 750 \\
$X$-displacement $E_{max}(e^{-})$ [mm] & 135& 150 & 65 \\
Polarity Switch & Yes & Yes & Yes \\
Goal &\makecell{Dump XFEL $e^{-}$ }& \makecell{Dump XFEL $e^{-}$ \\ and measure pairs }& \makecell{Measure pairs} \\
\hline

\end{tabular}
\caption{Characteristics of the spectrometer dipole magnets used in LUXE. The $Z$-coordinate to object corresponds to the distance along the beam-line between the end of the magnet and the object which is meant to intercept the majority of the particles deflected by the magnet. The $X$-displacement is the displacement created by the magnet at this $Z$-position}
\label{tab:TC:Magnets}

\end{center}
\end{table}

In order to reduce the costs associated with these magnets, it was decided to use a single type everywhere in the experiment. After reviewing the already existing magnets used in DESY, one magnet type was found to meet all the criteria expected, and more. These magnets are called TDC magnets and are currently being used in the FLASH accelerator~\cite{FlashMagnets} to drive the electron beam to a dump. The main characteristics of the TDC magnets are summarised in Table~\ref{tab:Magnets:TDC}. The magnet can also function with the default type of power supplies used everywhere else in the accelerator such that in case of failure it could be replaced very easily. Polarity switches would be added to all the dipole magnets power supplies to allow ramping up and down in the same fashion in order to remain within the same hysteresis cycles.

\begin{table}[htbp]
\begin{center}
\begin{tabular}{|l|c|c|c|c|}
	\hline
Characteristic  & Technical data\\
\hline
Nominal current [A] & 372 \\
Max power [kW] & 28.1 \\
Resistivity at 20$^\circ$C [m$\Omega$] & 200\\
Water cooling [l/min] & 14.4 \\
Field [T] & 1.6 \\
Magnetic height [mm] & 60\\
Magnetic length [mm] & 1300\\
Overall length [mm] & 1440 \\
Aperture [mm] & 570 \\
Weight [kg] & 6000\\
\hline
\end{tabular}
\caption{Characteristics of the TDC dipole magnets used in FLASH. The aperture is in the bending plane of the magnet.
\label{tab:Magnets:TDC}
}
\end{center}
\end{table}

\begin{figure}[ht]
   \centering
     
   \includegraphics*[height=4.4cm]{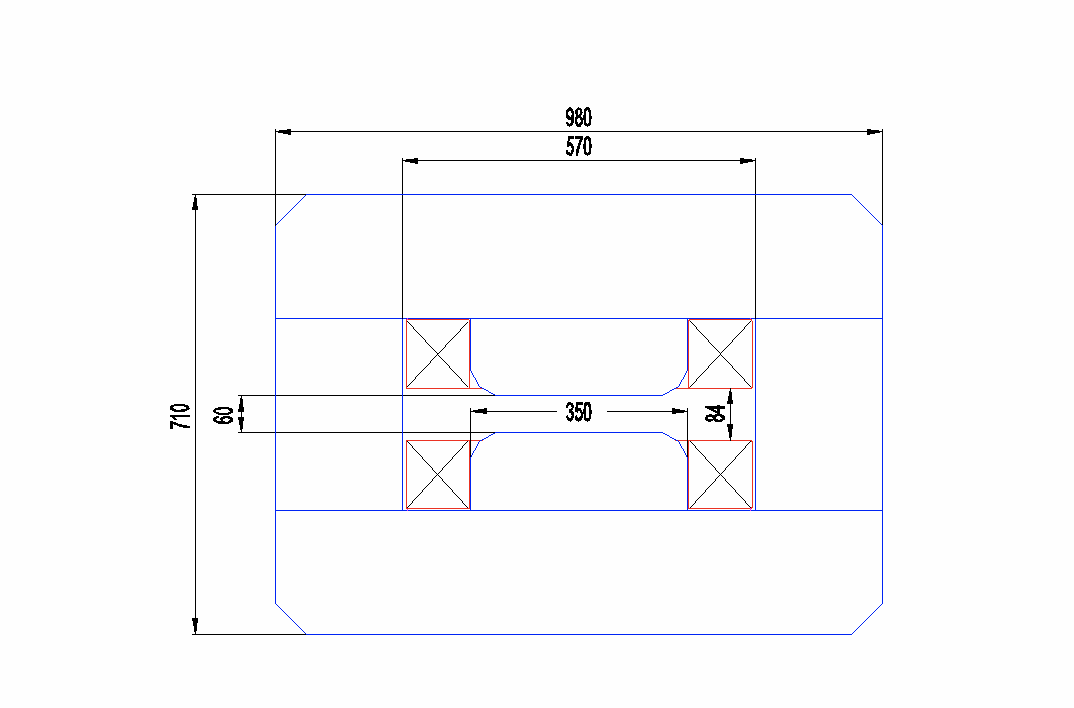}
   \includegraphics*[height=4.4cm]{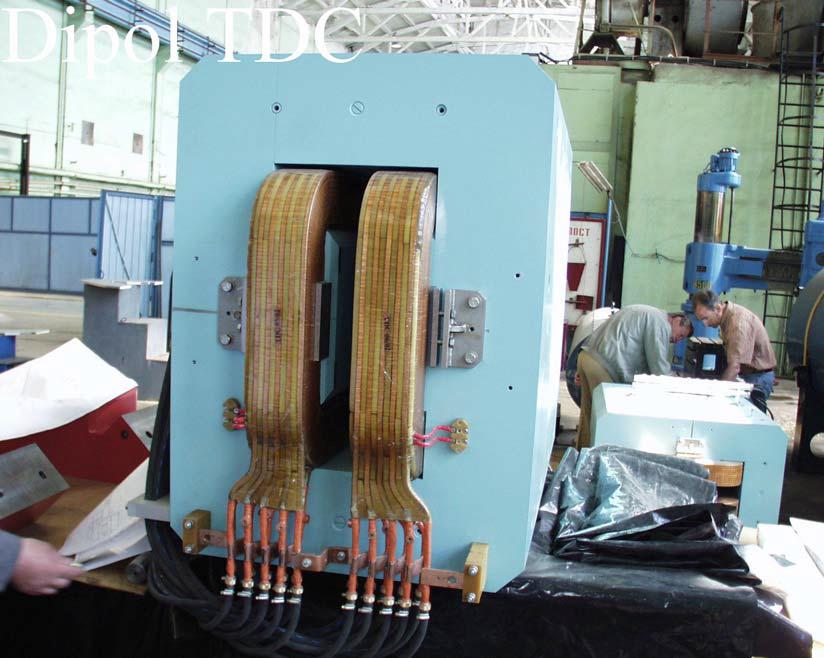}
   \includegraphics*[height=4.4cm]{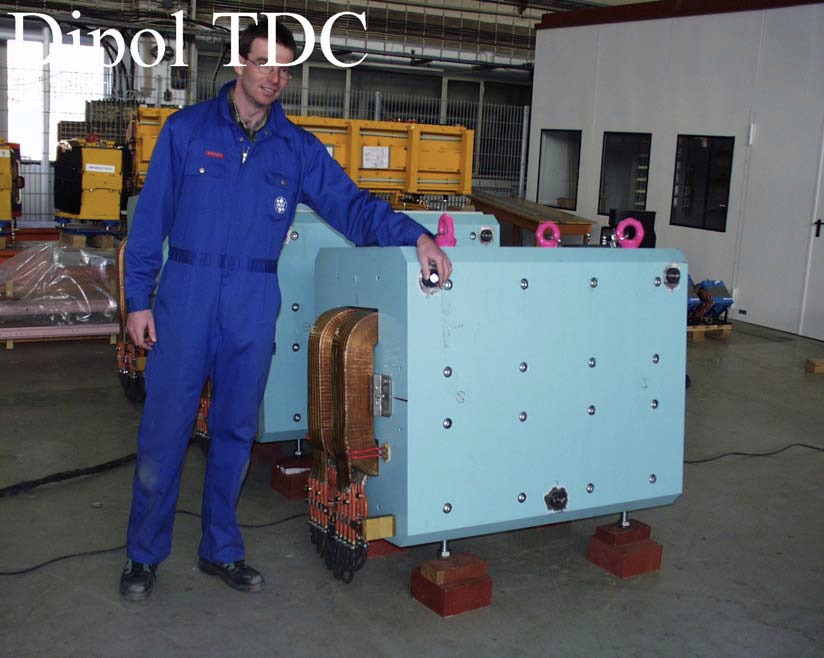}

   \caption{Sketch and pictures of the TDC magnet.
   \label{fig:magnets-TDC}}
\end{figure}

A sketch displaying the TDC characteristics, and some photos of the magnets are shown in Fig.~\ref{fig:magnets-TDC}.

\begin{figure}[ht]
   \centering
     
   \includegraphics*[height=8.2cm]{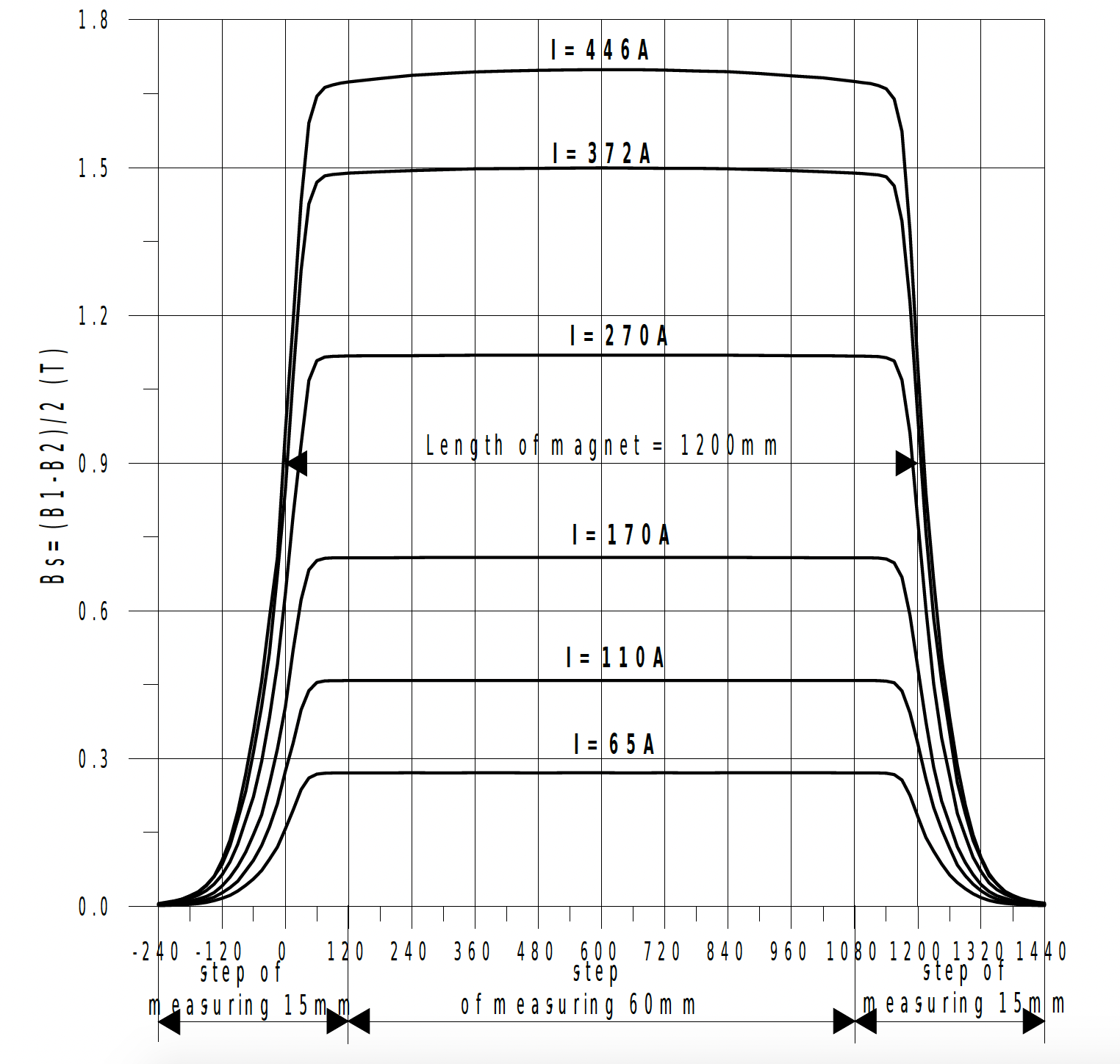}
   \caption{ Field values as a function of the distance in the TDC magnet.
   \label{fig:magnets-curve-TDC}}
\end{figure}

If needed, these magnets can also be used at different field values as shown in Fig.~\ref{fig:magnets-curve-TDC}.

\subsubsection{Vacuum Chamber and Beam Window}

\begin{figure}[ht]
   \centering
     
   \includegraphics*[height=6cm]{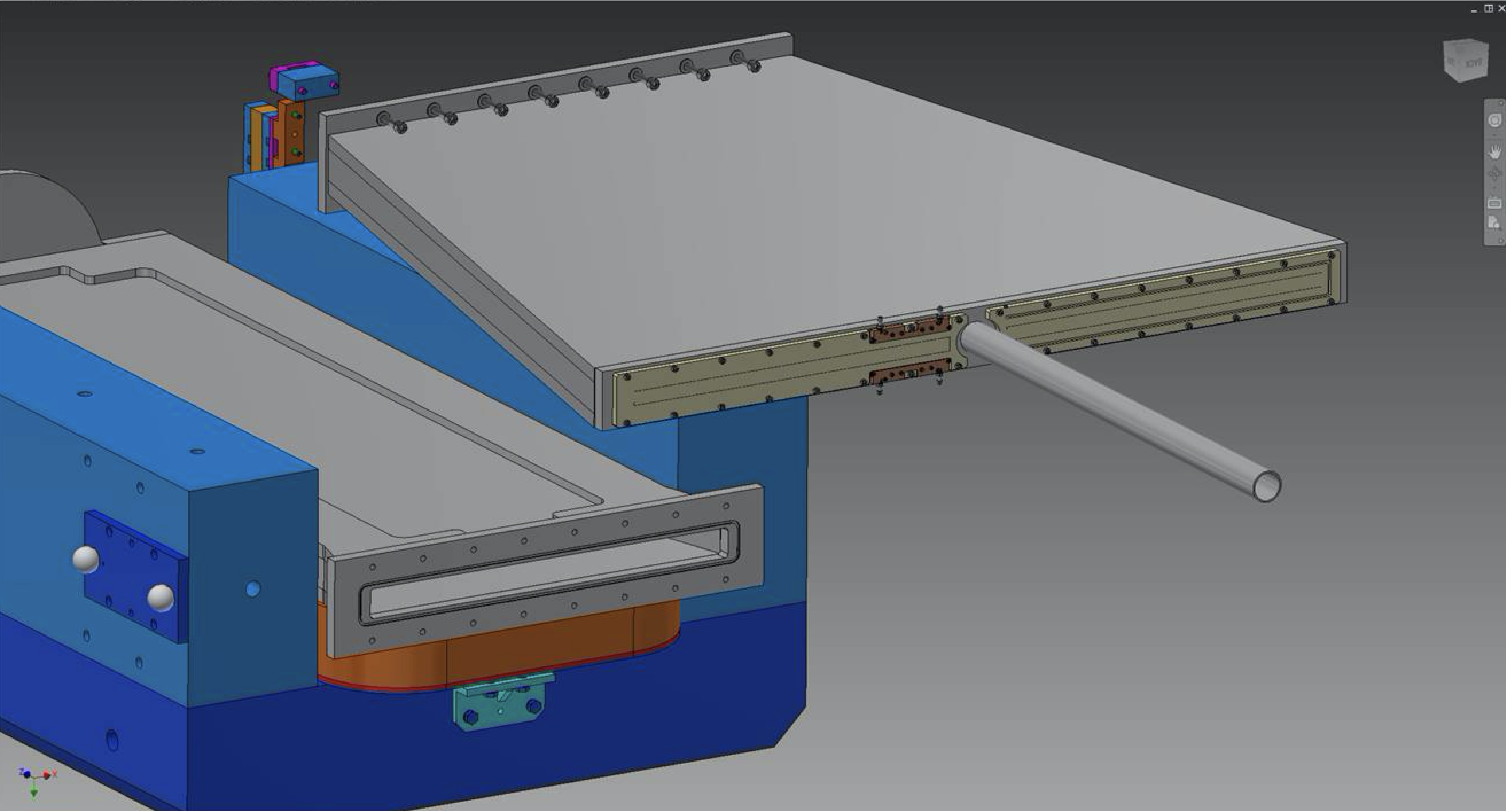}
   \includegraphics*[height=6cm]{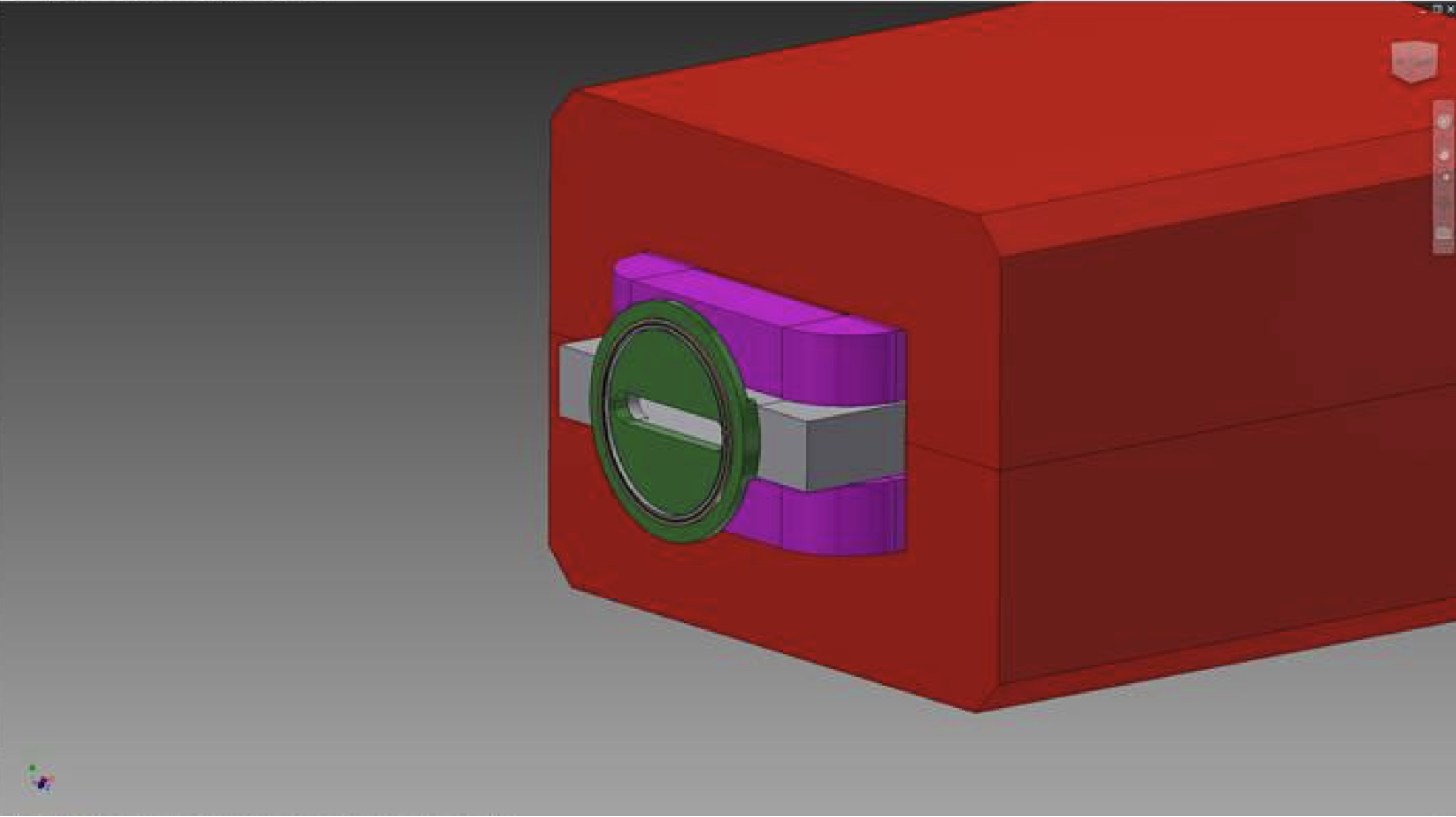}
   \includegraphics*[height=6cm]{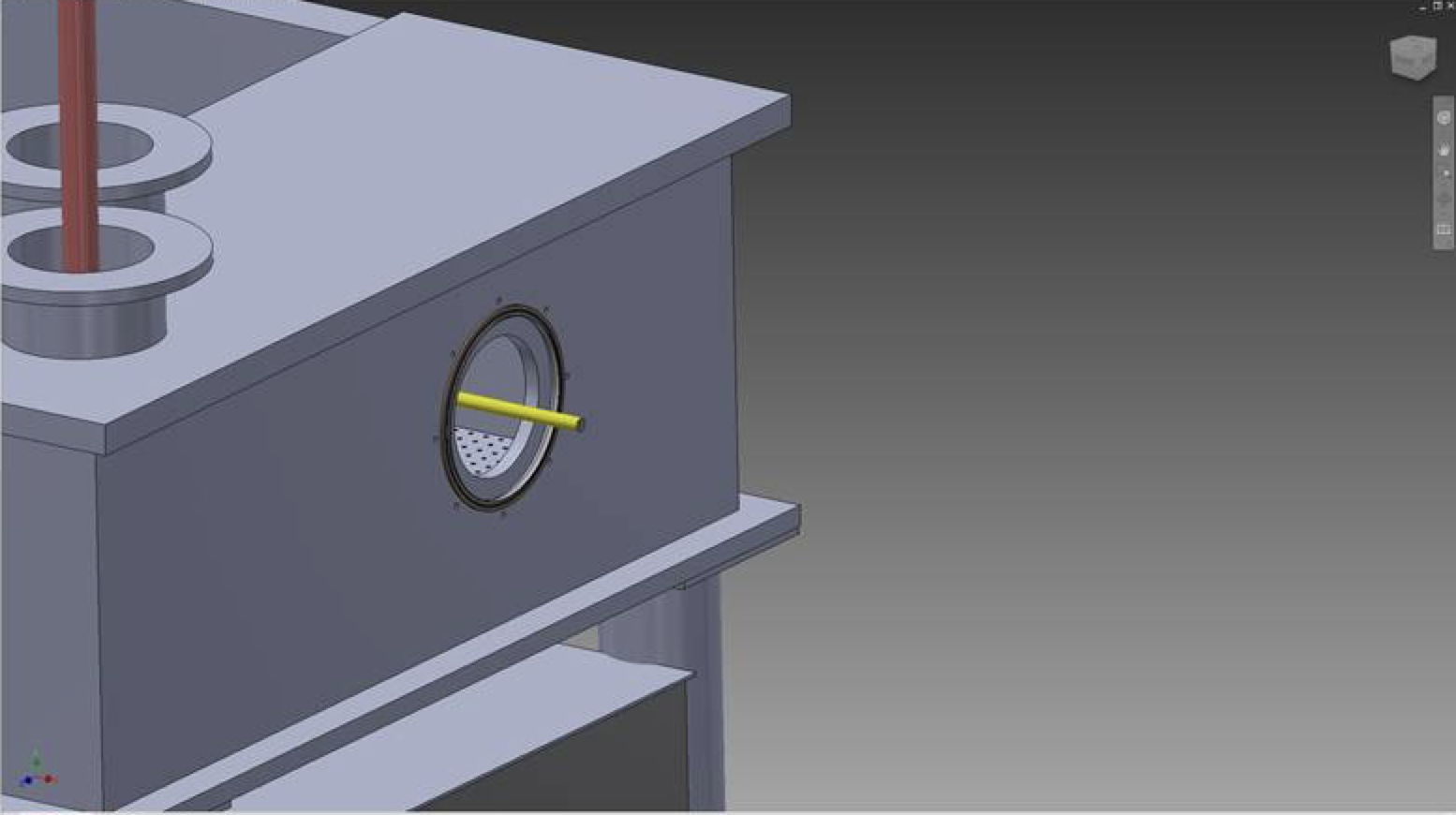}

   \caption{CAD drawing of the magnet vacuum chamber: (top) two parts of the chamber, (middle) bellows and connection link to the IP chamber, and (bottom) connection link in the IP chamber side.
   \label{fig:magnet:CAD-vacChamber}}
\end{figure}

Since two of the three magnets (IP and PDS) have to be used as precision spectrometers to measure $e^{+}e^{-}$ pairs, vacuum chambers terminated with windows to let the $e^{\pm}$ go through are mandatory. These chambers are separated in two different parts, a square chamber that will be mounted inside the magnet after opening it, and a trapezoidal part to allow the electrons to remain in vacuum after the end of the magnetic field for as long as possible. This latter part will be attached to the main square and vacuum sealed using an O-ring. At the interaction point the chamber will be attached to the IP chamber vacuum tank using a bellow and another O-ring. Figure~\ref{fig:magnet:CAD-vacChamber} shows  different views of the vacuum chamber inserted in the TDC magnet, and the connection to the IP chamber.

The beam-window will be made of a 300\,$\mu$m tick aluminum plate in order to sustain the thermal constraints due to interactions with the electron beam, but also to guarantee the mechanical resistance to the vacuum used in the experiment. The plate will also be equipped with a cooling-loop to dissipate the heat produced by the beam.

\begin{figure}[ht]
   \centering
     
   \includegraphics*[height=8cm]{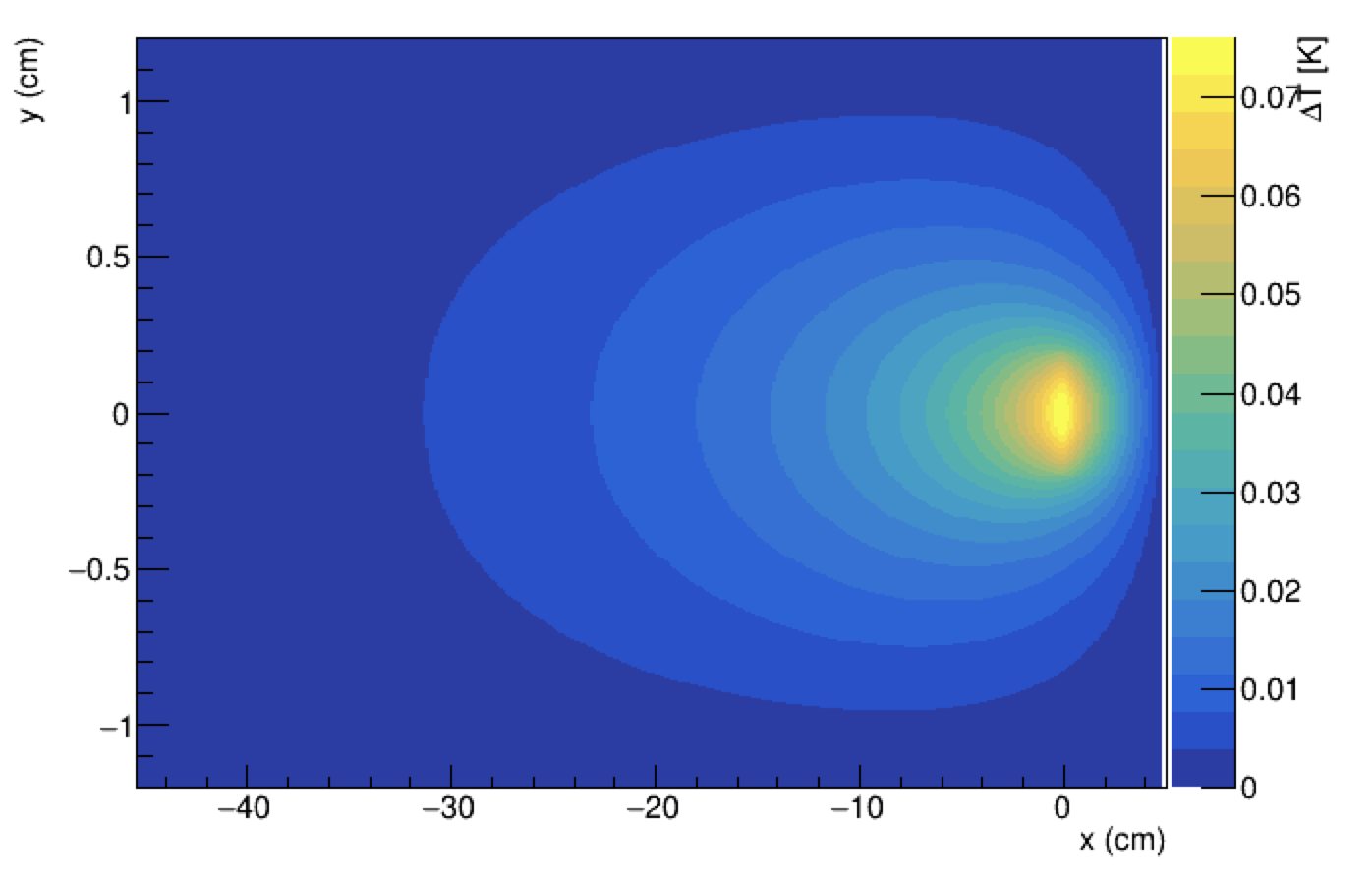}
   \includegraphics*[height=8cm]{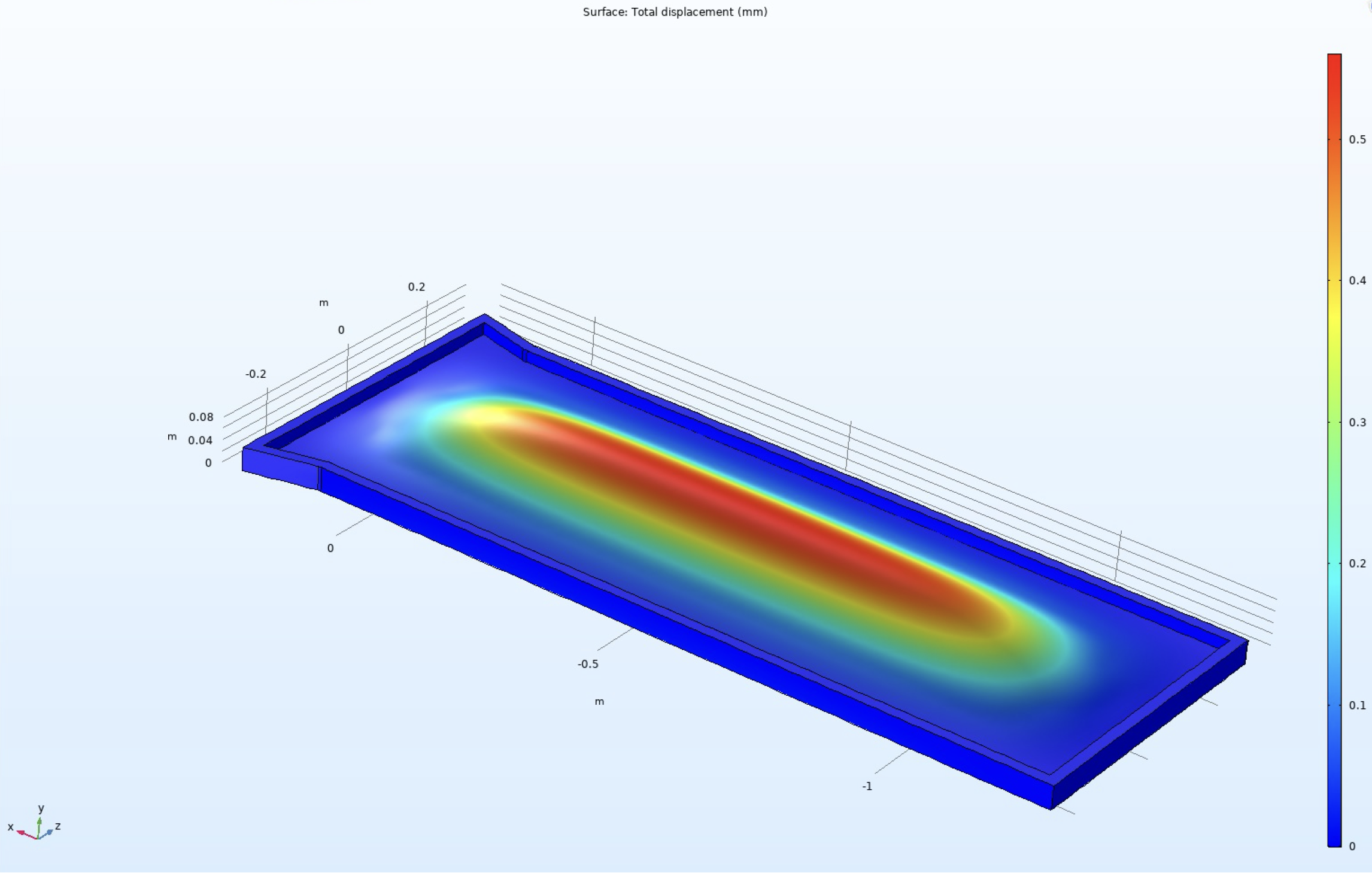}

   \caption{(top) Thermal dissipation of the electron beam through the 300\,$\mu$m aluminum beam window for 2\,h of data-taking assuming 10\,Hz of operation, obtained with FLUKA. (bottom) Mechanical stress of a 300\,$\mu$m aluminum beam window exposed to a vacuum of $10^{-5}$\,mbar, obtained with Comsol. The $x$ and $y$ axis are expressed in m, while the $z$ axis show the displacement in mm due to the pressure increase. The extreme values of the $z$ axis are bounded in red at 0.55\,mm for the maximal, and in blue at 0\,mm for the minimal.
   \label{fig:magnets-thermal-dissipation}}
\end{figure}

Simulations have been run using FLUKA~\cite{fluka} to estimate the total heat that will be dissipated by the electron beam in the window, as shown on Fig.~\ref{fig:magnets-thermal-dissipation}. It was estimated that without any cooling, the temperature of the plate would increase by 0.075\textdegree{}C  after 2 hours of data taking, assuming 10\,Hz of exposure. This shows that the beam can not damage the window, since to reach the melting point of the aluminum one would need up to about a year of continuous running. 

Simulations and experimentation have also been performed to estimate the mechanical resistance of the aluminum plate to vacuum deformation. It was found that the window would survive being exposed to $10^{-5}$\,mbar vacuum, which is the vacuum that would be used in the experiment, albeit with a deformation  found at the centre of the plate. This can be seen in Fig.~\ref{fig:magnets-thermal-dissipation}. It is expected from the simulations that the window will survive to lower values of the vacuum, however this was not checked experimentally because of a lack of turbo-pump to reach them.

\clearpage

\subsection{Interaction Chamber}


 The vacuum tank that will hold the IP chamber is being designed using the ALPS2~\cite{Bahre:2013ywa} main vacuum chamber as a model, but with upscaled dimensions ($2.4$\,m $\times$ 1.5\,m $\times$ 0.7\,m (L $\times$ W $\times$ H)). 
 
\begin{figure}[ht]
   \centering
   \includegraphics*[height=6cm]{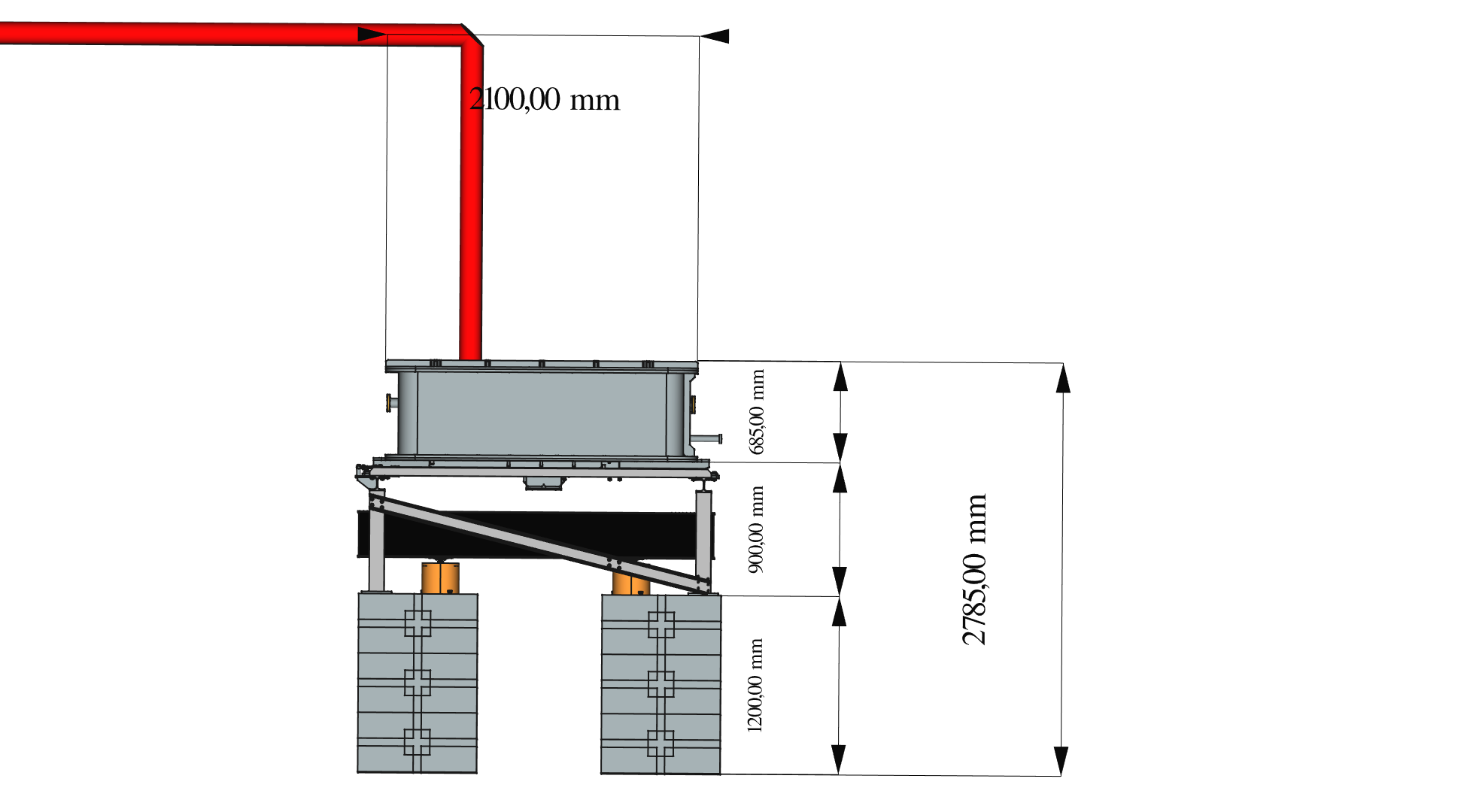}
      \includegraphics*[height=6cm]{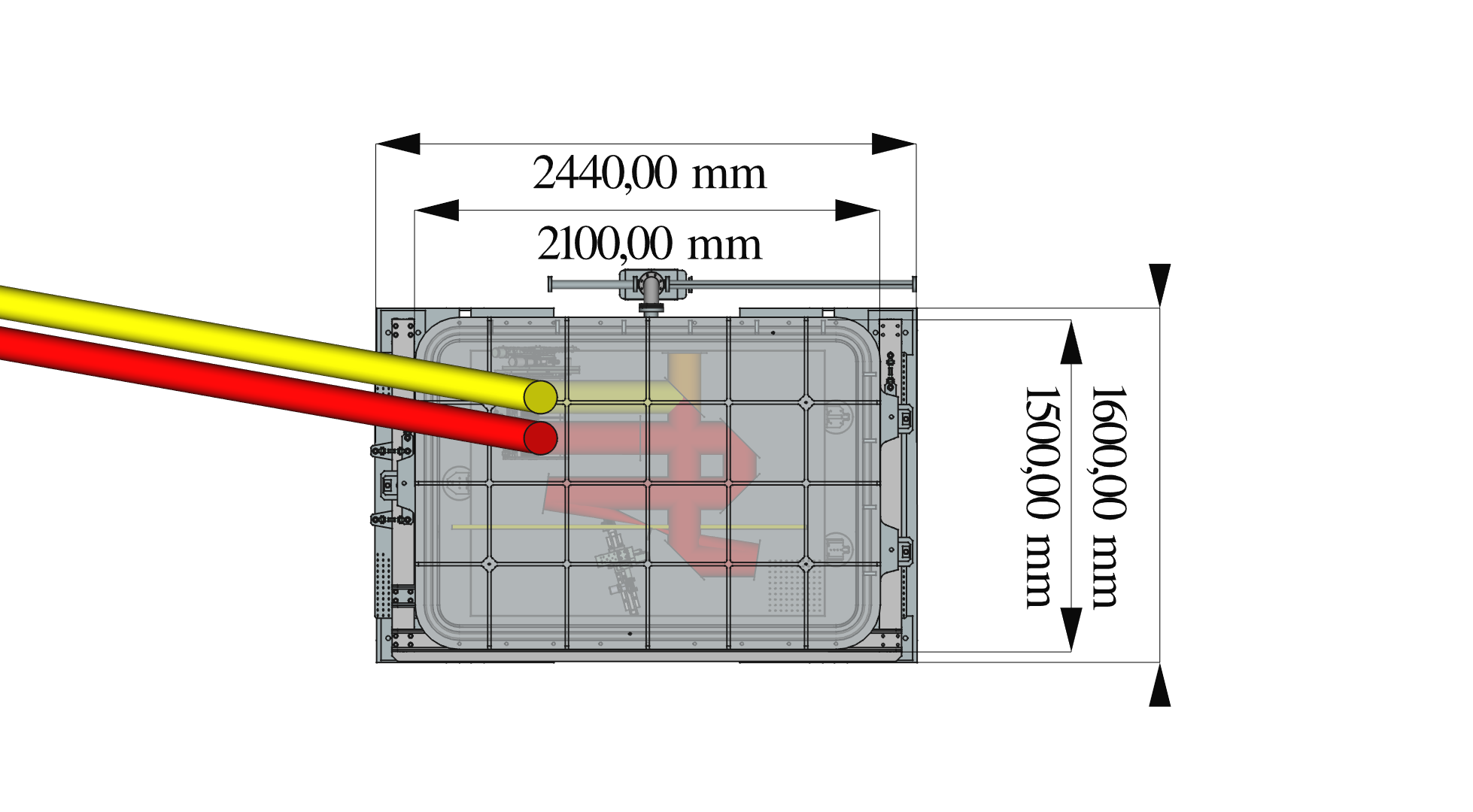}
      \includegraphics*[height=6cm]{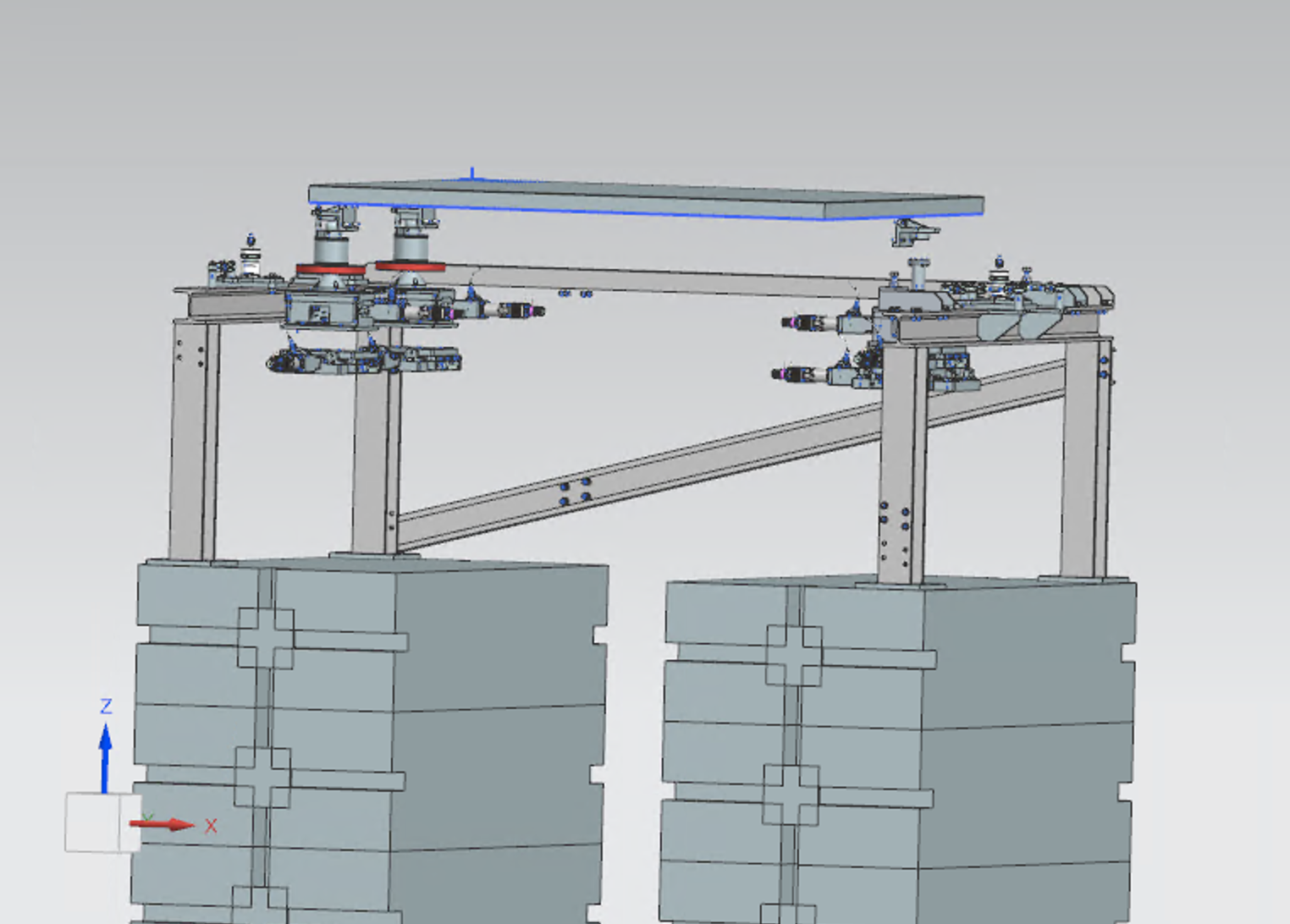}

   \caption{Conceptual design of the interaction chamber vacuum tank, (top) side view, (middle) top view. (Bottom) pictures of motors for  positioning the optical system precisely.
   \label{fig:IPChamber-layout}}
\end{figure}

 The breadboard for the optical elements is supported from the optical table located below the vacuum tank at three points. The supports are mechanically decoupled from the vacuum vessels by soft bellows. The breadboard will be movable on 3 axes using the EASY motor control system, as can be seen in Fig.~\ref{fig:IPChamber-layout}. 
The chamber and optical table will be mounted on standard concrete pillars to bring the chamber to the beam height.

The interaction chamber's internal structure, more specifically the optical elements that will populate it, is explained in detail in Chapter~\ref{chapt3}. The inner elements can also be seen in Fig.~\ref{fig:IPChamber-layout} (middle), where the box cover is displayed as transparent. The laser beam enters the box through the pipe shown in red, and exit it through the pipe shown in yellow.
 
Figure~\ref{fig:IPChamber-layout} shows the preliminary dimensions of the vacuum tank. As the vacuum tank is being designed to be operated at $10^{-5}$\,mbar, a rubber seal O-ring will be used. The chamber is designed such that it will experience minimal deformation when reaching the vacuum.

The vacuum will be reached thanks to a dedicated turbo molecular pump that will be connected to the tank using soft bellows to avoid parasitic vibrations. Some of the vacuum elements can be seen in Fig.~\ref{fig:IPChamber-layout} (middle) above the chamber.

Given the large volume of the chamber, it is estimated that the required level of vacuum will be reached in about 3\,h.  

\clearpage

\subsection{Beam Dumps}
\label{sec:beamDumps}

Three beam dumps will be required for a safe disposal of the high-energy particles (electrons or photons) that will be created in the experiment. In the following, the requirements for each of these dumps will be reviewed.

\subsubsection{$\gamma$-laser Electron Dump and Shielding}

For the LUXE Conceptual Design Report~\cite{luxecdr} a preliminary dump design for the $\gamma$-laser configuration was developed by machine experts from DESY. The main characteristics can be found in Section~3.4 of~\cite{luxecdr}, but they are summarised here for the sake of convenience. The dump was designed to sustain up to 10\,Hz of 1\,nC electron bunches at a maximum energy of 20\,GeV, resulting in a maximum dissipated thermal power of 200\,W. The same maximal thermal power requirement will be used to design the other dumps.

\begin{figure}[ht]
   \centering
   \includegraphics*[height=10cm]{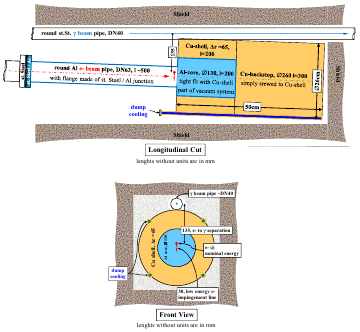}
   \caption{Conceptual design of the $\gamma$-laser electron dump made for the LUXE CDR.
   \label{fig:dump-layout}}
\end{figure}

The geometry of this preliminary dump is an aluminium cylinder with 13\,cm diameter and 20\,cm length, inserted in a copper cylinder of 26 cm diameter and 50\,cm length, as can be seen in Fig.~\ref{fig:dump-layout}. It has a total weight of 220\,kg. The dump is water cooled, and placed in a large (vertical, horizontal, depth: 4\,m, 3\,m, 1.5\,m) shielding made of concrete and neutrons absorbers (polyethylene and iron). From the radiation protection perspective a minimum of 0.5\,m of concrete must encapsulate the dump to contain as many  secondary particles created in the dump as possible.

\begin{figure}[ht]
   \centering
   \includegraphics*[height=4.2cm]{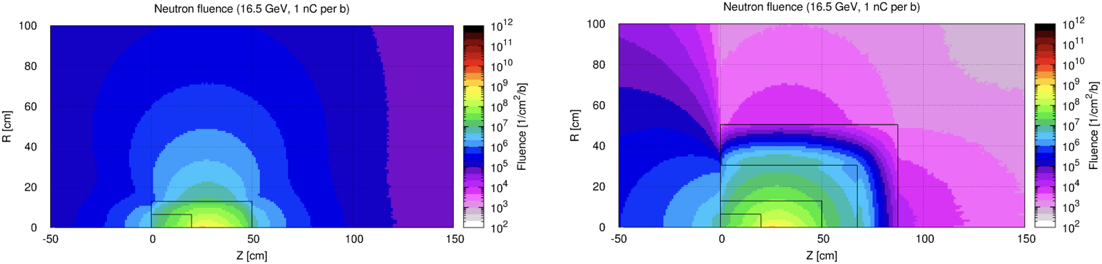}
   \caption{Flux of neutrons around the conceptual dump developed for the LUXE CDR obtained from FLUKA. (Left) No shielding around the dump is considered, (Right) 17.5\,cm of iron and 20\,cm of polyethylene are added.
   \label{fig:dump-neutronFluence}}
\end{figure}

This conceptual dump was implemented in different FLUKA and \geant simulations to study the leakage of particles in its vicinity. As shown in Fig.~\ref{fig:dump-neutronFluence}, adding iron and polyethylene around the dump reduces the flux of low energy neutrons by 3 orders of magnitude around the dump, while the flux of high energy neutrons (above 10\,MeV) is only reduced by a factor of about 2 to 3. The design of the dumps needed in the experiment will include an optimisation of the shielding to reduce the flux of particles leaking from it as much as possible, while remaining within an affordable cost. It is to be noticed that a similar performance has been obtained with both simulation tools.

\begin{figure}[ht]
   \centering
   \includegraphics*[height=6.2cm]{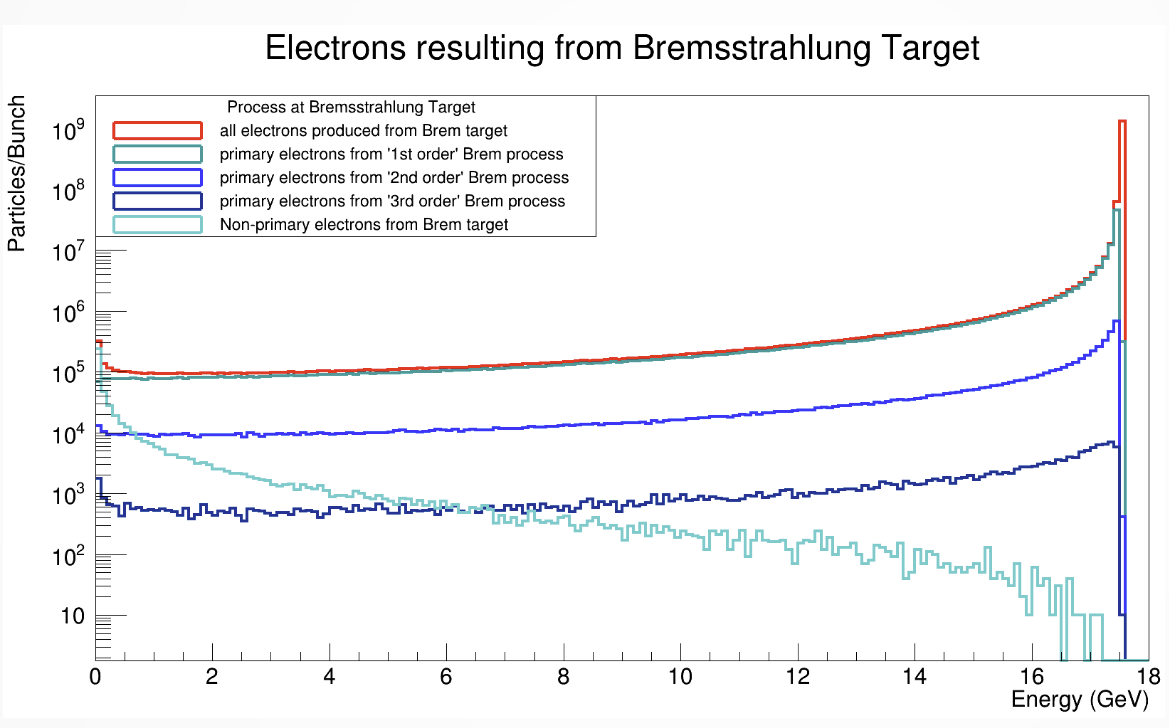}
   \caption{Energy spectrum of electrons reaching the electron dump in the $\gamma$-laser setup, when the accelerator is operated at 17.5 GeV.} 
   \label{fig:dump-Brem-input}
\end{figure}
 \begin{figure}[ht]
   \centering
   \includegraphics[height=10cm,trim={15cm 0 0 0},clip]{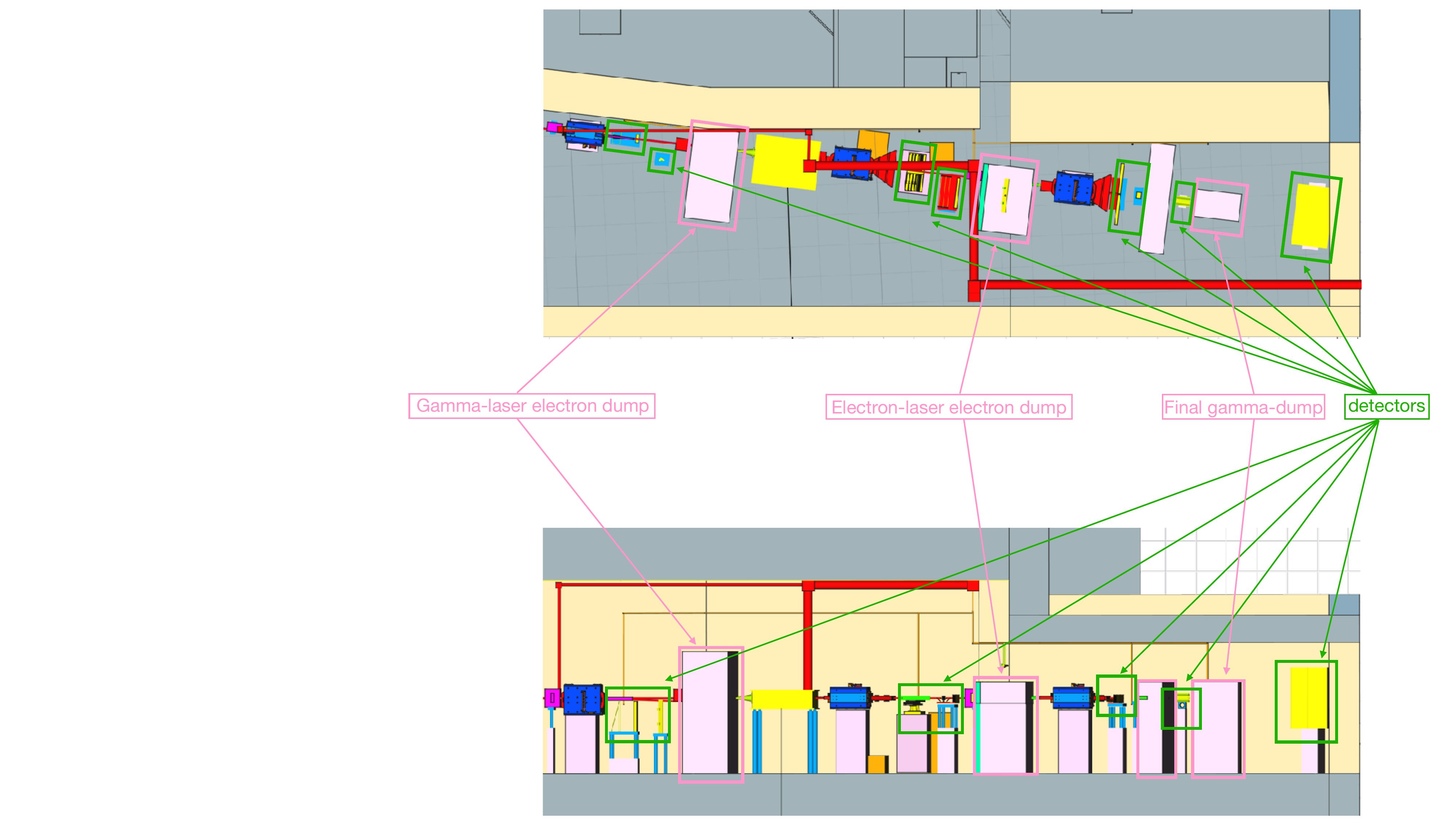}
   \caption{Position of the different dumps and detectors used in the particle flux study.}
   \label{fig:dump-positions}
\end{figure}

\begin{figure}[ht]
   \centering
   \includegraphics*[height=5.2cm]{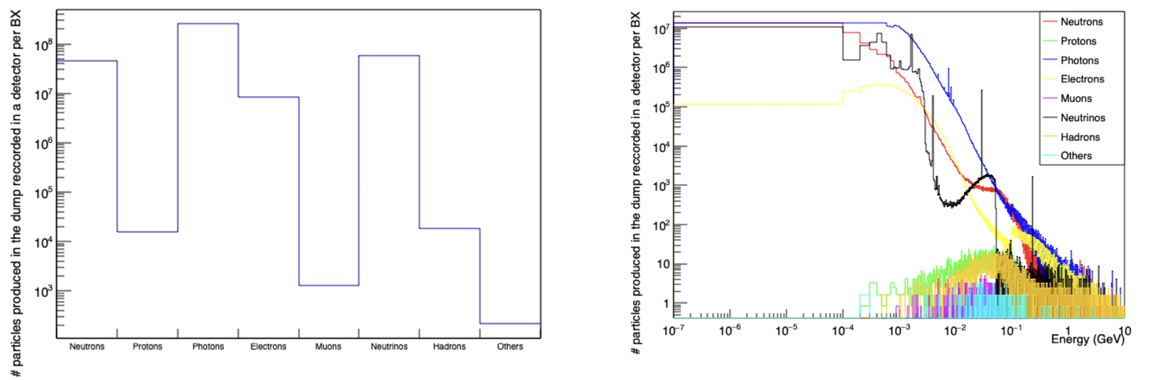}
   \caption{(Left) Number and type of particles obtained from the \geant simulation and detected in the detector site defined in Fig.~\ref{fig:dump-positions}  for the $\gamma$-laser electron dump. (Right) Energy spectrum of these same particles.
   \label{fig:dump-Brem-output}}
\end{figure}

The first dump of the experimental setup when viewed the beam, is a dump that will be used in the $\gamma$-laser mode. In this mode, a tungsten target is used to convert electrons into high energy photons. In order to remove the non-interacting electrons a dipole magnet (1.6\,T, 1.2\,m long) is used to divert them into the dump. This magnet is placed about 3\,m upstream of the dump and will displace the electrons by about 13\,cm when the XFEL will be operating at 16.5\,GeV.

The electron energy spectrum reaching the dump after the target is shown on Fig.~\ref{fig:dump-Brem-input}. This result was obtained with \geant. The energy spectrum appears continuous, where most of the electrons possess the beam energy. The number and types of particles originating from the dump, but detected in one of the LUXE detector locations (as shown in Fig.~\ref{fig:dump-positions}) are shown in Fig.~\ref{fig:dump-Brem-output}.

As it can be seen most of the particles created are low energetic photons and neutrons. But high energetic neutrons are also created which can be problematic for the electronics which are present in the cavern.

\subsubsection{$e$-laser Electron Beam Dump}

\begin{figure}[ht]
   \centering
   \includegraphics*[height=6.2cm]{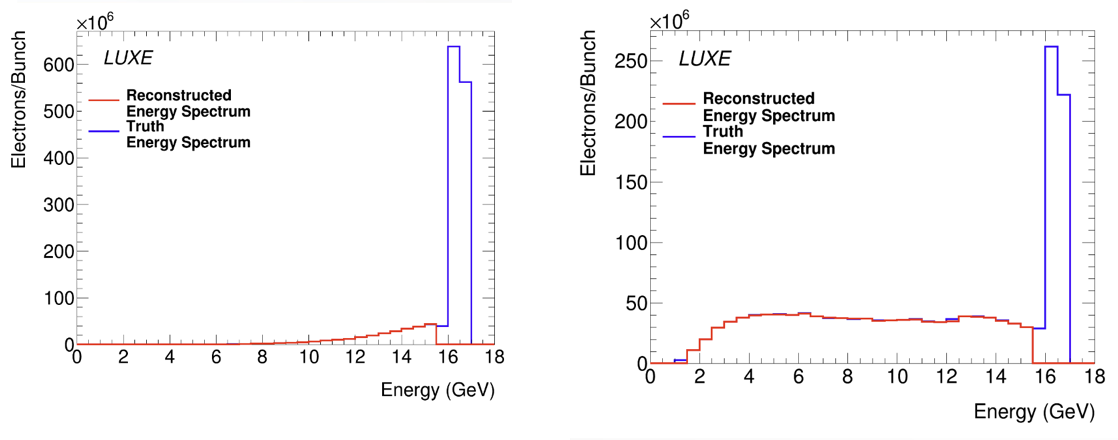}
   \caption{Energy spectrum of electrons reaching the electron dump in the $e$-laser setup. (Left) Phase-0 low $\xi$ run, (Right) Phase-1 high $\xi$ run.
   \label{fig:dump-eleLaser-input}}
\end{figure}


\begin{figure}[ht]
   \centering
   \includegraphics*[height=5.2cm]{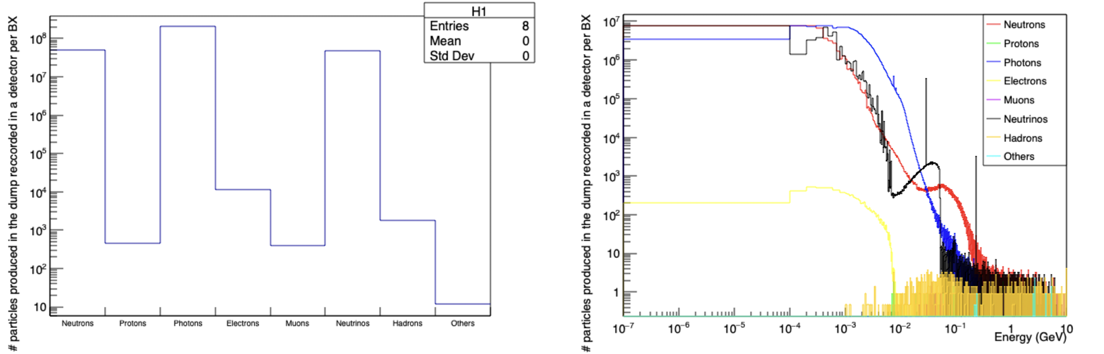}
   \caption{(Left) Number and type of particles obtained from the Geant4 simulation and detected in the detector locations defined in Fig.\,\ref{fig:dump-positions} for the electron-laser electron dump. (Right) Energy spectrum of these same particles.
   \label{fig:dump-eleLaser-output}}
\end{figure}

The second dump required for the experiment, is a dump that will be used in the $e$-laser data-taking mode. In this mode, the electrons from the \euxfel interact directly with the laser beam in the IP chamber. The non-interacting remaining electrons are dumped using the same type of dipole spectrometer magnet (1.6\,T, 1.2\,m long) used in the $\gamma$-laser data-taking mode. This dump is placed about 4.5\,m away from the magnet located after the IP, resulting in a deviation of about 1\,cm for the 16.5\,GeV electrons when the magnet is used at its nominal field. For non-linear Compton scattering studies, it is also envisioned to run the magnet at about 1.0\,T resulting in a deviation of about 10\,cm for 16.5\,GeV electrons

Figure~\ref{fig:dump-eleLaser-input} presents the energy spectrum of the electrons reaching the electron dump in the $e$-laser setup for two different phases of the data-taking. In the left side of the figure, the energy spectrum corresponds to the Phase-0 (40\,TW laser) at a moment where the laser beam is mildly focused (low $\xi$) at the interaction point. In this case it is clear that most of the electrons do not interact with the laser and are going to be dumped with an energy of 16.5\,GeV. On the right side of Fig.~\ref{fig:dump-eleLaser-input}, the energy spectrum corresponds to the Phase-1 (350\,TW laser), with a high laser focus (high $\xi$). In this case a large fraction of the electrons interact with the laser creating a continuum of electron energy with a large intensity that will need to be fully absorbed by the beam dump. To meet this technical challenge, the dump must be able to contain about 20 to 30 electromagnetic radiation lengths with energies ranging from 16.5\,GeV down to 2\,GeV after the magnet that will therefore create a horizontal spread. For this reason it has been proposed to create a dump with a wedge or step shape. The same type of behaviour exist for the $\gamma$-laser electron dump, but given the small tail of low energy electrons, it is for now thought to stop the low energetic electrons that will go through he magnetic field using more shielding material.

Figure~\ref{fig:dump-eleLaser-output} shows the number, type and energy spectrum of the particles that are created in this beam-dump. This figure was obtained using the \geant simulation of the LUXE experiment. As for the $\gamma$-laser dump most of the particles created are low energetic photons and neutrons, that can be caught using a dedicated shielding made with boronated concrete and polyethylene. While low energetic particles dominate the spectrum, care needs to be taken of the high energy neutrons that can be observed, and iron will have to be used to stop them in the shielding. 

In the current configuration the shielding has been designed to contain a notch to encompass as much as possible the secondary particles created in the dump.

\subsubsection{Final Photon Dump}

\begin{figure}[ht]
   \centering
   \includegraphics*[height=6.2cm]{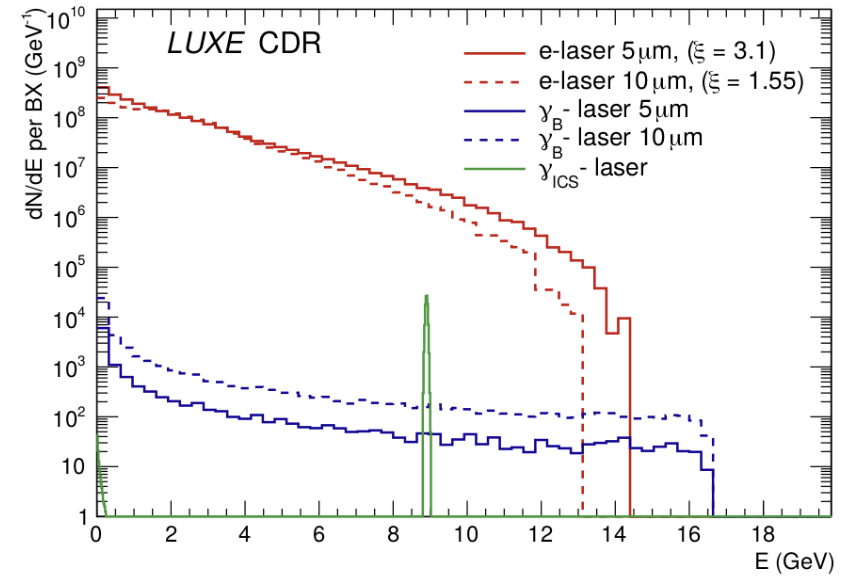}
   \caption{Energy spectrum of photons reaching the final photon dump for the different running modes.
   \label{fig:dump-finalPhoton-input}}
\end{figure}

\begin{figure}[ht]
   \centering
   \includegraphics*[height=5.2cm]{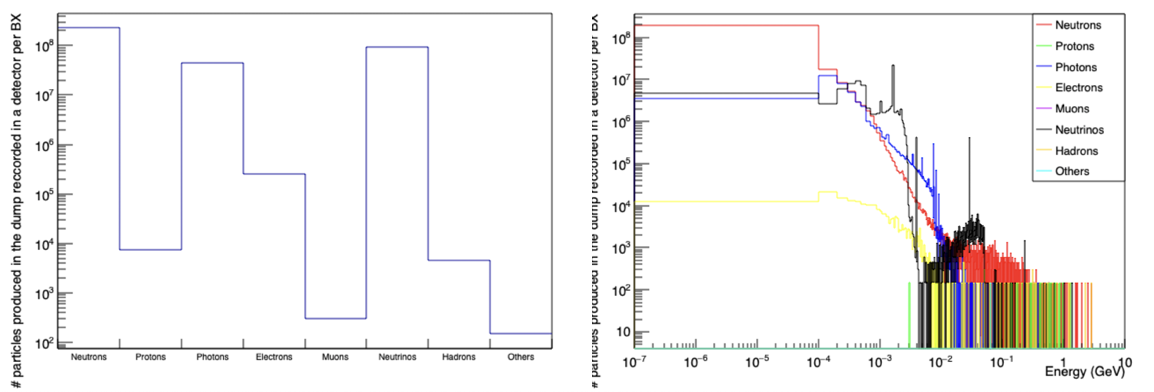}
   \caption{(Left) Number and type of particles obtained from the \geant simulation and detected in the detector site defined in Fig.~\ref{fig:dump-positions} for the final photon dump. (Right) Energy spectrum of these same particles.
   \label{fig:dump-finalPhoton-output}}
\end{figure}

In both data-taking modes, a large flux of high-energy photons is created in LUXE, either via bremsstrahlung or Compton scattering. These photons are created at the bremsstrahlung target or at the IP and travel further downstream the beam-line to a final photon dump. The absolute energy spectrum per bunch crossing of these photons is presented in Fig.~\ref{fig:dump-finalPhoton-input}.

Ideally, this photon-dump opens the possibility for searches for new axion like particles~\cite{LUXEBSM}. In the current \geant simulation implementation a 1\,m long and 50\,cm diameter lead  cylindric dump was used. Figure~\ref{fig:dump-finalPhoton-output} shows the number, particle types, and energy spectrum of the particles created in the dump. As with the two previously discussed dumps, neutrons and photons dominate the particles created, but with a softer energy spectrum. It is to be noted that in the current simulation, no shielding surrounds the beam dump. This will be modified in reality to cope with radiation-protection constraints. Finally, in~\cite{LUXEBSM} tungsten was suggested to be used, as the material is denser, presumably allowing better containment of secondary particles created in the dump, while optimising the sensitivity to new physics. This needs to be investigated.

\subsubsection{\euxfel Dump}

The main \euxfel dump is located in the XS1 access shaft in UG04 approximately 50\,m away from the annex where the experiment will be installed. While this dump is heavily shielded, it is the main reason why XS1 is exposed to radiation and cannot be accessed if the accelerator is running. In order to estimate the radiation level and particle flux, originating from this dump, a radiation (LB 6419\cite{Pandora-Detectors}) detector was installed by the DESY radiation protection group (D3) in the annex at the approximate position where the interaction chamber would be installed, but closer to the floor. 

\begin{figure}[ht]
   \centering
   \includegraphics*[height=4.2cm]{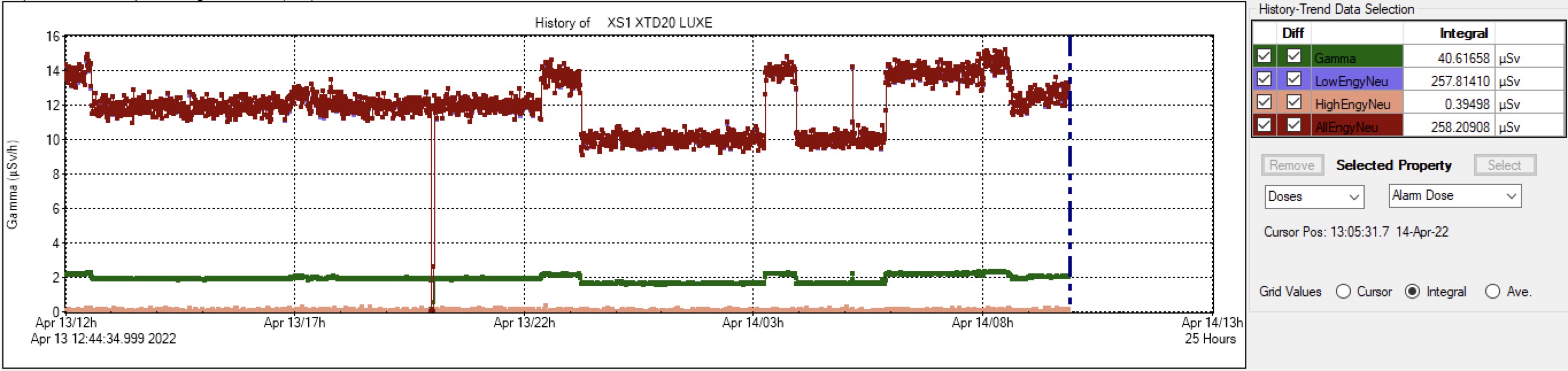}\vspace{1cm}
      \includegraphics*[height=4.2cm]{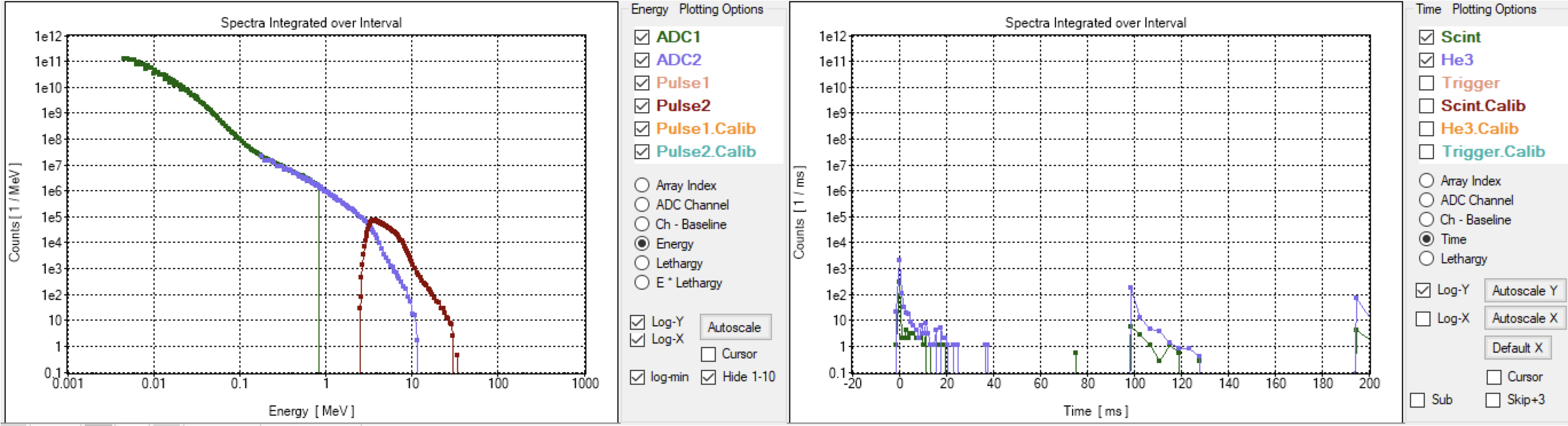}\vspace{1cm}
      \includegraphics*[height=4.2cm]{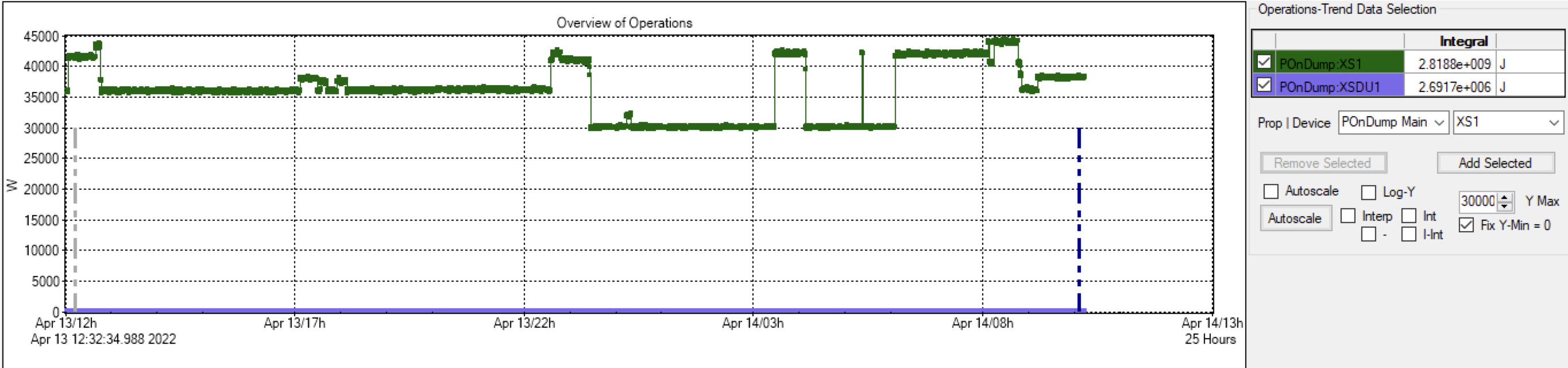}

   \caption{24\,h data of the LB 6419 detector located in the XS1 annex showing (top) the dose rate of neutrons, and photons, (middle) the energy spectrum of the particles detected, and (bottom) the energy dissipated on the main \euxfel dump.
   \label{fig:dump-XFEL-MainDumpReading-Flux}}
\end{figure}

Figure~\ref{fig:dump-XFEL-MainDumpReading-Flux} shows the data of this detector taken during 24~hours of regular operation of the \euxfel. The top plot shows the radiation levels (in $\mu$Sv) for $\gamma$, low energy neutrons ($<1\,$MeV), and high energy neutrons ($>1$\,MeV). The middle plot shows the energy spectrum of the particles that are detected. The bottom plot shows the power (in Watts) dissipated on the dump. One can clearly see a good correlation between the particles created and the energy dissipated on the dump, which indicates that the particles created and detected in the annex are indeed related to the dump.

Low energy neutrons represent the vast majority of the detected radiation originating from the dump, followed by photons.

It was estimated that when the dump dissipates 10\,kW of power, about 50\,$\mu$Sv/h of neutrons will be detected in the annex. Using the maximum conversion radiation type weighting factor for neutrons (which is a Gaussian centred at 20 for neutrons between 10\,keV and 1\,GeV), this would translate to about 22.5\,mGy of neutrons in a year. For photons the same exercise can be applied, the radiation type weighting factor is 1, which would translate to about 450\,mGy of  radiation due to photons in a year. These numbers are found to be small compared to the number that were obtained from the estimates (see Table~5.2 in the LUXE CDR~\cite{luxecdr}).


\clearpage

\section{Integration}
\subsection{Technical Infrastructure}
\subsubsection{Service Building} 

For the LUXE CDR~\cite{luxecdr}, it was originally assumed that the full experiment could be built inside the Osdorfer Born access-shaft building. The experiment would have been installed in underground level -3 (UG03), the laser and most services in level -2 (UG02), and the control room in the surface building. 

More recent discussion with the DESY radiation protection group have shown that this is in fact strongly disfavoured, as radiation safety regulations would severely restrict access to the laser and clean rooms and constitute a formidable obstacle to an efficient operation of the laser and of the related instrumentation. More details are found below, in the laser clean-room and service-room description. Constructing a new service building in the vicinity of the access shaft would therefore prove to be a better solution. However, such construction can be costly and could take a long time, which would be problematic for the experiment.

\begin{figure}[h!]
   \centering
   \includegraphics*[height=4.cm]{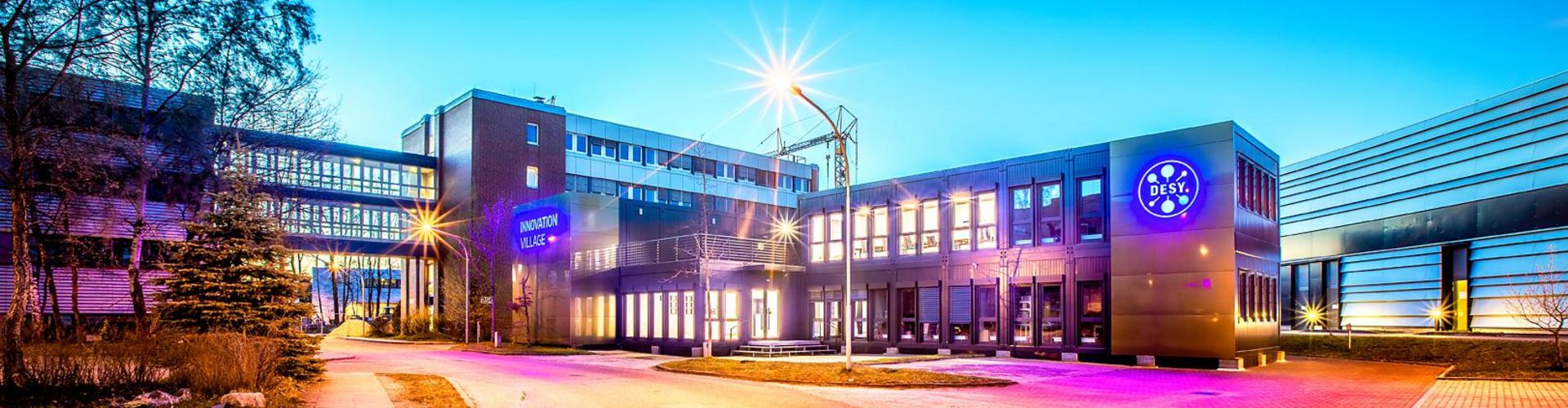}
   \caption{Picture of the DESY Innovation Village building.}   \label{fig:DESYInnovationVillage}
\end{figure}

In order to mitigate these points, the construction of a prefabricated building was investigated. On the DESY campus, a few example of such buildings have been erected. As an example, building 200, also known as the ''Innovation Village``, is a 1000\,m$^{2}$ multi-storey building, mixing different type of spaces, such as offices, laboratories, meeting-rooms, kitchen and toilets. This building was designed and constructed in about 1.5 years in 2018. A picture of the Innovation Village building can be seen in Fig.~\ref{fig:DESYInnovationVillage}.

Since the new service building will eventually host a high-intensity laser, floor stability and the possibility to add a laminar air flow based air-conditioning system with a filter fan unit above the work area are important requirements.

\begin{figure}[h!]
   \centering
   \includegraphics*[height=4.cm]{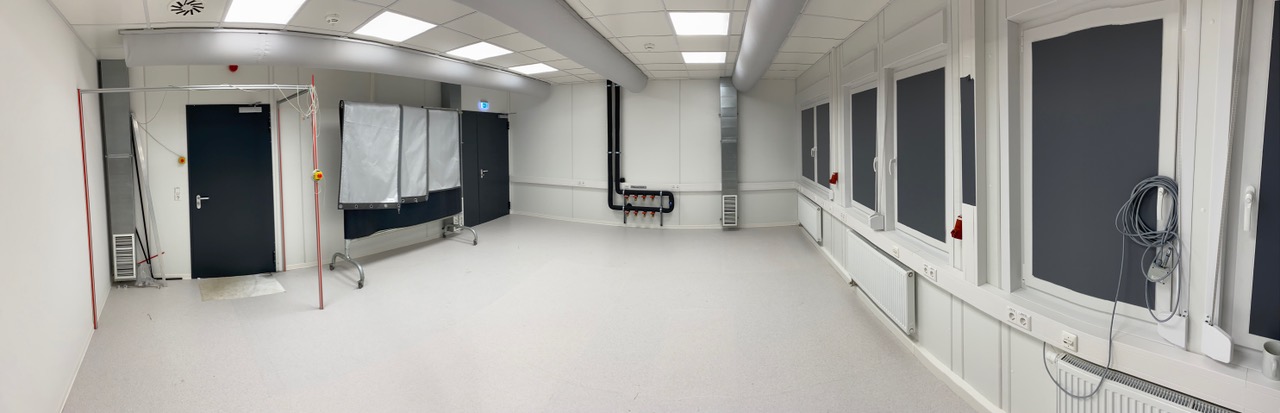}
   \caption{Picture of concrete floor lab-space in the Innovation Village building}.
   
\label{fig:DESYInnovationVillageLab}
\end{figure}

Similar laboratory space also exists in the Innovation Village building as shown in Fig.~\ref{fig:DESYInnovationVillageLab}.

\begin{figure}[h!]
   \centering
   \includegraphics*[height=8.cm]{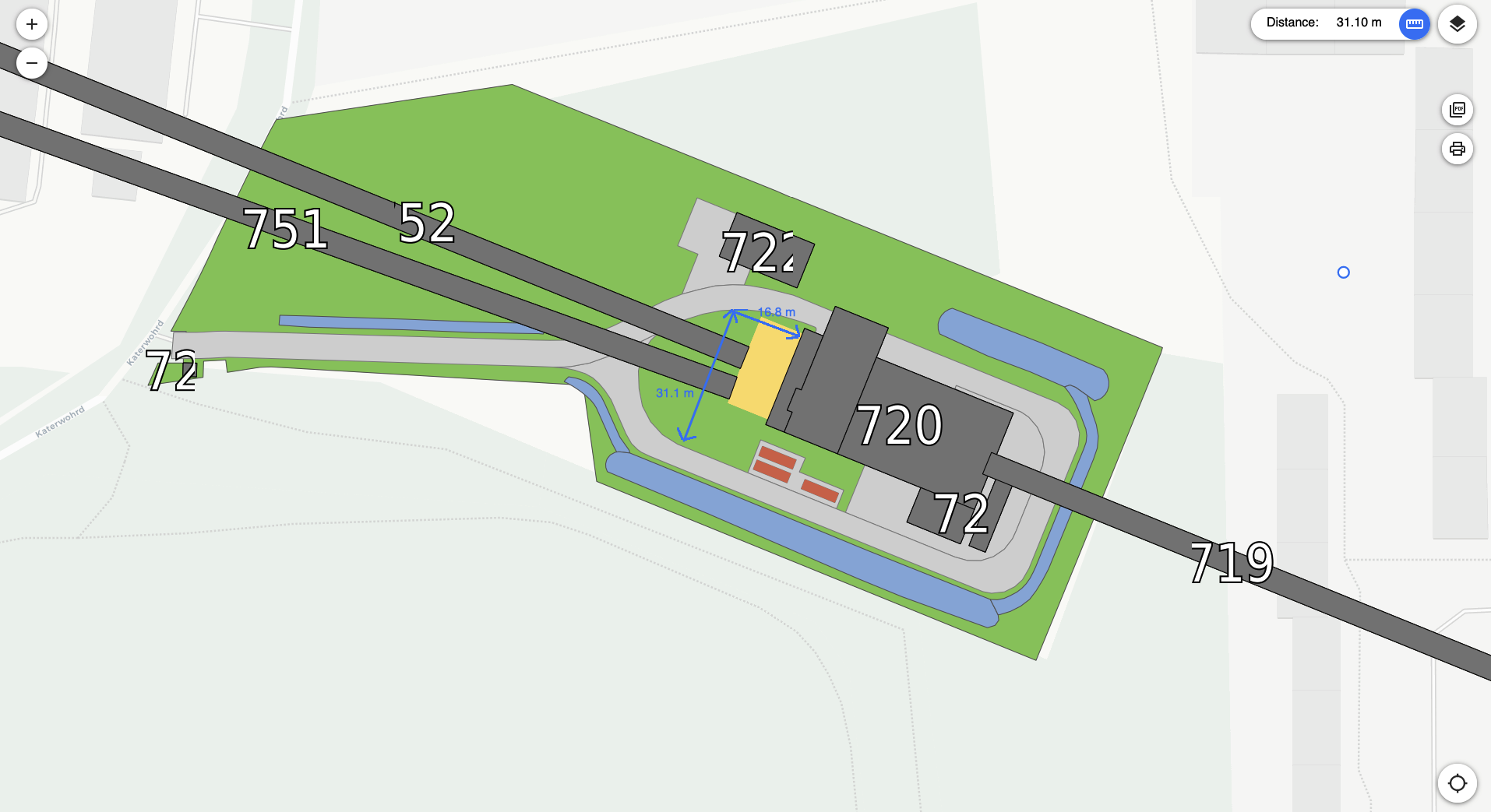}
   \caption{2D map of the direct vicinity of the XS1 access shaft where the new clean room and control room are currently planned to be installed}.
   \label{fig:LUXE_aerialmap_laser}
\end{figure}

Figure~\ref{fig:LUXE_aerialmap_laser} shows the footprint of the Osdorfer Born access shaft area. The space available directly above the XS1 tunnels and annex is about 17\,m long and 30\,m wide. 

\begin{figure}[h!]
   \centering
   \includegraphics*[height=5cm]{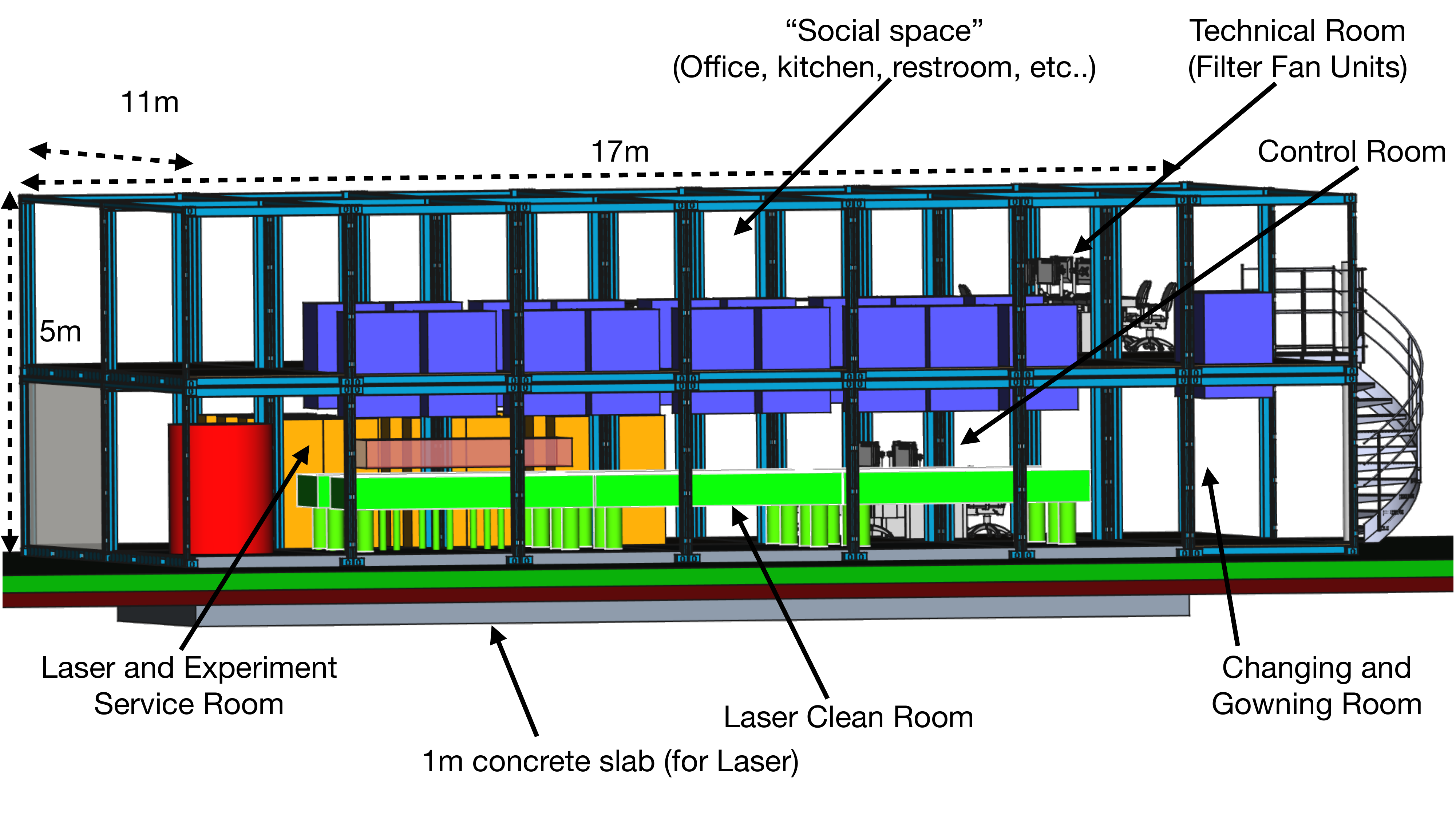}
  \includegraphics*[height=5cm]{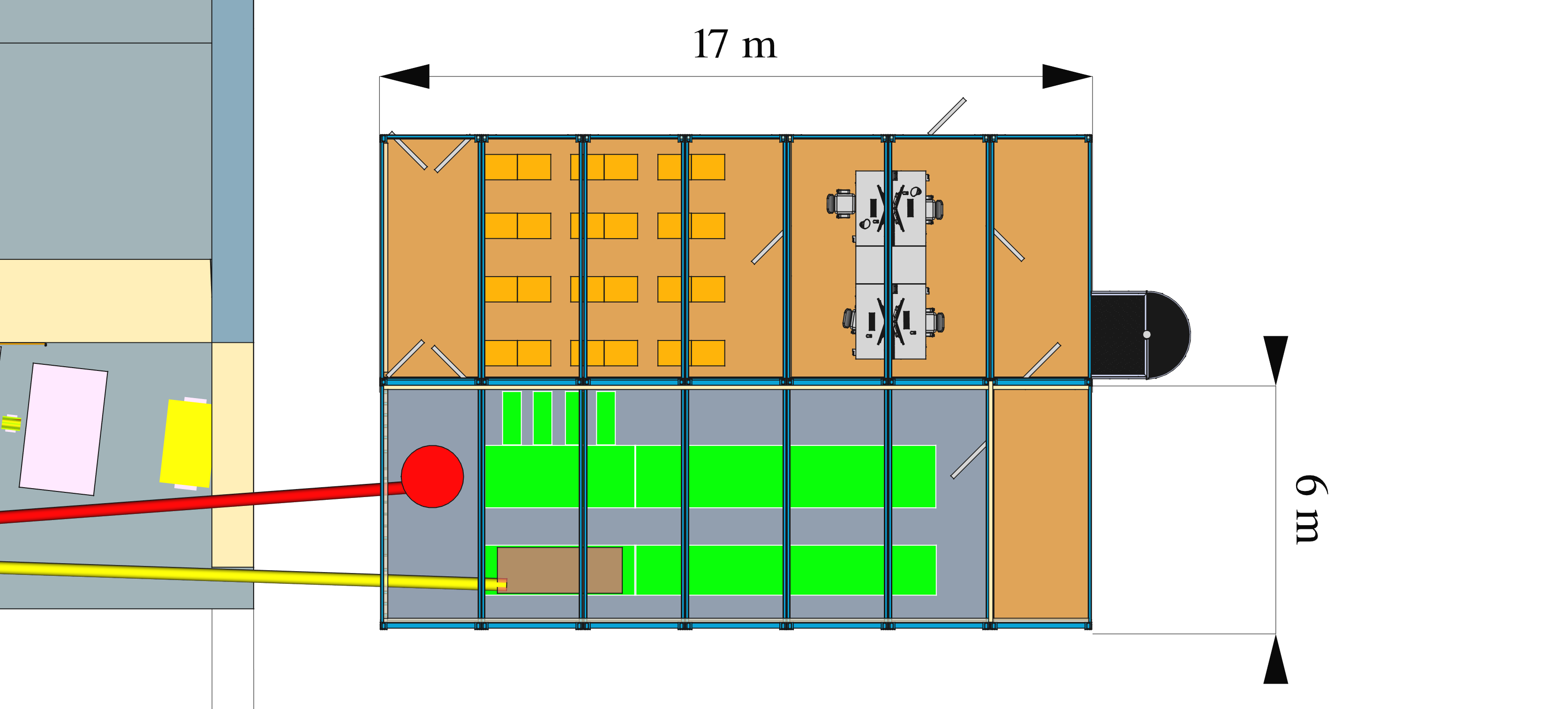}
  \includegraphics*[height=5cm]{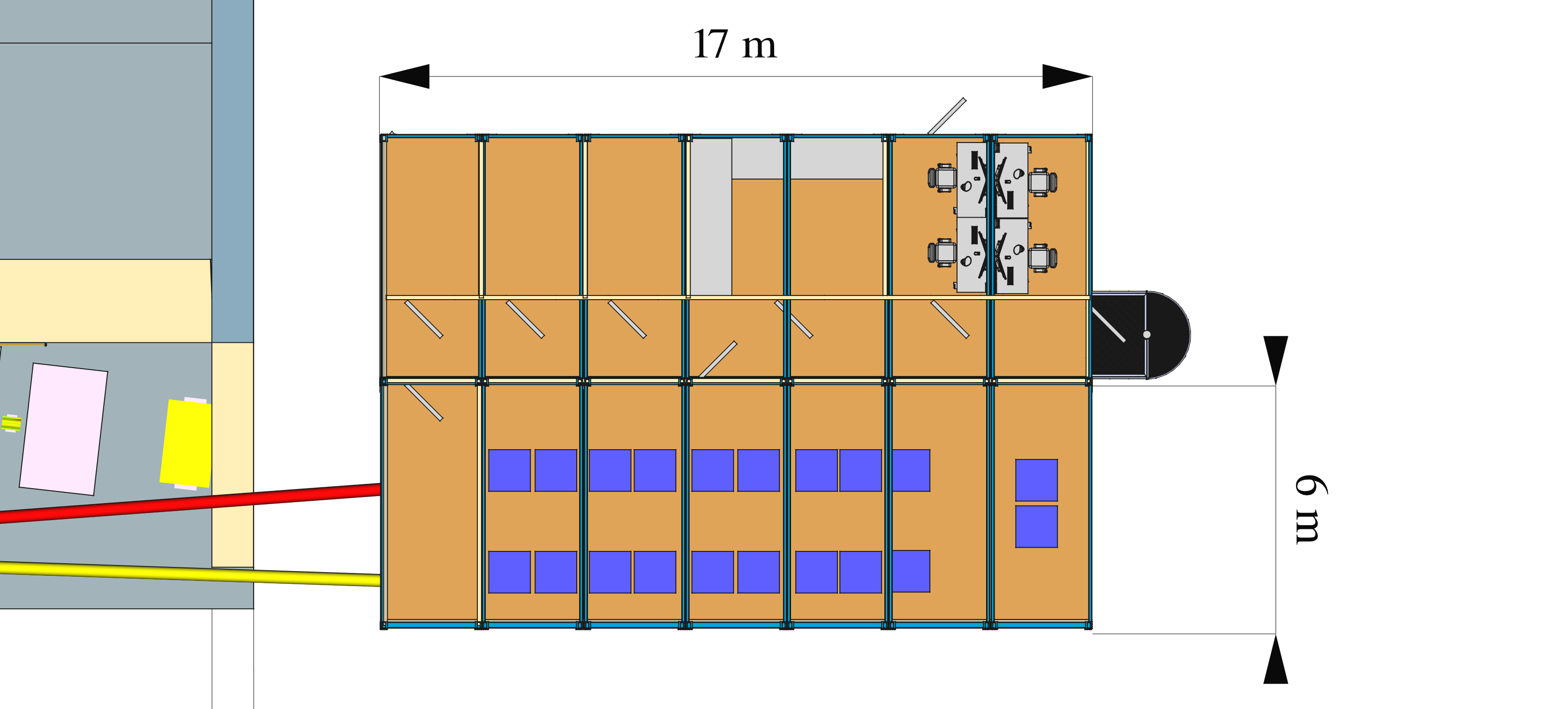}
  \caption{Conceptual CAD drawing of a conceptual new building located in front of the XS1 access-shaft. (Top) Side view with labels. The next two views show each floor: laser clean room, laser and experiment service room, and control room (ground floor); technical room for clean room, and social space (1st floor). To the left of the new building top views, the rightmost part of the XS1 building (compare with Fig.~\ref{fig:LUXE_CAD_Building}) as well as the two laser beampipes in red and yellow are visible} 
   \label{fig:LUXE_NewBuilding}
\end{figure}

Using standard size prefabricated elements (2.4\,m $\times$ 6\,m), it would be possible to construct a building that could potentially host all services needed by the experiment, as can be seen in Fig.~\ref{fig:LUXE_NewBuilding}. In this example, the building is 2 storeys high, and each floor is about 200\,m$^{2}$.

During the installation, commissioning, and data-taking phase, various activities will take place in the experimental area. Since the XS1 access shaft area is located about 2\,km away from the main DESY campus, it would be very convenient to build a control room and up to three offices on site. The control room will be equipped with PCs allowing up to four persons, shift leader,  detector expert, laser expert, and DAQ expert to monitor all components of the experiment. Ideally the control-room dimensions  should be about 40\,m$^{2}$. 
\subsubsection{Experimental Area}
\label{sec:TC:supportStructure}
\begin{figure}[ht]
   \centering
   \includegraphics*[height=8cm]{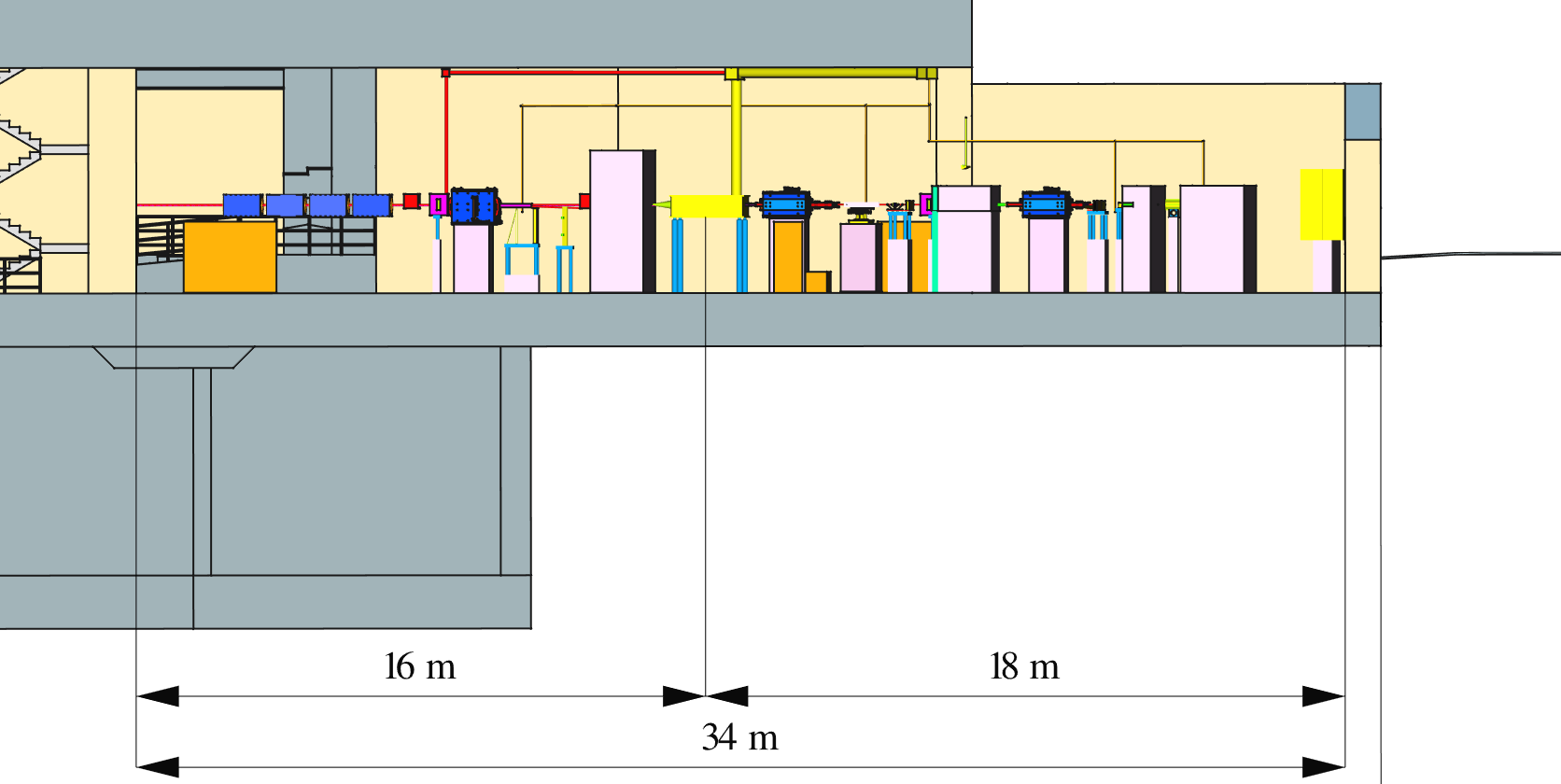} \hspace{2 cm} 
   \includegraphics*[height=8cm]{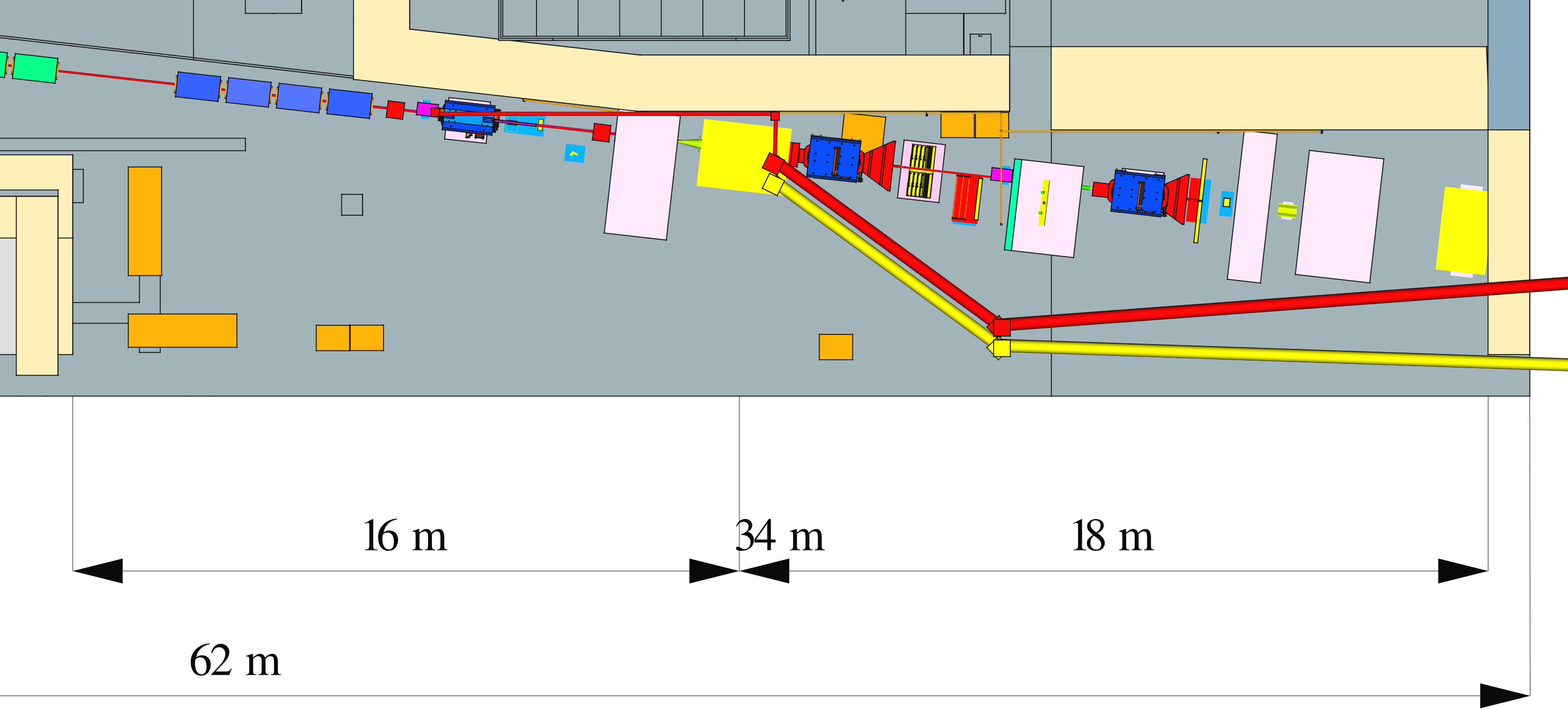}
   
   \caption{CAD drawing of the LUXE area in UG03 of the XS1 access shaft building: (top) sideview and (bottom) topview. The building is shown in beige and grey, while the different experimental elements are shown in colours.}
   \label{fig:LUXE_CAD_Building_detectors}
\end{figure}

As can be seen in Fig.~\ref{fig:LUXE_CAD_Building_detectors}, the \euxfel beam-pipes are located about 2.5\,m above the floor of XS1. In order to support and access the different experimental elements used in LUXE, support structures, shown in light blue, are required. To manipulate and access the different experimental elements, they will be accessed with movable bridges.

The support element used in the \euxfel for scientific instruments or beam elements consists of a concrete square pillar with 480\,mm side and 1850\,mm height. The elements have to be manufactured and are specifically adapted to each of the structural elements that needs to be supported in the experiment. For instance these pillars could be carved to host and shield the electronics that will need to be located closest to the experiment. About twelve of them will have to be built and installed. The support system which will be used by the different detectors will then be complemented with the kinematic mount system described in Chapter~\ref{chapt4}, devoted to the tracker detector.

\begin{figure}[ht]
   \centering
   \includegraphics*[height=12cm]{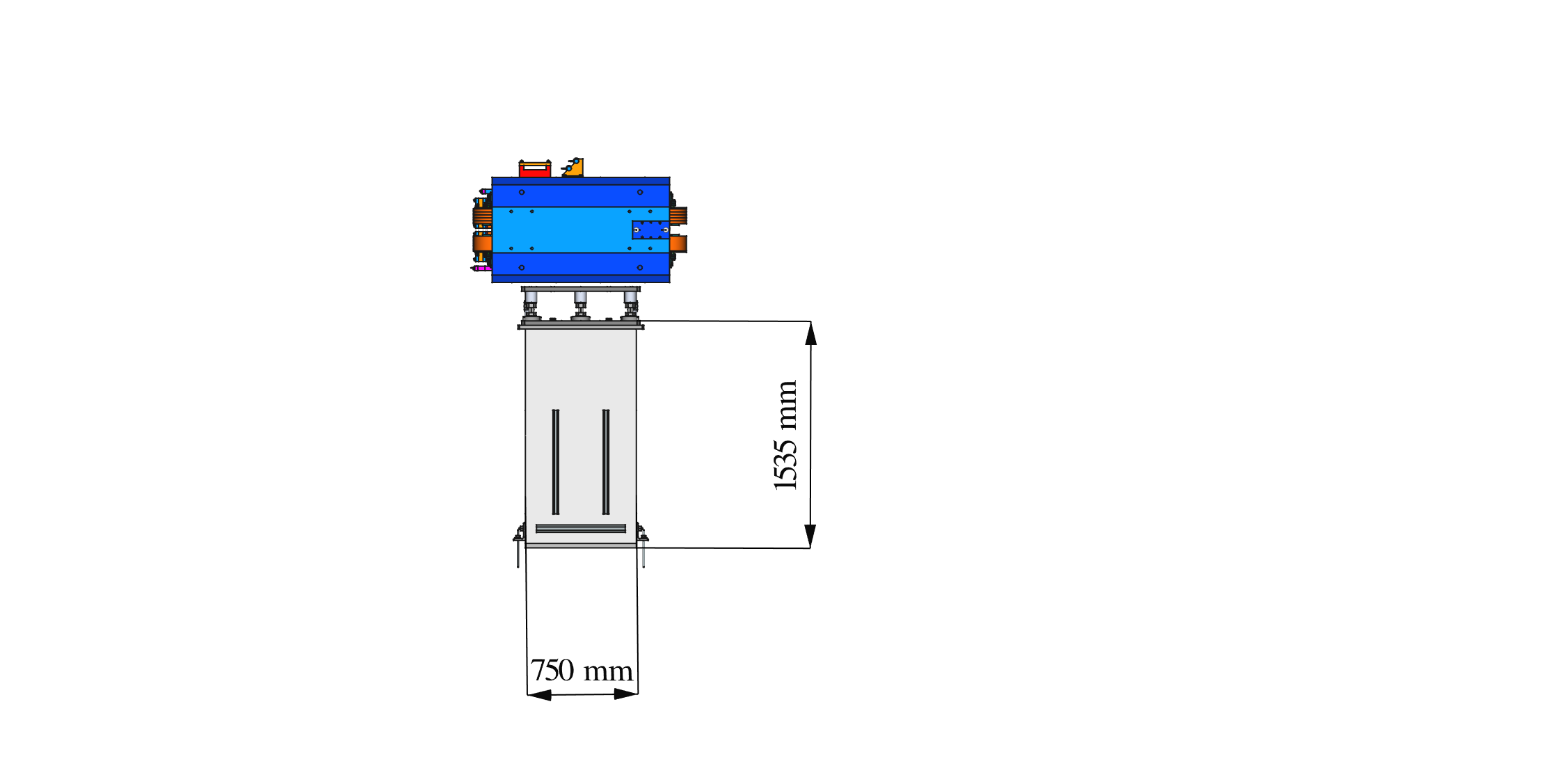}
   \caption{CAD drawing of the TDC magnet supporting structure.
   \label{fig:LUXE_CAD_MagnetSupport}}
\end{figure}

The magnet supports are the first elements that have been designed, and one example can be seen in Fig.~\ref{fig:LUXE_CAD_MagnetSupport}. It uses the default \euxfel supporting element and is surmounted by an adjustment plate which is attached to the magnet. Brackets attached with 12 screws allow the structure to be fixed to the floor. Since the TDC magnets have a mass of 6.5 T, the pillar side might have to be expanded up to 750\,mm to sustain the weight while limiting potential vibrations.  The magnets can either be attached vertically or horizontally, in order to allow bending electrons in both directions.

The different shieldings used in the experiment have been designed to be assembled from existing default hardened concrete blocks that can be reused for the experiment. Whenever necessary, lead shielding will also be used to absorb further primary or secondary particles coming from the beam. 

As discussed in Section~\ref{sec:beamDumps}, the beam dumps must be shielded to reduce the flux of secondary particles at the experimental instruments, or at other electronic devices located in the vicinity of the experiment, such as the control systems of the main \euxfel dump which are located in the annex tunnel.

\begin{figure}[ht]
   \centering
   \includegraphics*[height=8cm]{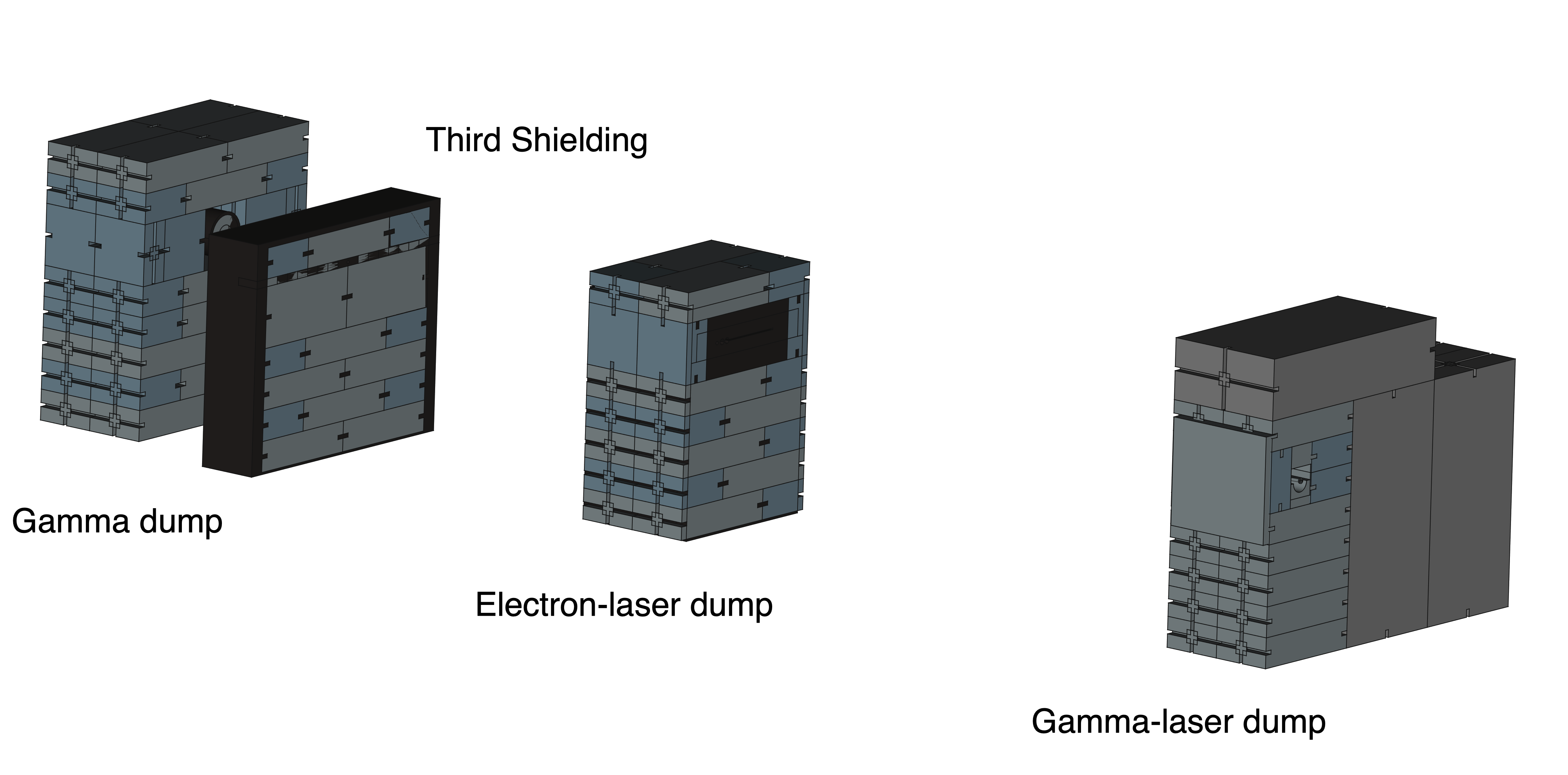}
   \caption{CAD drawing of the different shieldings used in the experiment made from standard concrete blocks that are available at DESY.
   \label{fig:LUXE_CAD_Shielding}}
\end{figure}

Four main shieldings are required in the experimental hall, as can be seen in Fig.~\ref{fig:LUXE_CAD_Shielding}.

The first shielding will host the $\gamma$-laser beam dump and will have dimensions of 4.8\,m $\times$ 4\,m $\times$ 1.6\,m (transverse and  horizontally to the beam, transverse and vertically to the beam, along the beam direction). This shielding is partially blocking the landing area of the access shaft. It was therefore designed to allow the extraction of the largest constituents using the access shaft crane, while leaving untouched the required amount of concrete blocks surrounding the dump. One of the reasons why this element is so massive compared to the other one is to protect the area from radiation and background co-linear to the extraction line. 

The second shielding will host the $e$-laser beam dump and will have dimensions of 2.4\,m $\times$ 3.2\,m $\times$ 1.6\,m (transverse and horizontally to the beam, transverse and vertically to the beam, along the beam direction). This shielding has been complemented with some layers of borated concrete with polyethylene aggregate to reduce the neutron yields as described in~\cite{PARK2014205}.

The third shielding will protect the detectors from primary particles that are back-scattered from the final photon dump, or from secondary particles created in this dump. Lead bricks will also be used in this shielding to stop the low energetic electrons that will be deflected by the dipole magnet of the photon spectrometer. The dimensions of this shielding are 2.4\,m $\times$ 3.2\,m $\times$ 1.6\,m (transverse and horizontally to the beam, transverse and vertically to the beam, along the beam direction). 

The fourth shielding will host the final photon beam dump and will have dimensions of 3.2\,m $\times$ 3.6\,m $\times$ 2\,m (transverse and horizontally to the beam, transverse and vertically to the beam, along the beam direction).

The shielding elements are currently rather massive and based on a preliminary study aiming to drastically reduce the particle flux around the sensitive detectors; further optimisation studies are planned to see what is required and technically feasible. 

\subsubsection{General Infrastructure and Services}

Movement of the different elements used by the detectors and the laser will benefit from the presence of an access shaft equipped with a crane sitting directly above the experiment. In the experimental area forklifts will be used for the transportation and installation of the experimental items.

\begin{figure}[ht]
  \centering
  \includegraphics*[height=8cm]{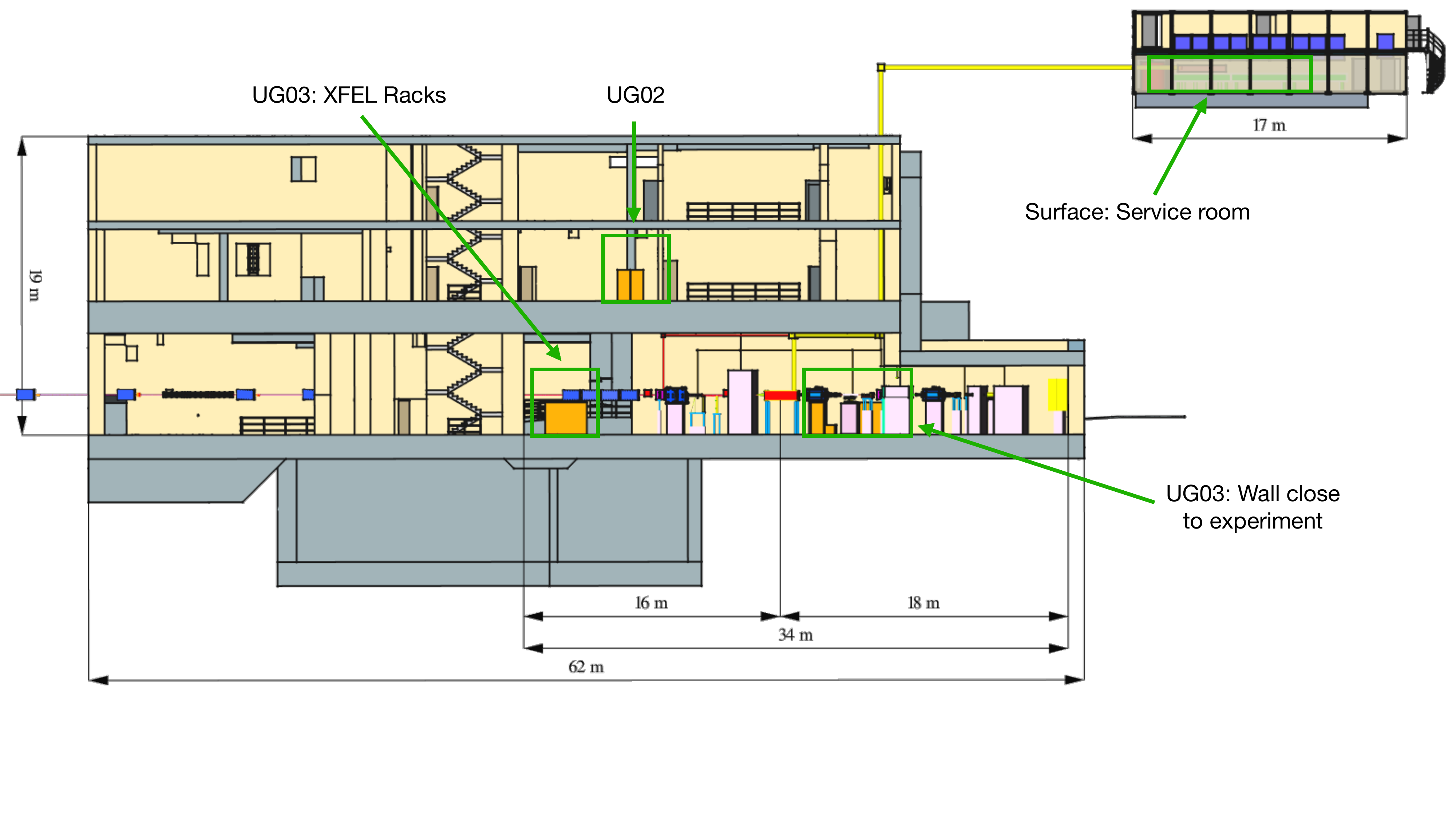}
  \caption{Placement of the different elements that could potentially host electronics racks for the experiments.
  \label{fig:LUXE_Infrastructure_Services}}
\end{figure}

Four locations are currently being investigated for the electronics racks used in the experiment. The locations are shown in Fig.~\ref{fig:LUXE_Infrastructure_Services}. The distance to the IP chamber, including some routing distance to these different sites are summarised in Table~\ref{tab:TC:lengthService}.

\begin{table}[htbp]
\begin{center}
\begin{tabular}{|l|c|}
	\hline
	
Area & Length \\	\hline

UG03: Side north wall & 7\,m \\
UG03: \euxfel rack & 16\,m \\
UG02  & 26\,m\\
Surface: service room& $\approx{}$50\,m\\
	\hline

\end{tabular}
\caption{Distance separating the IP chamber from the different sites where electronics racks could be installed for the experiment.\label{tab:TC:lengthService}
}
\end{center}
\end{table}

Only electronics, that need to be close to the detectors, will be installed in UG03, since the area can not be accessed more than a couple of time per months for a few hours. This will include the front-end readout of the ECal (see Chapter 5), the Cherenkov detector (Chapter 7), the gamma-profiler (Chapter 9) and gamma-flux monitor (Chapter 10), the readout system of the tracker (Chapter 4), and the cameras used in conjunction with the scintillating screens (Chapter 6).

\begin{figure}[ht]
  \centering
  \includegraphics*[height=8cm]{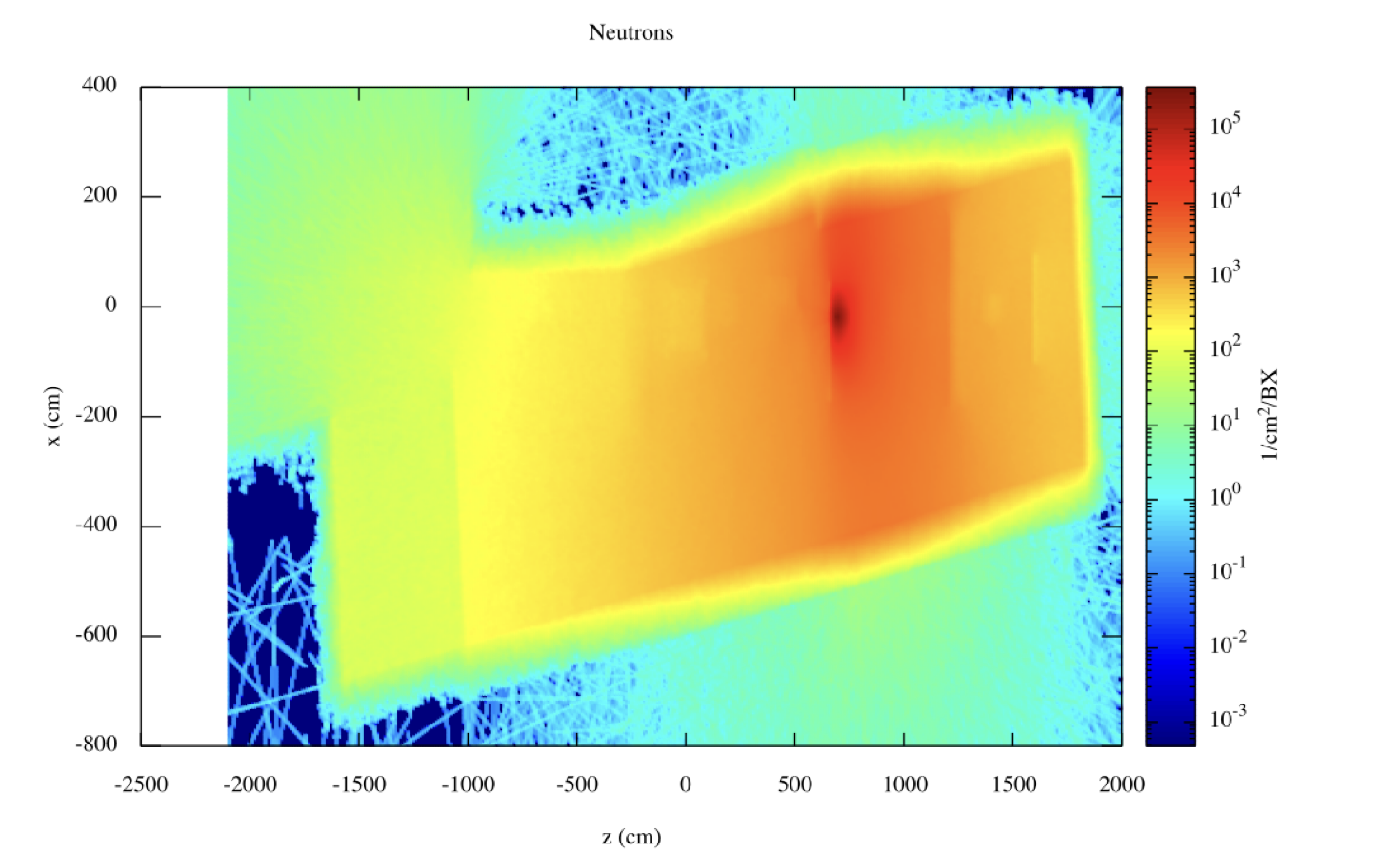}
  \caption{Radiation map of the experimental area in XS1-UG03, obtained with FLUKA and showing the neutrons fluence per bunch (BX) of the accelerator. The map is integrated over the 5\,m height of the annex.
  \label{fig:LUXE_RadiationMaps}}
\end{figure}

The exact placement of these elements is currently being investigated to minimise the amount of radiation that they will be exposed to. This is achieved using a dedicated FLUKA simulation of the experimental area to obtain radiation maps as shown in Fig.~\ref{fig:LUXE_RadiationMaps}. 

Preliminary studies performed with \geant  simulations~\cite{Electronics-Neutron-Study} have also shown that the amount of radiation expected at the bottom of the positron tracker supporting structure, or 2\,m bellow the beam-line level, would remain at an acceptable level. One would expect an integrated dose of about 1\,Gy per year from positrons, electrons and photons, while the neutron fluence would be of the order of $6\times{}10^{11}$\,cm$^{-2}$ per year. 

The rest of the readout chain and power supplies were originally thought to be installed in the UG02 room closest to the UG03 feed-through access shaft. However this is currently disfavoured due to the close location of the $e$-laser beam dump, which will induce a dose rate of up to 10\,$\mu$Sv per hour, closing the room for access during the data-taking of the experiment.

It is therefore currently planned to place the rest of the electronics, which mostly consists of the readout DAQ PCs and the power-supplies in two different locations. For the electronics that will need to be placed no more than 50\,m away from the detectors, mostly cameras readout and power over ethernet Cat6 cables used in the scintillating screens and in the laser diagnostics, installation is foreseen in the UG02 room where the laser was originally thought to be installed. The closest distance to this room from the IP is about 26\,m. The rest of the electronics will be placed directly in the new service building in the experiment control room, about 50\,m away from the IP.

In the LUXE CDR~\cite{luxecdr}, the type of services required by each detector system was estimated and is summarised in Table~\ref{tab:TC:detectorService}. The different services, signal cables, power cables, gas lines, will be installed in different cable trays that will be mounted on the wall of the experimental cavern.

\begin{table}[htbp]
\begin{tabular}{|l|c|c|c|c|}
	\hline
	
Detector             & Electrical consumption (kW)& Cooling & Gas       & N DAQ PCs\\   
	\hline
 
Scintillation screen & 2 & --    & --      & 4\\               
Cherenkov            & 2.6 & --   &  Air     & 4\\               
Tracker              & 11.5 & Water   & Nitrogen  & 2\\              
Calorimeter		     & 3 & Air     & Nitrogen  & 1\\               
Back-scattering	Calorimeter	 & 1.5 & --    & --      & 1\\  
Gamma Profiler       & 3.5 & --    & --      & 1\\
	\hline
             
\end{tabular}
\caption{Summary of the main service requirements needed by each detectors.
\label{tab:TC:detectorService}
}
\end{table}

The total electric power consumption expected by the experiment, which was estimated in the LUXE CDR~\cite{luxecdr}, is found to be of the order of 150\,kW in \phaseone and will grow to about 170\,kW in \phasetwo when the laser will be upgraded. The different contributions to these numbers are summarised in Table~\ref{tab:TC:detectorServiceConsumption}.


\begin{table}[htbp]
\begin{center}
\begin{tabular}{|l|c|c|c|c|}
	\hline
	Type                            &\phaseone &\phasetwo & Place\\
		\hline
Laser clean room AC	                & 10  & 10    &  Surface \\
Laser front-end	                    & 5	 & 5     &  Surface \\
Laser power amplifier+pump	        & 25  & 50    &  Surface \\
Laser pulse compressor	            & 3	 & 5     &  Surface \\
Laser additional flow box	        & 5	 & 5     &  Surface \\
Laser computers, controller, etc..	& 5	 & 5     &  Surface \\
	\hline

Laser beam-line (pumps)	            & 2	& 2      &  Surface \\ 
IP Box (pumps)	                    & 1	& 1      &  UG03 \\
Target chamber (pumps)	            & 2	    & 2       &  UG03 \\
Electron beam-line (pumps)      	& 2	& 2       &  UG03 \\
Magnet vacuum chamber pump	        & 2	& 2       &  UG03 \\
	\hline

Magnet power supplies	            & 60	& 60      &  UG02 \\
	\hline

Gamma profiler FE	                & 3 & 3  &  UG03 \\
Gamma profiler DAQ	                & 0.5  & 0.5  &  Surface \\
	\hline

Back-scattering Calo HV	            & 1 & 1  &  Surface \\
Back-scattering Calo DAQ	        & 0.5  & 0.5         &  Surface \\
	\hline

Cherenkov VME	                    & 1.6 & 1.6           &  Surface \\
Cherenkov DAQ	                    & 2 & 2         &  Surface \\
	\hline

Scintillation screen DAQ	        & 2 & 2        &  UG02 \\
	\hline

Calo DAQ	                     &   0.5   &  0.5         &  Surface \\
Calo HVPS	                     &   1.250  &  1.250        &  Surface \\
Calo FEB+VME	                 &   1.250  &  1.250         &  UG03 \\
	\hline

Tracker HV PS	                &    5.5 &	5.5        &  Surface \\
Tracker DAQ PC	                &    1&	1        &  Surface \\
Tracker RU+VME	                &    3&	3        &  UG03 \\
Tracker chiller 	            &    2&	2        &  UG03 \\
			\hline
Total in kW                      &    147&	174& \\
			\hline

\end{tabular}
\caption{LUXE scheduled power consumption (in kW) for the two running phases of the experiment. Estimated for the CDR.
\label{tab:TC:detectorServiceConsumption}
}
\end{center}
\end{table}

\clearpage
\subsubsection{Laser Clean-Room and Service-Room}
\label{sec:lasercleanroom}
\begin{figure}[h!]
   \centering
   \includegraphics*[height=8.cm]{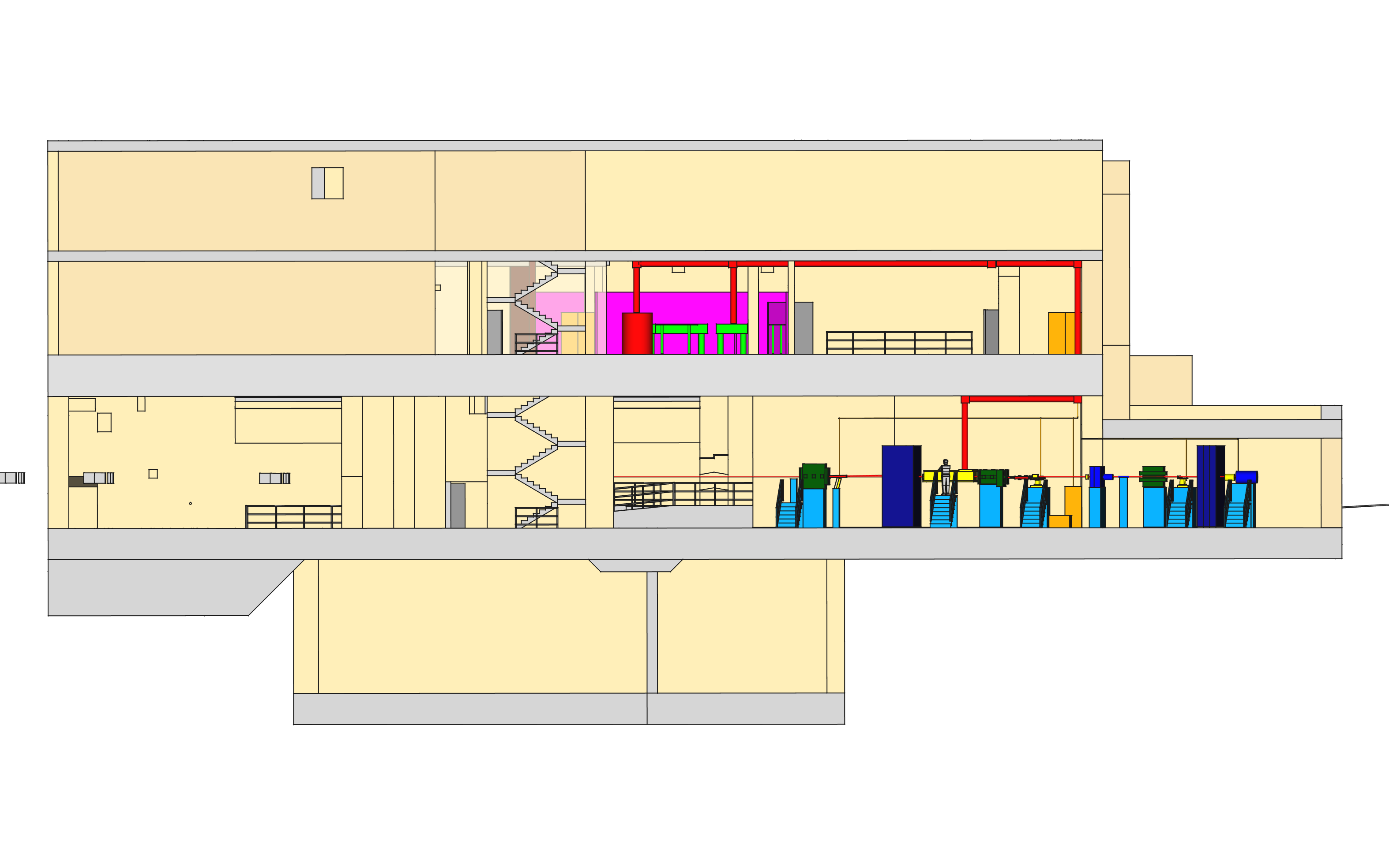}
   \caption{CAD drawing of the XS1 access shaft building, including the LUXE experiment and laser, as thought for the CDR. The building is shown in grey, while the different experimental elements are shown coloured.
   \label{fig:LUXE_Laser_CAD_CDR_Building}}
\end{figure}

\begin{figure}[h!]
   \centering
   \includegraphics*[height=8.cm]{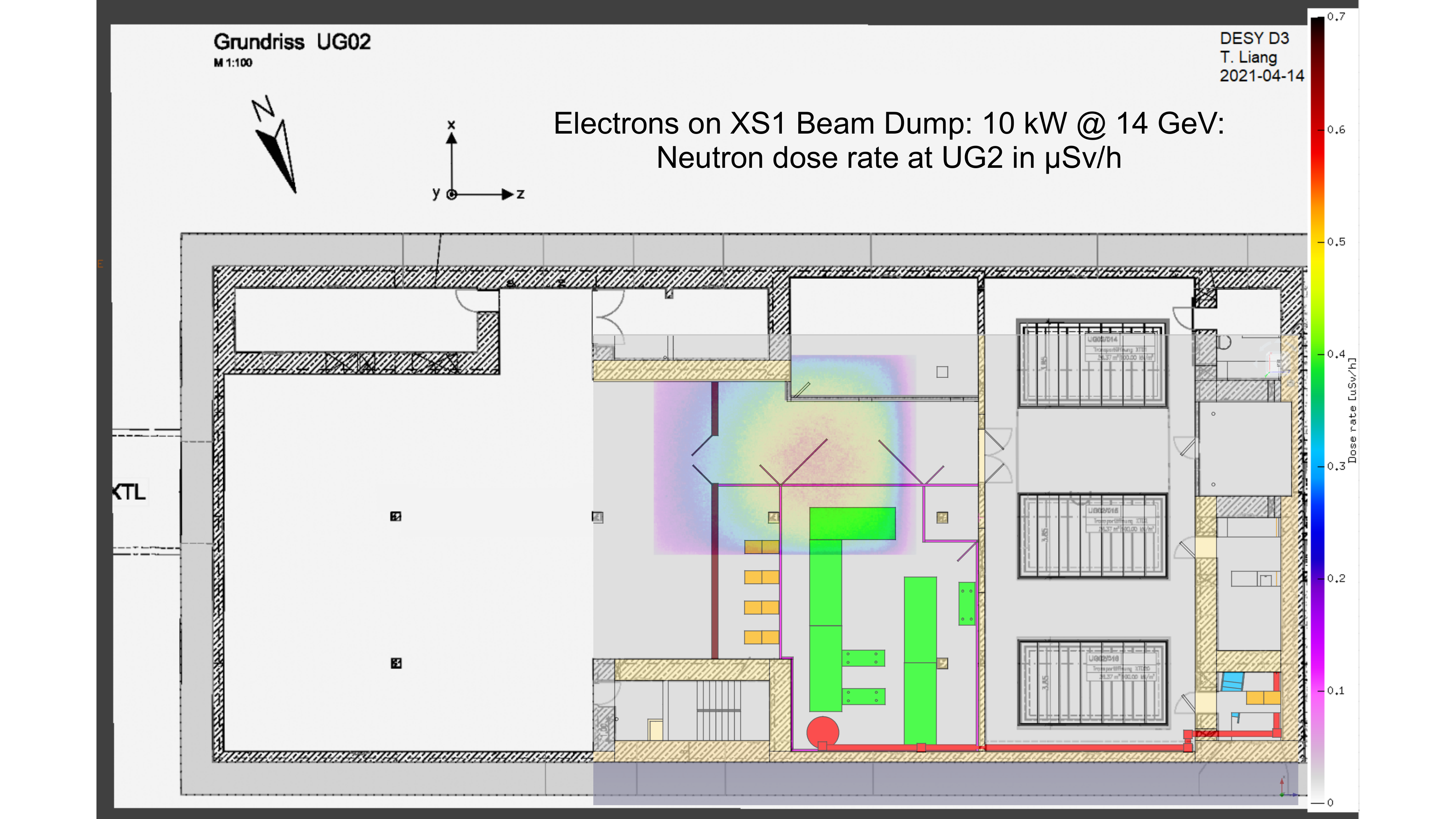}
   \caption{2D drawing of the XS1 access shaft building, in UG02 including the LUXE Laser. The building is shown in grey, while the different experimental elements are shown coloured. Also shown overlaid is the neutron dose rate map obtained from FLUKA simulation.
   \label{fig:LUXE_2D_Building}}
\end{figure}

\begin{figure}[h!]
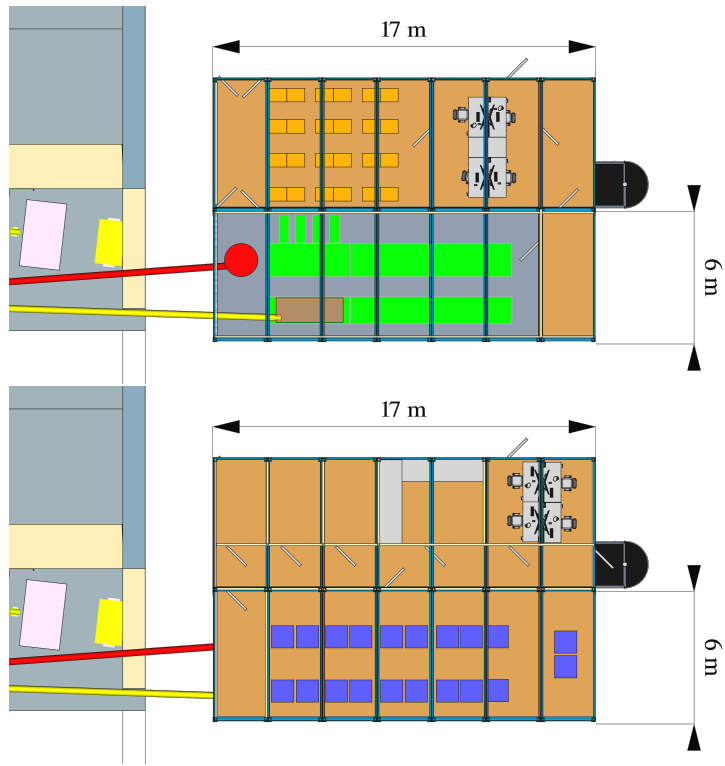

   \centering
   \includegraphics*[height=5.cm]{fig/TechnicalInfrastructure/XS1LaserControlRoomEGFloorTopView.png} 
   \includegraphics*[height=5.cm]{fig/TechnicalInfrastructure/XS1LaserControlRoom1stFloorTopView.png}
   \caption{CAD drawing of the new laser building and control room outside the XS1 access shaft building, the different experimental elements are shown coloured. (top) Laser clean room floor, (bottom) control room and service room for the laser and the experiment.
   \label{fig:LUXE_new_Laser}}
\end{figure}

The laser system will be installed in a clean-room of ISO-6 standard that will need to be constructed. The room was originally thought to be installed in the XS1 building in UG02, as can be seen in Fig.~\ref{fig:LUXE_Laser_CAD_CDR_Building}. However the presence of the main \euxfel dump located in UG04 directly below the space that was pre-selected for the installation, would result in a high neutron dose exposure during the \euxfel operation, as can be seen in the Fig.~\ref{fig:LUXE_2D_Building}. For this reason it is currently envisaged to install the laser infrastructure in the new temporary building that will be located outside the XS1 shaft, as already discussed and shown in Fig.~\ref{fig:LUXE_new_Laser}. In order to reduce vibrations that could be experienced by the laser the building would be constructed on a concrete base (1\,m thick). In order to reduce even further the dependence of the laser on vibrations, the optical tables used by the Laser in the room could be installed directly on the concrete floor with some feed-throughs.

The clean-room will have a surface of about 100\,m$^{2}$.
In order to control precisely the laser environment to a temperature of $21\pm 0.5 ^{\circ} $C and a humidity of $40 \pm 5 \%$, an air conditioning control system will be installed.

The space constraints required by the elements of a 350 TW laser for \phasetwo have also been taken into account in the design of the clean-room, such that enough space is available for both phases. Moreover, a gowning room attached to the main laser room is also planned. 

For \phaseone, it is currently considered that a new 40\,TW laser system will be bought and installed in the clean-room. The installation includes the assembly of components of the laser system such as front end, pre-amplifiers, main amplifier and compressor. In addition, a dedicated diagnostics station will also be installed in the laser room. All this is detailed in Chapter~\ref{chapt3}.

After the compressor, the high-power laser beam needs to be guided to the IP chamber. After the interaction, the attenuated part of the beam will be brought back to the laser clean-room for post diagnostic purposes. Furthermore, one laser beam is required for the inverse Compton scattering and another laser beam is required for measuring the spatio-temporal overlap of the laser and the electron beams.
For this purpose two vacuum pipes made of stainless steel with diameter 15-20\,cm and a length of about 40\,m, will be constructed. These pipes will pass through the 2\,m high concrete floor via a feed-through hole, already present, to the floor UG03, and will extend up to the surface, as illustrated in Fig.~\ref{fig:LUXE_CAD_Building}. The diameter is sufficient for the laser transport for both the 40\,TW and the 350\,TW laser.


For \phasetwo, a few upgrades have to be carried out to ensure a power of 350\,TW can be achieved. Firstly, a new amplification stage will be added. Secondly, as a consequence of the increased energy in the pulse, the size of the laser beam will be increased to 14\,cm, compared to 5\,cm for the 40\,TW laser, to keep the fluence below the damage threshold for the downstream optics. This implies that the compressor also will have to be upgraded to be larger.



A service room (area $20$\,m$^2$) is foreseen, placed adjacent to or above the laser clean-room. This will host all the power supplies and cooling units for the pump lasers in both phases. Moreover, this separate service room helps reduce the unnecessary heat generated through the power units in the laser area.  


\subsection{Installation}
\begin{figure}[ht]
   \centering
   \includegraphics*[height=4.2cm]{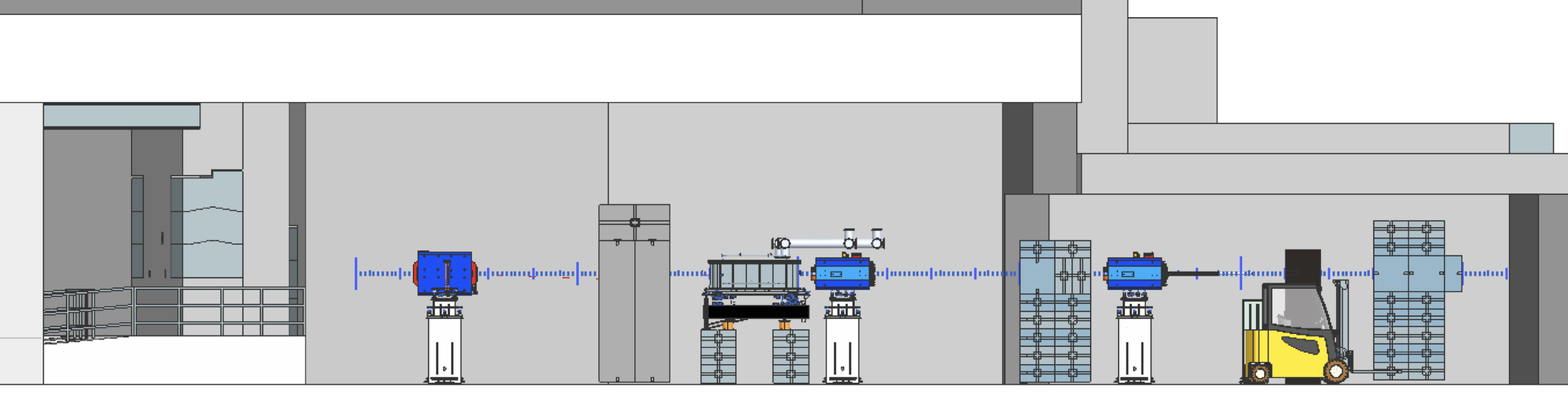}
   \includegraphics*[height=4.2cm]{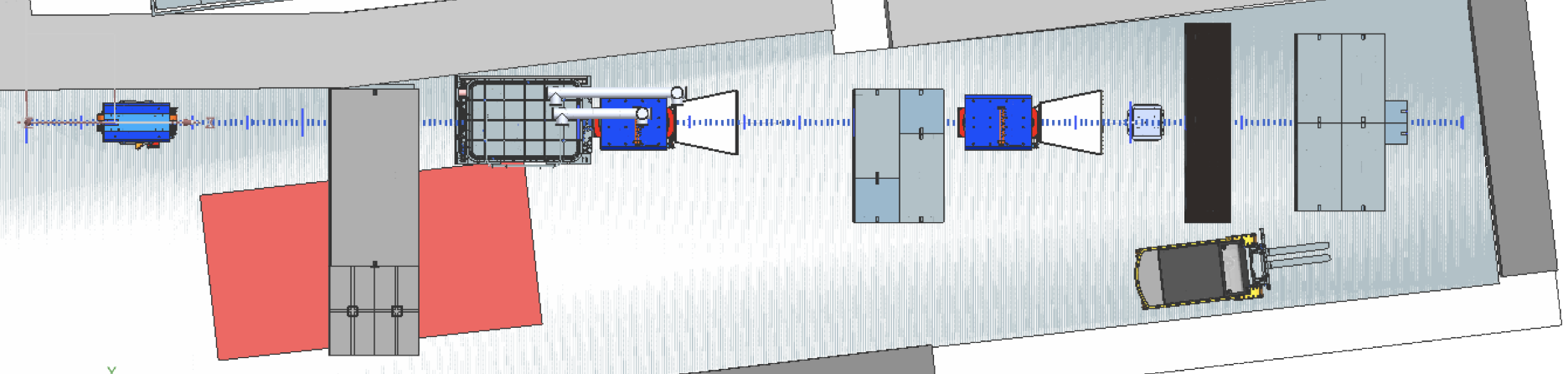}

   \caption{NX design of the experimental area. (Top) Side view, (bottom) top view.
   \label{fig:NX-IP}}
\end{figure}

The experimental area is implemented in the DESY CAD system (Siemens NX). The state of the current implementation can be seen on Fig.~\ref{fig:NX-IP}. The experimental area mostly contains the largest items, shielding, magnets with supporting structure, etc. This design work has been used to quantify the amount of material and work that will be needed for the installation of the experiment.

This preliminary work relies on few assumptions: 
\begin{itemize}
    \item All the big elements that would block the way of the forklift in the hall have to be installed with a certain sequence, which means that some items, even if they are not used for the reduced version of the experimental setup, such as that foreseen in bare-bones LUXE, must be available in order to be installed. This concerns specifically all the supporting pillars of all detector and experimental apparatus, the dipole magnets at the IP and used in the gamma spectrometer, and all the shielding.
    
    \item All the other elements (dumps, magnets, detectors, etc.) of bare-bones LUXE must be available.
    
    \item The design of the beam line, and more specifically the placement of each element in it is precisely known enough in advance such that the survey used for their positioning can happen as soon as possible.
    
    \item The planning is done taking into a account a regular running of the \euxfel, we do not consider any long shutdown of the facility after 2025/26, and consider 4 weeks usable in the winter and 2 weeks in the summer.
    
    \item If an activity is stopped and continued in a later shutdown, 1 week of overhead is added to compensate the time lost to restart the activity.
    
    \item The structure and the rules of access to the EuXFEL tunnels almost rule out any possibility to perform installation work of the TD20 transfer line and the experiment in parallel. Interference with work on the transfer line is however not considered. 
    
\end{itemize}

\begin{sidewaysfigure}[ht]
   \centering
   \includegraphics*[height=15cm]{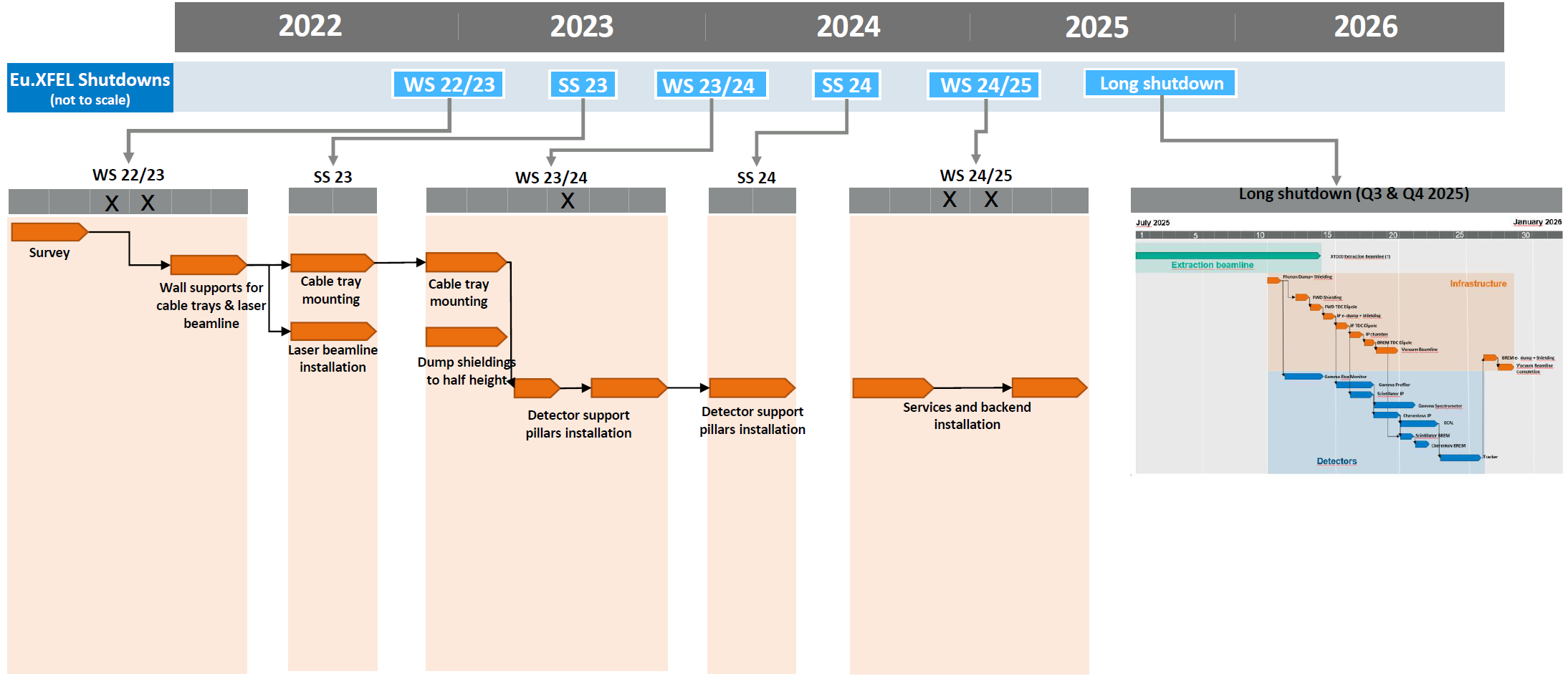}
   \caption{Preliminary Gantt installation schedule for LUXE.
   \label{fig:Gantt-Y1}}
\end{sidewaysfigure}

\begin{sidewaysfigure}[ht]
   \centering
   \includegraphics*[height=12cm]{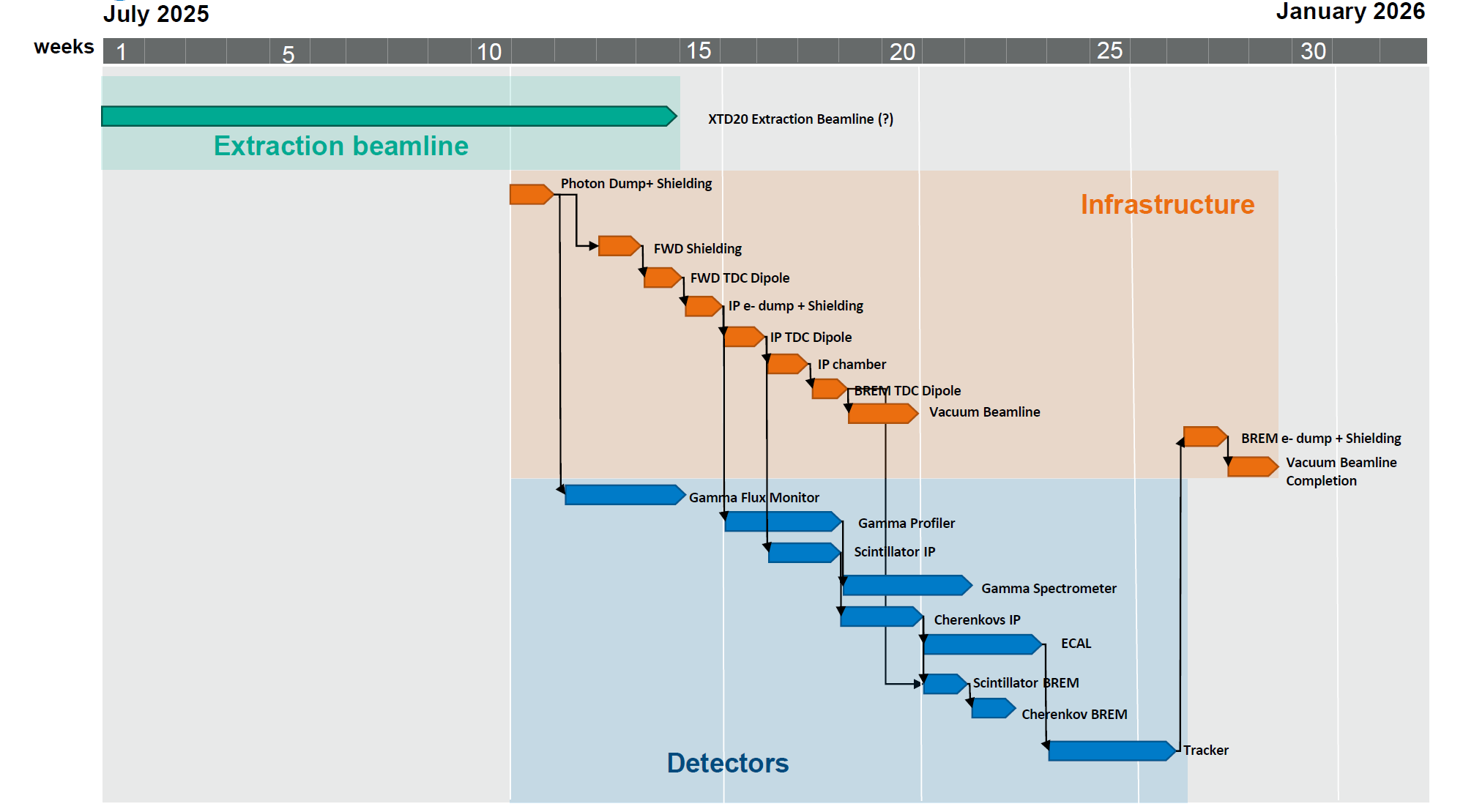}

   \caption{Preliminary Gantt installation schedule for LUXE, during a 6-month long shutdown of the \euxfel.
   \label{fig:Gantt-Y2}}
\end{sidewaysfigure}

Figures~\ref{fig:Gantt-Y1} and~\ref{fig:Gantt-Y2} show the preliminary installation schedule obtained with these assumptions. The years shown are only indicative.
LUXE could potentially be installed in about 3.5 year timescale, and this time could likely be shortened slightly if the long shutdown is factorised in.

\clearpage

\section{Safety and Radiation Protection}

Since the experiment will be built underground in the \euxfel complex, its safety concept must comply with all the safety requirements that were developed for the \euxfel and are described in details in Ref.~\cite{XFELTDR}. 

In particular, new supporting structures will be built with fire-resistant construction materials that can resist fire for 90 min. All cable and insulation material that will be used in the underground area will be halogen-free, contain flame-retardant materials and be in line with the DESY cable specification. The usage of gas when necessary will be monitored for gas-leaks and the environment where it is used will be controlled with oxygen deficiency monitoring devices, even if the amount of gas used in the experiment is expected to be low. The level of radiation created in the experiment will be monitored and should not exceed the limits set within the \euxfel complex. 
All sensitive equipment that could be damaged from failures of the main power supply will be equipped with batteries allowing to slowly turn it off in order to prevent damage.

In order to also comply with the DESY safety requirements, the concept shall detail mitigation strategies, escape paths, radiation protection monitoring devices, gas sensors, etc. Safety signs for laser protection, high voltage, etc.\ will be installed in the experimental areas whenever needed. Only specifically trained people will be allowed to access the different experimental elements. The concept is being built with the help of the different safety departments in DESY.



Finally, the work in the experimental area in UG03 will be severely restricted and limited to essential access. The access to the experimental equipment will only be possible during stops of the \euxfel, the area being interlocked to the electron beam and not accessible outside these breaks. 

In the laser clean-room, a daily access to the laser will be authorised for laser calibration and tuning purposes. Lasers of power 40 to 350 TW are classified as class 4. Such apparatus exceeds the maximum permissible exposure limit by a factor of more than $10^6$. Therefore, the required safety regulations will be followed. Protective personal equipment (PPE) and safety interlocks will be implemented as per legal regulations.
A laser interlock will be installed around the IP chamber such that it will be possible to open the chamber, while a reduced power laser could be running from the laser room, for alignment purposes. 
The highest level of protection and safety will be met in all places where the Laser will be running.
Safety officers will be appointed to restrict and control the access to the high power laser and interaction areas.



\newpage




\printbibliography
\end{refsection}

\chapter*{Acknowledgements}

We would like to acknowledge with thanks the following persons for their contribution in the earlier stage of this study: A. Fedotov, M. Gotskin, A. Mironov, S. Rykovanov, L. Shaimerdenova, M, Skakunov, A.I. Titov, O. Tolbanov, A. Tyazhev, A. Zarubin and A. Zhemchukov.







\backmatter



\end{document}